\documentclass[iop,revtex4]{emulateapj}
\usepackage{natbib}
\usepackage{threeparttable,lscape}
\usepackage{rotating}
\usepackage{hyperref}
\usepackage{color}
\usepackage{xcolor}

\definecolor{yaleblue}{rgb}{0.1,0.3,0.9}
\definecolor{ultramarine}{rgb}{0, 0, 150}
\hypersetup{colorlinks=true, linkcolor=black, urlcolor=blue, citecolor=yaleblue}

\shorttitle{Optical spectra of type II supernovae}
\shortauthors{Guti\'errez et al.}

\begin{document}

\title{Type II supernova spectral diversity I: Observations, sample characterization and spectral
line evolution\footnote{T\lowercase{his 
paper includes data gathered with the 6.5 m} M\lowercase{agellan} 
T\lowercase{elescopes located at} L\lowercase{as} C\lowercase{ampanas} O\lowercase{bservatory,} 
C\lowercase{hile; and the} G\lowercase{emini} O\lowercase{bservatory,} C\lowercase{erro} P\lowercase{achon,}
C\lowercase{hile} (G\lowercase{emini} P\lowercase{rogram} GS-2008B-Q-56). B\lowercase{ased on
observations collected at the} E\lowercase{uropean} O\lowercase{rganisation for} A\lowercase{stronomical}
R\lowercase{esearch in the} S\lowercase{outhern} H\lowercase{emisphere,} C\lowercase{hile} 
(ESO P\lowercase{rogrammes} 076.A-0156, 078.D-0048, 080.A-0516, \lowercase{and} 082.A-0526).}}

\author{Claudia P. Guti\'errez\altaffilmark{1,2,3,4},
Joseph P. Anderson\altaffilmark{3},
Mario Hamuy\altaffilmark{2,1},
Nidia Morrell\altaffilmark{5},
Santiago Gonz\'alez-Gaitan\altaffilmark{1,6,7},
Maximilian D. Stritzinger\altaffilmark{8},
Mark M. Phillips\altaffilmark{5},
Lluis Galbany\altaffilmark{9},
Gast\'on Folatelli\altaffilmark{10},
Luc Dessart\altaffilmark{11},
Carlos Contreras\altaffilmark{5},
Massimo Della Valle\altaffilmark{12,13},
Wendy L. Freedman\altaffilmark{14},
Eric Y. Hsiao\altaffilmark{15},
Kevin Krisciunas\altaffilmark{16},
Barry F. Madore\altaffilmark{17},
Jos\'e Maza\altaffilmark{2},
Nicholas B. Suntzeff\altaffilmark{16},
Jose Luis Prieto\altaffilmark{18,1},
Luis Gonz\'alez\altaffilmark{2},
Enrico Cappellaro\altaffilmark{19},
Mauricio Navarrete\altaffilmark{5},
Alessandro Pizzella\altaffilmark{20},
Maria T. Ruiz\altaffilmark{2},
R. Chris Smith\altaffilmark{21},
Massimo Turatto\altaffilmark{19}}

\altaffiltext{1}{Millennium Institute of Astrophysics, Casilla 36-D, Santiago, Chile }
\altaffiltext{2}{Departamento de Astronom\'ia, Universidad de Chile, Casilla 36-D, Santiago, Chile}
\altaffiltext{3}{European Southern Observatory, Alonso de C\'ordova 3107, Casilla 19, Santiago, Chile}
\altaffiltext{4}{Department of Physics and Astronomy, University of Southampton, Southampton, SO17 1BJ, UK}
\email{C.P.Gutierrez-Avendano@soton.ac.uk}
\altaffiltext{5}{Carnegie Observatories, Las Campanas Observatory, Casilla 601, La Serena, Chile}
\altaffiltext{6}{Center for Mathematical Modelling, University of Chile, Beauchef 851, Santiago, Chile}
\altaffiltext{7}{CENTRA, Instituto Superior T\'ecnico - Universidade de Lisboa, Portugal}
\altaffiltext{8}{Department of Physics and Astronomy, Aarhus University, Ny Munkegade 120, DK-8000 Aarhus C, Denmark}
\altaffiltext{9}{PITT PACC, Department of Physics and Astronomy, University of Pittsburgh, Pittsburgh, PA 15260, USA}
\altaffiltext{10}{Facultad de Ciencias Astron\'omicas y Geof\'isicas, Universidad Nacional de La Plata, Instituto de 
Astrof\'isica de La Plata (IALP), CONICET, Paseo del Bosque S\/N, B1900FWA La Plata, Argentina}
\altaffiltext{11}{Unidad Mixta Internacional Franco-Chilena de Astronom\'ia (CNRS UMI 3386), Departamento de Astronom\'ia, 
Universidad de Chile, Camino El Observatorio 1515, Las Condes, Santiago, Chile}
\altaffiltext{12}{INAF, Osservatorio Astronomico di Capodimonte, salita Moiariello 16, 80131 Napoli, Italy}
\altaffiltext{13}{International Center for Relativistic Astrophysics, Piazzale della Repubblica 2, I-65122 Pescara, Italy} 
\altaffiltext{14}{Department of Astronomy and Astrophysics, University of Chicago, 5640 South Ellis Avenue, Chicago, IL 60637, USA}
\altaffiltext{15}{Department of Physics, Florida State University, Tallahassee, FL 32306, USA}
\altaffiltext{16}{Department of Physics and Astronomy, Texas A\&M University, College Station, TX 77843, USA}
\altaffiltext{17}{Observatories of the Carnegie Institution for Science, Pasadena, CA 91101, USA}
\altaffiltext{18}{N\'ucleo de Astronom\'ia de la Facultad de Ingenier\'ia y Ciencias, Universidad Diego Portales, Av. Ej\'ercito 441, Santiago, Chile}
\altaffiltext{19}{INAF, Osservatorio Astronomico di Padova, Vicolo dell'Osservatorio 5, 35122 Padova, Italy}
\altaffiltext{20}{Dipartimento di Fisica e Astronomia - Universita’ di Padova, Vicolo dell’Osservatorio 3, I-35122 Padova, Italy}
\altaffiltext{21}{Cerro Tololo Inter-American Observatory, National Optical Astronomy Observatory, Casilla 603, La Serena, Chile}

\begin{abstract}
We present 888 visual-wavelength spectra of 122 nearby type II supernovae (SNe II) obtained between 
1986 and 2009, and ranging between 3 and 363 days post explosion. In this first paper, we outline our 
observations and data reduction techniques, together with a characterization based on the spectral 
diversity of SNe~II. A statistical analysis of the spectral matching technique is discussed as an 
alternative to non-detection constraints for estimating SN explosion epochs. The time evolution of 
spectral lines is presented and analysed in terms of how this differs for SNe of different photometric, 
spectral, and environmental properties: velocities, pseudo-equivalent widths, decline rates, magnitudes, 
time durations, and environment metallicity.
Our sample displays a large range in ejecta expansion velocities, from $\sim9600$ to $\sim1500$ km s$^{-1}$
at 50 days post explosion with a median H$_{\alpha}$ value of 7300 km s$^{-1}$. 
This is most likely explained through differing explosion energies.
Significant diversity is also observed in the absolute strength of spectral lines, characterised through their 
pseudo-equivalent widths. This implies significant diversity in both temperature evolution (linked to
progenitor radius) and progenitor metallicity between different SNe~II.
Around 60\% of our sample show an extra absorption component on the blue side of the H$_{\alpha}$ 
P-Cygni profile (``Cachito'' feature) between 7 and 120 days since explosion. Studying the nature of
Cachito, we conclude that these features at early times (before $\sim35$ days) are associated with 
\ion{Si}{2} $\lambda6355$, while past the middle of the plateau phase they are related to high velocity
(HV) features of hydrogen lines. 

\end{abstract}


\section{Introduction}

Supernovae (SNe) exhibiting prevalent Balmer lines in their spectra are known as Type II SNe
(SNe~II henceforth, \citealt{Minkowski41}). They are produced by the explosion of massive ($>8$ M$_\odot$)
stars, which have retained a significant part of their hydrogen envelope at the time of explosion. Red 
supergiant (RSG) stars have been found at the position of SN~II explosion sites in pre-explosion images
\citep[e.g.][]{VanDyk03, Smartt04, Smartt09, Maund05, Smartt15}, suggesting that they are the direct
progenitors of the vast majority of SNe~II. \\
\indent Initially SNe~II were classified according to the shape of the light curve: SNe with faster 
`linear' declining light curves were cataloged as SNe~IIL, while SNe with a plateau (quasi-constant 
luminosity for a period of a few months) as SNe~IIP  \citep{Barbon79}. Years later, two spectroscopic 
classes and one photometric were added within the SNe~II group: SNe~IIn and SNe~IIb, 
and SN~1987A-like, respectively. SNe~IIn show long-lasting narrow emission lines in their 
spectra \citep{Schlegel90}, attributed to interaction with circumstellar medium (CSM), 
while SNe~IIb are thought to be transitional objects, between SNe~II and SNe~Ib \citep{Filippenko93}.
On the other hand, the 1987A-like events, following the prototype of SN~1987A 
\citep[e.g.][]{Blanco87,Menzies87,Hamuy88,Phillips88,Suntzeff88}, are spectrotroscopically
similar to the typical SNe~II, however their light curves display a peculiar long rise to maximum
($\sim100$ days), which is consistent with a compact progenitor.
The latter three sub-types (IIn, IIb and 87A-like) are not included in the bulk
of the analysis for this paper.\\
\indent Although it has been shown that SNe~II\footnote{Throughout the remainder of the manuscript we use 
SN~II to refer to all SNe which would historically have been classified as SN~IIP or SN~IIL. 
In general we will differentiate these events by referring to their specific light curve or spectral
morphology, and we only return to this historical separation if clarification and comparison with
previous works is required.} are a continuous single population \citep[e.g.][]{Anderson14, Sanders15,
Valenti16}, a large spectral and photometric diversity is observed.
\citet{Pastorello04} and \citet{Spiro14} studied a sample of low luminosity SNe~II.
They show that these events present, in addition to low luminosities (M$_V\geq -15.5$ at peak), narrow spectral
lines. Later, \citep{Inserra13} analyzed a sample of moderately luminous SNe~II, finding 
that these SNe, in contrast to the low luminosity events, are relatively bright at peak (M$_V\leq -16.95$).\\
\indent In addition to these samples, many individual studies have been published showing spectral line 
identification, evolution and parameters such as velocities and pseudo-equivalent widths (pEWs) for 
specific SNe. Examples of very well studied SNe include SN~1979C \citep[e.g.][]{Branch81, Immler05}, SN~1980K
\citep[e.g.][]{Buta82,Dwek83,Fesen99}, SN~1999em \citep[e.g.][]{Baron00,Hamuy01,Leonard02b,Dessart06},
SN~1999gi \citep[e.g.][]{Leonard02a}, SN~2004et \citep[e.g.][]{Li05,Sahu06,Misra07,Maguire10}, SN~2005cs
\citep[e.g.][]{Pastorello06,Dessart08b,Pastorello09}, and SN~2012aw \citep[e.g.][]{Bose13,Dallora14,Jerkstrand14}.
The first two SNe (1979C and 1980K) are the prototypes of fast declining SNe~II (SNe~IIL), together with
unusually bright light curves and high ejecta velocities. On the other hand, the rest of the
objects listed are generally referred to as SNe~IIP as they display relatively slowly declining light curves.
For faint SNe similar to SN~2005cs, the expansion velocity and luminosity 
are even lower, probably due to low energy explosions (see \citealt{Pastorello09}).\\
\indent In recent years, the number of studies of individual SNe~II has continued to increase, however 
there are still only a handfull of statistical analyses of large samples 
\citep[e.g.][]{Patat94,Arcavi10,Anderson14,Gutierrez14,Faran14a,Faran14b,Sanders15,Pejcha15,Pejcha15a,Valenti16,
Galbany16,Muller17}.
Here we attempt to remedy this situation.  
The purpose of this paper is to present a statistical characterization of the 
optical spectra of SNe~II, as well as an initial analysis of their spectral features. 
We have analyzed 888 spectra of 122 SNe~II ranging between 3 and 363 days since explosion. We selected 
11 features in the photospheric phase with the aim of understanding the overall evolution of 
visual-wavelength spectroscopy of SNe~II with time.\\
\indent The paper is organized as follows. In section~\ref{data} we describe the data sample. The 
spectroscopic observations and data reduction techniques are presented in section~\ref{obs}.
In section~\ref{explo} the estimation of the explosion epoch is presented.
In section~\ref{prop} we describe the sample properties, while in section~\ref{line} we identify
spectral features. The spectral measurements are presented in section~\ref{measure} while 
the line evolution analysis and the conclusions are in section~\ref{ana} and~\ref{conc}, 
respectively.\\
\indent In Paper II, we study the correlations between different spectral and photometric parameters, 
and try to understand these in terms of the diversity of the underlying physics of the explosions and
their progenitors.

\section{Data sample}
\label{data}

\begin{figure}
\centering
\includegraphics[width=8.5cm]{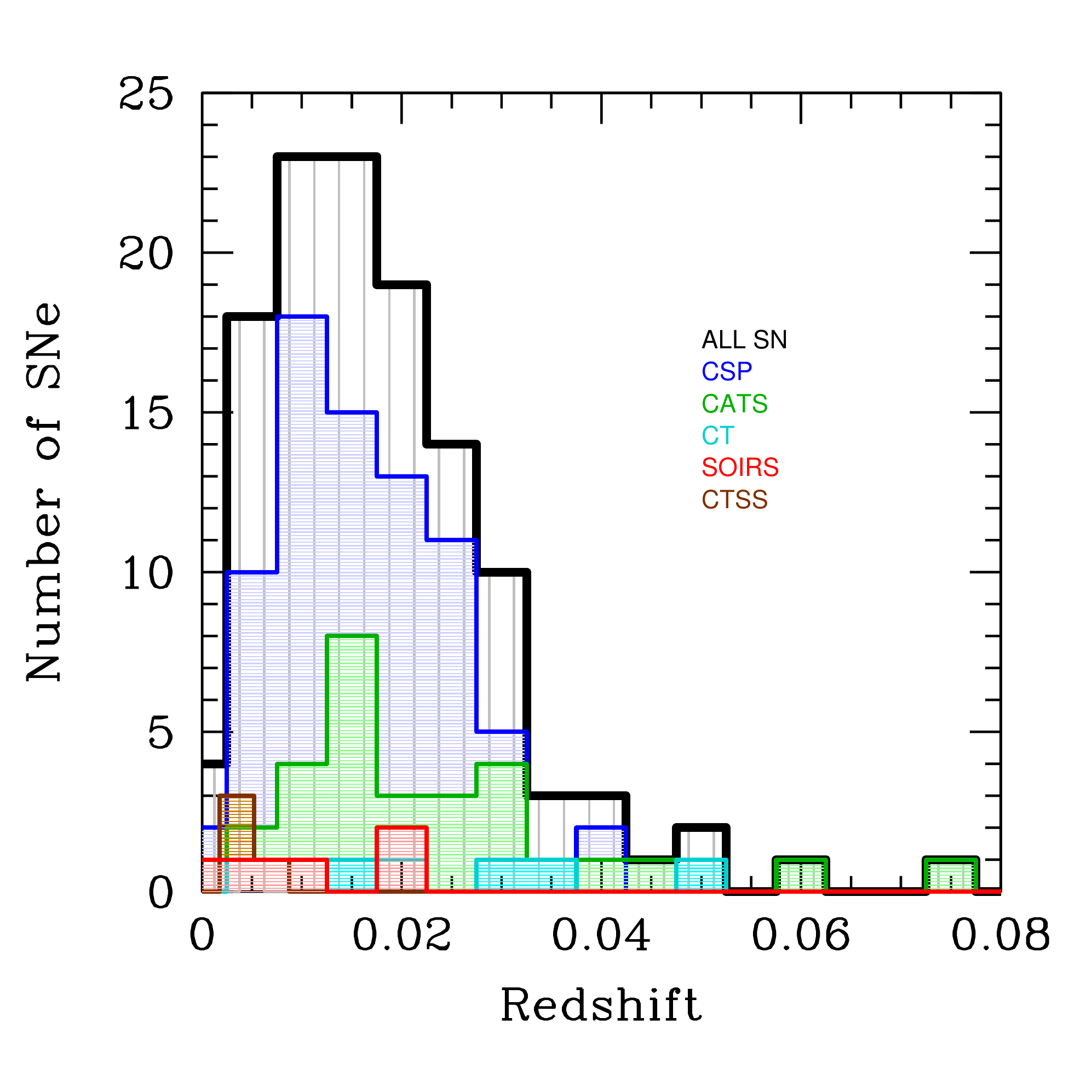}
\caption{Distribution of heliocentric redshifts for the 122 SN II in our sample.}
\label{snz}
\end{figure}

Our dataset was obtained between 1986 and 2009 from a variety of different sources.
This sample consists of 888 optical spectra of 122 SNe II\footnote{In the data release we include
eight spectra of the SN~2000cb, a SN~1987A-like event, which is not analyzed in this work.}, of which 
four were provided by the Cerro Tololo Supernova Survey (CTSS), seven were obtained by the 
Cal\'an/Tololo survey (CT, \citealt{Hamuy93}, PI: Hamuy 1989-1993), five by the Supernova Optical 
and Infrared Survey (SOIRS, PI: Hamuy, 1999-2000), 31 by the Carnegie Type II Supernova Survey 
(CATS, PI: Hamuy, 2002-2003) and 75 by the Carnegie Supernova Project (CSP-I, \citealt{Hamuy06}, 2004-2009).
These follow-up campaigns concentrated on obtaining well-sampled and high-cadence light curves and
spectral sequences of nearby SNe, based mainly on 2 criteria: 1) that the SN was brighter than V$\sim17$
mag at discovery, and 2) that those discovered SNe were classified as being relatively young, i.e.,  
less than one month from explosion. \\
\indent The redshift distribution of our sample is shown in Figure~\ref{snz}. 
The figure shows that the majority of the sample have a redshift $\leq0.03$. SN~2002ig has the highest
redshift in the sample with a value of 0.077, while the nearest SN (SN~2008bk) has a redshift of 0.00076.
The mean redshift value of the sample is 0.0179 and the median is 0.0152. 
The redshift information comes from the heliocentric recession velocity of each host galaxy as published
in the NASA/IPAC extragalactic Database (NED)\footnote{http://ned.ipac.caltech.edu}. These NED values
were compared with those obtained through the measurement of narrow emission lines observed within SN 
spectra and originating from host \ion{H}{2} regions. In cases of discrepancy between the two sources,
we give priority to our spectral estimations.  
Two of our objects (SN~2006Y and SN~2007ld) occur in unknown host galaxies. Their redshifts 
were obtained from the Asiago supernova catalog\footnote{http://sngroup.oapd.inaf.it} and from
the narrow emission lines within SN spectra originating from the underlying host galaxy, respectively.     
Table~\ref{t_info} lists the sample of SNe~II selected for this work, their host galaxy information, 
and the campaign to which they belong.\\
\indent From our SNe~II sample, SNe~IIn, SNe~IIb and SN~1987A-like events (SN~2006au and SN~2006V;
\citealt{Taddia12}) were excluded based on photometric information. Details of the SNe~IIn sample can
be found in \citet{Taddia13}, while those of the SNe~IIb in \citet{Stritzinger17} and \citet{Taddia17}.
The photometry of our sample in the $V-$band was published by 
\citet{Anderson14}. More recently, \citet{Galbany16} released the UBVRIz photometry of our sample 
obtained by CATS between 1986 and 2003. Around 750 spectra of $\sim100$ objects are published here for the 
first time. Now we briefly discuss each of the surveys providing SNe for our analysis.

\subsection{The Cerro Tololo Supernova Survey}

A total of 4 SNe~II (SN~1986L, SN~1988A, SN~1990E, and SN~1990K) were extensively observed at CTIO
by the Cerro Tololo SN program  (PIs: Phillips \& Suntzeff, 1986-2003). These SNe have been analyzed in 
previous works \citep[e.g][]{Schmidt93, Turatto93a, Cappellaro95, Hamuy01T}. 

\subsection{The Cal\'an/Tololo survey (CT)}

The Cal\'an/Tololo survey was a program of both discovery and follow-up of SNe.
A total of 50 SNe were obtained between 1989 and 1993. 
The analysis of SNe~Ia was published by \citet{Hamuy96}. Spectral and photometric details of 
six SNe~II were presented by \citet{Hamuy01T}. In this analysis we include these SNe~II and 
an additional object, SN~1993K.

\subsection{The Supernova Optical and Infrared Survey (SOIRS)}

The Supernova Optical and Infrared Survey carried out a program to obtain optical and IR photometry
and spectroscopy of nearby SNe ($z<0.08$). In the course of 1999-2000, 20 SNe were observed, six
of which are SNe~II. Details of these SNe were published by \citet{Hamuy01T, Hamuy01},  
\citet{Hamuy02L}, and \citet{Hamuy03}.

\subsection{The Carnegie Type II Supernova Survey (CATS)}

Between 2002 and 2003 the Carnegie Type II Supernova Survey observed 34 SNe~II. While optical spectroscopy
and photometry of these SNe~II have been previously used to derive distances \citep{Olivares08, Jones09}, 
the spectral observations have not been officially released until now.

\subsection{The Carnegie Supernova Project I (CSP-I)}

The Carnegie Supernova Project I (CSP-I)  was a five year follow-up program to obtain high quality optical 
and near infrared light curves and optical spectroscopy. The data obtained by the CSP-I between 2004 and 2009
consist of $\sim250$ SNe of all types, of which 75 correspond to SNe~II.  The first SN~Ia photometry data 
were published in \citet{Contreras10}, while their analysis was done by \citet{Folatelli10}. A second data  
release was provided by \citet{Stritzinger11}. A spectroscopy analysis of SNe~Ia was published by 
\citet{Folatelli13}. Recently, \citet{Stritzinger17} and \citet{Taddia17} published the 
photometry data release of stripped-envelope supernovae. The CSP-I spectral data for SNe~II are published 
here for the first time, while the complete optical and near-IR photometry will be published by Contreras et al. (in prep).

\section{Observations and data reduction}
\label{obs}

In this section we summarize our observations and the data reduction techniques. However, a 
detailed description of the CT methodology is presented in \citet{Hamuy93}, in the case of SOIRS is 
described in \citet{Hamuy01} and for CSP-I can be found in \citet{Hamuy06} and \citet{Folatelli13}.

\subsection{Observations}

\begin{figure}[h!]
\centering
\includegraphics[width=8.5cm]{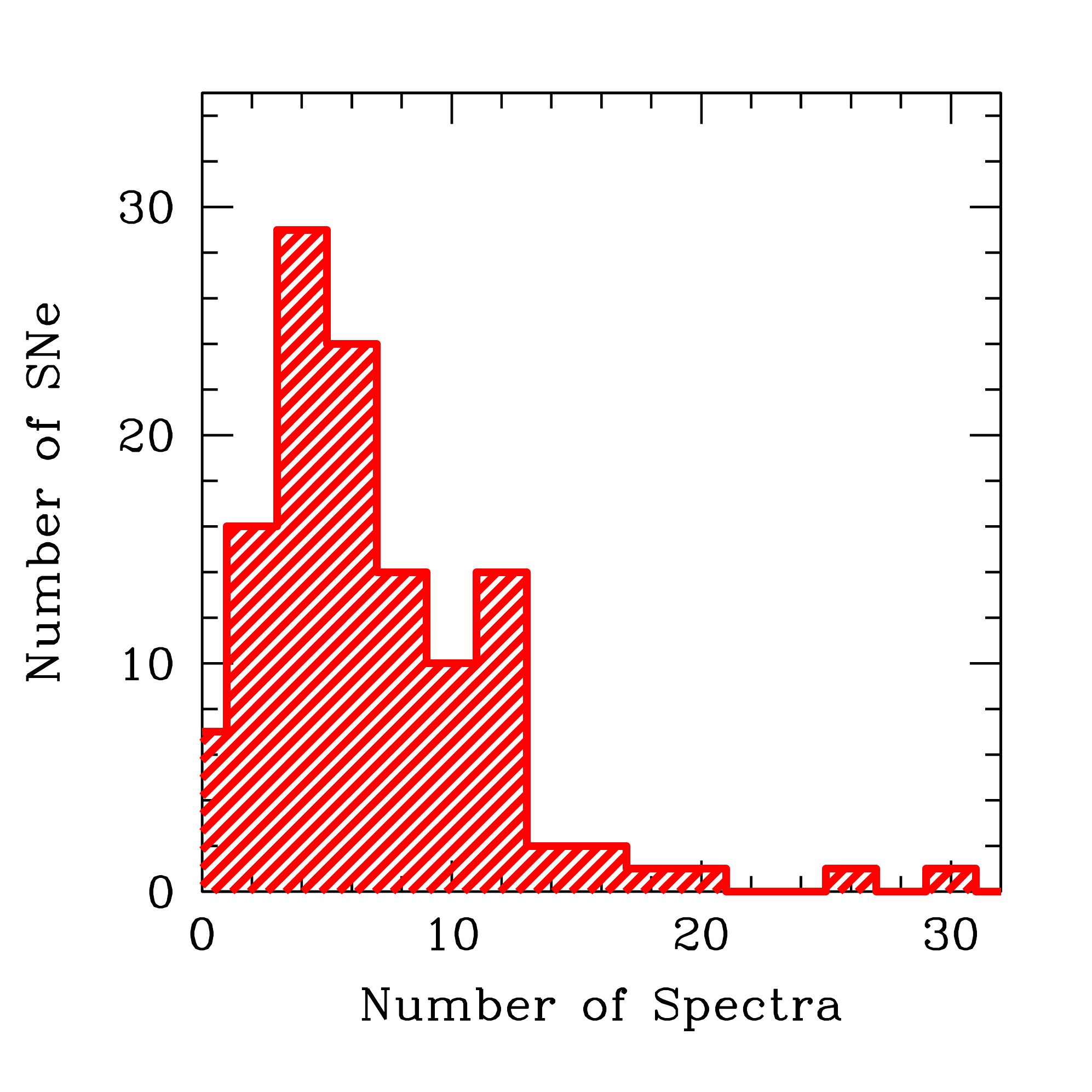}
\caption{Histogram of the number of spectra per SN. The distribution peaks at 4 spectra.}
\label{spec}
\end{figure}

The data presented here were obtained with a large variety of instruments and telescopes, as shown
in Table~\ref{t_spec}. The majority of the spectra were taken in long-slit spectroscopic mode with the 
slit placed along the parallactic angle. However, when the SN was located close to the host, it was necessary to pick a different
and more convenient angle to avoid contamination from the host. The majority of our spectra cover the range 
of $\sim3800$ to  $\sim9500$ \AA.
The observations were performed with the Cassegrain spectrographs at 1.5-m and 4.0-m telescopes at Cerro 
Tololo, with the Wide Field CCD Camera (WFCCD) at the 2.5m du Pont Telescope, the Low Dispersion Survey Spectrograph
(LDSS2; \citealt{Allington-Smith94}) on the Magellan Clay 6.5-m telescope and the Inamori Magellan Areal Camera and 
Spectrograph (IMACS; \citealt{Dressler11}) on the Magellan Baade 6.5-m telescope at 
Las Campanas Observatory. At La Silla, the observations were carried out with the ESO Multi-Mode Instrument 
(EMMI; \citealt{Dekker86}) in medium resolution spectroscopy mode (at the NTT)
and the ESO Faint Object Spectrograph and Camera (EFOSC; \citealt{Buzzoni84}) at the NTT and 3.6-m telescopes.
We also have 3 spectra for SN~2006ee obtained with the Boller \& Chivens CCD spectrograph at the Hiltner 2.4 m
Telescope of the MDM Observatory. 
Table~\ref{t_spec} displays a complete journal of the 888 spectral observations, listing for each spectrum
the UT and Julian dates, phases, wavelength range, FWHM resolution, exposure time,
airmass, and the telescope and instrument used. \\
\indent The distribution of the number of spectra per object for our sample is shown in
Figure~\ref{spec}. Seven SNe (SN~1993A, SN~2005dt, SN~2005dx, SN~2005es, SN~2005gz, SN2005me, SN~2008H)
only have one spectrum, while 90\% of the sample have between two and twelve spectra.
SN~1986L is the object with the most spectra (31), followed by SN~2008bk with 26.
On average we have 7 spectra per SN and a median of 6. There are 87 SNe~II for which we have
five or more spectra, 32 that have ten or more, and 6 objects with over 15 spectra 
(SN~1986L, SN~1993K, SN~2007oc, SN~2008ag, SN~2008bk and SN~2008if).
In the current work, 4\% of our obtained spectra are not used for analysis. 3\% correspond to spectra with low S/N that
does not allow useful extraction of our defined parameters, while 1\% are related with peculiarities in the spectra 
(see Section~\ref{prop} for more details). Despite this, these spectra are still included in the 
data release, and are noted in Table~\ref{t_spec}.

\subsection{Data reduction}

Spectral reduction was achieved in the same manner for all data, using 
IRAF and employing standard routines, including: bias subtraction, flat-fielding correction, one-dimensional (1D) spectral
extraction and sky subtraction, wavelength correction, and flux calibration. Telluric corrections have 
only been applied to data obtained after October 2004. \\
\indent In Appendix~\ref{spectra} (spectral series) we show plots with the spectral series for all SNe of our sample.

\section{Explosion epoch estimations}
\label{explo}

Before discussing the properties of our sample, in this section we outline our methods for estimating
explosion epochs. The non-detection of SNe on pre-discovery images with high cadence is the most accurate method for 
determining the explosion epoch for any given SN.
Explosion epochs based on non-detections  are set to the mid-point between  
SN discovery and non-detection. The representative uncertainty on this epoch is then (MJD$_{disc}-$MJD$_{non-det})/2$.
However within our sample (and for many other current SN search campaigns) many SNe do not have such 
accurate constraints from this method due to the low cadence of the observations. \\
\indent Over the last decade several tools have been published, enabling explosion epoch estimations through
matching of observed SN spectra to libraries of spectral templates. Programs such as the
Supernova Identification (SNID) code \citep{Blondin07}, the GEneric cLAssification TOol (Gelato)
\citep{Harutyunyan08}, and superfit \citep{Howell05} allow the user to estimate the type of supernova and 
its epoch by providing an observed spectrum. 
All perform classifications by comparison using different methods.
In our analysis we used only the first two methods: SNID
and Gelato. We find that Gelato gives a large percentage of their quality of fit to 
H$_{\alpha}$ P-Cygni profile. However, based on our analysis (see Section~\ref{ana}), the most significant
changes with time are observed in the blue part of the spectra (i.e. between 4000 and 6000 \AA). Moreover, according to
\citet{Gutierrez14}, the H$_{\alpha}$ P-Cygni profile shows a wide diversity and there is no clear, 
consistent evolution with time. 
In addition, SNID provides the possibility of adding additional templates to improve the accuracy of 
explosion epoch determinations. We take advantage of this attribute in the following sections by
adding new spectral templates, which aid in obtaining more accurate explosion epochs for our sample.\\
\indent While for many SNe this spectral matching is required to obtain a reliable explosion 
epoch, a significant fraction of our sample do have explosion epoch constraining SN non-detections
before discovery. In cases where the non-detection is $<20$ days before discovery, we use that information
to estimate our final values. In cases where this difference is larger than 20 days, we use the spectral 
matching technique. As a test of our methodology, for non-detection SNe we also estimate explosion 
epochs using spectral matching to check the latter's validity (see below for more details).

\subsection{SNID implementation}

To constrain the explosion epoch for our sample, we compare the first spectrum of each 
SN~II with a library of spectral templates provided by SNID and then, we choose
the best match. For each SN we examined multiple matches putting emphasis on the fit of 
the blue part of the spectrum between 4000 and 6000 \AA. This region contains many spectral 
lines that display a somewhat consistent evolution with time, unlike the dominant 
H$_{\alpha}$ profile at redder wavelengths. Explosion epoch errors from this spectral matching 
are obtained by taking the standard deviation of several good matches of the observed 
spectrum of our selected object with those from the SNID library.
H$_{\alpha}$ is the dominant feature in SN~II spectra, however its evolution and morphology
varies greatly between SNe in a manner that does not aid in the spectral matching technique.
We therefore ignore this wavelength region.\\ 
\indent The red part of the spectrum can be ignored during spectral matching in a variety 
of ways. 1) using the SNID options; or 2) checking only the match in the blue part.
For the former, SNID gives to the user the 
alternative to modifying some parameters. In our case, we can constrain the wavelength range using
\textit{wmin} and \textit{wmax}. Hence, the structure used is: \textit{``snid wmin=3500 
wmax=6000 spec.dat''}. For the latter, we just need to ignore visually the red part of the spectra
and explore the matches obtained by SNID until find a good fit in the blue part\footnote{Note that
the results obtained from the spectral matching are not altered if you use either all visible
wavelength spectrum or just the region between 4000 and 6000 Å.}.\\
\indent From the SNID library we use those template SNe that have well constrained explosion epochs,
meaning SNe~II with explosion epoch errors of less than five days (see Table~\ref{t_expl}). Specifically, 
we used SN~1999em \citep{Leonard02b}, SN~1999gi \citep{Leonard02a}, SN~2004et \citep{Li05}, SN~2005cs 
\citep{Pastorello06}, and SN~2006bp \citep{Dessart08b}. In the database of SNID there are a total of 166
spectra. However, these templates do not provide a good coverage of the
overall diversity of SNe~II within our sample/the literature. Most of the SNe in the library are
relatively `normal', with only one sub-luminous event (SN~2005cs). This means that any non-normal 
event within our sample will probably have poor constraints on its explosion epoch using these 
templates. For this reason we decided to use some of our own well-observed SNe~II to complement
the SNID database.

\subsection{New SNID templates}

We created a new set of spectral templates using our own SNe~II non-detection limits. SNe~II are included
as new SNID templates if they have errors on explosion epochs (through non-detection constraints) of less than 5 days.
Given this criterion, we included 22 SNe, which show significant spectral and photometric diversity. 
In this manner, the new SNID templates were constructed using $\sim150$ spectra and prepared using
the \textit{logwave} program included in the SNID packages. Adding our own template SNe to the SNID database
we can now use a total of 27 template SNe~II to estimate the explosion epoch.
Table~\ref{t_expl} shows the explosion epoch and the maximum dates in $V-$band for the reference SNe,
as well as the explosion epoch for our new templates. We note an important difference between our 
templates and previous ones in SNID: for the newer templates epochs are labelled with respect
to the explosion epoch, while for the older templates epochs are labelled with respect to maximum light 
(meaning that one then has to add the ``rise time'' to obtain the actual explosion date, see Table~\ref{t_expl}). \\

\subsection{Explosion epochs for the current sample}

\begin{figure*}
\centering
\includegraphics[width=6.5cm]{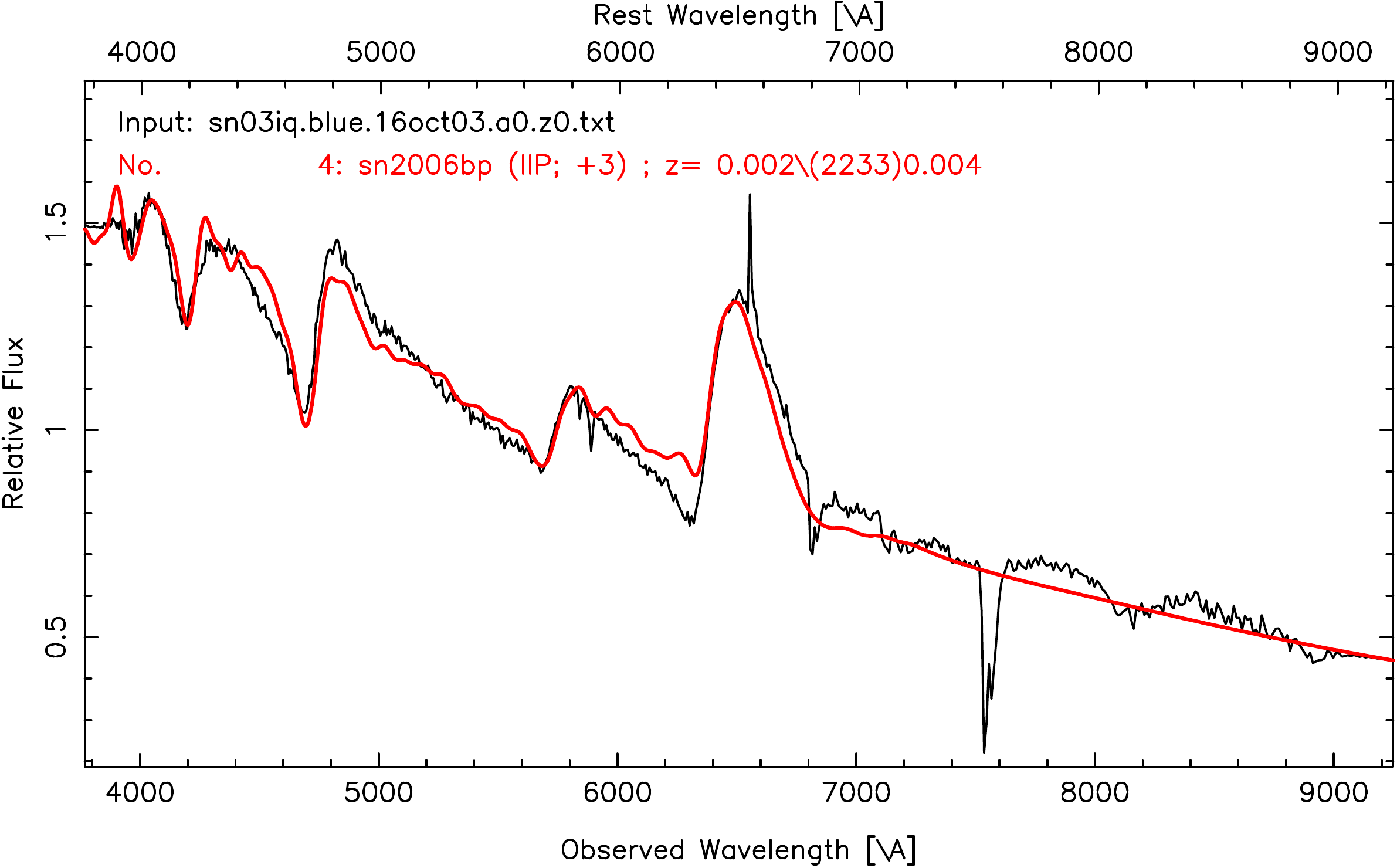}
\includegraphics[width=6.5cm]{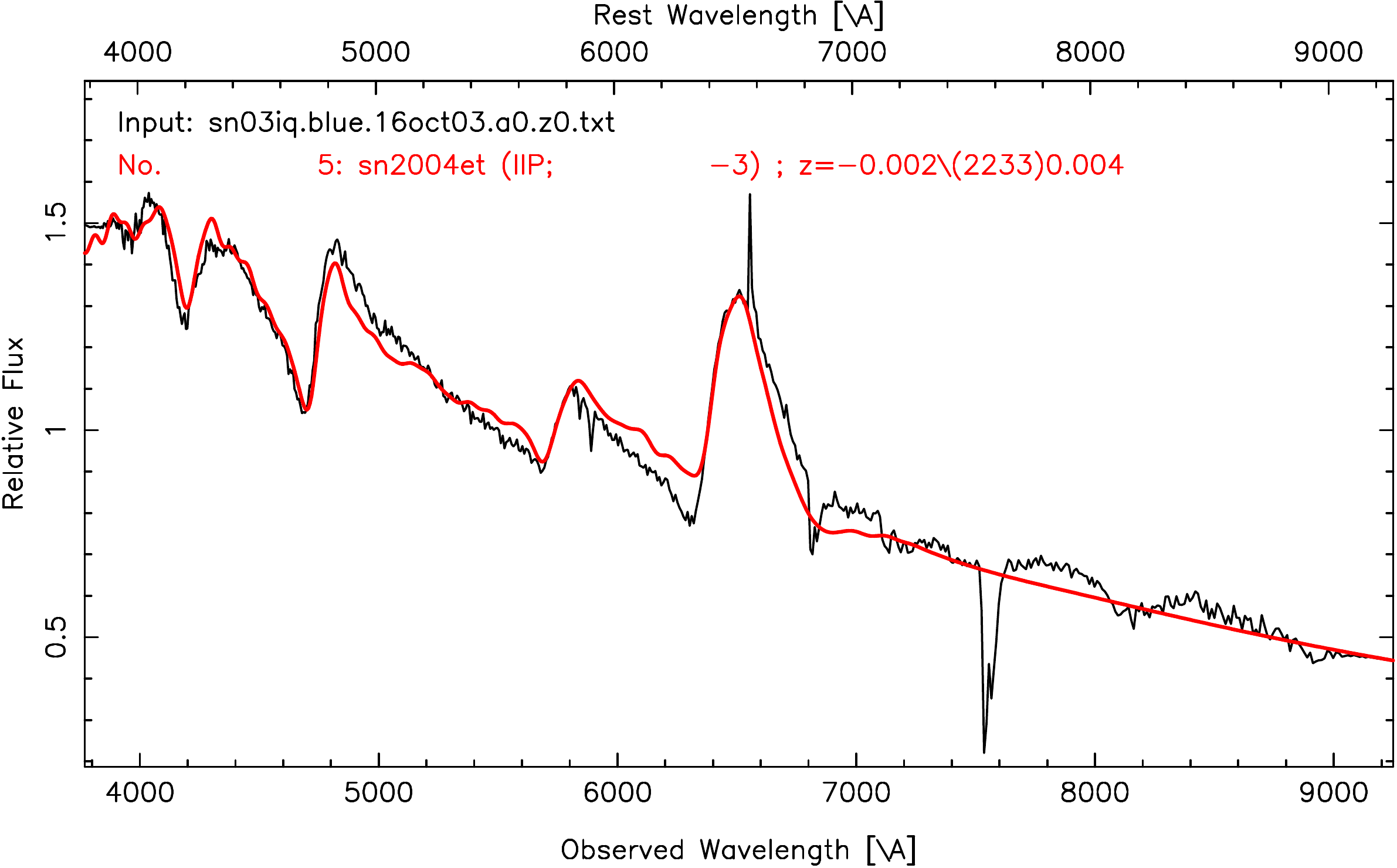}
\includegraphics[width=6.5cm]{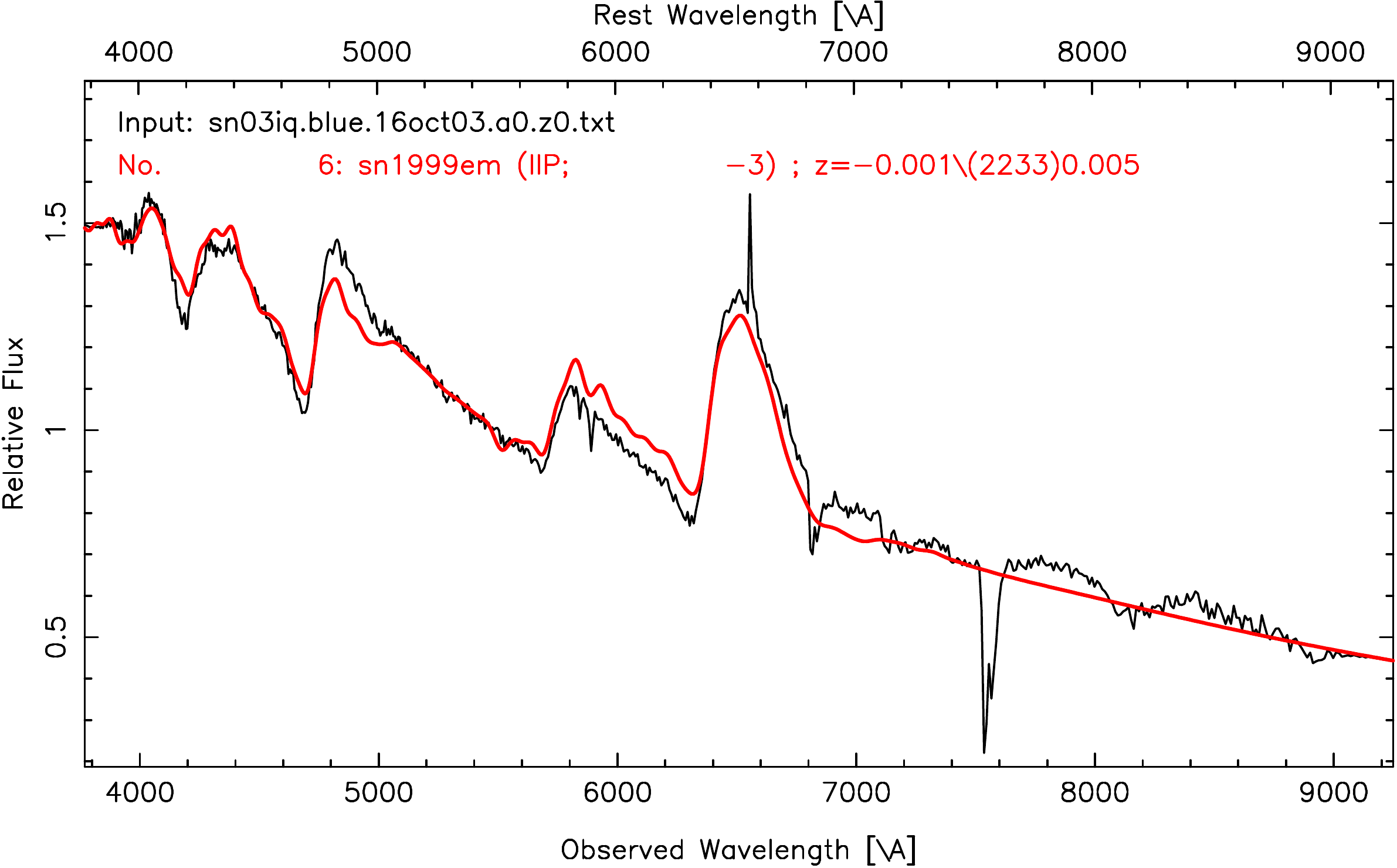}
\includegraphics[width=6.5cm]{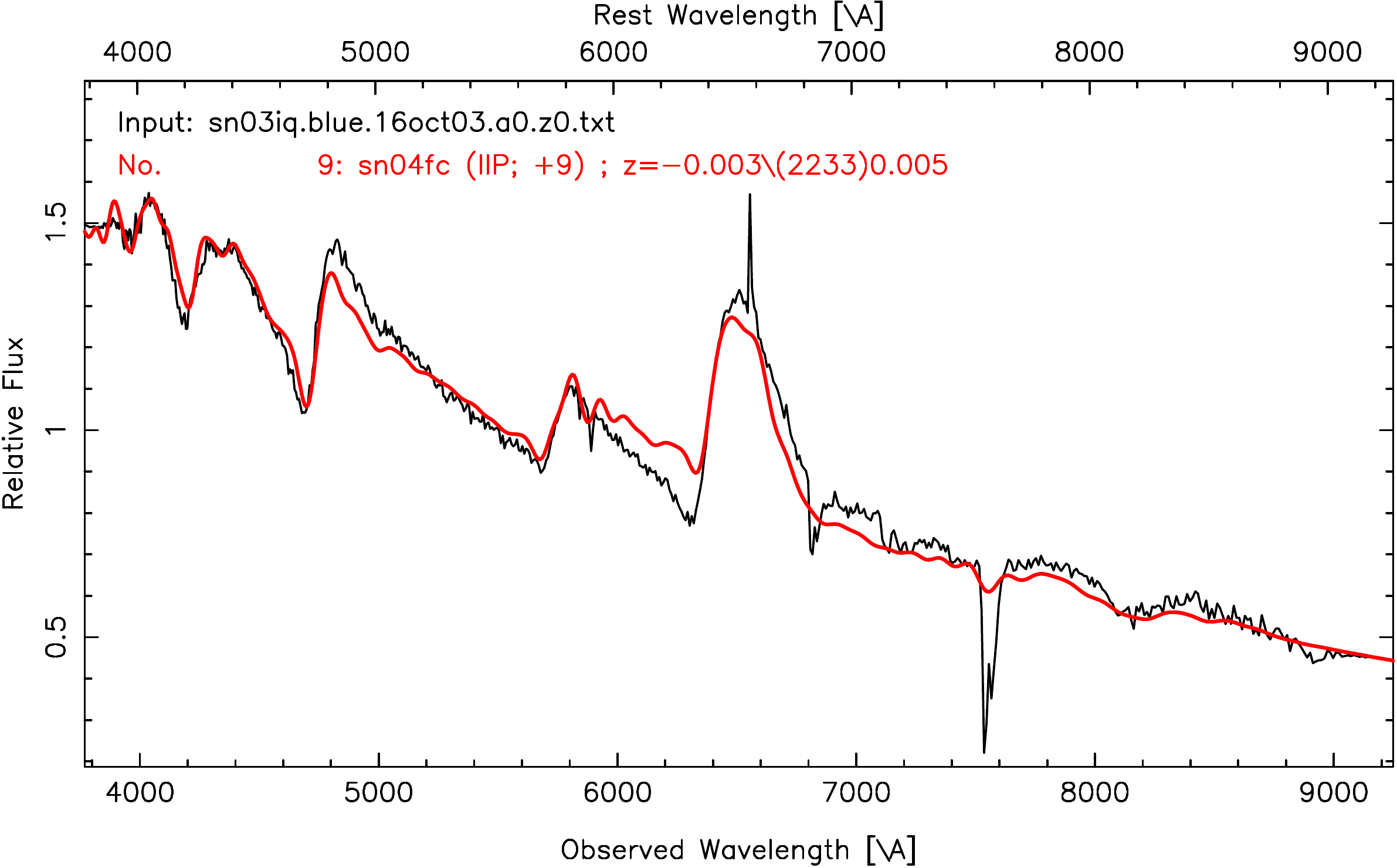}
\caption{Best spectral matching of SNe~2003iq using SNID. 
The plots show SN~2003iq compared with SN~2006bp, SN~2004et, SN~1999em and SN~2004fc
at 3, $-3$ and $-3$ and 9 days. As the first three SNe are included in the SNID database, they are
respect to the maximum, hence we have to add to them the days between the explosion and the $V-$band
maximum (see Table~\ref{t_expl}) to obtain the explosion epoch. On the other hand, SN~2004fc 
(included in this work) is respect to the explosion. Therefore, SN~2003iq has a good match with
these SNe at 12, 13, 7 and 9 days from explosion, respectively.
Taking the average, this means that this spectrum is at 10$\pm7$ days from explosion.}
\label{expl}
\end{figure*}

With the inclusion of these 22 SNe to SNID we estimated the explosion epoch for our full sample. 
An example of the best match is shown in Figure~\ref{expl}. We can see that first spectrum of 
SN~2003iq (October 16th) is best matched with SN~2006bp, SN~2004et, SN~1999em and SN~2004fc
12, 13, 7 and 9 days from explosion, respectively.
Taking the average, we conclude that the spectrum was obtained at 10$\pm7$ days since explosion. 
Table~\ref{t_info} shows the explosion epoch for each SN as well as the method
employed to derive it, while Table~\ref{table_explosion} shows all the details of spectral matching
and non-detection techniques.
Appendix~\ref{snid} (SNID matches) shows the plots with the best matches for each SN in our sample.

\begin{figure}[h!]
\centering
\includegraphics[width=8.5cm]{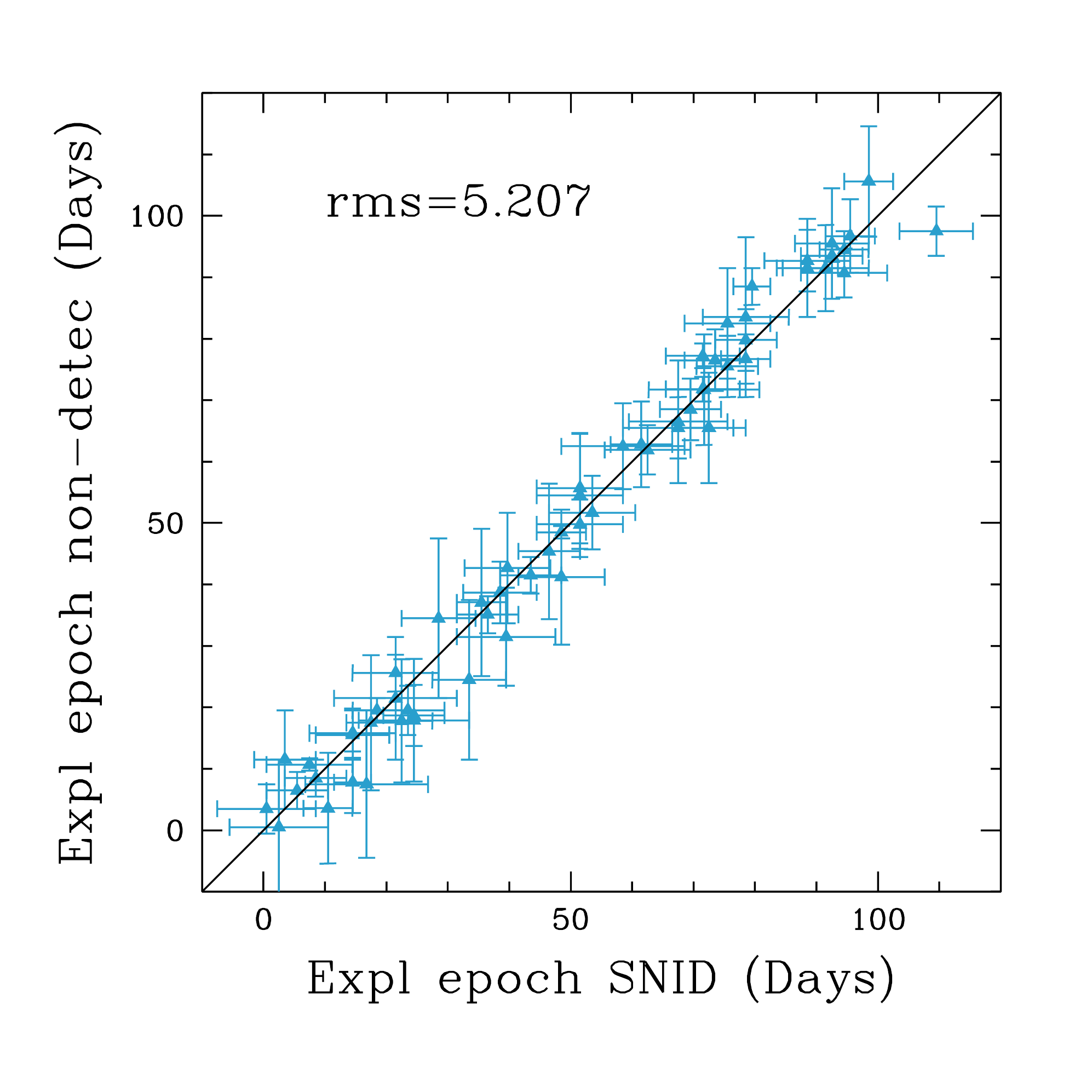}
\caption{Comparison between spectral matching and non-detection methods.}
\label{explosion}
\end{figure}

\indent To check the validity of spectral matching we compare the explosion epoch estimated 
with this technique and those with non-detections. These two estimations are displayed 
in Table~\ref{table_explosion}. 
From the second to the seventh column, the spectral matching details are shown (spectrum date,
best match found, days from maximum --from the SNID templates-- days from explosion, average,
and explosion date), while from eighth to tenth, those obtained from the non-detection (non-detection date,
discovery date and explosion date). The differences between both methods are presented in the last 
column. Such an analysis
was previously performed by \citet{Anderson14} where good agreement was found. With the use of 
our new templates we are able to improve the agreement between different explosion epoch constraining
methods, thus justifying their inclusion.
Figure~\ref{explosion} shows a comparison between both methods, where the mean absolute 
error between them diminishes from 4.2 \citep{Anderson14} to 3.9 days. 
Also the mean offset decreases from 1.5 days in \citet{Anderson14} to 0.5 days in this work.
Cases where explosion epochs have changed between \citet{Anderson14} and the current work 
are noted in Table~\ref{t_info}.
Nevertheless, although this method works well as a substitute for non-detections, exact constraints 
for any particular object are affected by any peculiarities inherent to the observed (or indeed template)
SN. For example, differences in the colour (and therefore temperature) evolution of events can
mimic differences in time evolution, while progenitor metallicity differences can delay/hasten the 
onset of line formation.
Further improvements of this technique can only be obtained by the inclusion of additional,
well observed SNeII in the future. \\

\section{Sample properties}
\label{prop}

\begin{figure}[h!]
\centering
\includegraphics[width=8.5cm]{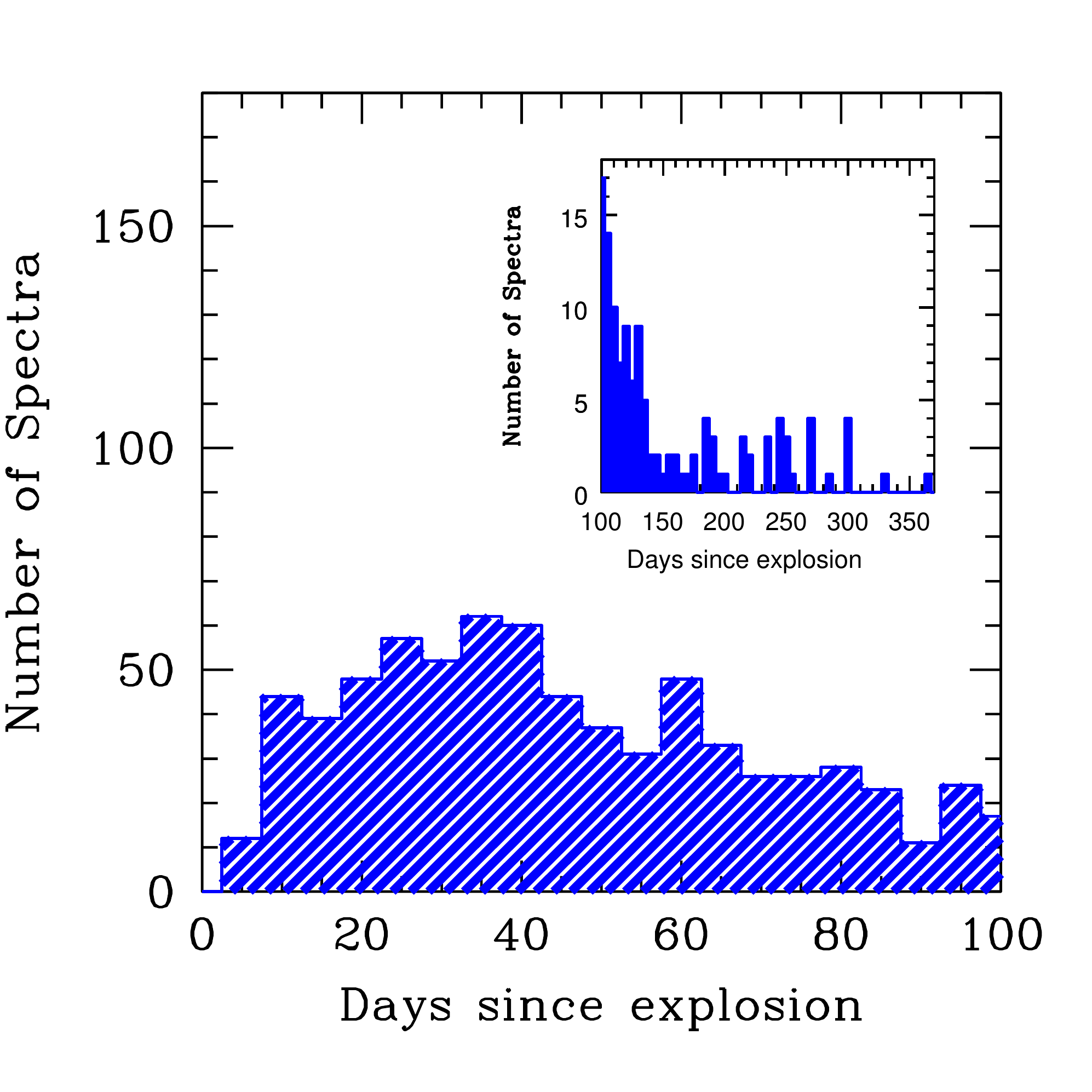}
\caption{Distribution of the number of spectra as a function of epoch from explosion. 
The inset on the right shows the same distribution between 100 and 370 days. }
\label{nspec}
\end{figure}

As mentioned in Section~\ref{data} we have 888 optical spectra of 122 SNe~II,
however due to low signal-to-noise ($S/N$) we remove 26 spectra of 12 SNe for our analysis.
We also remove nine spectra of SN~2005lw because they contain peculiarities that we expect are not
intrinsic to the SN (most probably defects resulting from the observing procedure or data
reduction). In total, we remove 35 spectra ($\sim4\%$).
Figure~\ref{nspec} shows the epoch distribution of our spectra 
since explosion to 370 days. One can see the majority (86\%) of the spectra
were observed between 0 and 100 days since explosion, with a total of 738 spectra.
Our earliest spectrum corresponds to SN~2008il at $3\pm3$ days and SN~2008gr at  $3\pm6$ days 
from explosion, while the oldest spectrum is at $363\pm9$ days for SN~1993K. 
53\% of the spectra were taken prior to 50 days, 3.8\% of which were observed before 10 days for 23 SNe.
Between $\sim30$ to 84 days there are 441 spectra of 114 SNe. 
There are 115 spectra older than 100 days and 27 older than 200 days, corresponding to
45 and 4 SNe, respectively. The average of spectra as a function of epoch from explosion is 60 days,
while its median is 46 days.

\begin{figure}
\centering
\includegraphics[width=8.5cm]{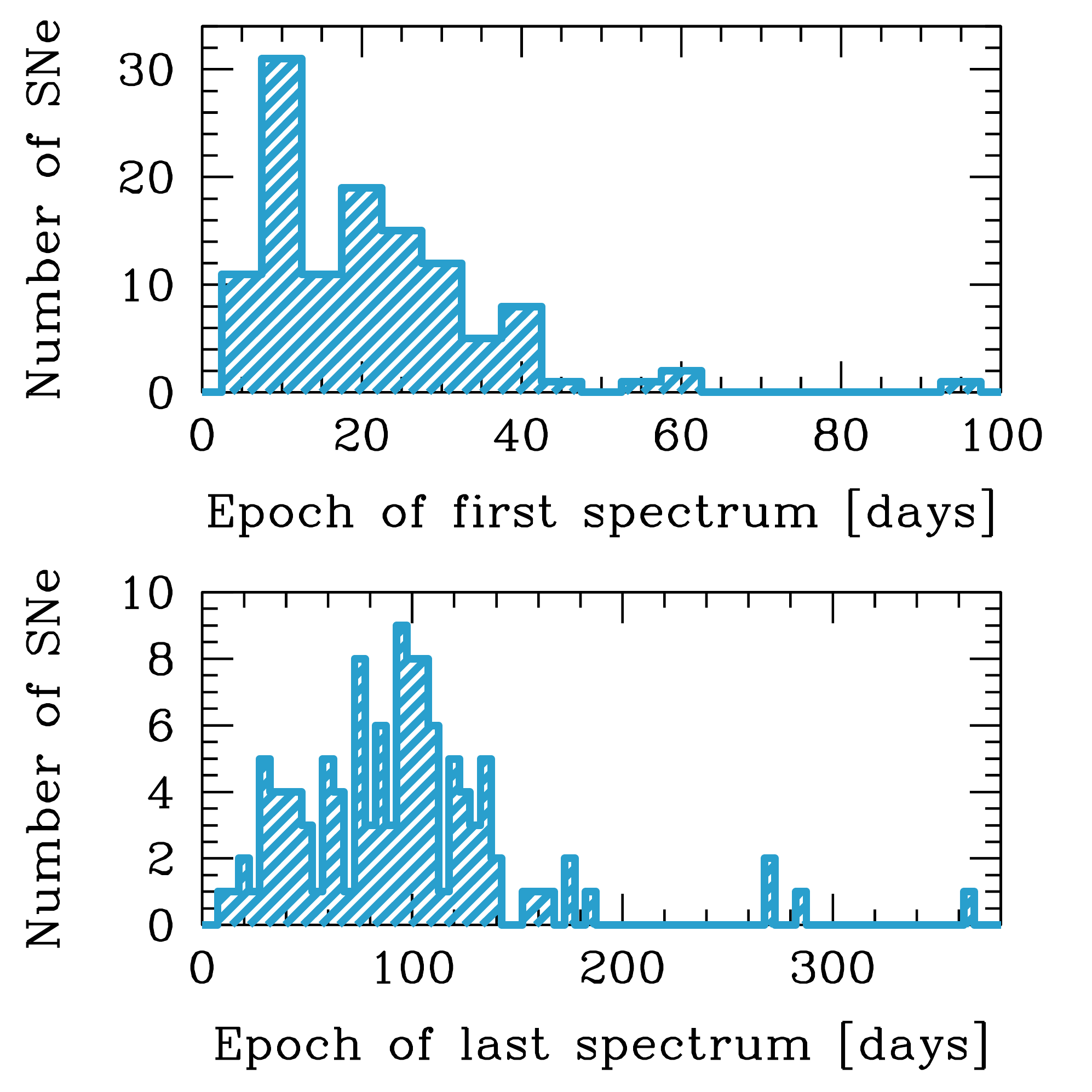}
\caption{\textit{Top:} Epoch from explosion of first spectrum. \textit{Bottom:} Epoch from explosion of last spectrum. }
\label{puspec}
\end{figure}

\indent Figure~\ref{puspec} shows the epoch distribution of the first and last spectrum for each SN in our sample.
The majority of SNe have their first spectra within 40 days from explosion. There are 31 SNe with their first
spectra around 10 days (the peak of the distribution). 
On the other hand, the peak of the distribution of the last spectrum is around 100 days. Almost all SNe have
their last spectra between 30 and 120 days, i.e., in the photospheric phase. There are 11 SNe with their last
spectrum after 140 days, while only 4 SNe (SN~1993K, 2003B, SN~2007it, SN~2008bk) have their last spectrum in 
the nebular phase ($\geq200$ days).

\begin{figure}[h!]
\centering
\includegraphics[width=8.5cm]{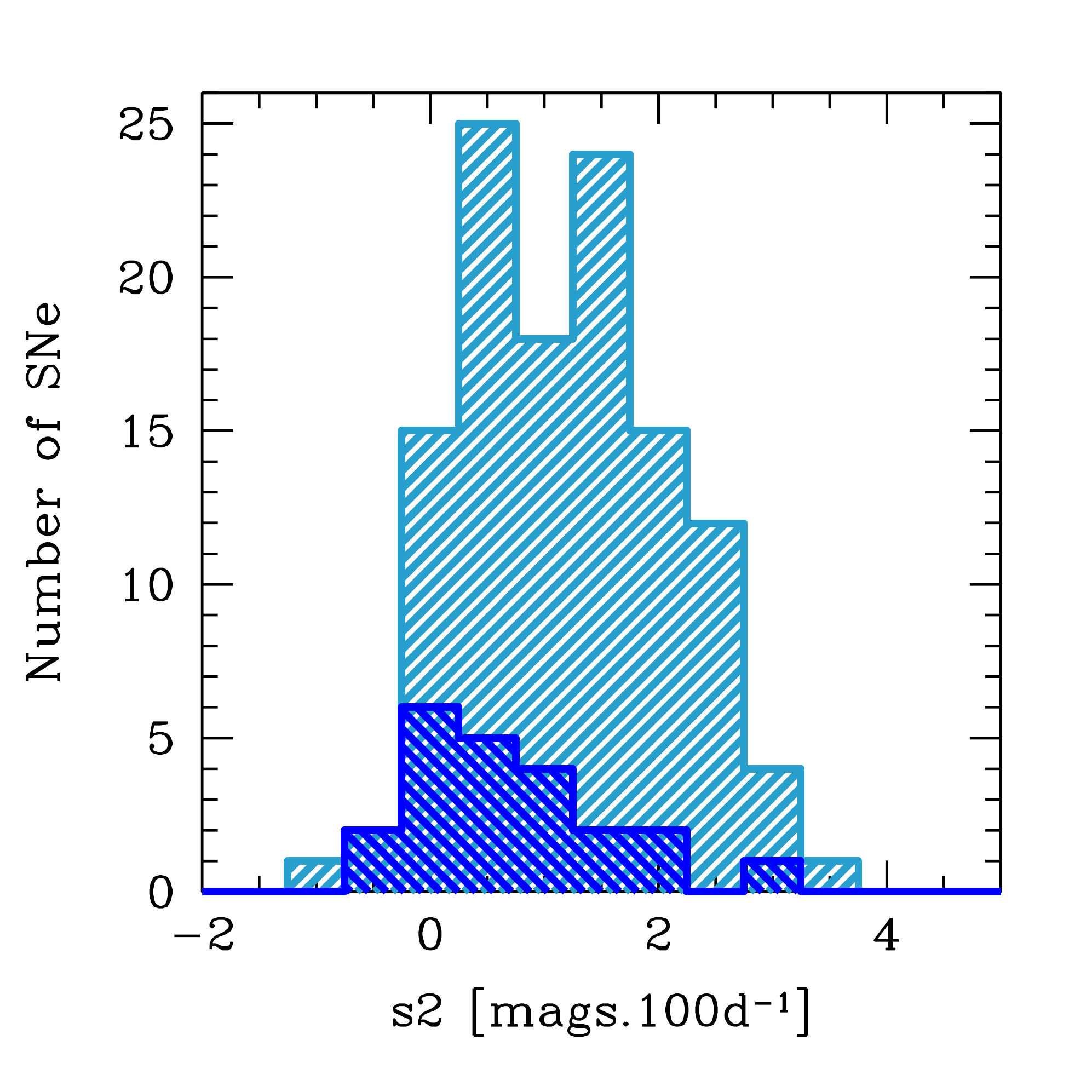}
\caption{Distribution of the plateau decline s$_2$ in $V-$band for 117 SNe of our sample.
The blue histogram presents the distribution of ``s$_2$" in $V-$band for 22 SNe~II used as a new 
template in SNID}.
\label{s2}
\end{figure}

\indent The photometric behaviour of our sample in terms of their plateau decline rate (s$_2$; defined in 
\citealt{Anderson14}) in the $V$ band 
is shown in Figure~\ref{s2}. For our sample of 117 SNe~II, we measure $s_2$ values ranging between $-0.76$
and 3.29 mag 100d$^{-1}$. Higher $s_2$ values mean that the SN has a faster declining
light curve.
We can see a continuum in the s$_2$  distribution, which shows that the majority of the SNe (83)
have a s$_2$ value between 0 and 2. There are 8 objects with s$_2$ values smaller than 0, while 3 SNe 
show a value larger than 3. The average of s$_2$ in our sample is 1.20. 
We are unable to estimate the s$_2$ value for 5 SNe as there is insufficient information from 
their light curves.
The s$_2$ distribution for the 22 SNe~II used as new templates in SNID is also shown in 
Figure~\ref{s2}.  Although the diversity
in the SNID templates increased with the inclusion of these SNe, the template distribution is still biased to 
low $s_2$ values.

\section{Spectral line identification}
\label{line}

\begin{figure}[h!]
\hspace{-0.3cm} 
\includegraphics[width=7cm,angle=270]{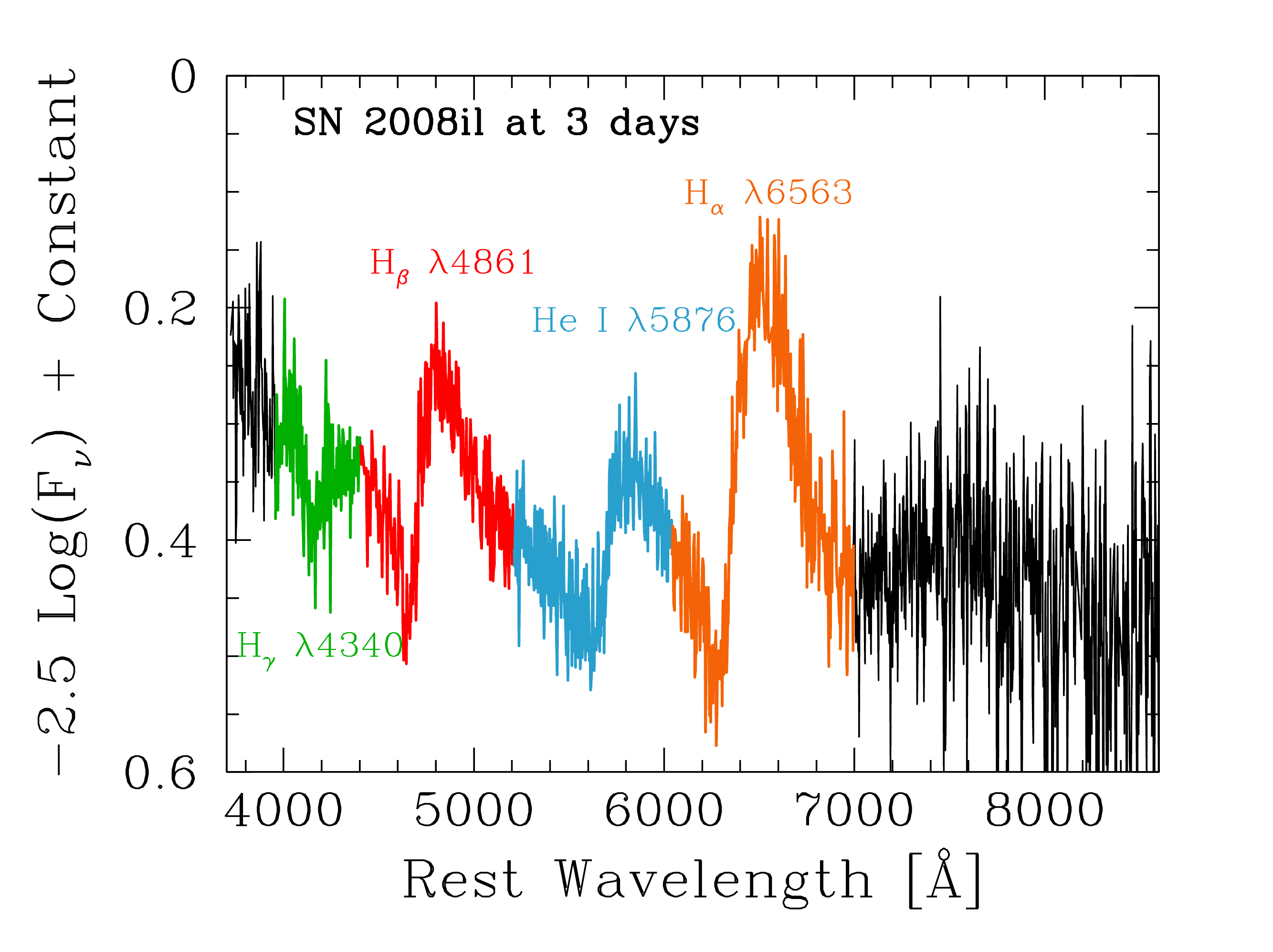}
\hspace{-0.3cm} 
\includegraphics[width=7cm,angle=270]{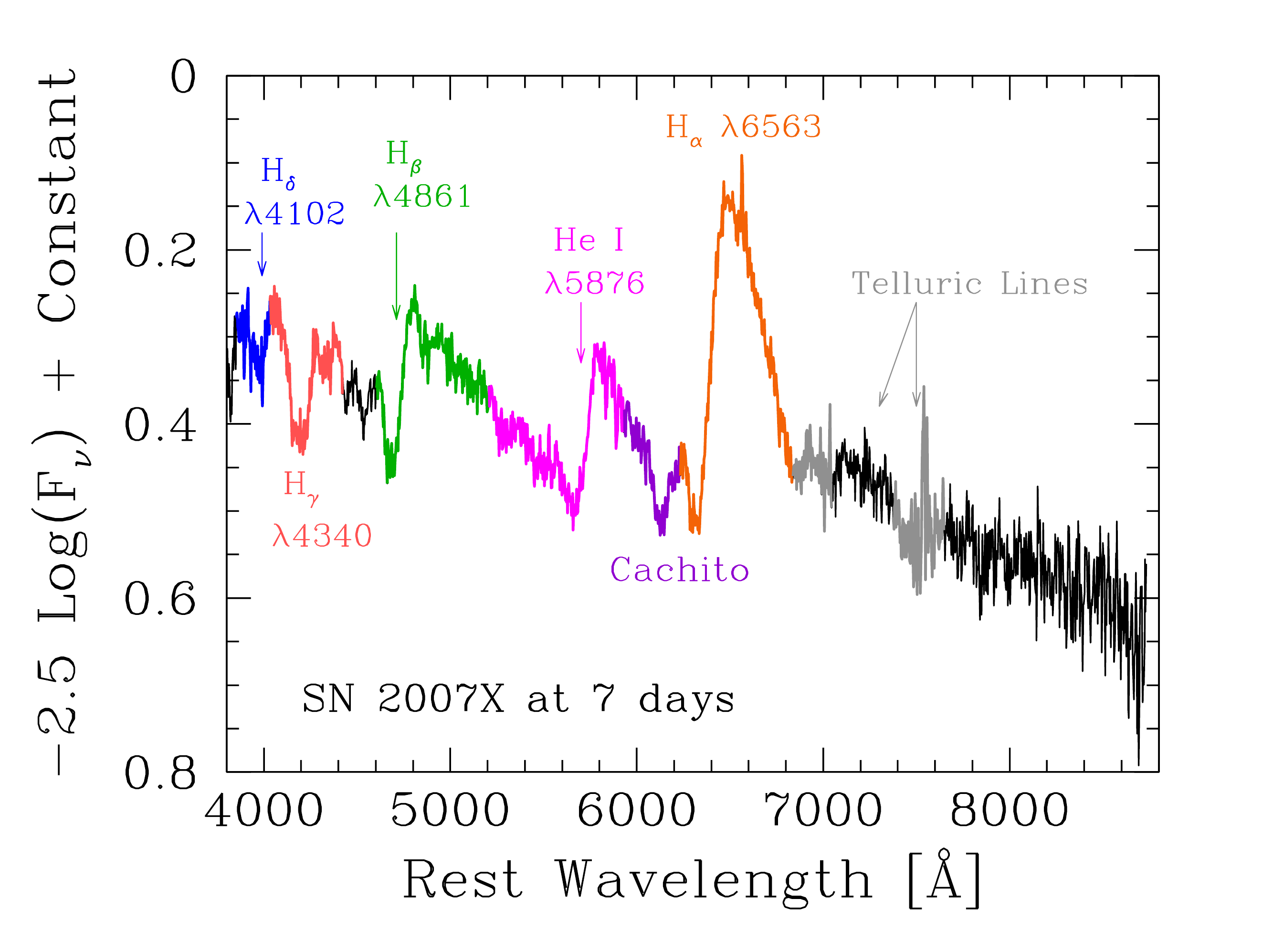}
\caption{Line identification in the early spectrum of SN~2008il (top) and SN~2007X (bottom).}
\label{linesear}
\end{figure}

We identified 20 absorption features within our photospheric spectra, in
the observed wavelength range of 3800 to 9500 \AA. Their identification was performed using the Atomic Spectra
Database\footnote{http://physics.nist.gov/asd3} and theoretical models \citep[e.g.][]{Dessart05, Dessart06, Dessart11}.
Early spectra exhibit lines of H$_{\alpha}$ $\lambda6563$,
H$_{\beta}$ $\lambda4861$, H$_{\gamma}$ $\lambda4341$, H$_{\delta}$ $\lambda4102$, and \ion{He}{1} 
$\lambda5876$, with the latter disappearing at $\sim20-25$ days past explosion. An extra absorption component 
on the blue side of H$_{\alpha}$ (hereafter ``Cachito''\footnote{Cachito is a Hispanic word that means
small piece of something (like a notch). We use this name to refer to the small absorption components blue ward of  H$_{\alpha}$,
giving its (until now) previously ambiguous nature.} is present in many SNe). 
That line has previously been attributed to high velocity (HV) features of hydrogen
or \ion{Si}{2} $\lambda6533$. Figure~\ref{linesear} shows the main lines in early spectra of SNe~II at 3 and 7 days from explosion.
We can see that SN~2008il shows the Balmer lines and \ion{He}{1}, while SN~2007X, in addition to these lines, 
also shows Cachito on the blue side of  H$_{\alpha}$.

\begin{figure}
\hspace*{-0.5cm} 
\vspace*{-1.3cm} 
\includegraphics[width=9.2cm,height=8.8cm]{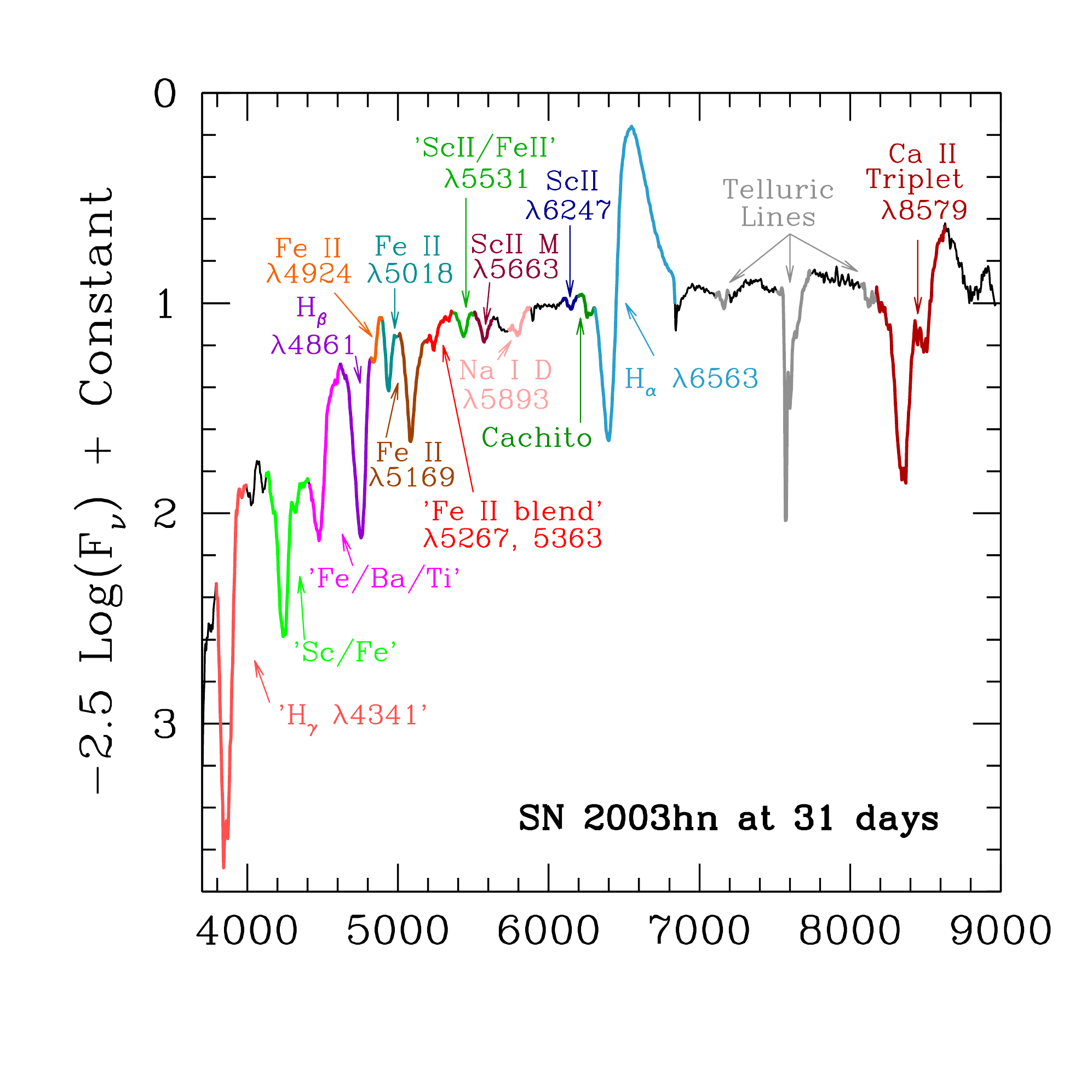}
\hspace*{-0.5cm} 
\vspace*{-1.3cm} 
\includegraphics[width=9.2cm,height=8.8cm]{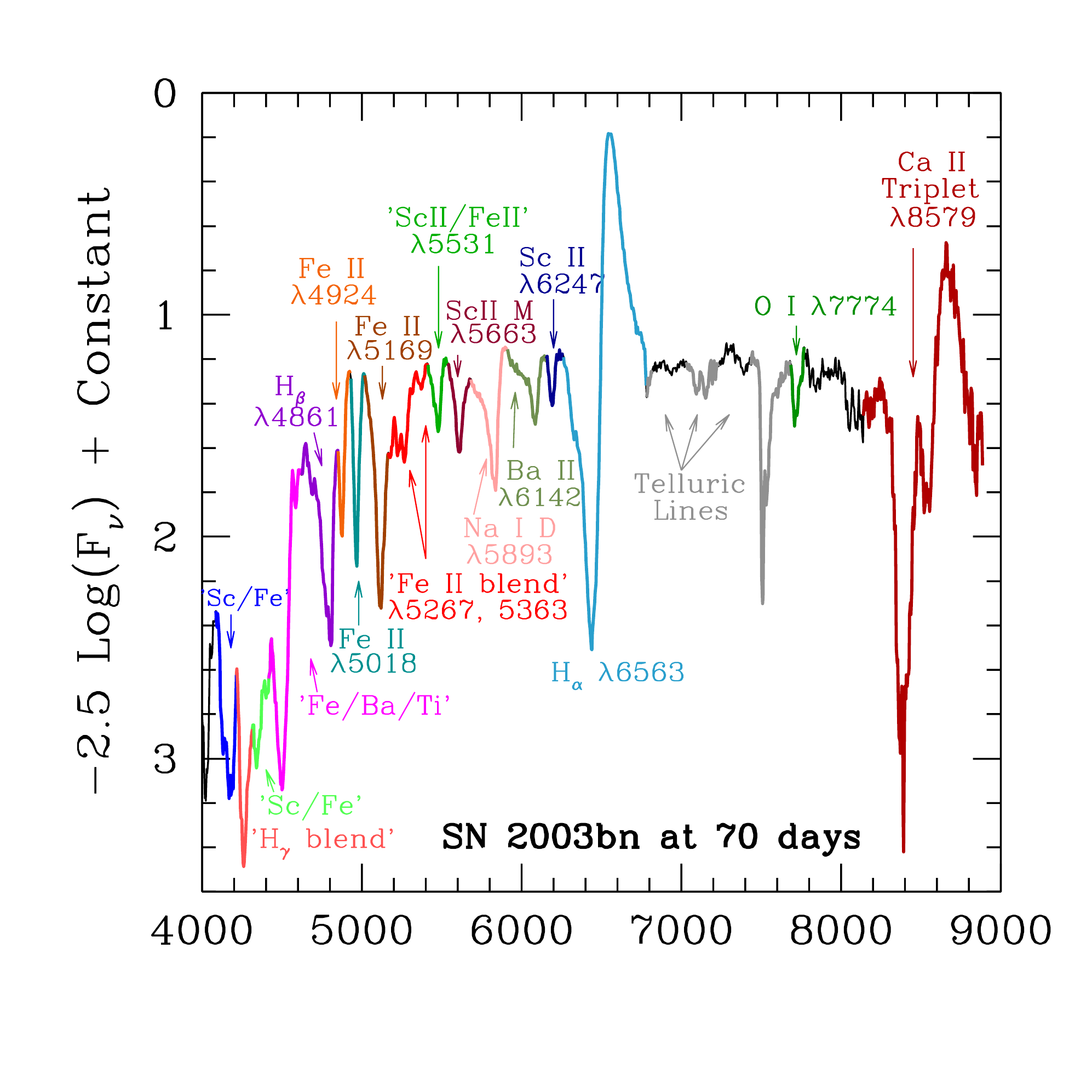}
\hspace*{-0.5cm} 
\includegraphics[width=9.2cm,height=8.8cm]{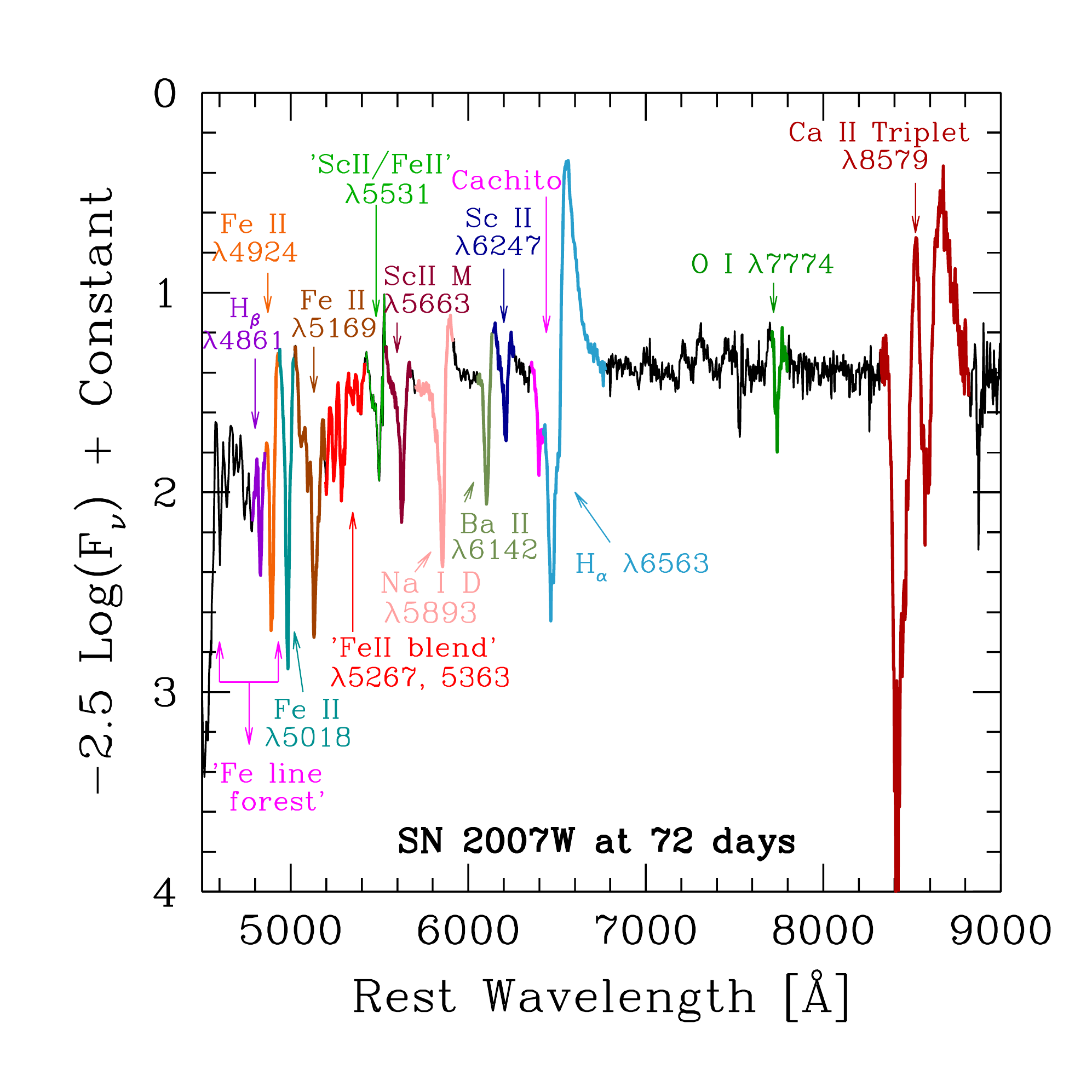}
\caption{Line identification in the photospheric phase for SNe~II 2003hn at 31 days (top), 
2003bn at 70 days (middle), and 2007W at 72 days (bottom).}
\label{linesph}
\end{figure}

In Figure~\ref{linesph} we label the lines present in the spectra of SNe~II during the photospheric phase
at 31, 70 and 72 days from explosion. Later than $\sim15$ days the iron group lines start to appear and dominate 
the region between 4000 and 6000 \AA. We can see Fe-group blends near $\lambda4554$, and between 5200 and 
5450 \AA\ (where we refer to the latter as ``Fe II blend'' throughout the rest of the text). Strong features such as
Fe II $\lambda4924$, \ion{Fe}{2} $\lambda5018$, \ion{Fe}{2} $\lambda5169$, \ion{Sc}{2}/\ion{Fe}{2}
$\lambda5531$, the \ion{Sc}{2} multiplet $\lambda5663$ (hereafter ``\ion{Sc}{2} M''), \ion{Ba}{2} $\lambda6142$, 
\ion{Sc}{2} $\lambda6247$, \ion{O}{1} $\lambda7774$, \ion{O}{1} $\lambda9263$ and the \ion{Ca}{2} triplet $\lambda\lambda8498,8662$ 
($\lambda8579$) are also present from $\sim20$ days to the end of the plateau. 
At 31 days, SN~2003hn shows all these lines, except  \ion{Ba}{2}, while at 70 and 72 days, SN~2003bn
and SN~2007W show all the lines. Unlike SN~2003bn, SN~2007W shows Cachito and the ``Fe line forest''\footnote{We label ``Fe line forest''
to that region around H$_{\beta}$ where a series of Fe-group (e.g. \ion{Fe}{2} $\lambda4629$, \ion{Sc}{2} $\lambda4670$,
\ion{Fe}{2} $\lambda4924$) absorption lines emerge.}. The Fe line forest is visible in a small fraction of 
SNe from 25-30 days (see the analysis
in section~\ref{ana}). As we can see there are significant differences between two different SNe at almost the same epoch.
Later we analyze and discuss how these differences can be understood in terms of overall diversity of SN~II properties.\\

\begin{figure}
\centering
\hspace*{-0.3cm} 
\includegraphics[width=9.2cm]{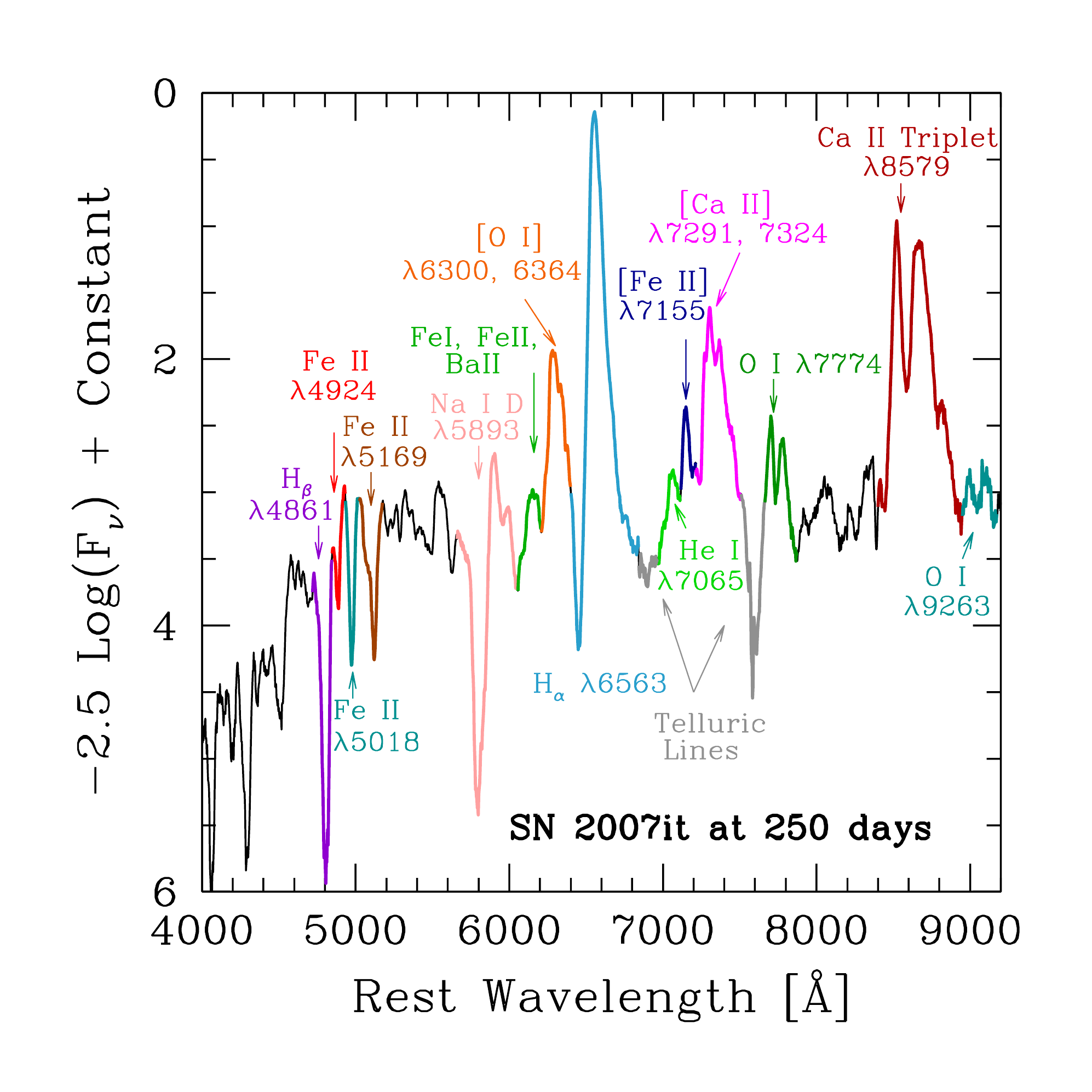}
\caption{Line identification in the nebular spectrum of SN~II 2007it at 250 days from explosion. }
\label{linesneb}
\end{figure}

\indent In the nebular phase, later than 200 days post explosion, the forbidden lines 
\ion{[Ca}{2]} $\lambda\lambda7291,$ 7323, \ion{[O}{1]} 
$\lambda\lambda6300,$ 6364 and \ion{[Fe}{2]} $\lambda7155$ emerge in the spectra. At this epoch
H$_{\alpha}$, H$_{\beta}$, \ion{Na}{1} D, the \ion{Ca}{2} triplet, \ion{O}{1} and the Fe group lines
between 4800 and 5500 \AA, and 6000-6500 \AA\ are also still present. Figure~\ref{linesneb}
shows a nebular spectrum of SN 2007it at 250 days from explosion.\\

\subsection{The H$_{\alpha}$ P-Cygni profile}
\label{halpha}

\indent H$_{\alpha}$ $\lambda6563$ is the dominant spectral feature in SNe~II. It is usually used
to distinguish different SN types using the initial spectral observation. This line is present 
from explosion until nebular phases, showing, in the majority of cases, a 
P-Cygni profile. Although the P-Cygni profile has an absorption and emission component, 
SNe display a huge diversity in the absorption feature. 

\begin{figure*}
\centering
\includegraphics[width=16cm]{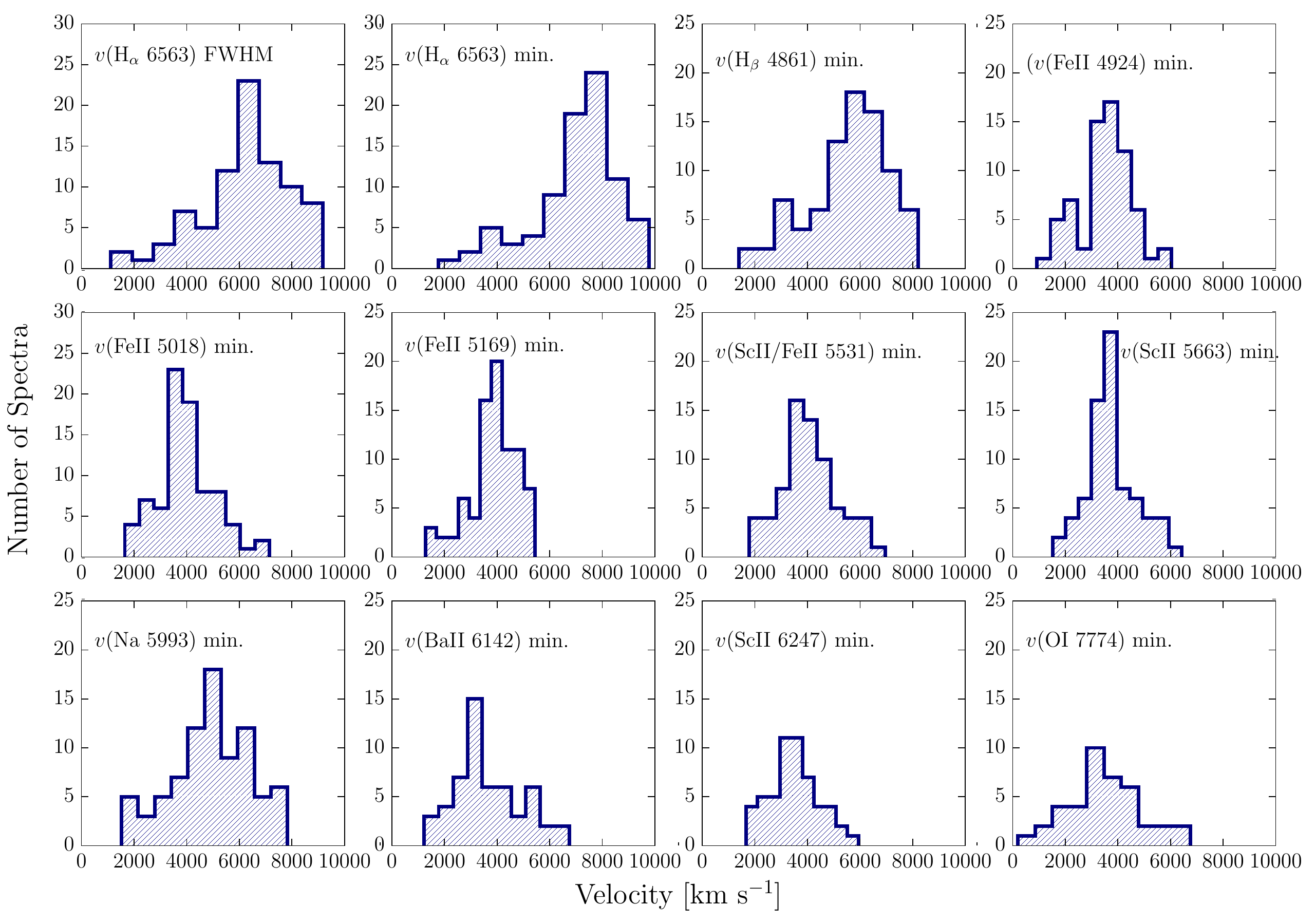}
\caption{Distribution of the expansion ejecta velocities for 11 optical features at 50 days. 
The first two panels show the H$_{\alpha}$ velocity obtained from the FWHM and from the minimum 
absorption flux. From third to twelfth panel are presented the H$_{\beta}$,
\ion{Fe}{2} $\lambda5018$, \ion{Fe}{2} $\lambda4924$, \ion{Fe}{2} $\lambda5169$, 
\ion{Sc}{2}/\ion{Fe}{2} $\lambda5531$, \ion{Sc}{2} $\lambda5663$, \ion{Na}{1} D $\lambda5893$,
\ion{Ba}{2} $\lambda6142$, \ion{Sc}{2} $\lambda6247$, \ion{O}{1} $\lambda7774$ velocities
obtained from the minimum absorption flux}.
\label{distvel}
\end{figure*}

\begin{figure}[h!]
\hspace*{-0.5cm} 
\includegraphics[width=9cm]{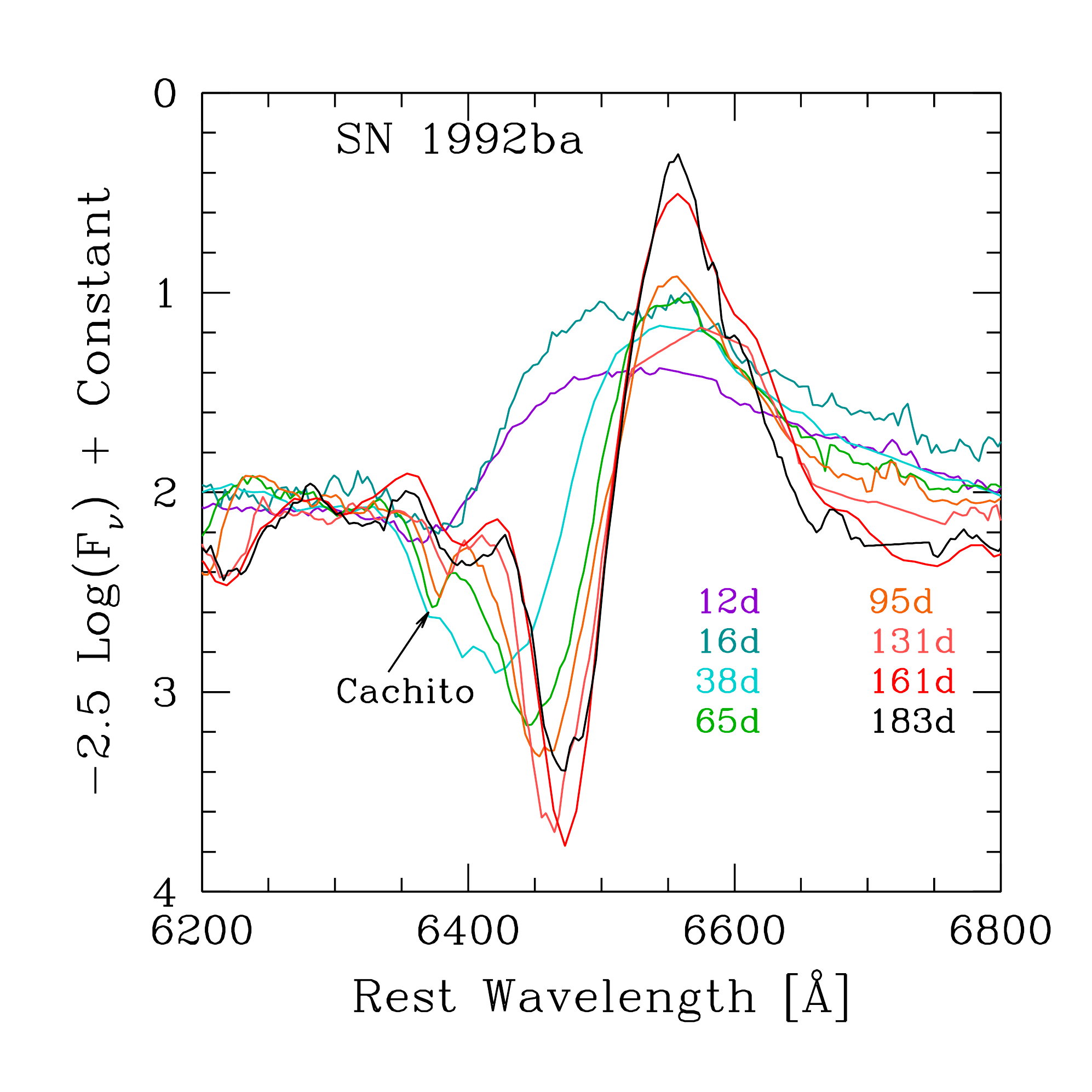}
\caption{H$_{\alpha}$ P-Cygni profile evolution in SN~1992ba. The epochs are labeled on the right.}
\label{Ha}
\end{figure}

\indent \citet{Gutierrez14} showed that SNe with little absorption of H$_{\alpha}$ (smaller 
absorption to emission ($a/e$) values) appear to have higher velocities, faster declining light curves
and tend to be more luminous. Here we show   
H$_{\alpha}$ displays a large range of velocities in the photospheric phase,
from 9500 km s$^{-1}$ to 1500 km s$^{-1}$ at 50 days
(see the first two panels in Figure~\ref{distvel}, which correspond to the H$_{\alpha}$ velocity derived from
the FWHM of the emission component and from the minimum flux of the absorption, respectively).\\
\indent The diversity of H$_{\alpha}$ in the photospheric phase is also observed through 
the blueshift of the emission peak at early times \citep{Dessart08a,Anderson14a} and the boxy profile 
\citep{Inserra11, Inserra12b}. The former is associated with differing density distributions of the
ejecta, while the latter with an interaction of the ejecta with a dense CSM.
In the nebular phase this shift in H$_{\alpha}$ emission peak has been
interpreted as evidence of dust production in the SN ejecta. Despite the fact that this 
is an important issue in SNe~II, only a few studies \citep[e.g.][]{Sahu06, 
Kotak09, Fabbri11} have focussed on these features. \\
\indent In Figure~\ref{Ha} we show an example of the evolution of H$_{\alpha}$ P-Cygni profile
in SN~1992ba. We can see in early phases a normal profile which evolves to a complicated 
profile around 65 days. Cachito on the blue side of H$_{\alpha}$ is present from 65 to 183 days.\\

\subsection{H$_{\beta}$, H$_{\gamma}$ and H$_{\delta}$ absorption features} 
\label{hbeta}

H$_{\beta}$ $\lambda4861$, H$_{\gamma}$ $\lambda4341$ and H$_{\delta}$ $\lambda4102$ 
like H$_{\alpha}$ are present from the first epochs. In earlier phases, these lines show a 
P-Cygni profile, however, from $\sim15$ days the spectra only display the absorption 
component, giving space to Fe group lines. The range of velocities of H$_{\beta}$, H$_{\gamma}$ 
and H$_{\delta}$ at 50 days post explosion vary from 8000 to 1000 km s$^{-1}$ (see Figure~\ref{distvel}). \\
\indent Although H$_{\delta}$ is a common line in SNe~II, we do not include a detailed analysis
of this line because in many cases the spectra are noisy in the blue part of the spectrum.
Besides, like other lines in the blue, this line is blended with Fe-group lines
later than 30 days.\\
\indent Around 30 days from explosion H$_{\gamma}$ starts to blend with other lines, such as \ion{Ti}{2} and 
\ion{Fe}{2}. Meanwhile in a few SNe, the H$_{\beta}$ absorption feature is surrounded by the Fe line forest.
Our later analysis shows that SNe displaying this behaviour are generally dimmer and lower velocity events 
(see Section~\ref{ana} for more details).

\subsection[\ion{He}{1}\ $\lambda5876$ and \ion{Na}{1} D $\lambda5893$]
{\ion{He}{1}\ $\lambda5876$ and \ion{Na}{1} D $\lambda5893$ }

\ion{He}{1} $\lambda5876$ is present in very early phases when the temperature of the ejecta is high enough 
to excite the ground state of helium.
As the temperature decreases, the \ion{He}{1} line starts to disappear due to low excitation of
\ion{He}{1} ions (around 15 days; \citealt{Roy11, Dessart10a}). At $\sim30$ days the \ion{Na}{1} D $\lambda5893$ absorption 
feature arises in the spectrum at a similar position where \ion{He}{1} was located. This new line 
evolves with time to a strong P-Cygni profile, displaying velocities between 8000 km s$^{-1}$ to 
1500 km s$^{-1}$ at 50 days from explosion (Figure~\ref{distvel}). \\
\indent In many SNe~II (or indeed SNe of all types), at these wavelengths one often observes 
narrow absorption features arising from slow-moving line of sight material from the interstellar medium, ISM (or possibly 
from circumstellar material, CSM). Such material can constrain the amount of foreground reddening suffered by SNe,
however we do not discuss this here.

\subsection{Fe-group lines}

When the SN ejecta has cooled sufficiently, \ion{Fe}{2} features start to dominate SNe~II spectra 
between 4000 to 6500 \AA. The first line that appears is \ion{Fe}{2} $\lambda5169$ on top of the emission 
component of H$_{\beta}$. With time \ion{Fe}{2} $\lambda5018$ and $\lambda4924$ emerge between H$_{\beta}$
and \ion{Fe}{2} $\lambda5169$. \ion{Fe}{2} $\lambda5169$ becomes a \ion{Fe}{2} blend
later than $\sim30-40$ days. At 
$\sim50$ days the 4000-5500 \AA\ region is completely filled with these lines and the continuum is diminished
due to \ion{Fe}{2} line-blanketing. The H$_{\gamma}$ and H$_{\delta}$ absorption features are blended with 
Fe-group lines, such as \ion{Fe}{2}, \ion{Ti}{2}, \ion{Sc}{2} and \ion{Sr}{2}. 
Between $\sim5400$ and 6500 \AA\ other metal lines appear in the spectra. Lines such as \ion{Sc}{2}/\ion{Fe}{2}
$\lambda5531$, \ion{Sc}{2} M, \ion{Ba}{2} $\lambda6142$ and \ion{Sc}{2} $\lambda6247$ get stronger with 
time.\\
\indent As we can see in Figure~\ref{distvel}, the Fe-group lines show a range of velocities between 
7000 km s$^{-1}$ to 500 km s$^{-1}$ at 50 days. The peak of the distribution of the \ion{Fe}{2} group lines velocities is around 
4000 km s$^{-1}$. In the case of \ion{Ba}{2}, the peak is lower (around 3000 km s$^{-1}$).\\
\indent Although \ion{Fe}{2} lines always appear at  late phases, 
few SNe show the iron line forest at 30 days. This feature appears earlier in low velocity/luminosity 
SNe (See section \ref{ana}). \\

\subsection[The Ca II NIR triplet]{The \ion{Ca}{2} NIR triplet}

The \ion{Ca}{2} NIR triplet is a strong feature in the spectra of SNe~II. This line appears at 
$\sim20-30$ days as an absorption feature, but with time it starts to show an emission component.
The \ion{Ca}{2} NIR triplet results in
a blend of $\lambda$8498 and $\lambda$8542 in the bluer part and a distinct component, 
$\lambda$8662 on the red part. In SNe~II with higher velocities these lines are blended producing a broad absorption 
and emission profile, however, in low-velocity SNe, we see two absorption components and one emission 
in the red part. The velocities of the \ion{Ca}{2} NIR triplet range between 9000 to
1000 km s$^{-1}$ at 50 days. In the nebular phase the \ion{Ca}{2} NIR triplet is also
present, however at this epoch it only exhibits the emission component.\\
\indent Although in the majority of our spectra we can not see \ion{Ca}{2} H \& K
$\lambda3945$, due to the poor signal to noise in this region, this line is present
in the photospheric phase of SNe~II. \\
\indent While the \ion{Ca}{2} NIR triplet is a prominent feature in SNe~II, we do not include
its analysis in the subsequent discussion, given that the overlap of lines makes a consistent comparison
of velocities and pseudo-equivalent widths (pEWs) difficult.

\subsection[O I lines]{\ion{O}{1} lines}
\label{oi}

The \ion{O}{1} $\lambda\lambda7772$, 7775 doublet (hereafter \ion{O}{1} $\lambda7774$) and
\ion{O}{1} $\lambda9263$ are the oxygen lines in the optical spectra of 
SNe~II. These lines are mainly driven by recombination and they appear when the 
temperature decreases sufficiently. 
The \ion{O}{1} $\lambda7774$ line is relatively strong and emerges around 20 days from 
explosion, however in the majority of cases it is contaminated by the telluric A-band 
absorption ($\sim7600-7630$ \AA), which hinders detailed analysis. \ion{O}{1} $\lambda9263$
is weaker and appears one month later than \ion{O}{1} $\lambda7774$. These lines are present
until the nebular phase and their expansion velocity at 50 days post explosion goes from $\sim7000$ km s$^{-1}$
to 500 km s$^{-1}$, as can be seen in Figure~\ref{distvel}. \\

\subsection[Cachito: Hydrogen High Velocity Features or the Si II $\lambda$6355 line?]
{Cachito: Hydrogen High Velocity (HV) Features or the Si II $\lambda$6355 line?}
\label{cachito}

\begin{figure}[h!]
\hspace*{-0.5cm} 
\includegraphics[width=9cm]{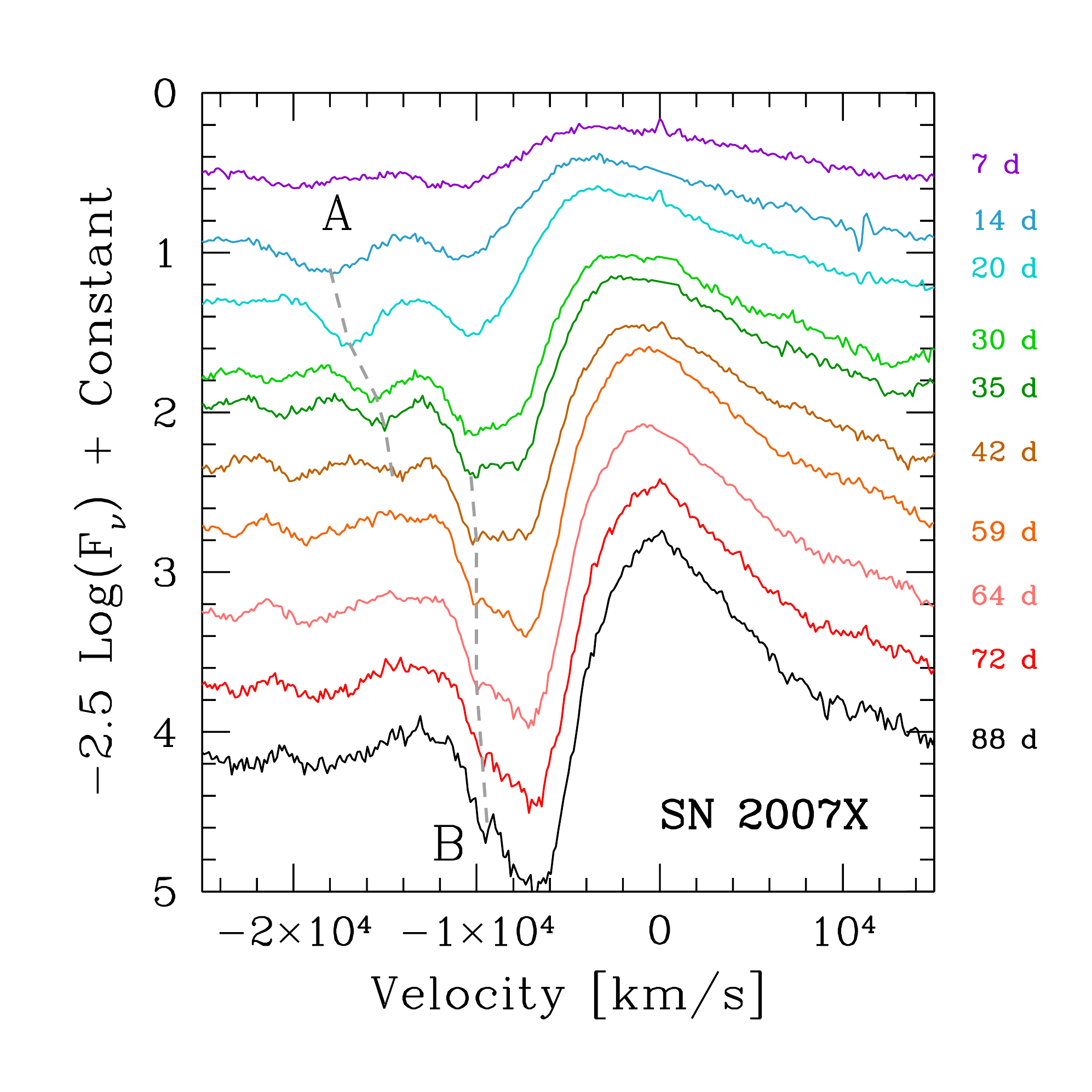}
\caption{H$_{\alpha}$ P-Cygni profile of the SN~2007X.
The epochs are labeled on the right. The dashed lines indicate the velocities for the A and B features, which we call `Cachito'.}
\label{cach}
\end{figure}

The extra absorption component on the blue side of H$_{\alpha}$ P-Cygni profile, called here 
``Cachito'', is seen in early phases in some SNe (e.g. SN~2005cs, \citealt{Pastorello06};
SN~1999em, \citealt{Baron00}) as well as in the plateau phase (e.g. SN~1999em, \citealt{Leonard02b}
SN~2007od, \citealt{Inserra11}). However, its shape and 
strength is completely different in the two phases.
\citet{Baron00} assigned the term ``complicated 
P-Cygni profle" to explain the presence of this component on the blue side of 
the Balmer series. They concluded that these features are due to velocity structures in the
expanding ejecta of the SNe~II. A few years later, \citet{Pooley02} and \citet{Chugai07} argued that this
extra component might originate from ejecta -- circumstellar (CS) interactions, while
\citet{Pastorello06} earmarked this feature as \ion{Si}{2} $\lambda6355$. \\
\indent In general, Cachito appears around 5-7 days between 6100 and 6300 \AA, 
and disappears at $\sim35$ days after explosion.
Later than 40 days the Cachito feature emerges closer to H$_{\alpha}$ (between 6250-6450 \AA) and  
it can be seen until 100-120 days.
Figure~\ref{cach} shows this component in SN~2007X. In early phases this feature is marked
with letter A and later with letter B. If attributed to H$_{\alpha}$ the derived velocities are
18000 km s$^{-1}$ and 10000 km s$^{-1}$, respectively.
A detailed analysis of this feature is presented in \ref{cacho}.

\subsection{Nebular Features}
\label{nebularf}

\begin{figure}[h!]
\centering
\includegraphics[width=9cm]{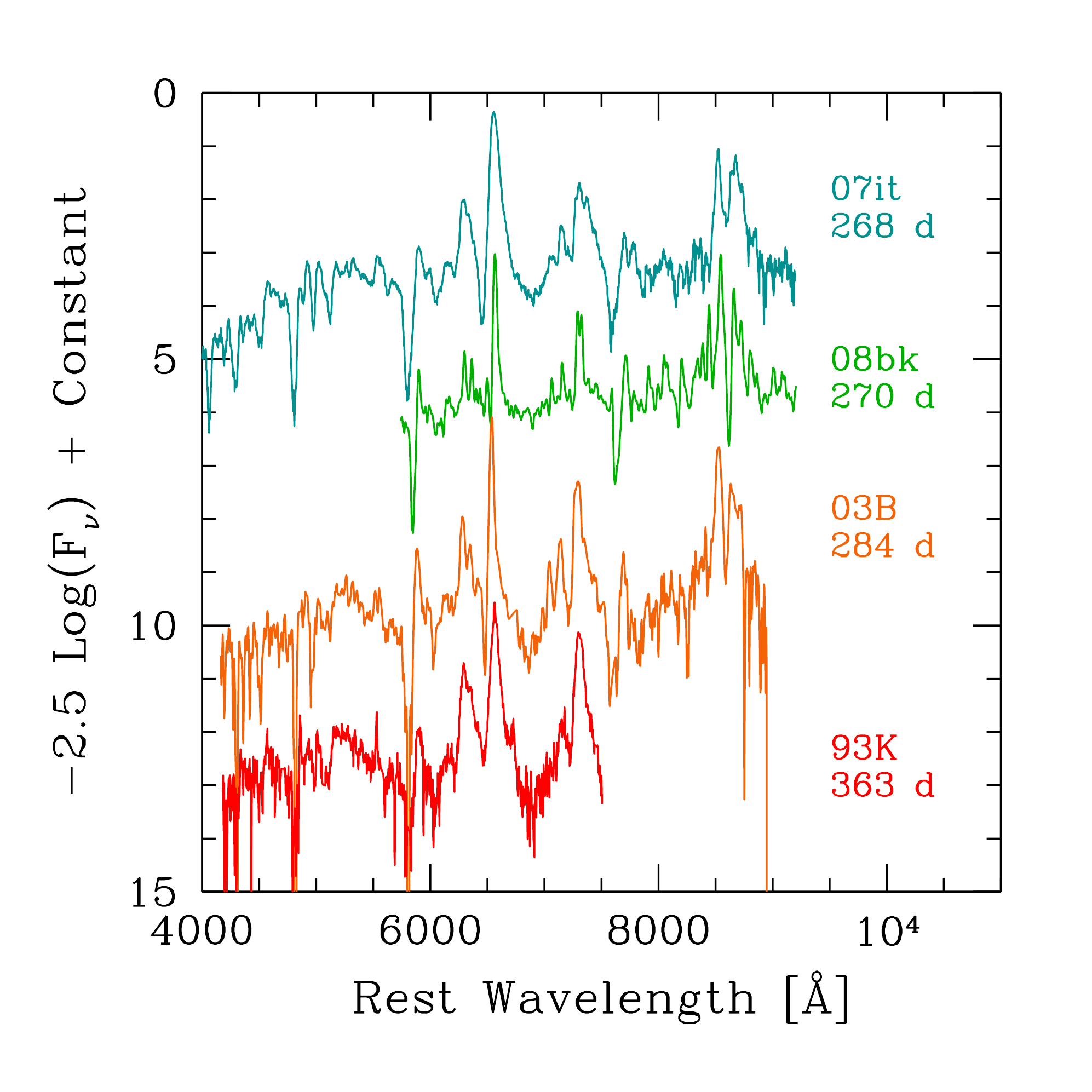}
\caption{Nebular spectra of seven different SNe of our sample. The spectra are organized 
according to epoch.}
\label{neb}
\end{figure}

As mentioned above, H$_{\alpha}$, H$_{\beta}$, the \ion{Ca}{2} NIR triplet, \ion{Na}{1}D ,
\ion{O}{1} and \ion{Fe}{2} are also present in the nebular phase (later than 200 days since explosion),
however in the case of the \ion{Ca}{2} NIR triplet, its appearance changes, passing from absorption and emission 
components to only emission components when the nebular phase starts.
The rest of the lines have the same behaviour but at much later epochs. The emergence of forbidden
emission lines signifies that the spectrum is now forming in regions of low density. At this phase, the ejecta has
become transparent, allowing us to peer into the inner layers of the rapidly expanding ejecta. 
Lines such as \ion{[Ca}{2]} $\lambda\lambda7291,$ 7323, 
\ion{[O}{1]} $\lambda\lambda6300,$ 6364 and H$_{\alpha}$ are the 
strongest features visible in the spectra.\\
\indent The \ion{[O}{1]} doublet observed at nebular times is one of the most important 
diagnostic lines of the helium-core mass \citep{Fransson87,Jerkstrand12}.
Usually the doublet is blended, however in SNe with low velocities these lines
can be resolved (see e.g. SN~2008bk). On the other hand, \ion{[Fe}{2]} $\lambda7155$ is easily 
detectable, but in most cases it is blended with \ion{[Ca}{2]} $\lambda\lambda7291,$ 7323
and \ion{He}{1} $\lambda7065$, which may hinder their analysis. In Figure~\ref{neb} we can see the
diversity found in the nebular spectra in our sample.\\

\section{Spectral measurements}
\label{measure}

As discussed previously, SNe~II spectra evolve from having a blue continuum with a few lines (Balmer series and \ion{He}{1})
to redder spectra with many lines: \ion{Fe}{2}, \ion{Ca}{2}, \ion{Na}{1} D, \ion{Sc}{2}, \ion{Ba}{2}, 
and \ion{O}{1}. To analyze the spectral properties of SNe~II, we measure 
the expansion velocities and pEWs of eleven features in the photospheric phase (see in Table~\ref{table_features} the features used), 
the ratio of absorption to emission ($a/e$) of H$_{\alpha}$ P-Cygni profile before 120 days, and the 
velocity decline rate of H$_{\beta}$.

\subsection{Expansion ejecta velocities}

The expansion velocities of the ejecta are commonly measured from the minimum flux of the absorption
component of the P-Cygni line profile. Using the Doppler relativistic equation and the rest wavelength of each line,
we can derive the velocity. To obtain the position of the minimum line flux (in wavelength), a Gaussian fitting was employed,
which  was performed with IRAF using the \textit{splot}
package. As the absorption component presents a wide diversity (e.g. asymmetries, flat shape, extra 
absorption components) we repeat the process many times (changing the pseudo-continuum),
and the mean of the measurements was taken
as the minimum flux wavelength. As our measurement error we take the standard deviation on the measurements. 
This error is added in quadrature to errors arising from the spectral resolution of our observations
(measured in \AA\ and converted to km s$^{-1}$)
and from peculiar velocities of host galaxies with respect to the Hubble flow (200 km s$^{-1}$).
This means that, in addition to the standard deviation error, which realizes the
width of the line and $S/N$, we take into account the spectral resolution, that in our case 
is the most dominant parameter to determine the error. \\
\indent The particular case of the H$_{\alpha}$ velocity was explored in \citet{Gutierrez14}. Due to the 
difficulty of measuring the minimum flux in a few SNe with little or extremely weak absorption component, 
we derive the expansion velocity of H$_{\alpha}$ using both the minimum flux of the absorption component and the
full-width-at-half-maximum (FWHM) of the emission line.\\
\indent In the case of \ion{O}{1} $\lambda$ 7774 where the telluric lines can affect our measurement of its 
absorption minimum, we only use SNe with a clear separation between the two features. This means
that the number of SNe with \ion{O}{1} measurements is significantly smaller (only 47 SNe) compared to the 
other measured features.

\subsection{Velocity decline rate}

To calculate the time derivative of the expansion velocity in SNe~II, we select the H$_{\beta}$ 
absorption line. It is present from the early spectra, it is easy to identify and it is relatively isolated.
To analyze quantitatively our sample, we introduce the $\Delta v($H$_{\beta})$ as the mean
velocity decline rate in a fixed phase range [t$_0$,t$_1$]:\\
\begin{center}
$\Delta v($H$_{\beta})=\frac{\Delta v_{abs}}{\Delta t}=\frac{v_{abs}(t_1)-v_{abs}(t_0)}{t_1-t_0}$.\\
\end{center}
This parameter was measured over the interval [+15,+30] d, [+15,+50] d, [+30,+50] d, [+30,+80] d, 
and [+50,+80] d.

\subsection{Pseudo-equivalent widths}

\begin{figure}
\hspace*{-0.5cm} 
\includegraphics[width=9cm]{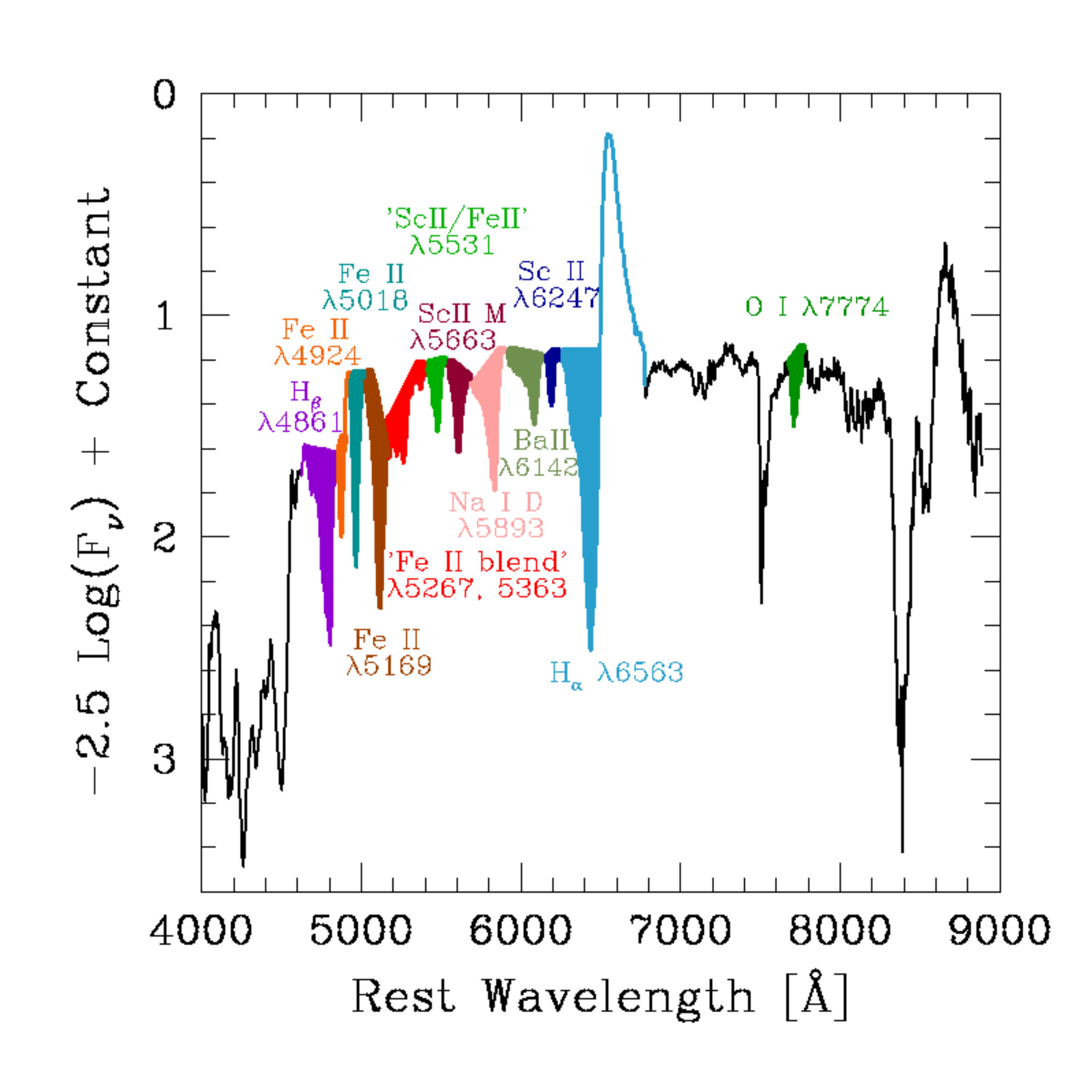}
\caption{Examples of pEWs used in this work for eleven features in the photospheric phase of SN~2003bn (at 70 days).}
\label{features}
\end{figure}

To quantify the spectral properties of SNe~II, another avenue for investigation is the measurement and 
characterization of spectral line pEWs. The prefix ``pseudo'' is used to indicate that the reference 
continuum level adopted does not represent the true underlying continuum level of the SN, given that 
in many regions the spectrum is formed from a superposition of many spectral lines. 
The pEW basically defines the strength of any given line 
(with respect to the pseudo-continuum) at any given time. The simplest and most often used method is to
draw a straight line across the absorption feature to mimic the continuum flux. 
Figure~\ref{features} shows an example of this technique applied to SN~2003bn. 
We do not include analysis of spectral lines where it is difficult to define the continuum level,
due to complicated line morphology, such as significant blending between lines.
For example, later than 20 - 25 days all absorption features bluer than H$_{\beta}$ are produced by blends 
of Fe-group lines plus other strong lines, such as \ion{Ca}{2} H \& K and H$_{\gamma}$.
On the other hand, the \ion{Ca}{2} NIR triplet $\lambda\lambda8498$, 8662 shows a profile that 
depends on the SN velocity (higher velocity SNe show a single broad absorption, while
low velocity SNe show two absorption characteristics). These attributes make a consistent 
analysis between SNe difficult, and therefore we do not include this line in our analysis.

\indent We measure the ratio of absorption
to emission ($a/e$) in H$_{\alpha}$ until 120 days. In the same way
the pEWs of the absorption lines are measured, we evaluate the pEWs for the emission in H$_{\alpha}$, thus we have:\\ 
\begin{center}
$a/e=\frac{pEW(H_{\alpha(abs)})}{pEW(H_{\alpha(emis)})}$.
\end{center}

\section{Line Evolution analysis}
\label{ana}

\begin{figure*}
\centering
\includegraphics[width=5.8cm]{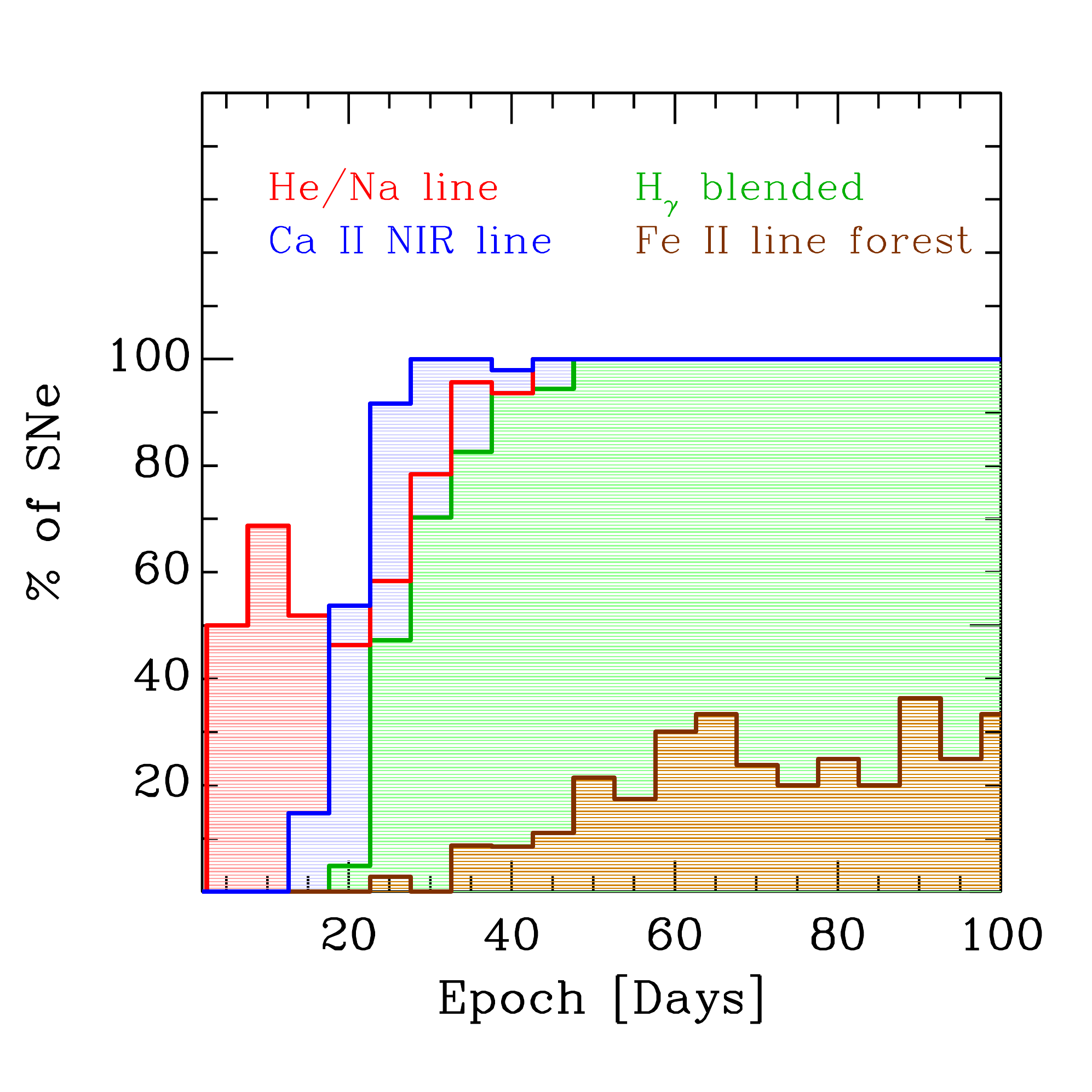}
\includegraphics[width=5.8cm]{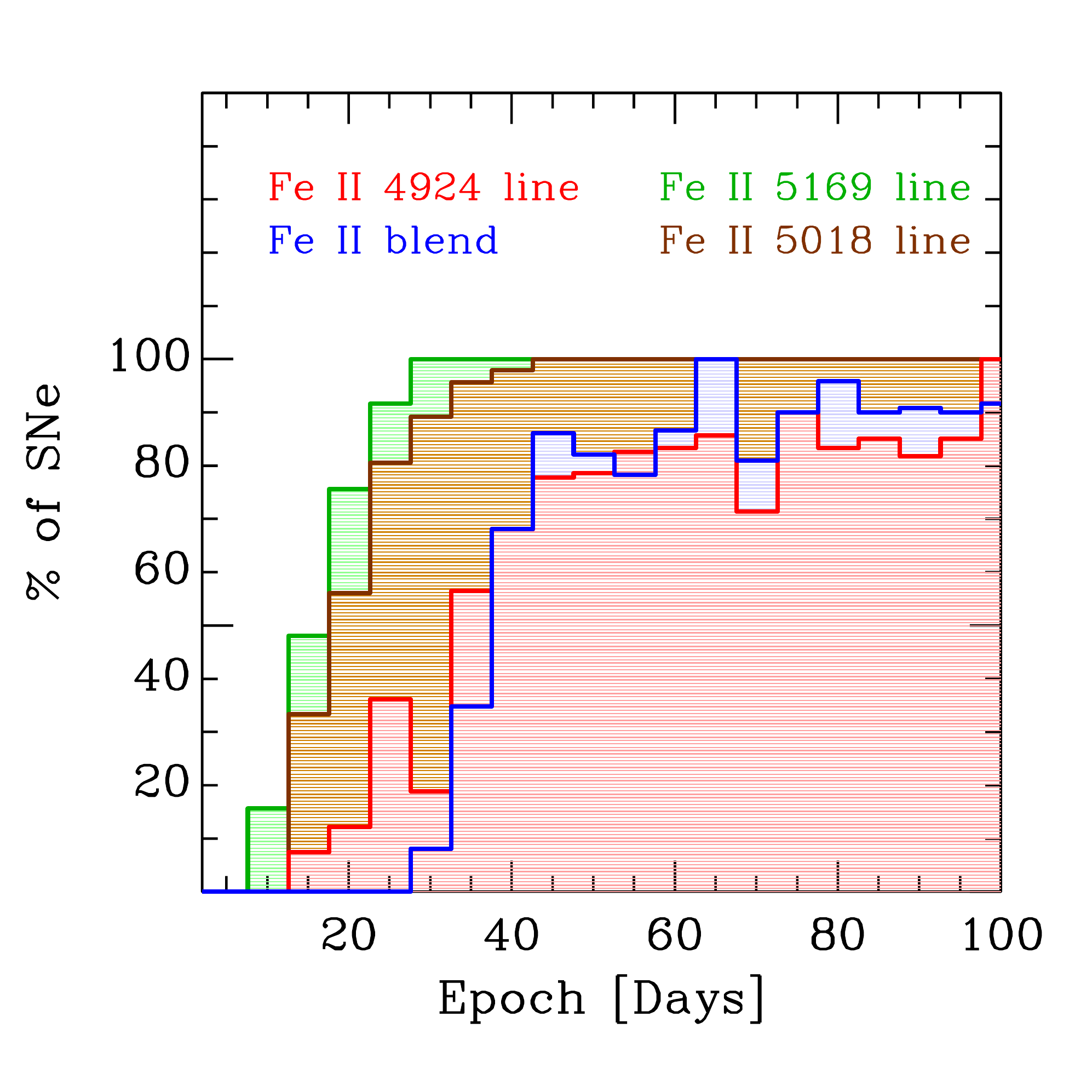}
\includegraphics[width=5.8cm]{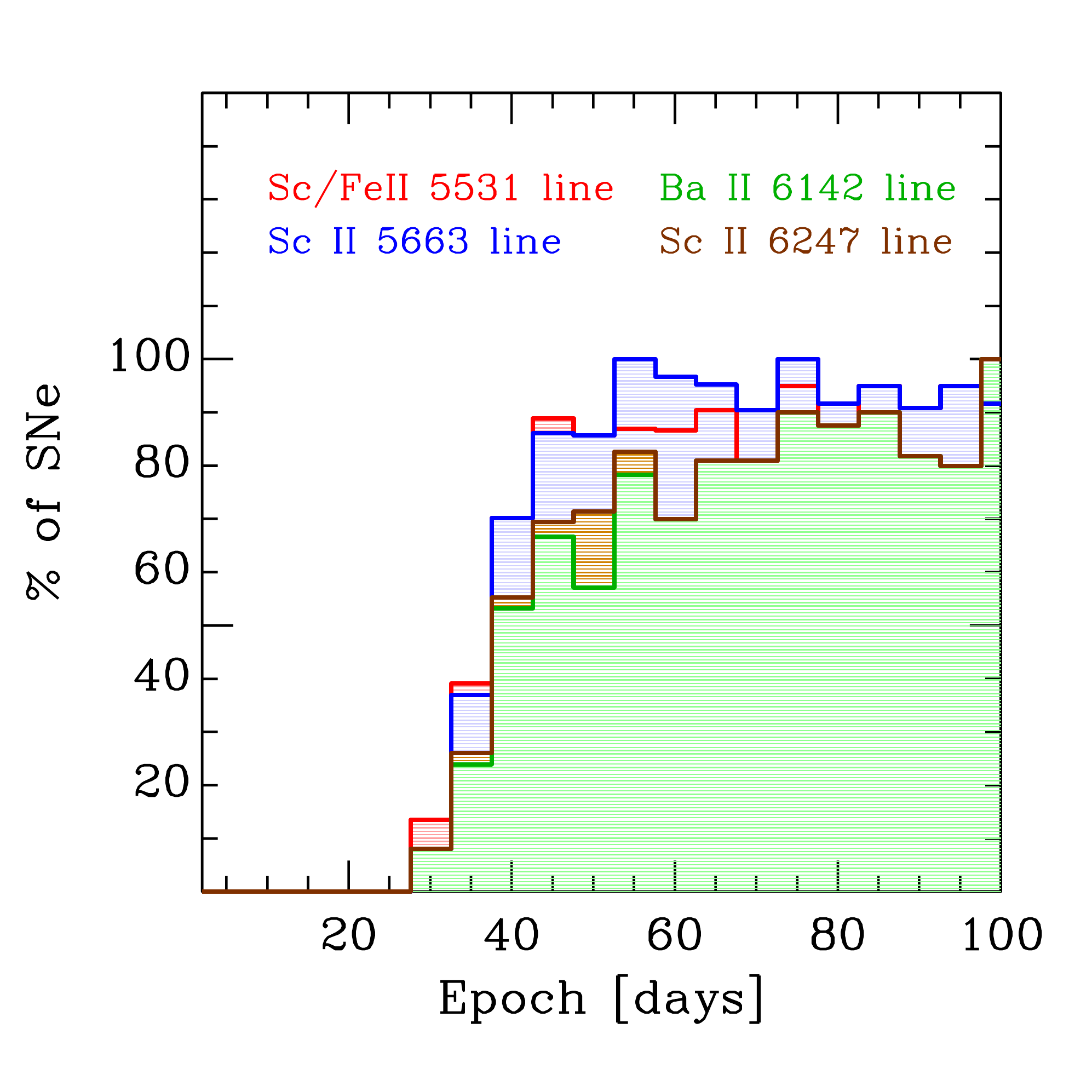}
\includegraphics[width=5.8cm]{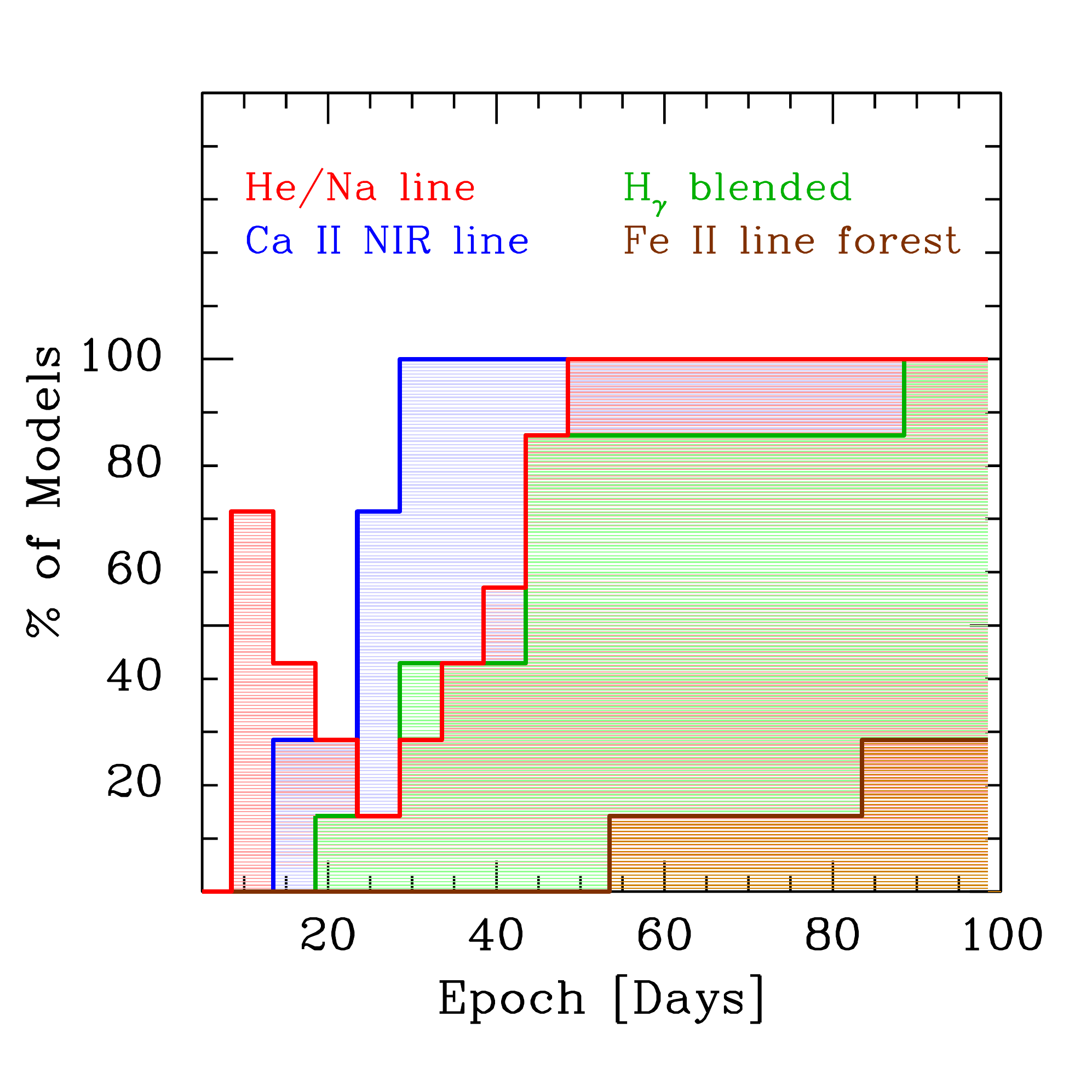}
\includegraphics[width=5.8cm]{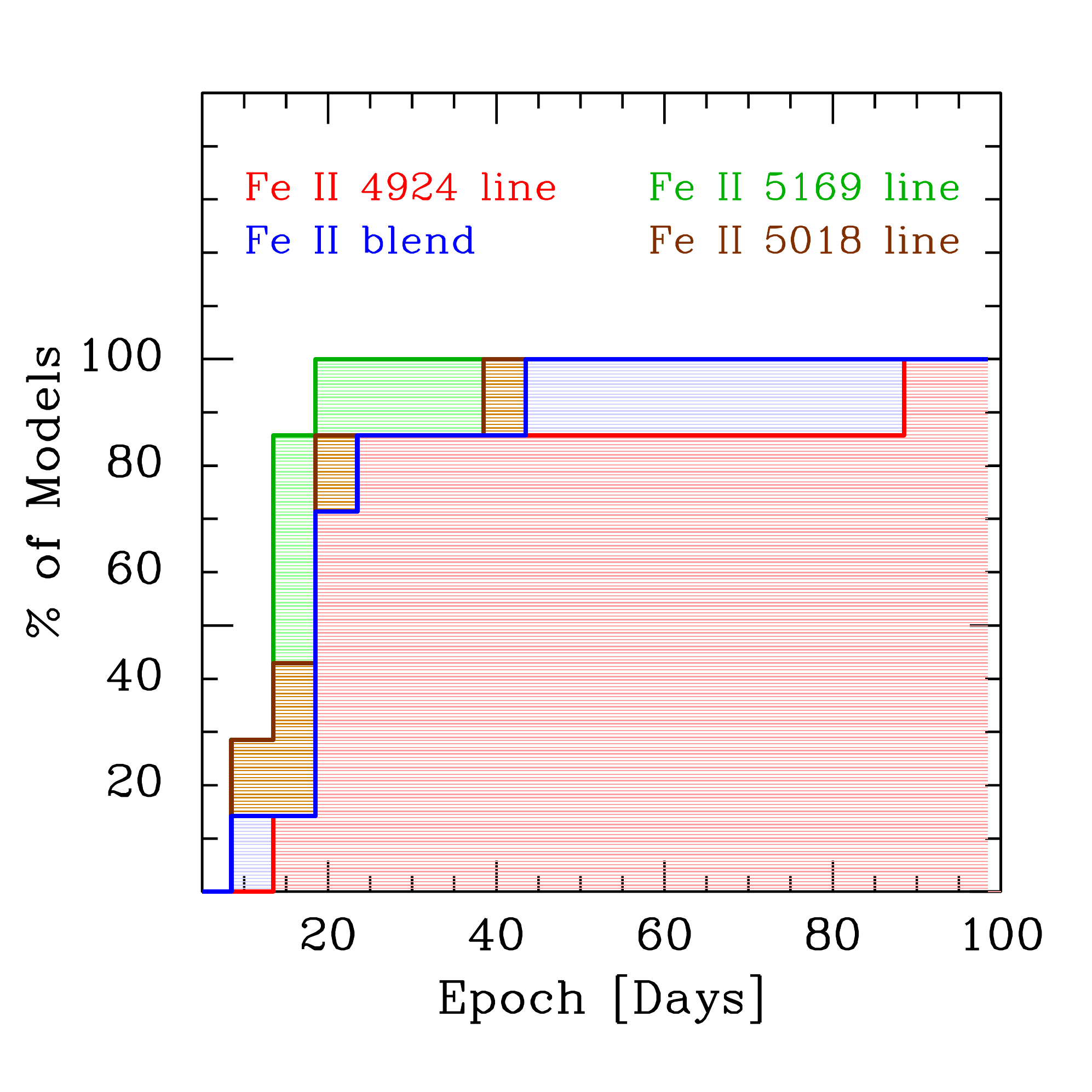}
\includegraphics[width=5.8cm]{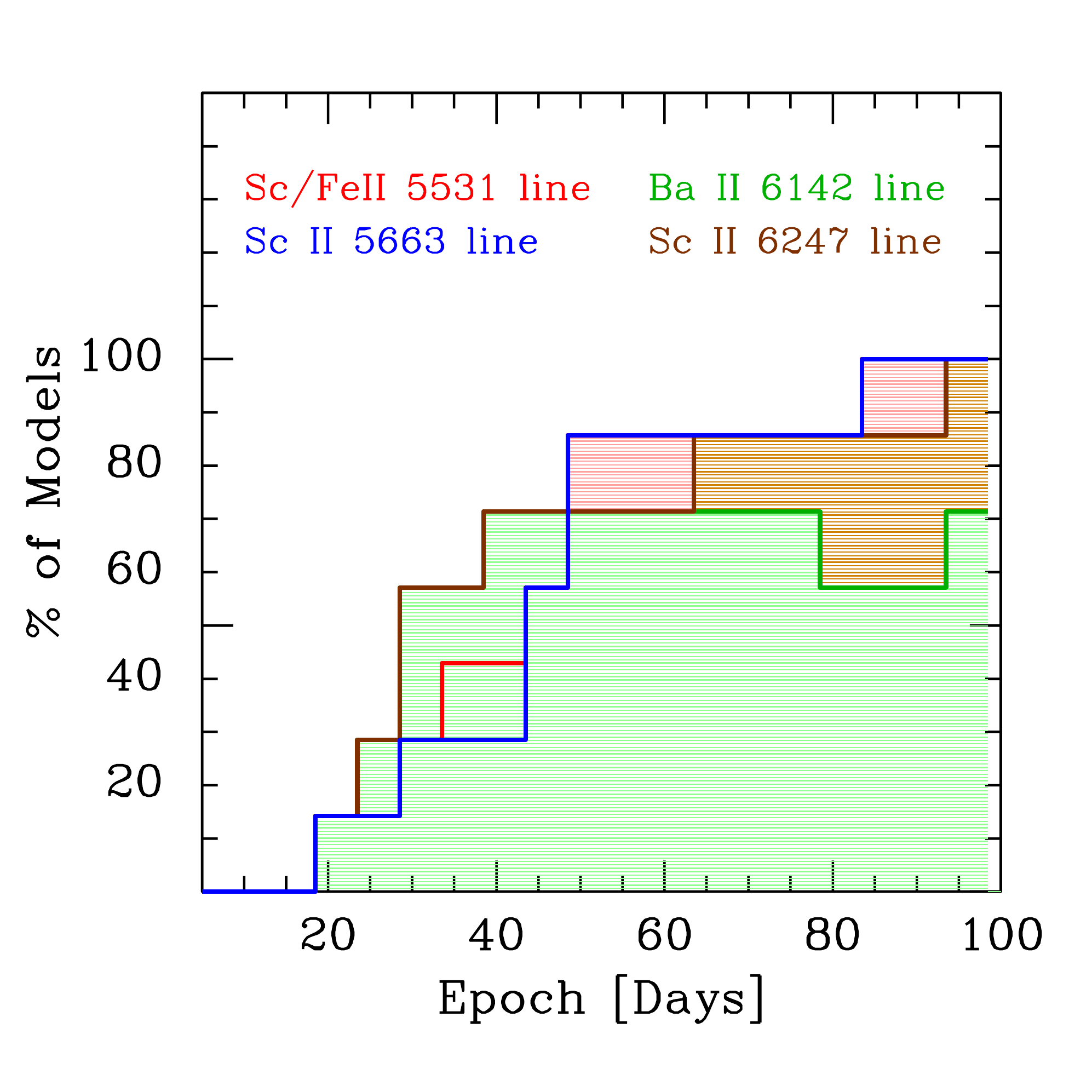}
\caption{Appearance of different lines in SNe~II between explosion and 100 days. \textbf{Top:} from the observed spectra.
\textbf{Bottom:} from synthetic spectra (see more details in Table~\ref{table_models}).}
\label{lines}
\end{figure*}

Here we study the time of appearance of different lines within different SNe, and make a comparison 
of those SNe with/without specific lines at different epochs. For all lines included in our analysis,
we search for their presence in each observed spectrum. Then, at any given epoch we obtain the percentage
of SNe that display each line. This enables an analysis of the overall line evolution of our sample and
whether the speed of this evolution changes between different SNe of different
light curve, spectral, and environment (metallicity) characteristics. \\
\indent In Figure~\ref{lines} we show the 
percentage of SNe displaying specific spectral features as a function of time.
As discussed previously, H$_{\alpha}$ and H$_{\beta}$ are permanently present in all the SNe~II spectra
from the first days, so we do not include them in the plot. We can see that:
\begin{itemize}
\item The feature located in the position of \ion{He}{1}/\ion{Na}{1} D is visible in all epochs, 
however around 15-25 days fewer SNe show the line respect to either earlier or later spectrum. We suggest that in this epoch the transition 
from \ion{He}{1} to \ion{Na}{1} D happens. Therefore, after 30 days we refer to this line as \ion{Na}{1} D.
It is present in 96\% of the spectra from $\sim35$ days. Later than 43 days it is present in all spectra. 
\item The \ion{Ca}{2} NIR triplet is present in $50\%$ of the sample at $\sim20$ days. Before 20 days it is present 
in $\sim12\%$ of the sample, while later than 25 days 
is visible in almost all the sample, but with one exception at 38 days. The latter is
SN~2009aj, which shows signs of CS interaction in the early phases. 
\item H$_{\gamma}$ blend with Fe-group lines starts at $\sim20$ days from explosion, growing dramatically
at 35-45 days. Only one spectrum at $\sim46$ days does not show the blend (SN~2008bp).  
\item The Fe-group lines start to appear at around 10 days (see Figure~\ref{lines}). The first line that emerges is 
\ion{Fe}{2} $\lambda5169$. We can see that few SNe exhibit the absorption feature before 
15 days, however later at 15 days around 50\% of SNe show the line and at 30 days all
objects have it. The next line that arises is \ion{Fe}{2} $\lambda5018$. This line is seen 
from 15 days, being present in all SNe later than 40 days. Meanwhile,
\ion{Fe}{2} $\lambda4924$ is seen in one spectrum at 13 days (SN~2008br).
From 30 days it is visible in more than 50\% of the spectra.
The \ion{Sc}{2}/\ion{Fe}{2} $\lambda5531$, \ion{Sc}{2} multiplet $\lambda5668$, \ion{Ba}{2} $\lambda6142$ and 
\ion{Sc}{2} $\lambda6246$ are detectable later than 30 days. The emergence of the \ion{Sc}{2}/\ion{Fe}{2} 
$\lambda5531$ and  \ion{Sc}{2} multiplet $\lambda5668$ happens at similar epochs, as well as \ion{Ba}{2} $\lambda6142$ and 
\ion{Sc}{2} $\lambda6246$. 
\end{itemize}

In order to further understand the differences in line-strength evolution of SNe~II, we separate the sample into those SNe that
do/do not display a certain spectral feature at some specific epoch. We then investigate whether these different samples also 
display differences in their light-curves and spectra. This is done by using the Kolmogorov-Smirnov (KS) test. 
Presented in Table~\ref{table_kstest} are all the results obtained with KS test: 
SNe with/without a given line as a function of $a/e$ and H$_{\alpha}$ velocity at t$_{tran+10}${\footnote{t$_{tran+10}$ is 
defined as the transition time (in $V-band$) between the initial and the plateau decline, plus 10 days. In other words,
{t$_{tran}$} marks the start of the recombination phase. (See A14 and \citep{Gutierrez14} for more details.)}},
M$_{max}$, s$_2$, and metallicity (derived from the ratio of 
H$_{\alpha}$ to [\ion{N}{2}] $\lambda6583$, henceforth M13 N2 diagnostic; \citealt{Marino13})
in a particular epoch. The values of the first four parameters can be found in Table~1 in \citet{Gutierrez14},
while the metallicity information was obtained from \citet{Anderson16}. 
We find that:
\begin{itemize}
\item SNe~II that never display the \ion{Fe}{2} line forest are distinctly different from those that do display the feature. 
Specifically, those that do show this feature have slow declining light curves (smaller s$_2$), are dimmer, and are found to
explode in higher metallicity regions within their hosts (see Table 4 for exact statistics).
\item There is less than a 2\% probability that those SNe~II where the \ion{He}{1} line is detected between 18 and 22 days 
post explosion arise from the same underlying parent population of $a/e$. This suggests that temperature differences between SNe~II
affect the morphology of the H$_{\alpha}$ feature.
\item \ion{Ba}{2} $\lambda$6142 and \ion{Sc}{2} $\lambda$6247 are both more
likely to be detected at around 40 days post explosion in dimmer SNe~II, with only a ~2\% probability that the two populations
(with and without these lines) are drawn from the same M$_{max}$ distribution.
\item Finally, when splitting the SNe~II sample into those that do and do not display: \ion{Sc}{2}/\ion{Fe}{2} $\lambda5531$, 
\ion{Sc}{2} multiplet $\lambda5668$, \ion{Ba}{2} $\lambda$6142 and \ion{Sc}{2} $\lambda$6247 at around 40 days post explosion we
find that there is only around a 1\% probability that the two samples are drawn from the same distribution of metallicity: 
those SNe that do not display these lines at this epoch are found to generally explode in regions of lower 
metallciity within their hosts 
\end{itemize}

\begin{figure*}
\centering
\includegraphics[width=\textwidth]{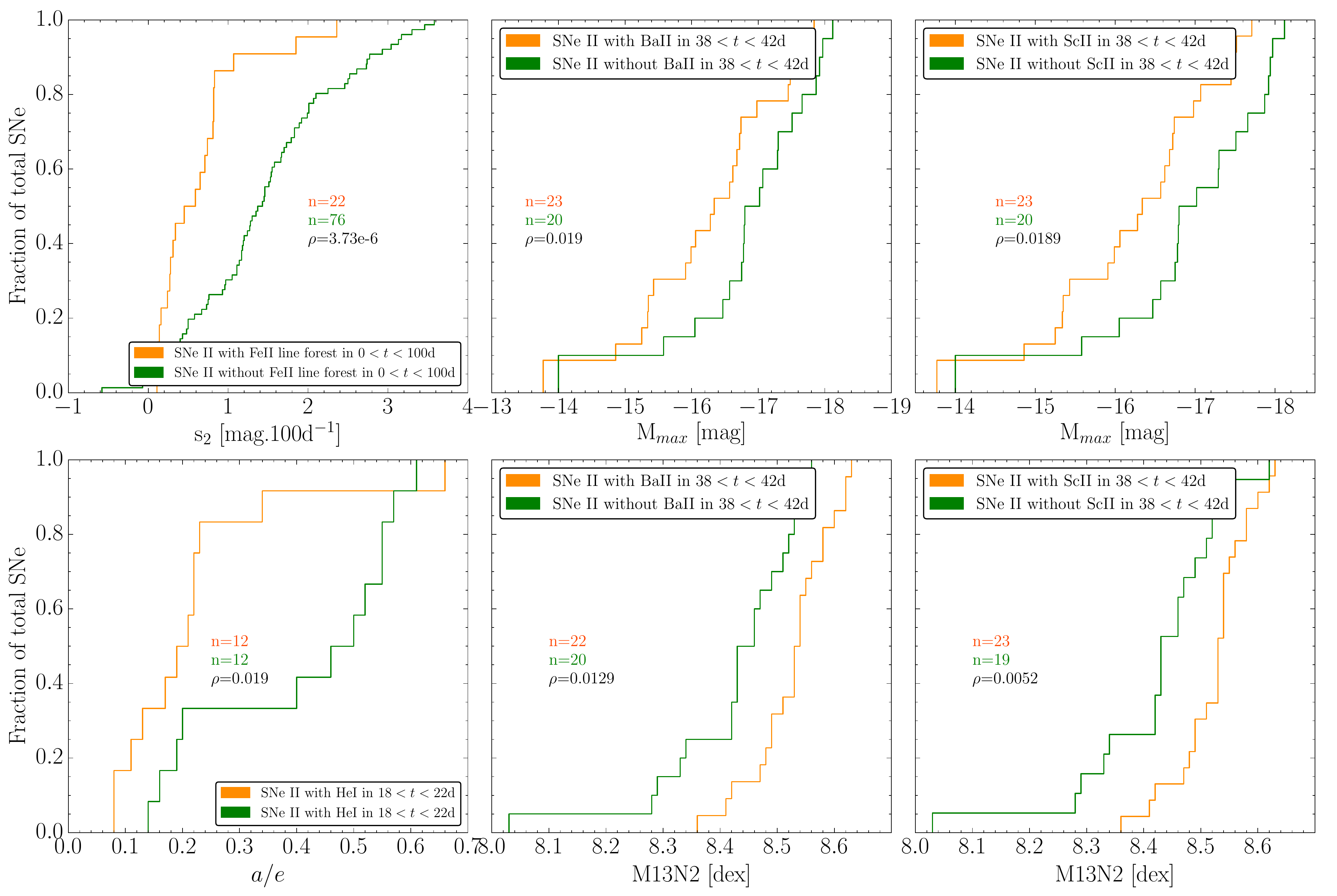}
\caption{Cumulative distributions of each SN with/without different lines as a function of spectral and 
photometric and environment properties. \textit{First panel}: SNe with/without \ion{Fe}{2} line forest between explosion and 100 days
as a function of s$_2$; \textit{second panel}: SNe with/without \ion{Ba}{2} between 38 and 42 days as a function of M$_{max}$;
\textit{third panel:} SNe with/without \ion{Sc}{2} between 38 and 42 days as a function of M$_{max}$; \textit{fourth panel:}
SNe with/without \ion{He}{1} between 18 and 22 days as a function of $a/e$; \textit{fifth panel:} SNe with/without \ion{Ba}{2} 
between 38 and 42 days as a function of M13N2 diagnostic; \textit{sixth panel:} SNe with/without \ion{Sc}{2} between 38 and 42 days as 
a function of M13N2 diagnostic.} 
\label{ksplot}
\end{figure*}

\indent Figure~\ref{ksplot} presents the cumulative distributions of the most significative findings
obtained with the KS-test analysis. 

\indent This analysis was also performed with synthetic spectra for seven different models from 
\citet{Dessart13a}. Four models (m15z2m3, m15z8m3, m15z8m3, m15z4m2) show differences in the metallicity, while
the rest of the properties are almost the same. The three remaining models have the same metallicity (solar metallicity),
however the other parameters are different: m15mlt1 has a bigger radius (twice times the radius of the two other models),
m15mlt3 has higher kinetic energy, while m12mlt3 displays a smaller final progenitor mass and less kinetic energy (1/5 E$_{kin}$
compared with the other models).
More details are shown in Table~\ref{table_models}. \\
\indent In general, the synthetic spectra show the same behaviour (in relation to the appearance of the lines) 
as observed spectra. However, some differences are found, the majority of which are probably related with the low area 
of parameter space covered by the models that currently exist as compared to the parameter space covered by real events.
The transition between \ion{He}{1} to \ion{Na}{1} D is more evident, and it happens between 18 and 40 days.
Although the transition in the models is unambiguously identified by knowing the optical depth of specific lines, in 
these synthetic spectra this happens a little bit later than in observed ones.
This suggests the temperature in specific models stays higher for a longer time than the average for observed SNe~II. 
It is also likely that the observed SNe~II span a smaller range in progenitor metallicity than the models 
(that go down to a tenth solar).
The \ion{Na}{1} D is visible in 100\% of the sample after 50 days, only 5 days later than
the observed spectra. 
\ion{Ca}{2} shows the same behaviour in both synthetic and observed spectra, however H$_{\gamma}$ 
is blended in all of the sample later than 90 days, unlike the observed spectra that show it from 45 days.
On the other hand, the \ion{Fe}{2} line forest is visible from 55 days, in contrast to 
the observed spectra that show this characteristic from 30 days.
This behaviour is only present in the spectra of higher metallicity model (2 times solar) and in the lower
explosion energy model. The iron lines (\ion{Fe}{2} $\lambda4924$, \ion{Fe}{2} $\lambda5018$,
\ion{Fe}{2} $\lambda5169$, and the
\ion{Fe}{2} blended) are present from $\sim10$ days. \ion{Fe}{2} $\lambda5169$ is visible in 50\% of the spectra
at $\sim15$ days, while \ion{Fe}{2} $\lambda4924$ is only visible in $\sim10$\%. From 20 days 
\ion{Fe}{2} $\lambda5169$ is present in all the synthetic spectra, 10 days before than in the observed ones. 
The behaviour of \ion{Fe}{2} $\lambda5018$ is similar in both synthetic and observed spectra, whereas 
\ion{Fe}{2} $\lambda4924$ starts faster in the models and it is visible in 85\% of the spectra from 30 days.
We can see differences in the \ion{Fe}{2} blend, which is visible in 100\% of the sample from 50 days
in the models, however in the observed spectra that never happens. More differences are also appreciable
between models and observation in \ion{Sc}{2}/\ion{Fe}{2} $\lambda5531$, the \ion{Sc}{2} multiplet
$\lambda5668$, \ion{Ba}{2} $\lambda6142$ and \ion{Sc}{2} $\lambda6246$. These lines in models arise from 20 days, 
but in the observations it occurs from 38-40 days. Nevertheless, the evolution of the distribution is similar from 50 days. 
In conclusion, while in general the models produce a time evolution of spectral lines that is quite similar to
the observations - supporting the robustness of the models - we observe small differences, suggesting 
a wider range of explosion and progenitor properties is required to explain the full diversity of observed SNe~II.

\subsection{Expansion velocity evolution}

\begin{figure*}
\includegraphics[width=18cm]{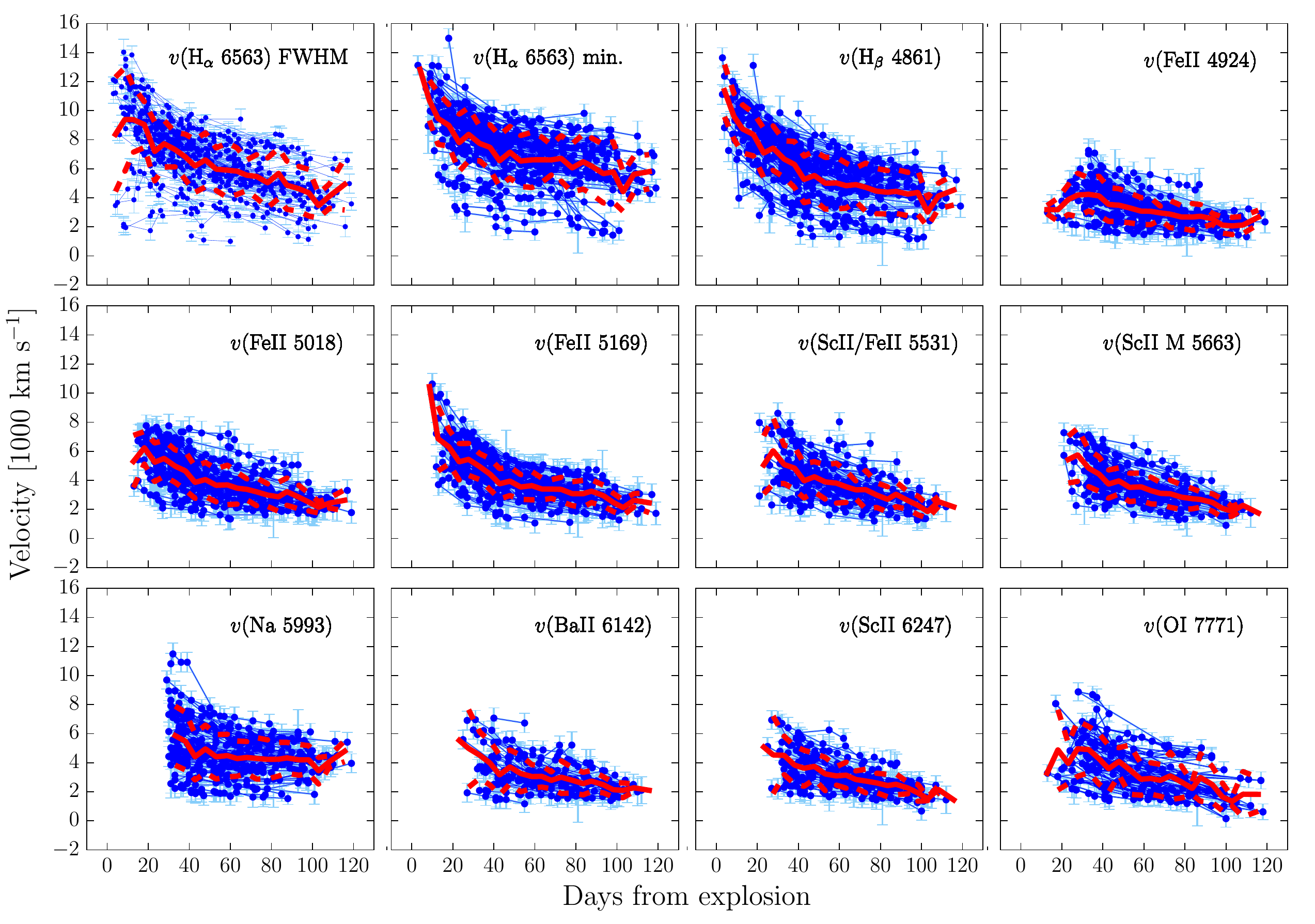}
\caption{Expansion velocity evolution for H$_{\alpha}$ (from the FWHM of emission and the minimum flux absorption),
H$_{\beta}$, \ion{Fe}{2} $\lambda4924$, \ion{Fe}{2} $\lambda5018$, \ion{Fe}{2} $\lambda5169$, 
\ion{Sc}{2}/\ion{Fe}{2}, \ion{Sc}{2} multiplet, \ion{Na}{1} D, \ion{Ba}{2}, \ion{Sc}{2} and \ion{O}{1}
from explosion to 120 days. The red solid line represents the mean velocity within each time bin,
while the dashed red lines indicate the standard deviation. Table~\ref{meanvel} presents these values.}
\label{vel}
\end{figure*}

\begin{figure}
\centering
\includegraphics[width=9cm]{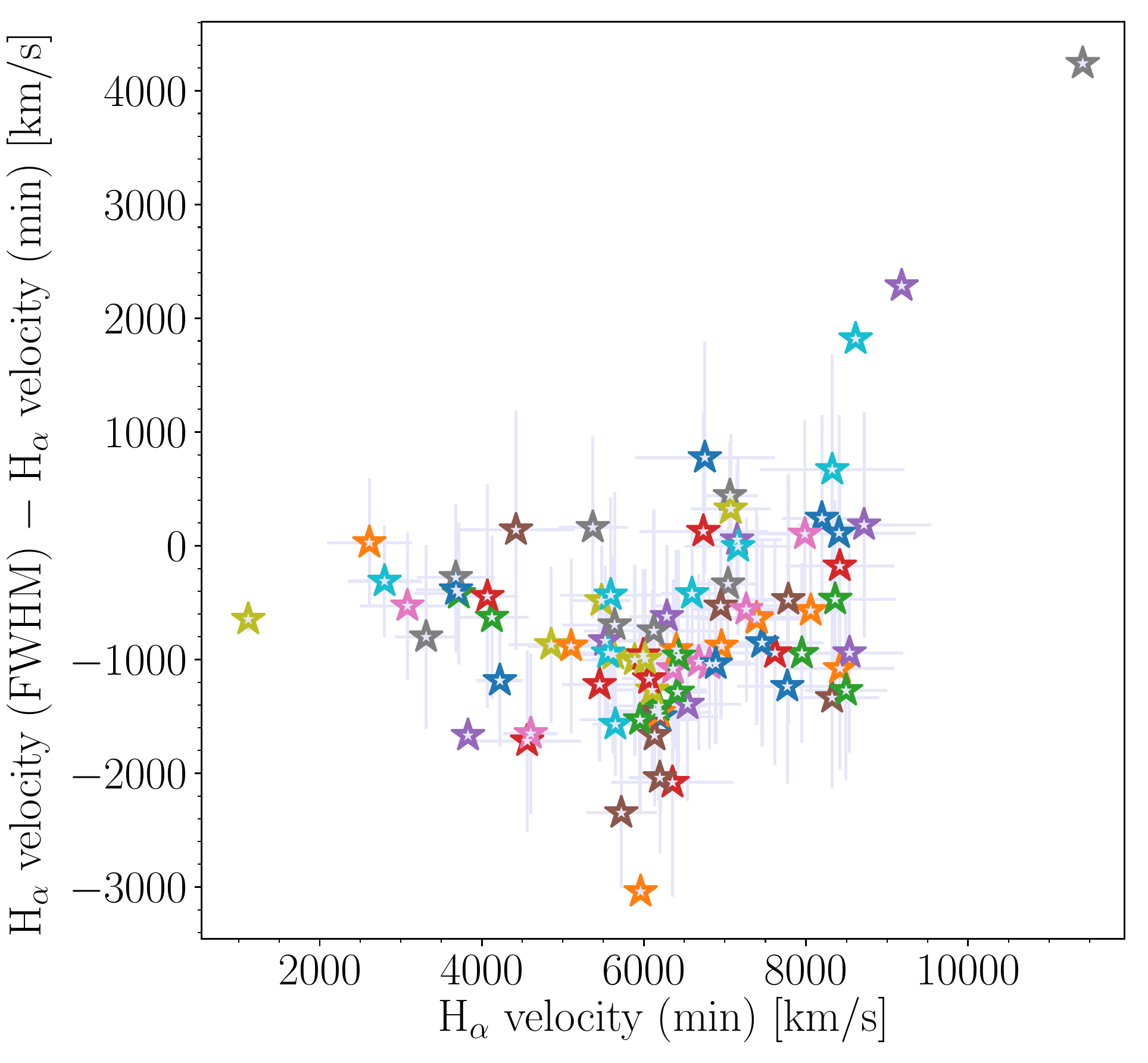}
\caption{Shifts of the H$_{\alpha}$ velocity obtained form the FWHM of the emission and from the
minimum of the absorption at 50 days post explosion.}
\label{hadist}
\end{figure}

Figure~\ref{vel} shows the velocity evolution of eleven spectral features as a
function of time. The first two panels of the plot show the expansion velocity of the
H$_{\alpha}$ feature: on the left the velocity derived from the FWHM and on the right that
derived from the minimum absorption flux. As we can see, the behaviour is similar, however the 
velocity obtained from the minimum absorption flux is offset between 10 and 20\%
to higher velocities. Figure~\ref{hadist} shows this shift at 50 days.
Velocities obtained from the minimum absorption flux are higher around $\sim1000$ km s$^{-1}$.
However, it is possible to see few SNe (with higher H$_{\alpha}$ velocities) showing higher 
values from the FWHM. Using the Pearson correlation test we find a weak correlation, with a value 
of $\rho=0.37$. SNe~II with narrower emission components display a larger offset between the velocity 
from the FWHM and that from the minimum of the absorption. In contrast, those SNe~II displaying the highest
FWHM velocities present comparatively lower minimum absorption velocities. 
We note also the presence of two outliers (extreme cases, the lowest and highest value). 
Figure~\ref{distvel} shows the velocity distribution for the eleven features at 50 days
post explosion. We can see that H$_{\alpha}$ shows higher velocities than the other lines, 
followed by H$_{\beta}$. The lowest velocities are presented by the iron-group lines.
In Figure~\ref{vel} it is possible to see that the H$_{\beta}$ expansion velocity shows the 
typical evolution for a homologous expansion and like H$_{\alpha}$, it is possible to see it 
from early phases. The iron lines display lower velocities than the Balmer lines.
So, the highest velocity in SNe~II is found in H$_{\alpha}$, which implies that
it is formed in the outer layers of the SN ejecta. Meanwhile, based on the lower velocities,
the iron-group lines form in the inner part, closer to the photosphere.
The \ion{O}{1} line does not show a strong evolution. As we can see, its velocity evolution is
almost flat. \\
\indent The lowest velocities are found in SN~2008bm, SN~2009aj and SN~2009au. However, these
SNe are distinct from the rest of the population. Unlike sub-luminous SNe~II (such as SN~2008bk and SN~1999br)
-- that also display low expansion velocities -- these events are relatively bright. They also show early signs
of CS interactions, e.g., narrow emission lines. By contrast, SN~2007ab, SN~2008if and SN~2005Z have the largest velocities.

\subsection{Velocity decline rate of H$_{\beta}$ analysis}

The velocity decline rate of SNe~II, denoted $\Delta v($H$_{\beta}$), has not been previously analyzed. We estimate  
$\Delta v($H$_{\beta})$ in five different epochs (outlined above) to understand
their behaviour. 
We find that SNe with a higher decline rate at early times
continue to show such behaviour at later times. 
The median velocity decline rate for our sample between 15 and 30 days is 105 km $s^{-1}$ d$^{-1}$, while
between 50 and 80 days is 29 km $s^{-1}$ d$^{-1}$. This results show an evident decrease in the velocity decline rate at two 
different intervals, which is consitent with homologous expansion. 



\subsection{pEWs evolution}

\begin{figure*}
\includegraphics[width=18cm]{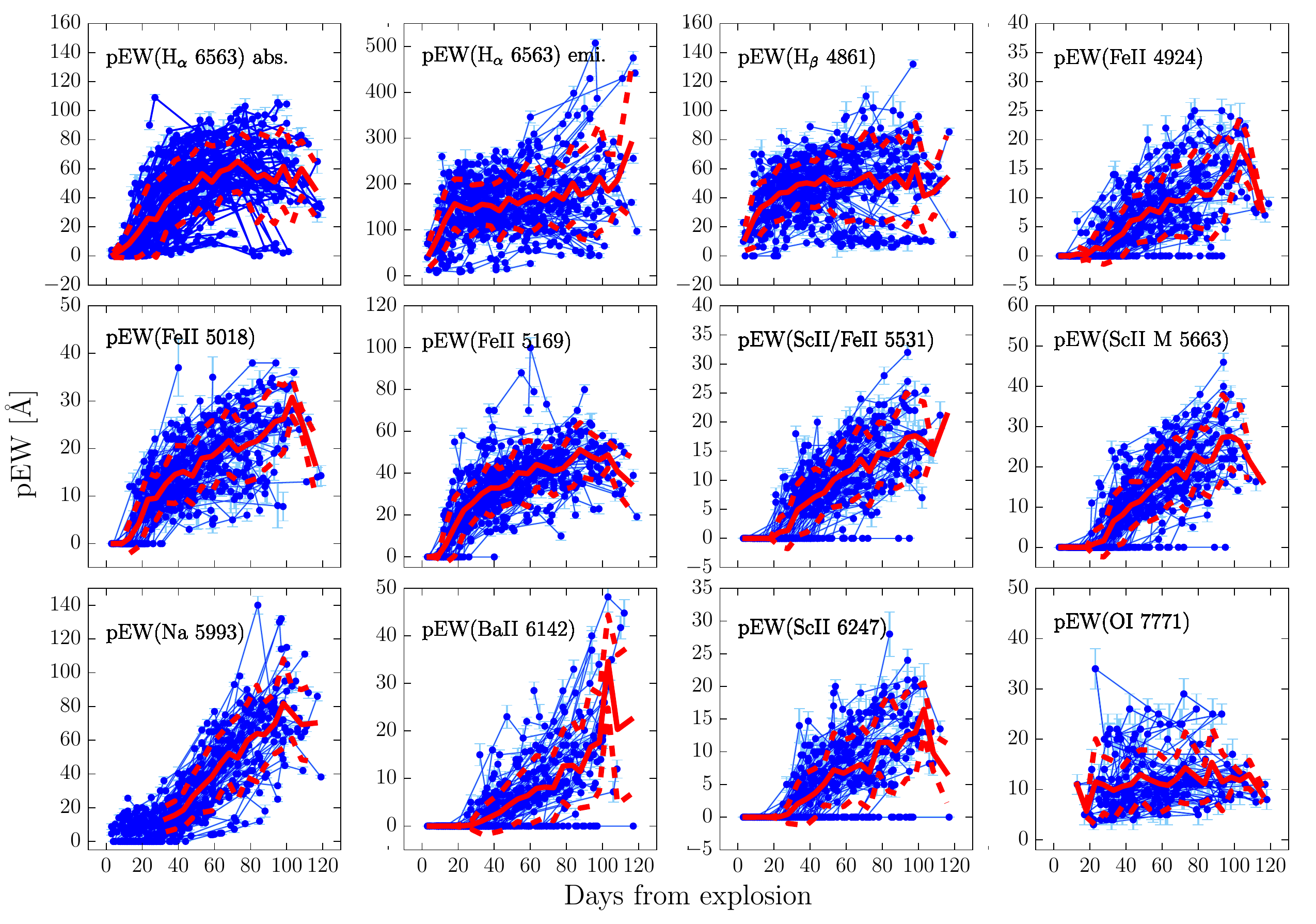}
\caption{pEWs evolution for H$_{\alpha}$ absorption, H$_{\alpha}$ emission, 
H$_{\beta}$, \ion{Fe}{2} $\lambda4924$, \ion{Fe}{2} $\lambda5018$, \ion{Fe}{2} $\lambda5169$,
\ion{Sc}{2}/\ion{Fe}{2}, \ion{Sc}{2} multiplet,
\ion{Na}{1} D, \ion{Ba}{2}, \ion{Sc}{2} and \ion{O}{1} from explosion to 120 days. 
The red sold line represents the mean pEW within each time bin,
while the dashed red lines indicate the standard deviation. Table~\ref{meanew} presents these values.}
\label{pews}
\end{figure*}

\begin{figure}
\hspace*{-0.3cm} 
\includegraphics[width=8.8cm]{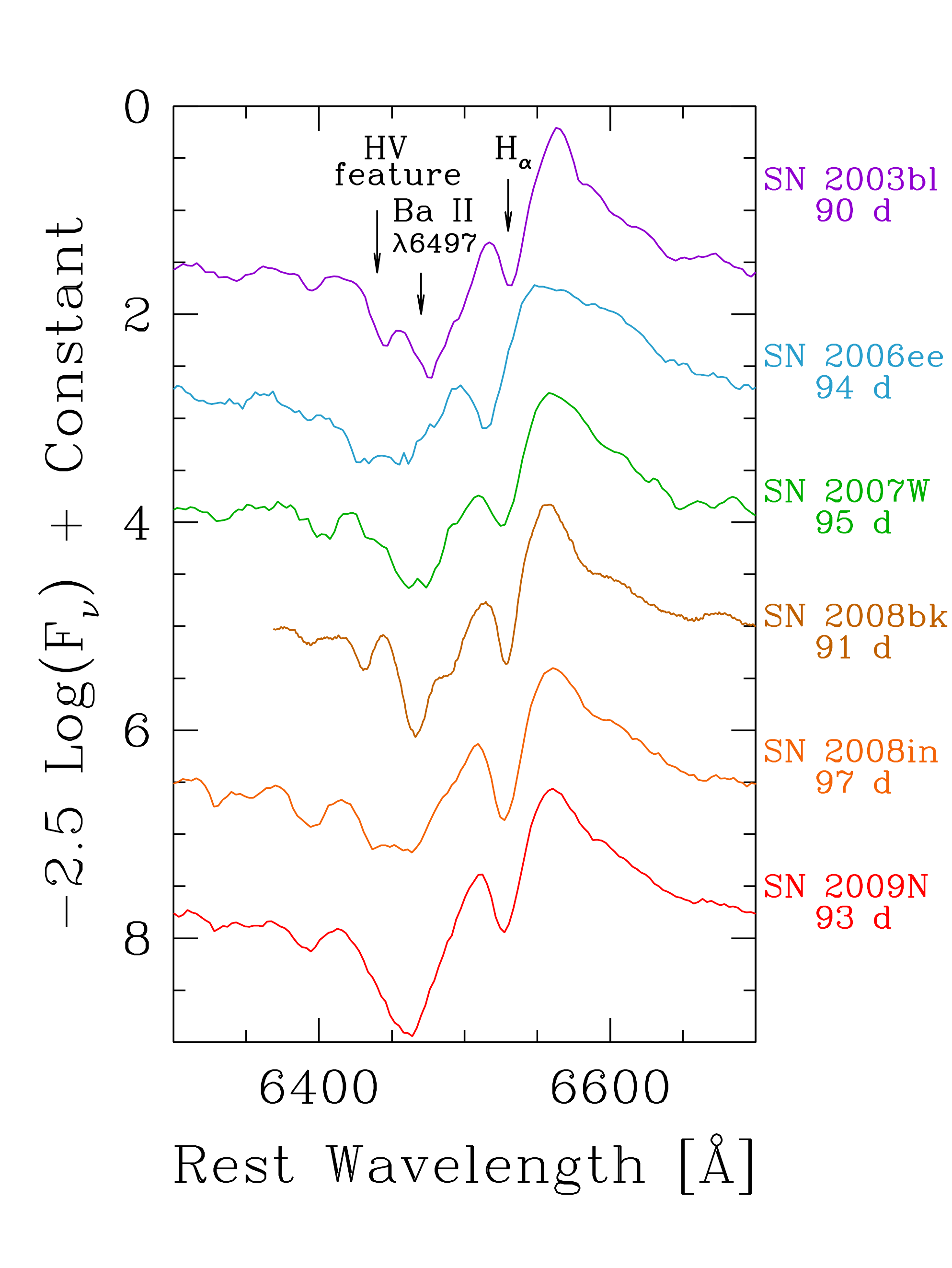}
\caption{H$_{\alpha}$ P-Cygni profile of low and intermediate velocity SNe~II:
2003bl, 2006ee, 2007W, 2008bk, 2008in and 2009N around 95 days post-explosion.  }
\label{ba2}
\end{figure}

The temporal evolution of pEWs for each of the eleven spectral features is shown in 
Figure~\ref{pews}. In general the pEWs increase quickly in the first 1-2 months 
then level off.
The first two panels show the pEW evolution of H$_{\alpha}$.
On the left is displayed the absorption, while on the right the emission component.
The absorption component monotonically increases from 0 increasing to $\sim100$ \AA, however in a few SNe
its evolution is different: from 70 days the pEW decreases significantly.
This behaviour is observed in low and intermediate velocity SNe
(e.g., SN~2003bl, SN~2006ee, SN~2007W, SN~2008bk, SN~2008in and SN~2009N). Generally, these SNe 
show a very narrow H$_{\alpha}$ P-cygni profile, and at around 70 days from explosion 
\ion{Ba}{2} $\lambda6497$ appears in the spectra as a dominant feature (see \citealt{Roy11,Lisakov17} for more details). 
In Figure~\ref{ba2} we can see the H$_{\alpha}$ P-Cygni profile with the 
presence of \ion{Ba}{2} $\lambda6497$, and the HV feature of hydrogen line (see section~\ref{cacho} for more details)
on the blue side of \ion{Ba}{2}. \\
\indent Figure~\ref{pews} also shows the H$_{\alpha}$ emission component evolution. An increment in the pEW
in the majority of SNe is appreciable. There are a couple cases (e.g. SN~2006Y), displaying a quasi-constant
evolution. The range of pEW of H$_{\alpha}$ emission goes up 400 \AA.  
In the case of H$_{\beta}$, we can see that from 60 days there are few SNe with low pEW values, which
show a quasi-constant evolution. SNe with this behaviour are those that show the \ion{Fe}{2} line forest.
The remaining SNe show an increase. The pEWs of iron-group lines grow with time, however there is
a group of SNe with pEW$=0$. This indicates that some specific SNe do not have the line yet.
For \ion{Sc}{2}/\ion{Fe}{2}, the \ion{Sc}{2} multiplet, \ion{Ba}{2}, and \ion{Sc}{2} this is more obvious. 
On the other hand, the \ion{O}{1} shows a quasi-constant behaviour and \ion{Na}{1} D a steady increase.
Comparing the values, we can see that the absorption of H$_{\alpha}$, H$_{\beta}$ and \ion{Na}{1} D have the highest values
(from 0 to $\sim120$), while \ion{Fe}{2} $\lambda4924$, \ion{Fe}{2} $\lambda5018$, \ion{Sc}{2}/\ion{Fe}{2}, the \ion{Sc}{2} multiplet,
\ion{Ba}{2}, \ion{Sc}{2} and \ion{O}{1} have the lowest ones (from 0 to $\sim50$).

\begin{figure}
\includegraphics[width=9cm]{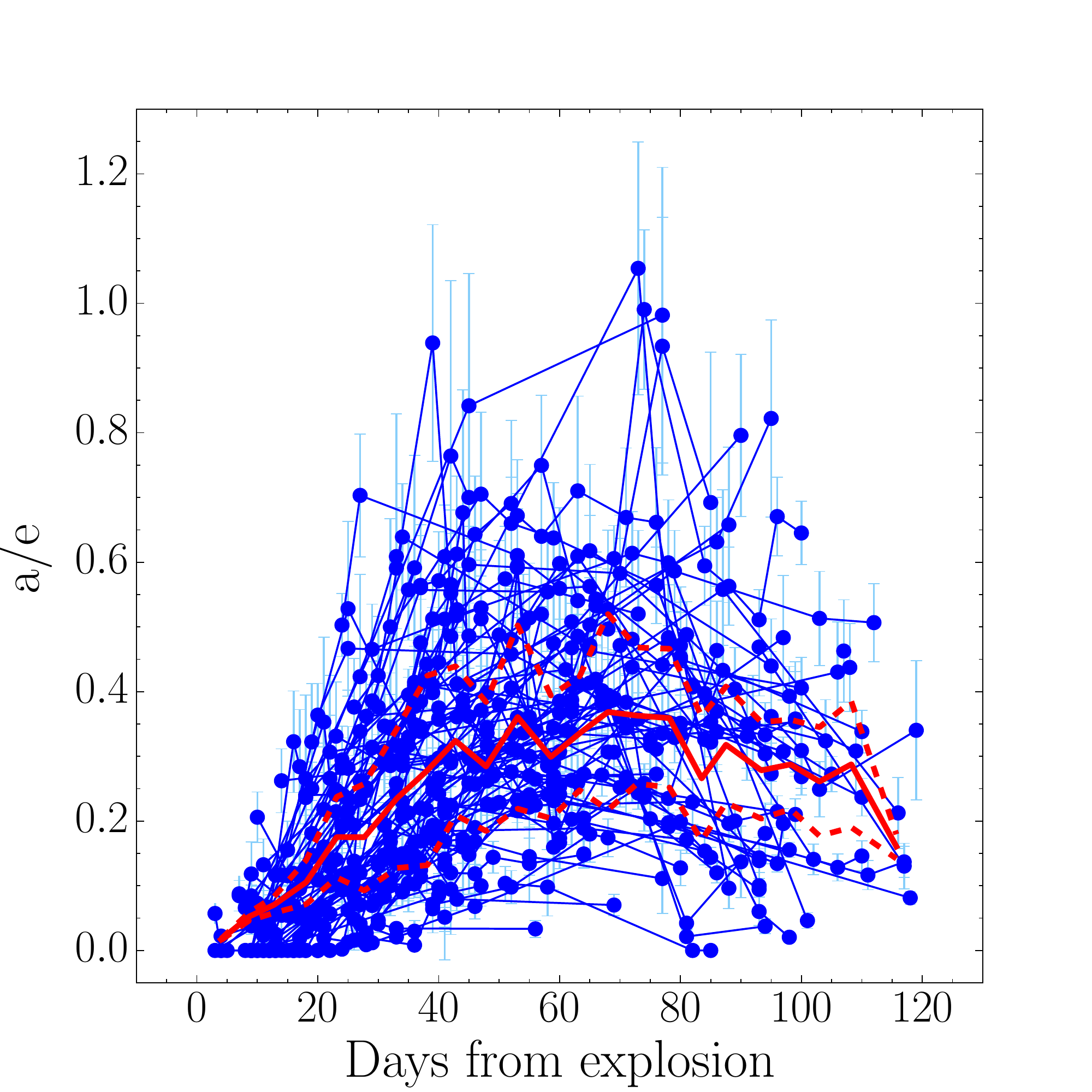}
\caption{Evolution of the ratio absorption to emission ($a/e$) of H$_{\alpha}$ between explosion and 120 days.}
\label{ae}
\end{figure}

The $a/e$ evolution is displayed in Figure~\ref{ae}. One can see an increase until $\sim60$ days 
and then, the quantity remains constant or slightly decreases.

\subsection{Cachito: Hydrogen HV features or \ion{Si}{2} line}
\label{cacho}

\begin{figure*}
\includegraphics[width=18cm]{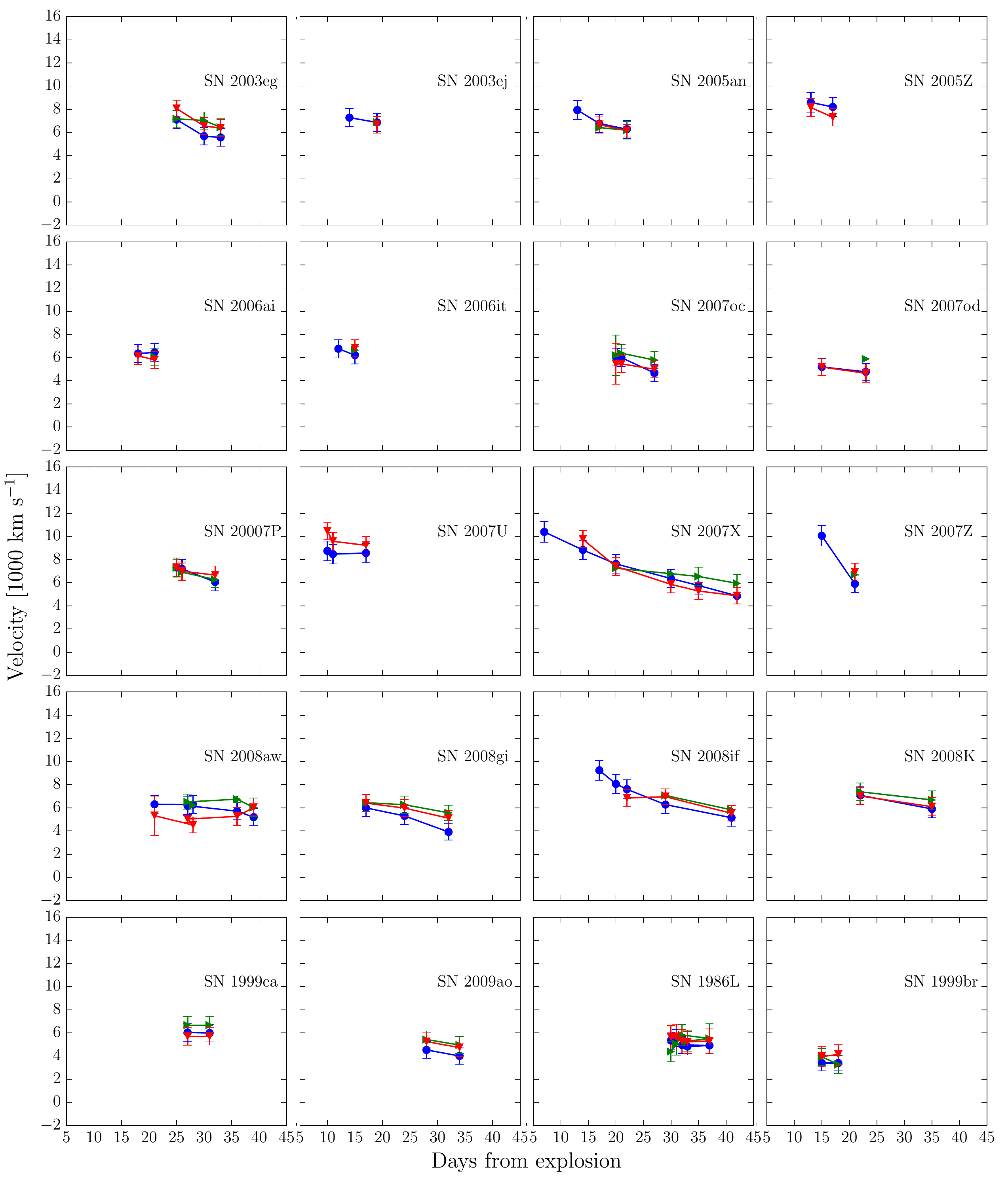}
\caption{Velocity evolution of Cachito (blue) at early phases
compared with \ion{Fe}{2} $\lambda$5018 (green) and \ion{Fe}{2} $\lambda$5169 (red). }
\label{velsi2}
\end{figure*}

\indent The nature of Cachito has recently been studied. Its presence on the blue side 
of H$_{\alpha}$ has given rise to multiple interpretations, such as HV features of hydrogen 
\citep[e.g.][]{Leonard02b, Baron00, Chugai07,Inserra11} or \ion{Si}{2} $\lambda6355$ 
\citep[e.g.][]{Pastorello06,Valenti13a,Tomasella13}.
Seventy SNe from our sample show Cachito in the photospheric phase,
between 7 and 120 days post-explosion, however its behaviour, shape and evolution is different
depending on the phase. To investigate the nature of Cachito we examine the following possibilities:
\begin{itemize}
\item If Cachito is produced by \ion{Si}{2} its velocity should be similar to those presented by 
other metal lines. 
\item If Cachito is related to HV features of hydrogen,  its velocity should be almost the same as those
obtained from H$_{\alpha}$ at early phases. In addition, if it is present, 
a counterpart should be visible on the blue side of H$_{\beta}$.
\end{itemize}
\indent Analyzing our sample we can detect Cachito in 50 SNe at early phases (before 40 days).
Because of the high temperatures at these epochs, the presence of \ion{Ba}{2} $\lambda6497$ is discarded.
Assuming that Cachito is produced by \ion{Si}{2}, we find that 60\% of SNe present a good match
with \ion{Fe}{2} $\lambda5018$ and \ion{Fe}{2} $\lambda5169$ velocities\footnote{Four SNe show a good
match with \ion{Si}{2} in very early phases, but between 30 and 40 days they do not show it.
They also show a different shape.}. Conversely, the rest of the sample shows velocities comparable to those
measured at very early phases for H$_{\alpha}$. Curiously the Cachito shape is different between
the two SN groups. In the former, the line is deeper and broader, while in the latter, the line is shallow. 
In Figure~\ref{velsi2} we present the velocity comparison for the former group, 
where a good agreement is found between Cachito, assumed as \ion{Si}{2} $\lambda6355$ (blue), and the iron lines,
\ion{Fe}{2} $\lambda5018$ (green) and \ion{Fe}{2} $\lambda5169$ (red). 

\begin{figure}
\includegraphics[width=9cm]{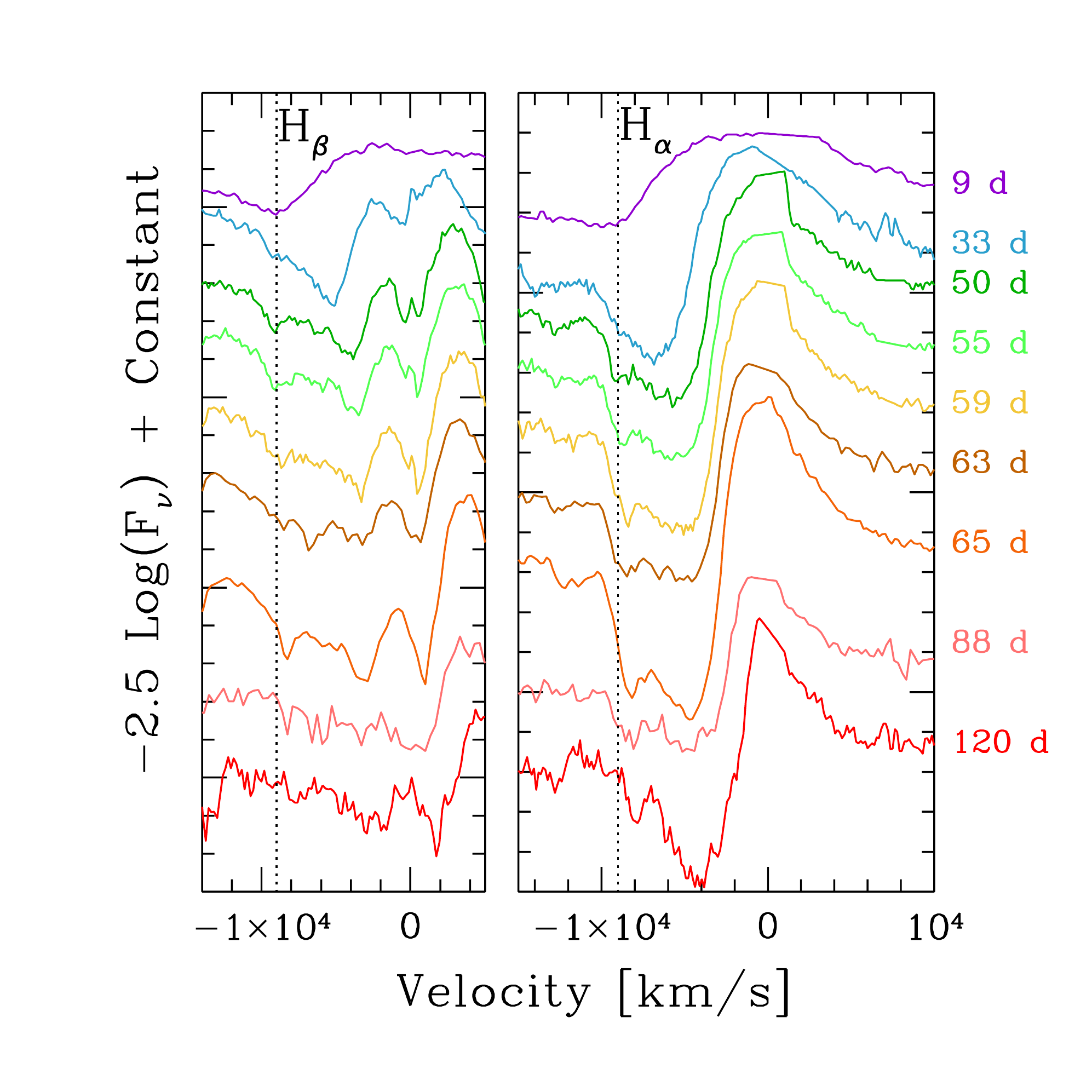}
\caption{Spectral evolution of H$_{\alpha}$ and H$_{\beta}$ lines of SN~2004fc. The dotted lines correspond to the 
HV features seen on the blue side of  H$_{\alpha}$ and H$_{\beta}$ from 50 to 120 days. We can see that the HV
features show a velocity evolution from $\sim9000$ to $\sim8000$ km s$^{-1}$.}
\label{hv04fc}
\end{figure}

\begin{figure*}
\centering
\includegraphics[width=18cm]{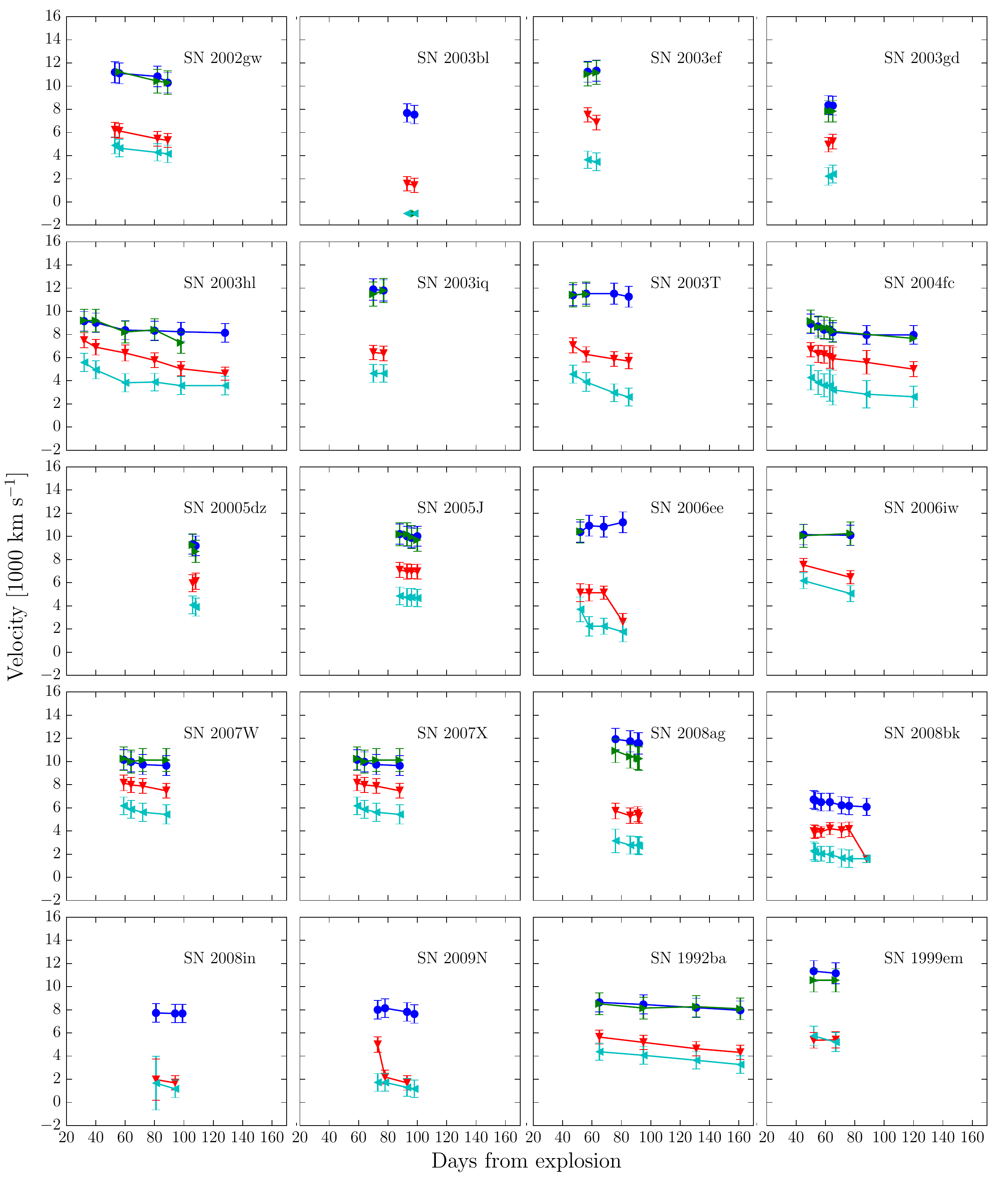}
\caption{Velocity evolution of Cachito in the plateau phase 
compared with the Balmer lines. In blue: HV of H$_{\alpha}$, in green: HV of H$_{\beta}$, in red: the H$_{\alpha}$
velocity and in cyan the H$_{\beta}$ velocity.}
\label{hvfs}
\end{figure*}

\indent Later than 40 days we detect Cachito in 43 SNe. Proceeding with the velocity comparison, we can discard
its identification as \ion{Si}{2} or \ion{Ba}{2} $\lambda6497$ (the latter, visible in few SNe from 60 days, see 
Figure~\ref{ba2}), which suggests that Cachito is associated to hydrogen. During the plateau it is 
possible to see Cachito as a shallow absorption feature only in H$_{\alpha}$
and/or as a narrow and deeper absorption on the blue side of both H$_{\alpha}$ and H$_{\beta}$
(see an example in Figure~\ref{hv04fc}). 
According to \citet{Chugai07}, the interaction between the SN ejecta and the RSG wind
should result in the emergence of these HV absorption features. They argue that the existence of a shallow
absorption feature is the result of the enhanced excitation of the outer unshocked ejecta, which is 
visible on the blue side of H$_{\alpha}$ (and \ion{He}{1} $\lambda$10830). At early times the H$_{\beta}$ Cachito 
feature is not predicted by \citet{Chugai07}, who argue that the optical depth is too low at the line forming region.
They also discuss that in addition to the HV shallow absorption,
a HV notch is formed in the cool dense shell (CDS) located behind the reverse shock.
Given the relatively high H$_{\alpha}$ optical depth of the CDS, a counterpart could be seen in H$_{\beta}$ as well.
We found that 63\% of the SNe with Cachito during the plateau show a counterpart in H$_{\beta}$
with the same velocity as that presented on H$_{\alpha}$, which favours the interpretation as CS
interaction. The HV notch of \ion{H}{1} is found in 27 SNe, however in the low velocity/luminosity SNe, it is only present
in H$_{\alpha}$. After 50 days the blue part of the spectrum (\textless 5000 \AA) is dominated 
by metal lines, which may hinder its detection.
Nonetheless, we argue that these can be HV \ion{H}{1}
because at least one low velocity/luminosity SN, SN~2006ee shows a Cachito feature on the blue side of both
H$_{alpha}$ and H$_{beta}$, at around 50 days with consistent velocities.
A summary of the analysis is displayed in Figure~\ref{hvfs}, where the H$_{\alpha}$ (red), HV H$_{\alpha}$ (blue), 
H$_{\beta}$ (cyan), and HV H$_{\beta}$ (green) velocity evolution is presented for 20 SNe.

\begin{figure}
\includegraphics[width=6.5cm,angle=270]{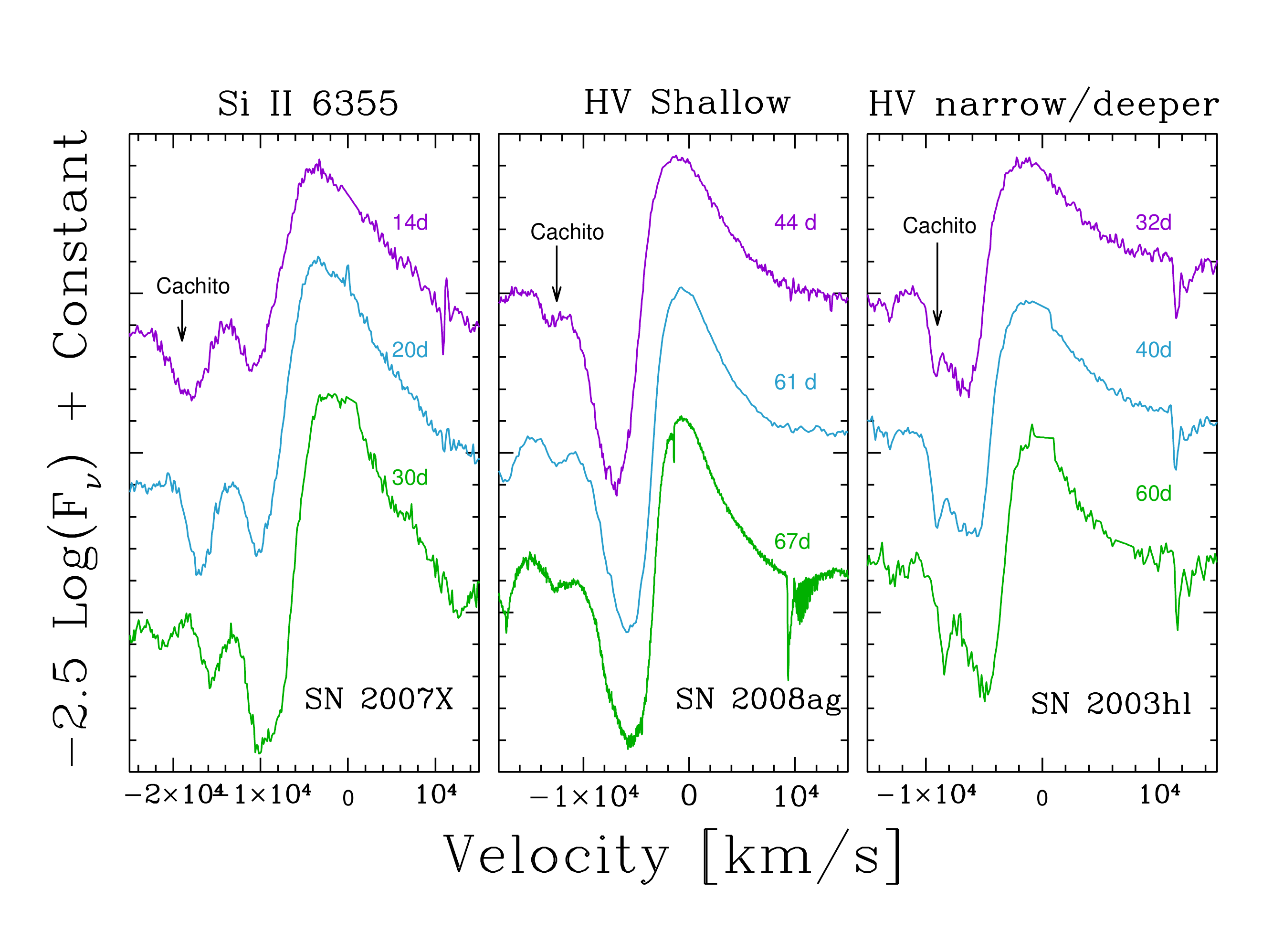}
\caption{Cachito's shape according to its nature. \textbf{On left panel:} The \ion{Si}{2} line in SN~2007X.
\textbf{In the middle panel:} HV features of \ion{H}{1} as a shallow absorption in SN~2008ag. \textbf{On right panel:}
HV features of \ion{H}{1} as narrow and deeper absorption component in SN~2003hl.}
\label{compc}
\end{figure}

\indent In addition to the 70 SNe where Cachito can be identified either with \ion{Si}{2} or HV features of 
\ion{H}{1}, we find six SNe~II that display
Cachito at certain epochs, however its exact properties do not align with the above interpretations 
(because of differences in shape and/or velocity). These are SN~2003bl, SN~2005an, SN~2007U, SN~2008br,
SN~2002gd and SN~2004fb. In summary 59\% of the full SNe sample show Cachito at some epoch,
while 41\% never show this feature.
Soon after shock-break out all SNe~II have extremely high temperature ejecta. Therefore, if we were able to obtain spectral 
sequences shortly after explosion, the \ion{Si}{2} feature
would always be observed. However, observationally this is not the case as there are many SNe~II within our sample 
without \ion{Si}{2} detections. This is simply an observational bias, due to the lack of data 
at very early times. Nevertheless, for SNe~II that stay hotter for longer
the probability of detecting \ion{Si}{2} becomes larger. We therefore speculate that SNe~II that have detected \ion{Si}{2}
at early times have larger radii, which leads to a slower cooling of the ejecta and hence facilitates \ion{Si}{2} detection.
Interestingly, when we split the sample into those SNe~II that do and do not display the 
\ion{Si}{2} line, those where the line is detected are found to have lower $a/e$ values, with only a 4\% chance that the
two populations are drawn from the same underlying distribution. This is also consistent with the previous finding that
those SNe~II with evident \ion{He}{1} detections at around 20 days post explosion are also found to have lower $a/e$ values,
suggesting that the value of $a/e$ is related to ejecta temperature evolution.\\
\indent In the case of those SNe~II displaying Cachito consistent with HV features, these are most likely produced by the interaction
of the SN ejecta with the RSG wind, where the exact shape and persistence of Cachito is related to the wind density \citet{Chugai07}.
In Figure~\ref{compc} one can observe the significant diversity in the different detection of Cachito.

\section{Conclusions}
\label{conc}

In this paper we have presented optical spectra of 122 nearby SNe~II observed between 1986
and 2009. A total of 888 spectra ranging between 3 and 363 days post explosion 
have been analyzed. The spectral matching technique was discussed as an
alternative to non-detection constraints for estimating SN explosion epochs.\\
\indent In order to quantify the spectral diversity we analyze the appearance of the
photospheric lines and their time evolution in terms of the $a/e$ and H$_{\alpha}$ velocity at 
the $B$-band transition time plus 10 days (t$_{tran+10}$; see \citealt{Gutierrez14} for more details), the magnitude at maximum (M$_{max}$),
the plateau decline (s$_2$), and metallicity (M13 N2). We analyzed the velocity decline rate of H$_{\beta}$, the $a/e$ 
evolution, the expansion ejecta velocities and the pEWs for eleven features: H$_{\alpha}$, H$_{\beta}$,
\ion{He}{1}/\ion{Na}{1} D, \ion{Fe}{2} $\lambda4924$, \ion{Fe}{2} $\lambda5018$,
\ion{Fe}{2} $\lambda5169$, \ion{Fe}{2} blend, 
\ion{Sc}{2}/\ion{Fe}{2}, \ion{Sc}{2} multiplet, \ion{Ba}{2}, \ion{Sc}{2}, and \ion{O}{1}. 
We find a large range in velocities and pEWs, which may be related with a diversity in the 
explosion energy, radius of the progenitor, and metallicity.
The evolution of line strengths was analyzed and compared to that of spectral models. SNe~II displaying differences
in spectral line evolution were also found to have other different spectral, photometric and environmental properties.
Finally, we discuss the detection and origin of Cachito on the blue side of H$_{\alpha}$. \\
\indent The main results obtained with our analysis are summarized as follows:
\begin{itemize}
\item The line evolution indicates differences in temperatures and/or metallicity. Thus,
SNe with slower temperature gradients show the appearance of the iron lines later, while SNe in 
environments with higher metallicities show them earlier. In fact, the \ion{Fe}{2} line
forest is present in faint SNe with low ejecta temperatures and/or in high metallicity environments.
Comparing this result with the synthetic spectra, we find that indeed this feature is only present
in higher metallicity (2 times solar) and lower
explosion energy models, which is consistent with our observations. 
\item SNe~II display a significant variety of expansion velocities, suggesting a large range in explosion
energies.
\item At early phases (before 25 days), SNe~II with a weak H$_{\alpha}$ absorption component show
\ion{He}{1} $\lambda5876$ and the \ion{Si}{2} $\lambda6355$ features. We speculate that this occurs because of
higher temperatures at these epochs.
\item Around 60\% of our SNe~II show the Cachito feature between 7 and 120 days since explosion.
When Cachito is detected less than 30 days post explosion then it is identified with \ion{Si}{2}. 
The epochs of early detection can thus inform us to the temperature evolution: SNe~II with 
\ion{Si}{2} detections at later epochs have higher temperatures, and this may be related to higher-radius 
progenitors. At later epochs, during the recombination phase, we suggest that Cachito is related to HV 
of hydrogen lines. Such HV features are most likely related to the interaction of the SN ejecta with the RSG wind.
\end{itemize}

All data analyzed in this work are available on http://csp.obs.carnegiescience.edu/, as well as the additional SNID templates 
(22 SNe), for the SNe~II comparison.

\acknowledgments
C.P.G., S.G.G. acknowledge support by projects IC120009 ``Millennium
Institute of Astrophysics (MAS)" and P10-064-F ``Millennium Center for Supernova
Science" of the Iniciativa Científica Milenio del Ministerio Economía,
Fomento y Turismo de Chile. C.P.G. acknowledges support from EU/FP7-ERC grant No. [615929].
M.D.S. is supported by the Danish Agency for Science and Technology
and Innovation realized through a Sapere Aude Level 2 grant and by a research grant (13261) from the VILLUM FONDEN.
We gratefully acknowledge support of the CSP by the NSF under grants AST–0306969, AST–0908886, AST–0607438, AST–1008343,
AST-1613472, AST-1613426 and AST-1613455.
This research has made use of the NASA/IPAC
Extragalactic Database (NED) which is operated by the Jet Propulsion 
Laboratory, California Institute of Technology, under contract with the
National Aeronautics.

\clearpage
\LongTables

\clearpage
\setcounter{table}{0}
\begin{landscape}
\clearpage
\centering
\tiny
\begin{longtable*}[t!]{lcccccccccccccccc}
\caption[SNe~II sample]{SN~II sample} 
\label{t_info} \\
\hline
& Host & Recession & Hubble & $E(B-V)_{\rm MW}$ & Discovery & Discovery  & Explosion & N of & \\
SN & Galaxy & velocity (km s$^{-1}$) & type & (mag) & date & Reference  & Epoch & spectra & Campaign\\
\hline
\hline
1986L	& NGC 1559			& 1305 		& SBcd  & 0.026 &   46711.1   & IAUC 4260  & 46708.0$^{n}$(3)  & 31	& CTSS\\
1988A	& NGC 4579			& 1517 		& SABb 	& 0.036	&   47179.0   & IAUC 4533  & 47177.2$^{n}$(2)  & 5	& CTSS\\
1990E	& NGC 1035			& 1241		& SAc 	& 0.022	&   47937.7   & IAUC 4965  & 47935.1$^{n}$(3)  & 5	& CTSS\\
1990K	& NGC 0150			& 1584	 	& SBbc 	& 0.013	&   48037.3   & IAUC 5022  & 48001.5$^{n}$(6)  & 9	& CTSS\\
1991al	& 2MASX J19422191–5506275	& 4575$^1$	& ?  	& 0.054	&   48453.7   & IAUC 5310  & 48442.5$^{s}$(8)* & 8	& CT\\
1992af	& ESO 340-G038			& 5541 		& S  	& 0.046	&   48802.8   & IAUC 5554  & 48798.8$^{s}$(8)* & 5	& CT\\
1992am	& MCG -01-04-039		& 14397$^1$ 	& S 	& 0.046	&   48829.8   & IAUC 5570  & 48813.9$^{s}$(6)* & 2	& CT\\
1992ba	& NGC 2082			& 1185  	& SABc  & 0.051	&   48896.2   & IAUC 5625  & 48884.9$^{s}$(7)  & 10	& CT\\
1993A	& 2MASX J07391822–6203095	& 8790$^1$  	& ? 	& 0.153	&   49004.6   & IAUC 5693  & 48995.5$^{n}$(9)  & 2	& CT\\
1993K	& NGC 2223			& 2724  	& SBbc  & 0.056	&   49075.5   & IAUC 5733  & 49065.5$^{n}$(9)  & 17	& CT\\
1993S	& 2MASX J22522390-4018432 	& 9903 		& S	& 0.014 &   49133.7   & IAUC 5812  & 49130.8$^{s}$(5)  & 4	& CT\\
1999br	& NGC 4900			& 960   	& SBc   & 0.021	&   51281.0   & IAUC 7141  & 51276.7$^{n}$(4)  & 8	& SOIRS\\
1999ca	& NGC 3120			& 2793  	& Sc    & 0.096 &   51296.0   & IAUC 7158  & 51277.5$^{s}$(7)* & 4	& SOIRS\\
1999cr	& ESO 576-G034			& 6069$^1$  	& S/Irr	& 0.086 &   51249.7   & IAUC 7210  & 51246.5$^{s}$(4)* & 5	& SOIRS\\
1999eg	& IC 1861			& 6708  	& SA0   & 0.104 &   51455.5   & IAUC 7275  & 51449.5$^{s}$(6)* & 2	& SOIRS\\
1999em	& NGC 1637			& 717   	& SABc  & 0.036 &   51481.0   & IAUC 7294  & 51476.5$^{n}$(5)  & 12	& SOIRS\\
2002ew	& NEAT J205430.50-000822.0	& 8975  	& ?	& 0.091 &   52510.8   & IAUC 7964  & 52500.6$^{n}$(10) & 7	& CATS\\
2002fa	& NEAT J205221.51+020841.9 	& 17988 	& ? 	& 0.088 &   52510.8   & IAUC 7967  & 52502.5$^{s}$(8)* & 6	& CATS\\
2002gd	& NGC 7537			& 2676 		& SAbc  & 0.059 &   52552.7   & IAUC 7986  & 52551.5$^{s}$(4)* & 12	& CATS\\
2002gw	& NGC 922			& 3084  	& SBcd  & 0.017 &   52560.7   & IAUC 7995  & 52553.5$^{s}$(8)* & 11	& CATS\\
2002hj	& NPM1G +04.0097		& 7080  	& ? 	& 0.102 &   52568.0   & IAUC 8006  & 52562.5$^{n}$(7)  & 7	& CATS\\
2002hx	& PGC 023727			& 9293  	& SBb   & 0.048 &   52589.7   & IAUC 8015  & 52582.5$^{n}$(9)  & 9	& CATS\\
2002ig	& SDSS J013637.22+005524.9	& 23100$^2$ 	& ? 	& 0.034 &   52576.7   & IAUC 8020  & 52570.5$^{s}$(5)* & 5	& CATS\\
210	& MCG +00-03-054		& 15420  	& ?     & 0.033 &   ?$^3$     & ?          & 52486.5$^{s}$(6)* & 6	& CATS\\
2003B	& NGC 1097			& 1272  	& SBb   & 0.024 &   52645.0   & IAUC 8042  & 52613.5$^{s}$(11)*& 9	& CATS\\
2003E	& MCG -4-12-004			& 4470$^1$  	& Sbc	& 0.043 &   52645.0   & IAUC 8044  & 52629.5$^{s}$(8)* & 8	& CATS\\
2003T	& UGC 4864			& 8373  	& SAab  & 0.028 &   52665.0   & IAUC 8058  & 52654.5$^{n}$(10) & 6	& CATS\\
2003bl	& NGC 5374			& 4377$^1$  	& SBbc	& 0.024 &   52701.0   & IAUC 8086  & 52696.5$^{s}$(4)* & 8	& CATS\\
2003bn	& 2MASX J10023529-2110531 	& 3828  	& ?	& 0.057 &   52698.0   & IAUC 8088  & 52694.5$^{n}$(3)  & 12	& CATS\\
2003ci	& UGC 6212			& 9111  	& Sb    & 0.053 &   52720.0   & IAUC 8097  & 52711.5$^{n}$(8)  & 7	& CATS\\
2003cn	& IC 849	        	& 5433$^1$  	&SABcd	& 0.019 &   52728.0   & IAUC 8101  & 52717.5$^{s}$(4)* & 5	& CATS\\
2003cx	& NEAT J135706.53-170220.0 	& 11100 	& ? 	& 0.083 &   52730.0   & IAUC 8105  & 52725.5$^{s}$(5)* & 6	& CATS\\
2003dq	& MAPS-NGP O4320786358      	& 13800 	& ? 	& 0.016 &   52739.7   & IAUC 8117  & 52731.5$^{n}$(8)  & 3	& CATS\\
2003ef	& NGC 4708			& 4440$^1$  	& SAab  & 0.041 &   52770.7   & IAUC 8131  & 52757.5$^{s}$(9)* & 6	& CATS\\
2003eg	& NGC 4727			& 4388$^1$ 	& SABbc & 0.046 &   52776.7   & IAUC 8134  & 52764.5$^{s}$(5)* & 5	& CATS\\
2003ej	& UGC 7820			& 5094  	& SABcd & 0.017 &   52779.7   & IAUC 8134  & 52775.5$^{n}$(5)  & 3	& CATS\\
2003fb	& UGC 11522			& 5262$^1$  	& Sbc   & 0.162 &   52796.0   & IAUC 8143  & 52772.5$^{s}$(10)*& 4	& CATS\\
2003gd	& M74	       			& 657   	& SAc   & 0.062 &   52803.2   & IAUC 8150  & 52755.5$^{s}$(9)* & 3	& CATS\\
2003hd	& MCG -04-05-010		& 11850 	& Sb    & 0.011 &   52861.0   & IAUC 8179  & 52855.9$^{s}$(5)* & 9	& CATS\\
2003hg	& NGC 7771			& 4281  	& SBa   & 0.065 &   52870.0   & IAUC 8184  & 52865.5$^{n}$(5)  & 5	& CATS\\
2003hk	& NGC 1085			& 6795  	& SAbc 	& 0.033 &   52871.6   & CBET 41    & 52866.8$^{s}$(4)* & 4	& CATS\\
2003hl	& NGC 772			& 2475  	& SAb   & 0.064 &   52872.0   & IAUC 8184  & 52868.5$^{n}$(5)  & 6	& CATS\\
2003hn	& NGC 1448			& 1170  	& SAcd  & 0.013 &   52877.2   & IAUC 8186  & 52866.5$^{n}$(10) & 9	& CATS\\
2003ho	& ESO 235-G58			& 4314  	& SBcd  & 0.034 &   52851.9   & IAUC 8186  & 52848.5$^{s}$(7)* & 5	& CATS\\
2003ib	& MCG -04-48-15			& 7446 		& Sb    & 0.043 &   52898.7   & IAUC 8201  & 52891.5$^{n}$(8)  & 5	& CATS\\
2003ip	& UGC 327			& 5403  	& Sbc   & 0.058 &   52913.7   & IAUC 8214  & 52896.5$^{s}$(4)  & 4	& CATS\\
2003iq	& NGC 772			& 2475  	& SAb   & 0.064 &   52921.5   & CBET 48    & 52919.5$^{n}$(2)  & 5	& CATS\\
2004dy  & IC 5090                       & 9352          & Sa    & 0.045 &   53242.5   & IAUC 8395  & 53240.5$^{n}$(2)  & 3	& CSP \\
2004ej	& NGC 3095			& 2723  	& SBc   & 0.061 &   53258.5   & CBET 78    & 53223.9$^{s}$(9)* & 9	& CSP\\
2004er	& MCG -01-7-24			& 4411  	& SAc   & 0.023 &   53274.0   & CBET 93    & 53271.8$^{n}$(2)  & 10	& CSP\\
2004fb	& ESO 340-G7			& 6100  	& S     & 0.056 &   53286.2   & IAUC 8420  & 53258.6$^{s}$(7)* & 4	& CSP\\
2004fc	& NGC 701			& 1831  	& SBc   & 0.023 &   53295.2   & IAUC 8422  & 53293.5$^{n}$(1)  & 10	& CSP\\  
2004fx	& MCG -02-14-3 			& 2673  	& SBc   & 0.090 &   53307.0   & IAUC 8431  & 53303.5$^{n}$(4)  & 10	& CSP\\
2005J	& NGC 4012			& 4183 	 	& Sb    & 0.025 &   53387.0   & IAUC 8467  & 53379.8$^{s}$(7)* & 11	& CSP\\
2005K	& NGC 2923			& 8204 	 	& ?   	& 0.035 &   53386.0   & IAUC 8468  & 53369.8$^{s}$(8)  & 2	& CSP\\
2005Z	& NGC 3363			& 5766 	 	& S     & 0.025 &   53402.0   & IAUC 8476  & 53396.7$^{n}$(6)  & 9	& CSP\\
2005af	& NGC 4945			& 563   	& SBcd  & 0.156 &   53409.7   & IAUC 8482  & 53320.8$^{s}$(17)*& 9	& CSP\\
2005an	& ESO 506-G11			& 3206  	& S0    & 0.083 &   53432.7   & CBET 113   & 53431.8$^{s}$(6)* & 7	& CSP\\
2005dk	& IC 4882			& 4708  	& SBb   & 0.043 &   53604.0   & IAUC 8586  & 53601.5$^{s}$(6)* & 7	& CSP\\
2005dn	& NGC 6861			& 2829  	& SA0   & 0.048 &   53609.5   & IAUC 8589  & 53602.6$^{s}$(6)* & 8	& CSP\\
2005dt	& MCG -03-59-6			& 7695  	& SBb  	& 0.025 &   53614.7   & CBET 213   & 53605.6$^{n}$(9)  & 1	& CSP\\
2005dw	& MCG -05-52-49			& 5269  	& Sab  	& 0.020 &   53612.7   & CBET 219   & 53603.6$^{n}$(9)  & 3	& CSP\\
2005dx	& MCG -03-11-9			& 8012  	& S   	& 0.021 &   53623.0   & CBET 220   & 53611.8$^{s}$(7)* & 1	& CSP\\
2005dz	& UGC 12717			& 5696  	& Scd   & 0.072 &   53623.7   & CBET 222   & 53619.5$^{n}$(4)  & 7	& CSP\\
2005es	& MCG +01-59-79 		& 11287 	& S 	& 0.076 &   53643.7   & IAUC 8608  & 53638.7$^{n}$(5)  & 1	& CSP\\
2005gz	& MCG -01-53-022 		& 8518 		& SBbc 	& 0.06 	&   53654.7   & IAUC 8616  & 53650.2$^{n}$(5)  & 1	& CSP \\
2005lw	& IC 672	       		& 7710  	& ?  	& 0.043 &   53719.0   & CBET 318   & 53716.8$^{s}$(10) & 14	& CSP\\
2005me	& ESO 244-31 			& 6726  	& SAc 	& 0.022 &   53728.2   & CBET 333   & 53717.9$^{s}$(10)*& 1	& CSP\\
2006Y	& anon	       			& 10074$^2$ 	& ? 	& 0.115 &   53770.0   & IAUC 8668  & 53766.5$^{n}$(4)  & 13	& CSP\\
2006ai	& ESO 005-G009			& 4571$^1$  	& SBcd 	& 0.113 &   53784.0   & CBET 406   & 53781.6$^{s}$(5)  & 12	& CSP\\
2006bc	& NGC 2397			& 1363  	& SABb  & 0.181 &   53819.1   & CBET 446   & 53815.5$^{n}$(4)  & 3	& CSP\\
2006be	& IC 4582			& 2145  	& S     & 0.026 &   53819.0   & CBET 449   & 53802.8$^{s}$(9)* & 4	& CSP\\
2006bl	& MCG +02-40-9			& 9708  	& ?   	& 0.045 &   53829.5   & CBET 597   & 53822.7$^{s}$(10)*& 3	& CSP\\
2006ee	& NGC 774			& 4620  	& S0    & 0.054 &   53966.0   & cbet 597   & 53961.9$^{n}$(4)  & 13	& CSP\\
2006it	& NGC 6956			& 4650  	& SBb   & 0.087 &   54009.5   & CBET 660   & 54006.5$^{n}$(3)  & 6	& CSP\\
2006iw	& 2MASX J23211915+0015329 	& 9226		&?	& 0.044 &   54011.5   & CBET 663   & 54010.7$^{n}$(1)  & 5	& CSP\\
2006ms	& NGC 6935			& 4543  	& SAa   & 0.031 &   54046.2   & CBET 725   & 54028.5$^{s}$(6)**& 4	& CSP\\
2006qr	& MCG -02-22-023		& 4350  	& SABbc & 0.040 &   54070.0   & CBET 766   & 54062.8$^{n}$(7)  & 8	& CSP\\
2007P	& ESO 566-G36 			& 12224 	& Sa   	& 0.036 &   54124.0   & CBET 819   & 54118.7$^{n}$(5)  & 6	& CSP\\
2007U	& ESO 552-65			& 7791  	& S   	& 0.046 &   54136.5   & CBET 835   & 54133.6$^{s}$(6)* & 7	& CSP\\
2007W	& NGC 5105			& 2902  	& SBc 	& 0.045 &   54146.5   & CBET 844   & 54130.8$^{s}$(7)* & 7	& CSP\\
2007X	& ESO 385-G32			& 2837  	& SABc 	& 0.060 &   54146.5   & CBET 844   & 54143.5$^{s}$(5)  & 12	& CSP\\
2007Z	& PGC 016993			& 5277 		& Sbc 	& 0.525	&   54148.7   & CBET 847   & 54135.6$^{s}$(5)  & 2	& CSP \\
2007aa	& NGC 4030			& 1465  	& SAbc  & 0.023 &   54149.7   & CBET 848   & 54126.7$^{s}$(8)* & 11	& CSP\\
2007ab	& MCG -01-43-2 			& 7056  	& SBbc 	& 0.235 &   54150.7   & CBET 851   & 54123.9$^{s}$(10) & 5	& CSP\\
2007av	& NGC 3279			& 1394  	& Scd   & 0.032 &   54180.2   & CBET 901   & 54173.8$^{s}$(5)* & 4	& CSP\\
2007bf	& UGC 09121			& 5327		& Sbc  	& 0.018	&   54285.0   & CBET 919   & 54191.5$^{n}$(7)  & 4	& CSP \\
2007hm	& SDSS J205755.65-072324.9	& 7540		&?	& 0.059	&   54343.7   & CBET 1050  & 54336.6$^{s}$(6)* & 7	& CSP\\
2007il	& IC 1704			& 6454  	& S  	& 0.042 &   54354.0   & CBET 1062  & 54349.8$^{n}$(4)  & 12	& CSP\\
2007it	& NGC 5530			& 1193  	& SAc	& 0.103 &   54357.5   & CBET 1065  & 54348.5$^{n}$(1)  & 11	& CSP\\
2007ld	& anon				& 7499$^1$	& ?	& 0.081 &   54379.5   & CBET 1098  & 54376.5$^{s}$(8)* & 7	& CSP\\
2007oc	& NGC 7418			& 1450 	 	& SABcd & 0.014 &   54396.5   & CBET 1114  & 54388.5$^{n}$(3)  & 17	& CSP\\
2007od	& UGC 12846			& 1734  	& Sm   	& 0.032 &   54407.2   & CBET 1116  & 54400.6$^{s}$(5)* & 14	& CSP\\
2007sq	& MCG -03-23-5			& 4579  	& SAbc 	& 0.183 &   54443.0   & CBET 1170  & 54422.8$^{s}$(6)* & 7	& CSP\\
2008F	& MCG -01-8-15			& 5506 		& SBa 	& 0.044 &   54477.5   & CBET 1207  & 54469.6$^{s}$(6)* & 2	& CSP\\
2008H	& ESO 499- G 005 		& 4287		& SAc  	& 0.057	&   54481.0   & CBET 1210  & 54432.8$^{s}$(8)  & 1	& CSP \\
2008K	& ESO 504-G5			& 7997 		& Sb 	& 0.035 &   54481.0   & CBET 1211  & 54475.5$^{s}$(6)* & 12	& CSP\\
2008M	& ESO 121-26			& 2267 		& SBc 	& 0.040 &   54480.7   & CBET 1214  & 54471.7$^{n}$(9)  & 12	& CSP\\
2008W	& MCG -03-22-7 			& 5757		& Sc 	& 0.086 &   54502.7   & CBET 1238  & 54483.8$^{s}$(8)* & 10	& CSP\\
2008ag	& IC 4729			& 4439  	& SABbc & 0.074 &   54499.5   & CBET 1252  & 54477.9$^{s}$(8)* & 18	& CSP\\
2008aw	& NGC 4939			& 3110  	& SAbc  & 0.036 &   54528.0   & CBET 1279  & 54517.8$^{n}$(10) & 12	& CSP\\
2008bh	& NGC 2642			& 4345  	& SBbc  & 0.020 &   54549.0   & CBET 1311  & 54543.5$^{n}$(5)  & 6	& CSP\\
2008bk	& NGC 7793			& 227   	& SAd   & 0.017 &   54550.7   & CBET 1315  & 54540.9$^{s}$(8)* & 26	& CSP\\
2008bm	& CGCG 071-101			& 9563  	& Sc  	& 0.023 &   54554.7   & CBET 1320  & 54522.8$^{s}$(6)  & 4	& CSP\\
2008bp	& NGC 3095			& 2723  	& SBc 	& 0.061 &   54558.7   & CBET 1326  & 54551.7$^{n}$(6)  & 5	& CSP\\
2008br	& IC 2522			& 3019  	& SAcd 	& 0.083 &   54564.2   & CBET 1332  & 54555.7$^{n}$(9)  & 4	& CSP\\
2008bu	& ESO 586-G2			& 6630  	& S   	& 0.376 &   54574.0   & CBET 1341  & 54566.8$^{s}$(7)  & 5	& CSP\\
2008ga	& LCSB L0250N			& 4639 		& ? 	& 0.582 &   54734.0   & CBET 1526  & 54711.5$^{s}$(7)  & 3	& CSP\\
2008gi	& CGCG 415-004			& 7328 		& Sc 	& 0.060 &   54752.0   & CBET 1539  & 54742.7$^{n}$(9)  & 6	& CSP\\
2008gr	& IC 1579			& 6831 		& SBbc	& 0.012 &   54768.7   & CBET 1557  & 54769.6$^{s}$(6)* & 5	& CSP\\
2008hg	& IC 1720			& 5684 		& Sbc 	& 0.016 &   54785.5   & CBET 1571  & 54779.8$^{n}$(5)  & 6	& CSP\\
2008ho	& NGC 922			& 3082 		& SBcd	& 0.017 &   54796.5   & CBET 1587  & 54792.7$^{n}$(5)  & 3	& CSP\\
2008if	& MCG -01-24-10			& 3440 		& Sb 	& 0.029 &   54812.7   & CBET 1619  & 54807.8$^{n}$(5)  & 20	& CSP\\
2008il	& ESO 355-G4			& 6276 		& SBb 	& 0.015 &   54827.7   & CBET 1634  & 54825.6$^{n}$(3)  & 3	& CSP\\
2008in	& NGC 4303			& 1566 		& SABbc	& 0.020 &   54827.2   & CBET 1636  & 54825.4$^{n}$(2)* & 13	& CSP\\
2009N	& NGC 4487			& 1034 		& SABcd	& 0.019 &   54856.3   & CBET 1670  & 54846.8$^{s}$(5)  & 13	& CSP\\
2009W	& SDSS J162346.79+114423	& 5100		& ?  	& 0.065 &   54865.0   & CBET 1683  & 54816.9$^{s}$(9)  & 1	& CSP \\
2009aj	& ESO 221- G 018		& 2883		& Sa  	& 0.130	&   54887.0   & CBET 1704  & 54880.5$^{n}$(7)  & 12	& CSP \\
2009ao	& NGC 2939			& 3339 		& Sbc 	& 0.034 &   54895.0   & CBET 1711  & 54890.7$^{n}$(4)  & 7	& CSP\\
2009au	& ESO 443-21			& 2819 		& Scd 	& 0.081 &   54902.0   & CBET 1719  & 54897.5$^{n}$(4)  & 10	& CSP\\
2009bu	& NGC 7408			& 3494 		& SBc	& 0.022 &   54916.2   & CBET 1740  & 54901.9$^{s}$(8)* & 6	& CSP\\
2009bz	& UGC 9814			& 3231 		& Sdm	& 0.035 &   54920.0   & CBET 1748  & 54915.8$^{n}$(4)  & 5	& CSP\\
\hline
\hline
\end{longtable*}
\setcounter{table}{0}
\tiny
\begin{list}{}{}
\item $^1$ Measured using our own spectra.\\
$^2$ Taken from the Asiago supernova catalog: http://graspa.oapd.inaf.it/ \citep{Barbon99}.\\
$^3$ The CATS survey performed the follow up of SN 210, which was discovered by the SN Factory \citet{Wood-Vasey04} and was never
reported to the International Astronomical Union (IAU) to provide an official designation.\\
$^{s}$ Explosion epoch estimation through spectral matching.\\
$^{n}$ Explosion epoch estimation from SN non-detection.\\
$^*$ Cases where explosion epochs have changed between \citet{Anderson14} and the current work.\\
\textit{Observing campaigns:} CTSS=Cerro Tololo Supernova Survey; CT=Cal\'an/Tololo Supernova Program; SOIRS=Supernova Optical and
Infrared Survey; CATS=Carnegie Type II Supernova Survey; CSP=Carnegie Supernova Project.
\item In the first column the SN name, followed by its host galaxy are listed. In column 3 we list the host
galaxy heliocentric recession velocity. These are taken from the Nasa Extragalactic Database (NED: http://ned.ipac.caltech.edu/)
unless indicated by a superscript (sources in table notes). In columns 4 and 5 we list the host galaxy morphological Hubble types (from NED)
and the reddening due to dust in our Galaxy \citep{Schlafly11} taken from NED. In column 6, 7 and 8 we list the discovery date, their reference
and the explosion epochs. The number of spectra and the the observing campaign from which each SN was taken are given in column 9 and 10, 
and acronyms are listed in the table notes.\\
\end{list}
\small
\clearpage
\end{landscape}

\clearpage
\setcounter{table}{1}
\begin{table*}
\centering
\tiny
\begin{threeparttable}
\caption{Reference SNe~II}
\label{t_expl}
\begin{tabular}[t]{ccccccccccc}
\hline
SN &  Explosion date & V-Maximum date & Days from Explosion to V-maximum & Reference\\ 
\hline
\hline
1999em	& 2451475.6 (5) & 2451485.5 &   5     &\citealt{Leonard02b} \\
1999gi	& 2451518.3 (3) & 2451530.0 &   12    &\citealt{Leonard02a} \\
2004et	& 2453270.5 (3) & 2453286.6 &   16    &\citealt{Li05, Sahu06} \\
2005cs	& 2453547.6 (1) & 2453553.6 &   6     &\citealt{Pastorello06} \\
2006bp	& 2453833.4 (1)	& 2453842.0 &   9     &\citealt{Dessart08b} \\
\hline
1988A	& 2447177.2 (2) & \nodata   & \nodata & This work \\
1990E	& 2447935.1 (3) & \nodata   & \nodata & This work \\
1999br	& 2451276.7 (4) & \nodata   & \nodata & This work \\
2003bn	& 2452694.5 (3) & \nodata   & \nodata & This work \\
2003iq	& 2452919.5 (2) & \nodata   & \nodata & This work \\
2004er	& 2453271.8 (2) & \nodata   & \nodata & This work \\
2004fc	& 2453293.5 (1) & \nodata   & \nodata & This work \\
2004fx	& 2453303.5 (4) & \nodata   & \nodata & This work \\
2005dz	& 2453619.5 (4) & \nodata   & \nodata & This work \\
2006bc	& 2453815.5 (4) & \nodata   & \nodata & This work \\
2006ee	& 2453961.9 (4) & \nodata   & \nodata & This work \\
2006it	& 2454006.5 (3) & \nodata   & \nodata & This work \\
2006iw	& 2454010.7 (1) & \nodata   & \nodata & This work \\
2006Y	& 2453766.5 (4) & \nodata   & \nodata & This work \\
2007il	& 2454349.8 (4) & \nodata   & \nodata & This work \\
2007it	& 2454348.5 (1) & \nodata   & \nodata & This work \\
2007oc	& 2454388.5 (3) & \nodata   & \nodata & This work \\	
2008il	& 2454825.6 (3) & \nodata   & \nodata & This work \\
2008in	& 2454825.4 (2) & \nodata   & \nodata & This work \\
2009ao	& 2454890.7 (4) & \nodata   & \nodata & This work \\
2009au	& 2454897.5 (4) & \nodata   & \nodata & This work \\
2009bz	& 2454915.8 (4) & \nodata   & \nodata & This work \\
\hline
\hline
\end{tabular}
\setcounter{table}{0}
\begin{list}{}{}
\item Columns: (1) SN name; (2) Julian date of the explosion epoch; (3) Julian date of the $V$-band maximum;
(4) Days from Explosion to $V-$band maximum; (5) References.\\
The first five SNe are included in SNID and are used as templates in this work.
Their respective references are presented in the column 4.
The rest of SNe showed after the line are taken from this work as new SNID templates.
\end{list}
\label{parameters}
\end{threeparttable}
\end{table*}
\small                      

\clearpage
\setcounter{table}{2}
\begin{table*}[h!]
\centering
\small 
\caption[Spectral features used in the statistical analysis]{Spectral features used to the statistical analysis in the photospheric and nebular phases}
\label{table_features}
\begin{tabular}{lcccccccccccccccc}
\hline
Feature name & Rest wavelength$^{\star}$ [\AA] & Blue-ward limit range$^{\clubsuit}$ [\AA] & Red-ward limit range$^{\clubsuit}$ [\AA]\\
\hline
\hline
H$_{\alpha}$ 		& 6563	& 6000 - 6300	& 6900 - 7100	\\
H$_{\beta}$		& 4861	& 4400 - 4700	& 4800 - 4900	\\
\ion{Fe}{2}		& 4924	& 4800 - 4900   & 4900 - 4950	\\
\ion{Fe}{2}		& 5018	& 4900 - 4500   & 5000 - 5050	\\
\ion{Fe}{2}		& 5169	& 5000 - 5050   & 5100 - 5300	\\
\ion{Na}{1}		& 5893	& 5500 - 6000   & 5800 - 6000	\\
\ion{Sc}{2}		& 5531	& 5400 - 5450   & 5500-  5550   \\
\ion{Sc}{2}/\ion{Fe}{2}	& 5663	& 5500 - 5550   & 5600 - 5700	\\
\ion{Ba}{2}		& 6142	& 6000 - 6050   & 6100 - 6150	\\
\ion{Sc}{2}		& 6247	& 6150 - 6170   & 6250 - 6270   \\
\ion{O}{1}		& 7774	& 7630 - 7650   & 7750 - 7780	\\
\hline
\hline
\end{tabular}
\small
\begin{list}{}{}
\item $^{\star}$ The rest wavelengths are weighted averages of the strongest spectral lines that give rise to each absorption feature.\\
$^{\clubsuit}$ These limits are necessary in order to account for variations in spectral feature width and expansion
velocity among SNe.
\end{list}
\label{table_features}
\end{table*} 


\clearpage
\setcounter{table}{3}
\begin{table*}[h!]
\centering
\small
\caption{KS-Test values}
\label{table_kstest}
\begin{tabular}[t]{ccccccccccc}
\hline
Feature name                    & $v$(H$_{\alpha}$) & $a/e$ & M$_{max}$ & $s_2$ & M13N2 & Epoch [days] \\
\hline
\hline
\ion{Fe}{2} line forest		& 5.19		& 9.80		& \textbf{2.17}	& \textbf{$3.73\times10^{-4}$} 	& \textbf{4.38}	& 0 - 100 \\
H$_{\gamma}$ blended		& 29.75		& 58.33		& 31.21		& 55.32				& 16.10		& 23 - 27 \\
\ion{He}{1} 5876		& 18.62	        & \textbf{1.91}	& 30.94		& 25.73				& 46.18 	& 18 - 22 \\
\ion{Ca}{2} IR triplet		& 87.73		& 91.97   	& 94.47	    	& 53.30   			& 98.82 	& 18 - 22 \\
\ion{Fe}{2} 4924		& \textbf{2.82} & 16.84		& \textbf{1.09} & \textbf{4.80}			& 99.40		& 28 - 32 \\
\ion{Fe}{2} 5018		& 15.68		& 90.80		& 99.02		& 68.84				& 61.53 	& 18 - 22 \\
\ion{Fe}{2} 5169		& 60.76		& 35.15		& 74.88		& 50.83				& 20.30 	& 13 - 17 \\
\ion{Fe}{2} multiplet		& 21.38		& 26.75		& \textbf{1.00}	& \textbf{0.25}			& 99.28 	& 33 - 37 \\
\ion{Sc}{2}/\ion{Fe}{2} 5531	& 75.60		& 89.60		& 30.34	    	& 45.20	   			& \textbf{1.84} & 38 - 42 \\
\ion{Sc}{2} multiplet 5663	& 63.54		& 63.54		& 30.34		& 80.10	    			& \textbf{0.79} & 38 - 42 \\
\ion{Ba}{2} 6142		& 45.75		& 83.58	   	& \textbf{1.90}	& 57.43	   			& \textbf{1.29} & 38 - 42 \\
\ion{Sc}{2} 6247		& 45.76		& 83.58   	& \textbf{1.89}	& 57.42	   			& \textbf{0.52}	& 38 - 42 \\
\hline 
\end{tabular}
\begin{list}{}{}
\item Percentage obtained using a KS Test to verify if two distributions (with and without each line ) are drawn from
the same parent population as a function of $v$(H$_{\alpha}$), $a/e$, M$_{max}$, $s_2$, and M13N2 in an particular epoch.
This epoch is shown in the last column of the table. 
\end{list}
\label{table_kstest}
\end{table*} 


\clearpage
\setcounter{table}{4}
\begin{table*}[h!]
\centering
\normalsize
\begin{threeparttable}
\caption{Model properties}
\label{table_models}
\begin{tabular}[t]{ccccccccccc}
\hline
Model &     Z       & M$_{final}$ & R$_*$       & E$_{kin}$\\
      & [Z$_\odot$] & [M$_\odot$] & [R$_\odot$] & [B] \\ 
\hline
\hline
m15z2m3	& 0.1	& 14.92	& 524	& 1.35 	 \\
m15z8m3	& 0.4	& 14.76	& 611	& 1.27	 \\
m15z8m3	& 1.0   & 14.09	& 768	& 1.27	 \\
m15z4m2	& 2.0	& 12.60	& 804   & 1.24	 \\
m15mlt1	& 1.0	& 14.01	& 1107	& 1.24	 \\
m15mlt3	& 1.0	& 14.08	& 501	& 1.34	 \\
m12mlt3	& 1.0	& 10.50	& 500	& 0.25	 \\
\hline
\hline
\end{tabular}
\setcounter{table}{0}
\small
\begin{list}{}{}
\item Summary of model properties used in this work.\\
Columns: (1) Model name; (2) Metallicity; (3) Final progenitor mass; (4) Progenitor radius; (5) Kinetic energy  
\end{list}
\label{table_models}
\end{threeparttable}
\end{table*}


\clearpage
\appendix

\LongTables
\renewcommand{\thetable}{A\arabic{table}}
\setcounter{table}{0}
\tiny
\tiny
\centering
\tablecolumns{11} 
\tablewidth{0pc} 
\begin{longtable*}[t]{ccccccccccc}
\caption{Spectroscopic observations information}
\label{t_spec}\\
\hline
UT Date &  JD &  Phase &  Tel. &  Inst. &  Wavelength &  Resol. &  Exp. &  Air-  \\
        &     & (days) &       &        &  Range (AA) &  (AA)   &  (s)  & mass  \\ 
 (1)    & (2) &  (3)   &  (4)  &  (5)   &  (6)        &  (7)    &  (8)  & (9) \\[0.3ex]
\hline
\hline 
\multicolumn{9}{c}{SN 1986L} \\[0.3ex] 
1986-10-09 & 2446712.50 & 4   &  \nodata  &  \nodata   & 3681-7728 &  \nodata   & 1000     &  1.21    \\[0.3ex]
1986-10-11 & 2446714.50 & 6   &  \nodata  &  \nodata   & 3730-7168 &  \nodata   & 1556     &  1.20    \\[0.3ex]
1986-10-12 & 2446715.50 & 7   &  \nodata  &  \nodata   & 3800-7322 &  \nodata   & 2000     &  1.21    \\
\end{longtable*}
\begin{list}{}{}
\item Note that up to 1999 we do not have acess to the telescope, intrument and resolution information. Between 2002 and 2003 the resolution information is not available. 
\item Columns: (1) UT date of the observation; (2) Julian date of the observation; (3) Phase in days
since explosion; (4) Telescope code – 3P6: ESO 3.6-m Telescope; BAA: Las Campanas Magellan I 6.5-m Baade Telescope; CLA: Las
Campanas Magellan II 6.5-m Clay Telescope; DUP: Las Campanas 2.5-m du Pont Telescope Telescope; NTT: New Technology Telescope;
(5) Instrument code – BC: Boller \& Chivens spectrograph; EF: ESO Faint Object Spectrograph and Camera (EFOSC-2);
EM: ESO Multi-Mode Instrument (EMMI); IM: Inamori Magellan Areal Camera and Spectrograph (IMACS), 
LD: Low Dispersion Survey Spectrograph (LDSS); WF: Wide Field Reimaging CCD Camera (WFCCD); (6) Wavelength
range covered; (7) Spectral resolution in \AA\ as estimated from arc-lamp lines; (8) Total exposure time;
(9) Airmass at the middle of the observation.\\[0.3ex]
$^{\star}$ Spectra with low S/N\\[0.3ex]
$^{\Box}$ Peculiar SN\\[0.3ex]
$^{\clubsuit}$ Spectra with defects resulting from the observing procedure or data reduction.
\end{list}
\small


\LongTables
\clearpage
\renewcommand{\thetable}{A\arabic{table}}
\setcounter{table}{1}
\begin{landscape}
\centering
\scriptsize
\tablecolumns{11} 
\tablewidth{0pc} 
\begin{longtable*}[t!]{ccccccccccc}
\caption[Explosion epoch estimations]{Explosion epoch estimations comparison}
\label{table_explosion}\\
\hline
 SN &  Spect. date &  Best Match &  Days from &  Days from & Average  & Explosion date & Non Detection date & Discovery date &
Explosion date & Difference  \\
 &  JD &    &  maximum &   explosion &  (Using match) &  (MJD) &  (MJD) &  (MJD) &  (MJD) &  (days) \\[0.3ex]
\hline	
\hline
1968L  & 46715.5  & 2006bp  & $-2$    & 7       & 7       & 46708.5 (5)  & 46705.5  & 46710.5  & 46708.0 (3)  & 0       \\[0.3ex]
       &          & 1999em  & $-4$    & 6       &         &              &          &          &              &         \\[0.3ex]
       &          &         &         &         &         &              &          &          &              &         \\[0.3ex]       
1988A  & 47188.5  & 1999em  & +5      & 15      & 17      & 47171.5 (6)  & 47175.5  & 47179.0  & 47177.2 (2)  & $-6$    \\[0.3ex] 
       &          & 2006bp  & +7      & 16      &         &              &          &          &              &         \\
\hline
\hline
\end{longtable*}
\setcounter{table}{3}
\begin{list}{}{}
\item Columns: (1) SN name; (2) Reduced Julian date of the spectrum used to the match (JD − 2400000); (3) Best match obtained with SNID; 
(4) Days from maximum of the template used to the match; (5) Days from explosion of the template used to the match;
(6) Average obtained from the days from explosion; (7) Explosion date obtained with the matching technique; 
(8) Non-detection date of the SN; (9) Discovery date of the SN; (10) Explosion date obtained from non-detection and discovery date;
(11) Difference in days between the explosion date from matching technique and non-detection. 
\end{list}
\small
\clearpage
\end{landscape}

\clearpage
\renewcommand{\thetable}{A\arabic{table}}
\setcounter{table}{2}
\vspace{-4cm} 
\begin{landscape}
\tiny
\begin{table*}
\centering
\caption{Mean velocity values and the standard deviation for our sample.}
\label{meanvel}
\hspace{-2cm} 
\begin{tabular}{lcccccccccccccccc}
Epoch	&  H$_{\alpha}$    &  H$_{\alpha}$   &  H$_{\beta}$	& Fe II $\lambda$4924 & Fe II $\lambda$5018   & Fe II $\lambda$5169  & Fe II/Sc II  &  Sc II Mult. & Na I D       &  Ba II       &      Sc II   &  O I\\[0.5ex]		
(Days)  &    (km s$_{-1}$) &    (km s$_{-1}$) &  (km s$_{-1}$)  &    (km s$_{-1}$)    &     (km s$_{-1}$)    &    (km s$_{-1}$)   &    (km s$_{-1}$)   &    (km s$_{-1}$)       &    (km s$_{-1}$) &    (km s$_{-1}$)  &    (km s$_{-1}$)     &    (km s$_{-1}$)    \\
\hline
\hline
4	&  $8369\pm3930$  &  $12845\pm950$  &  $11379\pm1800$ &	 \nodata	& \nodata	  &  \nodata         &  \nodata         &  \nodata	  & \nodata	   & \nodata	     &  \nodata	       &  \nodata       \\
8.6	&  $9399\pm3564$  &  $10702\pm1382$ &  $9605\pm1574$  &	 \nodata	& \nodata	  &  $10447\pm1050$  &  \nodata         &  \nodata	  & \nodata	   & \nodata	     &  \nodata	       &  \nodata       \\
12.8	&  $9384\pm2240$  &  $9468\pm1787$  &  $8748\pm1653$  &	 $3187\pm281$   &  $5298\pm1791$  &  $6871\pm2234$   &  \nodata         &  \nodata	  & \nodata	   & \nodata	     &  \nodata	       &  $3280\pm1680$  \\
18.1	&  $9044\pm1576$  &  $8987\pm1430$  &  $8364\pm1478$  &	 $3183\pm796$   &  $6237\pm1117$  &  $6300\pm1174$   &  \nodata         &  \nodata	  & \nodata	   & \nodata	     &  \nodata	       &  $4888\pm2762$  \\
23.1	&  $7191\pm2027$  &  $7798\pm1966$  &  $7083\pm1690$  &	 $3882\pm1025$  &  $5241\pm1328$  &  $5274\pm1254$   &  $5057\pm1995$	&  $5426\pm1612$  & \nodata	   &  $5539\pm3200$  &  $5076\pm420$   &  $3951\pm1396$  \\
27.7	&  $7728\pm1606$  &  $8369\pm1825$  &  $7426\pm1527$  &	 $4193\pm1499$  &  $5506\pm1229$  &  $5440\pm1098$   &  $6033\pm2276$	&  $5747\pm1826$  & \nodata	   &  $4984\pm2790$  &  $4537\pm2678$  &  $4961\pm1900$  \\
33.1	&  $7319\pm1190$  &  $7745\pm1194$  &  $6668\pm1260$  &	 $4240\pm1139$  &  $4974\pm1114$  &  $4942\pm892$    &  $5085\pm1307$	&  $4832\pm1267$  & $5865\pm2029$  &  $4521\pm1595$  &  $4504\pm1323$  &  $4908\pm1586$  \\
38.1	&  $6815\pm1563$  &  $7478\pm1548$  &  $6297\pm1582$  &	 $4135\pm1346$  &  $4641\pm1324$  &  $4428\pm1065$   &  $4835\pm1458$	&  $4427\pm1181$  & $5479\pm1977$  &  $3979\pm1513$  &  $3837\pm1301$  &  $4274\pm1737$  \\
42.8	&  $6188\pm1807$  &  $6551\pm1745$  &  $5267\pm1600$  &	 $3544\pm1158$  &  $3857\pm1097$  &  $3760\pm1045$   &  $3969\pm1188$	&  $3739\pm968$   & $4411\pm1596$  &  $3164\pm1367$  &  $3621\pm1238$  &  $3587\pm1605$  \\
47.8	&  $6616\pm1864$  &  $7145\pm1772$  &  $5541\pm1688$  &	 $3493\pm1130$  &  $4049\pm1181$  &  $3938\pm990$    &  $4243\pm1035$	&  $3956\pm899$   & $4952\pm1634$  &  $3712\pm1165$  &  $3747\pm888$   &  $4090\pm947$   \\
53.1	&  $5975\pm1457$  &  $6535\pm1755$  &  $5004\pm1785$  &	 $3307\pm1027$  &  $3654\pm1135$  &  $3537\pm851$    &  $3888\pm1181$	&  $3507\pm978$   & $4408\pm1531$  &  $3258\pm1258$  &  $3227\pm1146$  &  $3520\pm1690$  \\
58.6	&  $5907\pm1883$  &  $6615\pm1900$  &  $5025\pm1774$  &	 $3086\pm924$   &  $3682\pm1358$  &  $3631\pm973$    &  $3803\pm1371$	&  $3552\pm935$   & $4491\pm1467$  &  $3047\pm815$   &  $3092\pm638$   &  $2861\pm1099$  \\
63.3	&  $5836\pm1597$  &  $6619\pm1565$  &  $4836\pm1548$  &	 $3074\pm919$   &  $3455\pm1093$  &  $3401\pm758$    &  $3553\pm1073$	&  $3294\pm918$   & $4284\pm1213$  &  $3062\pm735$   &  $3145\pm898$   &  $2919\pm1077$  \\
68	&  $5556\pm1053$  &  $6613\pm1038$  &  $4909\pm1146$  &	 $2963\pm464$   &  $3378\pm722$	  &  $3397\pm639$    &  $3342\pm609$	&  $3099\pm752$   & $4359\pm1098$  &  $2785\pm581$   &  $2843\pm593$   &  $2786\pm1095$  \\
72.8	&  $5473\pm1496$  &  $6720\pm1564$  &  $4725\pm1665$  &	 $2875\pm936$   &  $3203\pm952$	  &  $3374\pm828$    &  $3352\pm1315$	&  $3074\pm935$   & $4296\pm1426$  &  $3004\pm1123$  &  $2738\pm813$   &  $3091\pm1631$  \\
78.2	&  $5037\pm1706$  &  $6061\pm1660$  &  $4460\pm1553$  &	 $2685\pm784$   &  $2980\pm778$	  &  $3078\pm820$    &  $2981\pm852$	&  $2841\pm850$   & $4229\pm1344$  &  $2792\pm877$   &  $2607\pm668$   &  $2682\pm1109$  \\
83.5	&  $5687\pm1722$  &  $6490\pm1883$  &  $4386\pm1492$  &	 $2679\pm929$   &  $2863\pm1000$  &  $3074\pm919$    &  $2686\pm861$	&  $2734\pm700$   & $4240\pm1156$  &  $2564\pm776$   &  $2428\pm773$   &  $2232\pm424$   \\
87.5	&  $4871\pm1664$  &  $6197\pm1930$  &  $4448\pm1514$  &	 $2732\pm807$   &  $3217\pm937$   &  $3253\pm785$    &  $3041\pm855$	&  $2679\pm532$   & $4335\pm1062$  &  $2735\pm724$   &  $2517\pm722$   &  $2986\pm16344$ \\
93.3	&  $4627\pm1541$  &  $5666\pm2016$  &  $4261\pm1493$  &	 $2555\pm372$   &  $2841\pm811$   &  $2946\pm718$    &  $2607\pm610$	&  $2478\pm596$   & $4200\pm1059$  &  $2484\pm554$   &  $2256\pm398$   &  $2493\pm12364$ \\
98.2	&  $4349\pm1602$  &  $5788\pm2073$  &  $4372\pm1505$  &	 $2122\pm550$   &  $2498\pm603$   &  $2476\pm631$    &  $2276\pm644$	&  $2069\pm629$   & $4203\pm971$   &  $2139\pm501$   &  $1929\pm580$   &  $1655\pm950$   \\	
103	&  $3466\pm836$   &  $4422\pm1353$  &  $3014\pm993$   &	 $2067\pm593$   &  $2141\pm393$   &  $2119\pm525$    &  $1844\pm471$	&  $1864\pm334$   & $3445\pm915$   &  $101\pm506$    &  $1514\pm124$   &  $1335\pm321$   \\	
108.2	&  $4114\pm836$   &  $5625\pm1226$  &  $4128\pm885$   &	 $2128\pm594$   &  $2430\pm346$   &  $2625\pm457$    &  $2531\pm344$	&  $2278\pm242$   & $4073\pm388$   &  $274\pm259$    &  $2161\pm249$   &  $1833\pm1555$  \\	
115.7	&  $4927\pm1763$  &  $5805\pm1176$  &  $4536\pm1025$  &	 $2667\pm484$   &  $2623\pm725$   &  $2451\pm679$    &  $2170\pm1860$	&  $1748\pm210$   & $4793\pm722$   &  $100\pm108$    &  $1441\pm220$   &  $1823\pm1160$  \\	
\hline
\hline
\end{tabular}
\hspace{-2cm} 
\begin{list}{}{}
\item Columns: (1) Epoch; (2) Velocity of H$_{\alpha}$ from FWHM of emission component; (3) Velocity of H$_{\alpha}$ from the minimum flux of the absorption component; (4) Velocity of H$_{\beta}$; (5) Velocity of Fe II $\lambda$4924;
(6) Velocity of Fe II $\lambda$5018; (7) Velocity of Fe II $\lambda$5169; (8) Velocity of Fe II/Sc II; (9) Velocity of Sc II Multiplet; (10) Velocity of Na I D; (11) Velocity of Ba II; (12) Velocity of ScII;
and (13) Velocity of O I. \\	
\end{list}             
\end{table*}
 
\clearpage
\end{landscape}

\clearpage
\renewcommand{\thetable}{A\arabic{table}}
\setcounter{table}{3}
\vspace{-4cm} 
\begin{landscape}
\tiny
\begin{table*}
\centering
\caption{Mean pEW values and the stardard deviations for our sample.}
\label{meanew}
\begin{tabular}{lcccccccccccccccc}
Epoch	&  H$_{\alpha}$&  H$_{\alpha}$   &  H$_{\beta}$	& Fe II $\lambda$4924 & Fe II $\lambda$5018   & Fe II $\lambda$5169  & Fe II/Sc II  &  Sc II Mult. & Na I D       &  Ba II       &      Sc II   &  O I\\[0.5ex]		
(Days)        &    (\AA)     &      (\AA)      &     (\AA)    &    (\AA)         &     (\AA)         &    (\AA)         &    (\AA)       &    (\AA)       &    (\AA)       &    (\AA)       &    (\AA)     &    (\AA)    \\
\hline
\hline	               
4	&  $0.8\pm1.8$	  &  $47.3\pm28.7$    &  $11.9\pm8.3$   &  $0\pm0$       &  $0\pm0$       &  $0\pm0$        &  $0\pm0$       &  $0\pm0$        &  \nodata	       &  $0\pm0$        &  $0\pm0$       &  \nodata        \\
8.6	&  $4.34\pm5.4$	  &  $83.2\pm46.4$    &  $24.8\pm19.2$  &  $0\pm0$       &  $0\pm0$       &  $0.1\pm0.6$    &  $0\pm0$       &  $0\pm0$        &  \nodata	       &  $0\pm0$        &  $0\pm0$       &  \nodata        \\
12.8	&  $8.6\pm10.9$	  &  $121.2\pm61.2$   &  $33.3\pm17.6$  &  $0.3\pm1.3$   &  $1.2\pm3.2$   &  $5.5\pm8.6$    &  $0\pm0$       &  $0\pm0$        &  \nodata	       &  $0\pm0$        &  $0\pm0$       &  $11\pm1.1$     \\
18.1	&  $16.6\pm17.3$  &  $157.2\pm53.7$   &  $37.4\pm14.8$  &  $0.2\pm0.8$   &  $4.2\pm4.8$   &  $14.5\pm14.0$  &  $0\pm0$       &  $0\pm0$        &  \nodata	       &  $0\pm0$        &  $0\pm0$       &  $5.67\pm1.1$   \\
23.1	&  $25.7\pm22.1$  &  $147.8\pm62.2$   &  $43.3\pm18.8$  &  $1.4\pm2.5$   &  $9.3\pm5.5$   &  $22.4\pm10.9$  &  $1.1\pm2.5$   &  $1.02\pm3.21$  &  \nodata	       &  $0.1\pm0.3$    &  $0.1\pm0.30$  &  $11.5\pm8.6$   \\
27.7	&  $25.0\pm25.9$  &  $142.7\pm55.2$   &  $43.8\pm16.7$  &  $1.1\pm2.5$   &  $9.6\pm4.5$   &  $25.5\pm9.5$   &  $1.4\pm3.1$   &  $1.61\pm3.99$  &  \nodata	       &  $0.3\pm1.0$    &  $0.5\pm1.40$  &  $10.9\pm5.6$   \\
33.1	&  $36.6\pm20.4$  &  $155.7\pm44.5$   &  $49.2\pm13.6$  &  $2.8\pm3.6$   &  $12.5\pm4.0$  &  $30.2\pm8.4$   &  $4.9\pm4.4$   &  $6.48\pm6.30$  &  $13.3\pm7.6$   &  $1.1\pm2.7$    &  $2.1\pm3.17$  &  $9.8\pm5.8$    \\
38.1	&  $42.5\pm23.3$  &  $152.9\pm44.2$   &  $50.2\pm16.4$  &  $3.8\pm3.8$   &  $14.5\pm6.0$  &  $33\pm13.5$    &  $6.0\pm5.0$   &  $8.09\pm7.14$  &  $15.6\pm8.3$   &  $2.0\pm3.4$    &  $2.8\pm4.00$  &  $10.6\pm3.9$   \\
42.8	&  $46.1\pm22.7$  &  $142.0\pm64.0$   &  $49.2\pm15.9$  &  $6.1\pm4.2$   &  $15.1\pm5.4$  &  $32.9\pm9.2$   &  $7.1\pm4.8$   &  $10.5\pm5.87$  &  $18.7\pm10.3$  &  $2.8\pm3.9$    &  $3.7\pm3.77$  &  $11.0\pm5.5$   \\
47.8	&  $48.1\pm21.6$  &  $169\pm61.7$     &  $54.2\pm18.3$  &  $6.4\pm4.1$   &  $14.2\pm5.8$  &  $34.6\pm9.6$   &  $7.9\pm3.3$   &  $11.5\pm4.83$  &  $26.5\pm12.5$  &  $4.3\pm4.7$    &  $5.1\pm3.26$  &  $13.0\pm4.9$   \\
53.1	&  $56.6\pm21.7$  &  $156.9\pm55.2$   &  $48.8\pm20.3$  &  $8.1\pm5.5$   &  $17.9\pm5.9$  &  $40.2\pm13.5$  &  $10.1\pm4.7$  &  $14.0\pm7.53$  &  $32.4\pm15.2$  &  $5.6\pm4.7$    &  $7.2\pm5.64$  &  $11.9\pm5.9$   \\
58.5	&  $50.7\pm24.9$  &  $169.7\pm78.6$   &  $49.6\pm26.4$  &  $7.5\pm4.7$   &  $18.3\pm7.4$  &  $39.8\pm15.7$  &  $11.2\pm5.5$  &  $16.0\pm8.35$  &  $38.3\pm20.0$  &  $6.4\pm6.2$    &  $6.7\pm5.09$  &  $11.1\pm5.8$   \\
63.3	&  $58.1\pm18.4$  &  $173.3\pm70.1$   &  $49.5\pm25.1$  &  $9.7\pm5.7$   &  $19.8\pm5.6$  &  $44.1\pm11.0$  &  $11.8\pm5.3$  &  $18.2\pm7.63$  &  $46.0\pm17.5$  &  $7.9\pm6.8$    &  $7.4\pm4.78$  &  $10.8\pm4.6$   \\
68.0	&  $60.2\pm17.3$  &  $163.3\pm42.1$   &  $53.1\pm27.0$  &  $9.4\pm6.0$   &  $21.6\pm7.0$  &  $42.8\pm9.6$   &  $13.6\pm6.1$  &  $19.5\pm7.79$  &  $52.2\pm17.1$  &  $8.2\pm6.3$    &  $8.0\pm5.88$  &  $12.1\pm4.5$   \\
72.8	&  $65.2\pm20.8$  &  $179.5\pm71.8$   &  $56.3\pm32.2$  &  $9.6\pm6.3$   &  $19.5\pm6.3$  &  $41\pm11.7$    &  $11.5\pm5.9$  &  $17.3\pm8.15$  &  $49.4\pm24.5$  &  $8.1\pm6.9$    &  $6.9\pm5.61$  &  $14.3\pm7.1$   \\
78.2	&  $60.0\pm21.4$  &  $167.0\pm71.7$   &  $46.9\pm26.1$  &  $11.6\pm6.6$  &  $20.8\pm7.8$  &  $42.4\pm10.1$  &  $14.7\pm5.2$  &  $22.9\pm7.55$  &  $59.7\pm21.8$  &  $12.6\pm7.56$  &  $11.4\pm4.7$  &  $12.7\pm5.6$   \\
83.5	&  $53.8\pm31.1$  &  $202.2\pm86.1$   &  $52.4\pm27.9$  &  $10.4\pm5.9$  &  $21.5\pm7.8$  &  $47.0\pm13.5$  &  $14.3\pm5.2$  &  $21.3\pm8.00$  &  $63.9\pm28.6$  &  $12.8\pm9.41$  &  $11.4\pm7.4$  &  $11.1\pm3.7$   \\
87.5	&  $56.1\pm26.9$  &  $176.4\pm95.0$   &  $55.2\pm28.8$  &  $10.5\pm7.6$  &  $23.0\pm7.5$  &  $51.3\pm12.8$  &  $14.9\pm6.2$  &  $21.6\pm9.26$  &  $63.2\pm19.0$  &  $11.4\pm7.67$  &  $10.1\pm6.5$  &  $15.4\pm6.9$   \\
93.3	&  $50.9\pm28.8$  &  $182.9\pm107.3$  &  $47.0\pm26.7$  &  $13.3\pm8.1$  &  $25.6\pm8.2$  &  $48.2\pm11.1$  &  $17.3\pm7.7$  &  $27.3\pm10.6$  &  $69.7\pm17.7$  &  $16.6\pm12.4$  &  $12.4\pm6.5$  &  $11.5\pm5.8$   \\
98.2	&  $61.6\pm28.3$  &  $214.4\pm117.6$  &  $62.7\pm29.3$  &  $14.9\pm4.7$  &  $26.2\pm7.0$  &  $46.4\pm9.9$   &  $17.7\pm6.2$  &  $27.6\pm7.81$  &  $81.8\pm26.4$  &  $17.6\pm10.2$  &  $12.9\pm6.7$  &  $12.4\pm2.7$   \\
103.0	&  $48.2\pm25.9$  &  $184.8\pm80.3$   &  $41.2\pm24.8$  &  $19.1\pm4.7$  &  $30.7\pm3.9$  &  $48.8\pm6.4$   &  $16.8\pm6.6$  &  $26.4\pm9.00$  &  $76.0\pm14.9$  &  $34.6\pm9.78$  &  $16.5\pm4.0$  &  $11.8\pm1.6$   \\
108.2	&  $60.1\pm19.0$  &  $208.9\pm56.1$   &  $43.8\pm15.5$  &  $15.9\pm4.0$  &  $25.7\pm2.6$  &  $41\pm10.2$    &  $15.0\pm0.0$  &  $21.5\pm4.11$  &  $69.2\pm20.8$  &  $20.3\pm15.7$  &  $9.8\pm2.7$   &  $13.0\pm2.8$   \\
115.7	&  $46.2\pm18.6$  &  $287.5\pm158.8$  &  $53.4\pm29.4$  &  $7.9\pm0.9$   &  $16.8\pm6.2$  &  $35.1\pm12.9$  &  $21.2\pm3.5$  &  $16.3\pm4.5$   &  $70.2\pm23.1$  &  $22.4\pm15.7$  &  $6.7\pm4.5$   &  $8.9\pm1.5$    \\
\hline
\hline
\end{tabular}
\hspace{-2cm} 
\begin{list}{}{}
\item Columns: (1) SN name; (2) pEW of H$_{\alpha}$ absorption component; (3) pEW of H$_{\alpha}$ emission component; (4) pEW of H$_{\beta}$; (5) pEW of Fe II $\lambda$4924;
(6) pEW of Fe II $\lambda$5018; (7) pEW of Fe II $\lambda$5169; (8) pEW of Fe II/Sc II; (9) pEW of Sc II Multiplet; (10) pEW of Na I D; (11) pEW of Ba II; (12) pEW of ScII; 
(13) pEW of O I. \\	
\end{list}             
\end{table*} 
\clearpage
\end{landscape}

\section{Spectral series}
\label{spectra}

\begin{figure*}
\centering
\includegraphics[width=5.5cm]{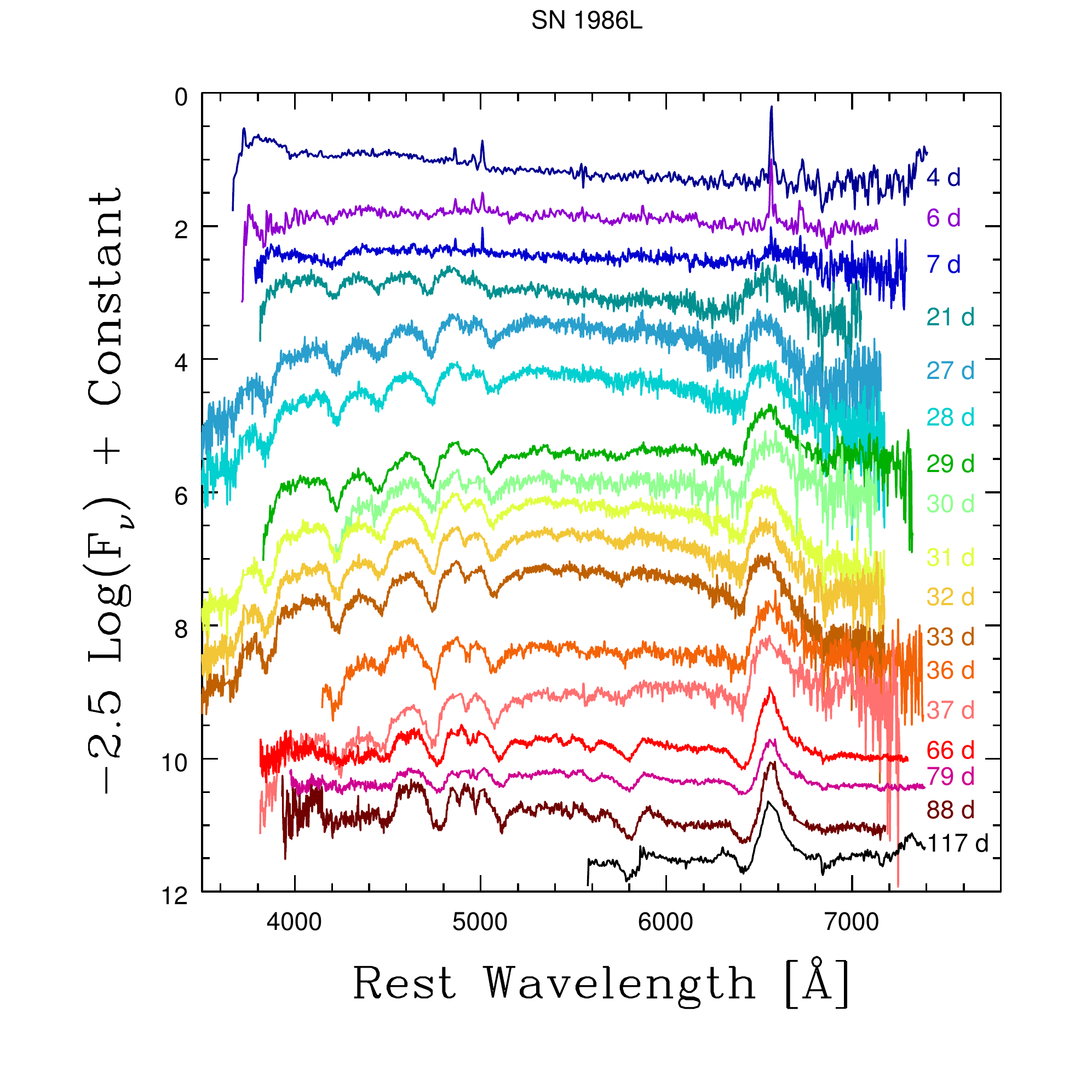}
\includegraphics[width=5.5cm]{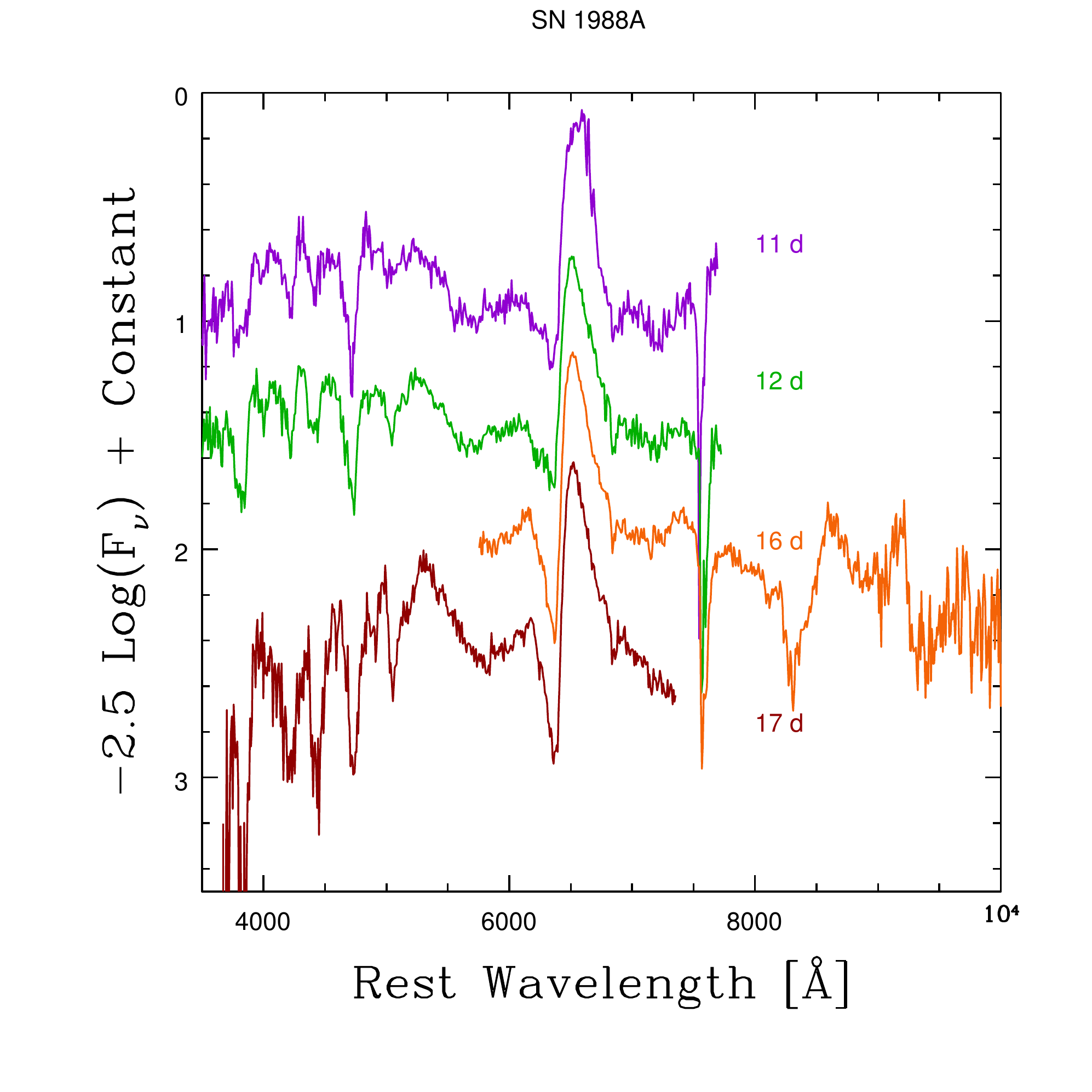}
\includegraphics[width=5.5cm]{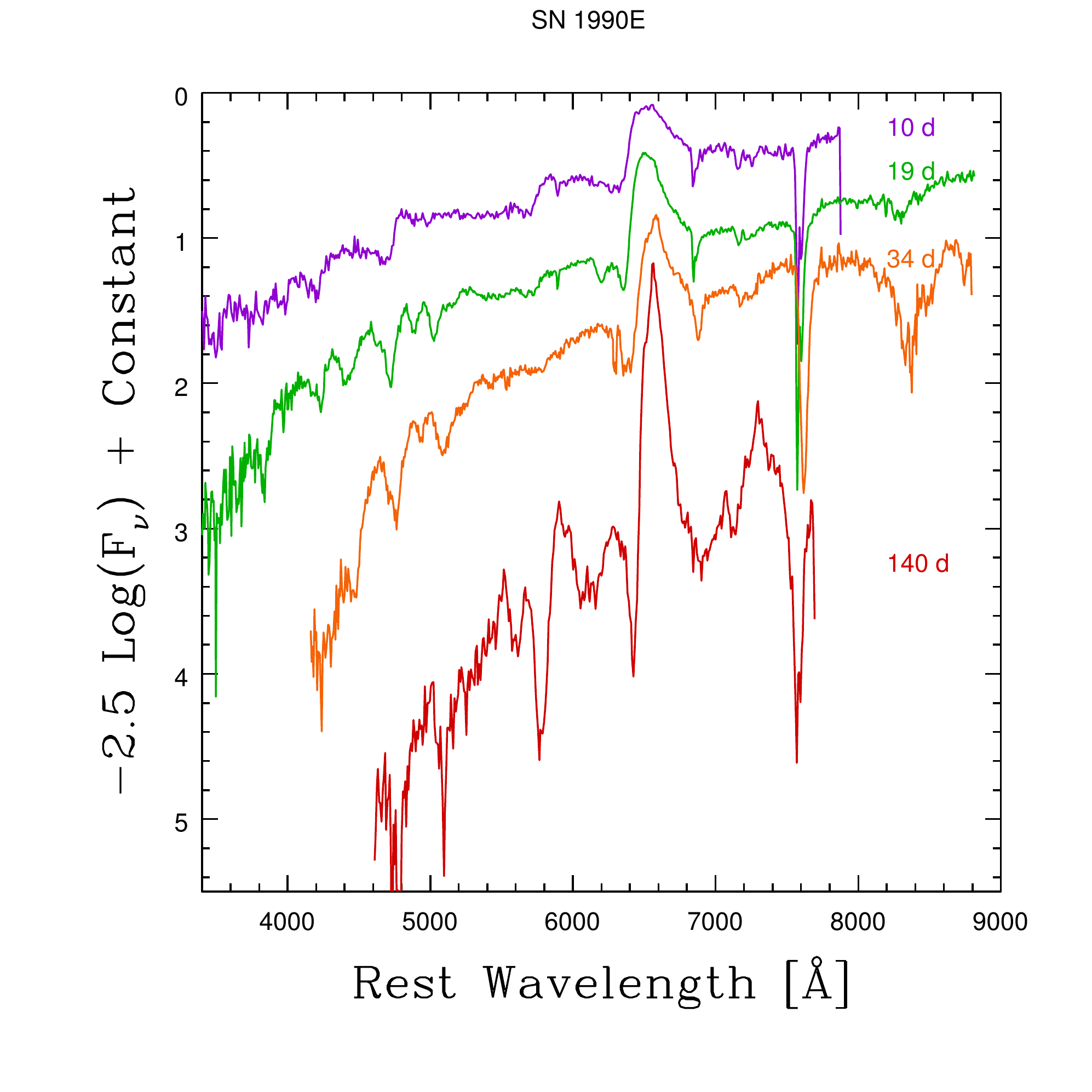}
\includegraphics[width=5.5cm]{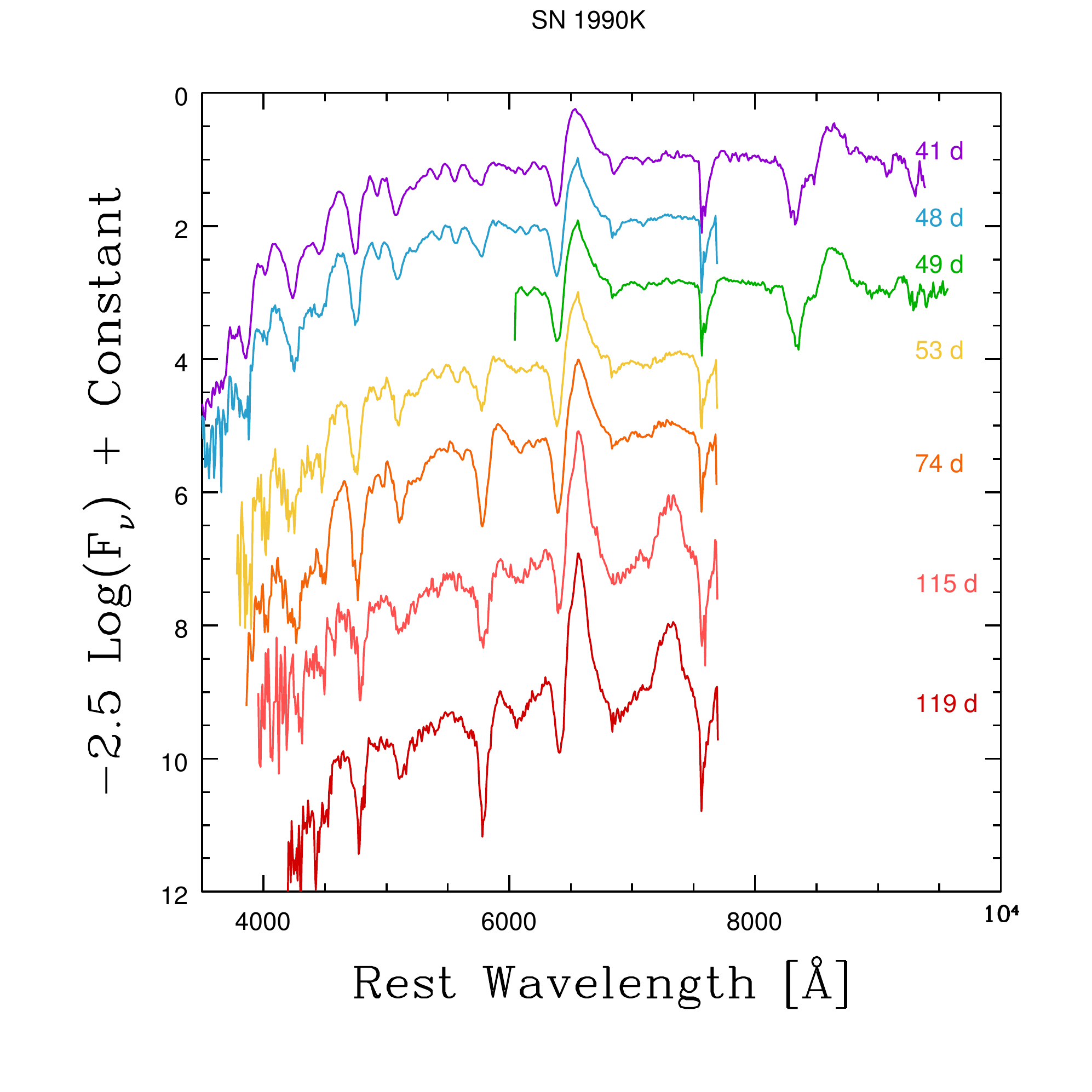}
\includegraphics[width=5.5cm]{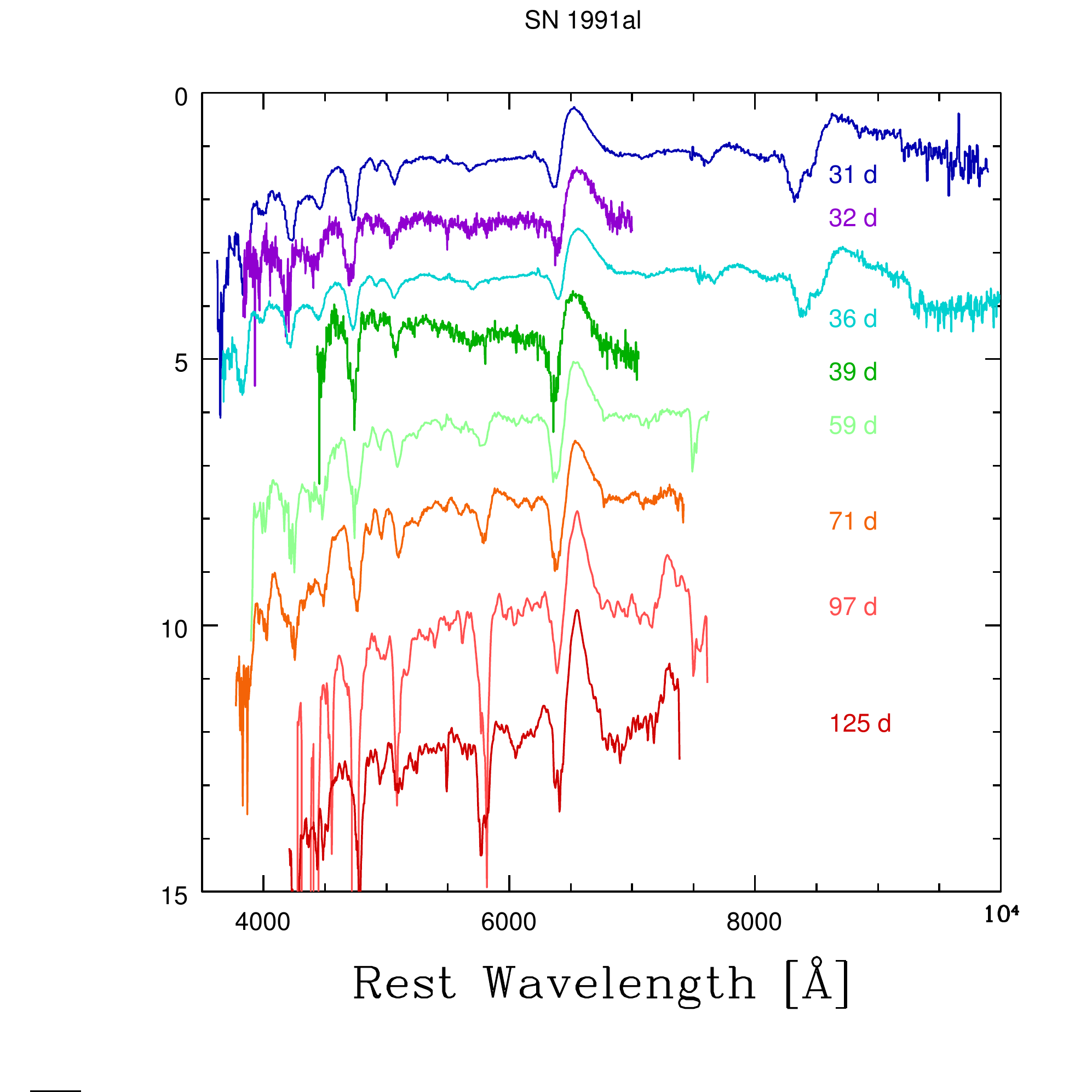}
\includegraphics[width=5.5cm]{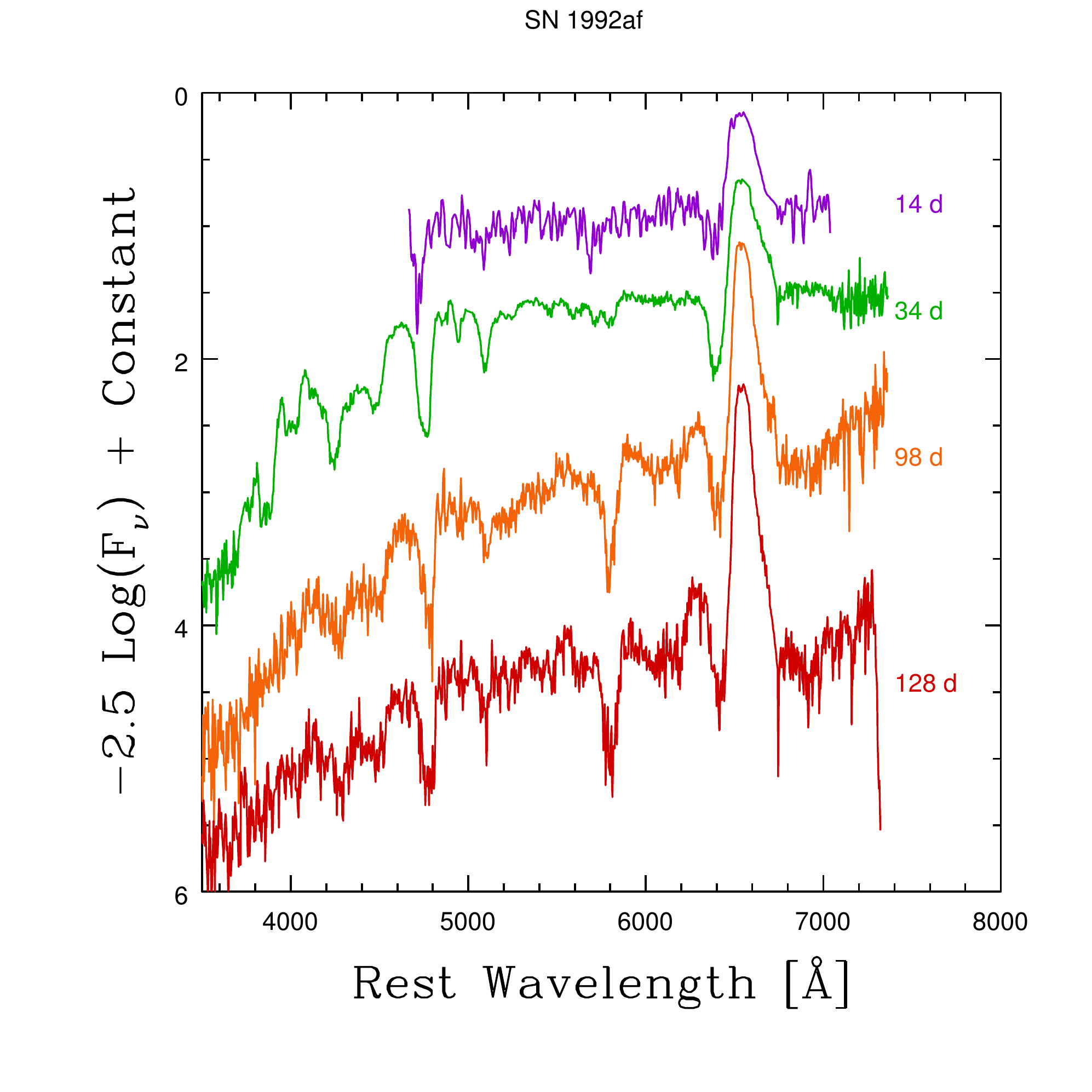}
\includegraphics[width=5.5cm]{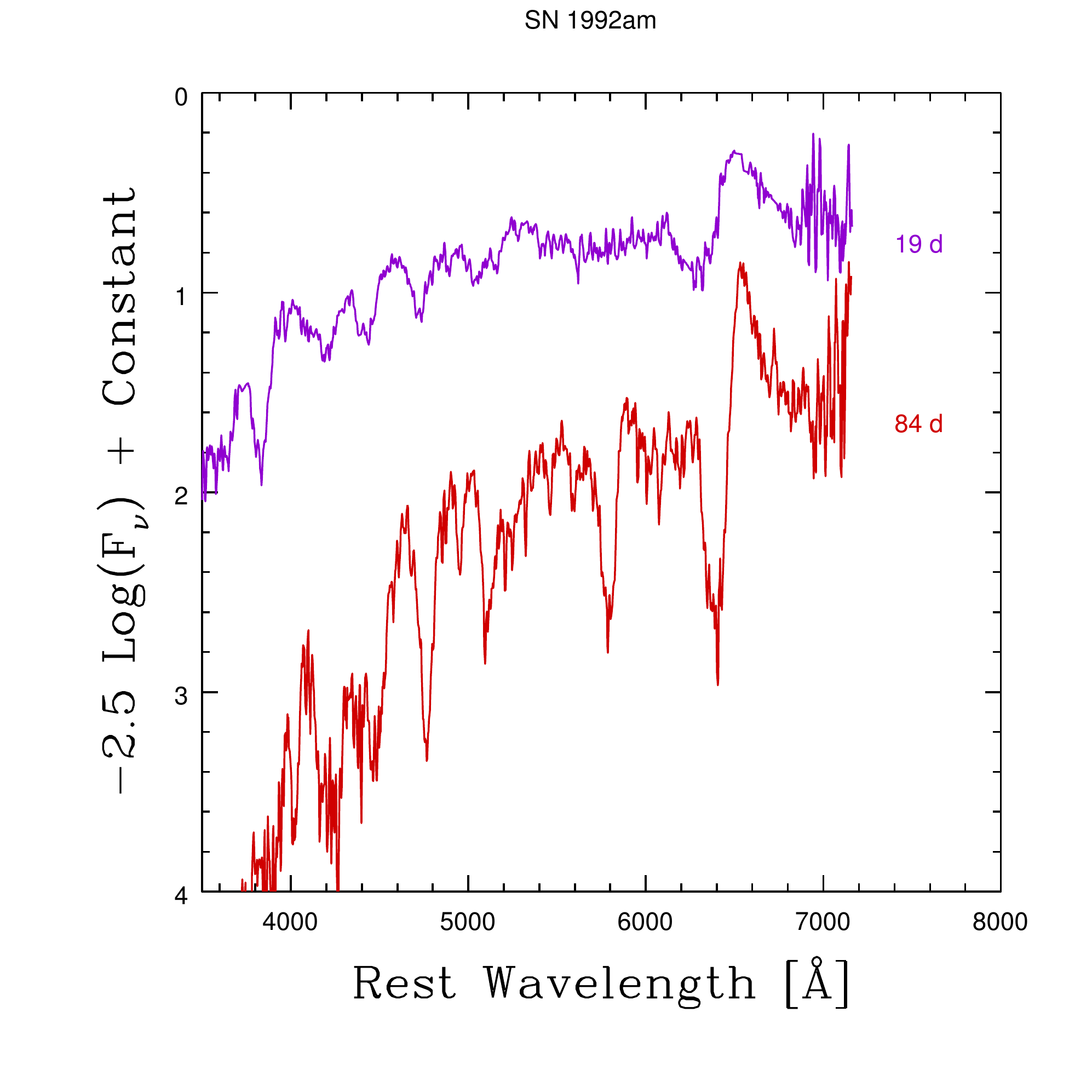}
\includegraphics[width=5.5cm]{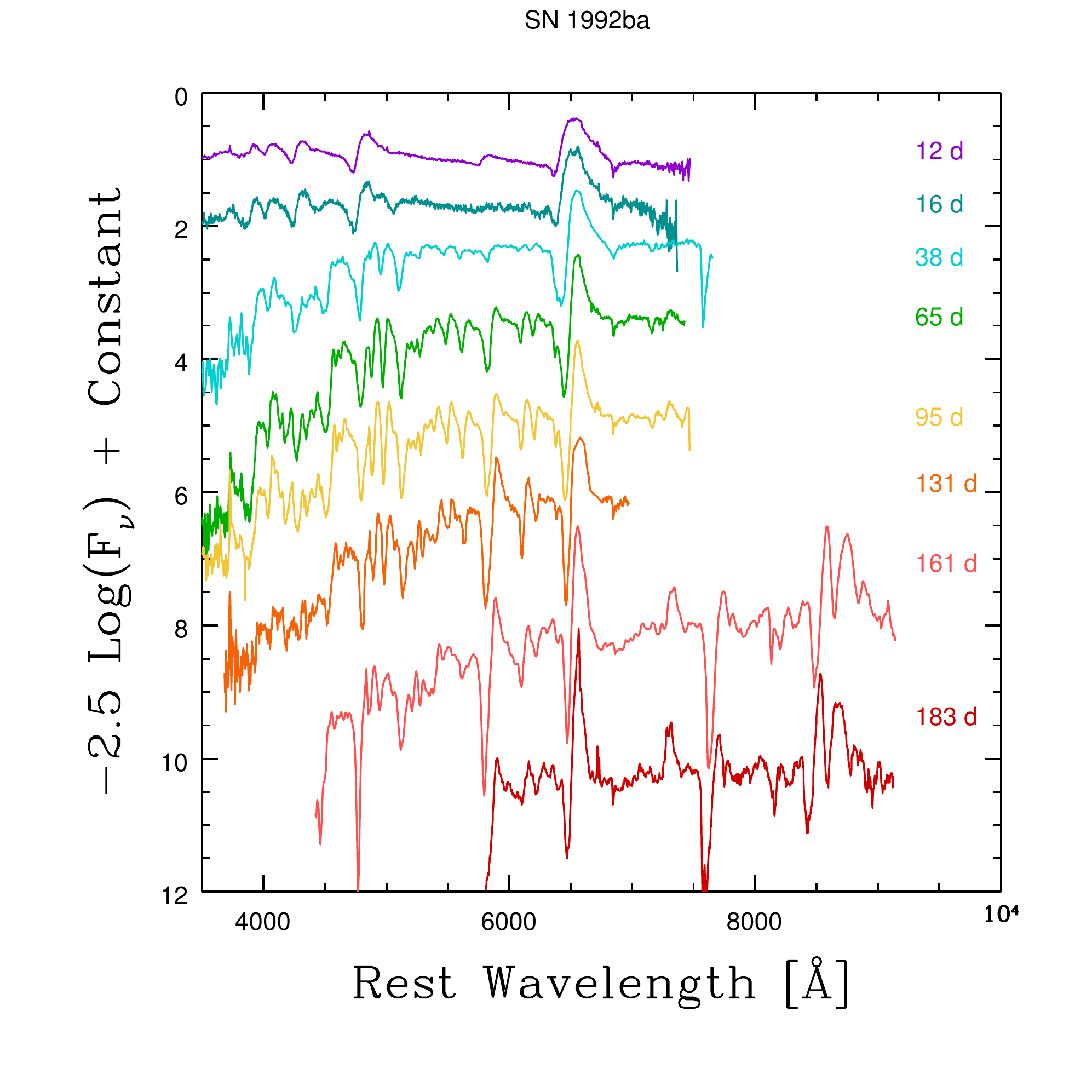}
\includegraphics[width=5.5cm]{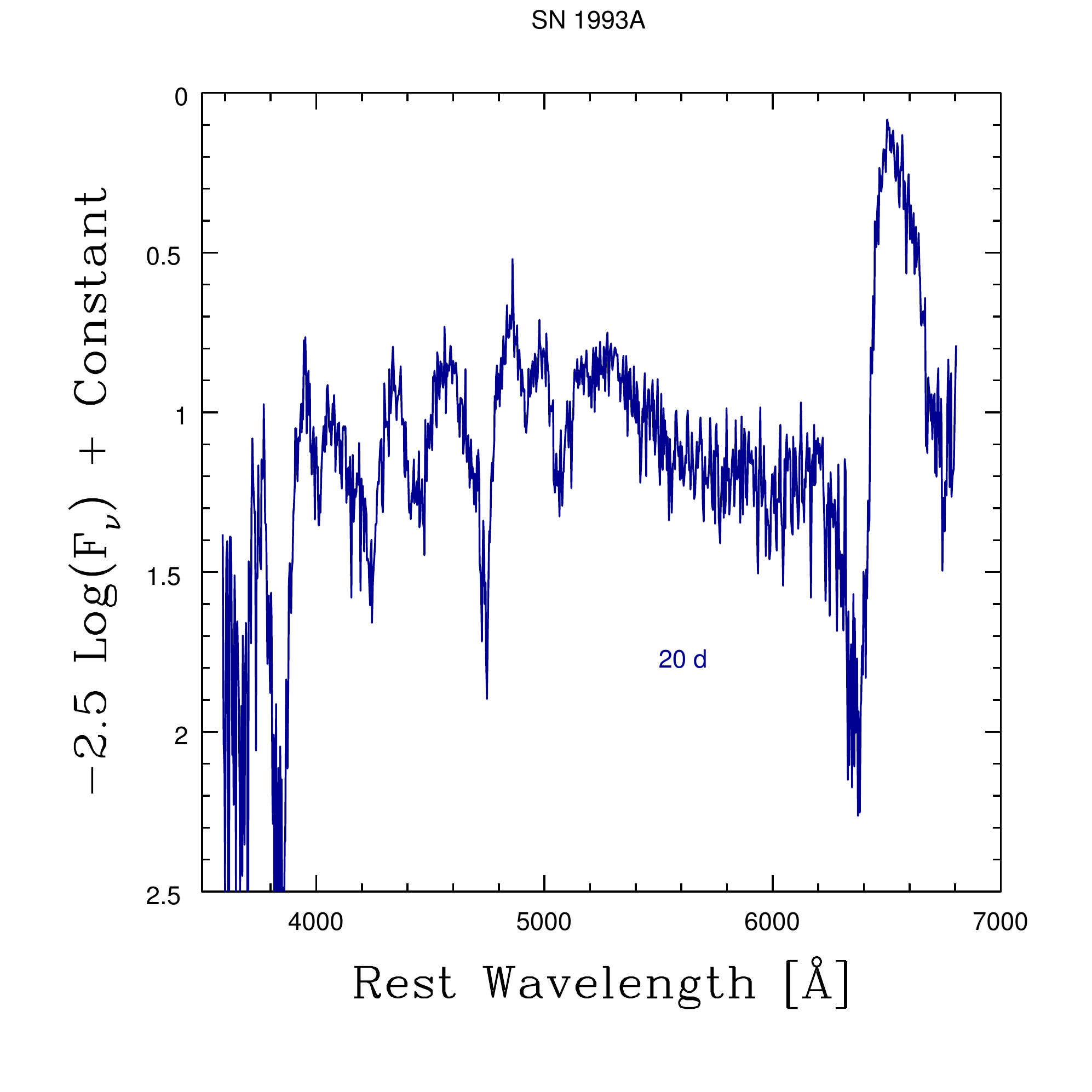}
\includegraphics[width=5.5cm]{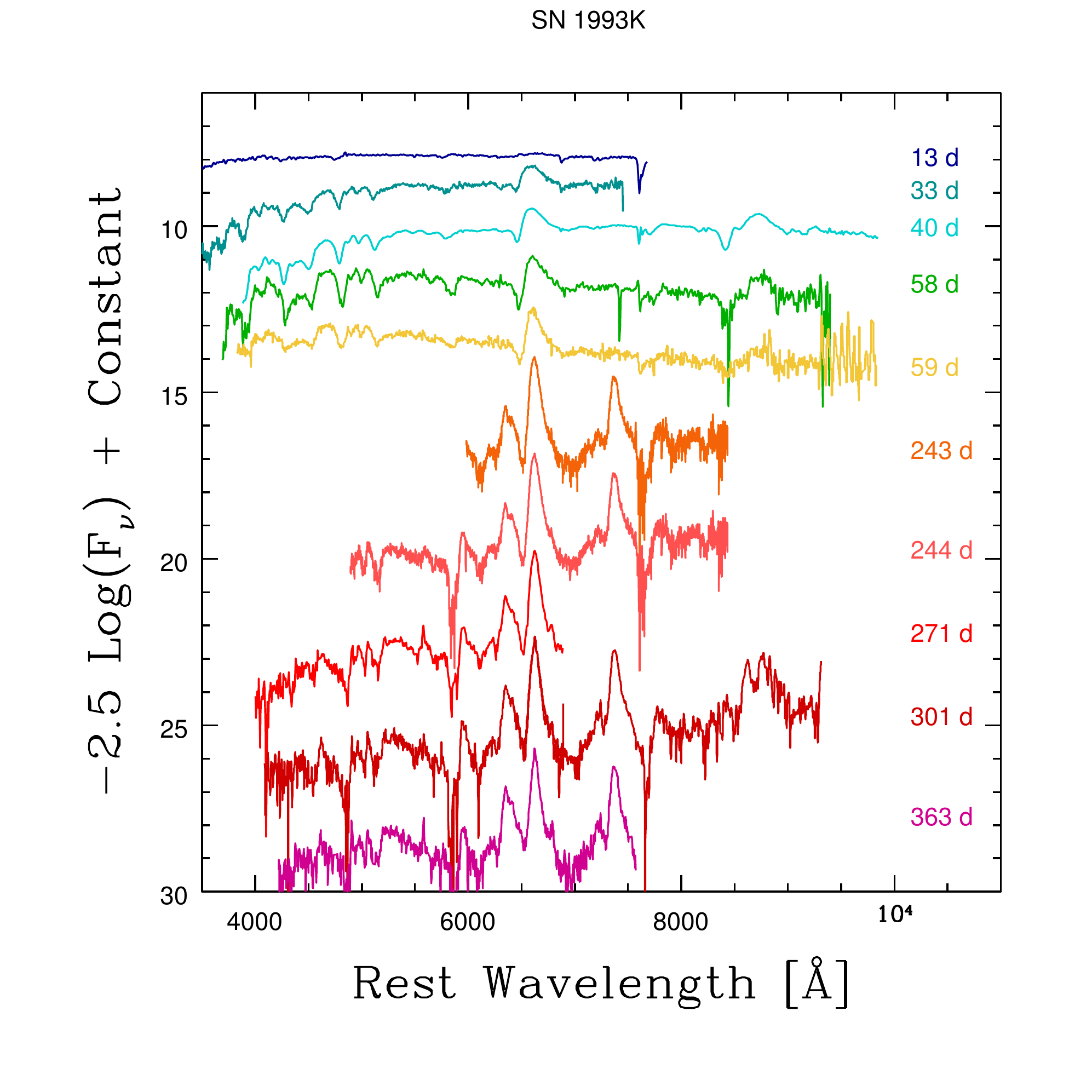}
\includegraphics[width=5.5cm]{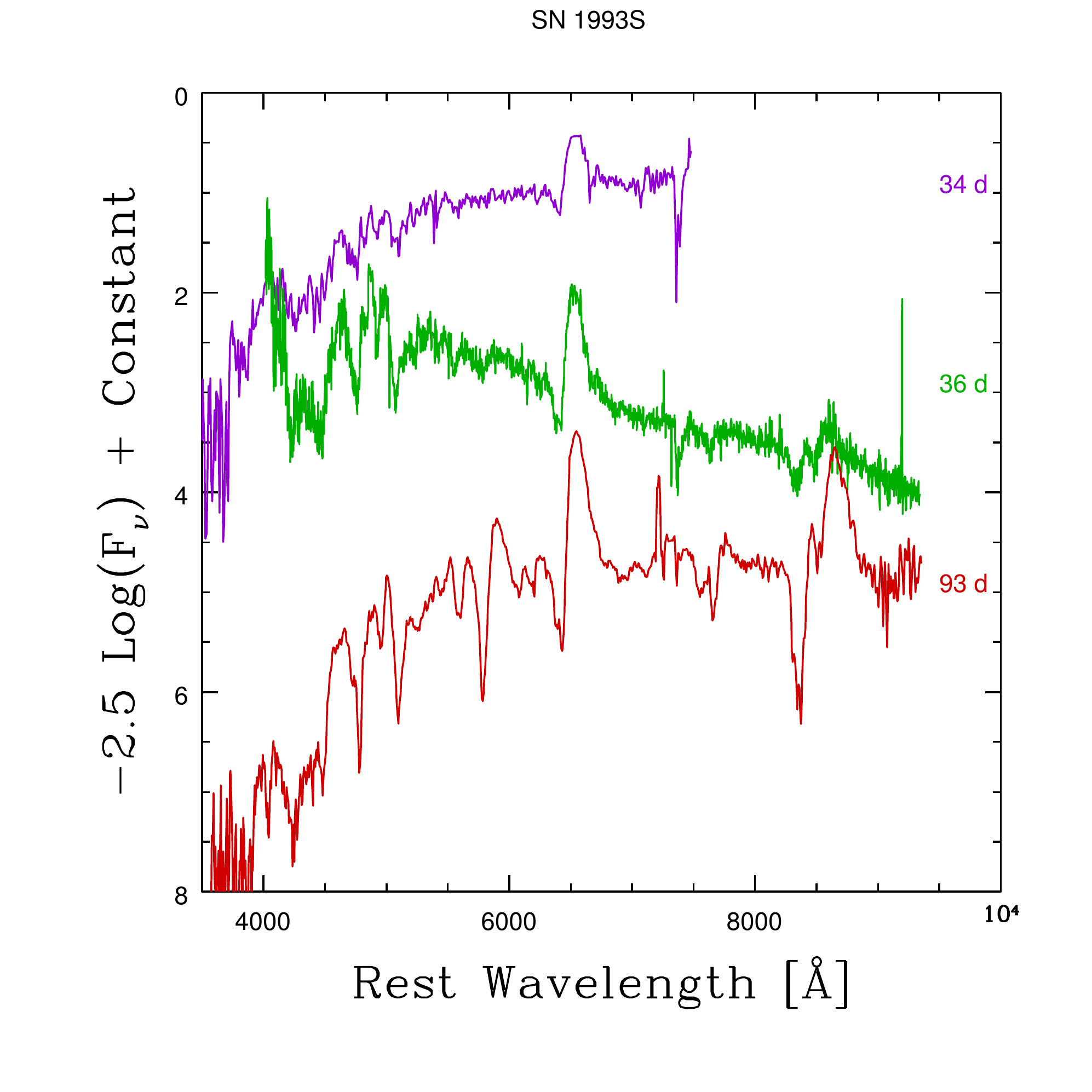}
\includegraphics[width=5.5cm]{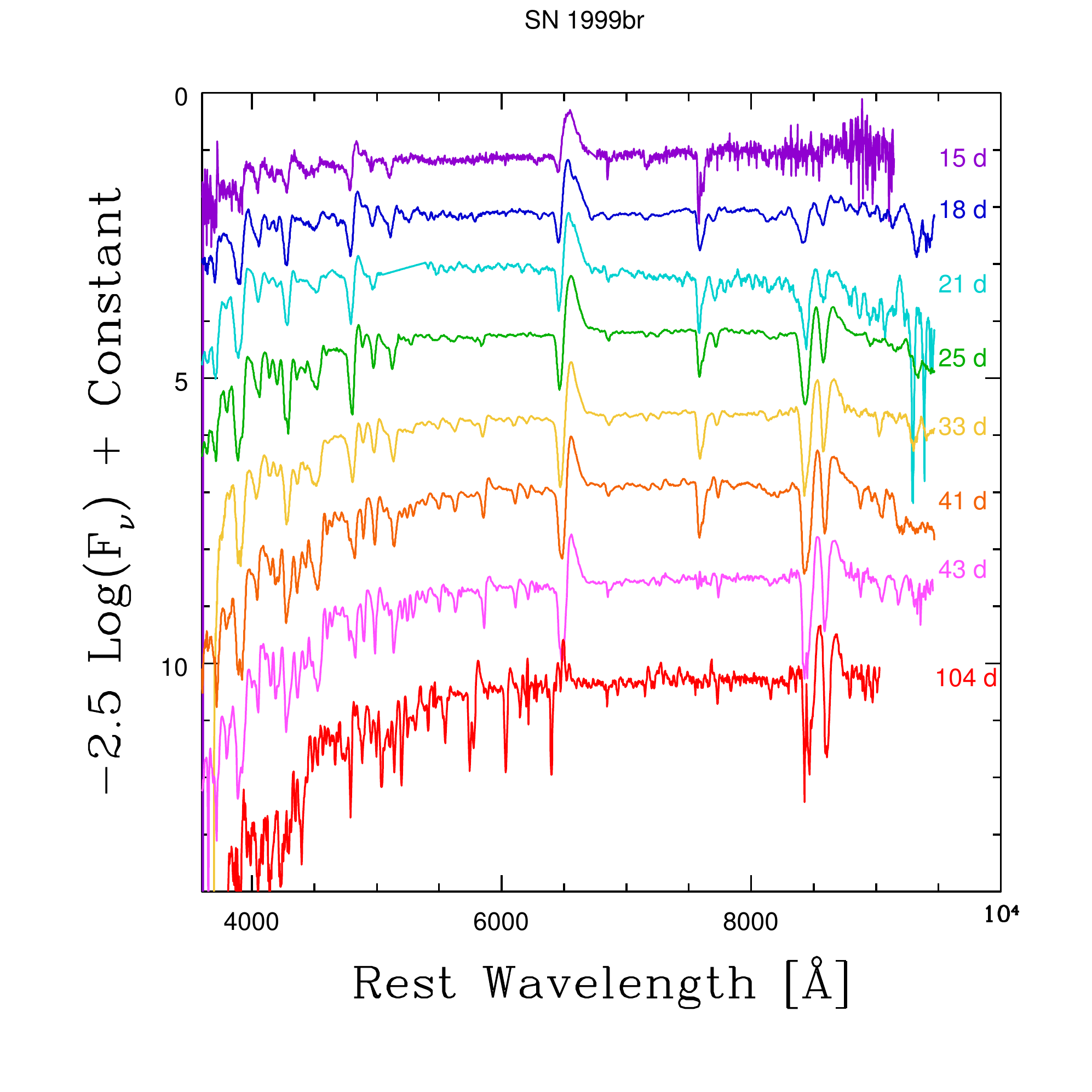}
\caption{Examples of SNe~II spectra: SN~1986L, SN~1988A, SN~1990E, SN~1990K, SN~1991al, SN~1992af, SN~1992am, SN~1992ba, SN~1993A, SN~1993K, SN~1993S, SN~1999br.}
\label{example}
\end{figure*}

\begin{figure*}[h!]
\centering
\includegraphics[width=5.5cm]{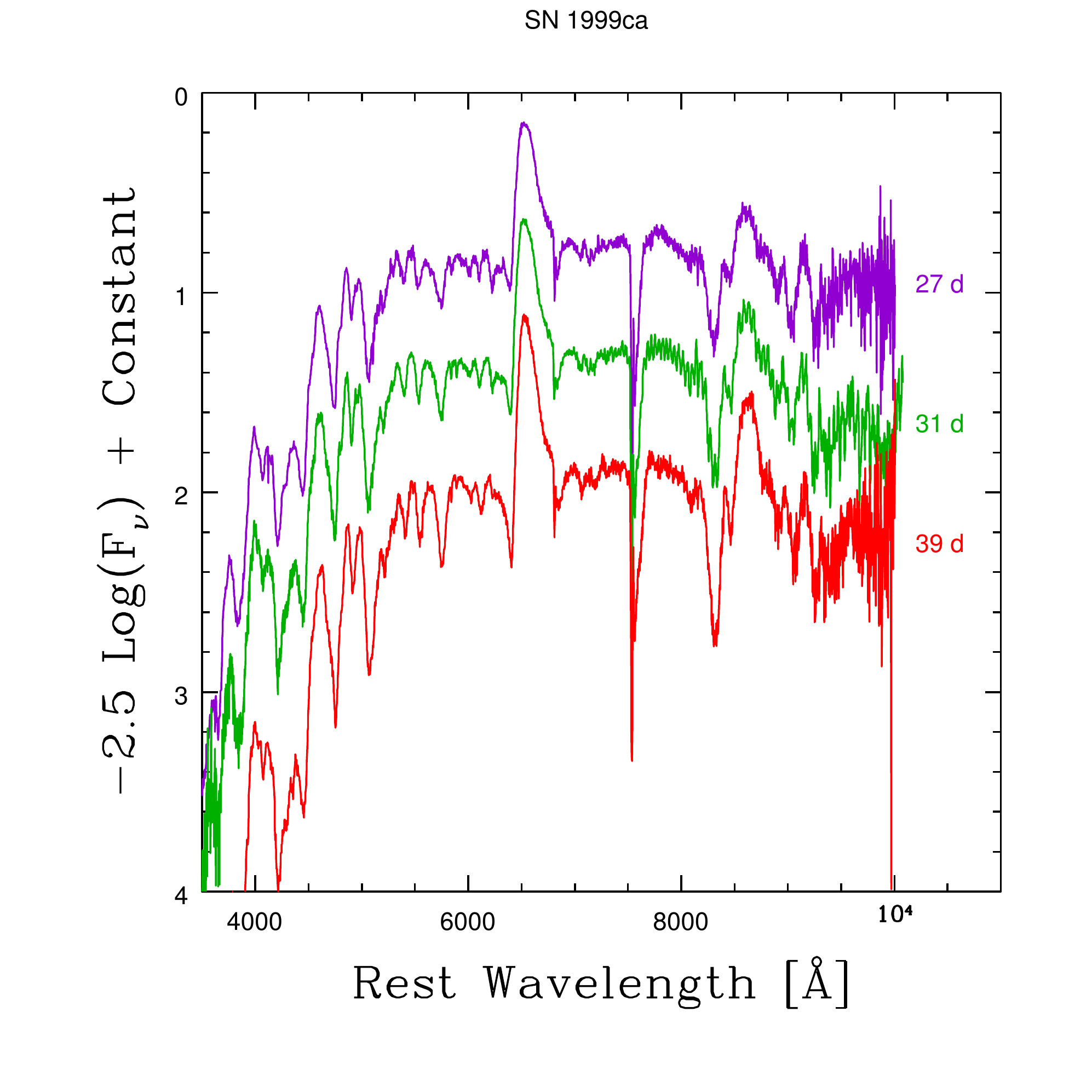}
\includegraphics[width=5.5cm]{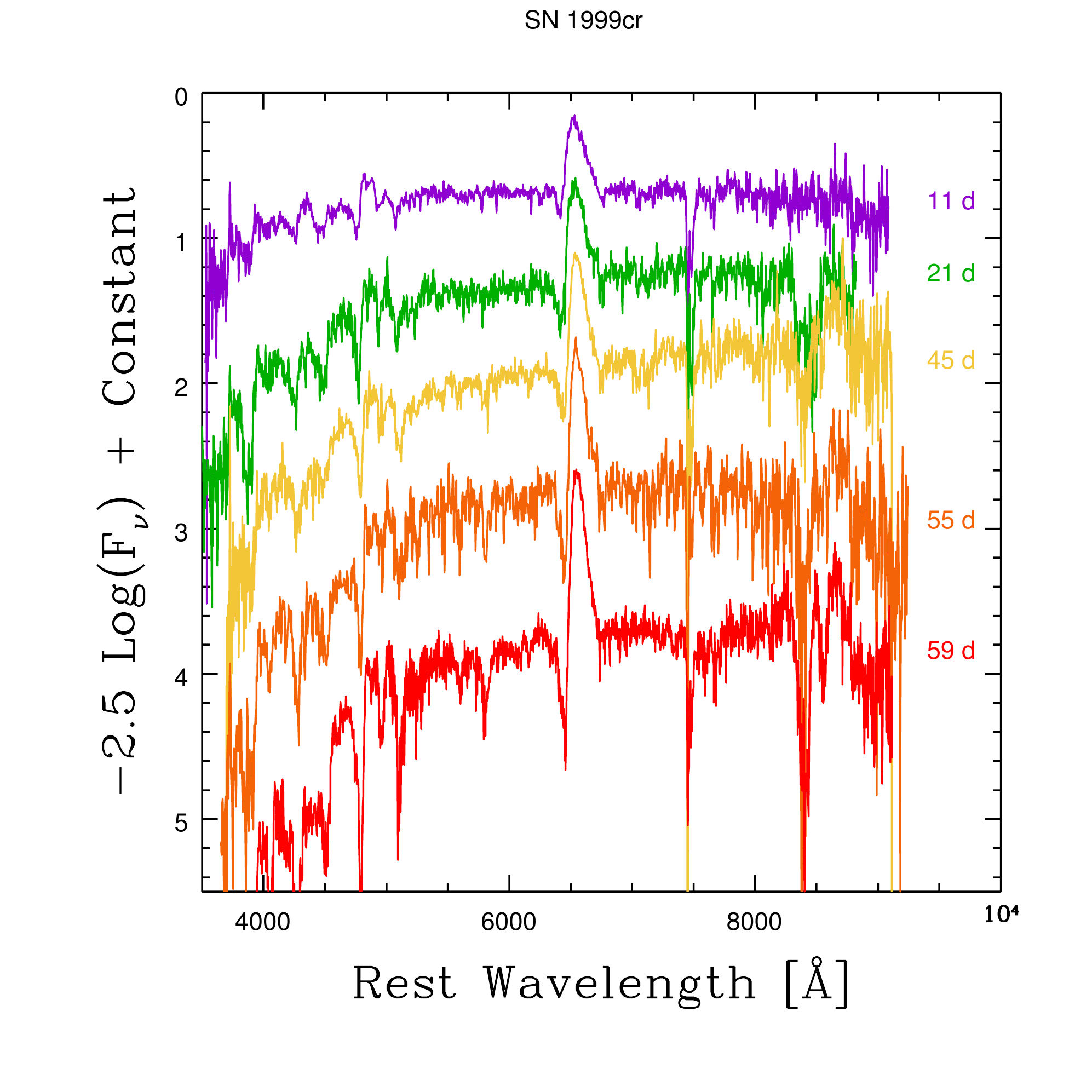}
\includegraphics[width=5.5cm]{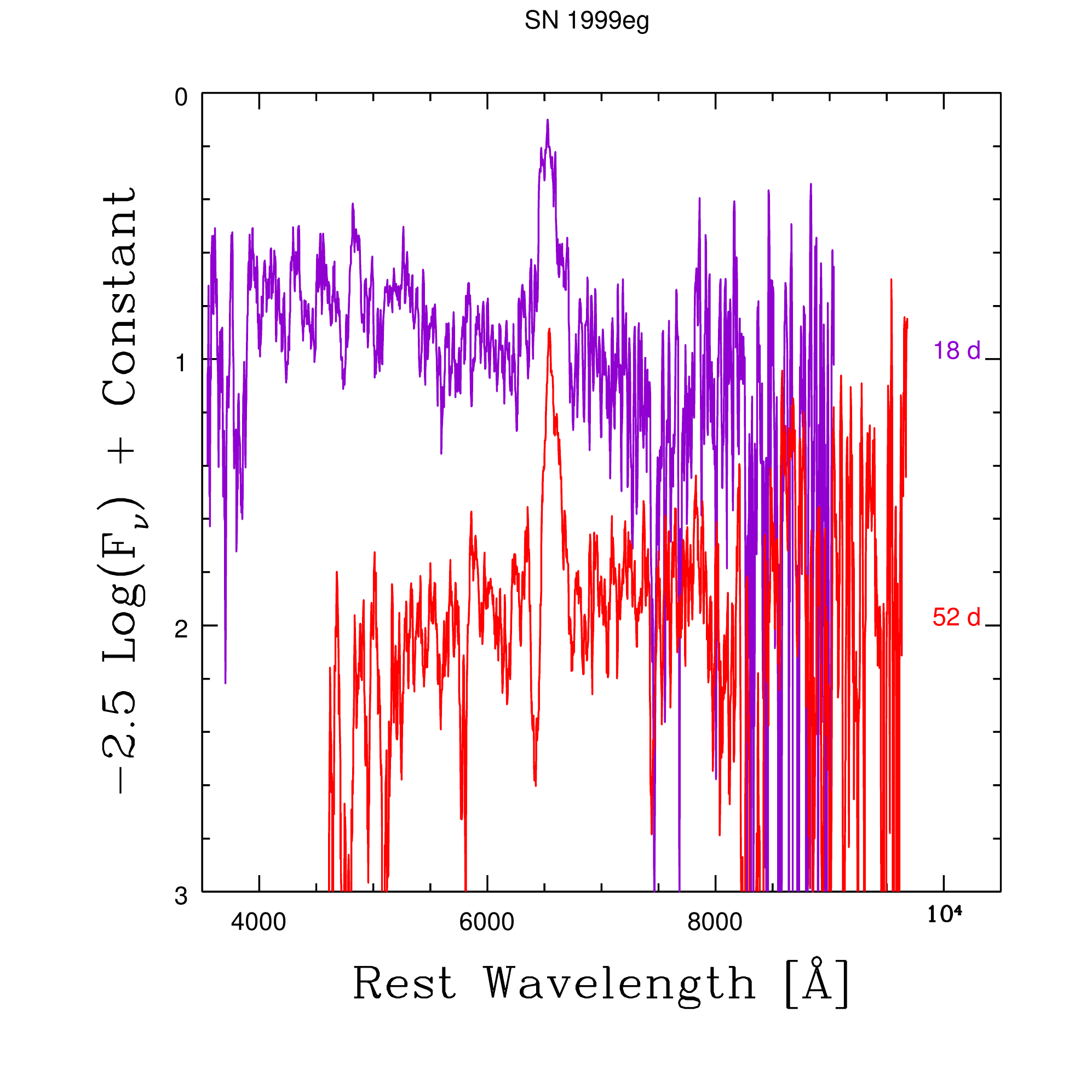}
\includegraphics[width=5.5cm]{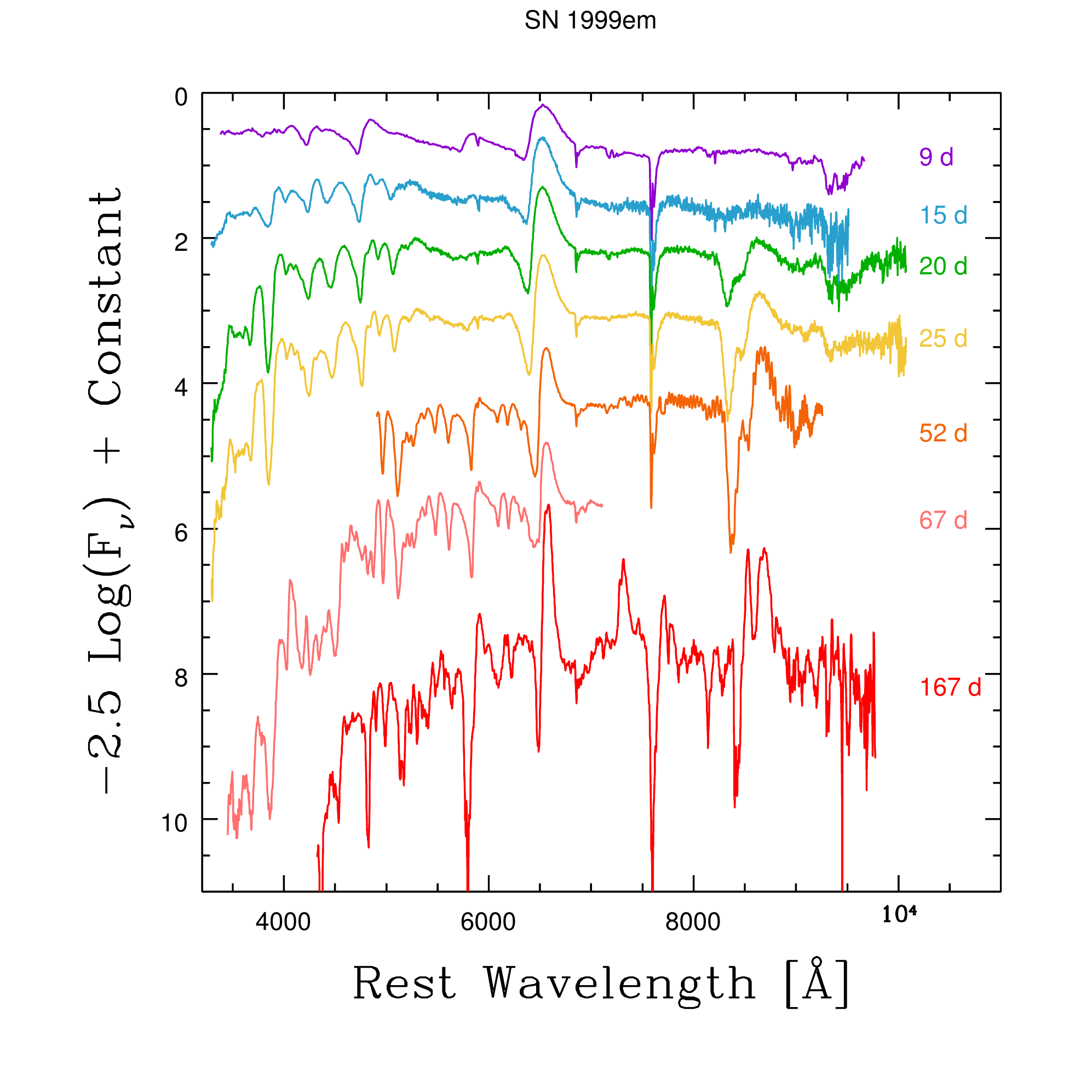}
\includegraphics[width=5.5cm]{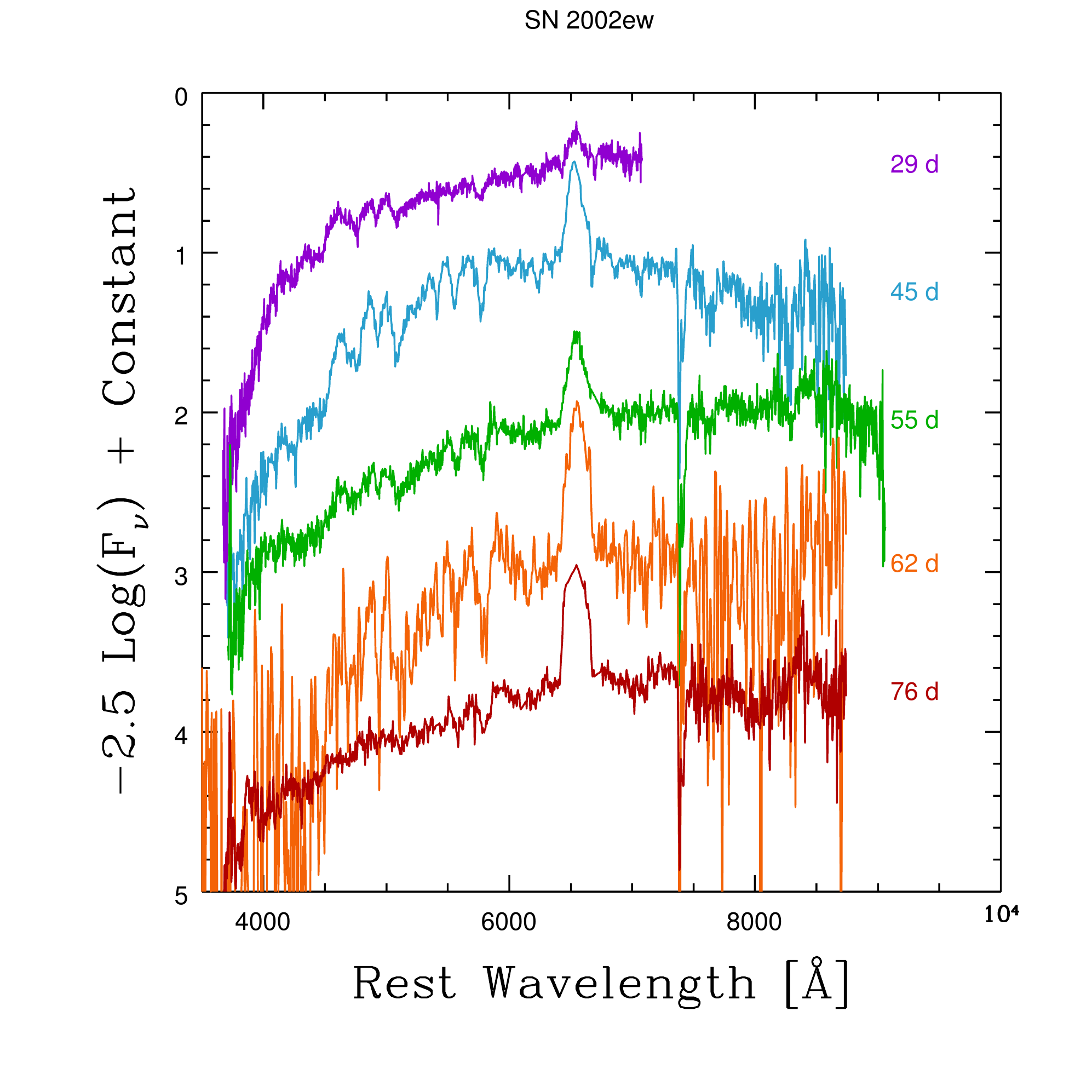}
\includegraphics[width=5.5cm]{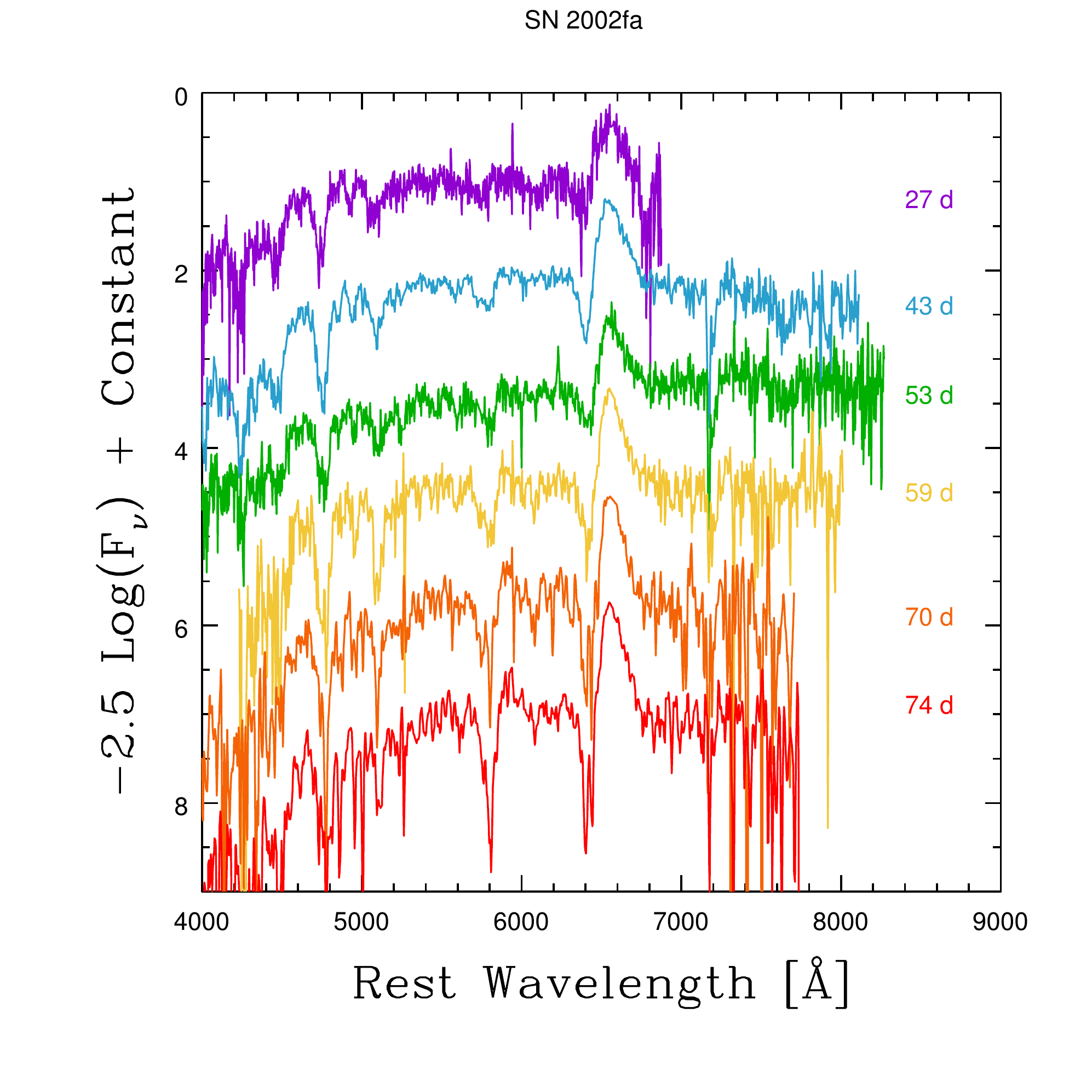}
\includegraphics[width=5.5cm]{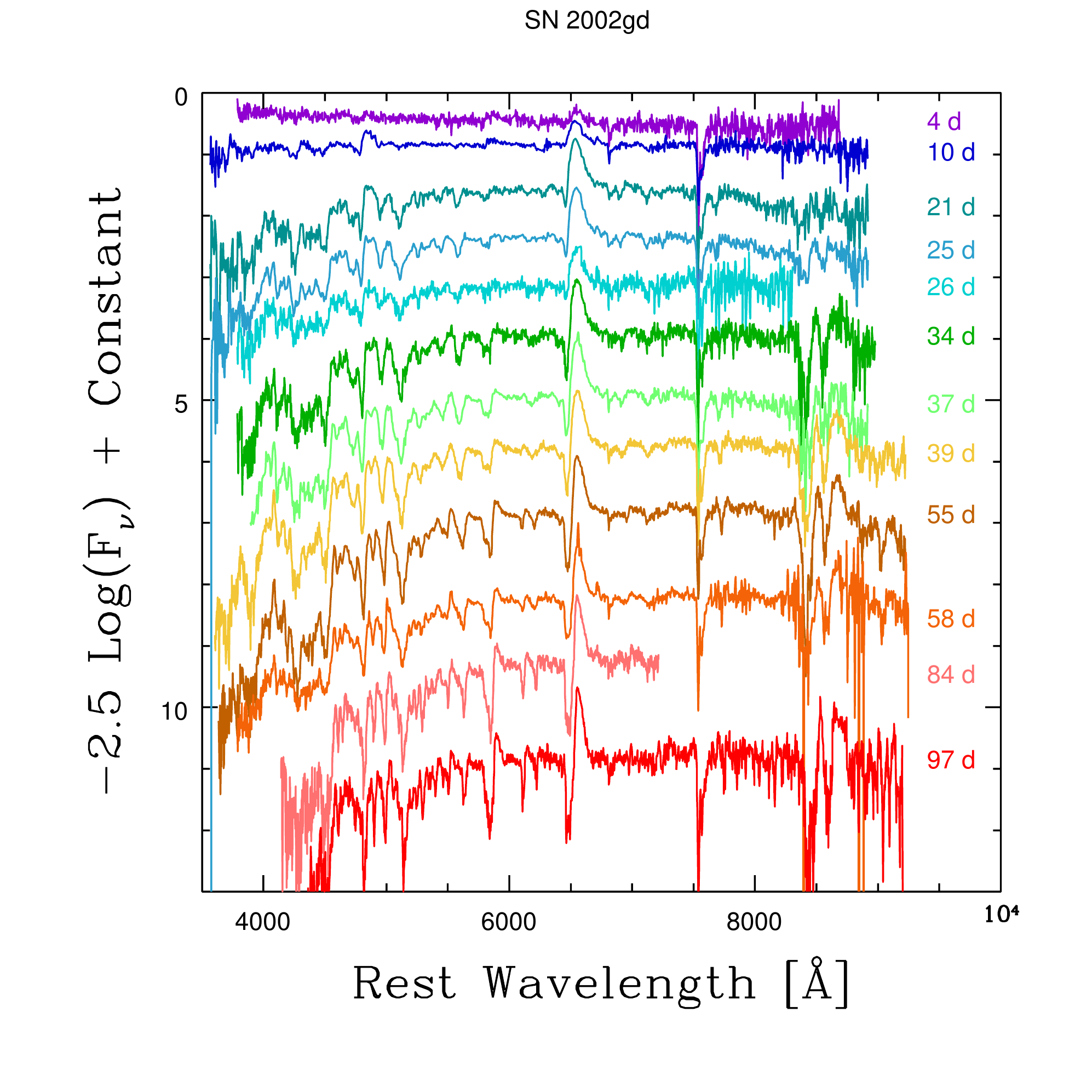}
\includegraphics[width=5.5cm]{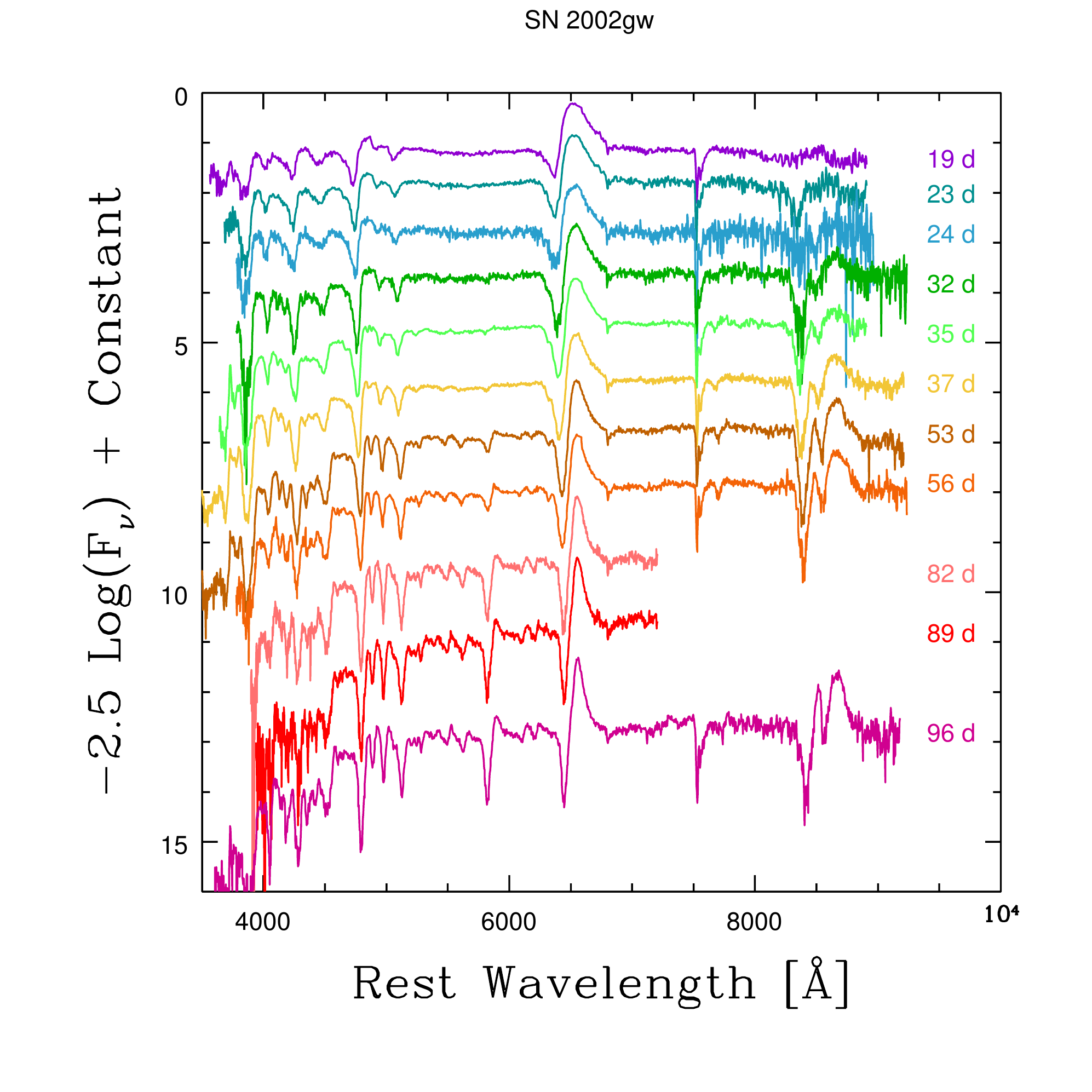}
\includegraphics[width=5.5cm]{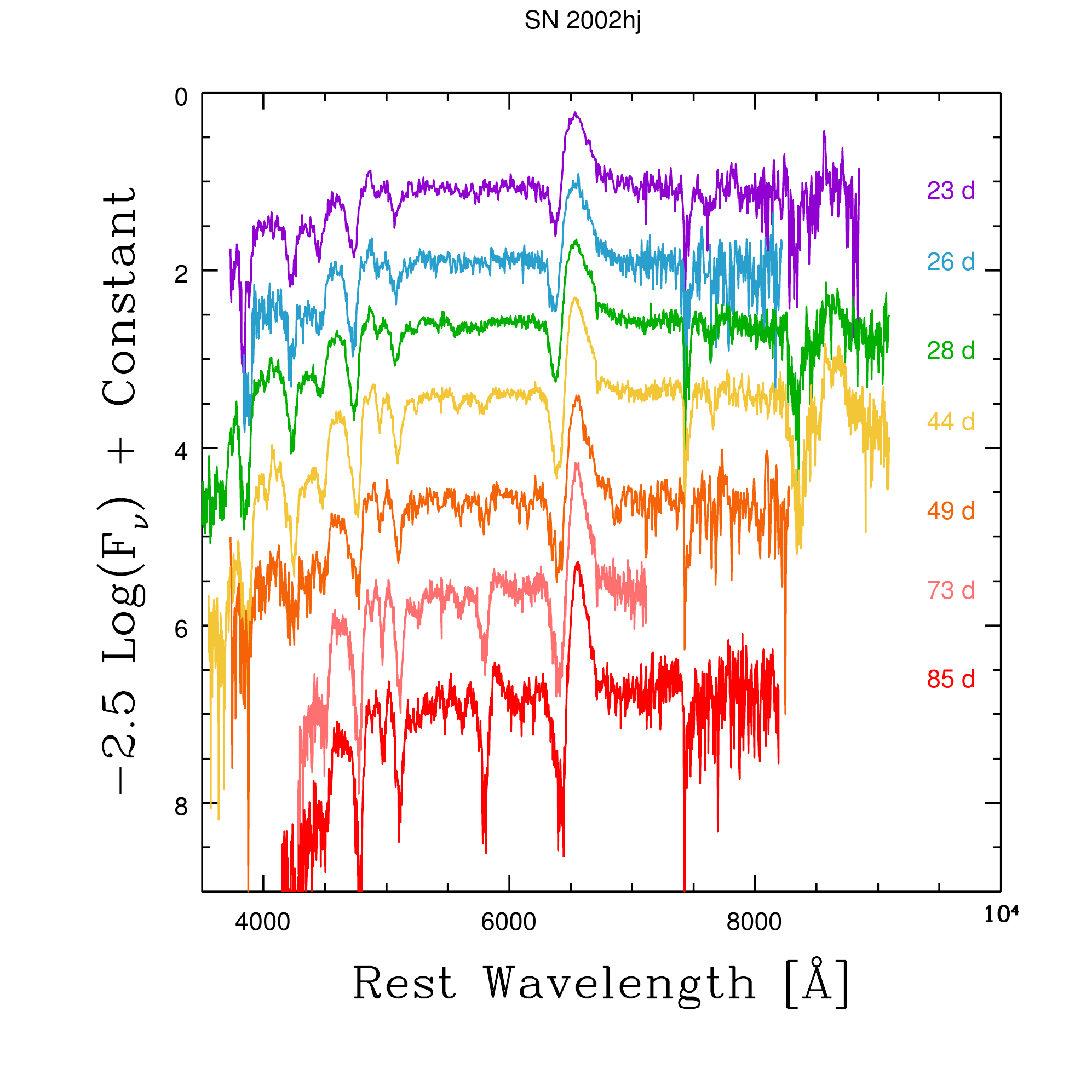}
\includegraphics[width=5.5cm]{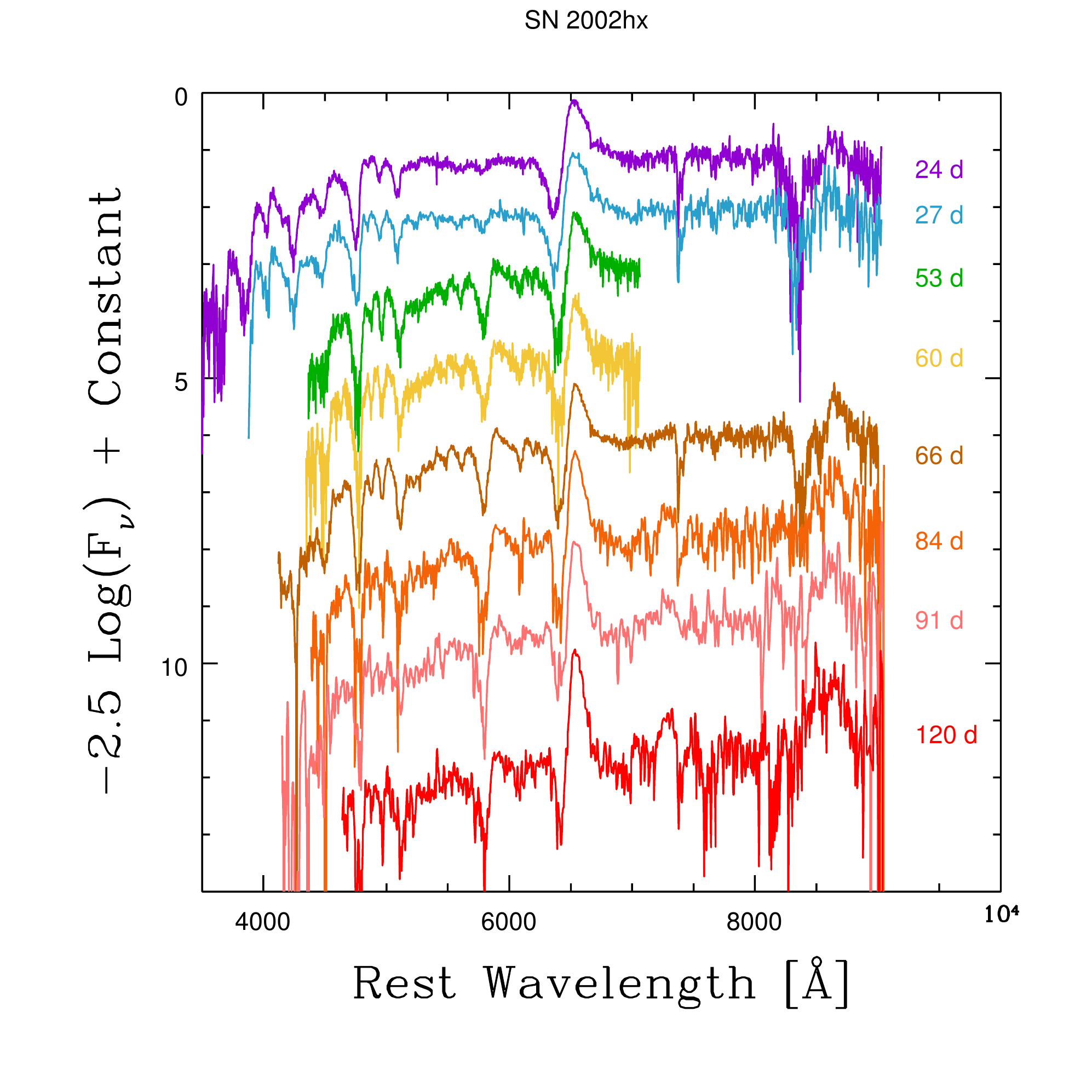}
\includegraphics[width=5.5cm]{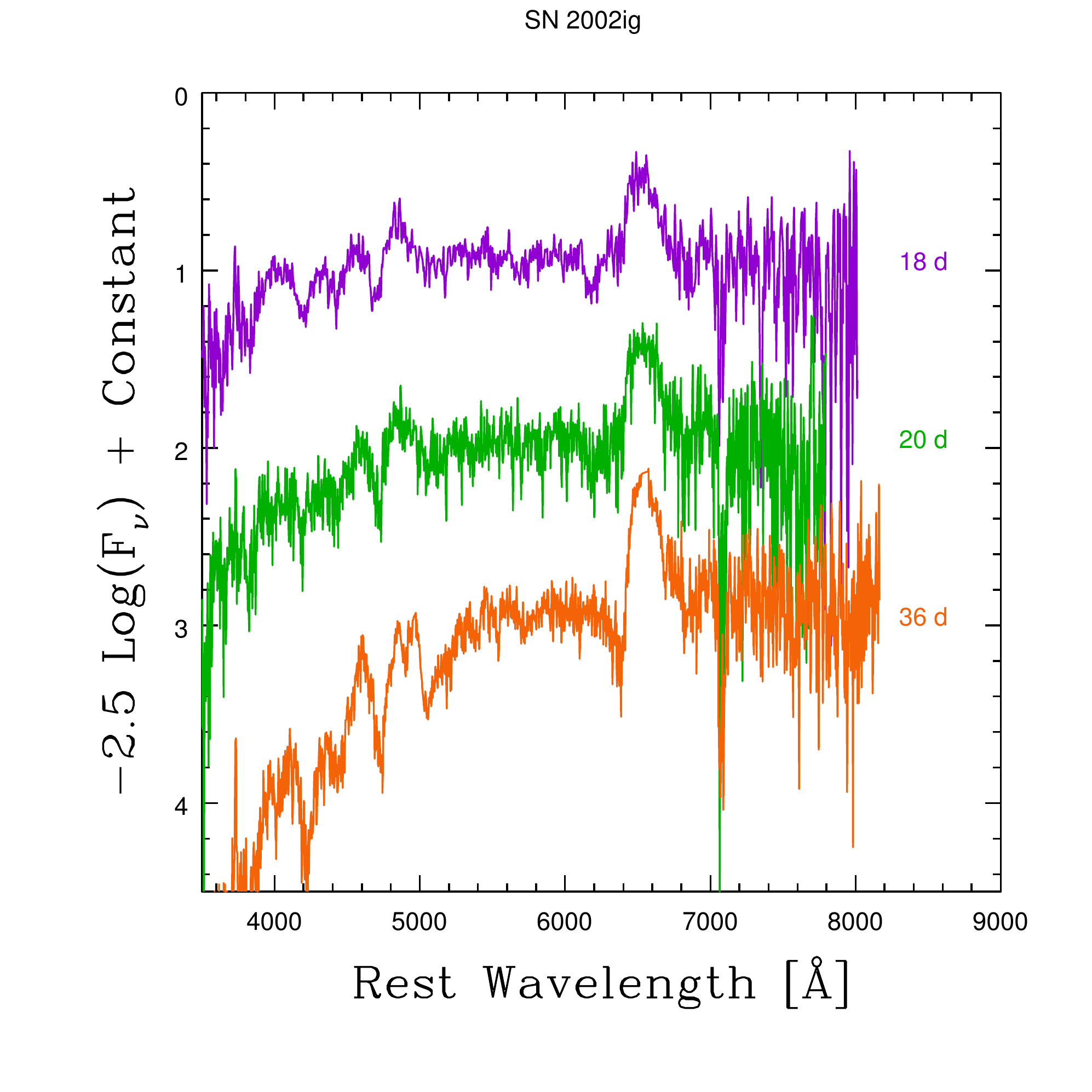}
\includegraphics[width=5.5cm]{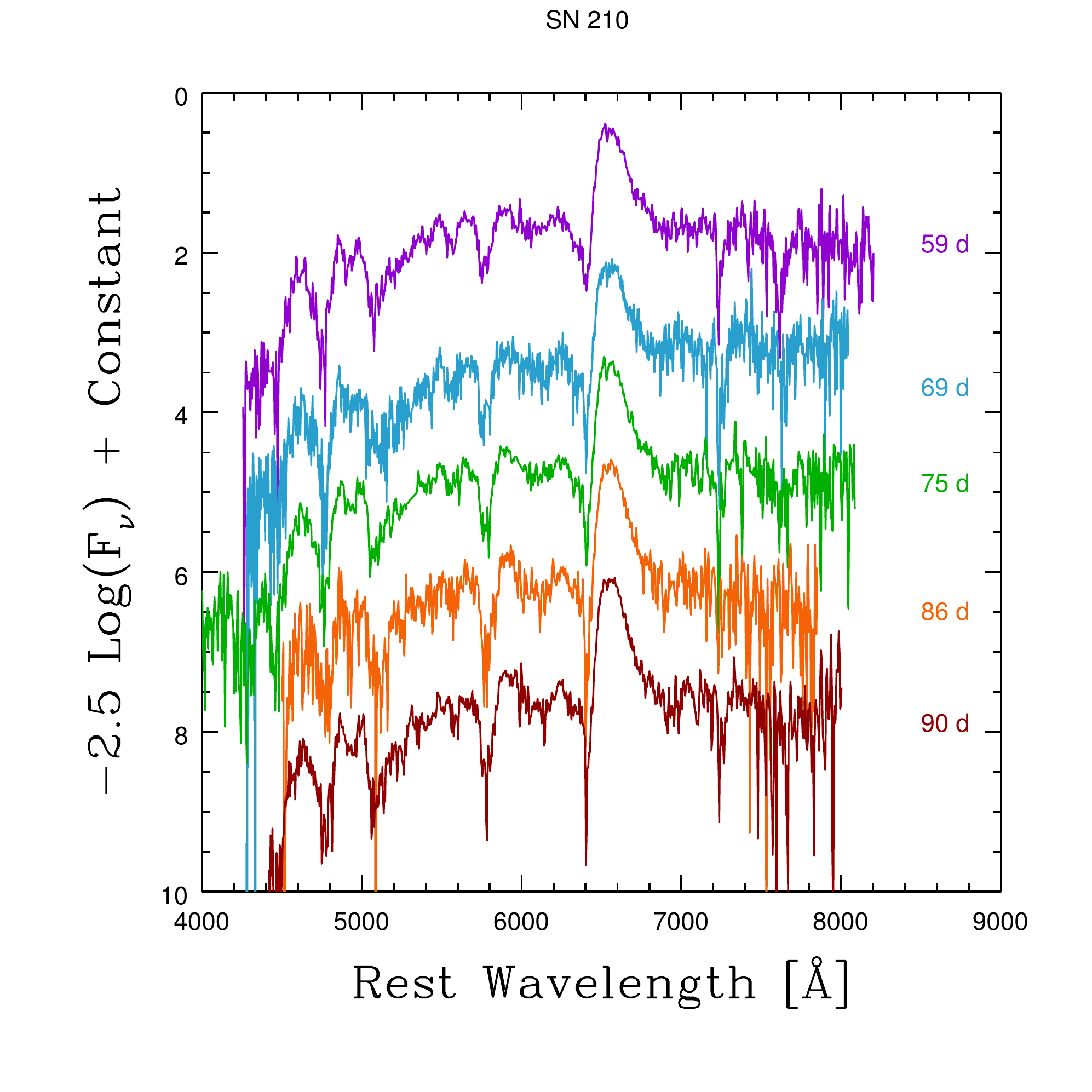}
\caption{Examples of SNe~II spectra: SN~1999ca, SN~1999cr, SN~1999eg, SN~1999em, SN~2002ew, SN~2002fa, SN~200gd, SN~200gw, SN~2002hj, SN~2002hx, SN~2002ig, SN~210.}
\label{example}
\end{figure*}

\begin{figure*}[h!]
\centering
\includegraphics[width=5.5cm]{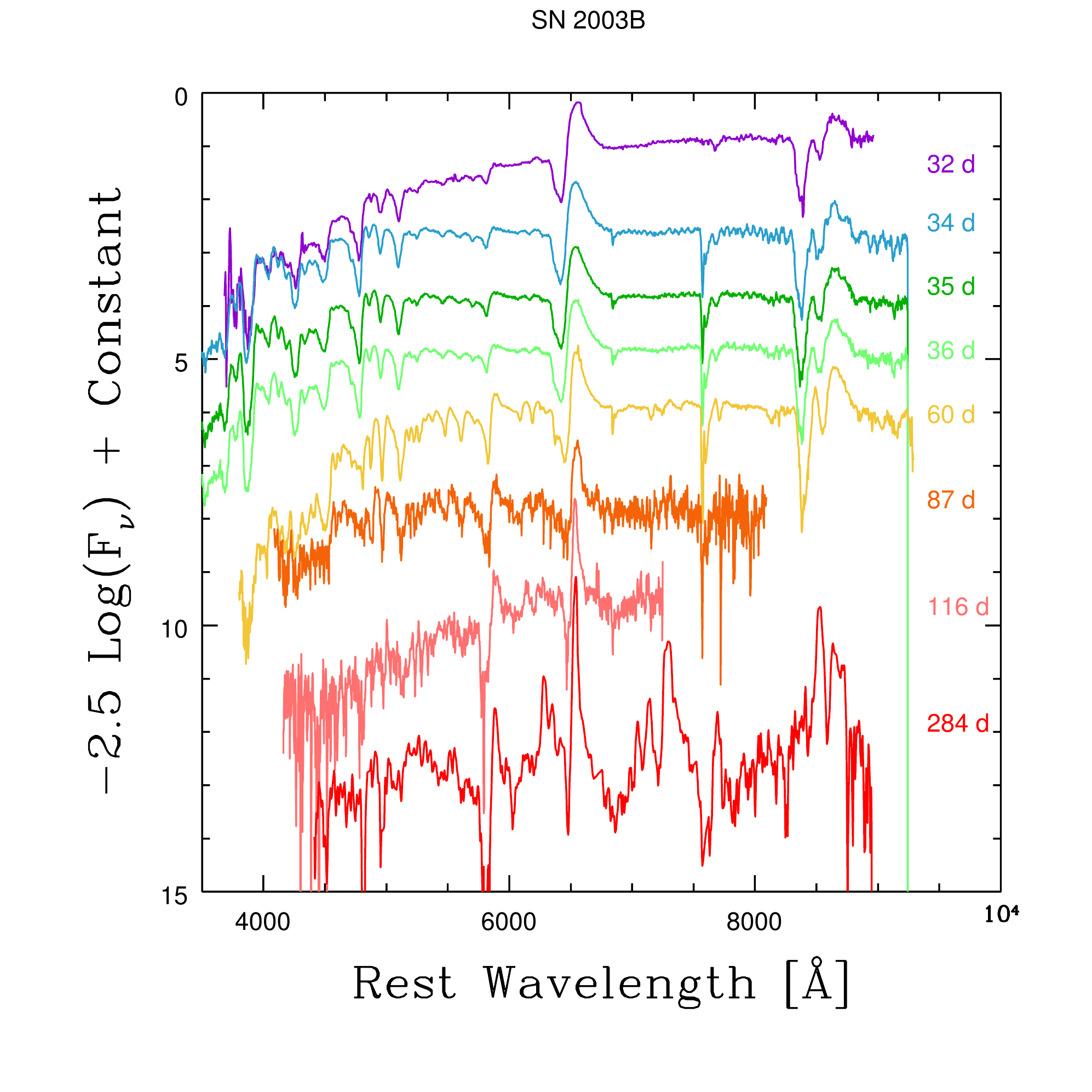}
\includegraphics[width=5.5cm]{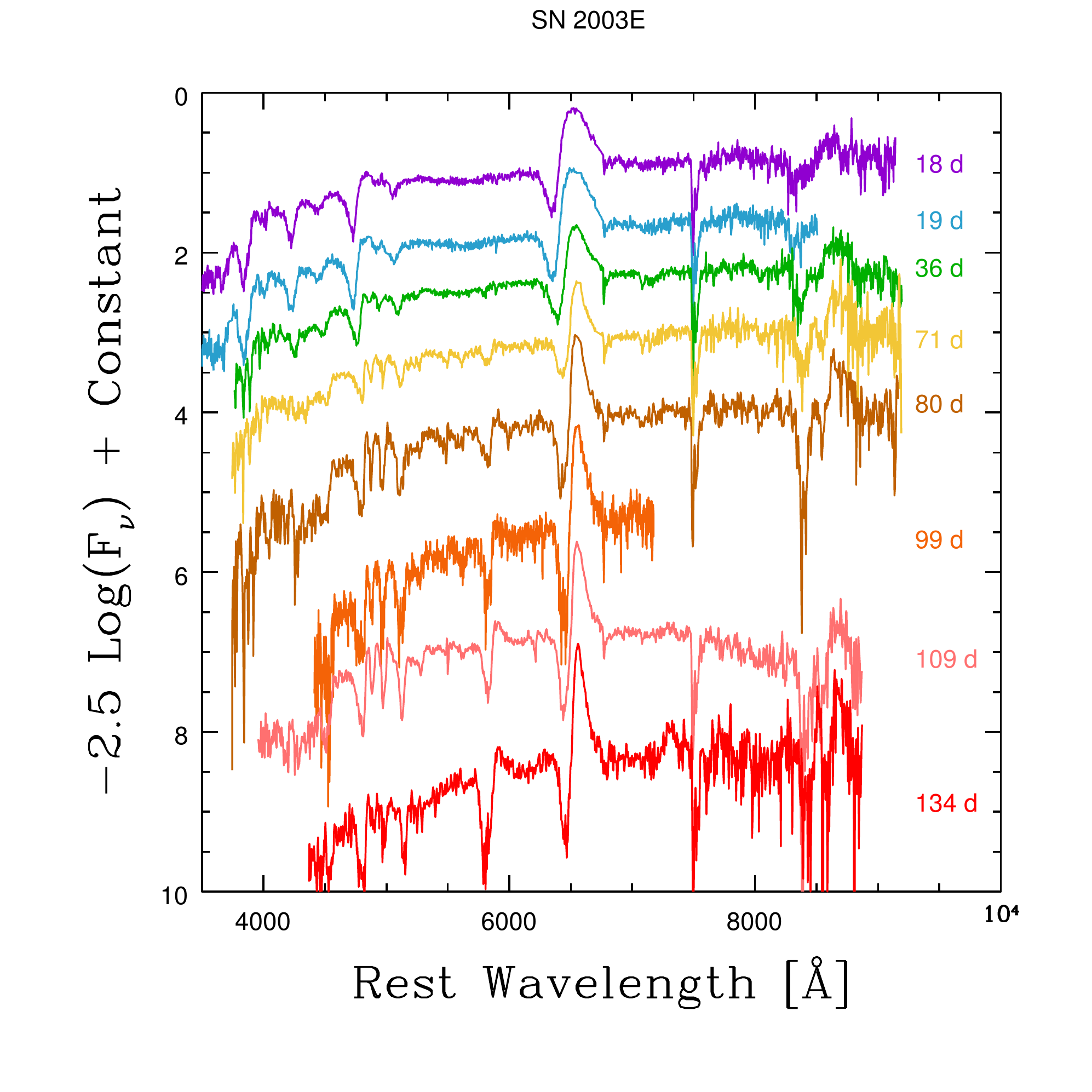}
\includegraphics[width=5.5cm]{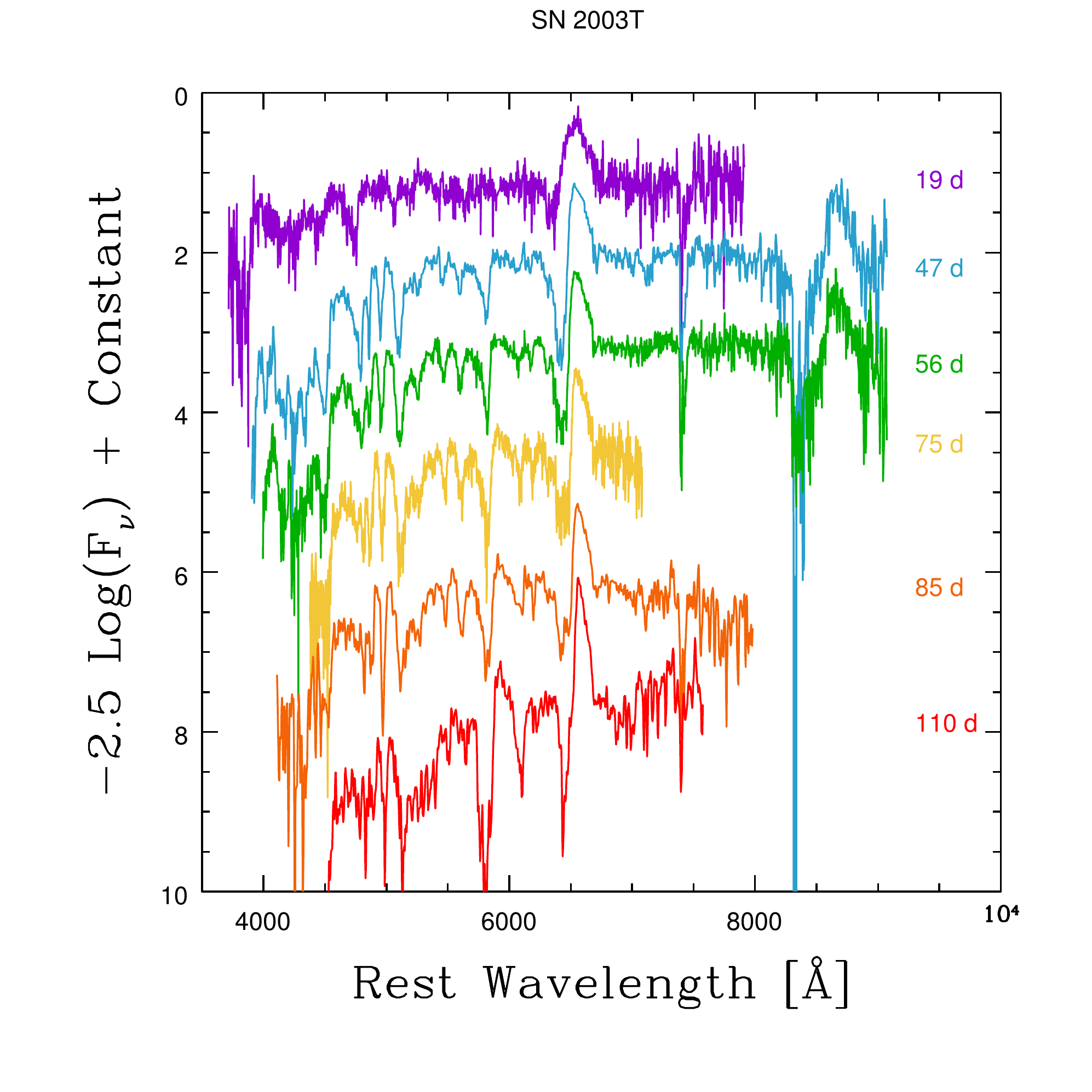}
\includegraphics[width=5.5cm]{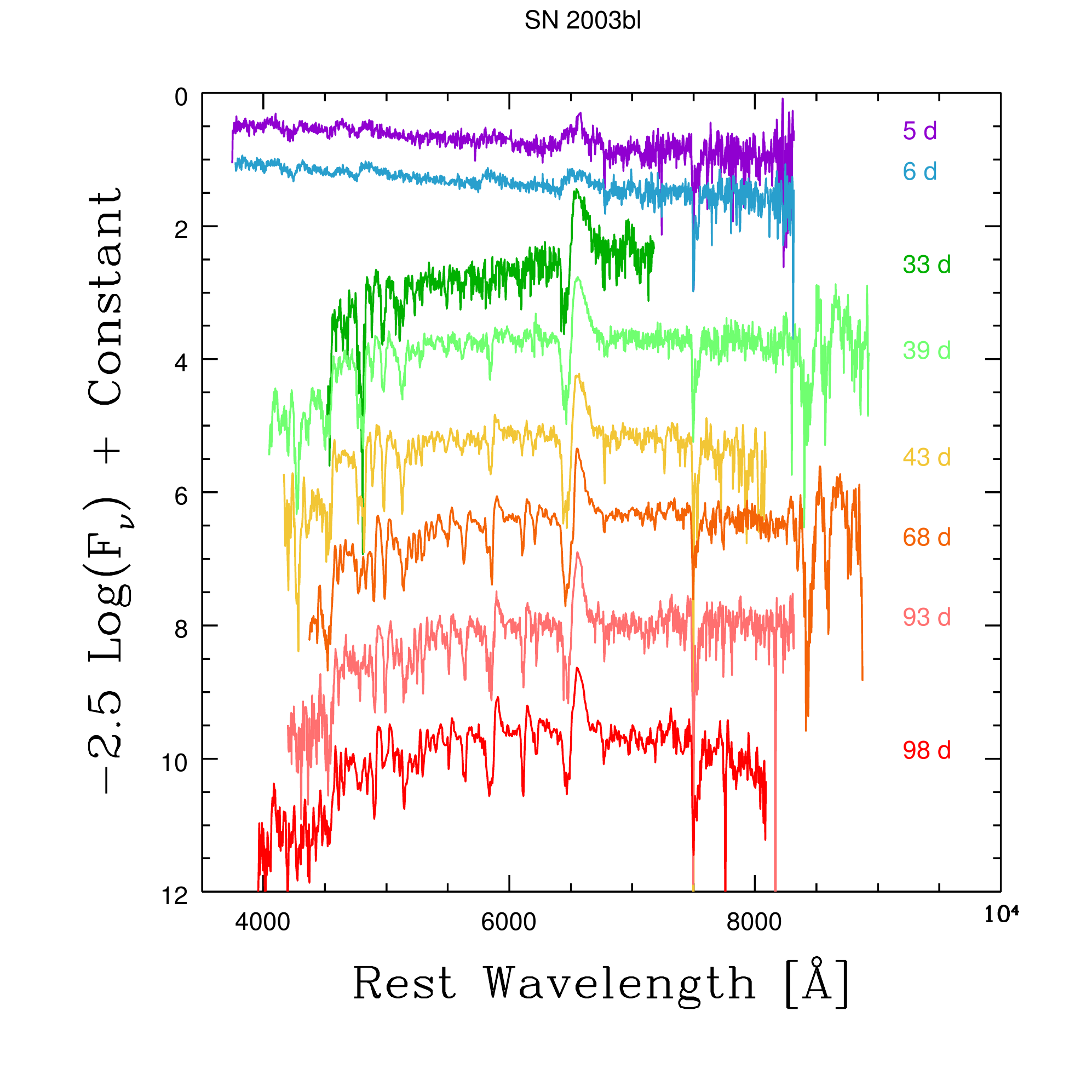}
\includegraphics[width=5.5cm]{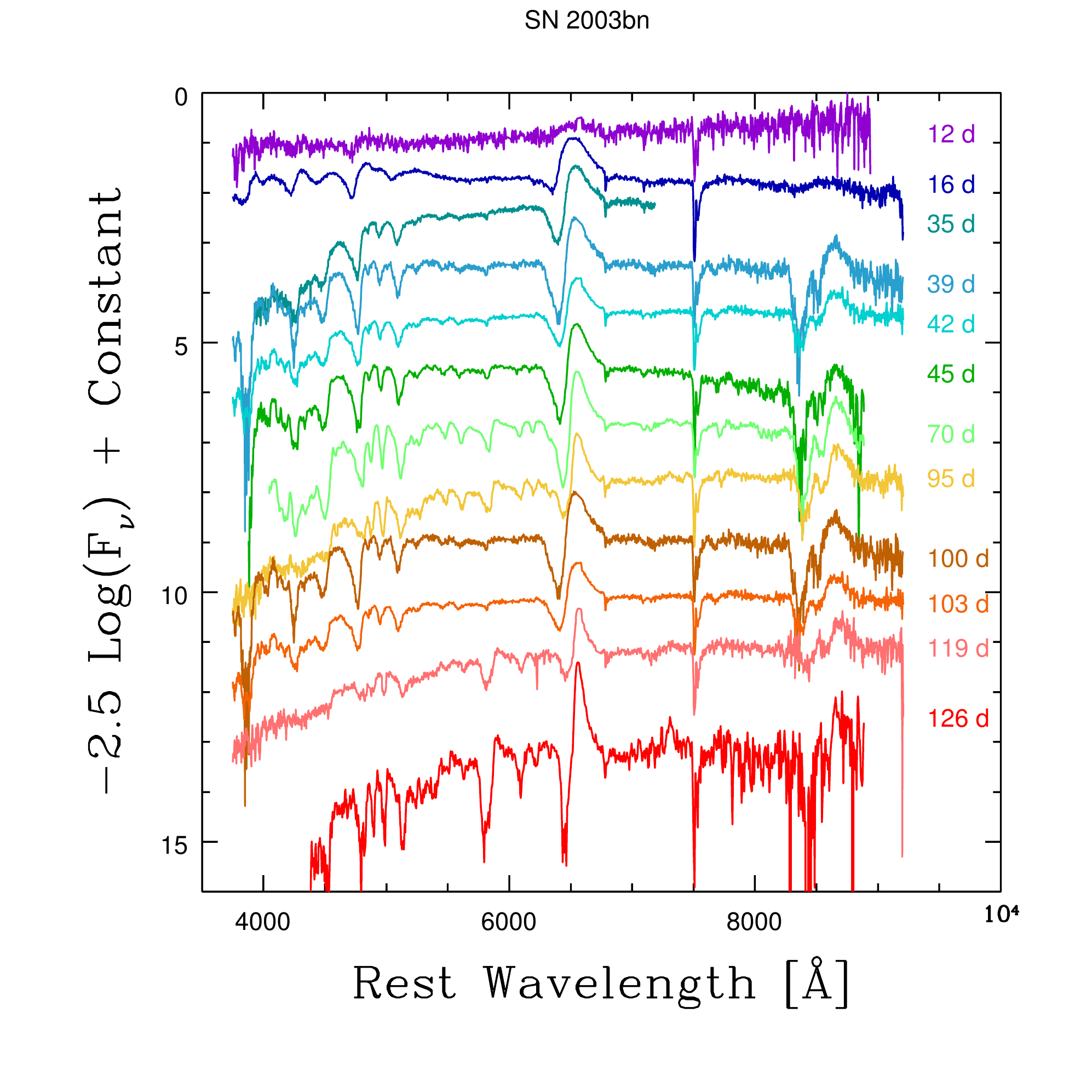}
\includegraphics[width=5.5cm]{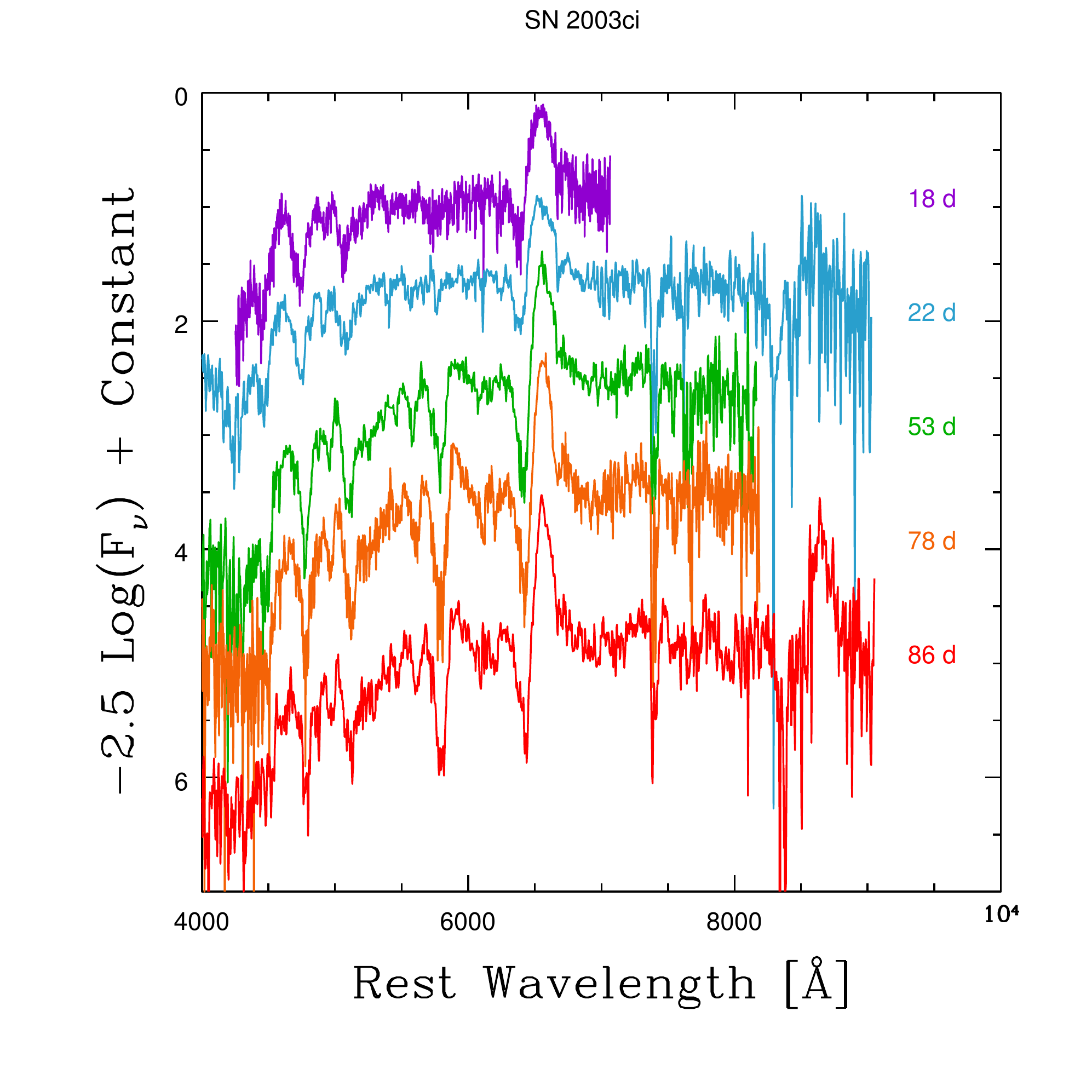}
\includegraphics[width=5.5cm]{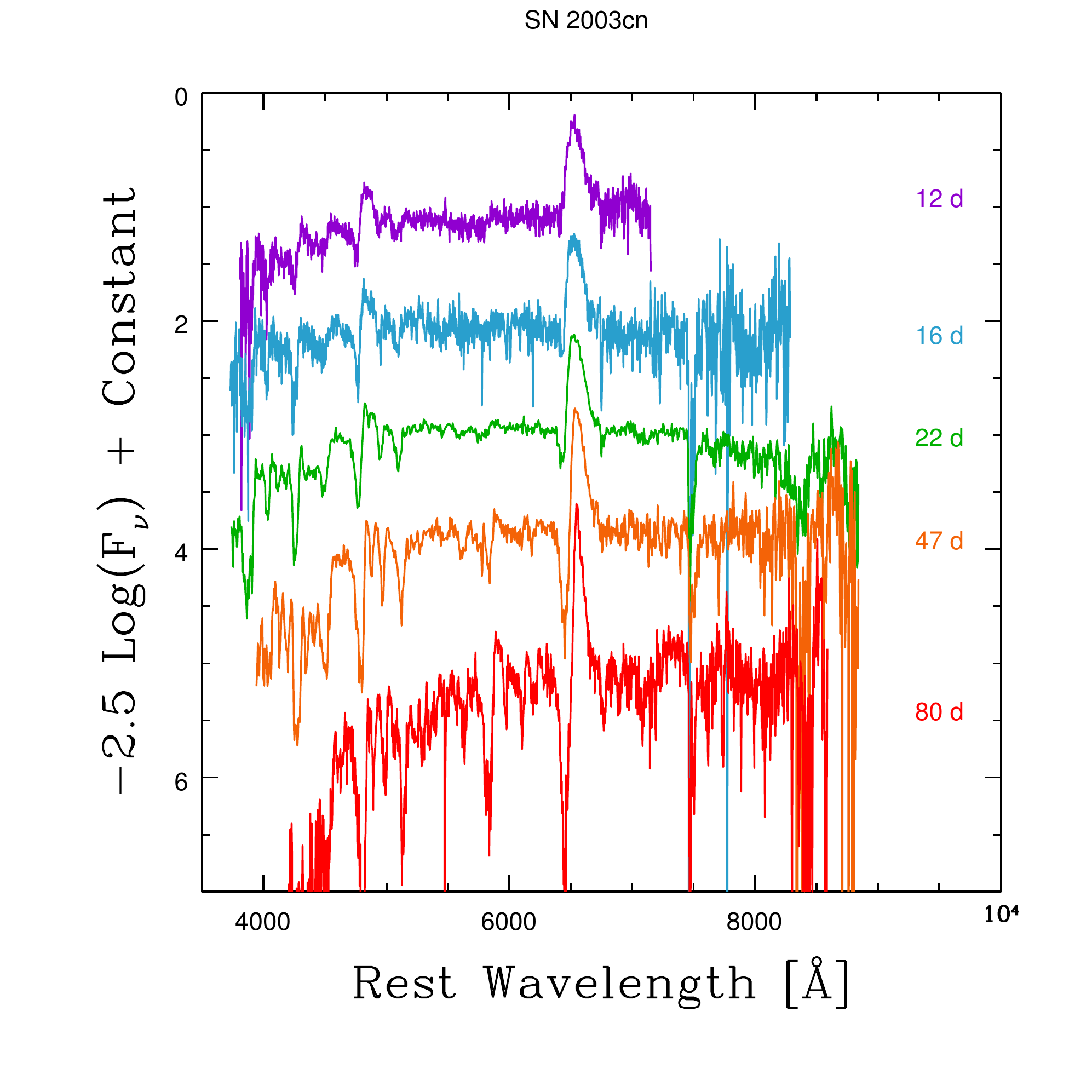}
\includegraphics[width=5.5cm]{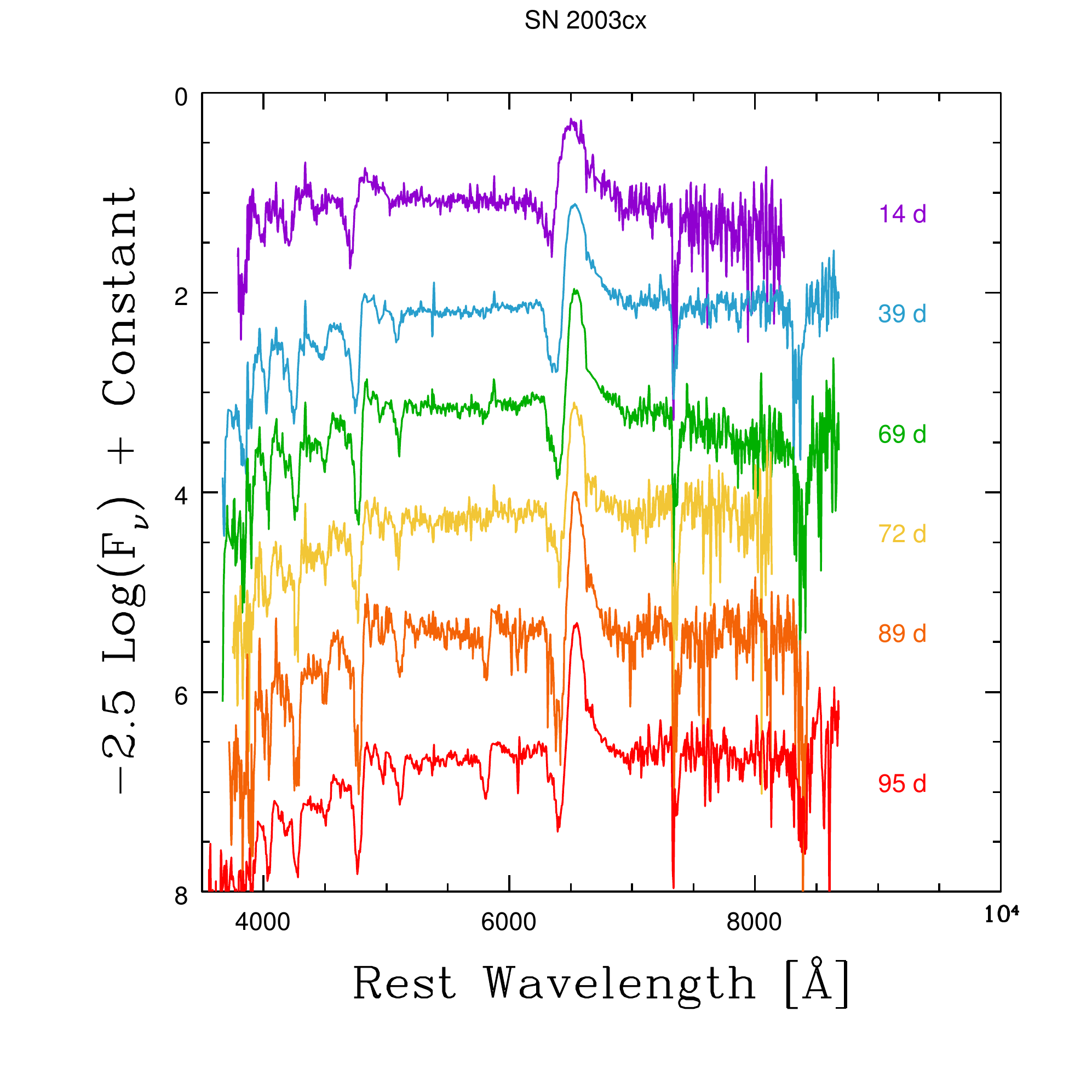}
\includegraphics[width=5.5cm]{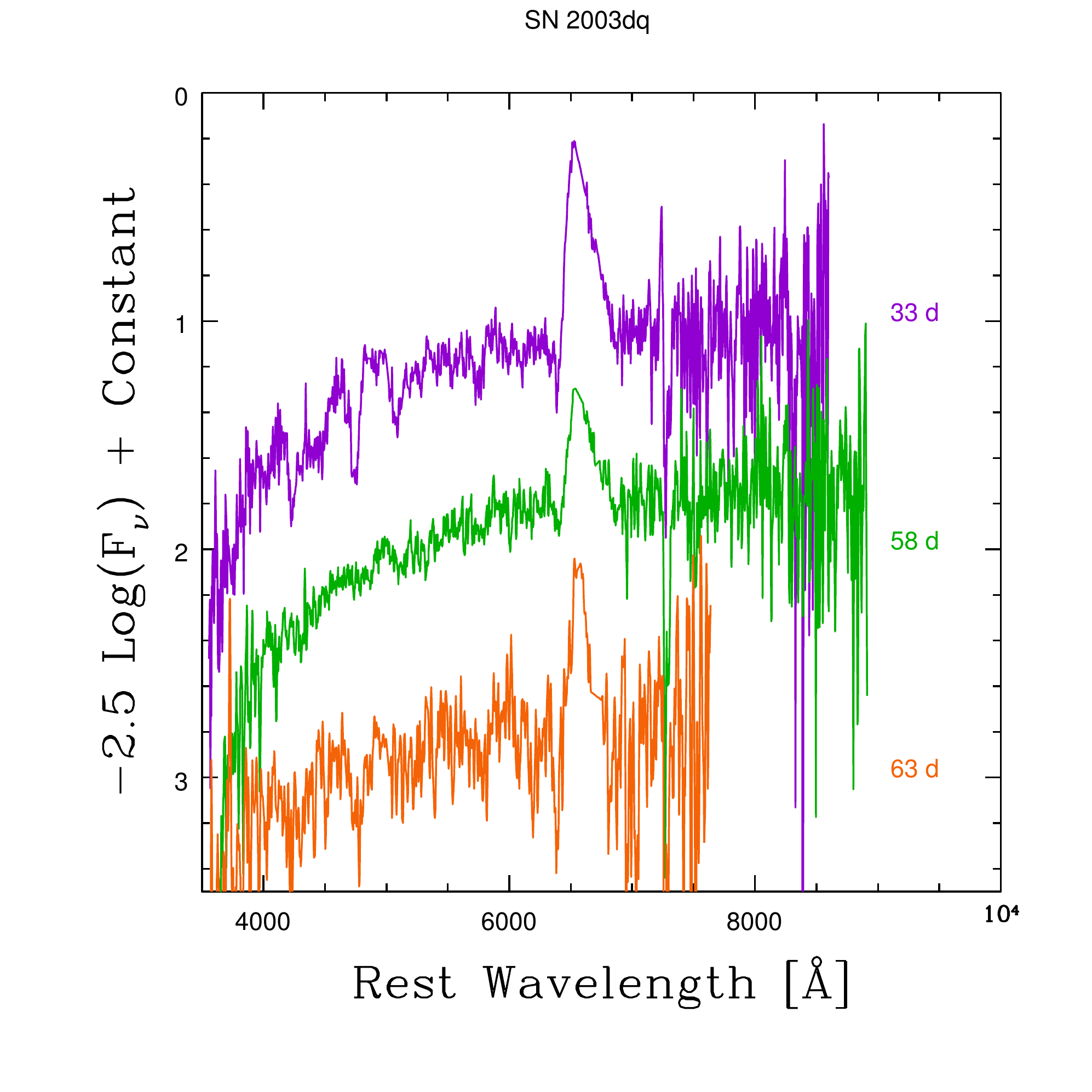}
\includegraphics[width=5.5cm]{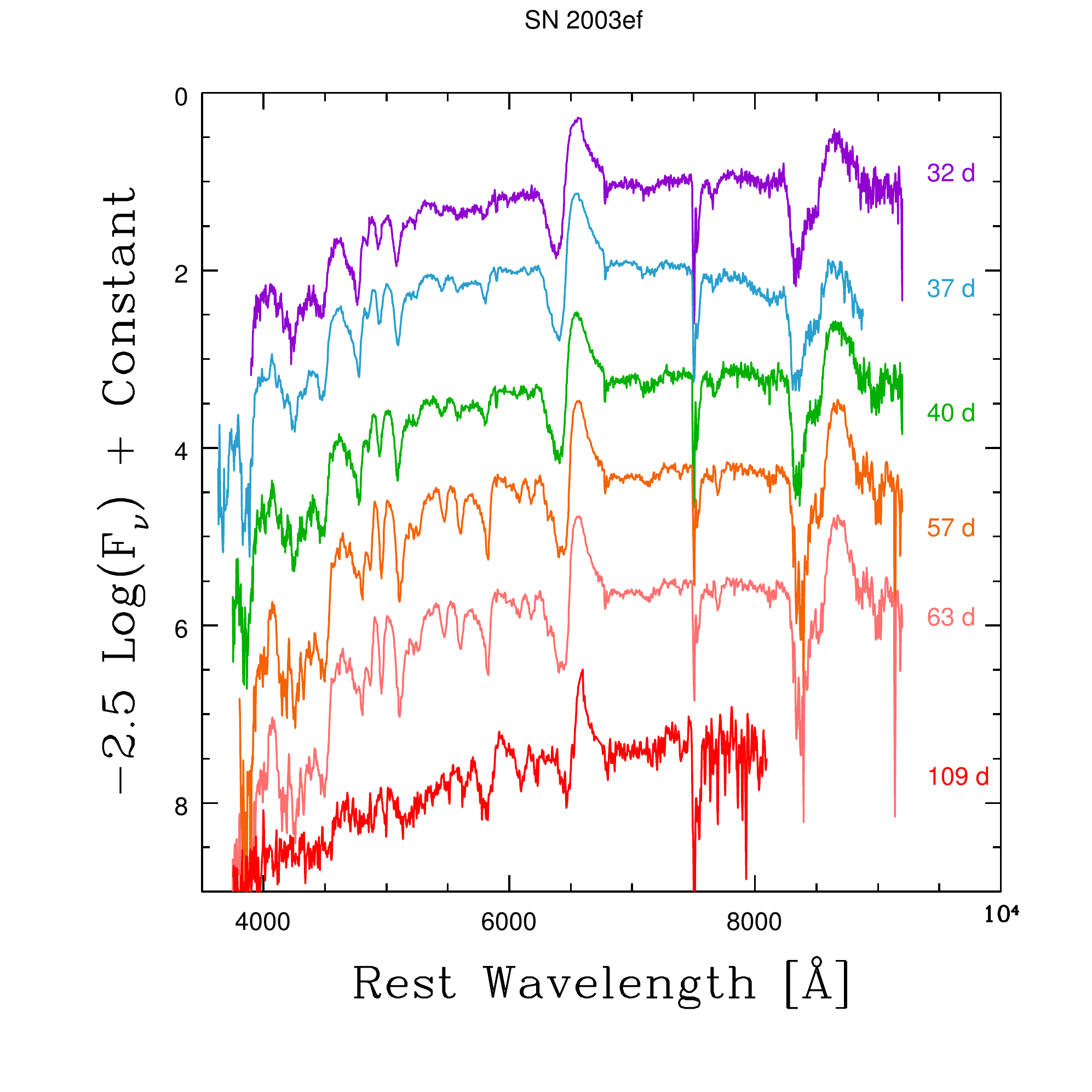}
\includegraphics[width=5.5cm]{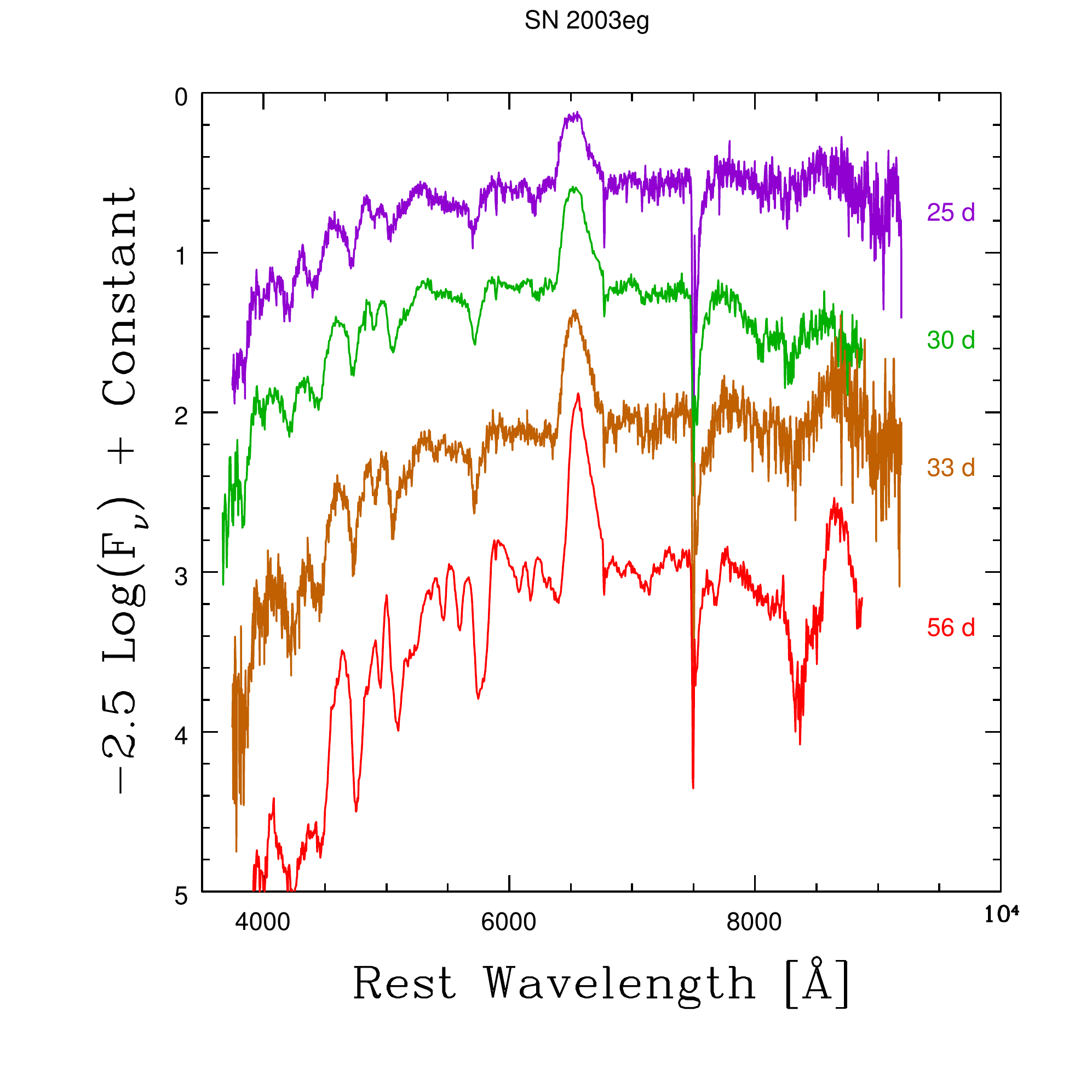}
\includegraphics[width=5.5cm]{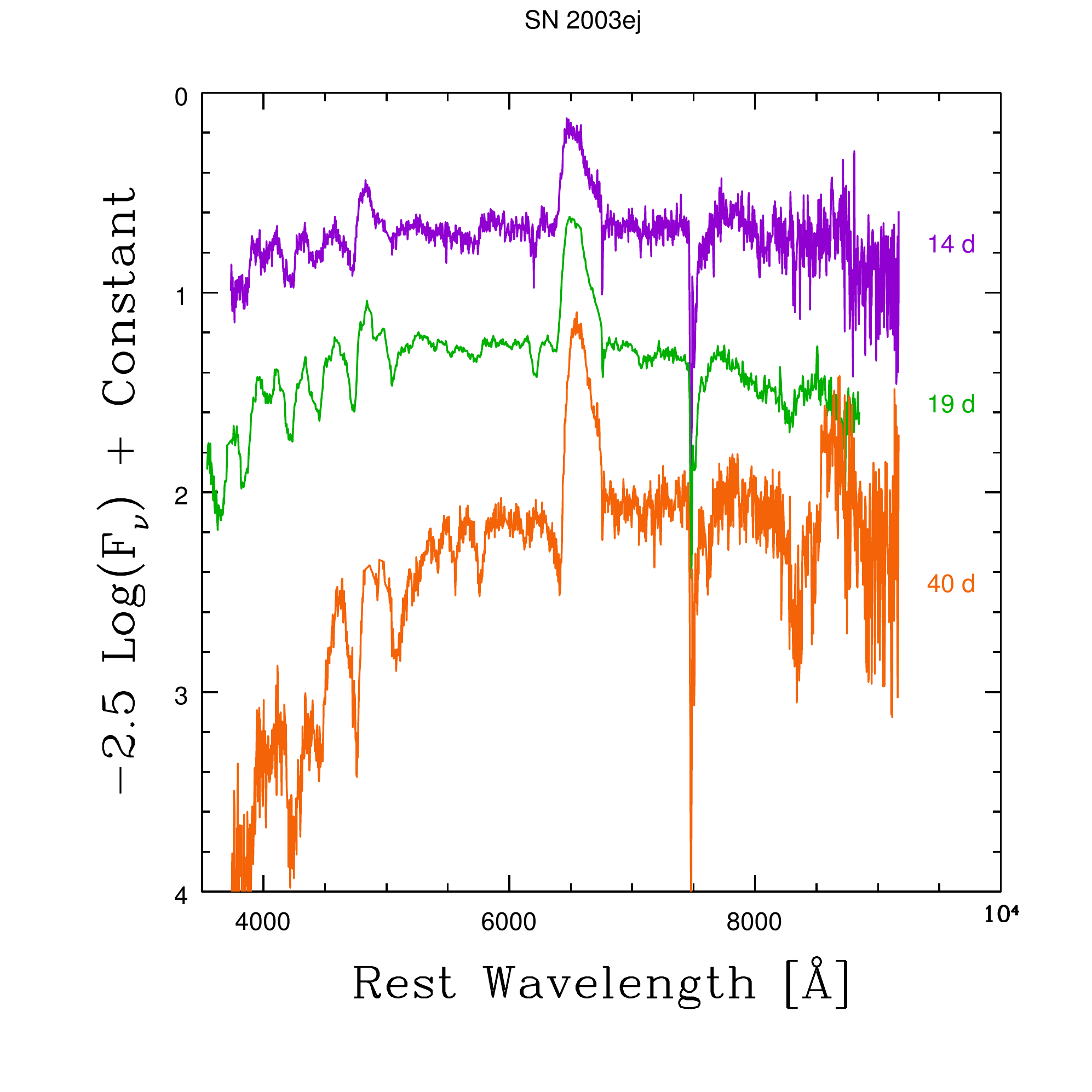}
\caption{Examples of SNe~II spectra: SN~2003B, SN~2003E, SN~2003T, SN~2003bl, SN~2003bn, SN~2003ci, SN~2003cn, SN~2003cx, SN~2003dq, SN~2003ef, SN~2003eg, SN~2003ej.}
\label{example}
\end{figure*}

\begin{figure*}[h!]
\centering
\includegraphics[width=5.5cm]{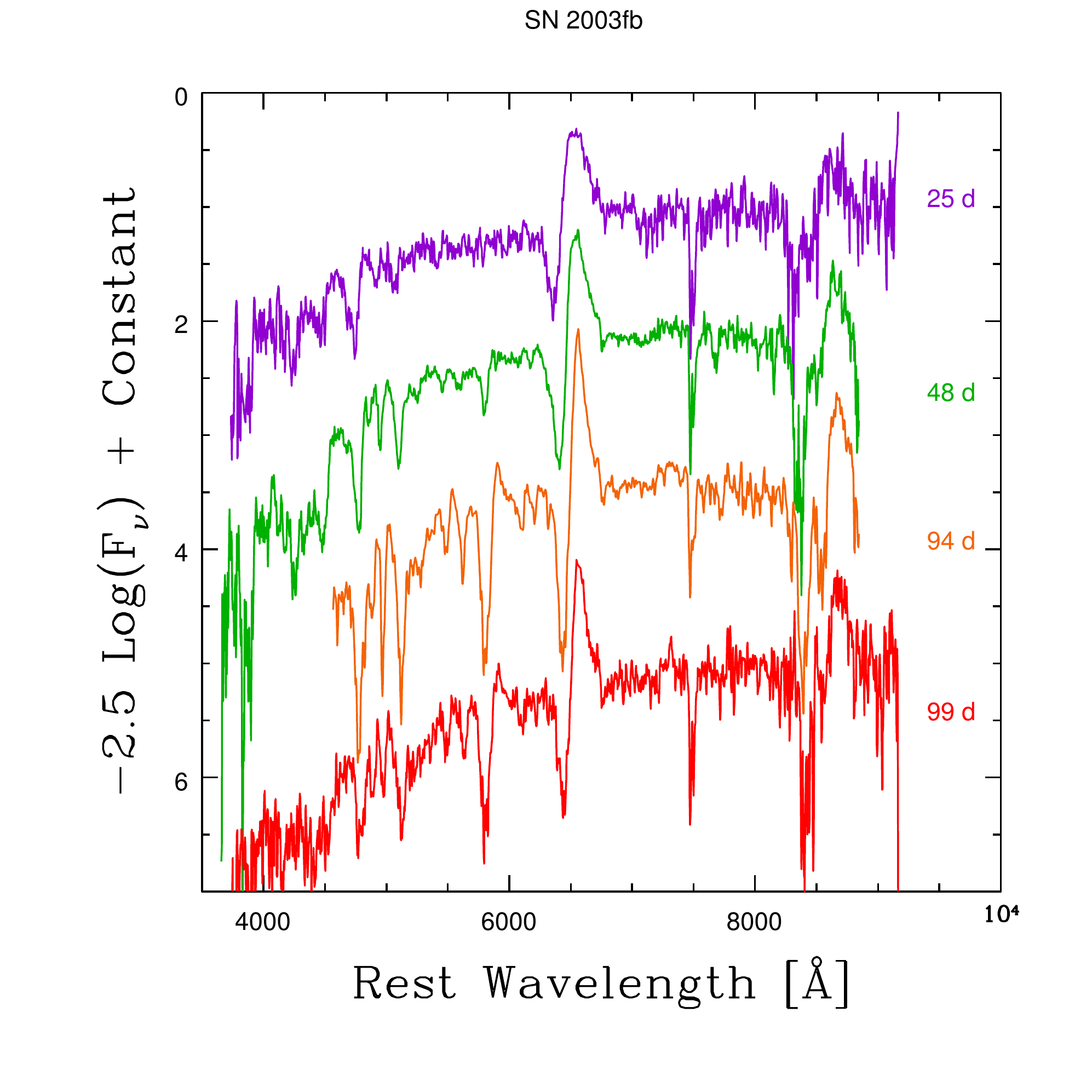}
\includegraphics[width=5.5cm]{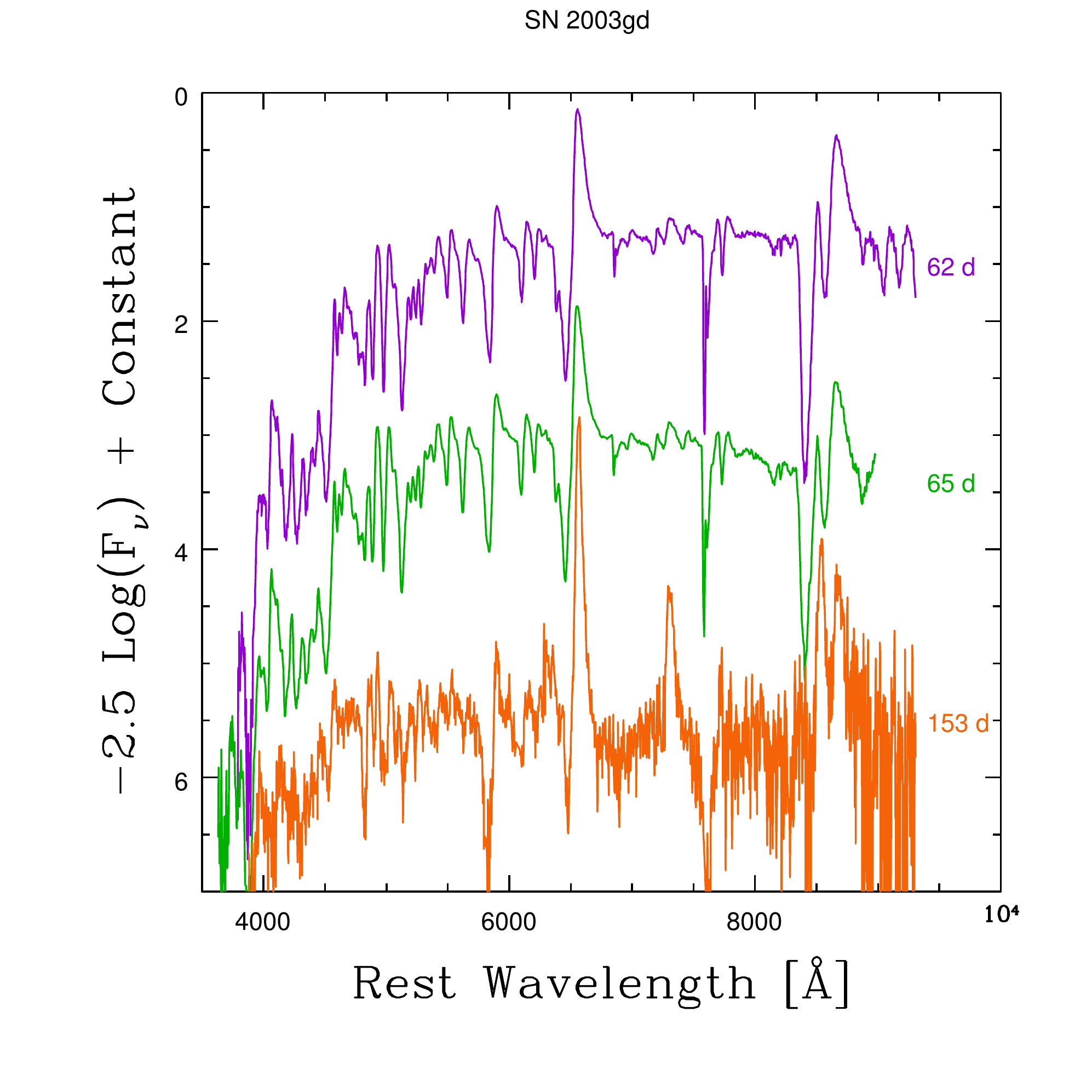}
\includegraphics[width=5.5cm]{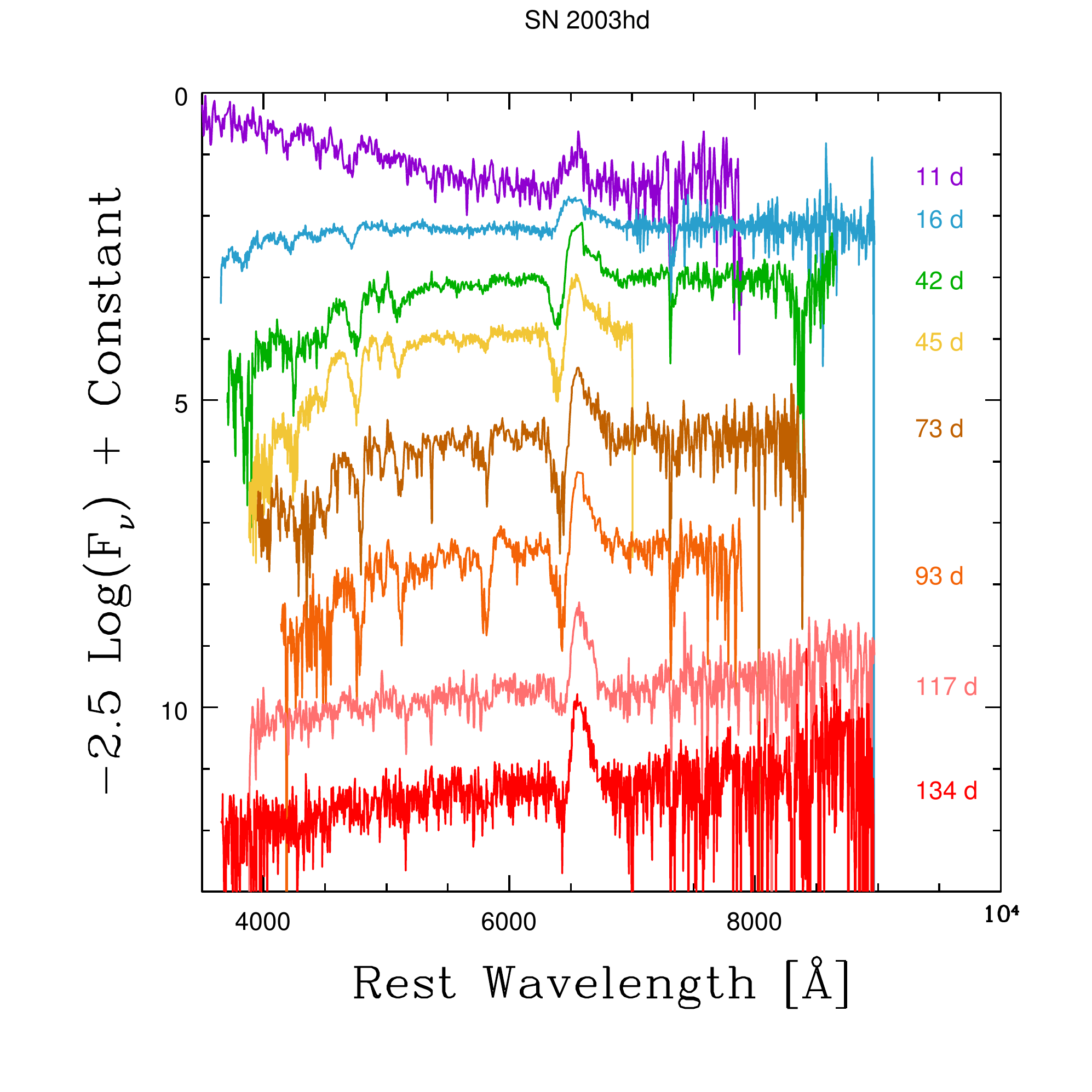}
\includegraphics[width=5.5cm]{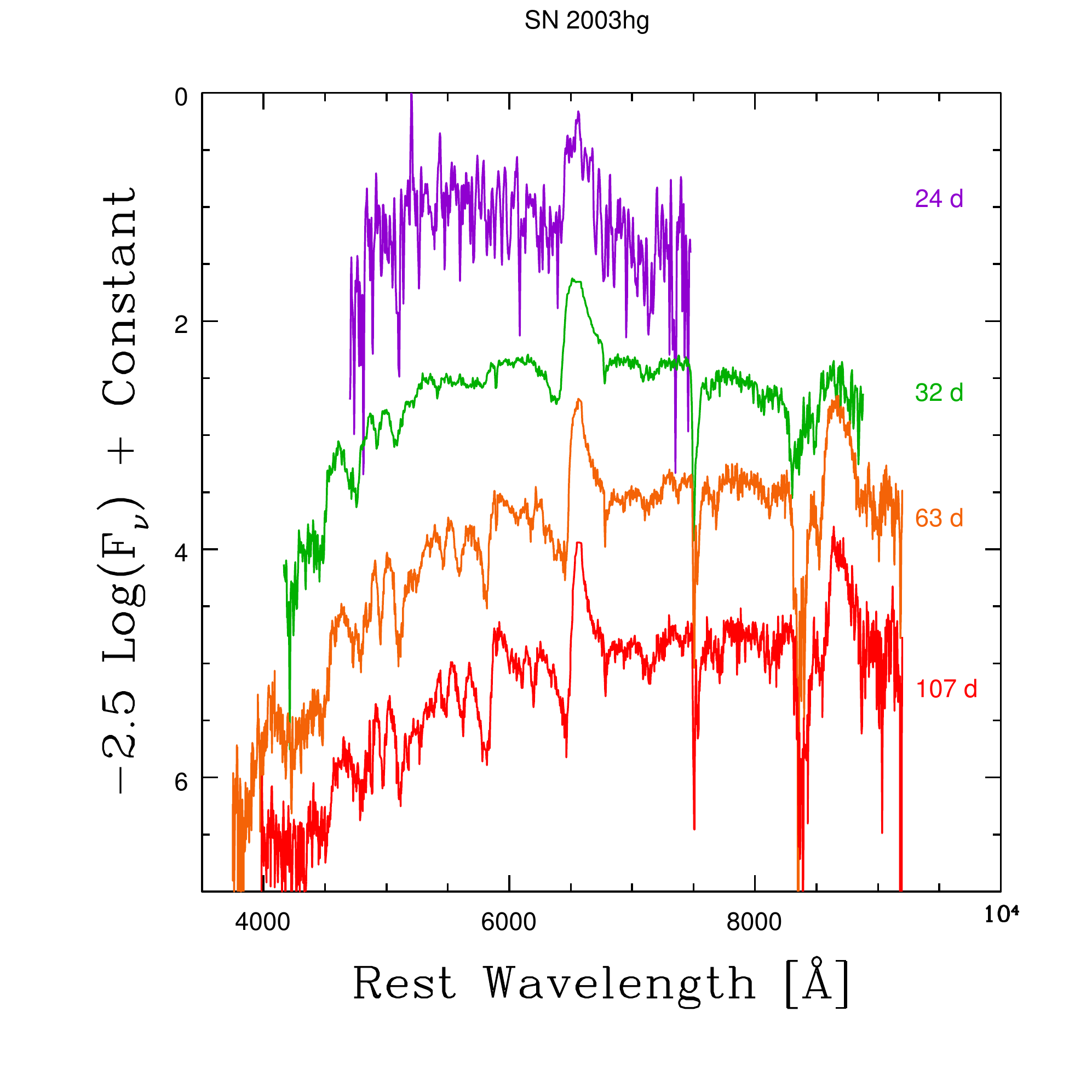}
\includegraphics[width=5.5cm]{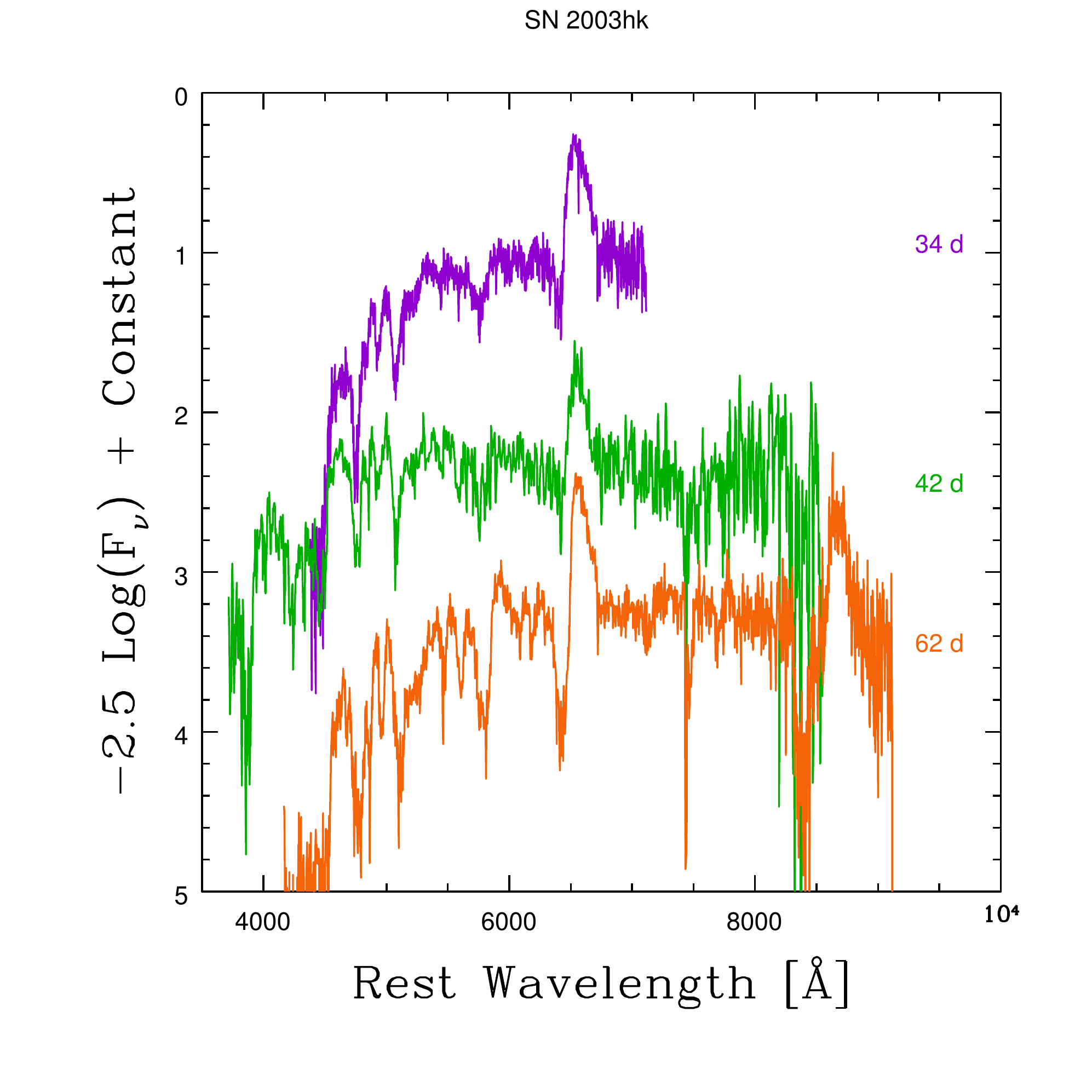}
\includegraphics[width=5.5cm]{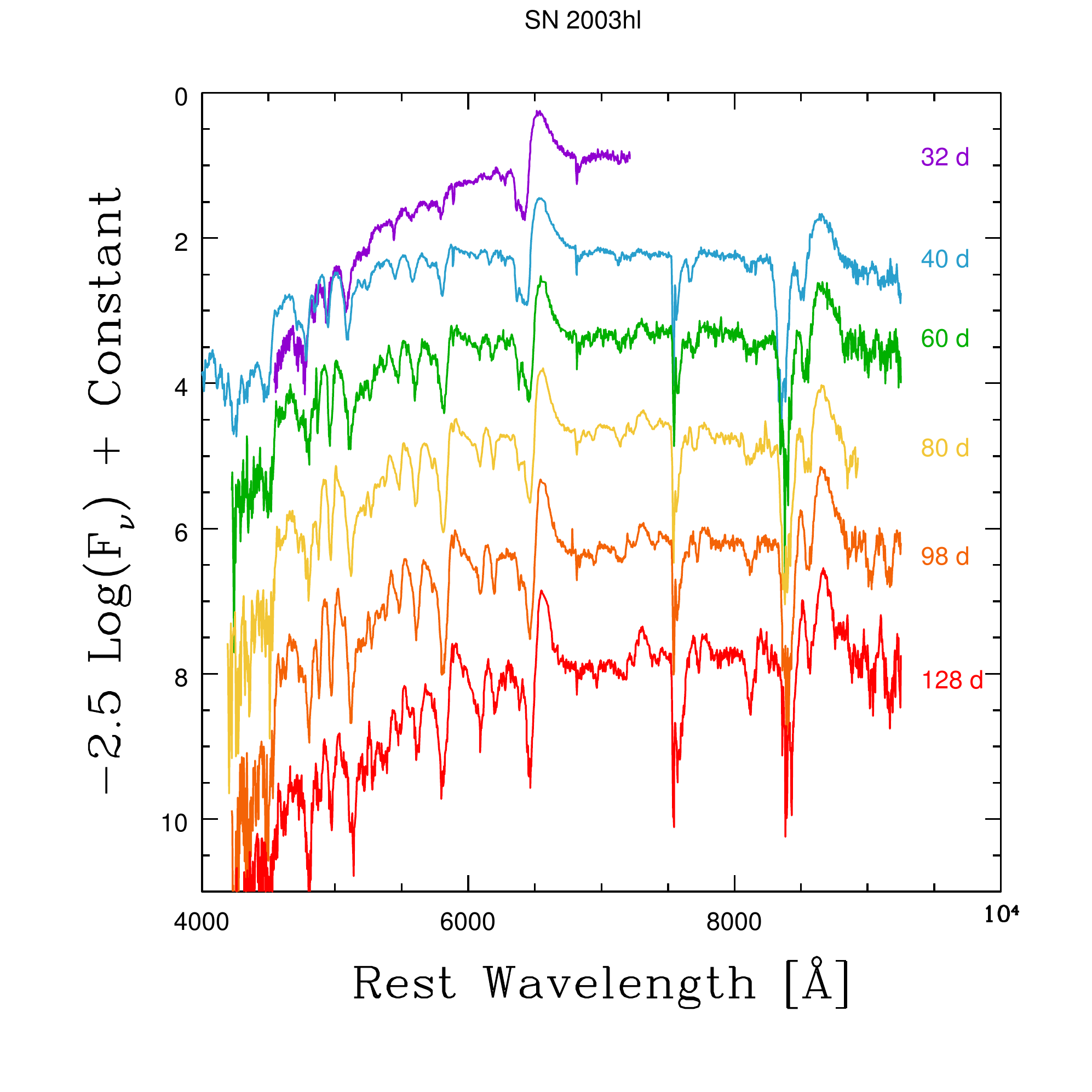}
\includegraphics[width=5.5cm]{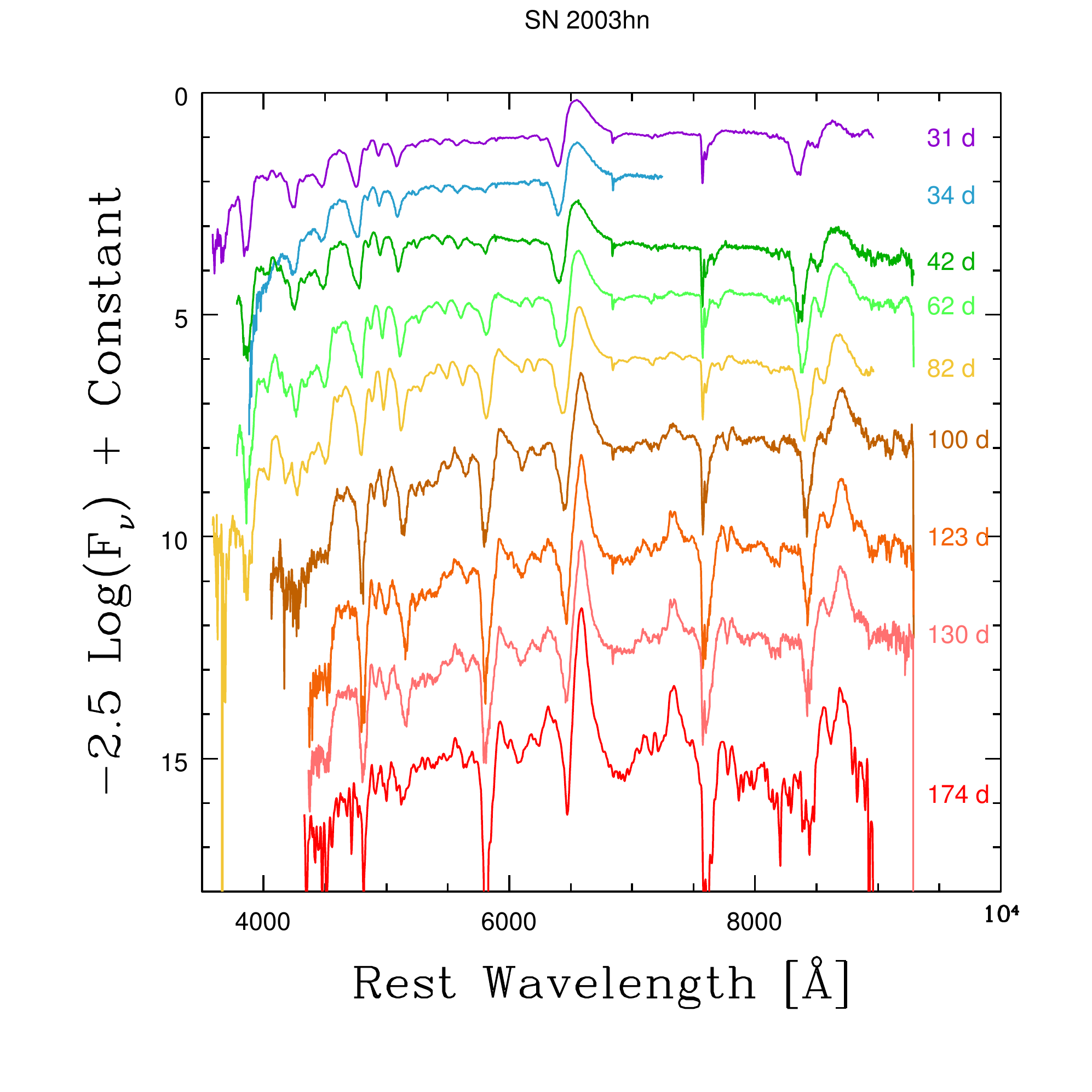}
\includegraphics[width=5.5cm]{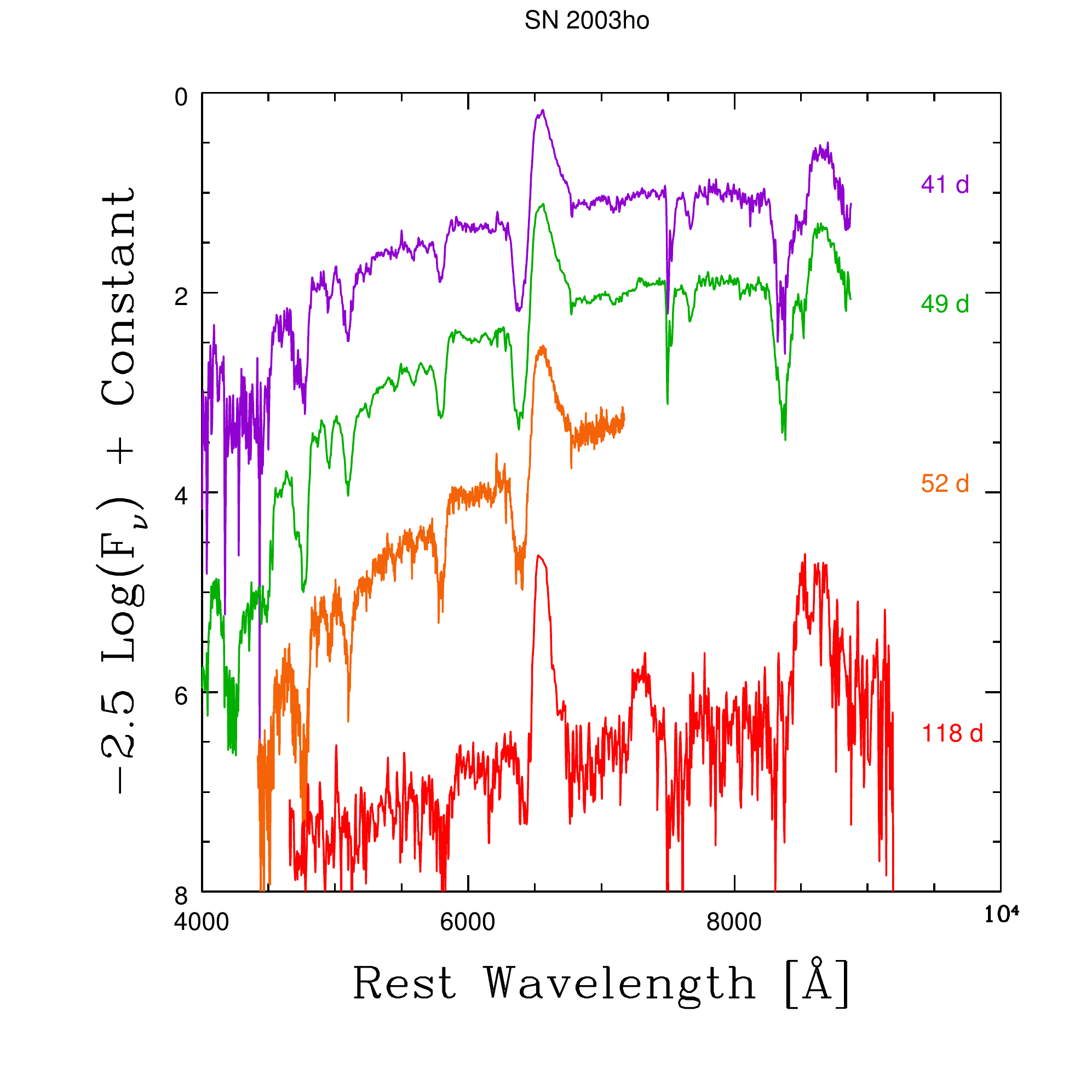}
\includegraphics[width=5.5cm]{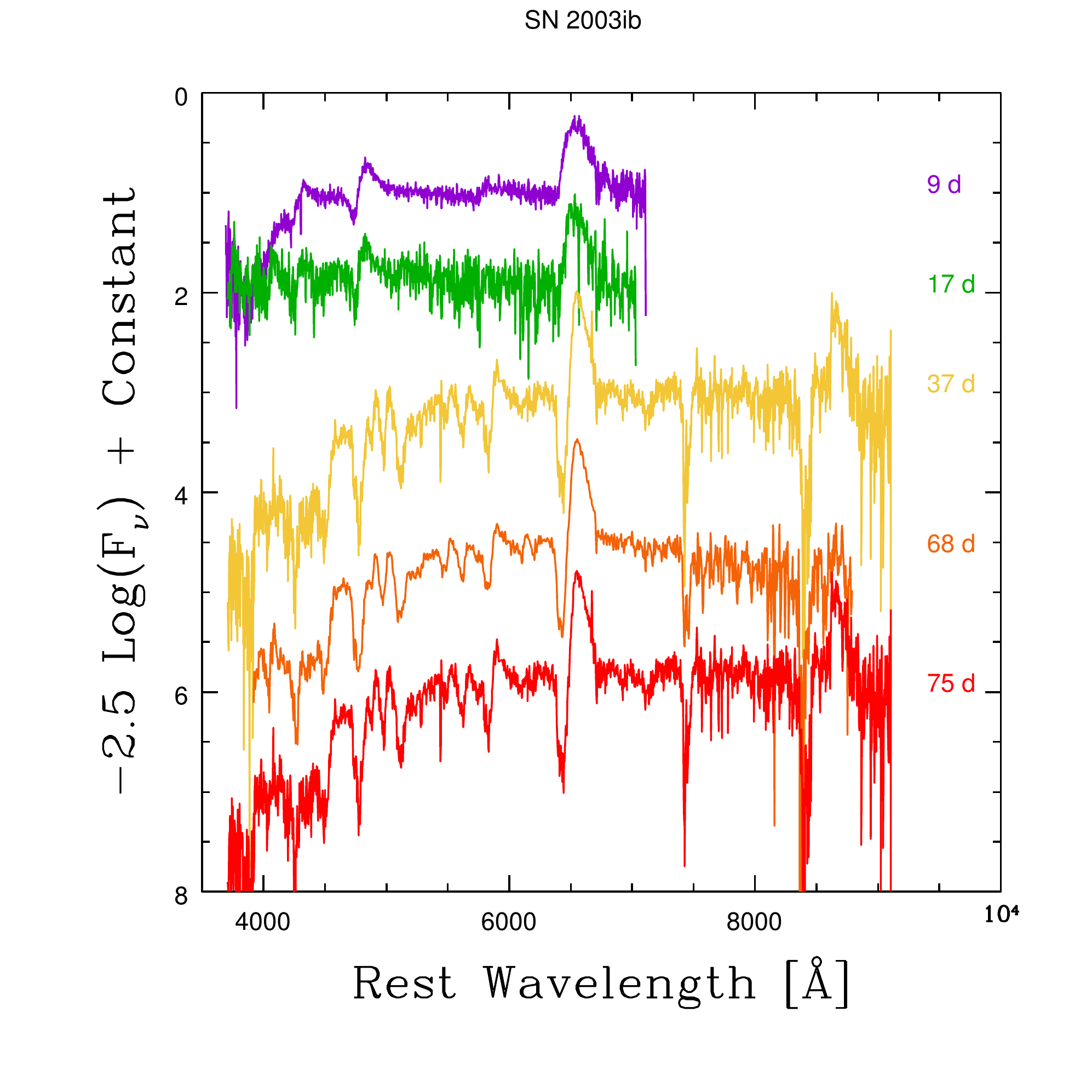}
\includegraphics[width=5.5cm]{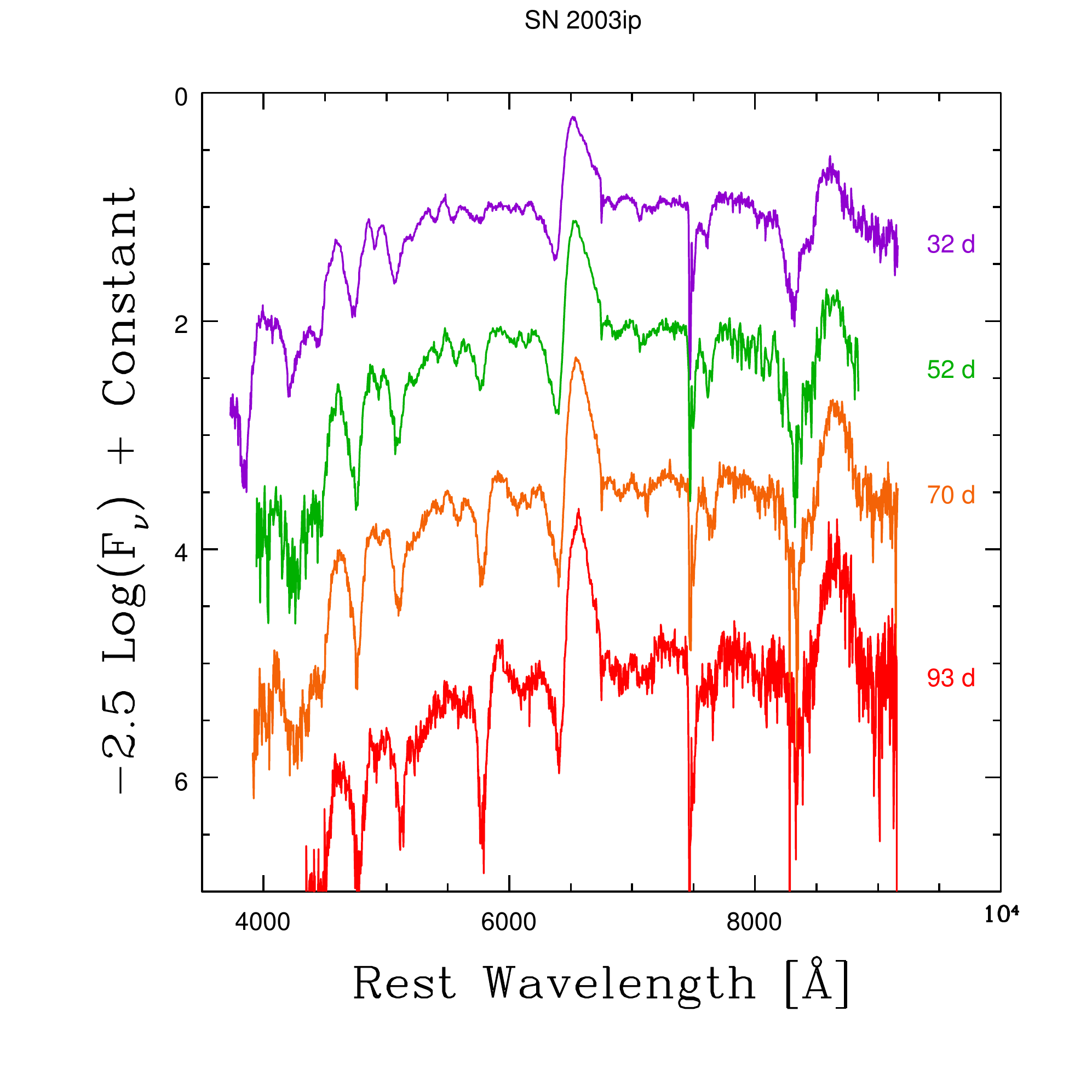}
\includegraphics[width=5.5cm]{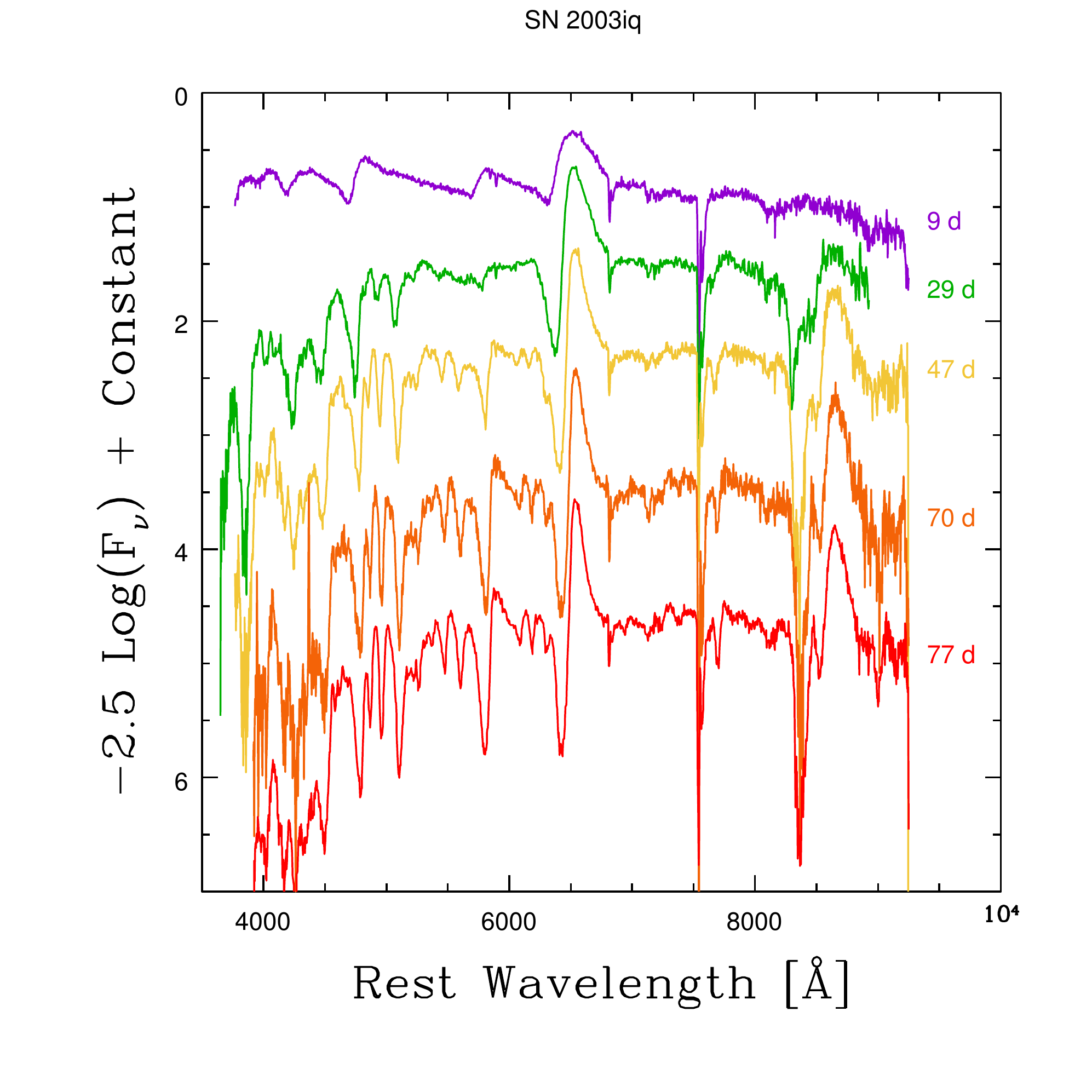}
\includegraphics[width=5.5cm]{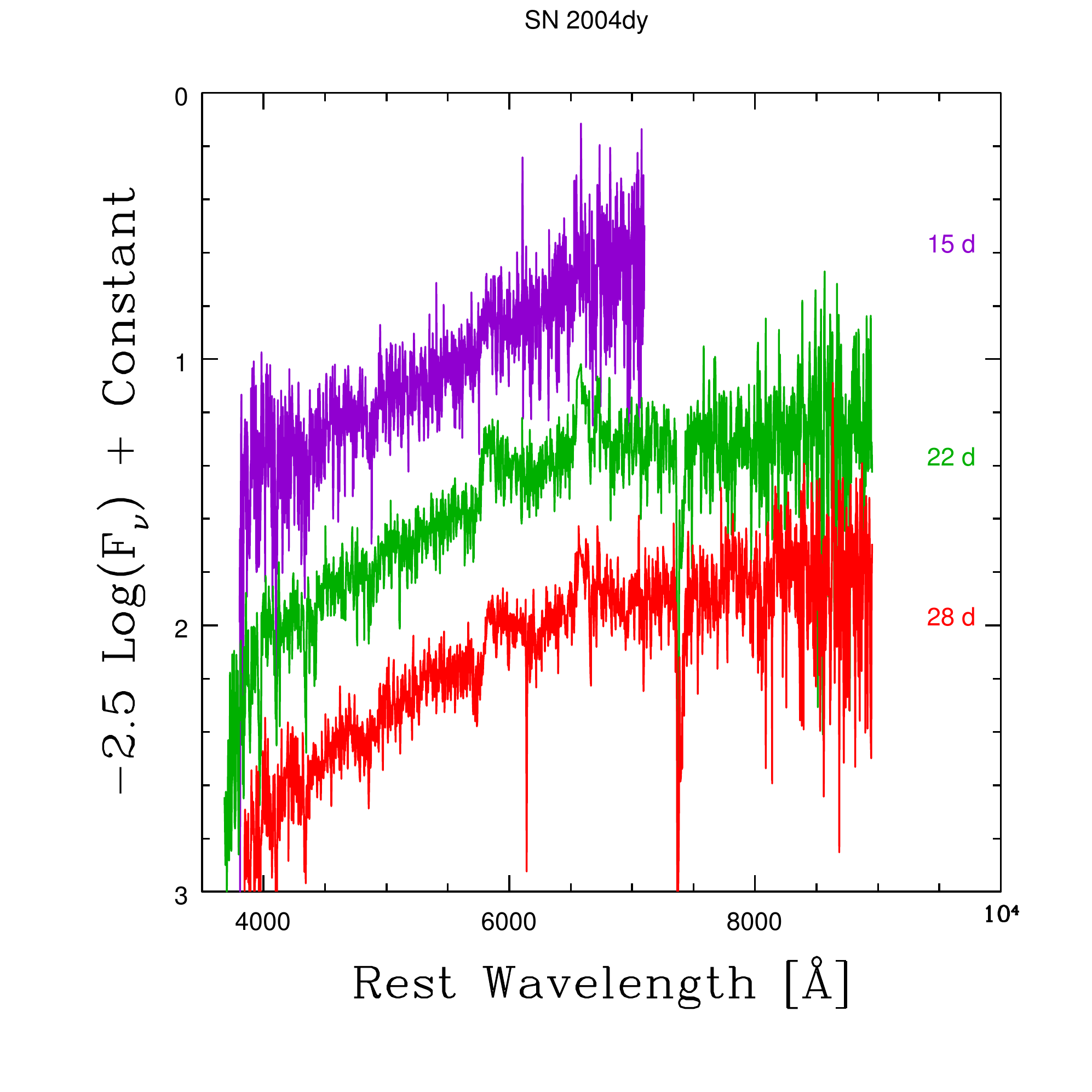}
\caption{Examples of SNe~II spectra: SN~2003fb, SN~2003gd, SN~2003hd, SN~2003hg, SN~2003hk, SN~2003hl, SN~2003hn, SN~2003ho, SN~2003ib, SN~2003ip, SN~2003iq, SN~2004dy.}
\label{example}
\end{figure*}

\begin{figure*}[h!]
\centering
\includegraphics[width=5.5cm]{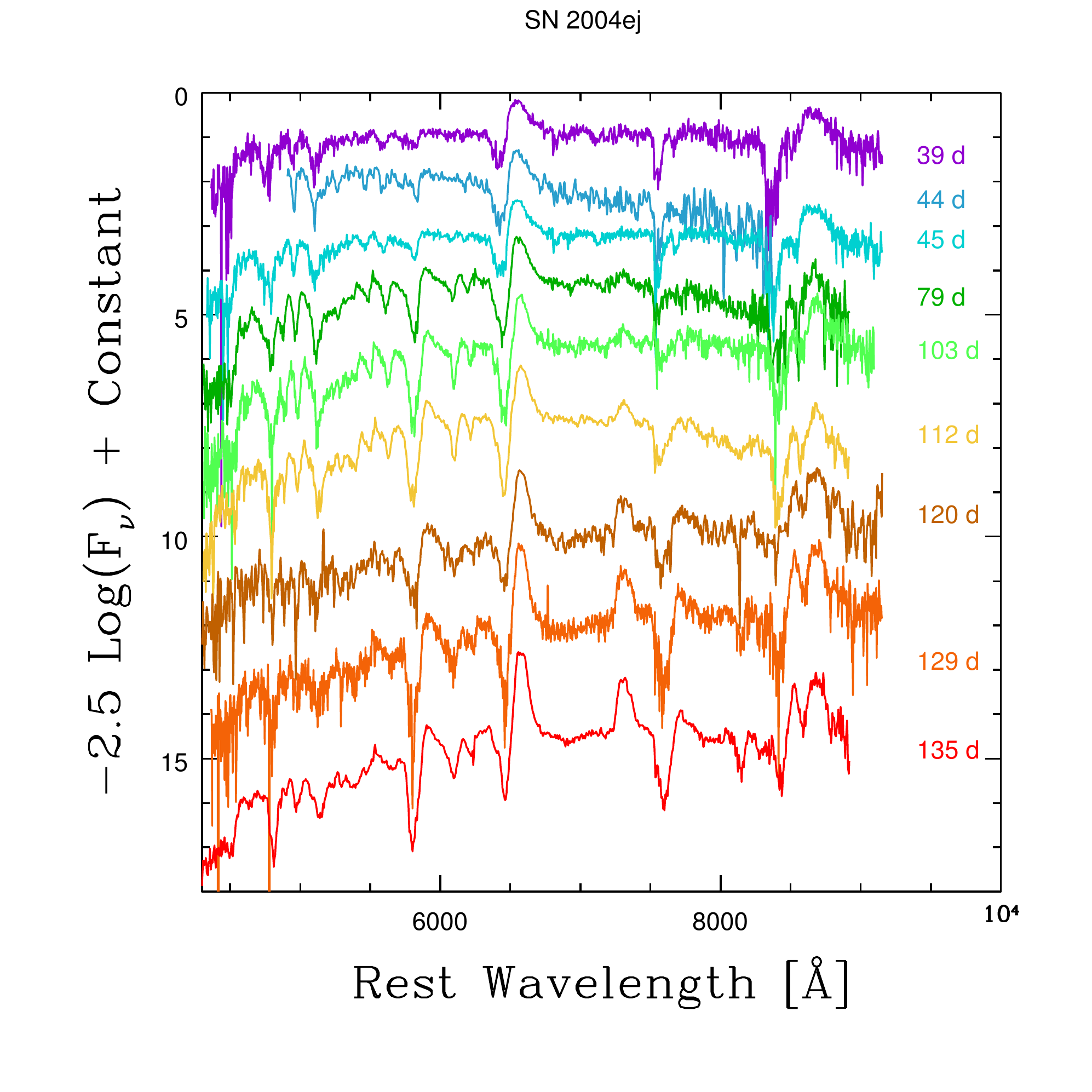}
\includegraphics[width=5.5cm]{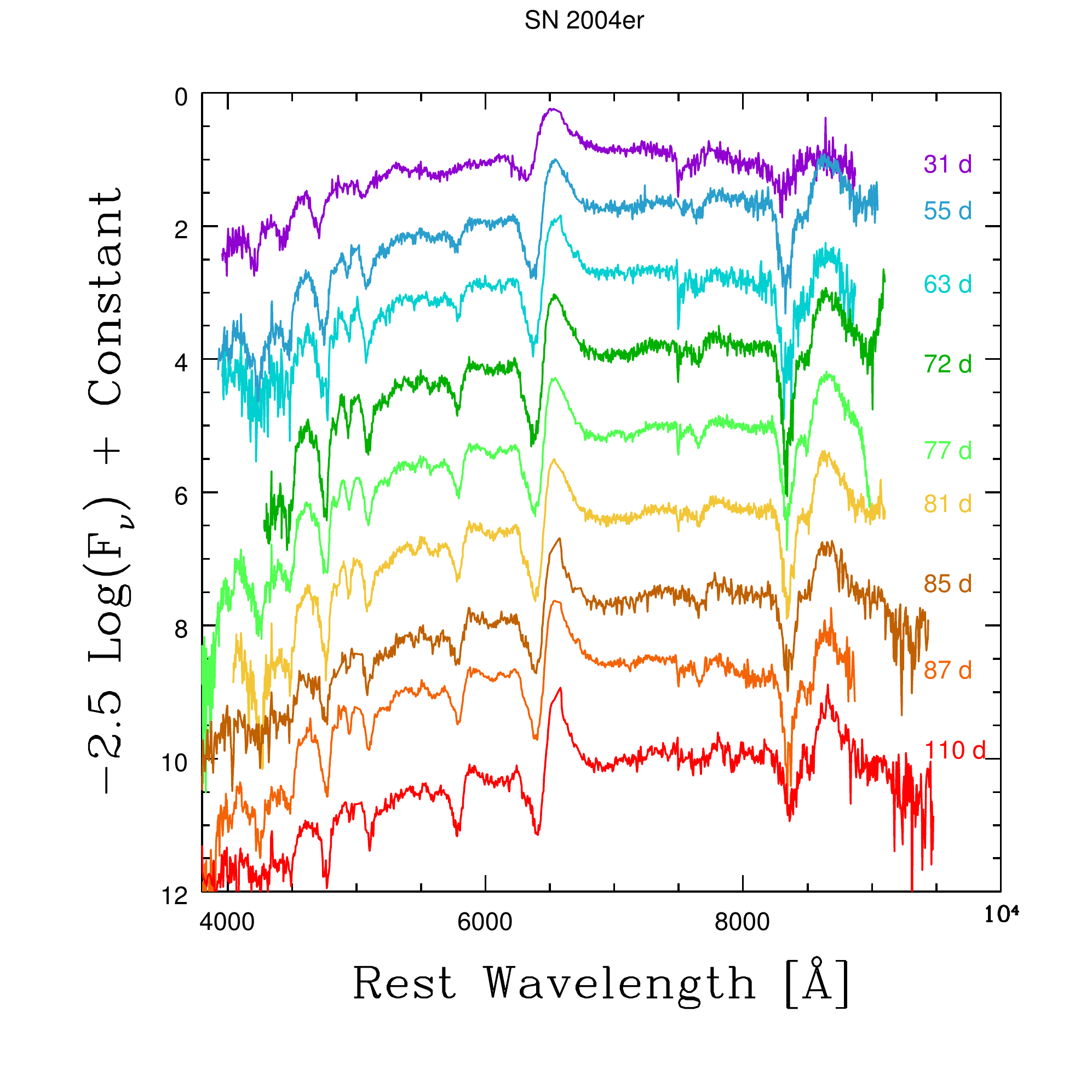}
\includegraphics[width=5.5cm]{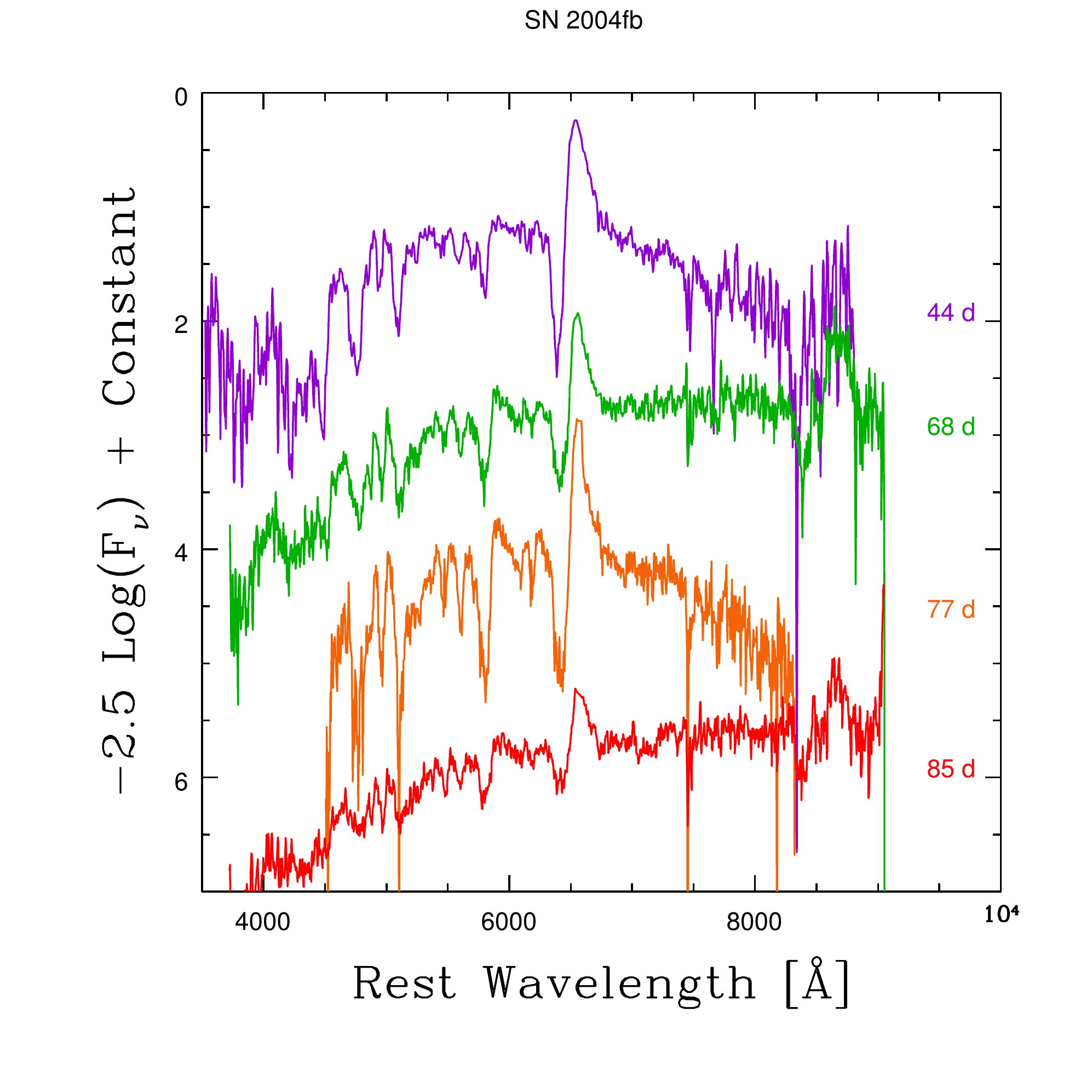}
\includegraphics[width=5.5cm]{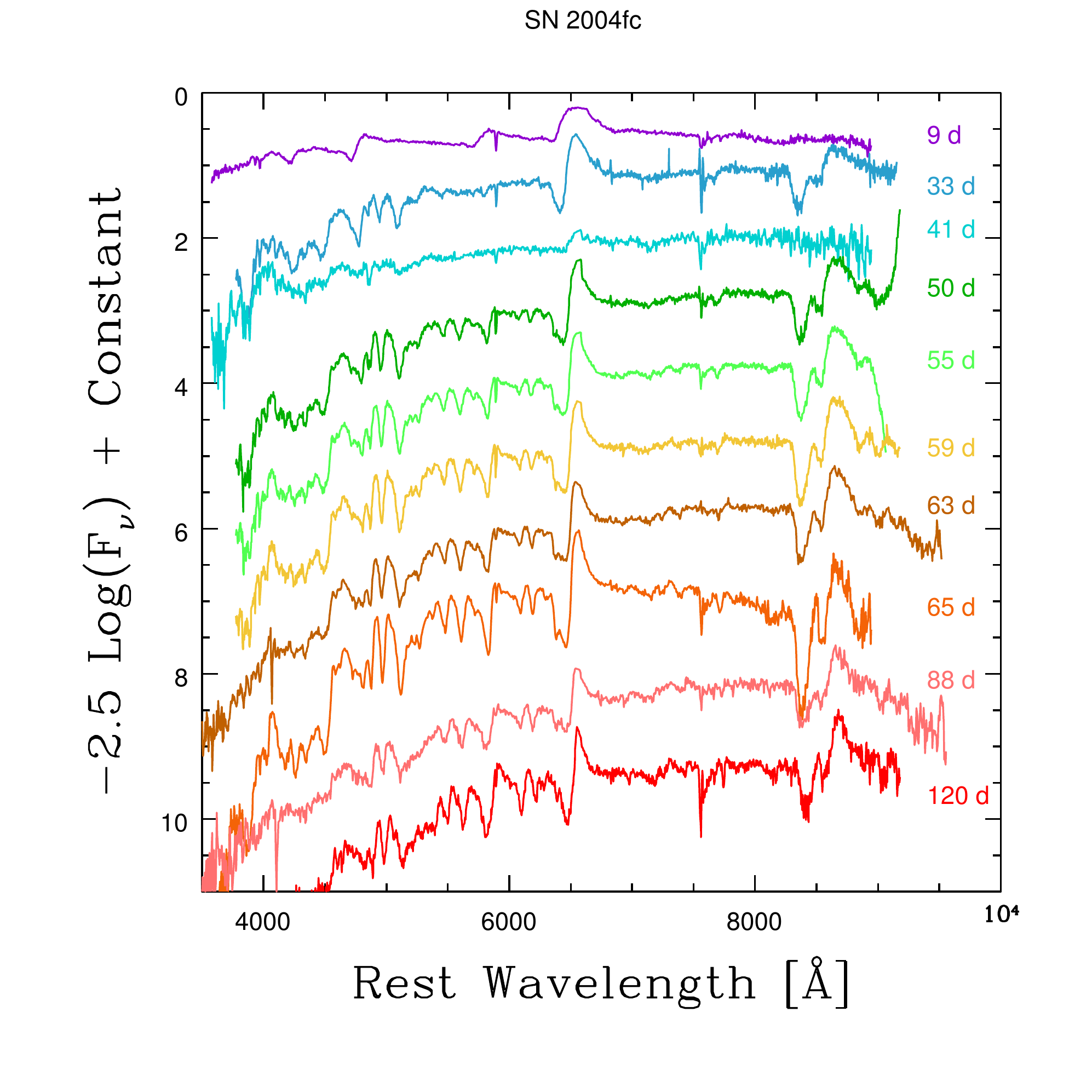}
\includegraphics[width=5.5cm]{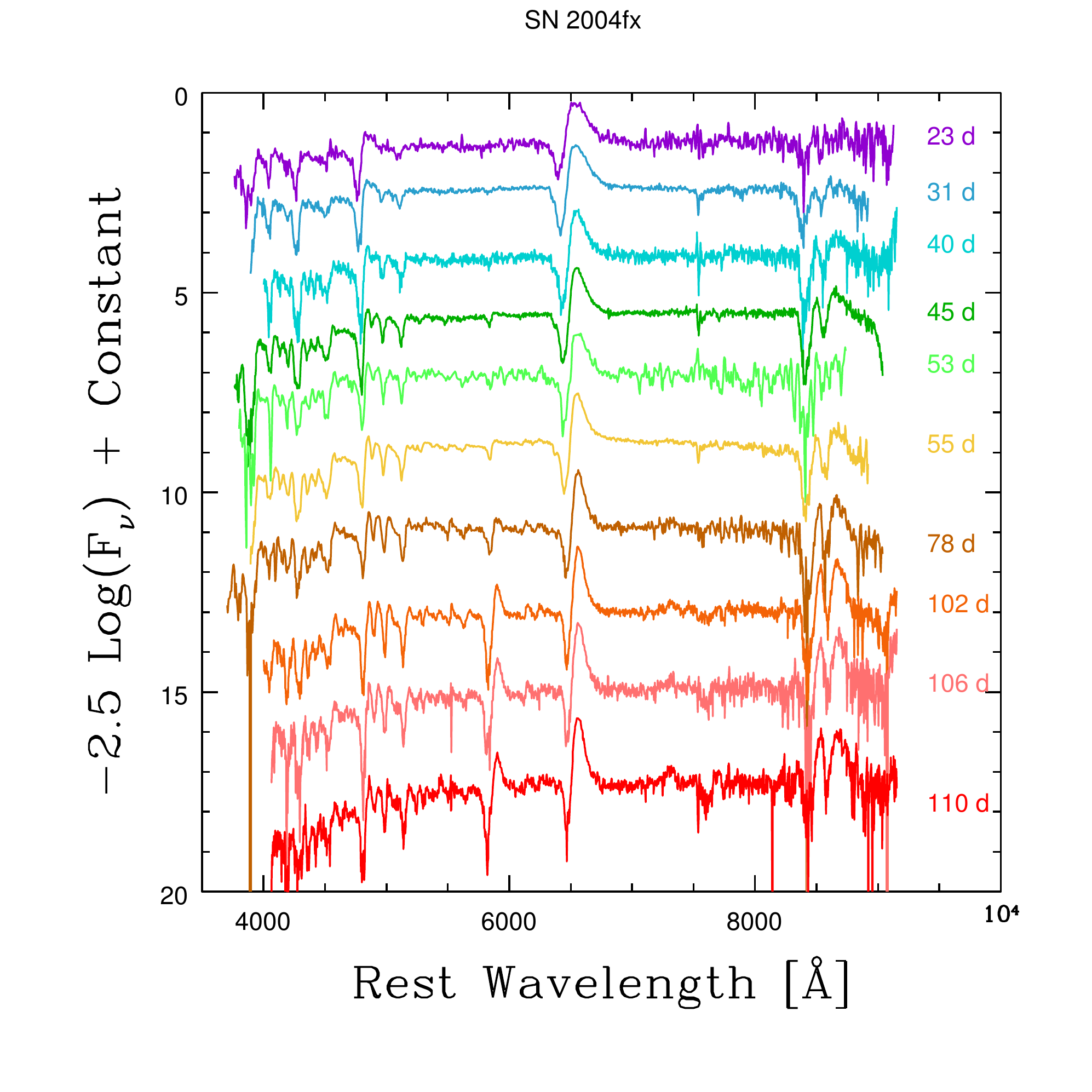}
\includegraphics[width=5.5cm]{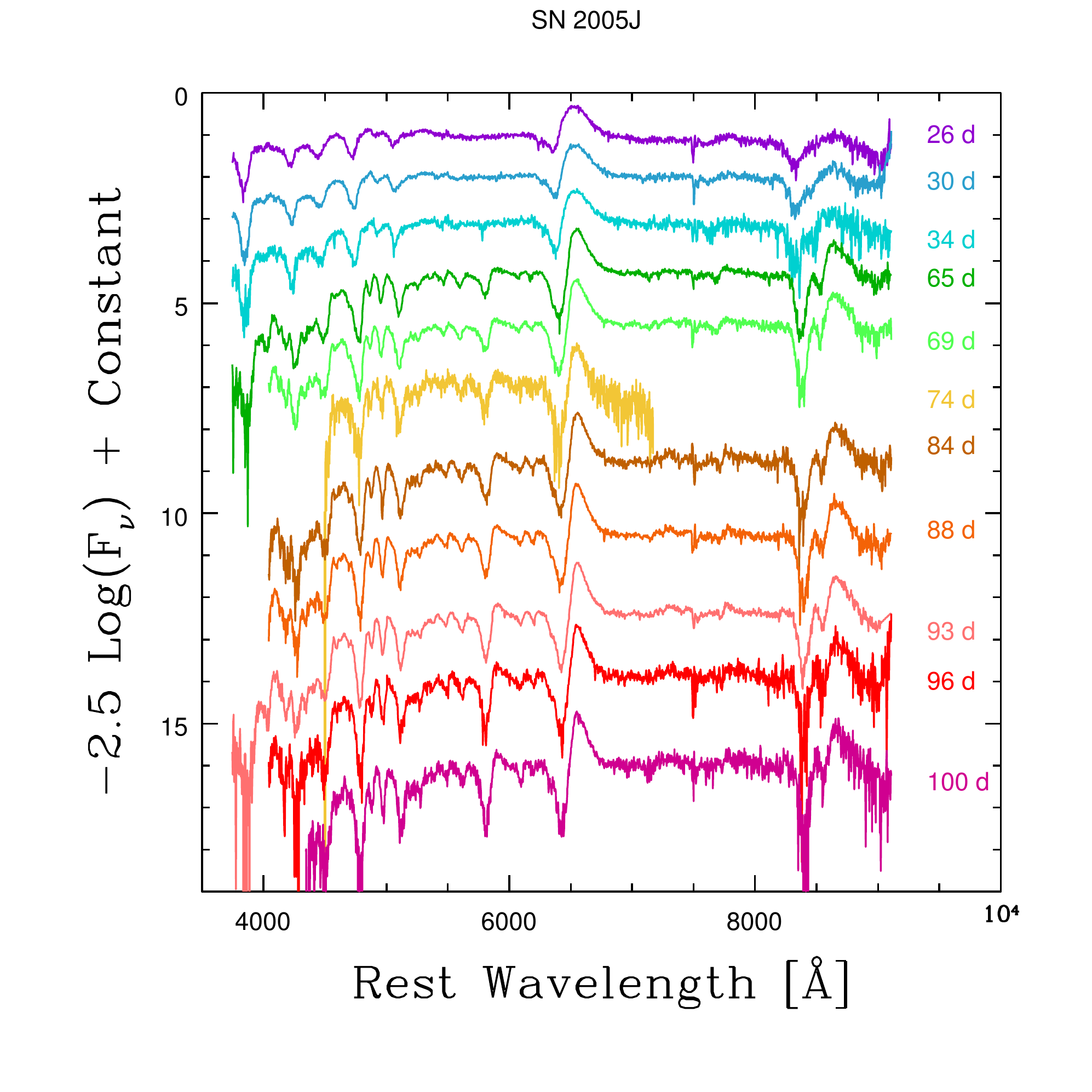}
\includegraphics[width=5.5cm]{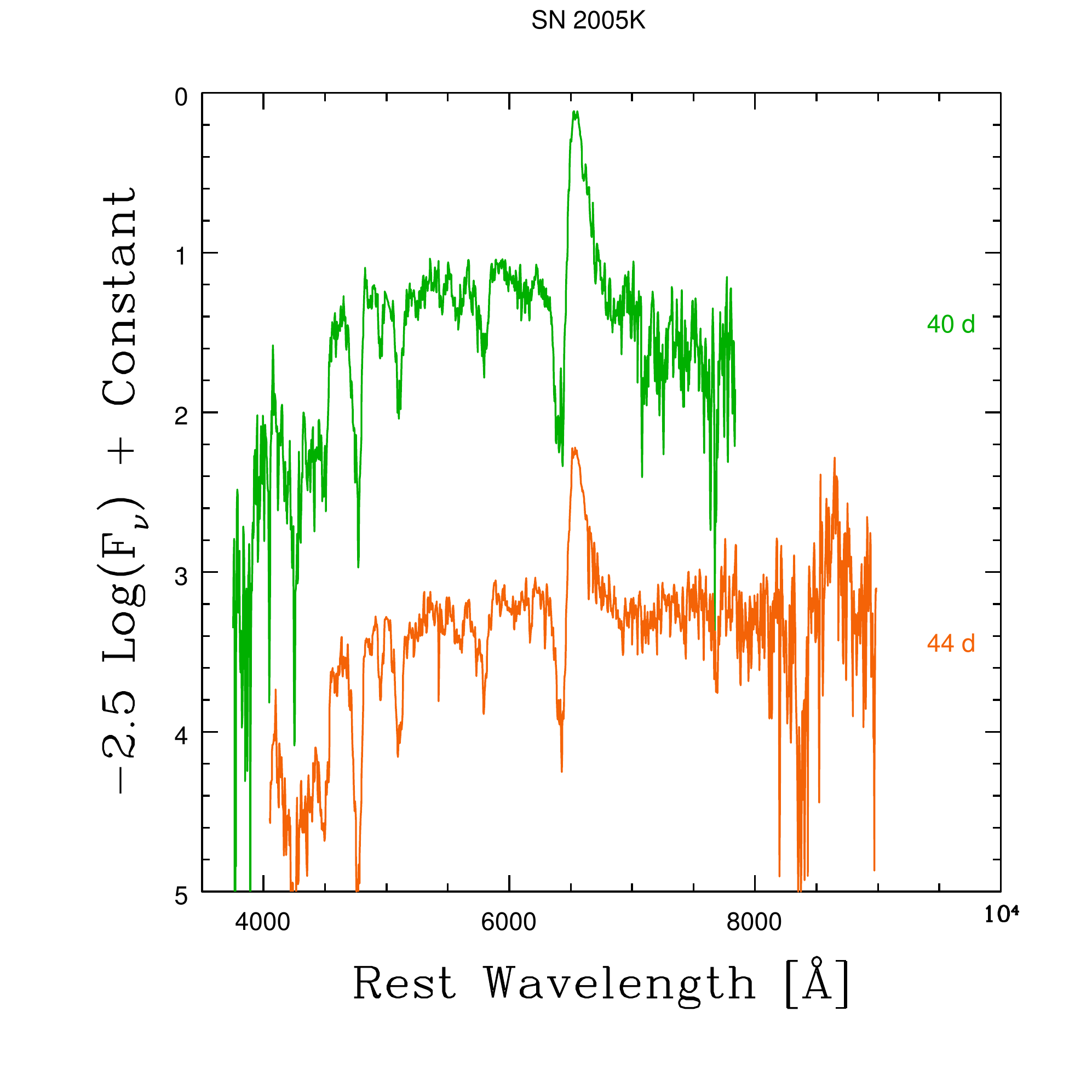}
\includegraphics[width=5.5cm]{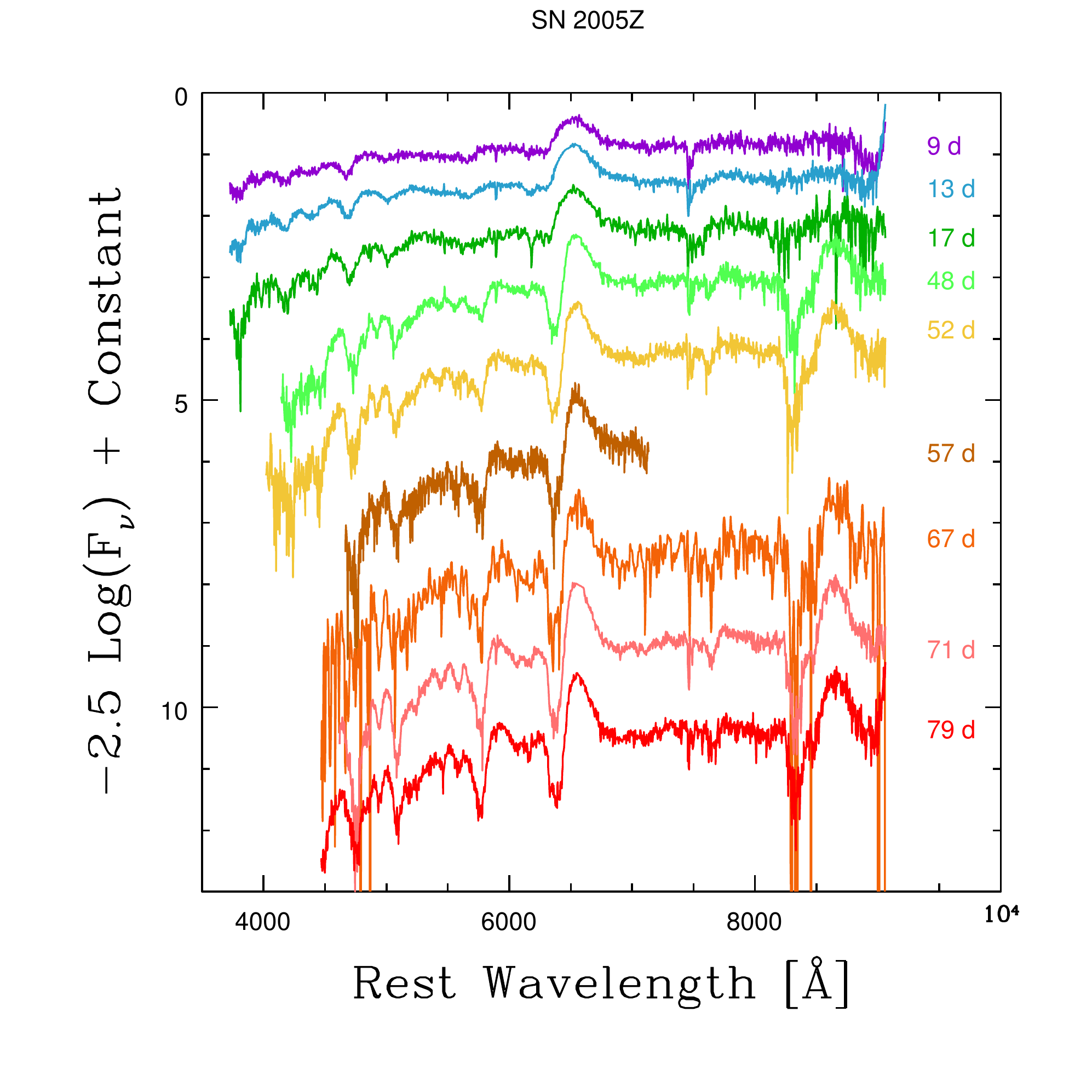}
\includegraphics[width=5.5cm]{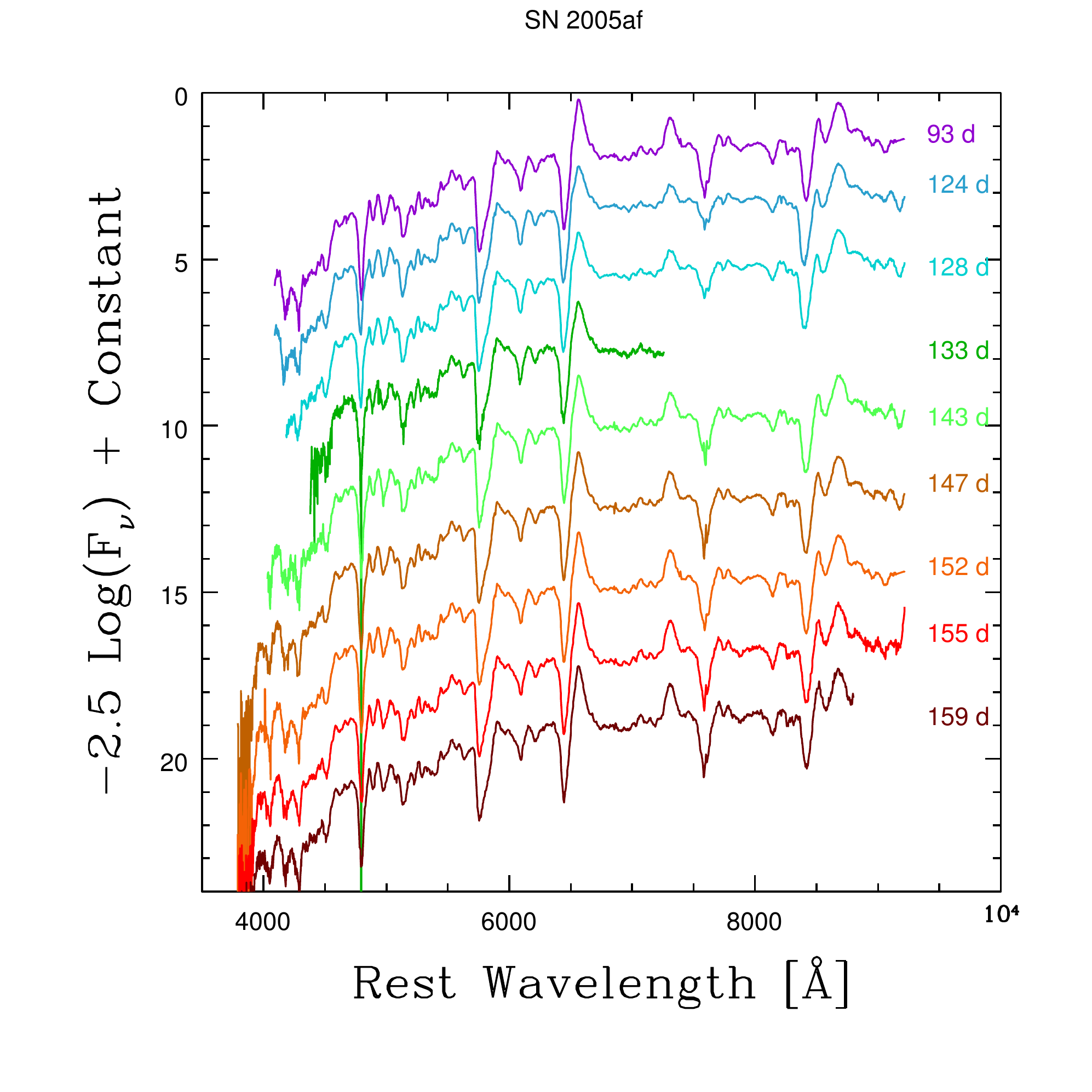}
\includegraphics[width=5.5cm]{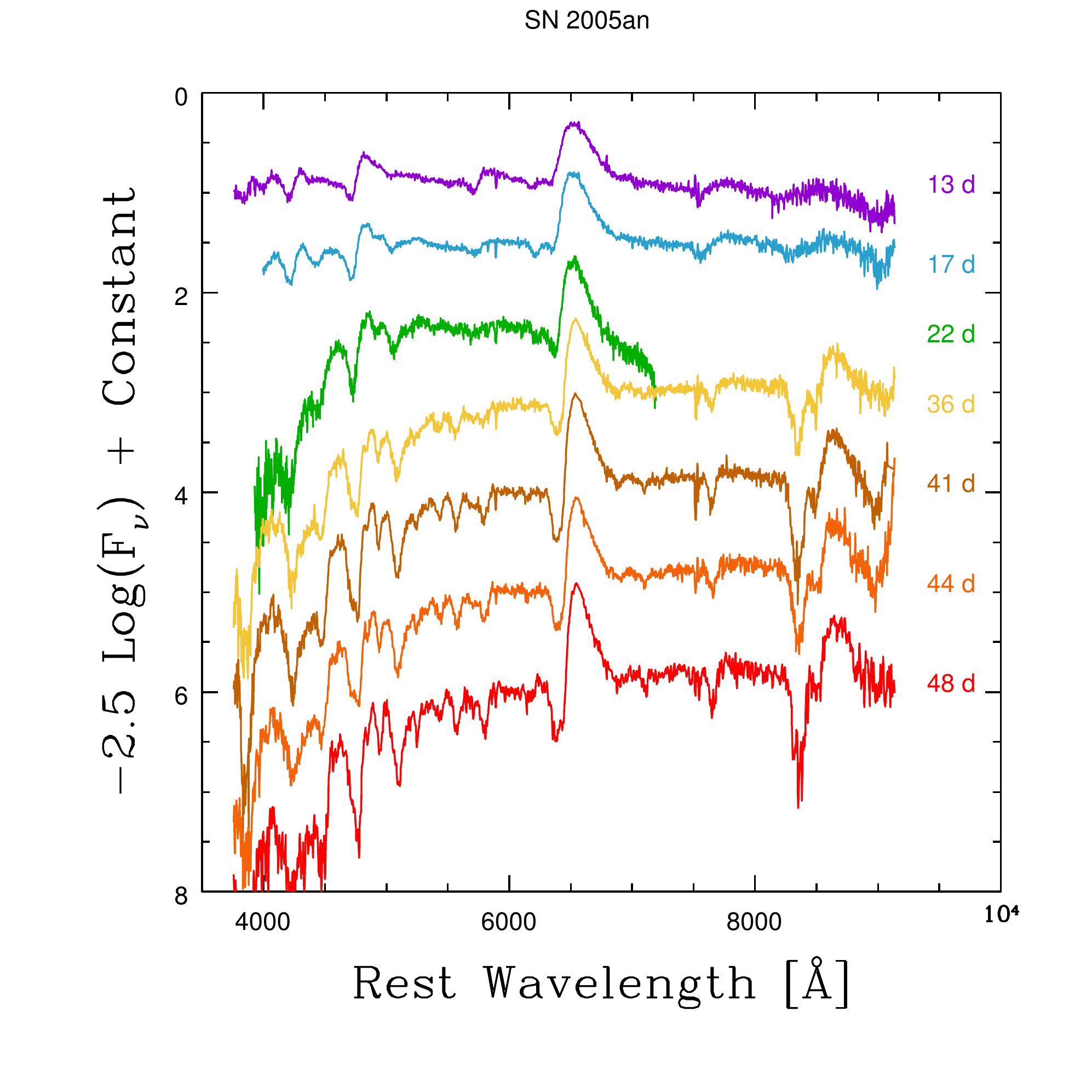}
\includegraphics[width=5.5cm]{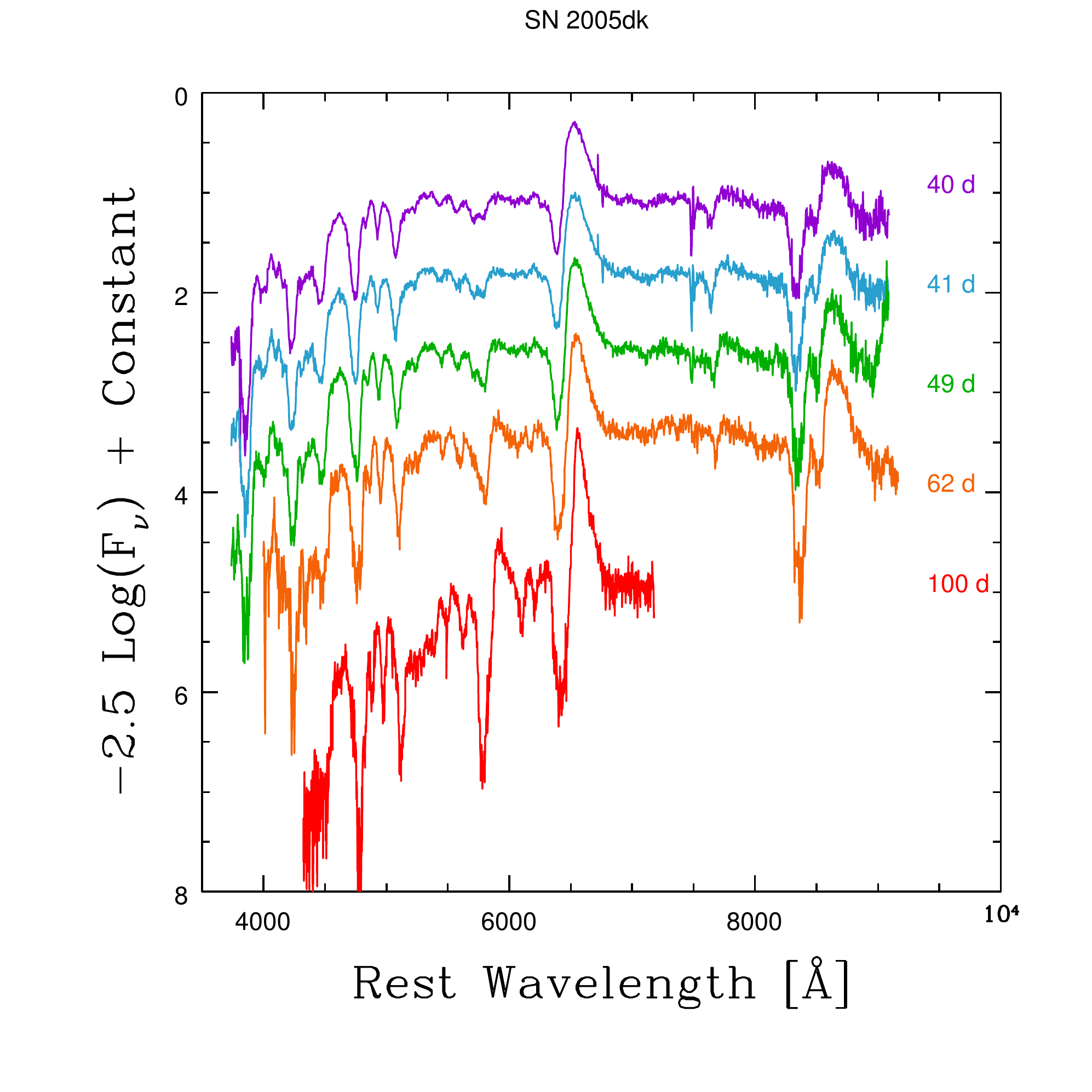}
\includegraphics[width=5.5cm]{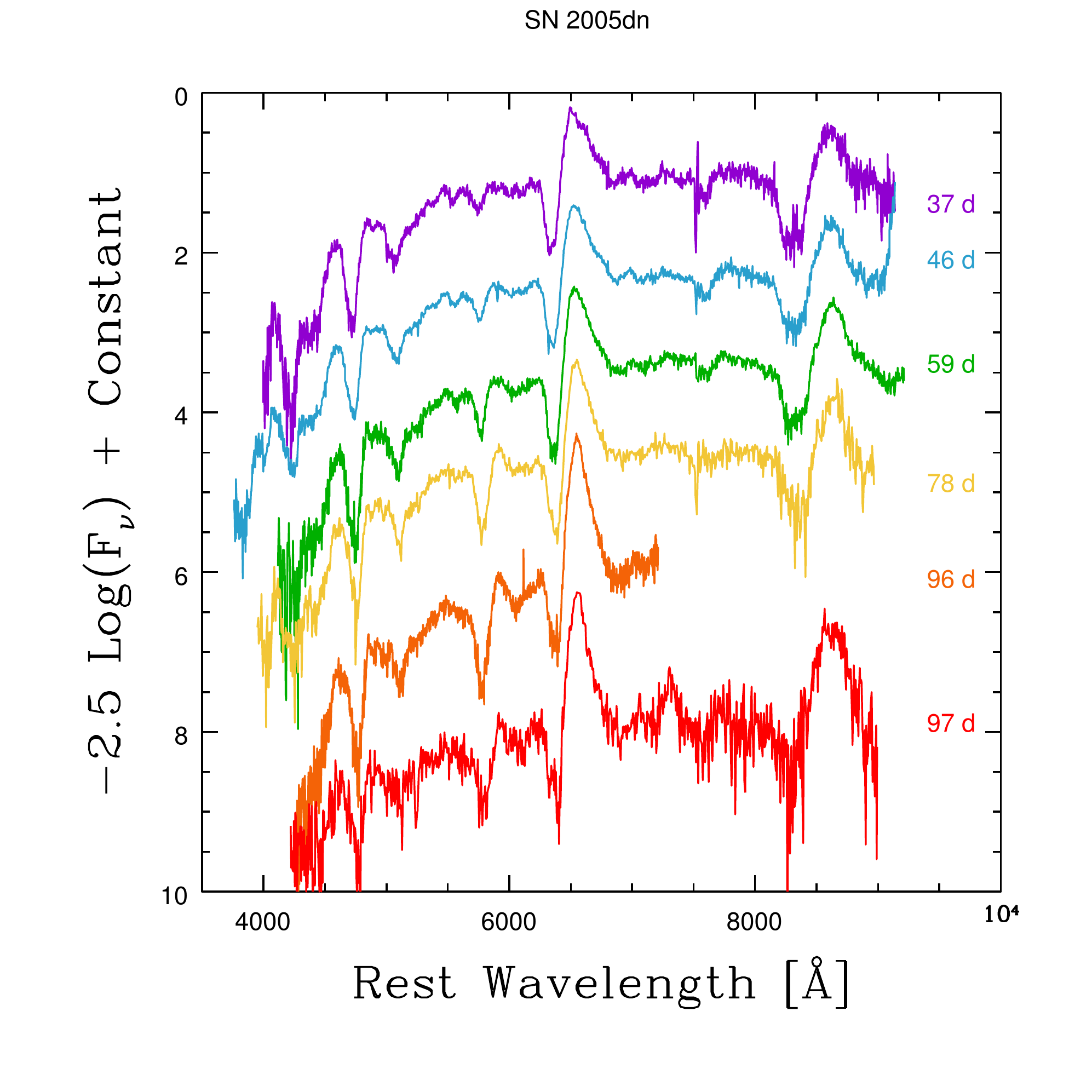}
\caption{Examples of SNe~II spectra: SN~2004ej, SN~2004er, SN~2004fb, SN~2004fc, SN~2004fx, SN~2005J, SN~2005K, SN~2005Z, SN~2005af, SN~2005an, SN~2005dk, SN~2005dn.}
\label{example}
\end{figure*}

\begin{figure*}[h!]
\centering
\includegraphics[width=5.5cm]{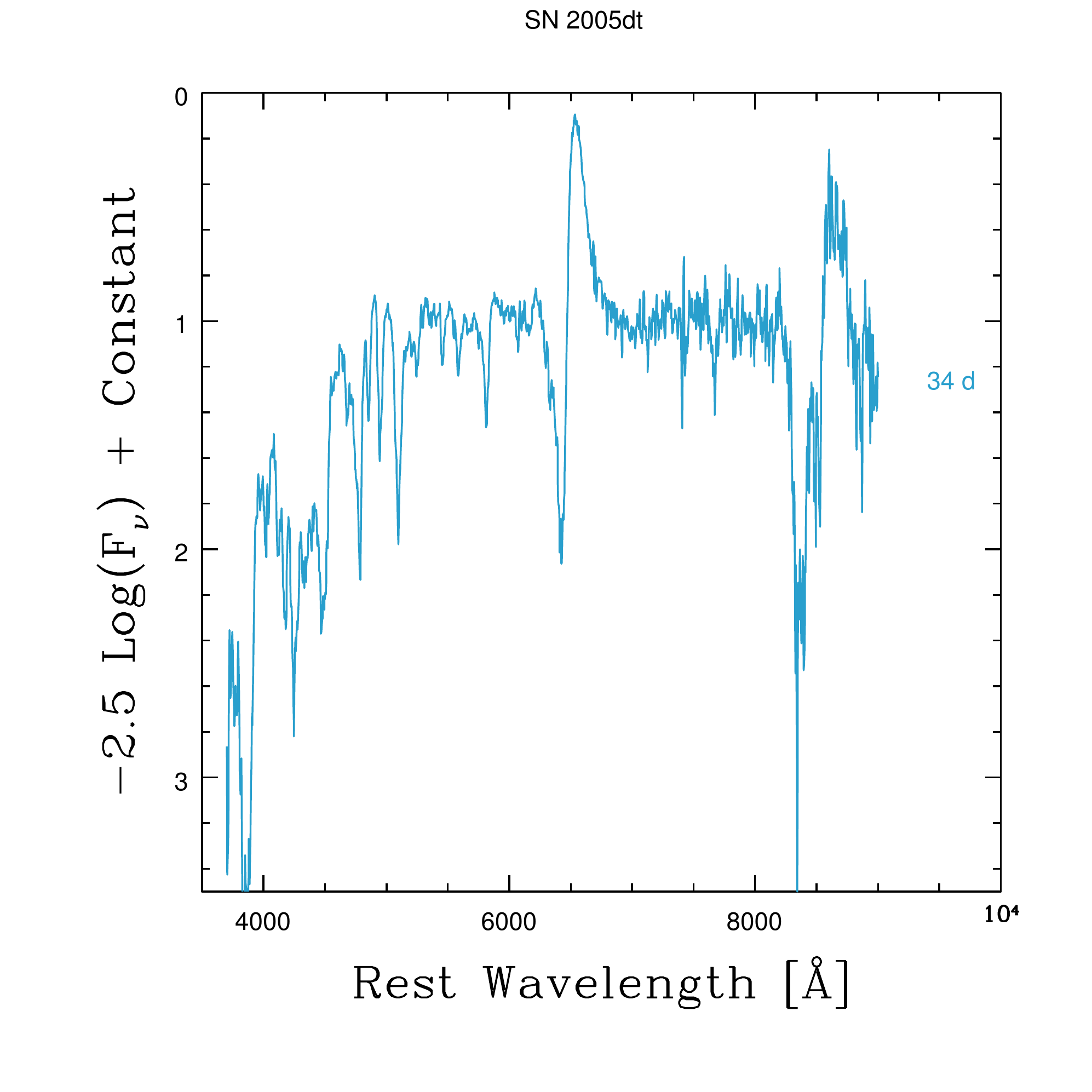}
\includegraphics[width=5.5cm]{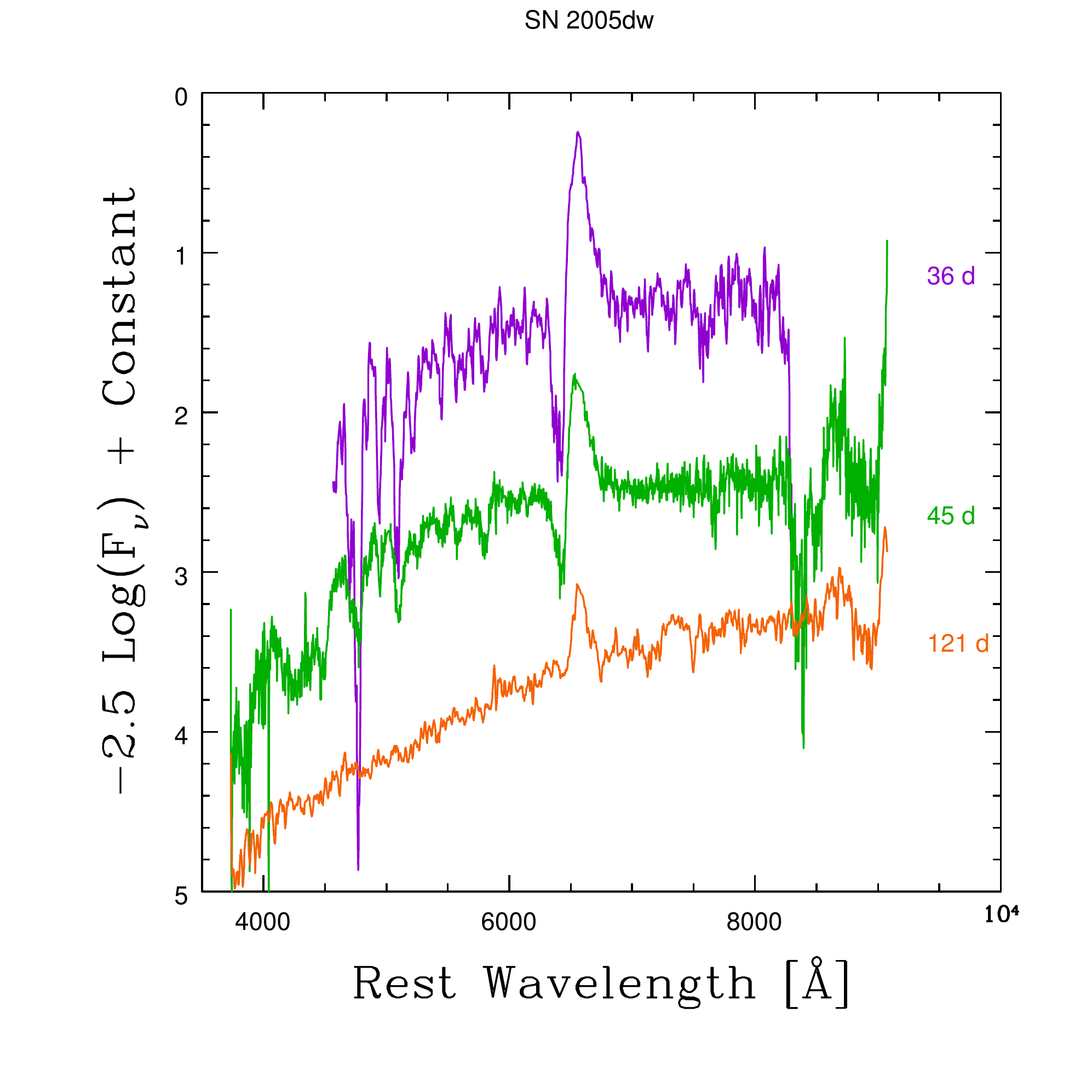}
\includegraphics[width=5.5cm]{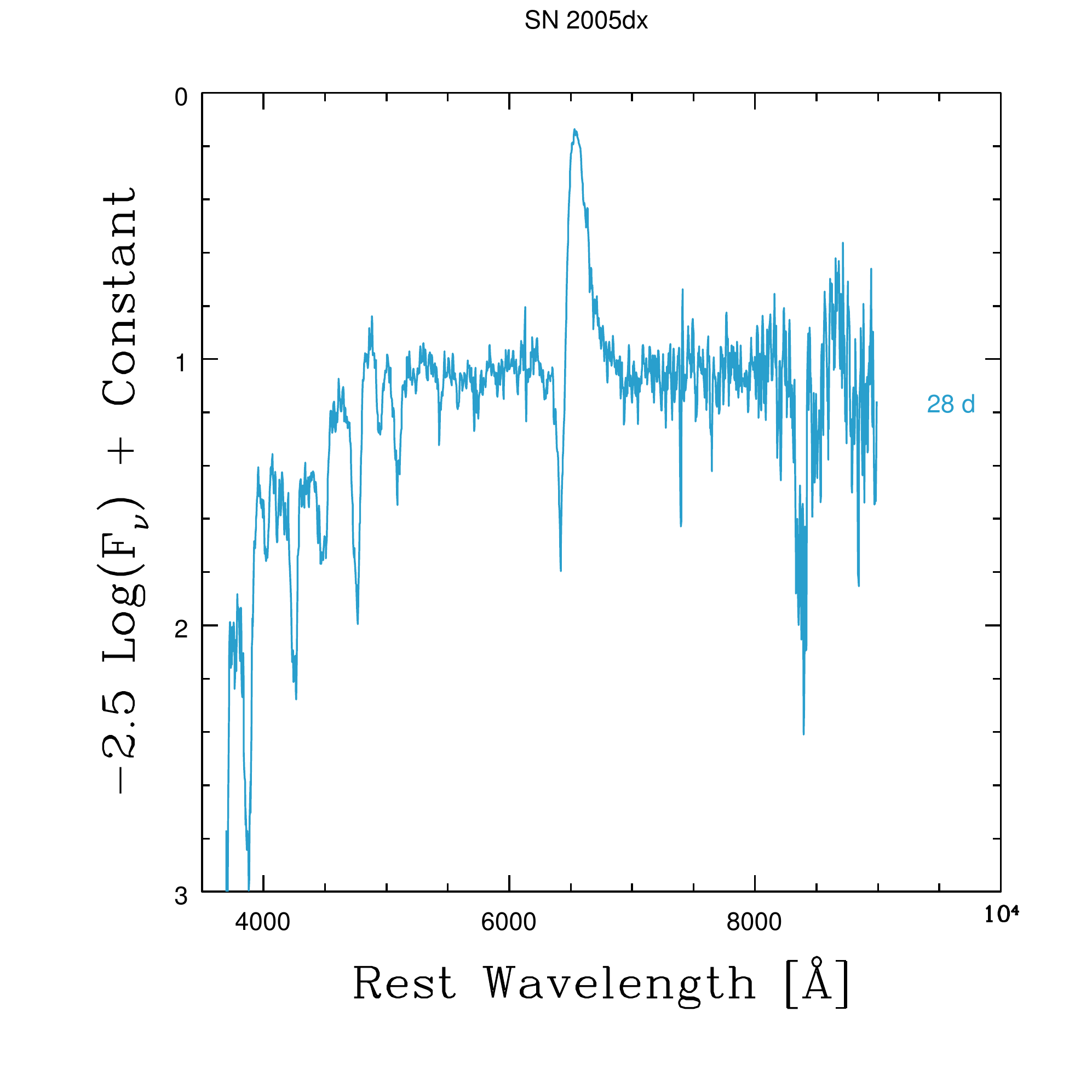}
\includegraphics[width=5.5cm]{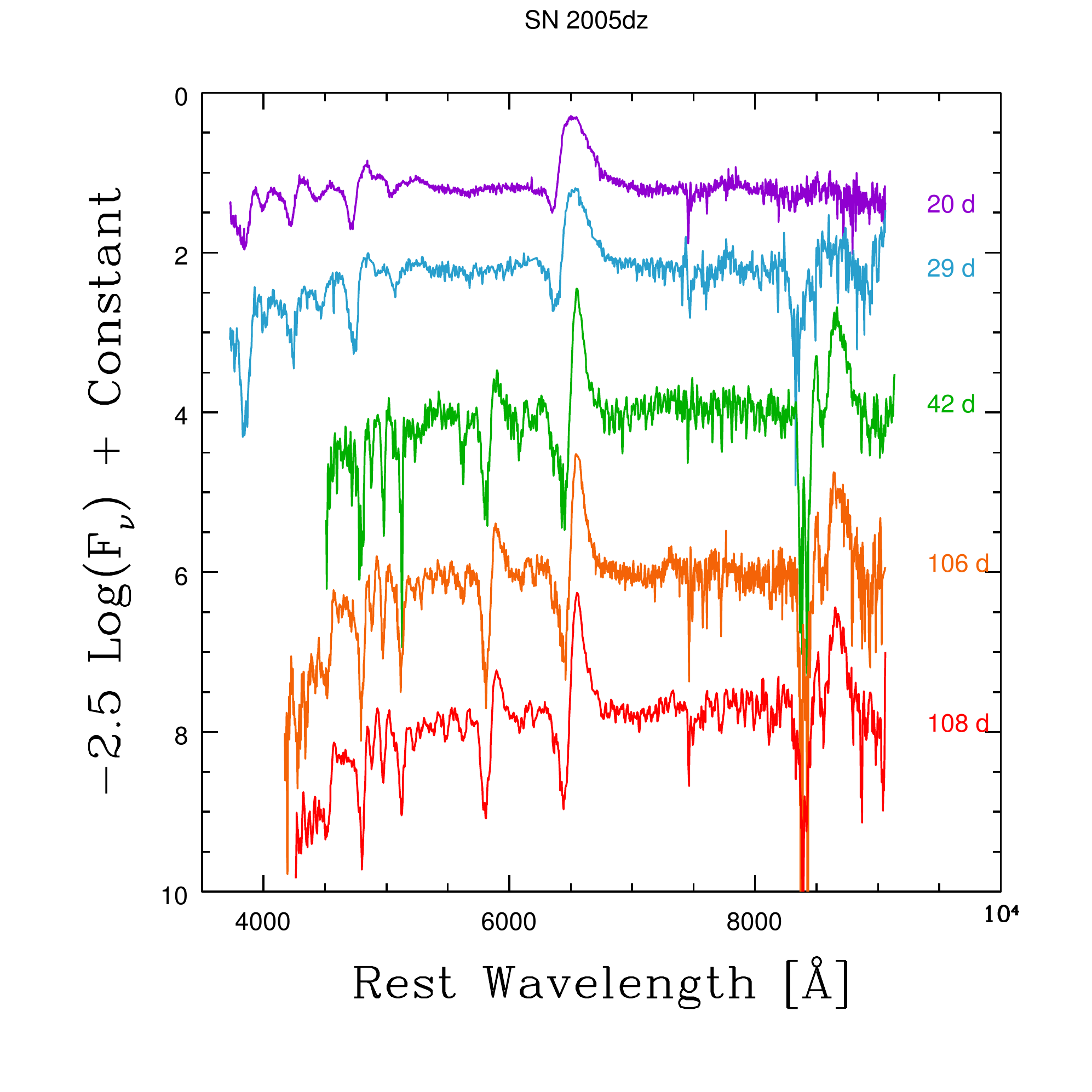}
\includegraphics[width=5.5cm]{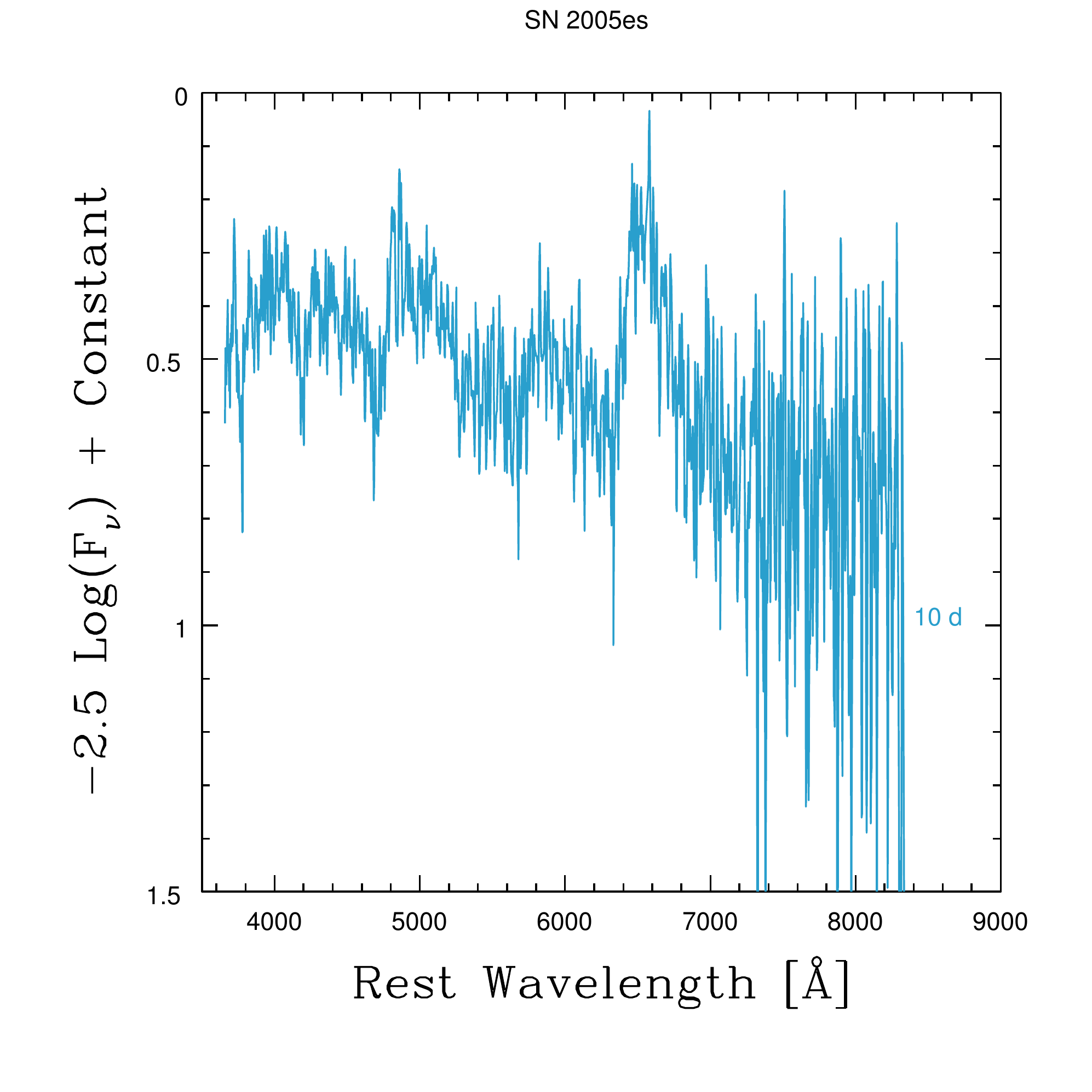}
\includegraphics[width=5.5cm]{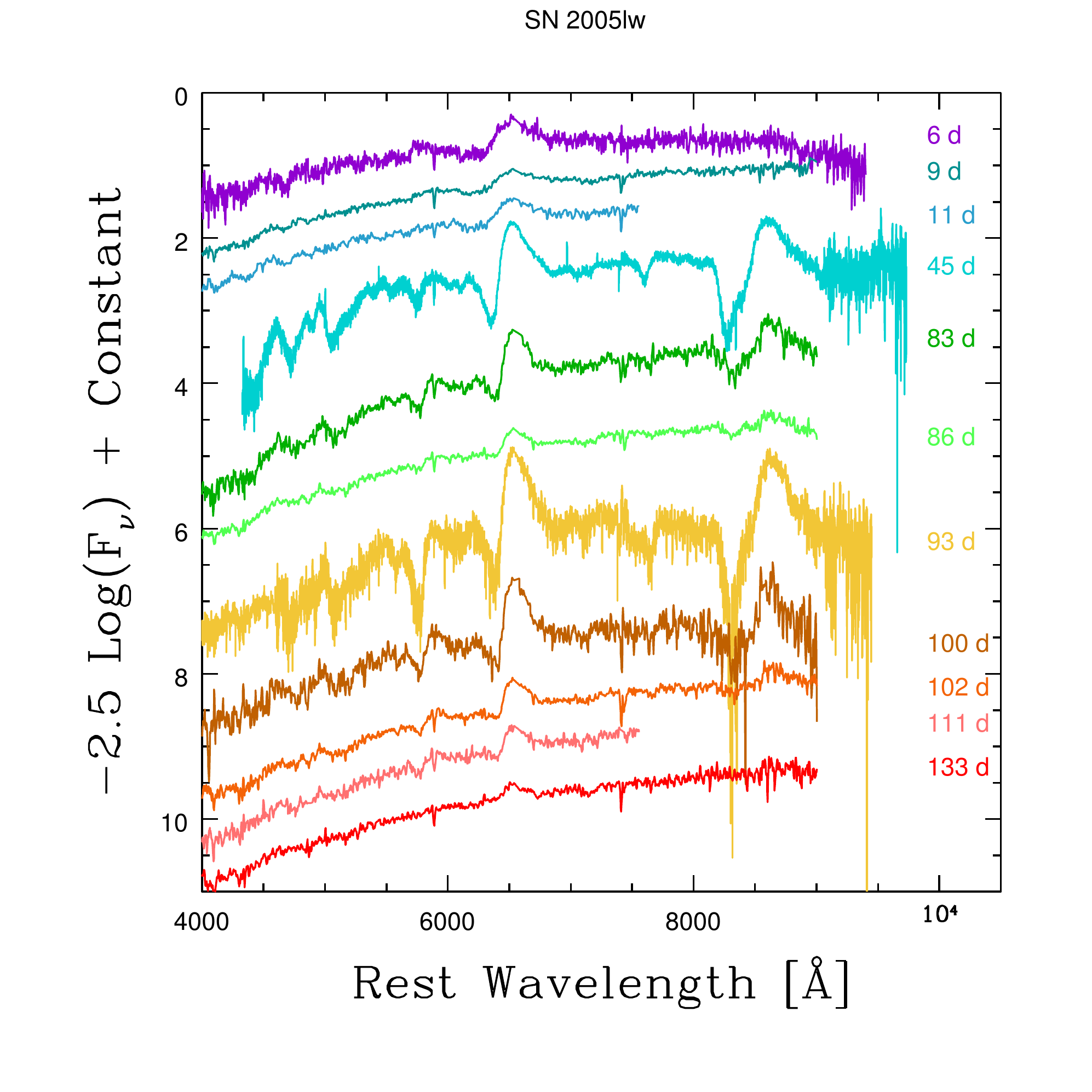}
\includegraphics[width=5.5cm]{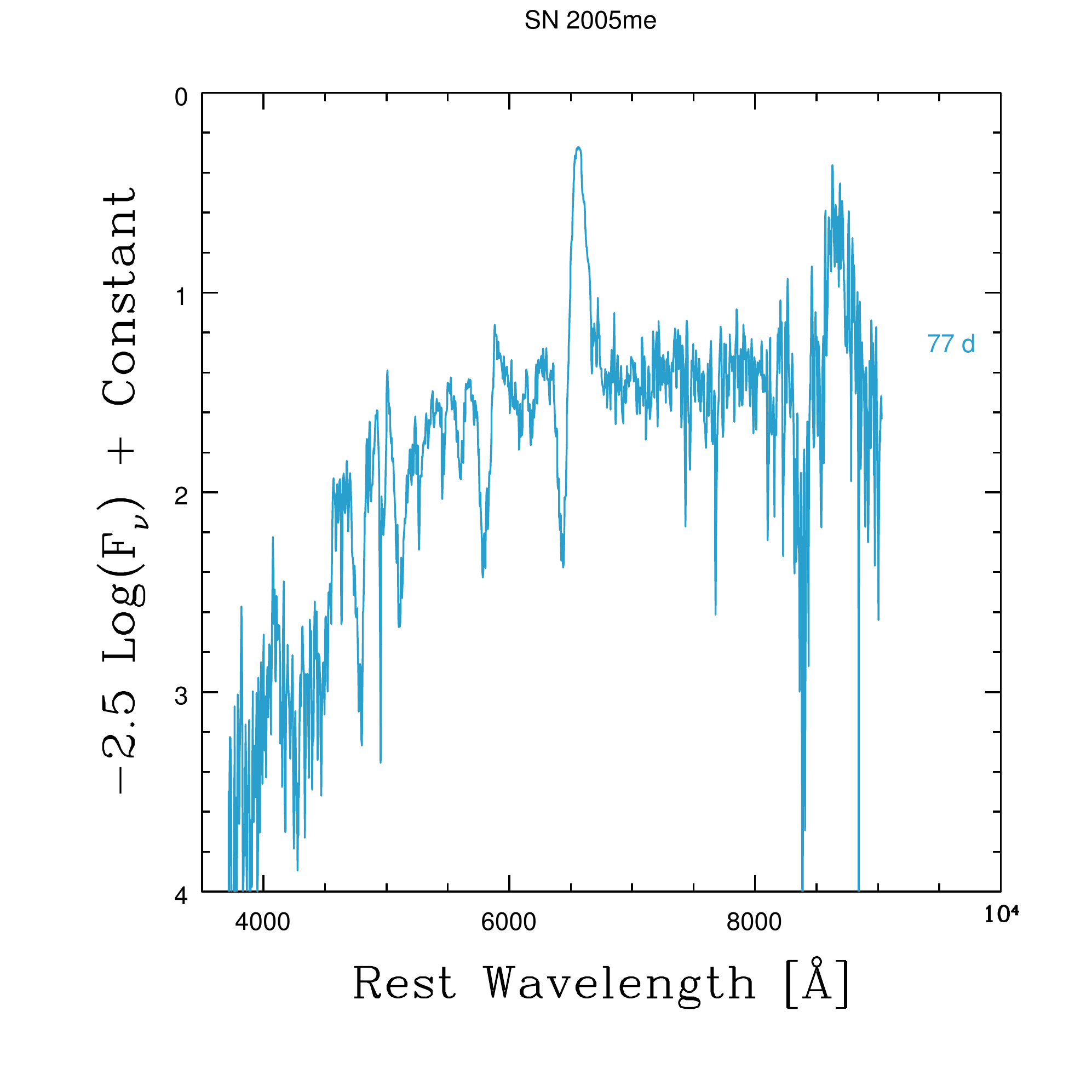}
\includegraphics[width=5.5cm]{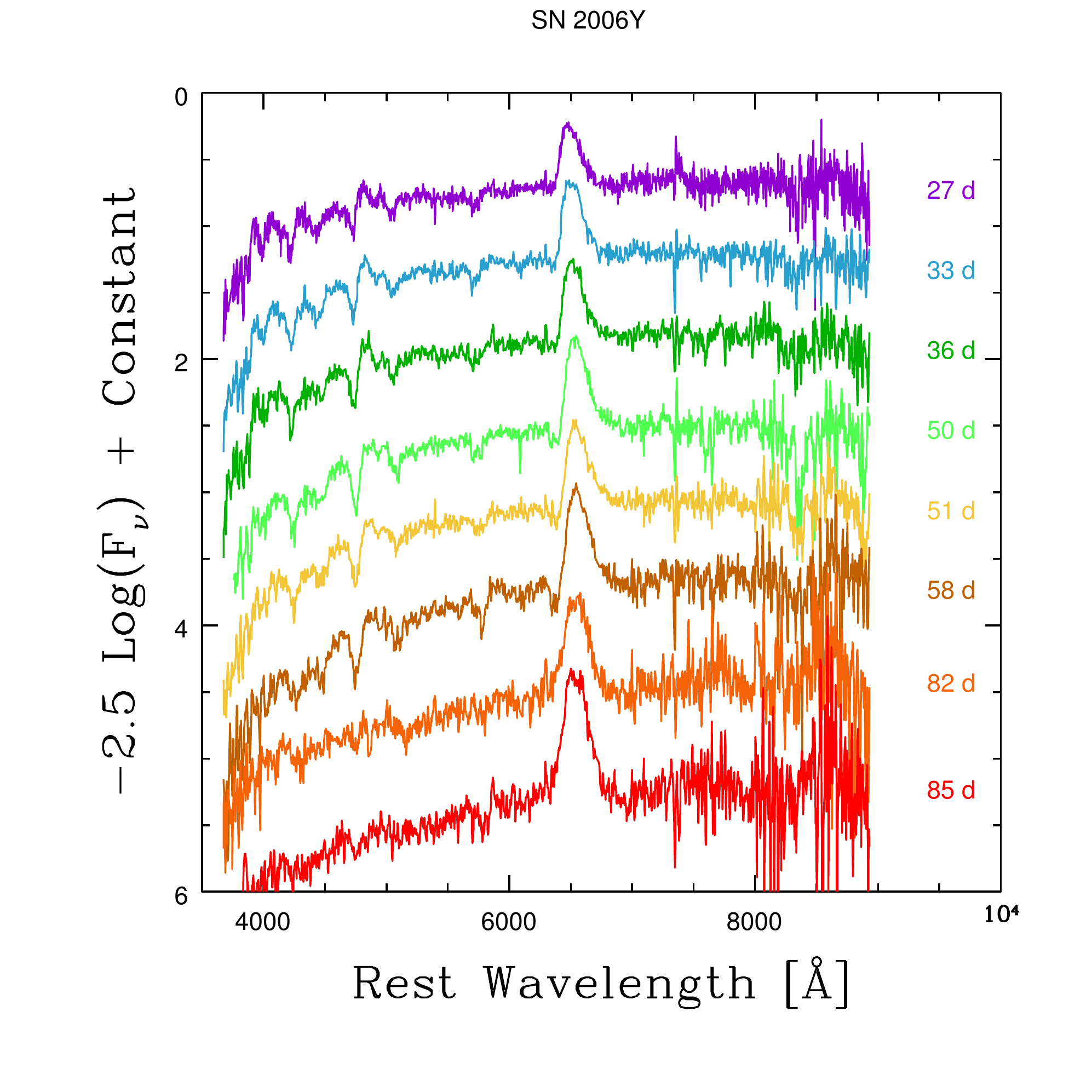}
\includegraphics[width=5.5cm]{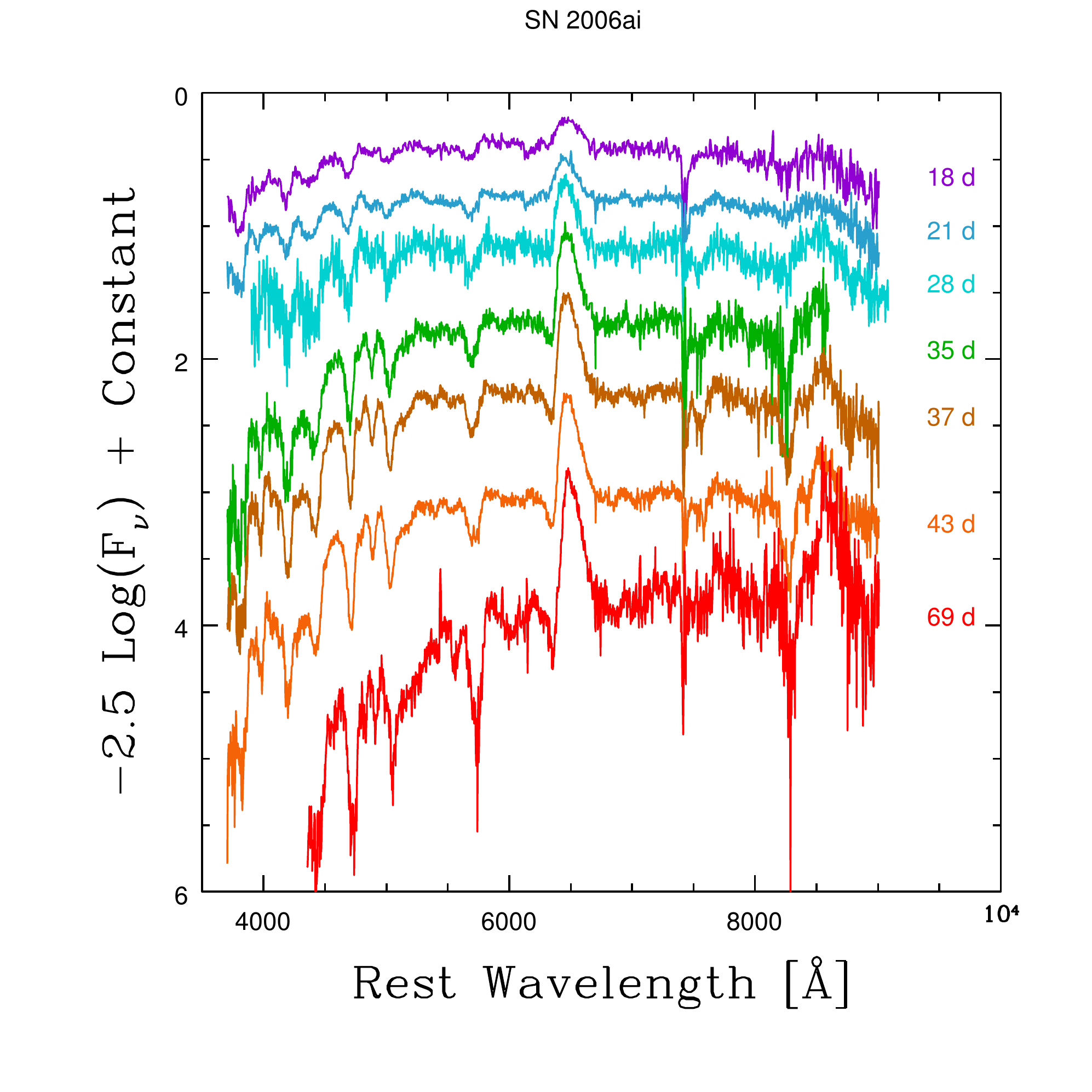}
\includegraphics[width=5.5cm]{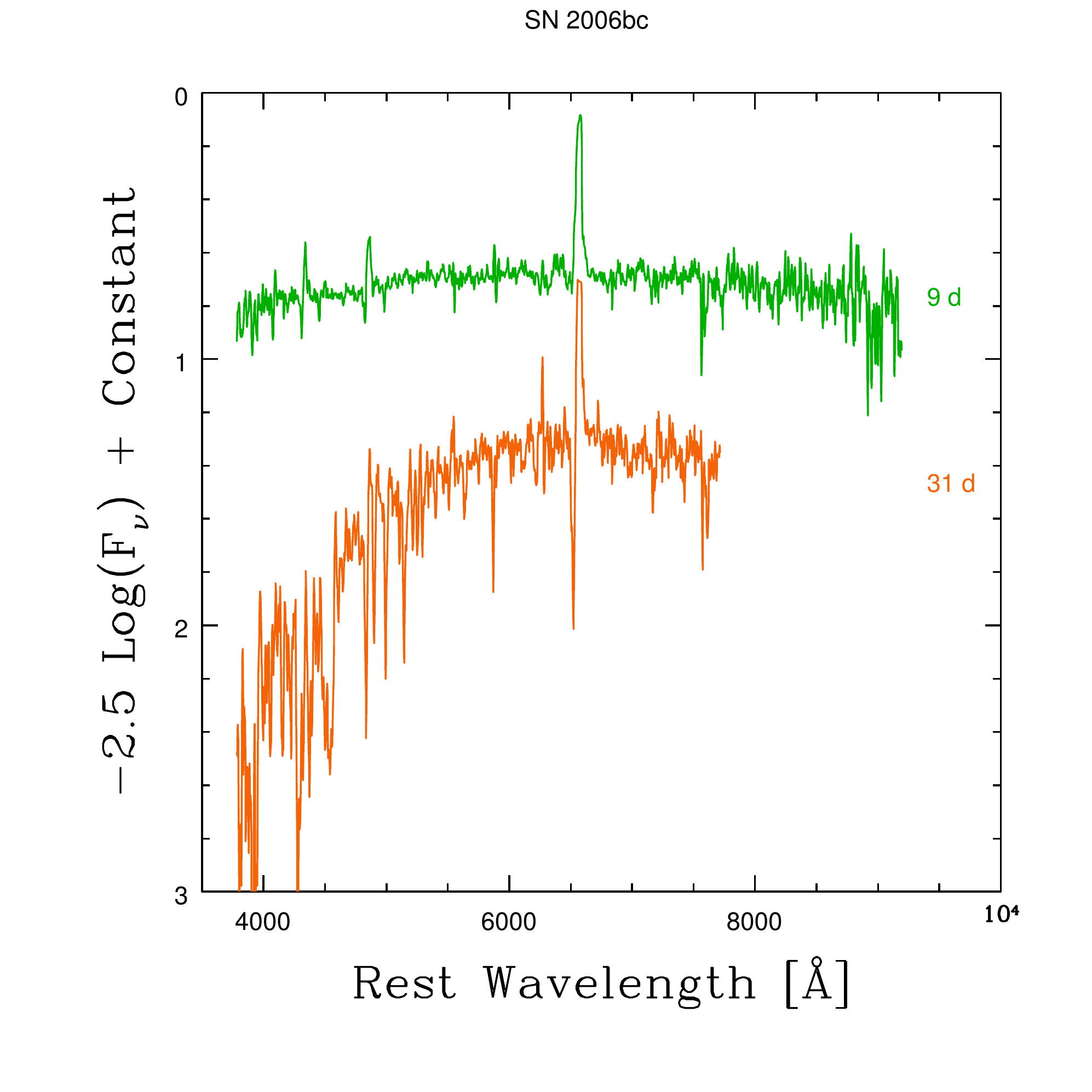}
\includegraphics[width=5.5cm]{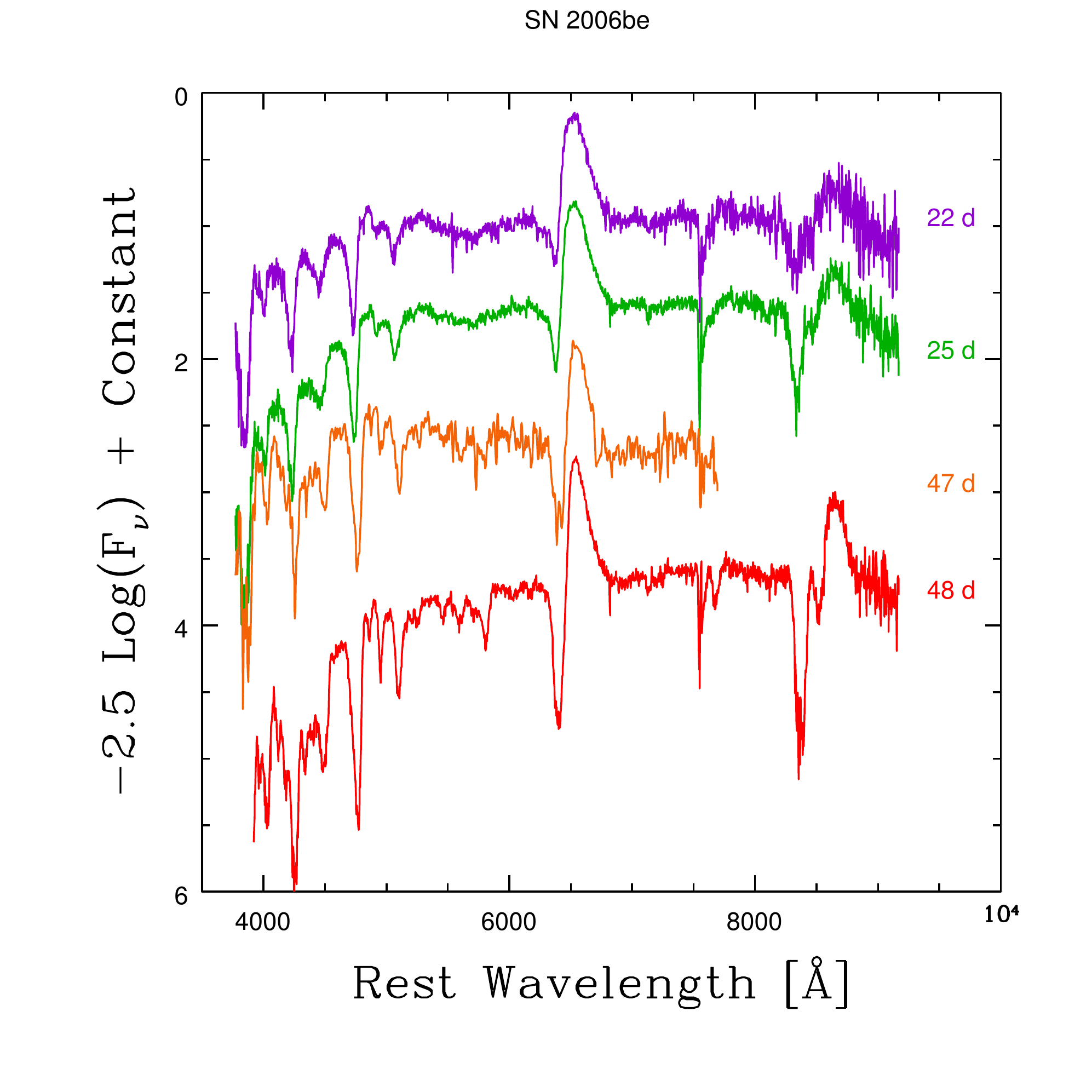}
\includegraphics[width=5.5cm]{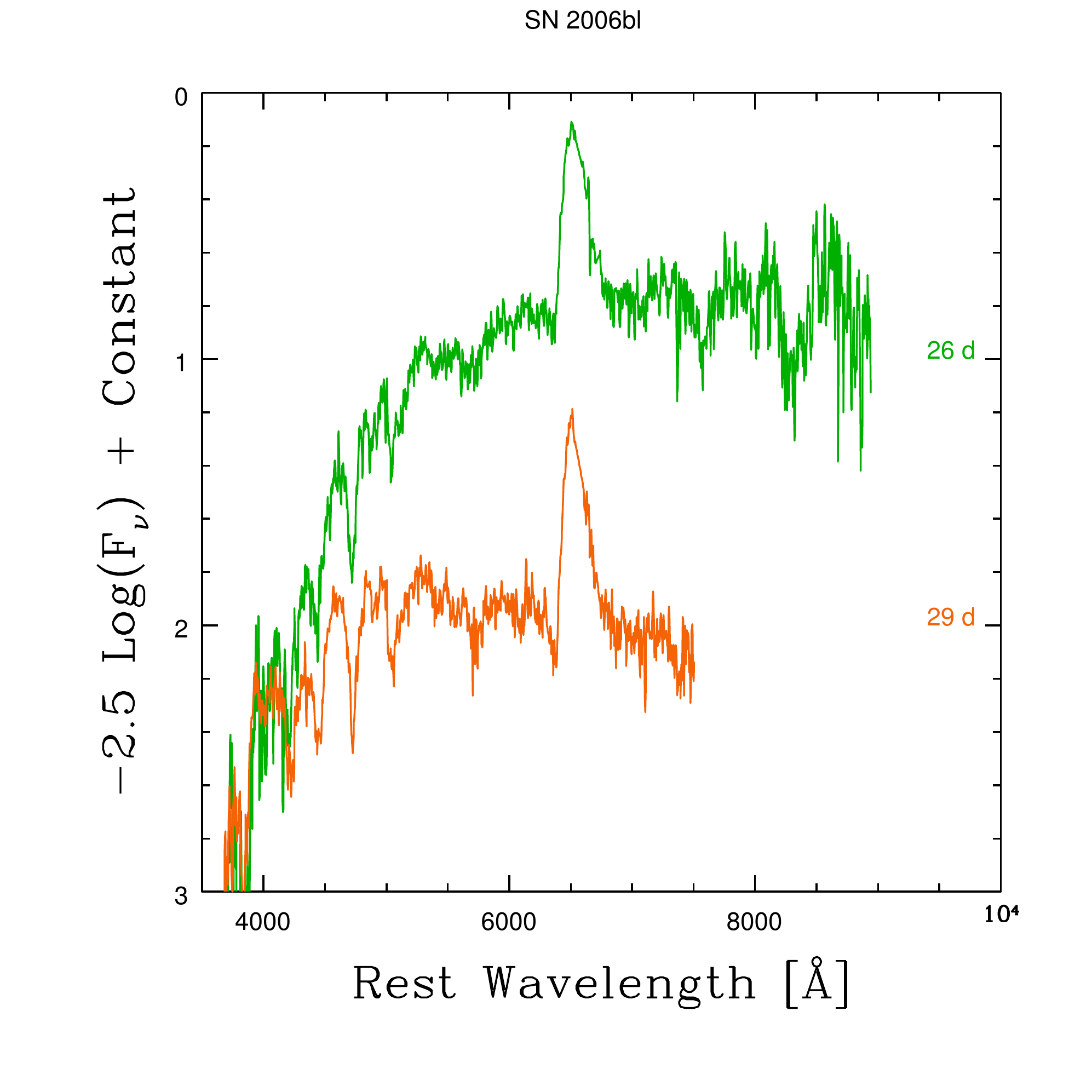}
\caption{Examples of SNe~II spectra: SN~2005dt, SN~2005dw, SN~2005dx, SN~2005dz, SN~2005es, SN~2005lw, SN~2005me, SN~2006Y, SN~2006ai, SN~2006be, SN~2006be, SN~2006bl.}
\label{example}
\end{figure*}

\begin{figure*}[h!]
\centering
\includegraphics[width=5.5cm]{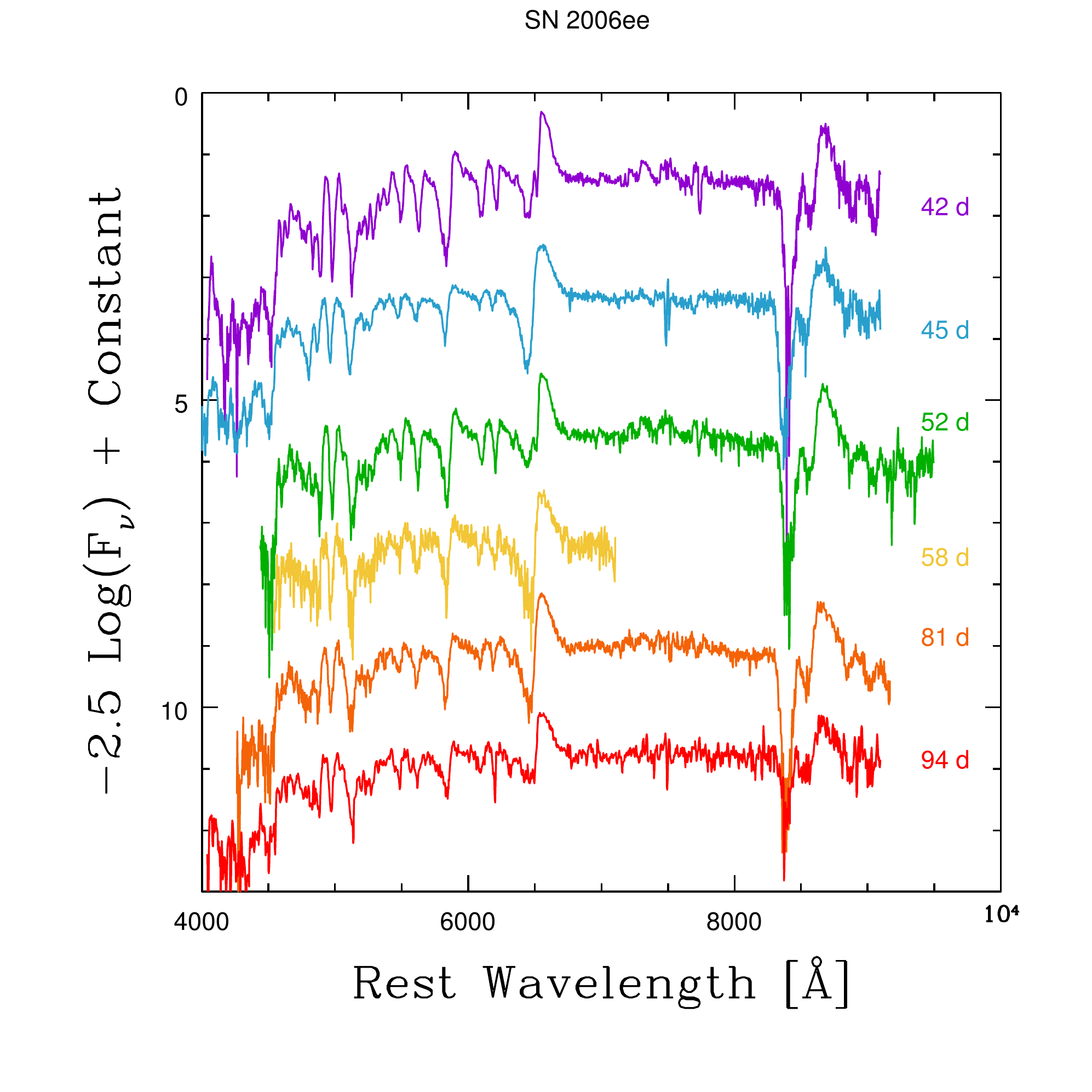}
\includegraphics[width=5.5cm]{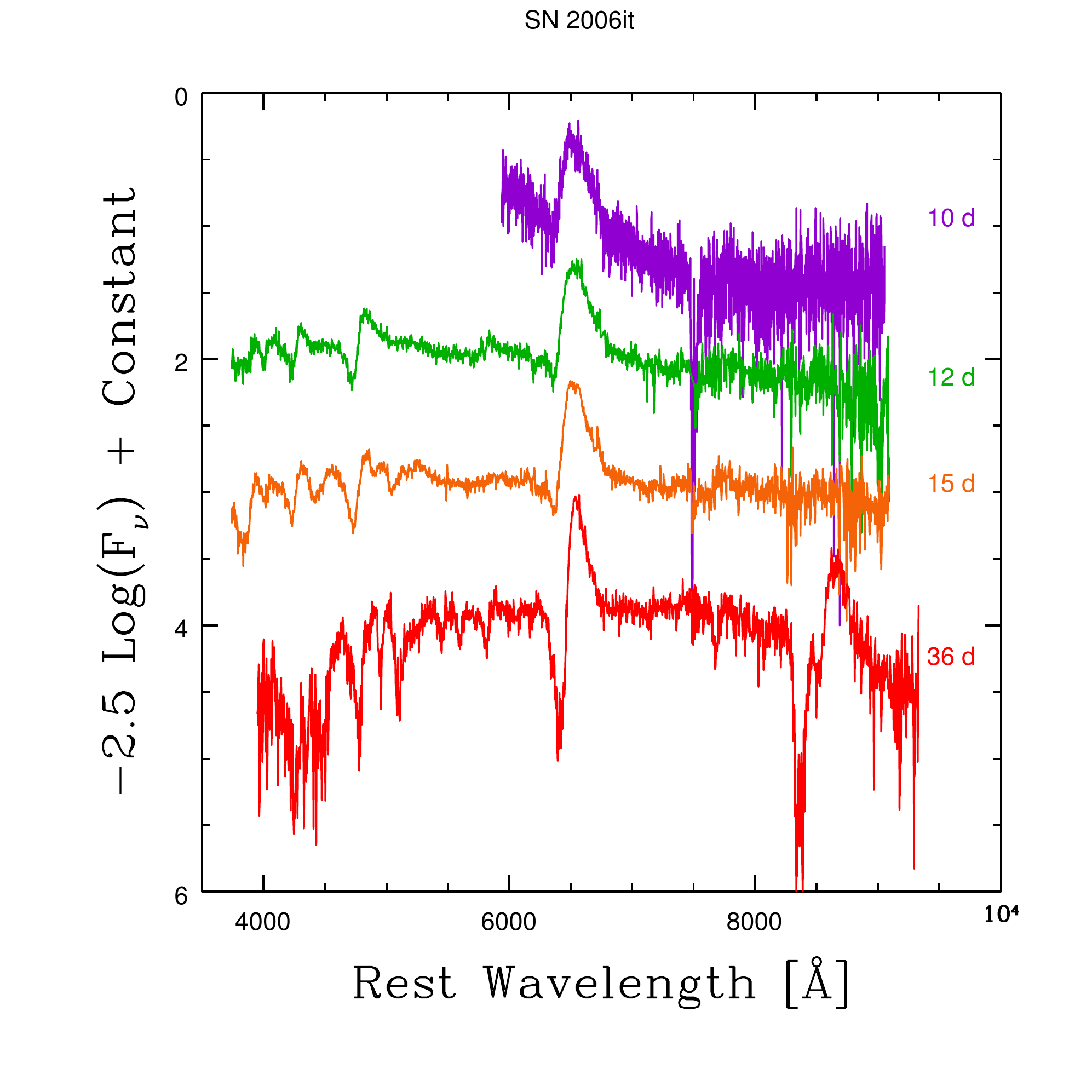}
\includegraphics[width=5.5cm]{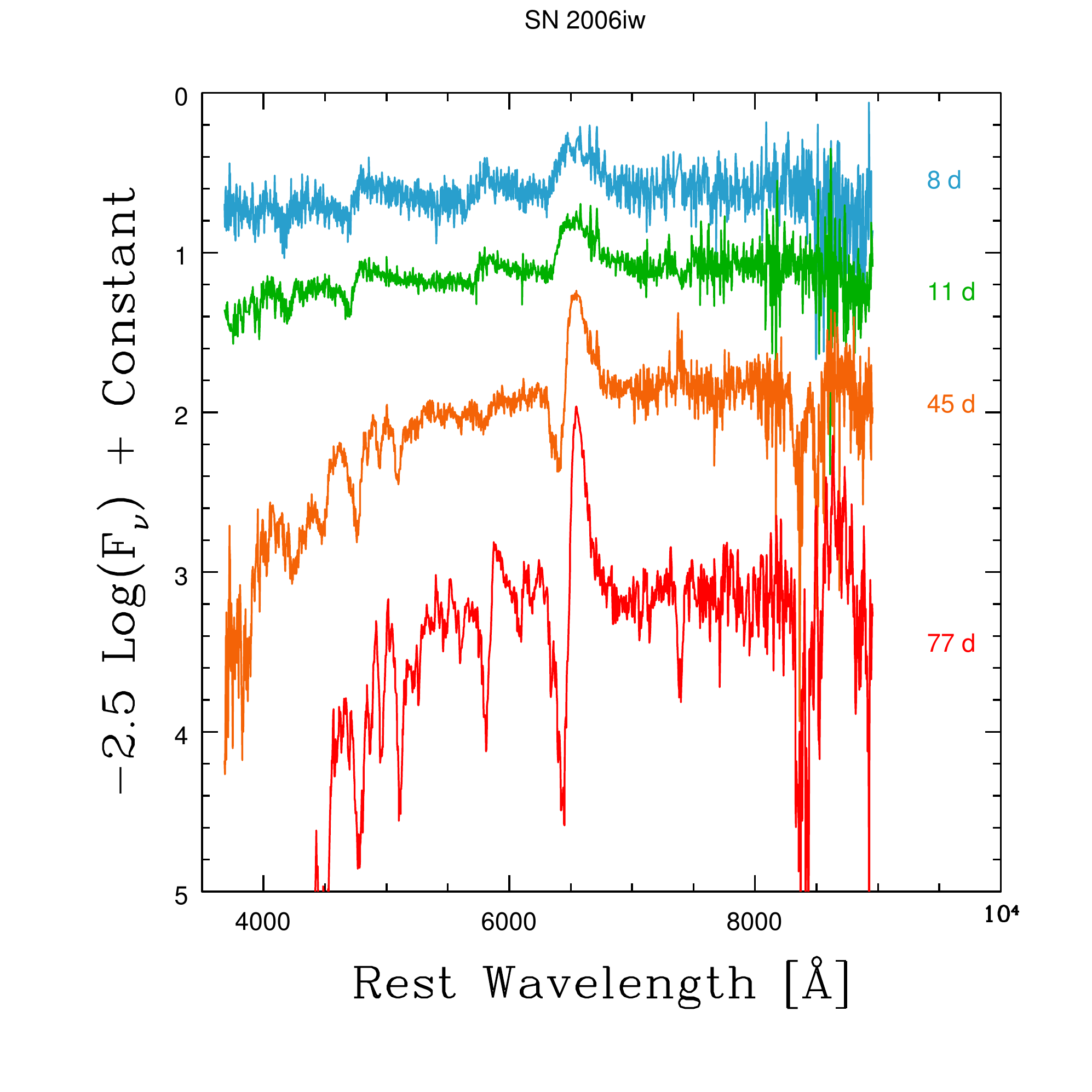}
\includegraphics[width=5.5cm]{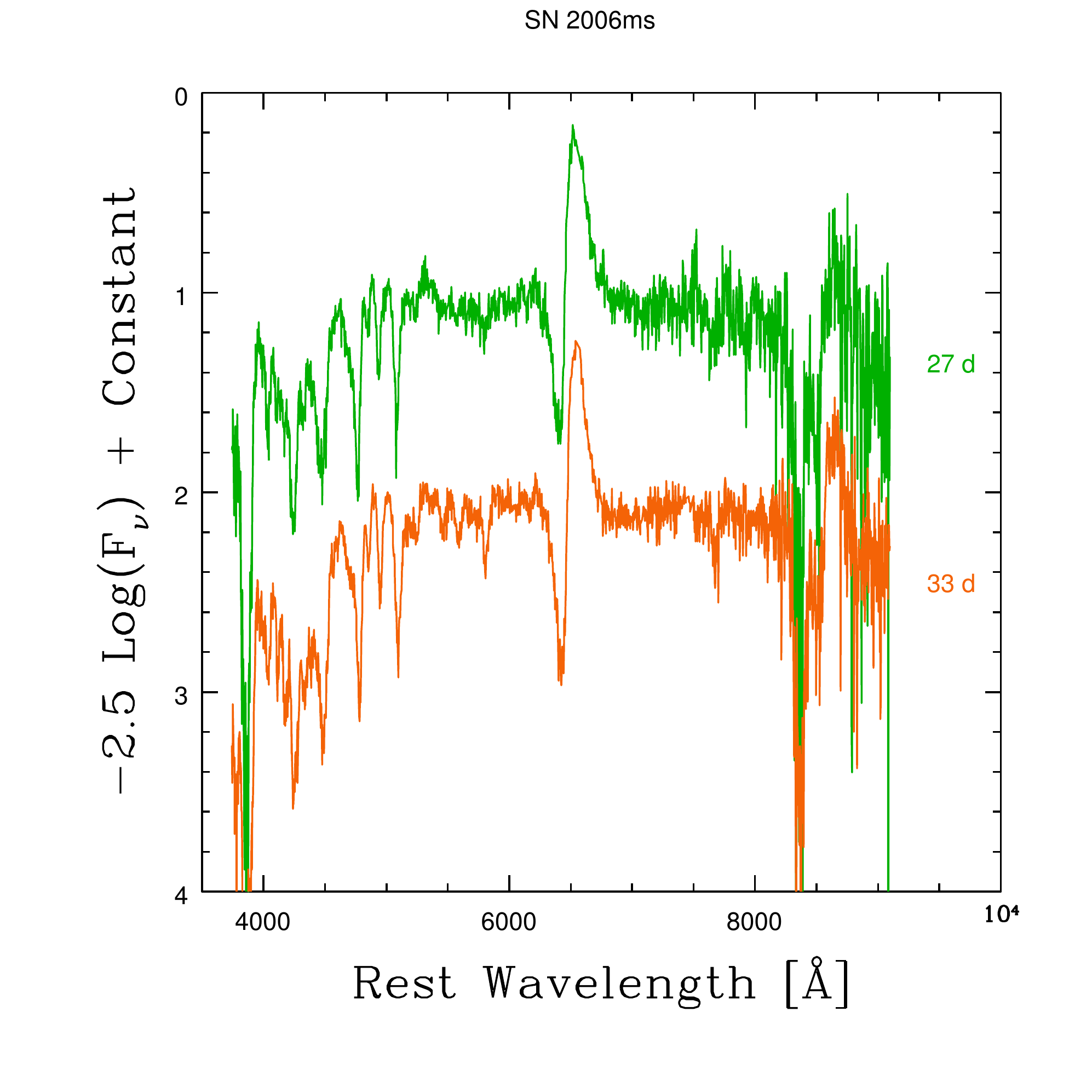}
\includegraphics[width=5.5cm]{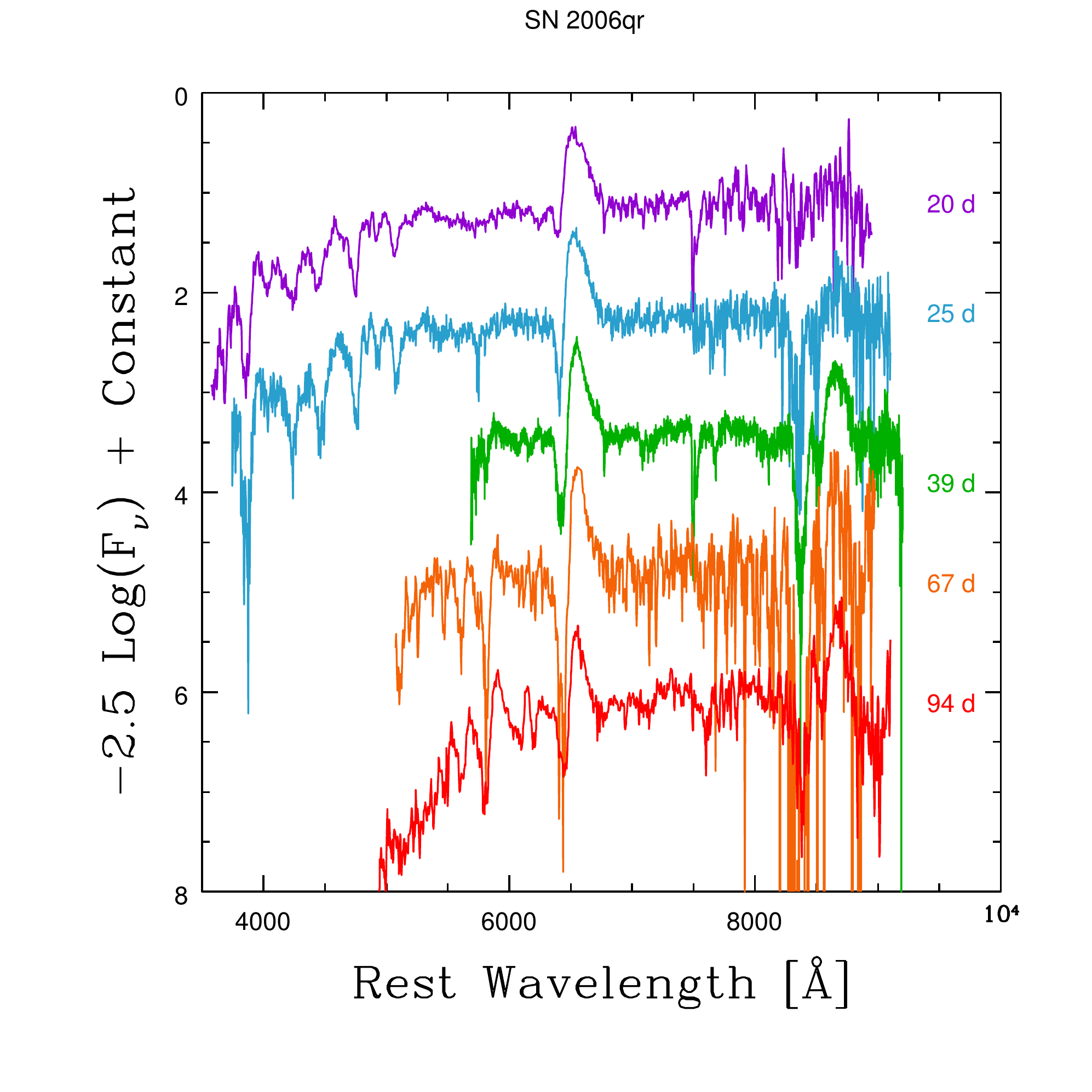}
\includegraphics[width=5.5cm]{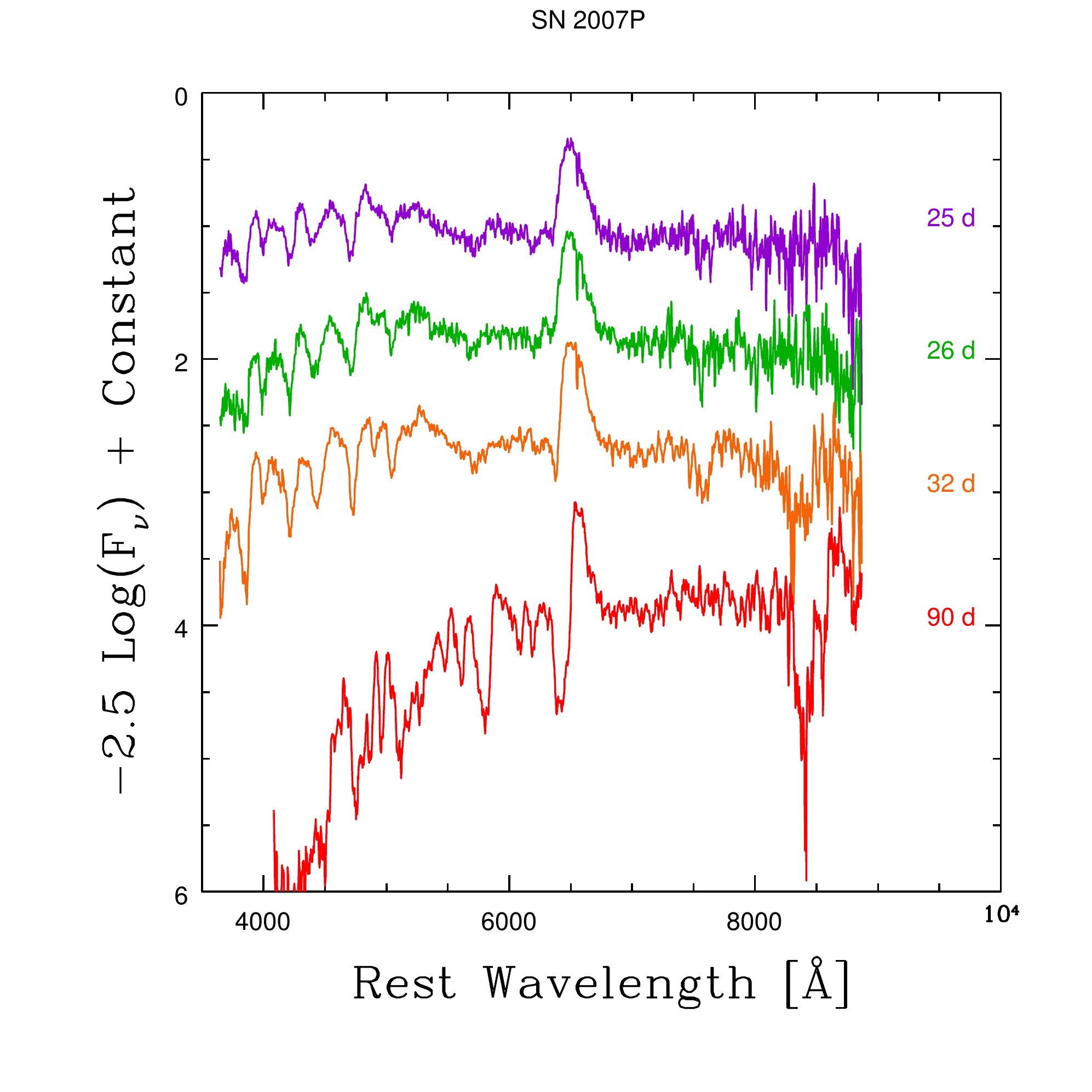}
\includegraphics[width=5.5cm]{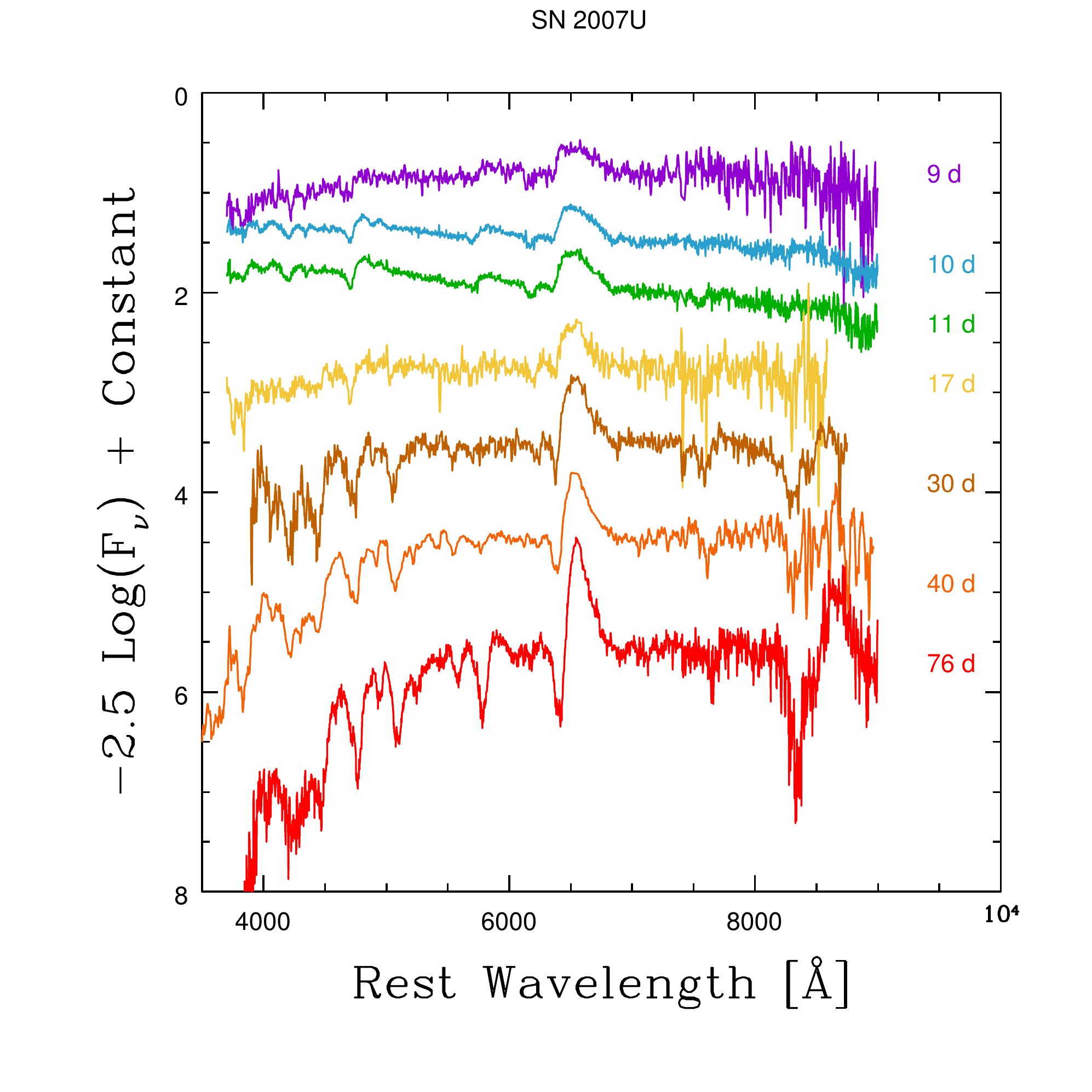}
\includegraphics[width=5.5cm]{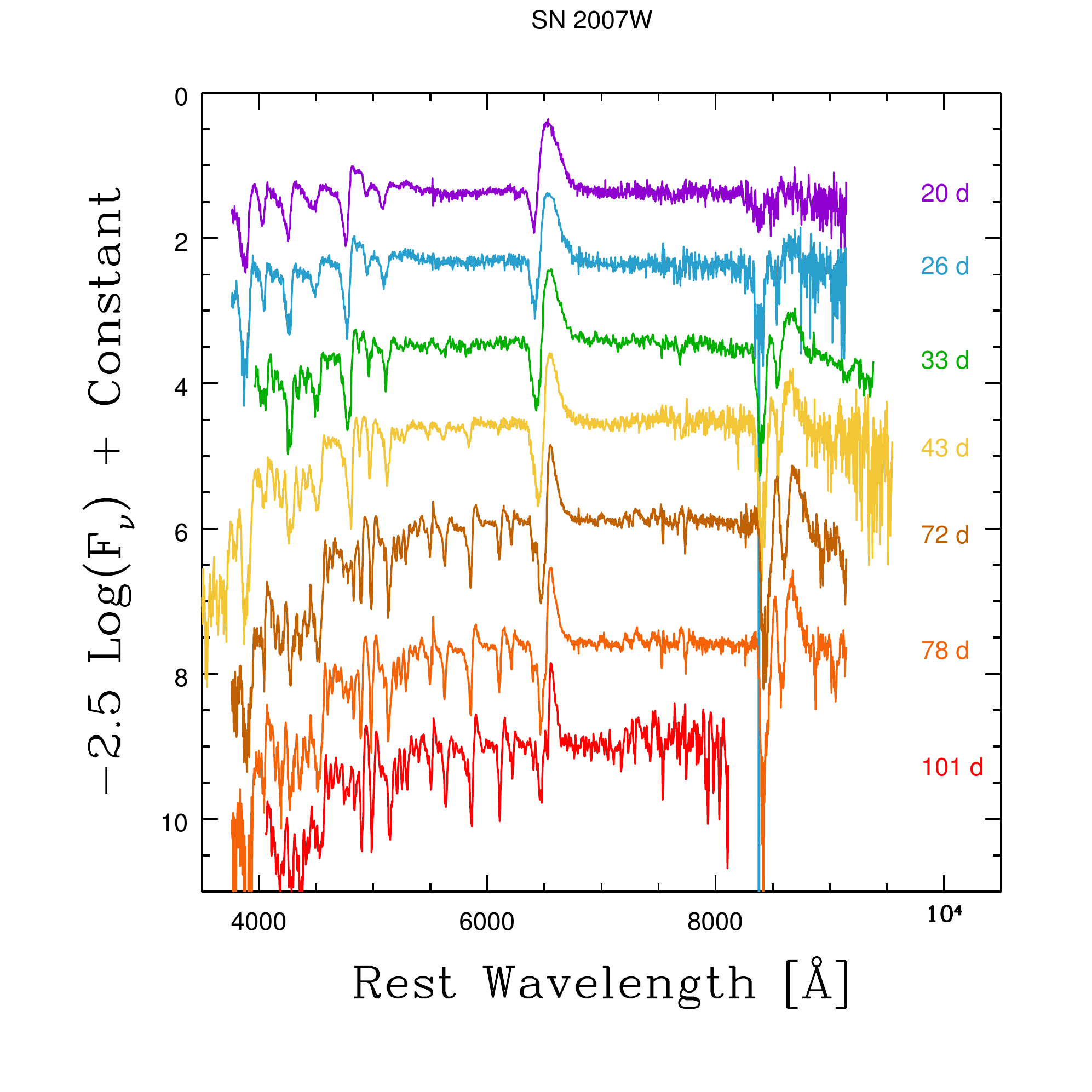}
\includegraphics[width=5.5cm]{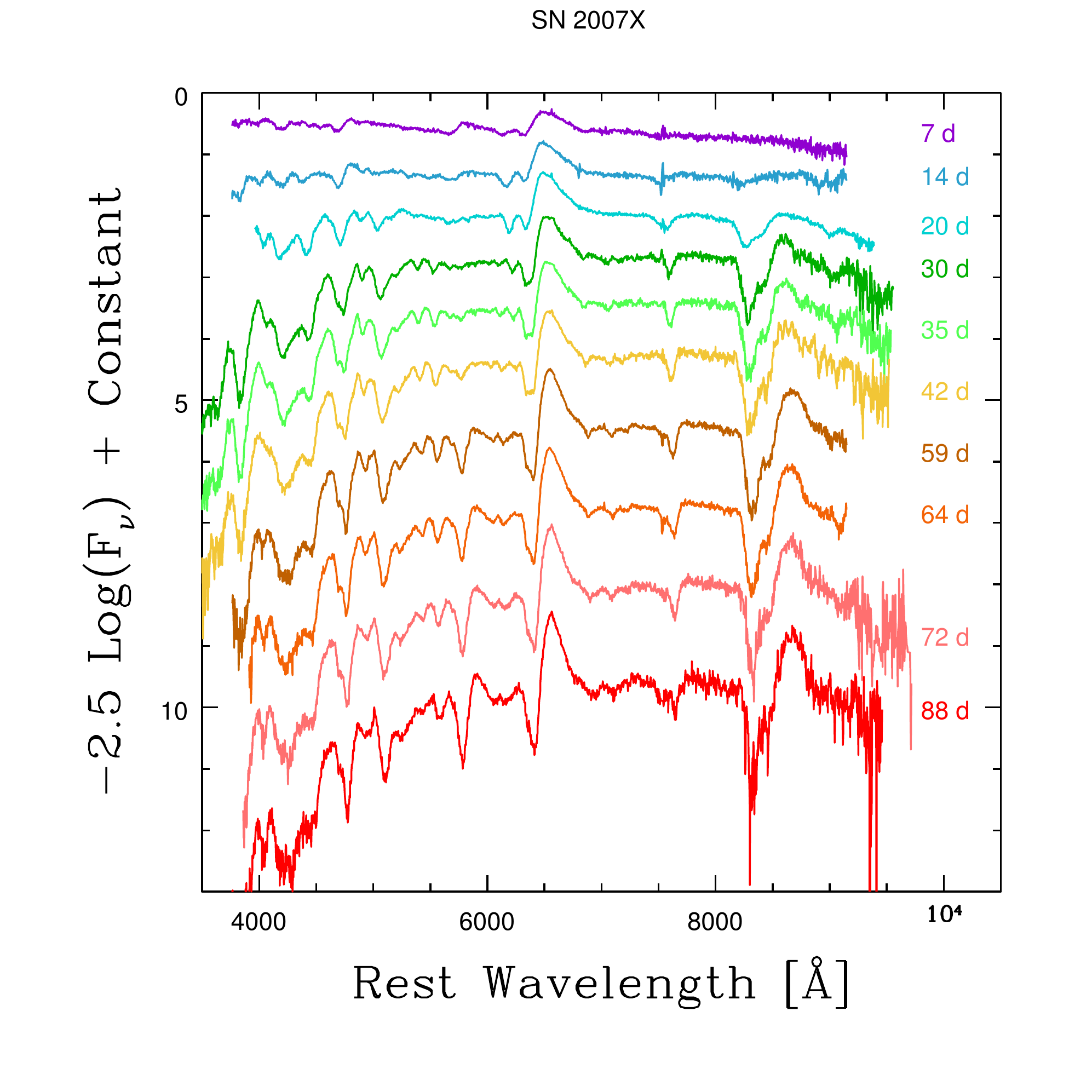}
\includegraphics[width=5.5cm]{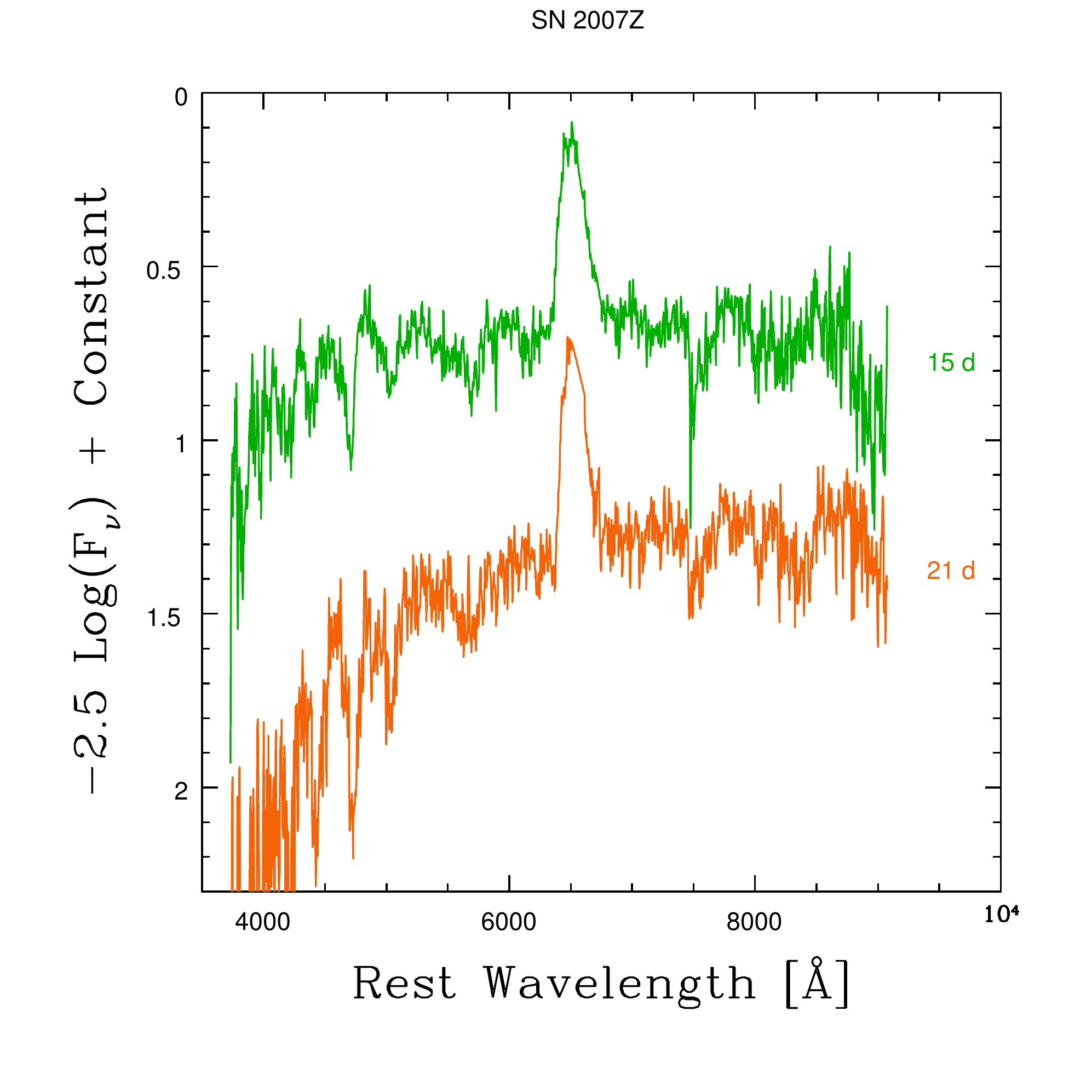}
\includegraphics[width=5.5cm]{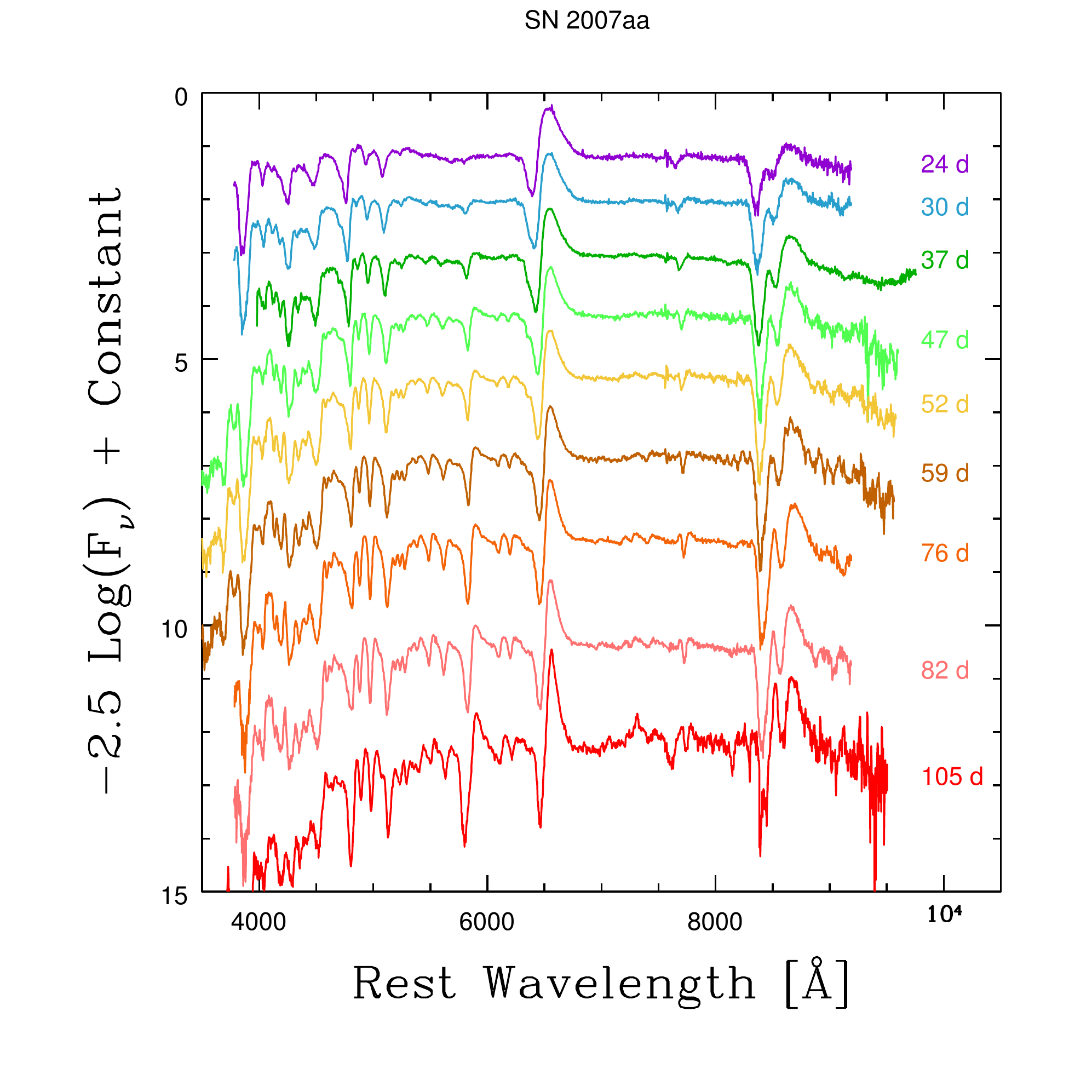}
\includegraphics[width=5.5cm]{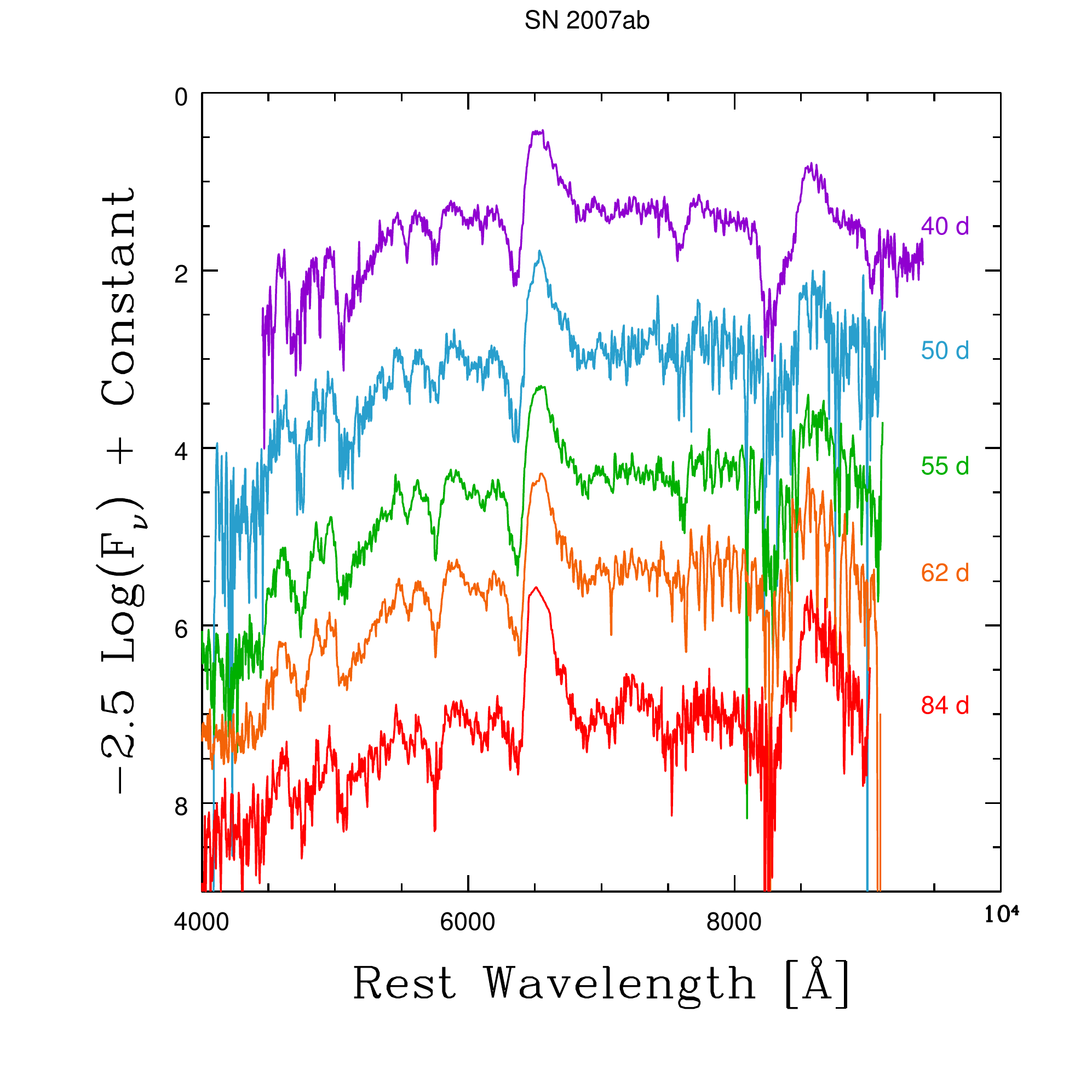}
\caption{Examples of SNe~II spectra: SN~2006ee, SN~2006it, SN~2006iw, SN~2006ms, SN~2006qr, SN~2007P, SN~2007U, SN~2007W, SN~2007X, SN~2007Z, SN~2007aa, SN~2007ab.}
\label{example}
\end{figure*}

\begin{figure*}[h!]
\centering
\includegraphics[width=5.5cm]{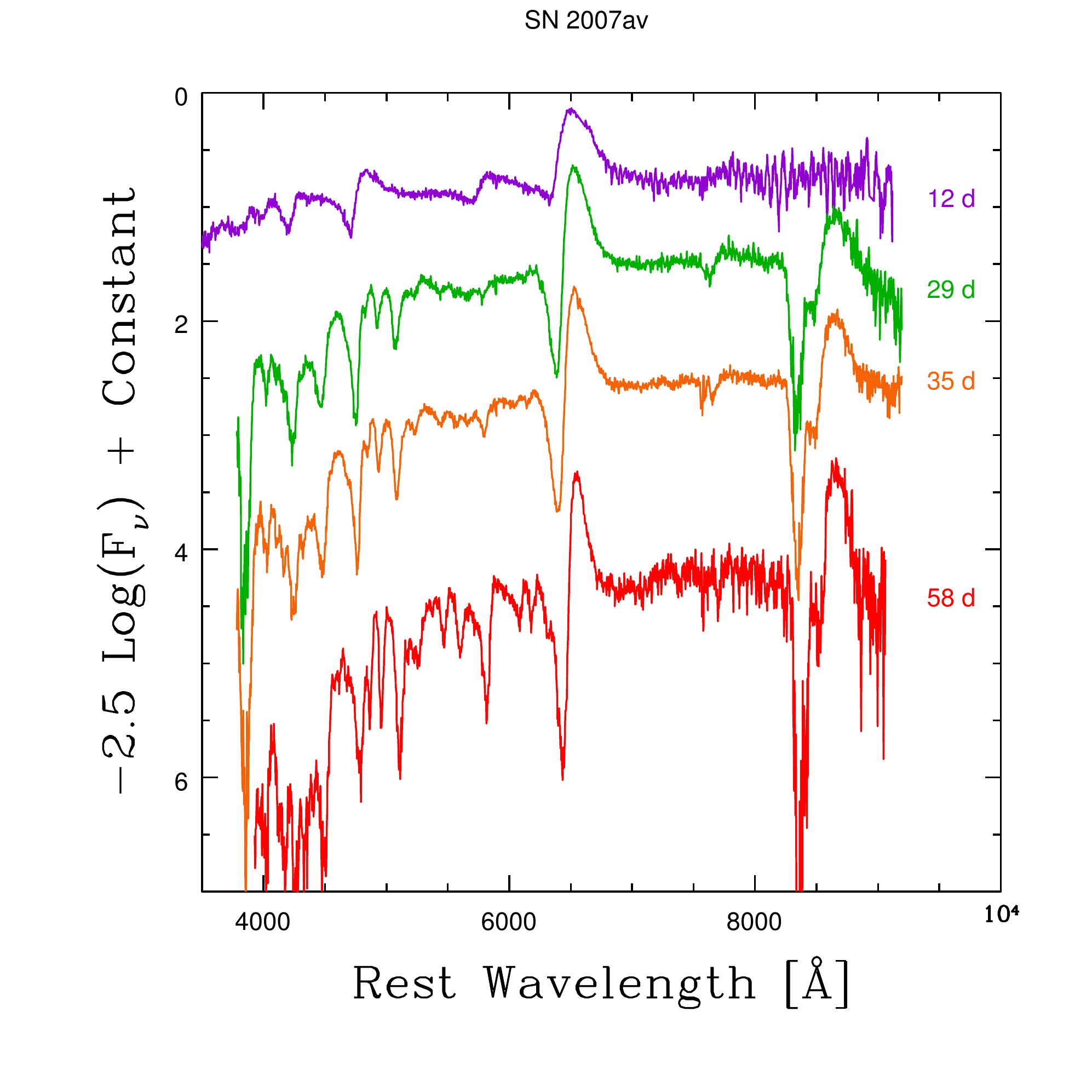}
\includegraphics[width=5.5cm]{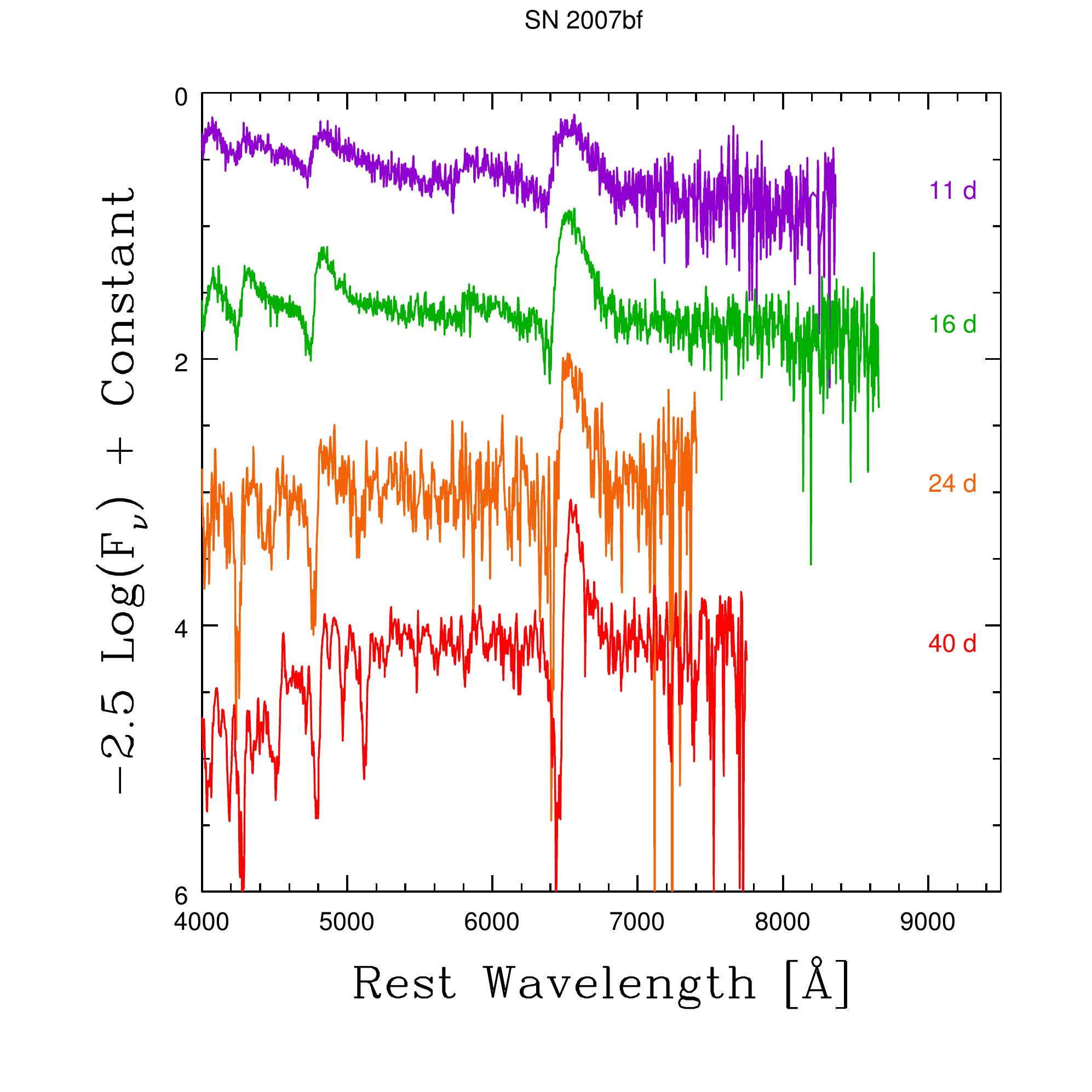}
\includegraphics[width=5.5cm]{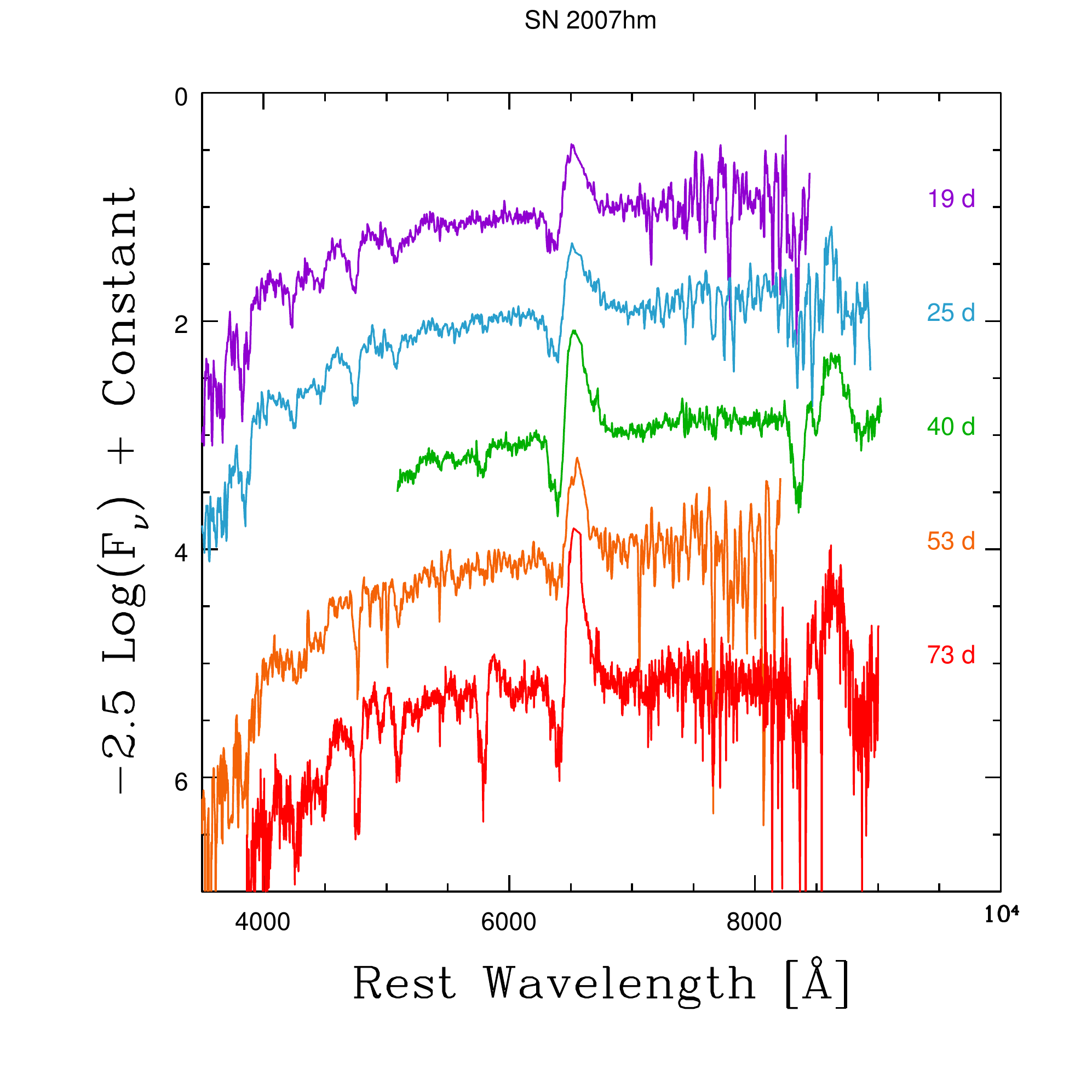}
\includegraphics[width=5.5cm]{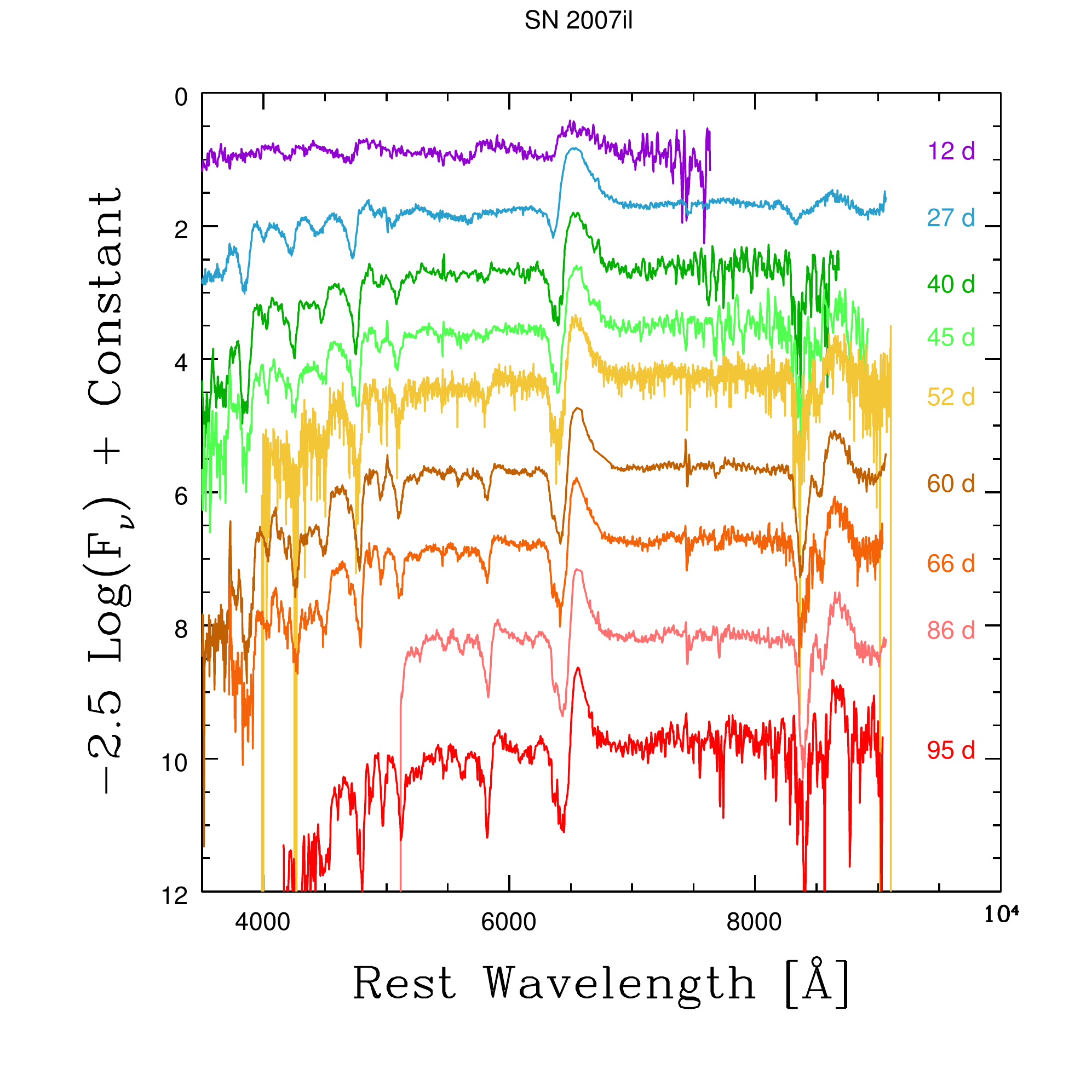}
\includegraphics[width=5.5cm]{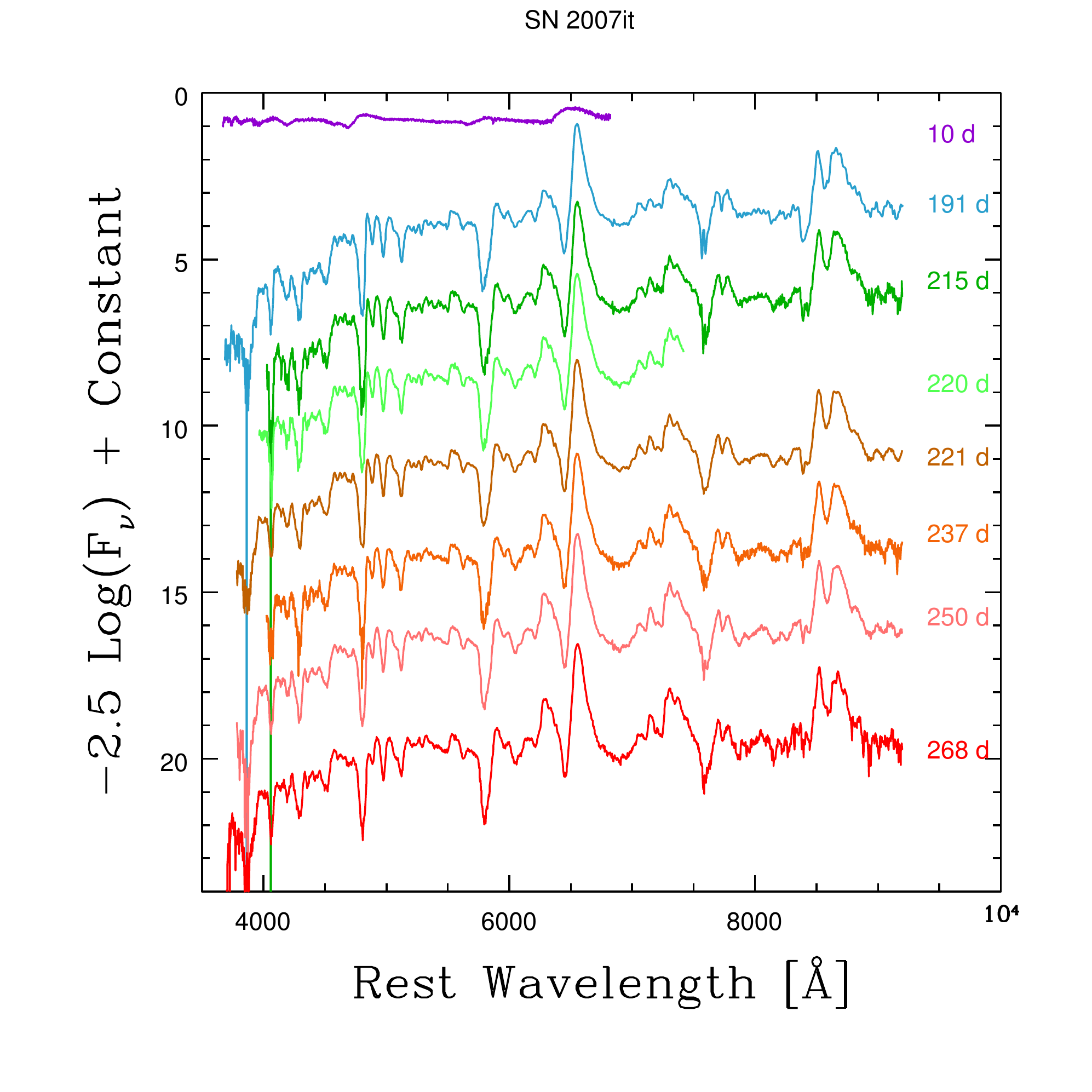}
\includegraphics[width=5.5cm]{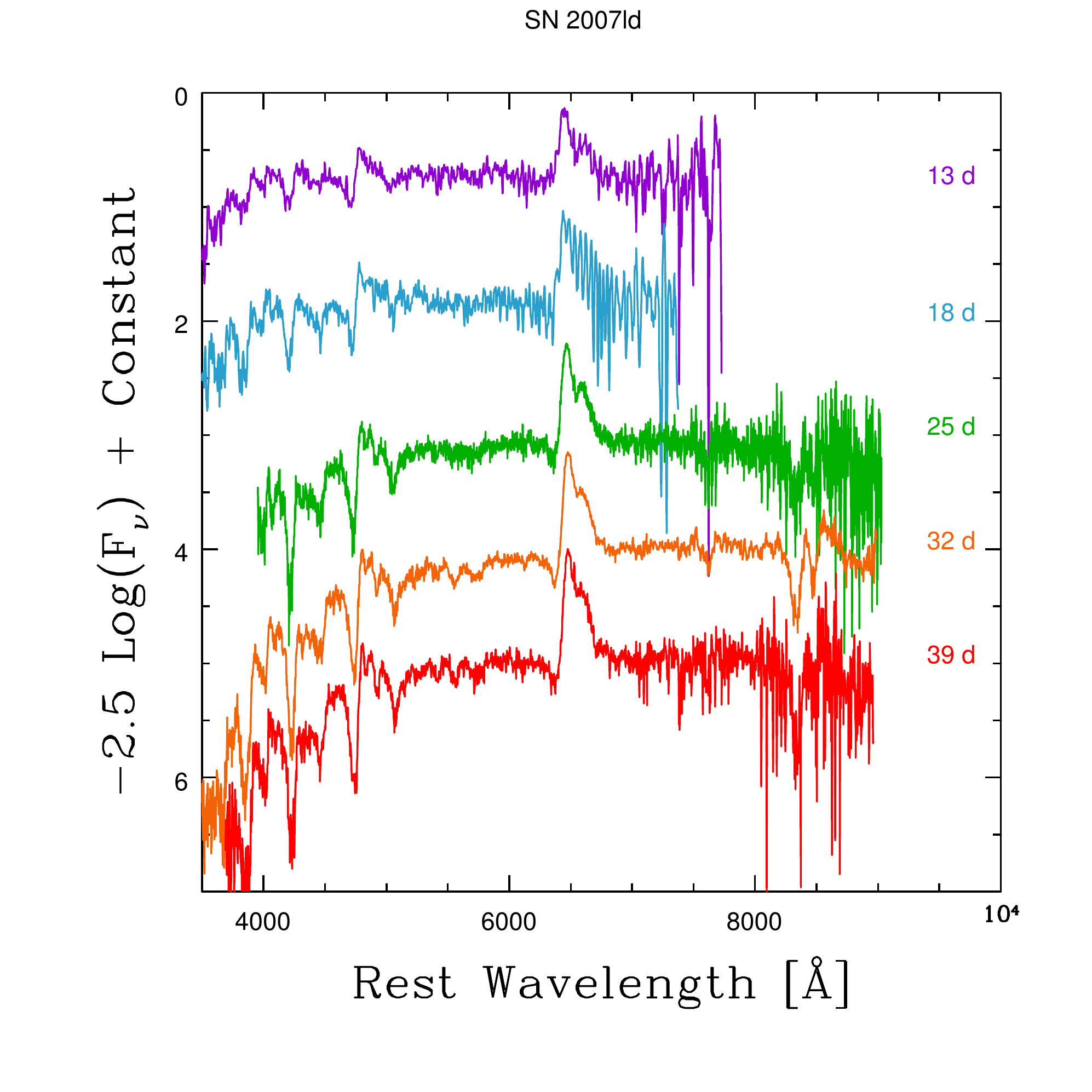}
\includegraphics[width=5.5cm]{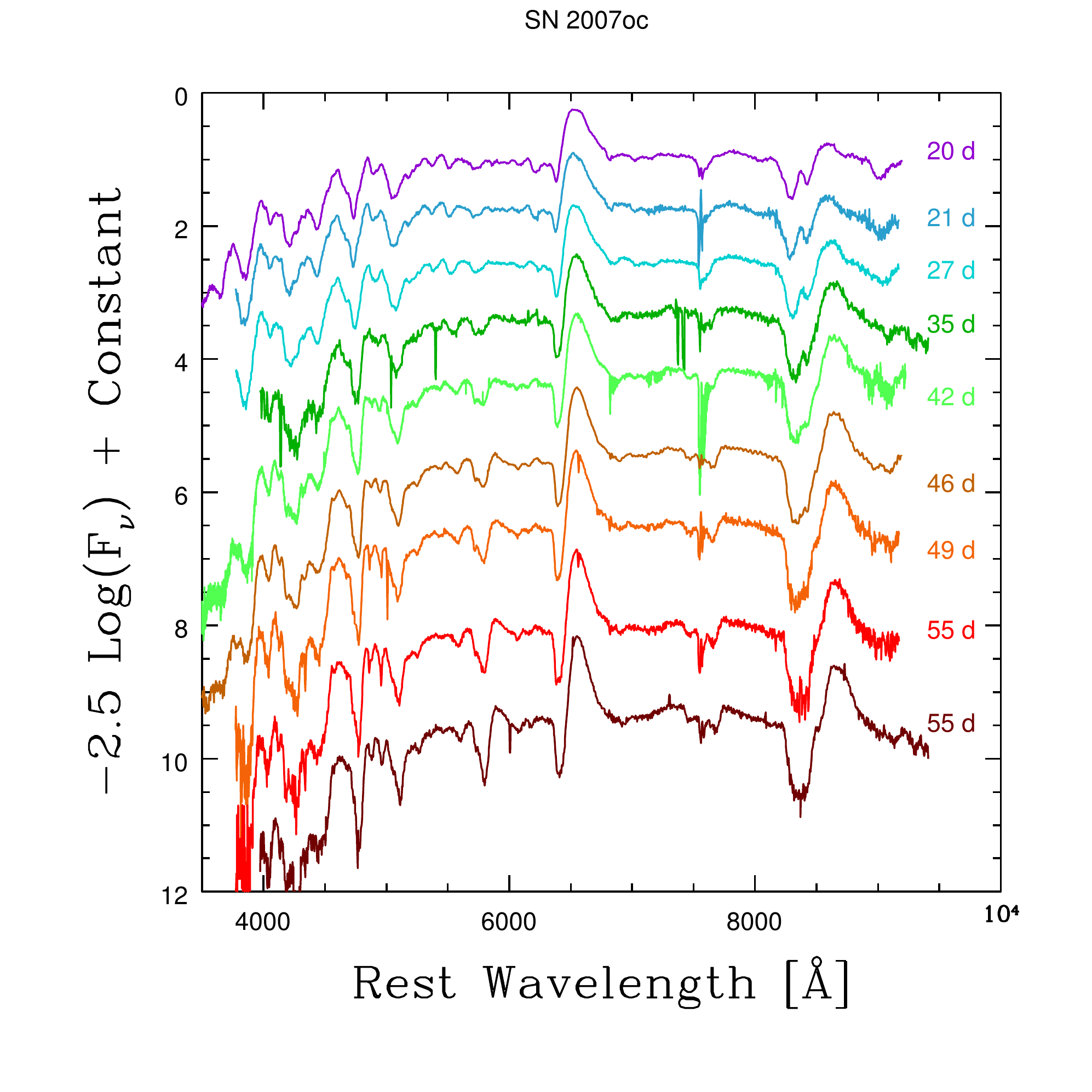}
\includegraphics[width=5.5cm]{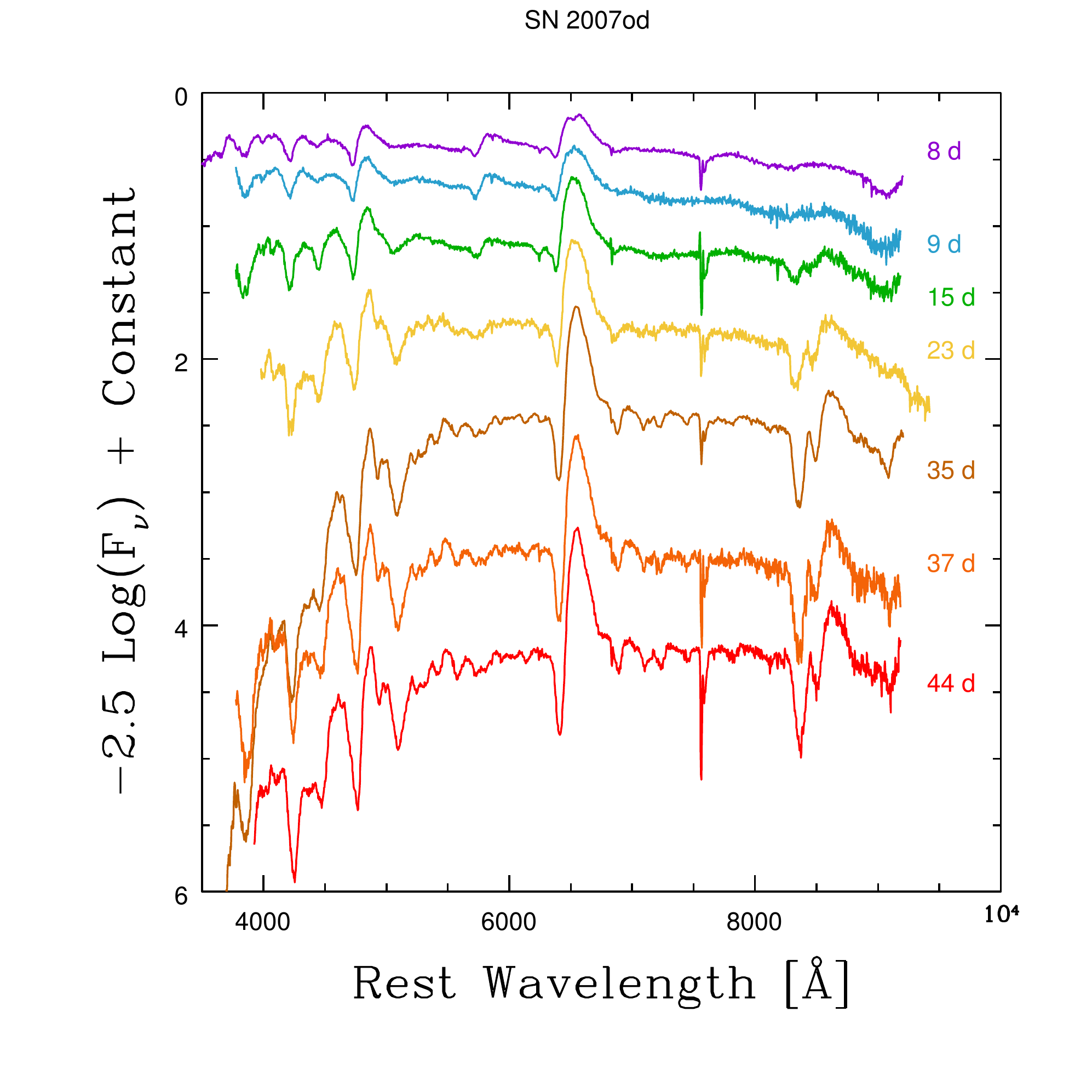}
\includegraphics[width=5.5cm]{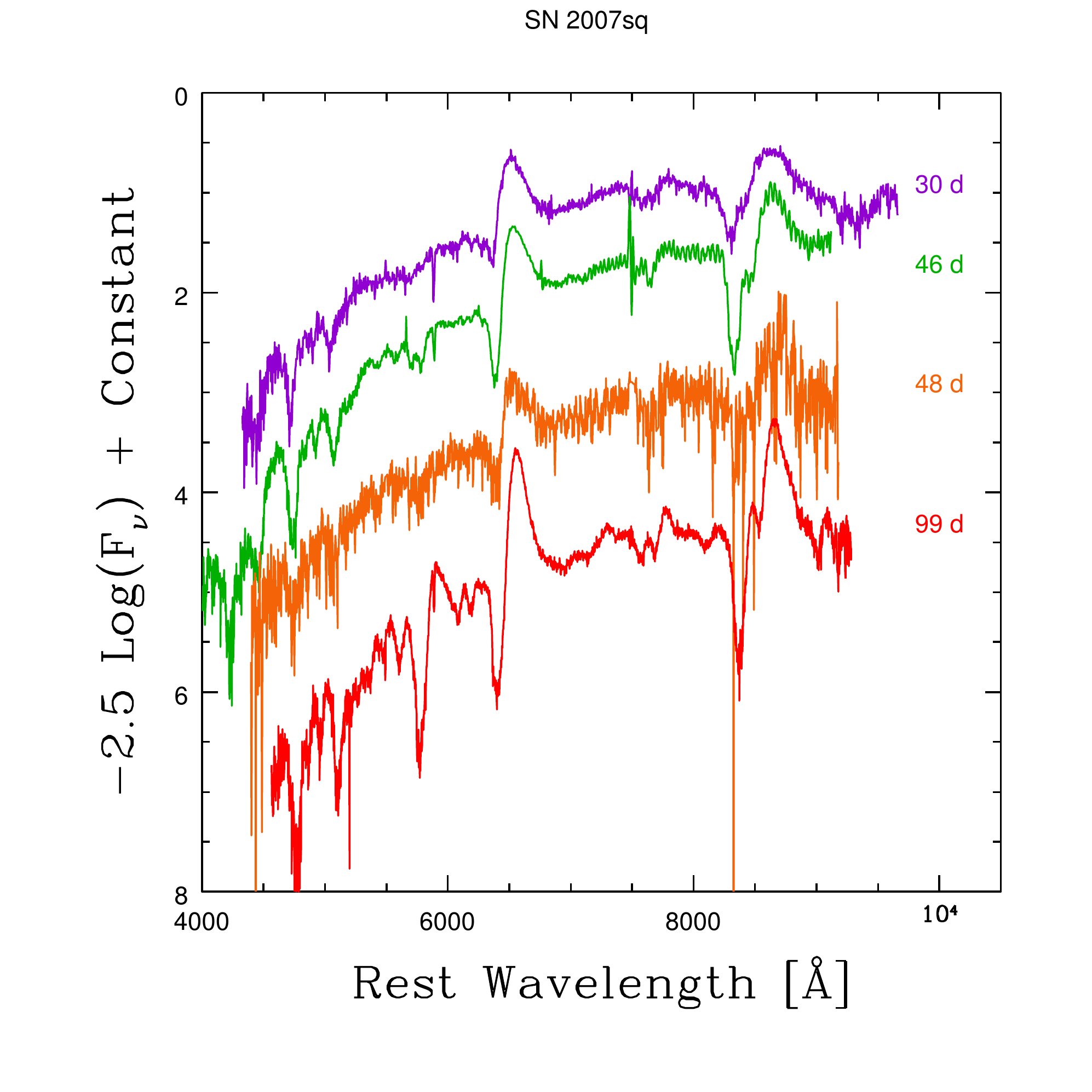}
\includegraphics[width=5.5cm]{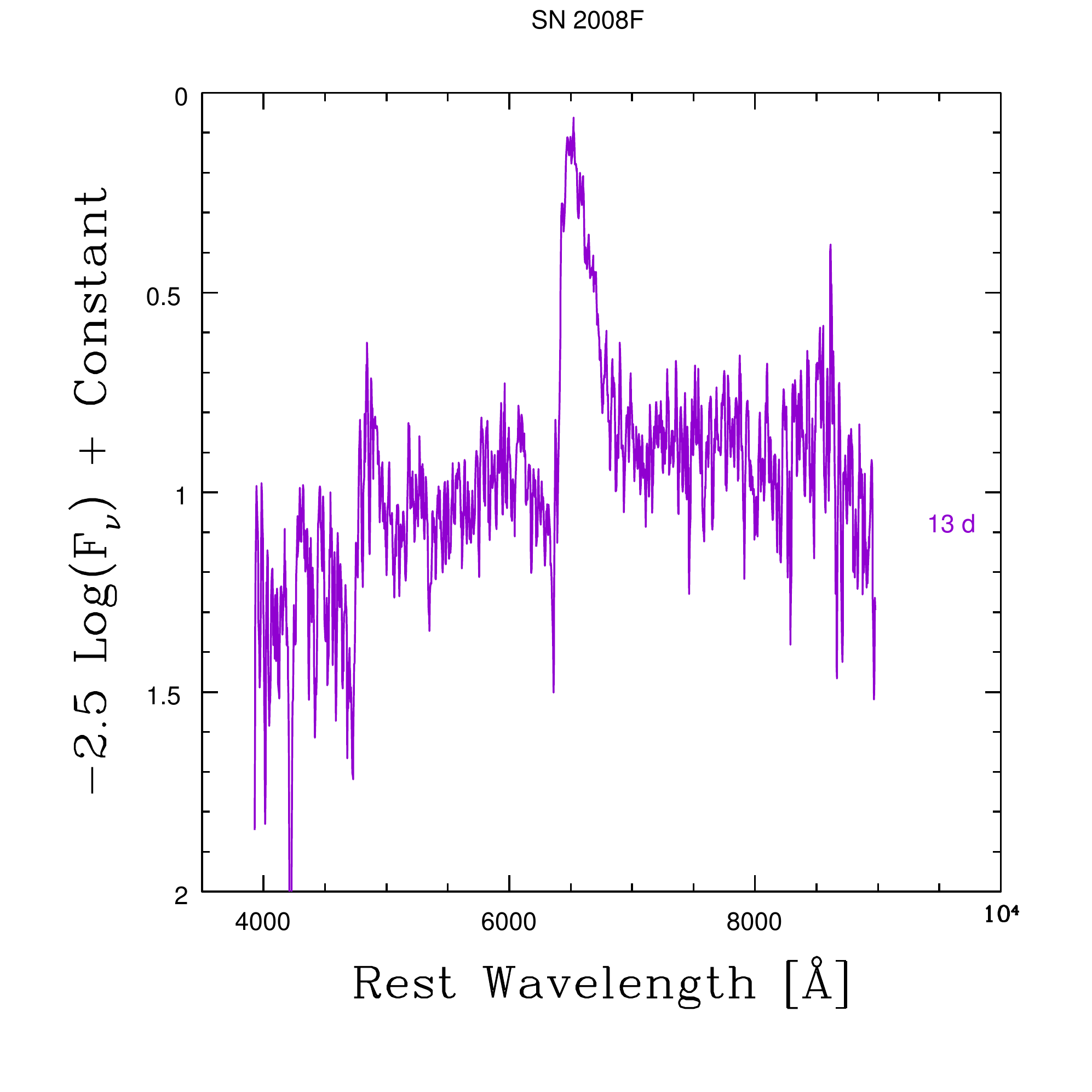}
\includegraphics[width=5.5cm]{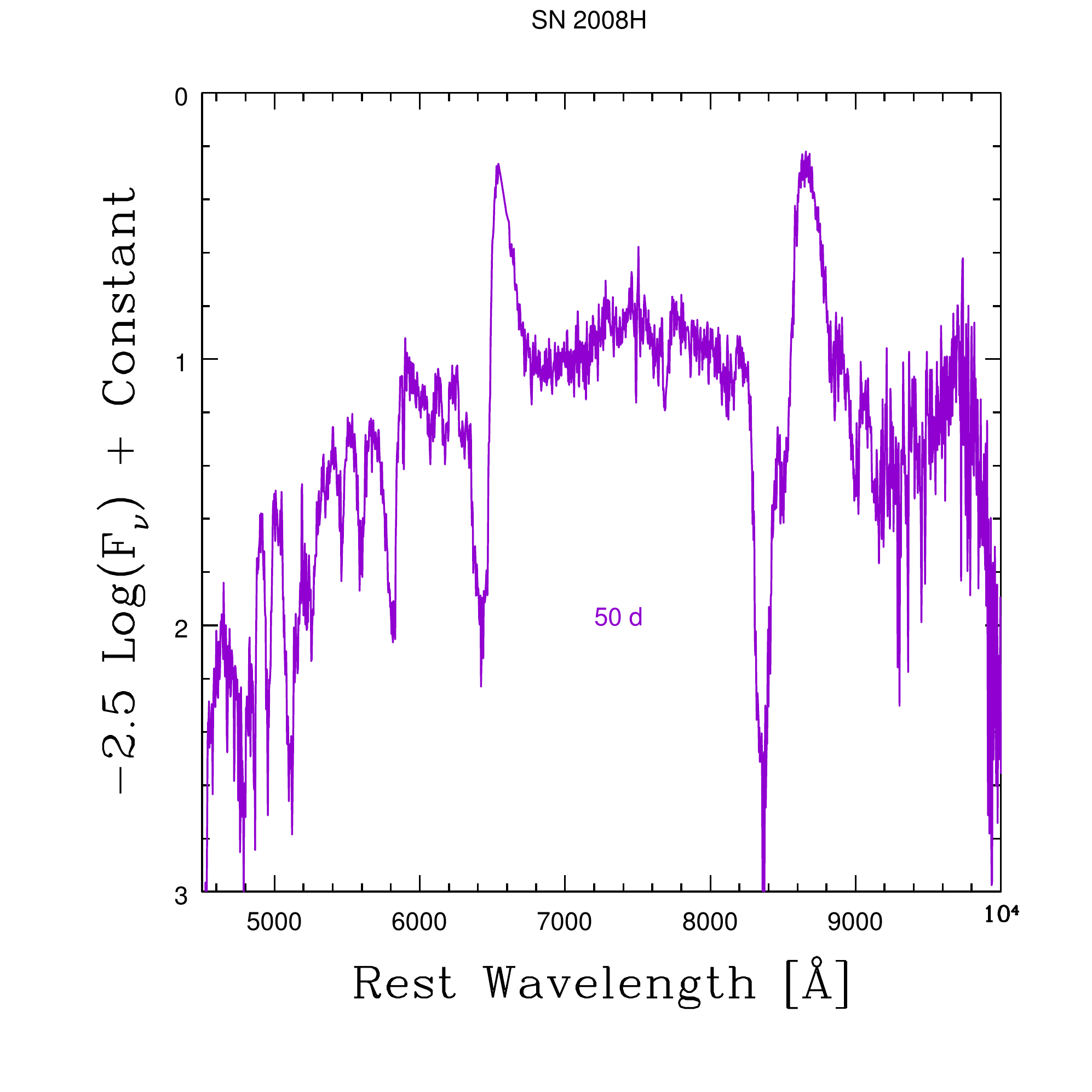}
\includegraphics[width=5.5cm]{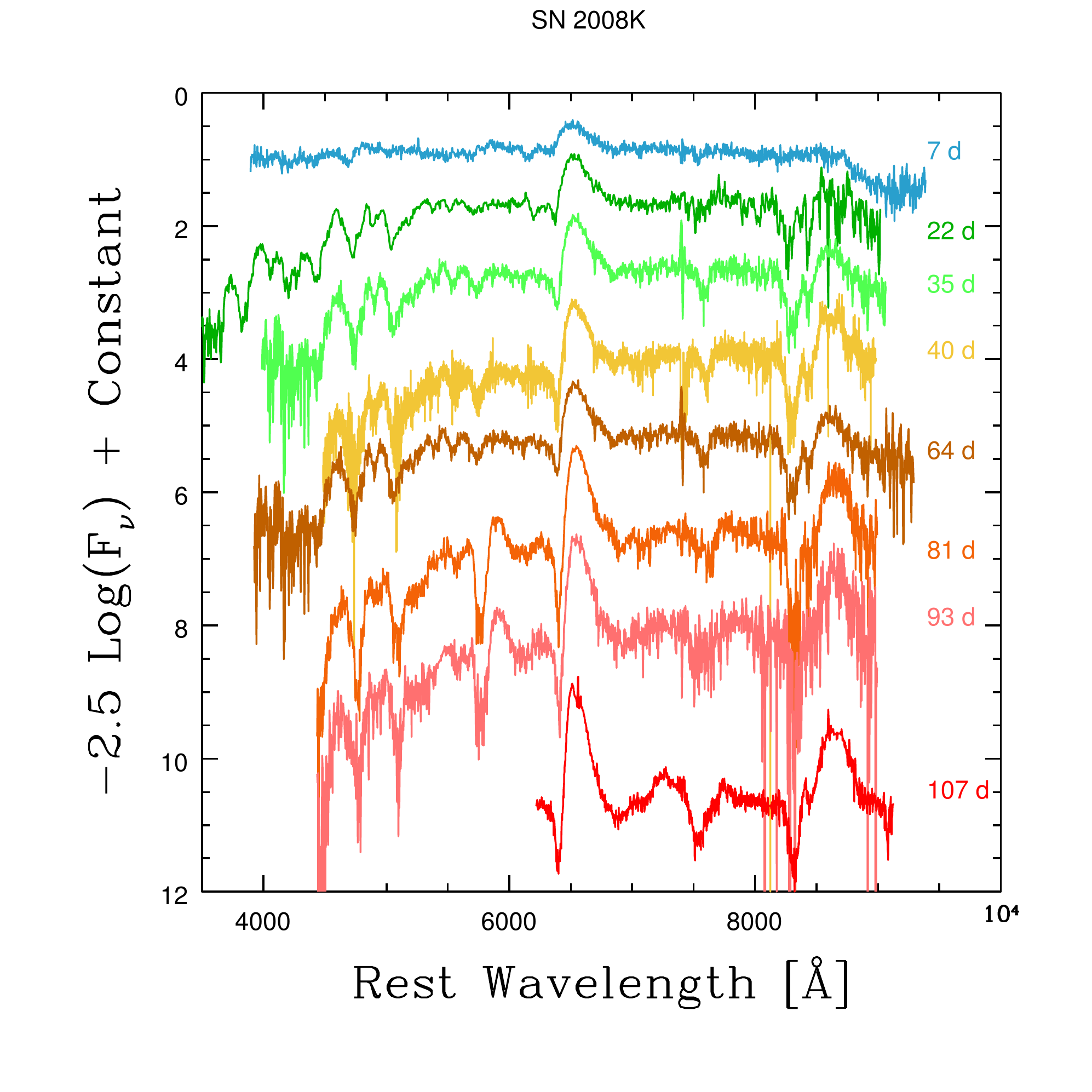}
\caption{Examples of SNe~II spectra: SN~2007av, SN~2007bf, SN~2007hm, SN~2007il, SN~2007it, SN~2007ld, SN~2007oc, SN~2007od, SN~2007sq, SN~2008F, SN~2008H, SN~2008K.}
\label{example}
\end{figure*}

\begin{figure*}[h!]
\centering
\includegraphics[width=5.5cm]{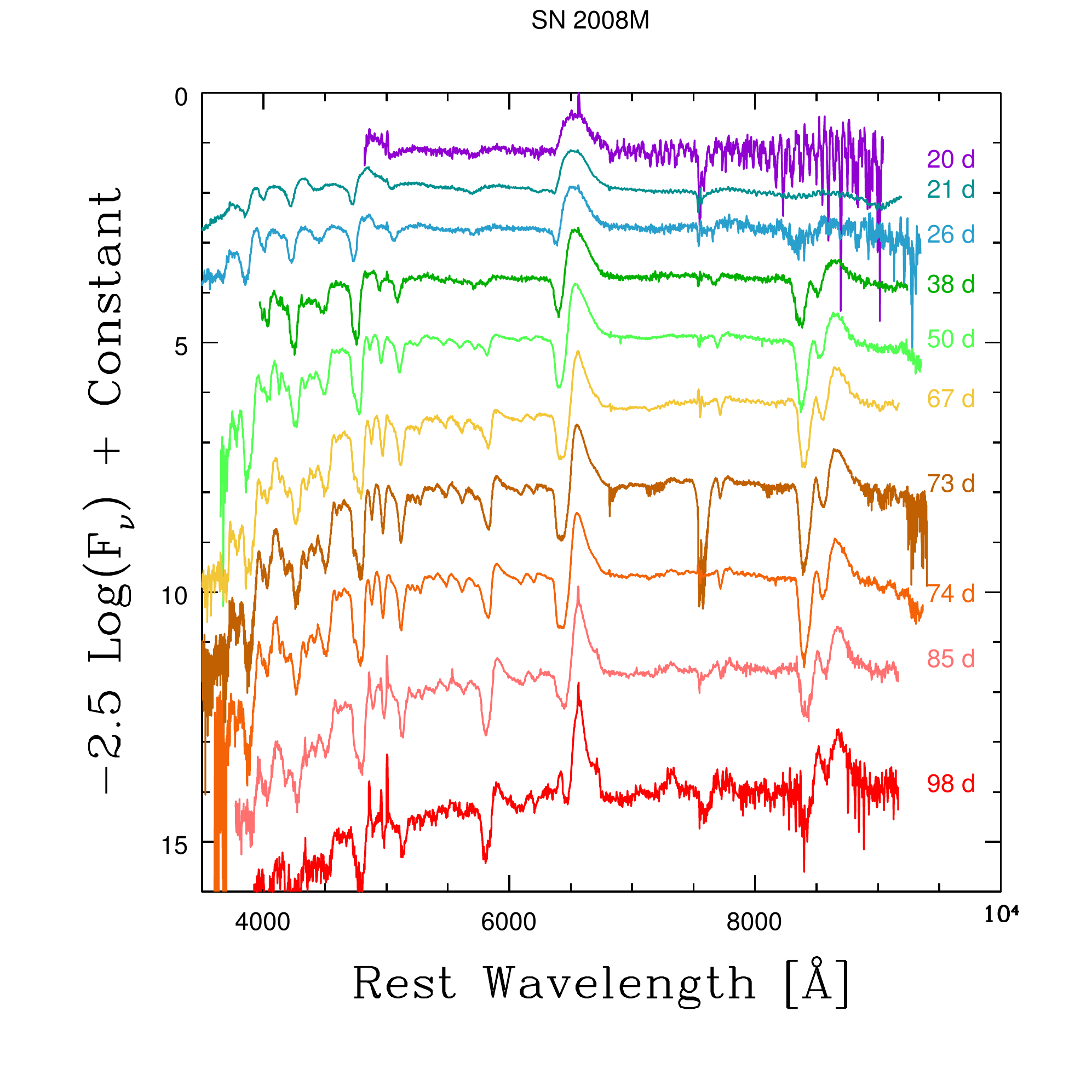}
\includegraphics[width=5.5cm]{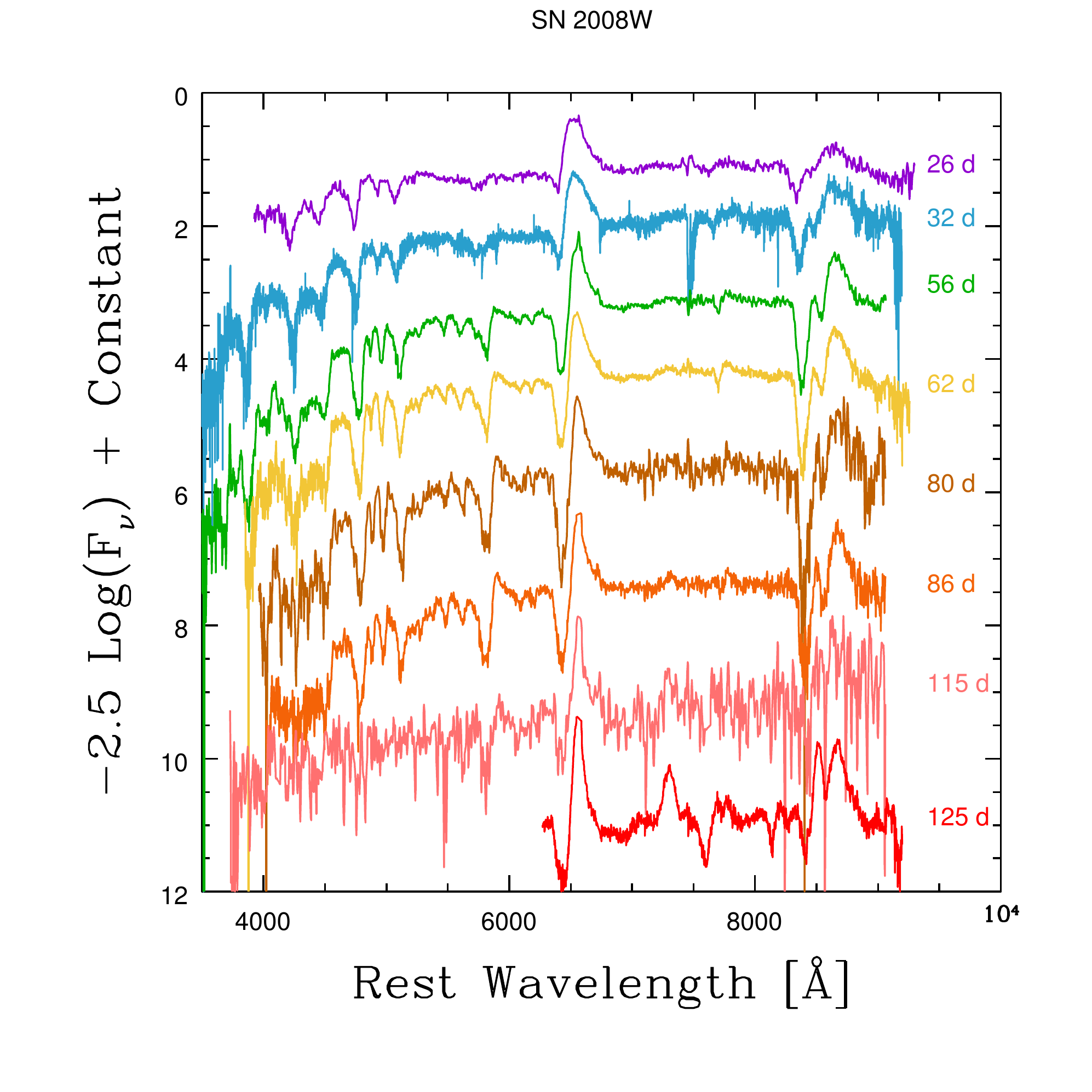}
\includegraphics[width=5.5cm]{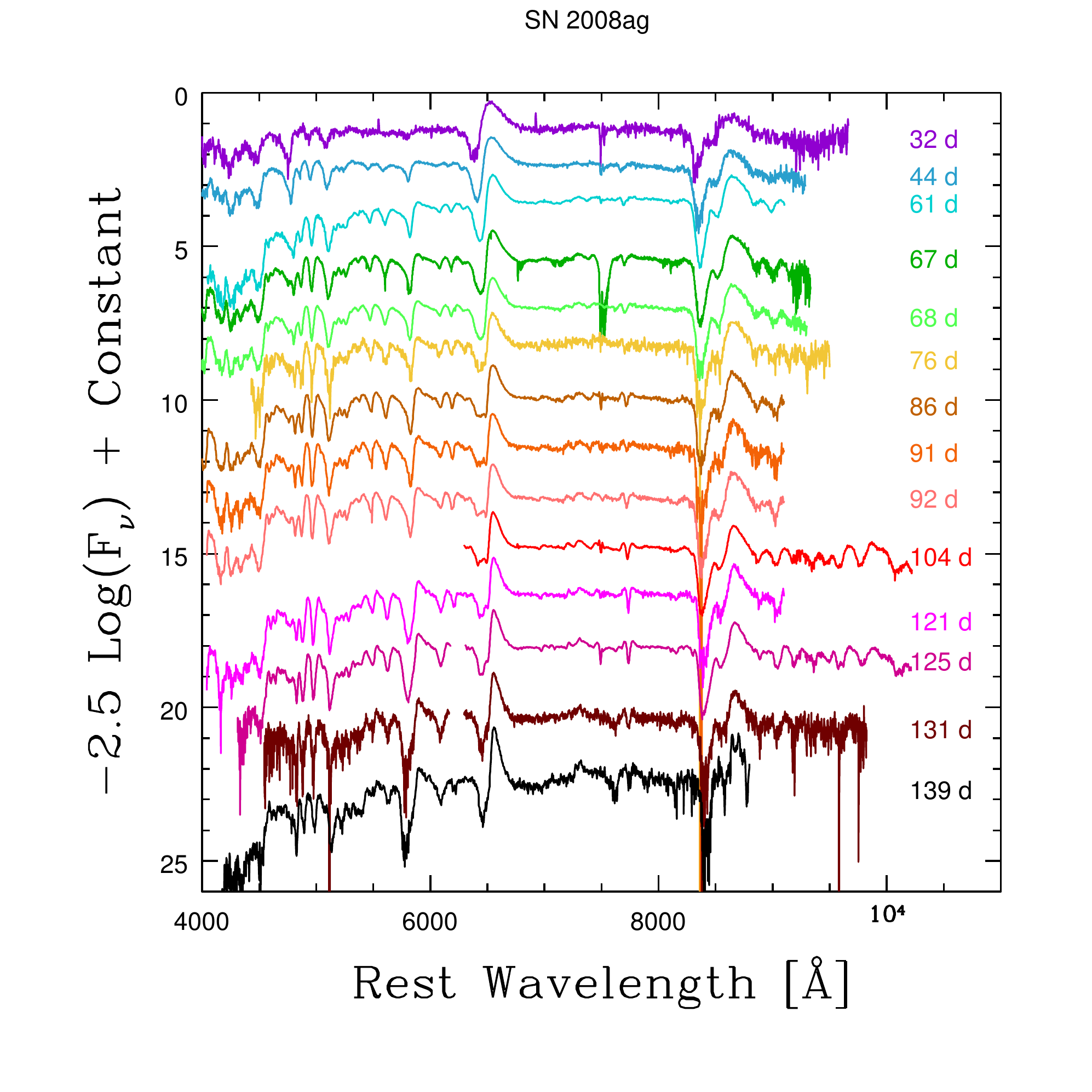}
\includegraphics[width=5.5cm]{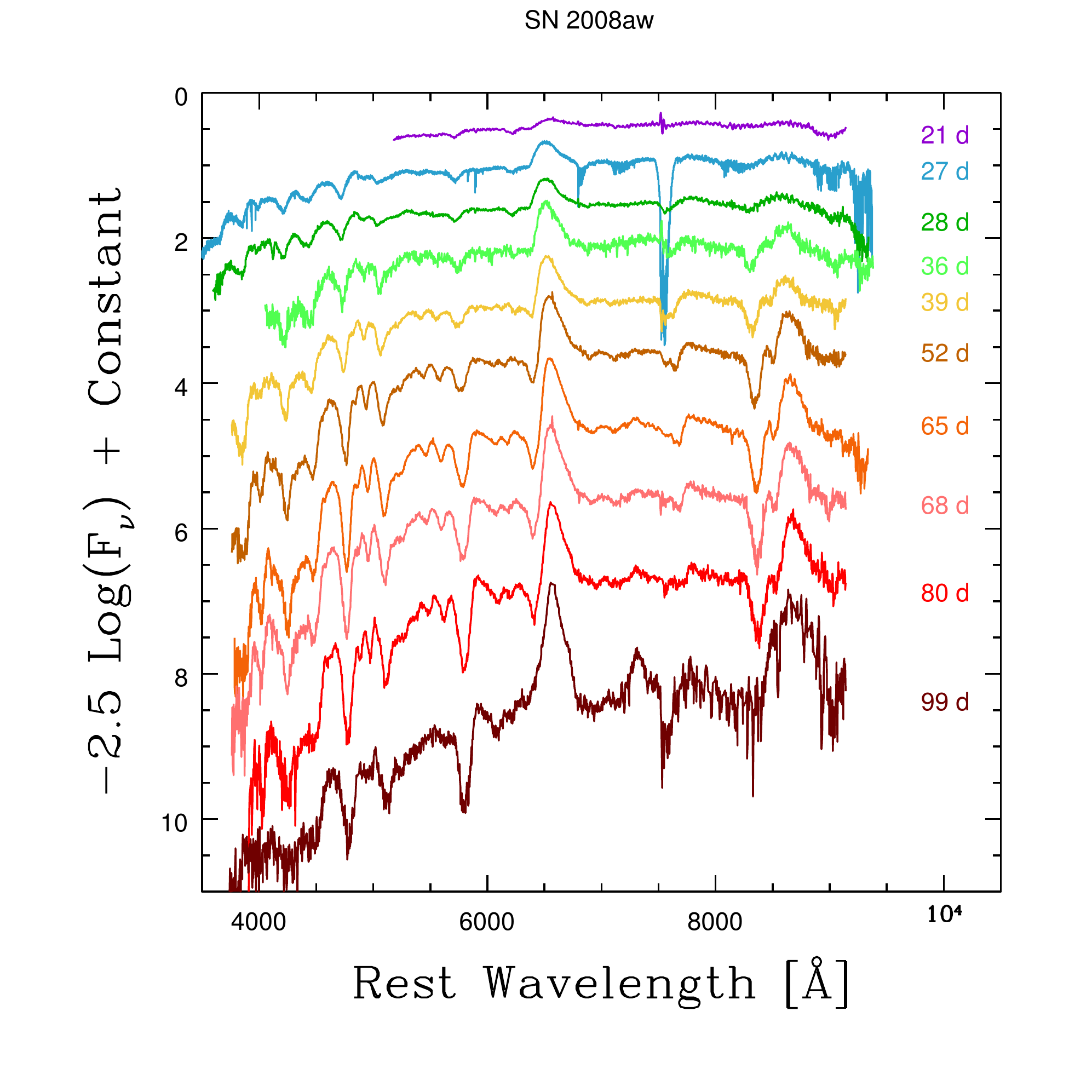}
\includegraphics[width=5.5cm]{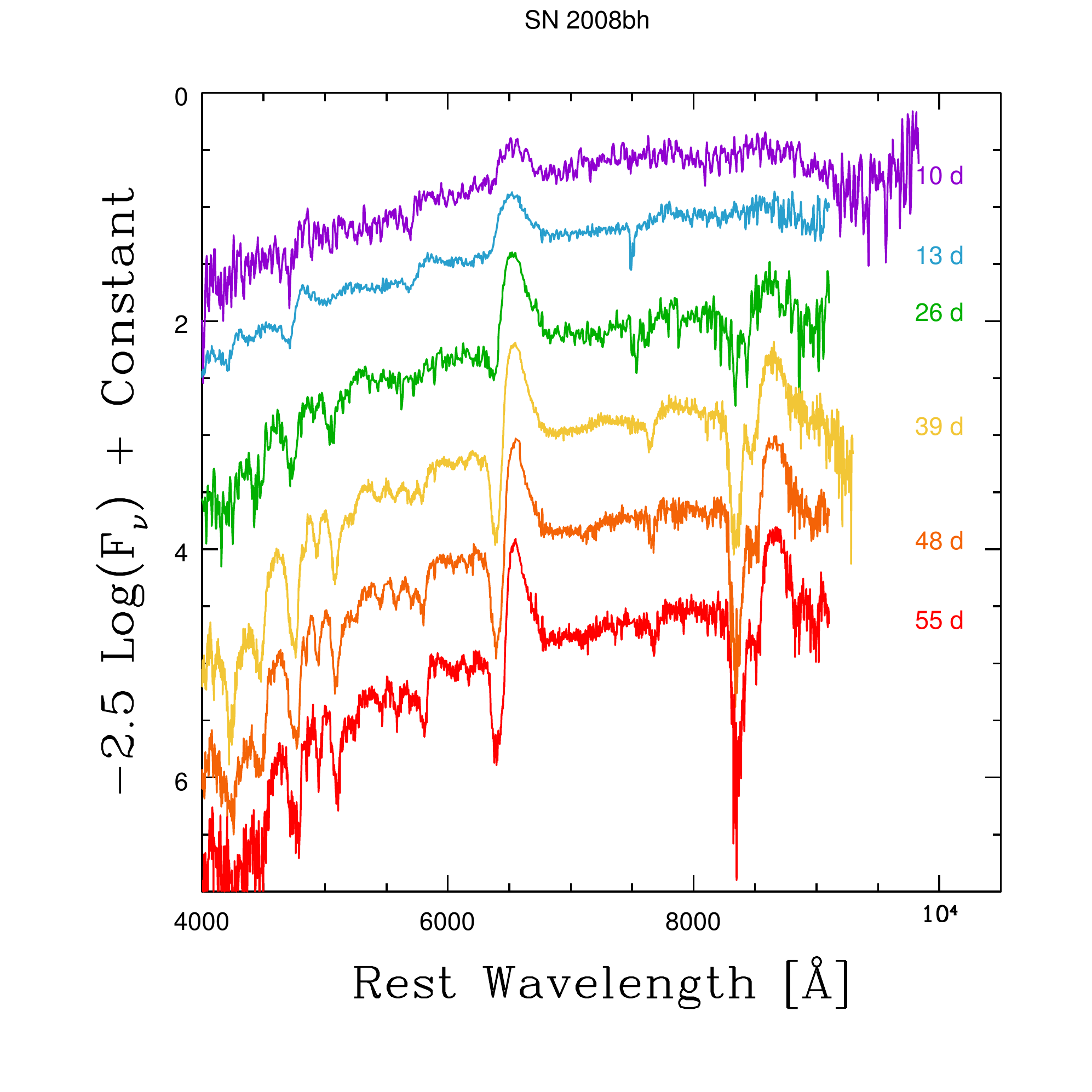}
\includegraphics[width=5.5cm]{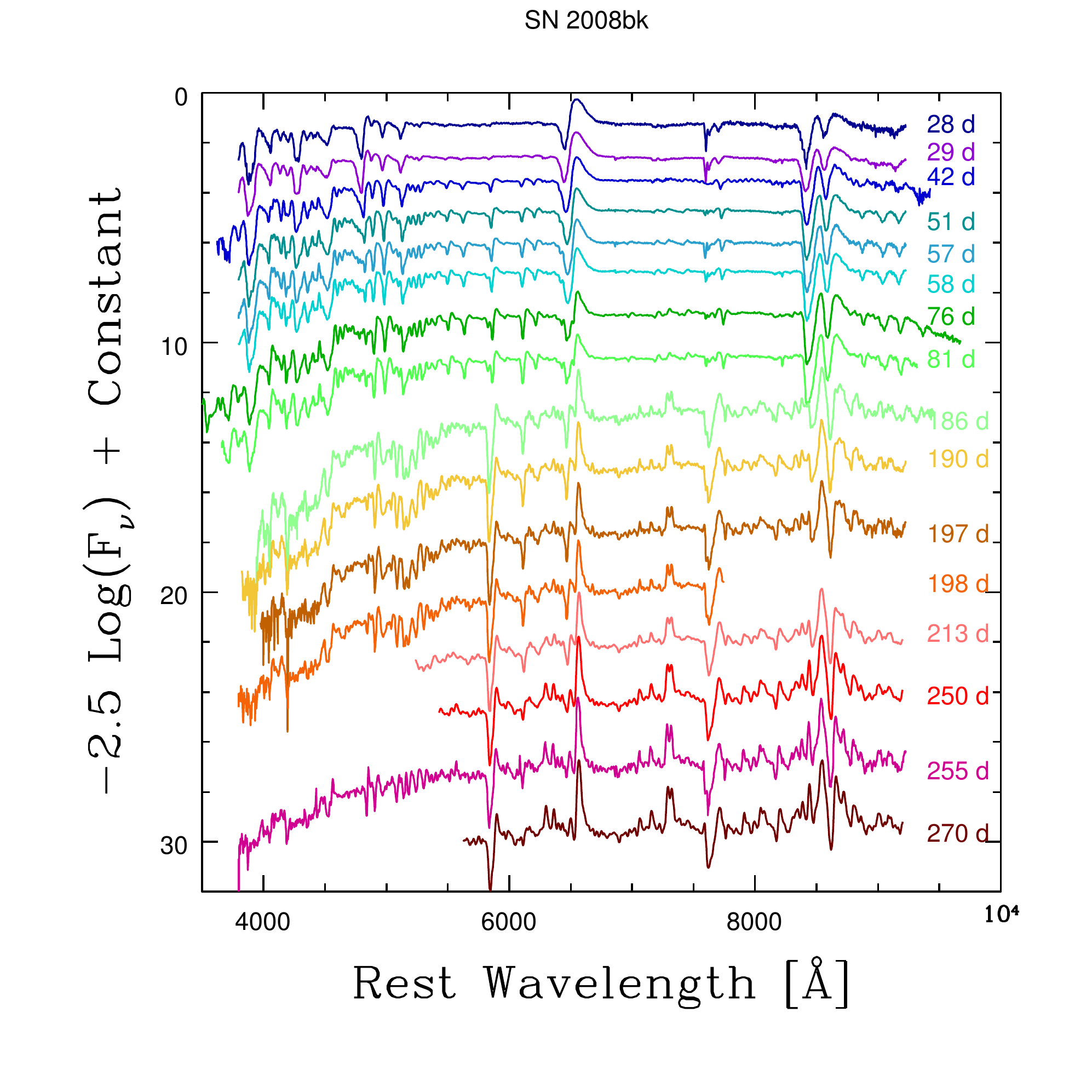}
\includegraphics[width=5.5cm]{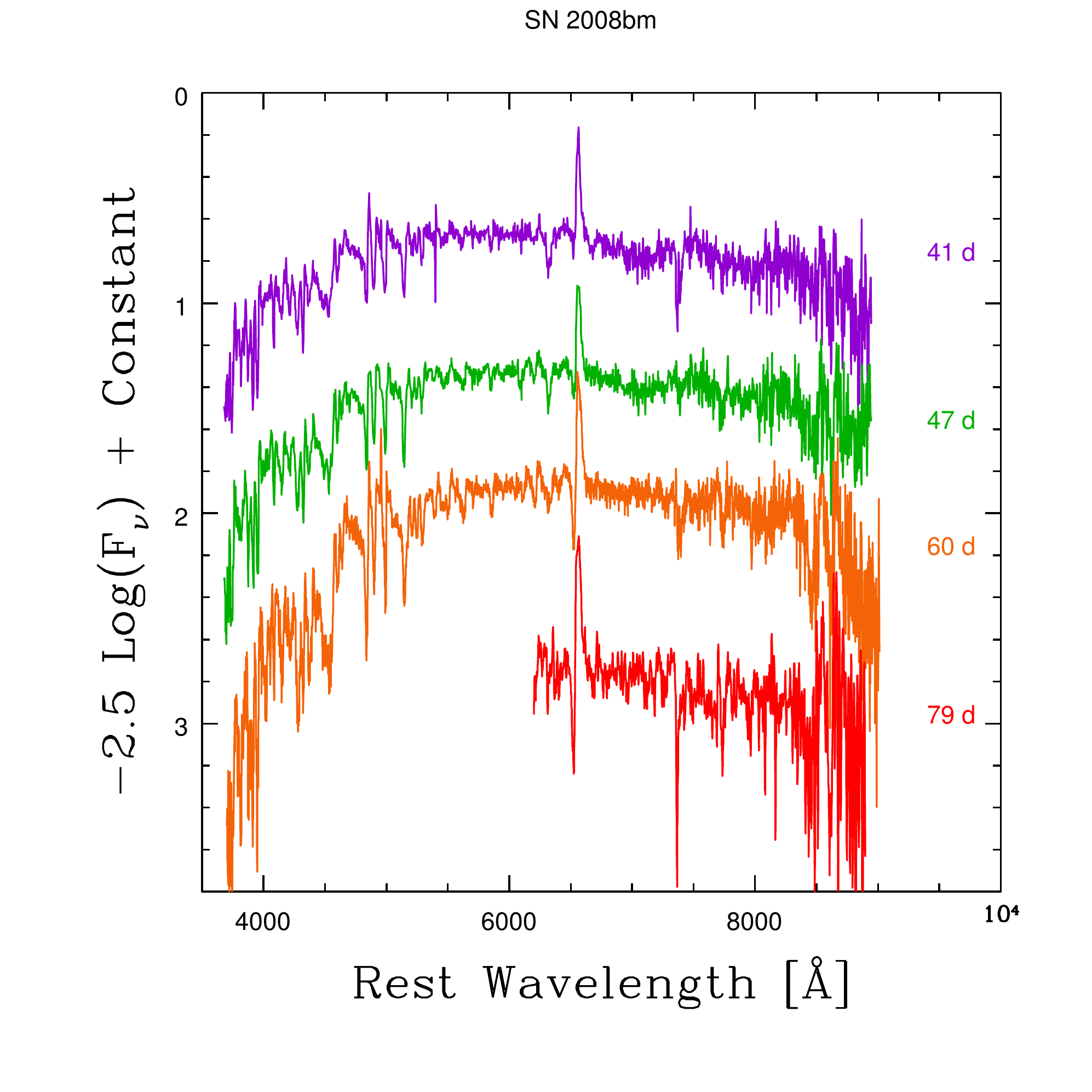}
\includegraphics[width=5.5cm]{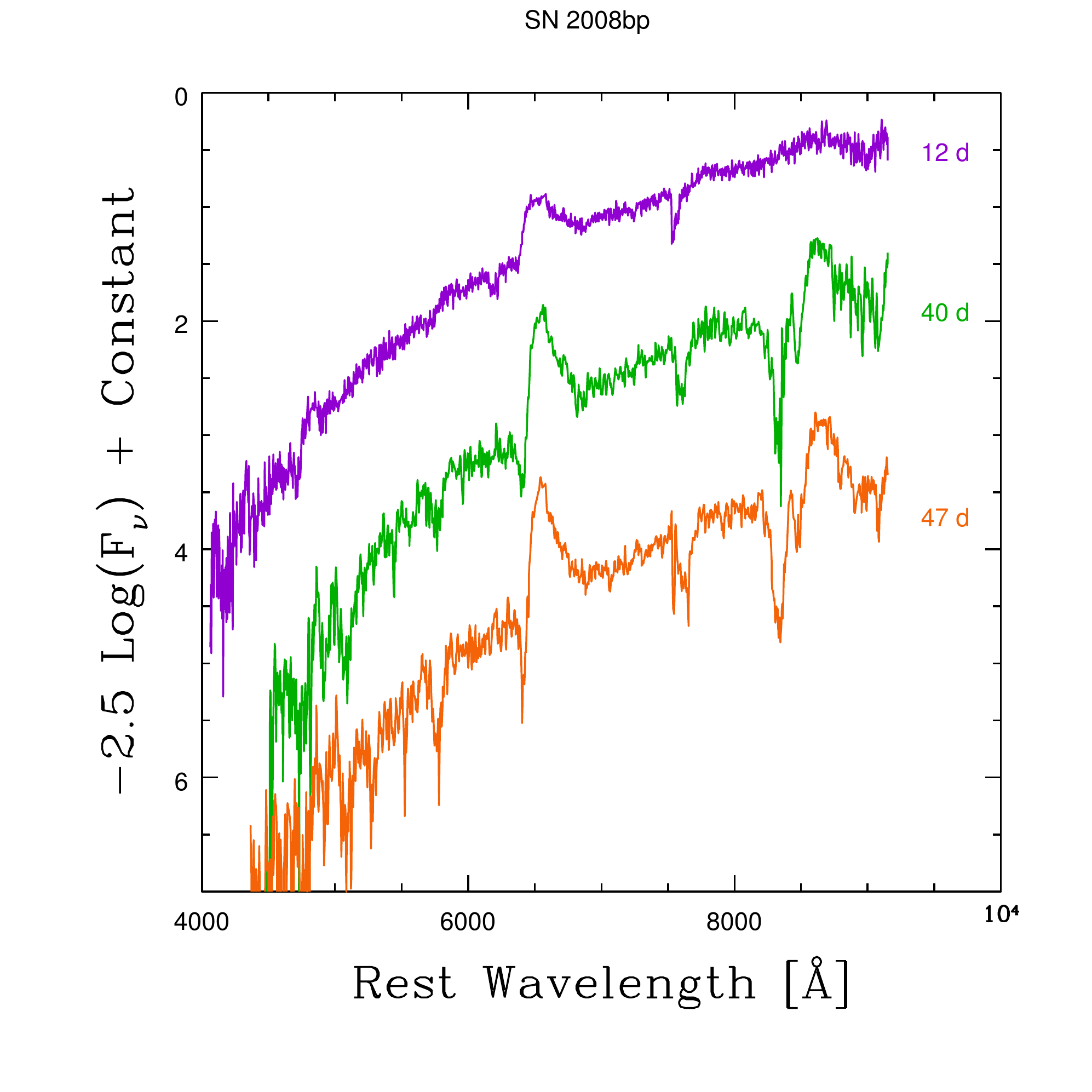}
\includegraphics[width=5.5cm]{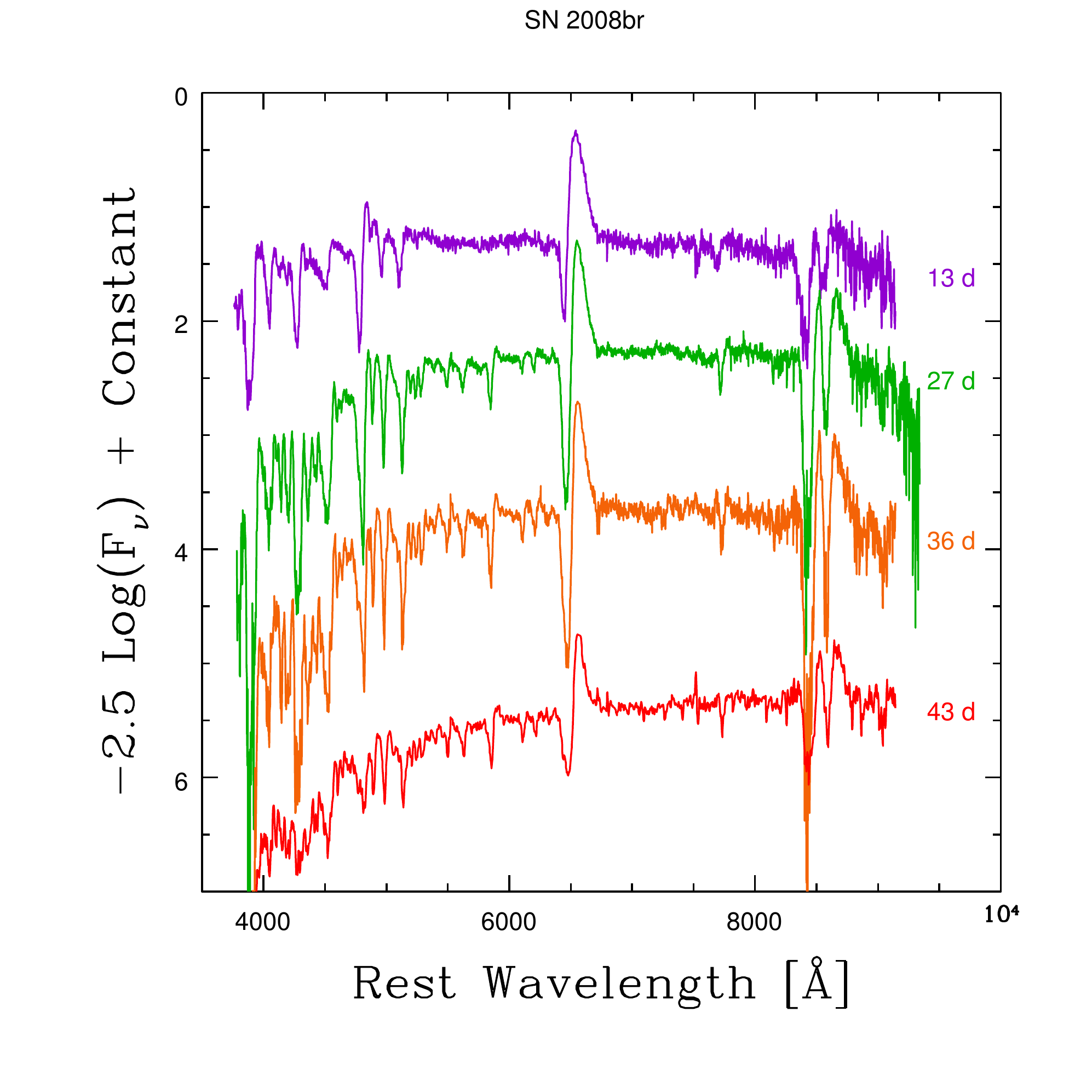}
\includegraphics[width=5.5cm]{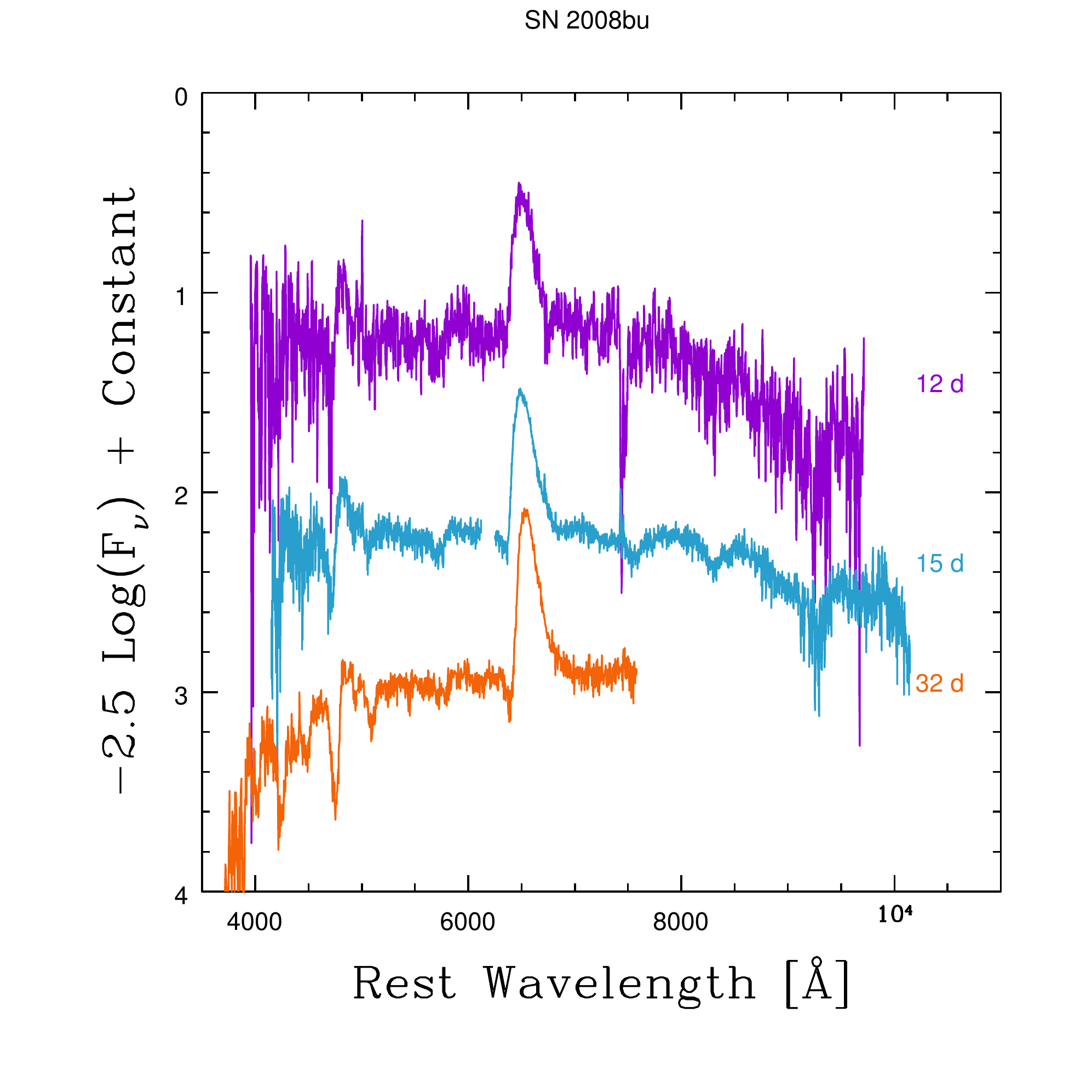}
\includegraphics[width=5.5cm]{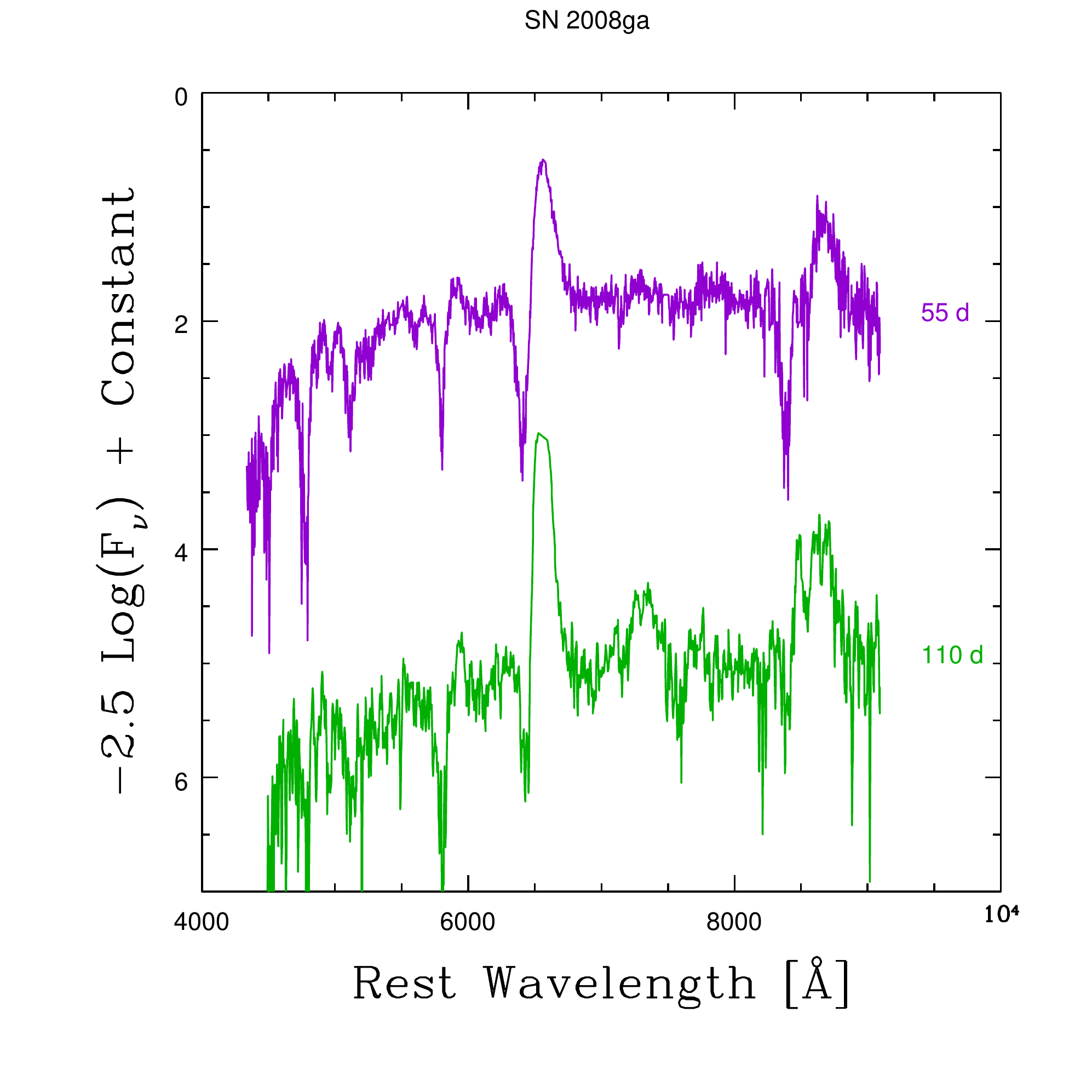}
\includegraphics[width=5.5cm]{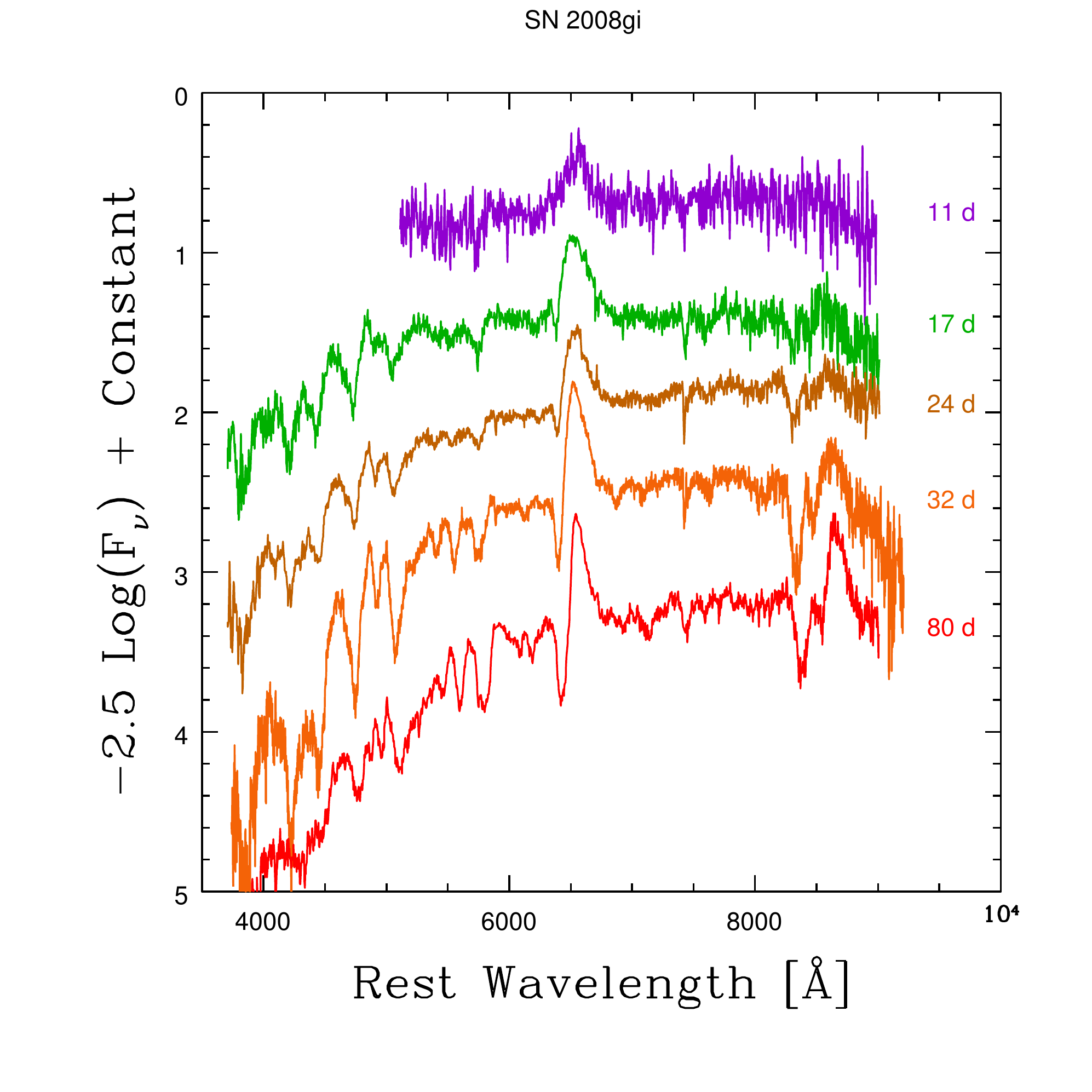}
\caption{Examples of SNe~II spectra: SN~2008M, SN~2008W, SN~2008ag, SN~2008aw, SN~2008bh, SN~2008bk, SN~2008bm, SN~2008bp, SN~2008br, SN~2008bu, SN~2008ga, SN~2008gi.}
\label{example}
\end{figure*}

\begin{figure*}[h!]
\centering
\includegraphics[width=5.5cm]{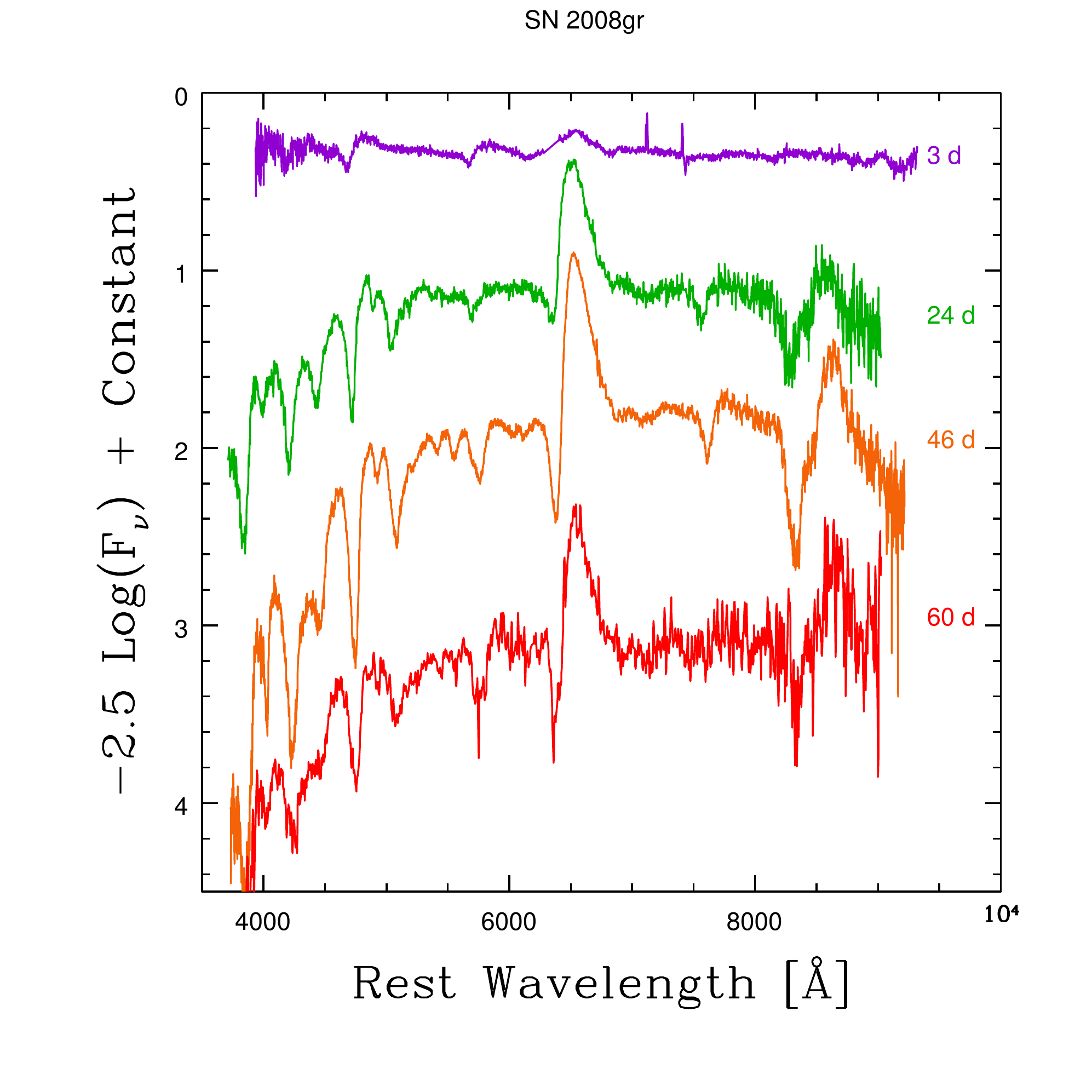}
\includegraphics[width=5.5cm]{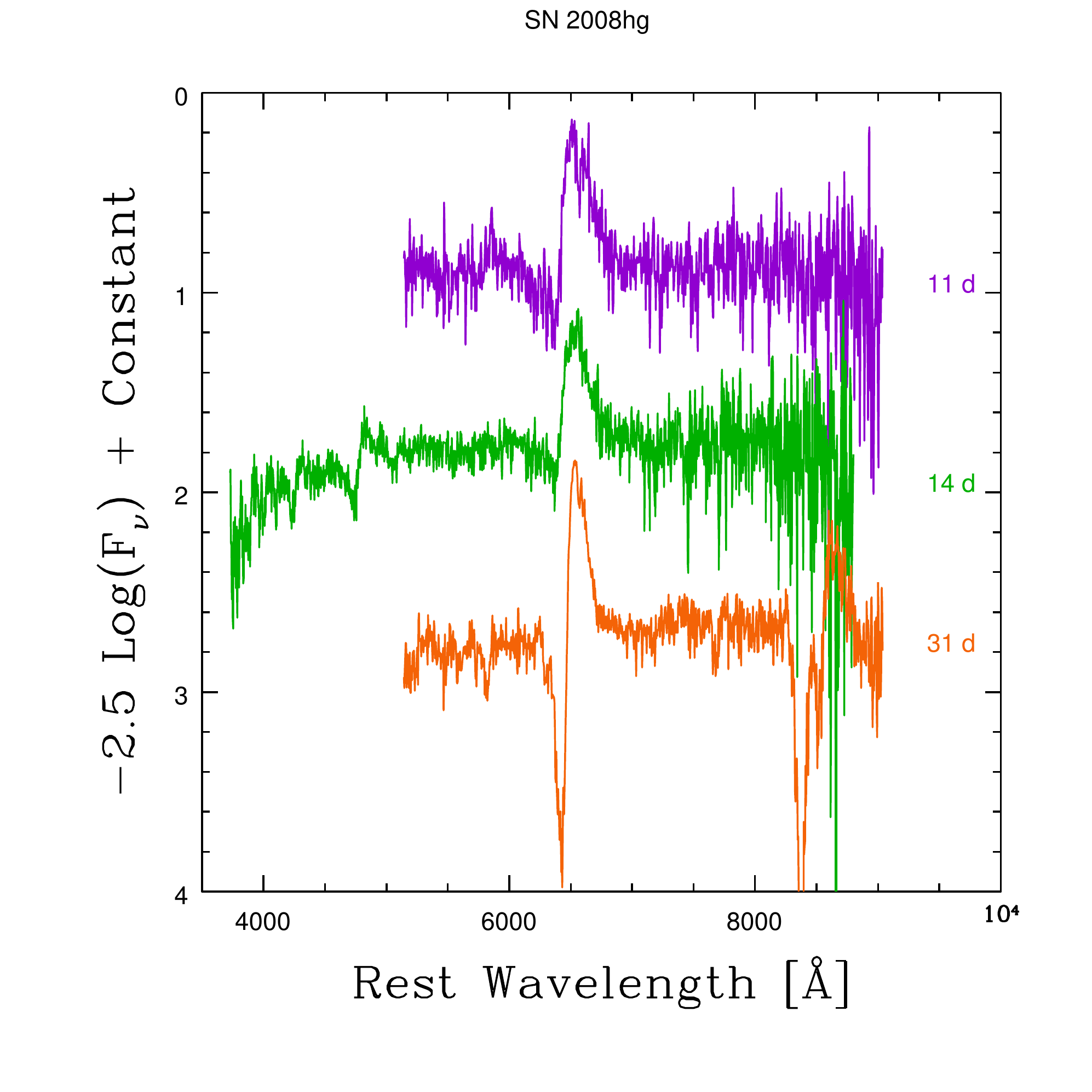}
\includegraphics[width=5.5cm]{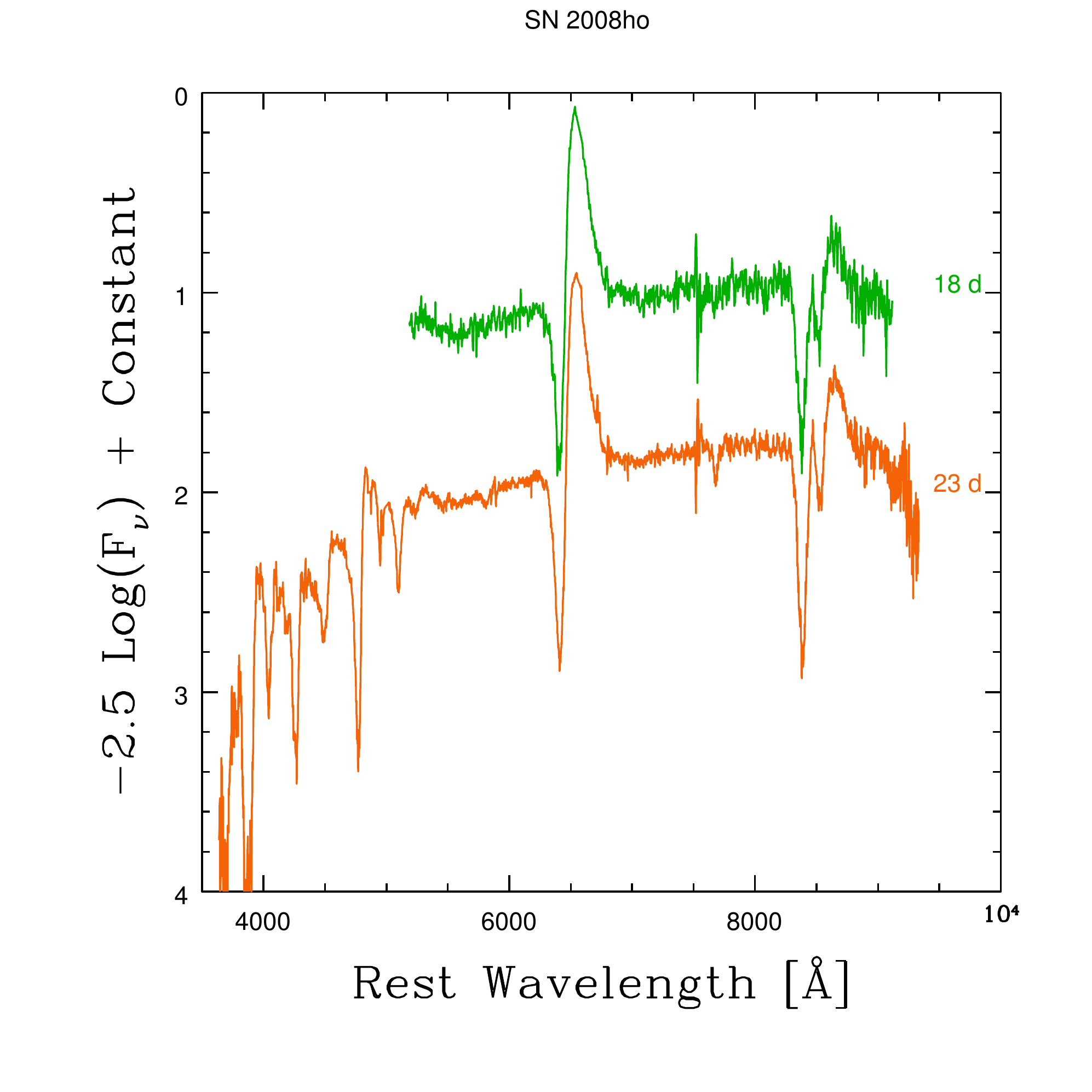}
\includegraphics[width=5.5cm]{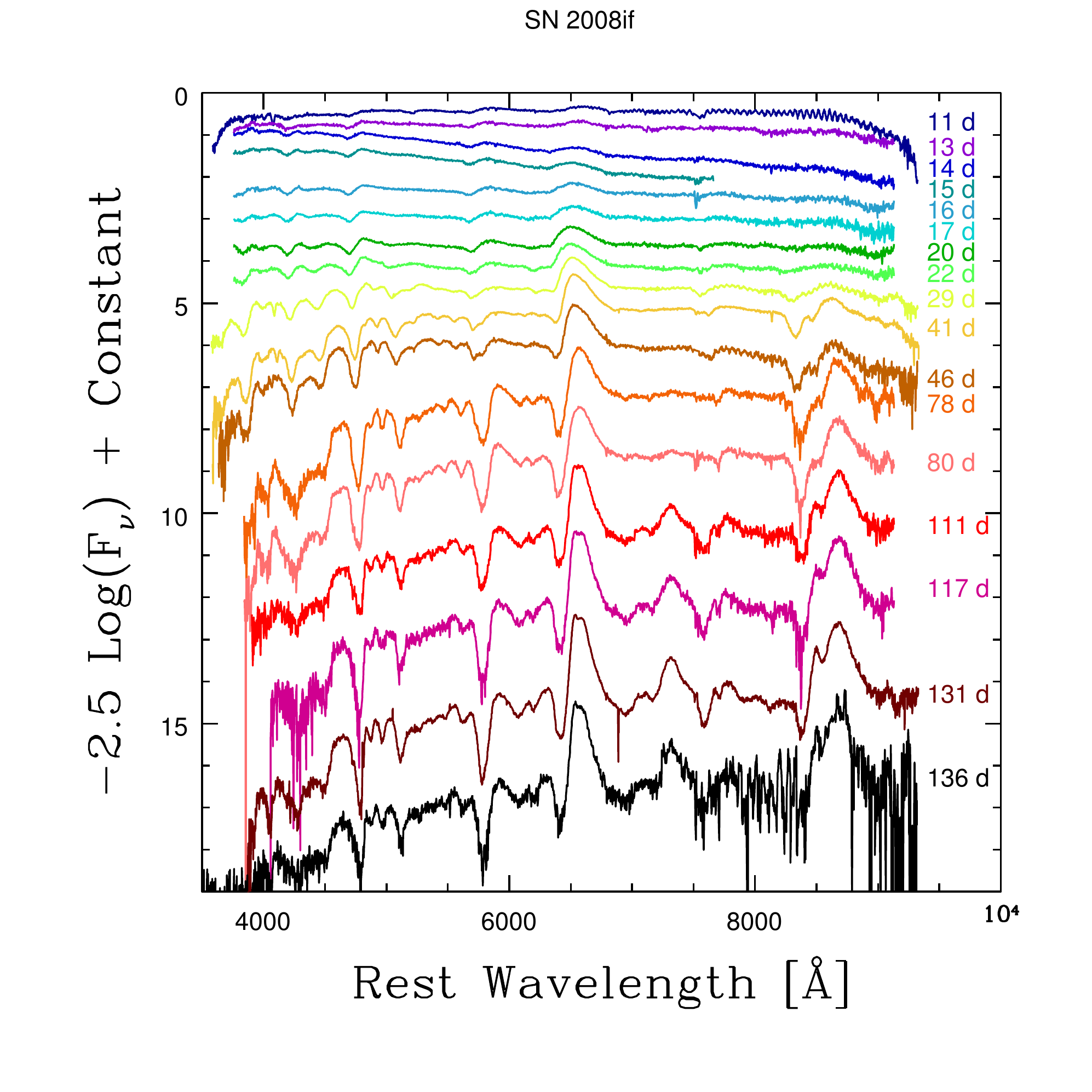}
\includegraphics[width=5.5cm]{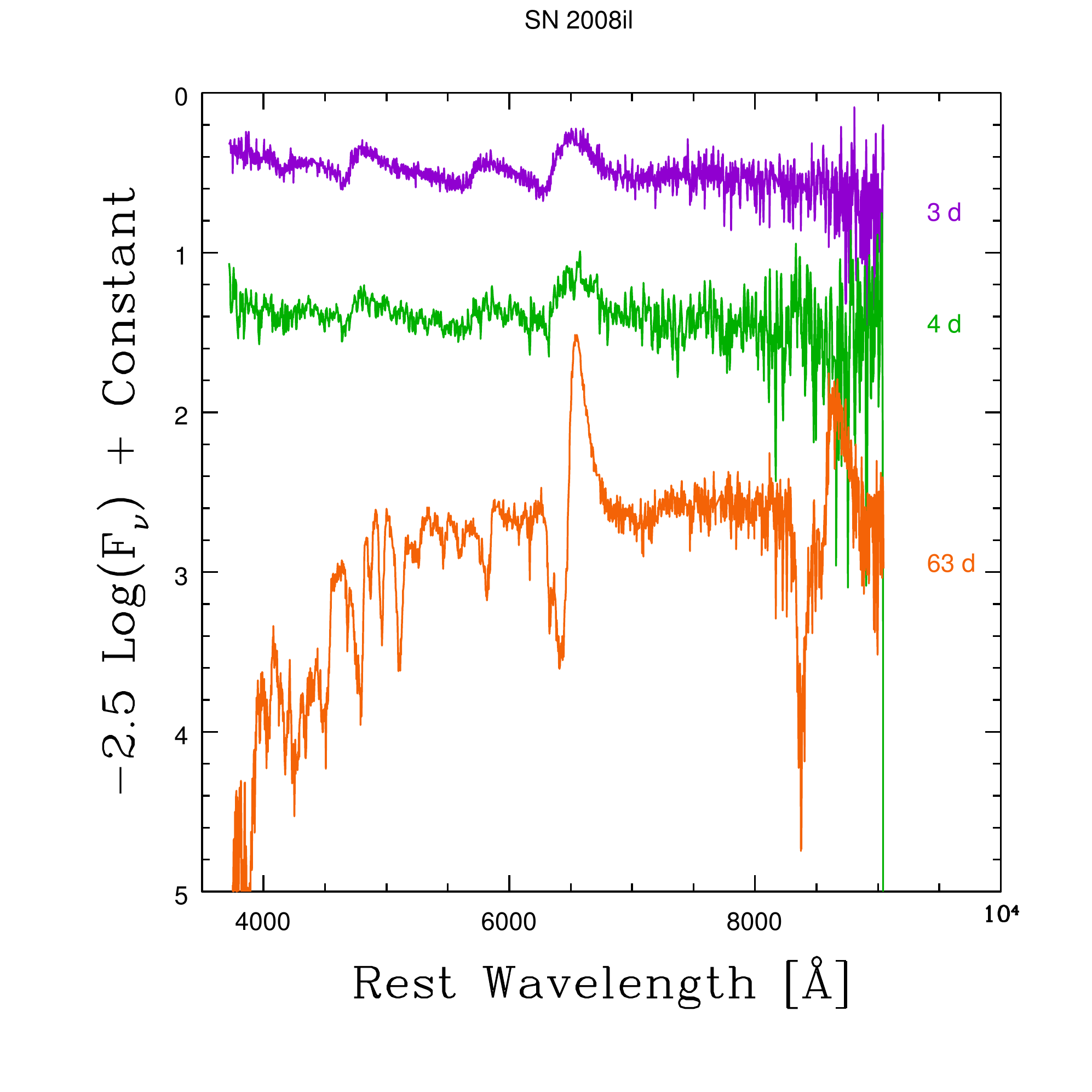}
\includegraphics[width=5.5cm]{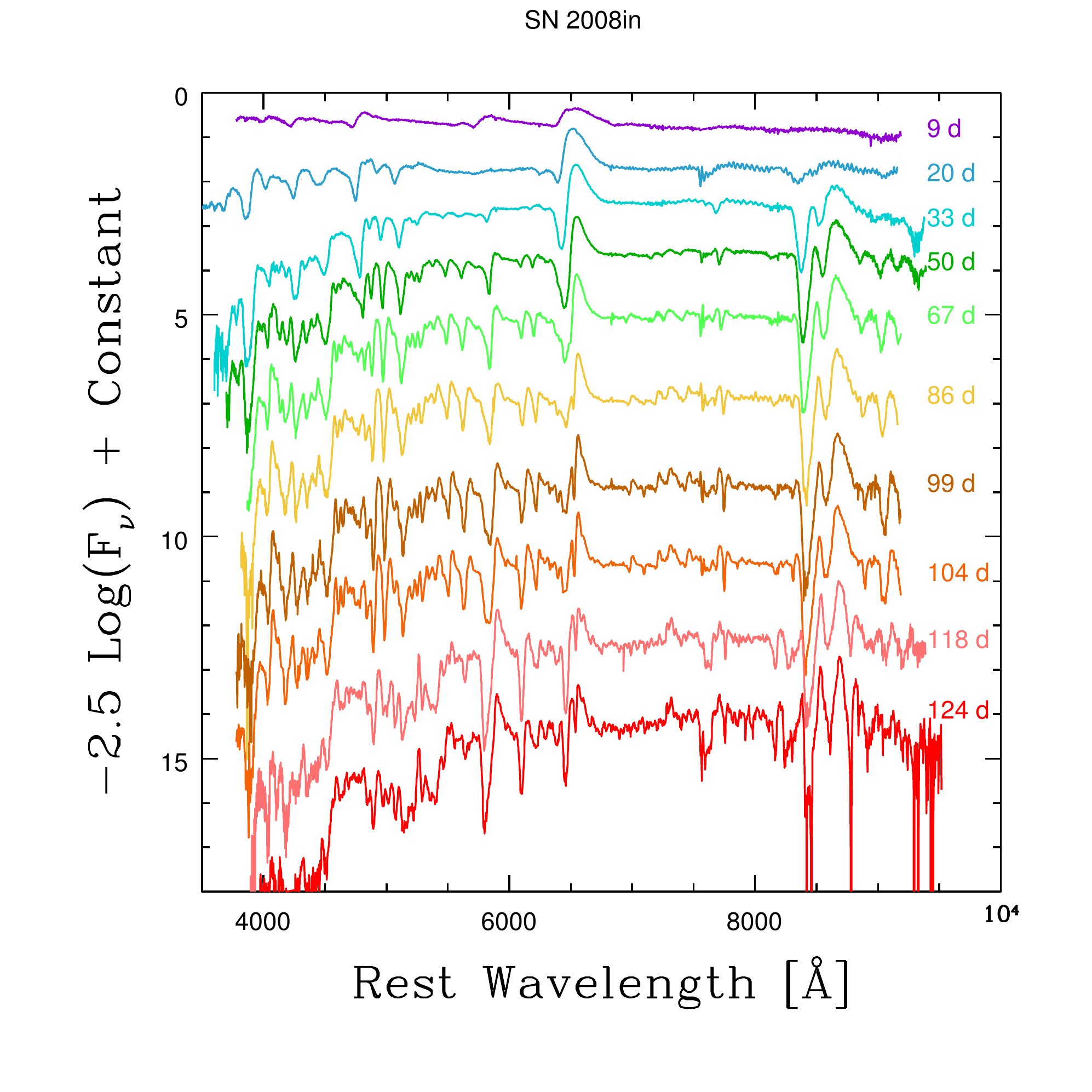}
\includegraphics[width=5.5cm]{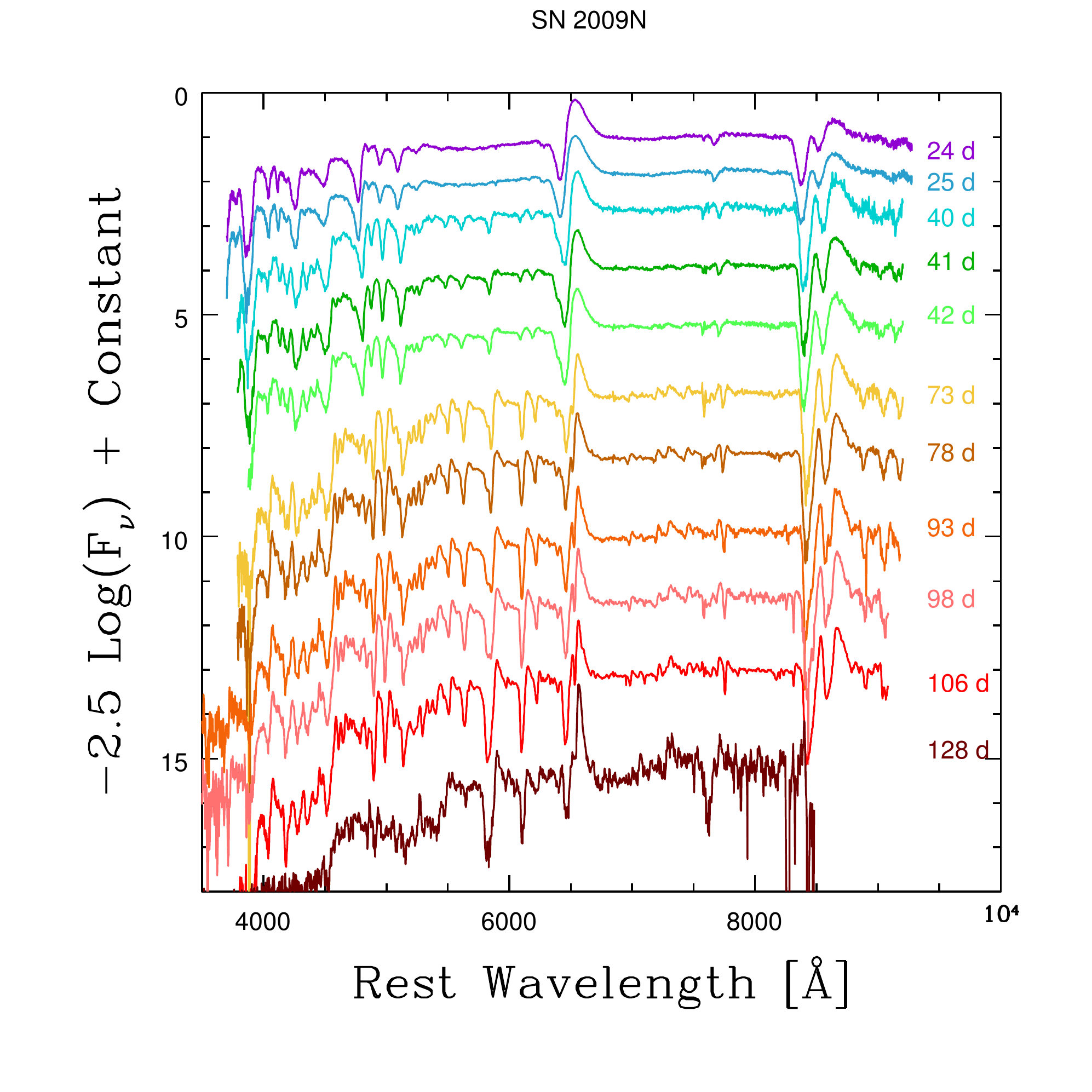}
\includegraphics[width=5.5cm]{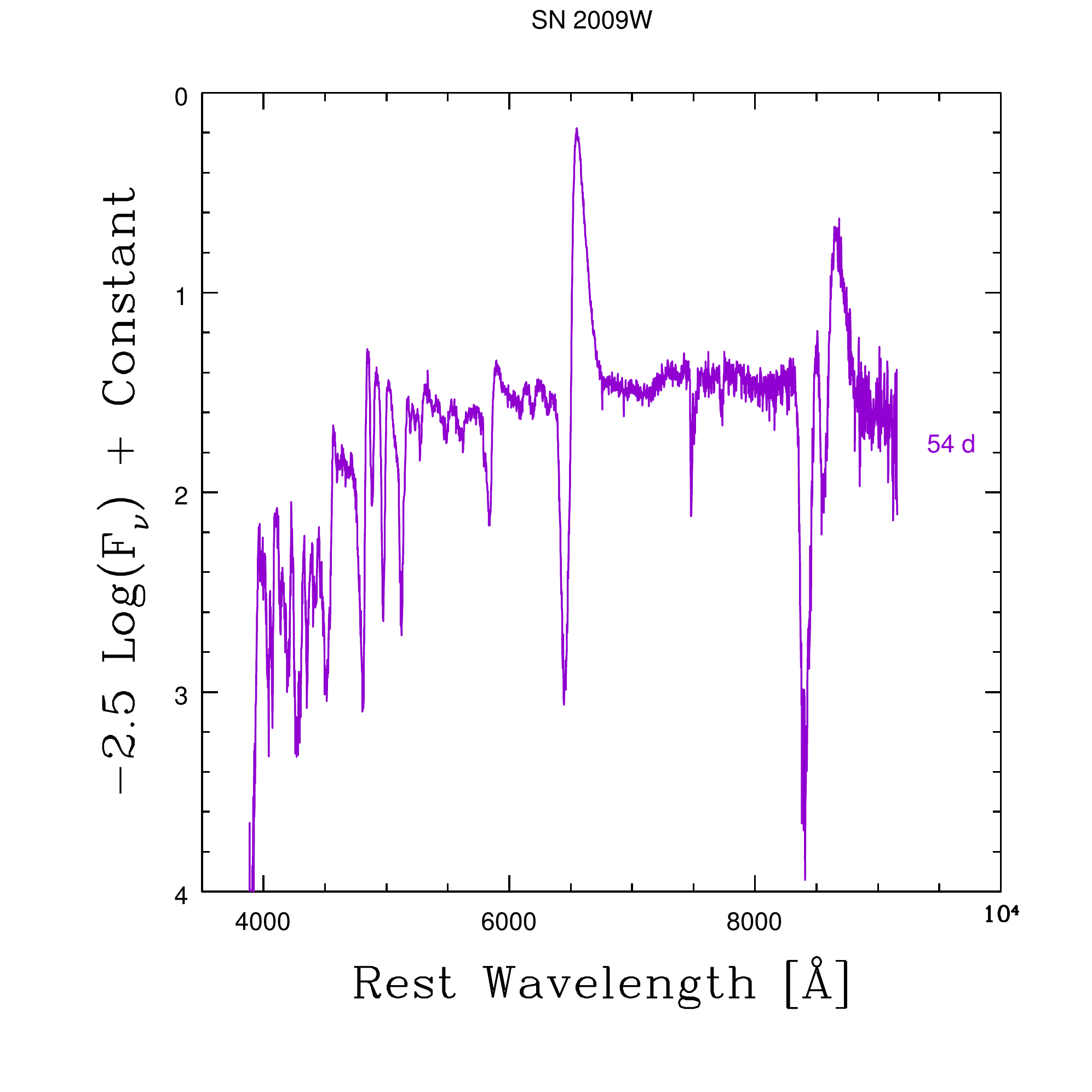}
\includegraphics[width=5.5cm]{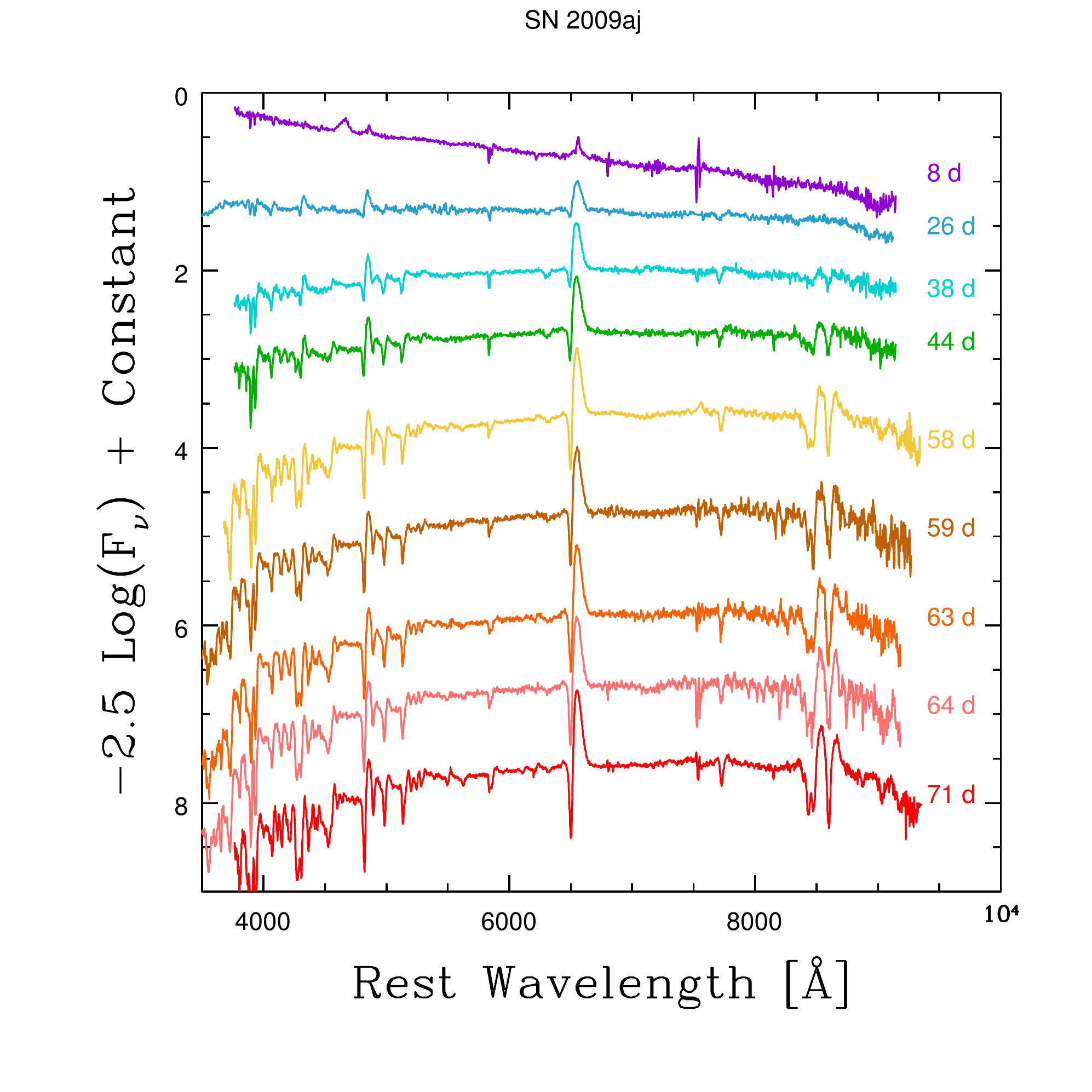}
\includegraphics[width=5.5cm]{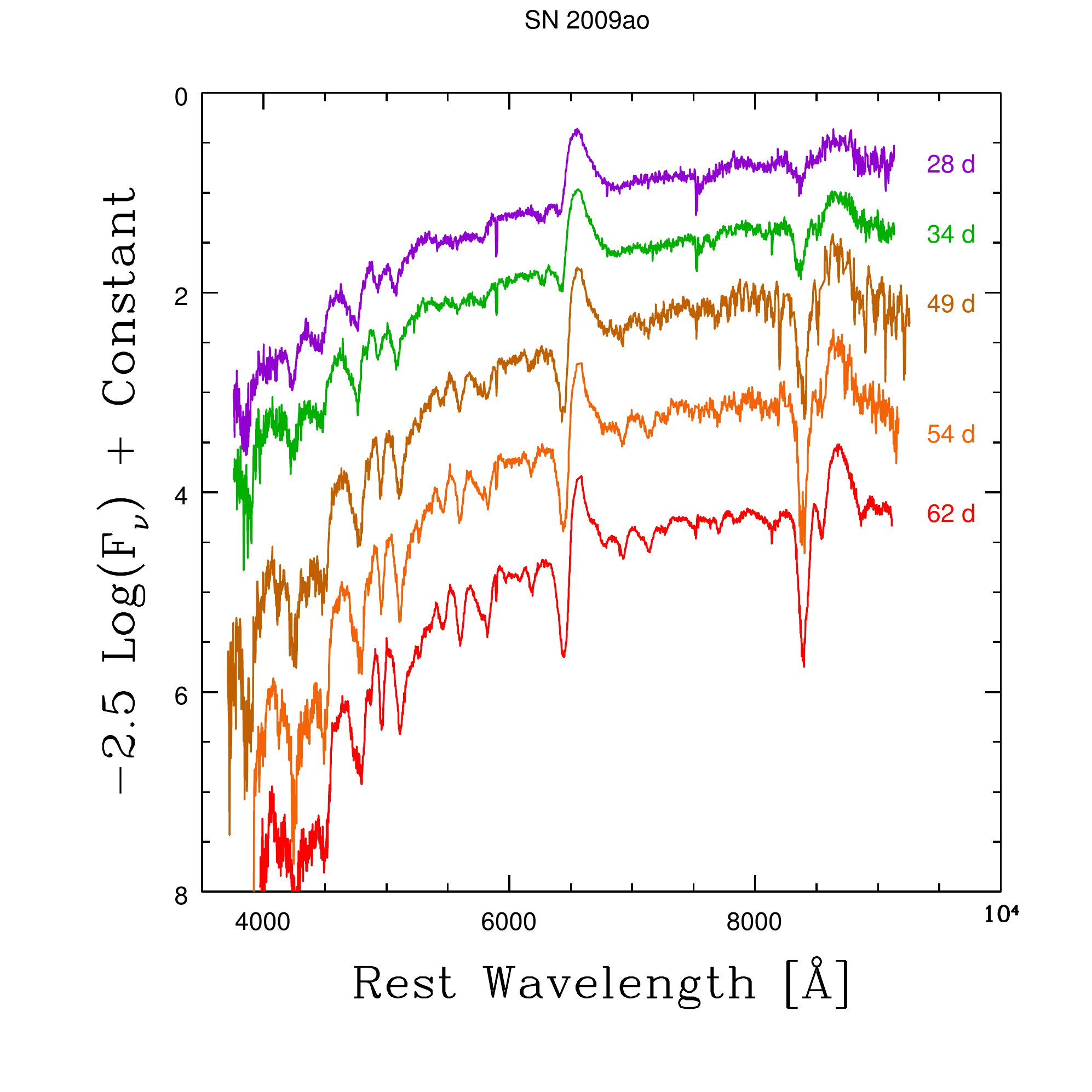}
\includegraphics[width=5.5cm]{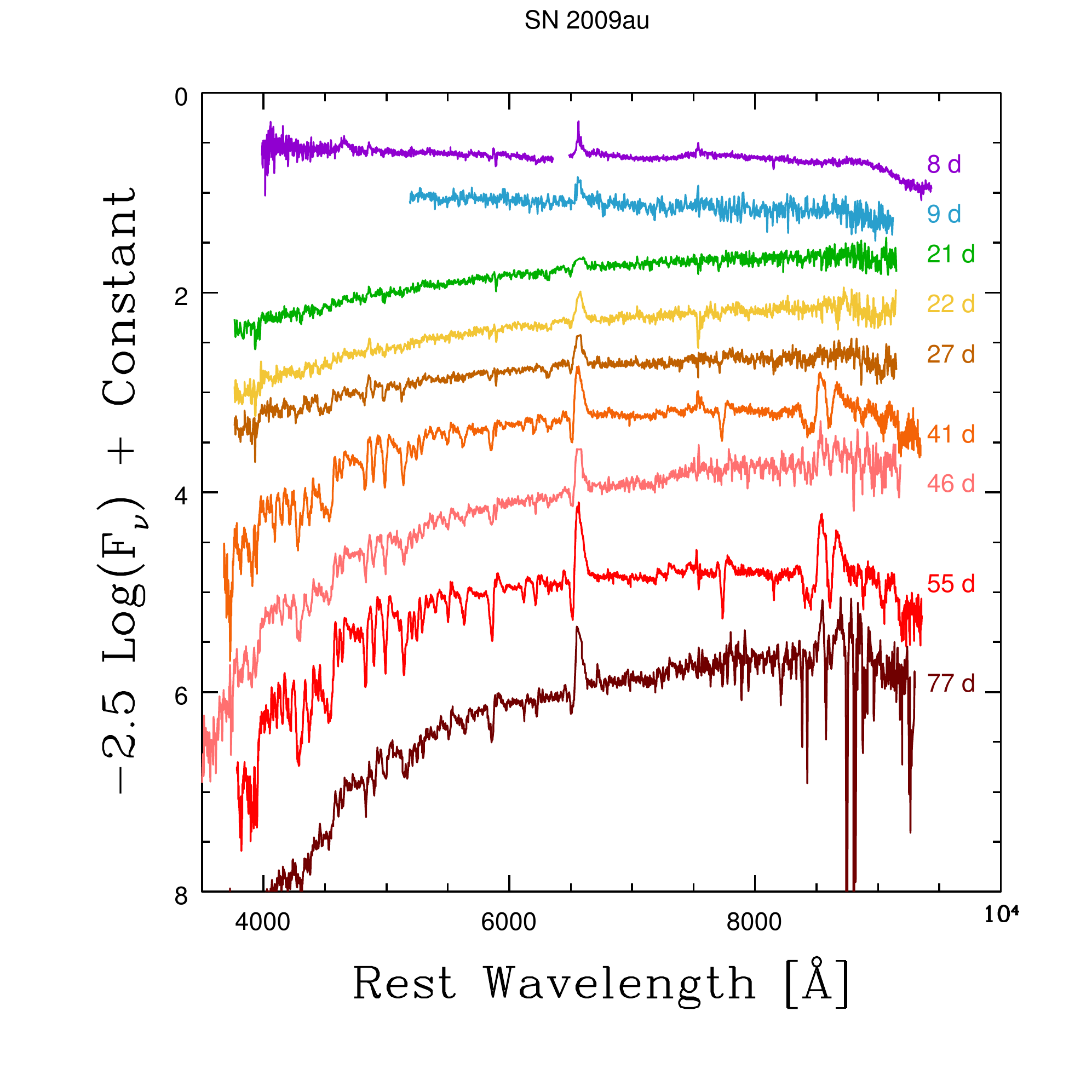}
\includegraphics[width=5.5cm]{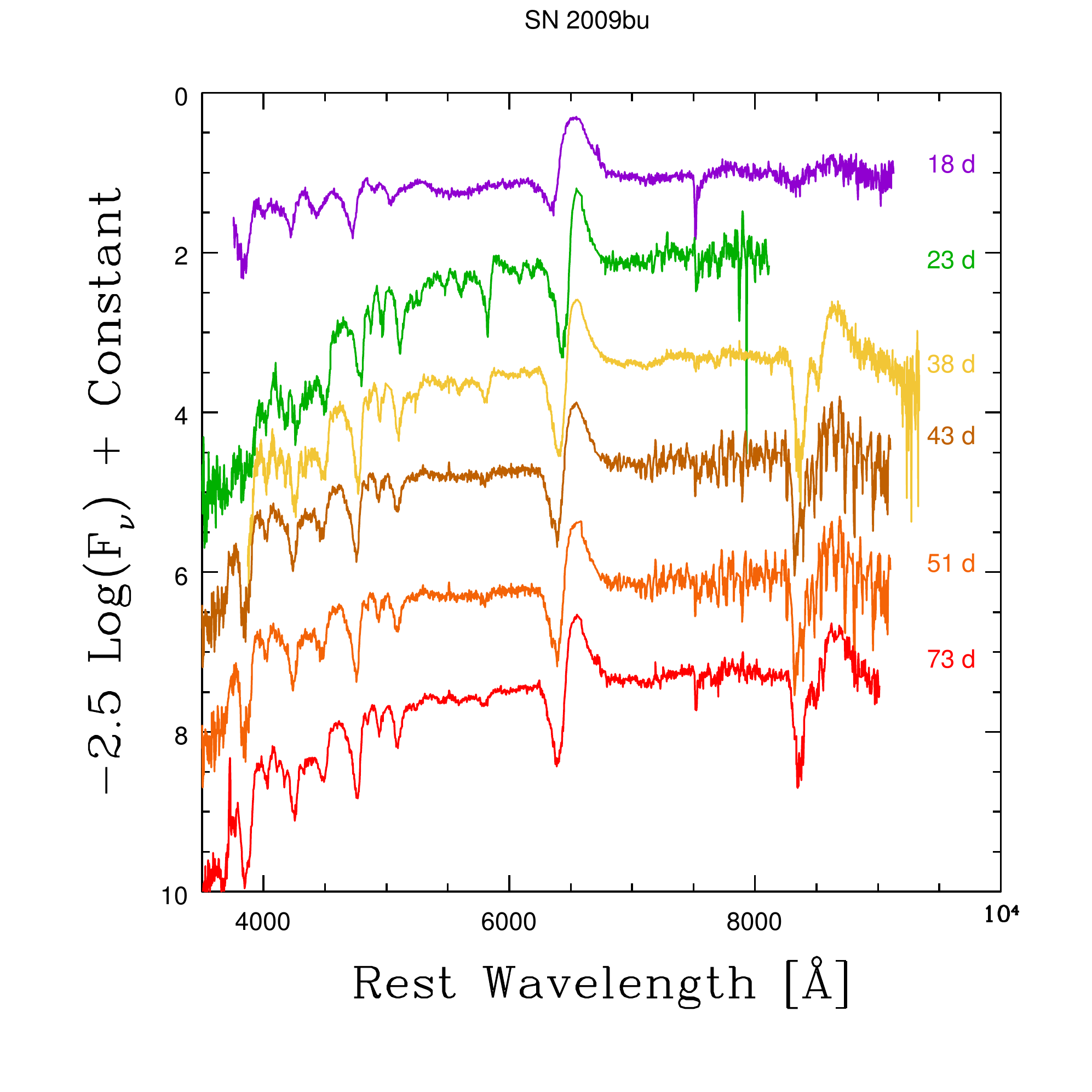}
\caption{Examples of SNe~II spectra: SN~2008gr, SN~2008hg, SN~2008ho¸ SN~2008if, SN~2008il, SN~2008in, SN~2009N, SN~2009W, SN~2009aj, SN~2009ao, SN~2009au, SN~2009bu.}
\label{example}
\end{figure*}

\begin{figure*}[h!]
\centering

\includegraphics[width=5.5cm]{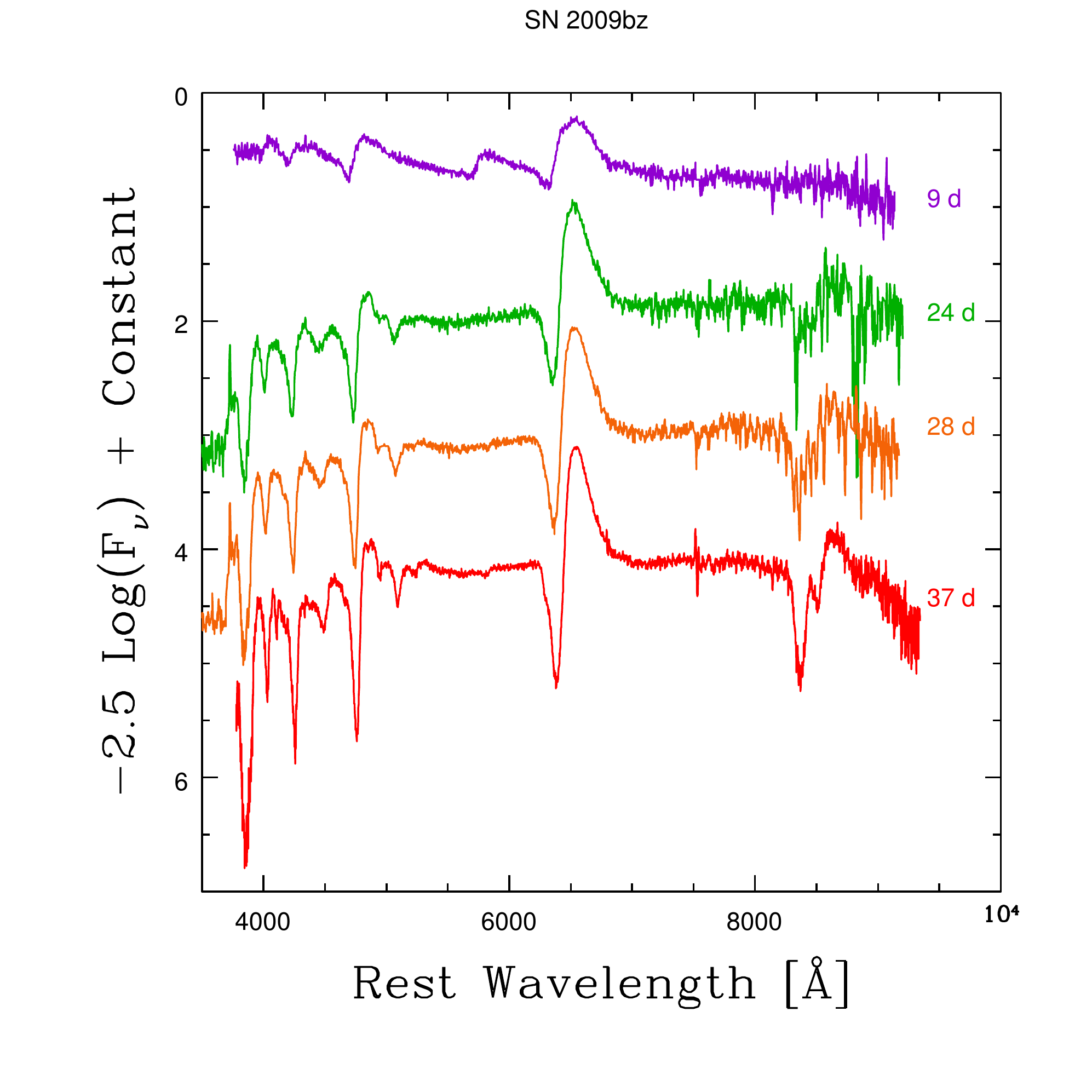}
\caption{Examples of SNe~II spectra: SN~2009bz.}
\label{example}
\end{figure*}

\clearpage

\section{SNID matches}
\label{snid}

\begin{figure}[h!]
\centering
\includegraphics[width=4.4cm]{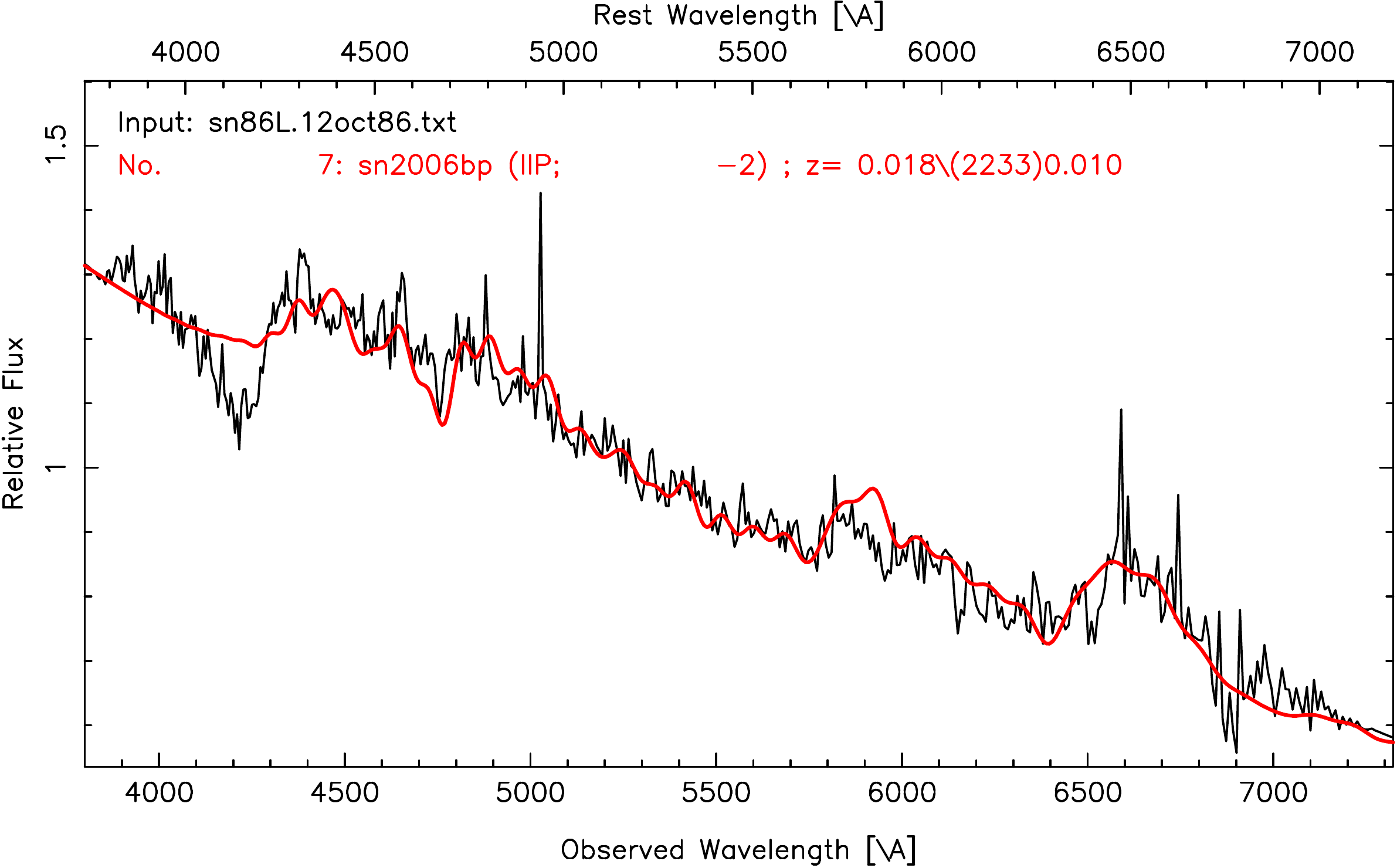}
\includegraphics[width=4.4cm]{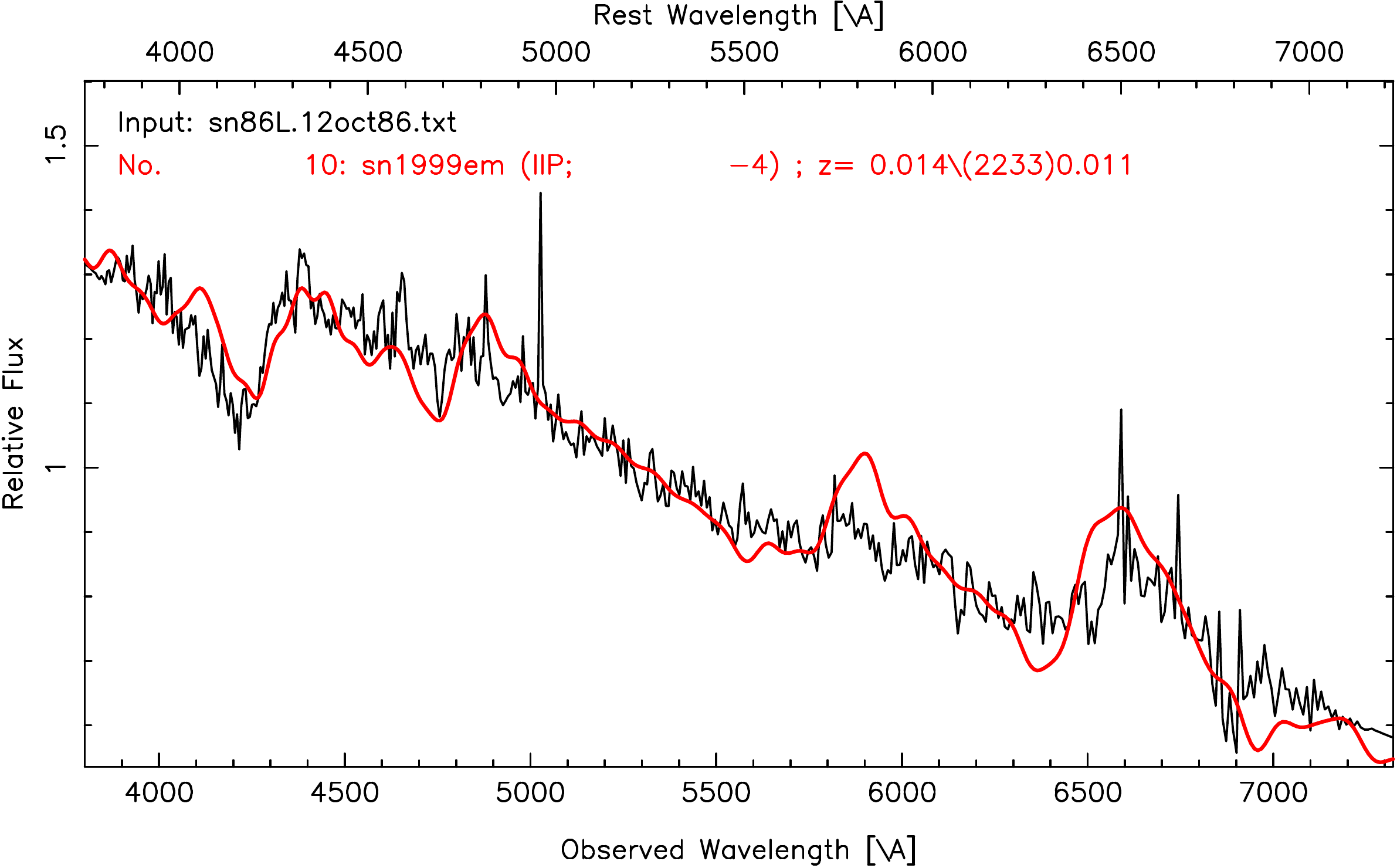}
\caption{Best spectral matching of SN~1986L using SNID. The plots show SN~1986L compared with 
SN~2006bp and SN~1999em at 6 and 7 days from explosion.}
\end{figure}

\begin{figure}[h!]
\centering
\includegraphics[width=4.4cm]{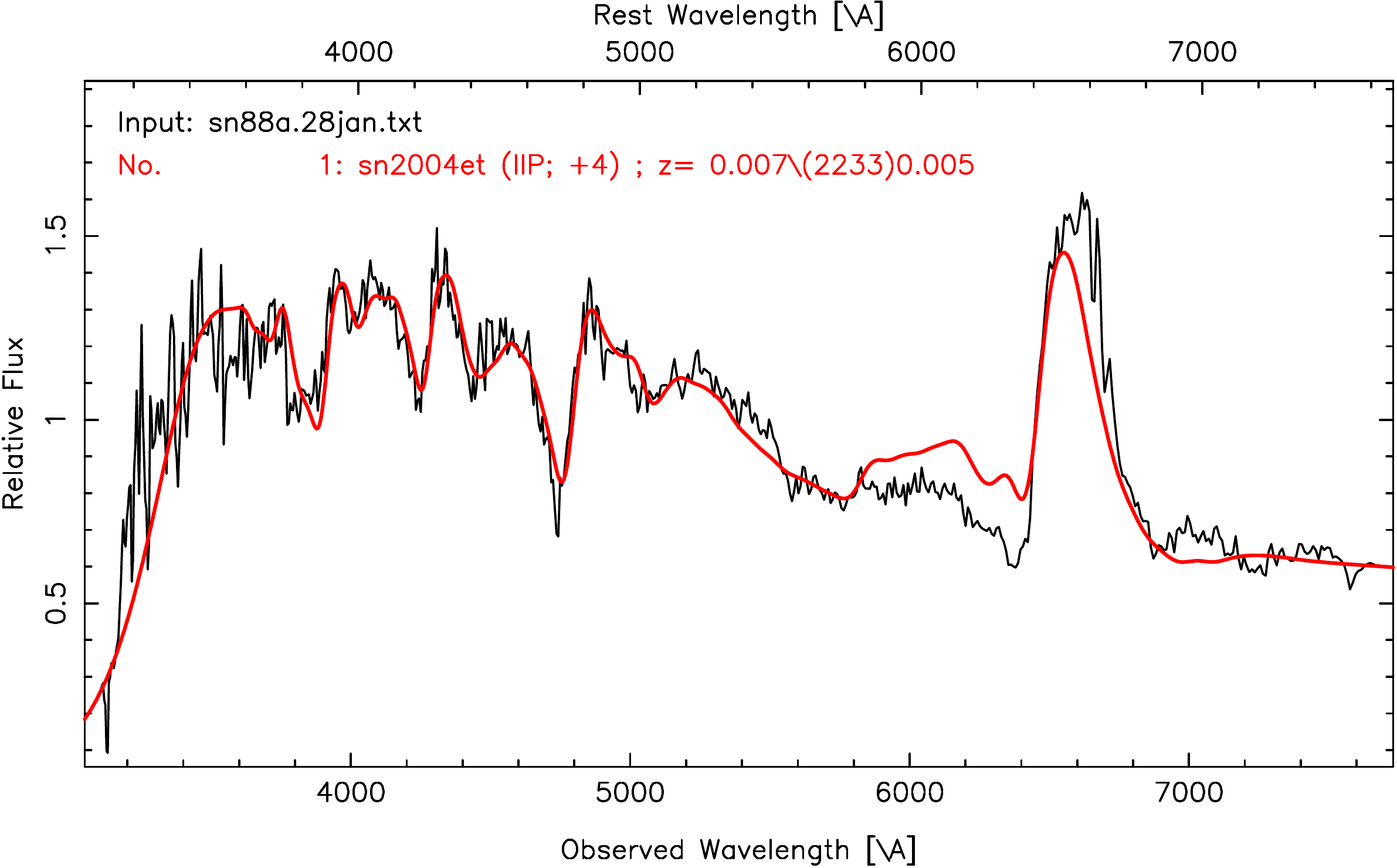}
\includegraphics[width=4.4cm]{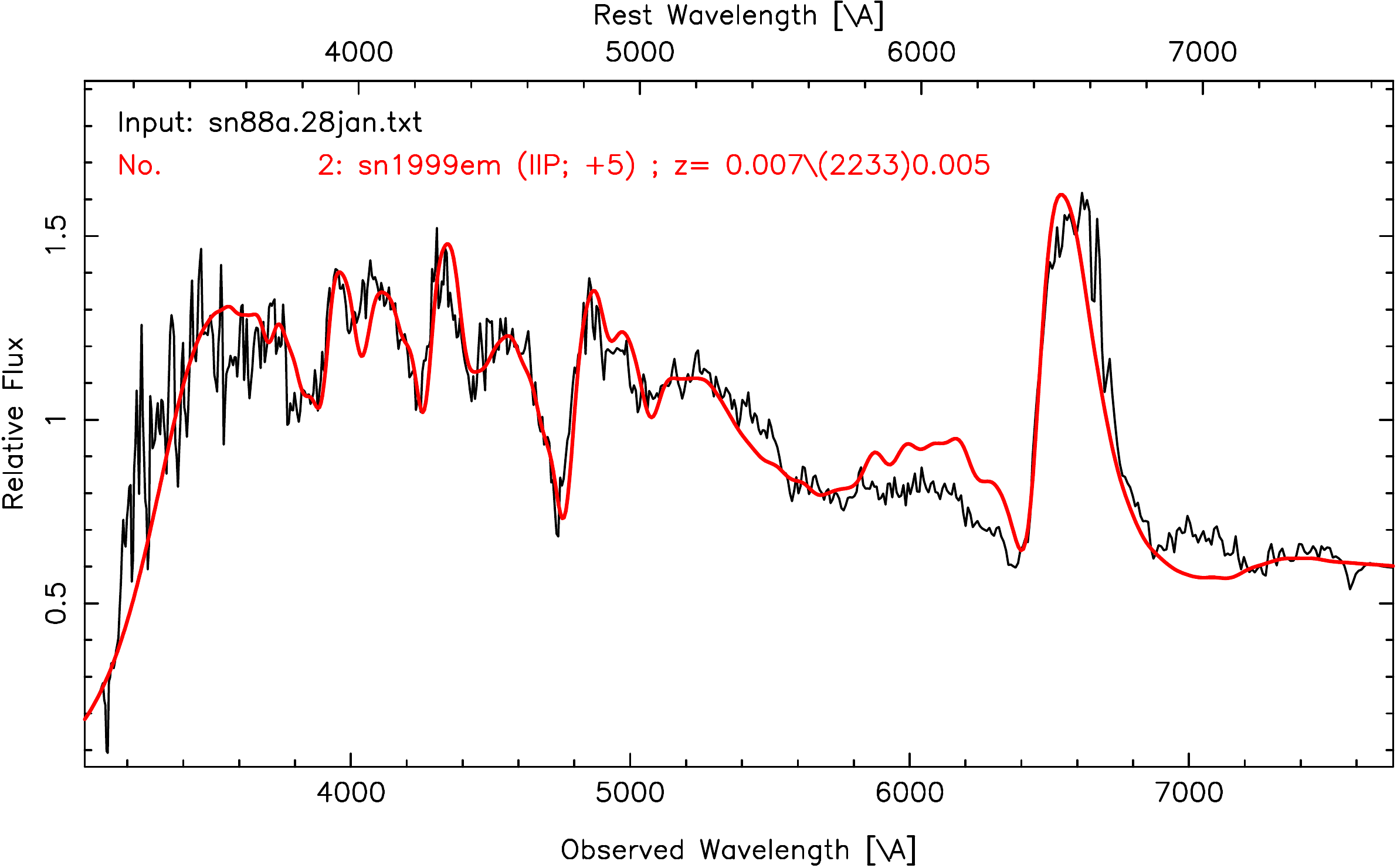}
\includegraphics[width=4.4cm]{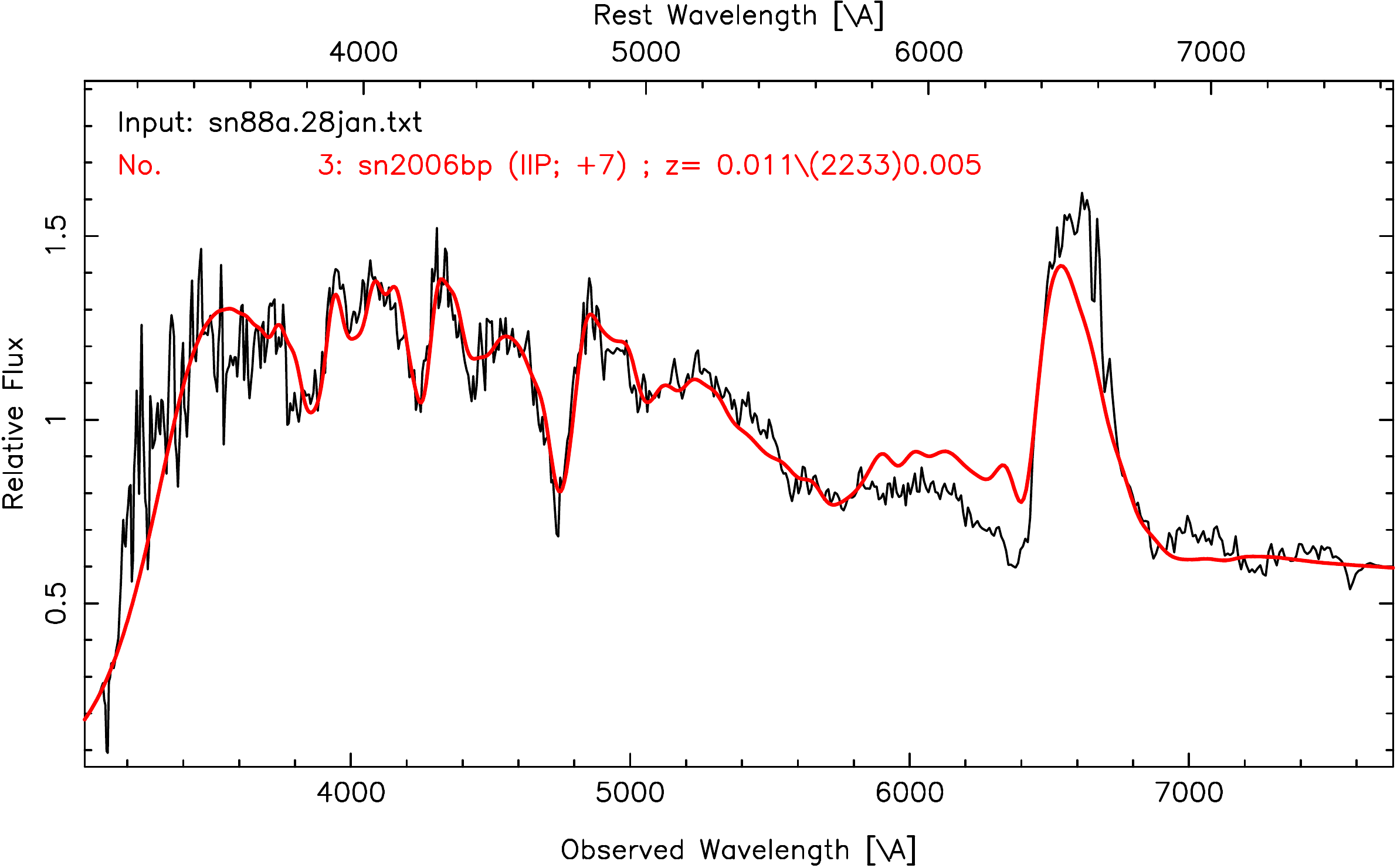}
\caption{Best spectral matching of SN~1988A using SNID. The plots show SN~1988A compared with 
SN~2004et, SN~1999em, SN~2006bp at 20, 15 and 16 days from explosion.}
\end{figure}

\begin{figure}[h!]
\centering
\includegraphics[width=4.4cm]{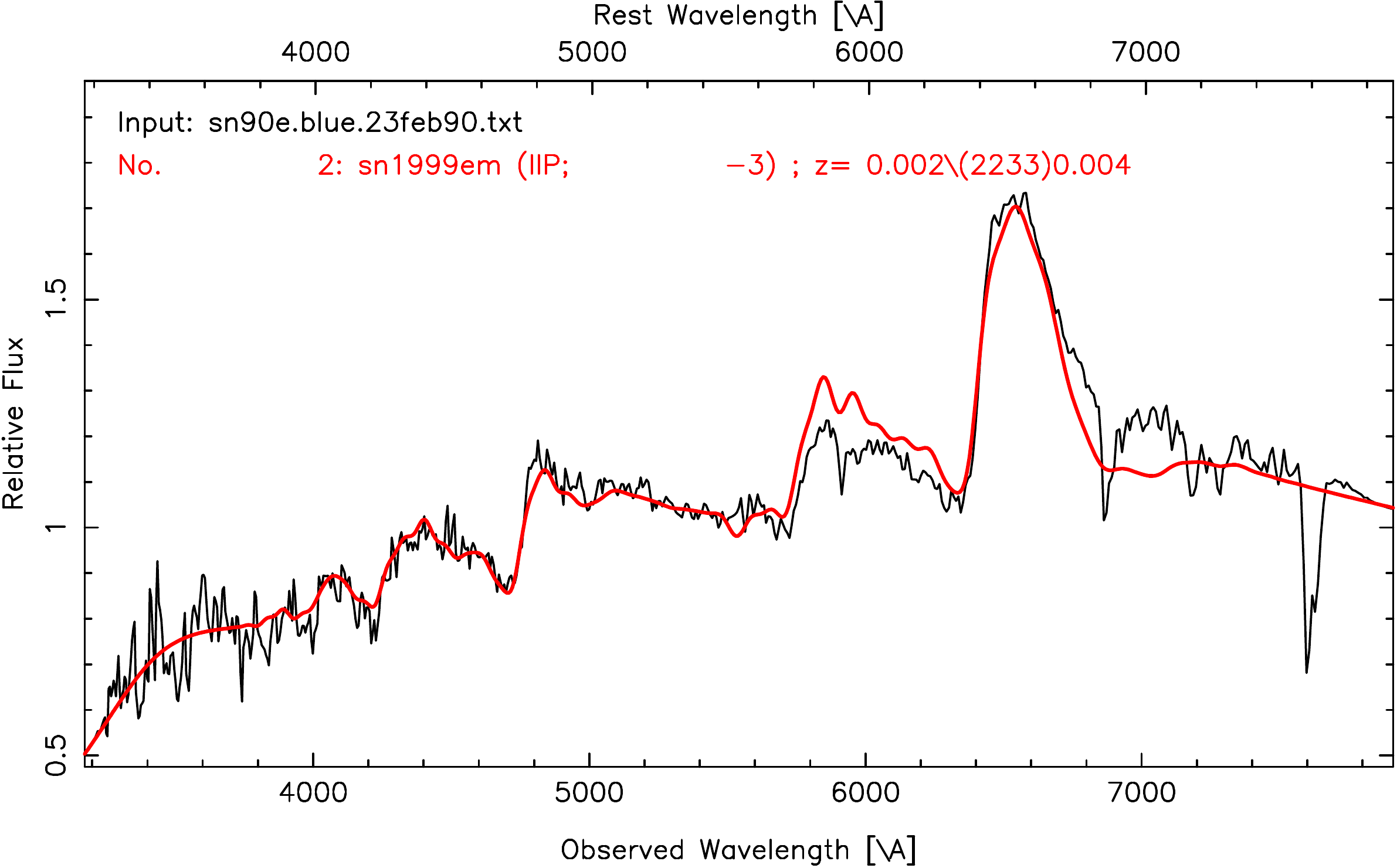}
\includegraphics[width=4.4cm]{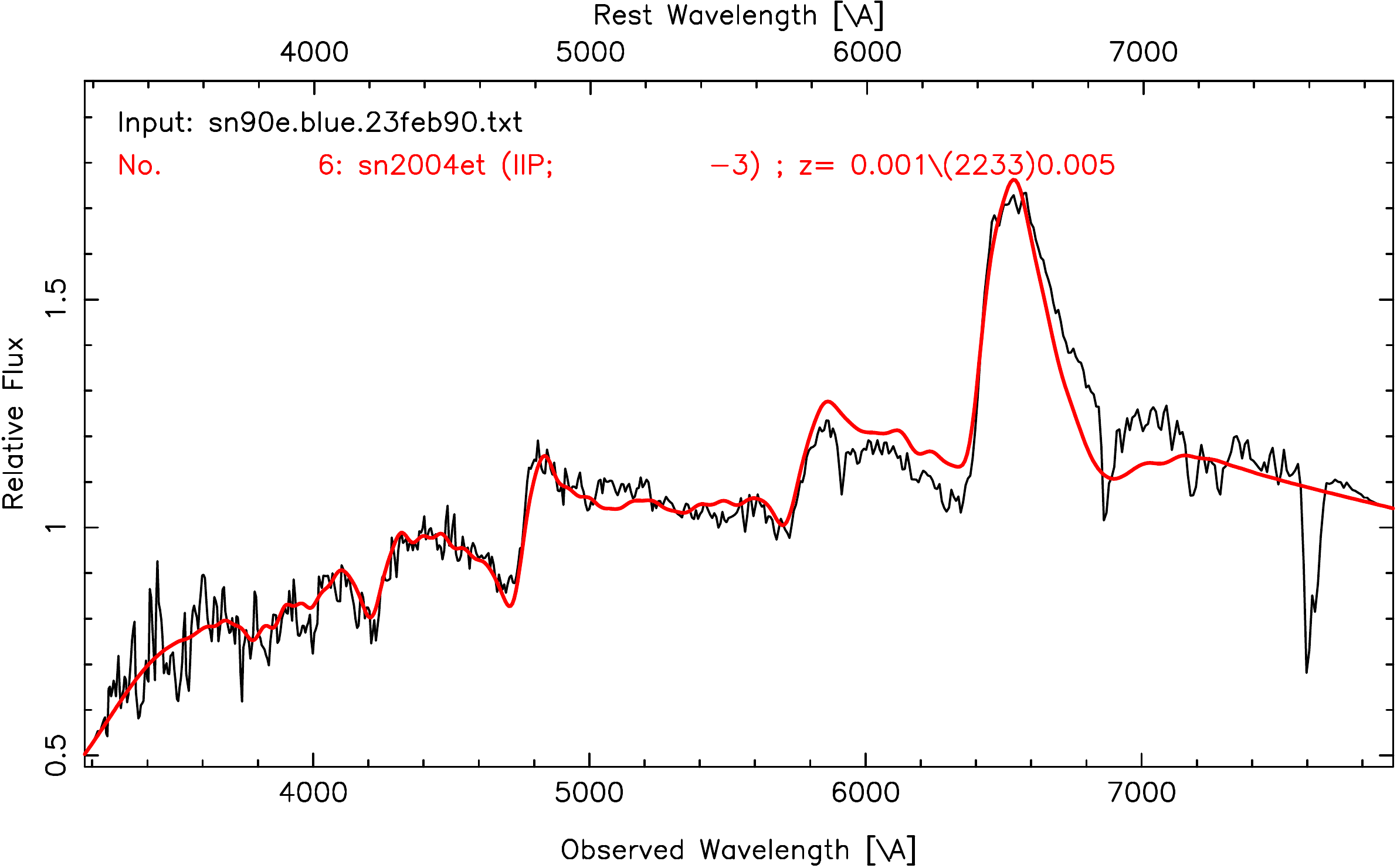}
\includegraphics[width=4.4cm]{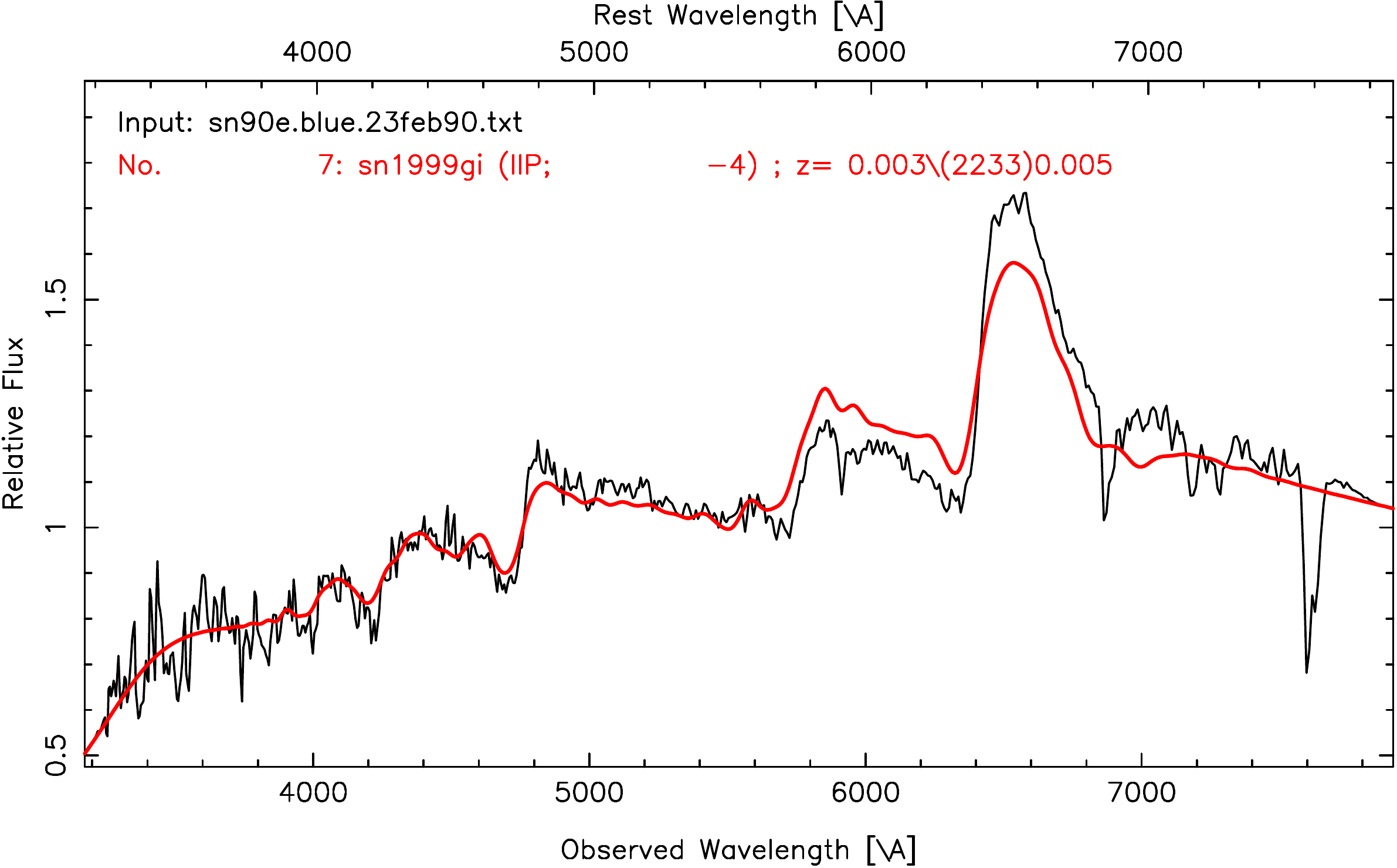}
\includegraphics[width=4.4cm]{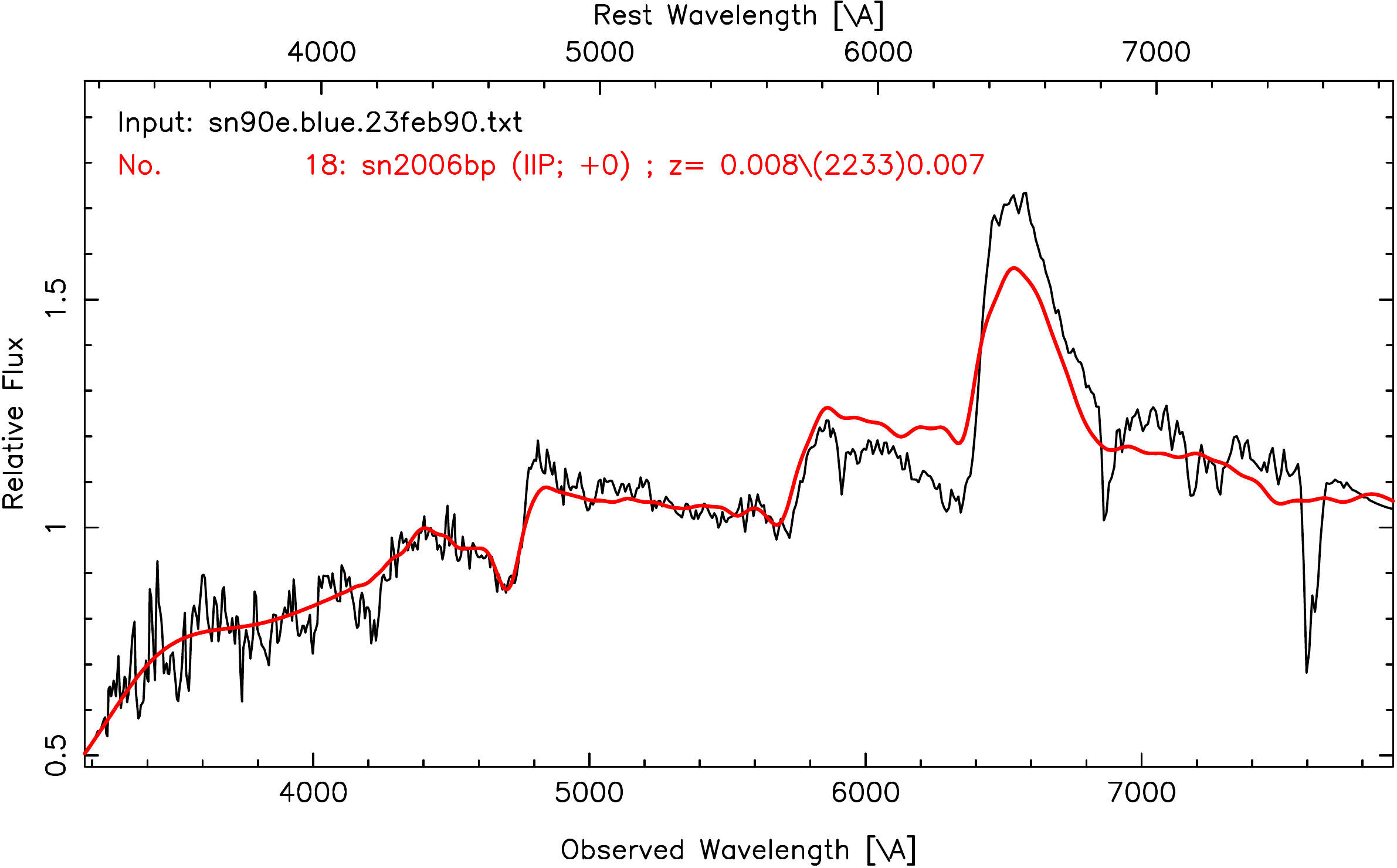}
\caption{Best spectral matching of SN~1990E using SNID. The plots show SN~1990E compared with 
SN~1999em, SN~2004et, SN~1999gi and SN2006bp at 7, 13, 8 and 9 days from explosion.}
\end{figure}

\clearpage

\begin{figure}[h!]
\centering
\includegraphics[width=4.4cm]{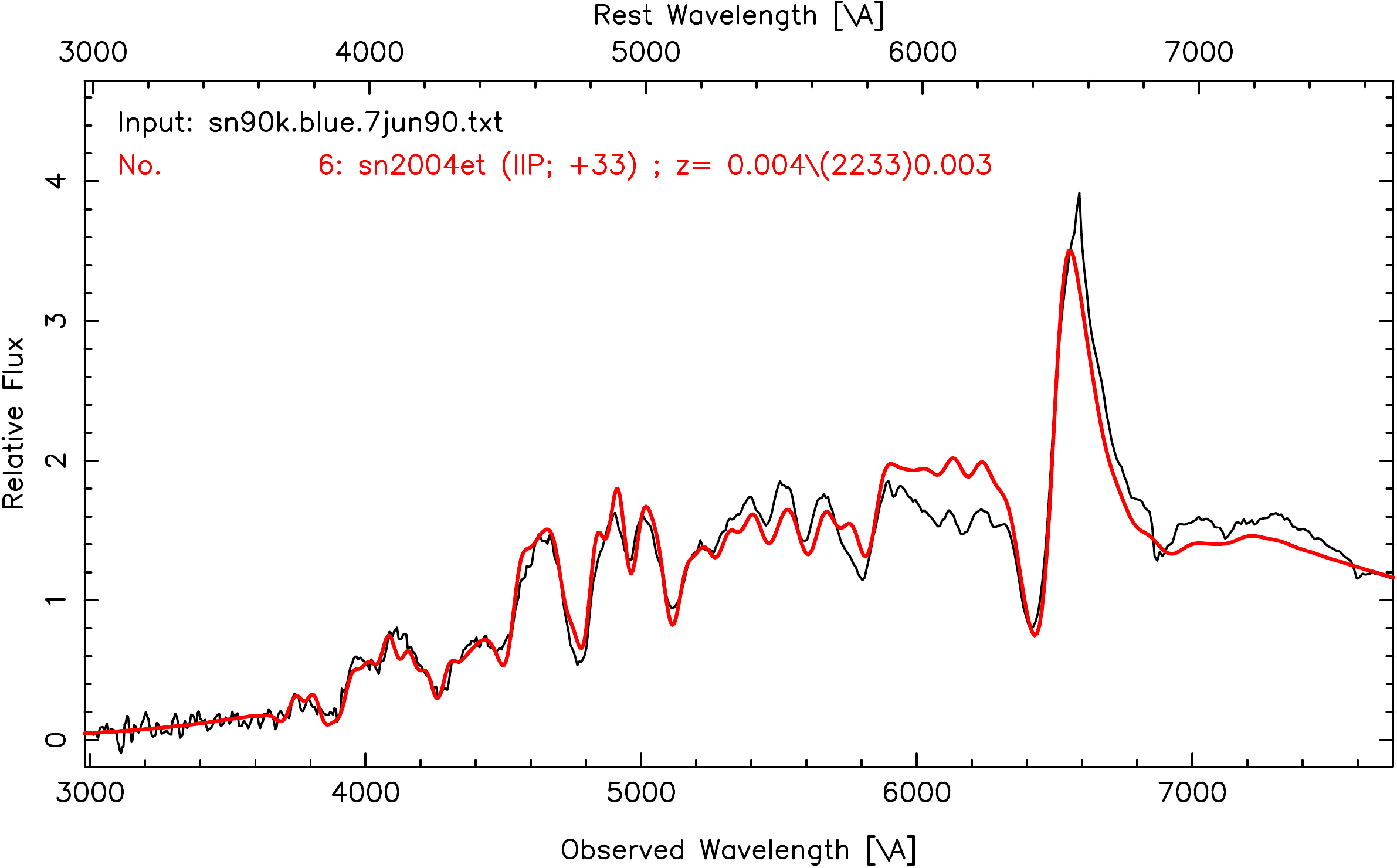}
\includegraphics[width=4.4cm]{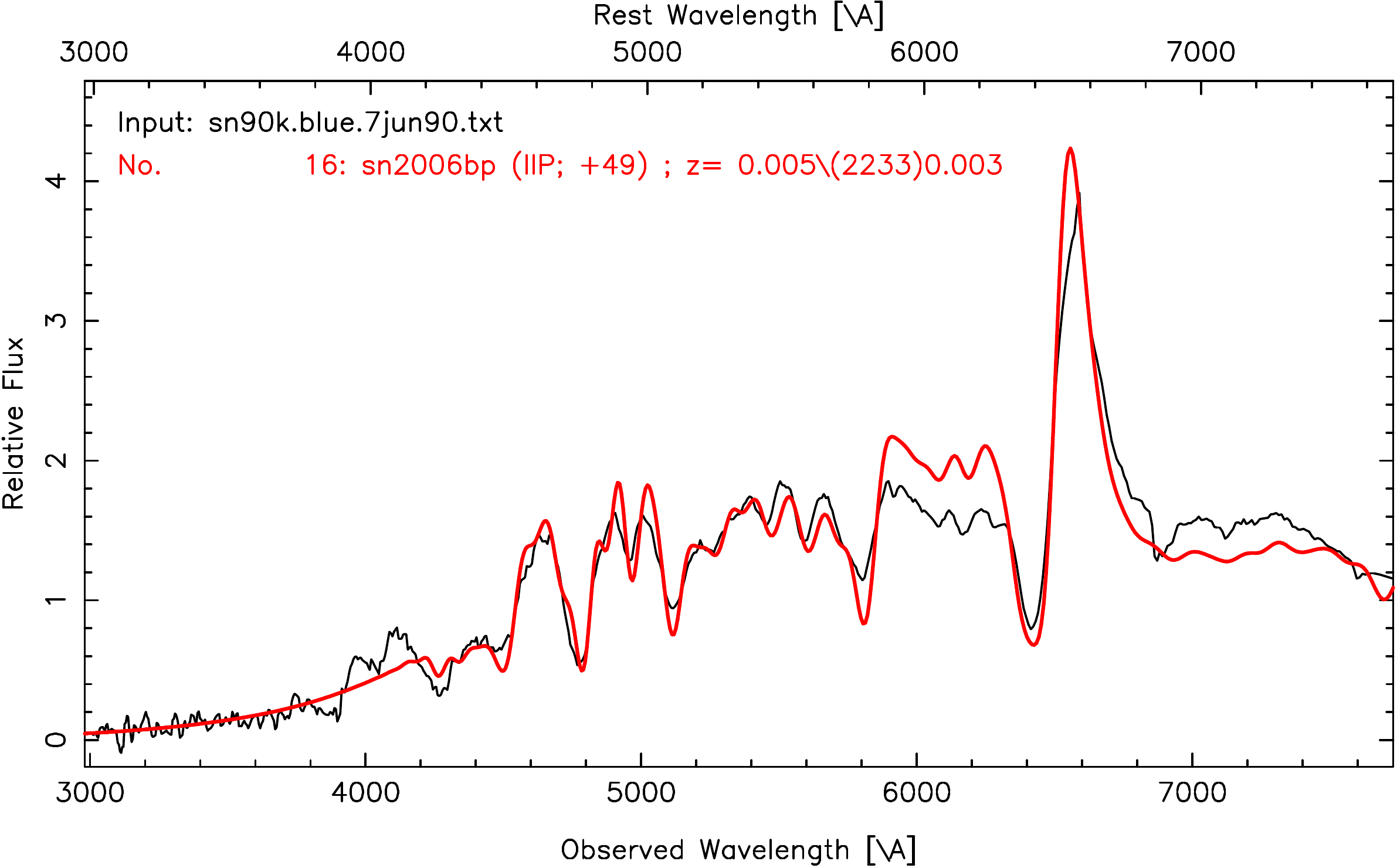}
\includegraphics[width=4.4cm]{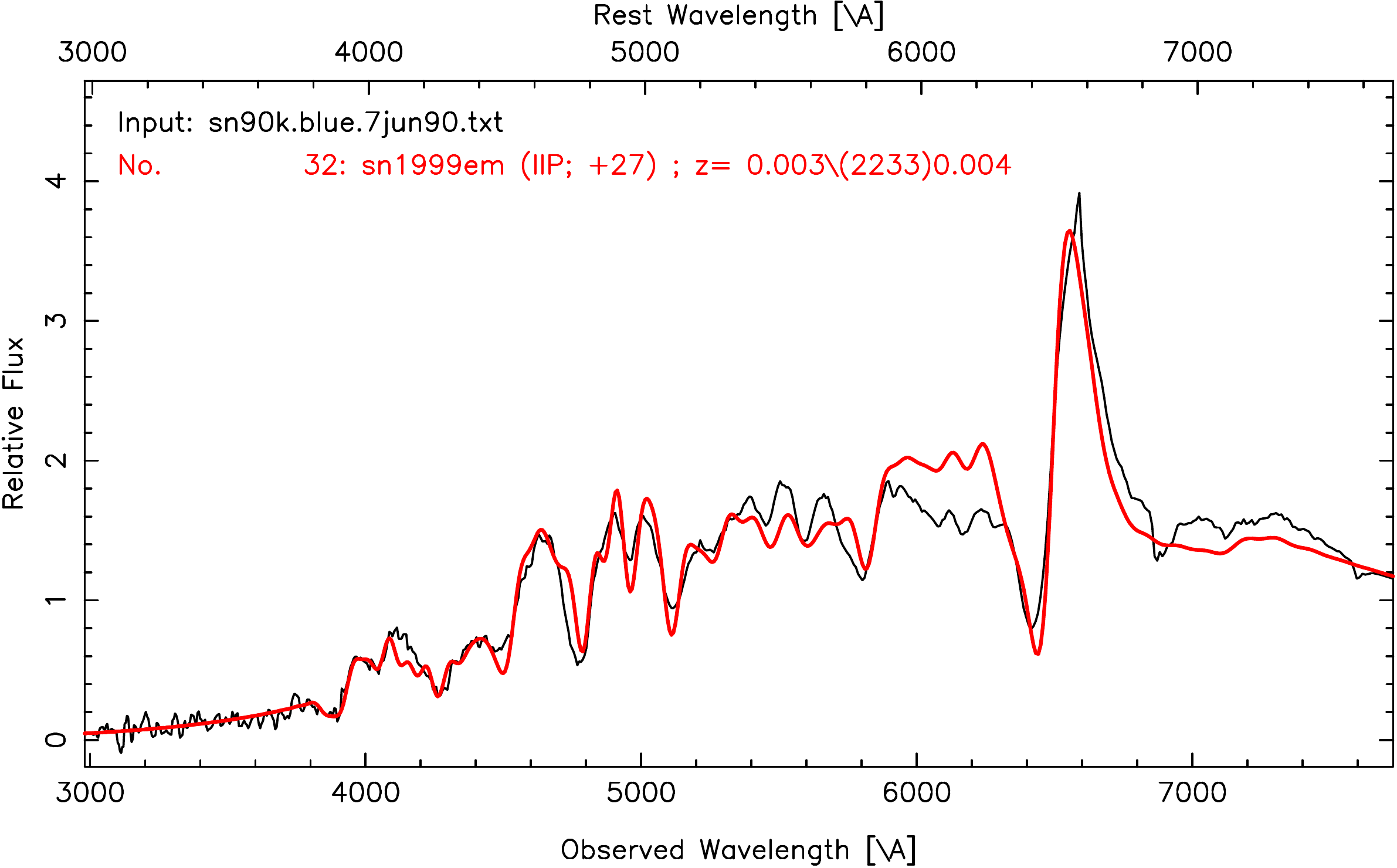}
\caption{Best spectral matching of SN~1990K using SNID. The plots show SN~1990K compared with 
SN~2004et, SN~2006bp , 1999em at 49, 58, and 37 days from explosion.}
\end{figure}

\begin{figure}[h!]
\centering
\includegraphics[width=4.4cm]{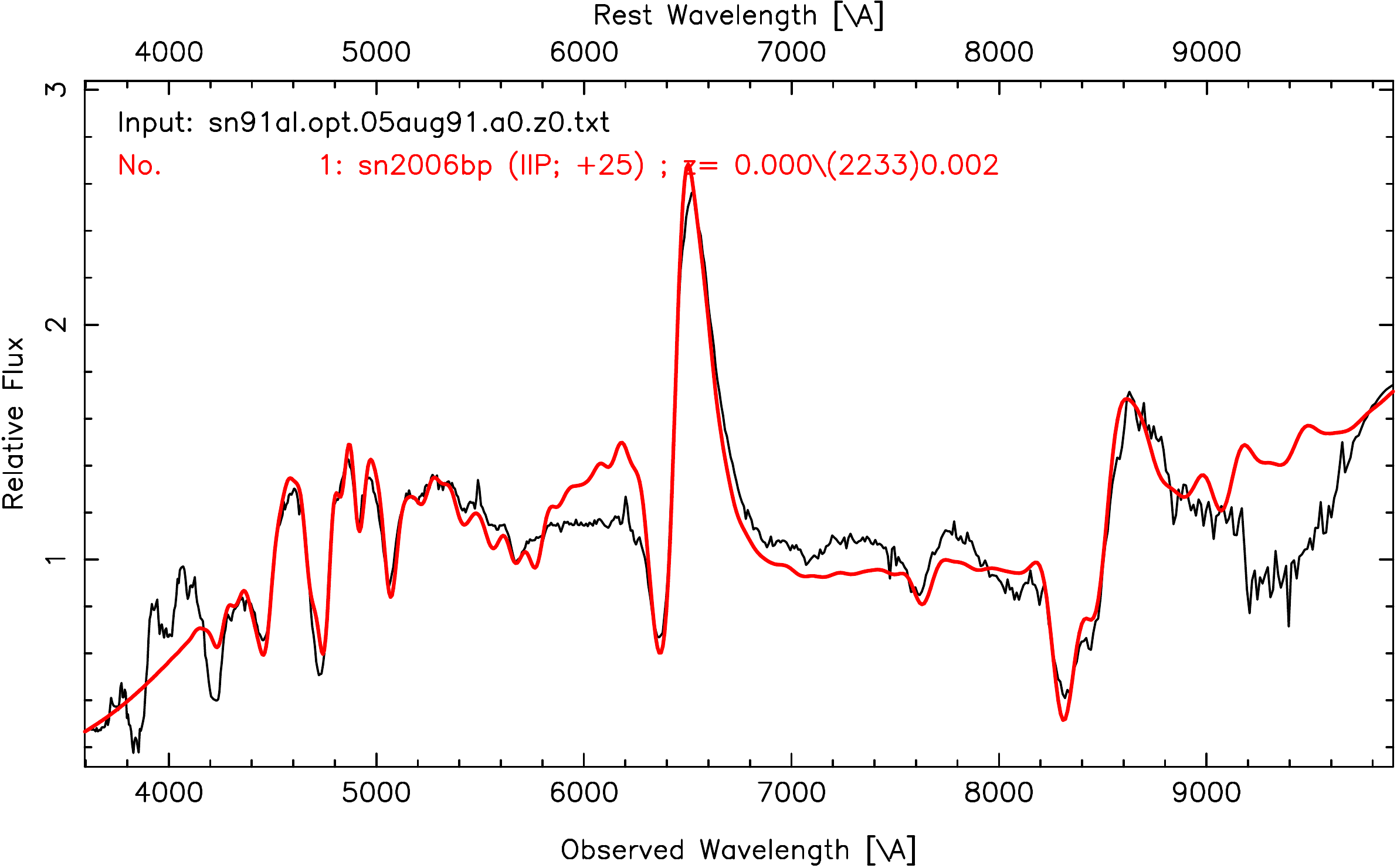}
\includegraphics[width=4.4cm]{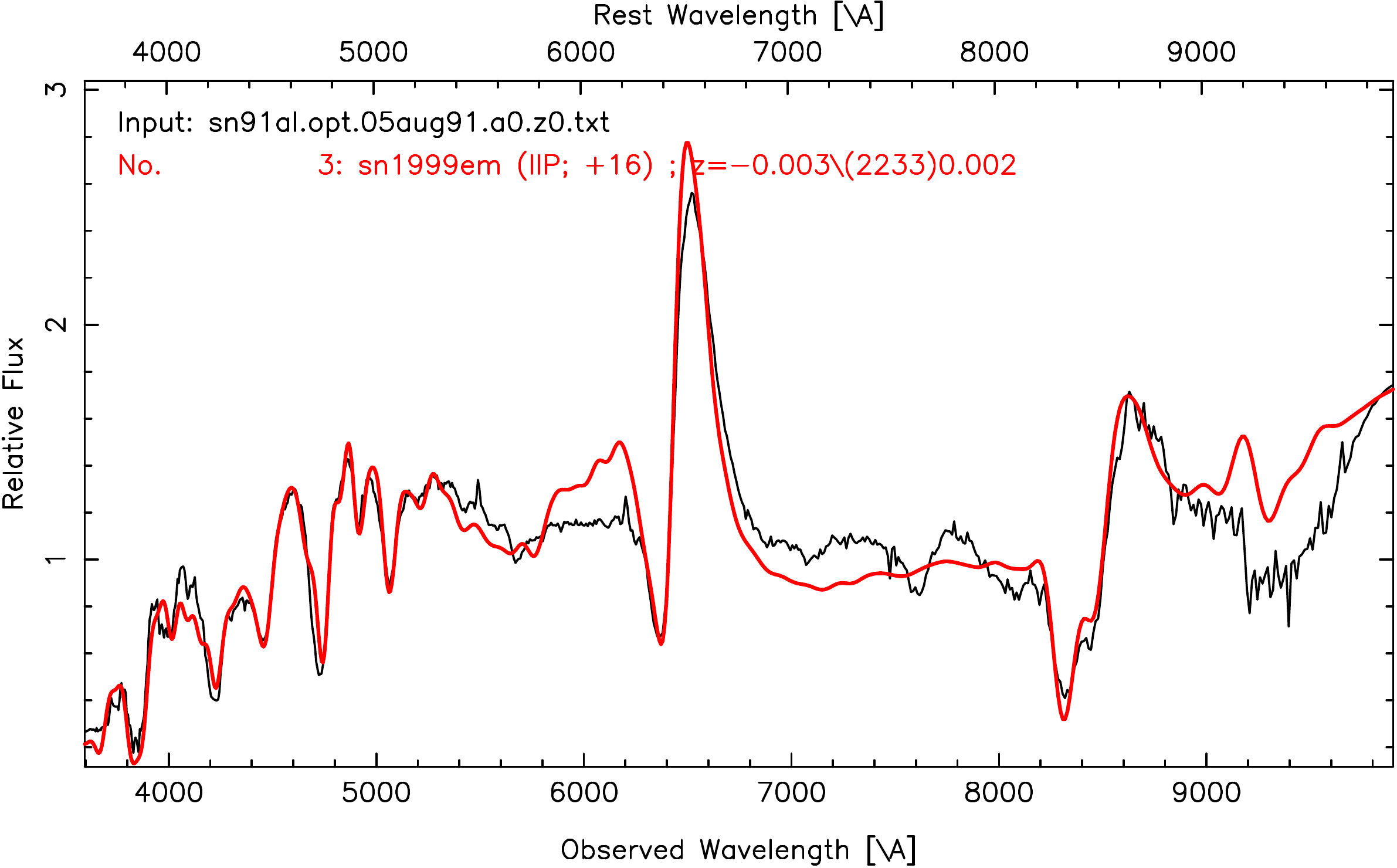}
+\includegraphics[width=4.4cm]{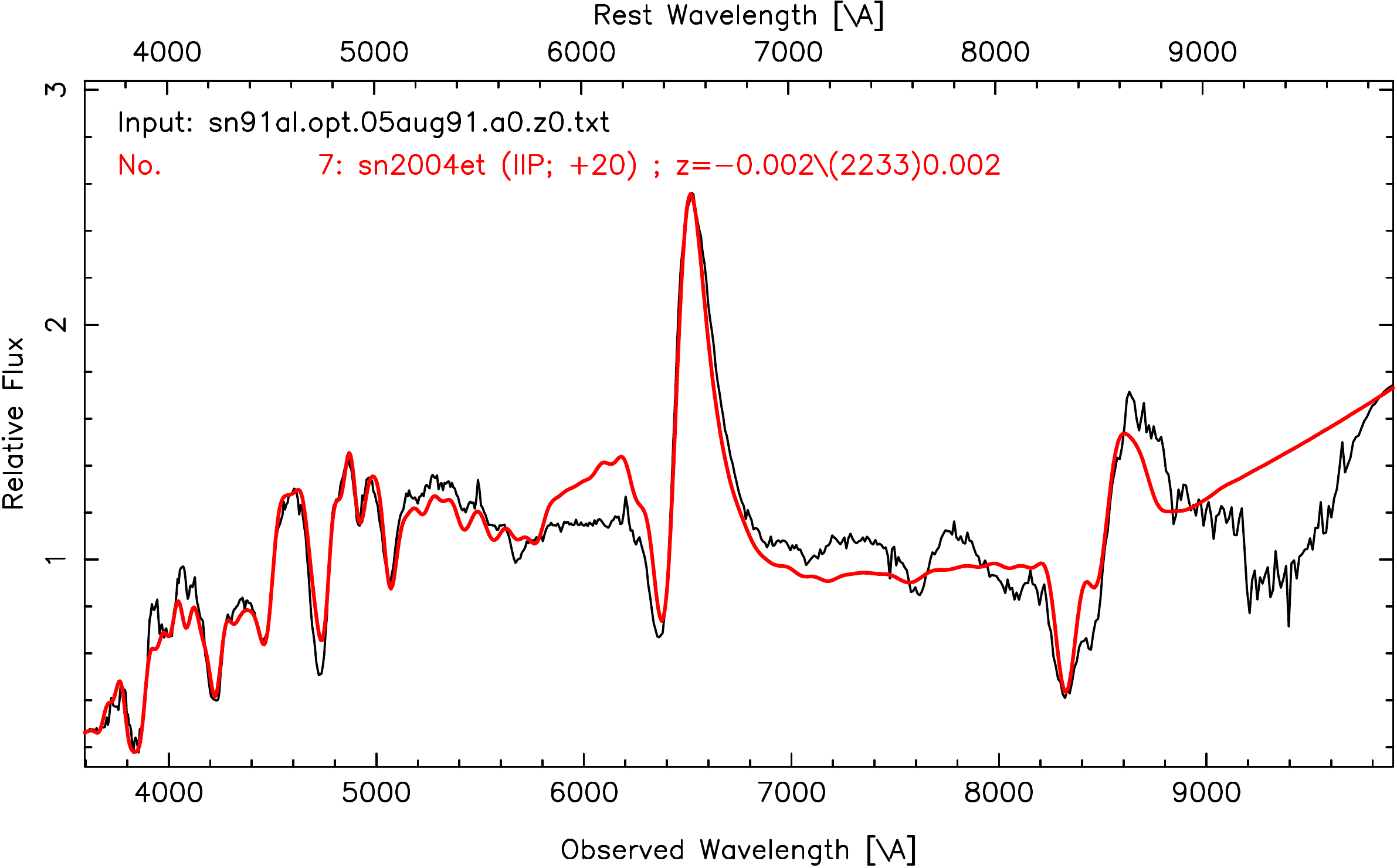}
\includegraphics[width=4.4cm]{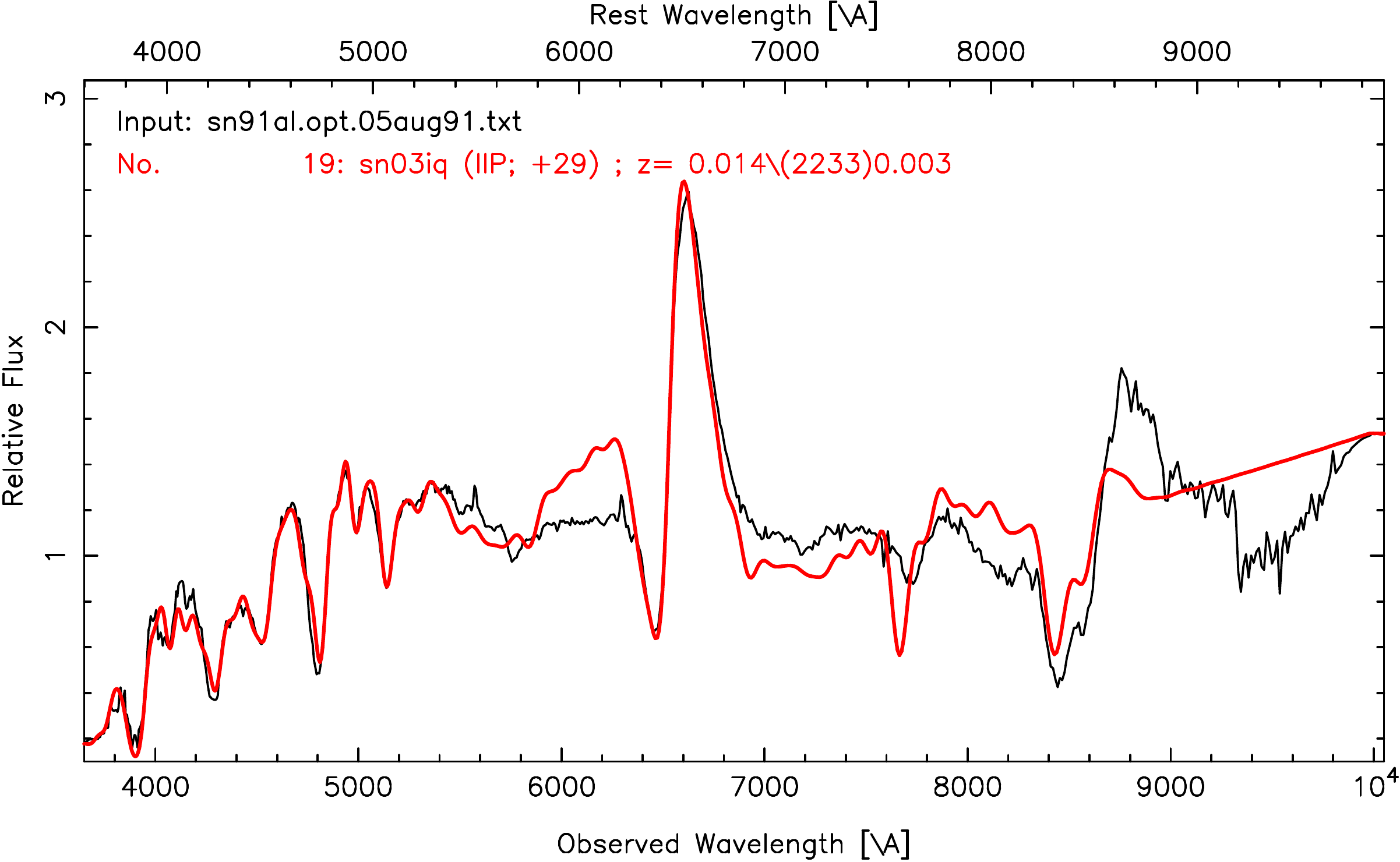}
\caption{Best spectral matching of SN~1991al using SNID. The plots show SN~1991al compared with 
SN~2006bp, SN~1999em, SN004et, and SN~2003iq at 34, 26, 36, and 29 days from explosion.}
\end{figure}

\clearpage

\begin{figure}[h!]
\centering
\includegraphics[width=4.4cm]{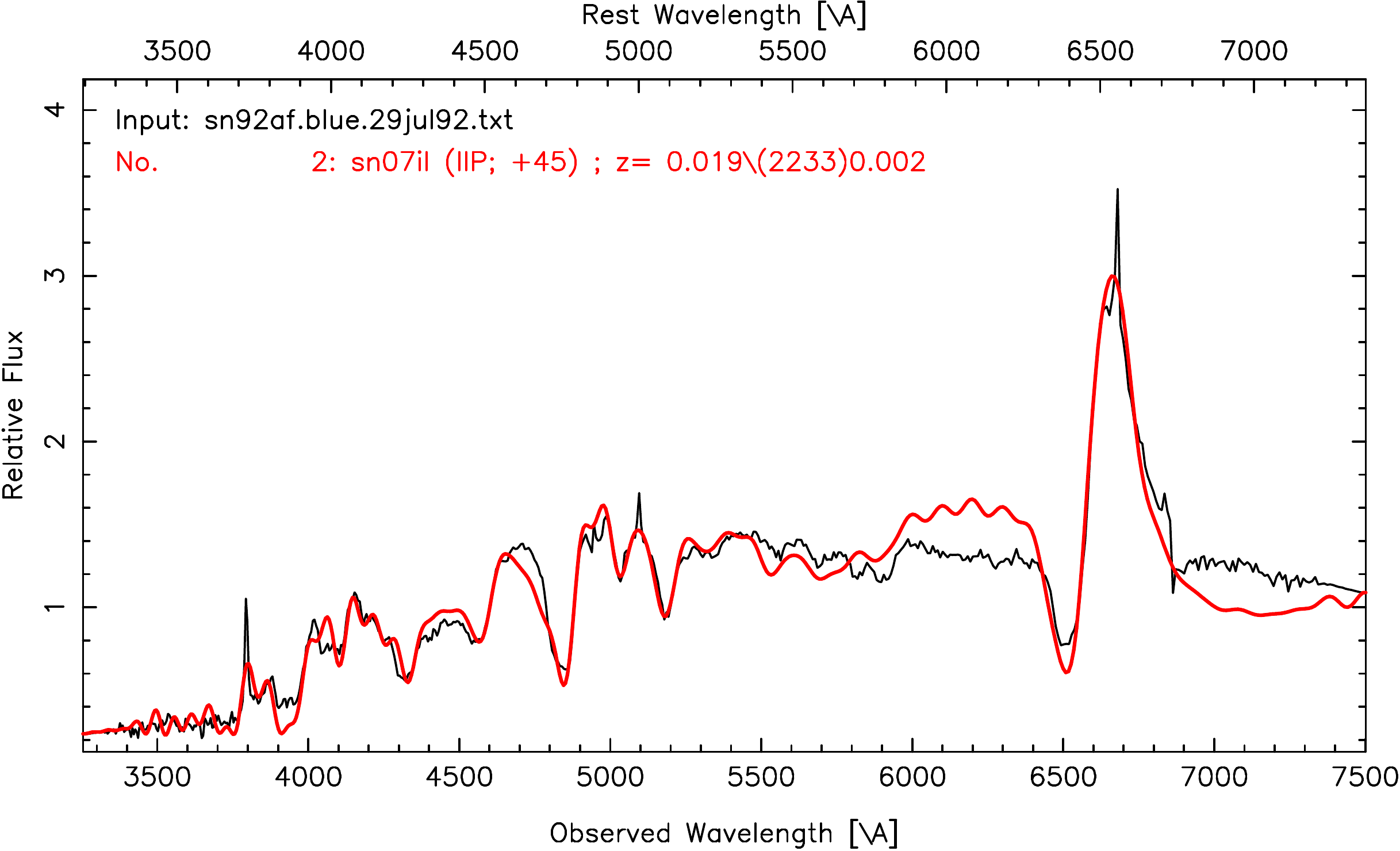}
\includegraphics[width=4.4cm]{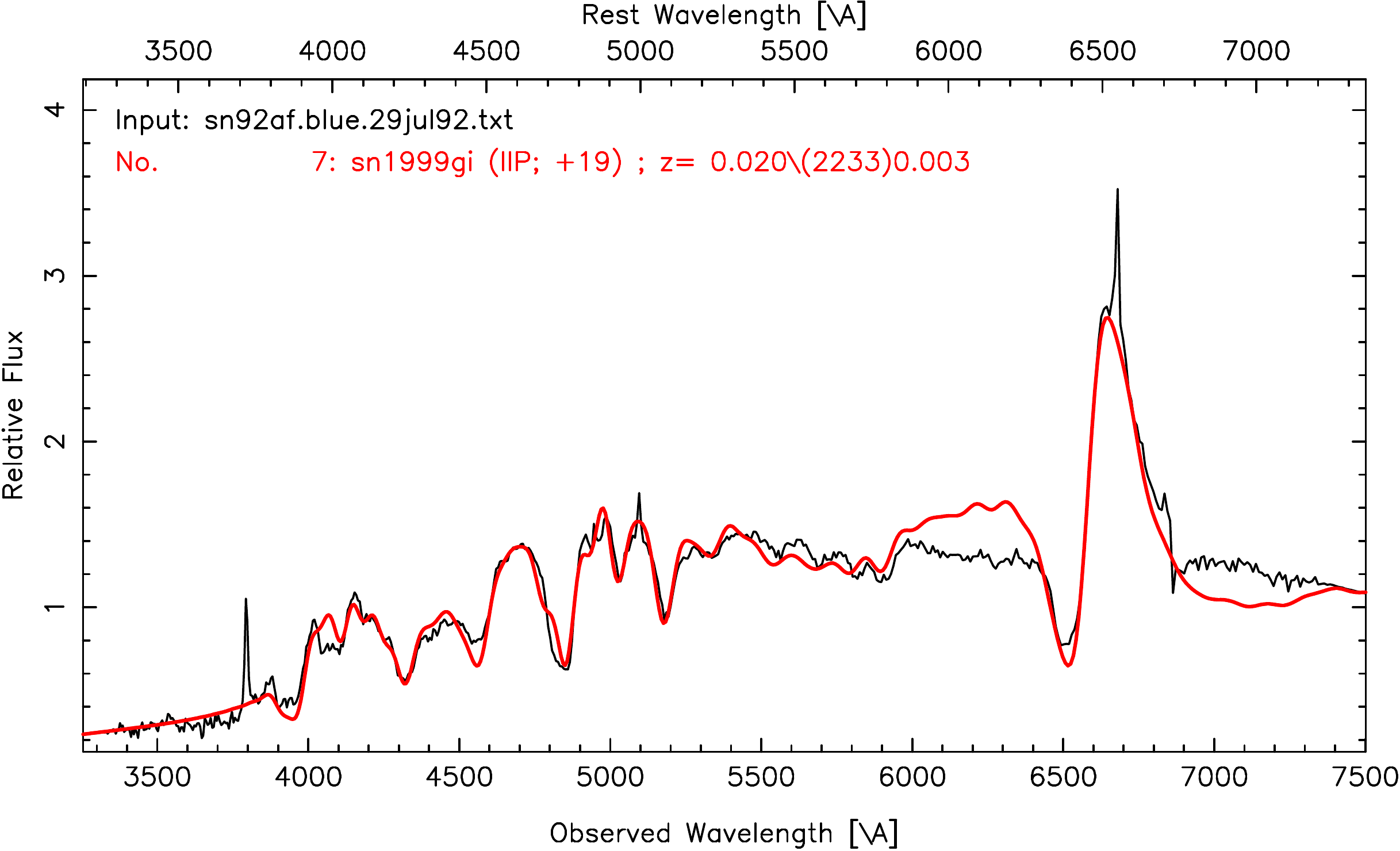}
\includegraphics[width=4.4cm]{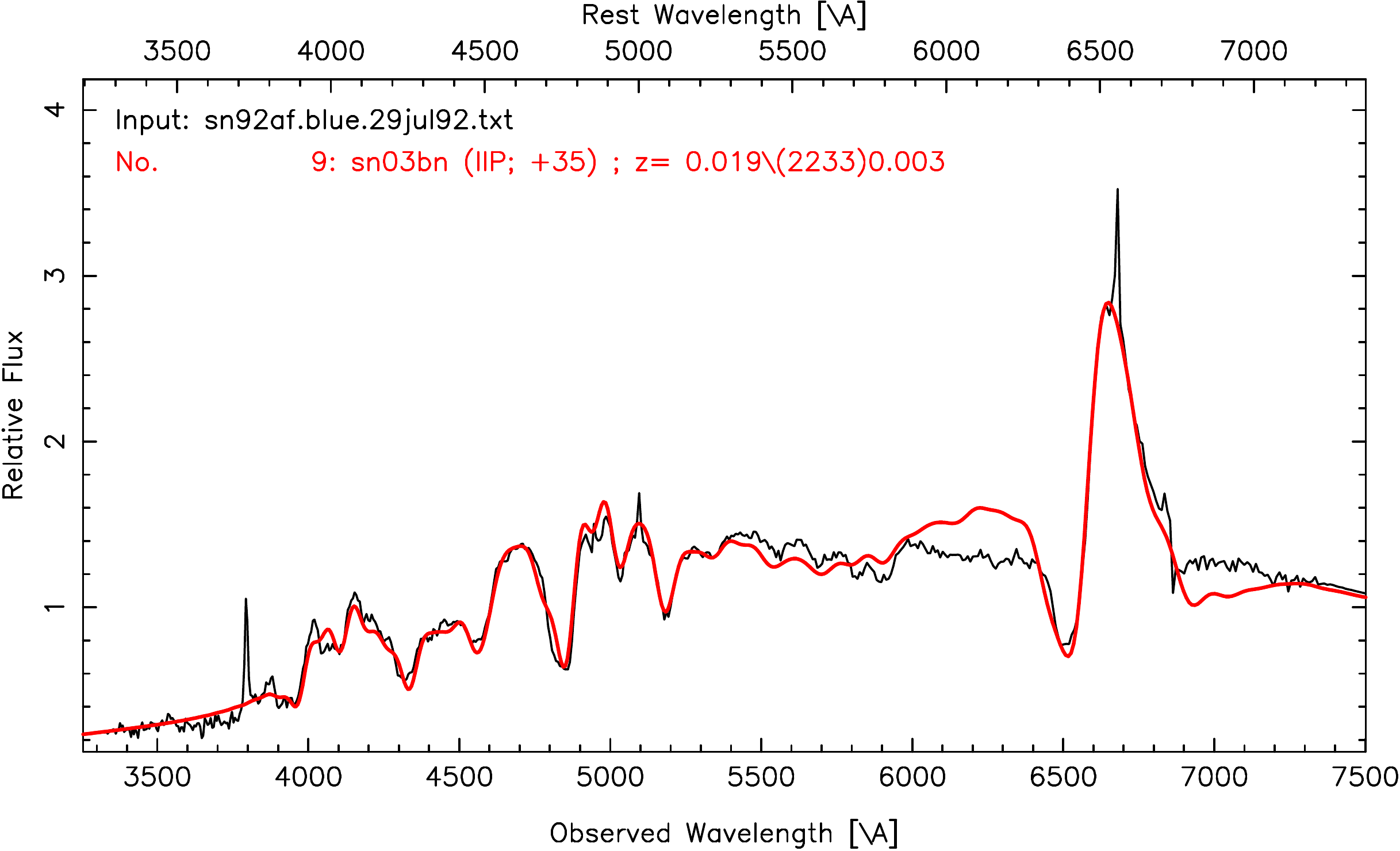}
\includegraphics[width=4.4cm]{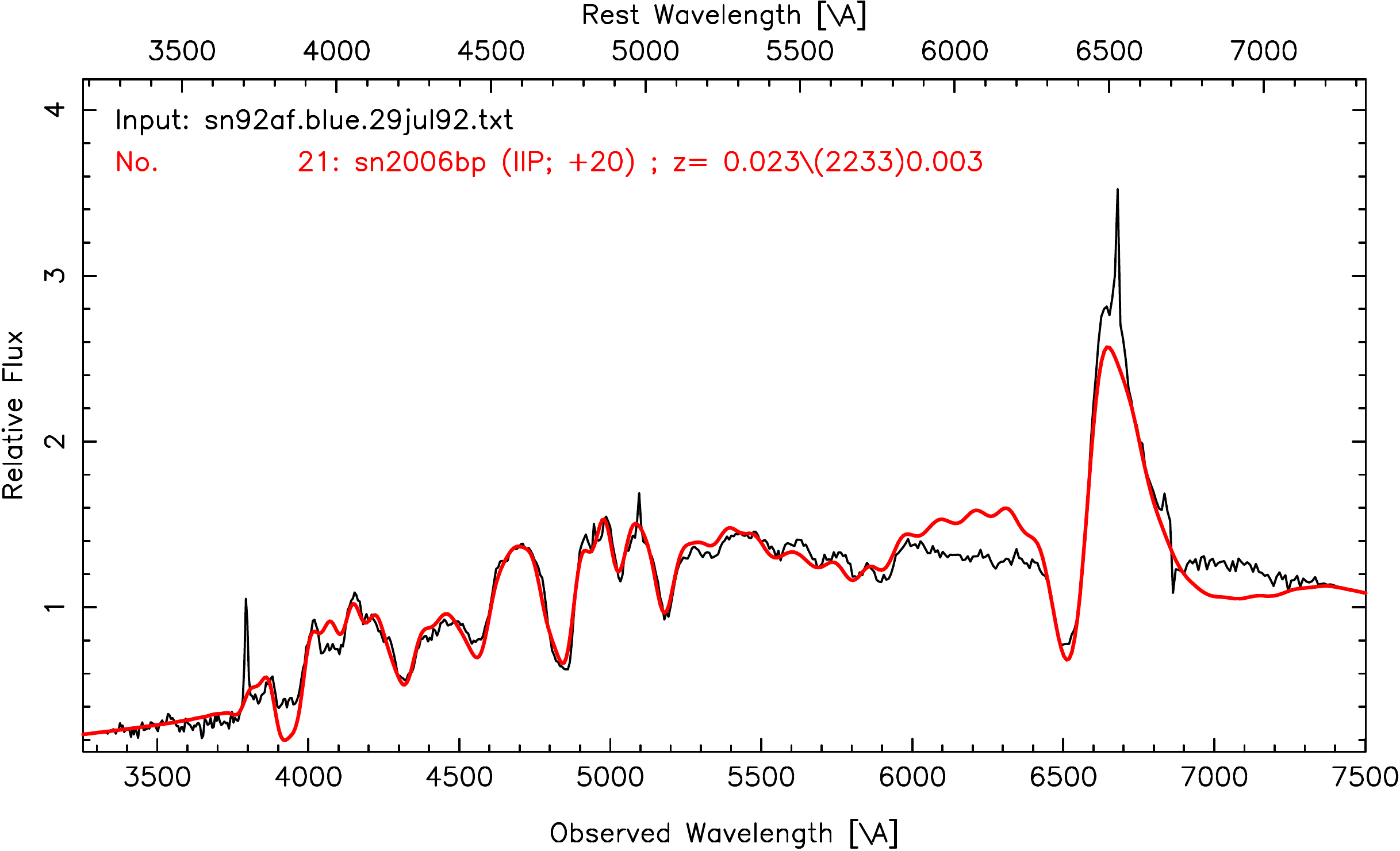}
\includegraphics[width=4.4cm]{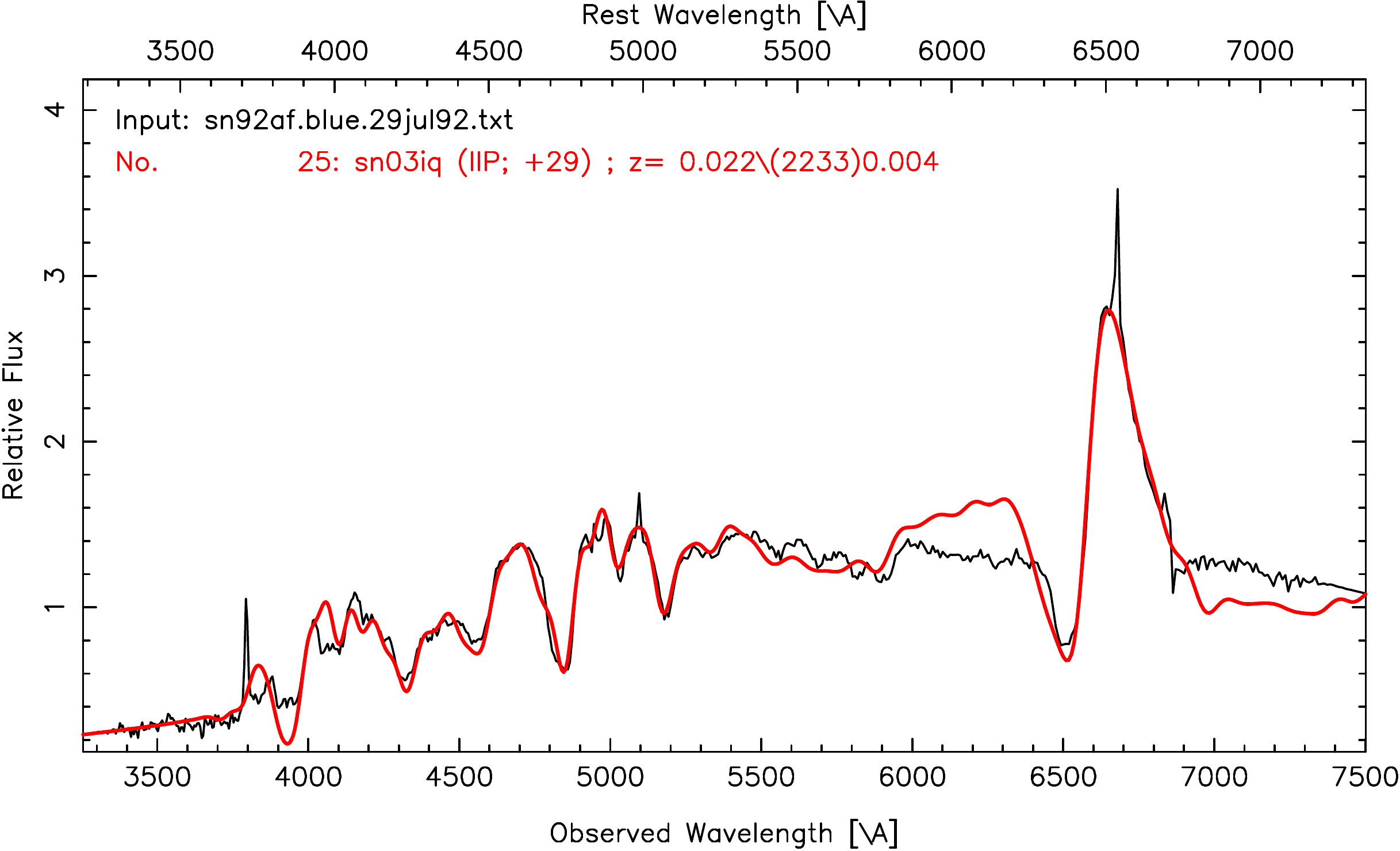}
\includegraphics[width=4.4cm]{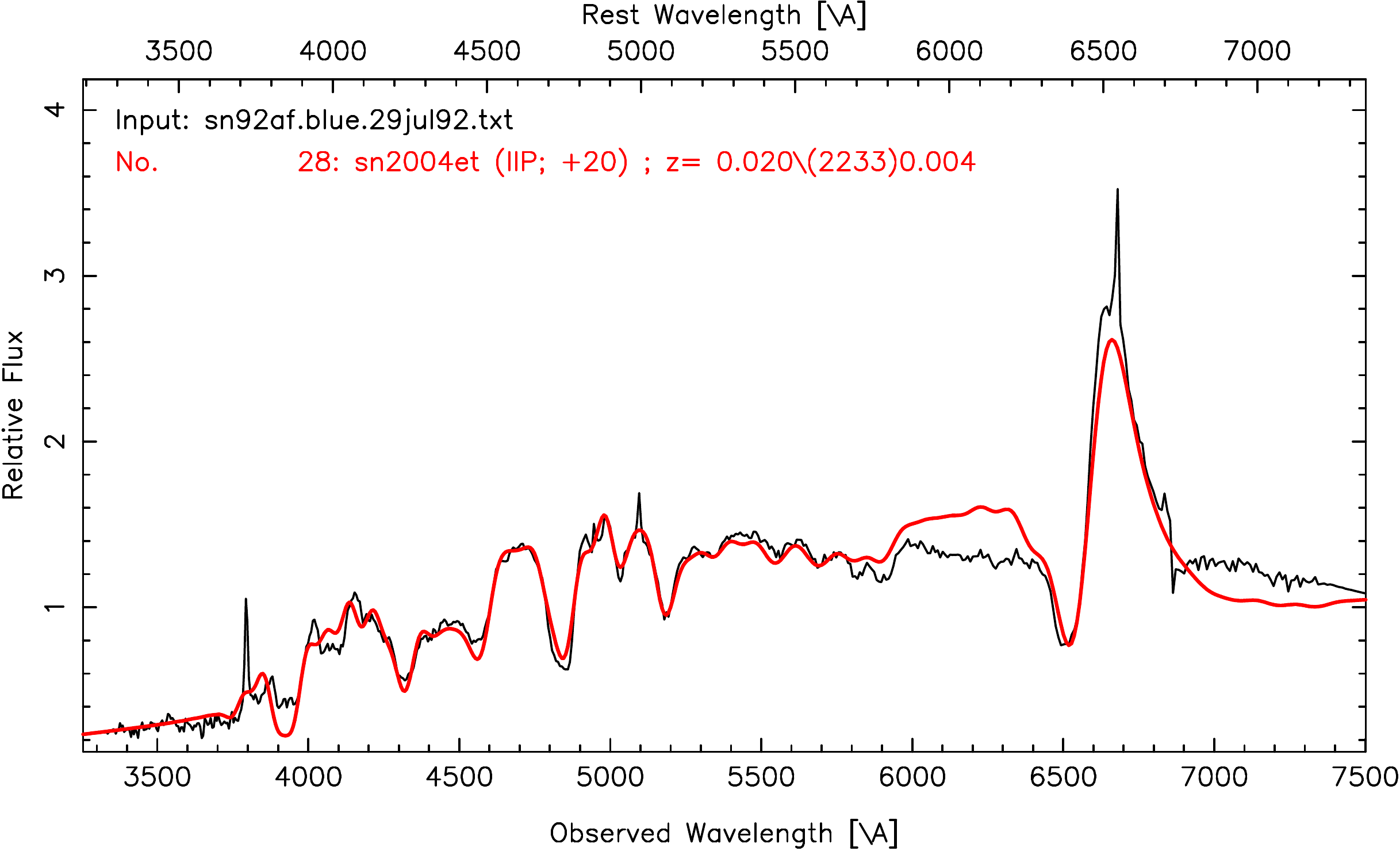}
\caption{Best spectral matching of SN~1992af using SNID. The plots show SN~1992af compared with 
SN~2007il, SN~199gi, SN~2003bn, SN~2006bp, SN~2003iq, and SN~2004et at 45, 31, 35, 29, 29, and 36 days from explosion.}
\end{figure}

\begin{figure}[h!]
\centering
\includegraphics[width=4.4cm]{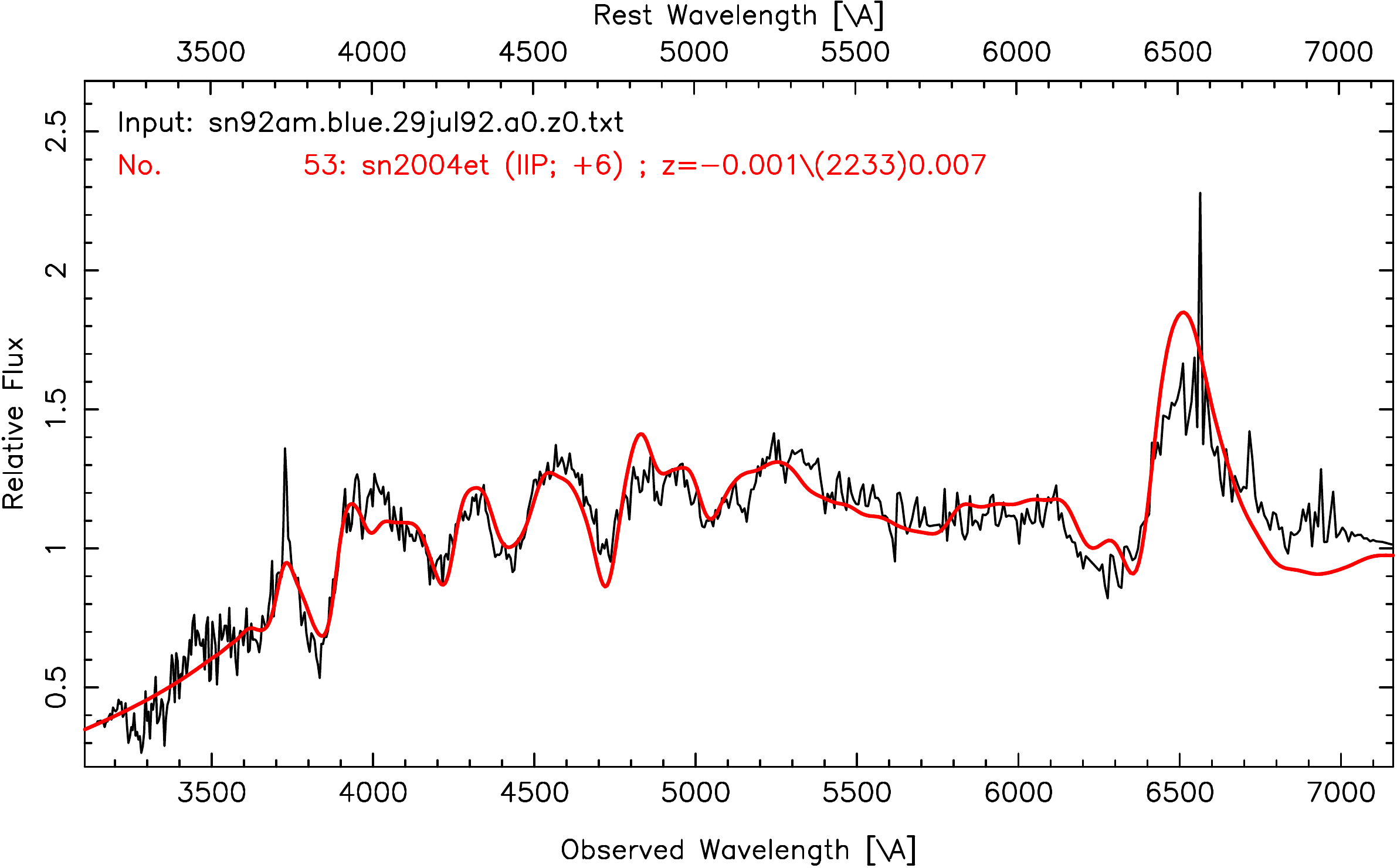}
\includegraphics[width=4.4cm]{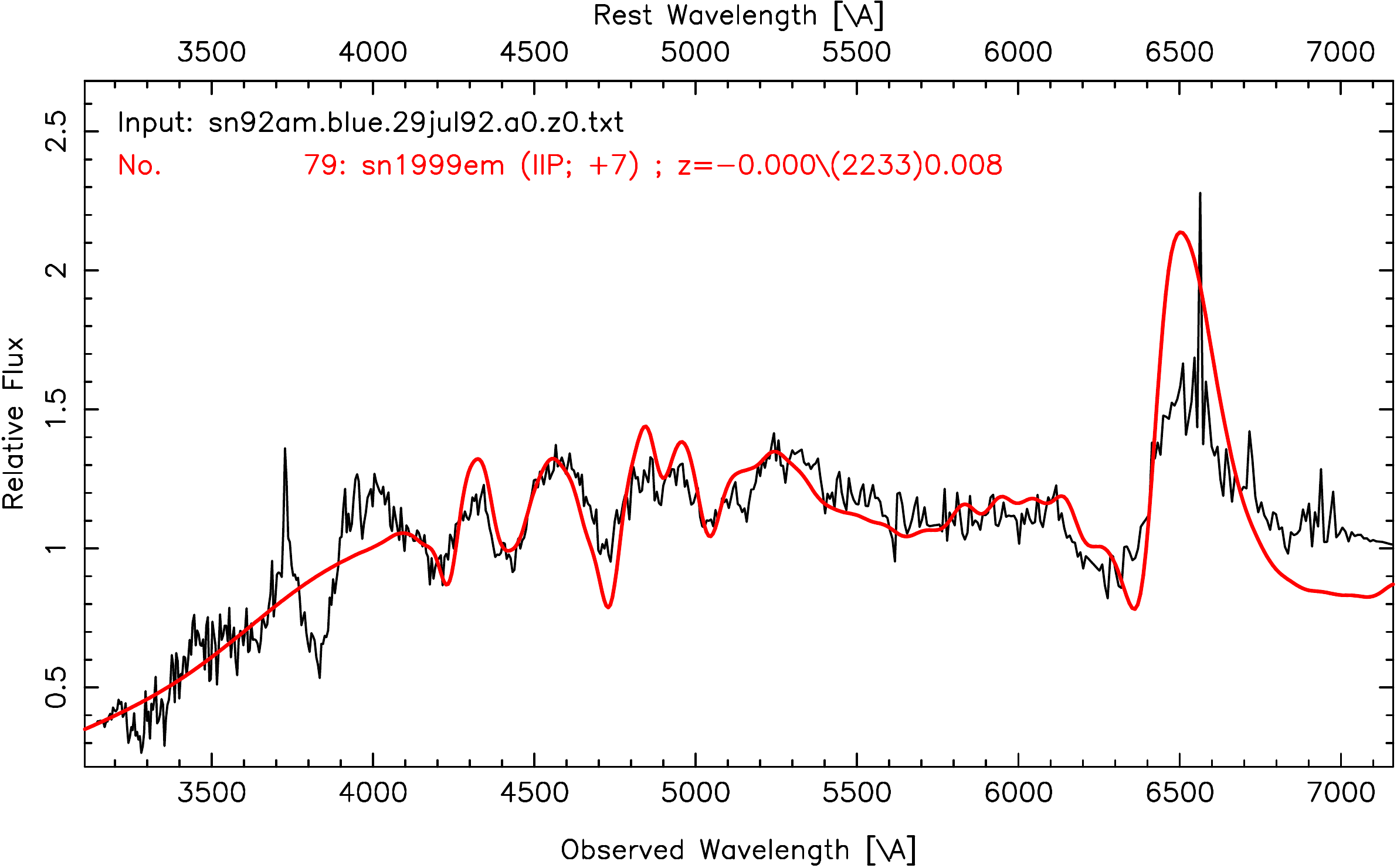}
\caption{Best spectral matching of SN~1992am using SNID. The plots show SN~1992am compared with 
SN~2004et and SN~1999em at 21 and 17 days from explosion.}
\end{figure}

\begin{figure}[h!]
\centering
\includegraphics[width=4.4cm]{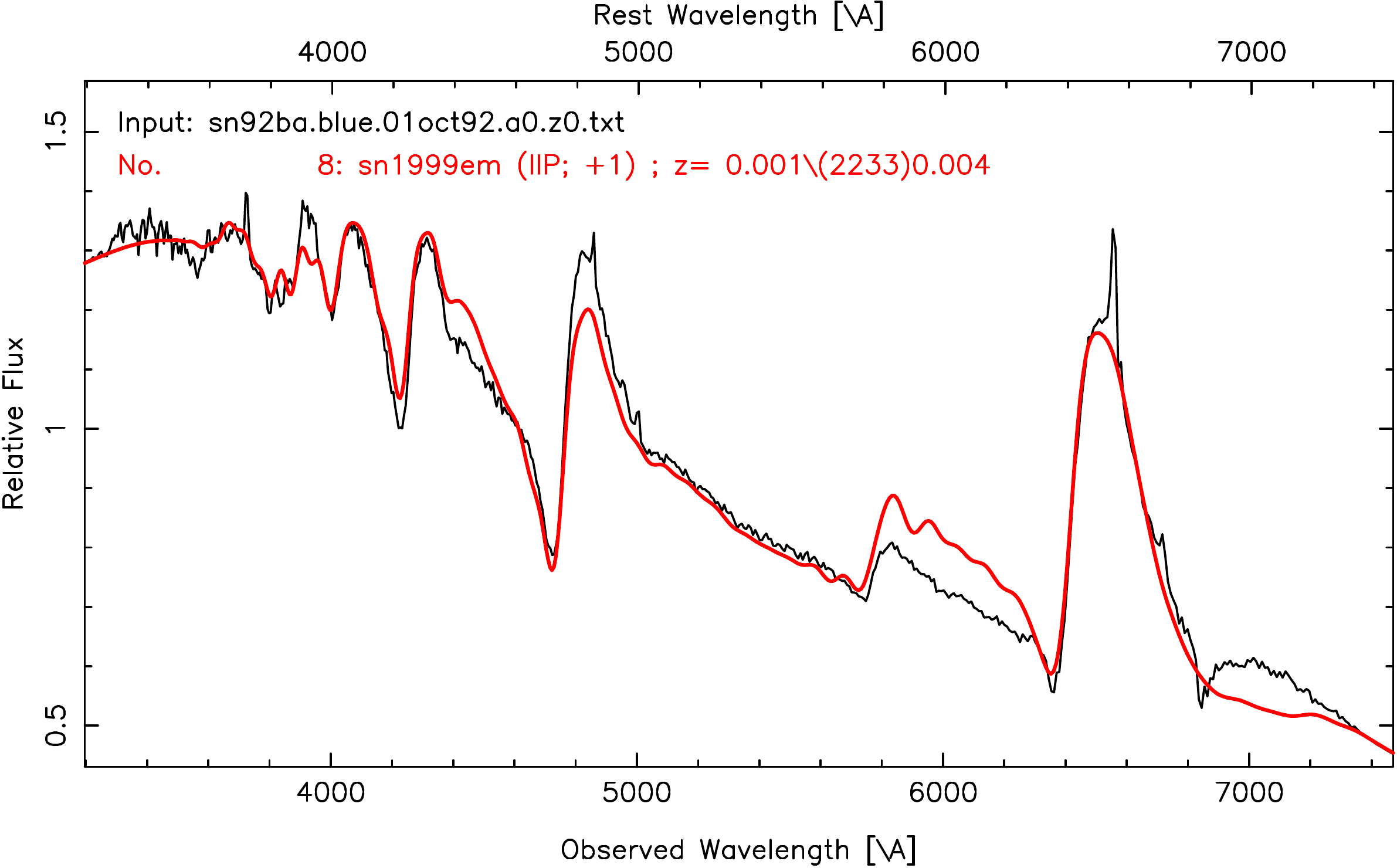}
\includegraphics[width=4.4cm]{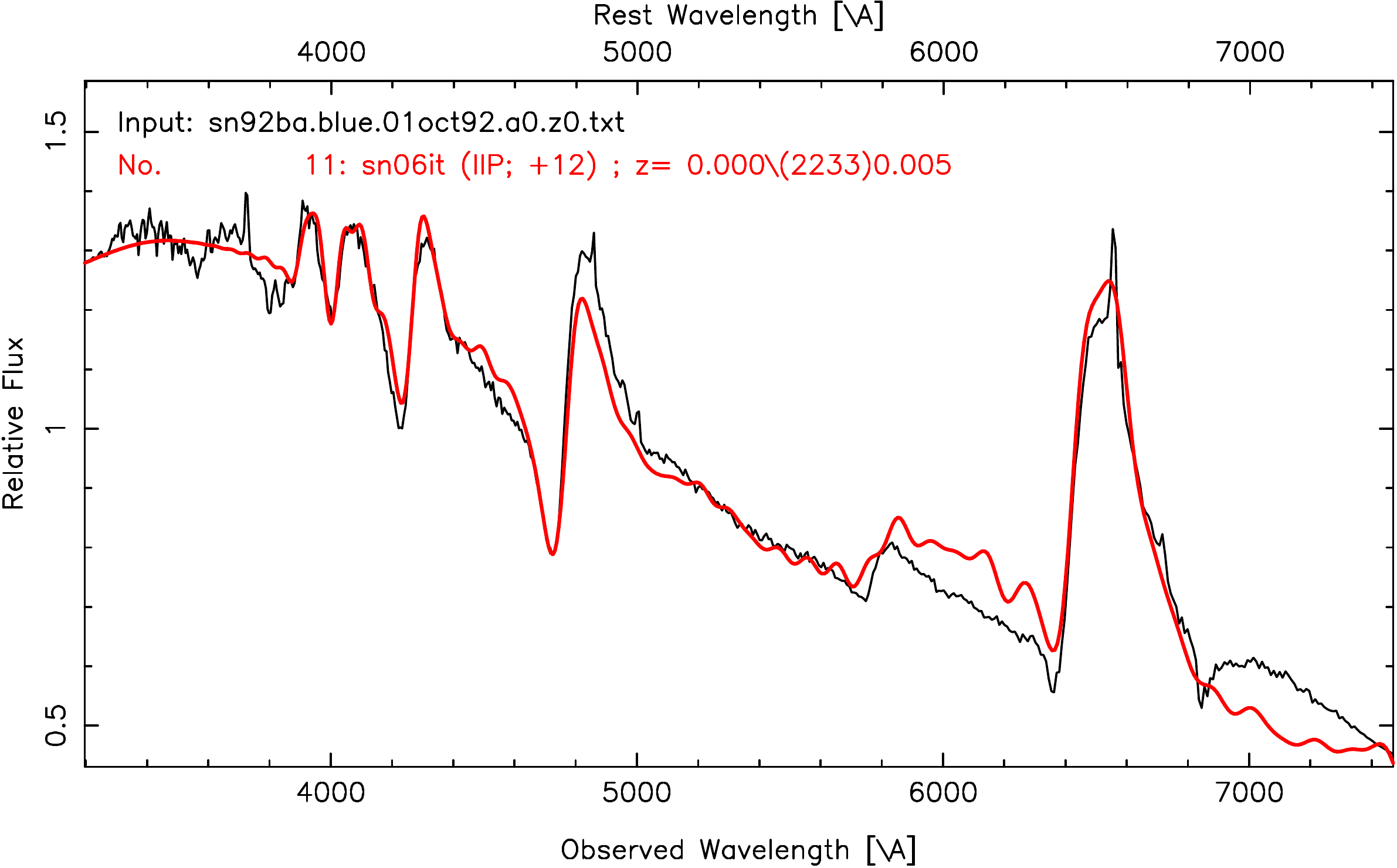}
\caption{Best spectral matching of SN~1992ba using SNID. The plots show SN~1992ba compared with 
SN~1999em and SN~2006it at 11 and 12 days from explosion.}
\end{figure}

\clearpage

\begin{figure}[h!]
\centering
\includegraphics[width=4.4cm]{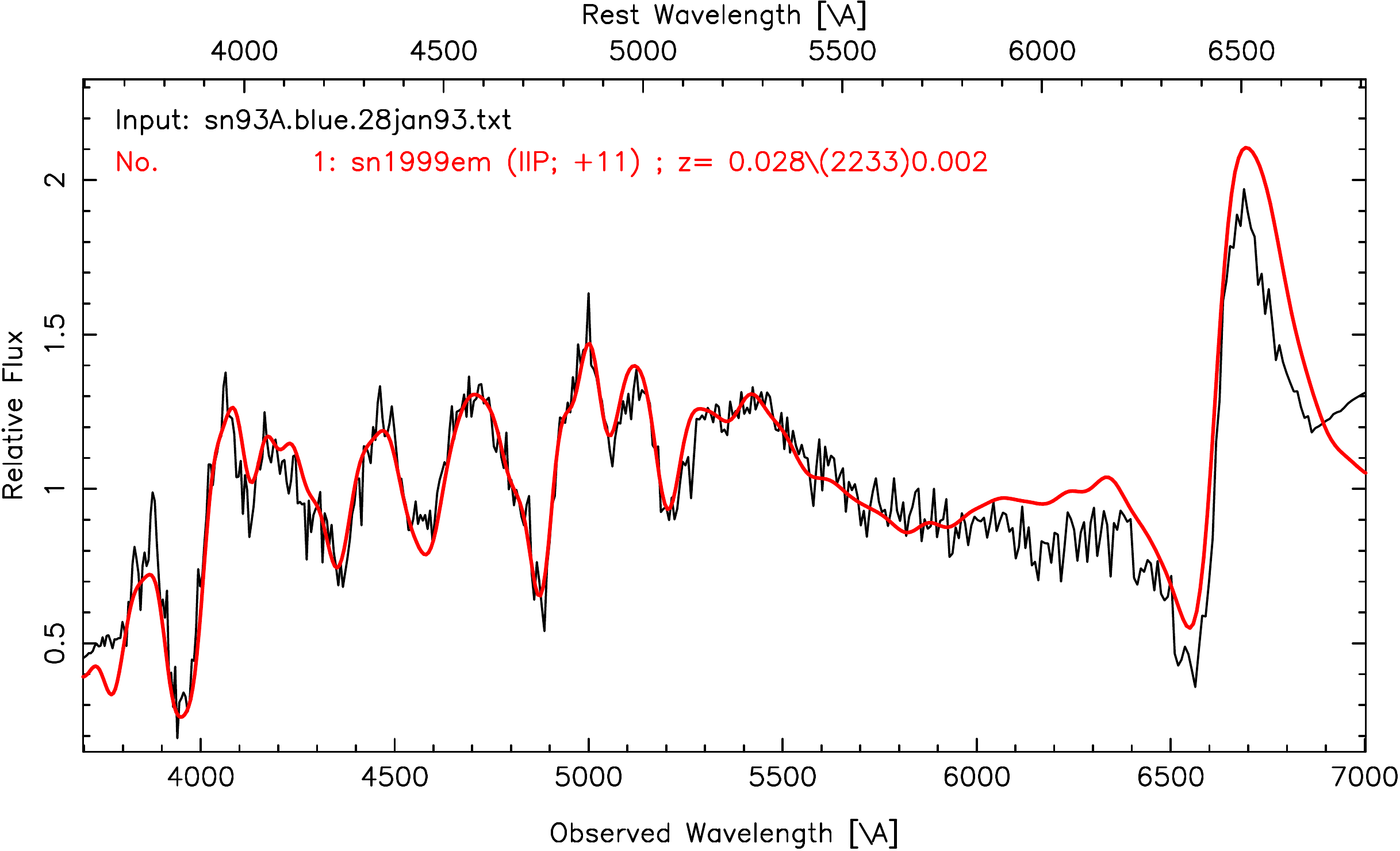}
\includegraphics[width=4.4cm]{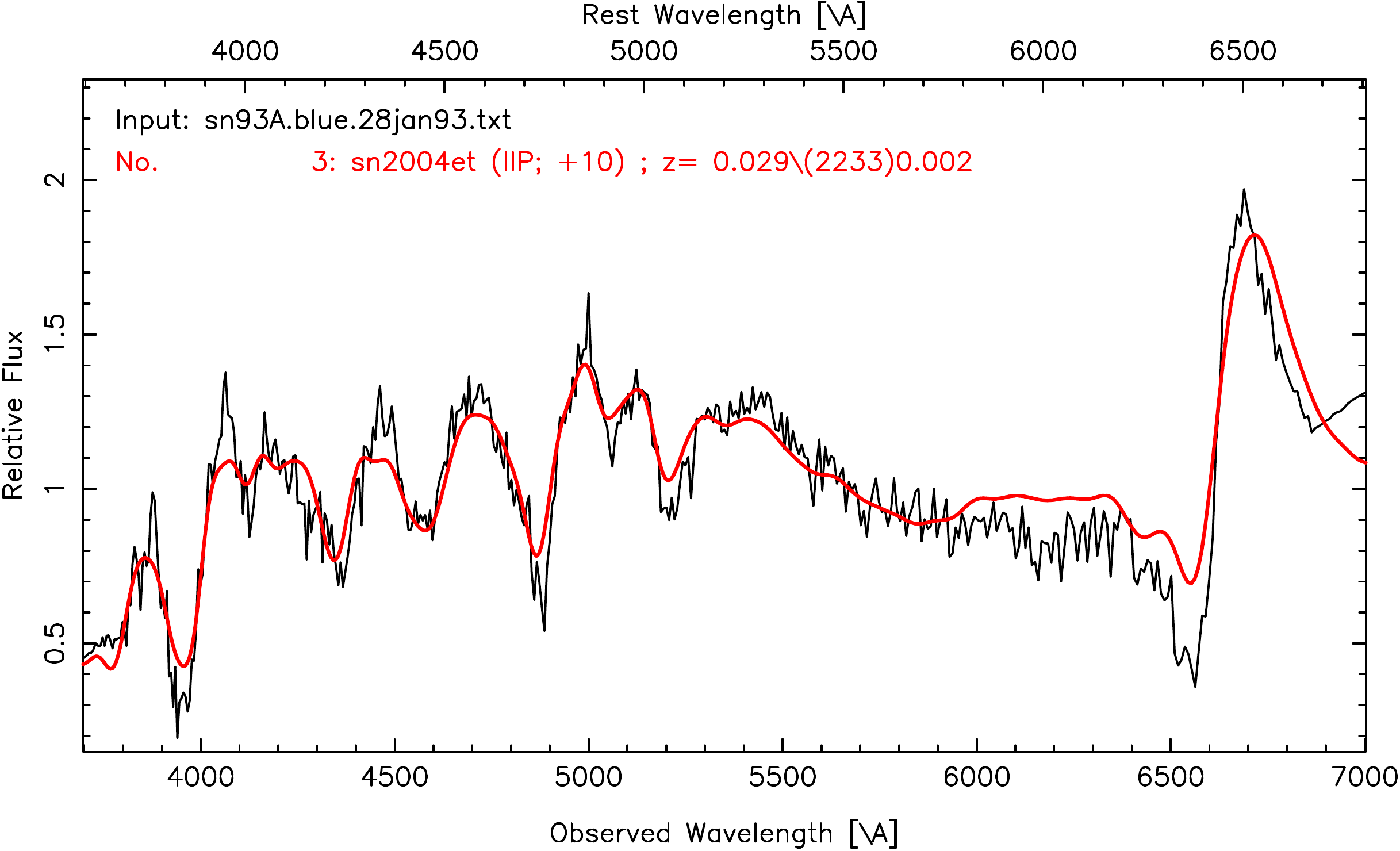}
\includegraphics[width=4.4cm]{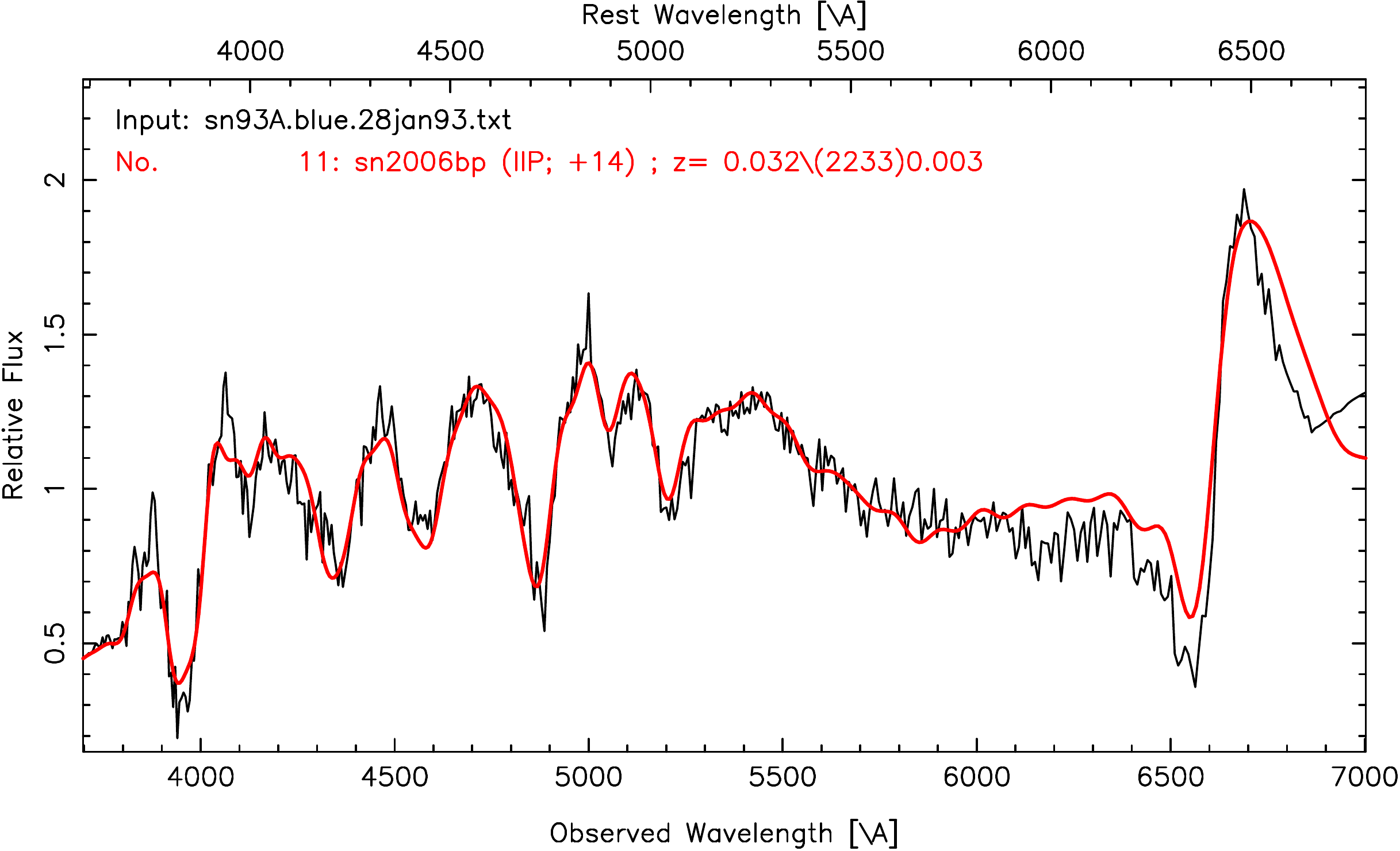}
\caption{Best spectral matching of SN~1993A using SNID. The plots show SN~1993A compared with 
SN~1999em, SN~2004et, SN~2006bb at 21, 36, and 23 days from explosion.}
\end{figure}

\begin{figure}[h!]
\centering
\includegraphics[width=4.4cm]{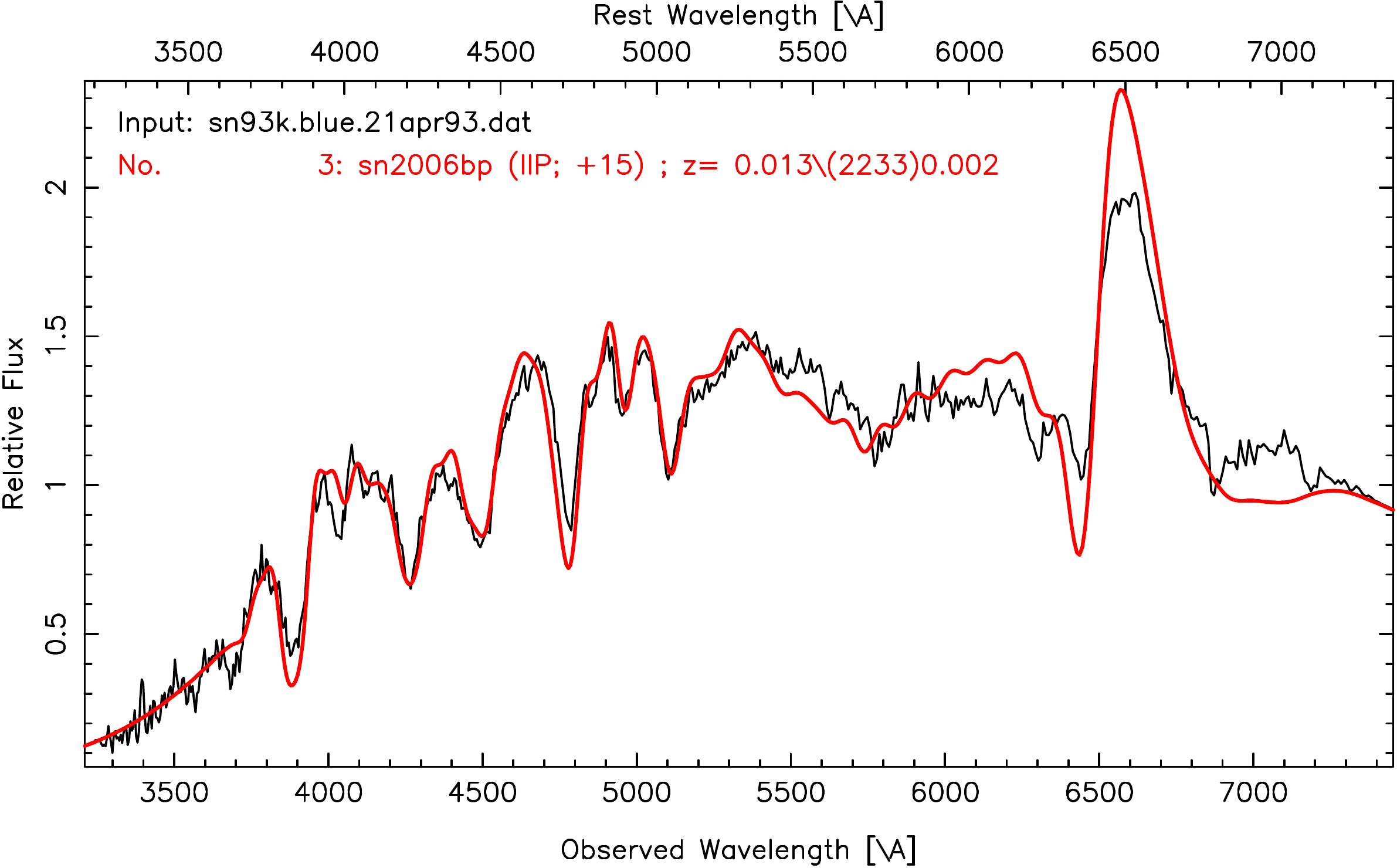}
\includegraphics[width=4.4cm]{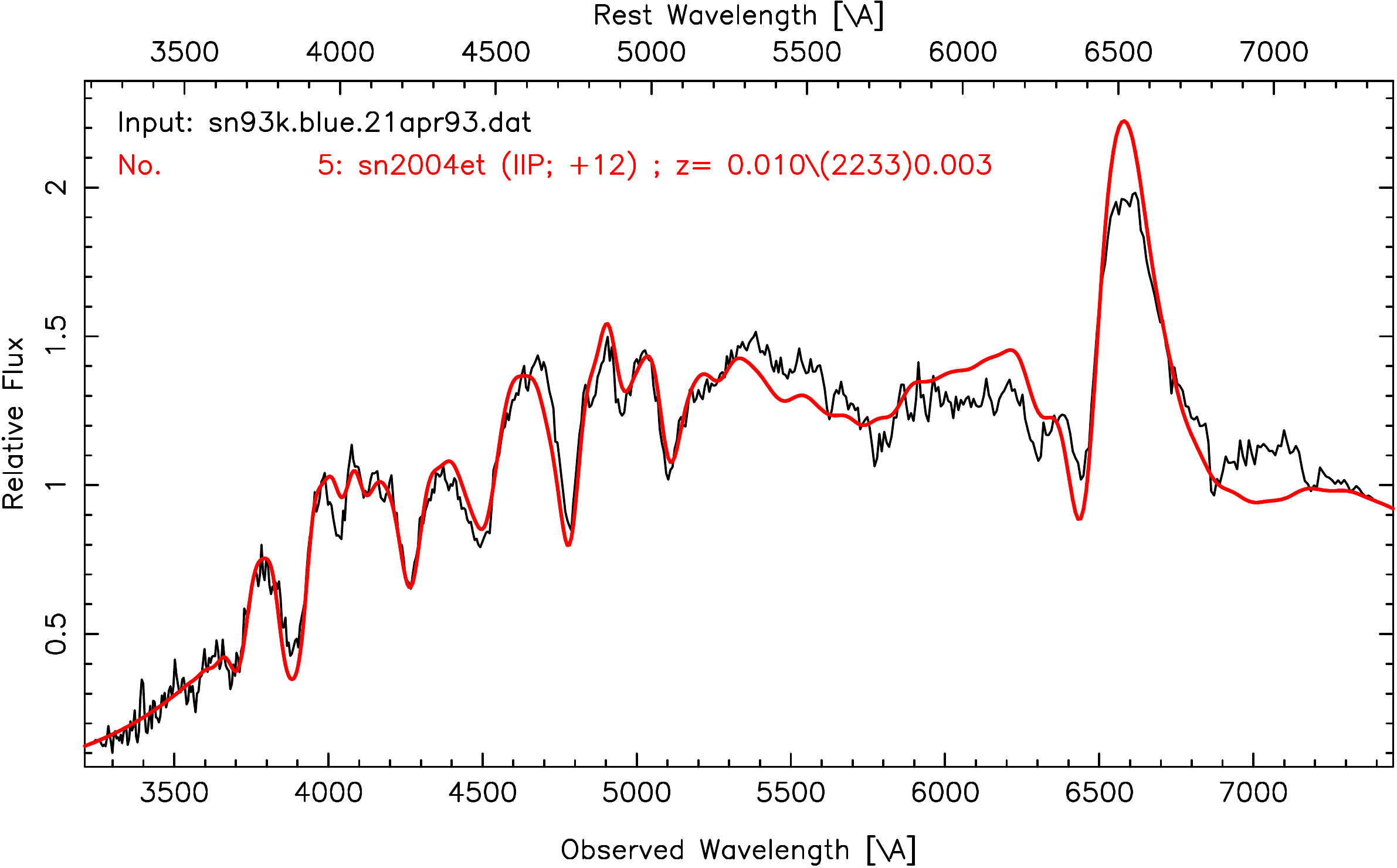}
\includegraphics[width=4.4cm]{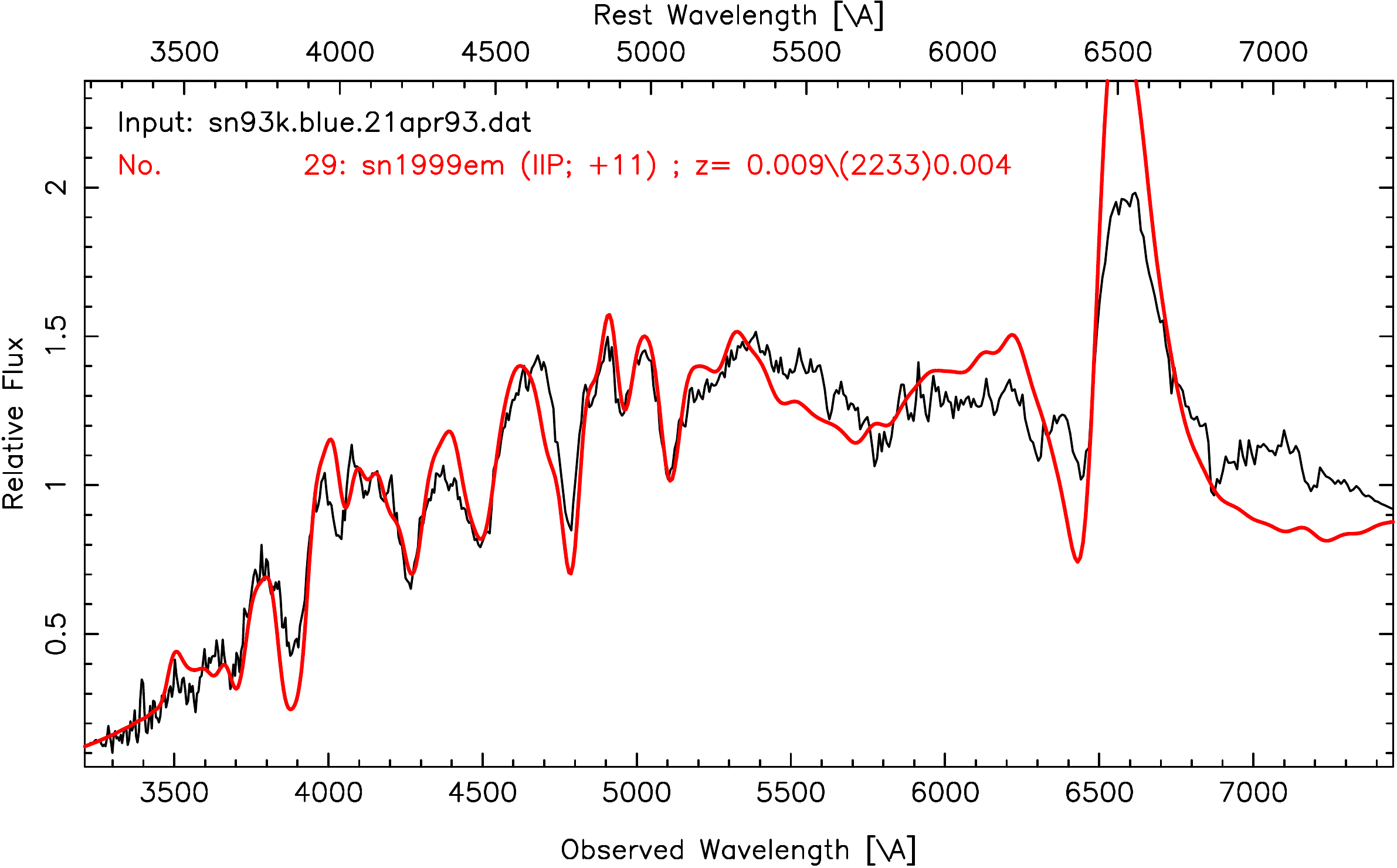}
\includegraphics[width=4.4cm]{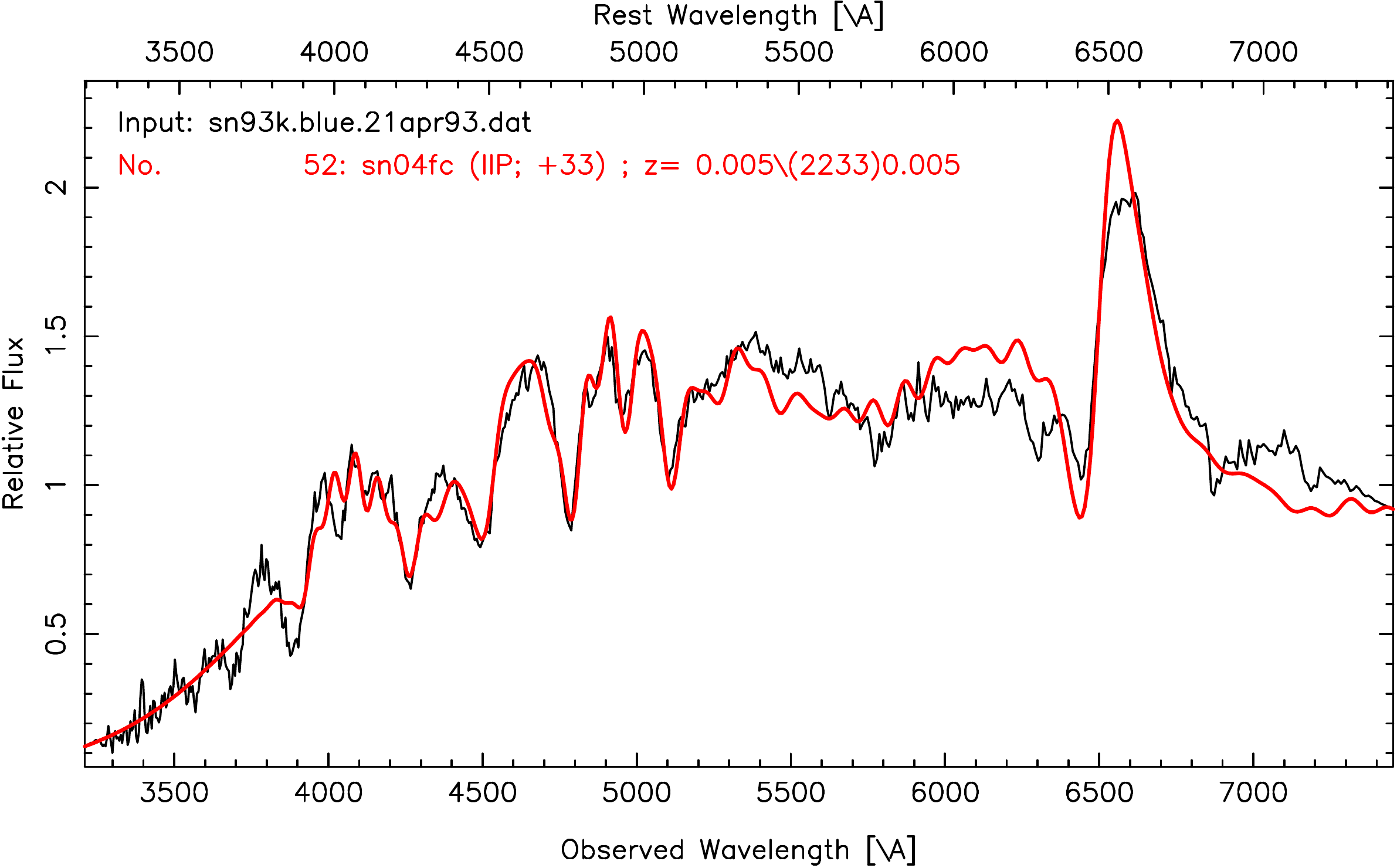}
\caption{Best spectral matching of SN~1993K using SNID. The plots show SN~1993K compared with 
SN~2006bp, SN~2004et, SN~1999em, and SN~2004fc at 24, 28, 21, and 33 days from explosion.}
\end{figure}

\begin{figure}[h!]
\centering
\includegraphics[width=4.4cm]{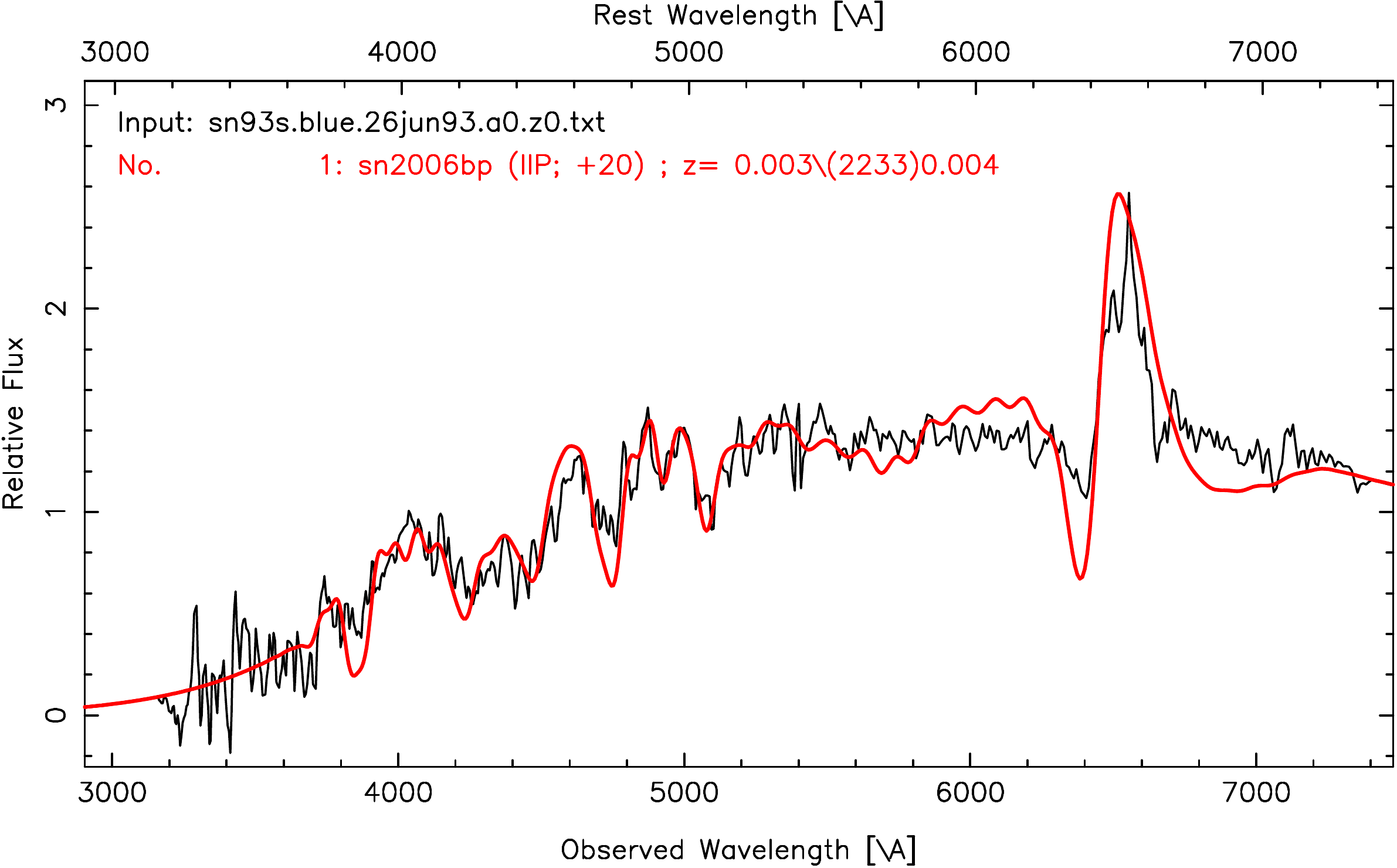}
\includegraphics[width=4.4cm]{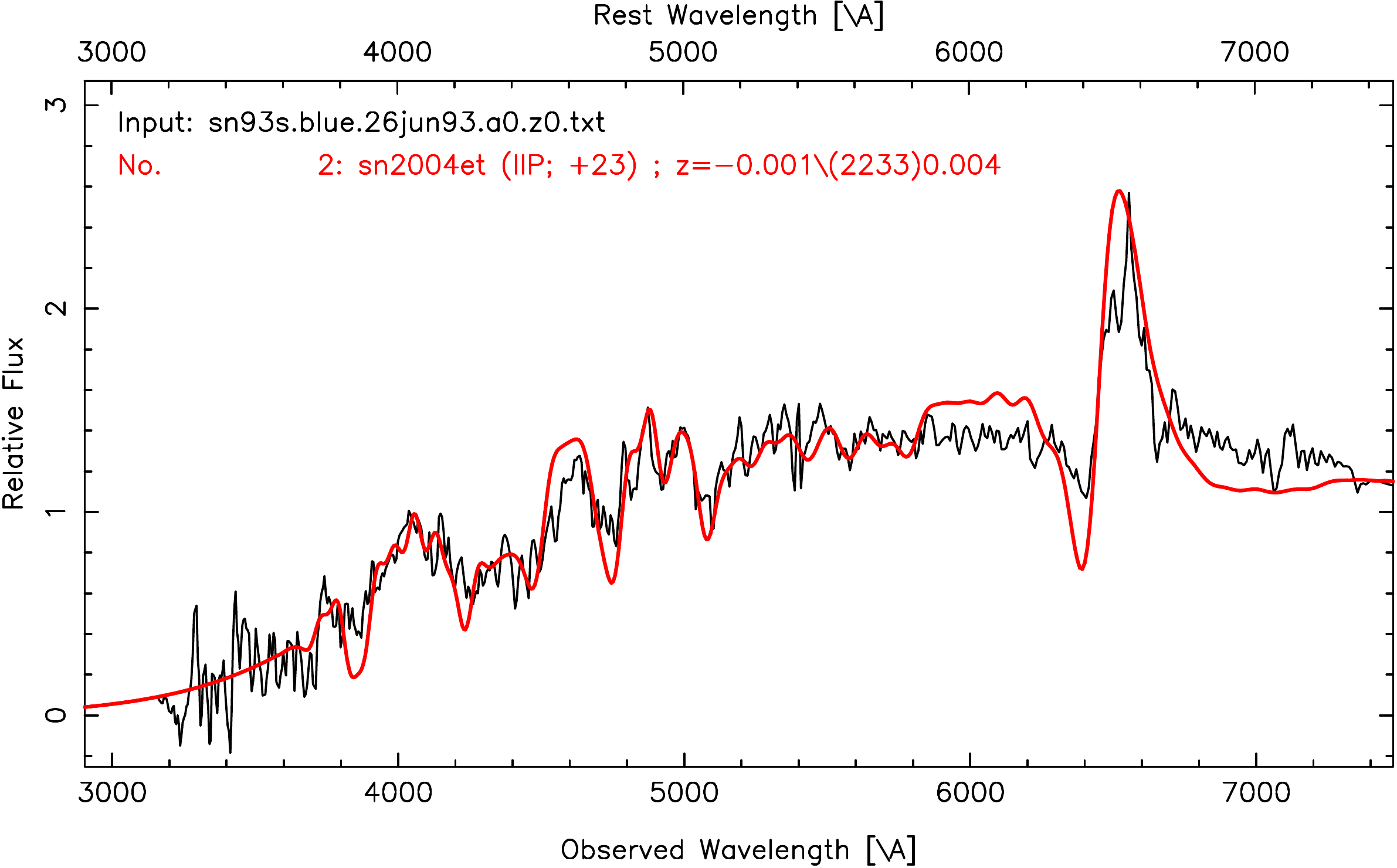}
\caption{Best spectral matching of SN~1993S using SNID. The plots show SN~1993S compared with 
SN~2006bp and SN~2004et at 29 and 39 days from explosion.}
\end{figure}

\clearpage

\begin{figure}[h!]
\centering
\includegraphics[width=4.4cm]{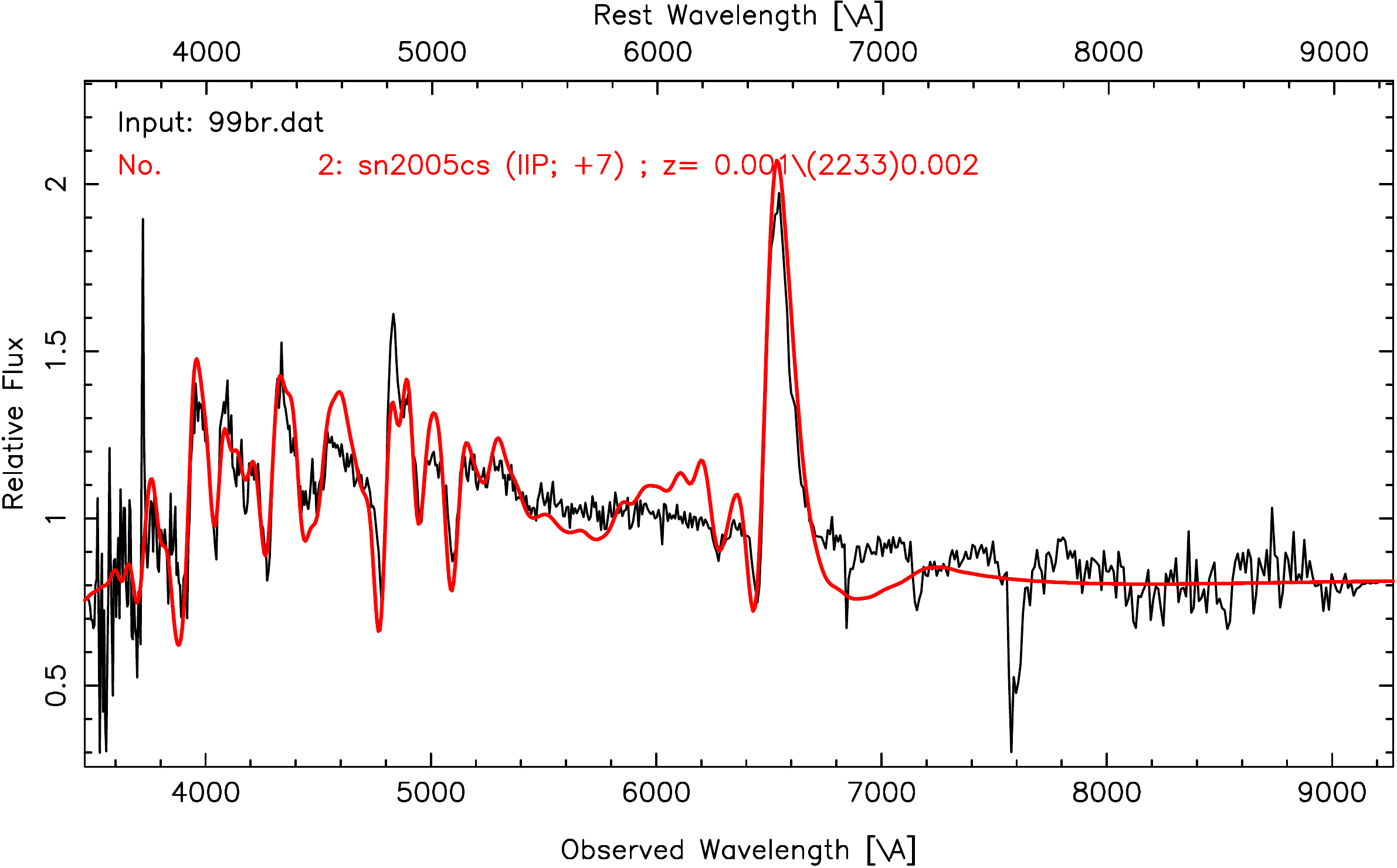}
\caption{Best spectral matching of SN~1999br using SNID. The plots show SN~1999br compared with 
SN~1999br at 13 days from explosion.}
\end{figure}

\begin{figure}[h!]
\centering
\includegraphics[width=4.4cm]{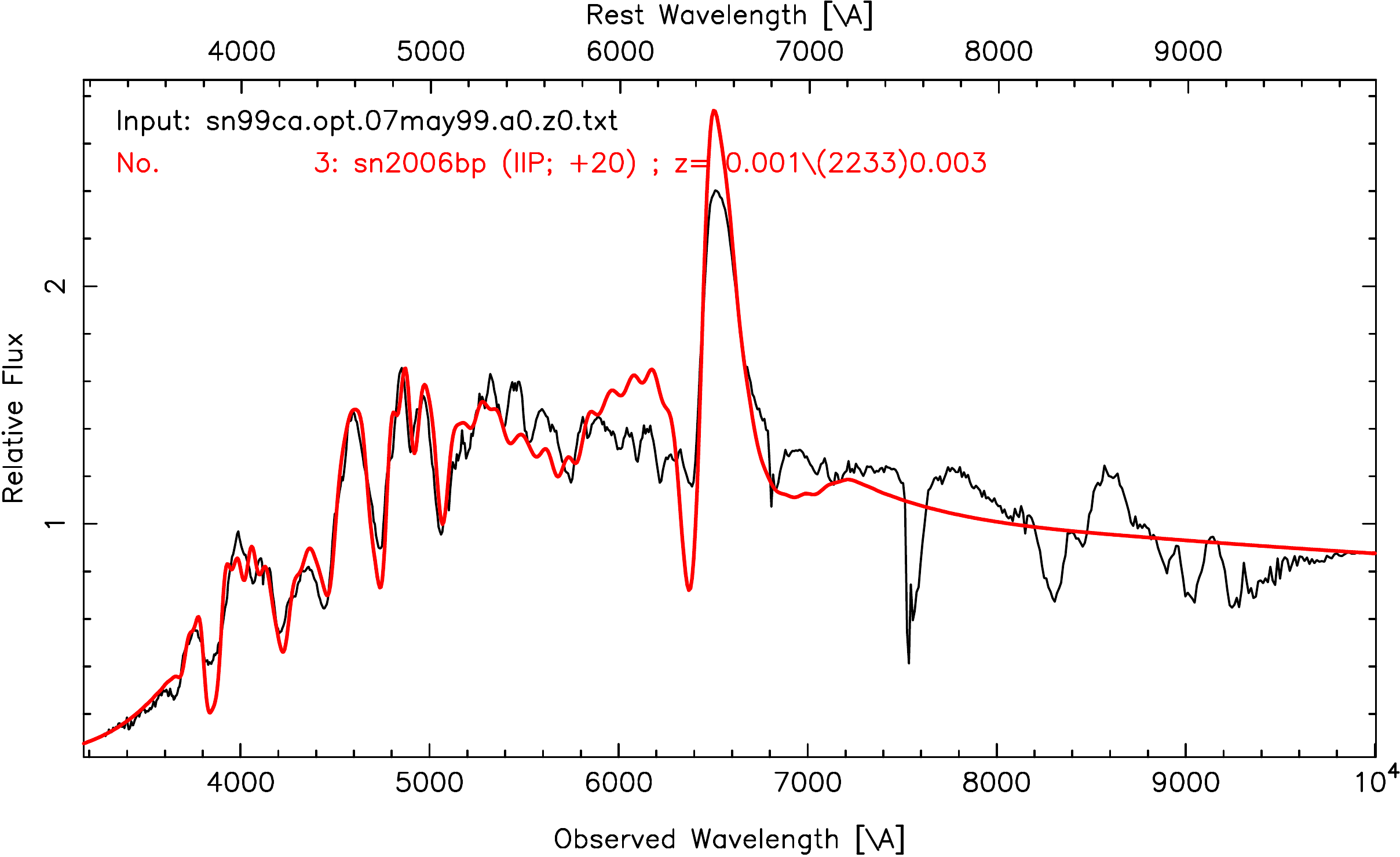}
\includegraphics[width=4.4cm]{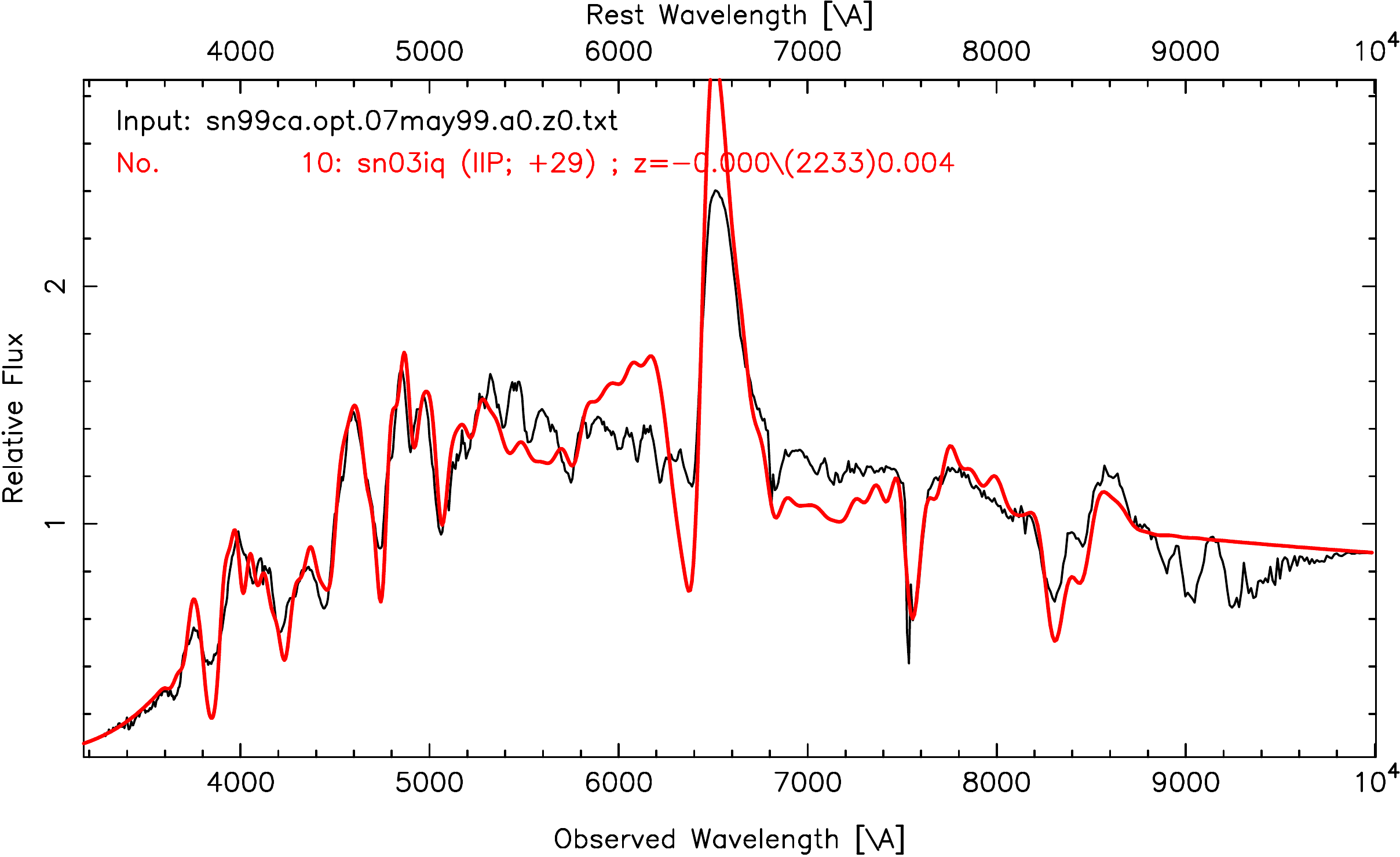}
\includegraphics[width=4.4cm]{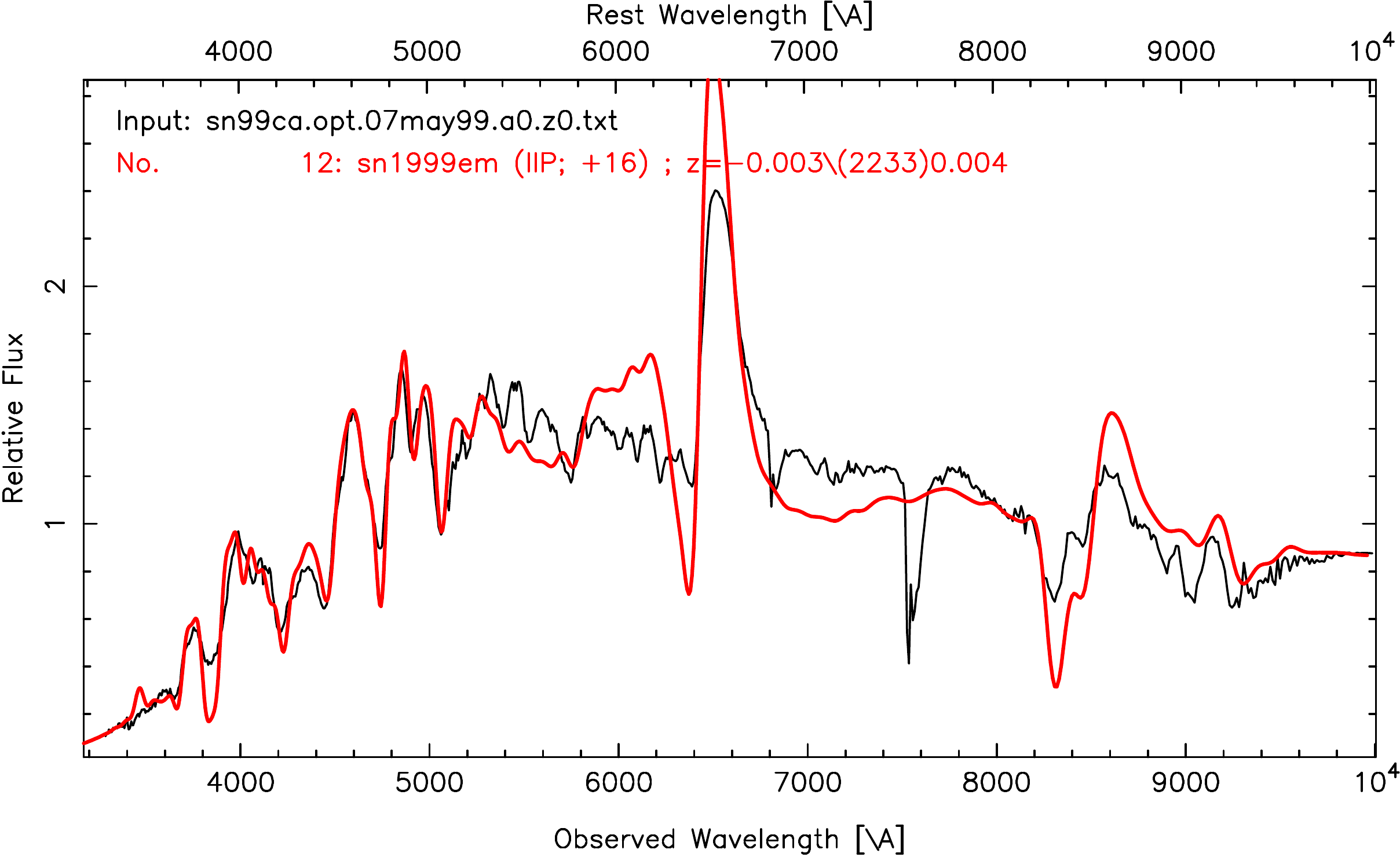}
\includegraphics[width=4.4cm]{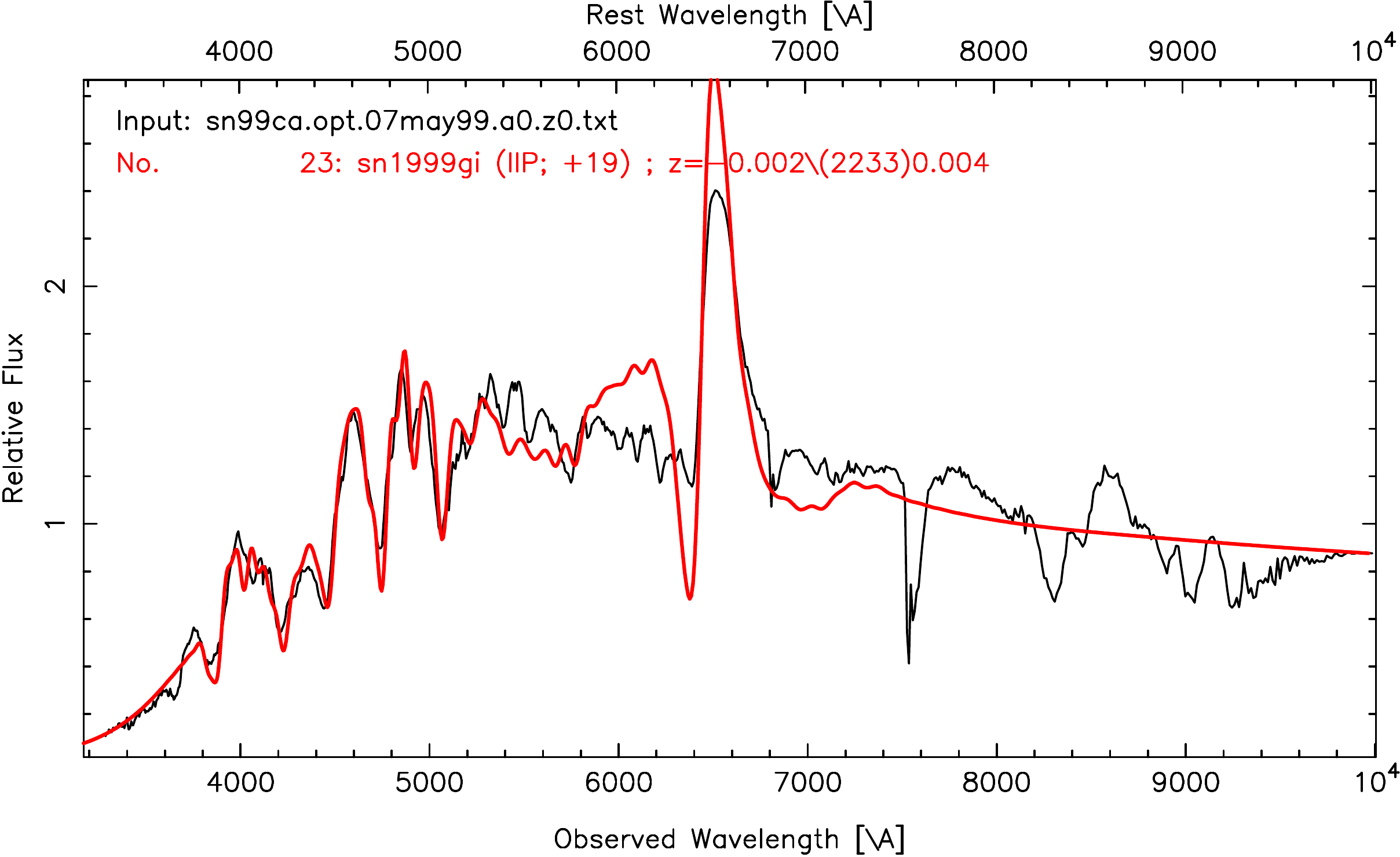}
\includegraphics[width=4.4cm]{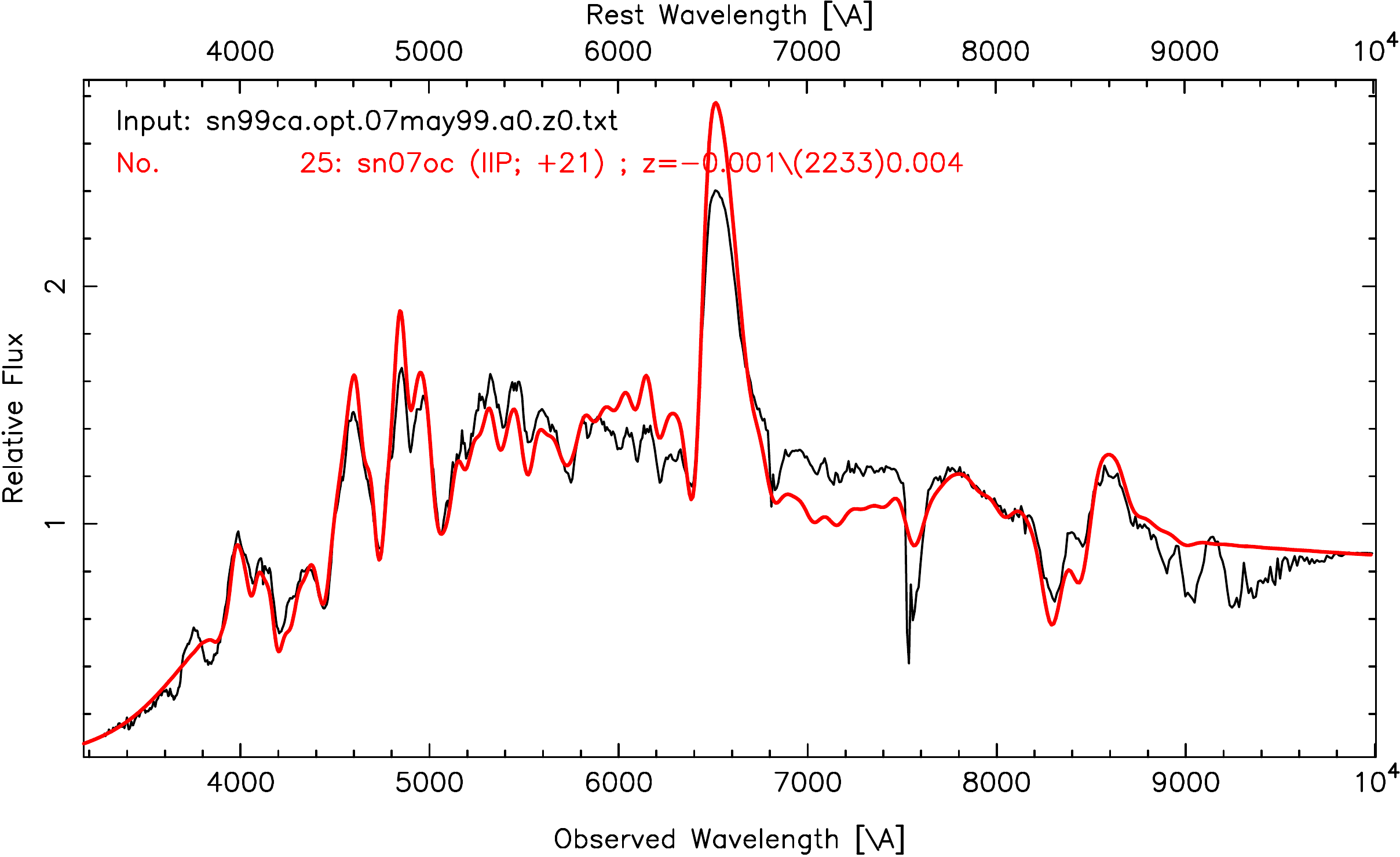}
\caption{Best spectral matching of SN~1999ca using SNID. The plots show SN~1999ca compared with 
SN~2006bp, SN~2003iq, SN~1999em, SN~1999gi, and 2007oc at 29, 29, 26, 31 and 21 days from explosion.}
\end{figure}

\begin{figure}[h!]
\centering
\includegraphics[width=4.4cm]{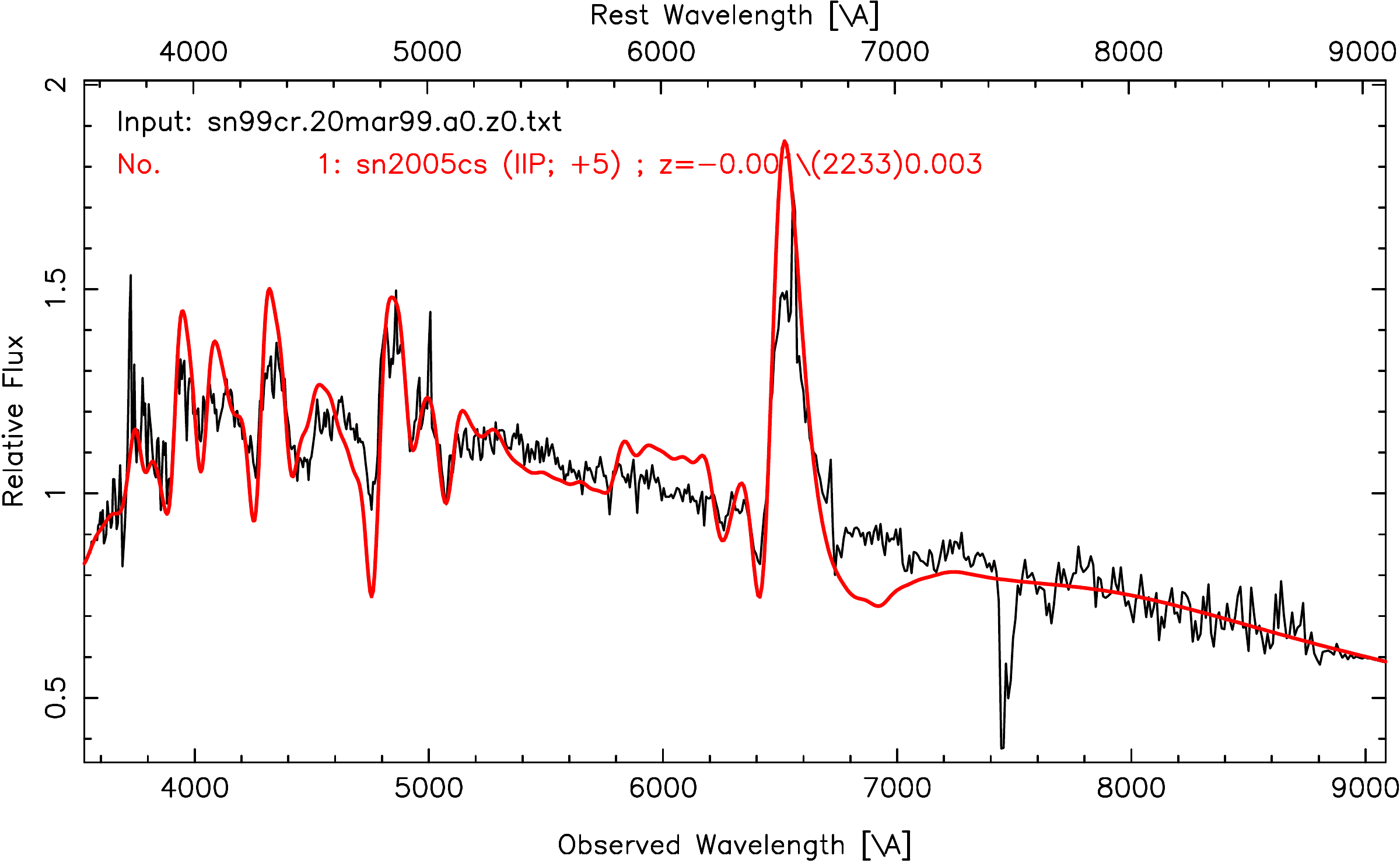}
\caption{Best spectral matching of SN~1999cr using SNID. The plots show SN~1999cr compared with 
SN~2005cs at 11 days from explosion.}
\end{figure}

\clearpage 

\begin{figure}[h!]
\centering
\includegraphics[width=4.4cm]{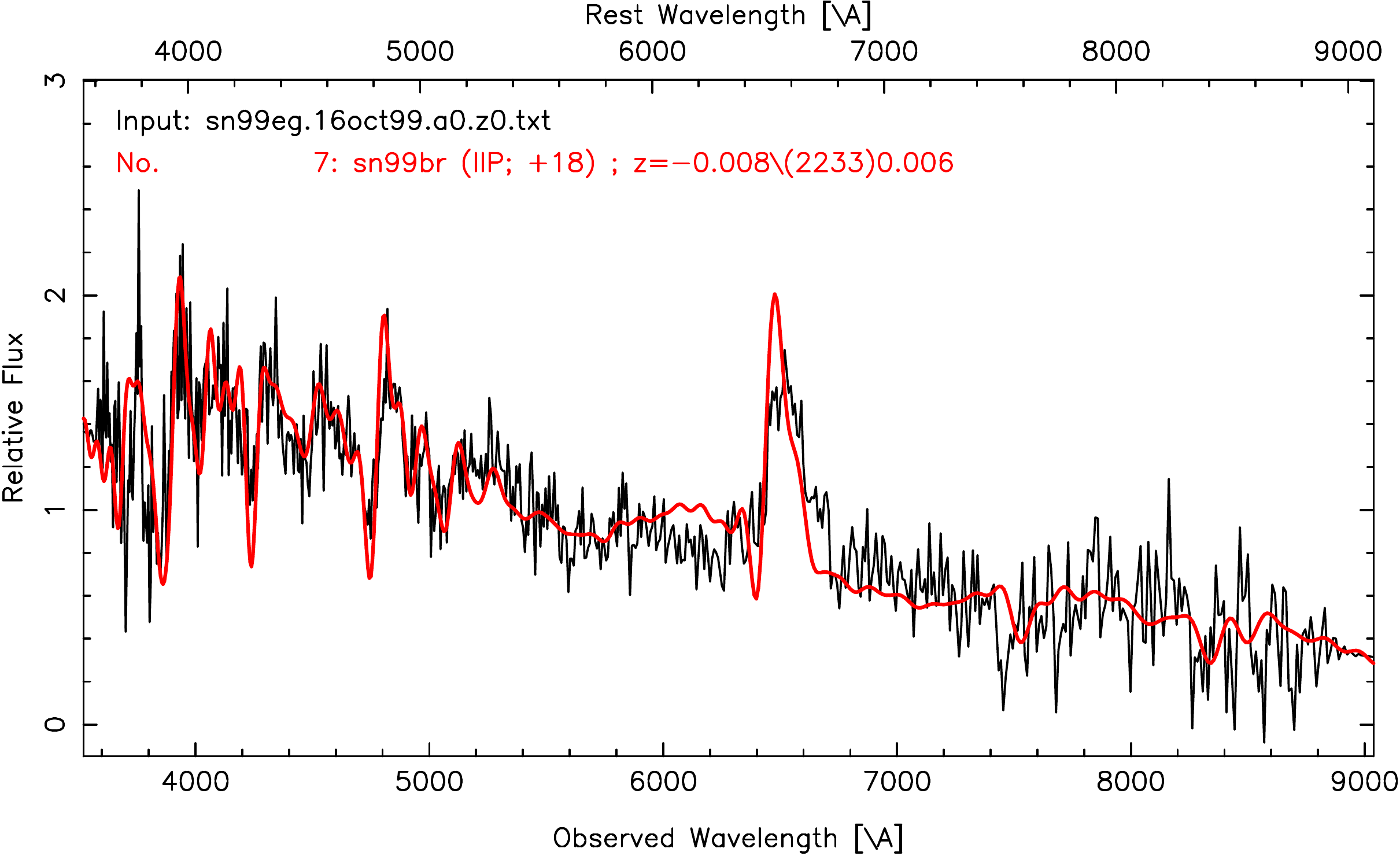}
\includegraphics[width=4.4cm]{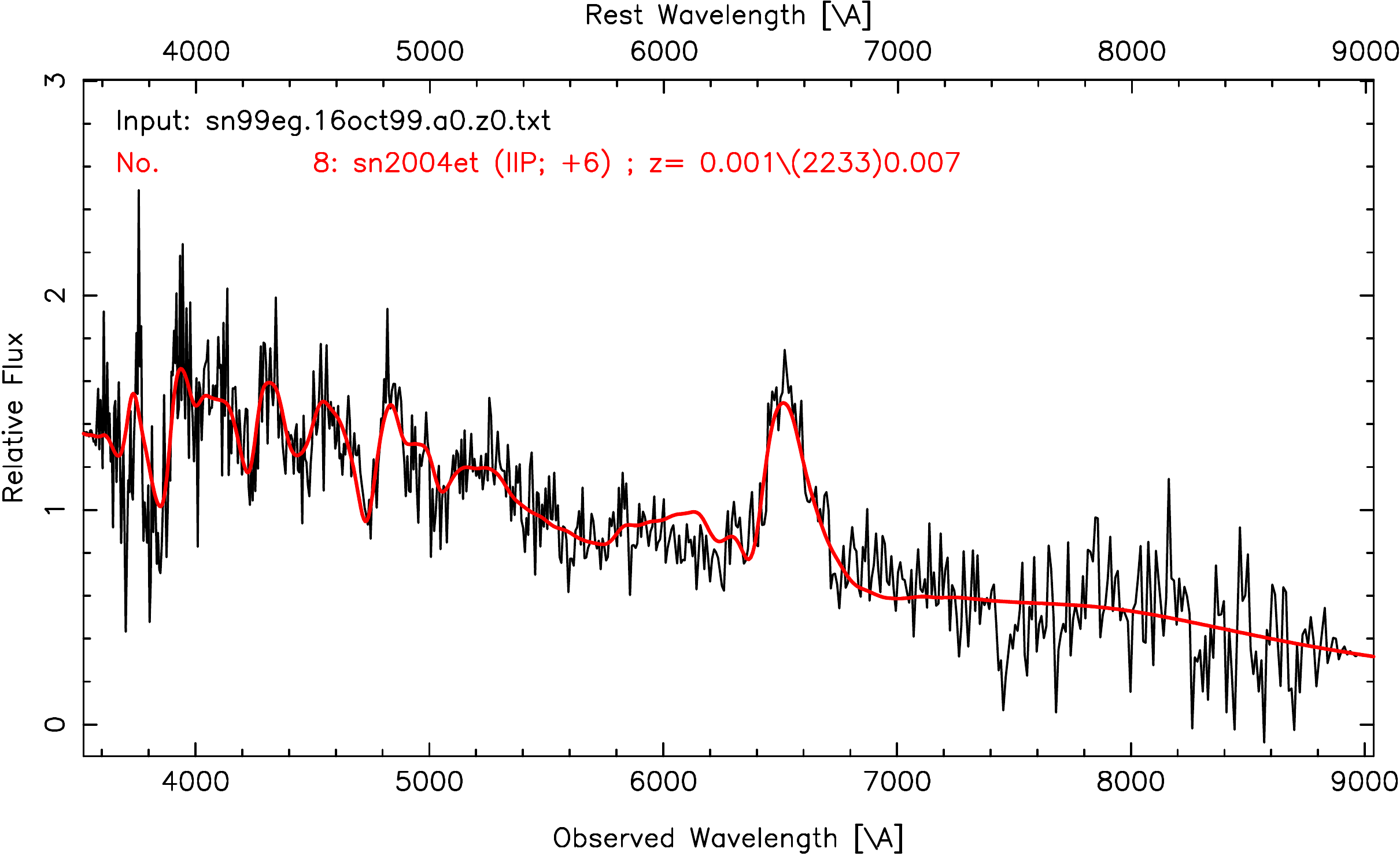}
\includegraphics[width=4.4cm]{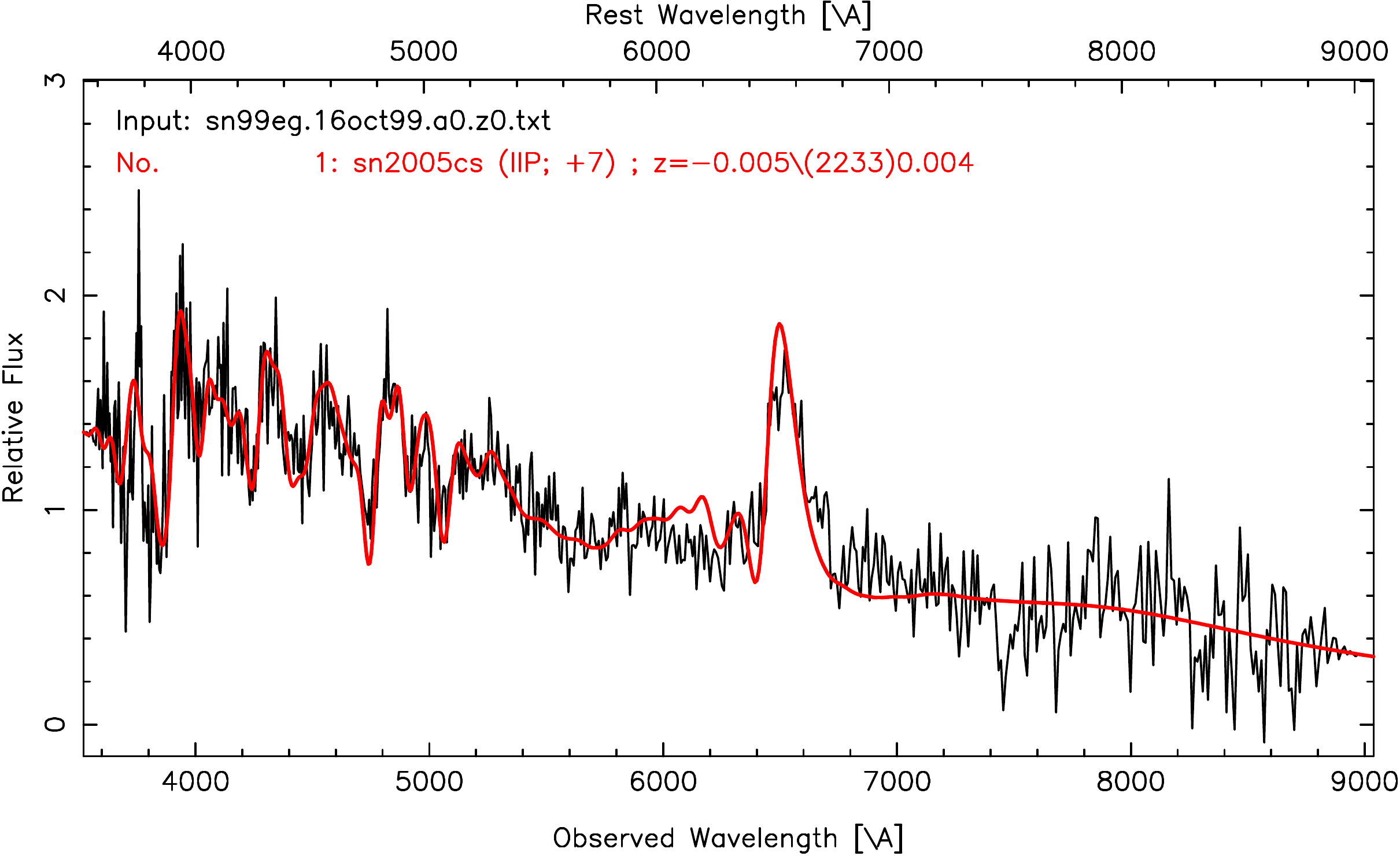}
\caption{Best spectral matching of SN~1999eg using SNID. The plots show SN~1999eg compared with 
SN~1999br, SN~2004et, and SN~2005cs at 18, 22, and 13 days from explosion.}
\end{figure}

\begin{figure}[h!]
\centering
\includegraphics[width=4.4cm]{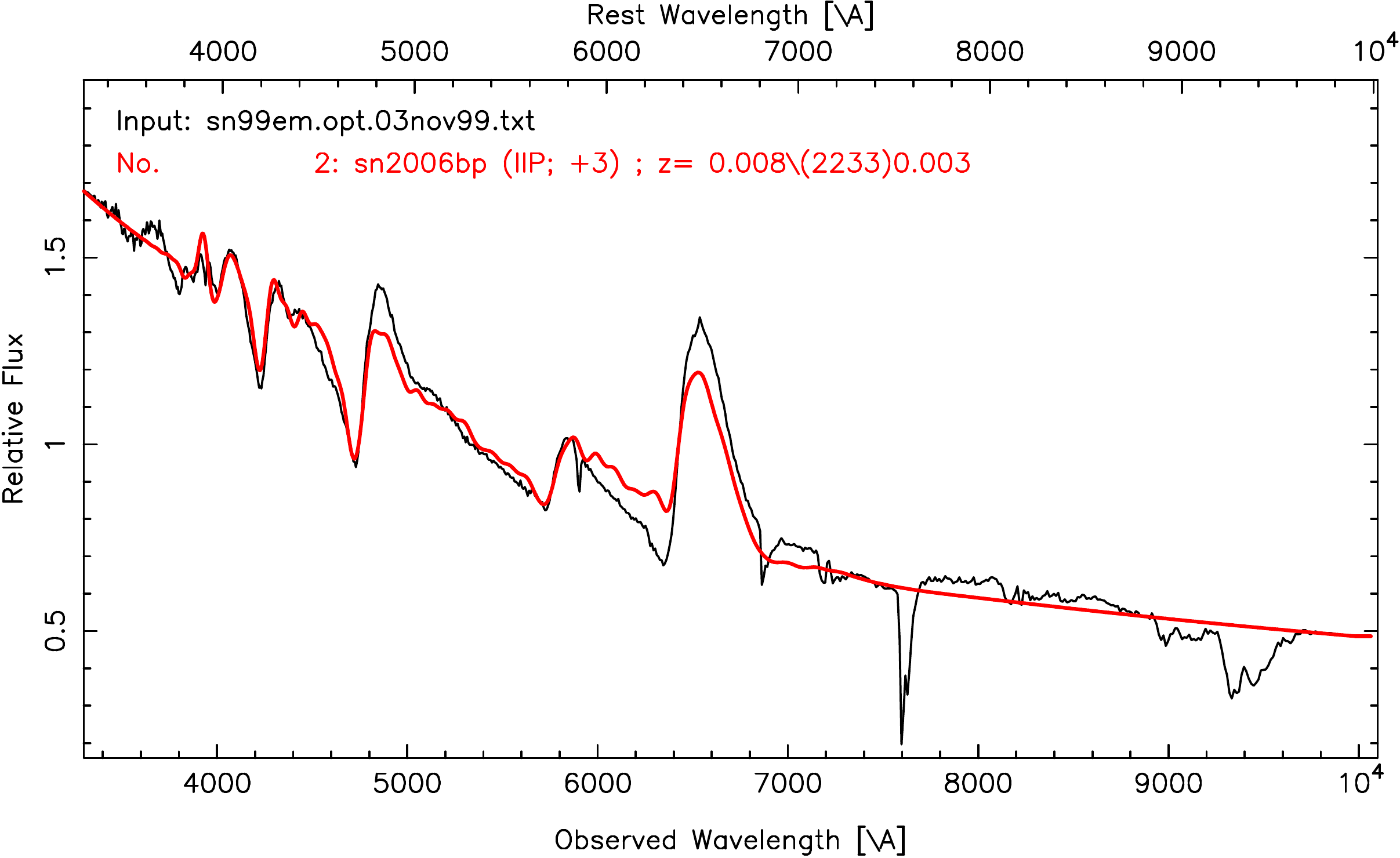}
\includegraphics[width=4.4cm]{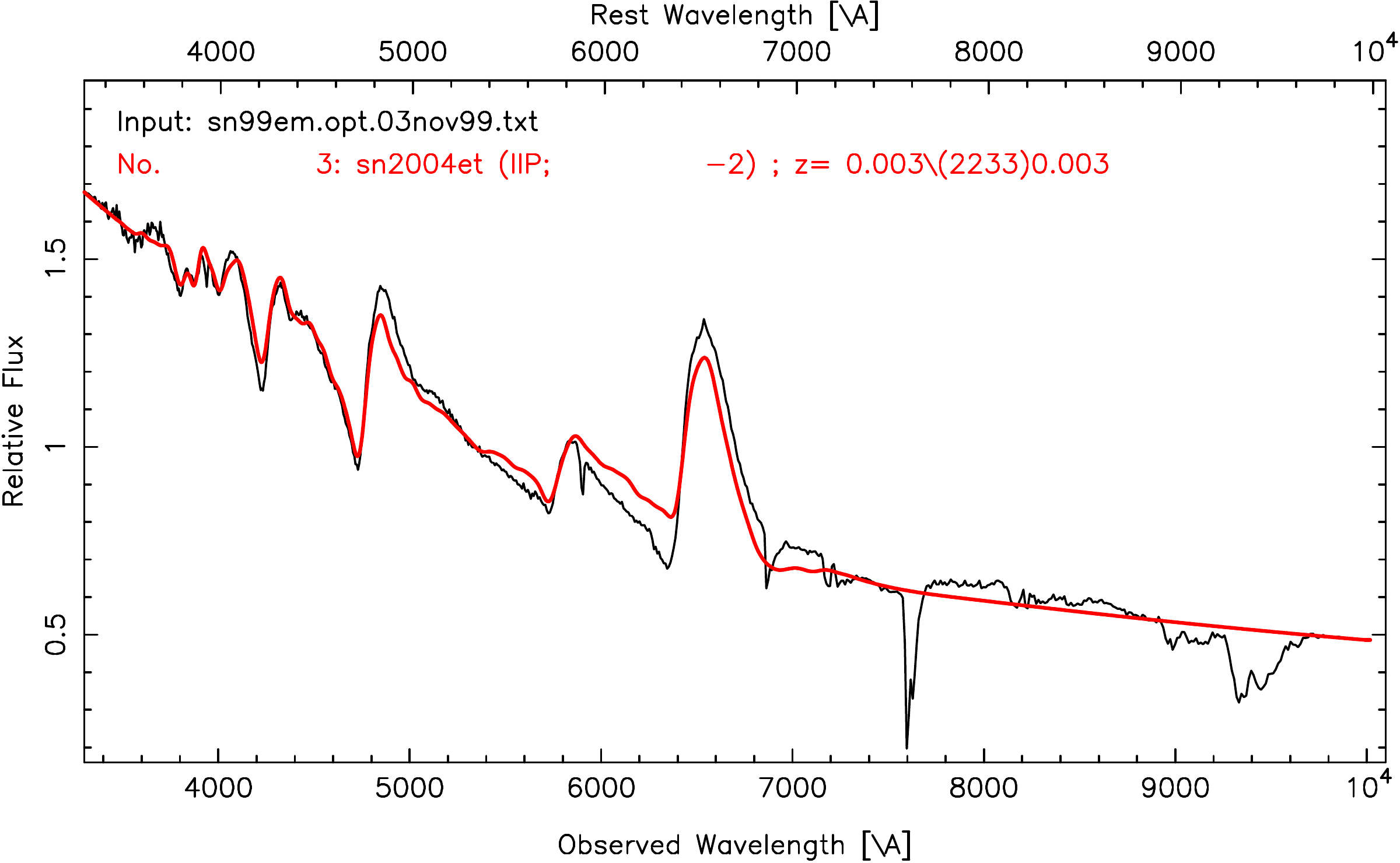}
\caption{Best spectral matching of SN~1999em using SNID. The plots show SN~1999em compared with 
SN~2006bp and SN~2004et at 12 and 13 days from explosion.}
\end{figure}

\begin{figure}[h!]
\centering
\includegraphics[width=4.4cm]{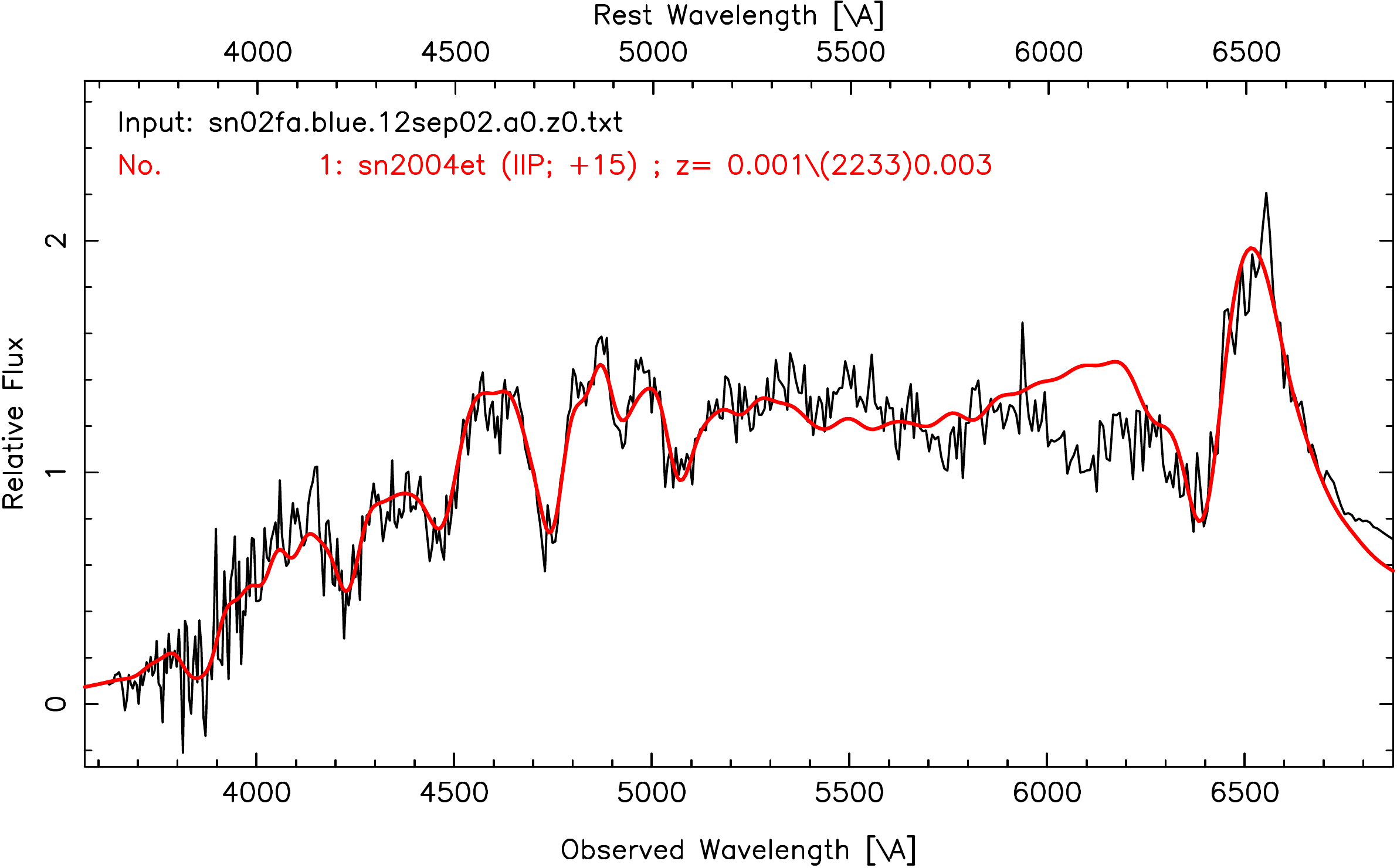}
\includegraphics[width=4.4cm]{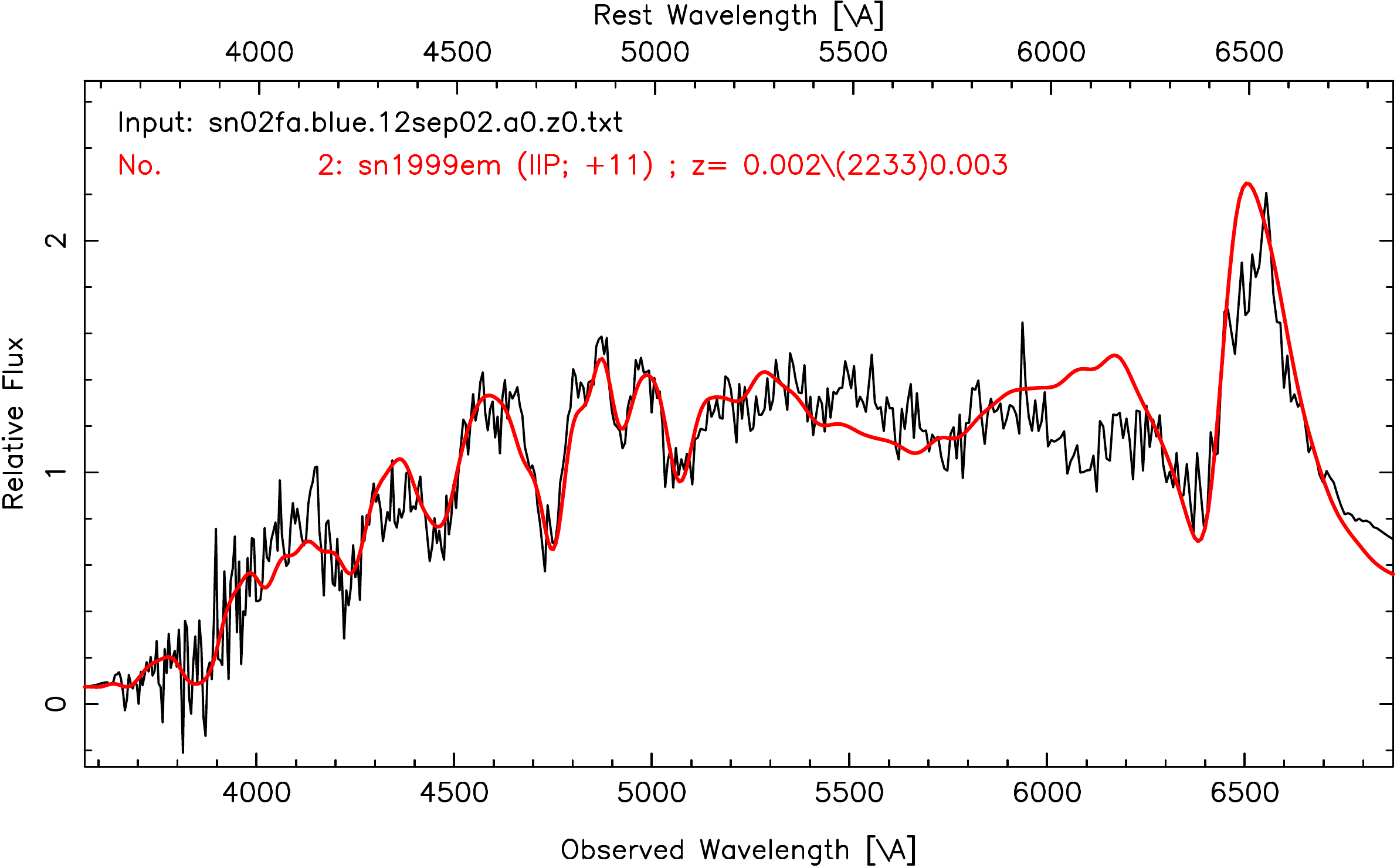}
\includegraphics[width=4.4cm]{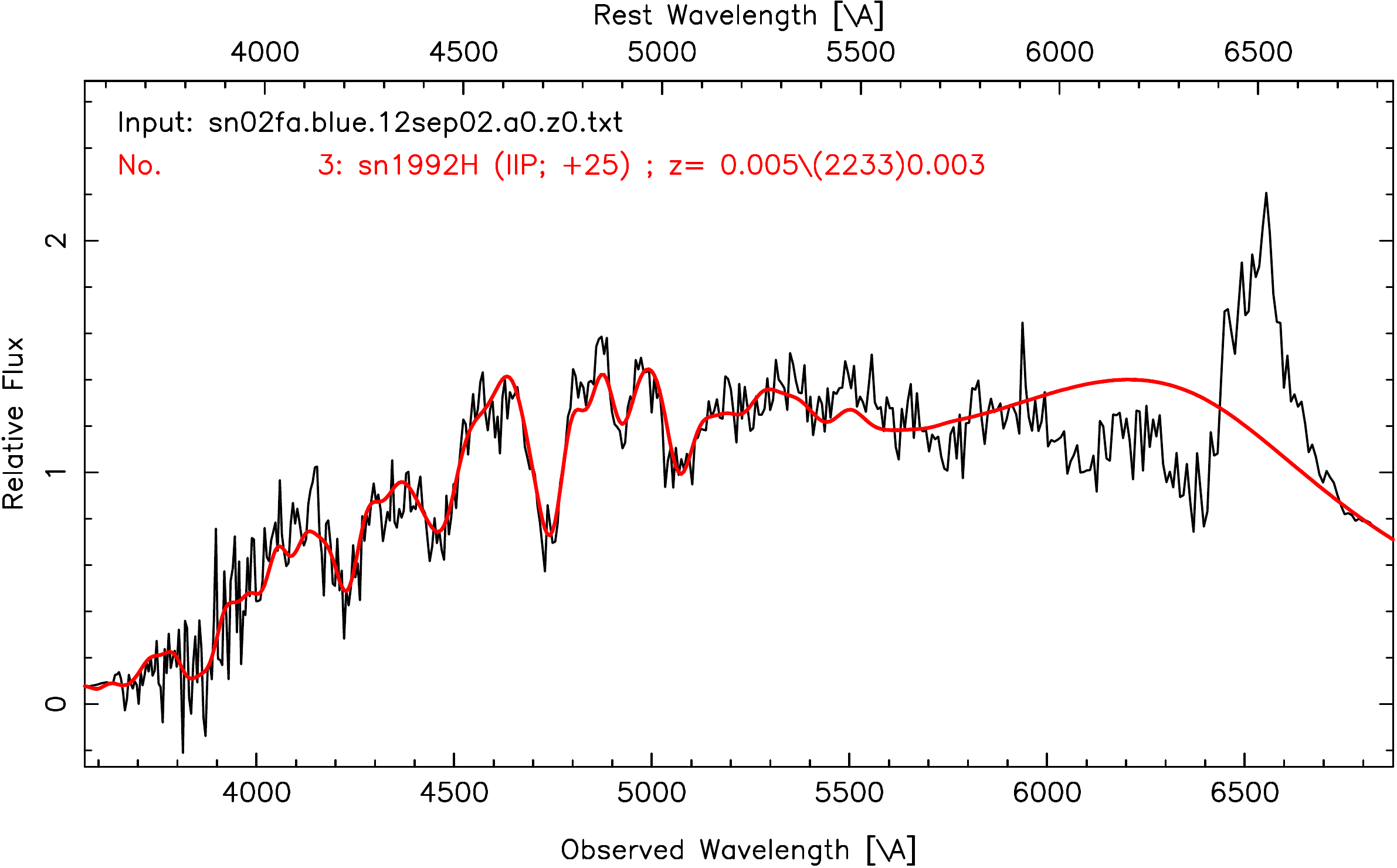}
\includegraphics[width=4.4cm]{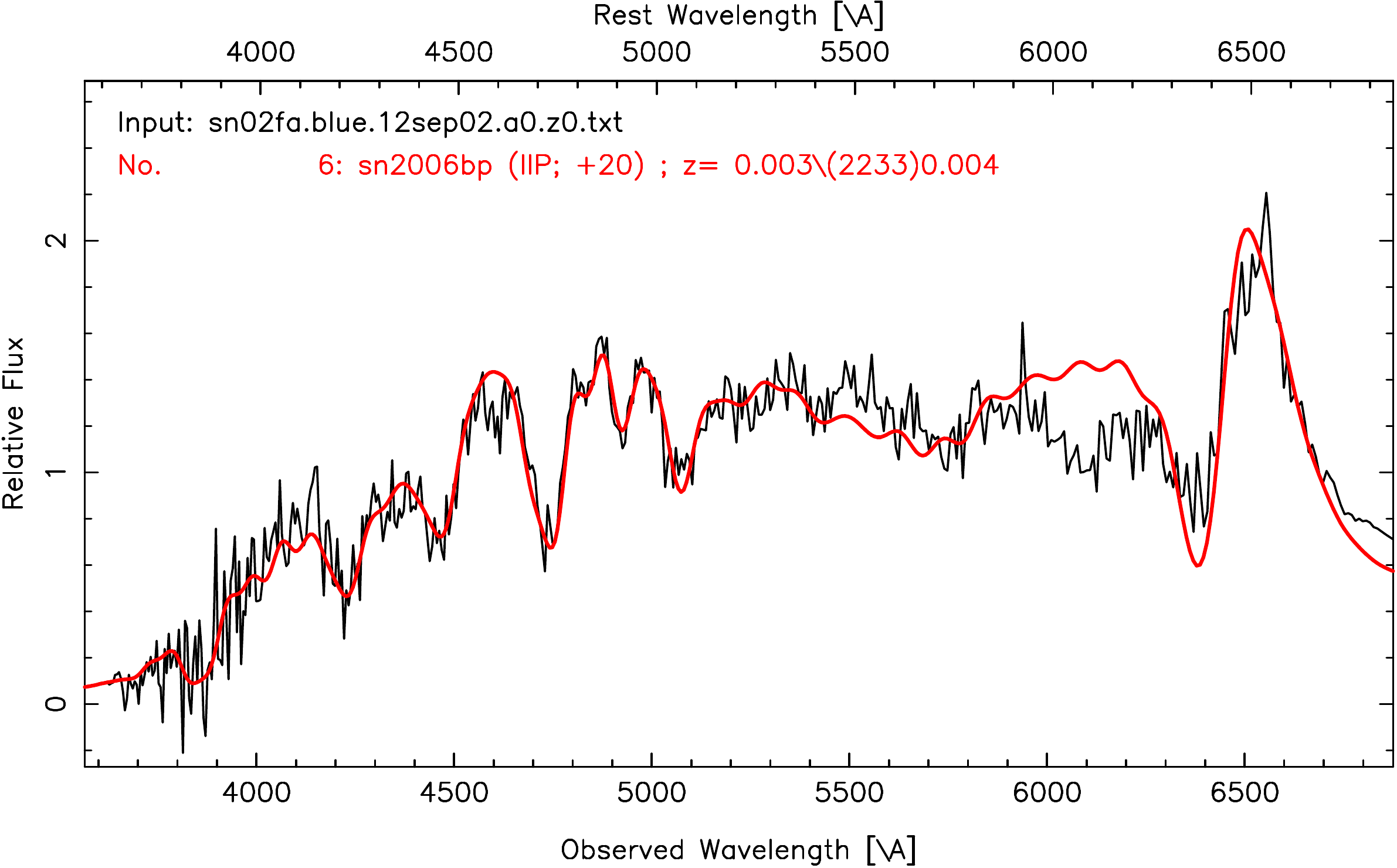}
\includegraphics[width=4.4cm]{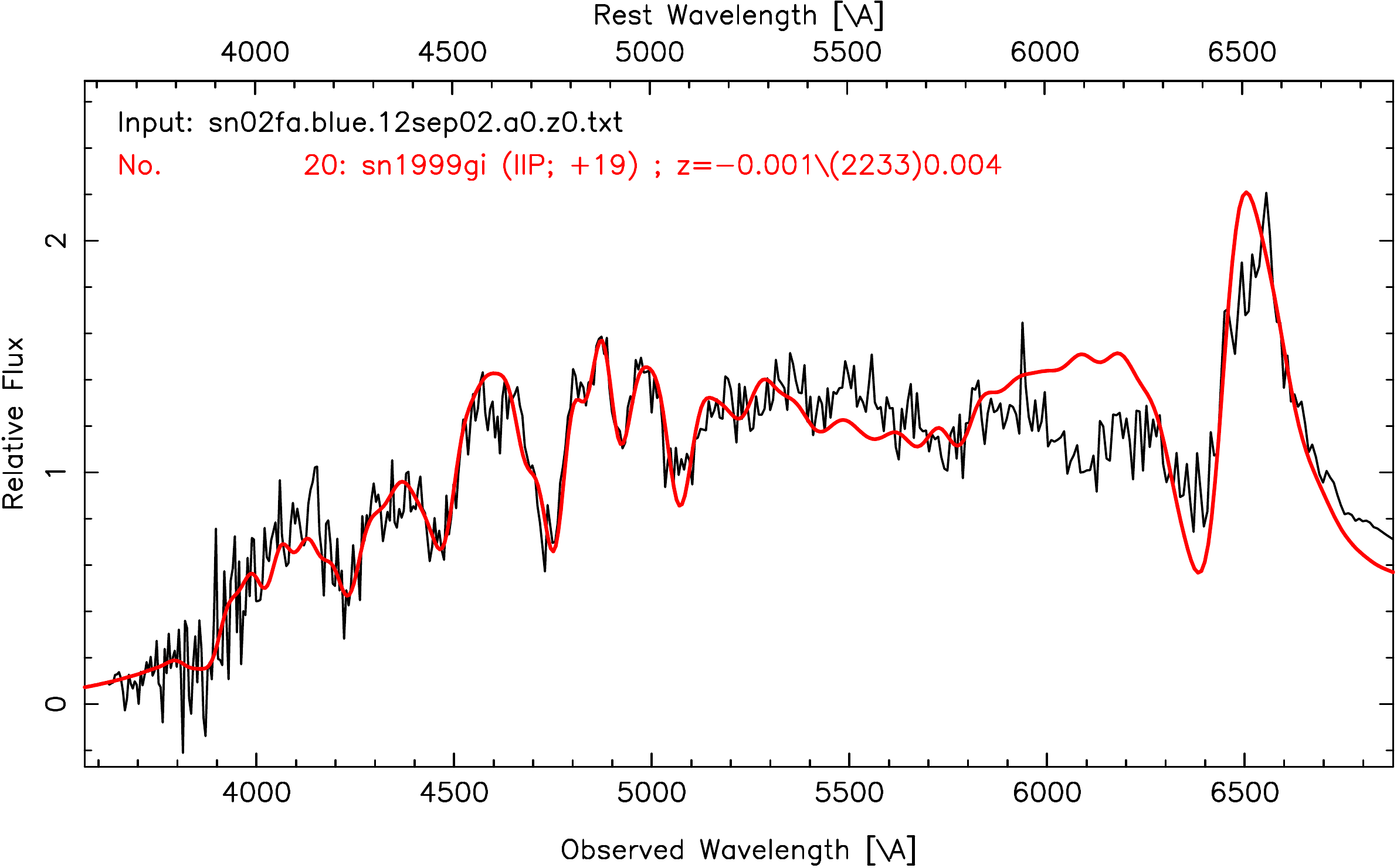}
\caption{Best spectral matching of SN~2002fa using SNID. The plots show SN~2002fa compared with 
SN~2004et, SN~1999em, SN~1992H, SN~2006bp, and SN~1999gi at 31, 21, 30, 29, and 31 days from explosion.}
\end{figure}

\clearpage

\begin{figure}[h!]
\centering
\includegraphics[width=4.4cm]{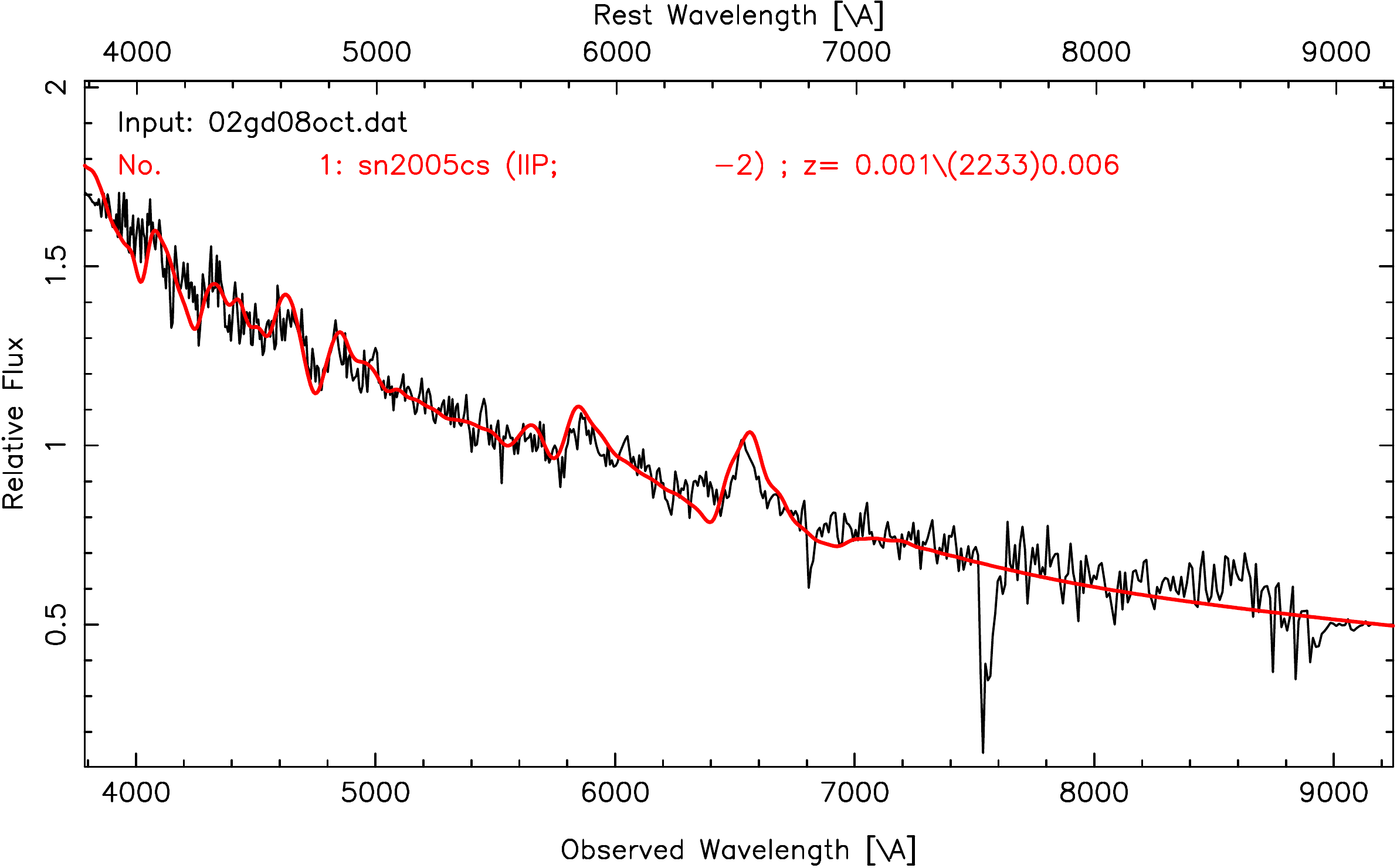}
\caption{Best spectral matching of SN~2002gd using SNID. The plots show SN~2002gd compared with 
SN~2005cs at 4 days from explosion.}
\end{figure}

\begin{figure}[h!]
\centering
\includegraphics[width=4.4cm]{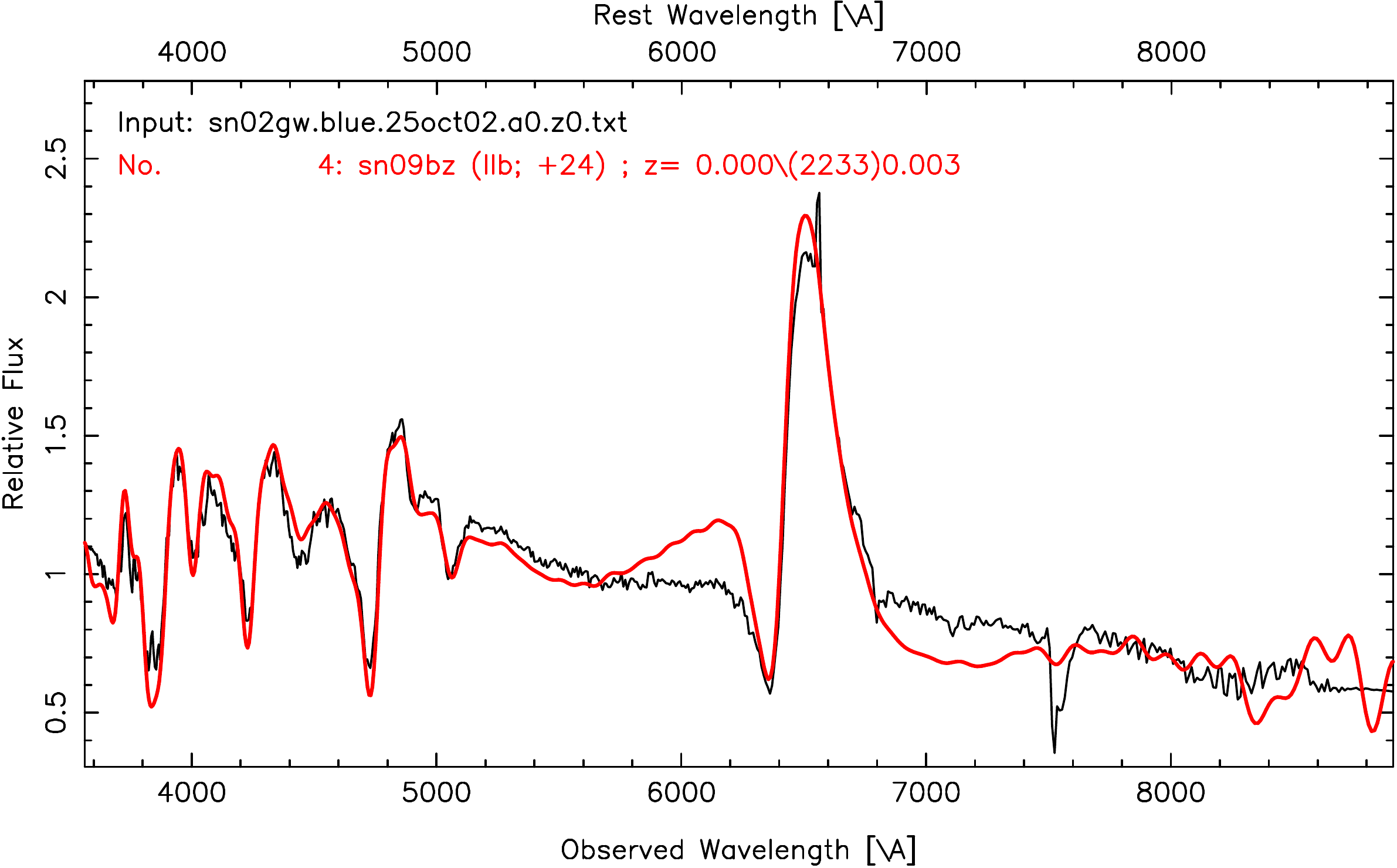}
\includegraphics[width=4.4cm]{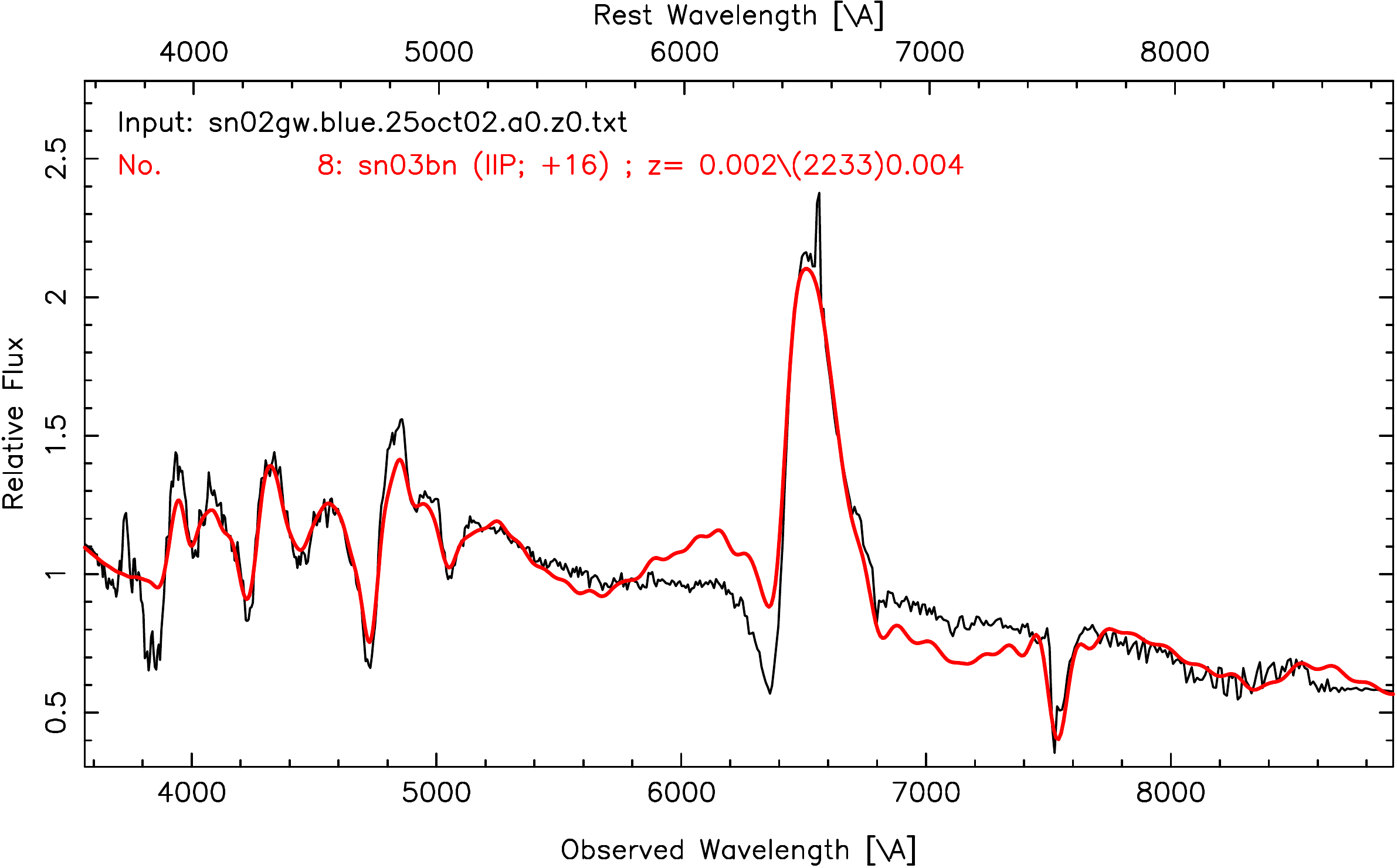}
\includegraphics[width=4.4cm]{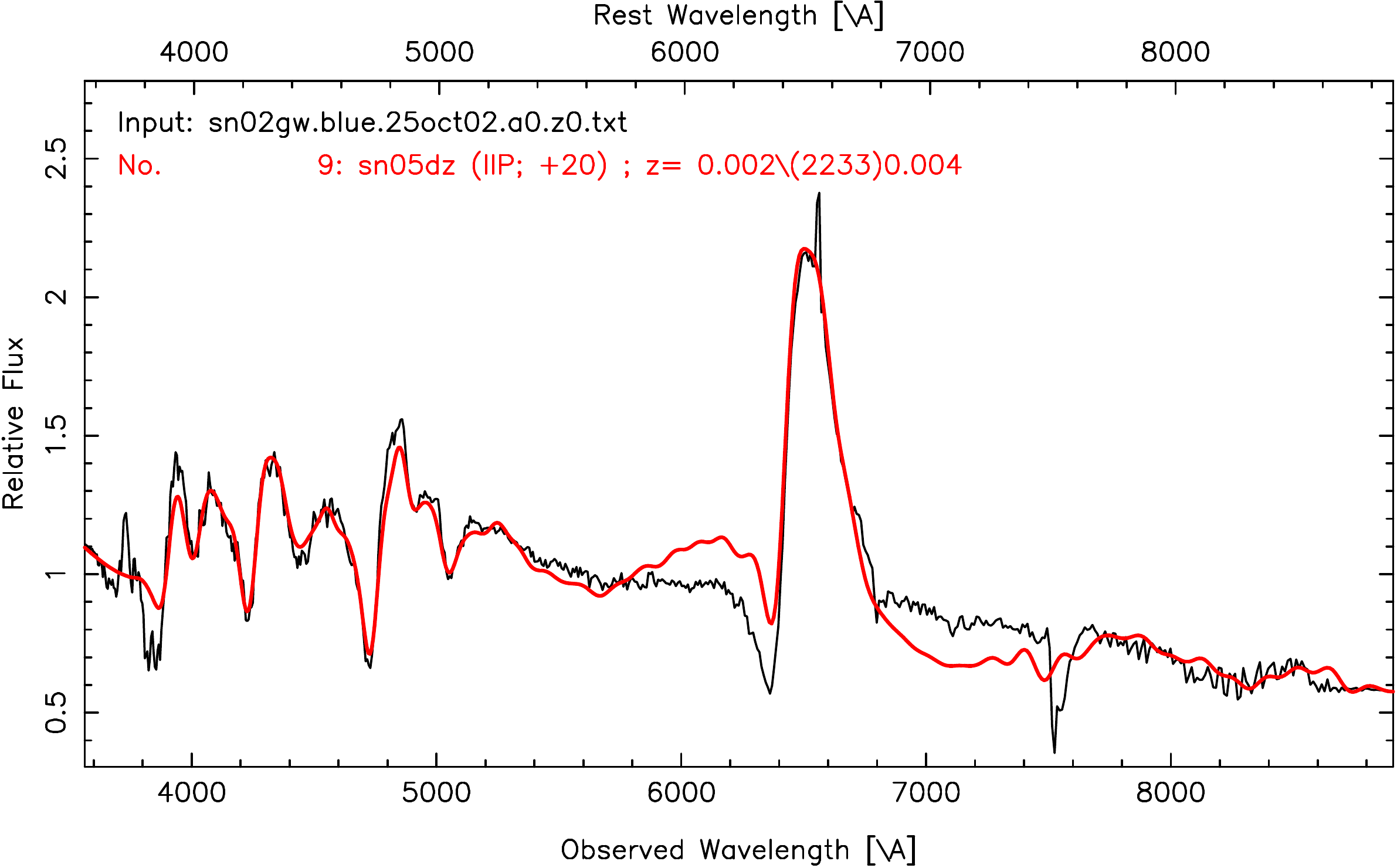}
\includegraphics[width=4.4cm]{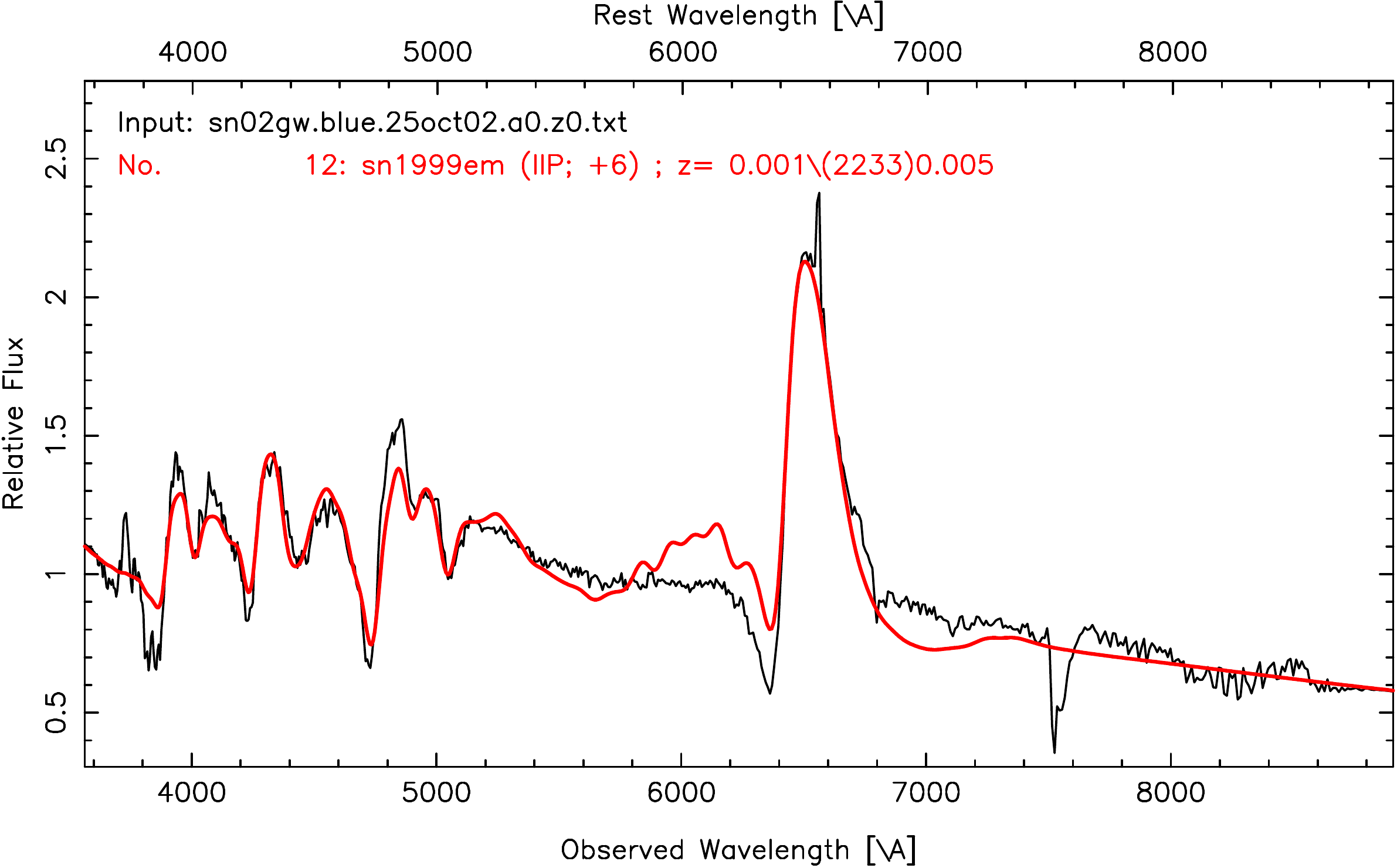}
\caption{Best spectral matching of SN~2002gw using SNID. The plots show SN~2002gw compared with 
SN~2009bz, SN~2003bn, SN~2005dz, and SN~1999em at 24, 16, 20, and 16 days from explosion.}
\end{figure}

\clearpage

\begin{figure}[h!]
\centering
\includegraphics[width=4.4cm]{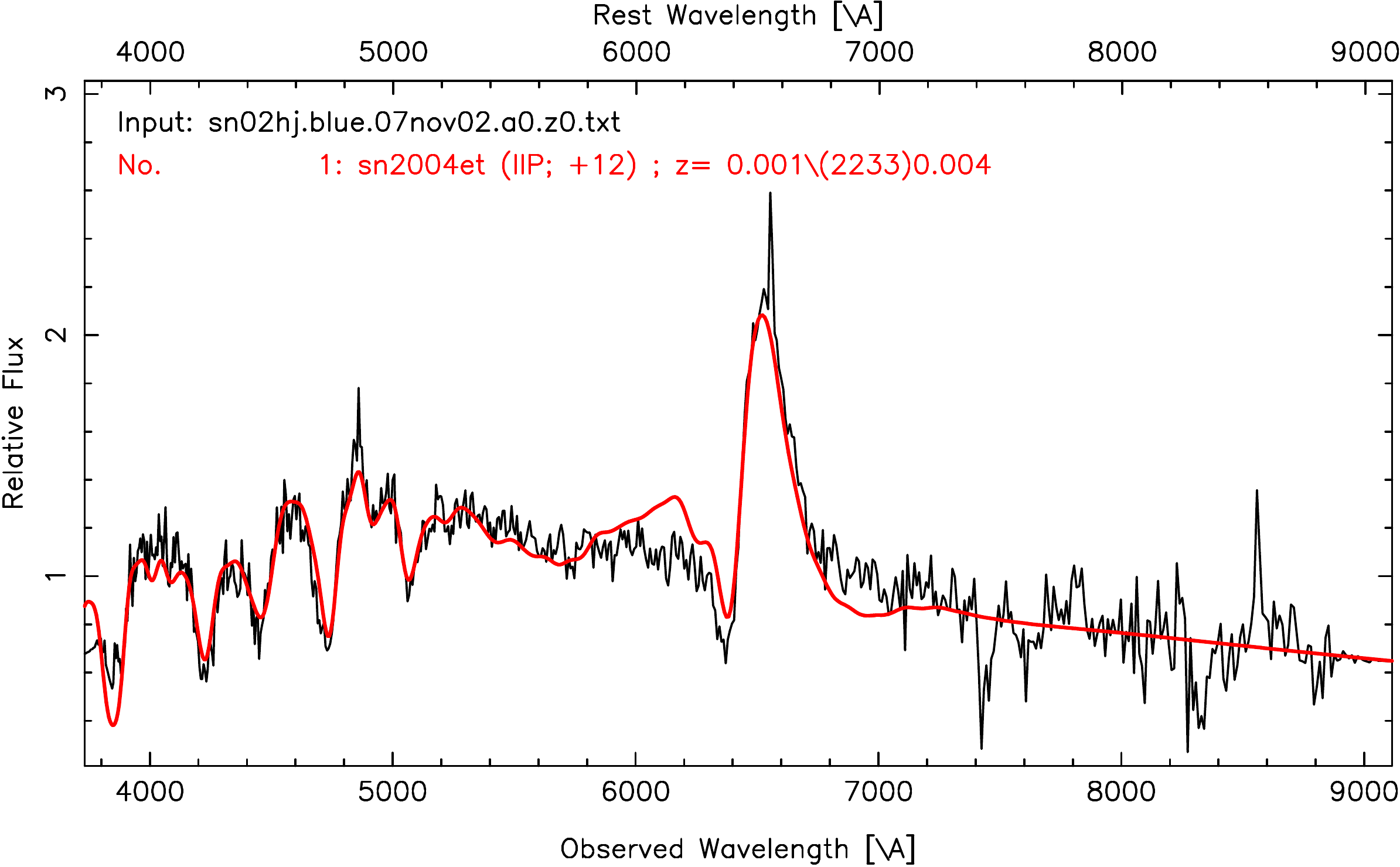}
\includegraphics[width=4.4cm]{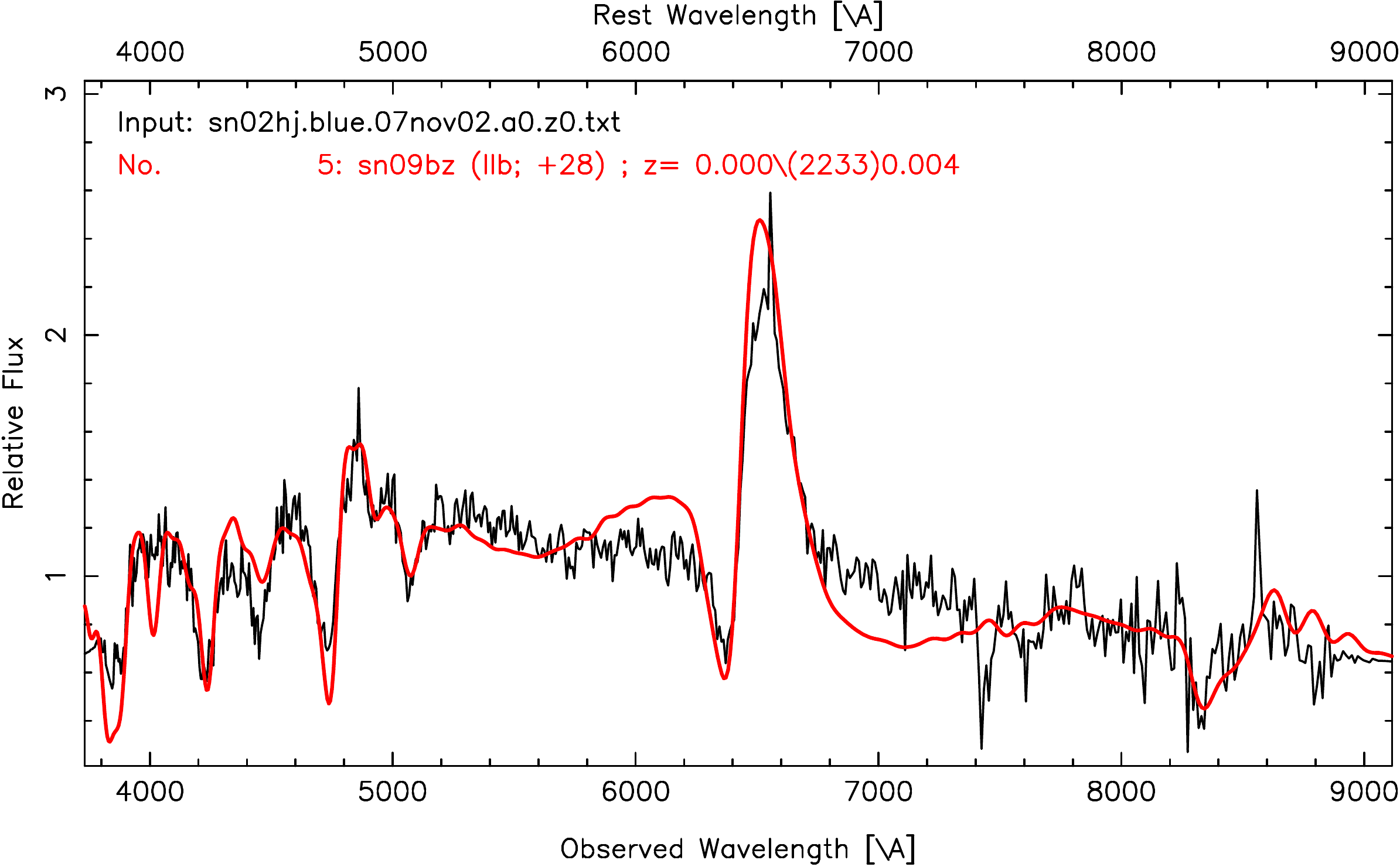}
\includegraphics[width=4.4cm]{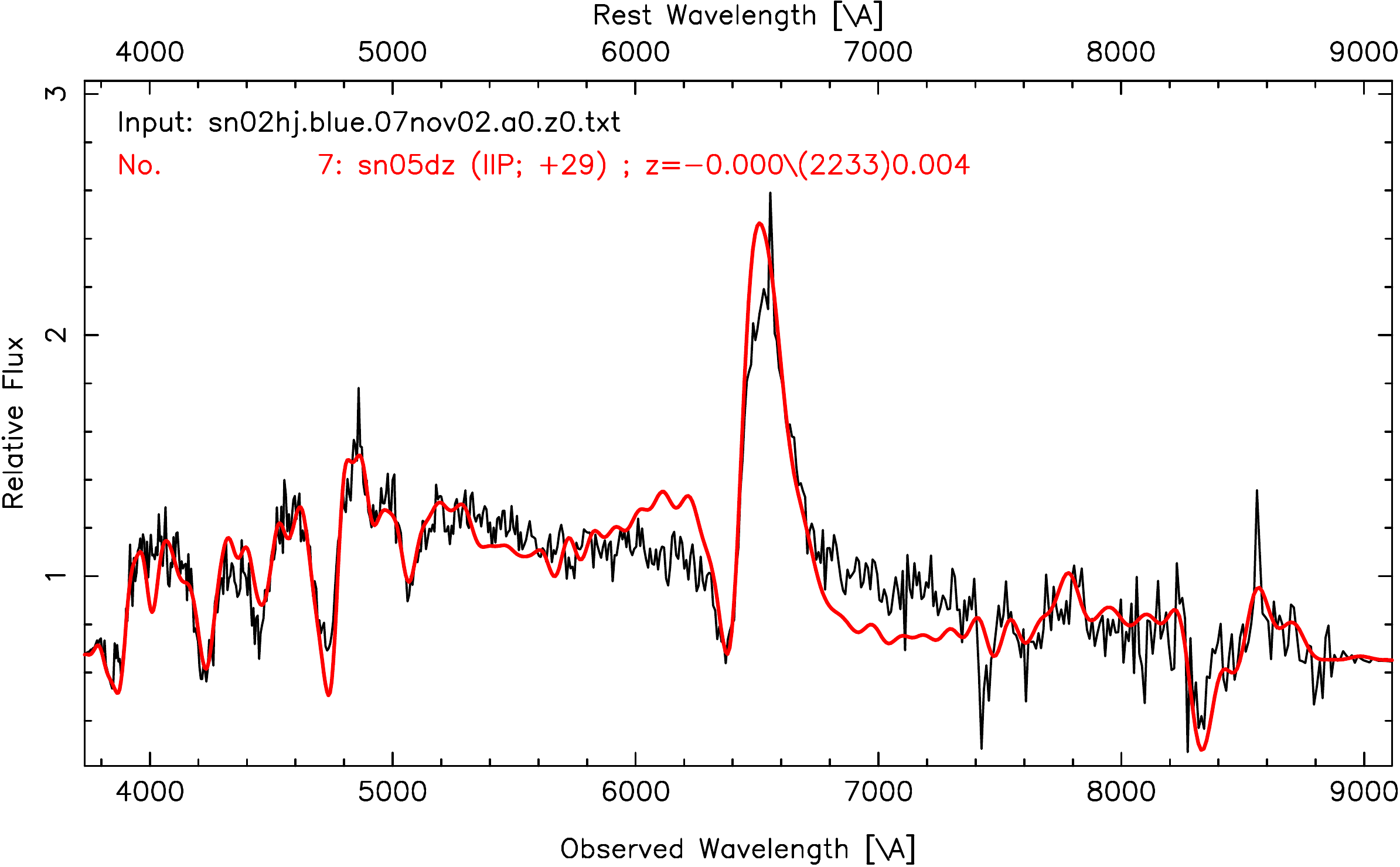}
\includegraphics[width=4.4cm]{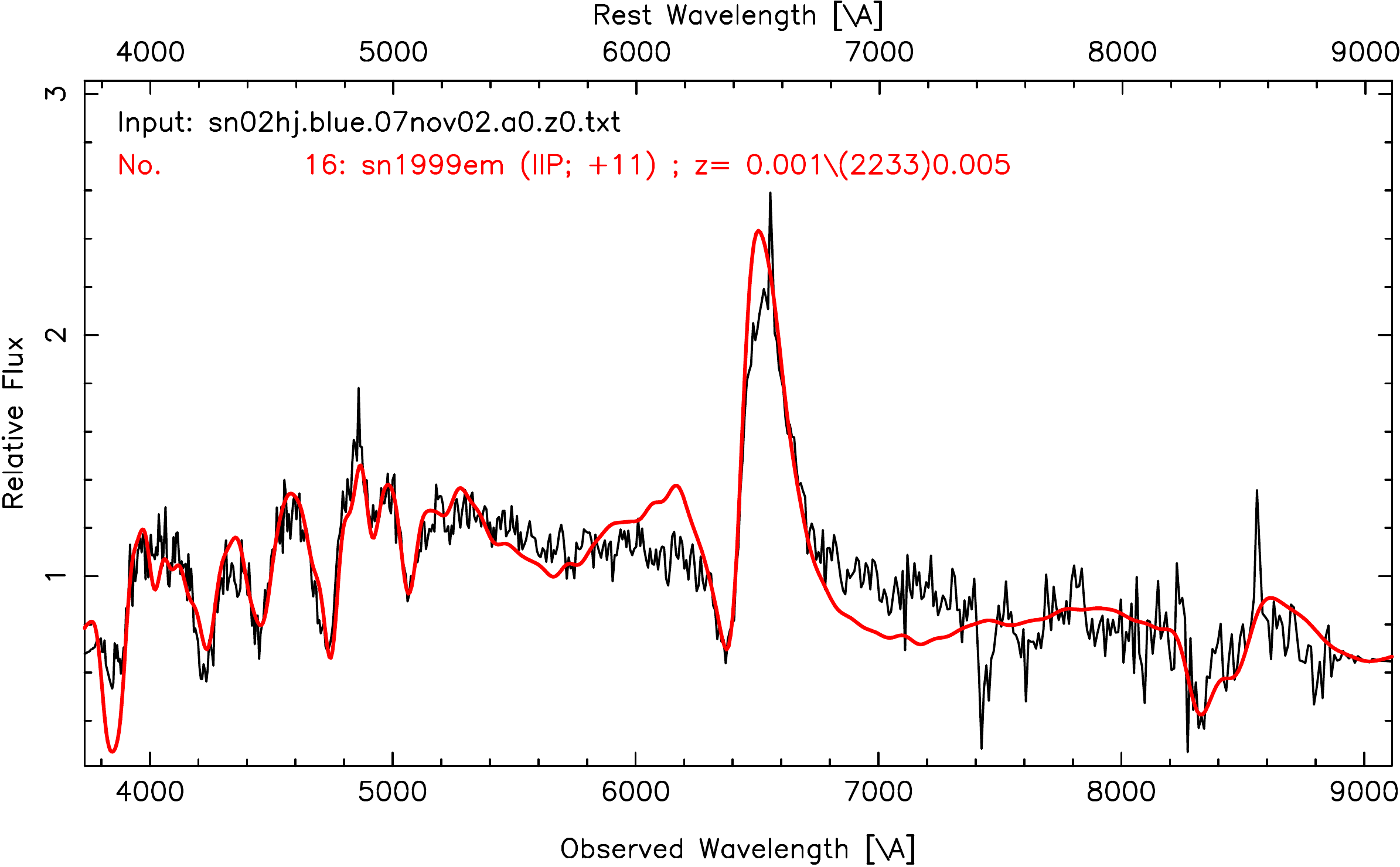}
\includegraphics[width=4.4cm]{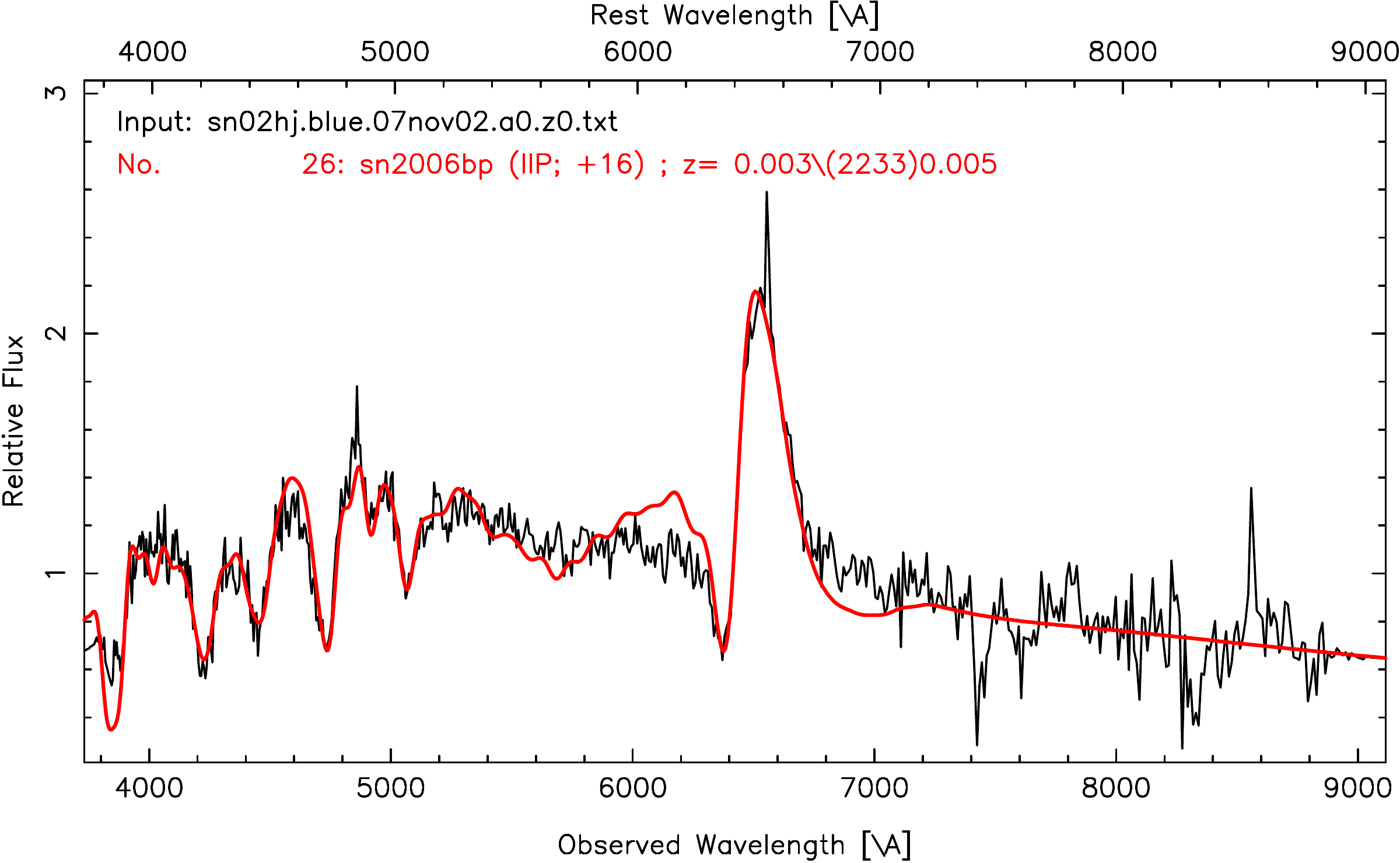}
\includegraphics[width=4.4cm]{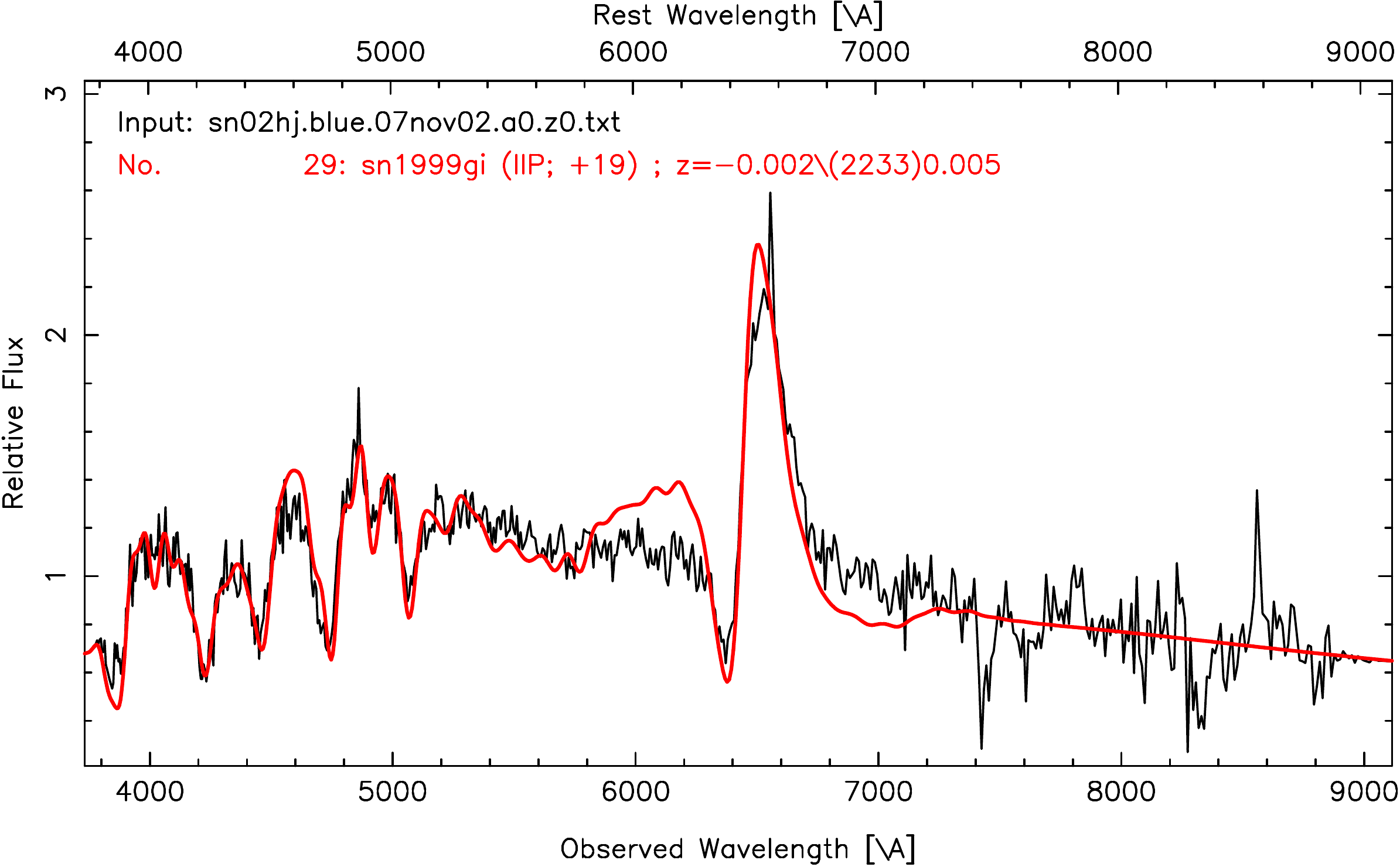}
\includegraphics[width=4.4cm]{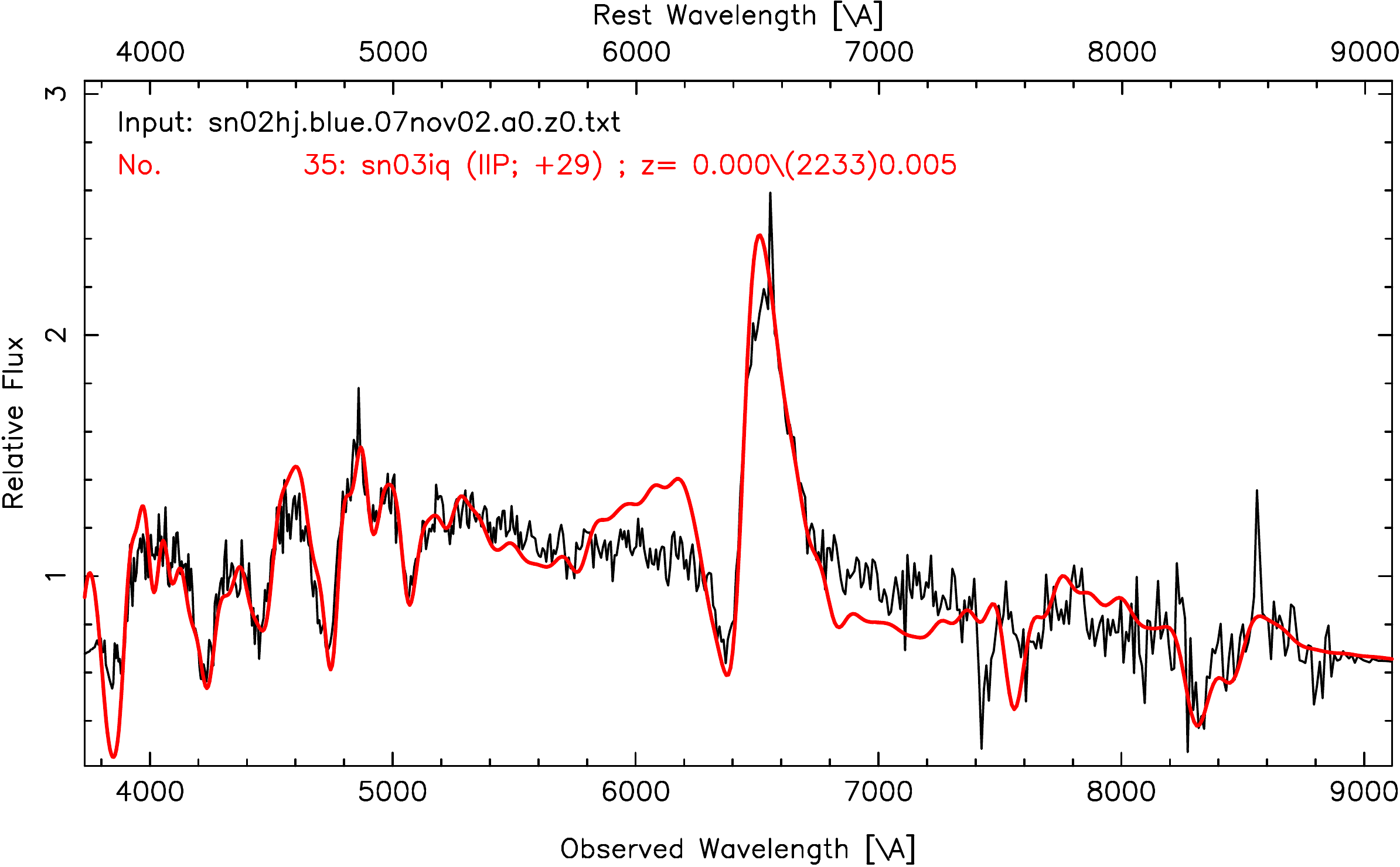}
\caption{Best spectral matching of SN~2002hj using SNID. The plots show SN~2002hj compared with 
SN~2004et, SN~2009bz, SN~2005dz, SN~1999em, SN~2006bp, SN~1999gi, and SN~2003iq at 28, 28, 29, 21, 25, 31, and 29 days from explosion.}
\end{figure}

\clearpage

\begin{figure}[h!]
\centering
\includegraphics[width=4.4cm]{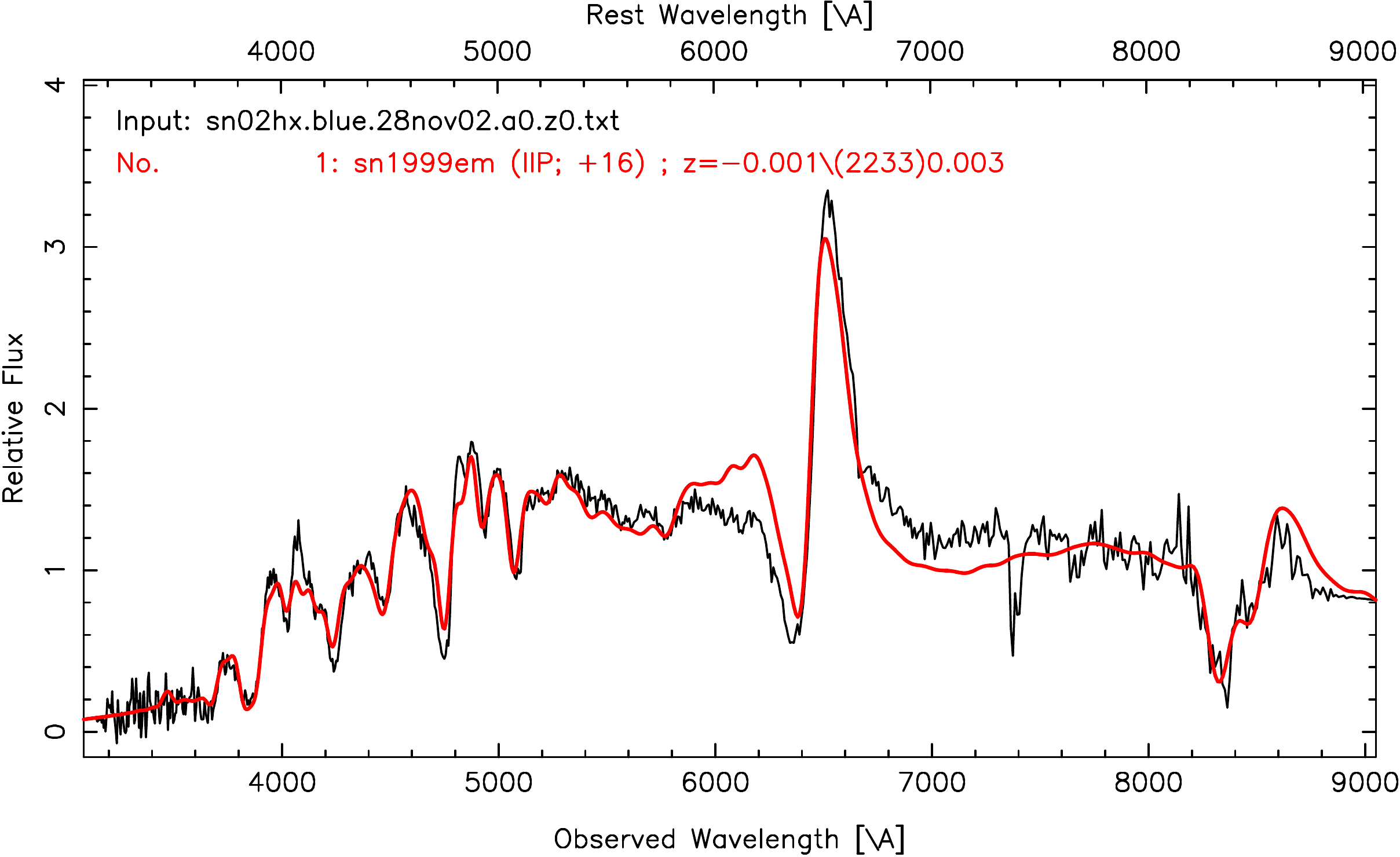}
\includegraphics[width=4.4cm]{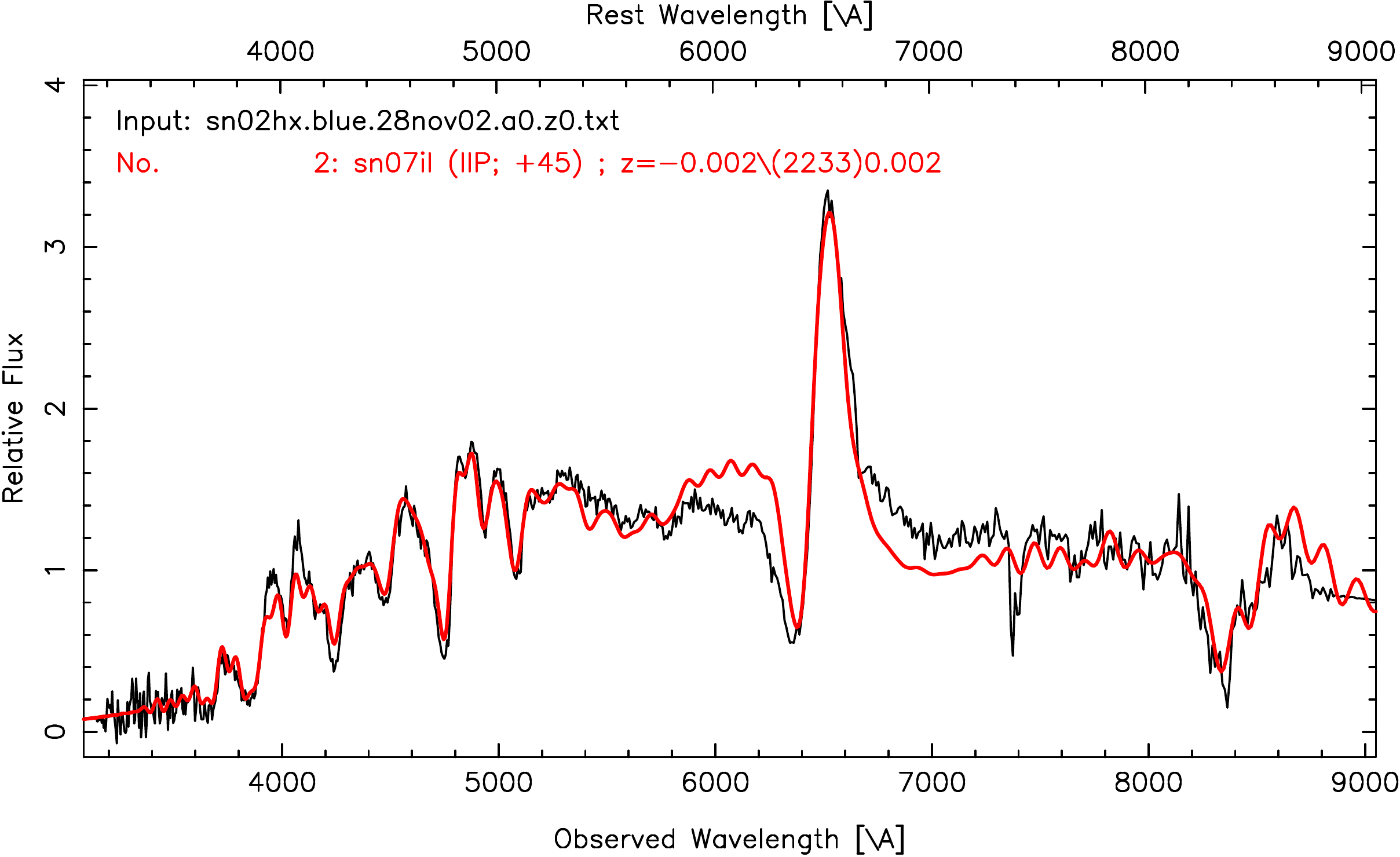}
\includegraphics[width=4.4cm]{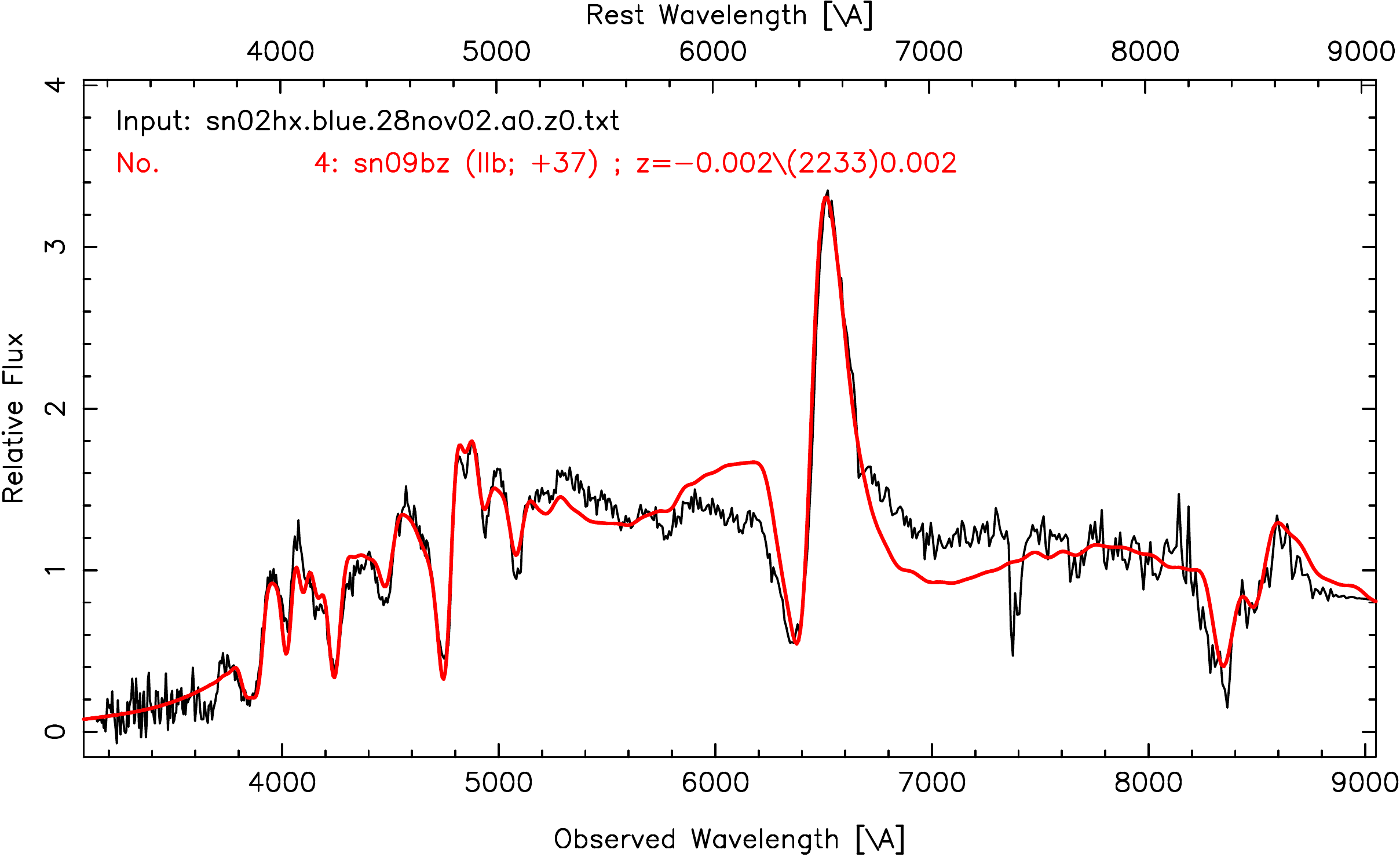}
\includegraphics[width=4.4cm]{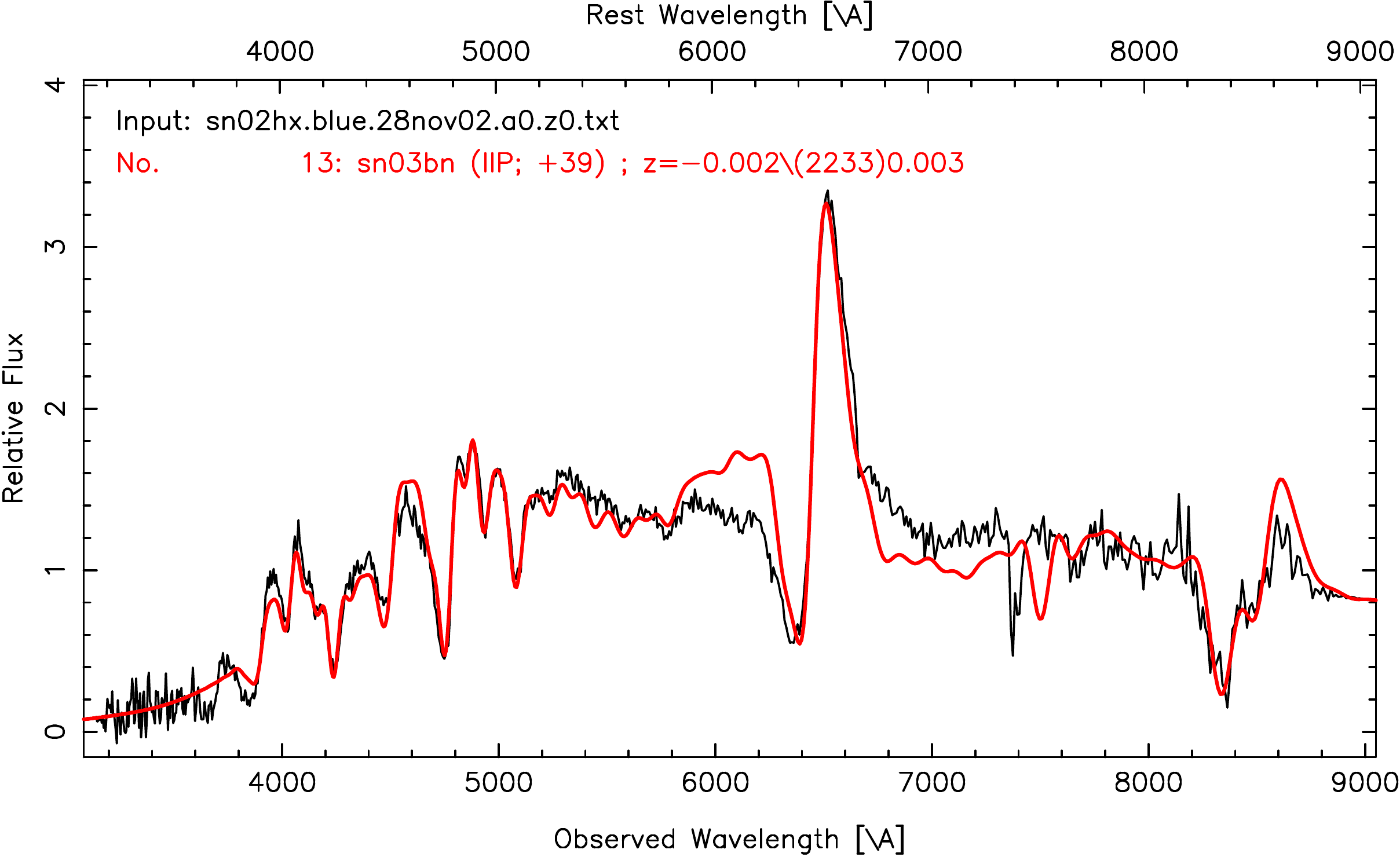}
\includegraphics[width=4.4cm]{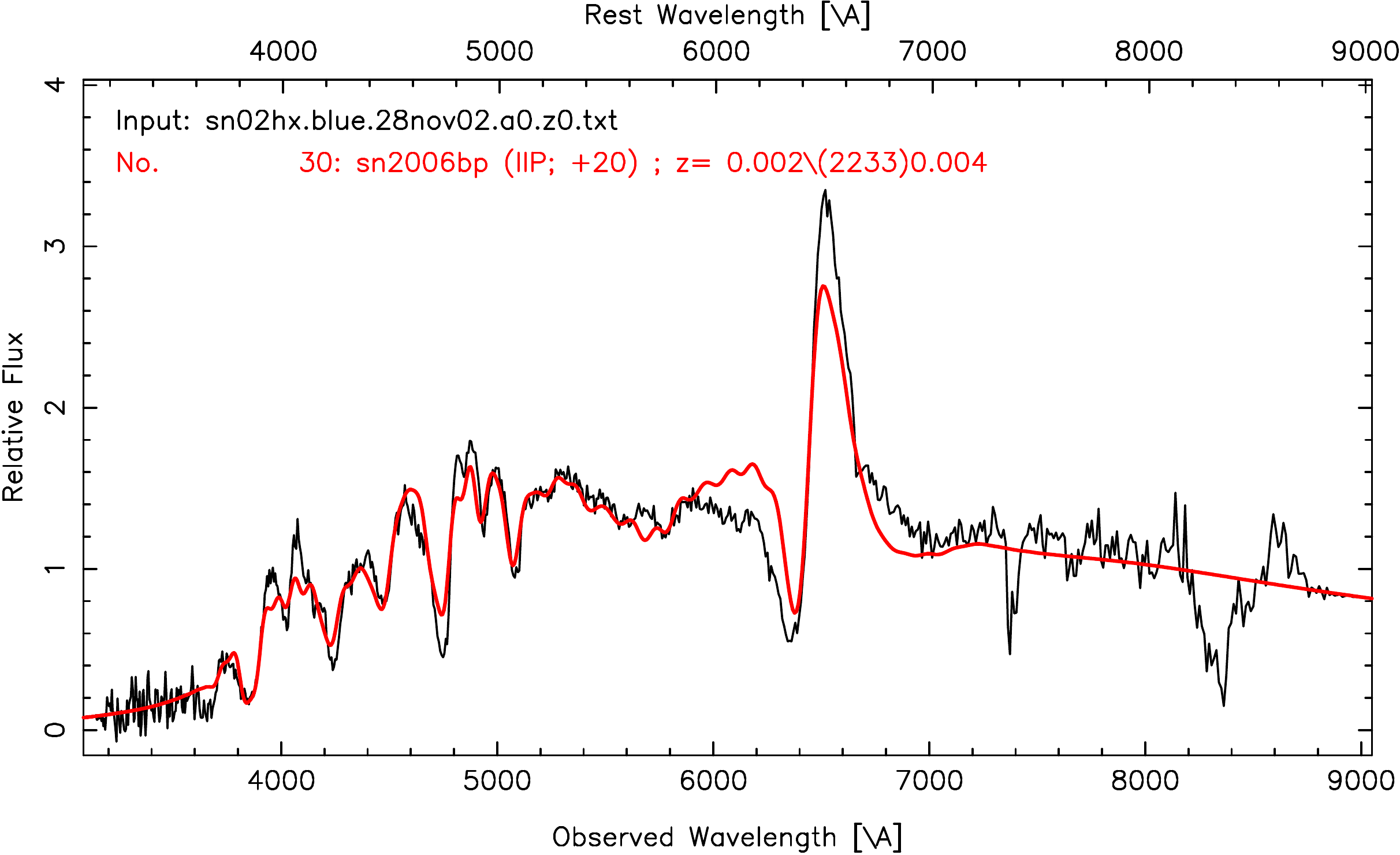}
\includegraphics[width=4.4cm]{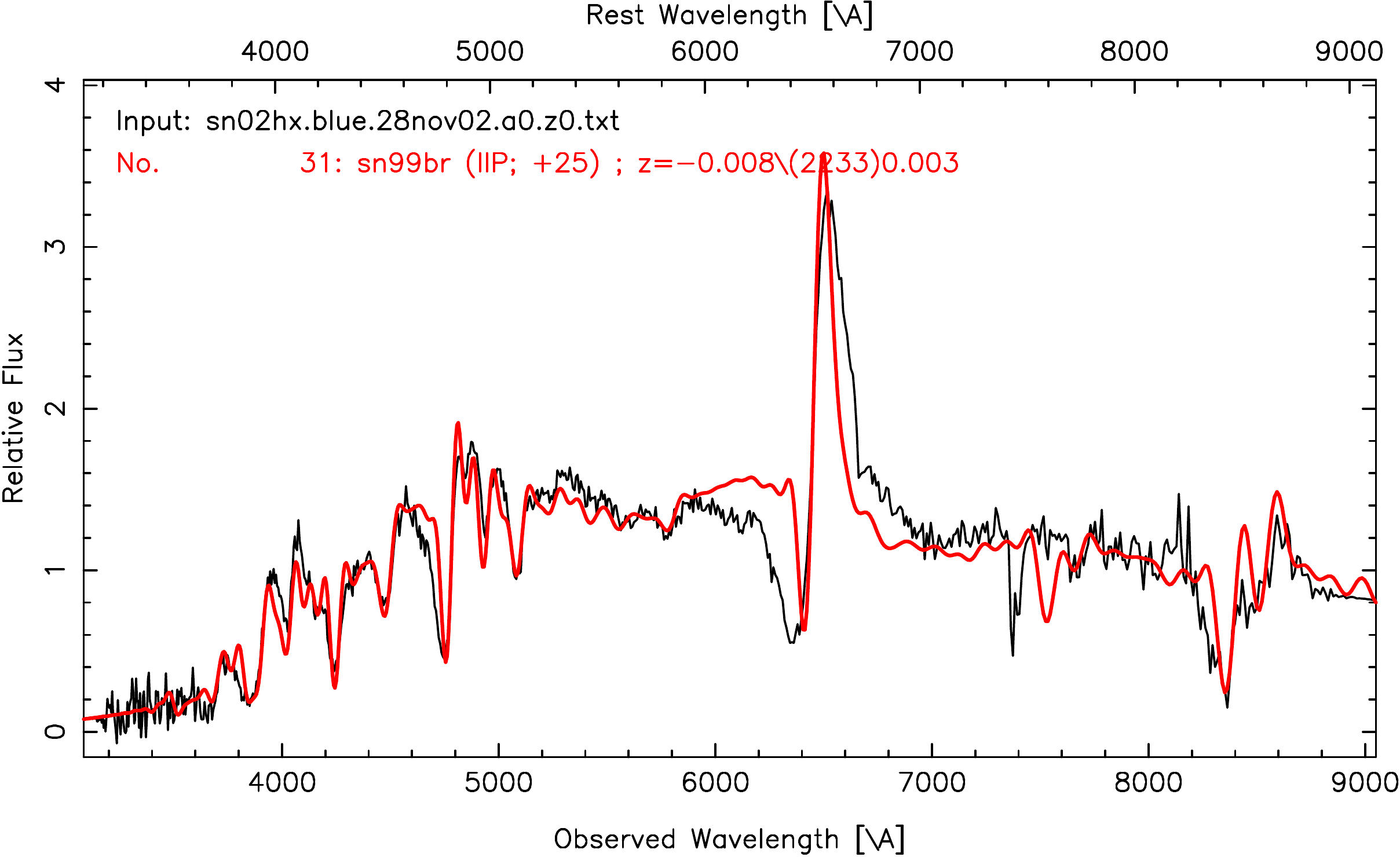}
\includegraphics[width=4.4cm]{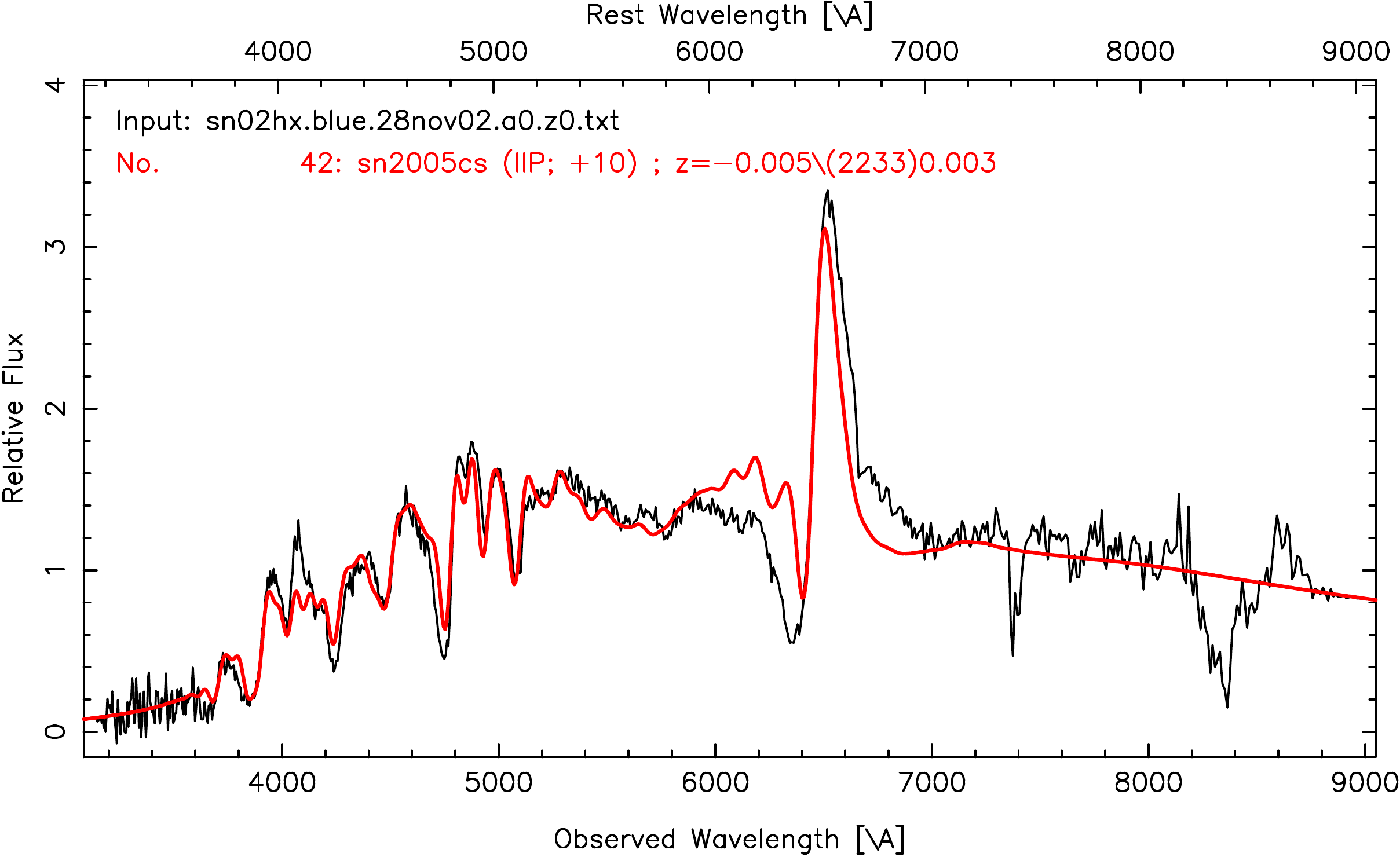}
\caption{Best spectral matching of SN~2002hx using SNID. The plots show SN~2002hx compared with 
SN~1999em, SN~2007il, SN~2009bz, SN~2003bn, SN~2006bp, SN~1999br, and SN~2005cs at 26, 45, 37, 39, 29, 25, and 16 days from explosion.}
\end{figure}

\begin{figure}[h!]
\centering
\includegraphics[width=4.4cm]{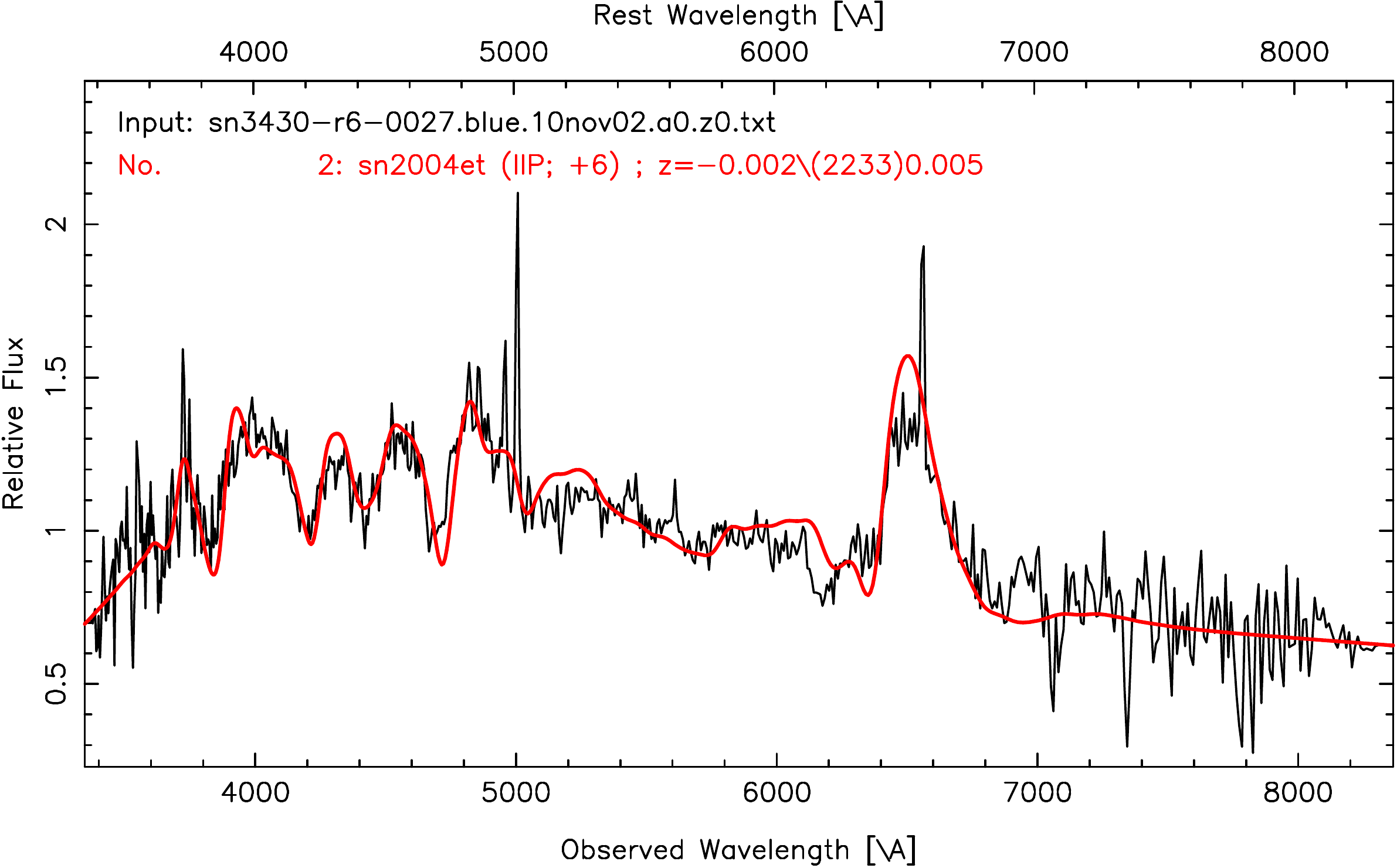}
\includegraphics[width=4.4cm]{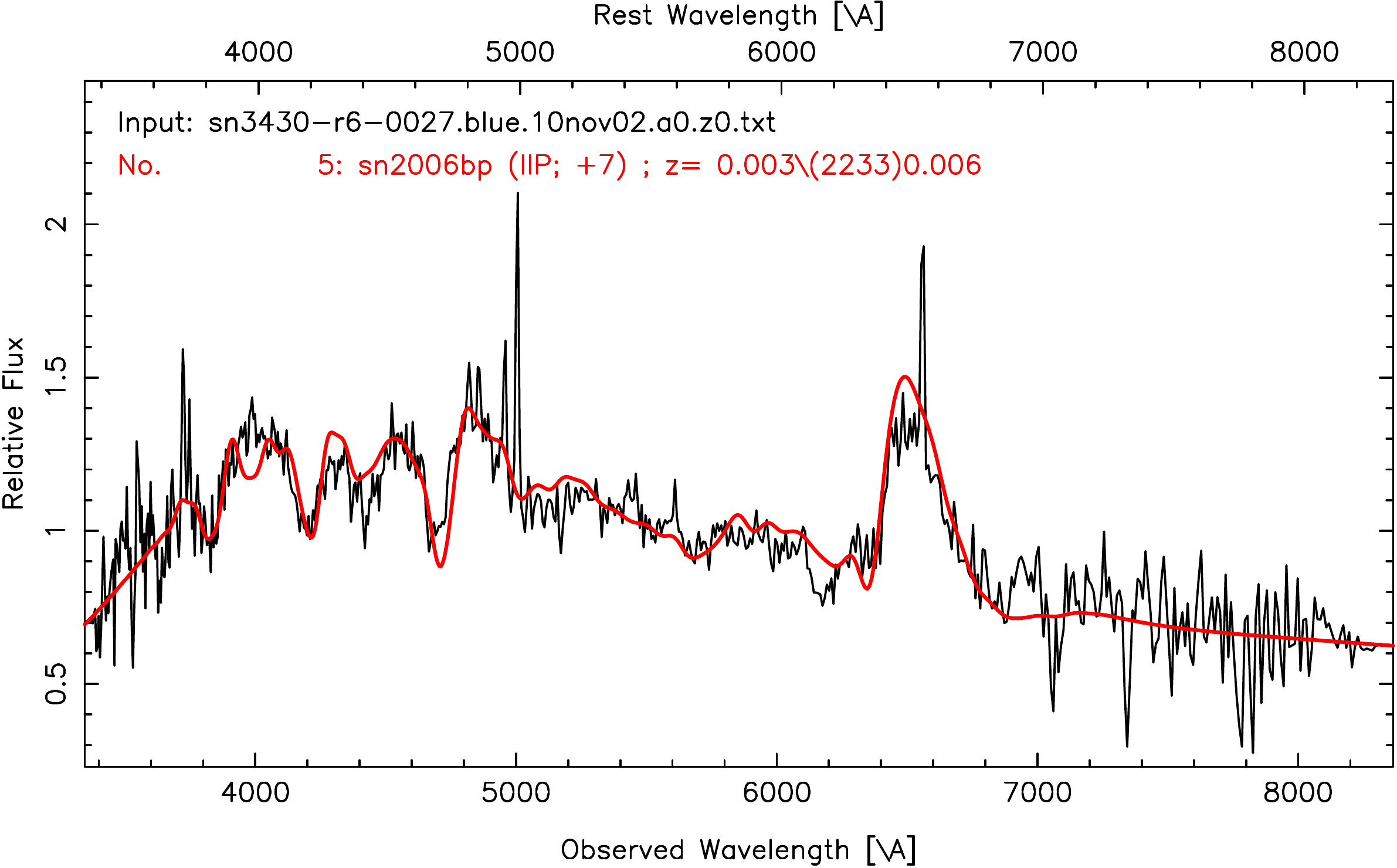}
\includegraphics[width=4.4cm]{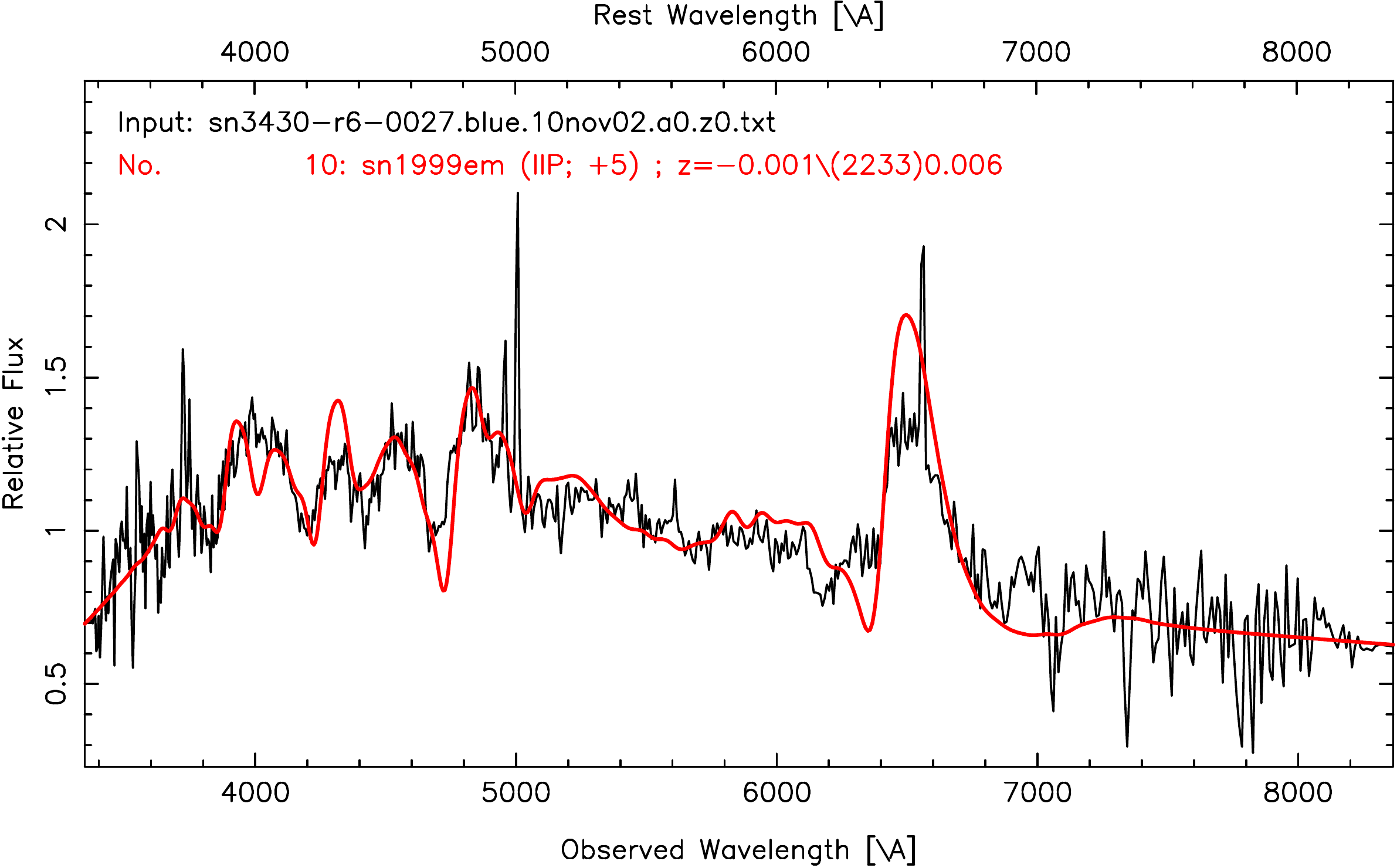}
\caption{Best spectral matching of SN~2002ig using SNID. The plots show SN~2002ig compared with 
SN~2004et, SN~2006bp, and SN~1999em at 22, 16, and 15 days from explosion.}
\end{figure}

\clearpage

\begin{figure}[h!]
\centering
\includegraphics[width=4.4cm]{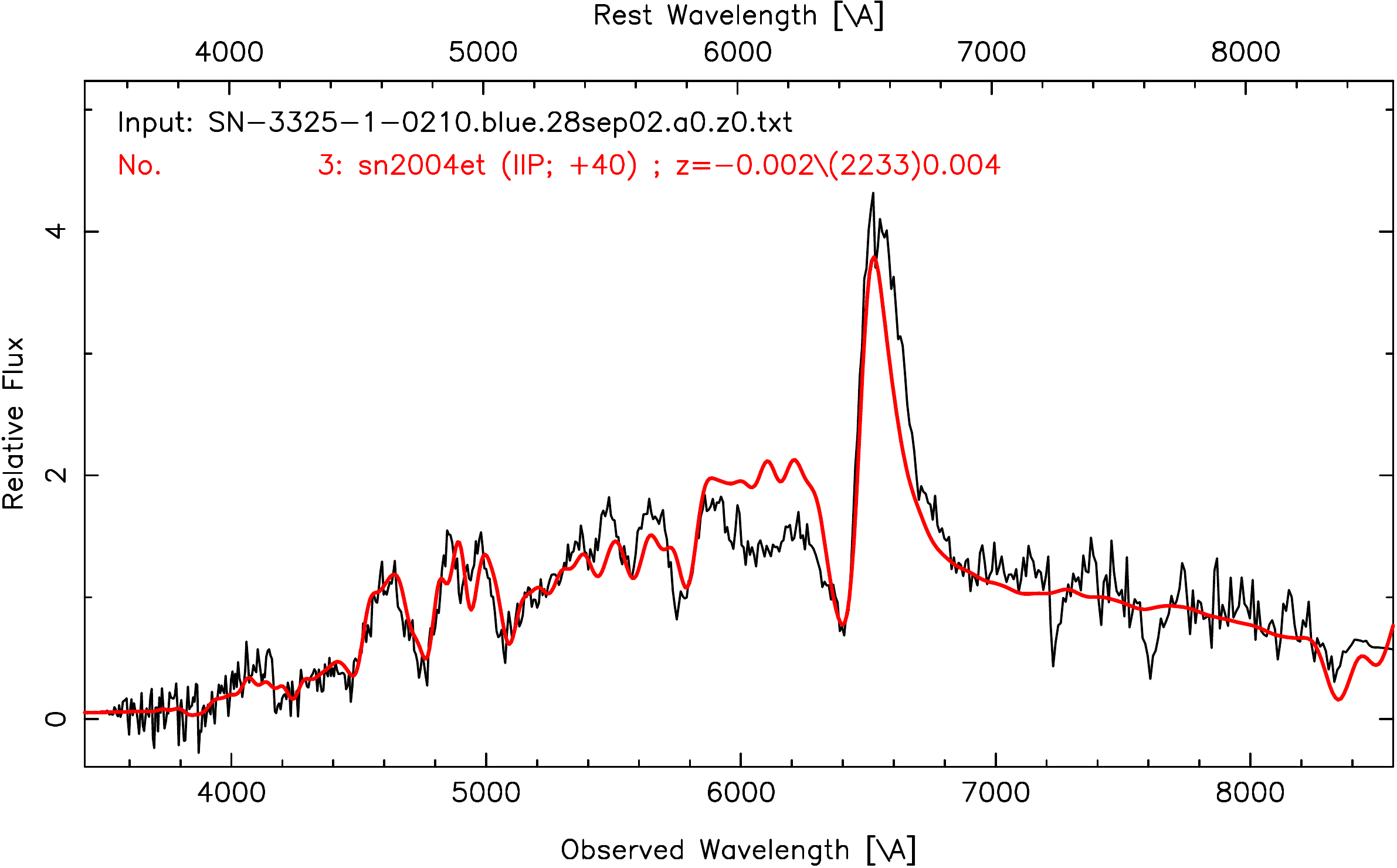}
\includegraphics[width=4.4cm]{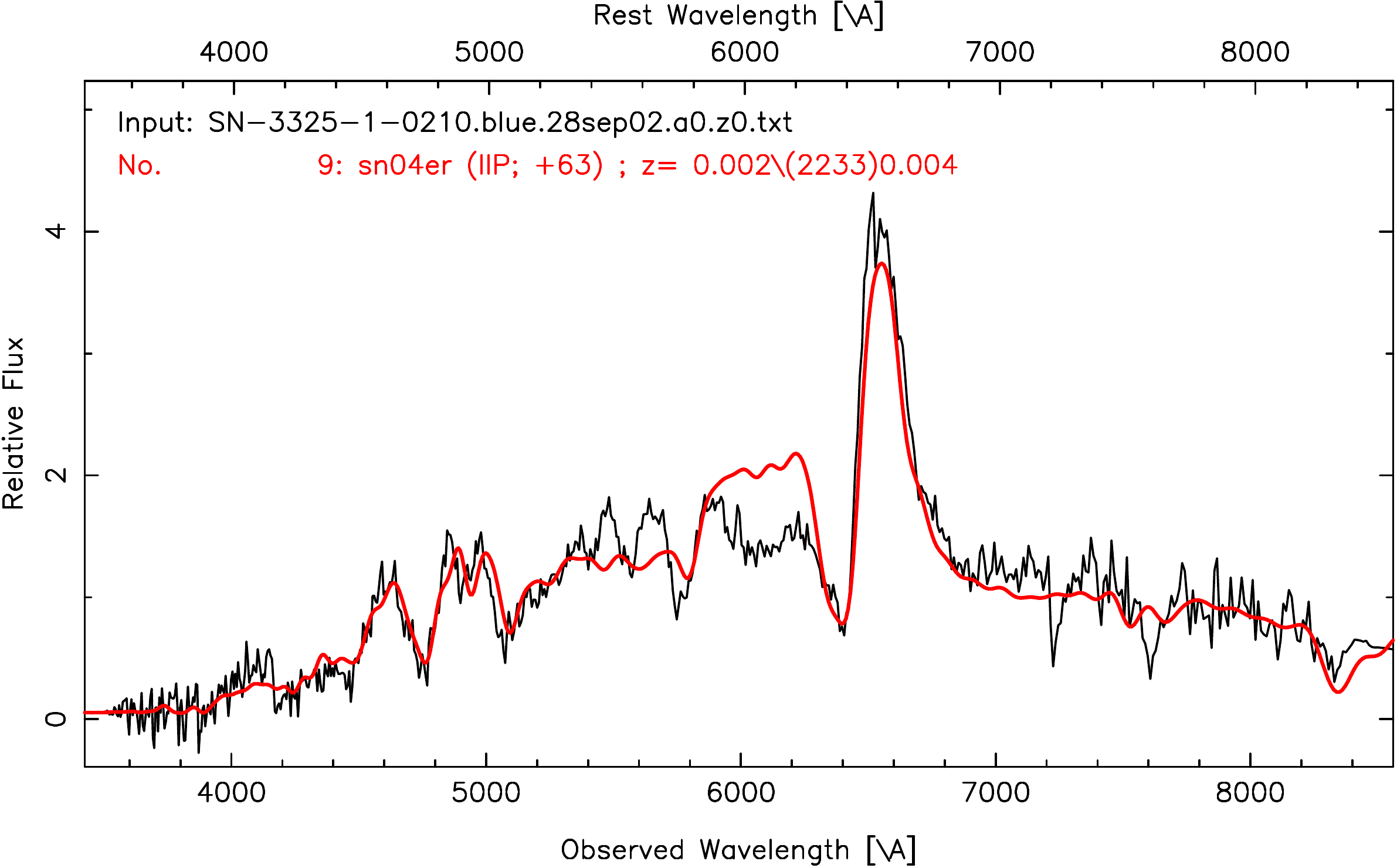}
\includegraphics[width=4.4cm]{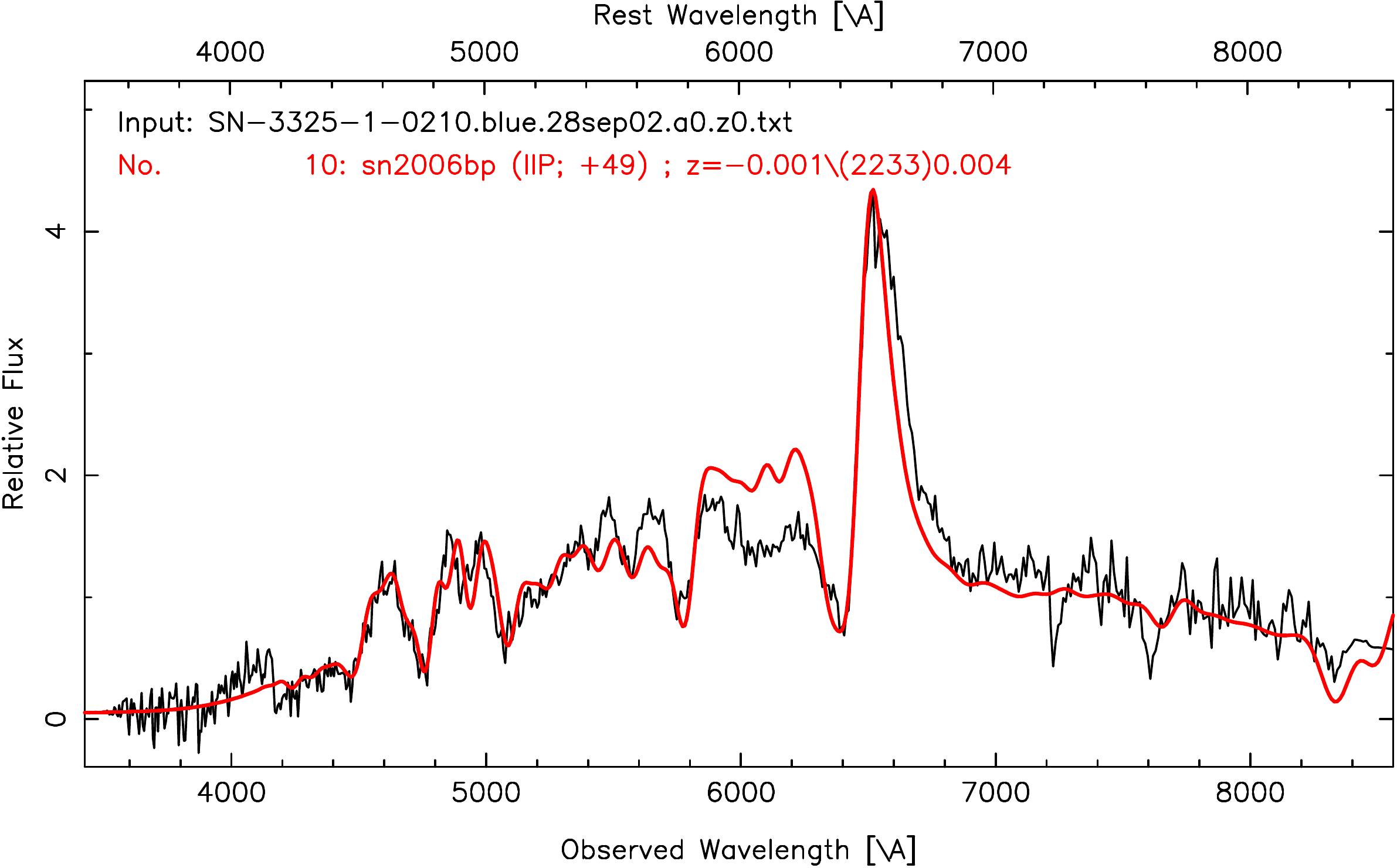}
\caption{Best spectral matching of SN~210 using SNID. The plots show SN~210 compared with 
SN~2004et, SN~2004er, and SN~2006bp at 56, 63, and 58 days from explosion.}
\end{figure}

\begin{figure}[h!]
\centering
\includegraphics[width=4.4cm]{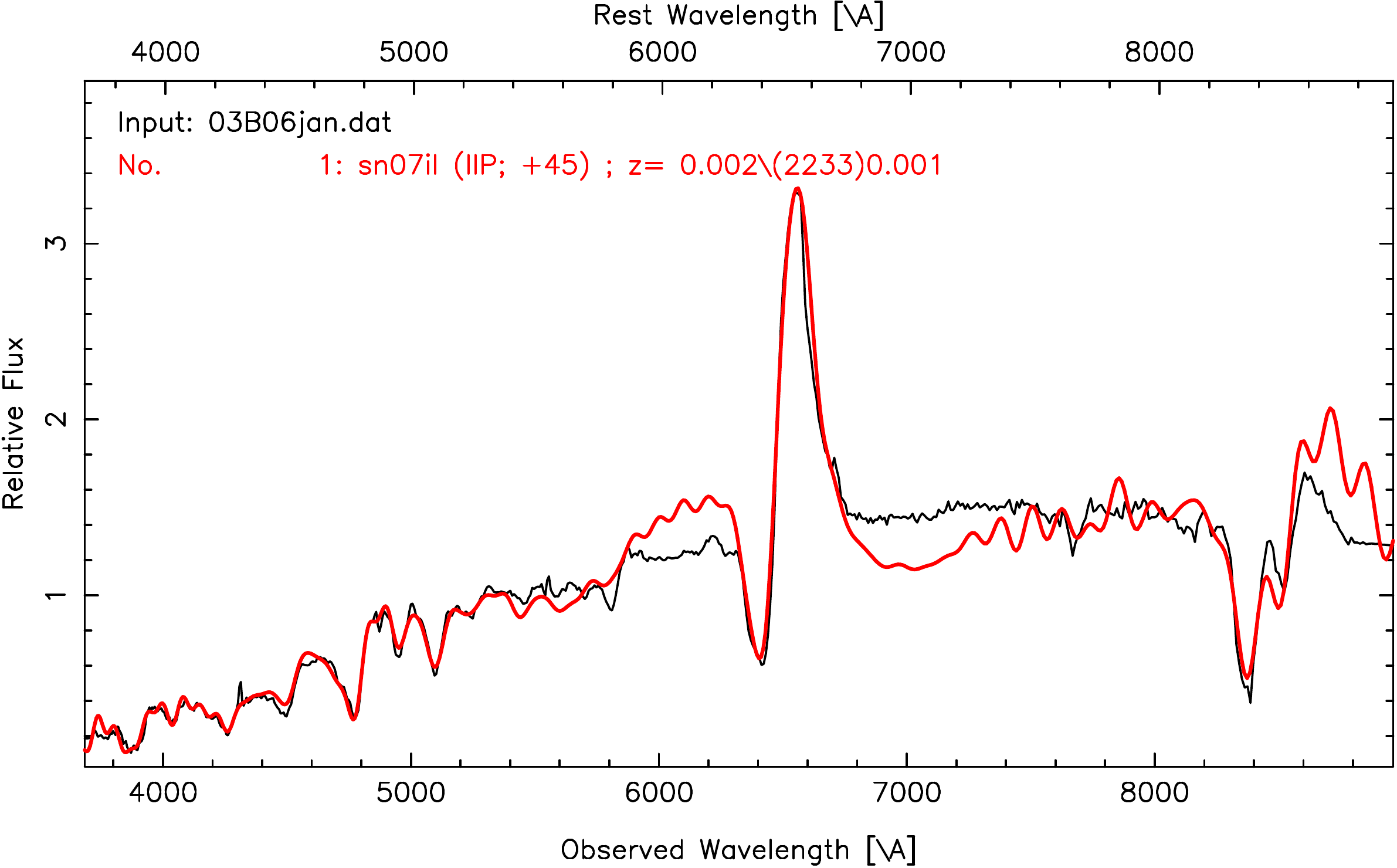}
\includegraphics[width=4.4cm]{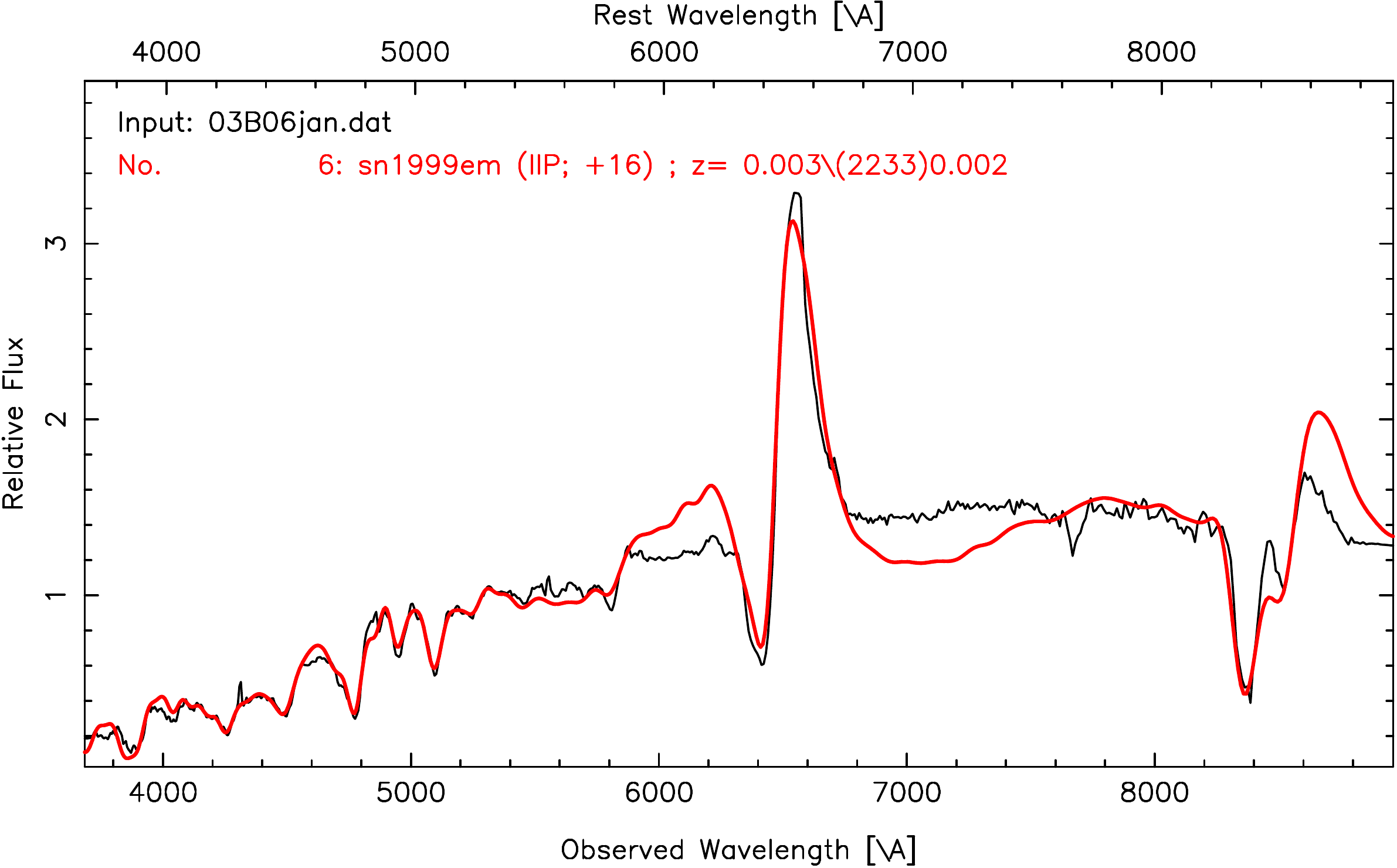}
\includegraphics[width=4.4cm]{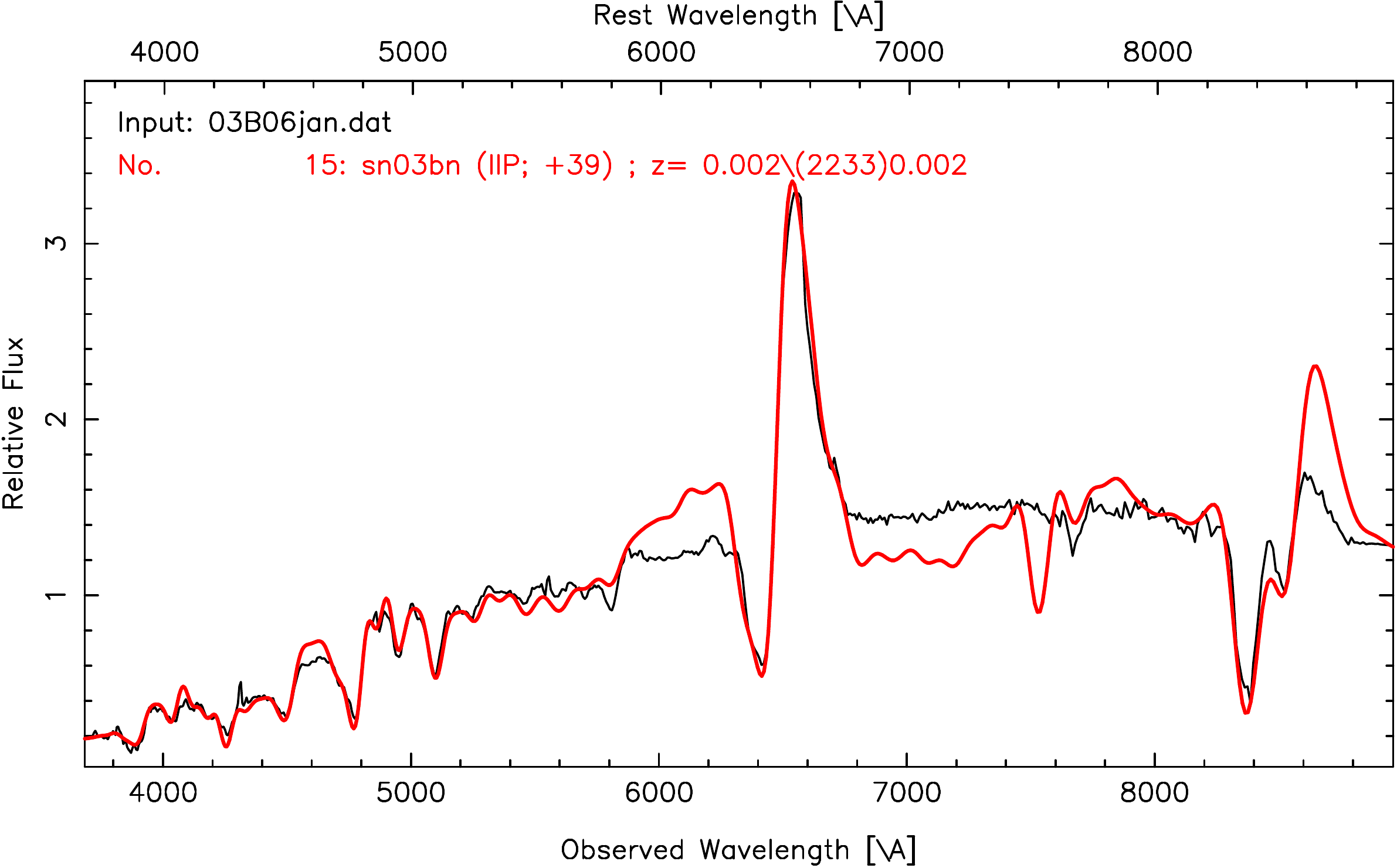}
\includegraphics[width=4.4cm]{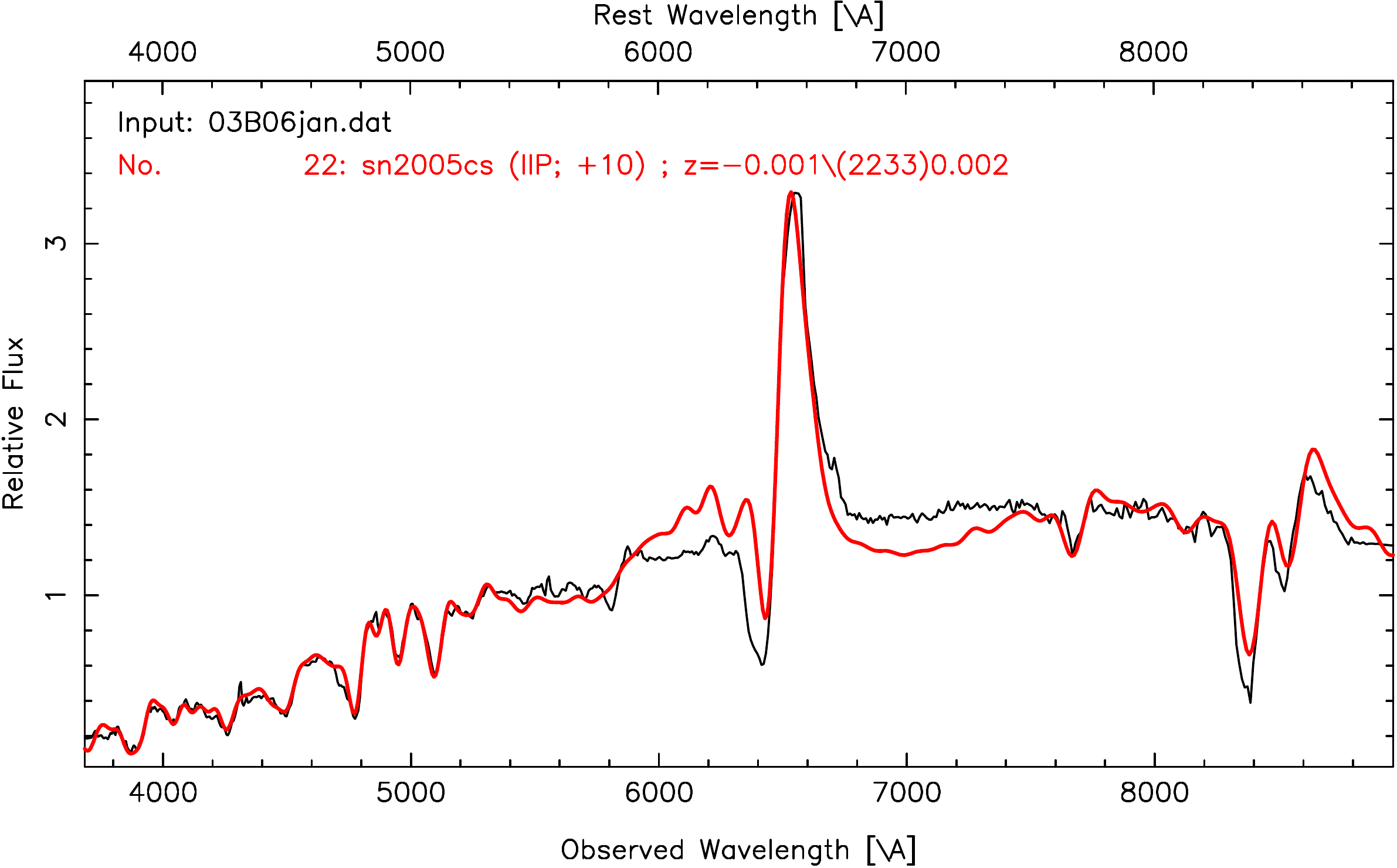}
\includegraphics[width=4.4cm]{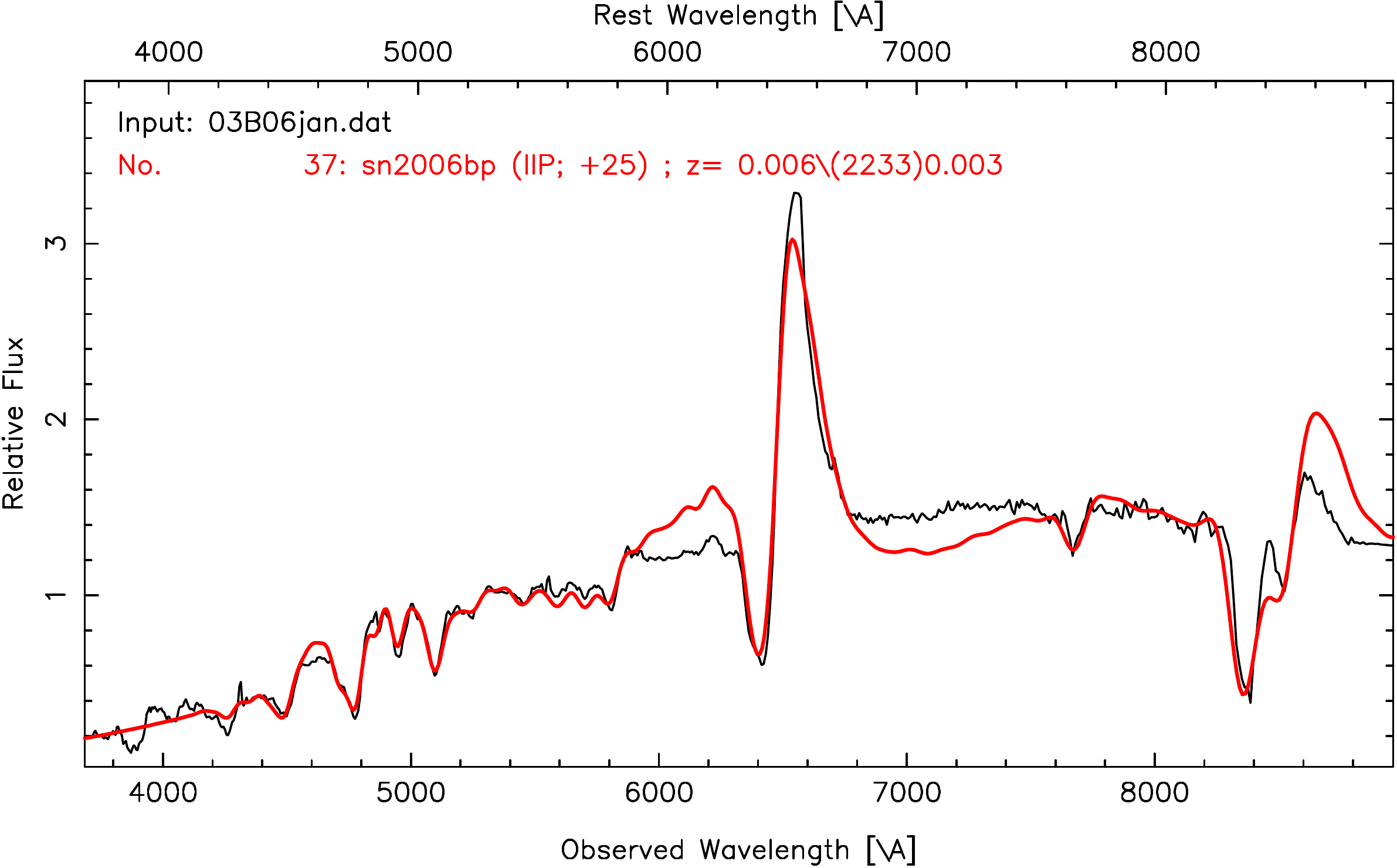}
\caption{Best spectral matching of SN~2003B using SNID. The plots show SN~2003B compared with 
SN~2007il, SN~1999em, SN~2003bn, SN~2005cs, and SN~2006bp at 45, 26, 39, 16, and 34 days from explosion.}
\end{figure}

\clearpage

\begin{figure}[h!]
\centering
\includegraphics[width=4.4cm]{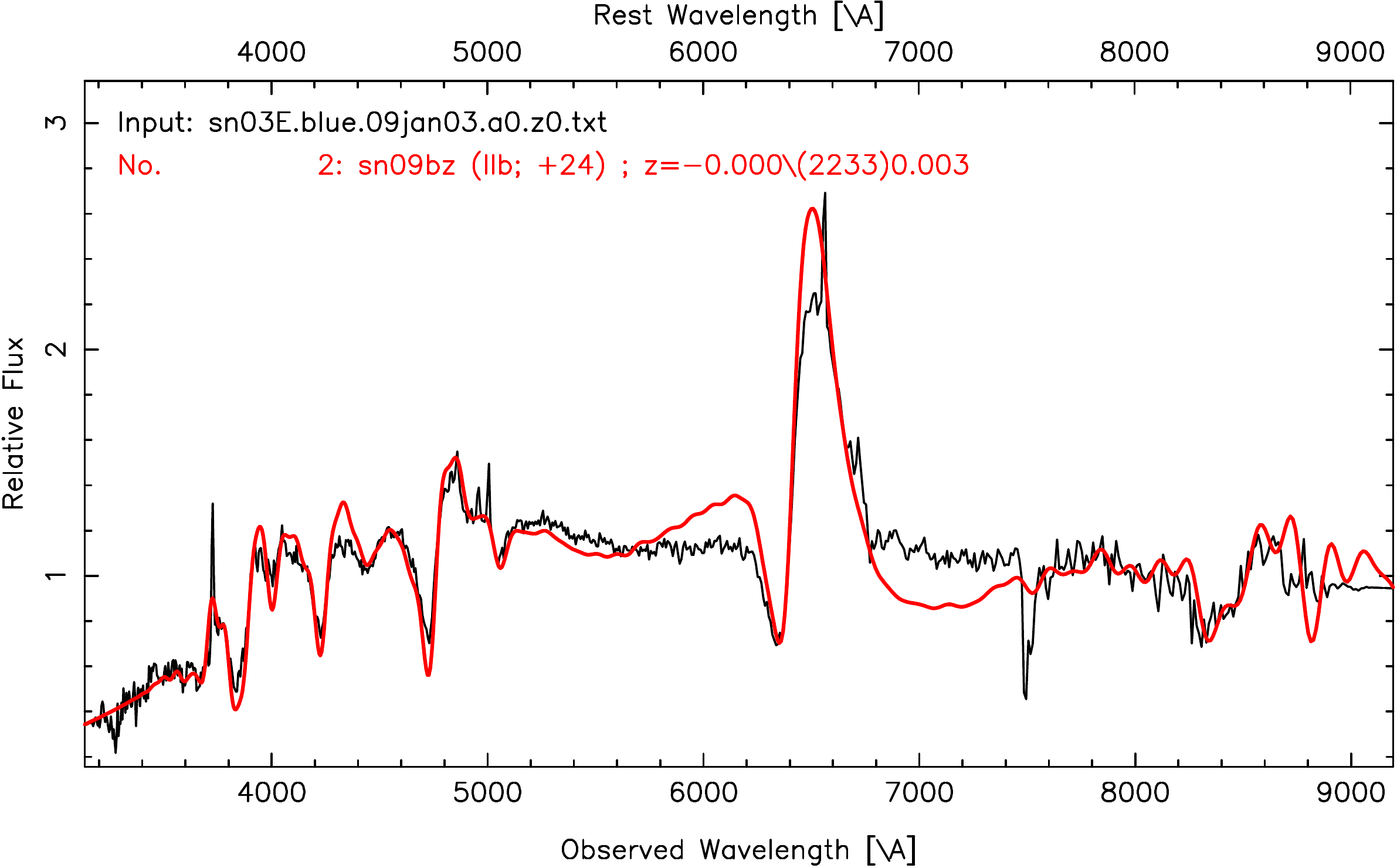}
\includegraphics[width=4.4cm]{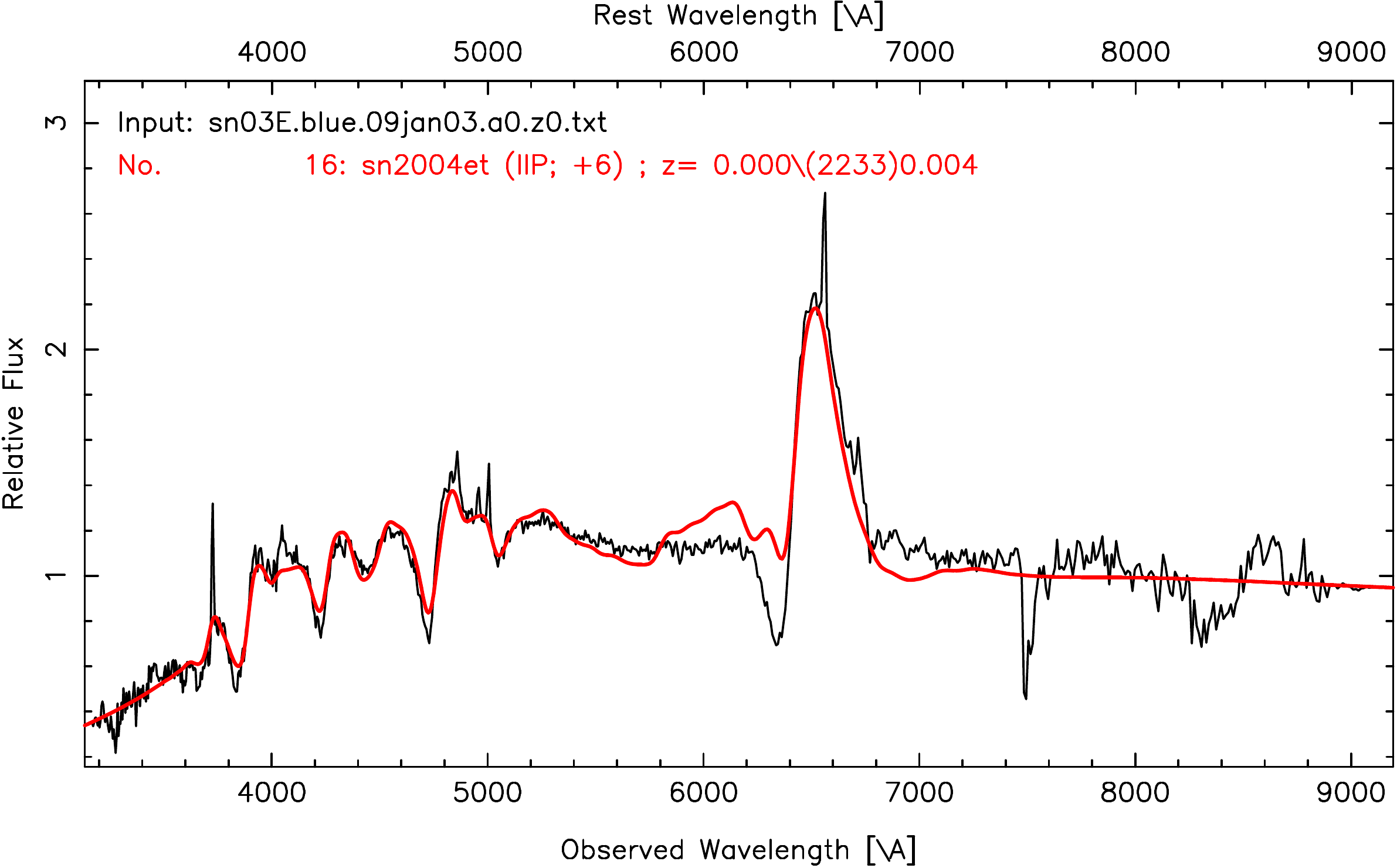}
\includegraphics[width=4.4cm]{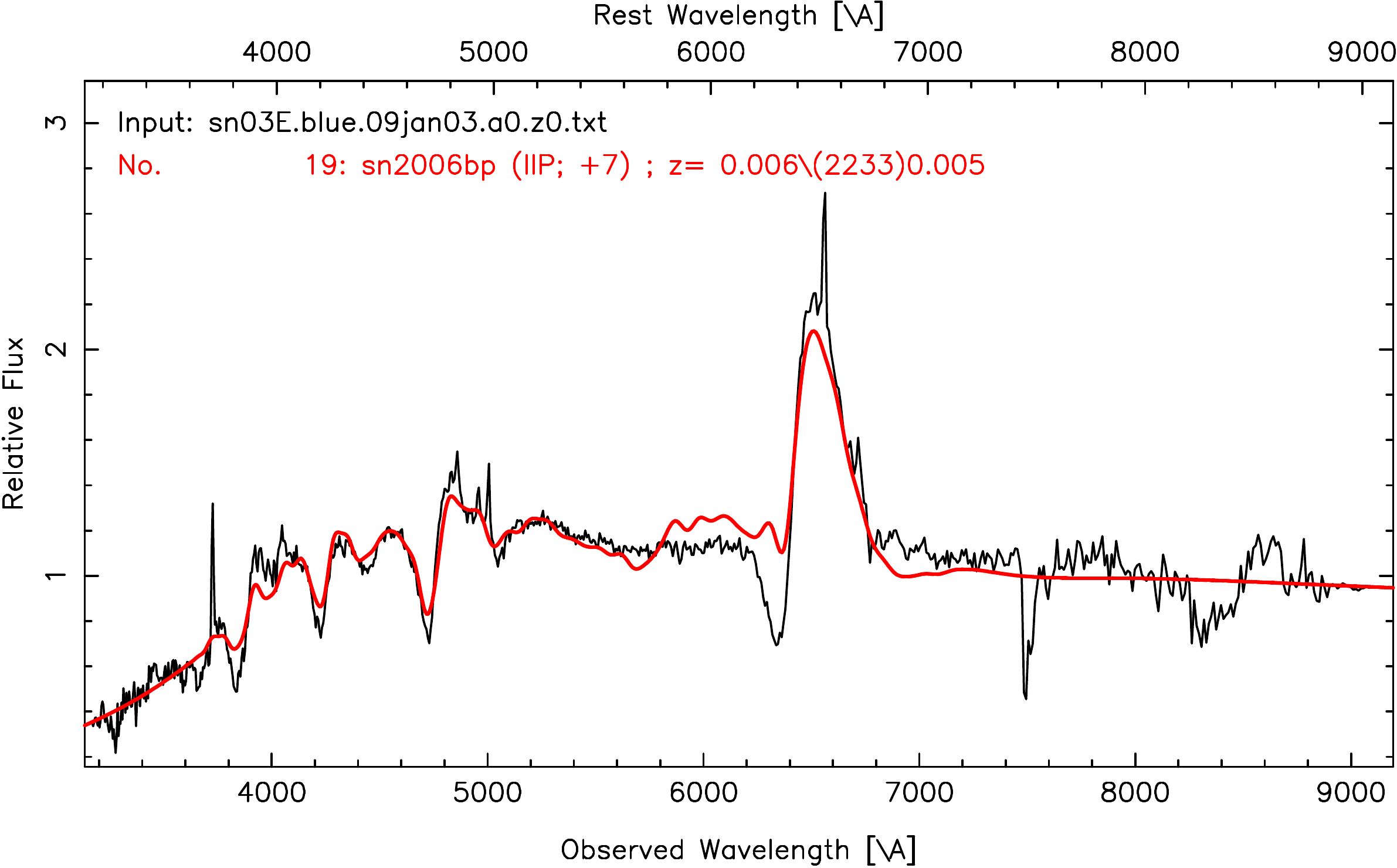}
\includegraphics[width=4.4cm]{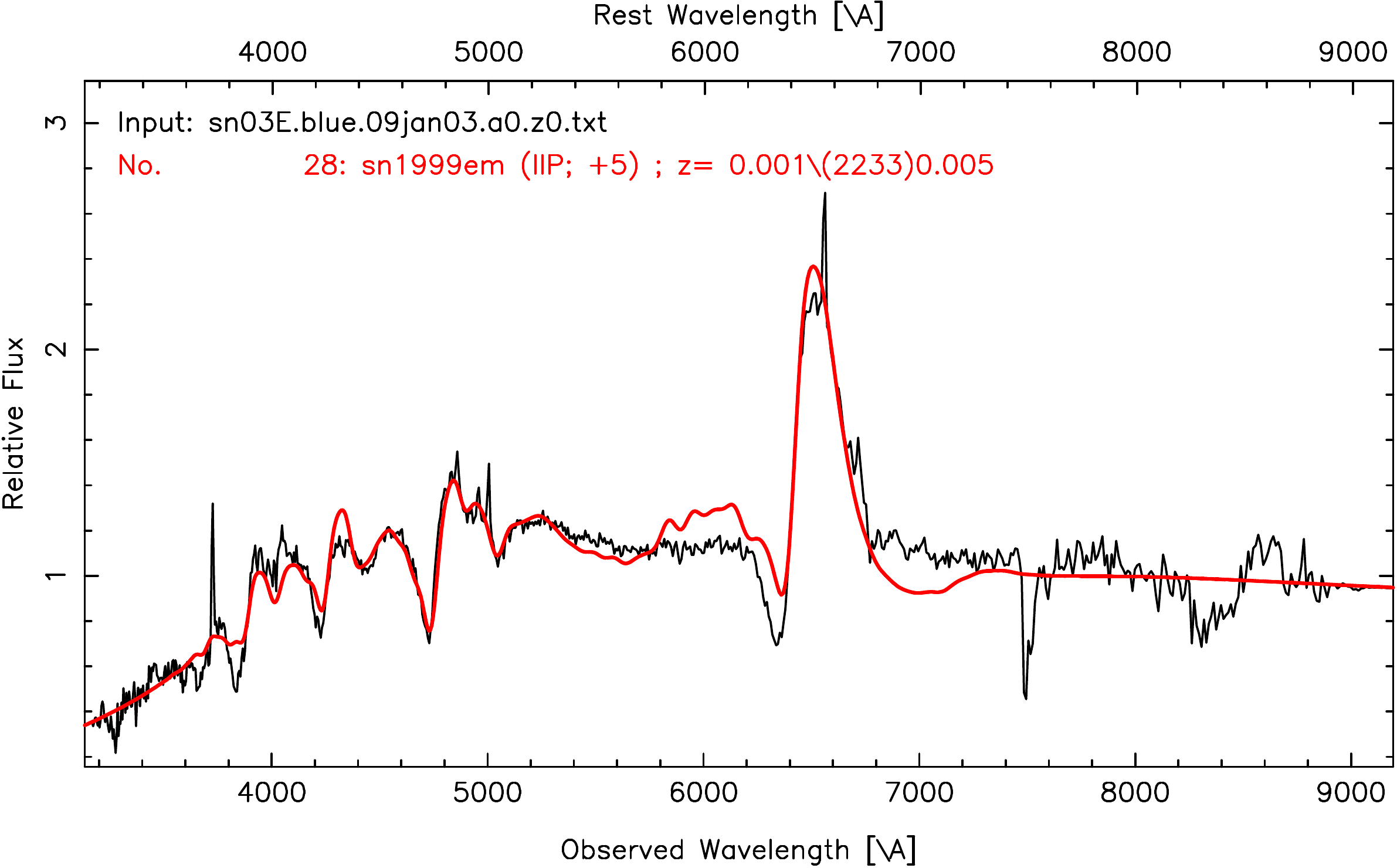}
\includegraphics[width=4.4cm]{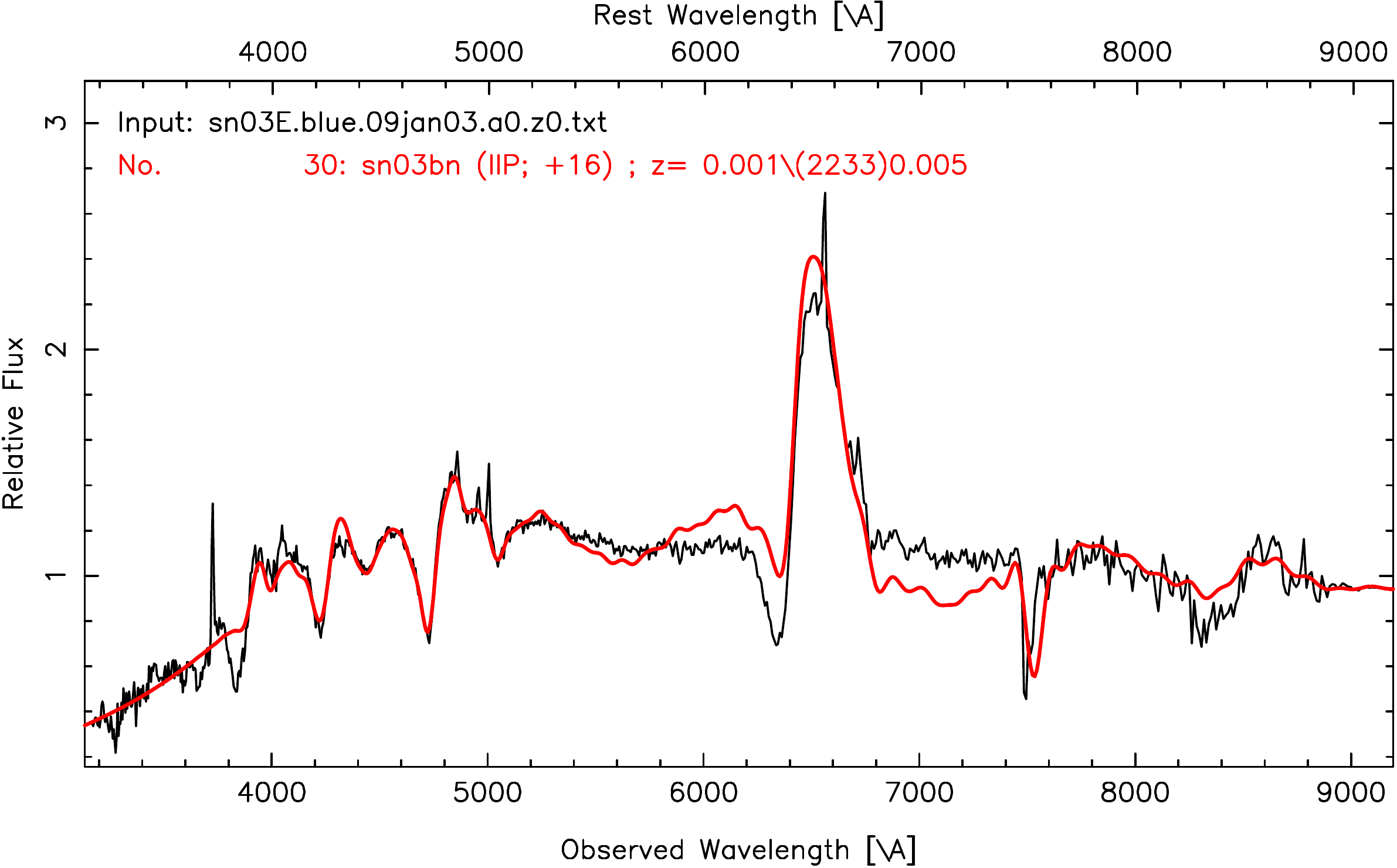}
\caption{Best spectral matching of SN~2003E using SNID. The plots show SN~2003E compared with 
SN~2009bz, SN~2004et, SN~2006bp, SN~1999em, and SN~2003bn at 24, 22, 16, 15, and 16 days from explosion.}
\end{figure}

\begin{figure}[h!]
\centering
\includegraphics[width=4.4cm]{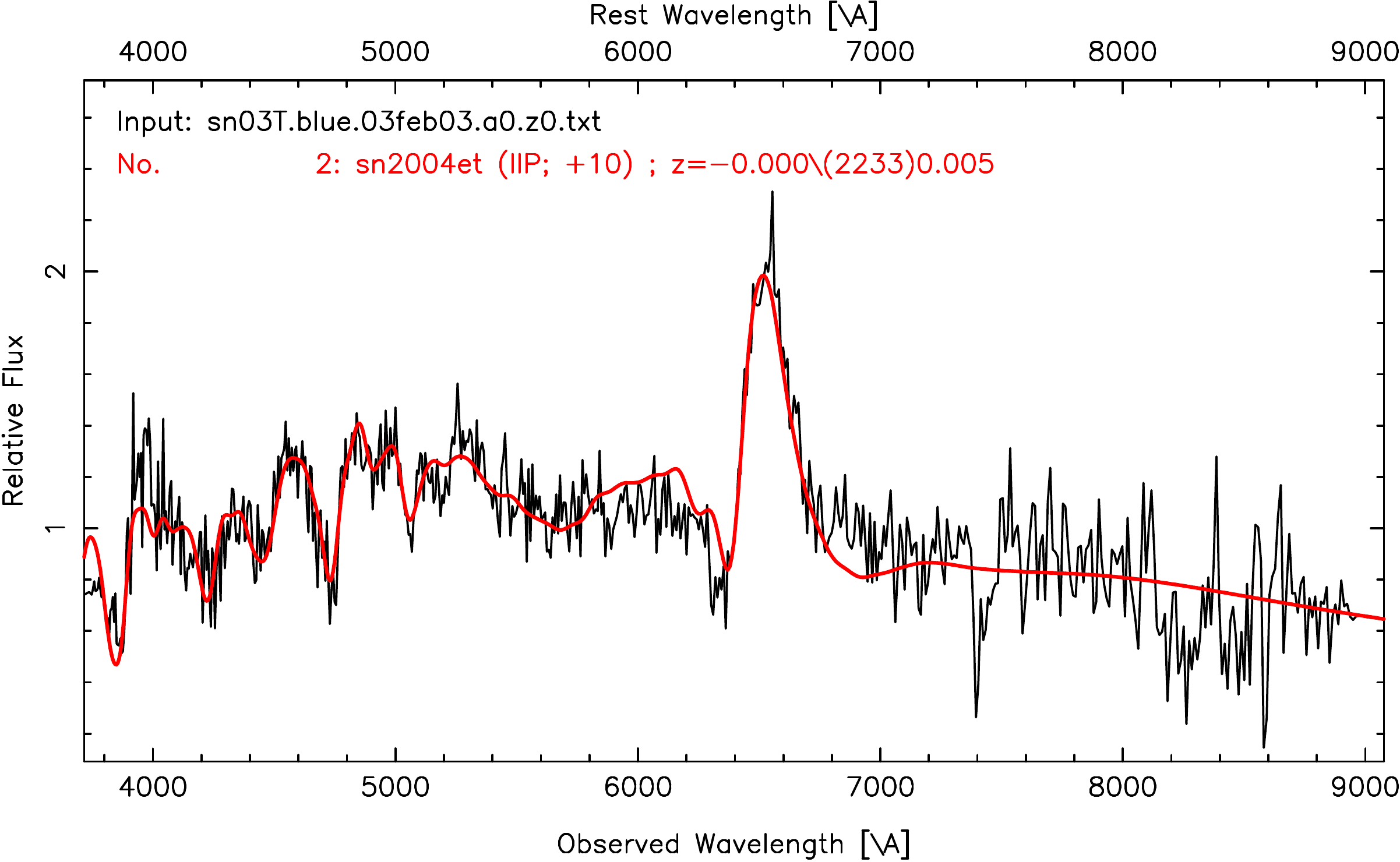}
\includegraphics[width=4.4cm]{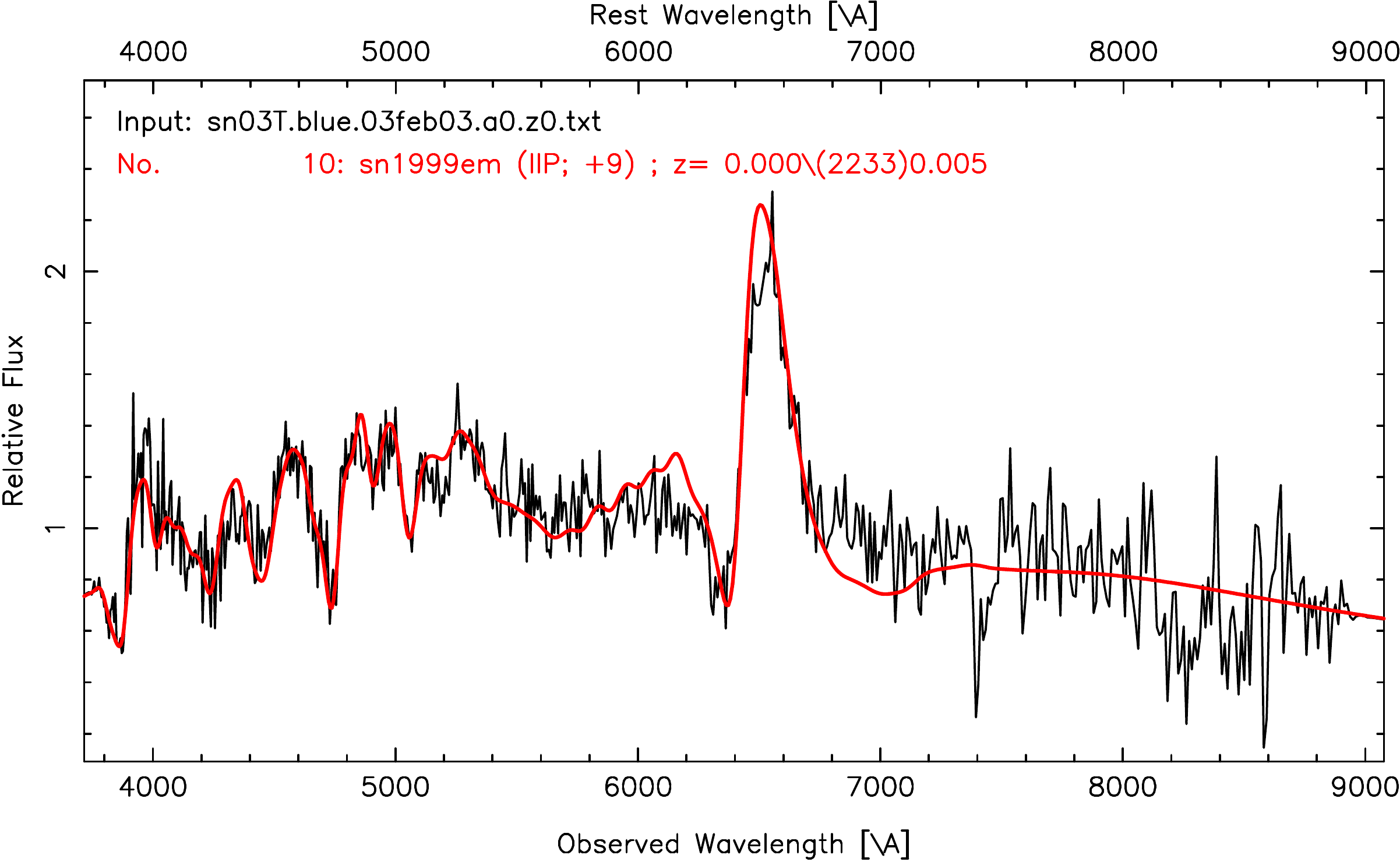}
\includegraphics[width=4.4cm]{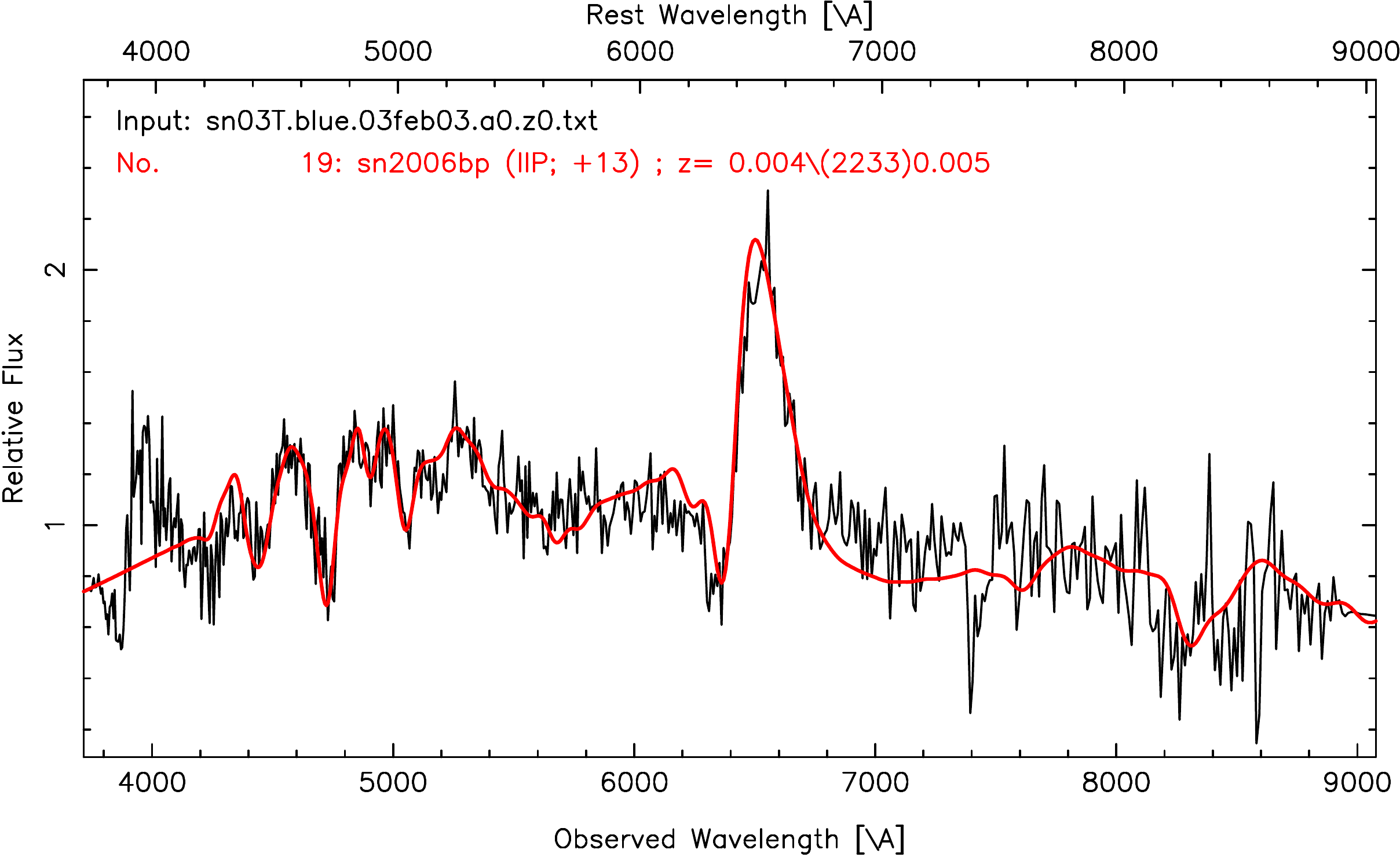}
\caption{Best spectral matching of SN~2003T using SNID. The plots show SN~2003T compared with 
SN~2004et, SN~1999em, and SN~2006bp at 26, 19, and 22 days from explosion.}
\end{figure}

\begin{figure}[h!]
\centering
\includegraphics[width=4.4cm]{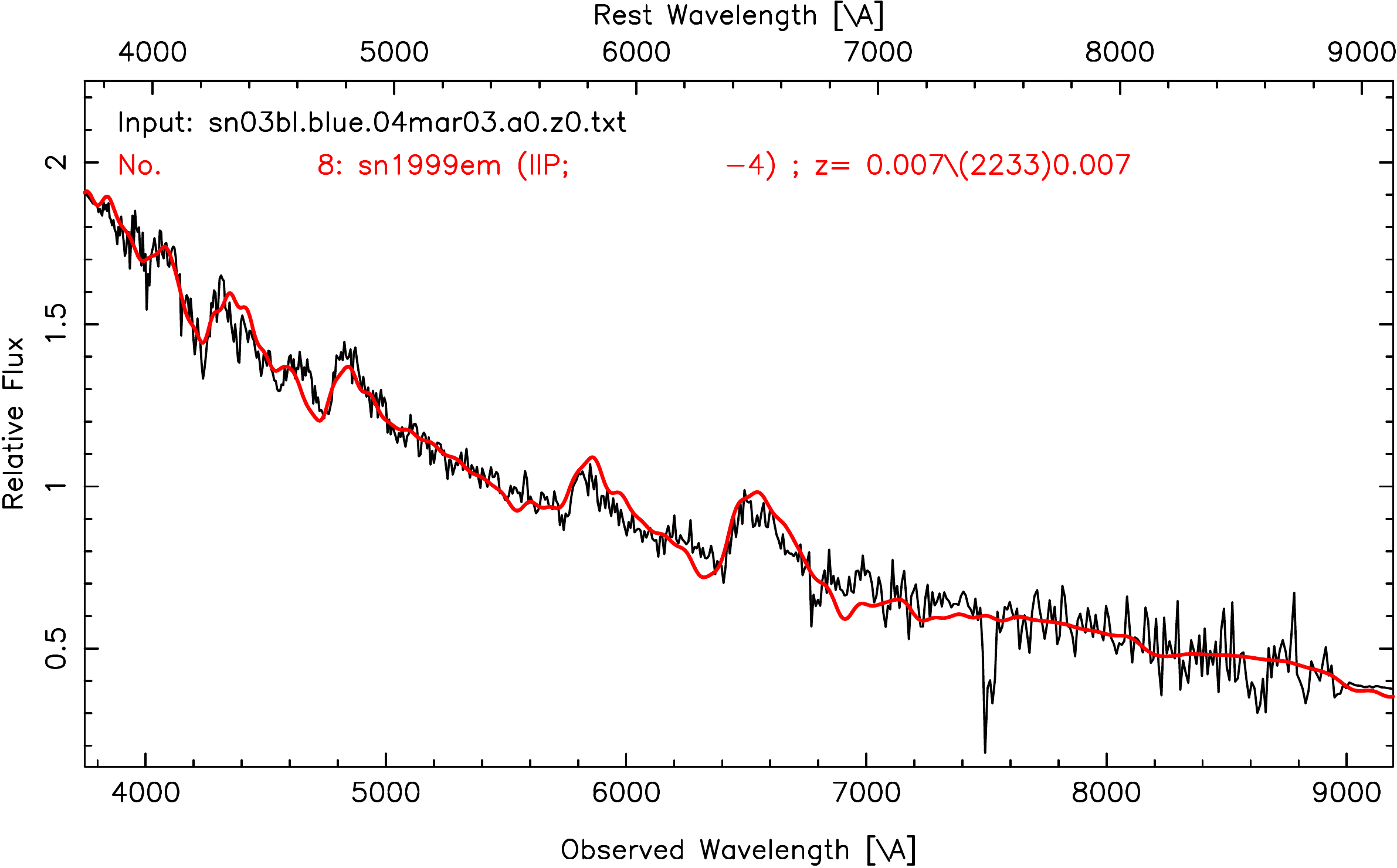}
\includegraphics[width=4.4cm]{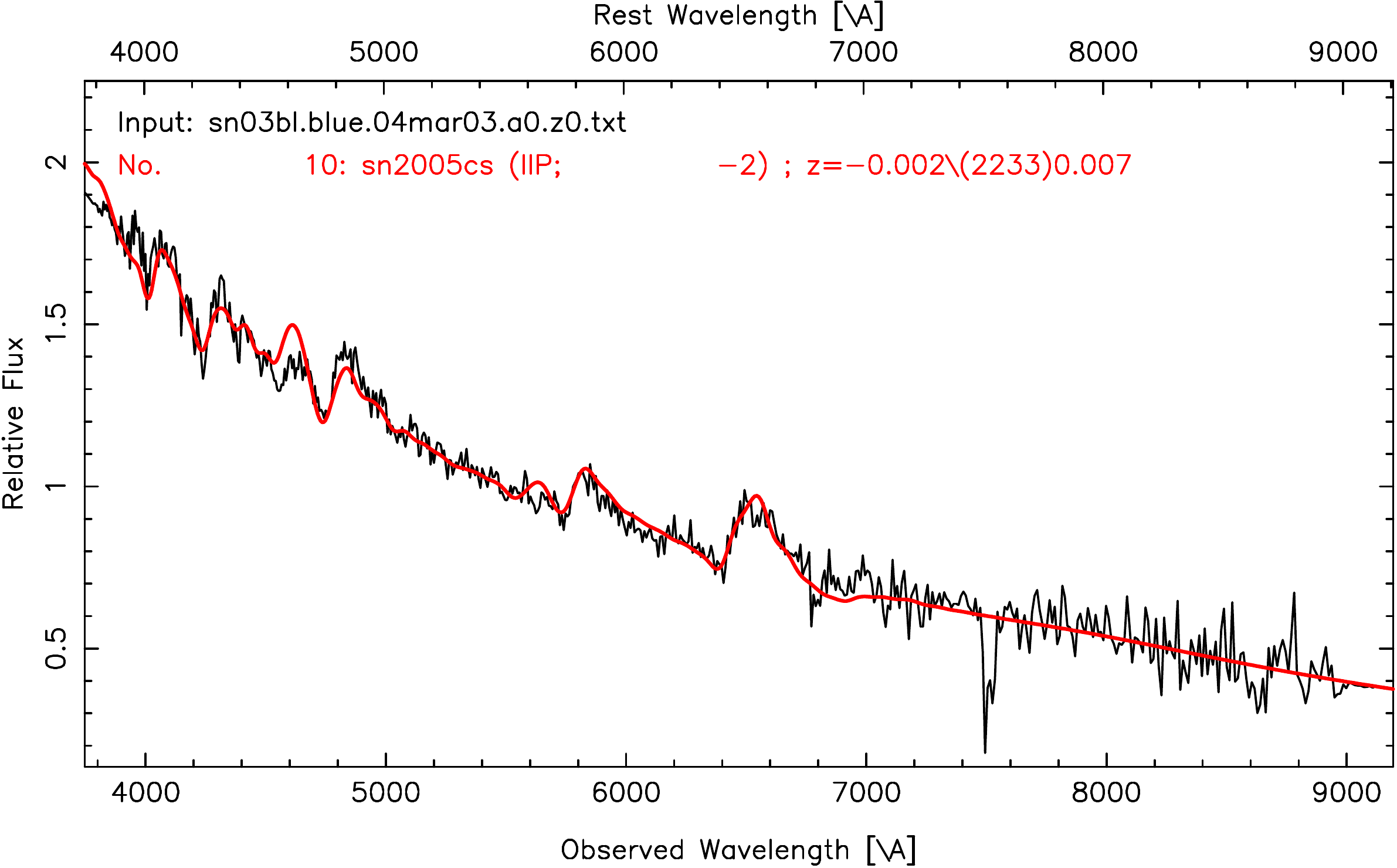}
\includegraphics[width=4.4cm]{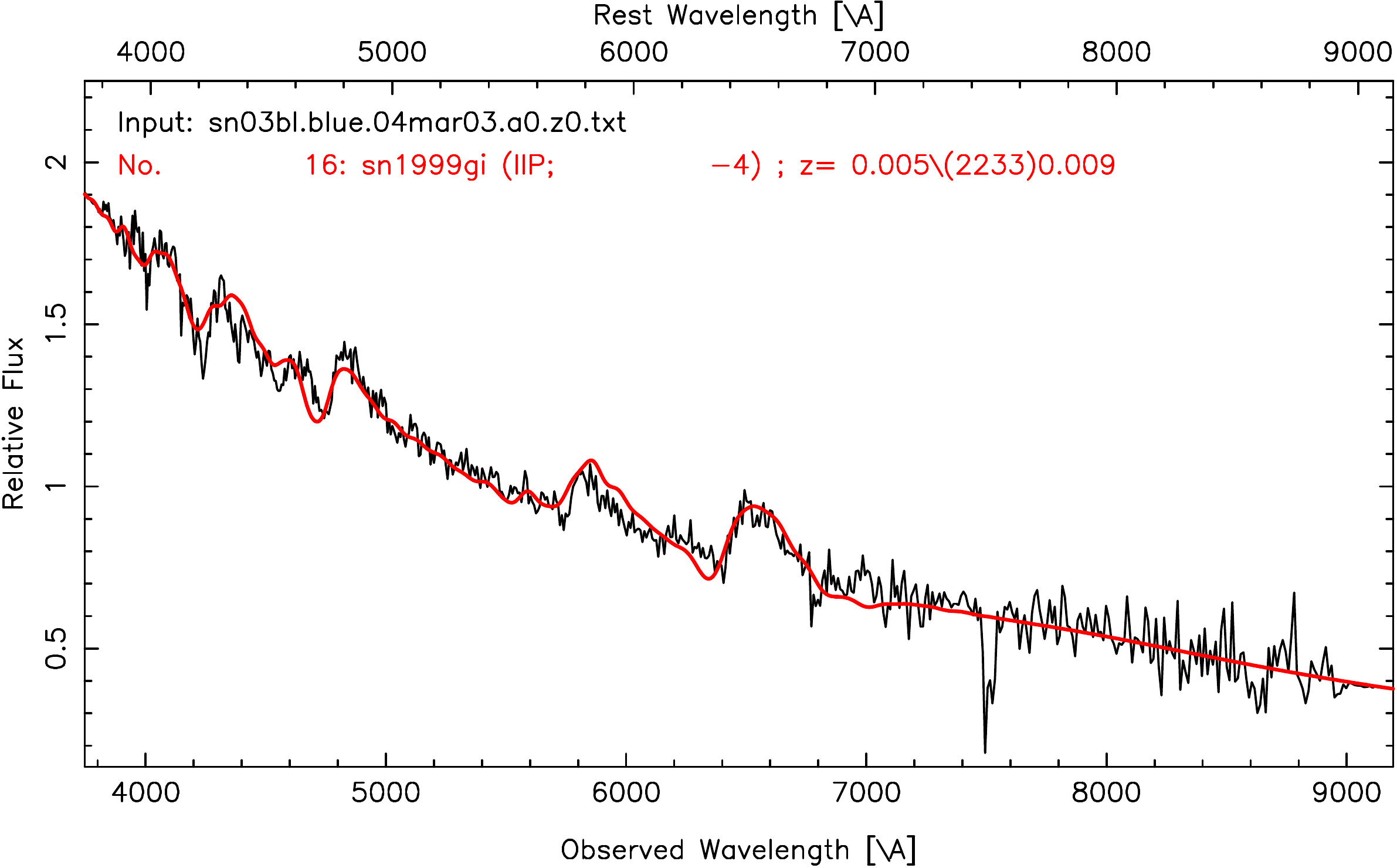}
\caption{Best spectral matching of SN~2003bl using SNID. The plots show SN~2003bl compared with 
SN~1999em, SN~2005cs, and SN~1999gi at 6, 4, and 8 days from explosion.}
\end{figure}

\clearpage

\begin{figure}[h!]
\centering
\includegraphics[width=4.4cm]{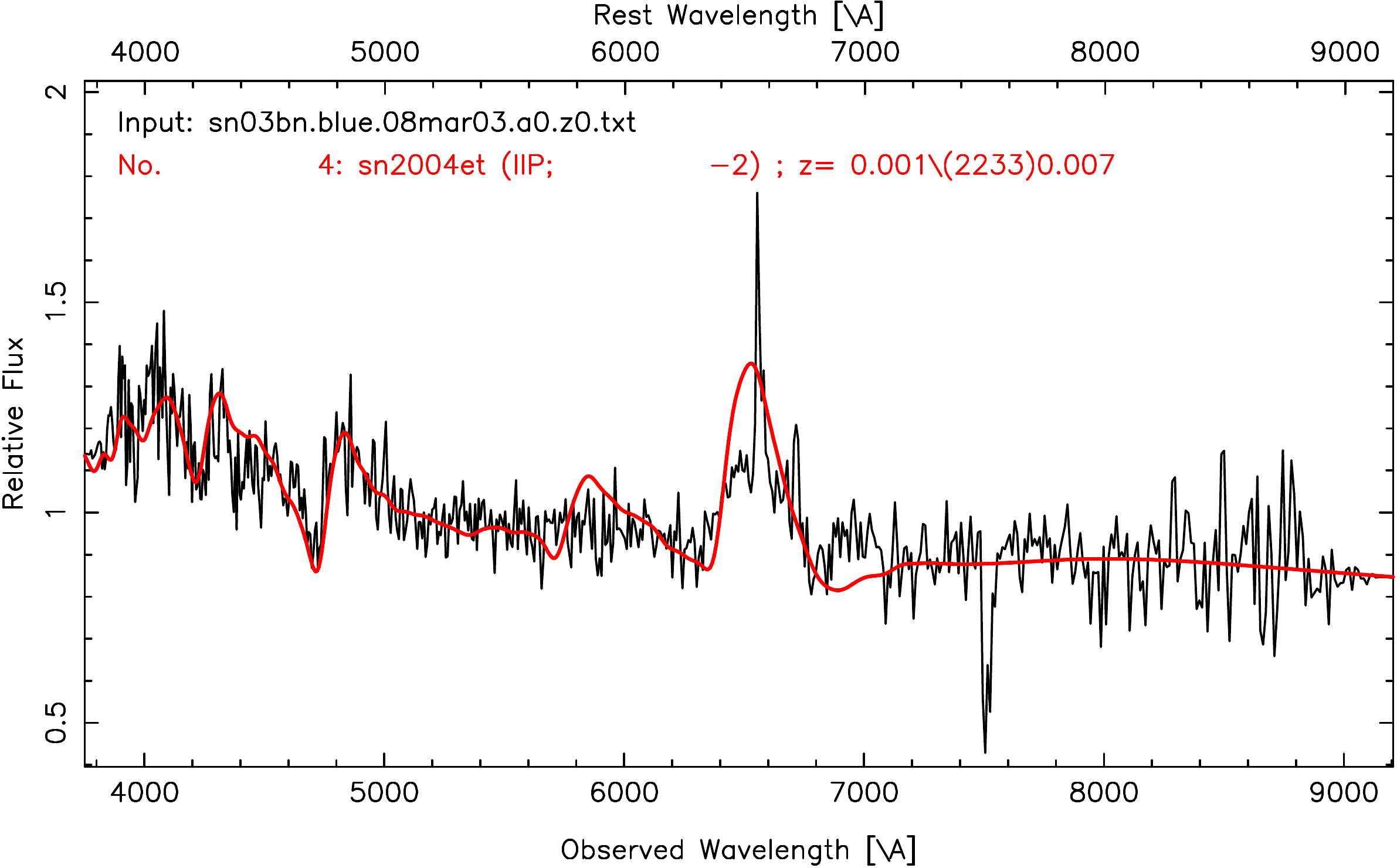}
\includegraphics[width=4.4cm]{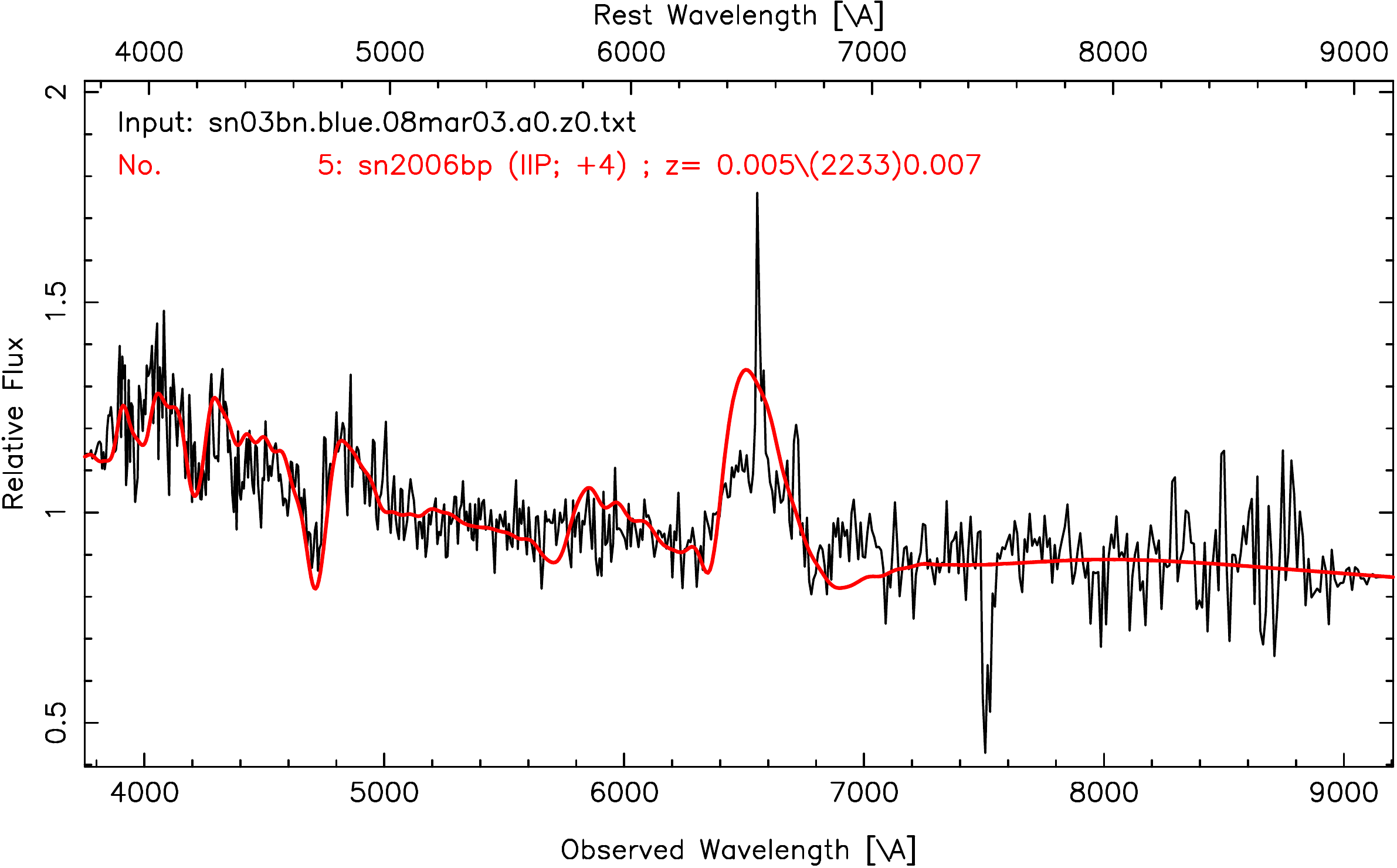}
\includegraphics[width=4.4cm]{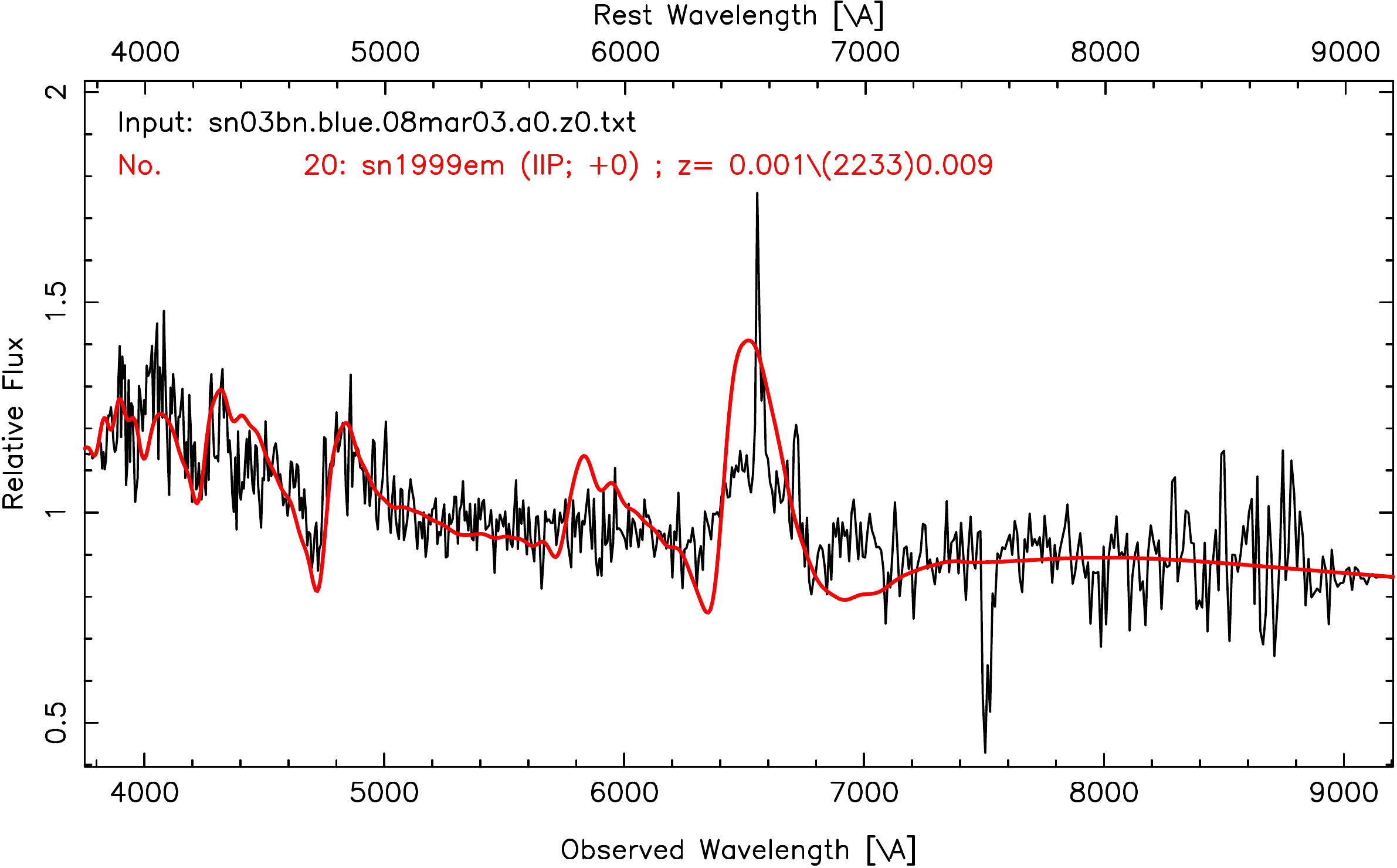}
\caption{Best spectral matching of SN~2003bn using SNID. The plots show SN~2003bn compared with 
SN~2004et, SN~2006bp, and SN~1999em at 14, 13, and 10 days from explosion.}
\end{figure}

\begin{figure}[h!]
\centering
\includegraphics[width=4.4cm]{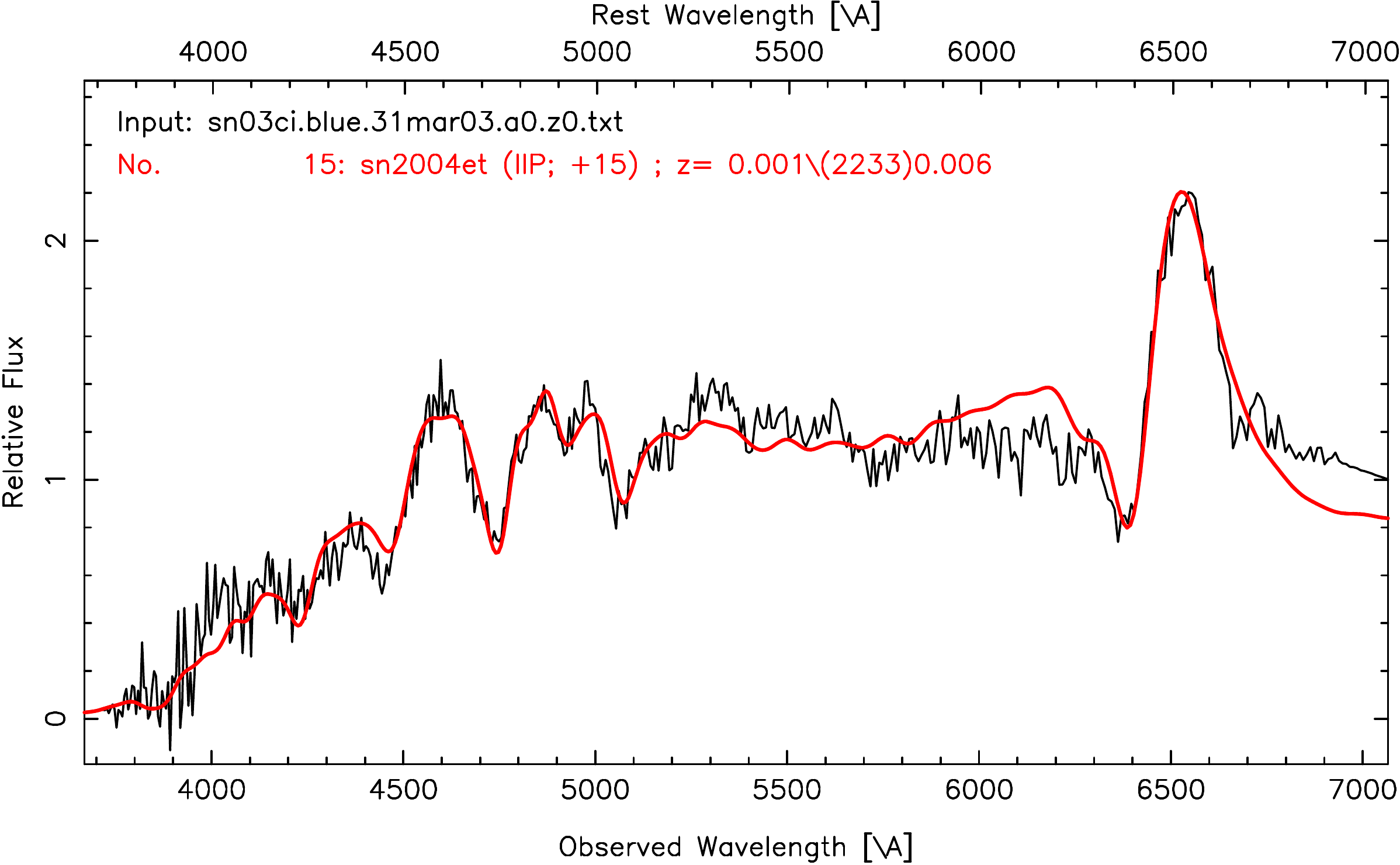}
\includegraphics[width=4.4cm]{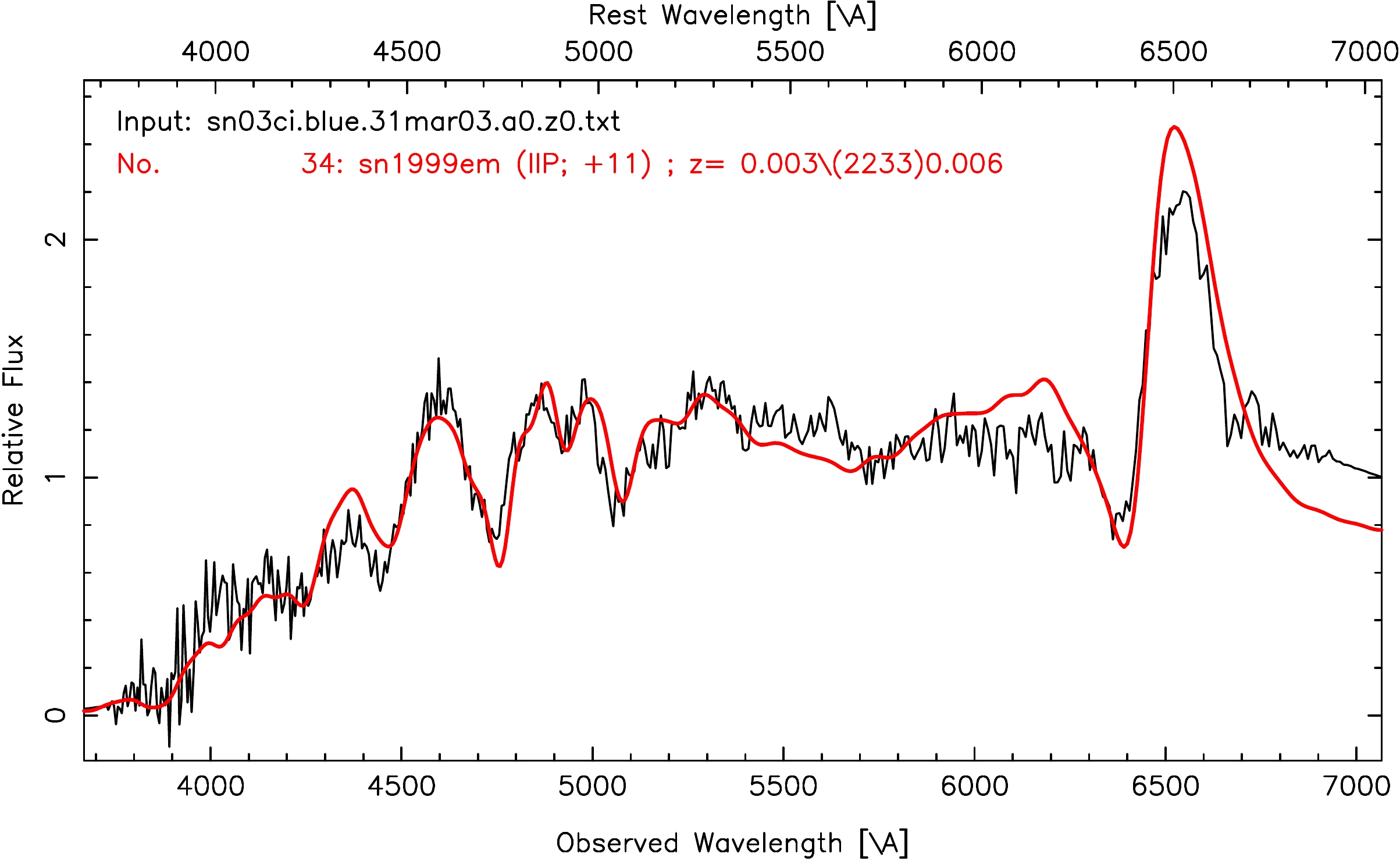}
\includegraphics[width=4.4cm]{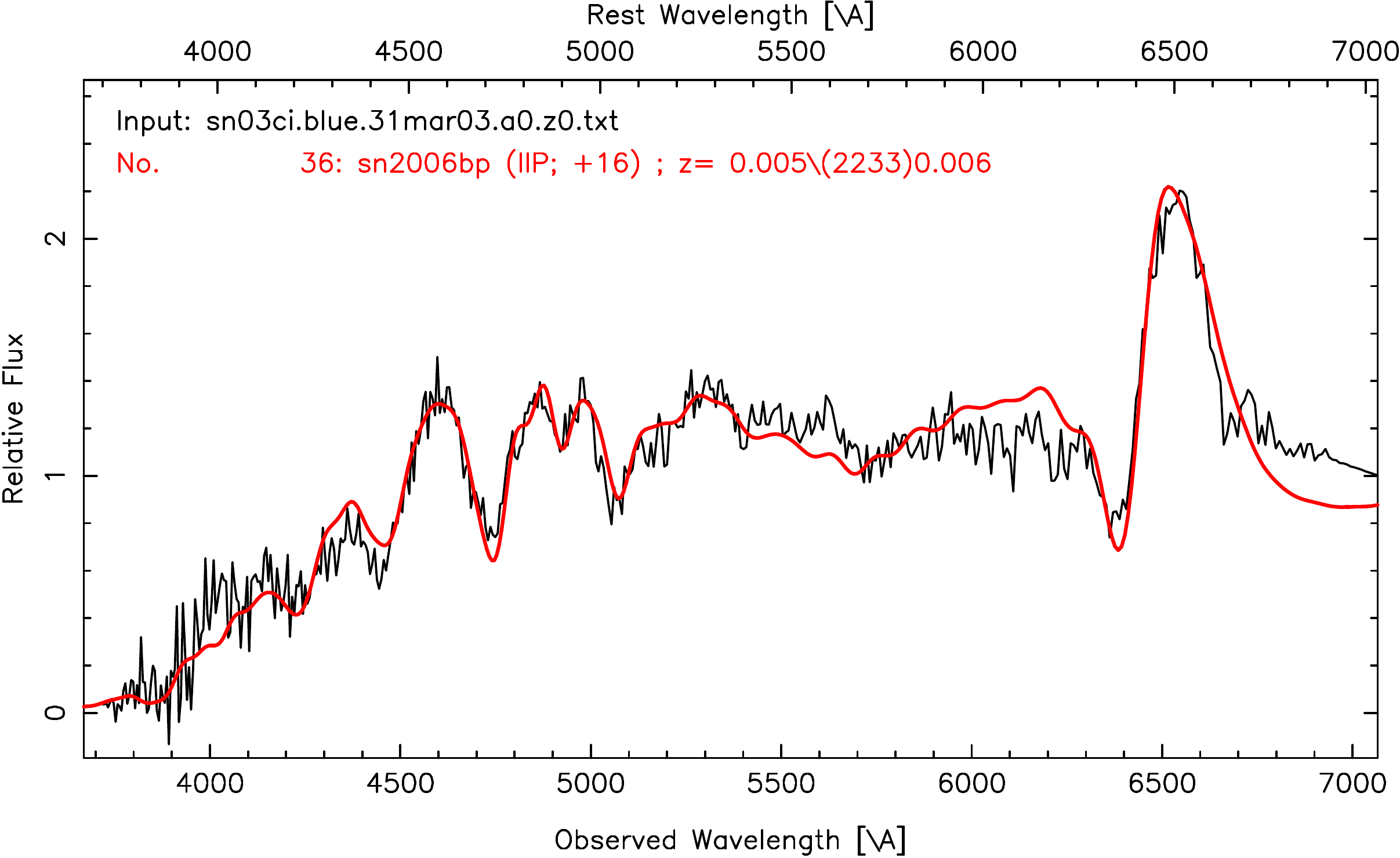}
\caption{Best spectral matching of SN~2003ci using SNID. The plots show SN~2003ci compared with 
SN~2004et, SN~1999em, and SN~2006bp at 31, 21, and 25 days from explosion.}
\end{figure}

\begin{figure}[h!]
\centering
\includegraphics[width=4.4cm]{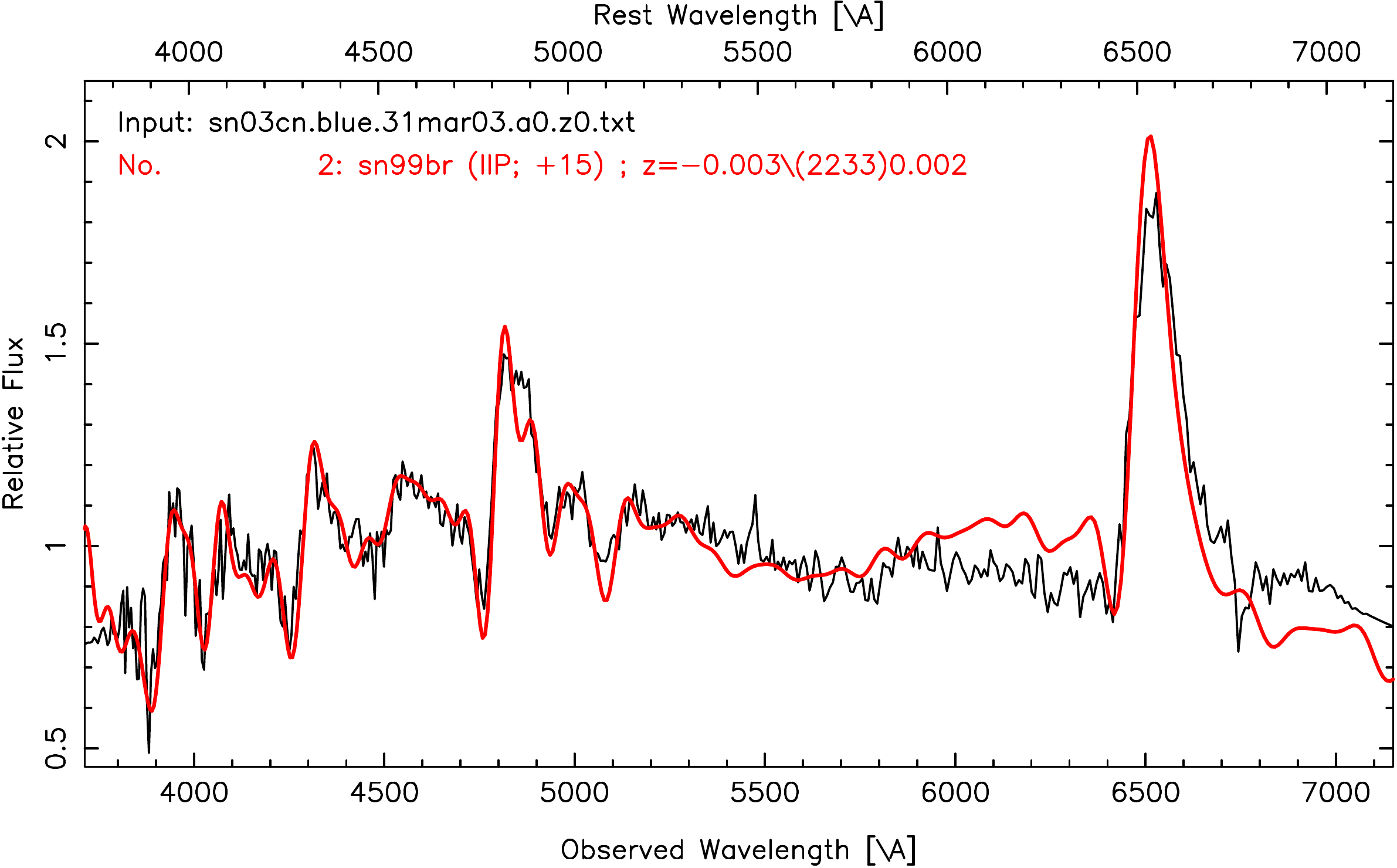}
\includegraphics[width=4.4cm]{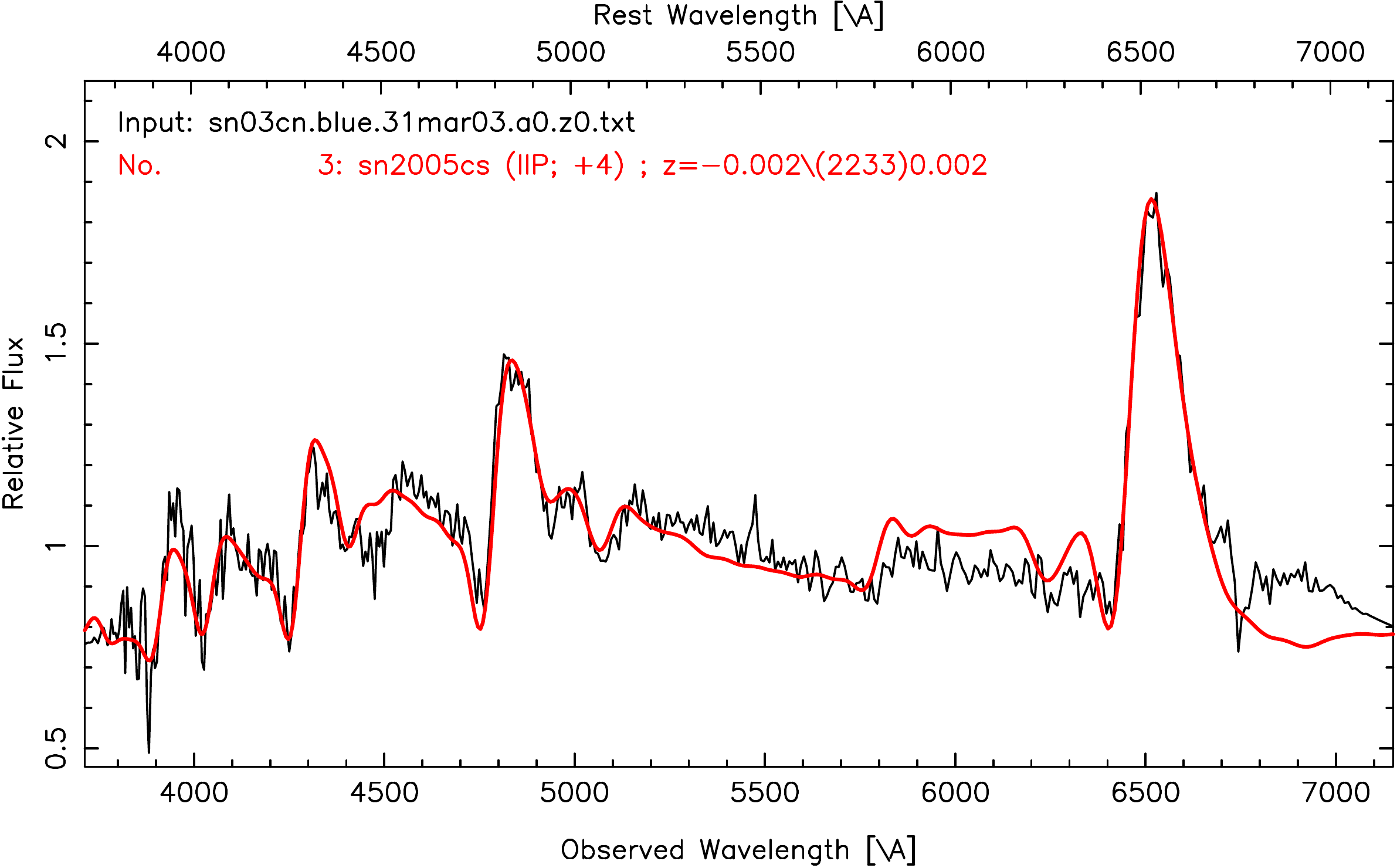} 
\caption{Best spectral matching of SN~2003cn using SNID. The plots show SN~2003cn compared with 
SN~1999br and SN~2005cs at 15 and 10 days from explosion.}
\end{figure}

\clearpage

\begin{figure}[h!]
\centering
\includegraphics[width=4.4cm]{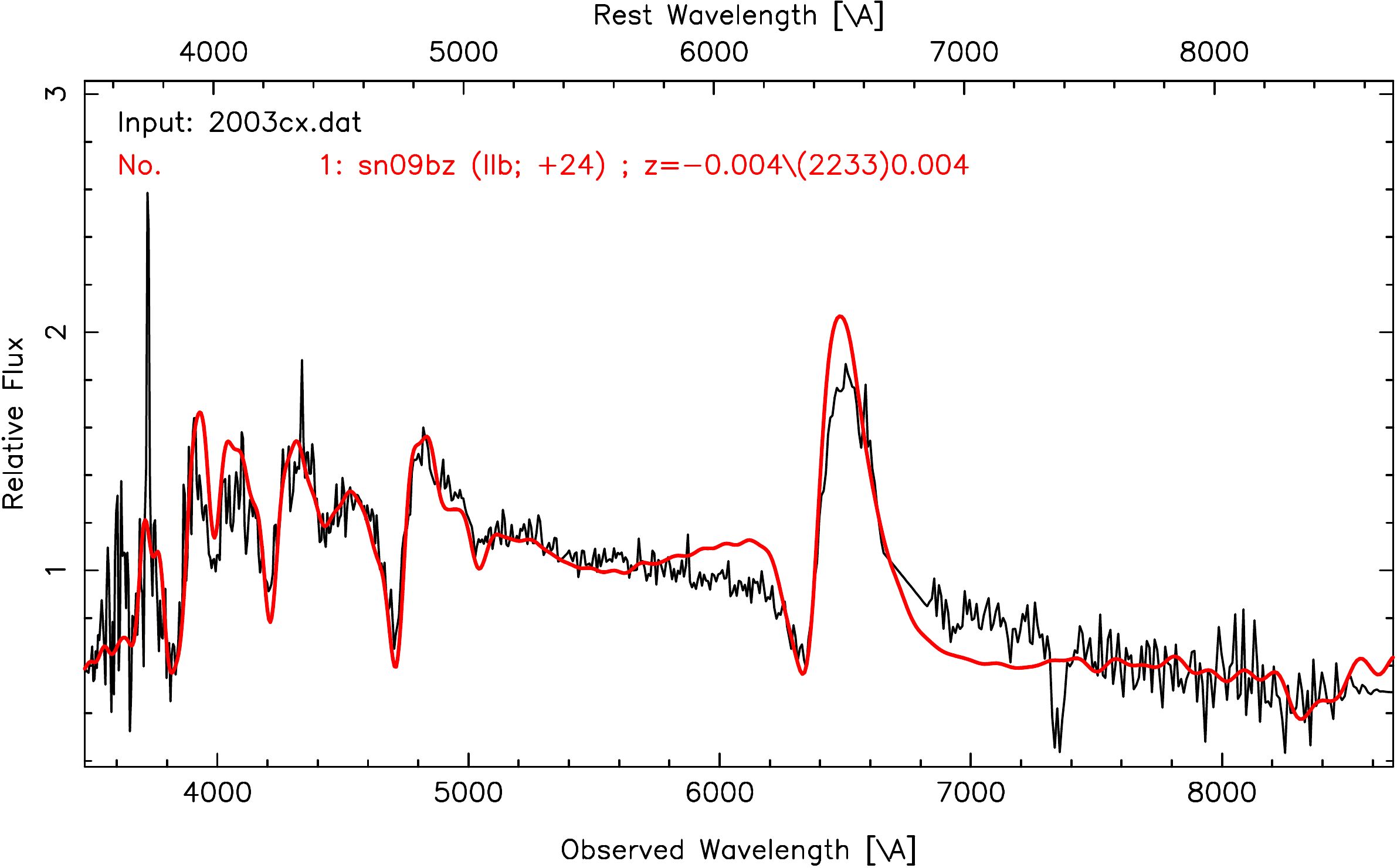}
\includegraphics[width=4.4cm]{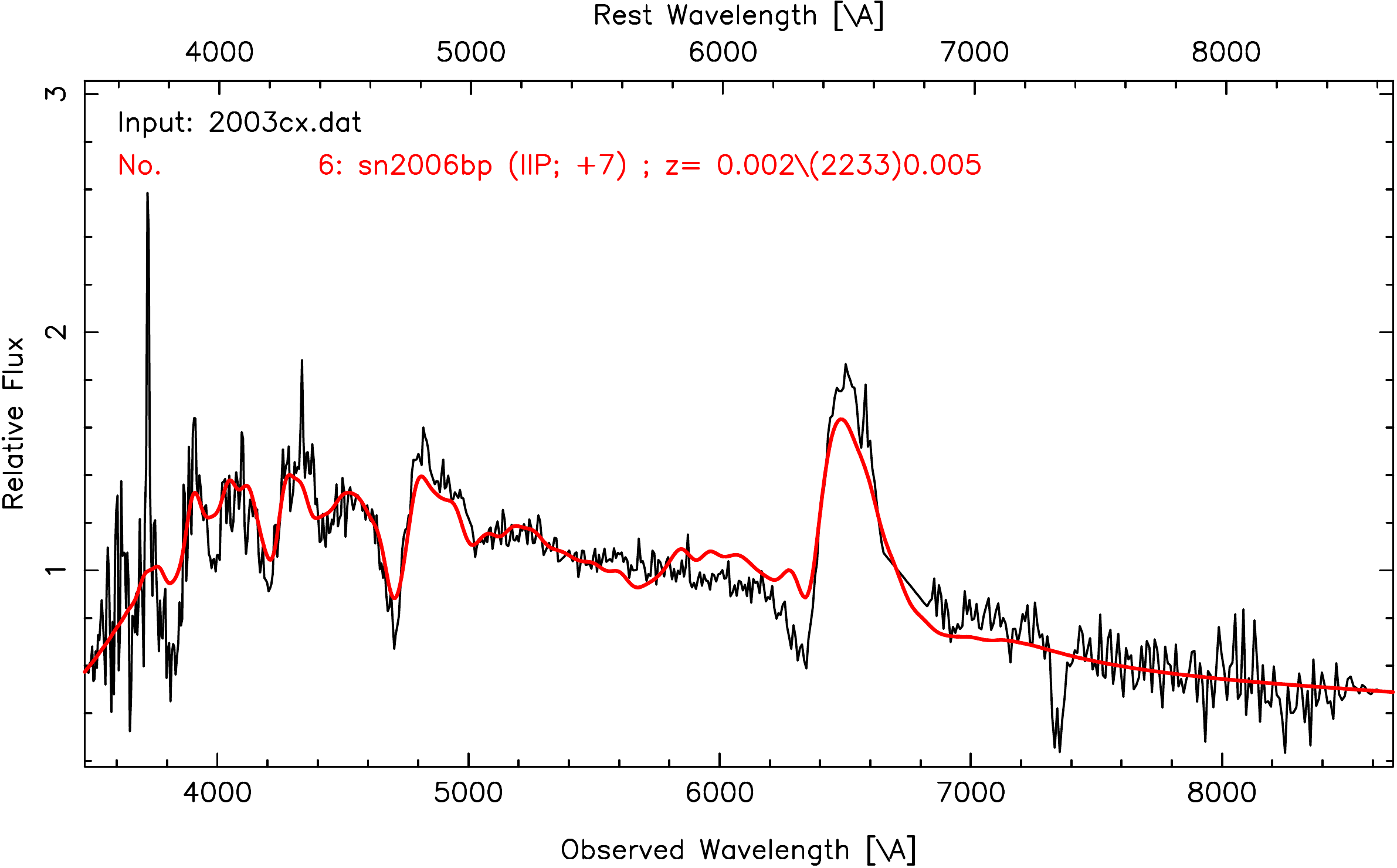}
\includegraphics[width=4.4cm]{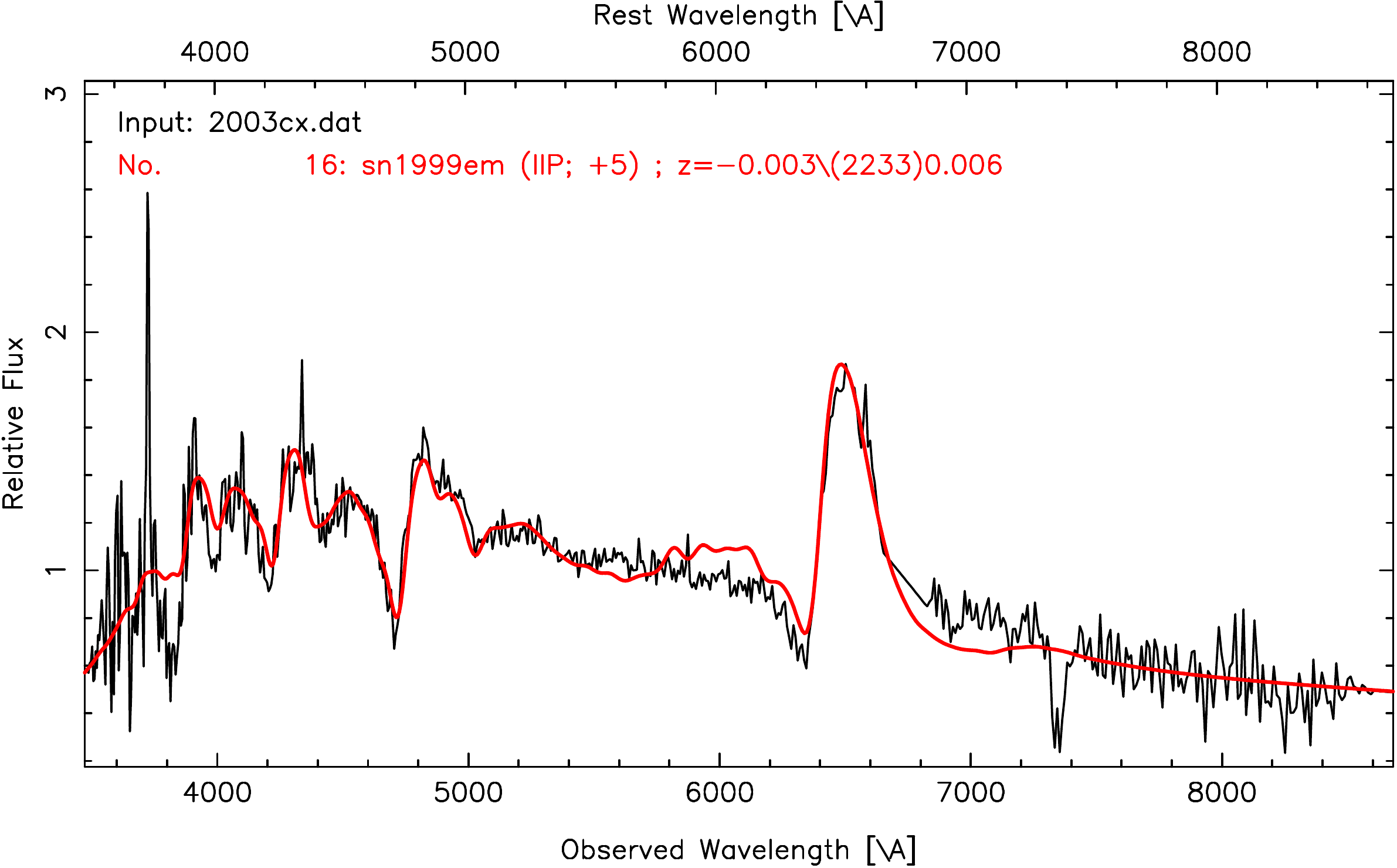}
\includegraphics[width=4.4cm]{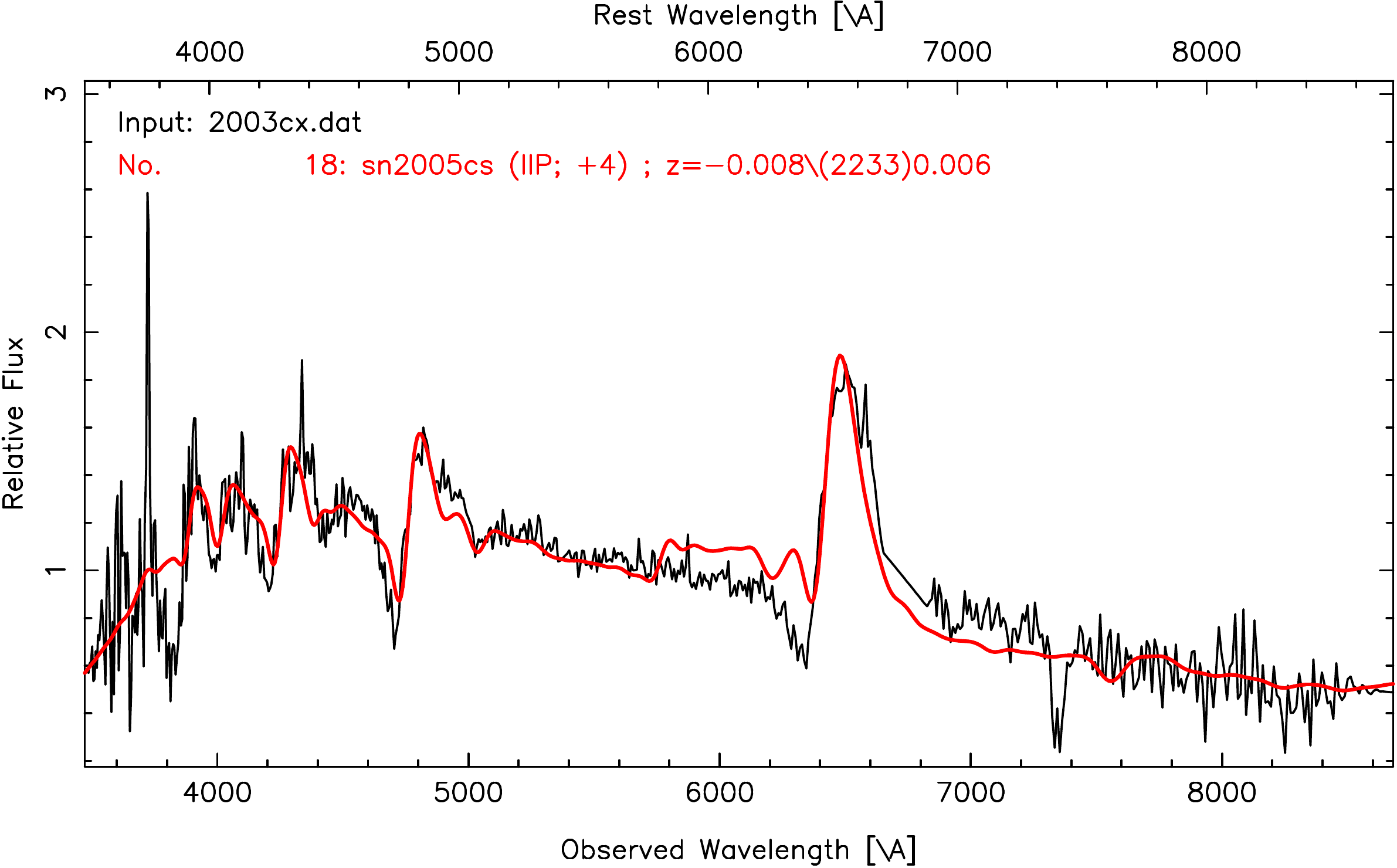}
\caption{Best spectral matching of SN~2003cx using SNID. The plots show SN~2003cx compared with 
SN~2009bz, SN~2006bp, SN~1999em, and SN~2005cs at 24, 16, 15, and 10 days from explosion.}
\end{figure}

\begin{figure}[h!]
\centering
\includegraphics[width=4.4cm]{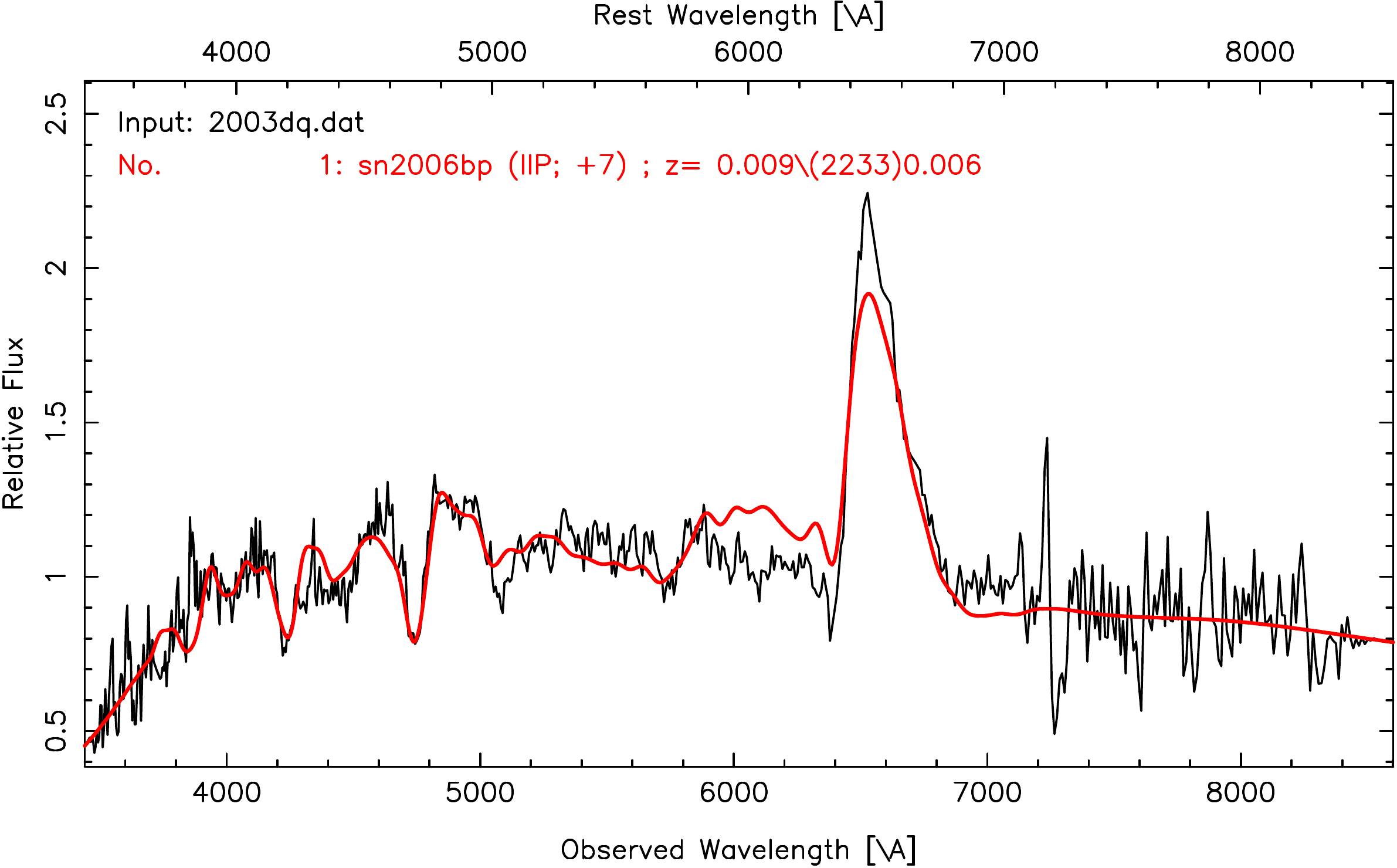}
\includegraphics[width=4.4cm]{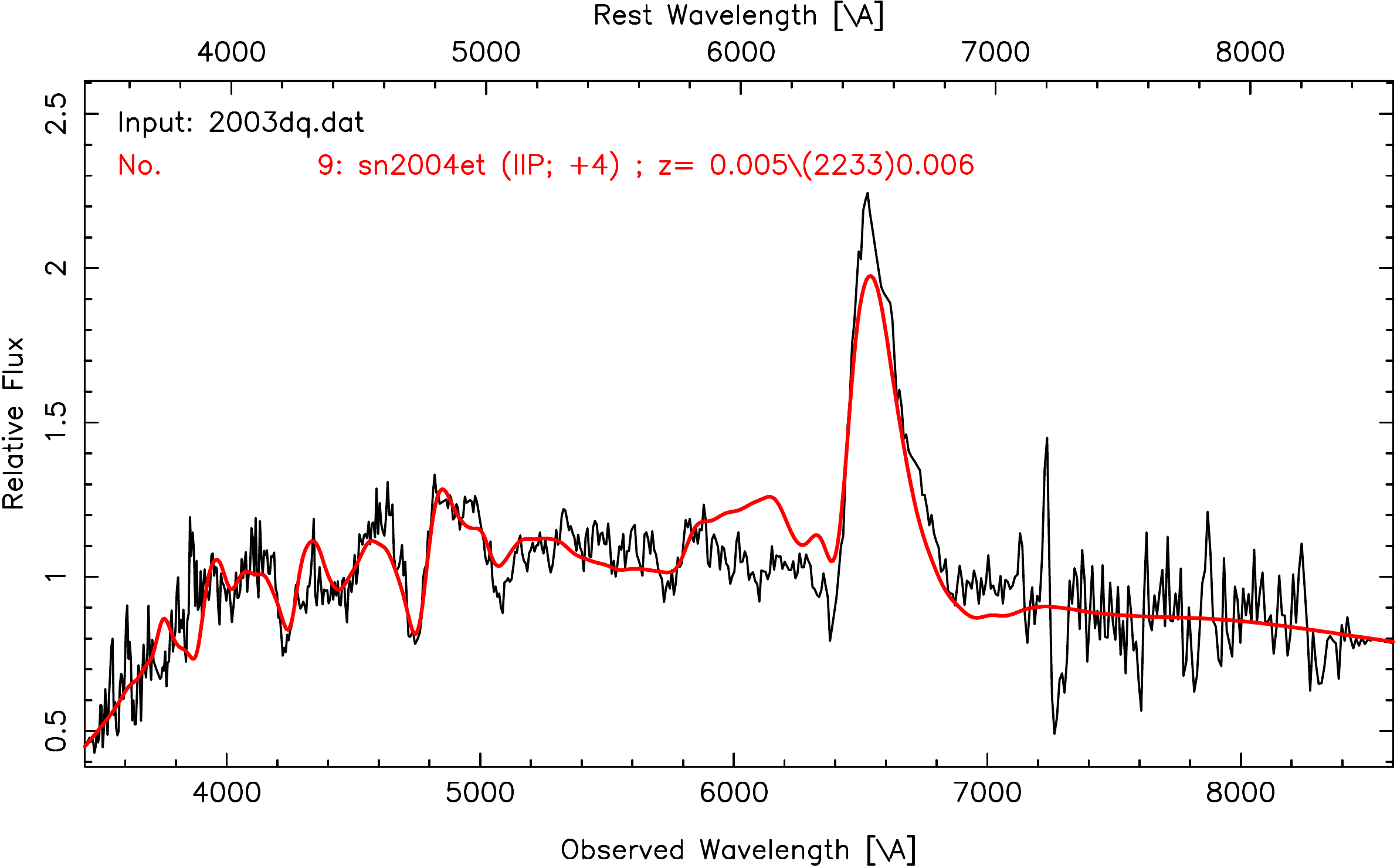}
\includegraphics[width=4.4cm]{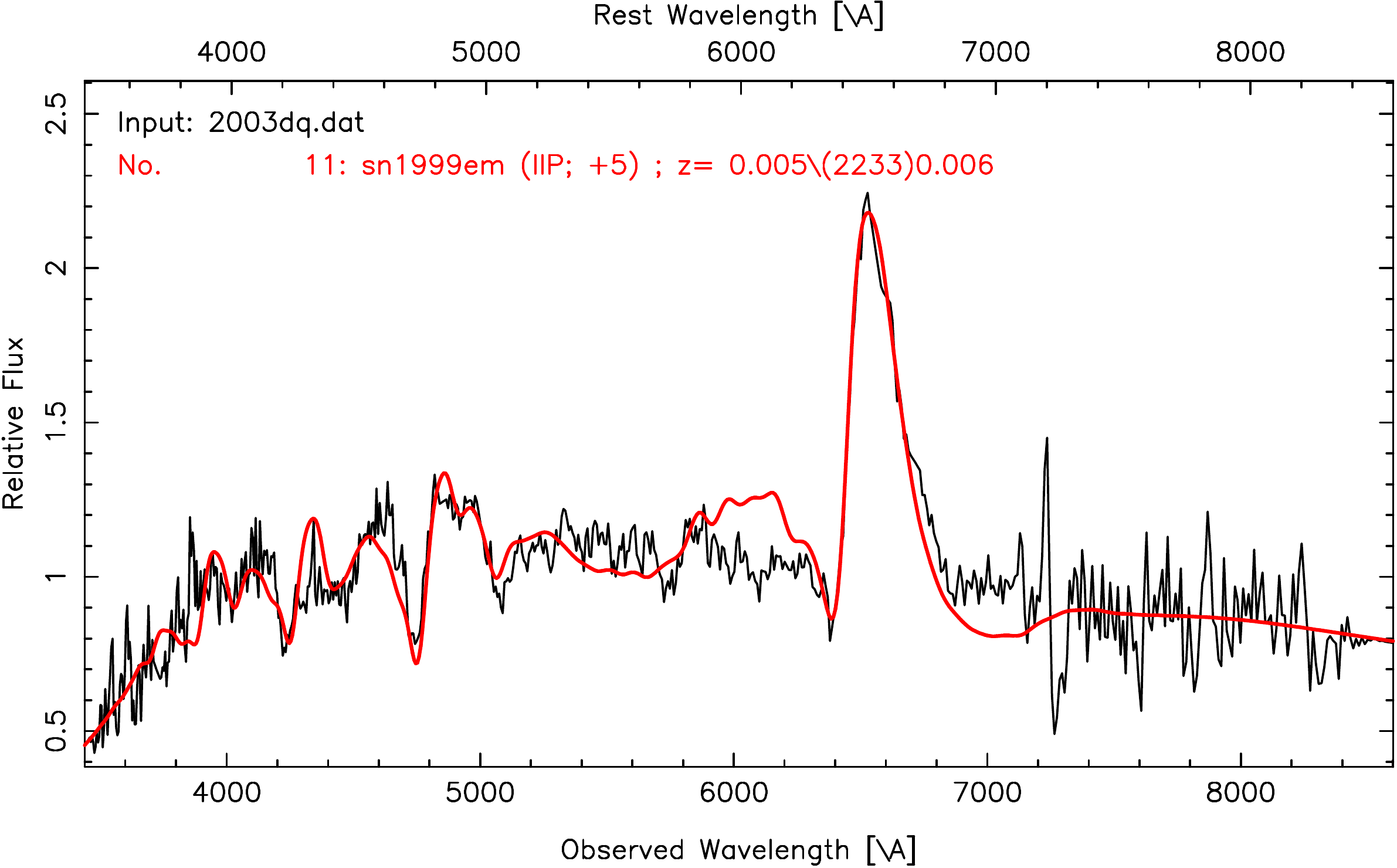}
\includegraphics[width=4.4cm]{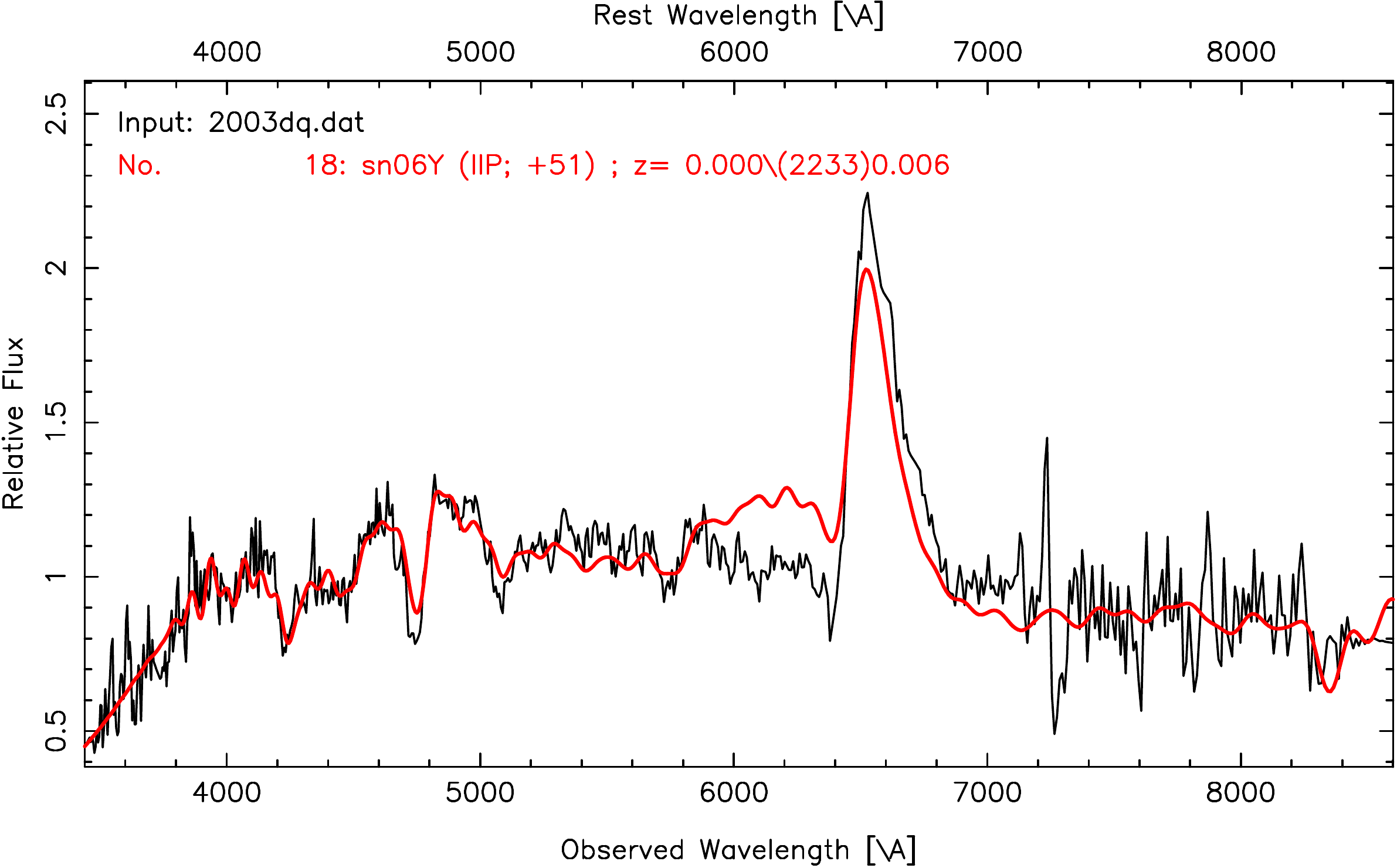}
\caption{Best spectral matching of SN~2003dq using SNID. The plots show SN~2003dq compared with 
SN~2006bp, SN~2004et, SN~1999em, and SN~2006Y at 16, 20, 15, 51 days from explosion.}
\end{figure}

\clearpage

\begin{figure}[h!]
\centering
\includegraphics[width=4.4cm]{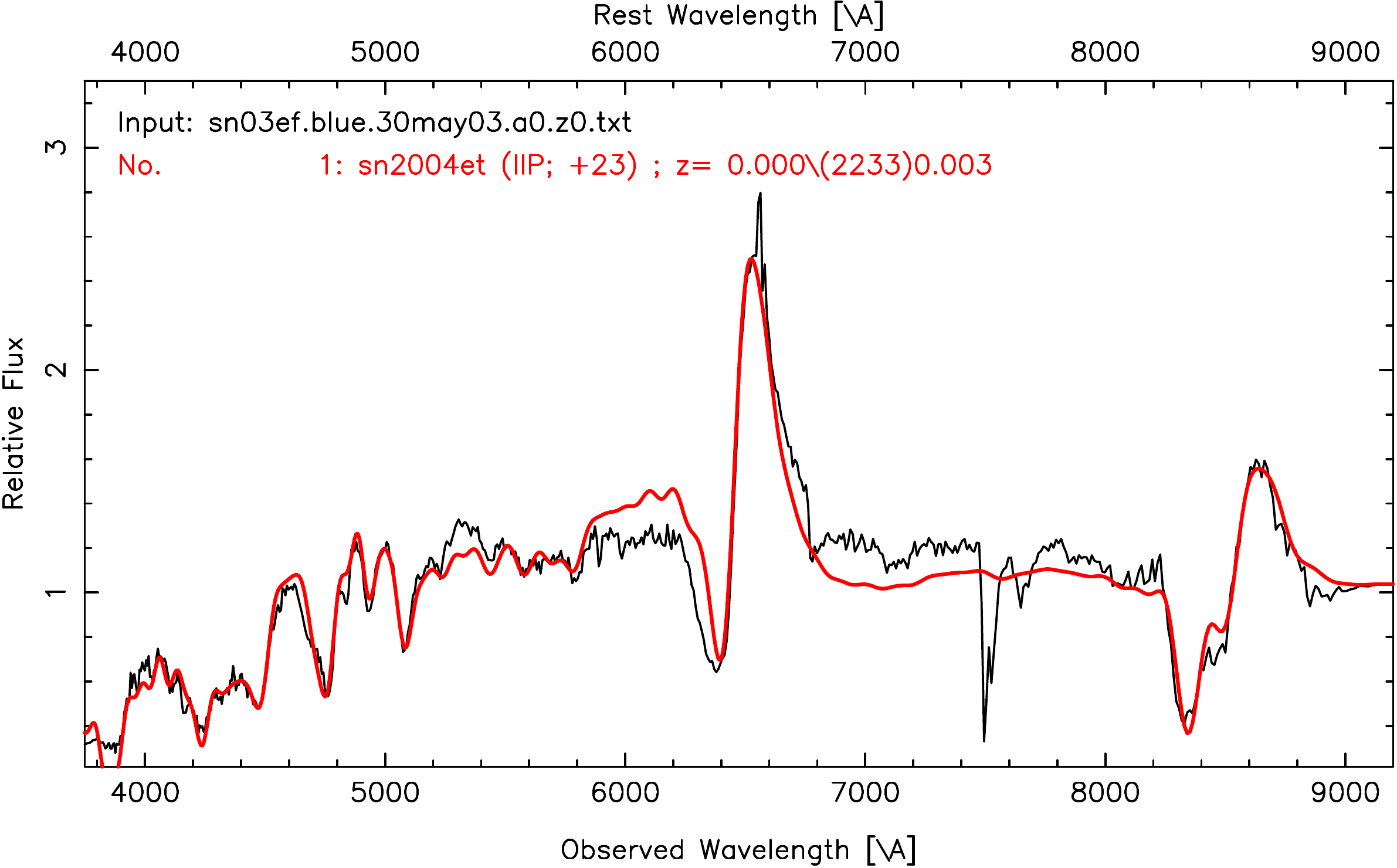}
\includegraphics[width=4.4cm]{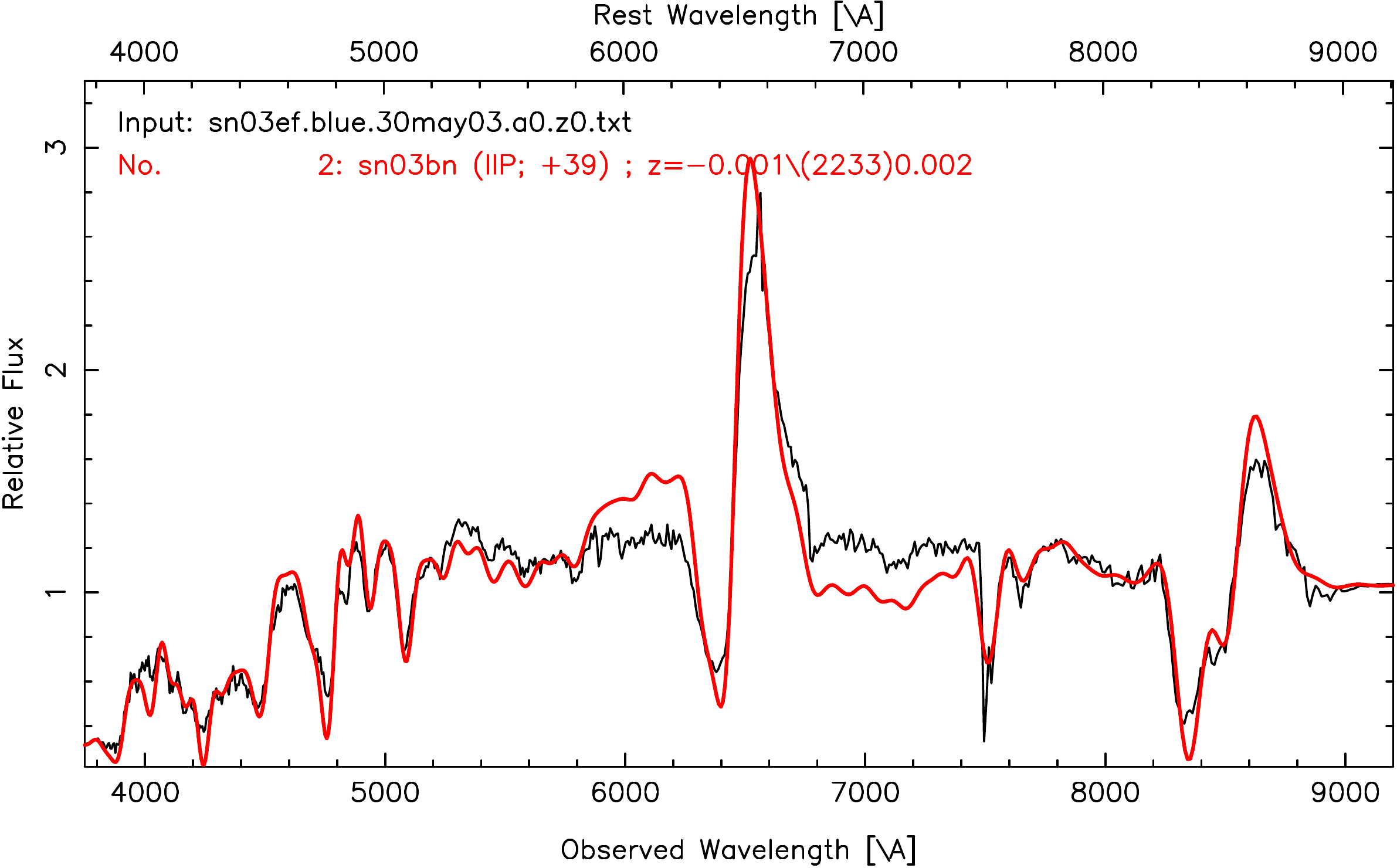}
\includegraphics[width=4.4cm]{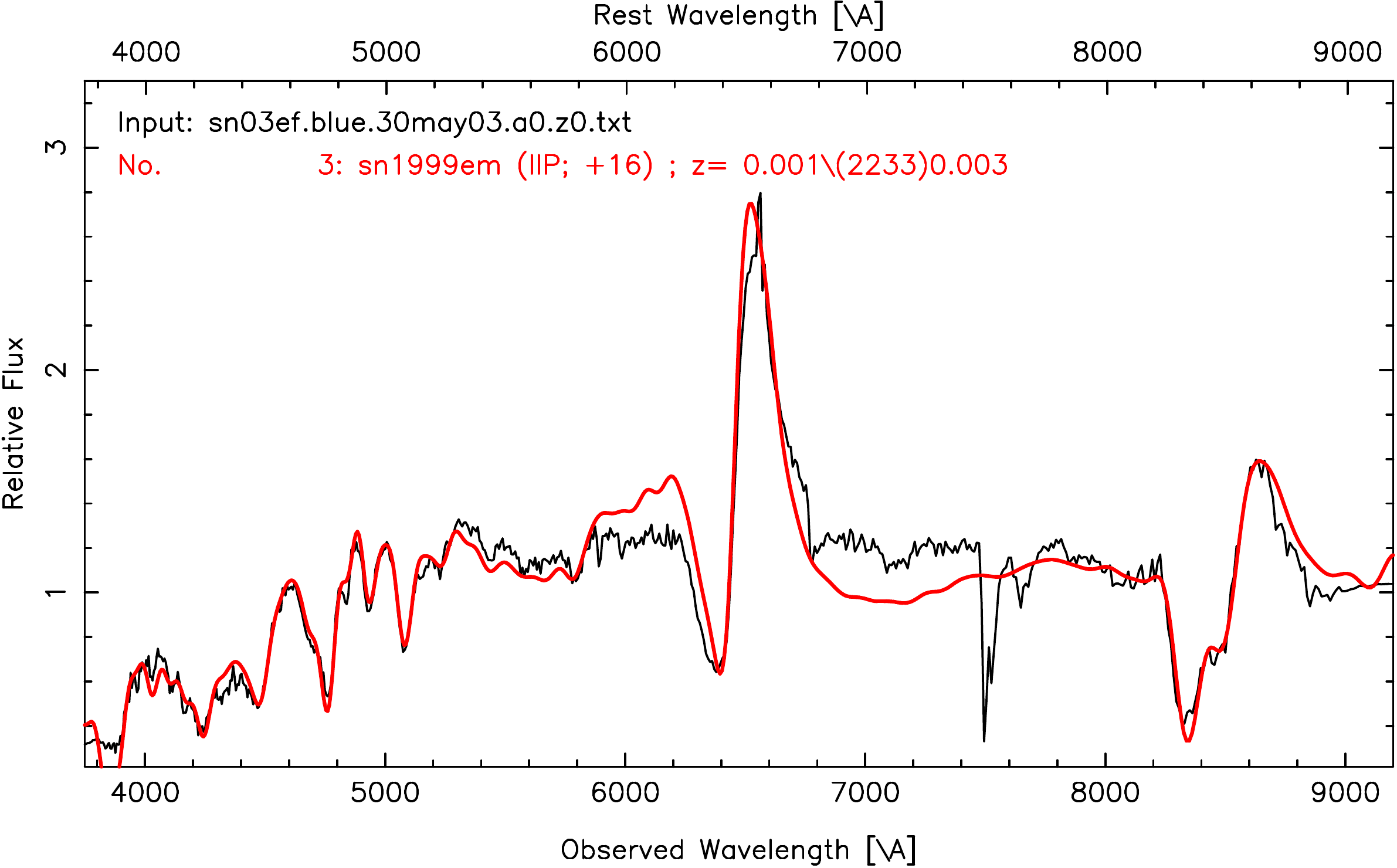}
\includegraphics[width=4.4cm]{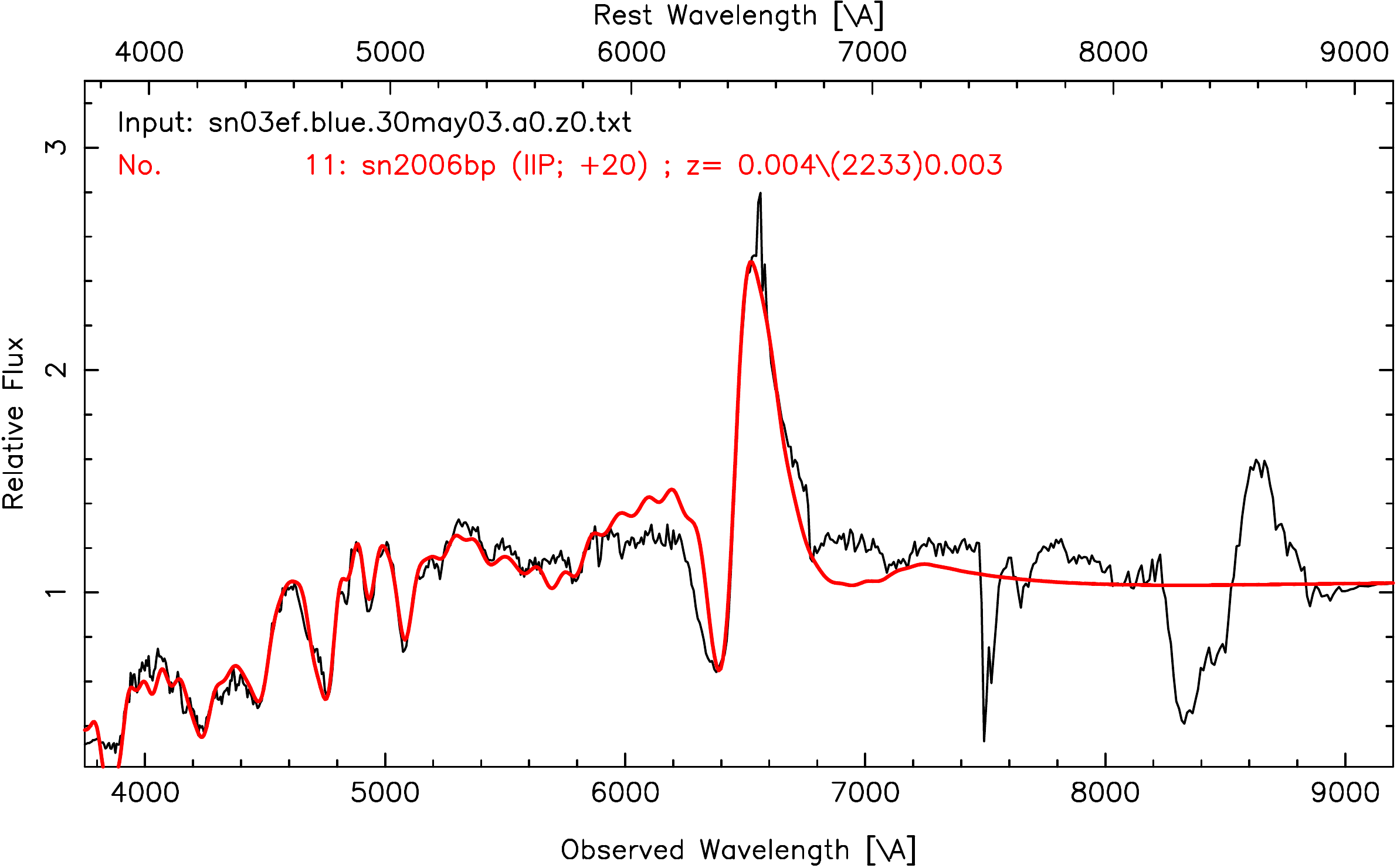}
\includegraphics[width=4.4cm]{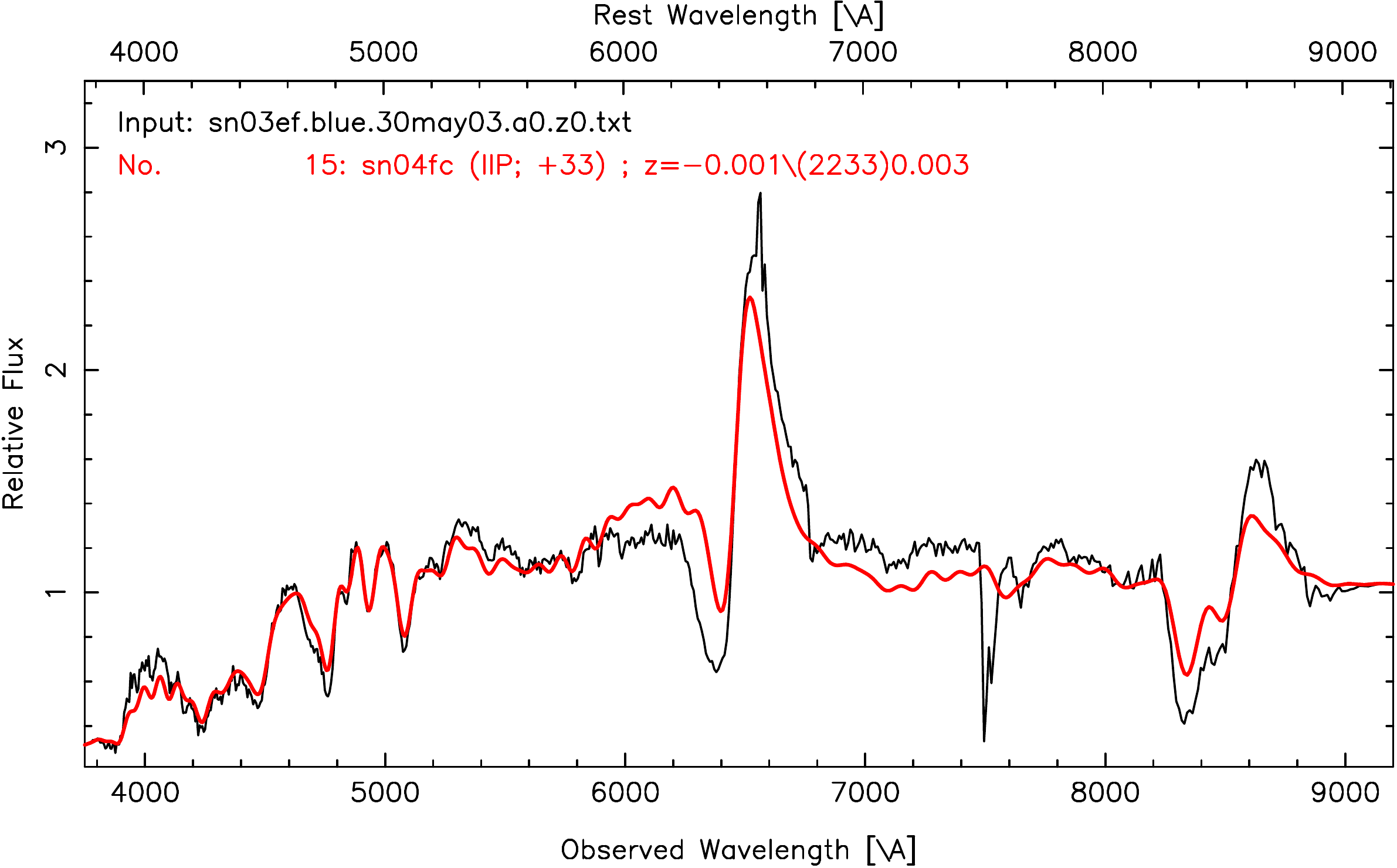}
\includegraphics[width=4.4cm]{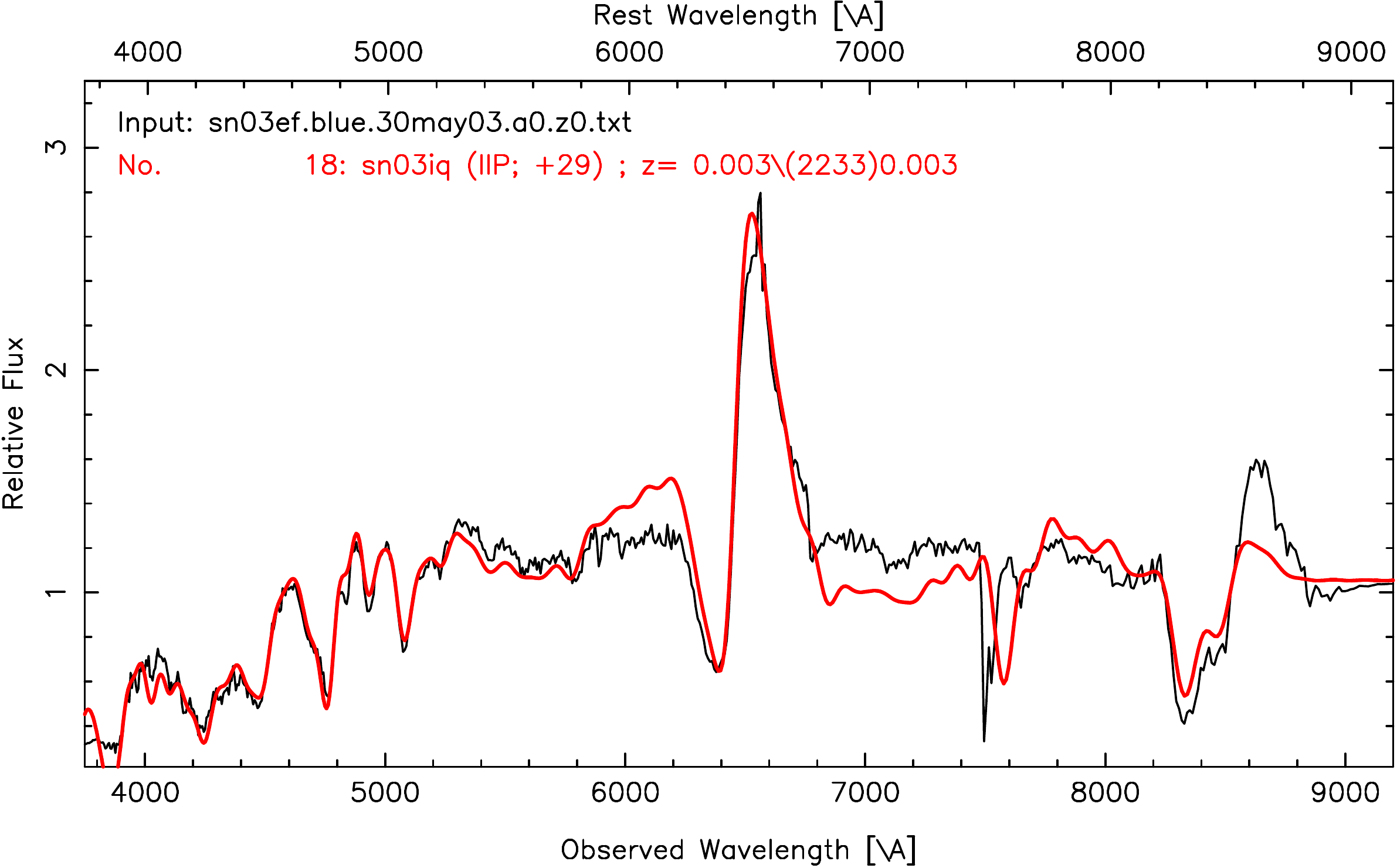}
\caption{Best spectral matching of SN~2003ef using SNID. The plots show SN~2003ef compared with 
SN~2004et, SN~2003bn, SN~1999em, SN~2006bp, SN~2004fc, and SN~2003iq at 39, 39, 26, 29, 33 days from explosion.}
\end{figure}

\begin{figure}[h!]
\centering
\includegraphics[width=4.4cm]{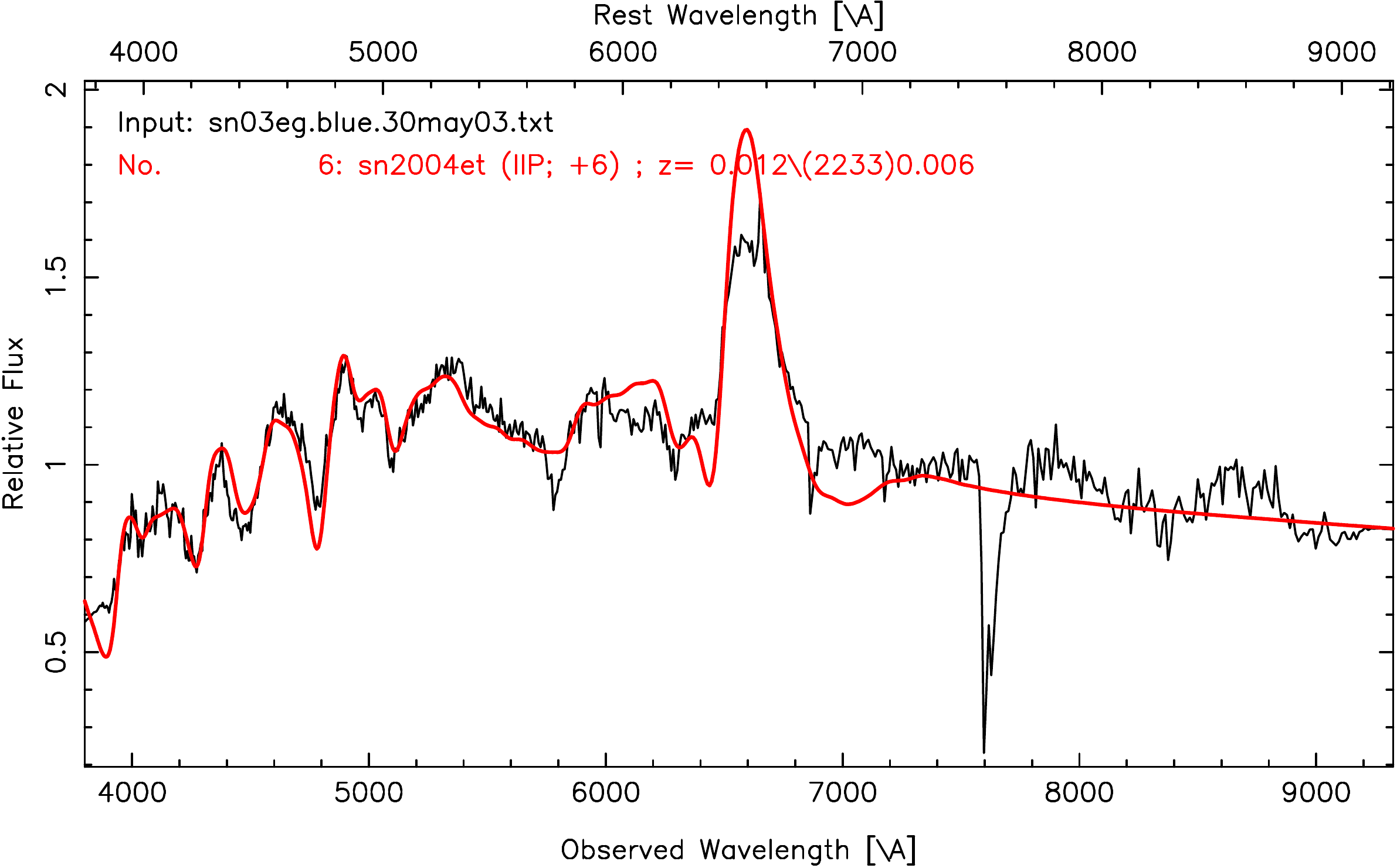}
\includegraphics[width=4.4cm]{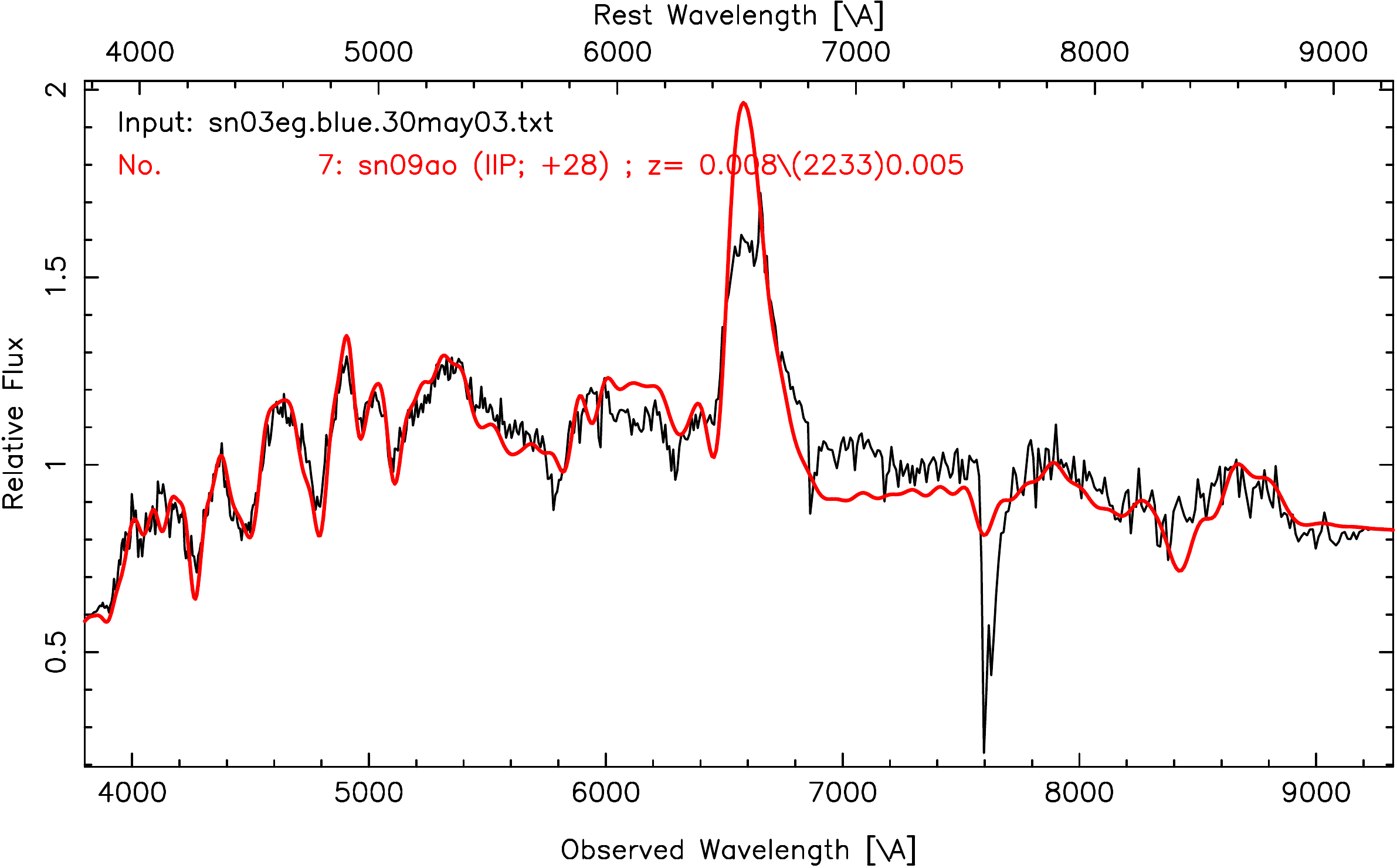}
\caption{Best spectral matching of SN~2003eg using SNID. The plots show SN~2003eg compared with 
SN~2004et, and SN~2009ao at 22 and 28 days from explosion.}
\end{figure}

\begin{figure}[h!]
\centering
\includegraphics[width=4.4cm]{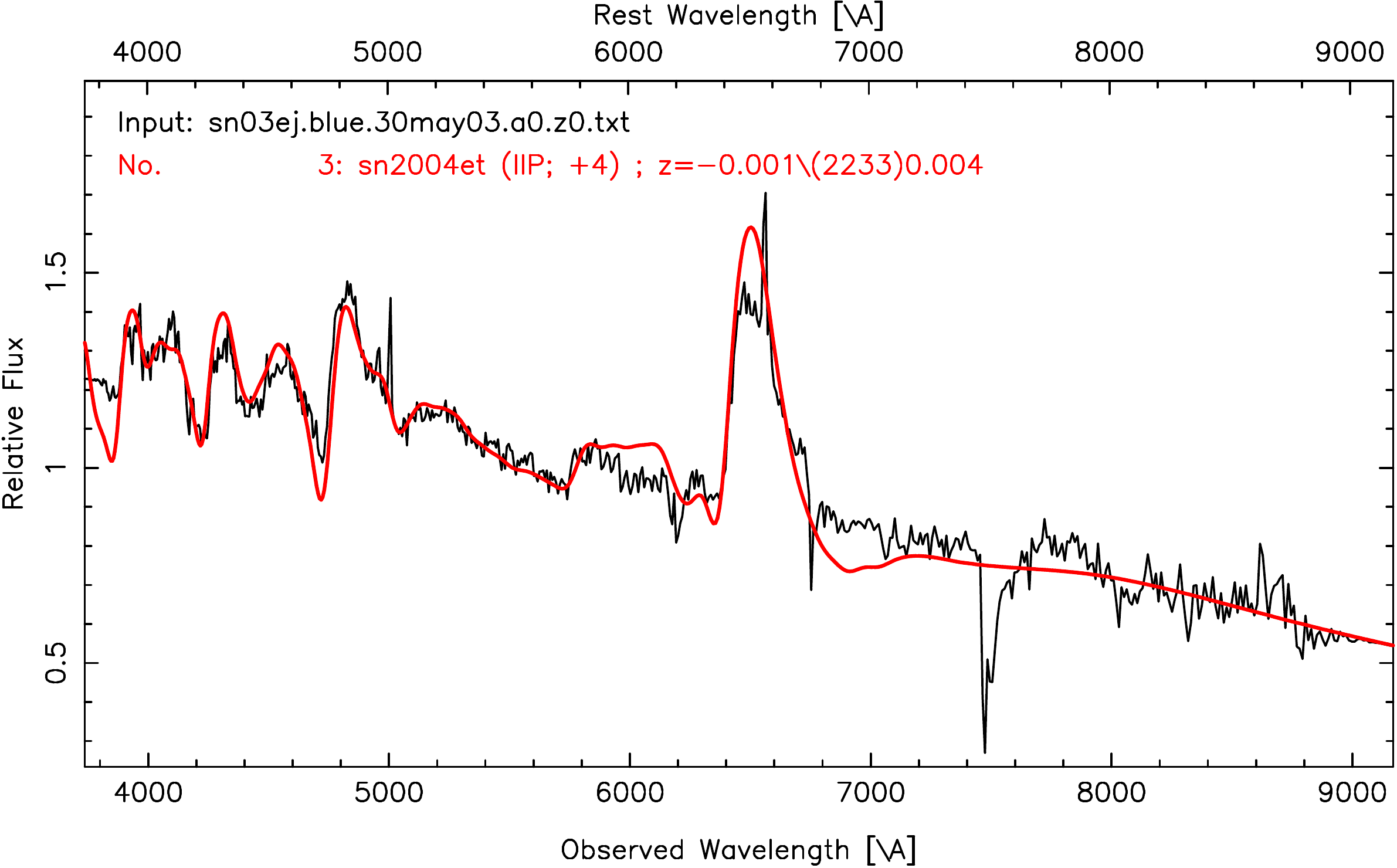}
\includegraphics[width=4.4cm]{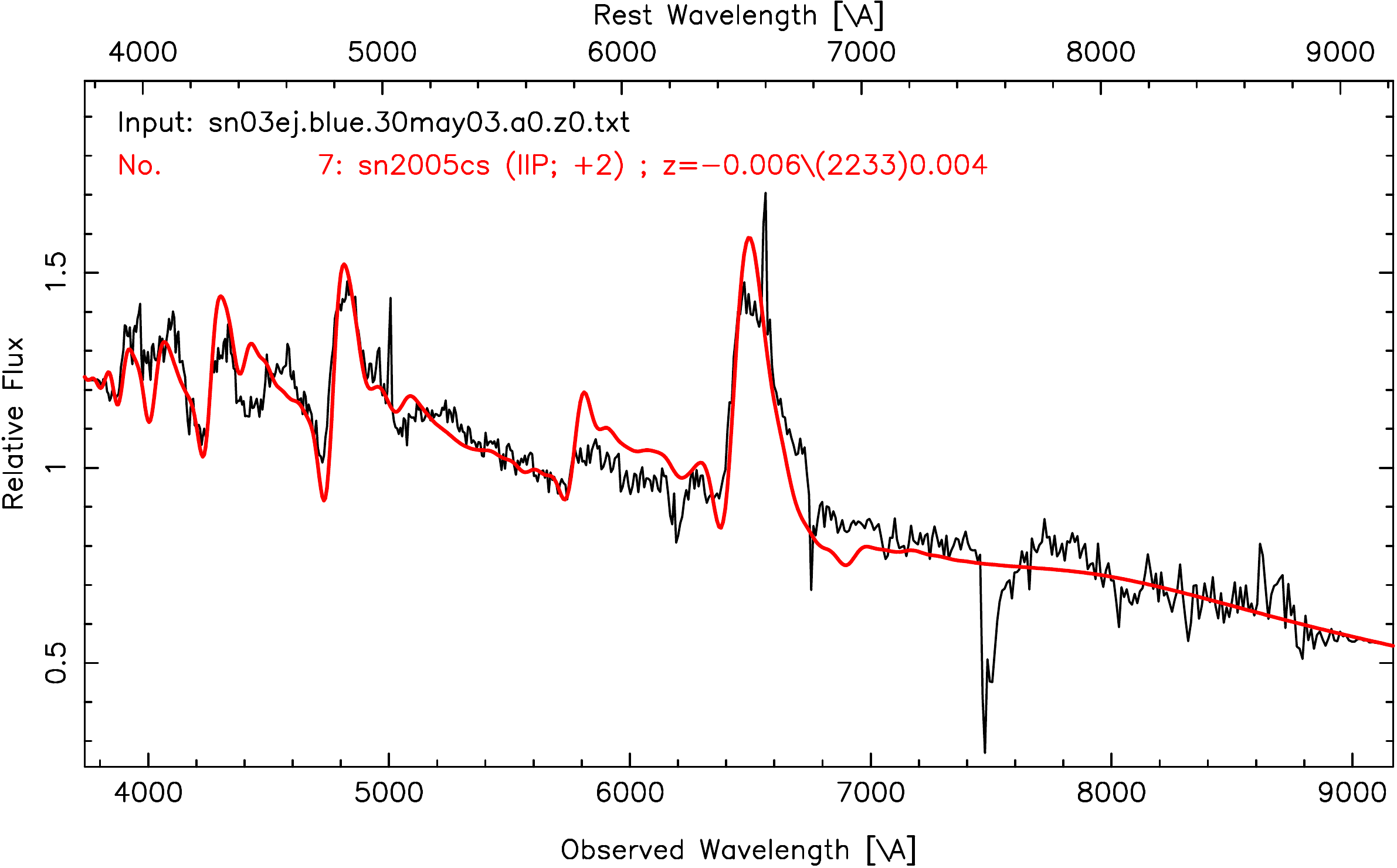}
\includegraphics[width=4.4cm]{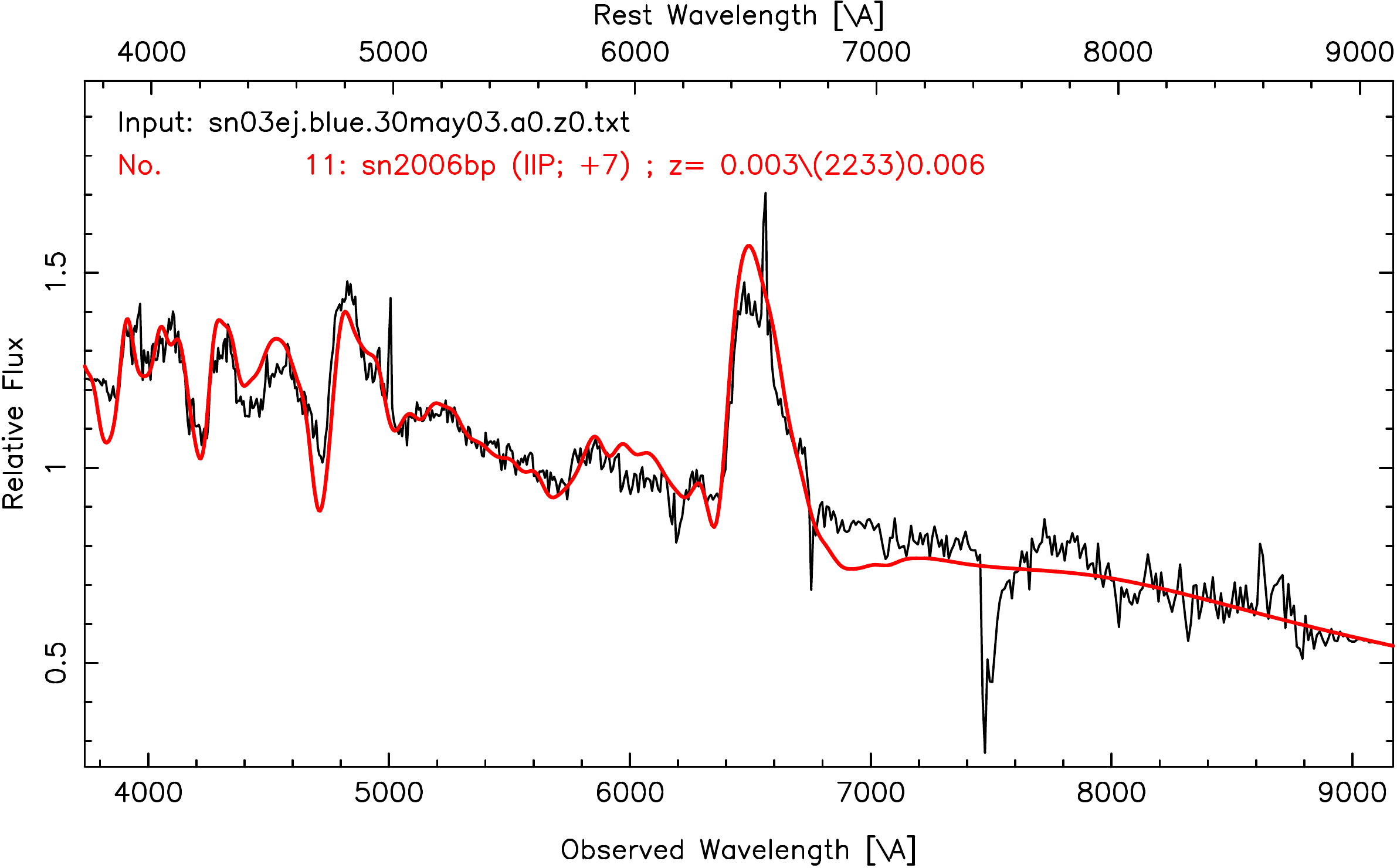}
\caption{Best spectral matching of SN~2003ej using SNID. The plots show SN~2003ej compared with 
SN~2004et, SN~2005cs, and SN~2006bp at 20, 8, and 16 days from explosion.}
\end{figure}

\clearpage

\begin{figure}[h!]
\centering
\includegraphics[width=4.4cm]{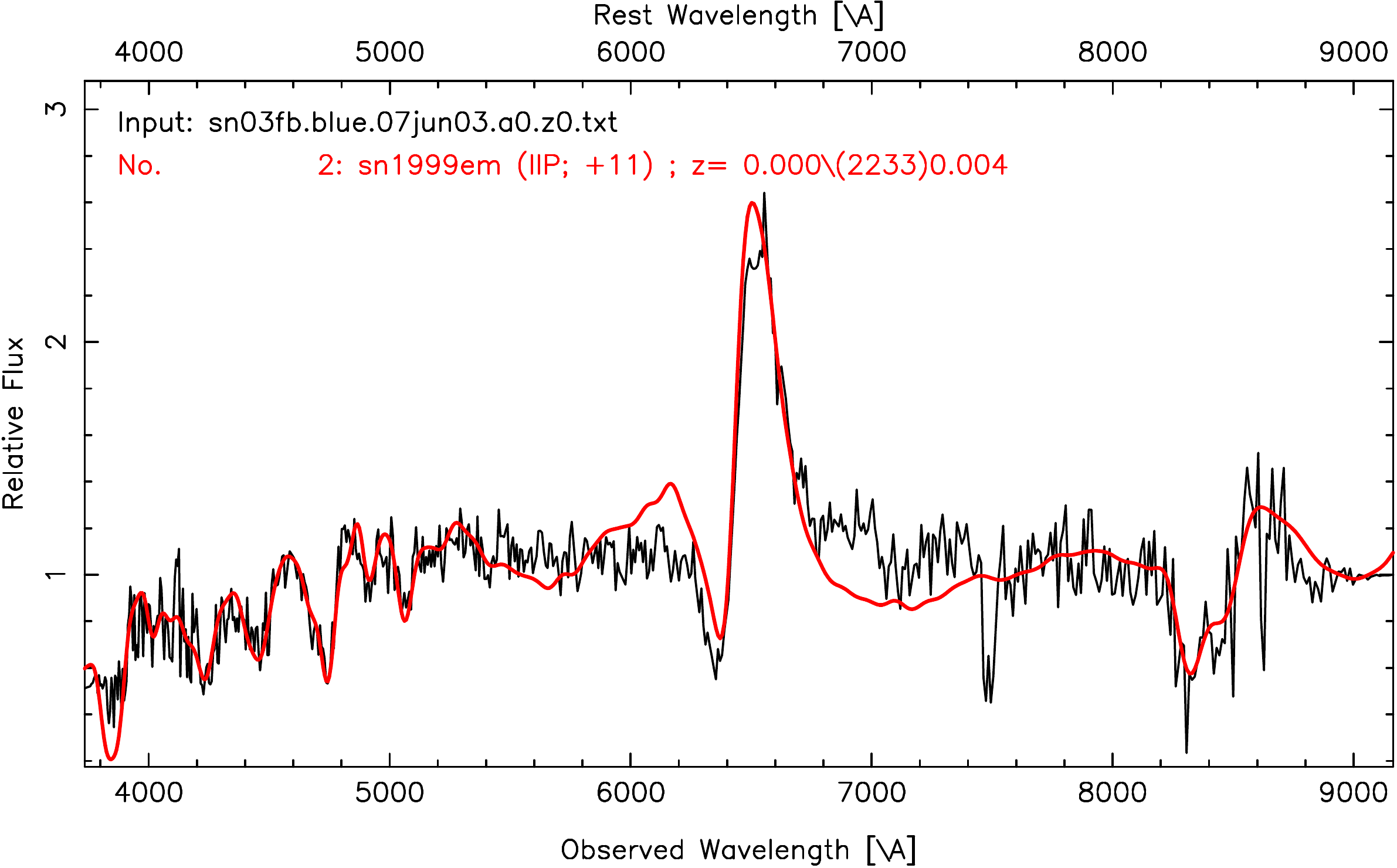}
\includegraphics[width=4.4cm]{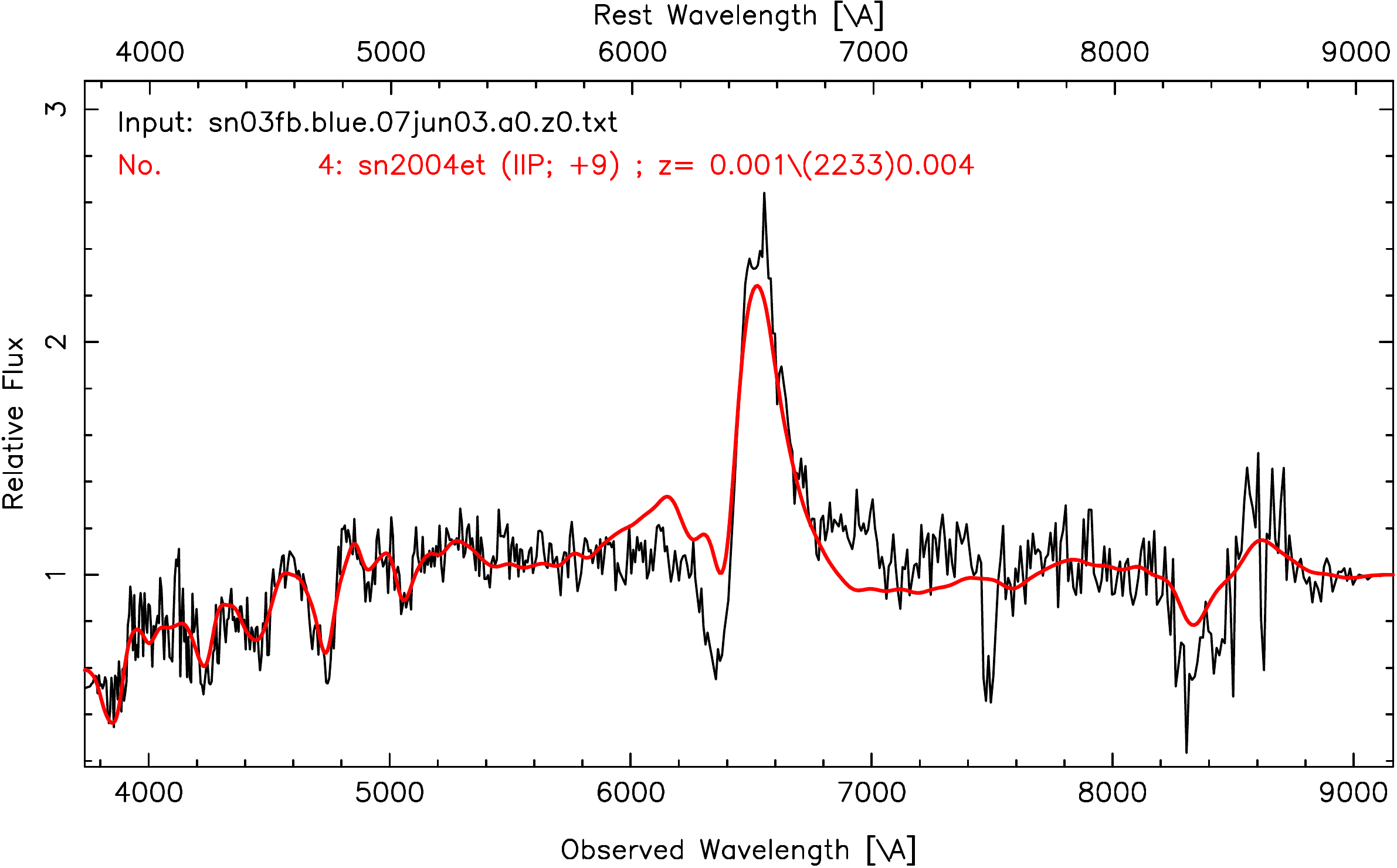}
\includegraphics[width=4.4cm]{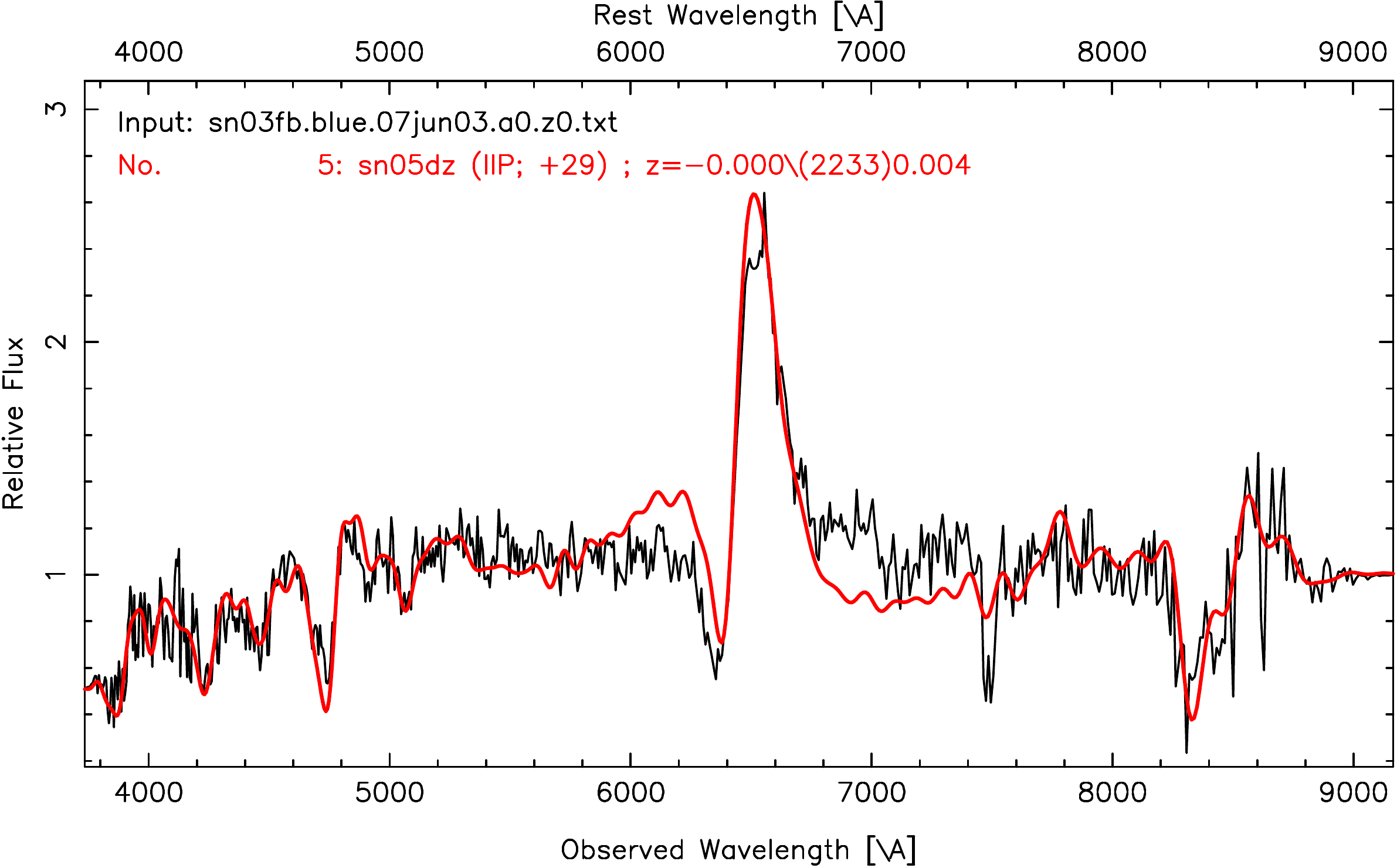}
\includegraphics[width=4.4cm]{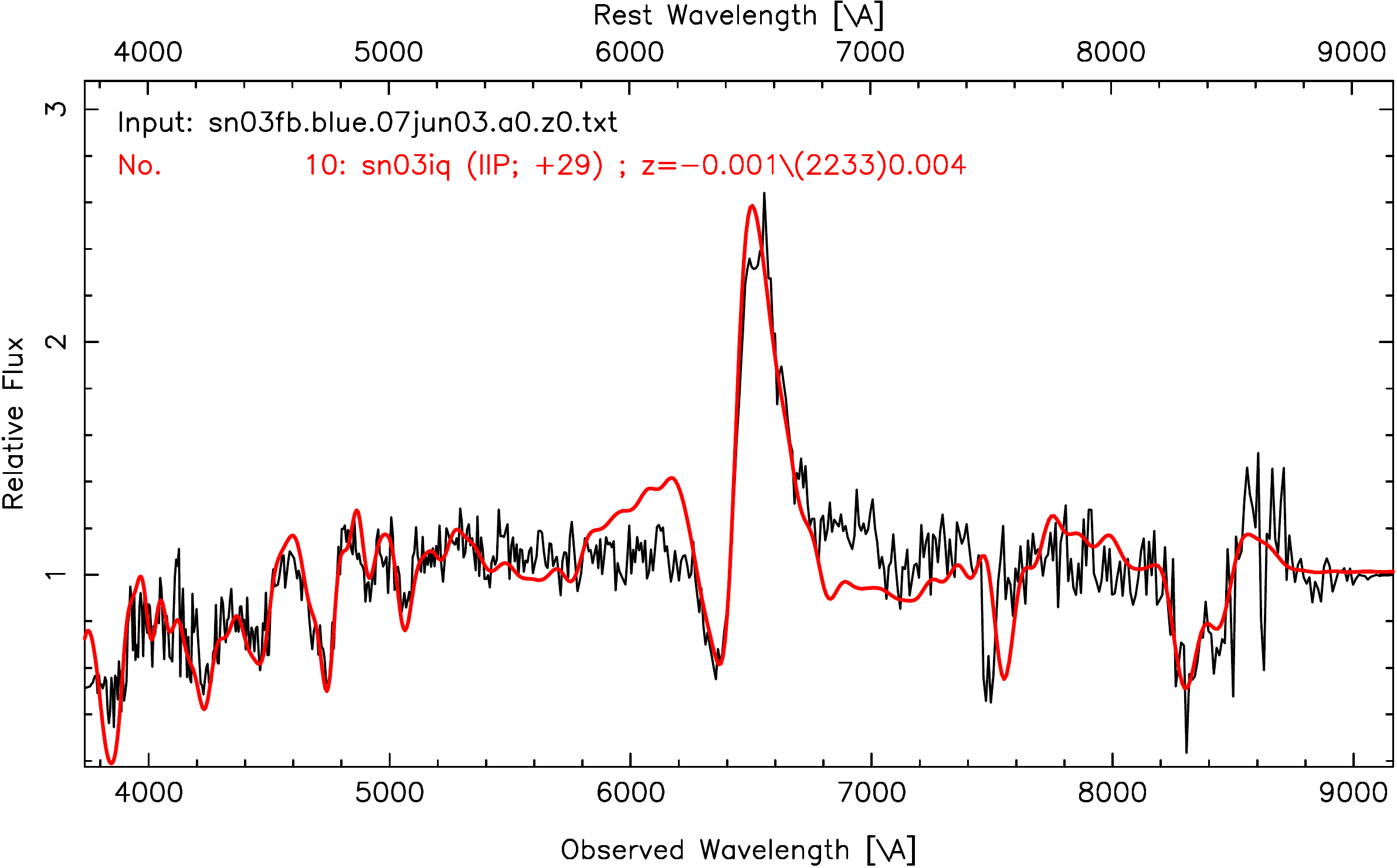}
\includegraphics[width=4.4cm]{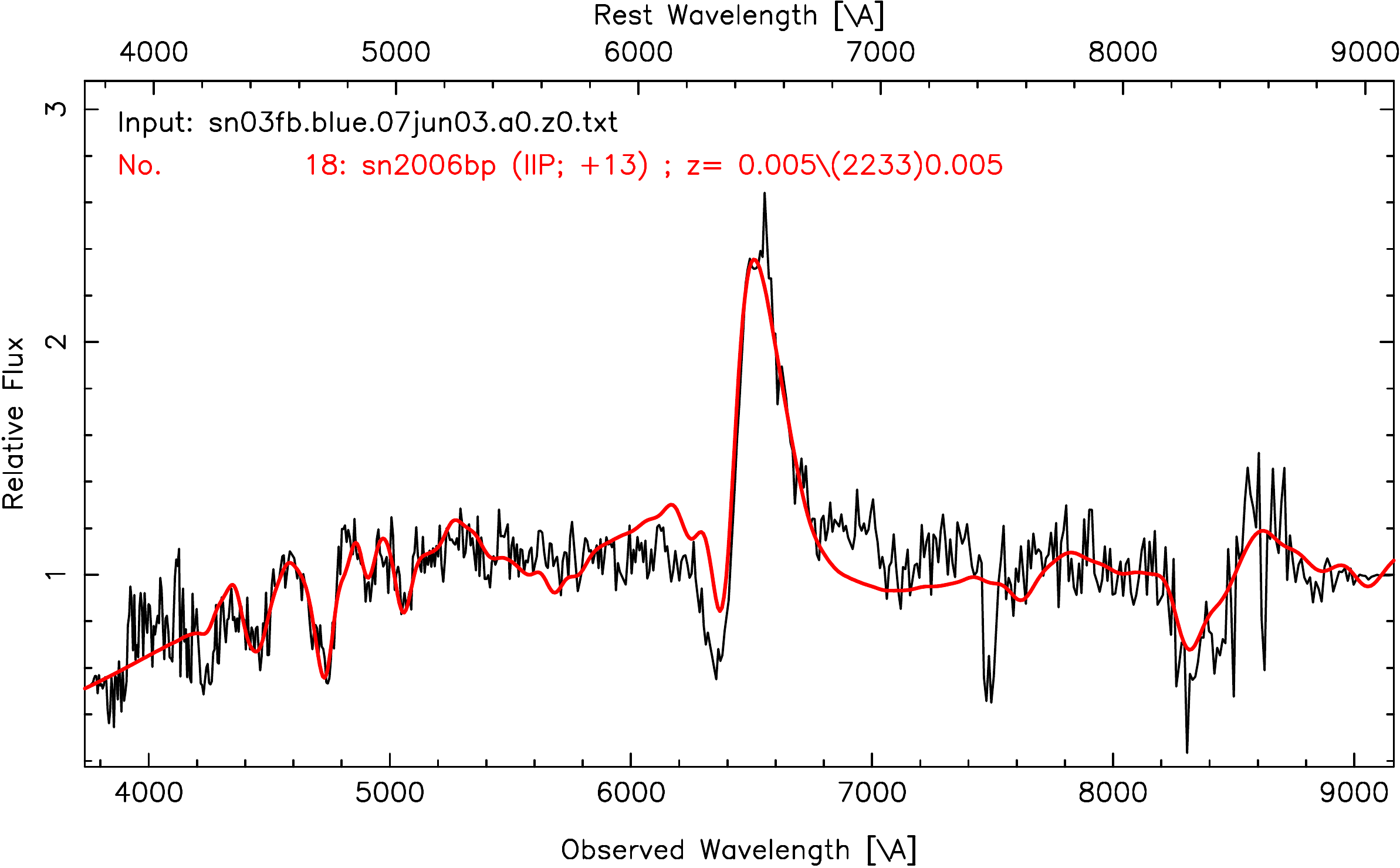}
\caption{Best spectral matching of SN~2003fb using SNID. The plots show SN~2003fb compared with 
SN~1999em, SN~2004et, SN~2005dz, SN~2003iq, and SN~2006bp at 21, 25, 29, 29, and 22 days from explosion.}
\end{figure}

\begin{figure}[h!]
\centering
\includegraphics[width=4.4cm]{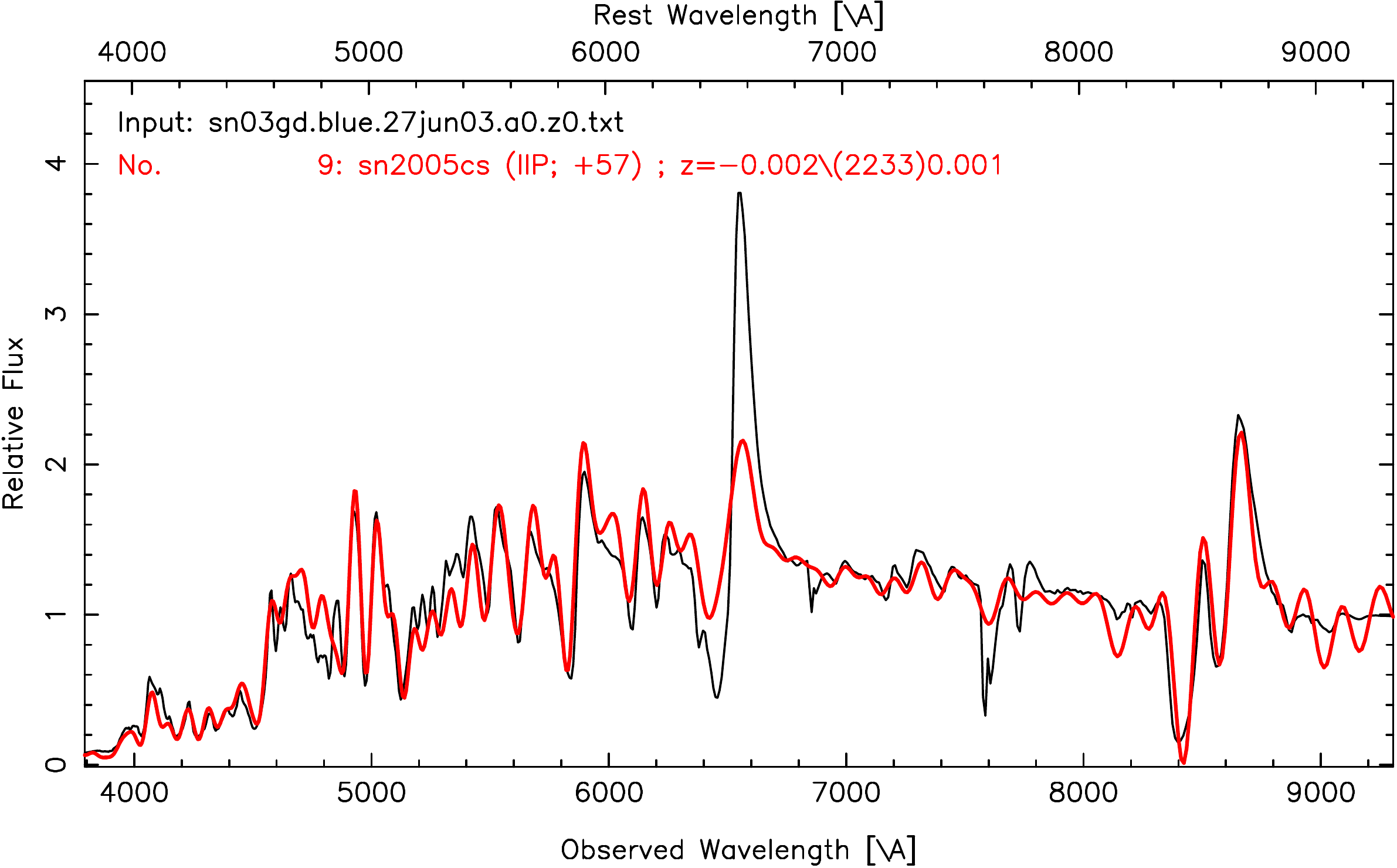}
\includegraphics[width=4.4cm]{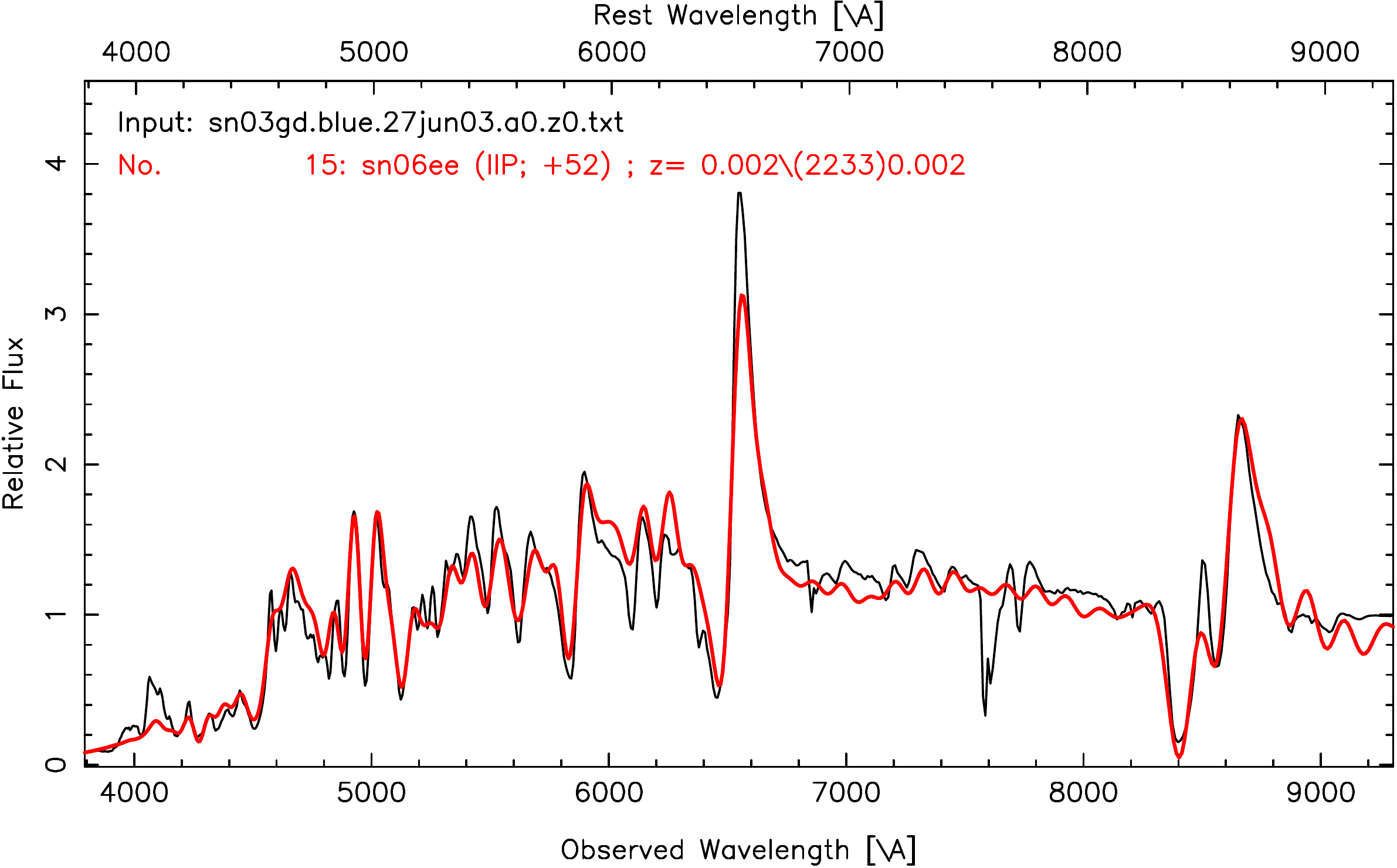}
\includegraphics[width=4.4cm]{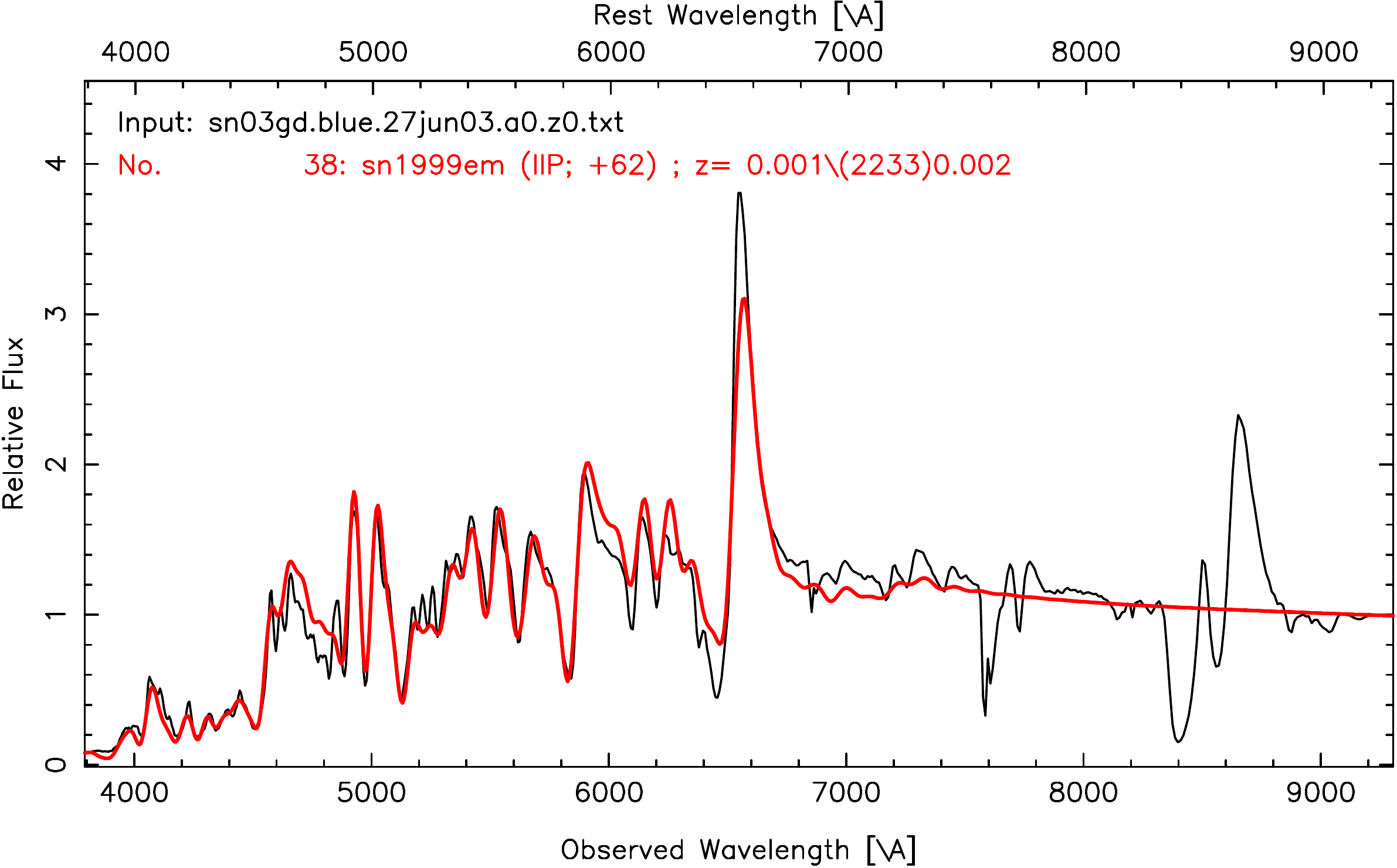}
\caption{Best spectral matching of SN~2003gd using SNID. The plots show SN~2003gd compared with 
SN~2005cs, SN~2006ee, and SN~1999em at 63, 52, and 72 days from explosion.}
\end{figure}

\begin{figure}[h!]
\centering
\includegraphics[width=4.4cm]{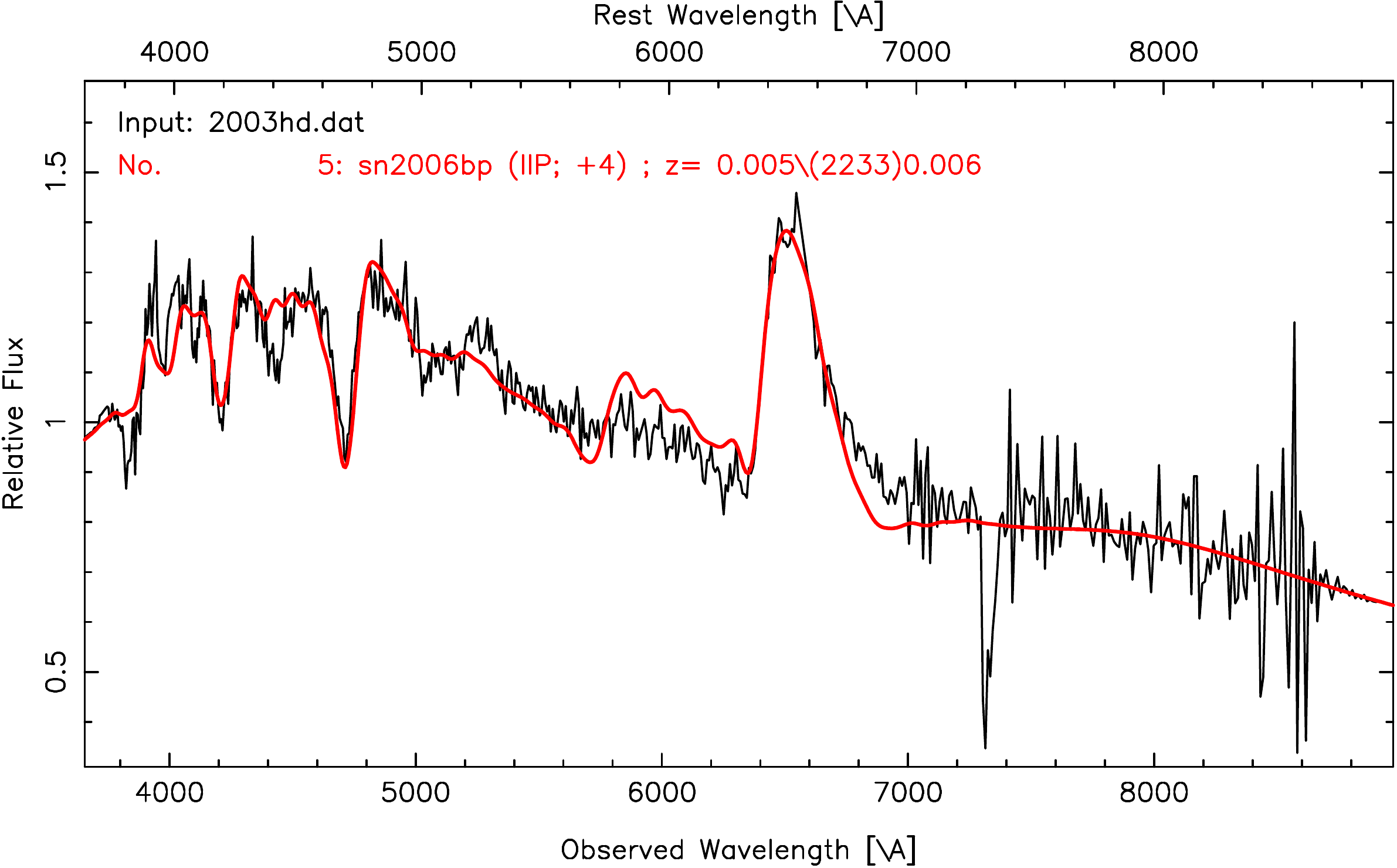}
\includegraphics[width=4.4cm]{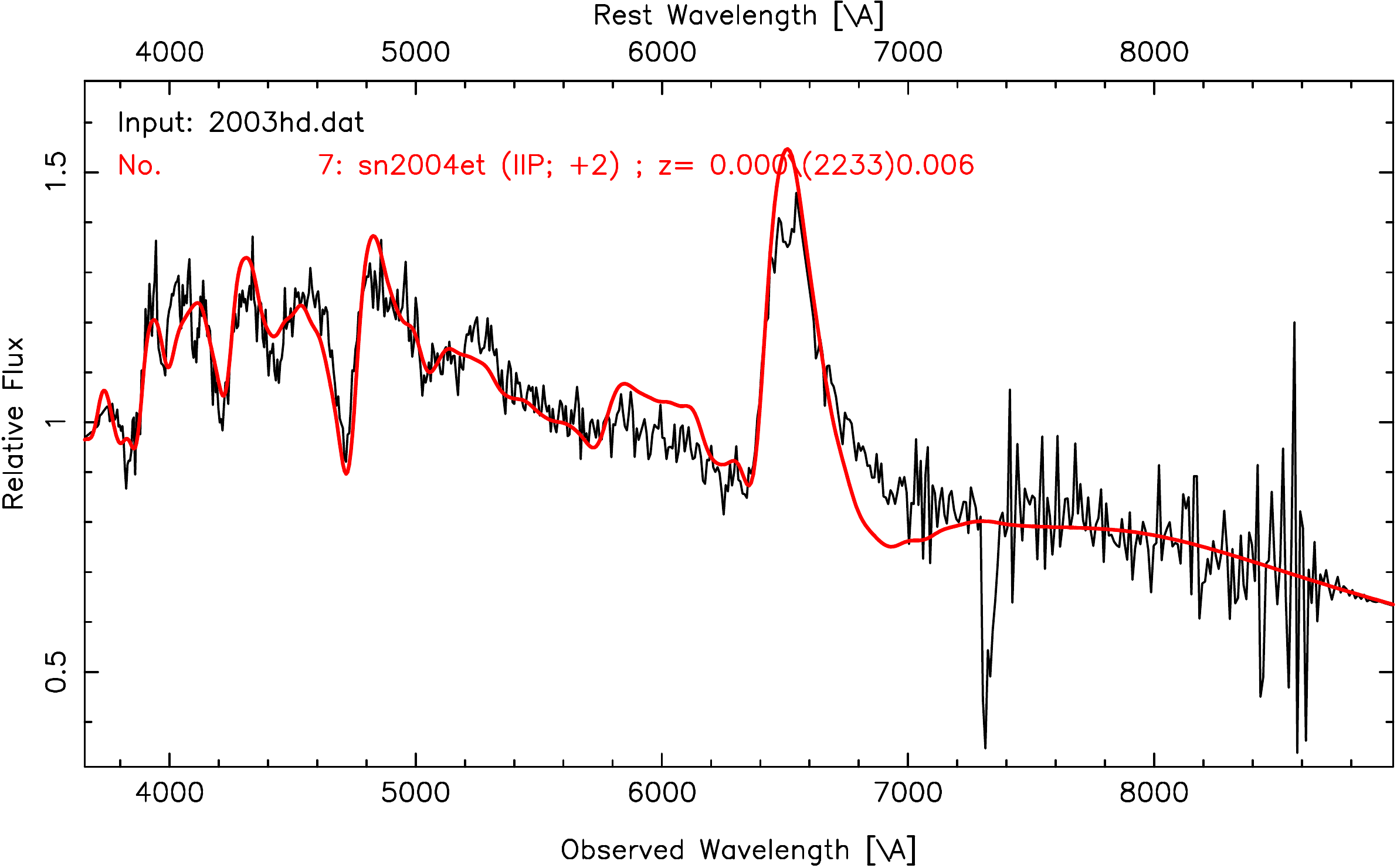}
\caption{Best spectral matching of SN~2003hd using SNID. The plots show SN~2003hd compared with 
SN~2006b and SN~2004et at 13 and 18 days from explosion.}
\end{figure}

\clearpage

\begin{figure}
\centering
\includegraphics[width=4.4cm]{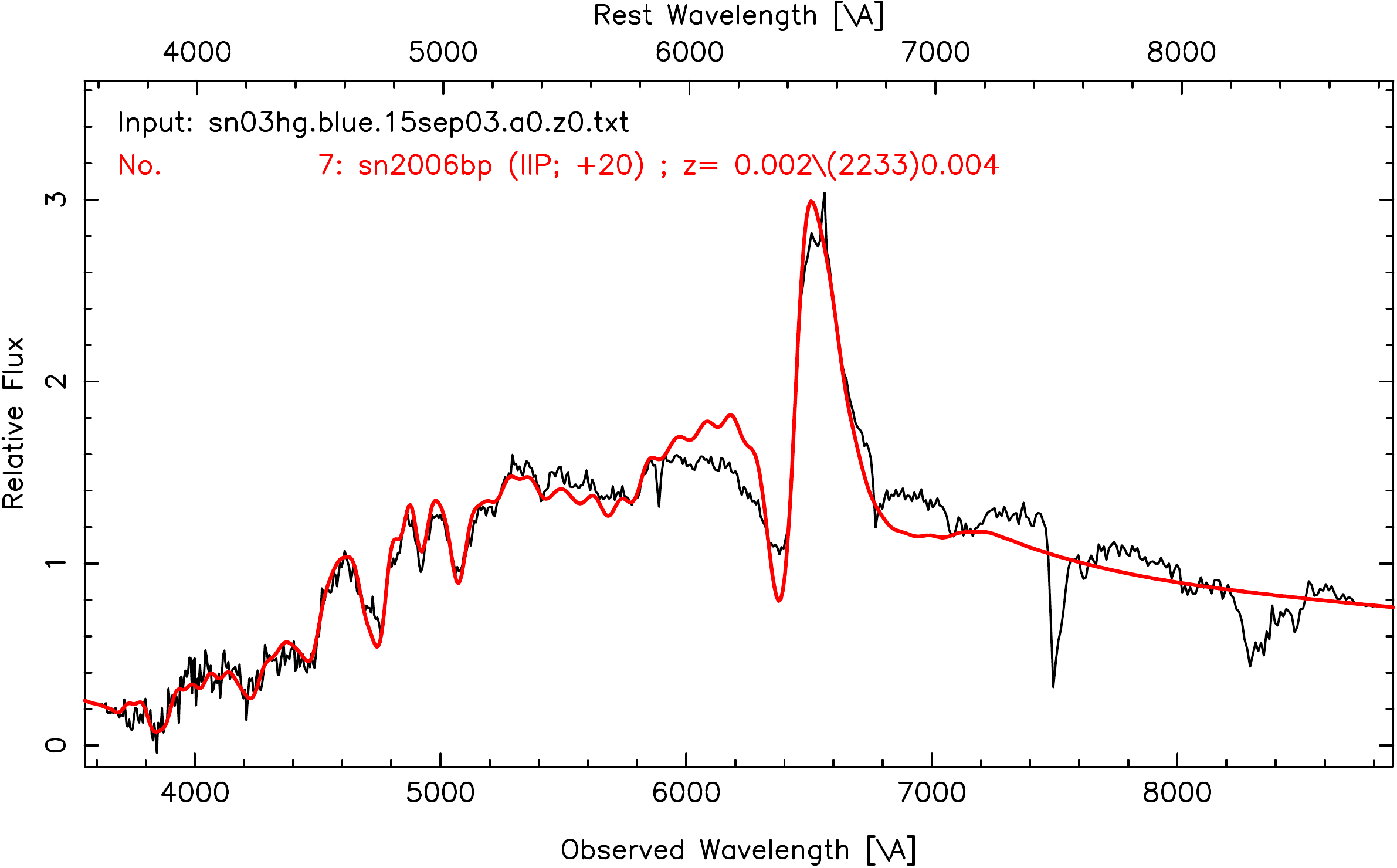}
\includegraphics[width=4.4cm]{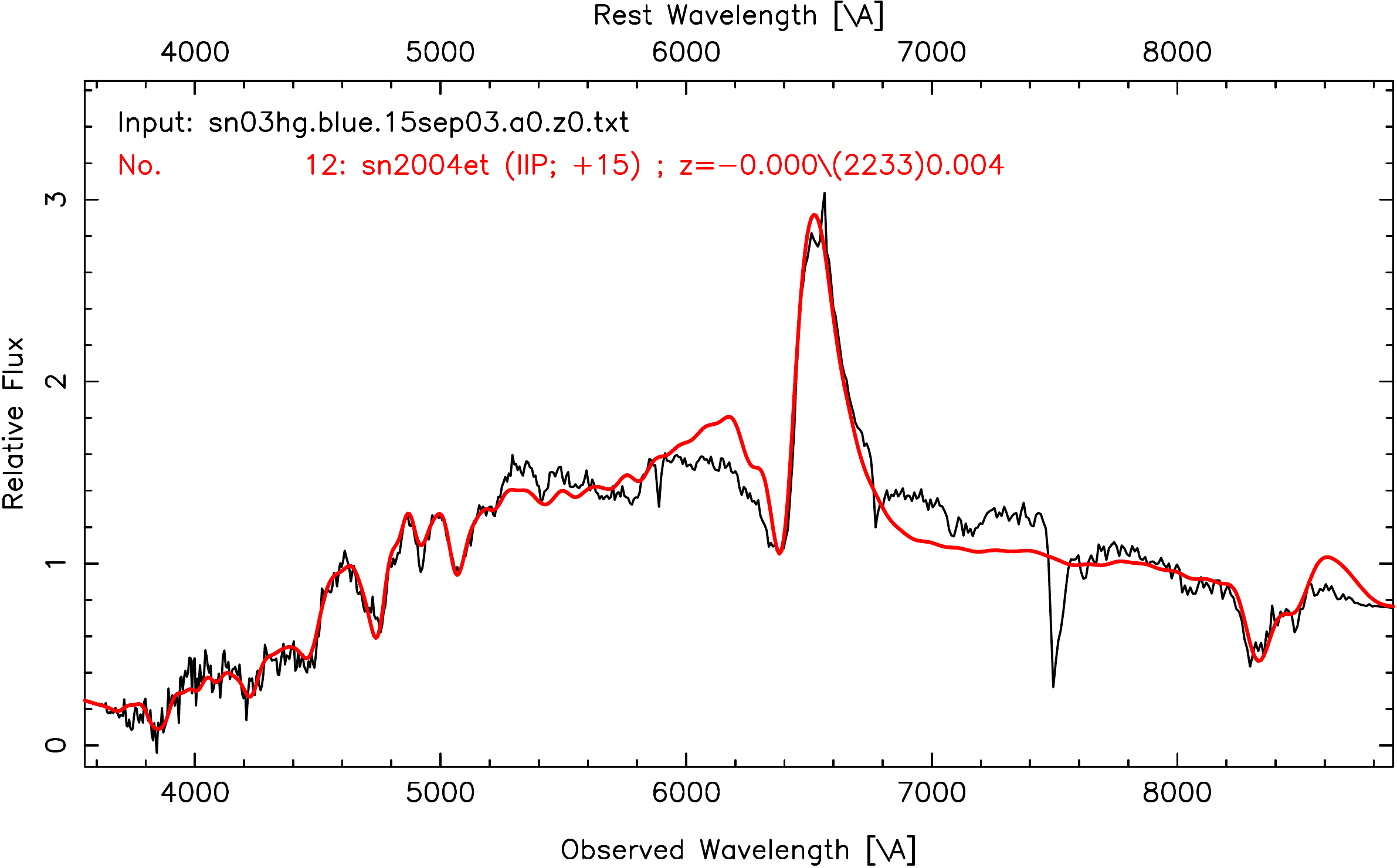}
\includegraphics[width=4.4cm]{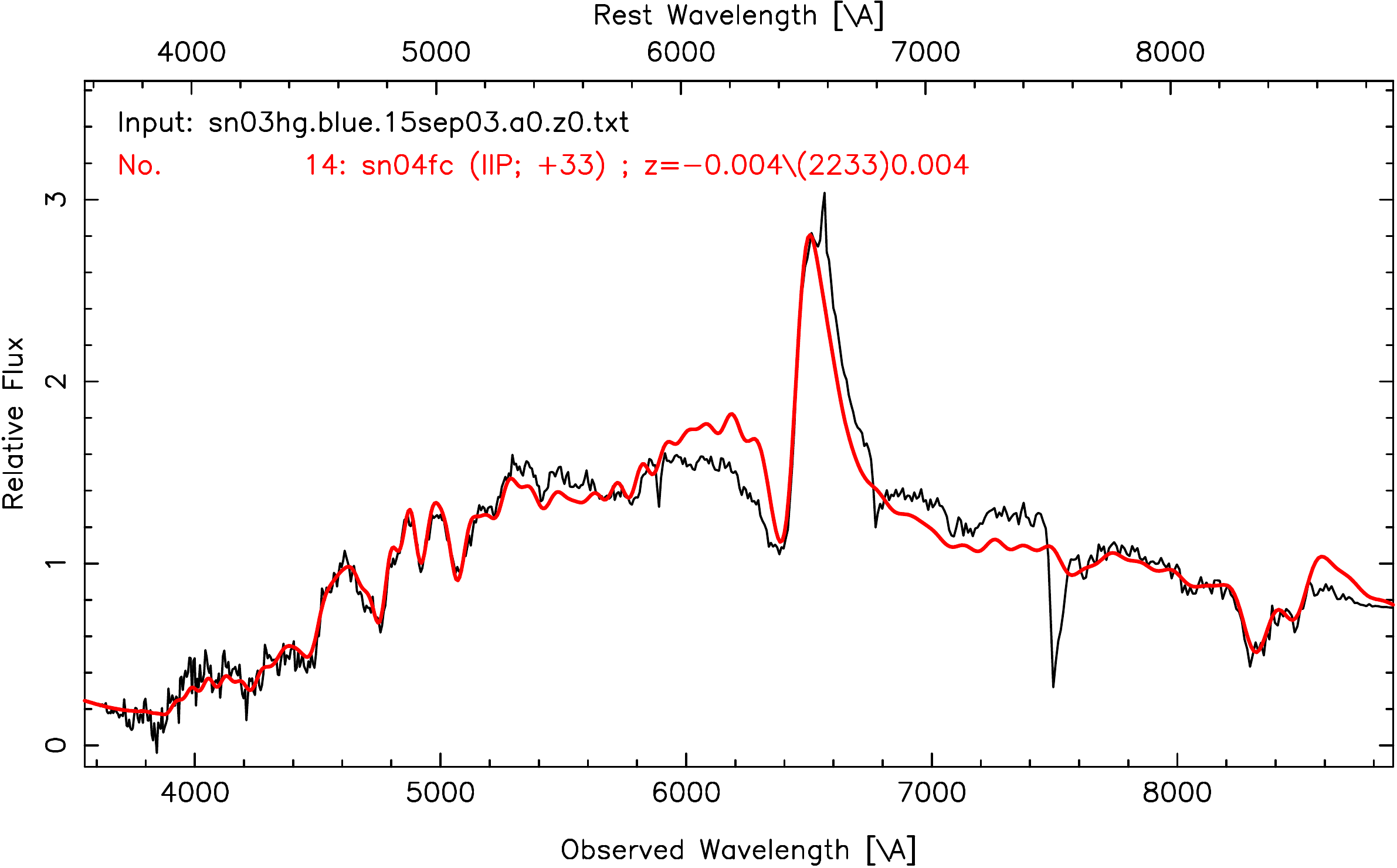}
\includegraphics[width=4.4cm]{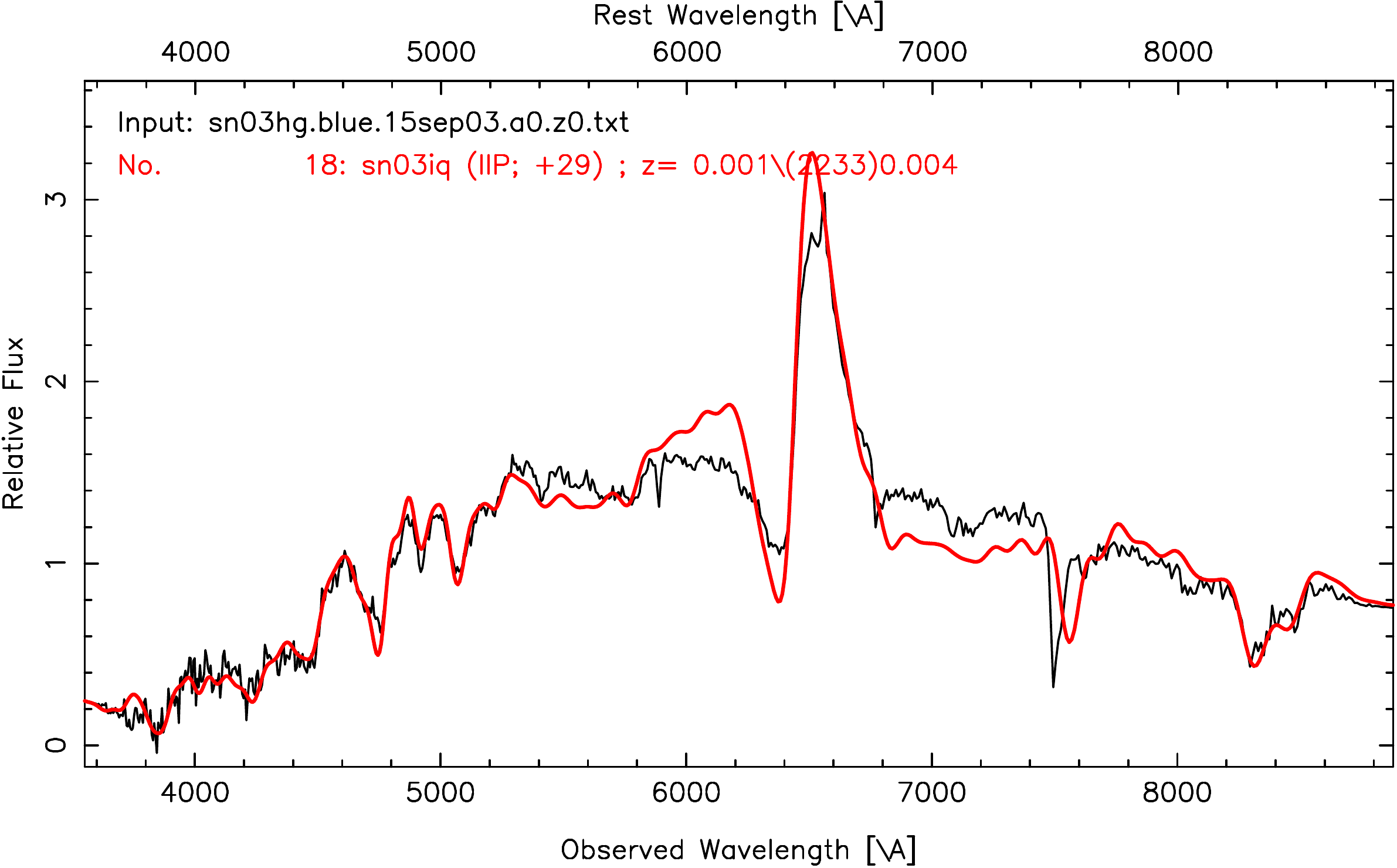}
\includegraphics[width=4.4cm]{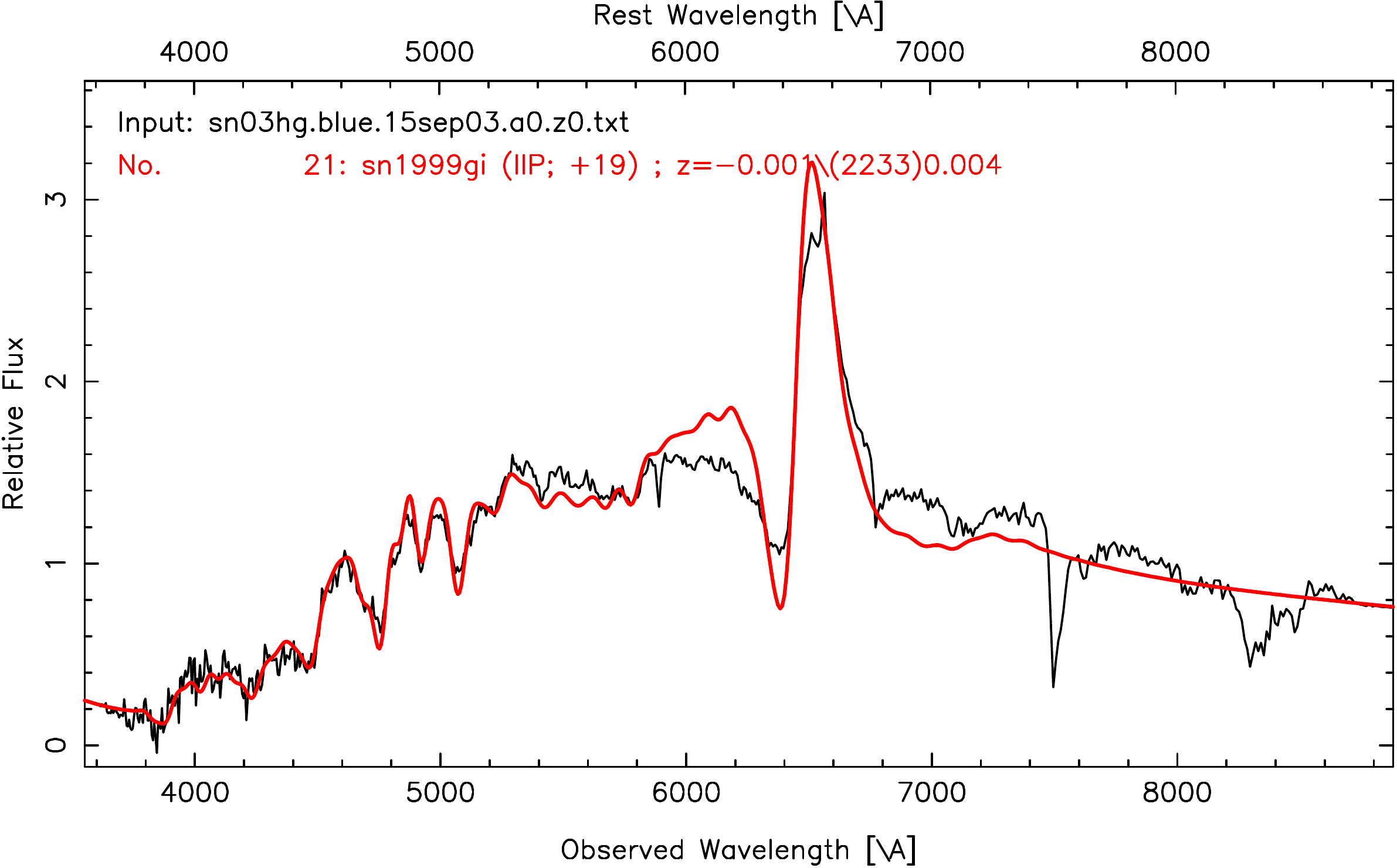}
\caption{Best spectral matching of SN~2003hg using SNID. The plots show SN~2003hg compared with 
SN~2006bp, SN~2004et, SN~2004fc, SN~2003iq, and SN~1999gi at 29, 31, 33, 29, and 31 days from explosion.}
\end{figure}

\begin{figure}
\centering
\includegraphics[width=4.4cm]{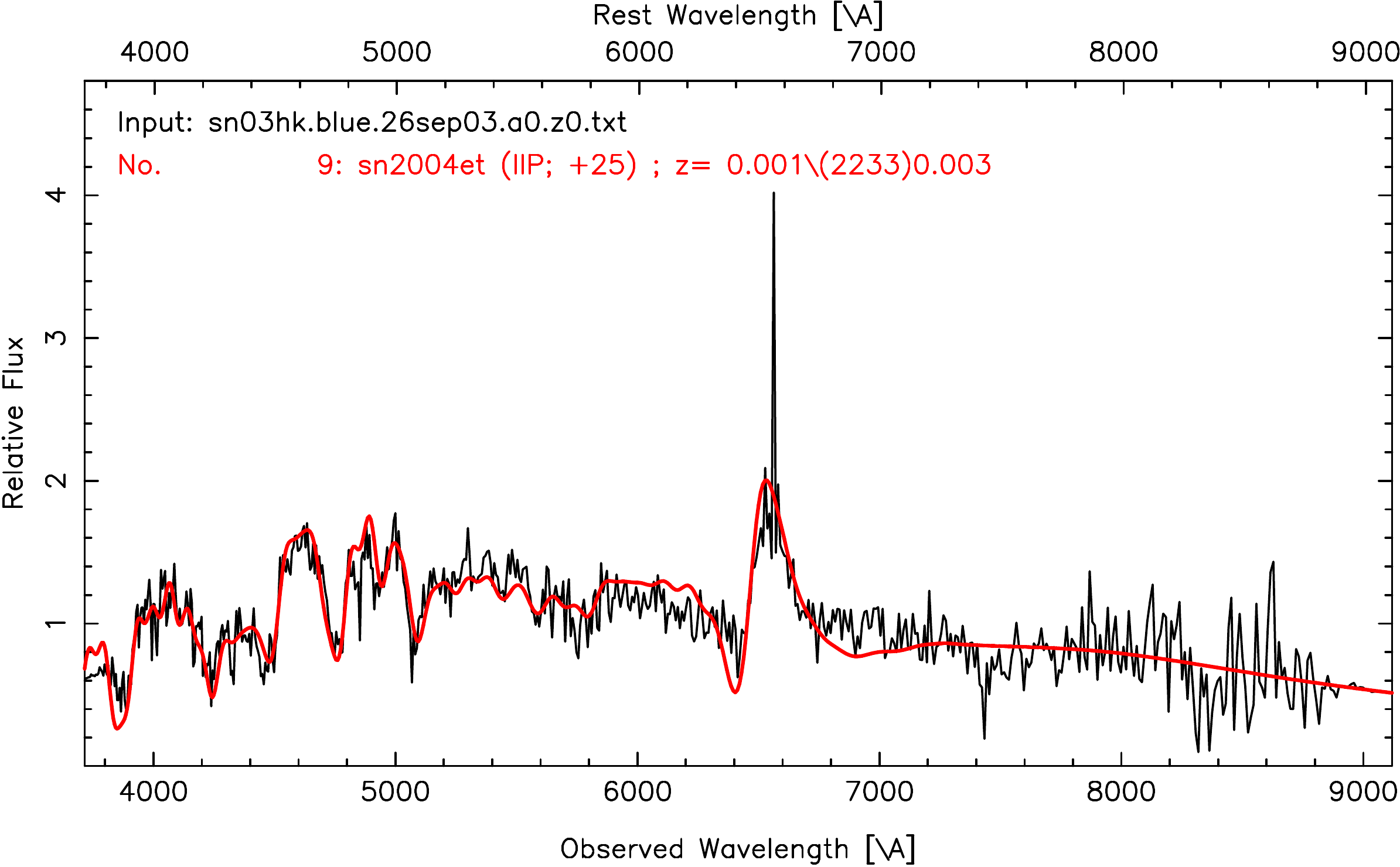}
\includegraphics[width=4.4cm]{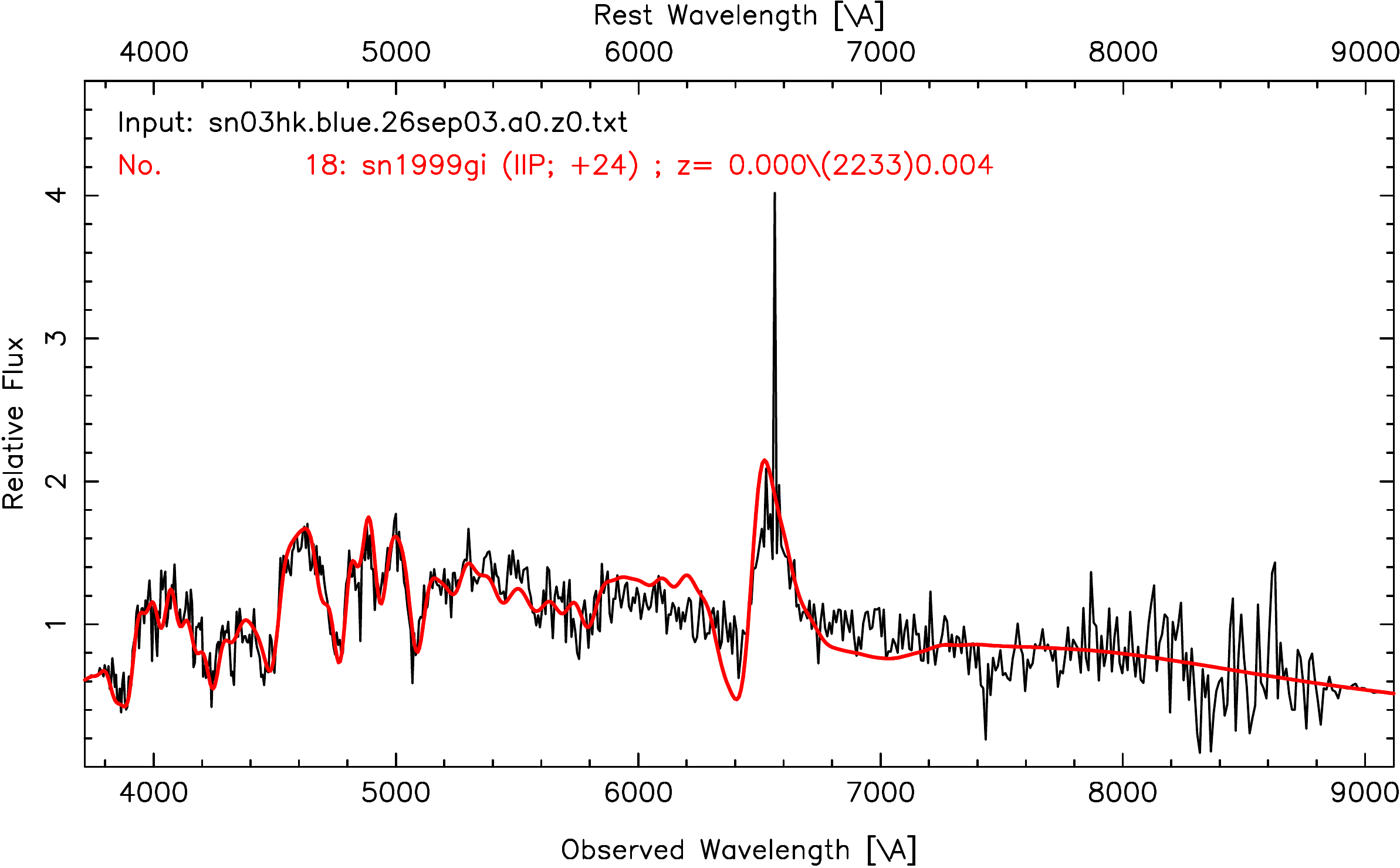}
\includegraphics[width=4.4cm]{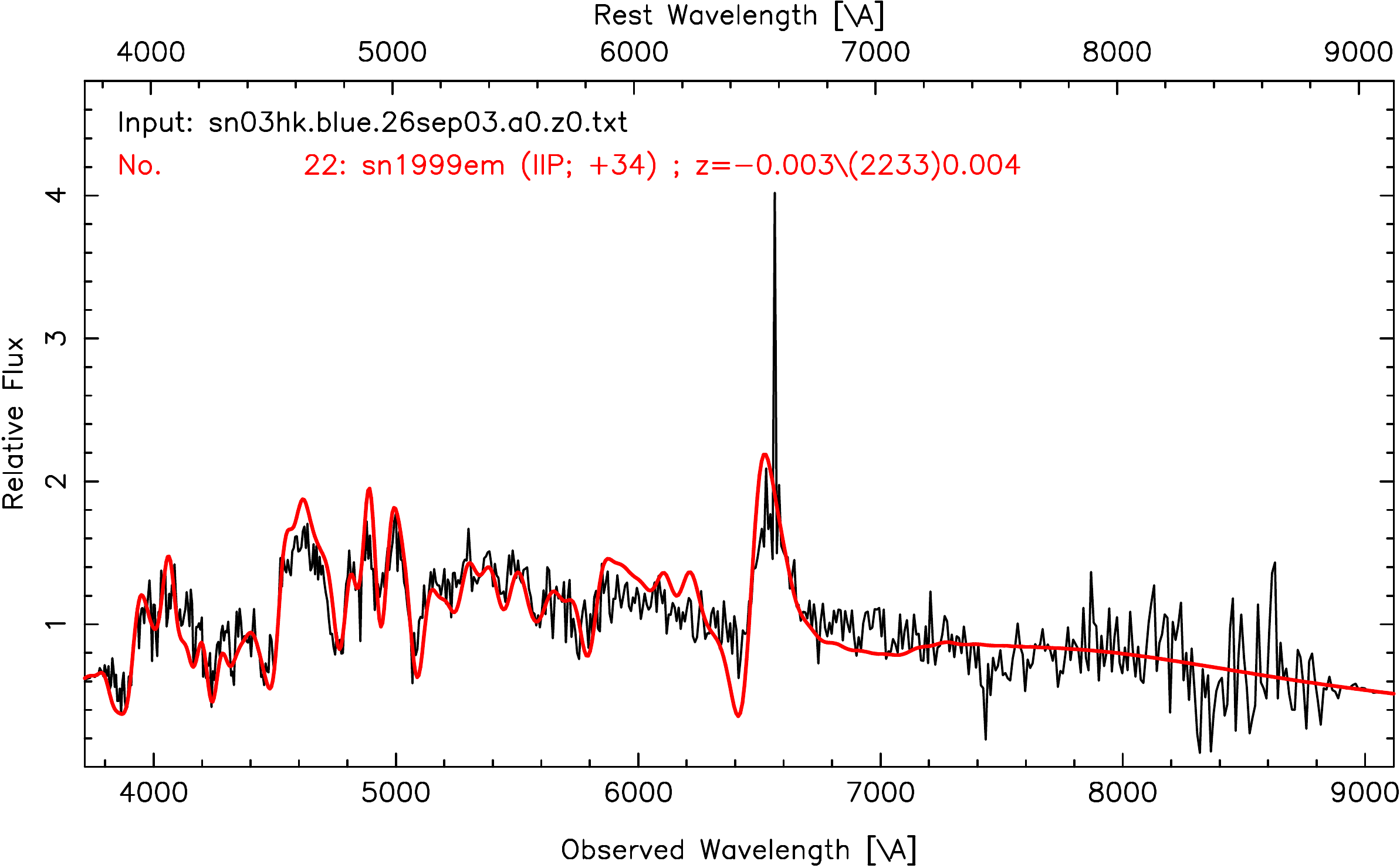}
\caption{Best spectral matching of SN~2003hk using SNID. The plots show SN~2003hk compared with 
SN~2004et, SN~1999gi, and SN~1999em at 41, 41, and 44 days from explosion.}
\end{figure}

\clearpage

\begin{figure}[h!]
\centering
\includegraphics[width=4.4cm]{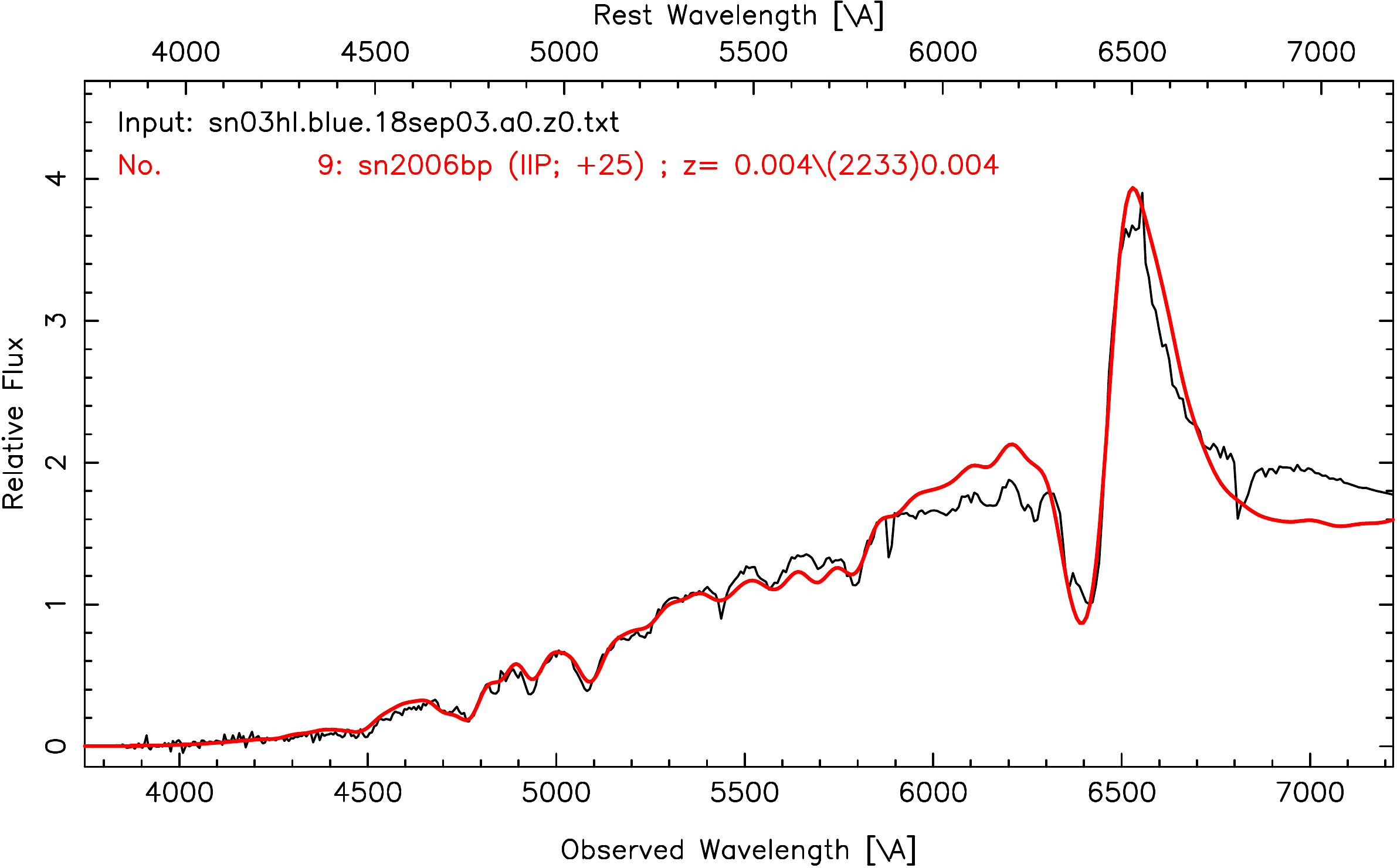}
\includegraphics[width=4.4cm]{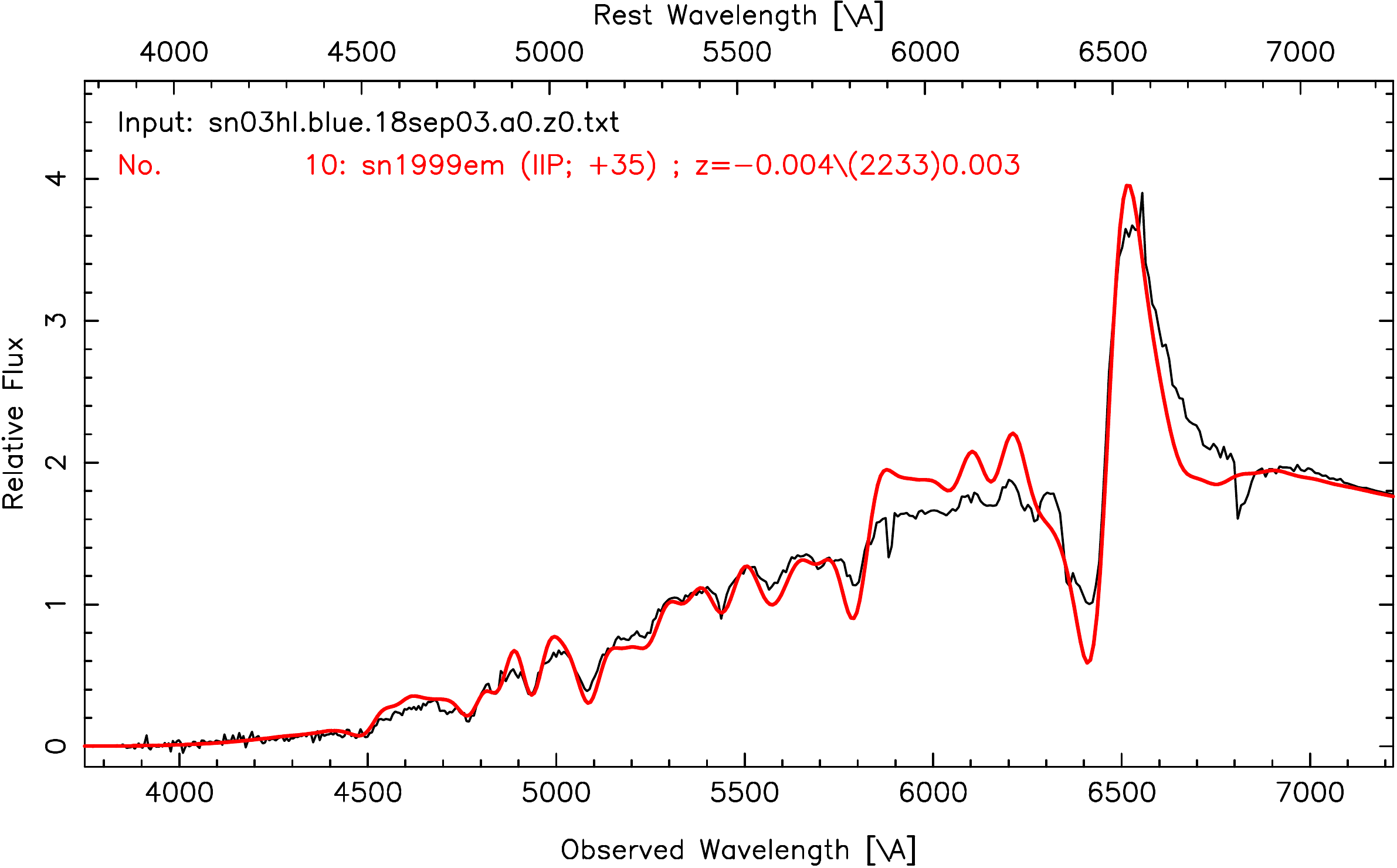}
\includegraphics[width=4.4cm]{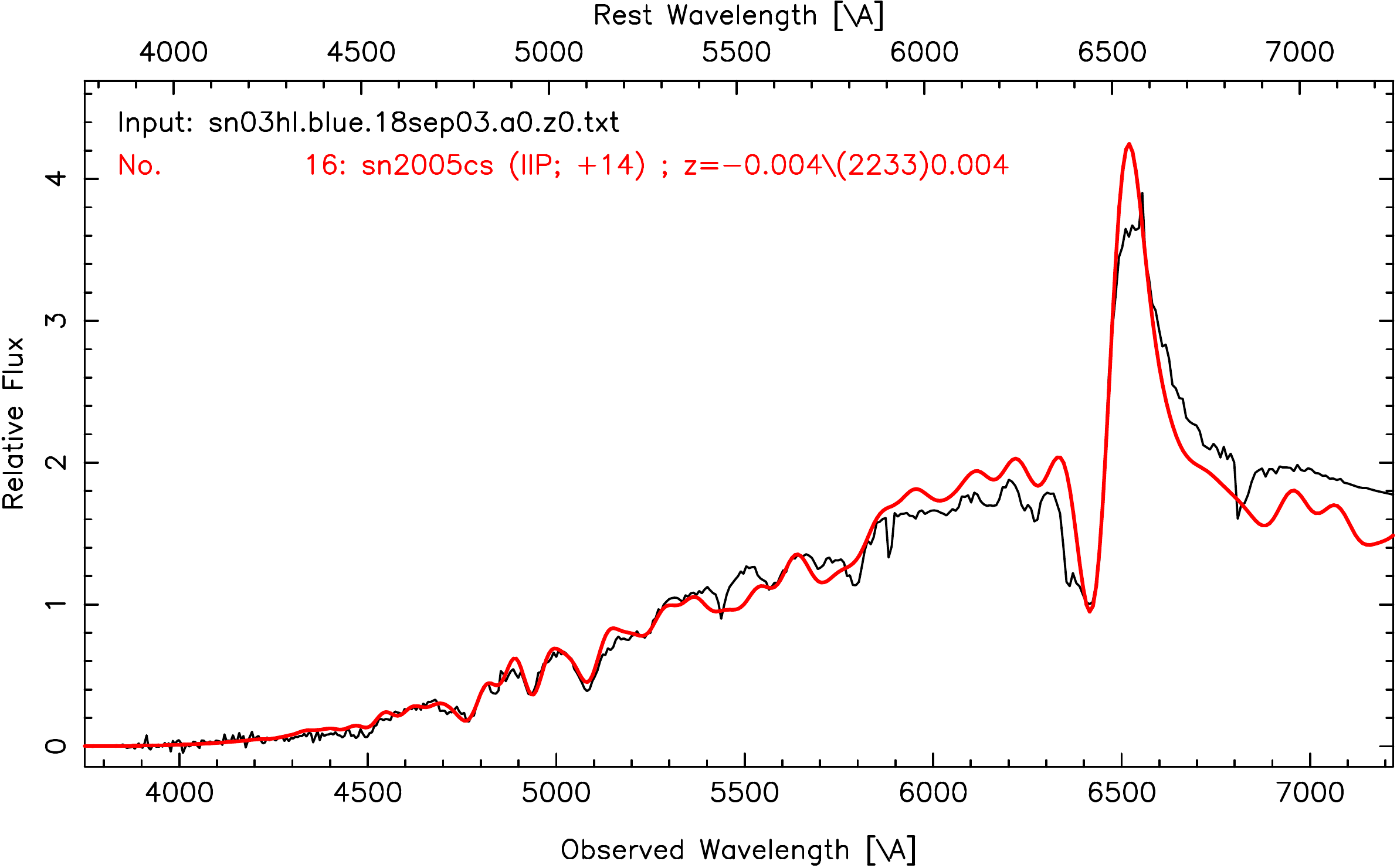}
\caption{Best spectral matching of SN~2003hl using SNID. The plots show SN~2003hl compared with 
SN~2006bp, SN~1999em, and SN~2005cs at 34, 45, and 20 days from explosion.}
\end{figure}

\begin{figure}[h!]
\centering
\includegraphics[width=4.4cm]{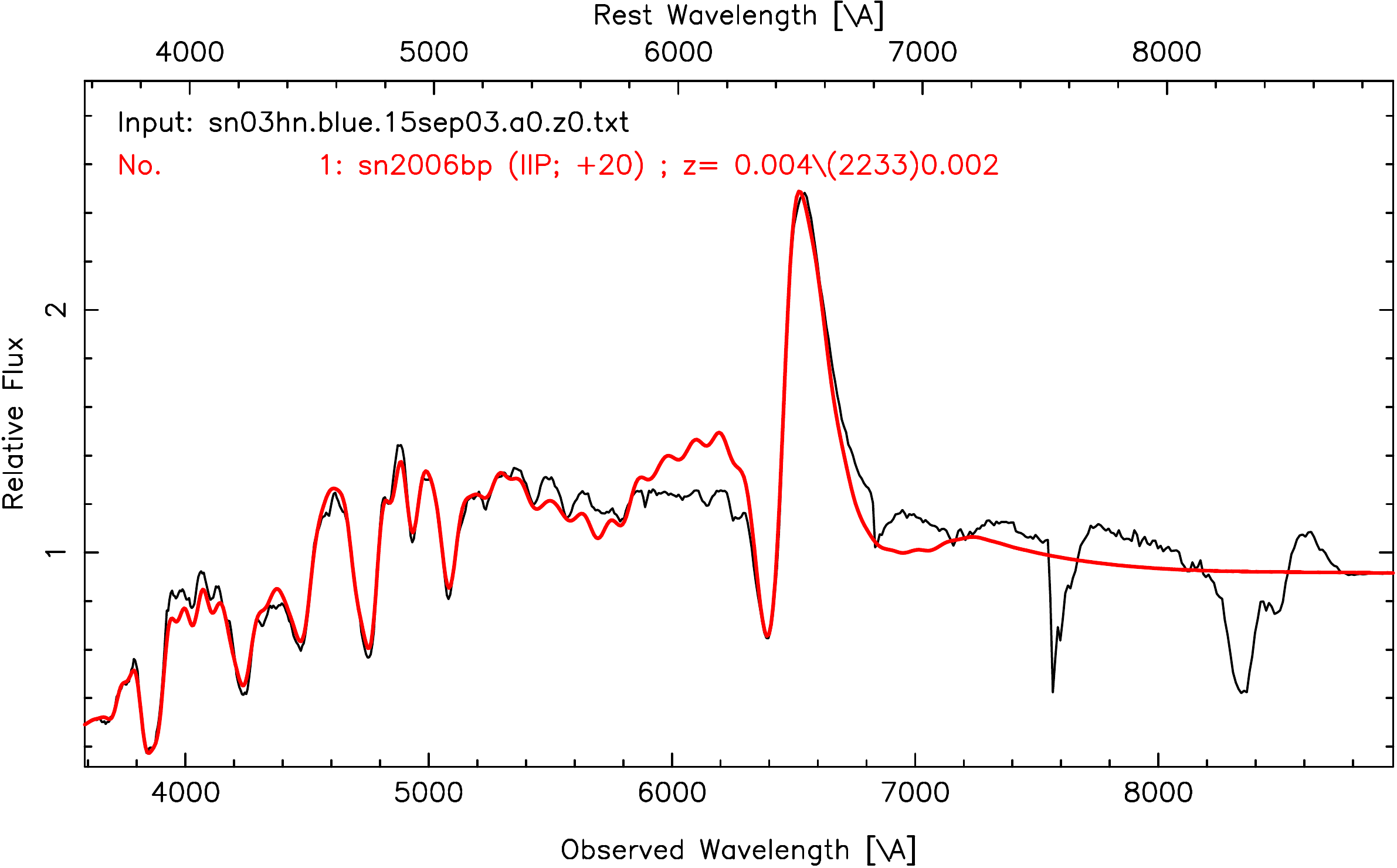}
\includegraphics[width=4.4cm]{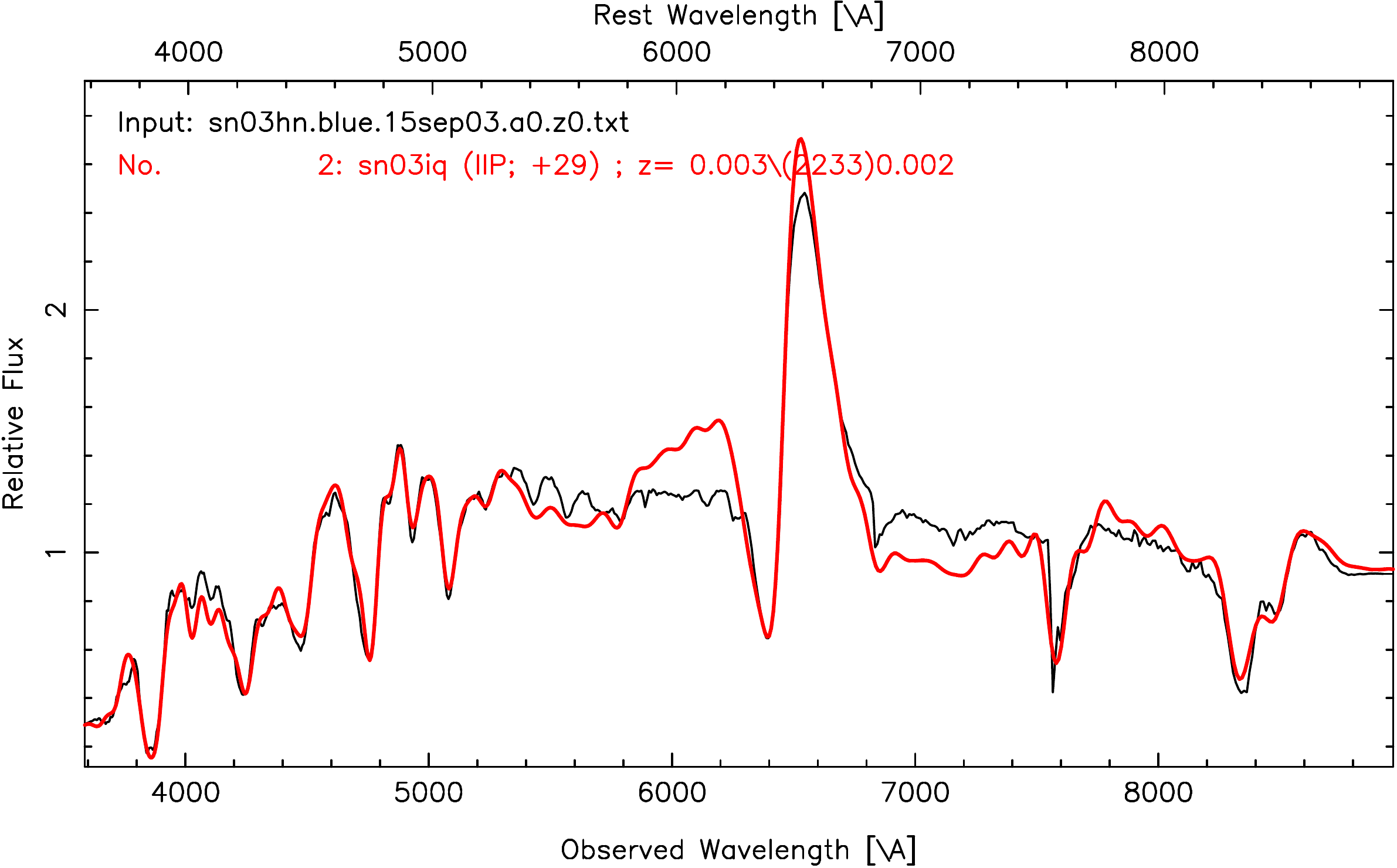}
\includegraphics[width=4.4cm]{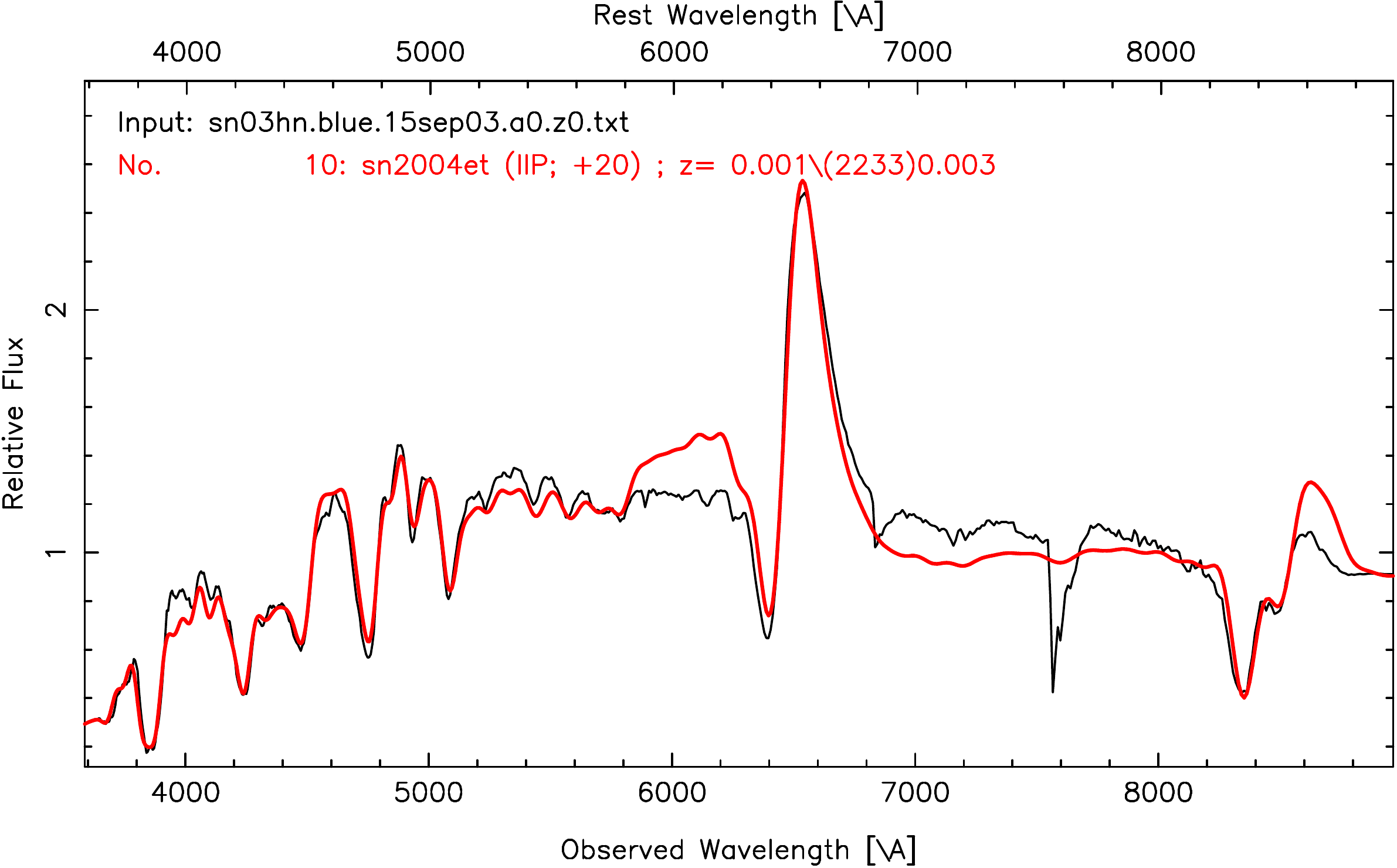}
\caption{Best spectral matching of SN~2003hn using SNID. The plots show SN~2003hn compared with 
SN~2006bp, SN~2003iq, and SN~2004et at 29, 29, and 36 days from explosion.}
\end{figure}

\begin{figure}[h!]
\centering
\includegraphics[width=4.4cm]{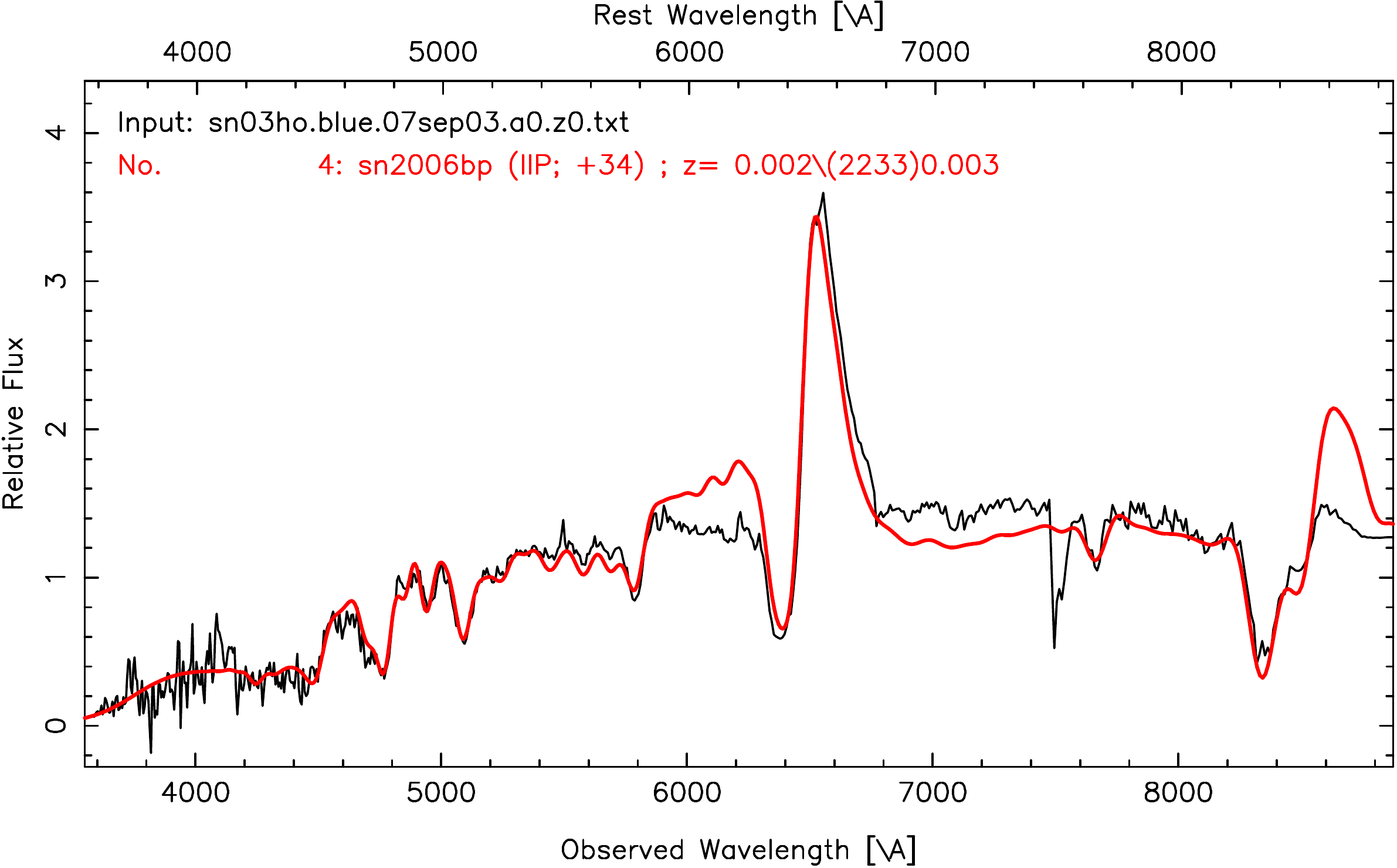}
\includegraphics[width=4.4cm]{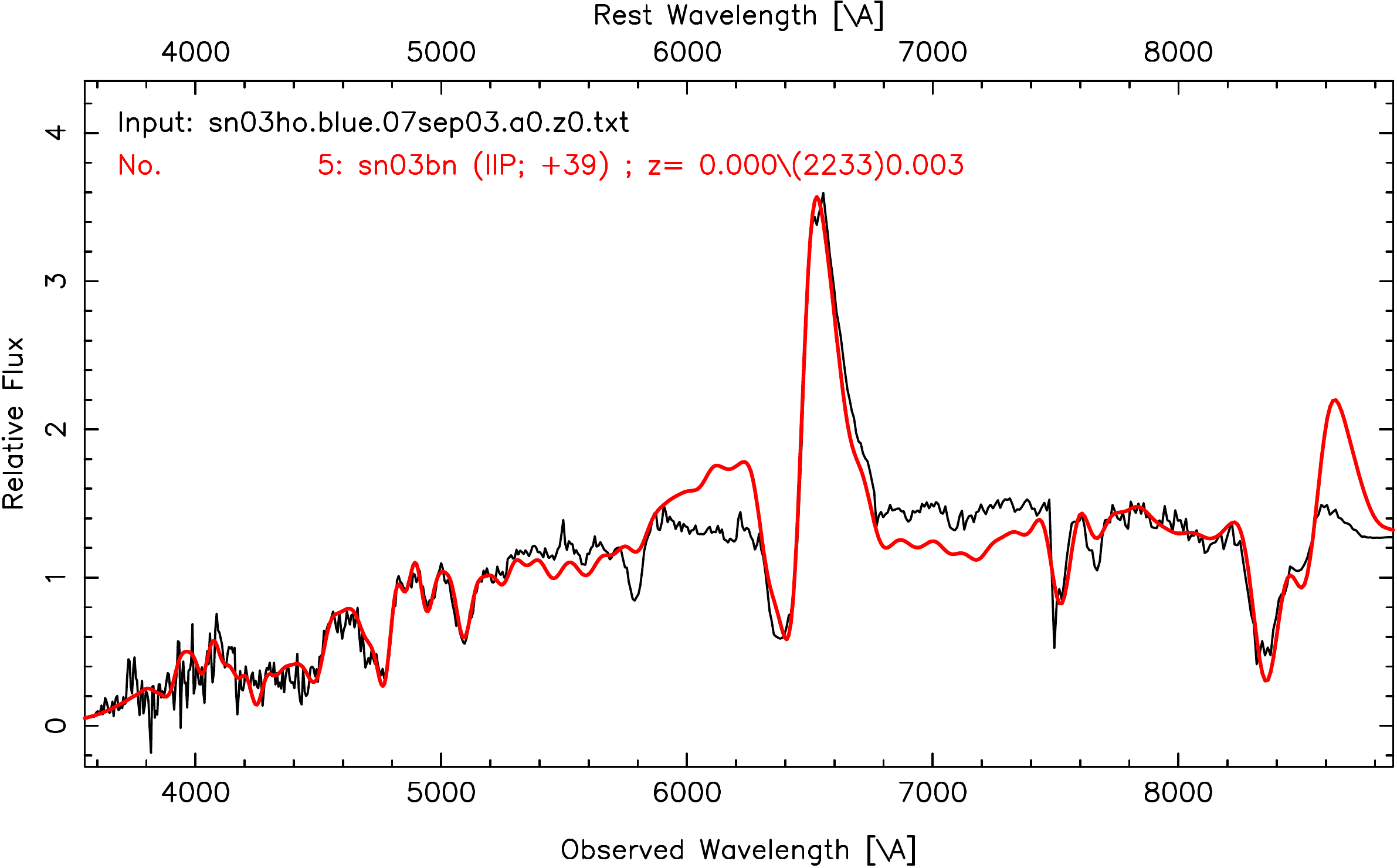}
\includegraphics[width=4.4cm]{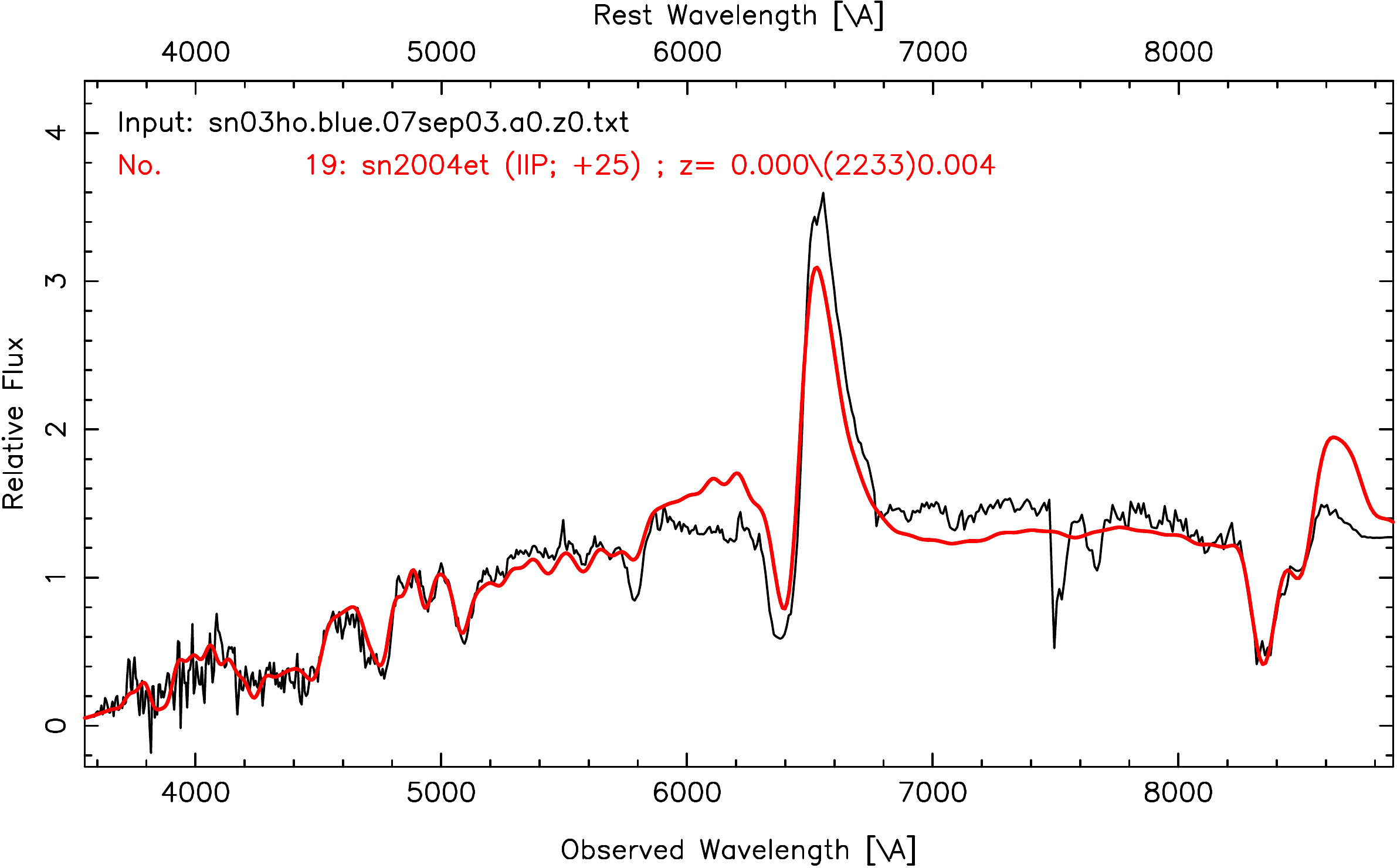}
\caption{Best spectral matching of SN~2003ho using SNID. The plots show SN~2003ho compared with 
SN~2006bp, SN~2003bn, and SN~2004et at 43, 39, and 41 days from explosion.}
\end{figure}

\begin{figure}[h!]
\centering
\includegraphics[width=4.4cm]{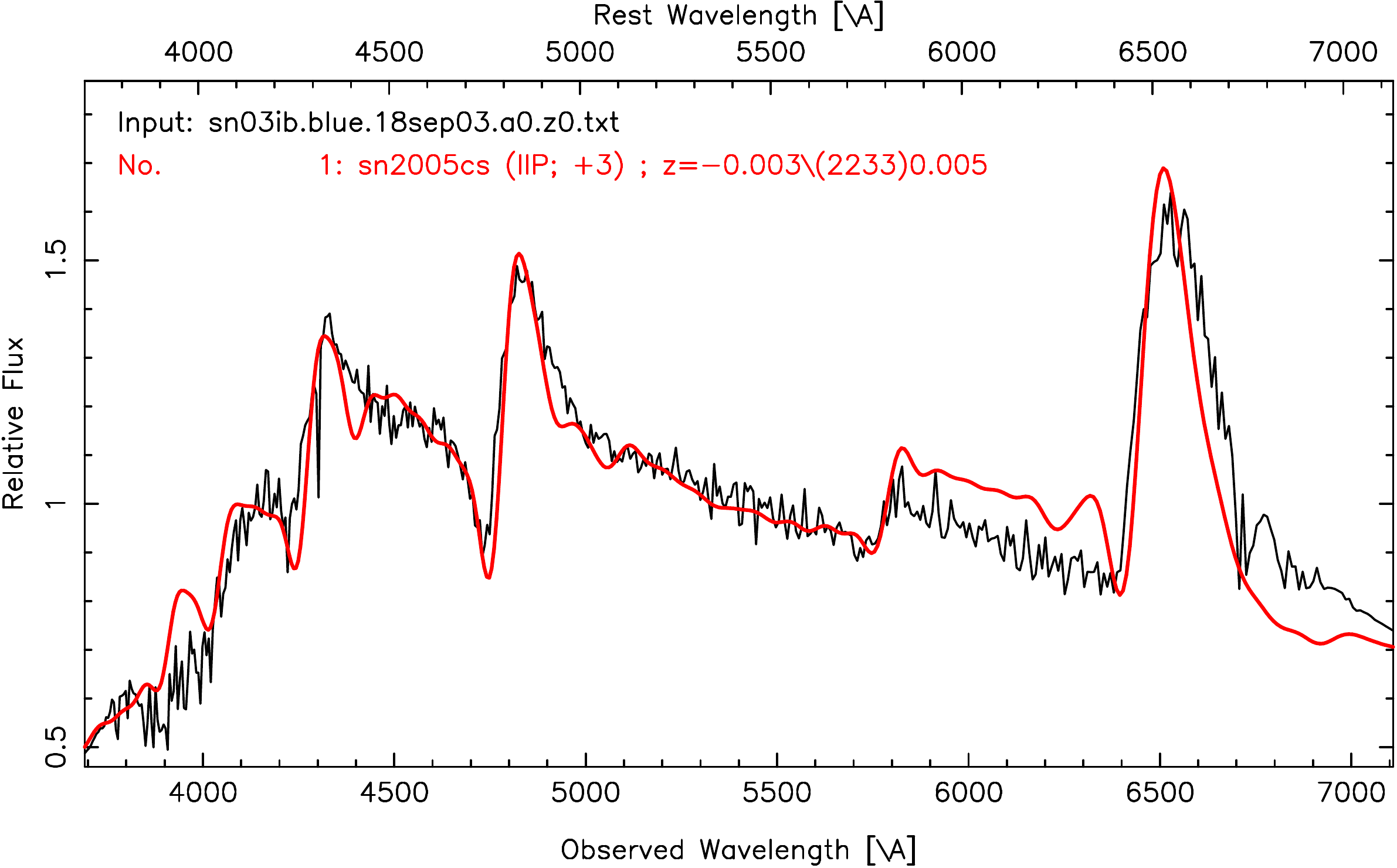}
\includegraphics[width=4.4cm]{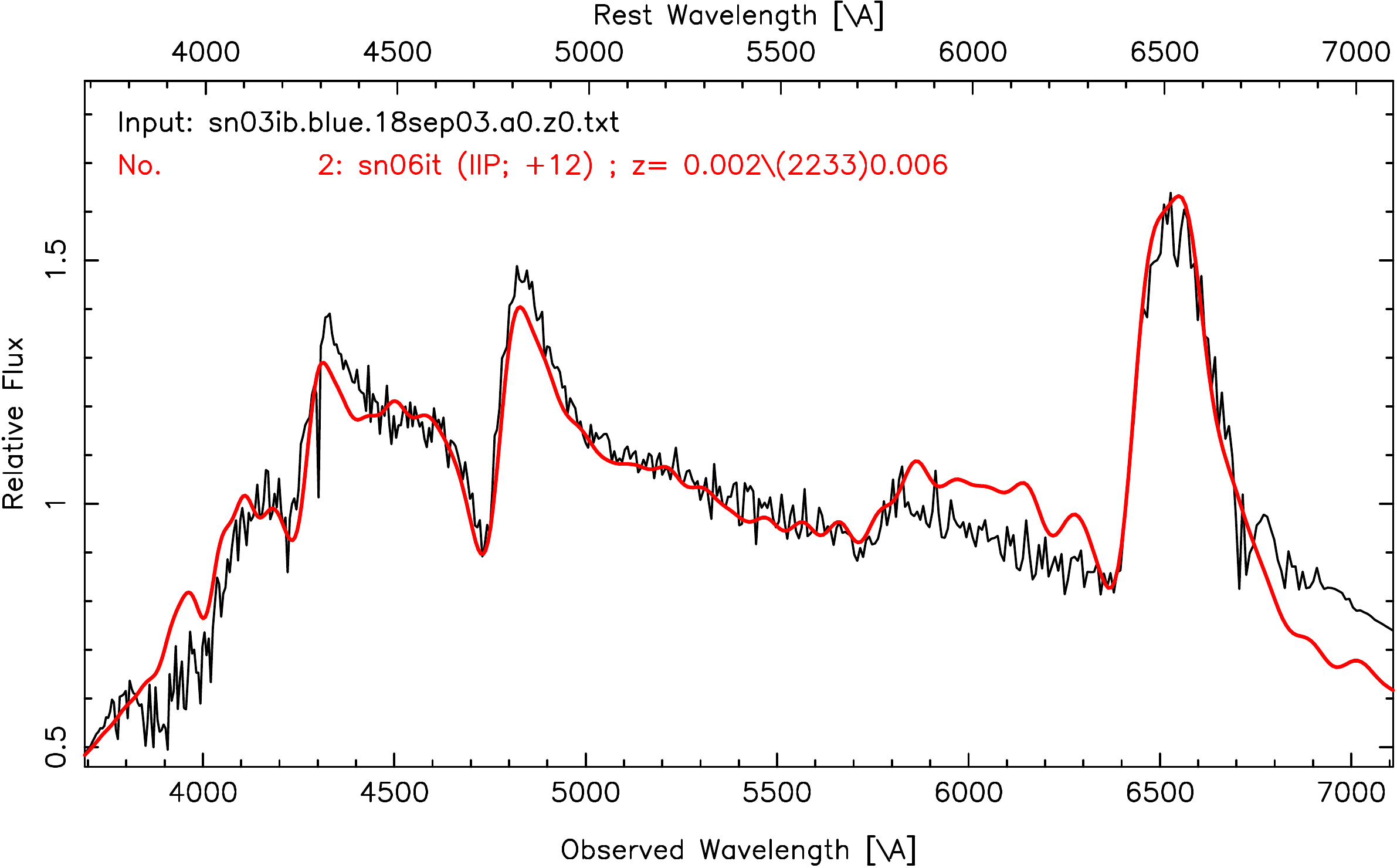}
\includegraphics[width=4.4cm]{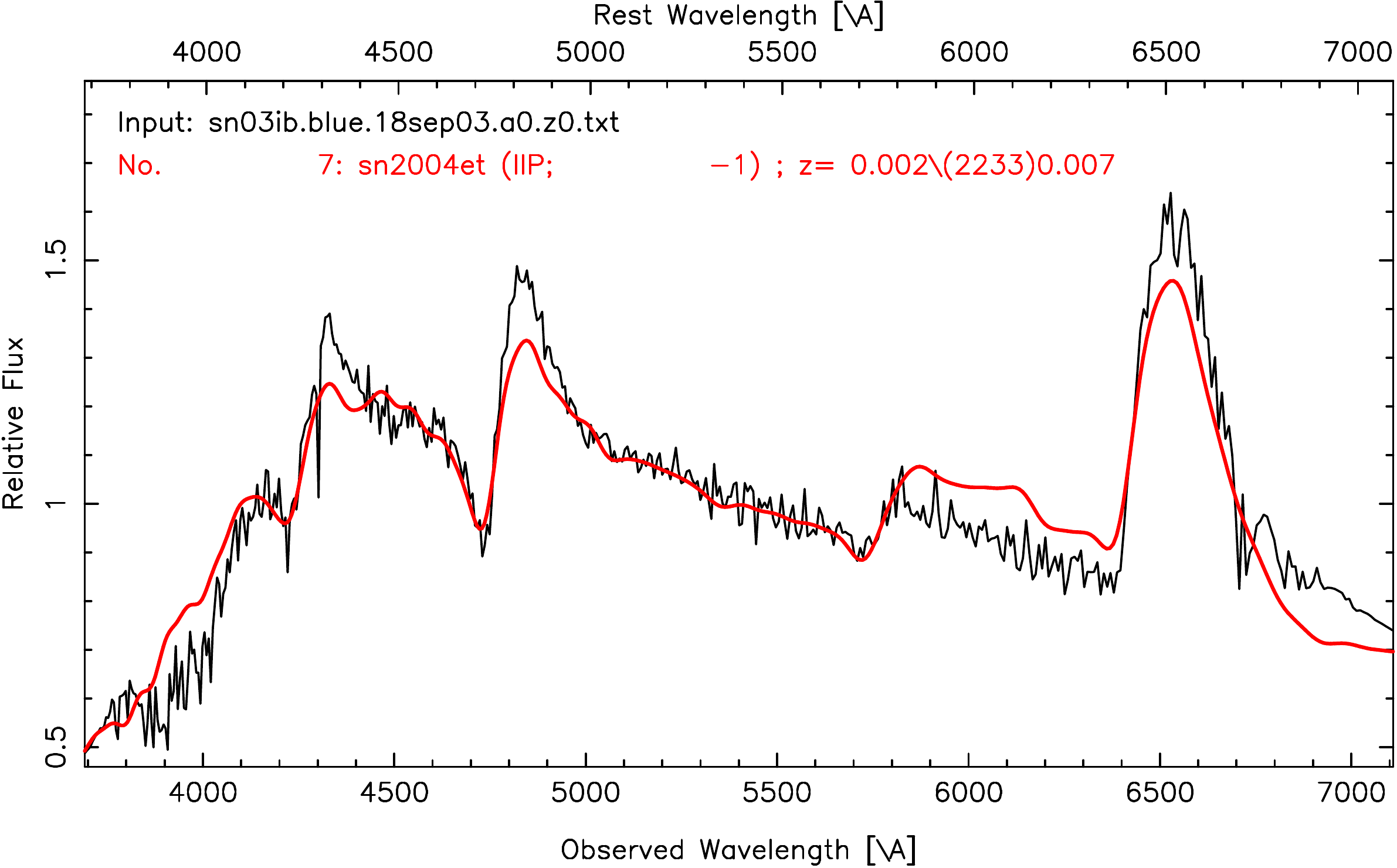}
\includegraphics[width=4.4cm]{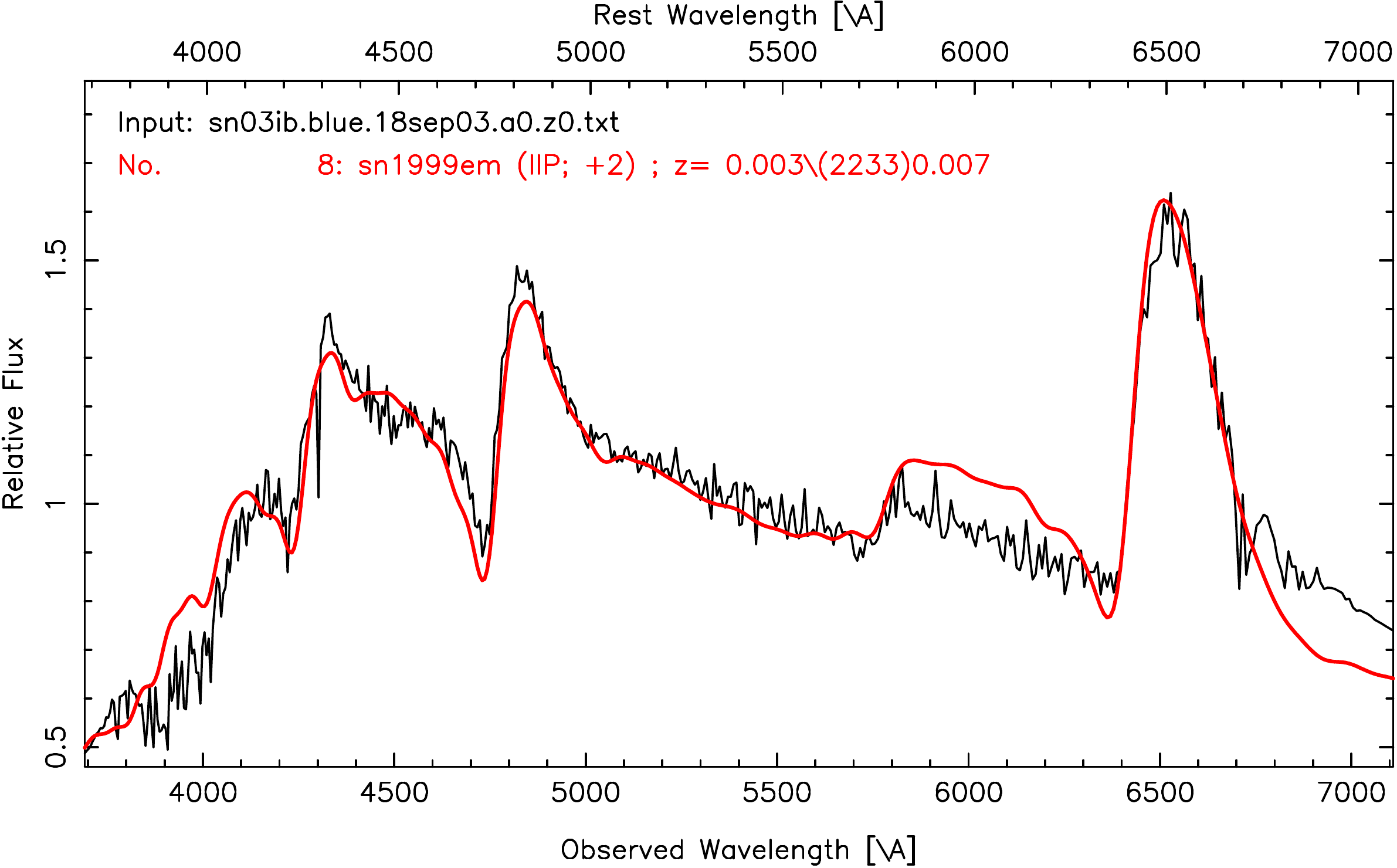}
\caption{Best spectral matching of SN~2003ib using SNID. The plots show SN~2003ib compared with 
SN~2005cs, SN~2006it, SN~2004et, and SN~1999em at 9, 12, 15, and 12 days from explosion.}
\end{figure}

\begin{figure}[h!]
\centering
\includegraphics[width=4.4cm]{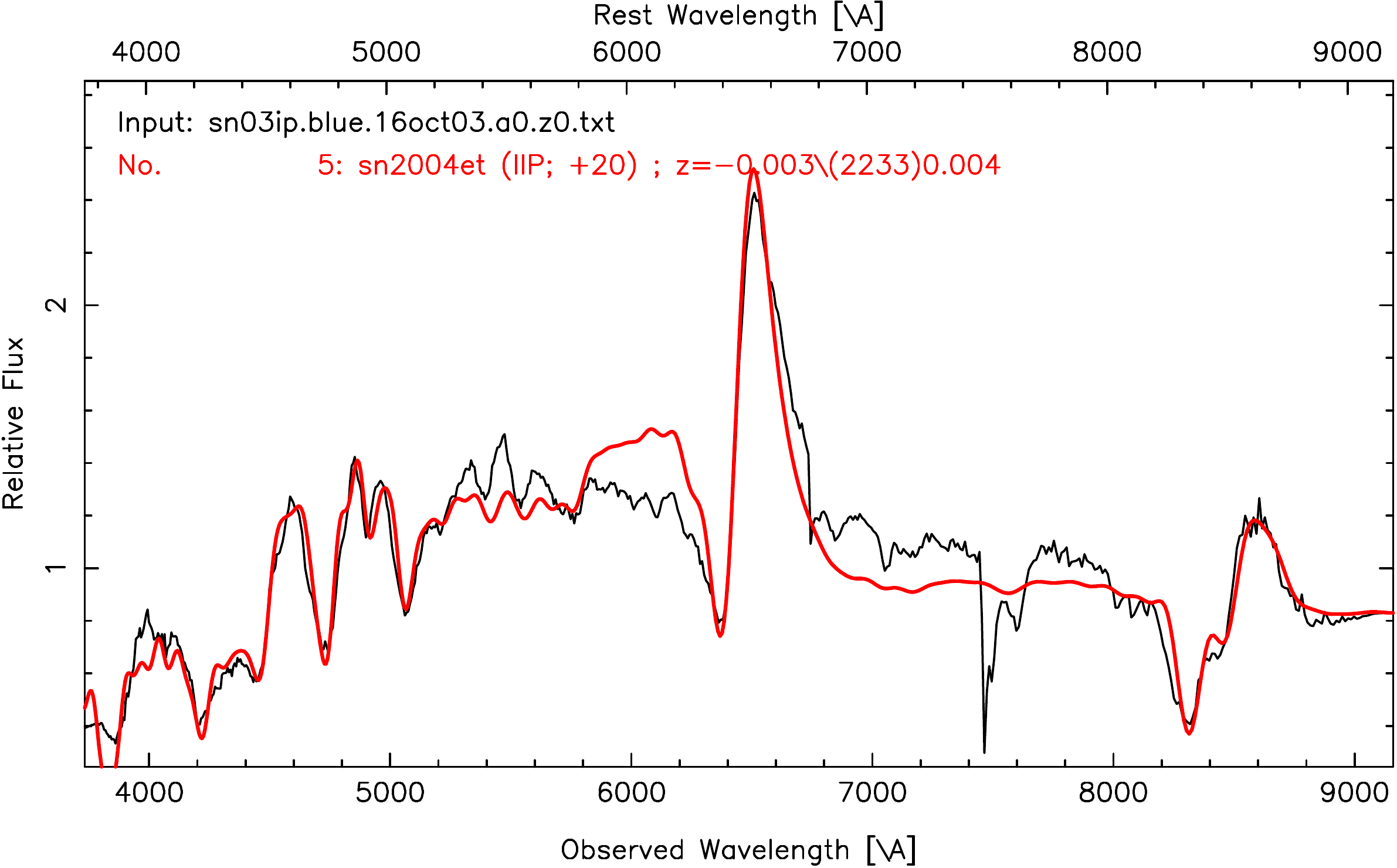}
\includegraphics[width=4.4cm]{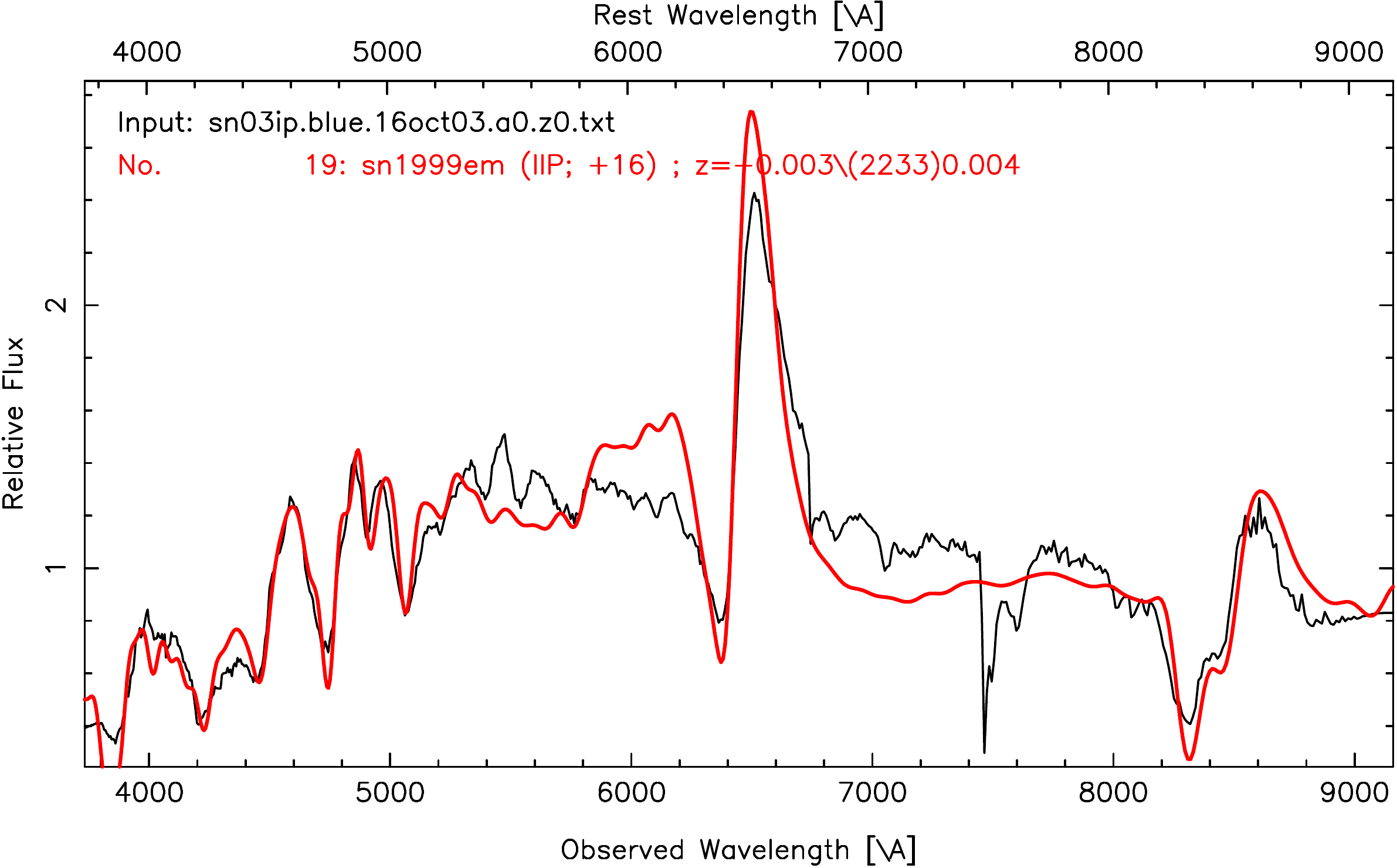}
\caption{Best spectral matching of SN~2003ip using SNID. The plots show SN~2003ip compared with 
SN~2004et and SN~1999em at 36 and 26 days from explosion.}
\end{figure}

\clearpage

\begin{figure}[h!]
\centering
\includegraphics[width=4.4cm]{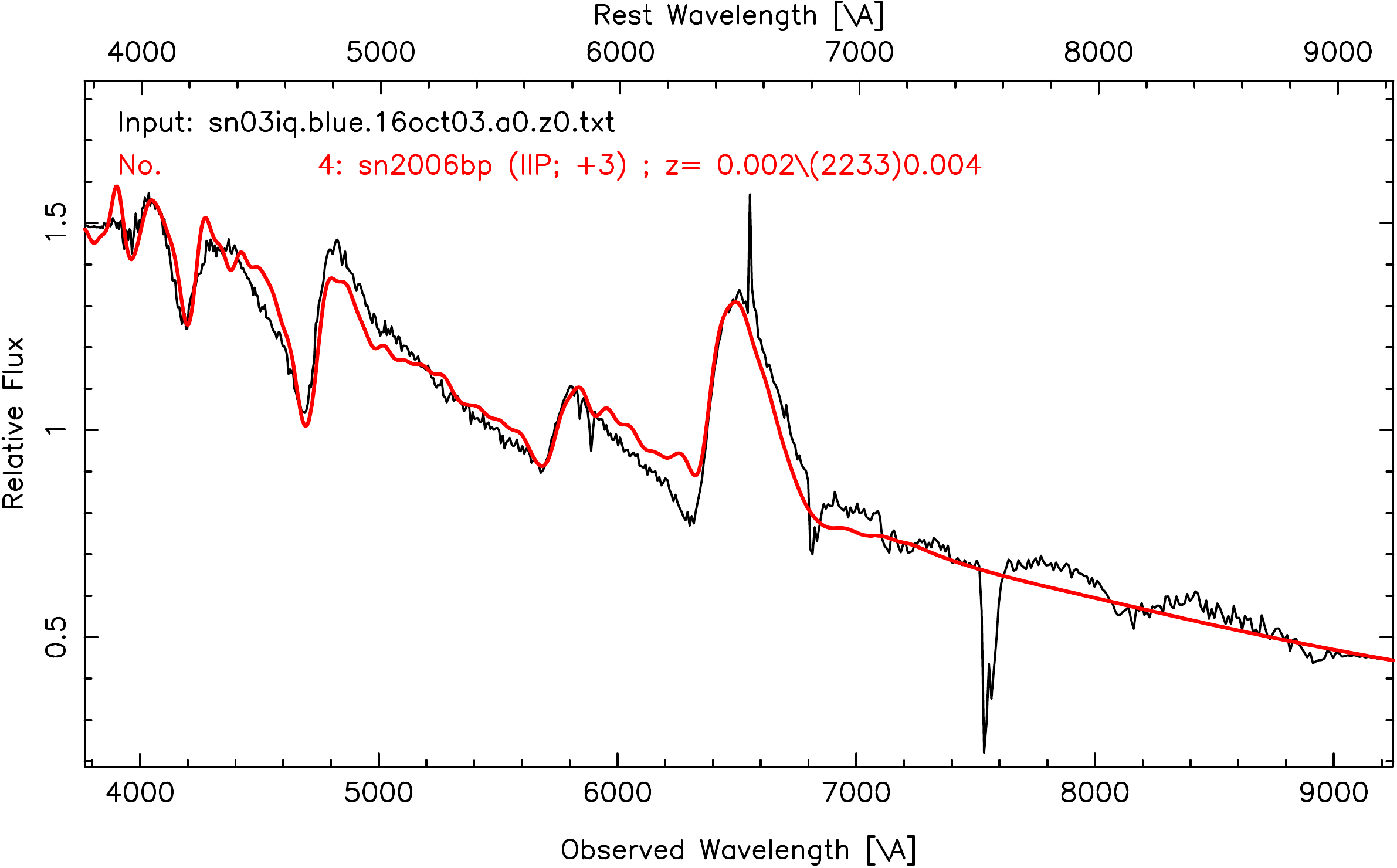}
\includegraphics[width=4.4cm]{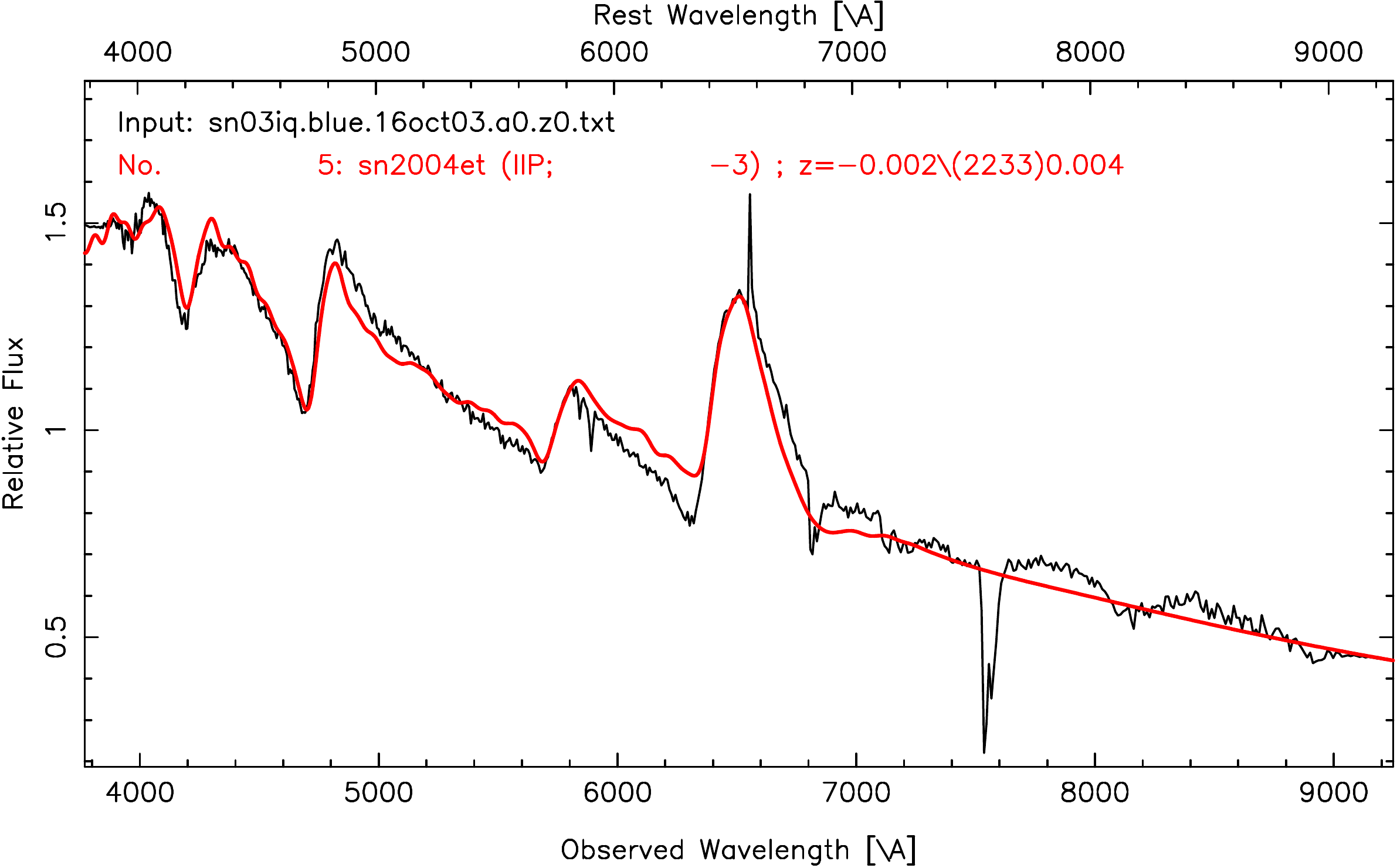}
\includegraphics[width=4.4cm]{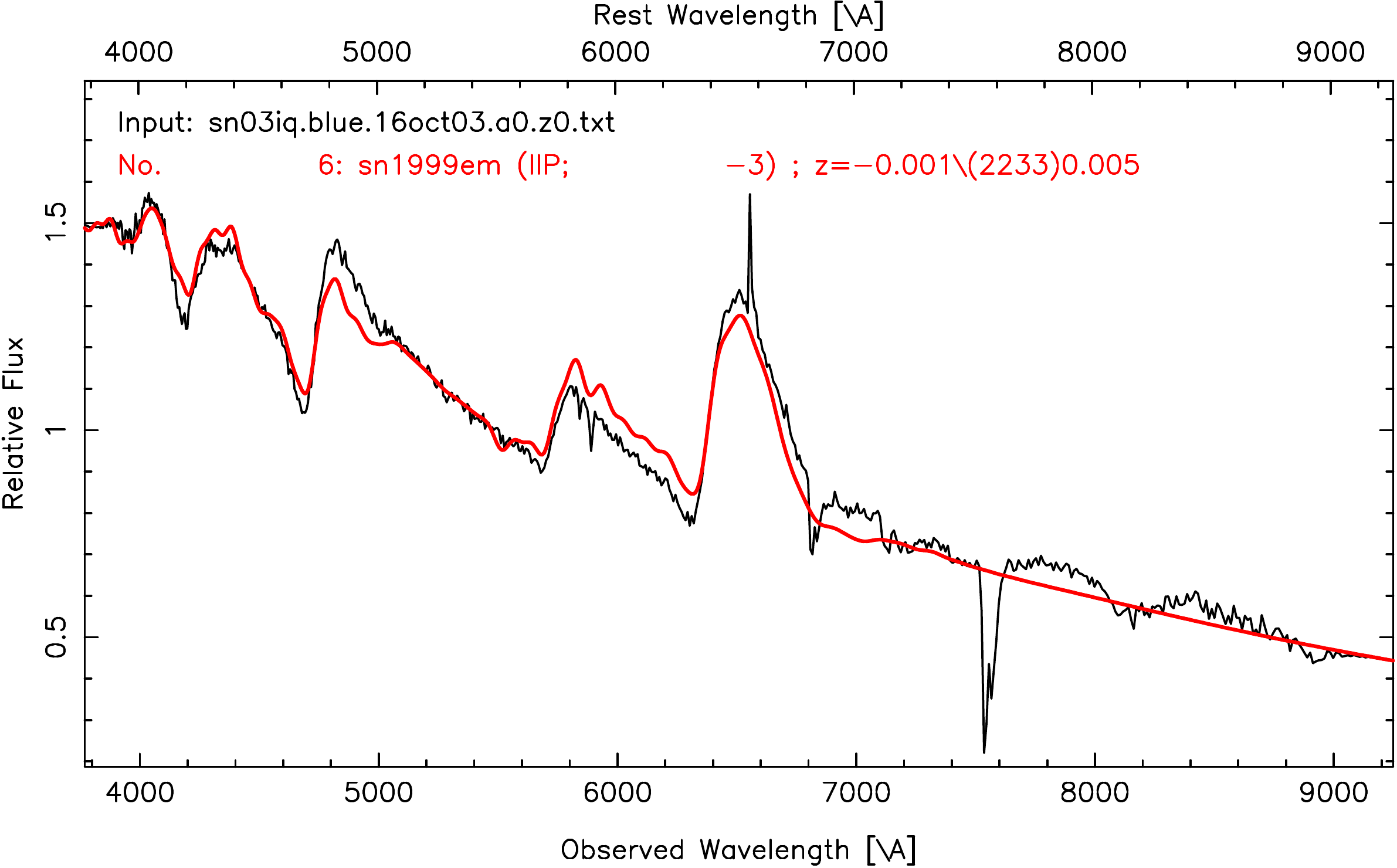}
\includegraphics[width=4.4cm]{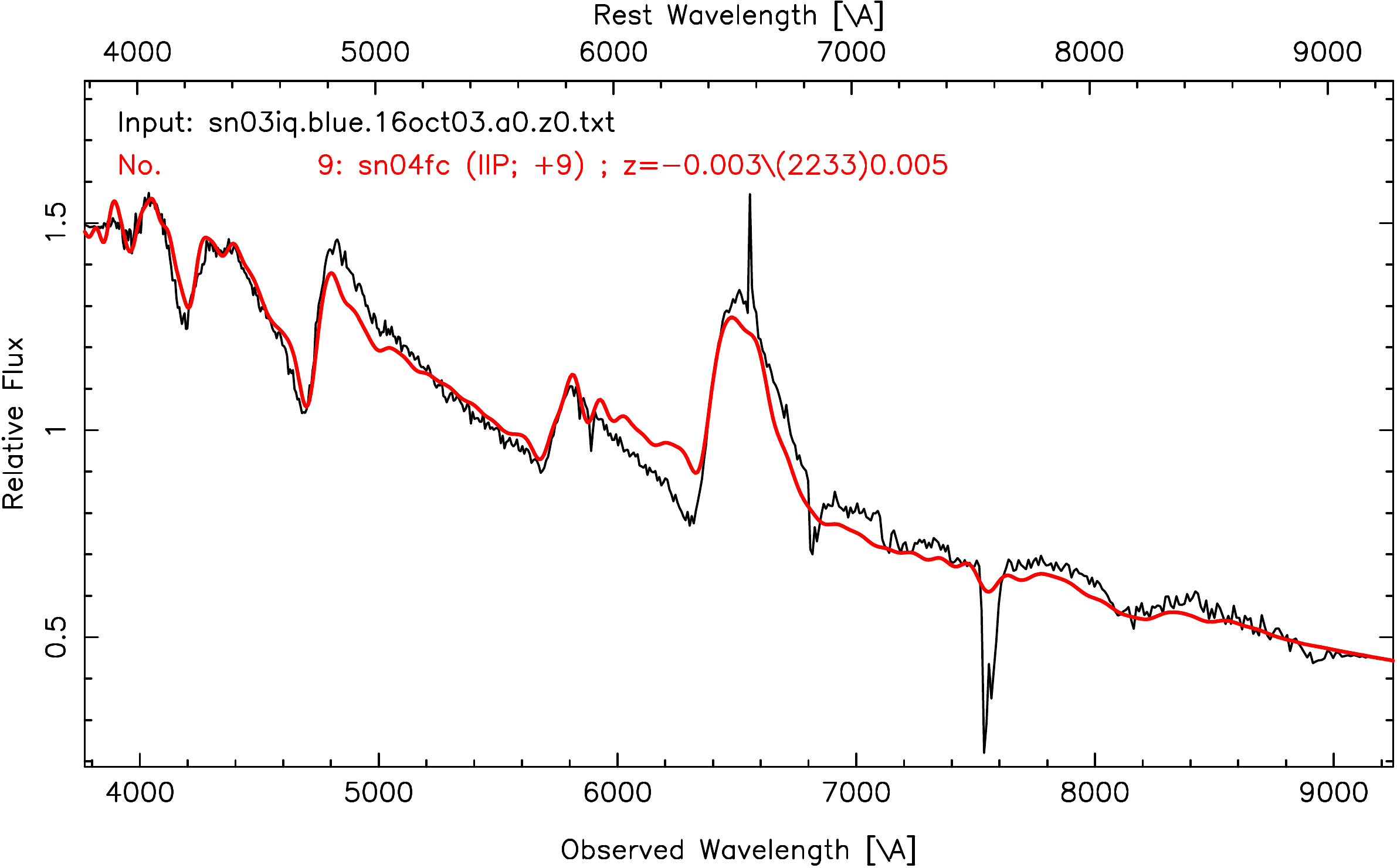}
\caption{Best spectral matching of SN~2003iq using SNID. The plots show SN~2003iq compared with 
SN~2006bp, SN~2004et, SN~1999em, and SN~2004fc at 13, 12, 7, and 9 days from explosion.}
\end{figure}

\begin{figure}[h!]
\centering
\includegraphics[width=4.4cm]{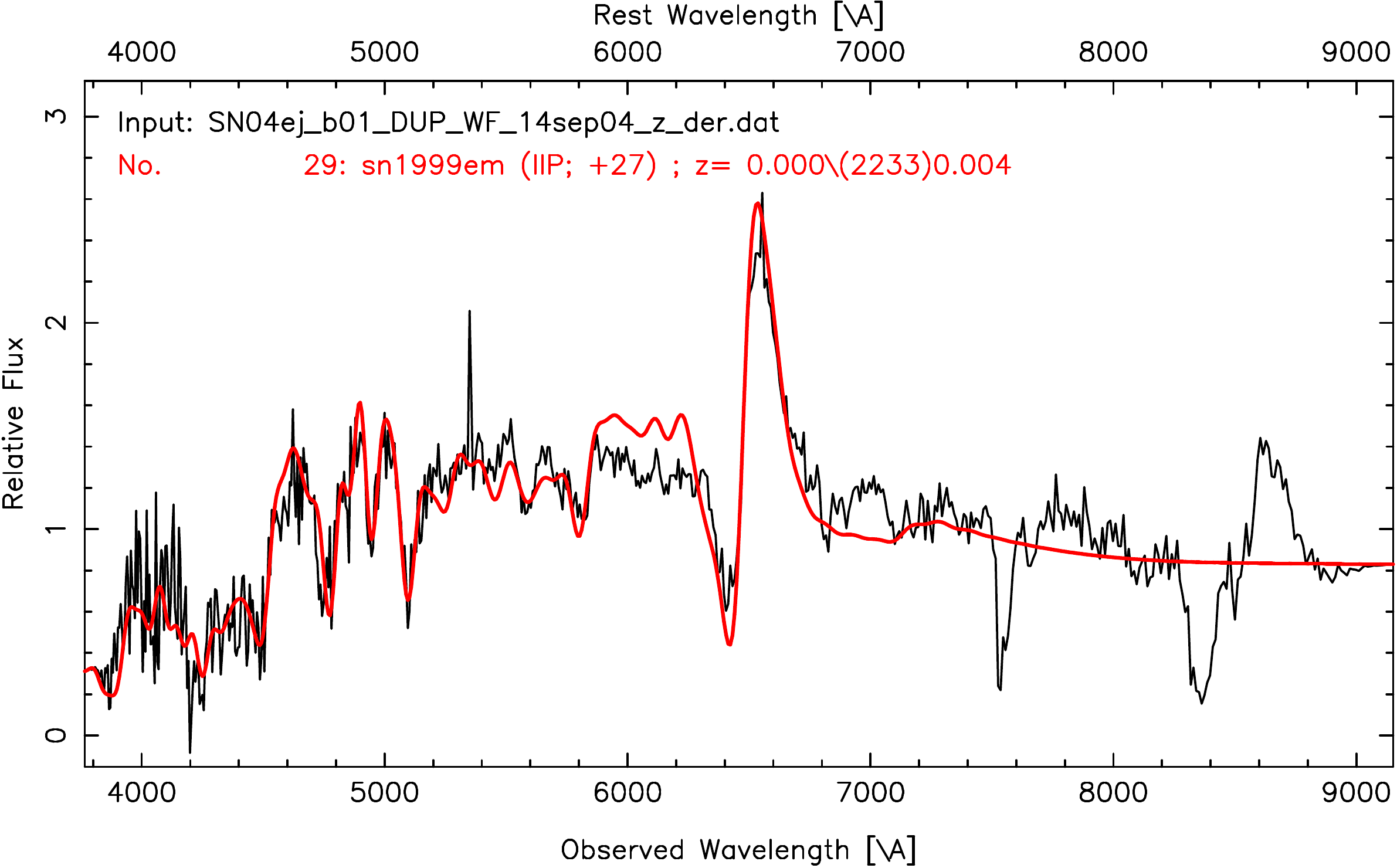}
\includegraphics[width=4.4cm]{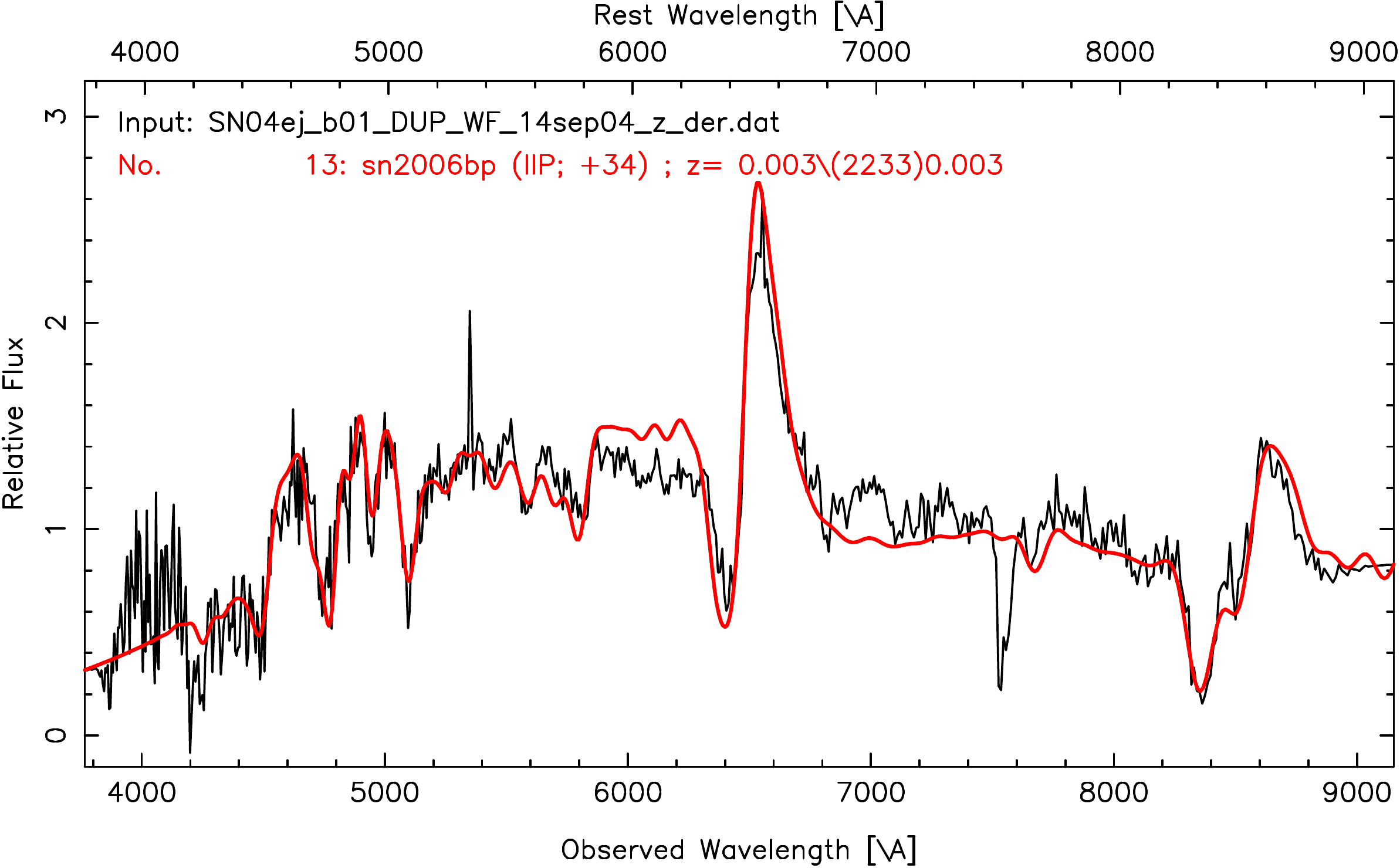}
\includegraphics[width=4.4cm]{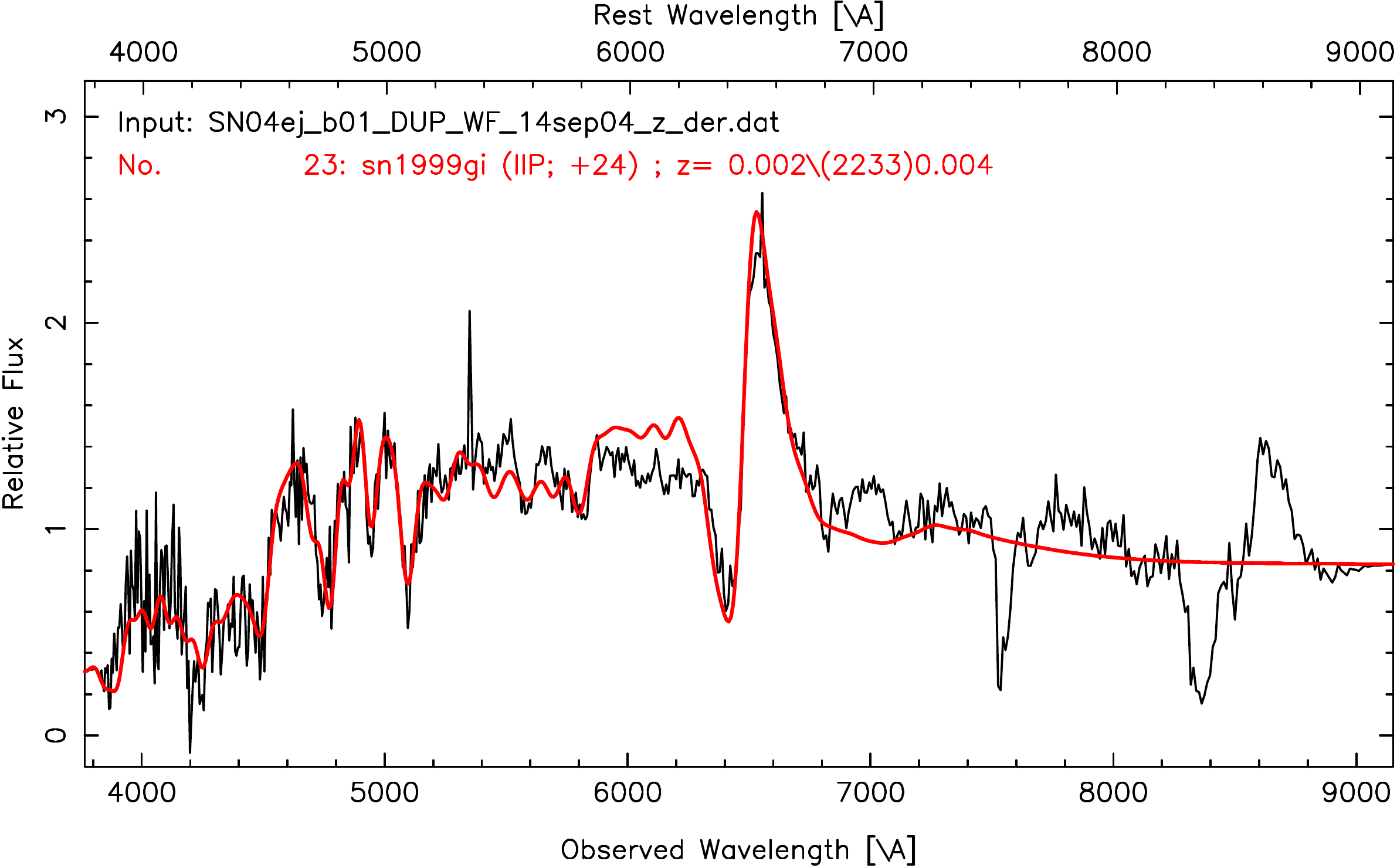}
\includegraphics[width=4.4cm]{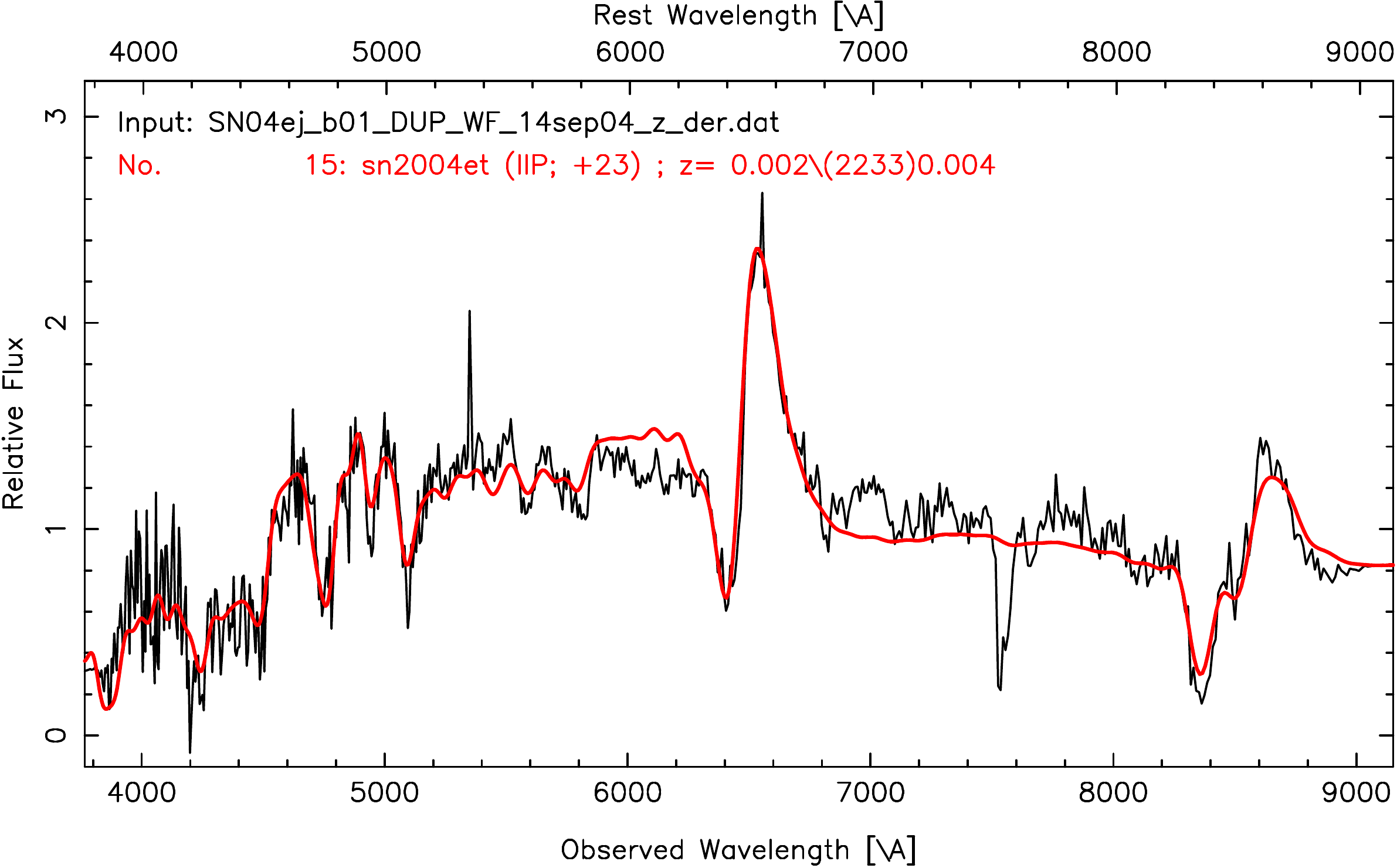}
\caption{Best spectral matching of SN~2004ej using SNID. The plots show SN~2004ej compared with 
SN1999em~, SN~2006bp, SN~1999gi, and SN~2004et at 37, 43, 36, and 39 days from explosion.}
\end{figure}

\clearpage

\begin{figure}[h!]
\centering
\includegraphics[width=4.4cm]{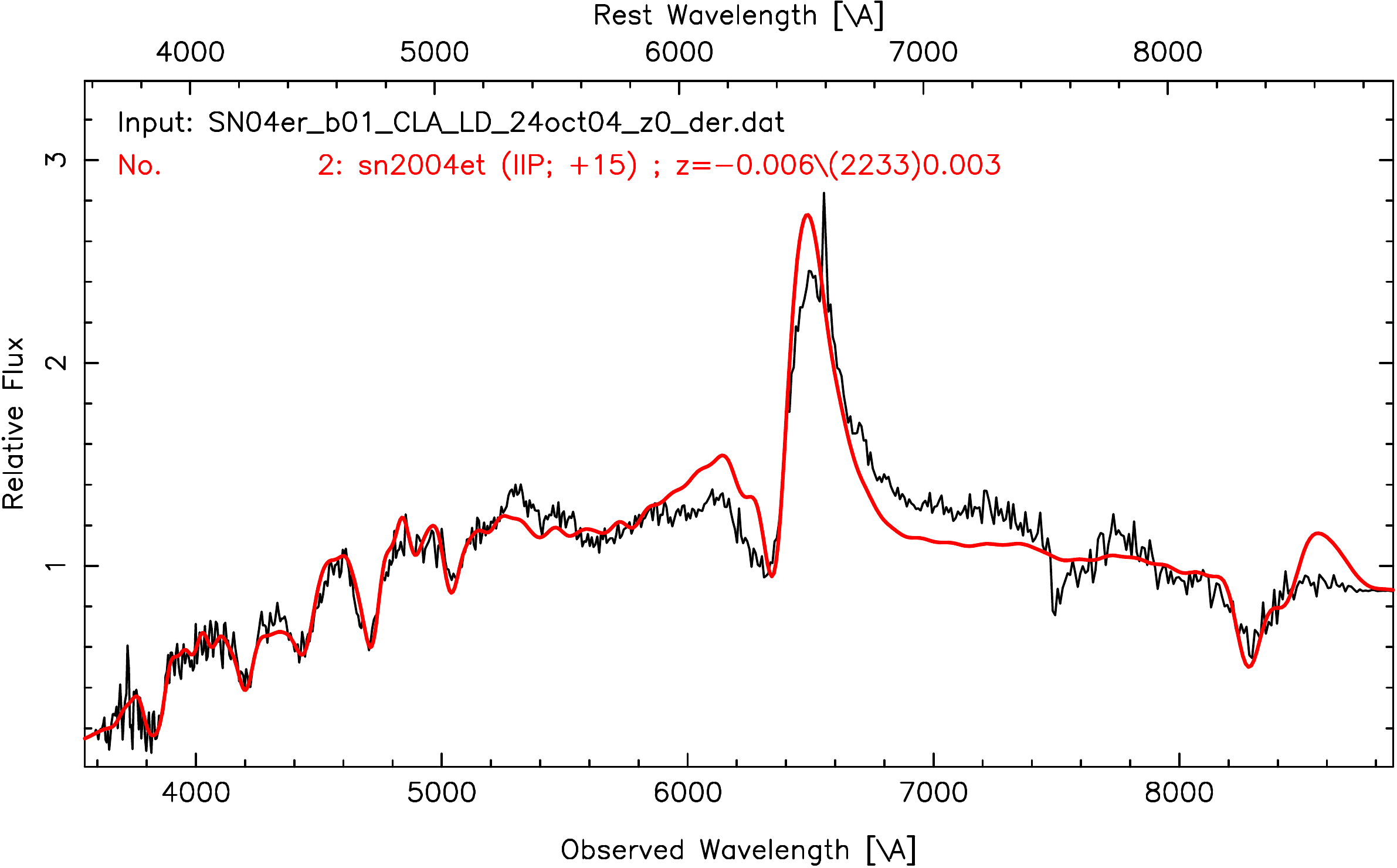}
\includegraphics[width=4.4cm]{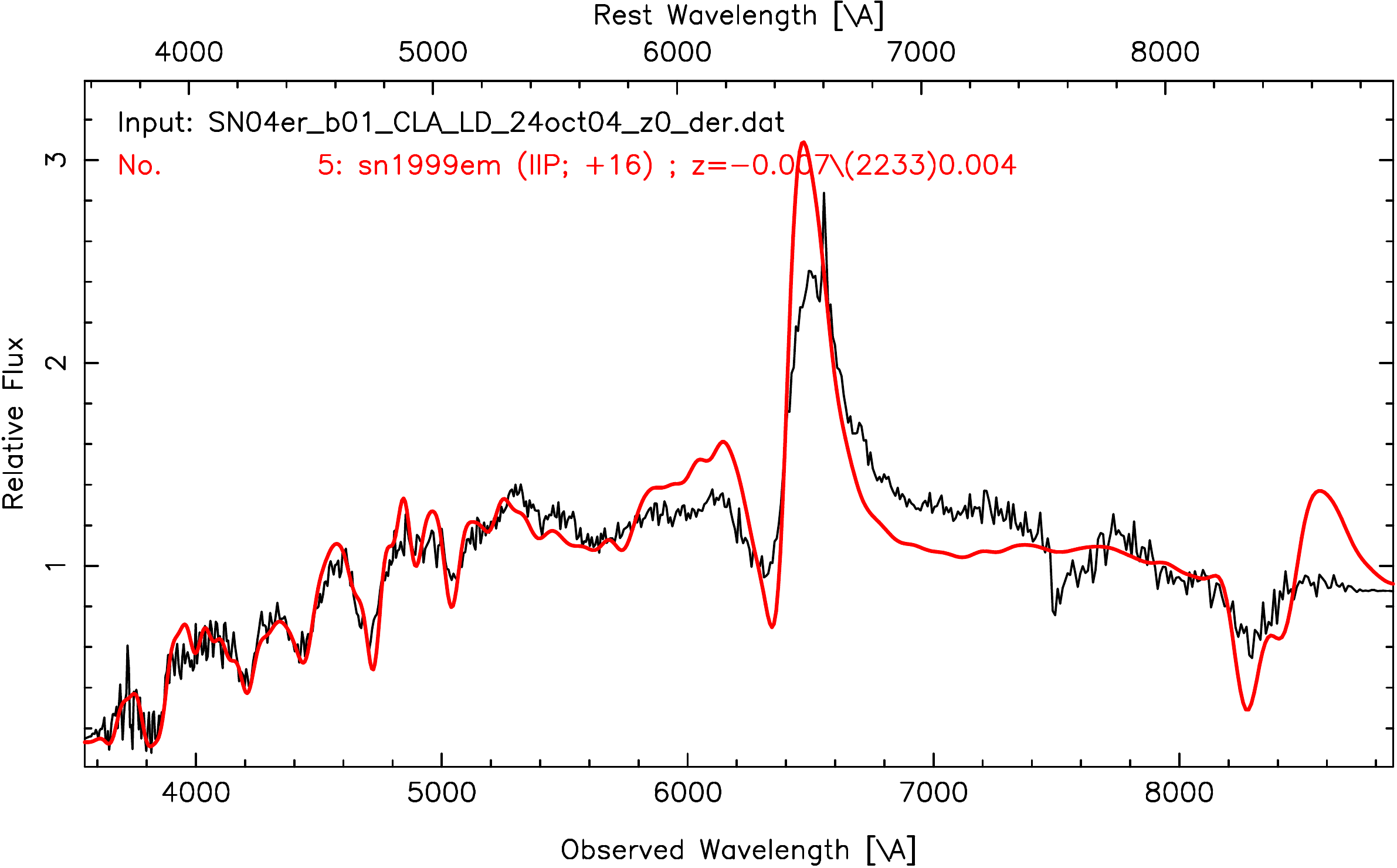}
\includegraphics[width=4.4cm]{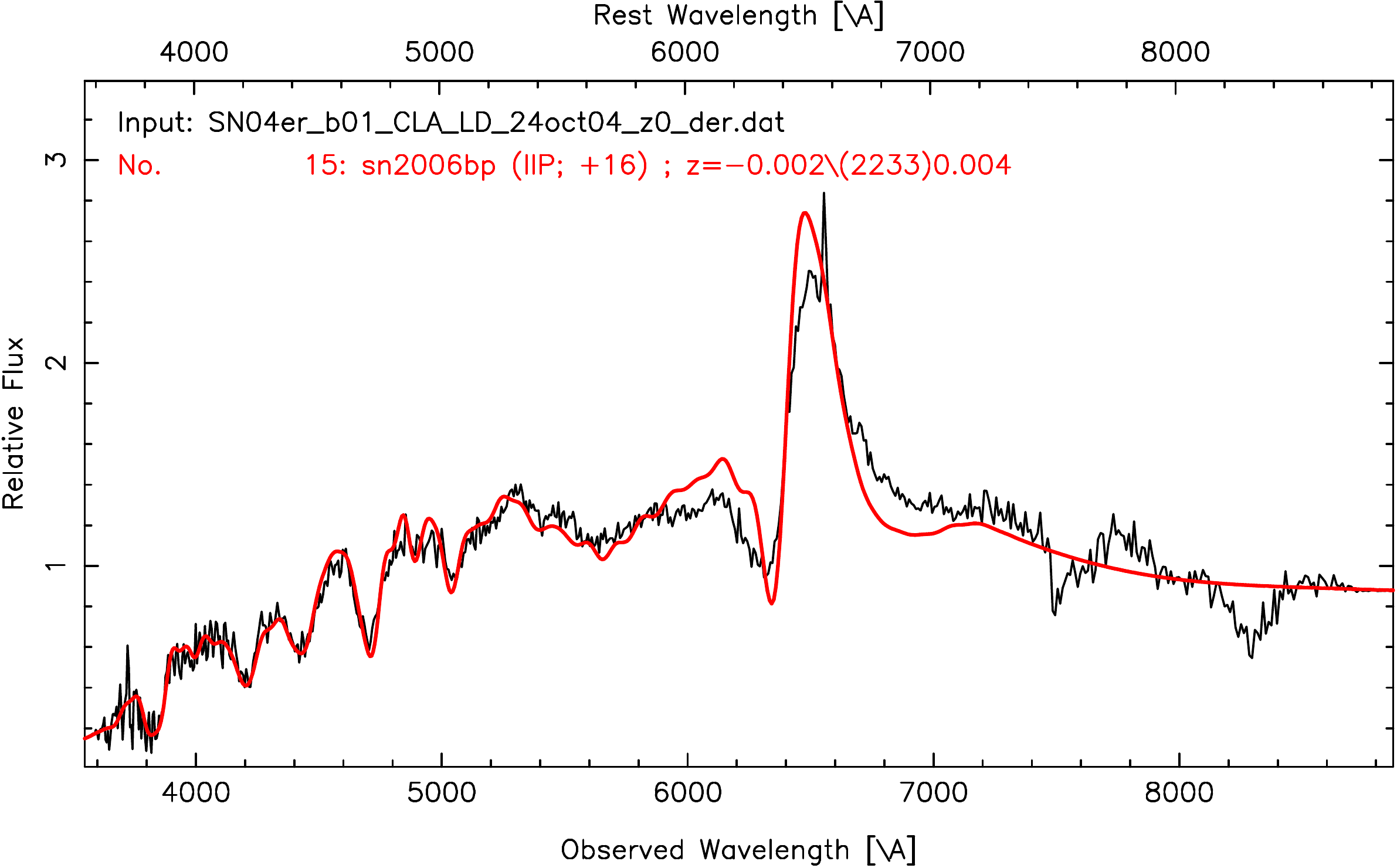}
\caption{Best spectral matching of SN~2004er using SNID. The plots show SN~2004er compared with 
SN~2004et, SN~1999em, and SN~2006bp at 31, 36, and 25 days from explosion.}
\end{figure}

\begin{figure}[h!]
\centering
\includegraphics[width=4.4cm]{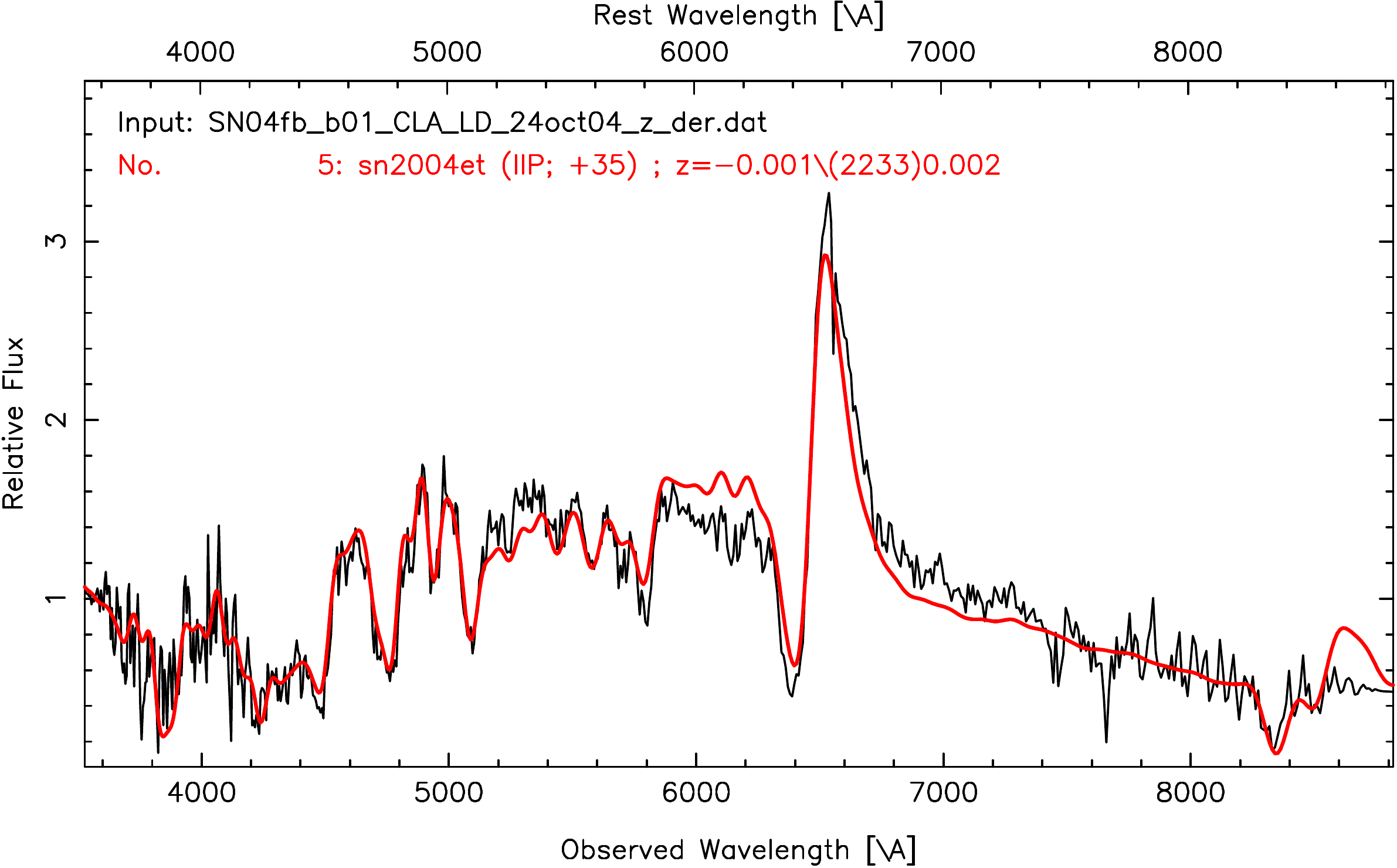}
\includegraphics[width=4.4cm]{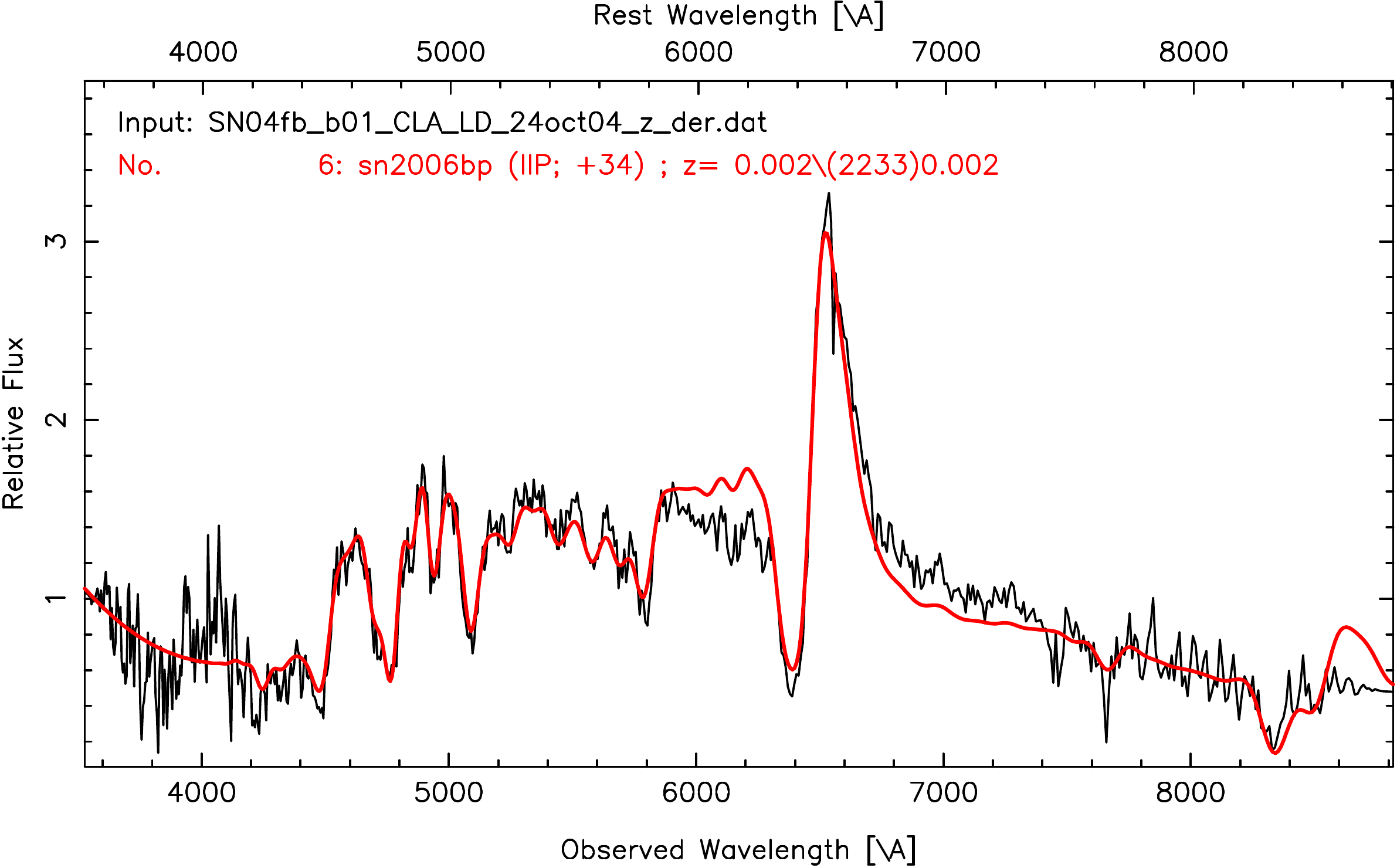}
\includegraphics[width=4.4cm]{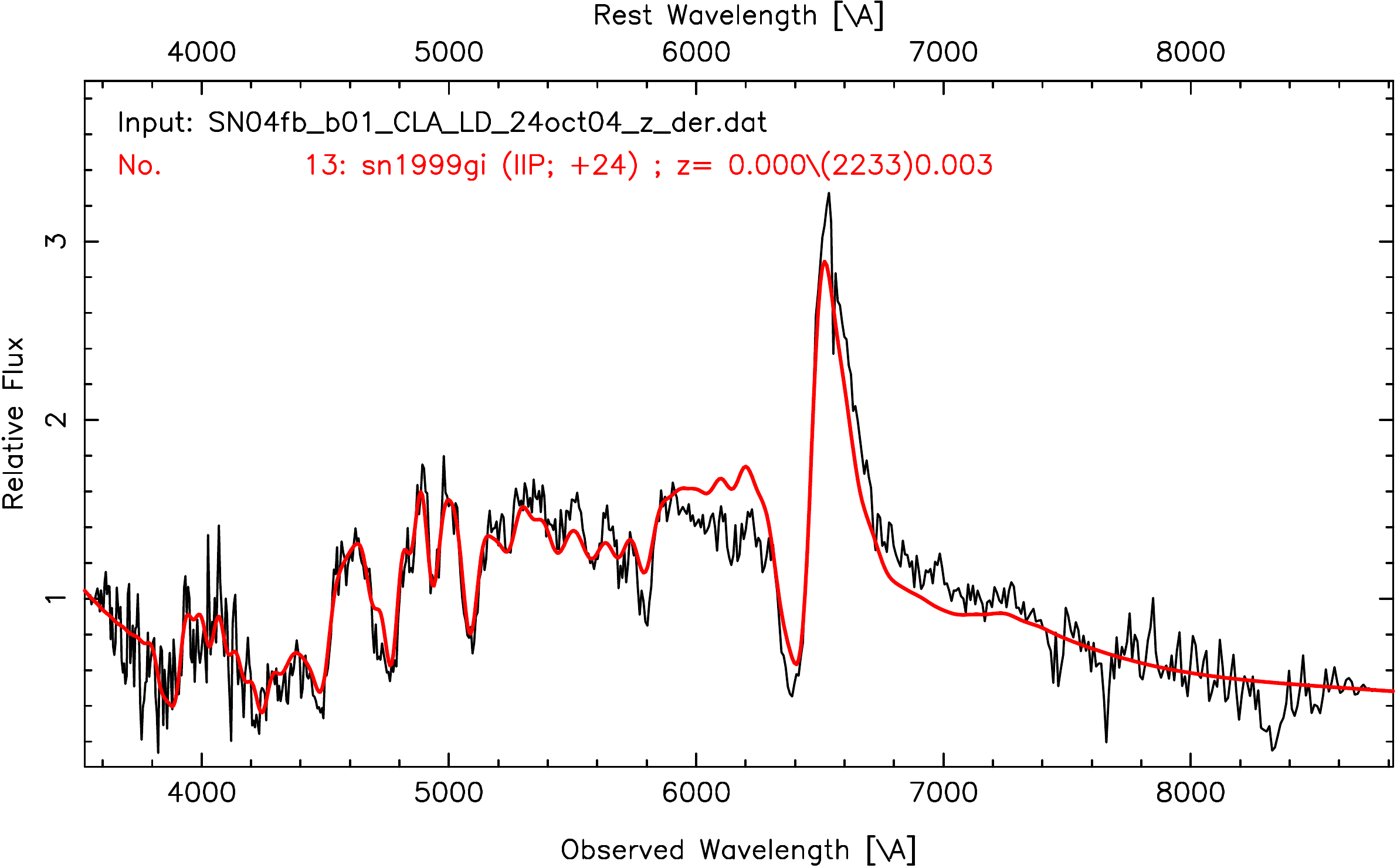}
\includegraphics[width=4.4cm]{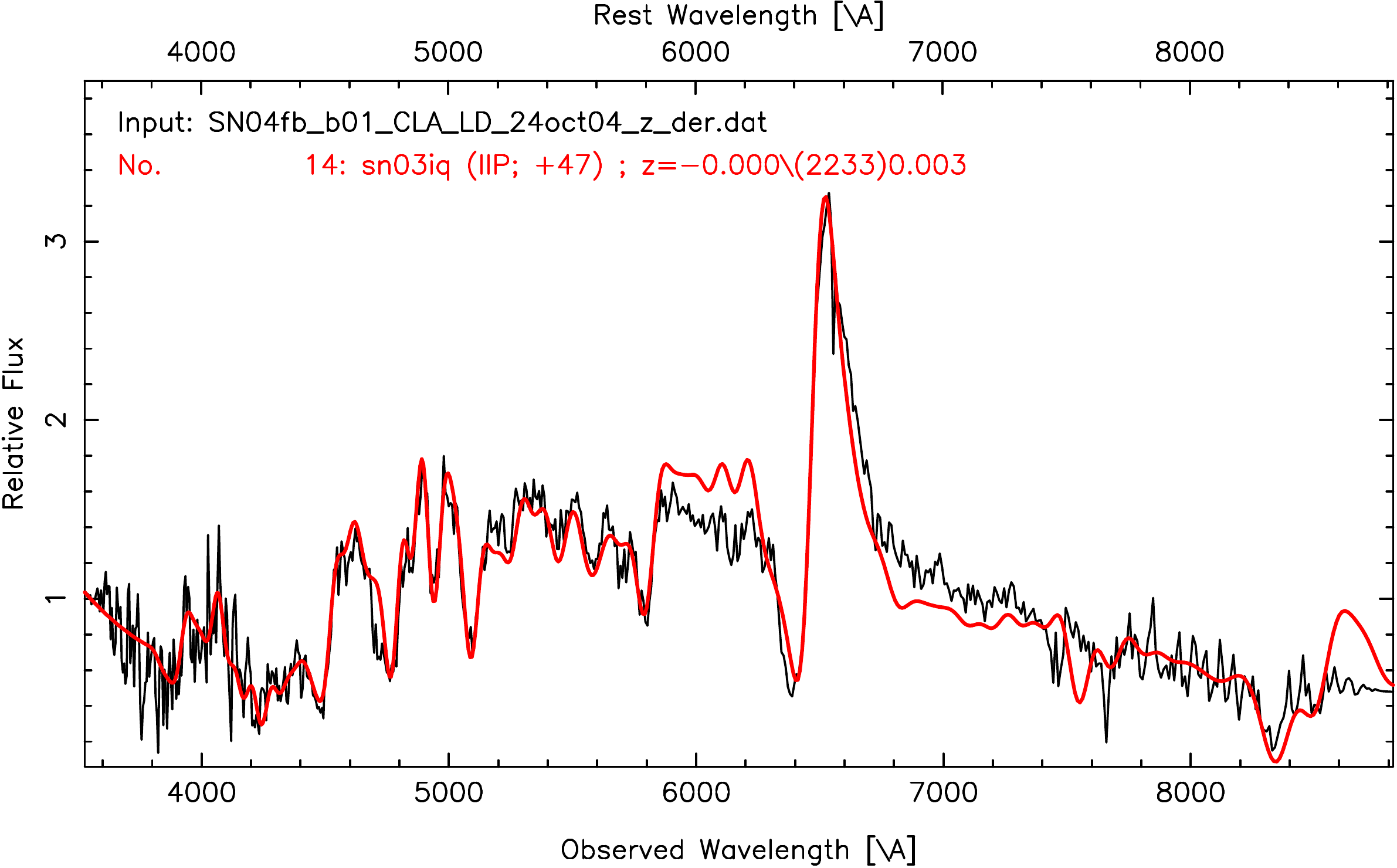}
\includegraphics[width=4.4cm]{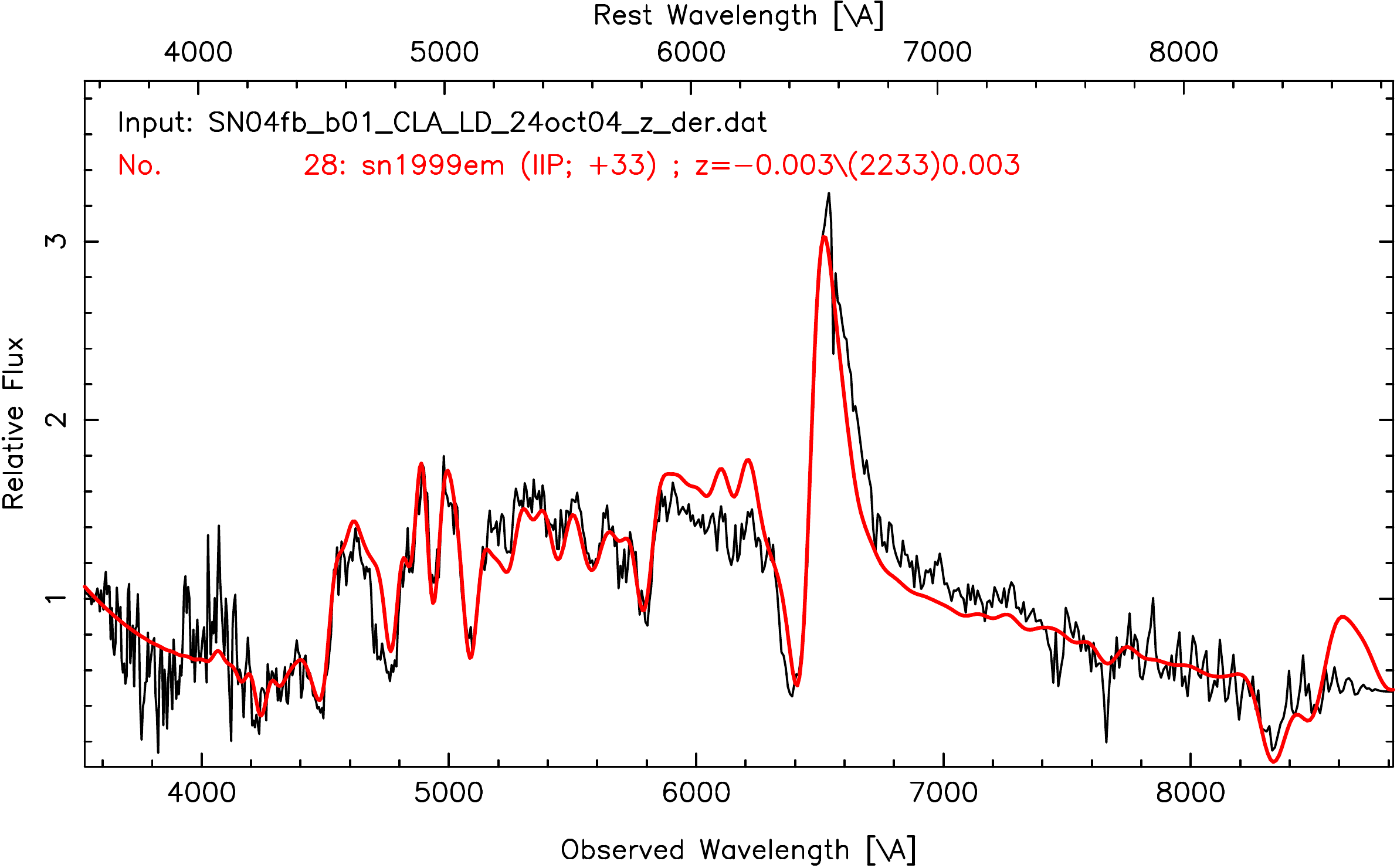}
\caption{Best spectral matching of SN~2004fb using SNID. The plots show SN~2004fb compared with 
SN~2004et, SN~2006bp, SN~1999gi, SN~2003iq, and SN~1999em at 51, 43, 36, 47, and 43 days from explosion.}
\end{figure}

\clearpage

\begin{figure}
\centering
\includegraphics[width=4.4cm]{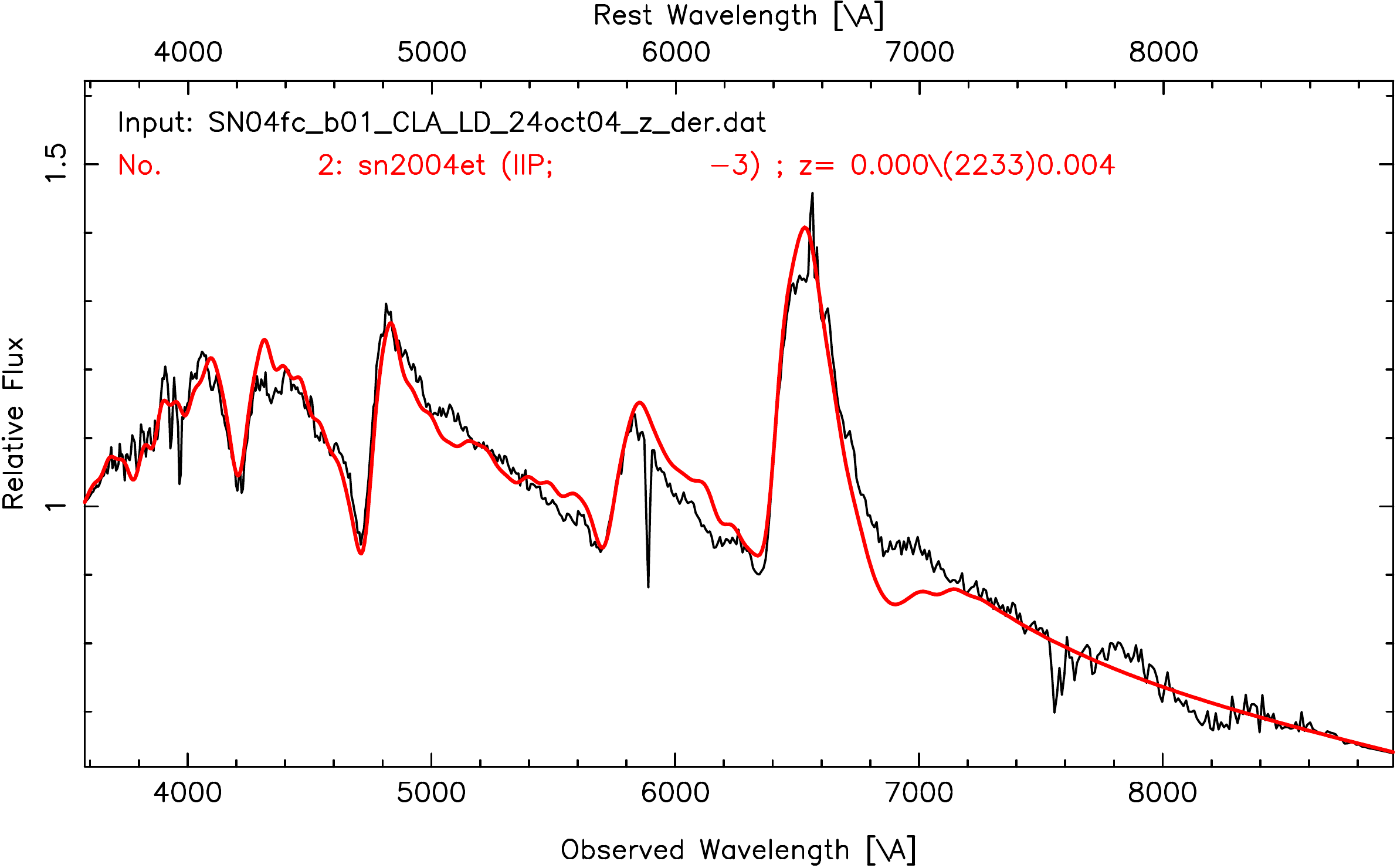}
\includegraphics[width=4.4cm]{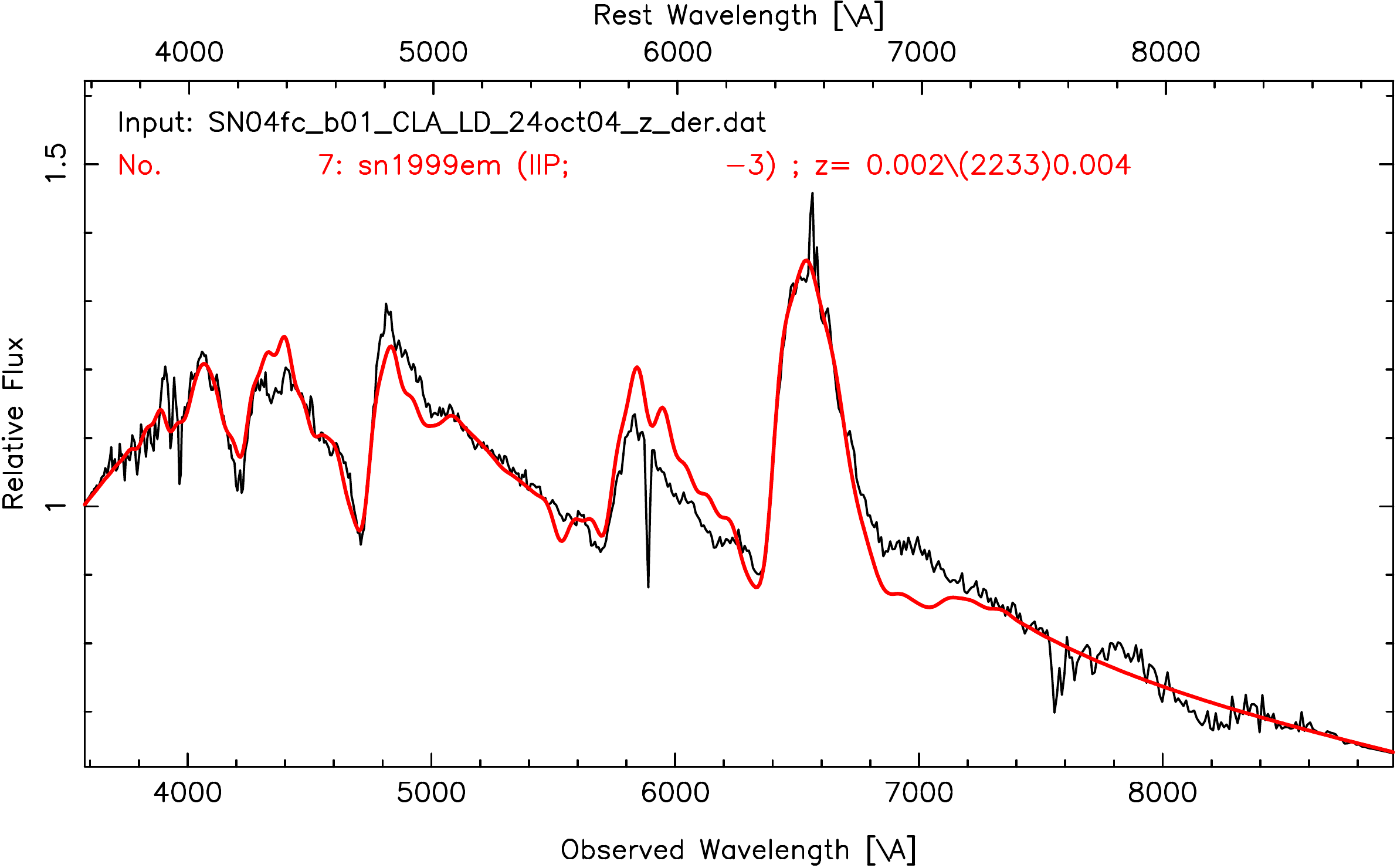}
\includegraphics[width=4.4cm]{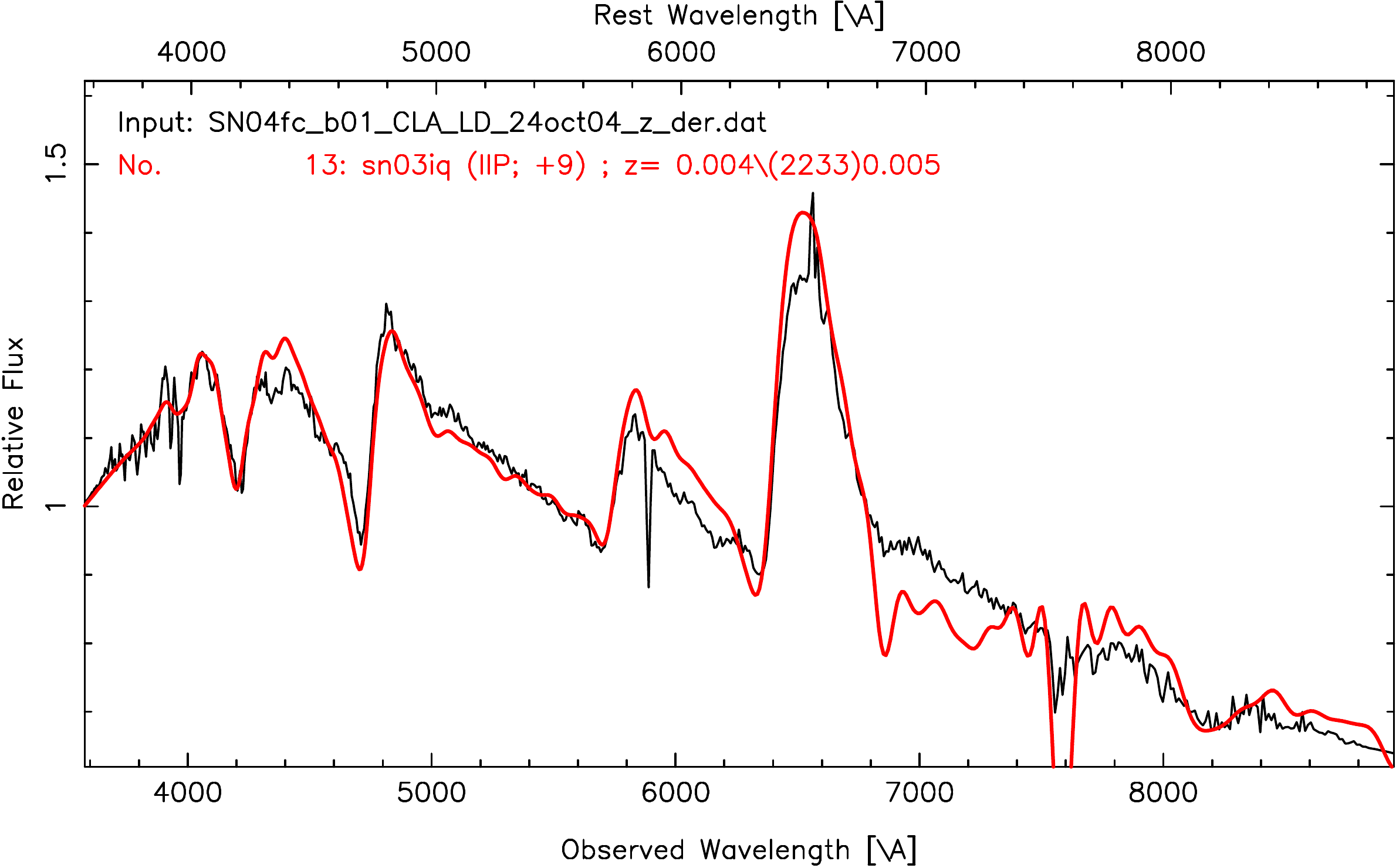}
\includegraphics[width=4.4cm]{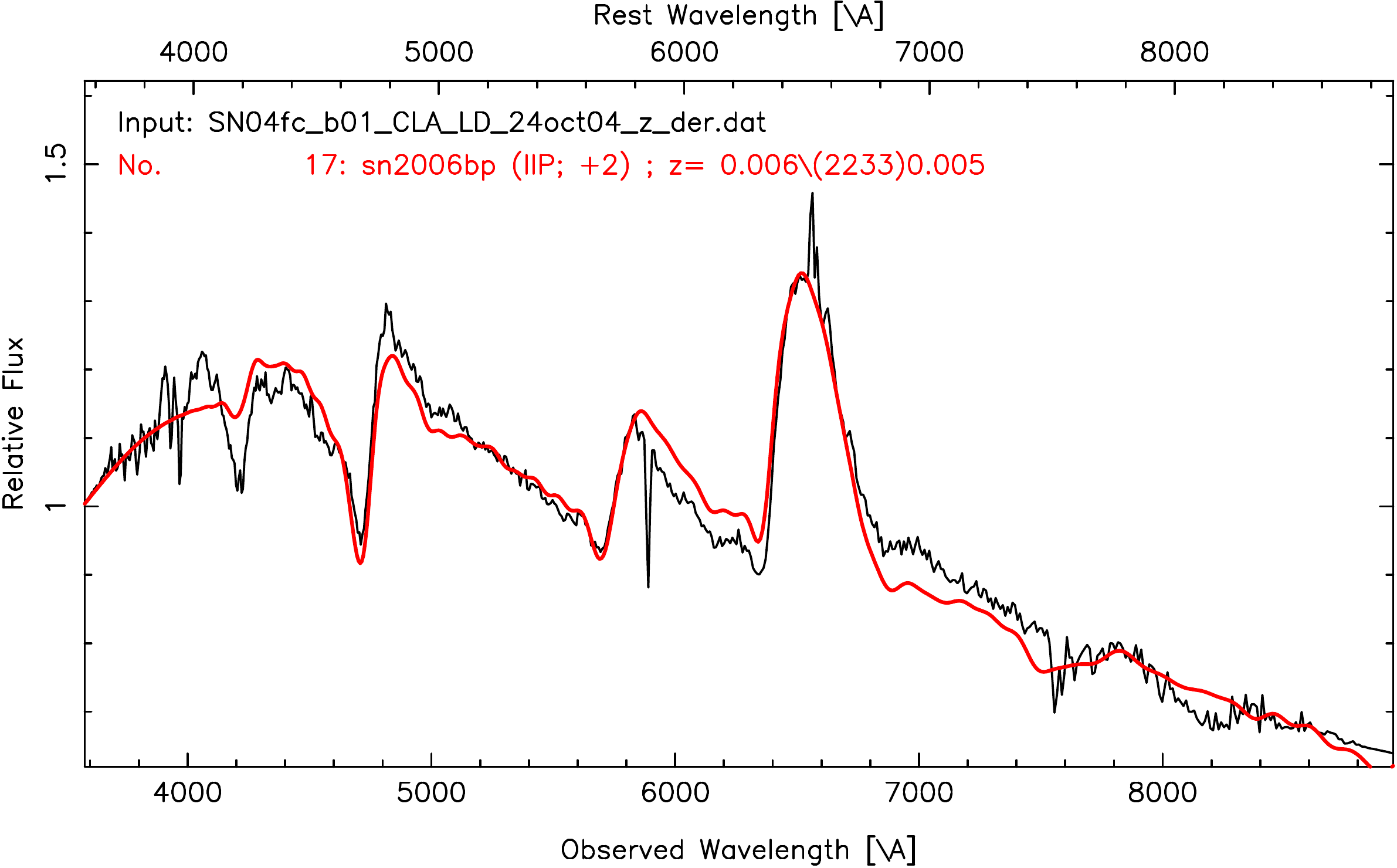}
\caption{Best spectral matching of SN~2004fc using SNID. The plots show SN~2004fc compared with 
SN~2004et, SN~1999em, SN~2003iq, and SN~2006bp at 13, 7, 9, and 11 days from explosion.}
\end{figure}

\begin{figure}
\centering
\includegraphics[width=4.4cm]{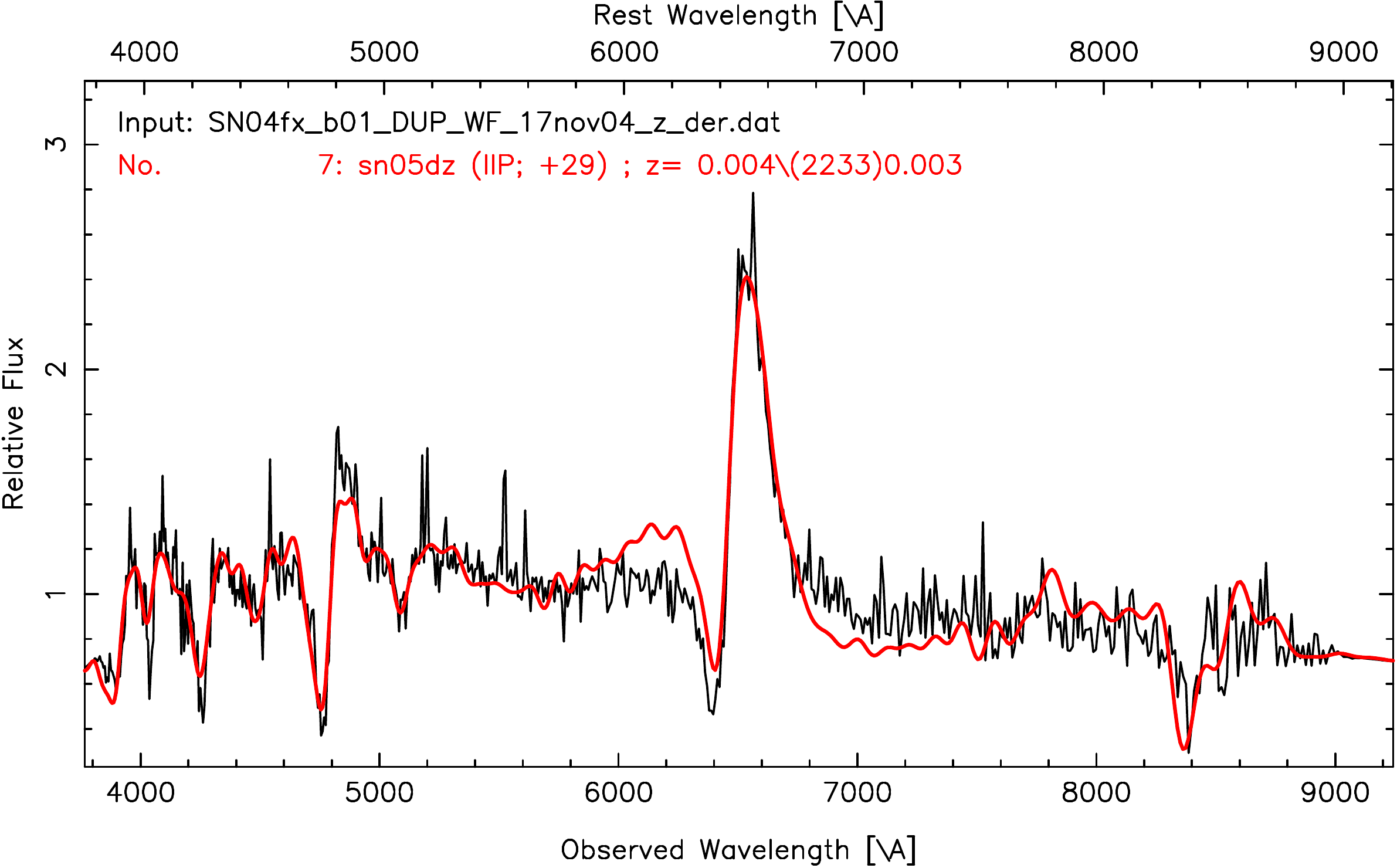}
\includegraphics[width=4.4cm]{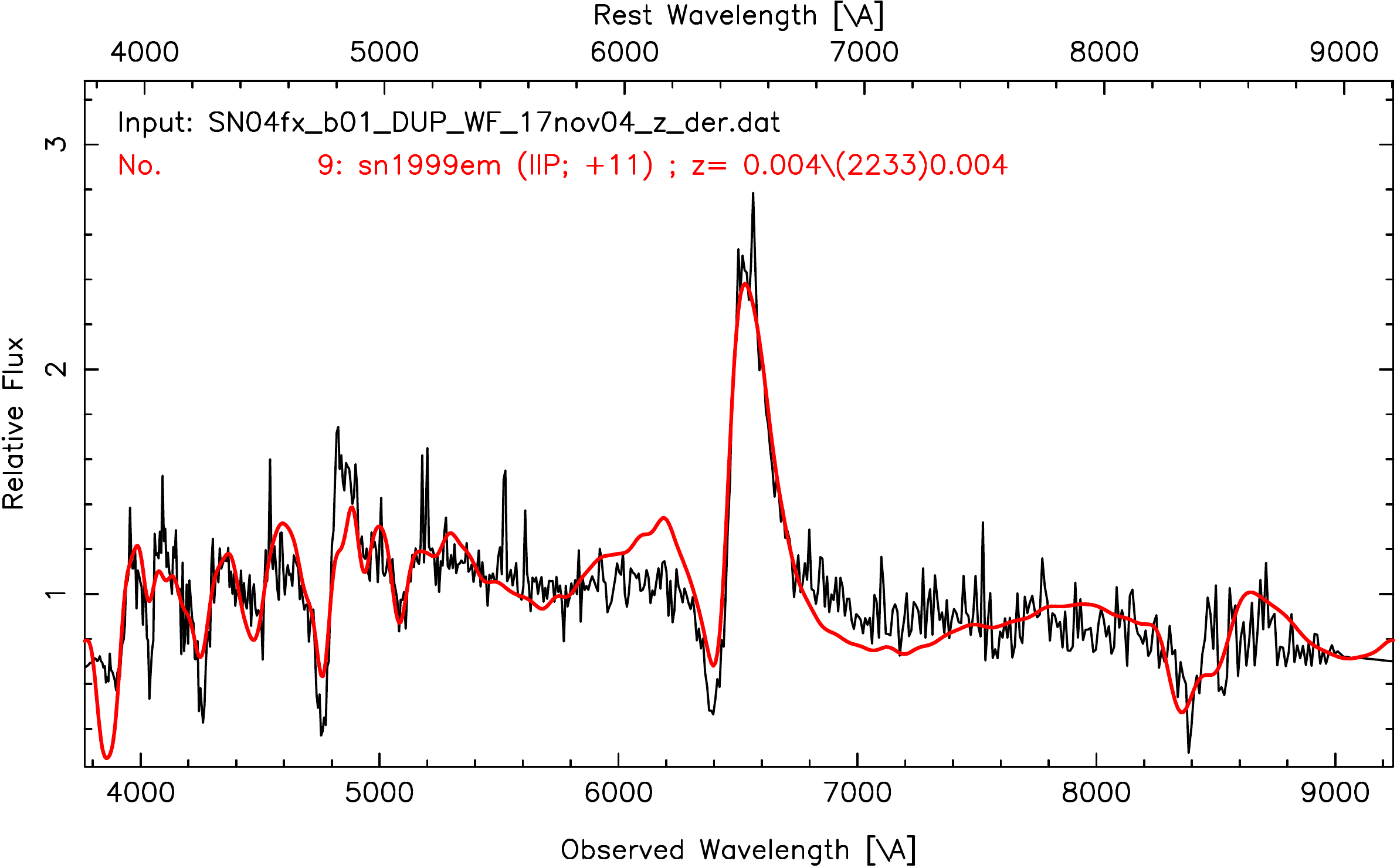}
\includegraphics[width=4.4cm]{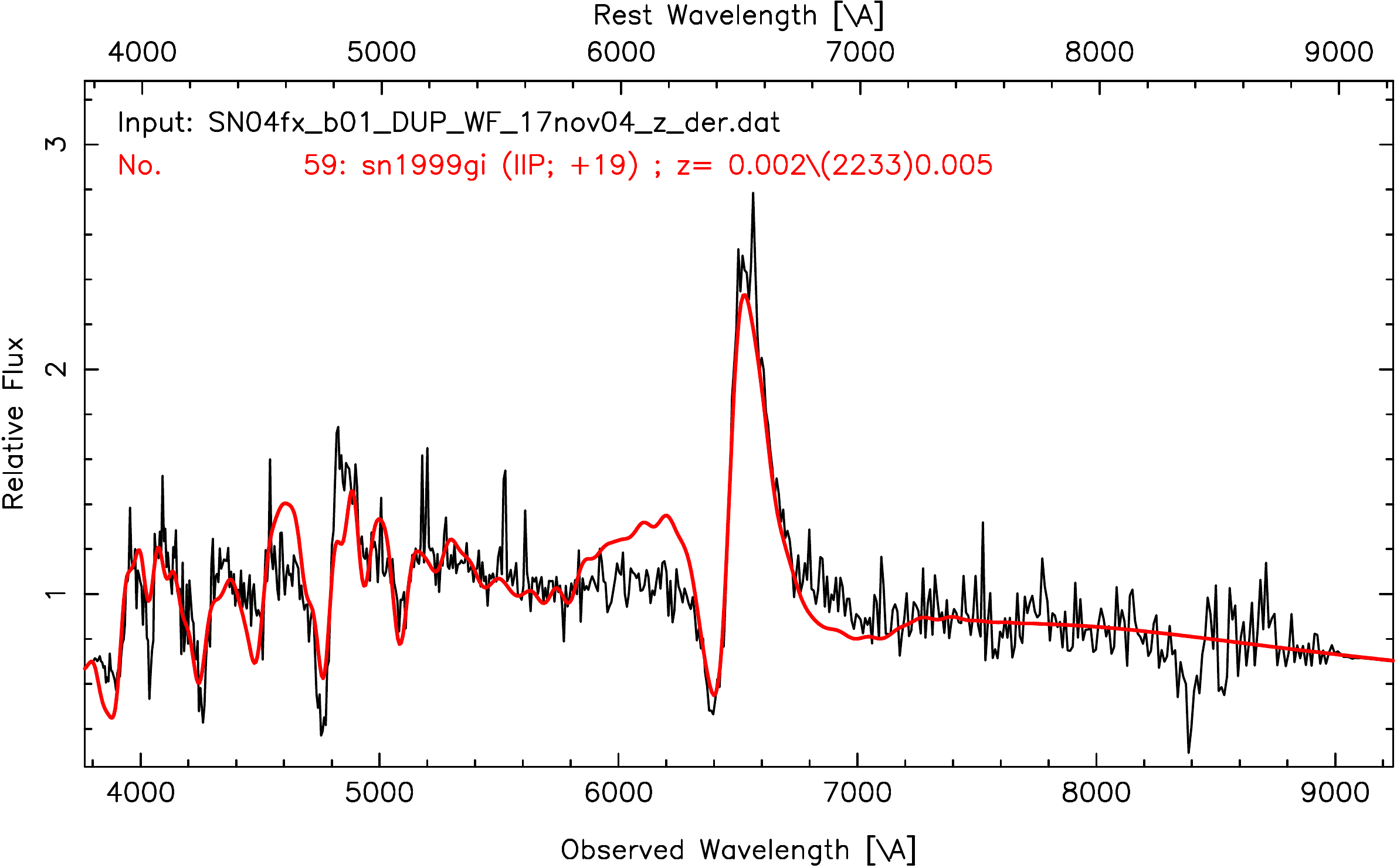}
\includegraphics[width=4.4cm]{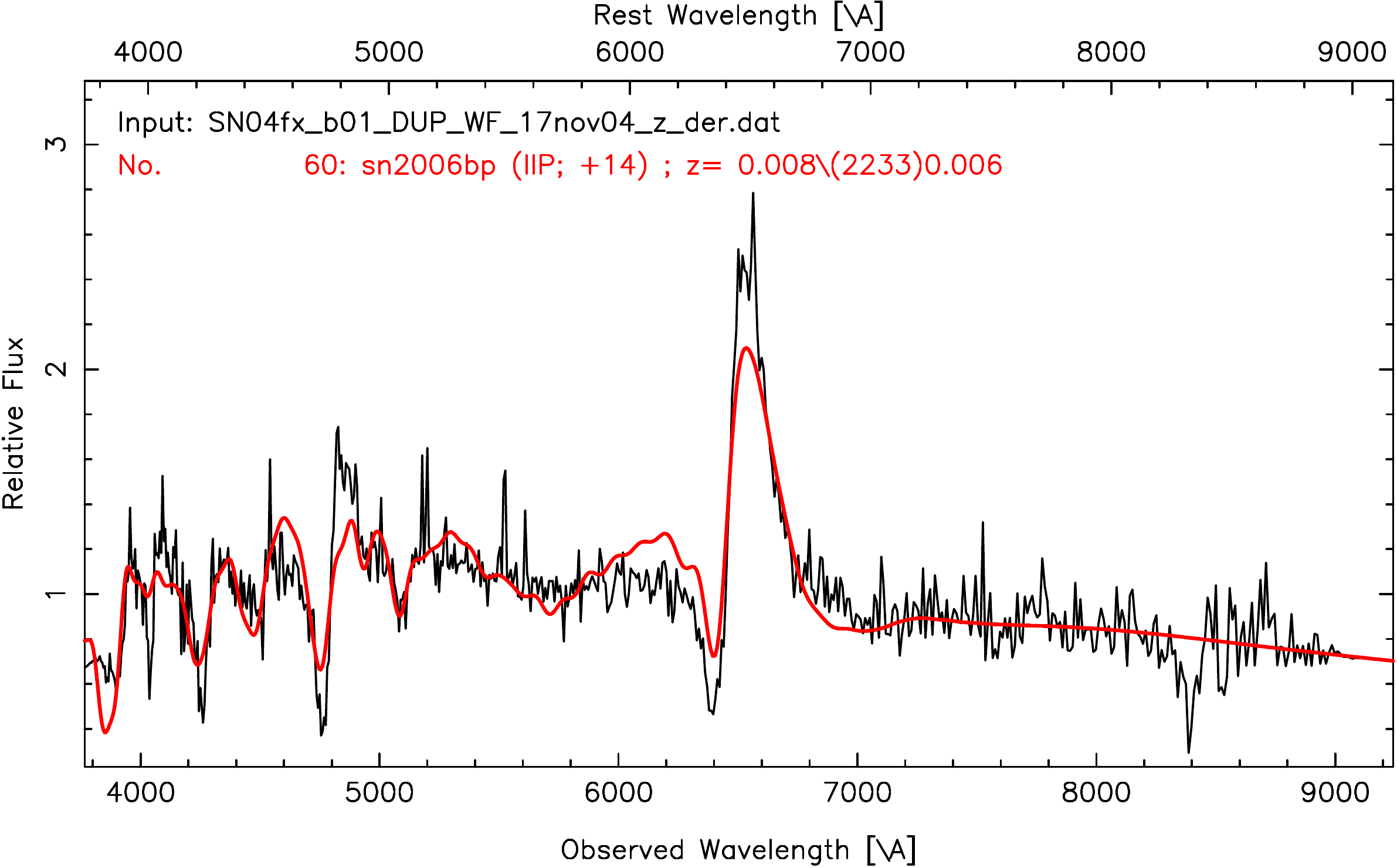}
\caption{Best spectral matching of SN~2004fx using SNID. The plots show SN~2004fx compared with 
SN~2005dz, SN~1999em, SN~1999gi, and SN~2006bp at 29, 21, 31, and 23 days from explosion.}
\end{figure}

\clearpage

\begin{figure}
\centering
\includegraphics[width=4.4cm]{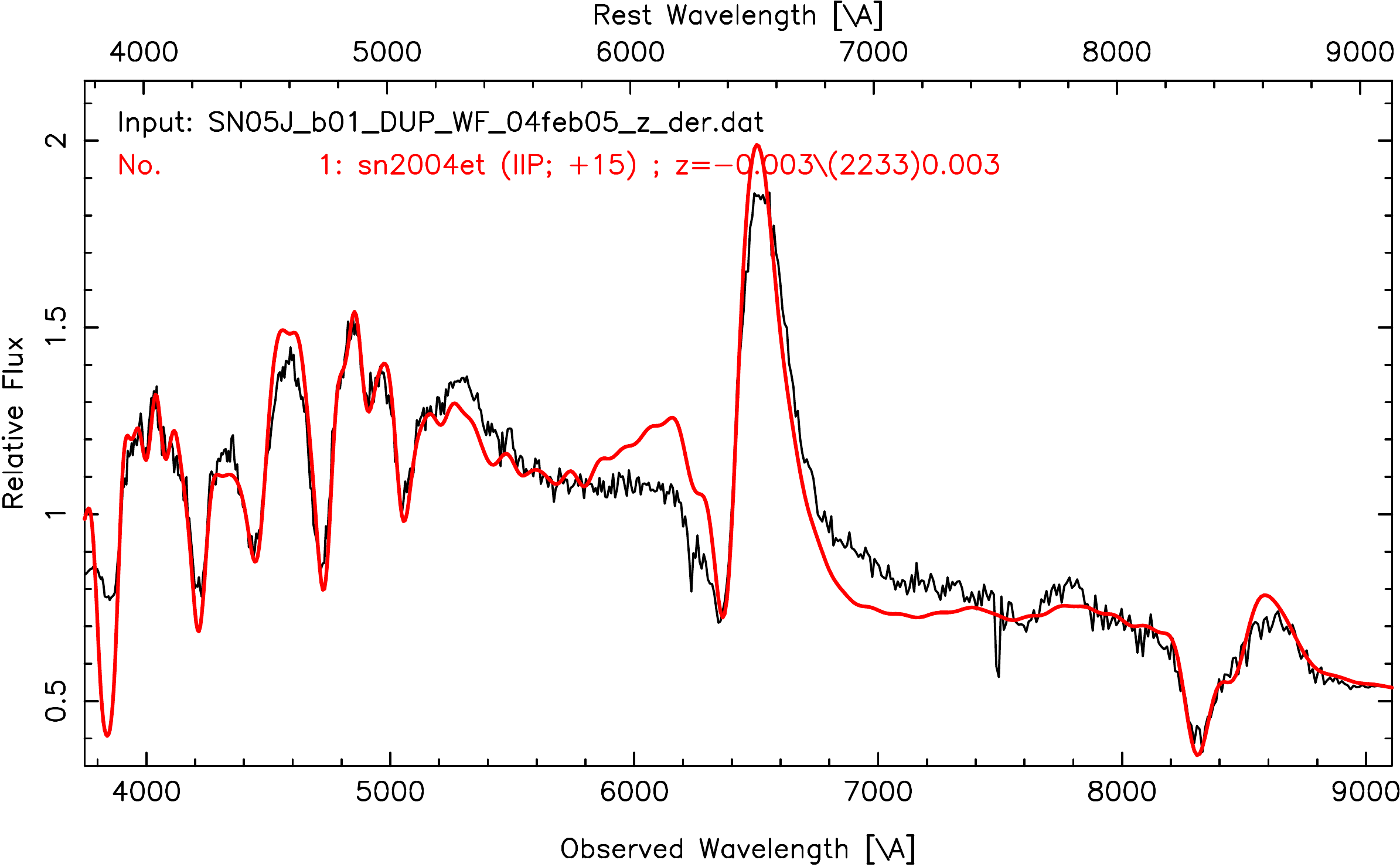}
\includegraphics[width=4.4cm]{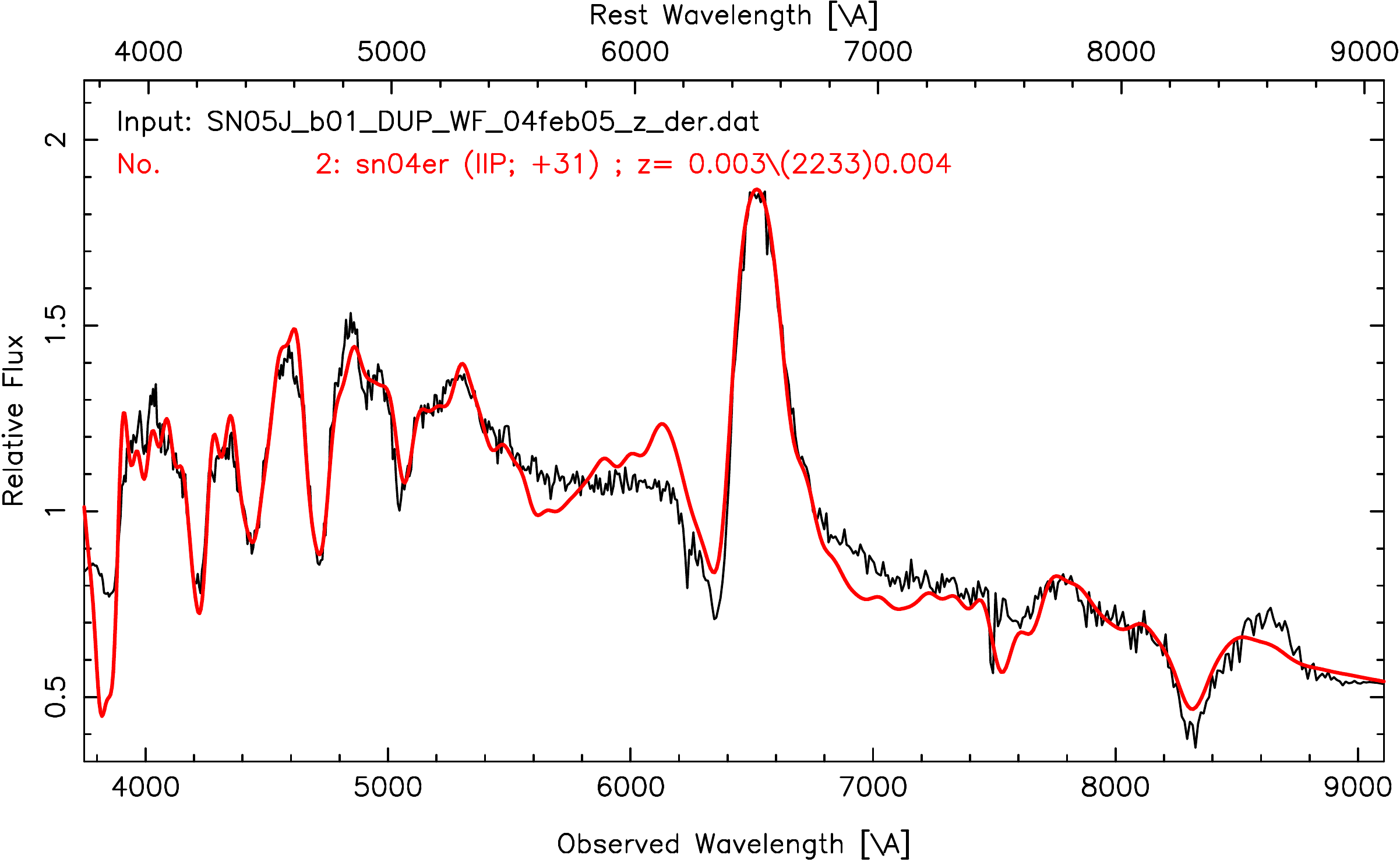}
\includegraphics[width=4.4cm]{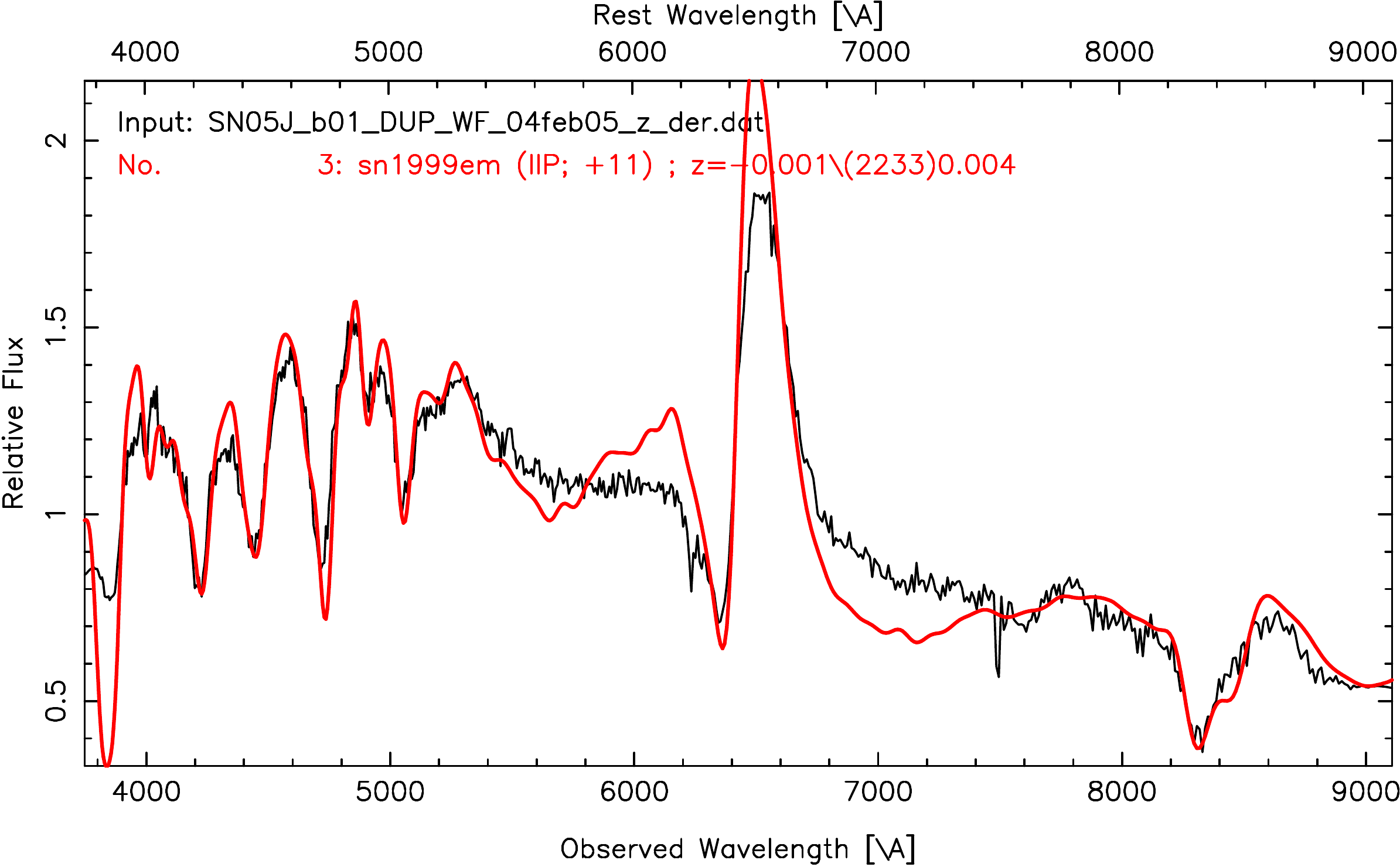}
\includegraphics[width=4.4cm]{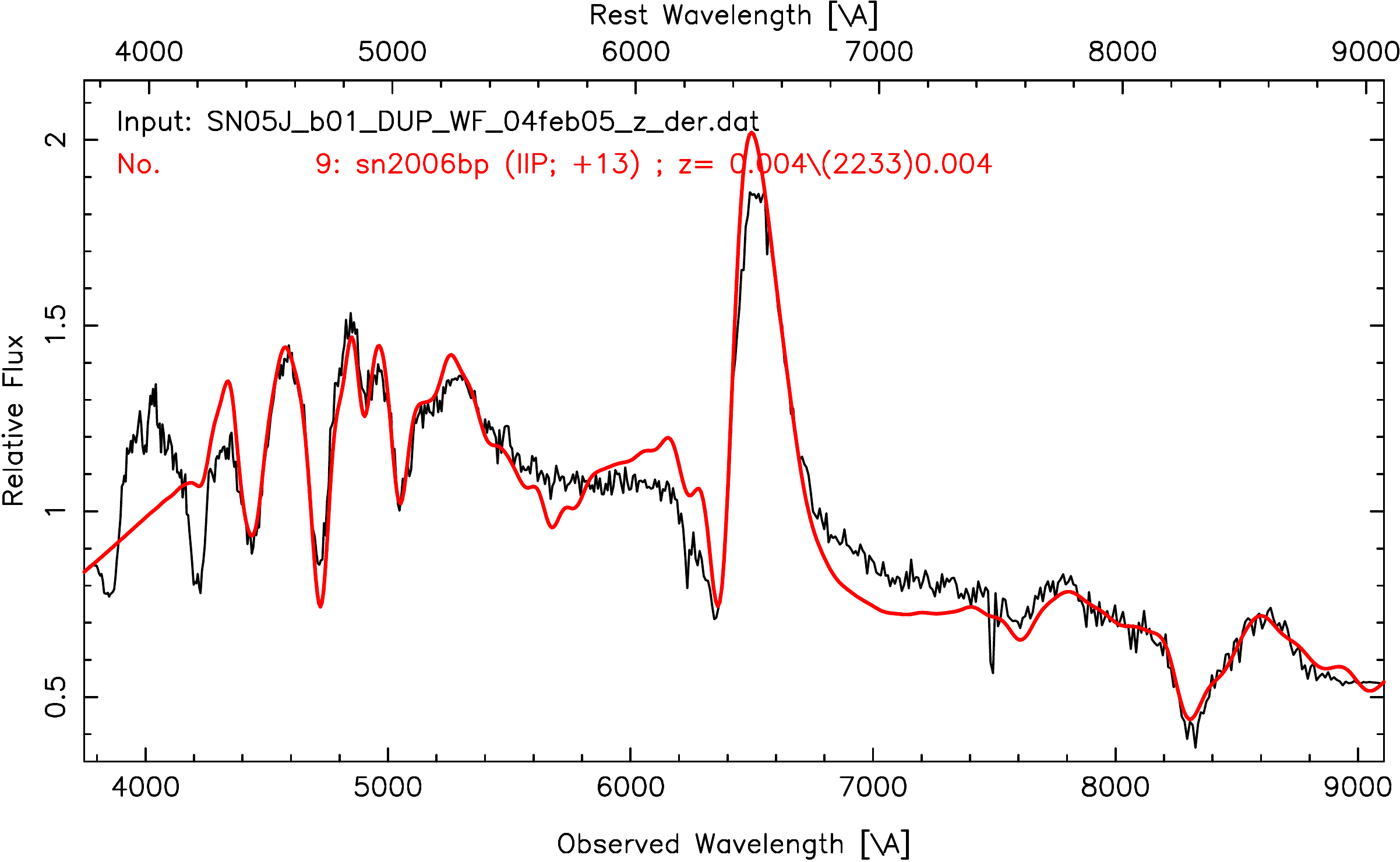}
\caption{Best spectral matching of SN~2005J using SNID. The plots show SN~2005J compared with 
SN~2004et, SN~2004er, SN~1999em, and SN~2006bp at 31, 31, 21, and 22 days from explosion.}
\end{figure}

\begin{figure}
\centering
\includegraphics[width=4.4cm]{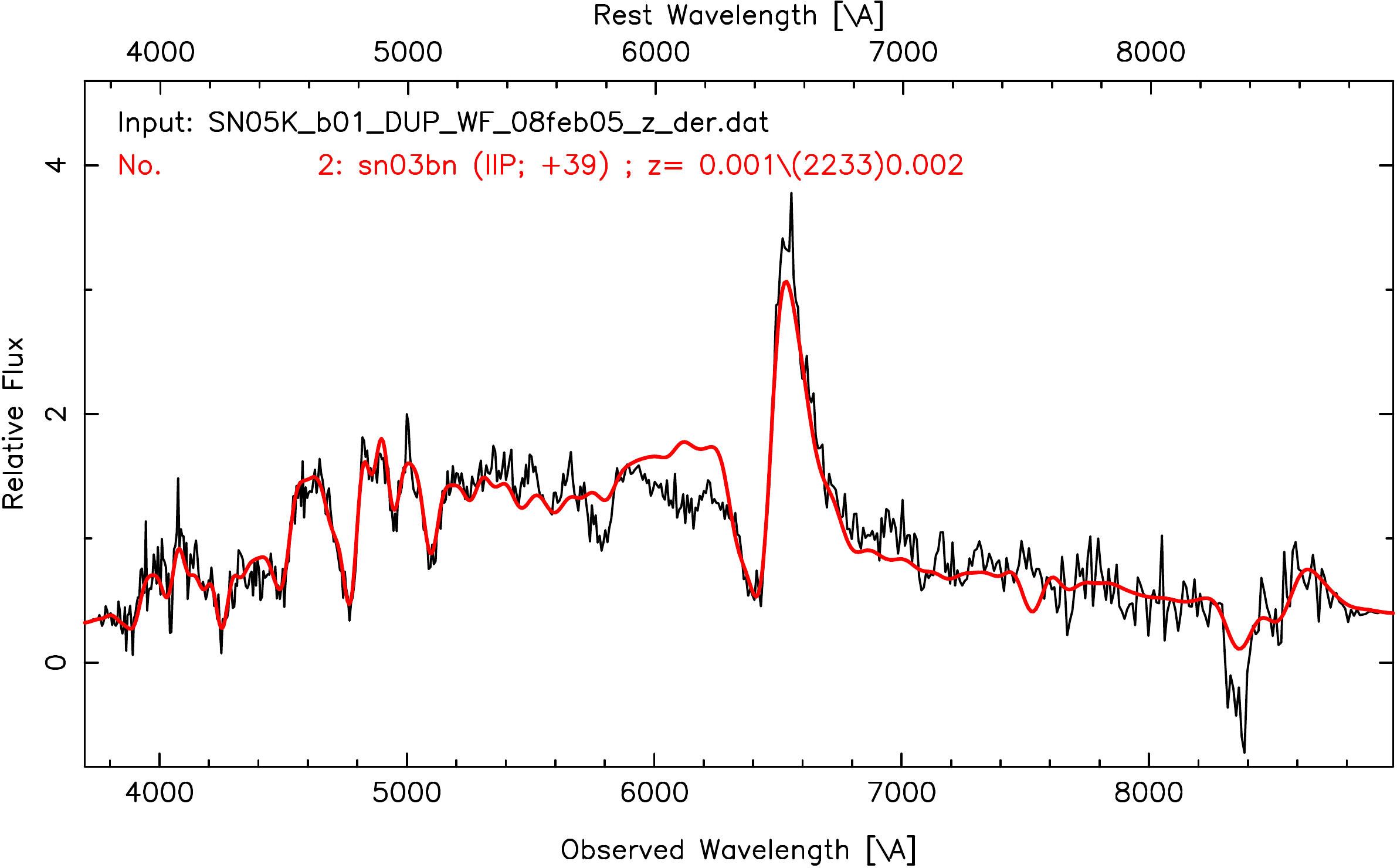}
\includegraphics[width=4.4cm]{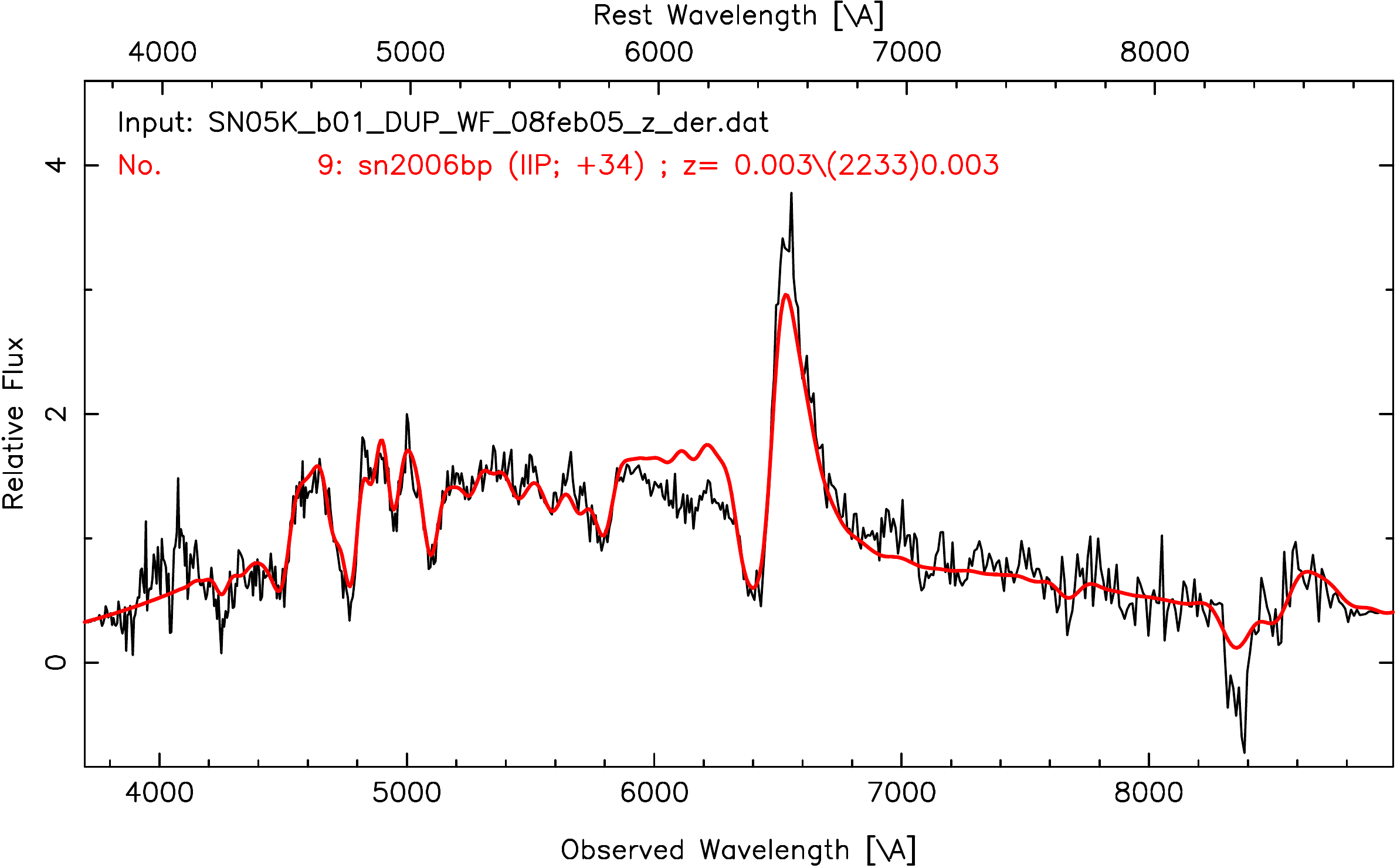}
\includegraphics[width=4.4cm]{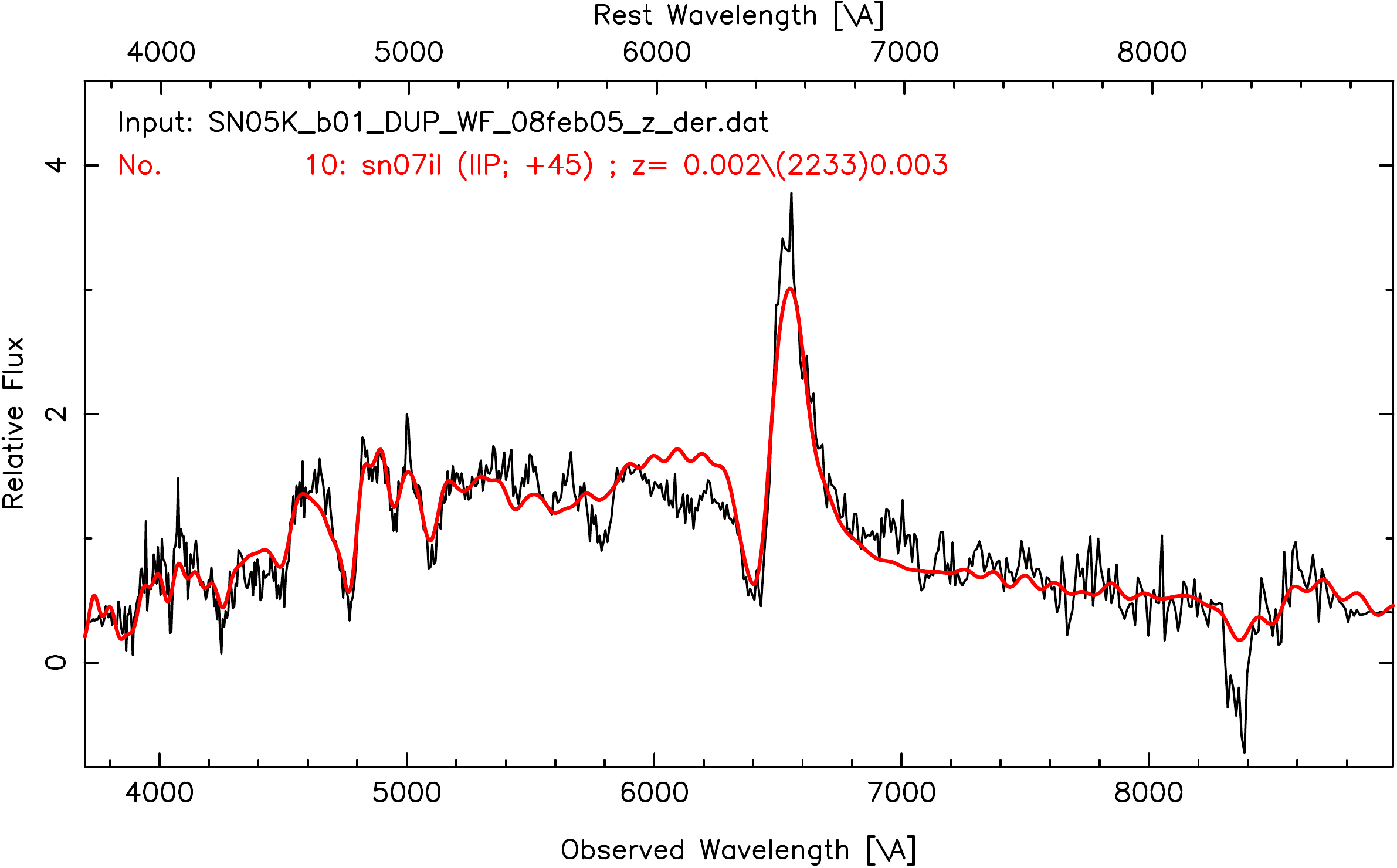}
\includegraphics[width=4.4cm]{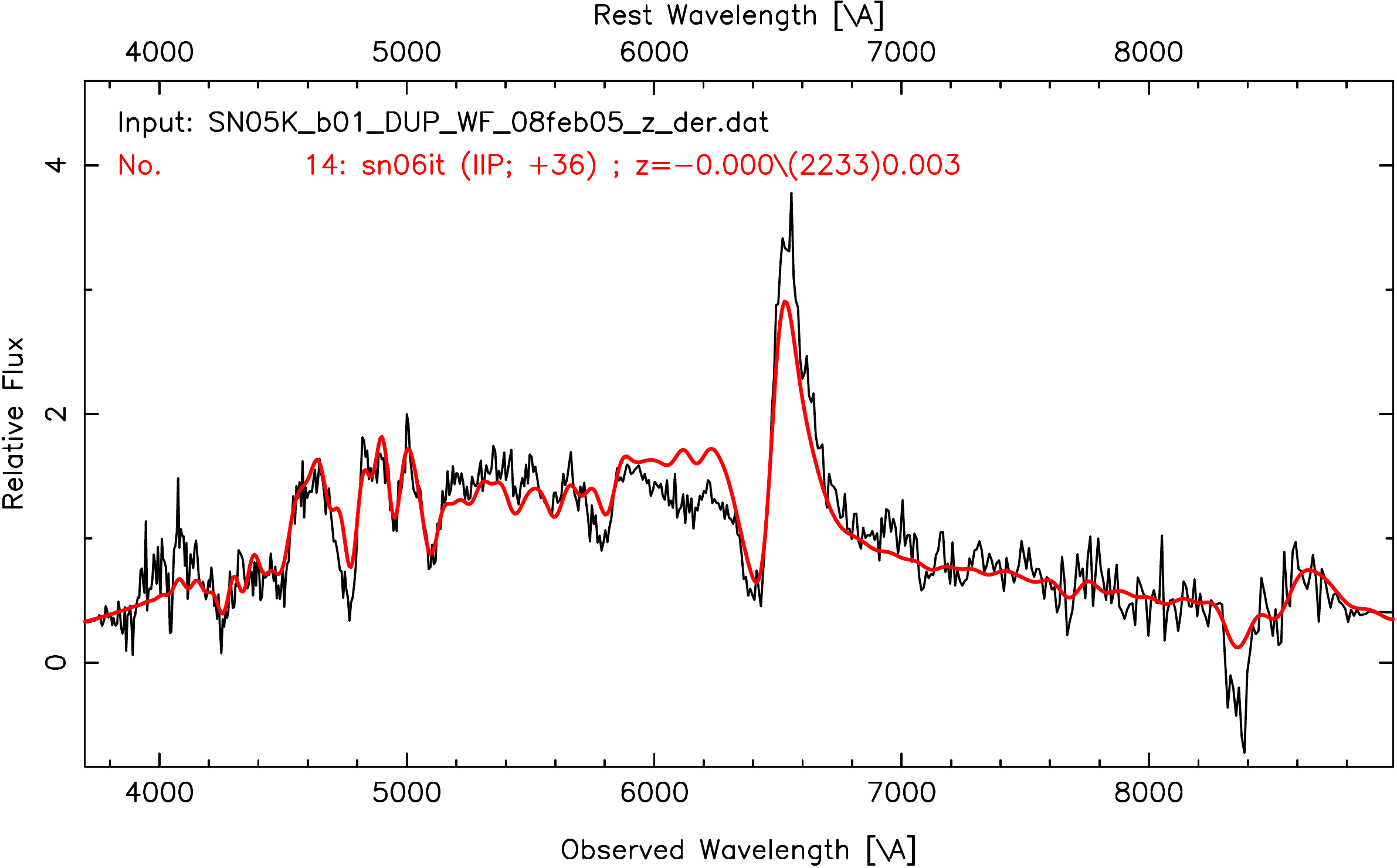}
\includegraphics[width=4.4cm]{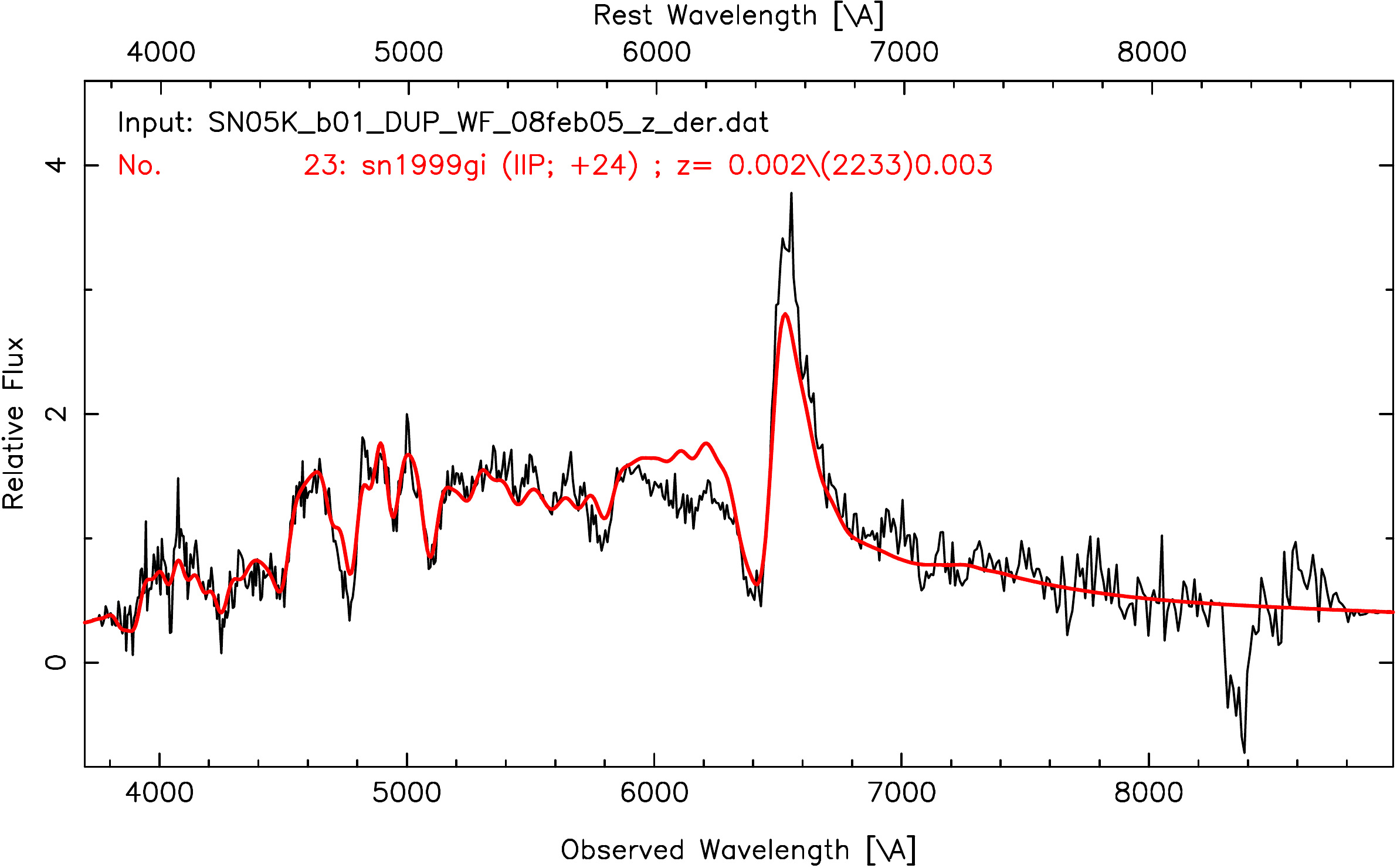}
\caption{Best spectral matching of SN~2005K using SNID. The plots show SN~2005K compared with 
SN~2003bn, SN~2006bp, SN~2007il, SN~2006it, and SN~1999gi at 39, 43, 45, 36, and 36 days from explosion.}
\end{figure}

\clearpage

\begin{figure}
\centering
\includegraphics[width=4.4cm]{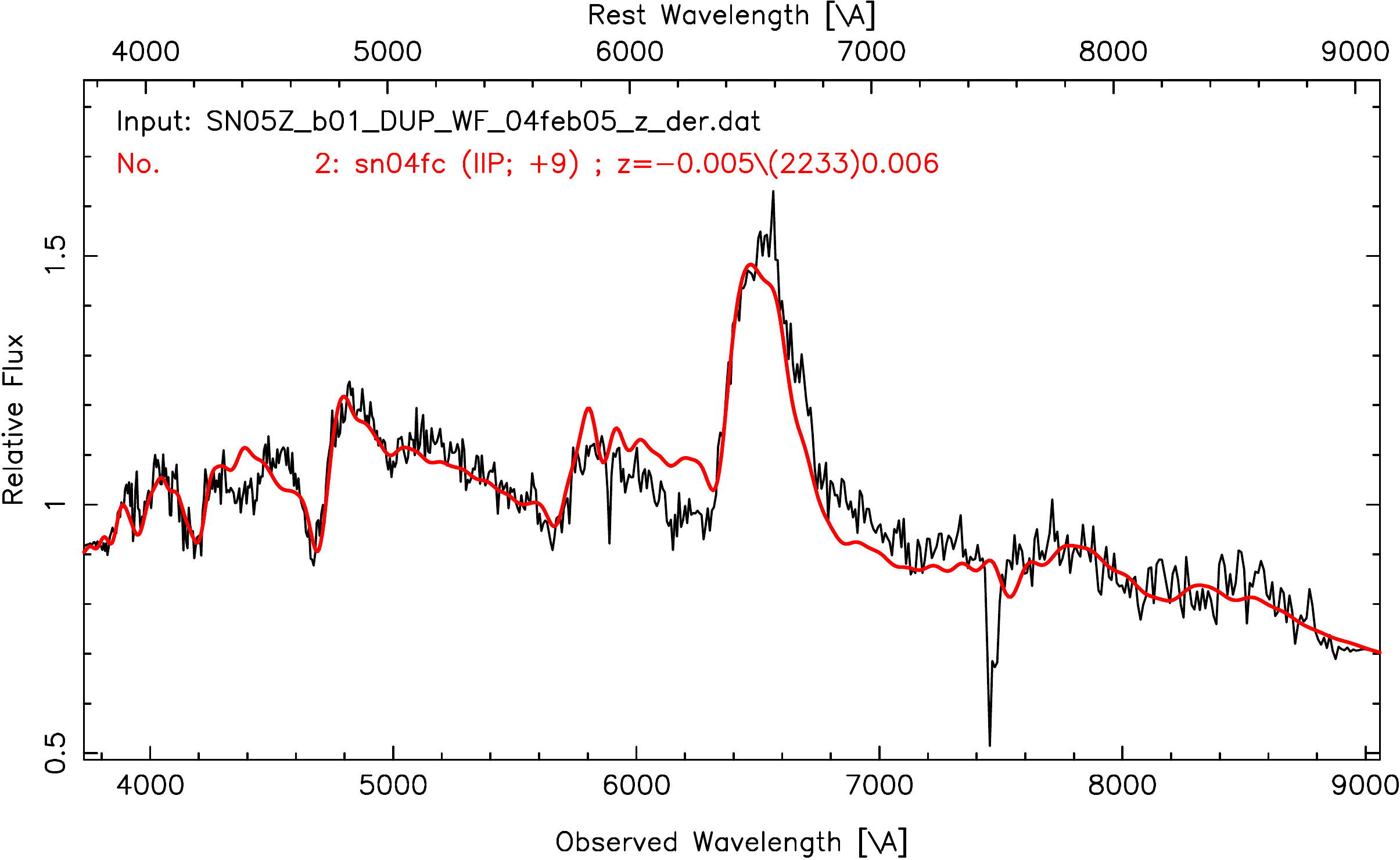}
\includegraphics[width=4.4cm]{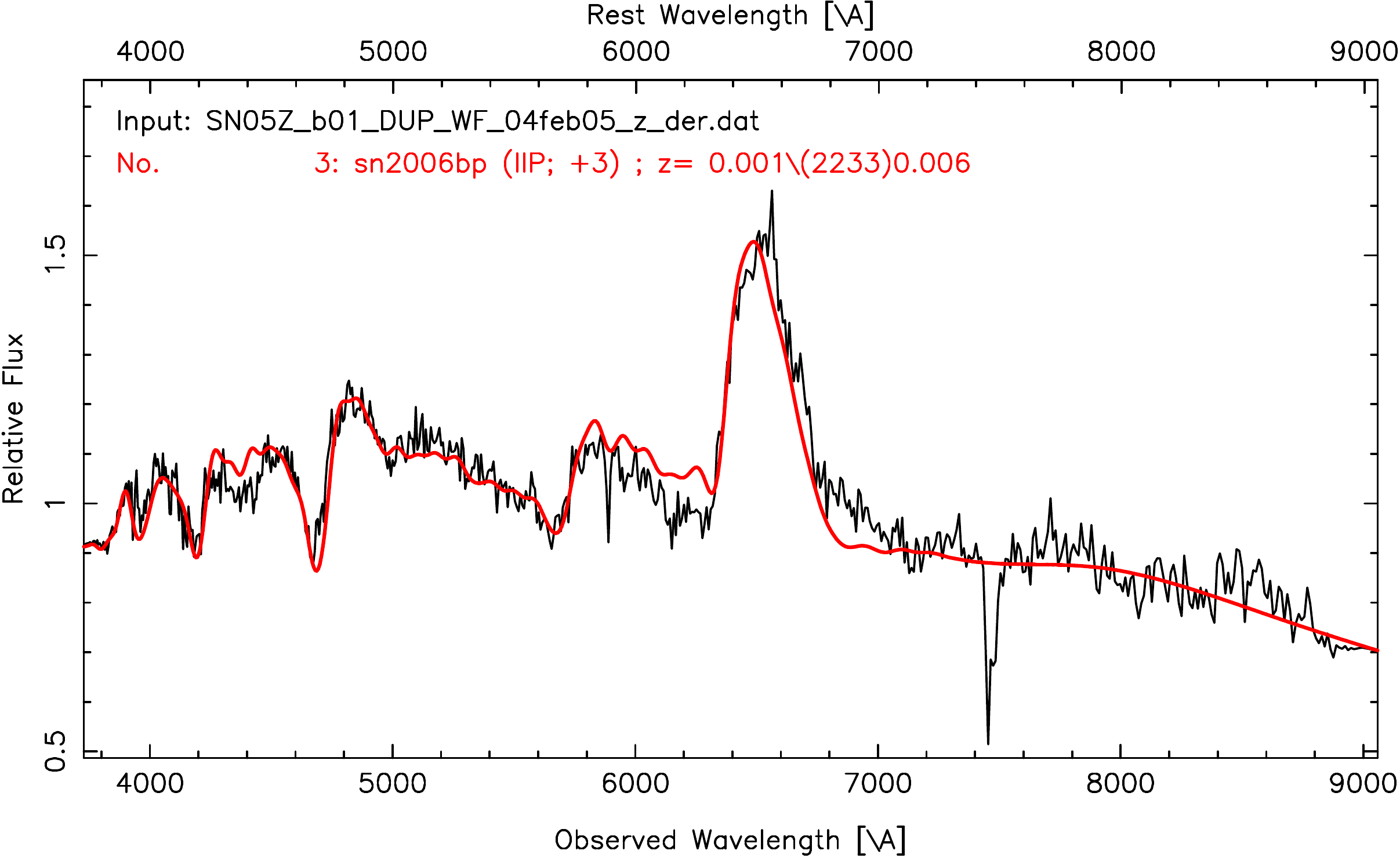} 
\includegraphics[width=4.4cm]{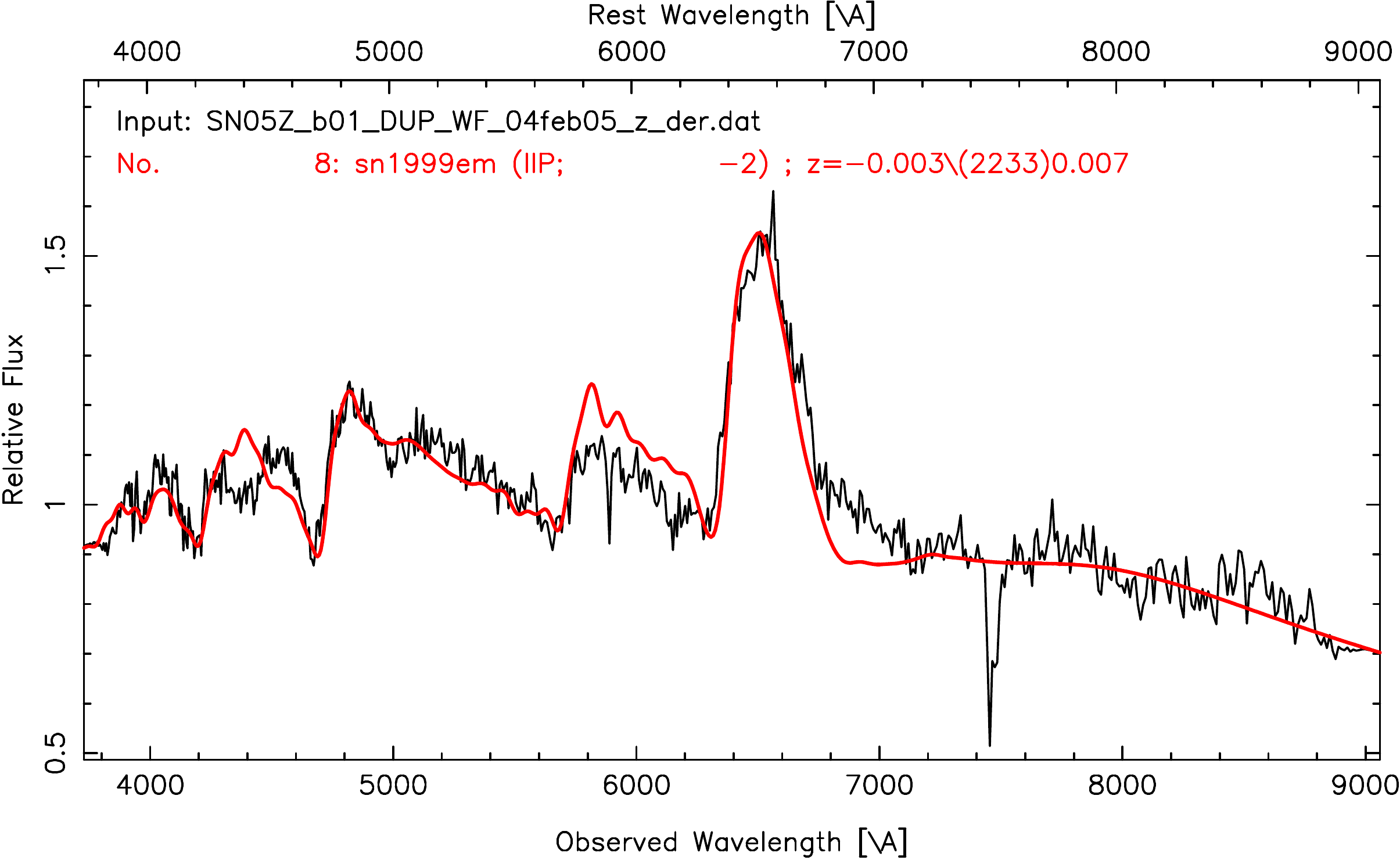}
\includegraphics[width=4.4cm]{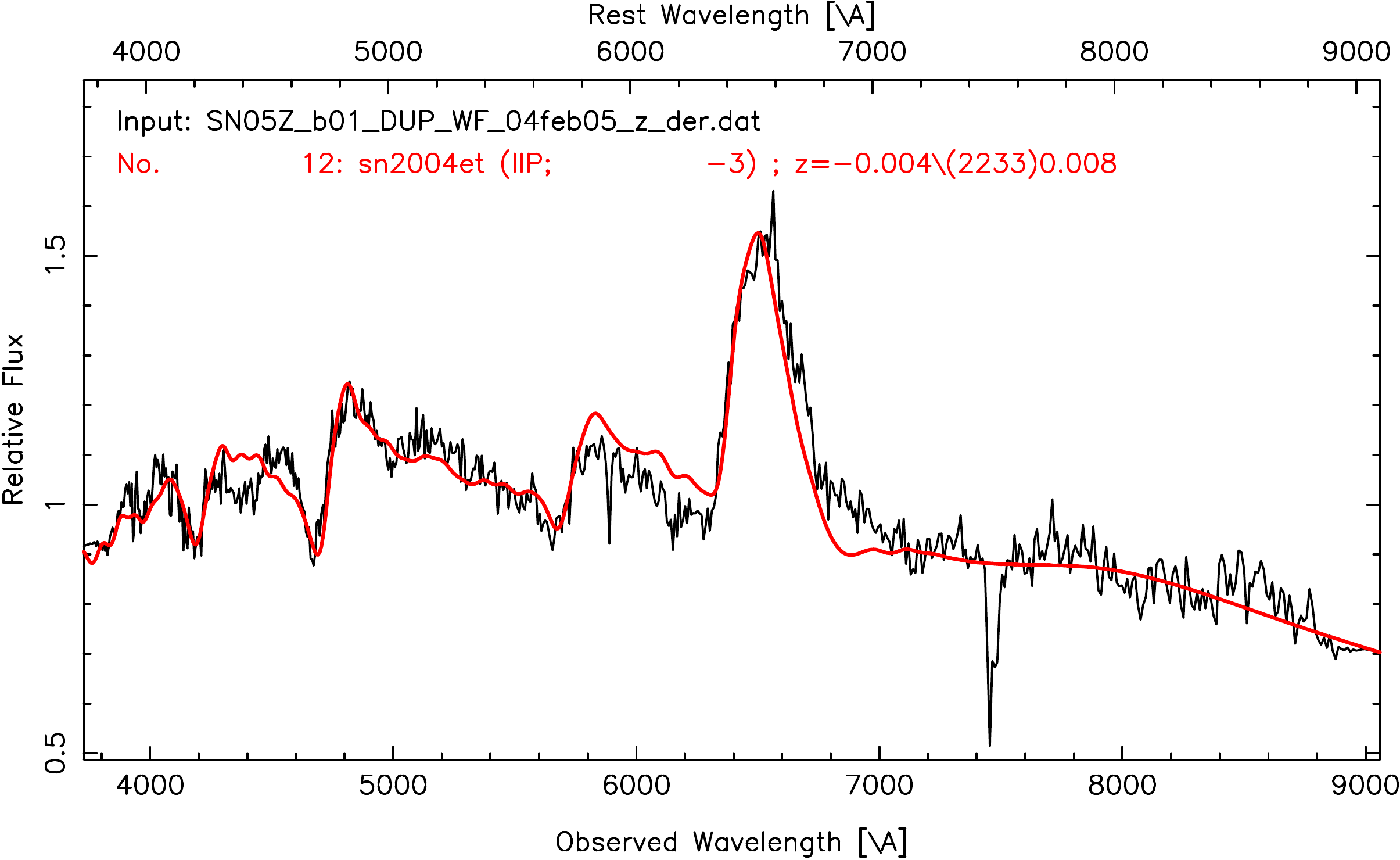}
\caption{Best spectral matching of SN~2005Z using SNID. The plots show SN~2005Z compared with 
SN~2004fc, SN~2006bp, SN~1999em, and 2004et at 9, 11, 8 and 13 days from explosion.}
\end{figure}

\begin{figure}
\centering
\includegraphics[width=4.4cm]{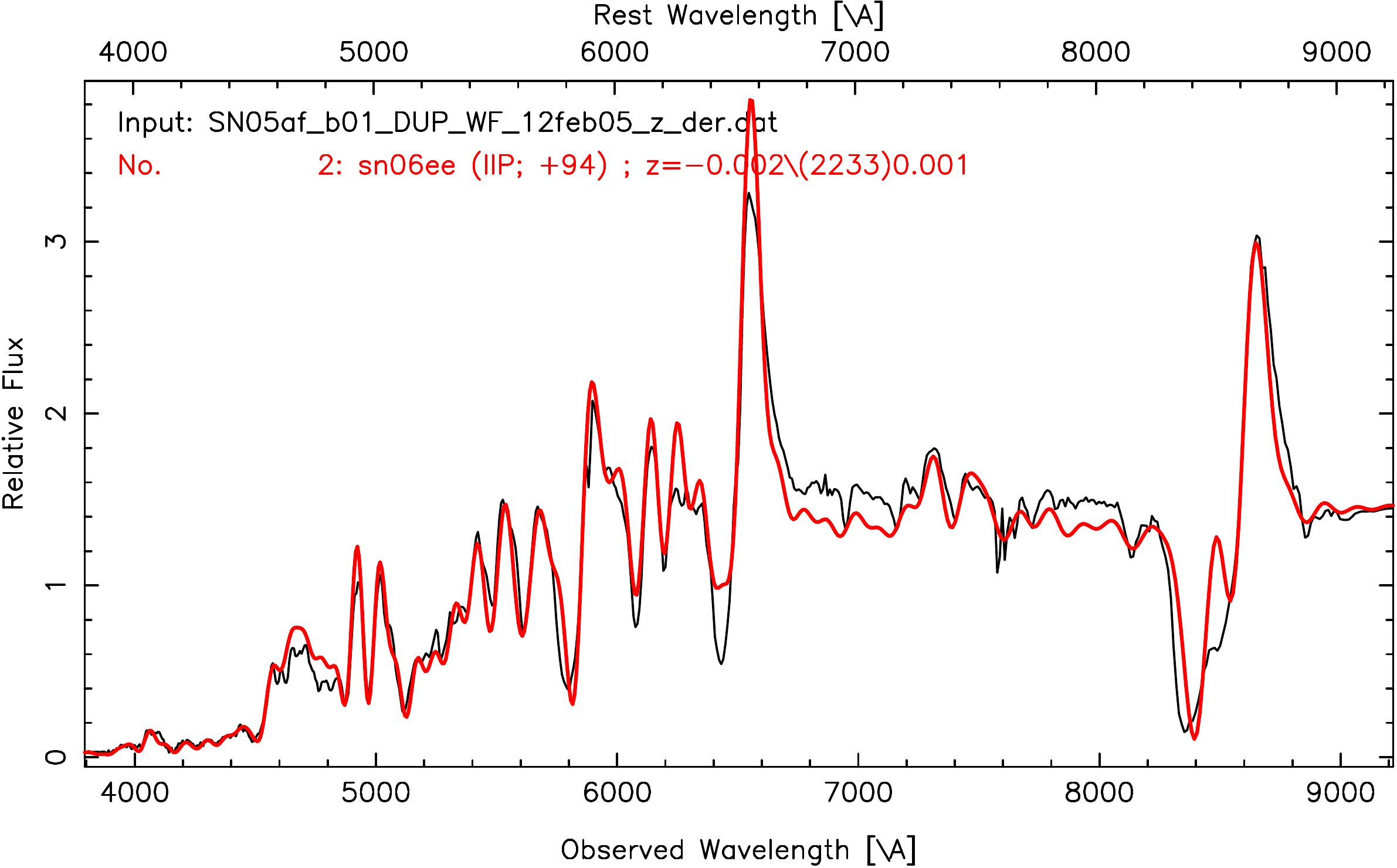}
\includegraphics[width=4.4cm]{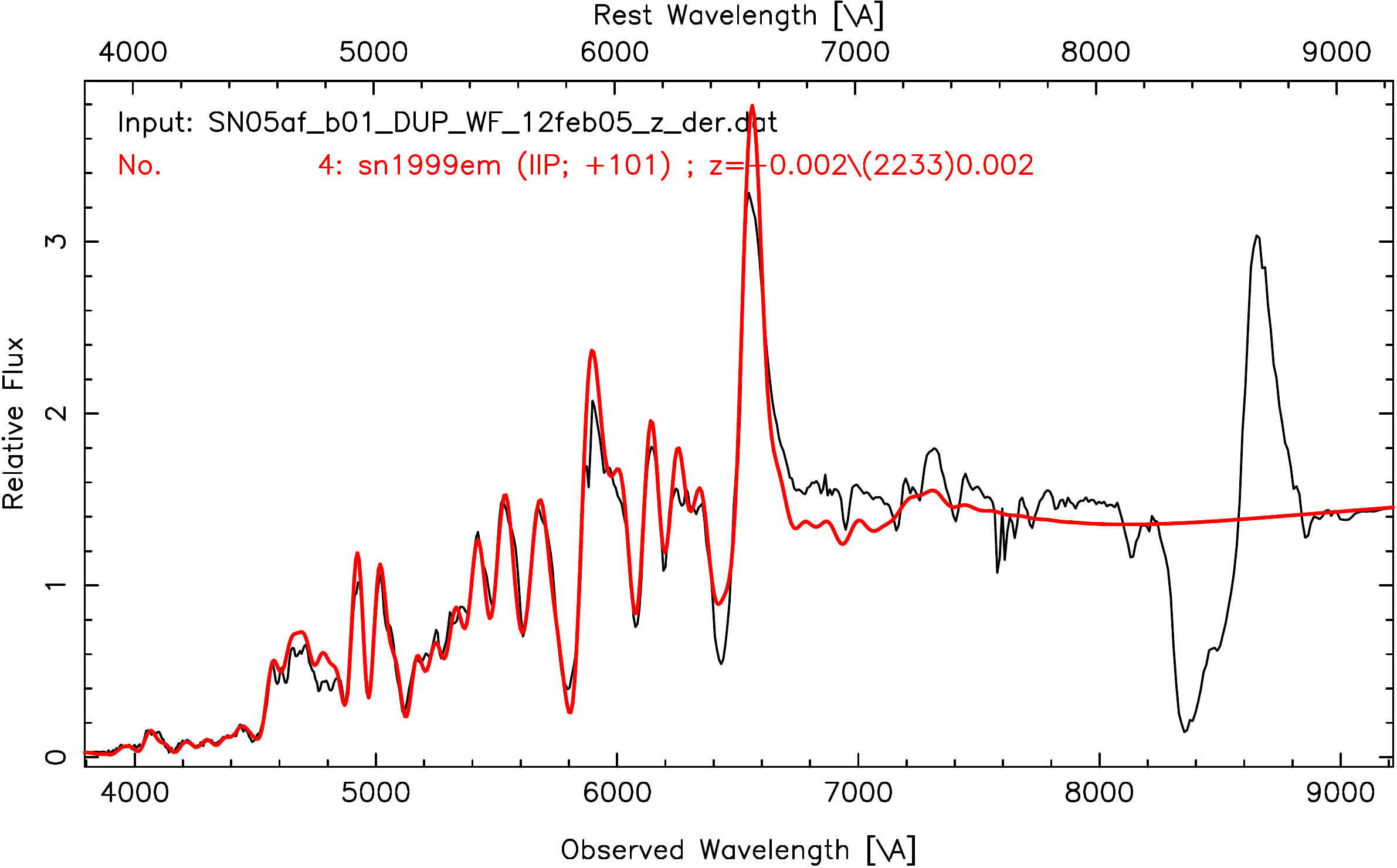}
\includegraphics[width=4.4cm]{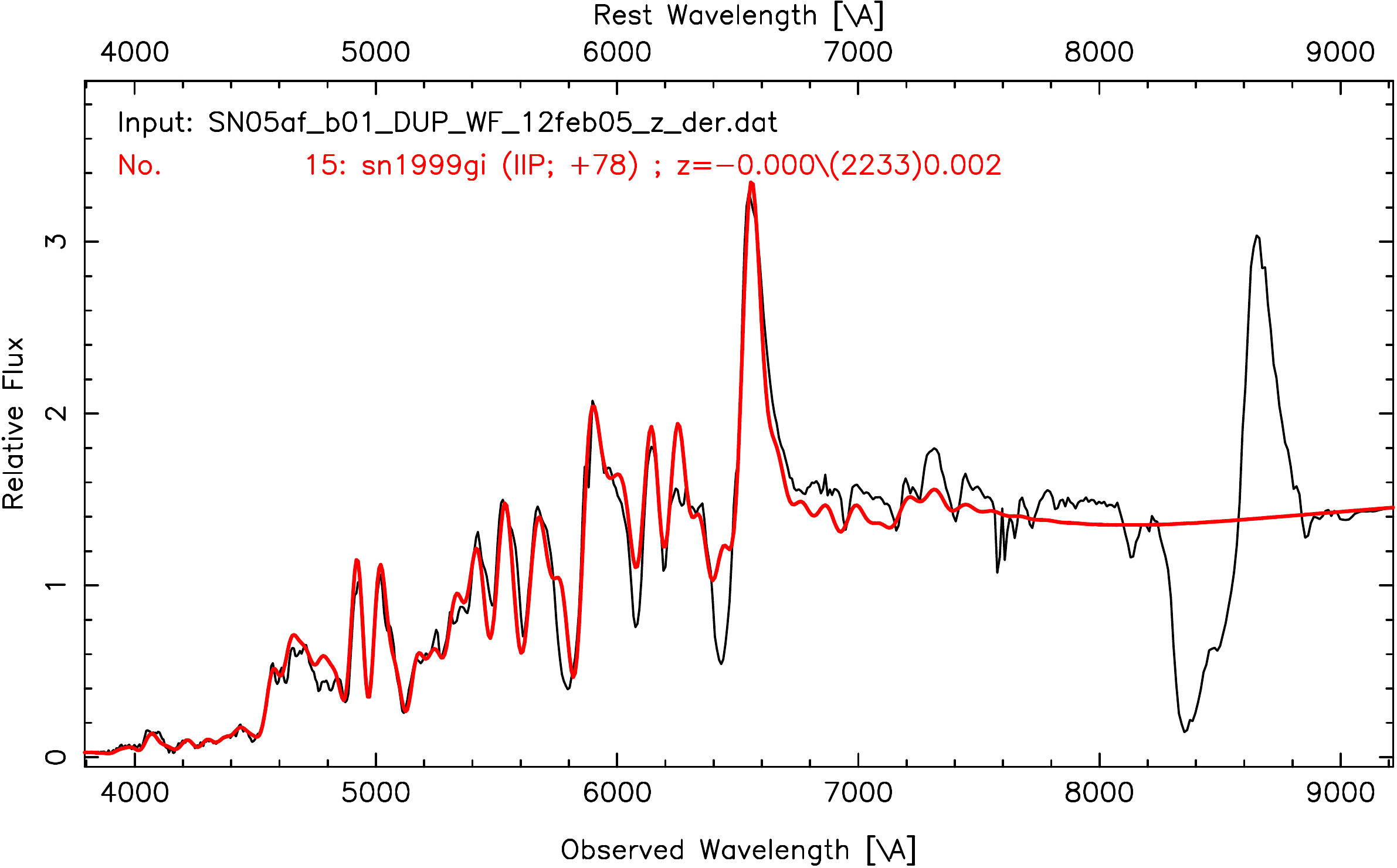}
\includegraphics[width=4.4cm]{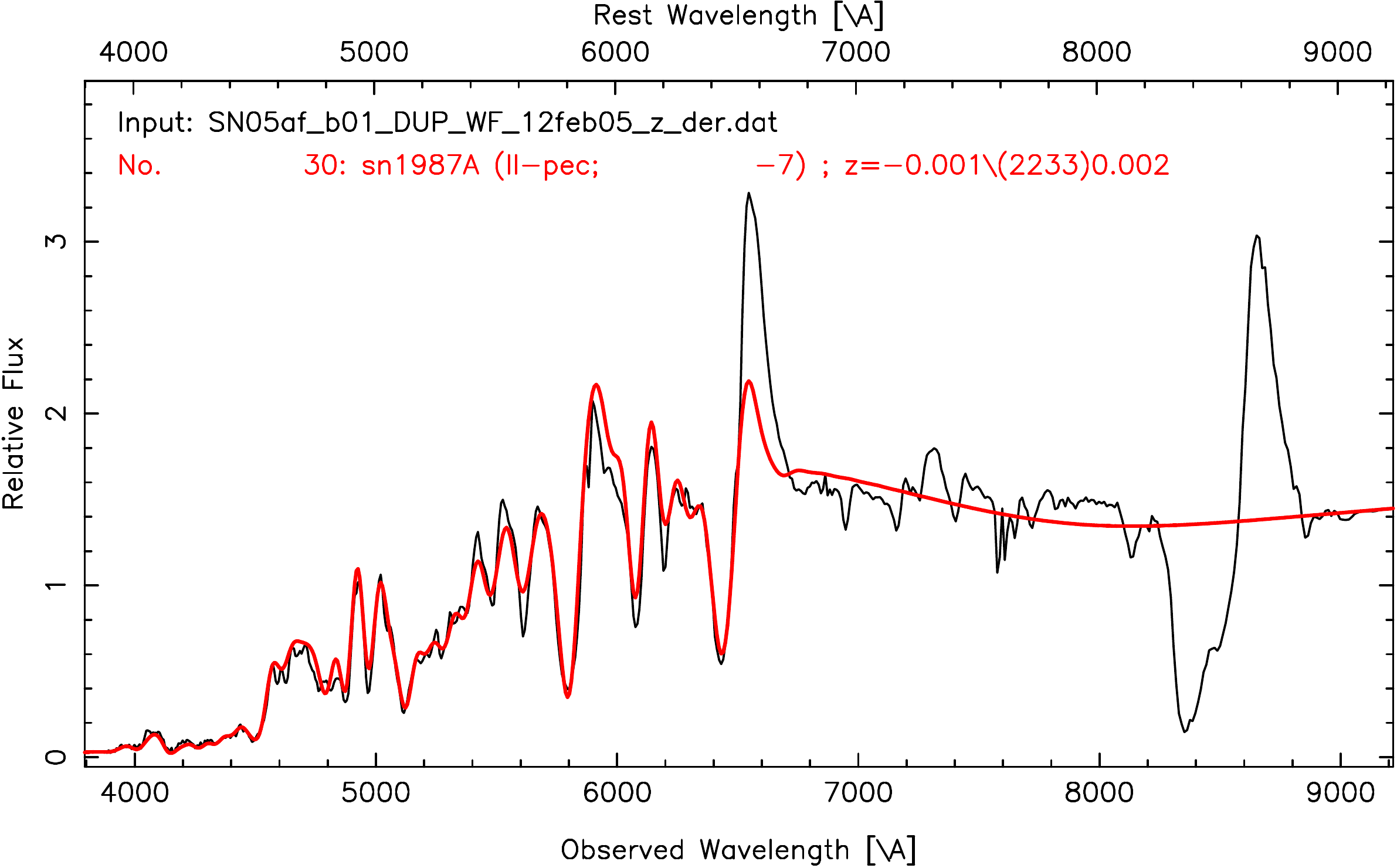}
\caption{Best spectral matching of SN~2005af using SNID. The plots show SN~2005af compared with 
SN~2006ee, SN~1999em, SN~1999gi, and 1987A at 94, 111, 90, and 78 days from explosion.}
\end{figure}

\clearpage

\begin{figure}
\centering
\includegraphics[width=4.4cm]{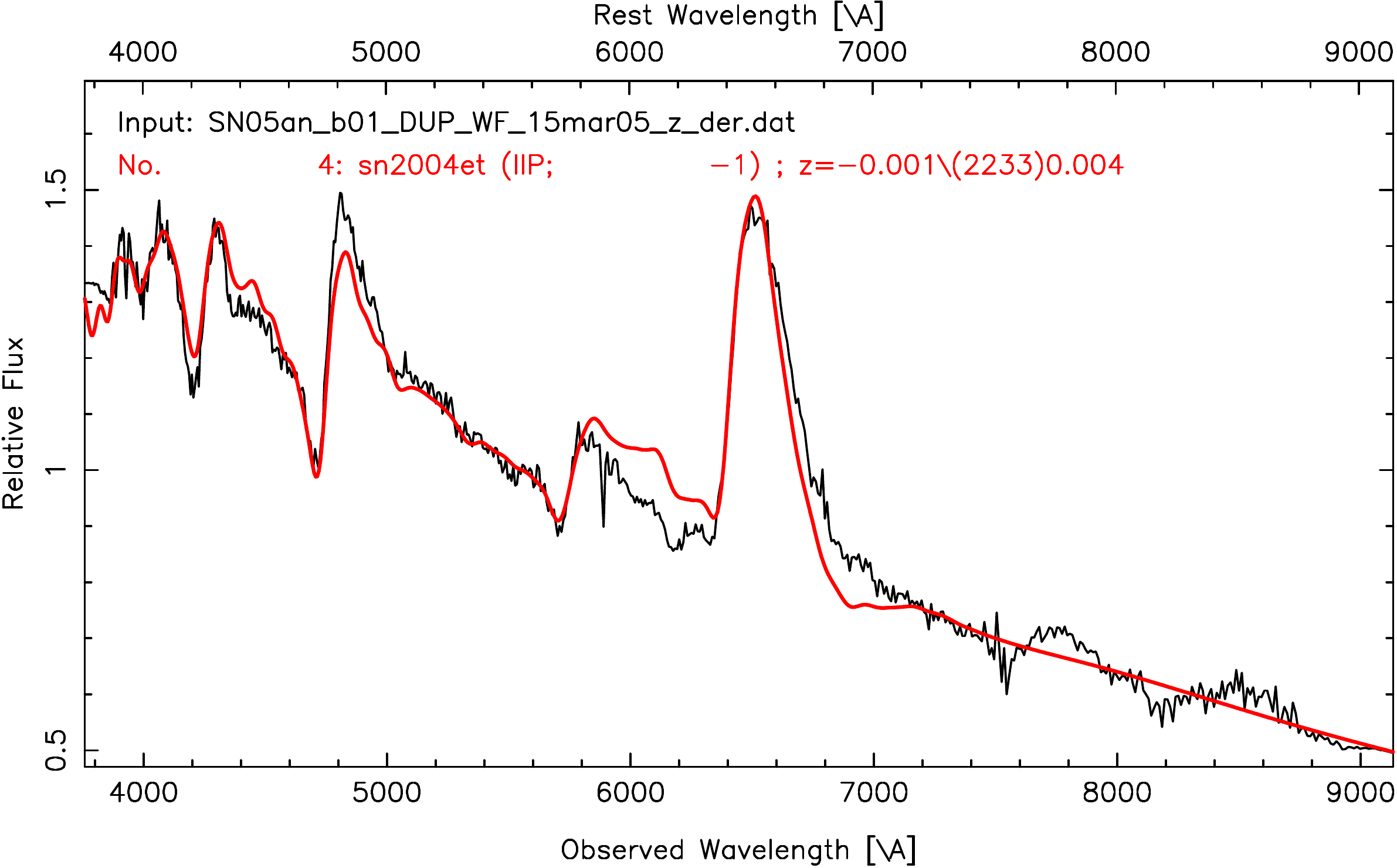}
\includegraphics[width=4.4cm]{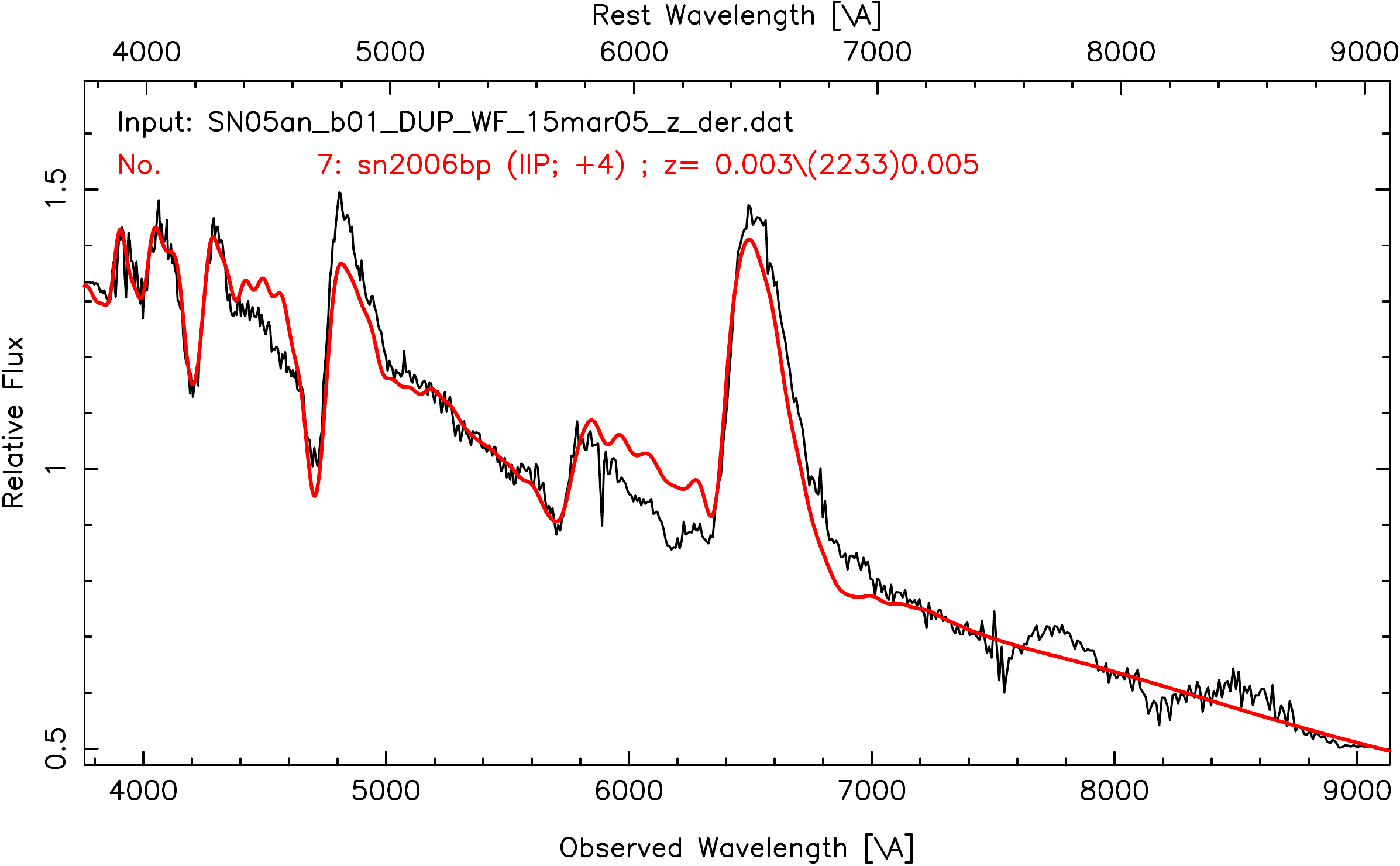}
\includegraphics[width=4.4cm]{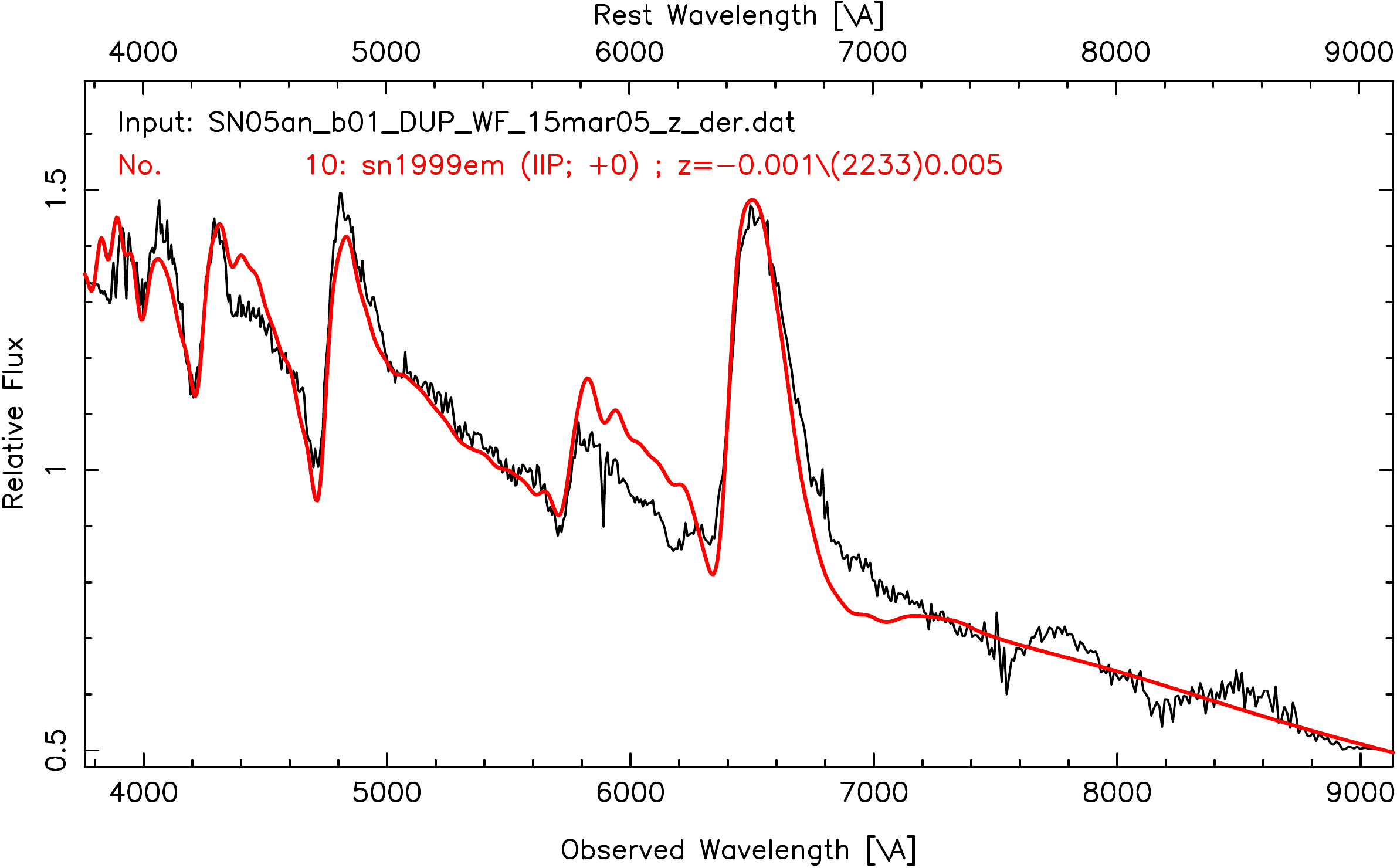}
\includegraphics[width=4.4cm]{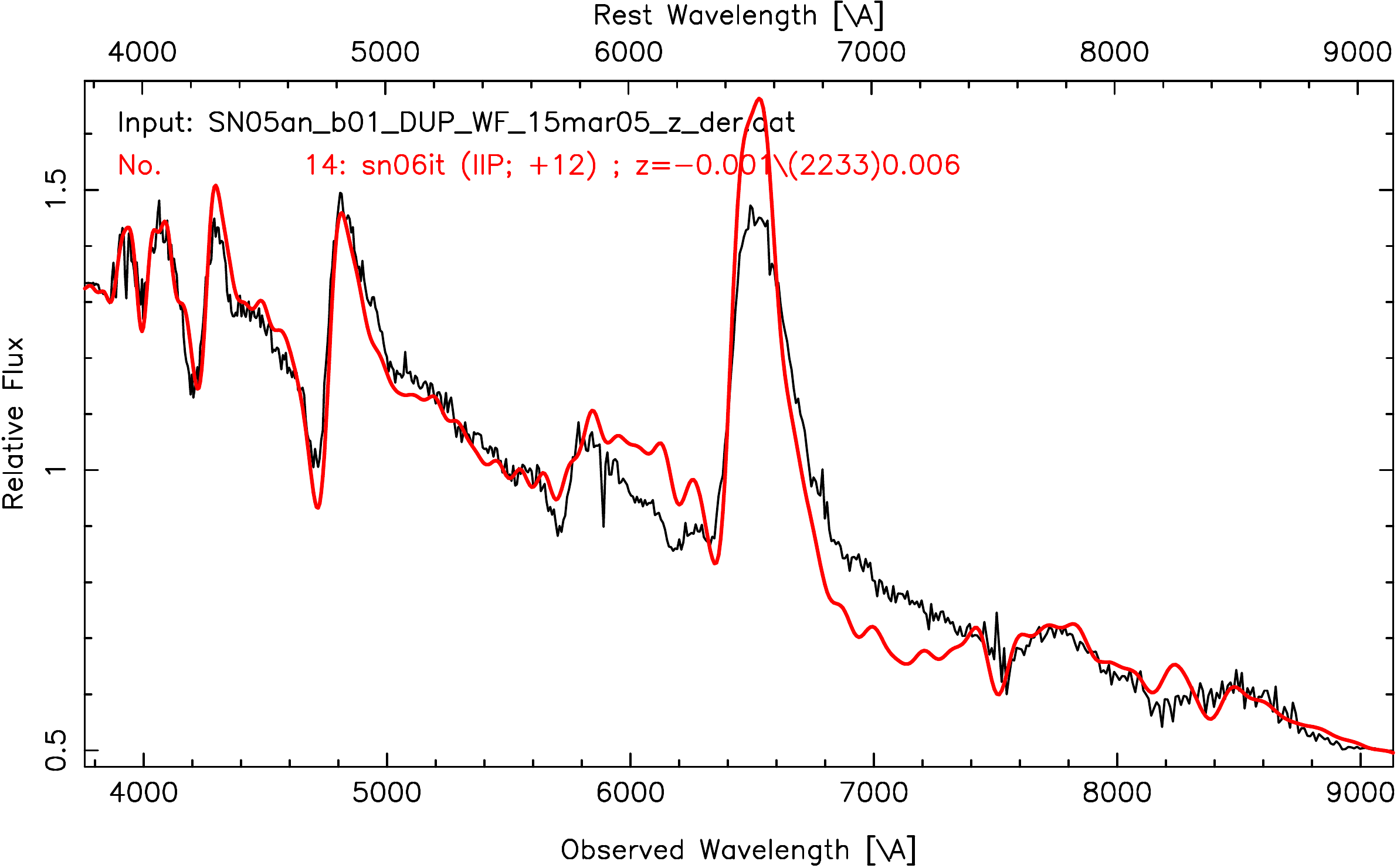}
\caption{Best spectral matching of SN~2005an using SNID. The plots show SN~2005an compared with 
SN~2004et, SN~2006bp, SN~1999em, and SN~2006it at 15, 13, 10, and 12 days from explosion.}
\end{figure}

\begin{figure}
\centering
\includegraphics[width=4.4cm]{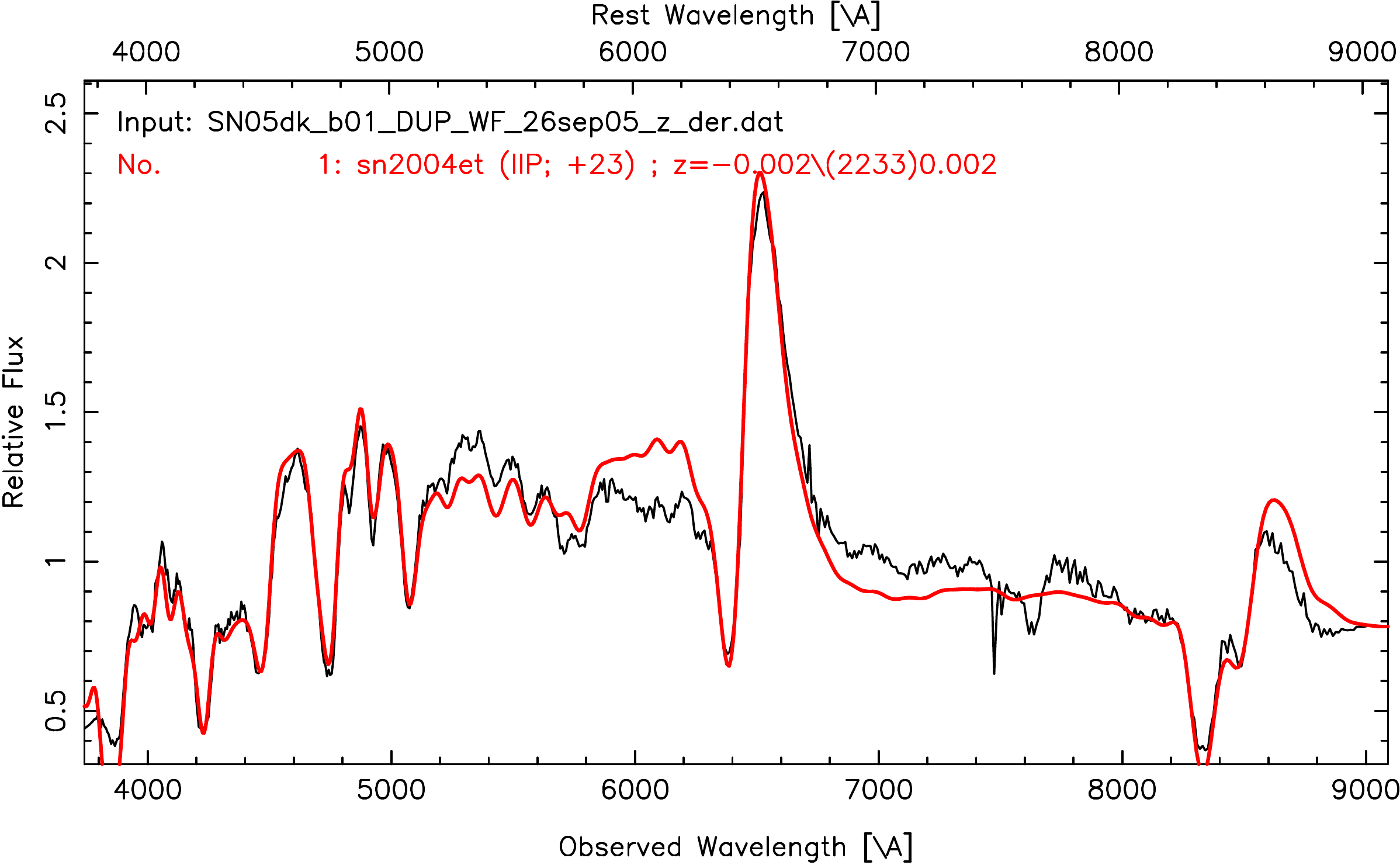}
\includegraphics[width=4.4cm]{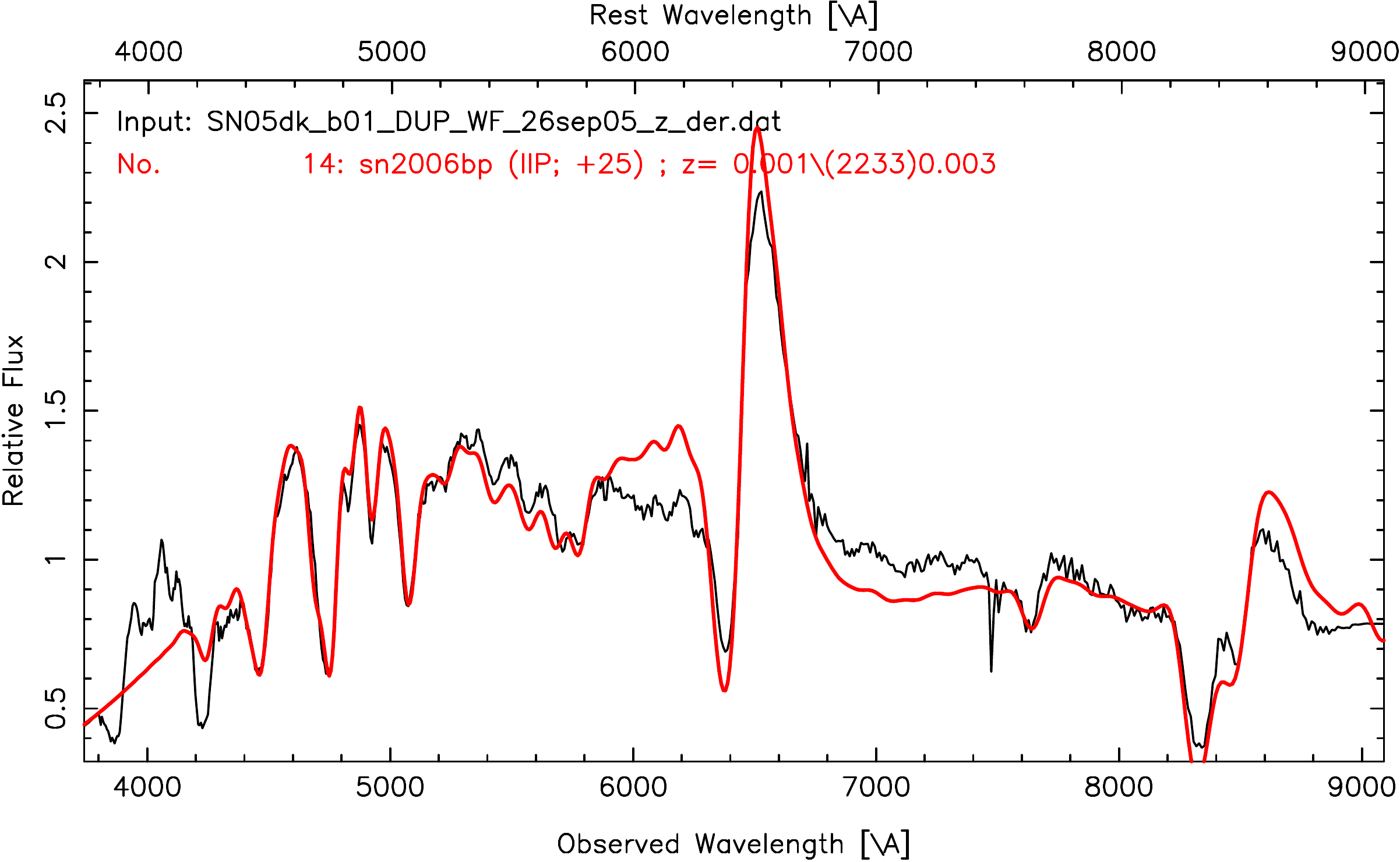}
\includegraphics[width=4.4cm]{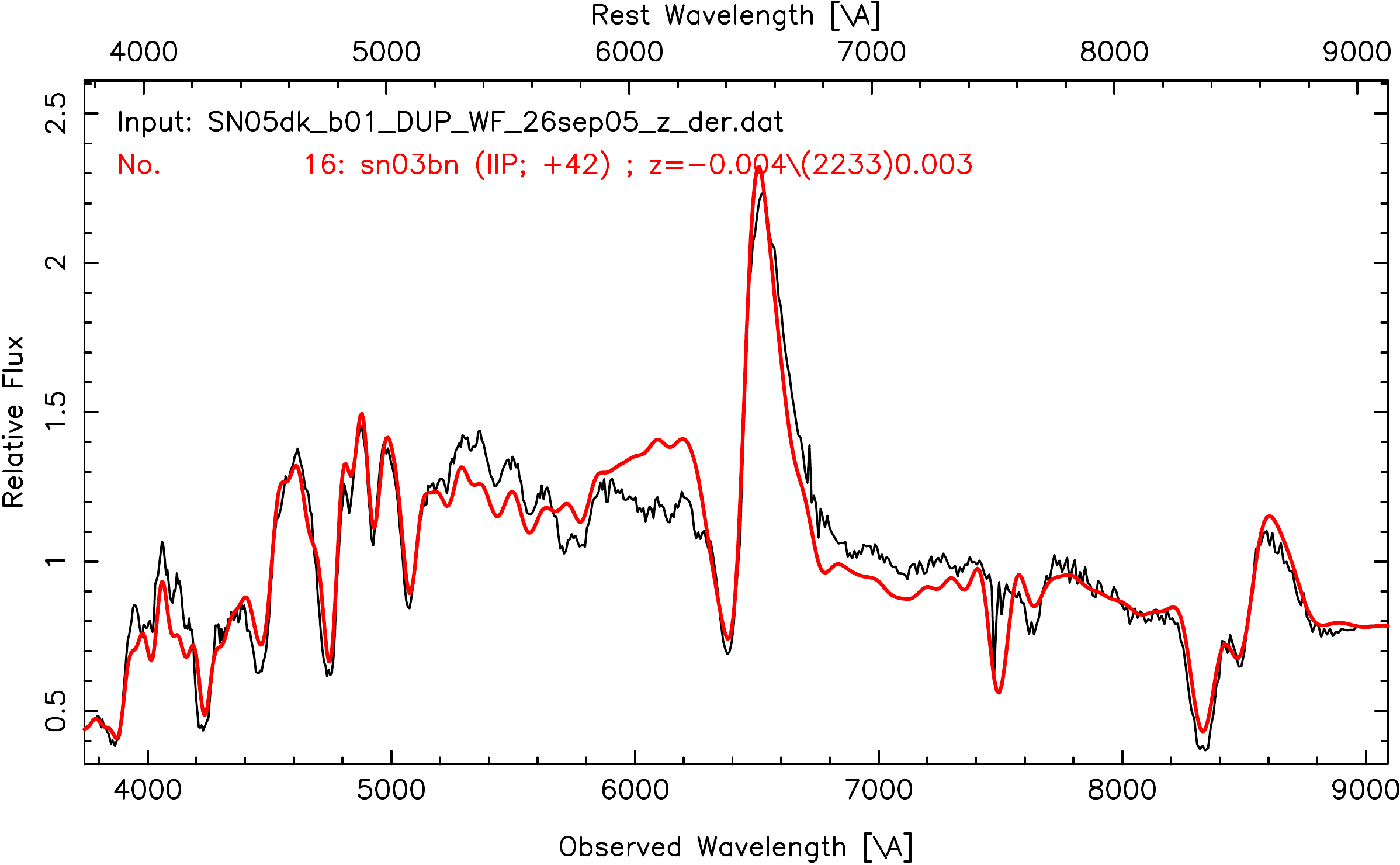}
\includegraphics[width=4.4cm]{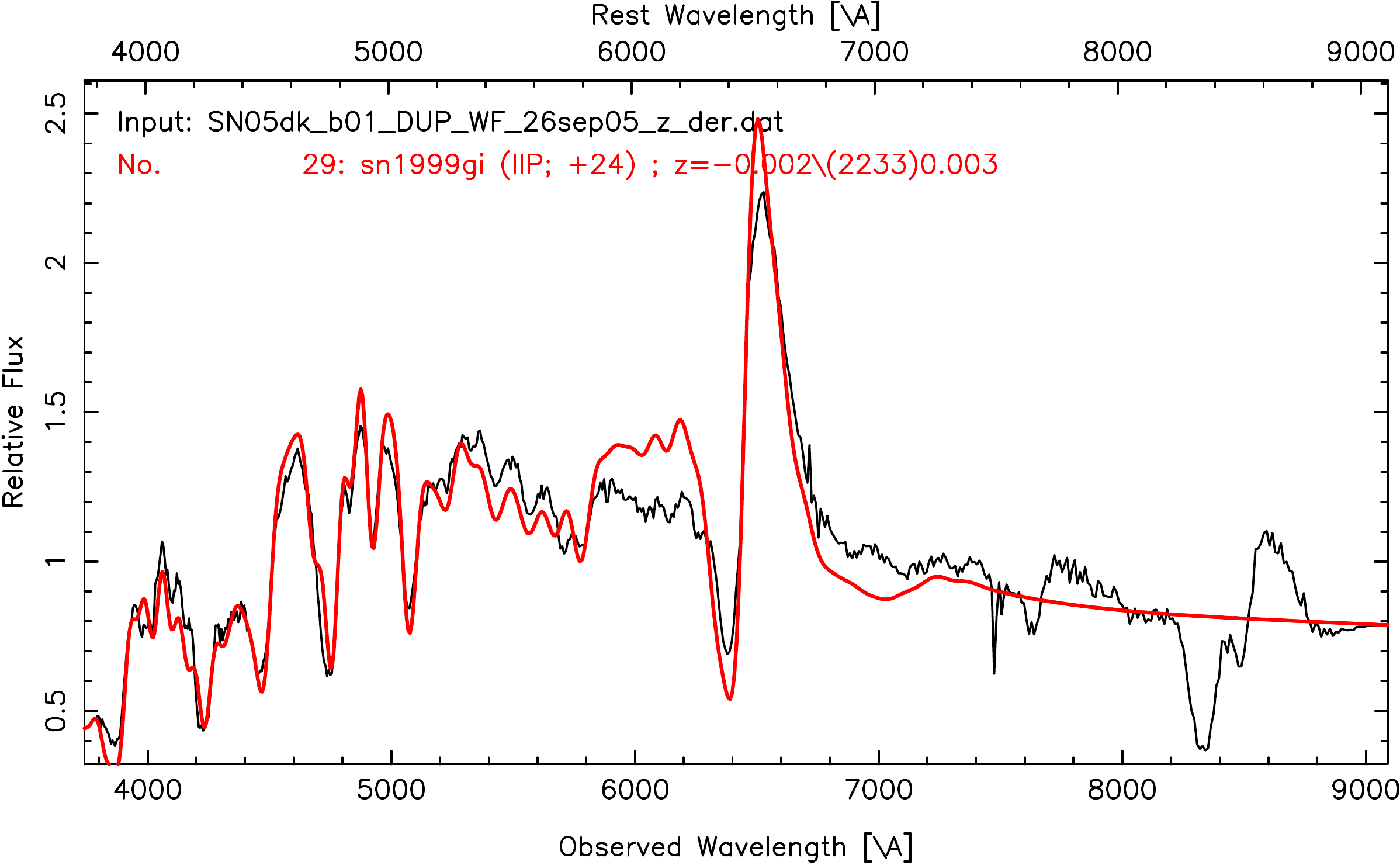}
\caption{Best spectral matching of SN~2005dk using SNID. The plots show SN~2005dk compared with 
SN~2004et, SN~2006bp, SN~2003bn, and SN~1999gi at 39,34, and 36 days from explosion.}
\end{figure}

\clearpage

\begin{figure}
\centering
\includegraphics[width=4.4cm]{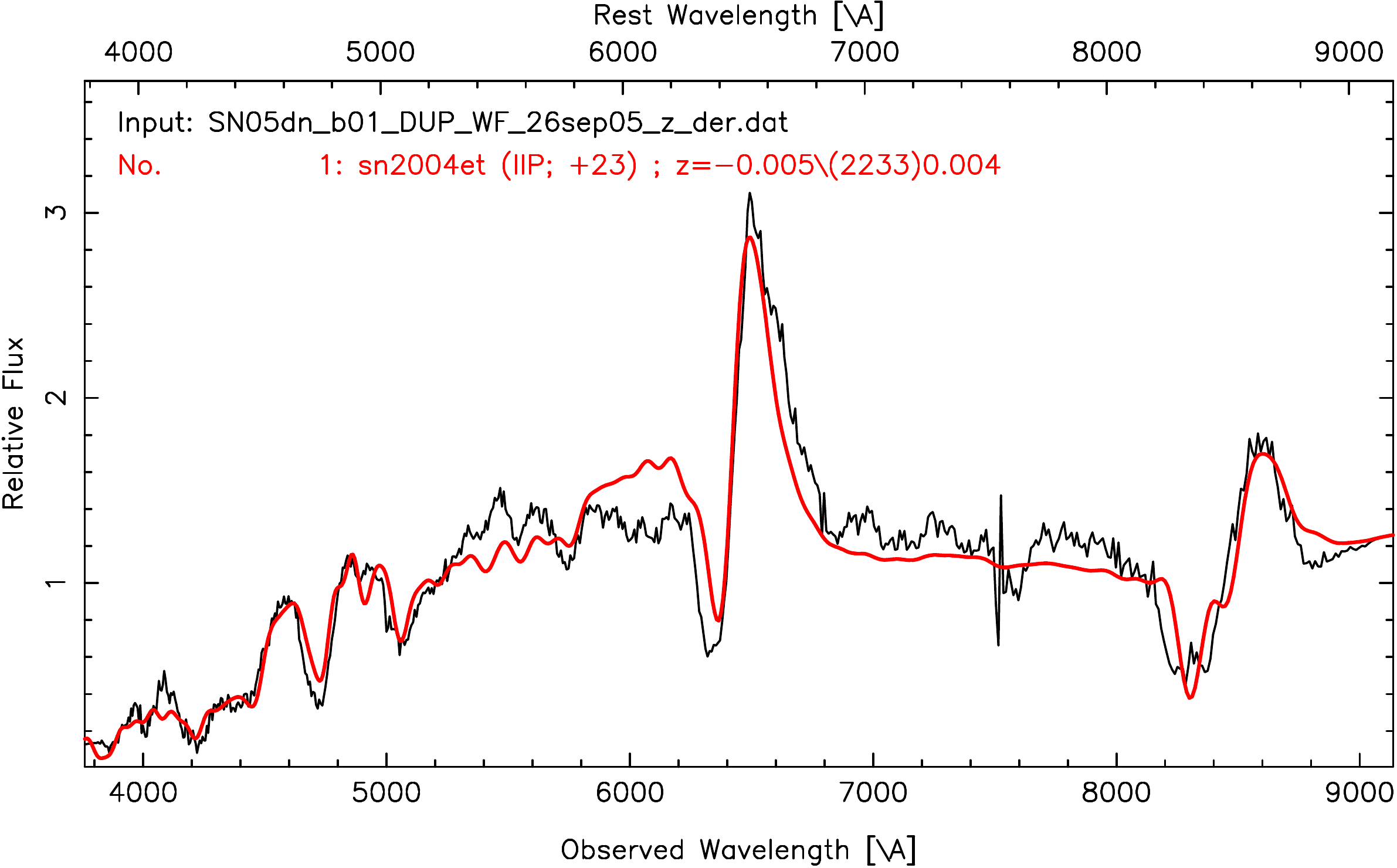}
\includegraphics[width=4.4cm]{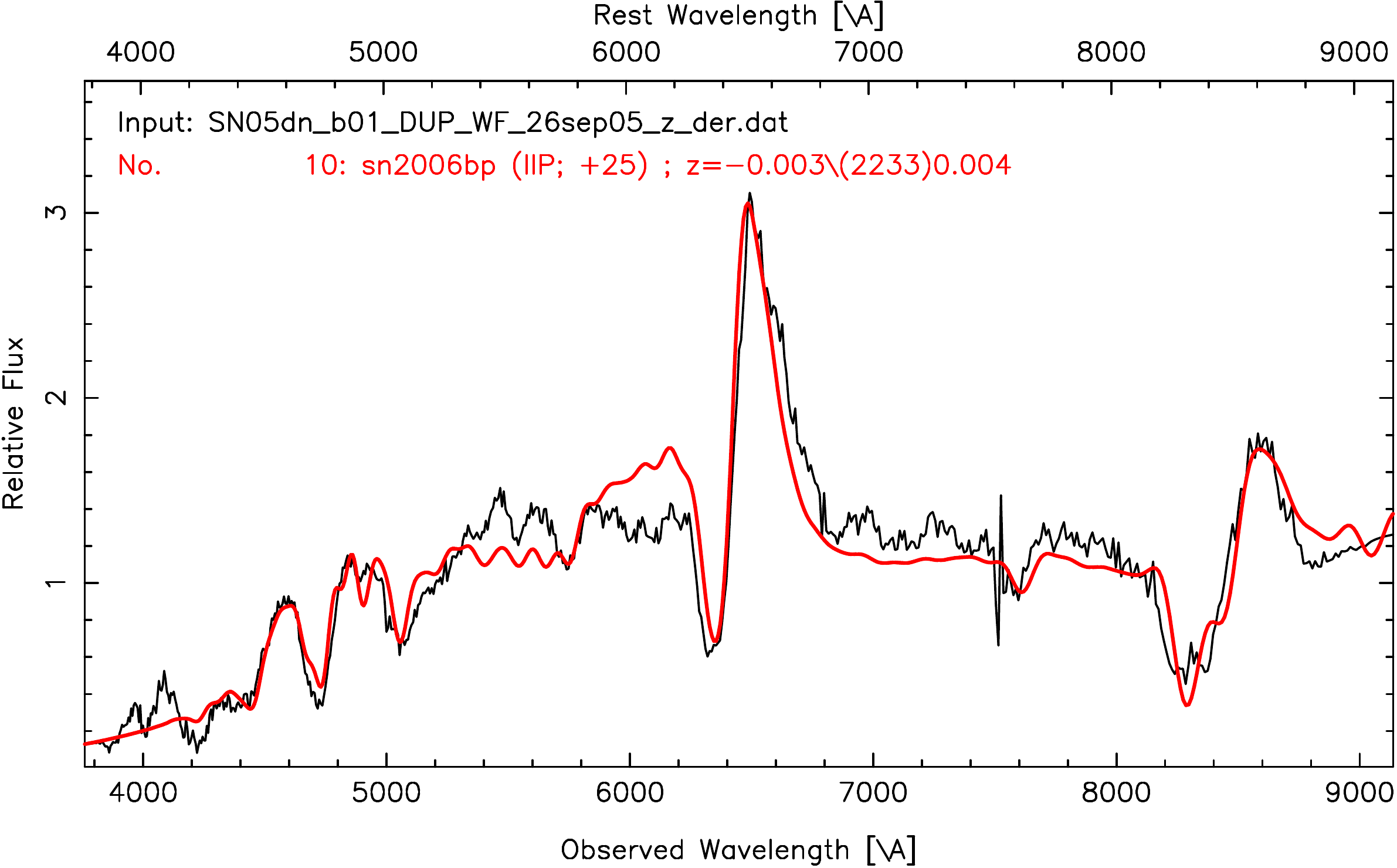}
\caption{Best spectral matching of SN~2005dn using SNID. The plots show SN~2005dn compared with 
SN~2004et and SN~2006bp at 39 and 34 days from explosion.}
\end{figure}

\begin{figure}
\centering
\includegraphics[width=4.4cm]{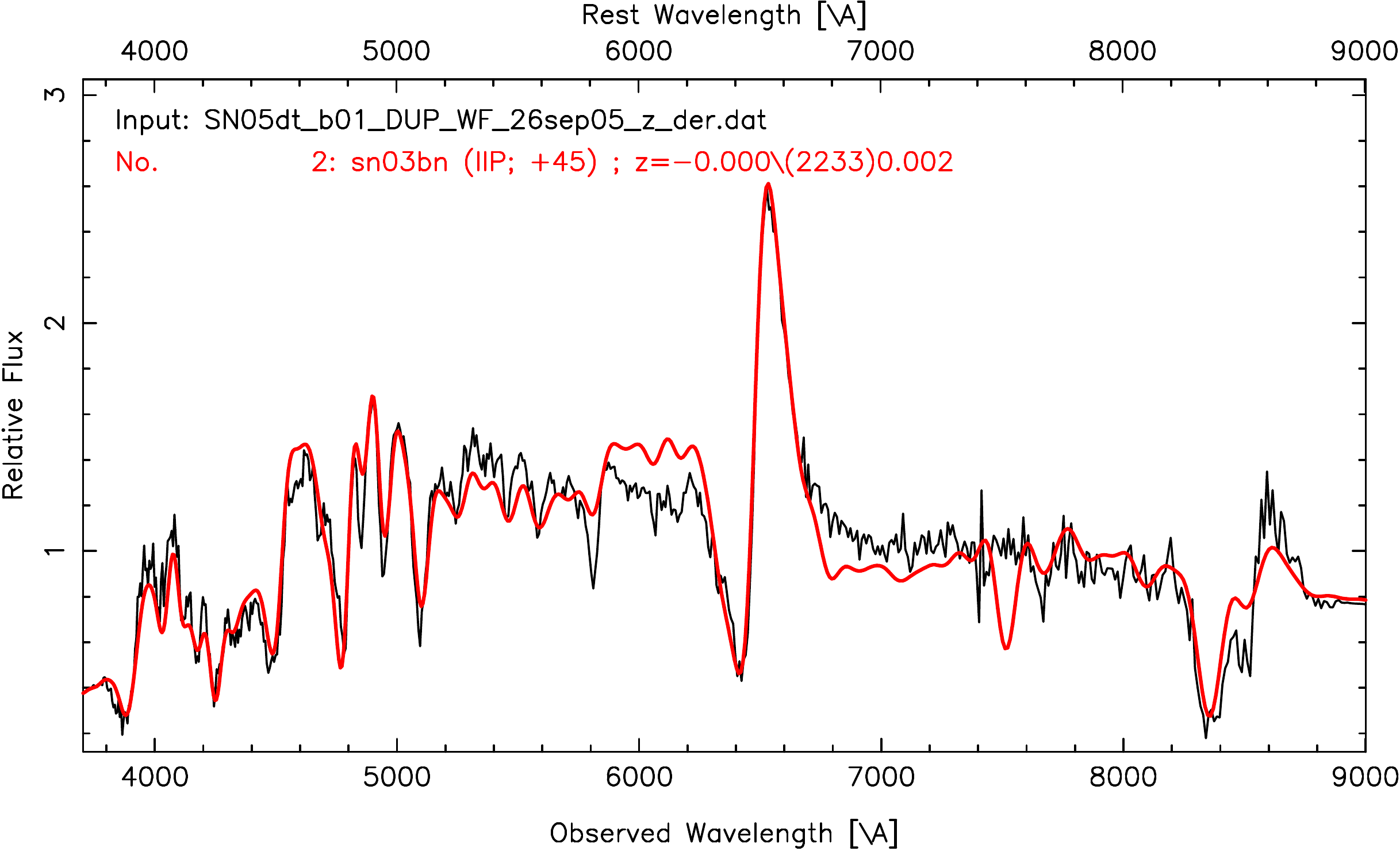} 
\includegraphics[width=4.4cm]{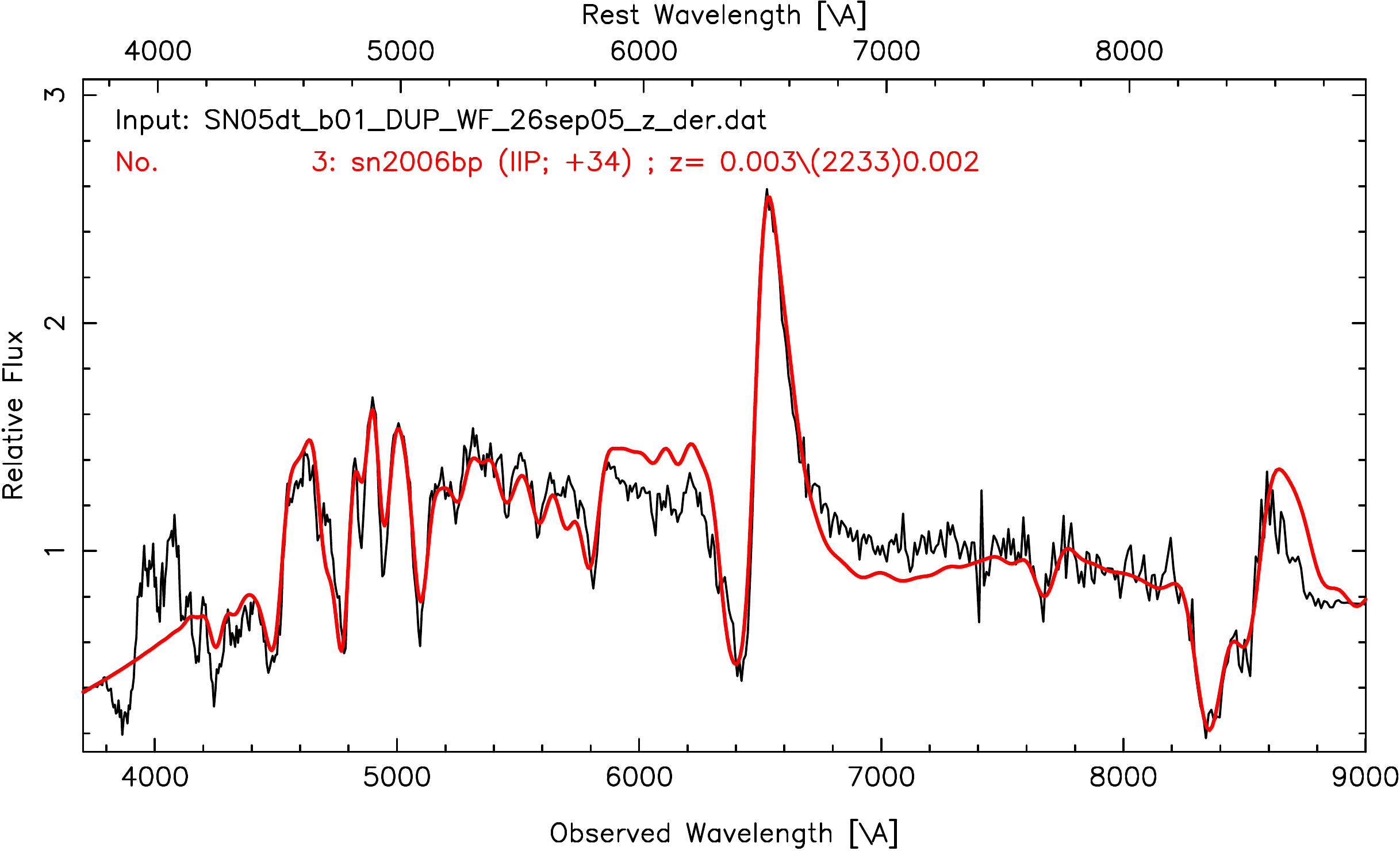}
\includegraphics[width=4.4cm]{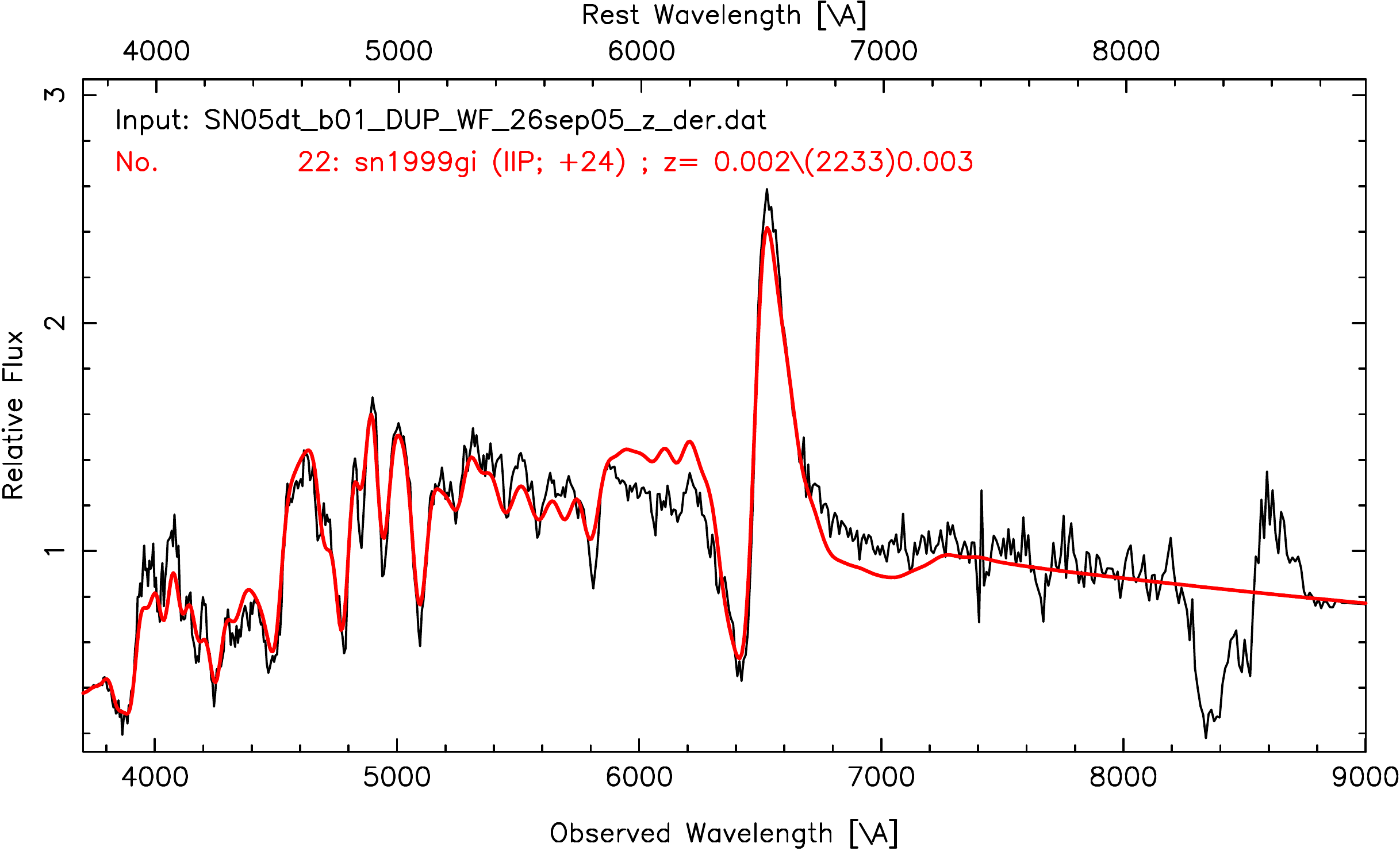}
\caption{Best spectral matching of SN~2005dt using SNID. The plots show SN~2005dt compared with 
SN~2003bn, SN~2006bp, and SN~1999gi at 45, 43, and 36 days from explosion.}
\end{figure}

\begin{figure}
\centering
\includegraphics[width=4.4cm]{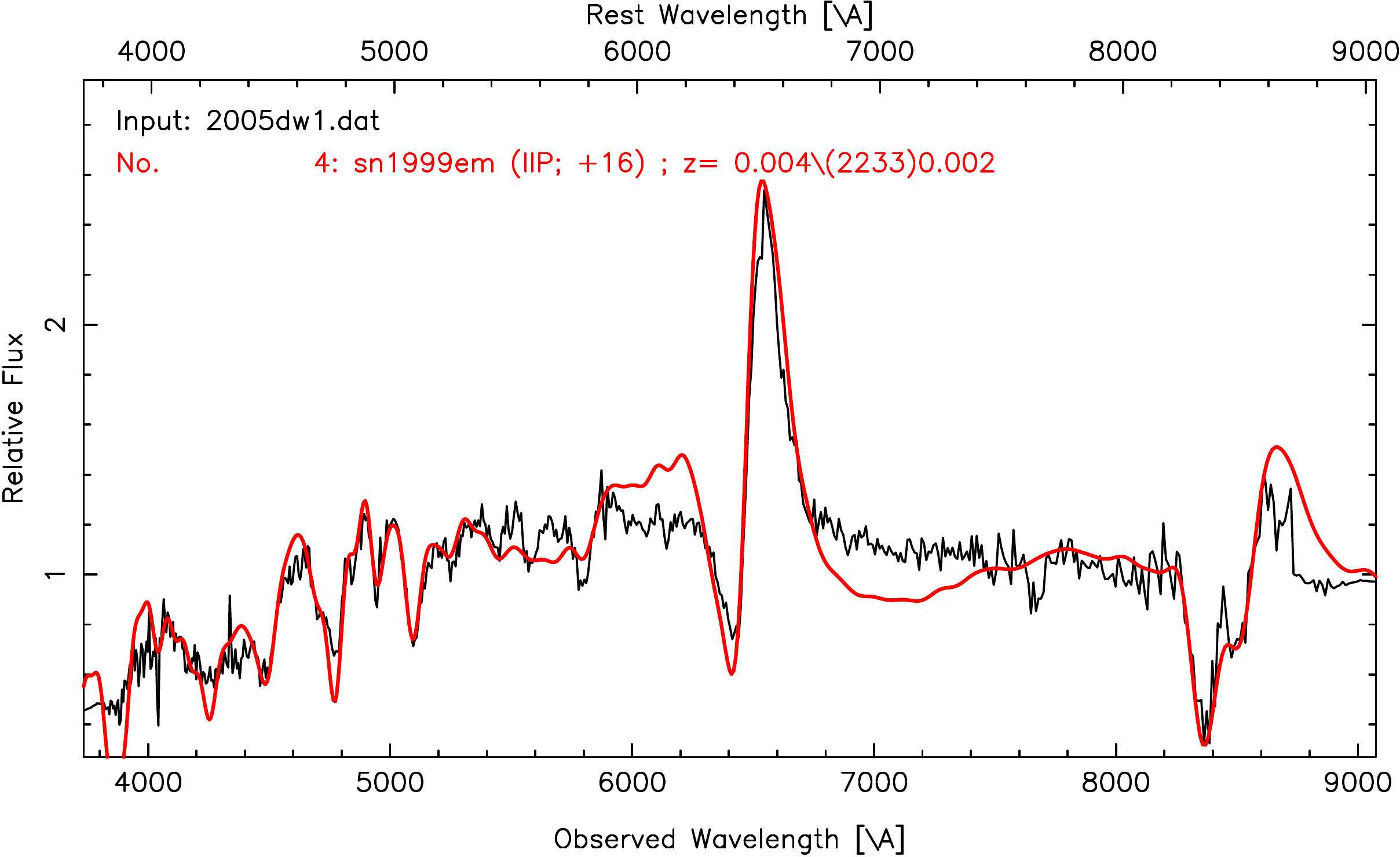}
\includegraphics[width=4.4cm]{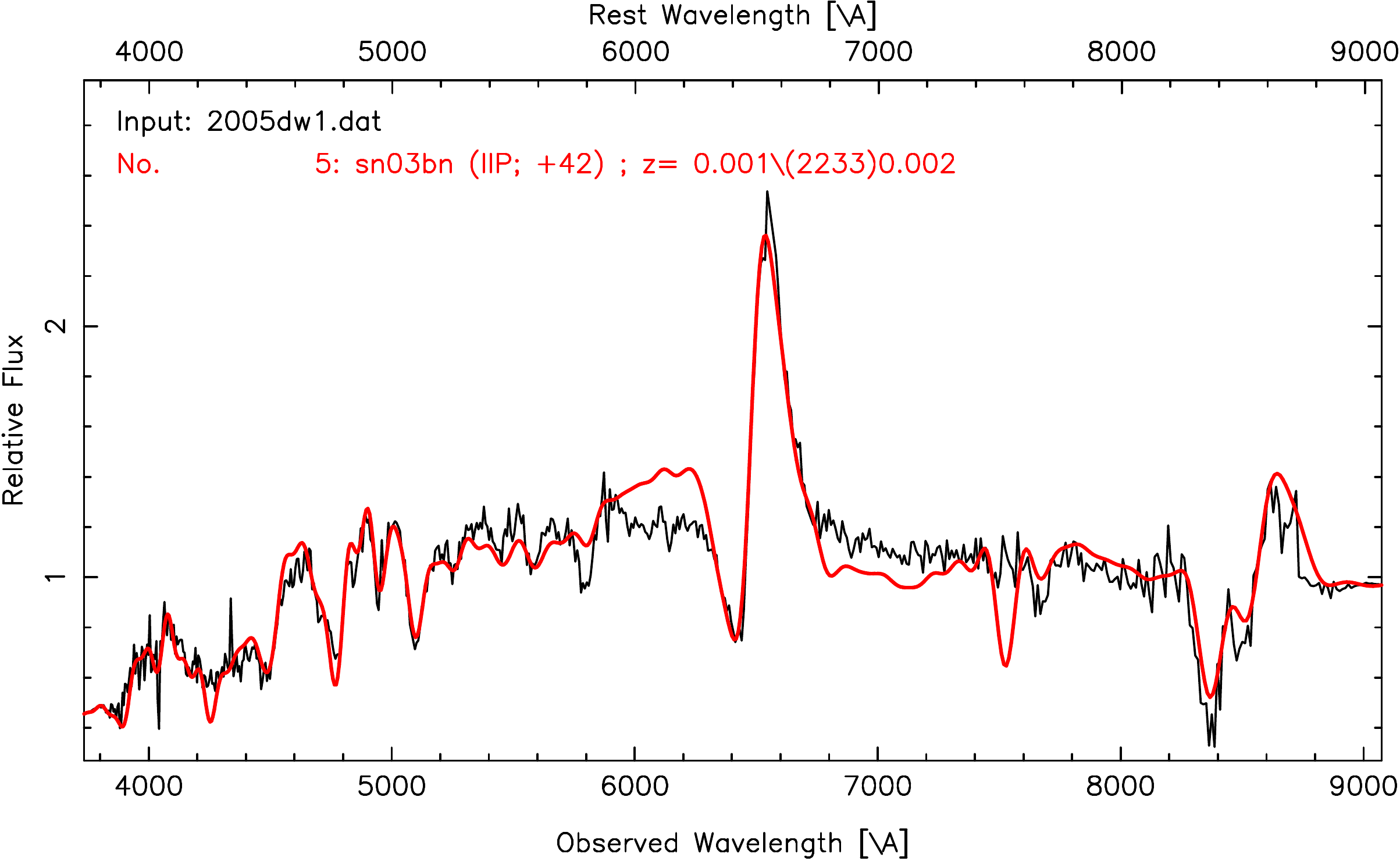}
\includegraphics[width=4.4cm]{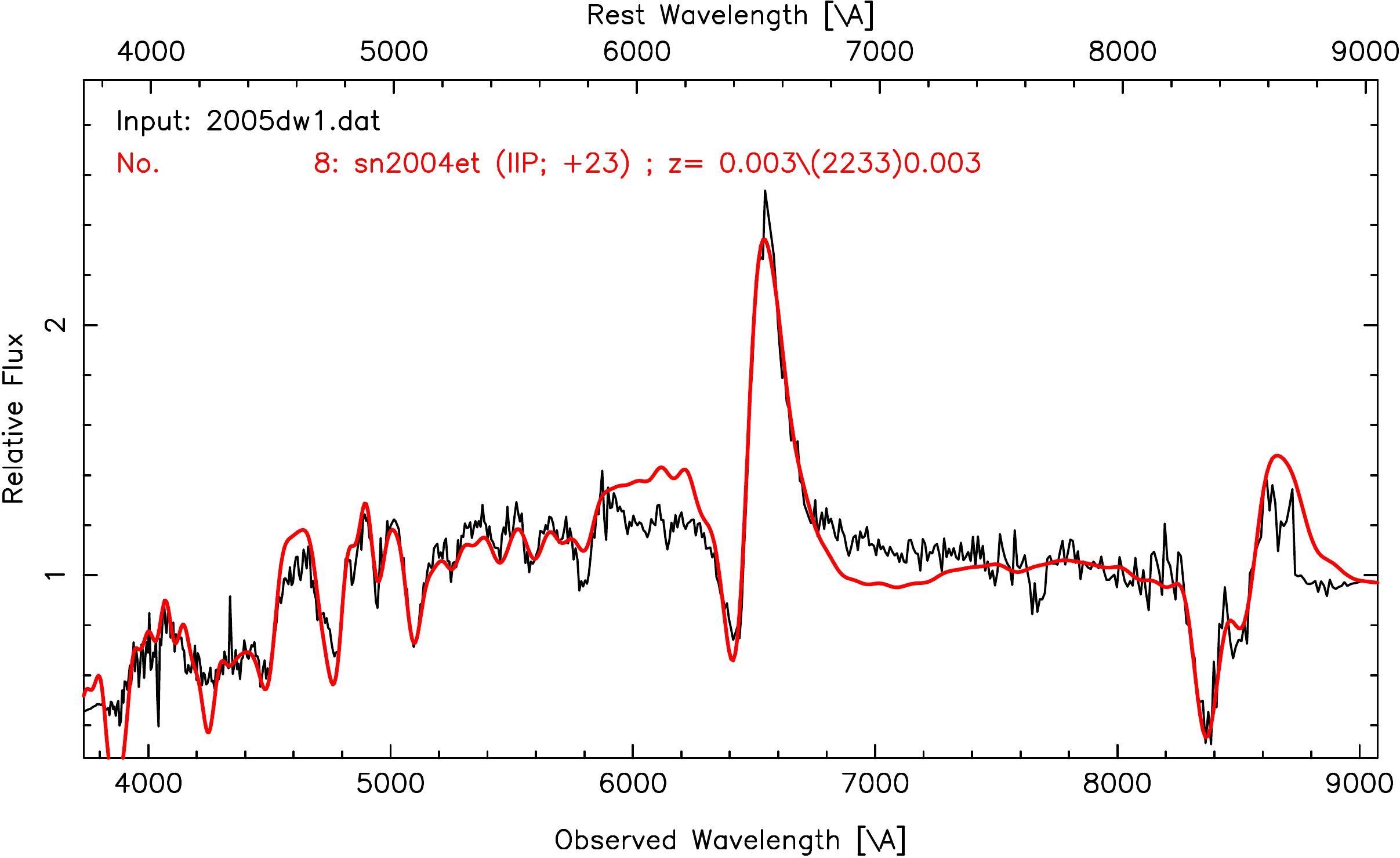}
\includegraphics[width=4.4cm]{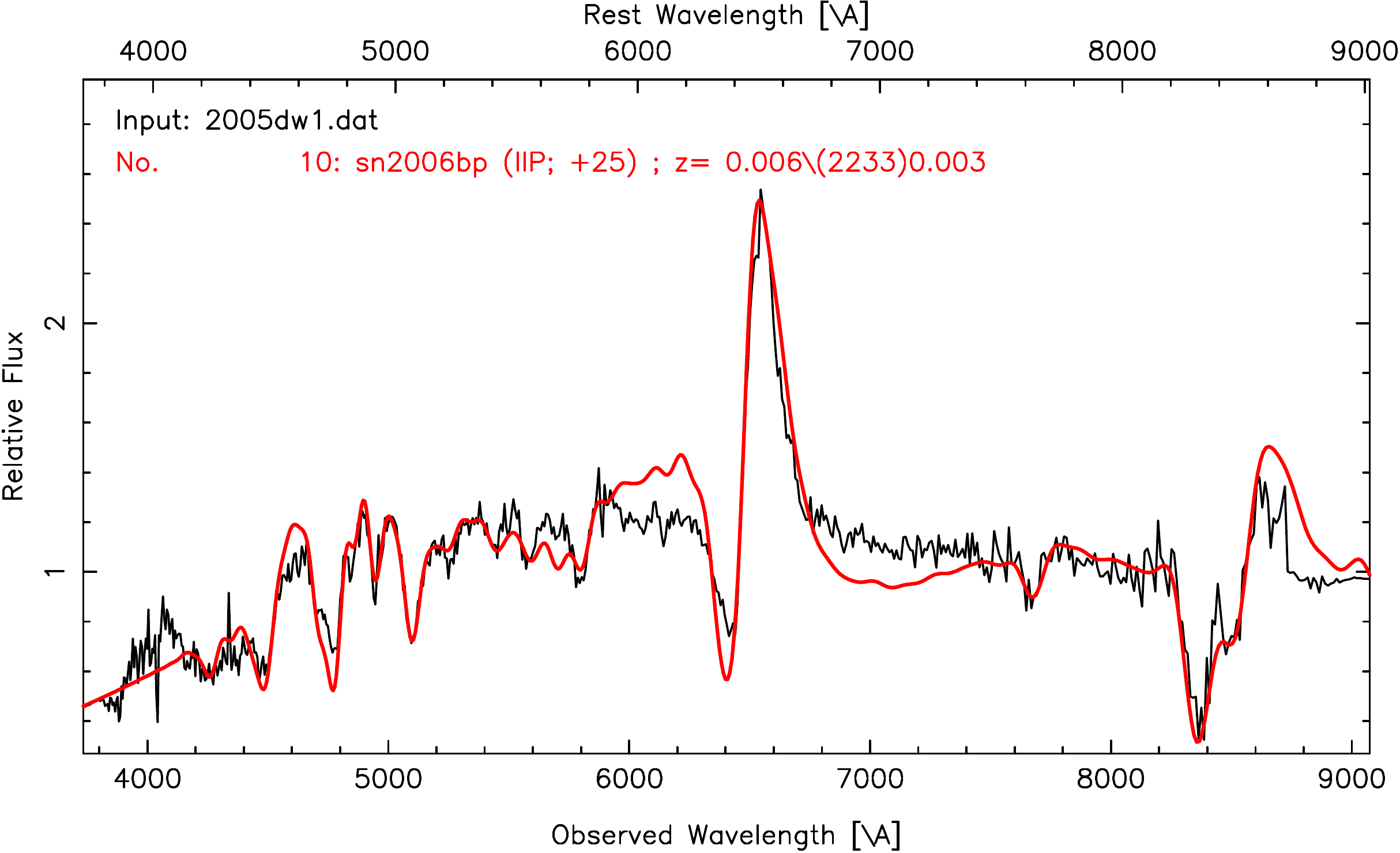}
\includegraphics[width=4.4cm]{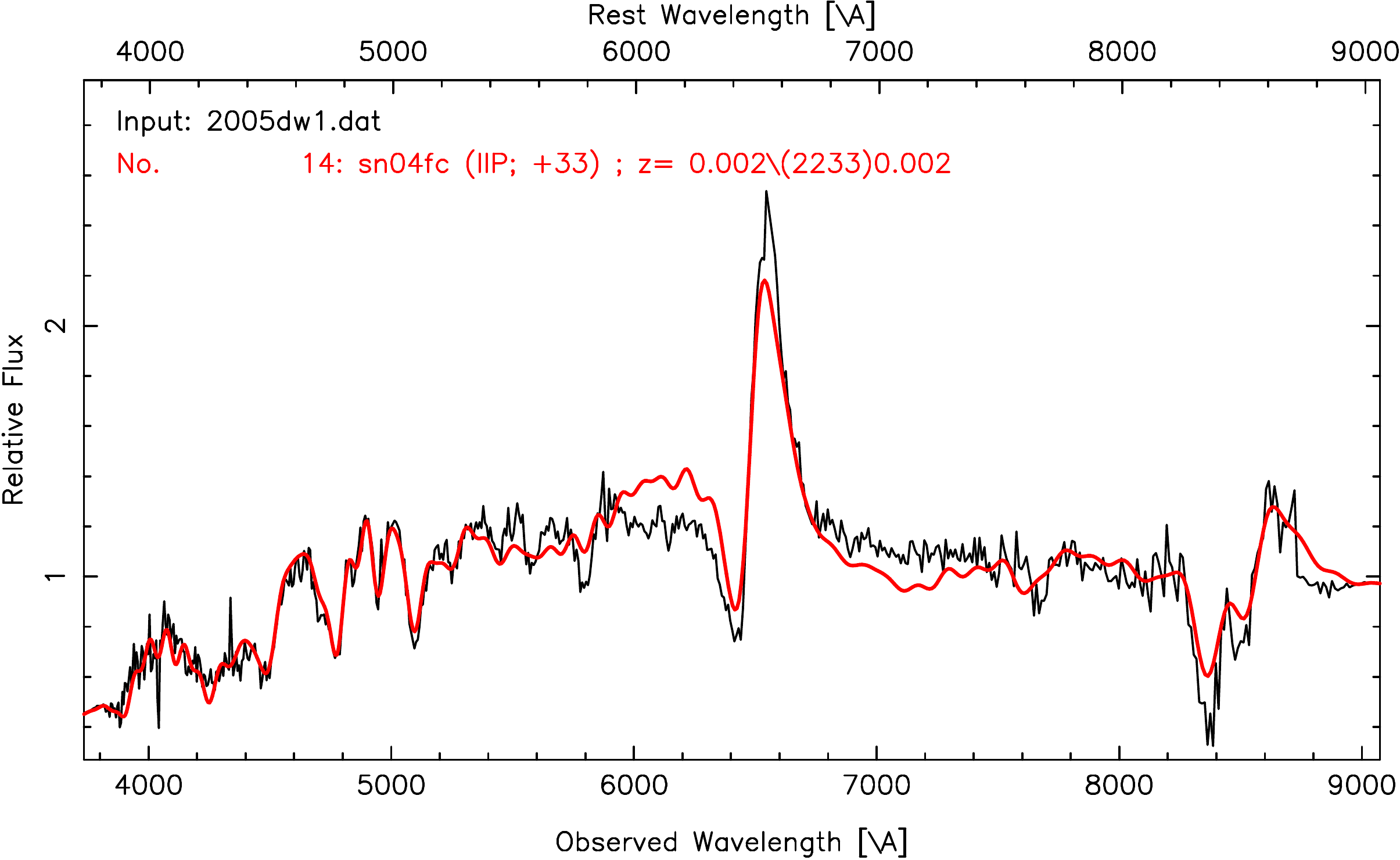}
\includegraphics[width=4.4cm]{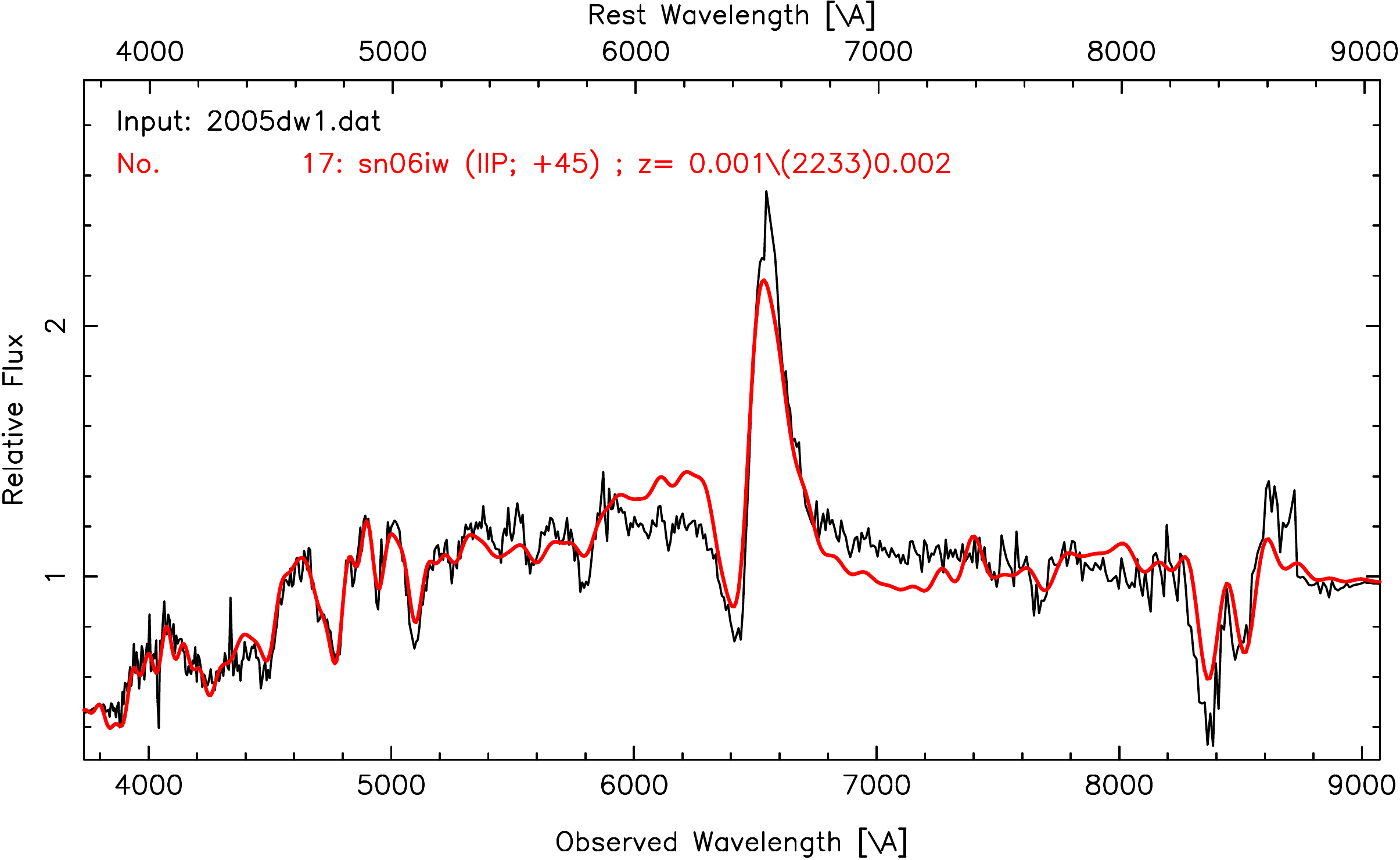}
\includegraphics[width=4.4cm]{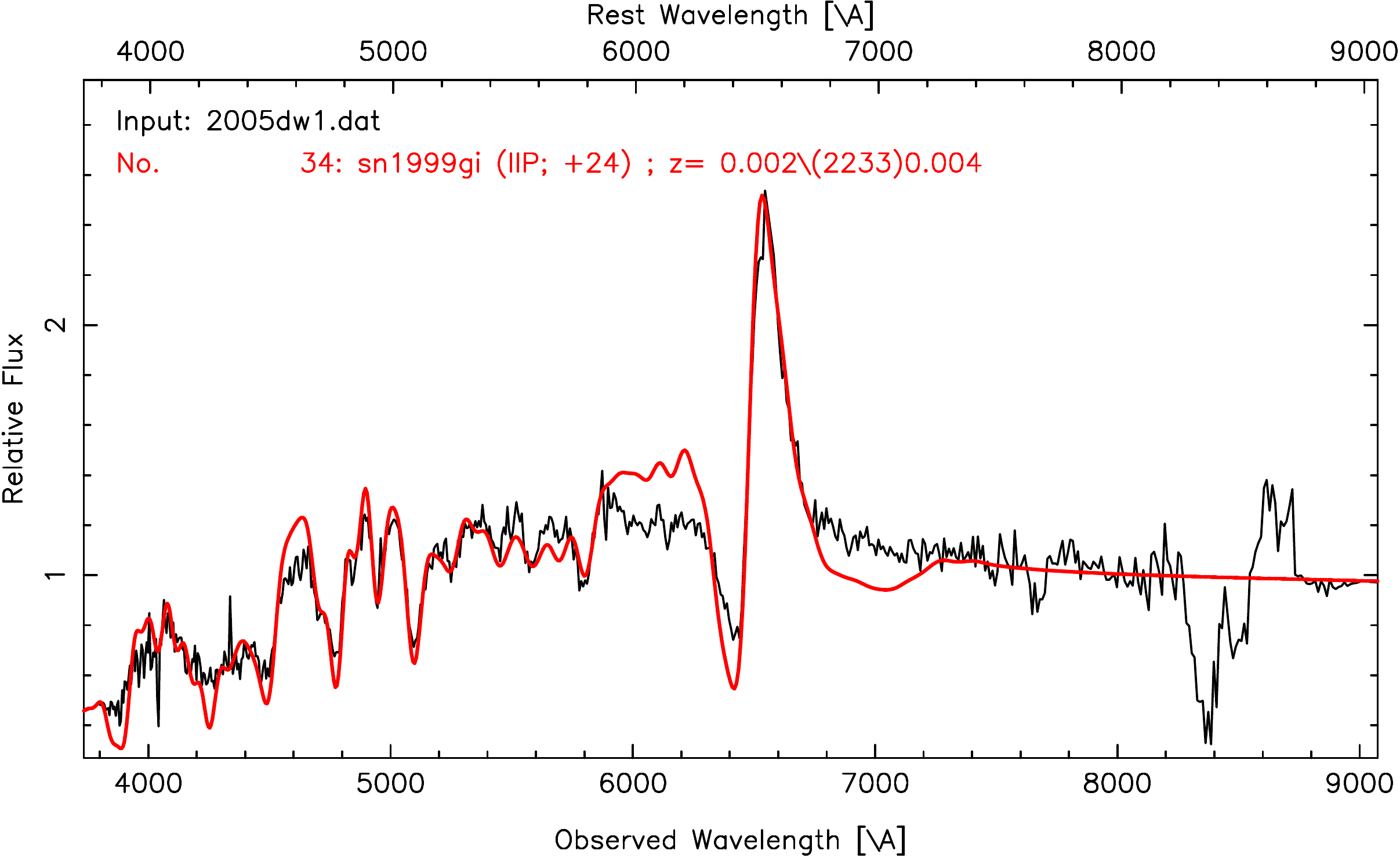}
\caption{Best spectral matching of SN~2005dw using SNID. The plots show SN~2005dw compared with 
SN~1999em, SN~2003bn, SN~2004et, SN~2006bp, SN~2004fc, SN~2006iw, and SN~1999gi at 36, 42, 39, 34, 33, 45, and 36 days from explosion.}
\end{figure}

\begin{figure}
\centering
\includegraphics[width=4.4cm]{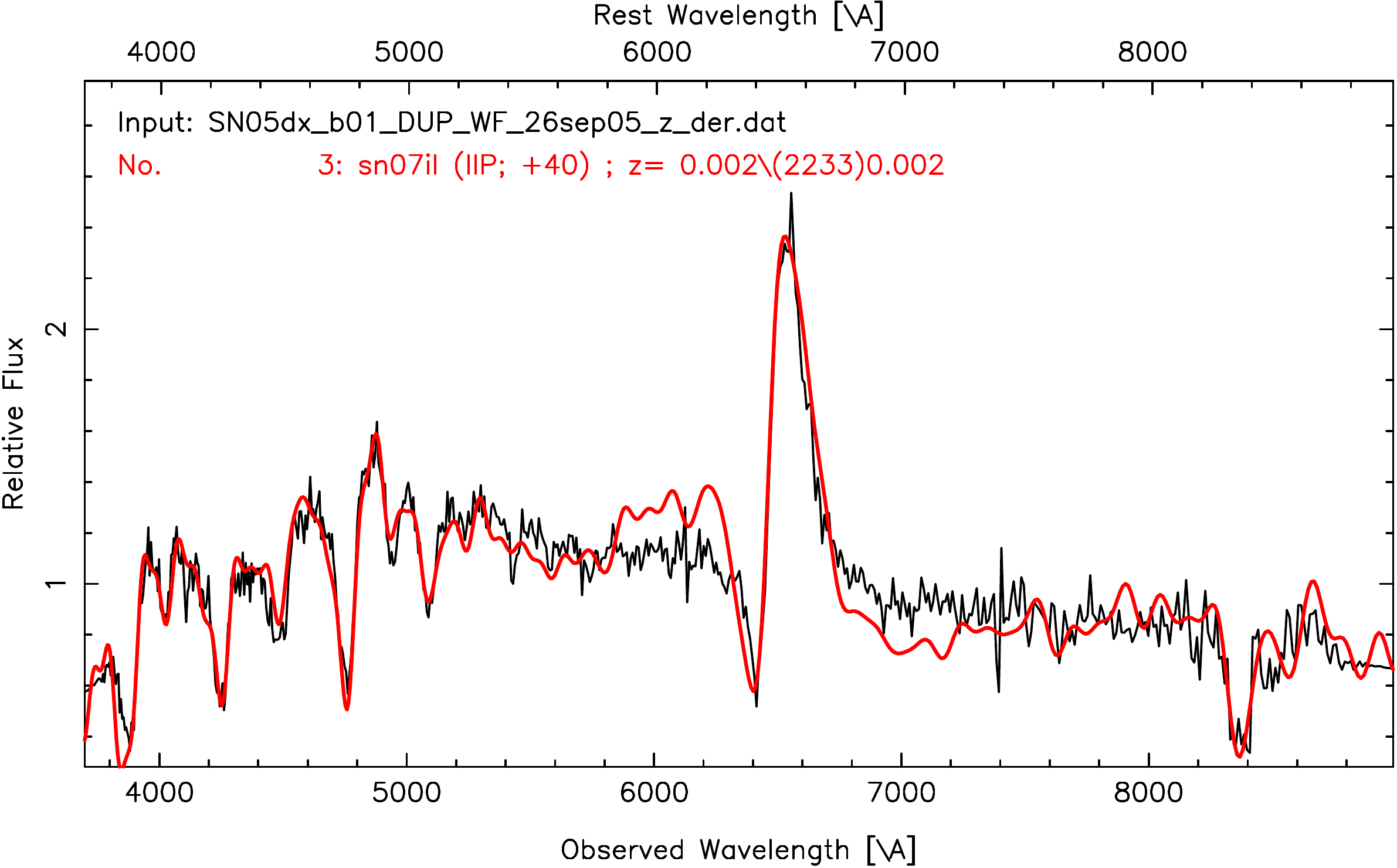}
\includegraphics[width=4.4cm]{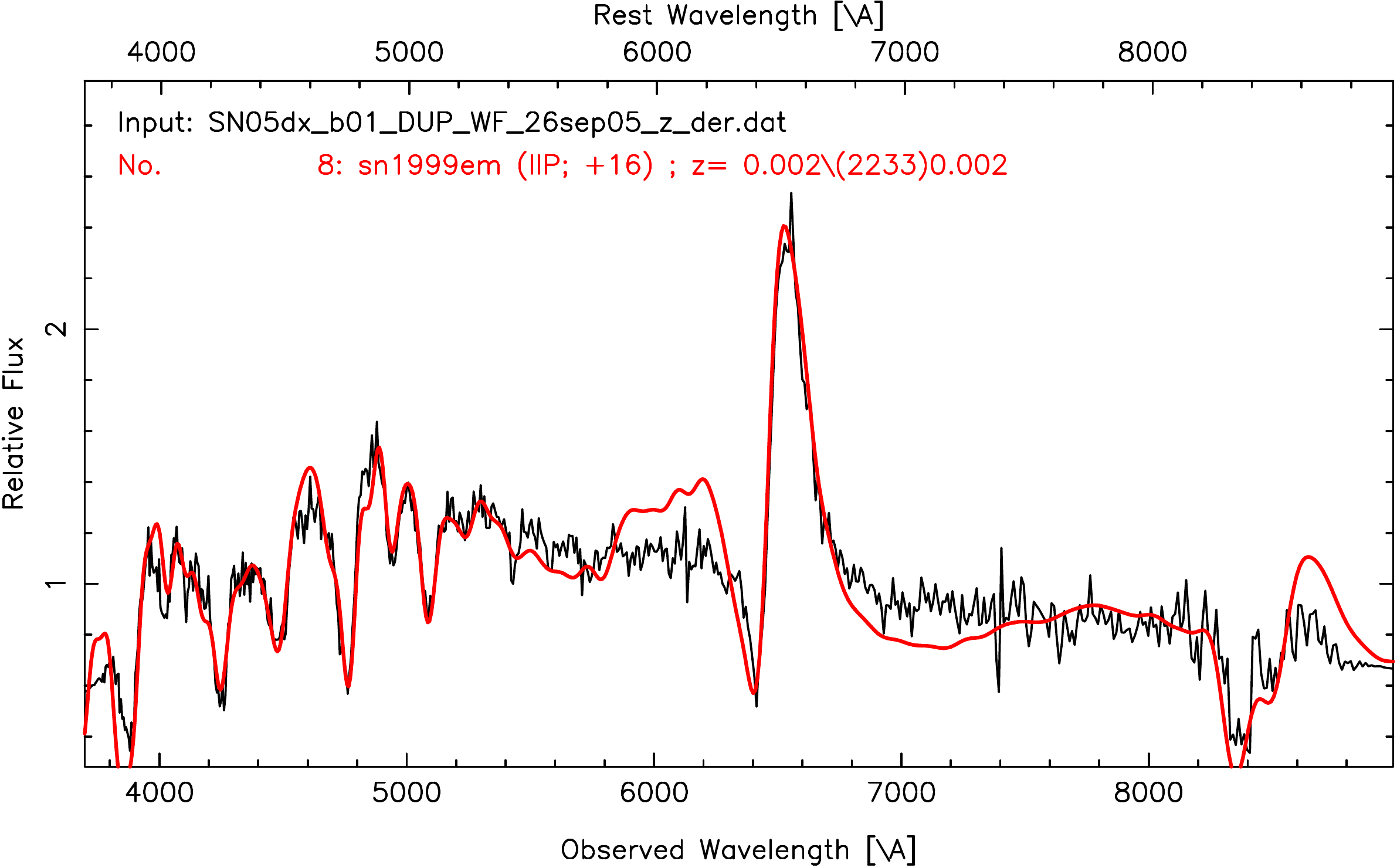}
\includegraphics[width=4.4cm]{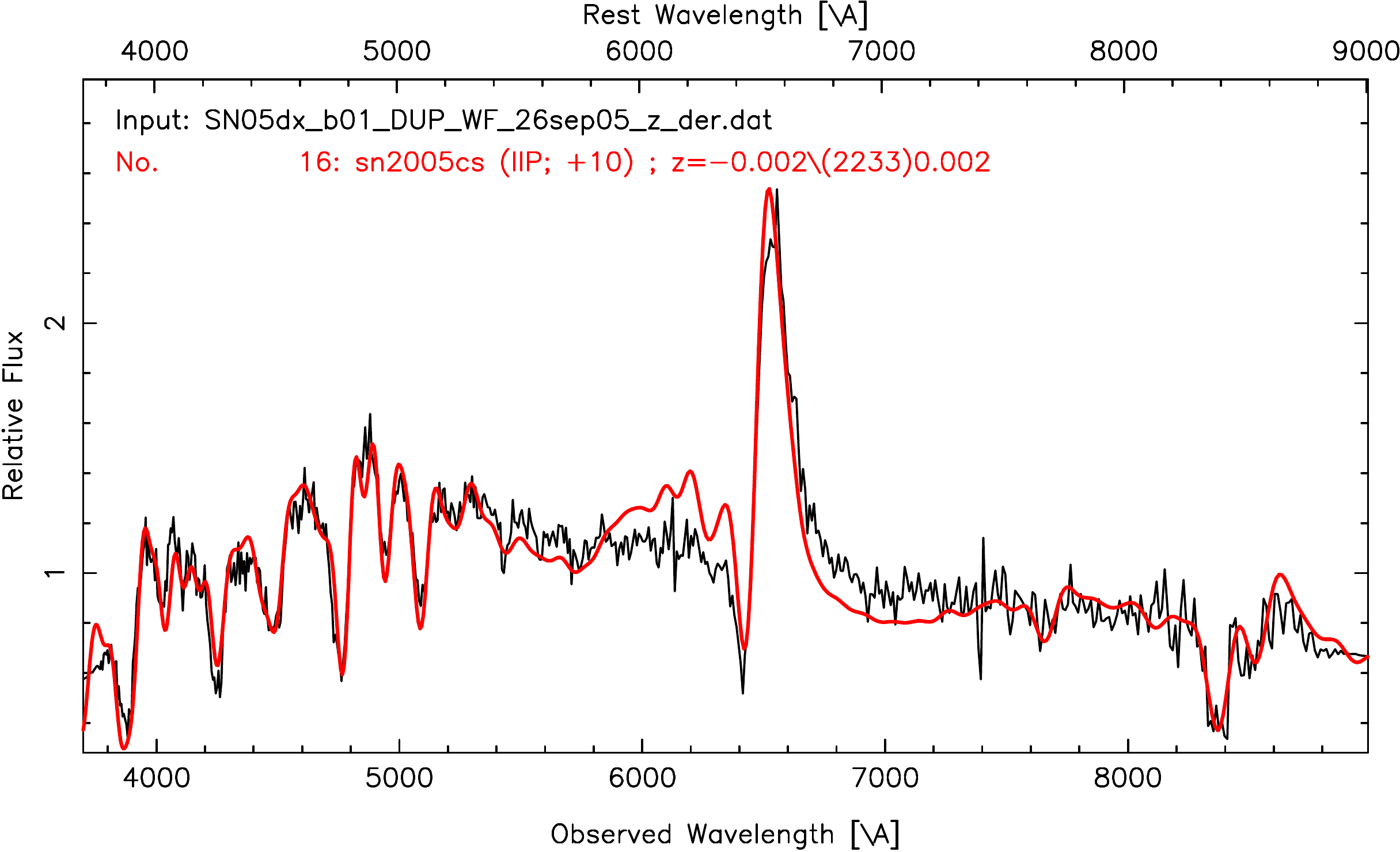}
\includegraphics[width=4.4cm]{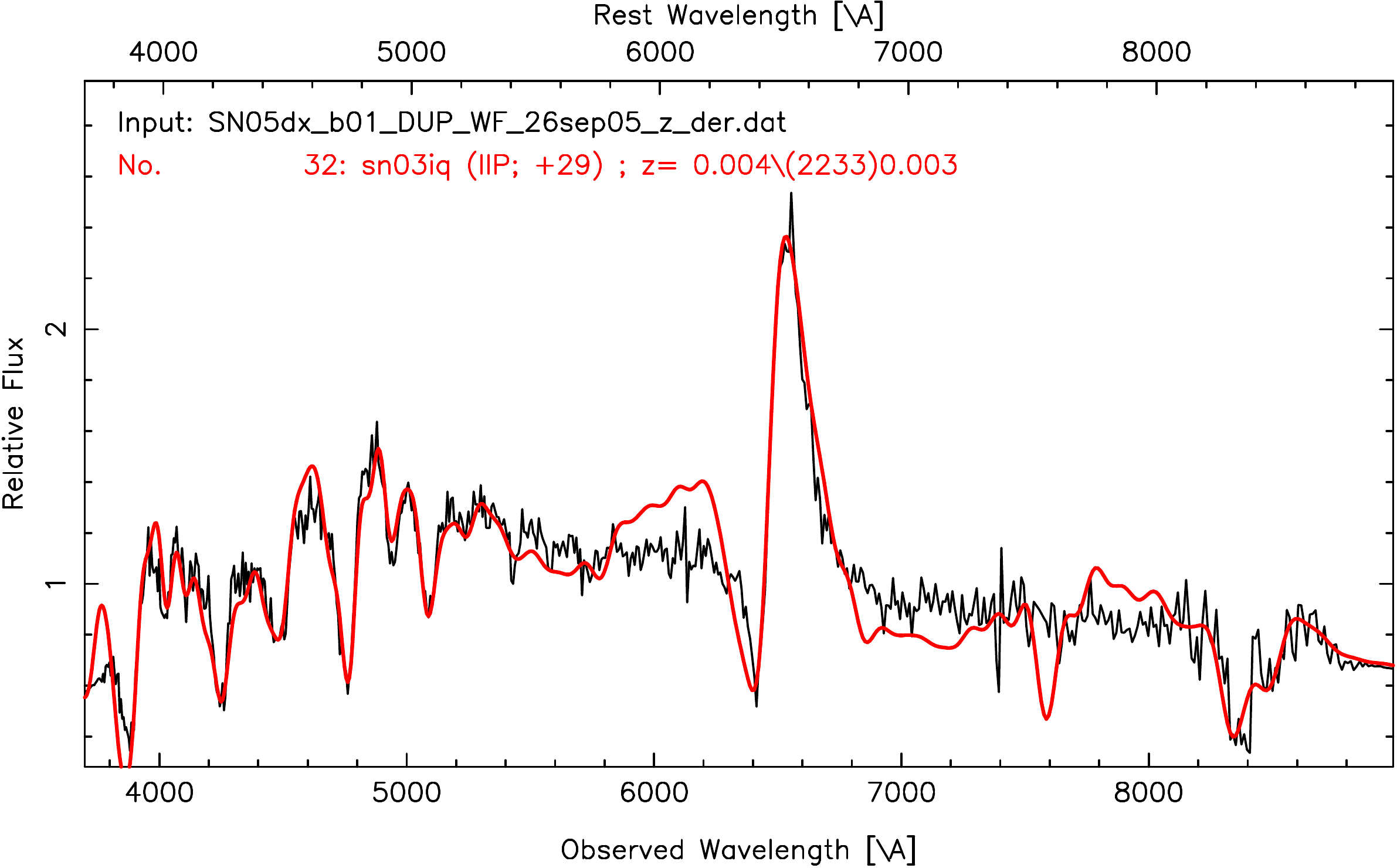}
\caption{Best spectral matching of SN~2005dx using SNID. The plots show SN~2005dx compared with 
SN~2007il, SN~1999em, SN~2005cs, and SN~2003iq at 40, 26, 16, and 29 days from explosion.}
\end{figure}

\begin{figure}
\centering
\includegraphics[width=4.4cm]{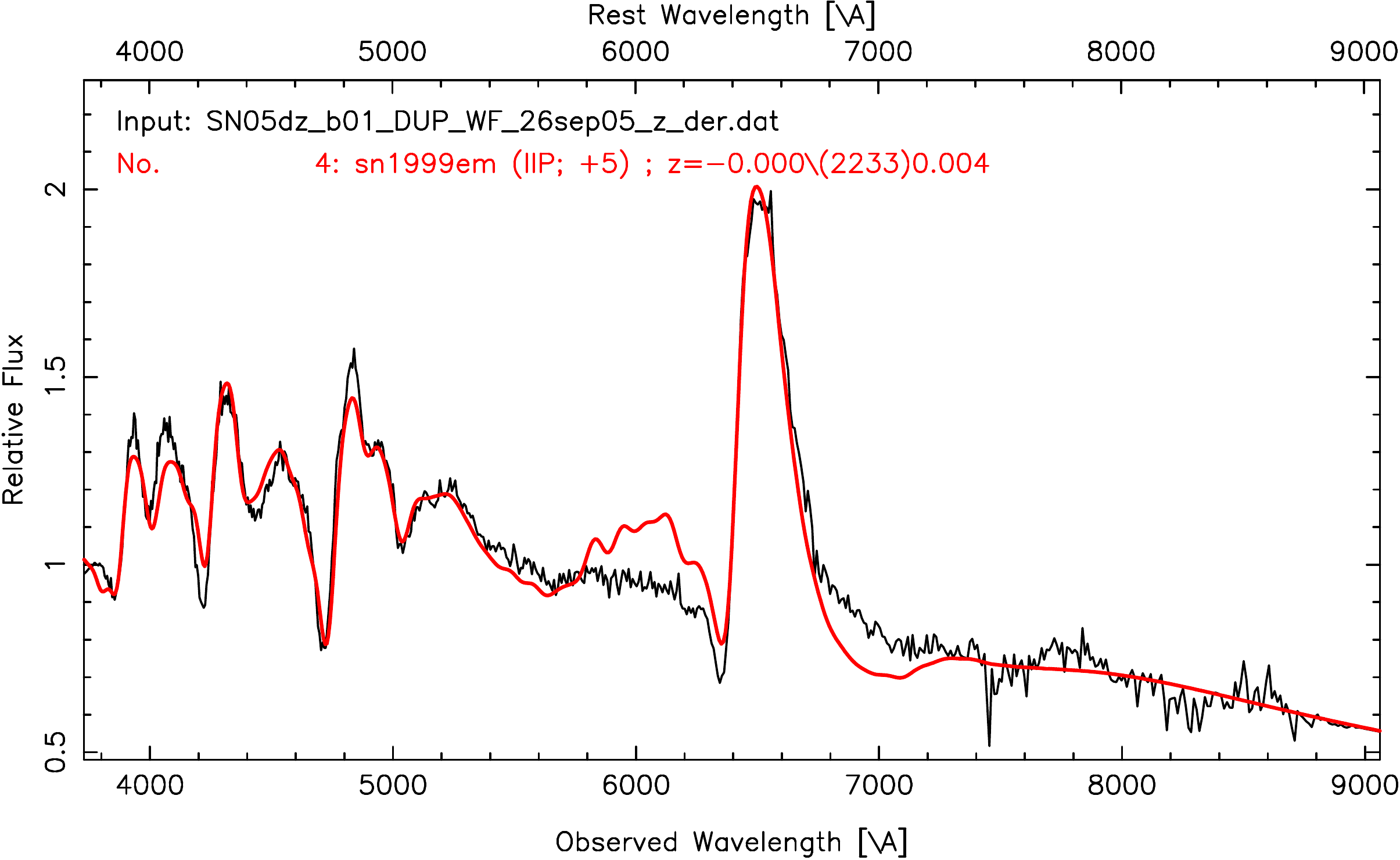}
\includegraphics[width=4.4cm]{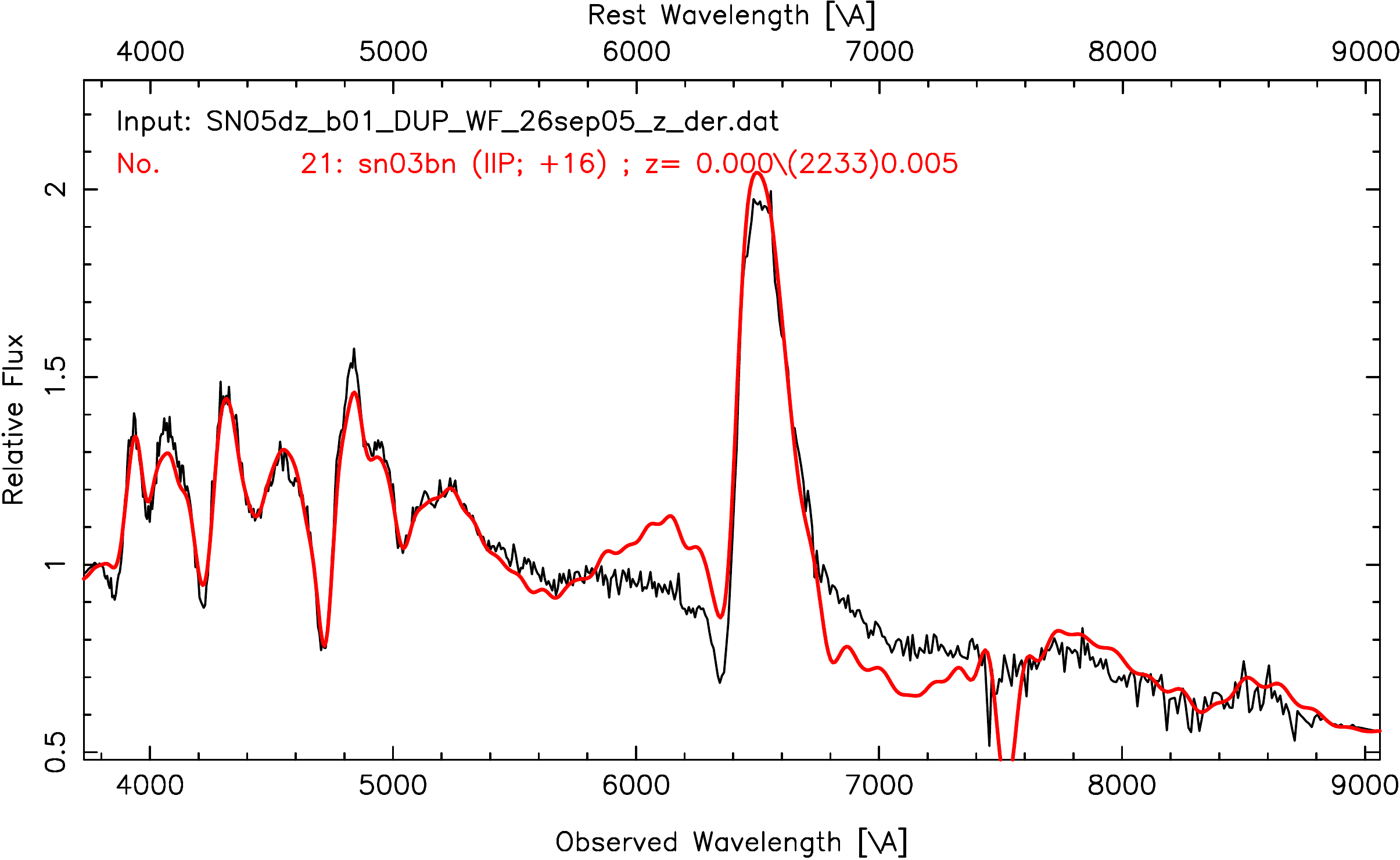}
\caption{Best spectral matching of SN~2005dz using SNID. The plots show SN~2005dz compared with 
SN~1999em and SN~2003bn at 15 and 16 days from explosion.}
\end{figure}

\clearpage

\begin{figure}
\centering
\includegraphics[width=4.4cm]{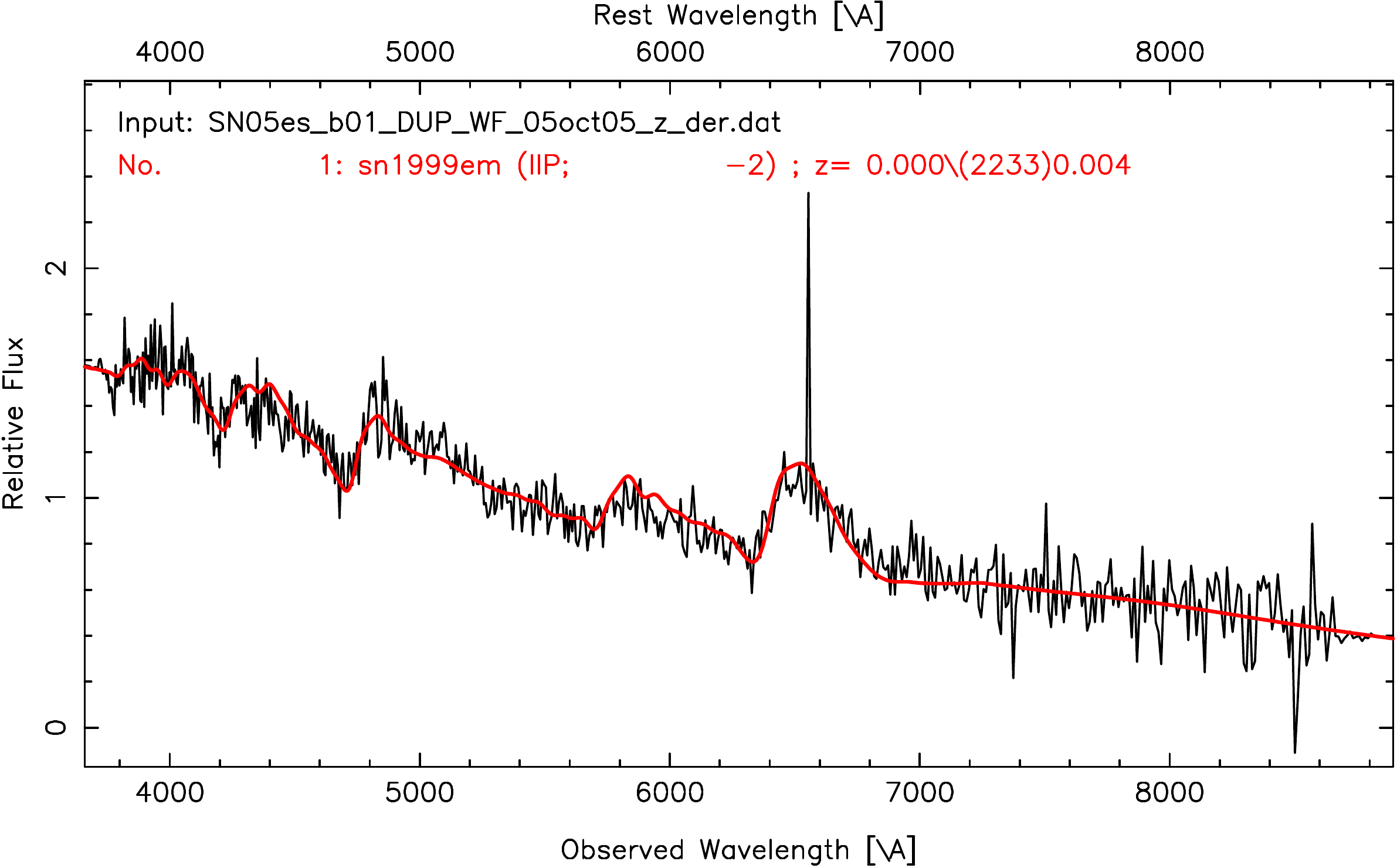}
\includegraphics[width=4.4cm]{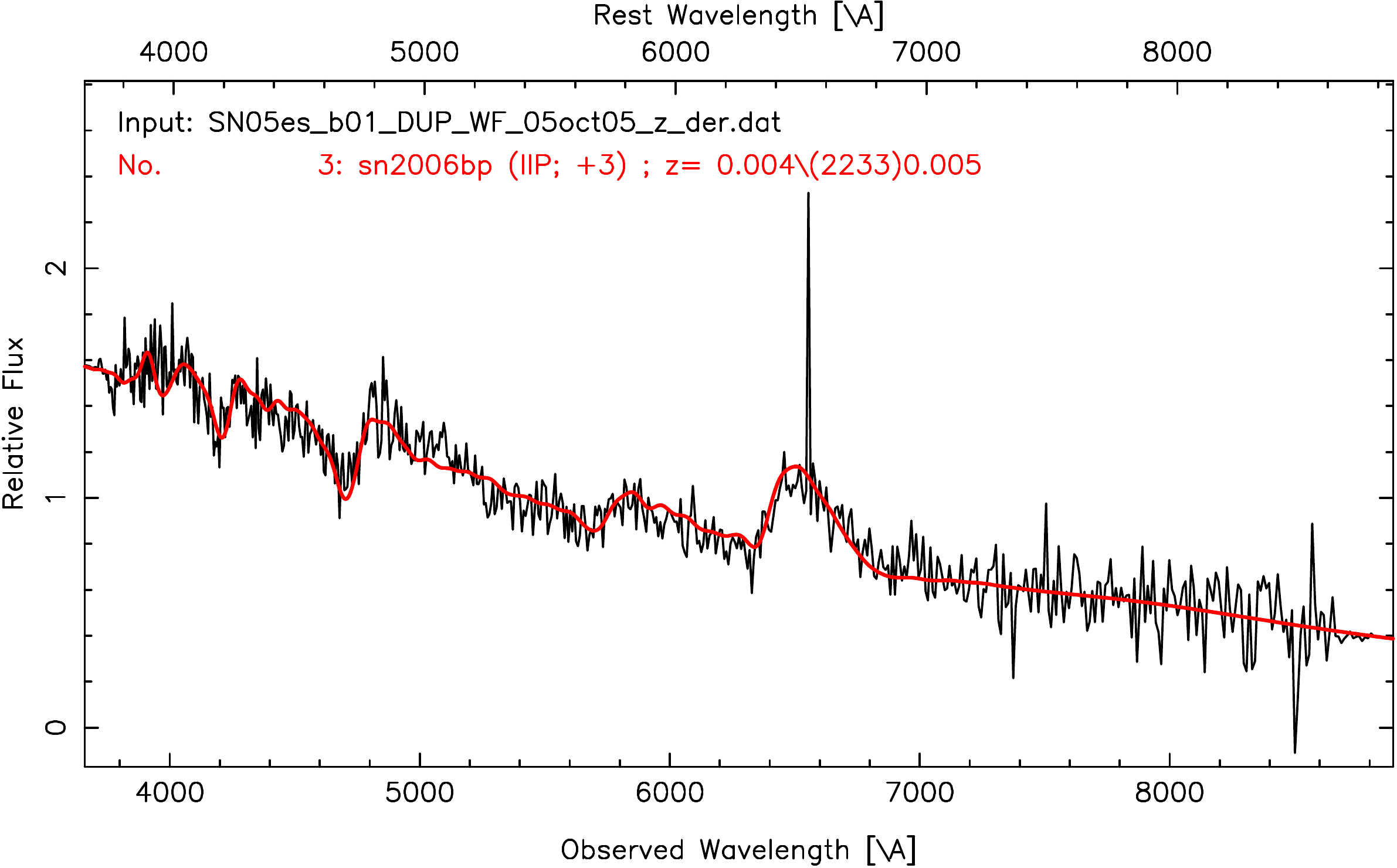}
\includegraphics[width=4.4cm]{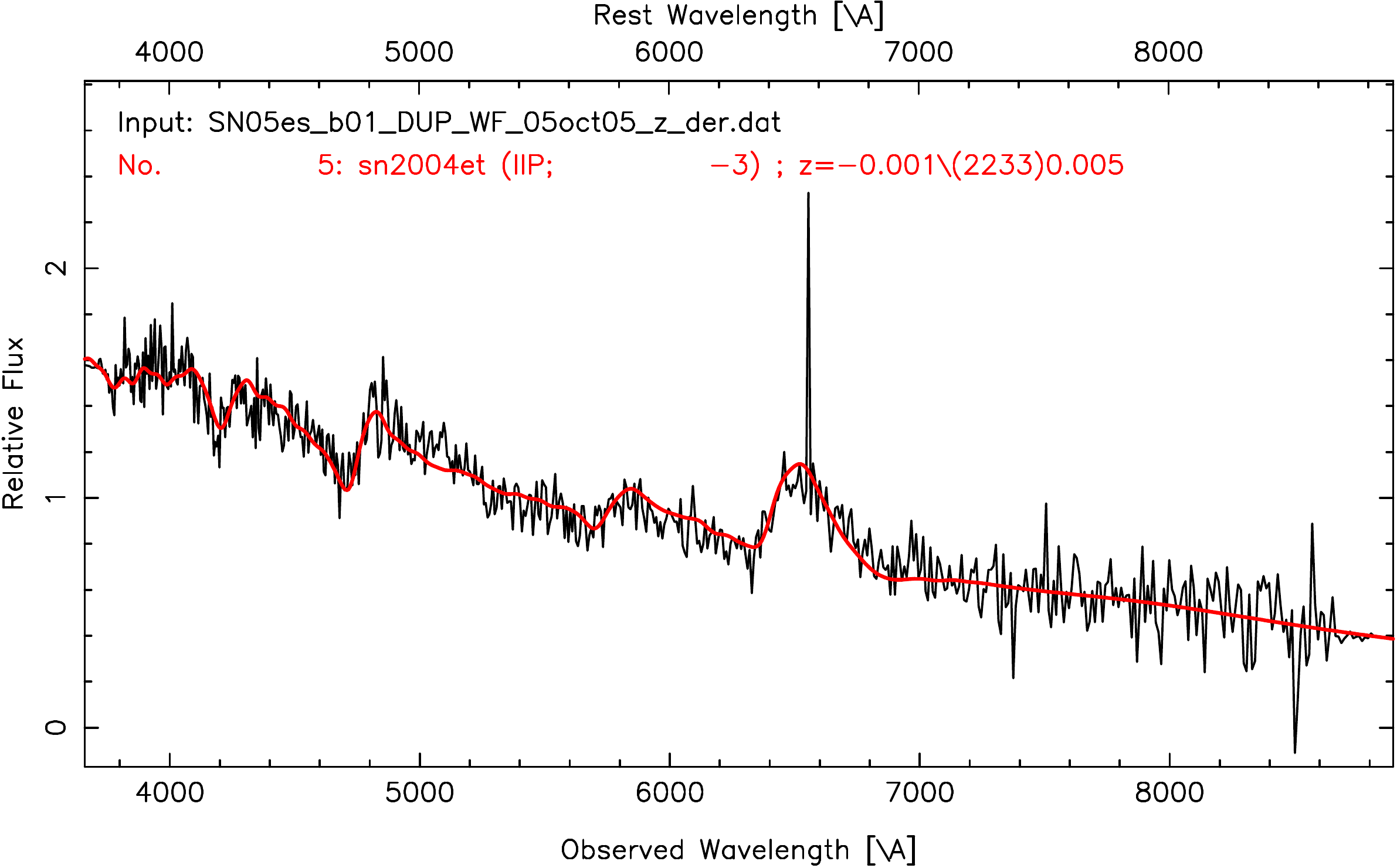}
\includegraphics[width=4.4cm]{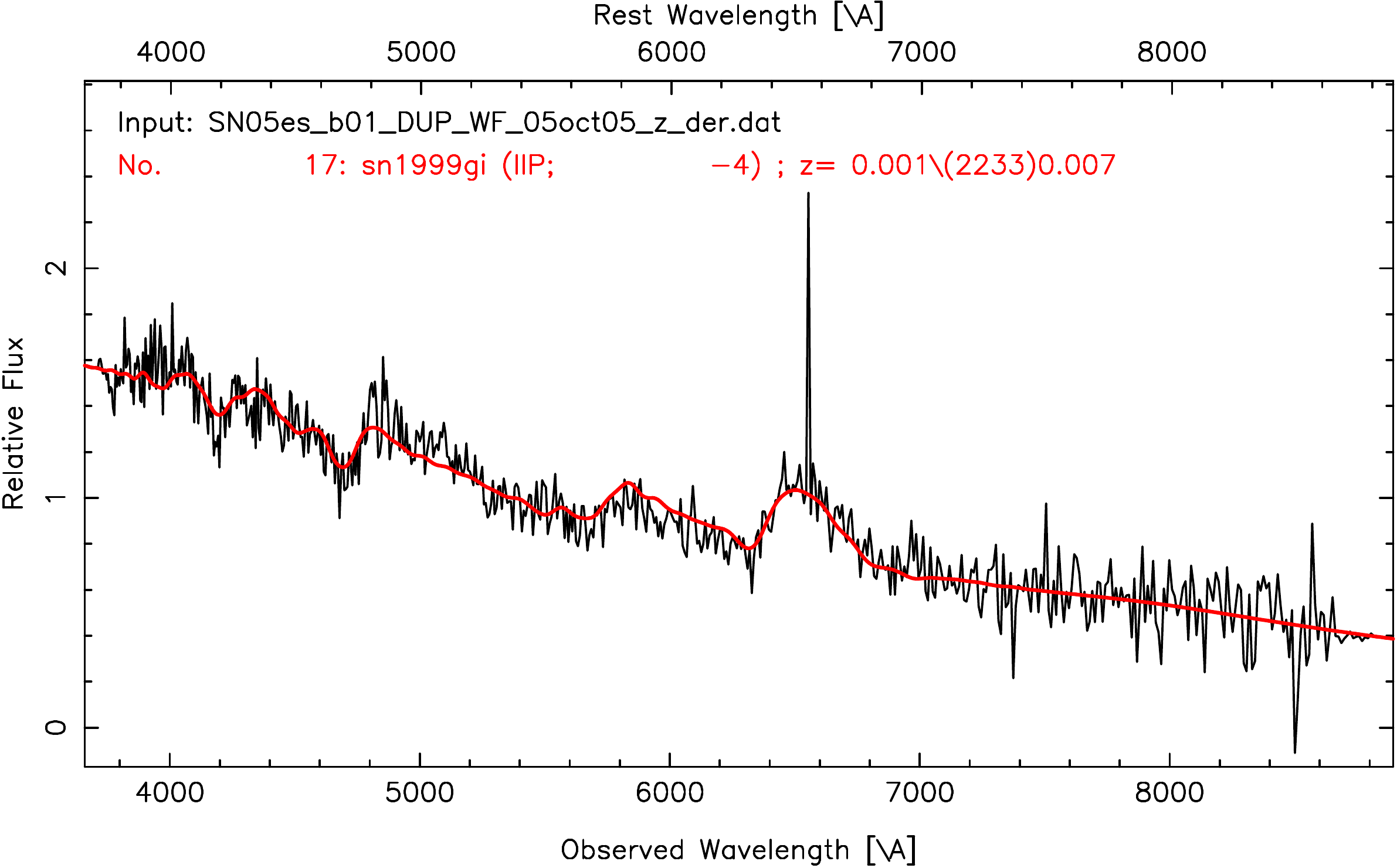}
\includegraphics[width=4.4cm]{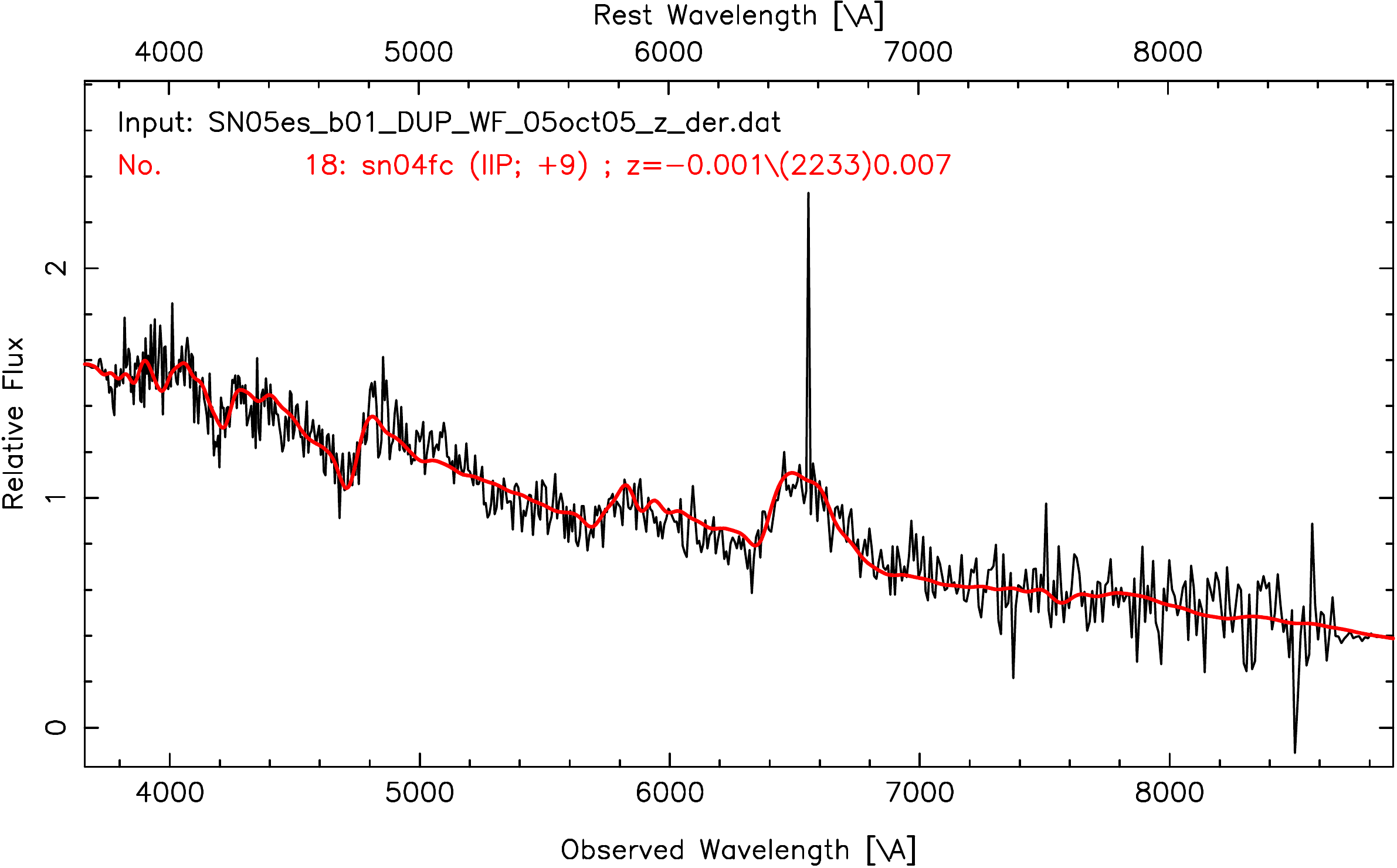}
\includegraphics[width=4.4cm]{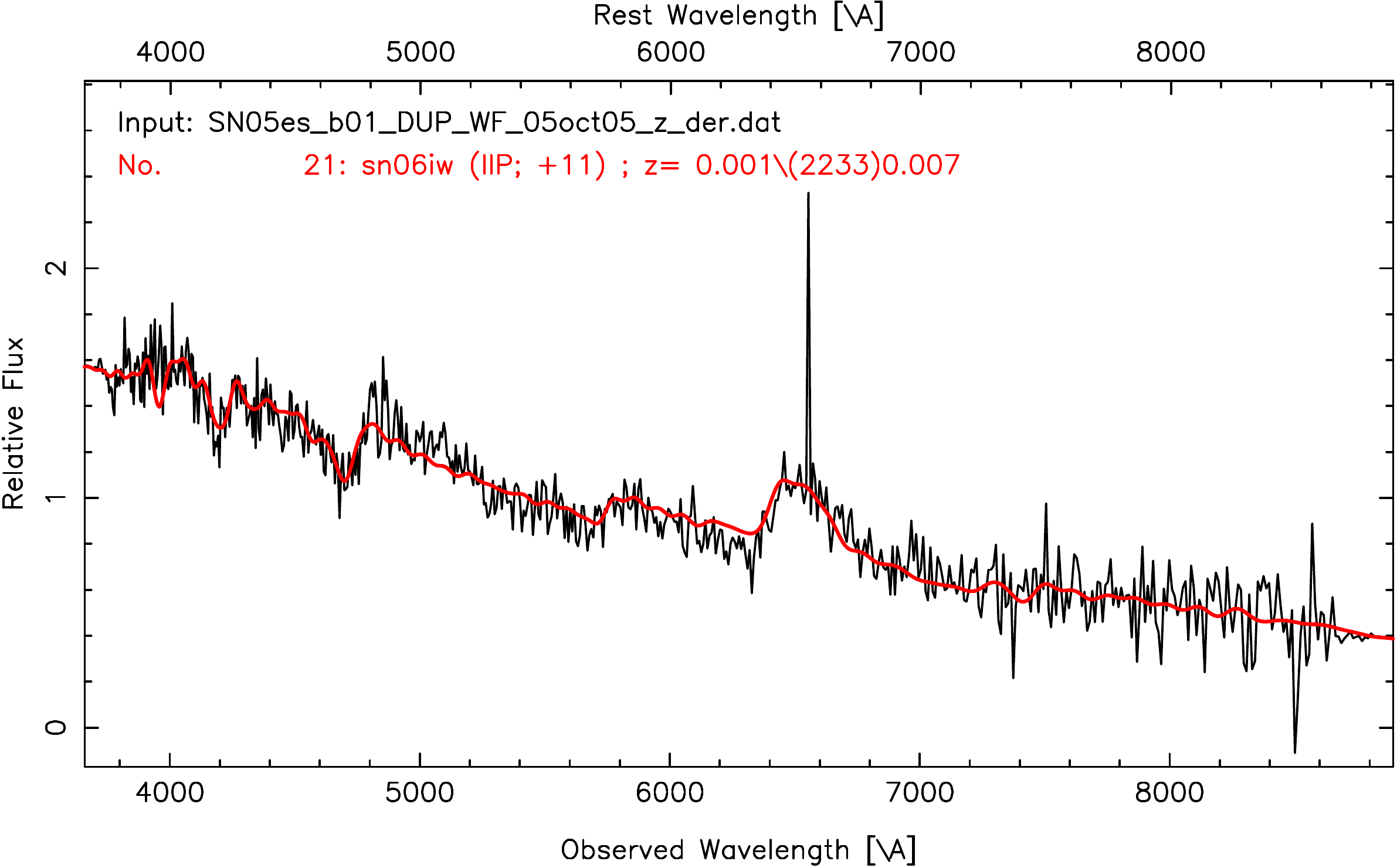}
\caption{Best spectral matching of SN~2005es using SNID. The plots show SN~2005es compared with 
SN~1999em, SN~2006bp, SN~2004et, SN~1999gi, SN~2004fc, and SN~2006iw at 8, 12, 13, 8, 9, and 11 days from explosion.}
\end{figure}


\begin{figure}
\centering
\includegraphics[width=4.4cm]{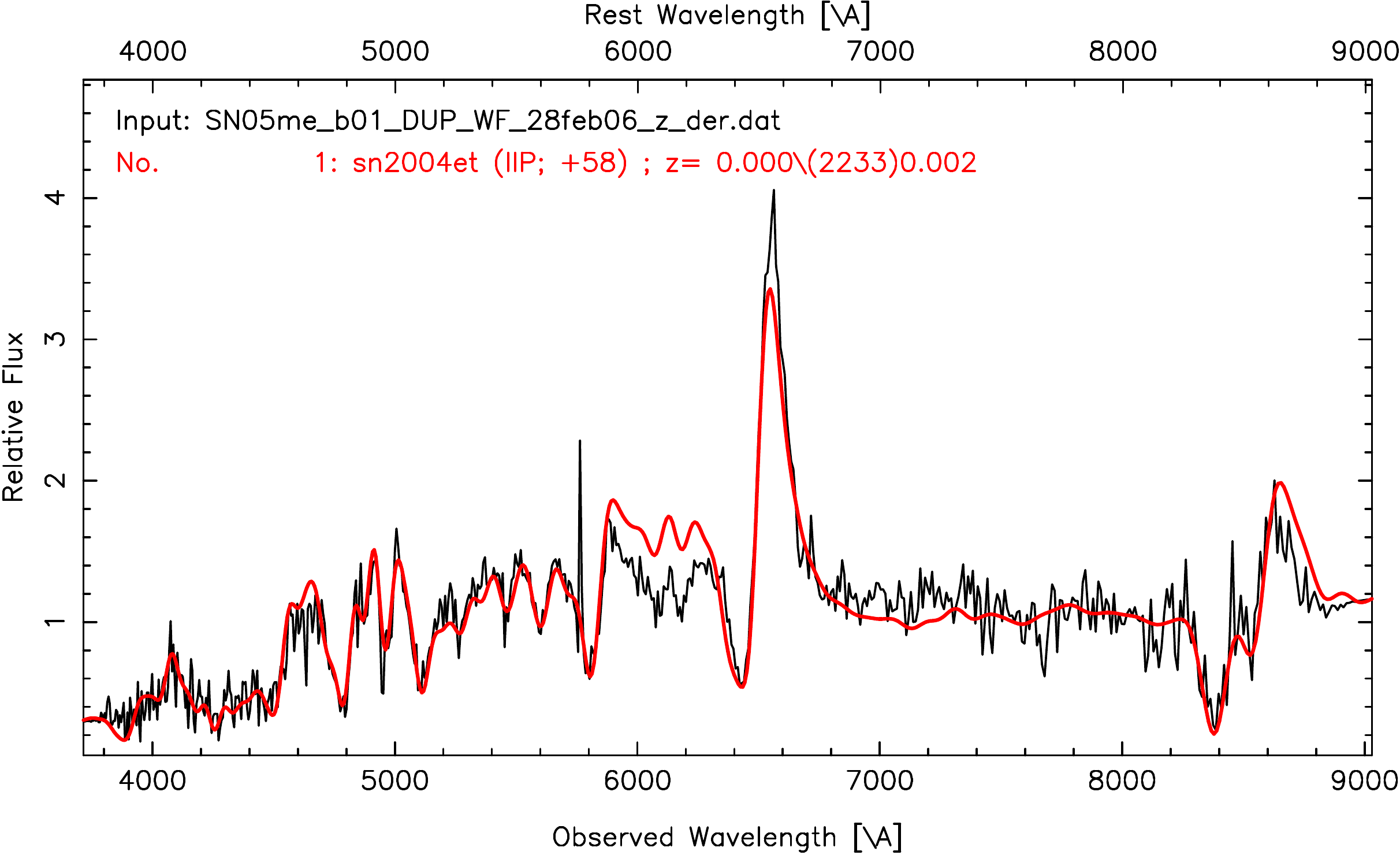}
\includegraphics[width=4.4cm]{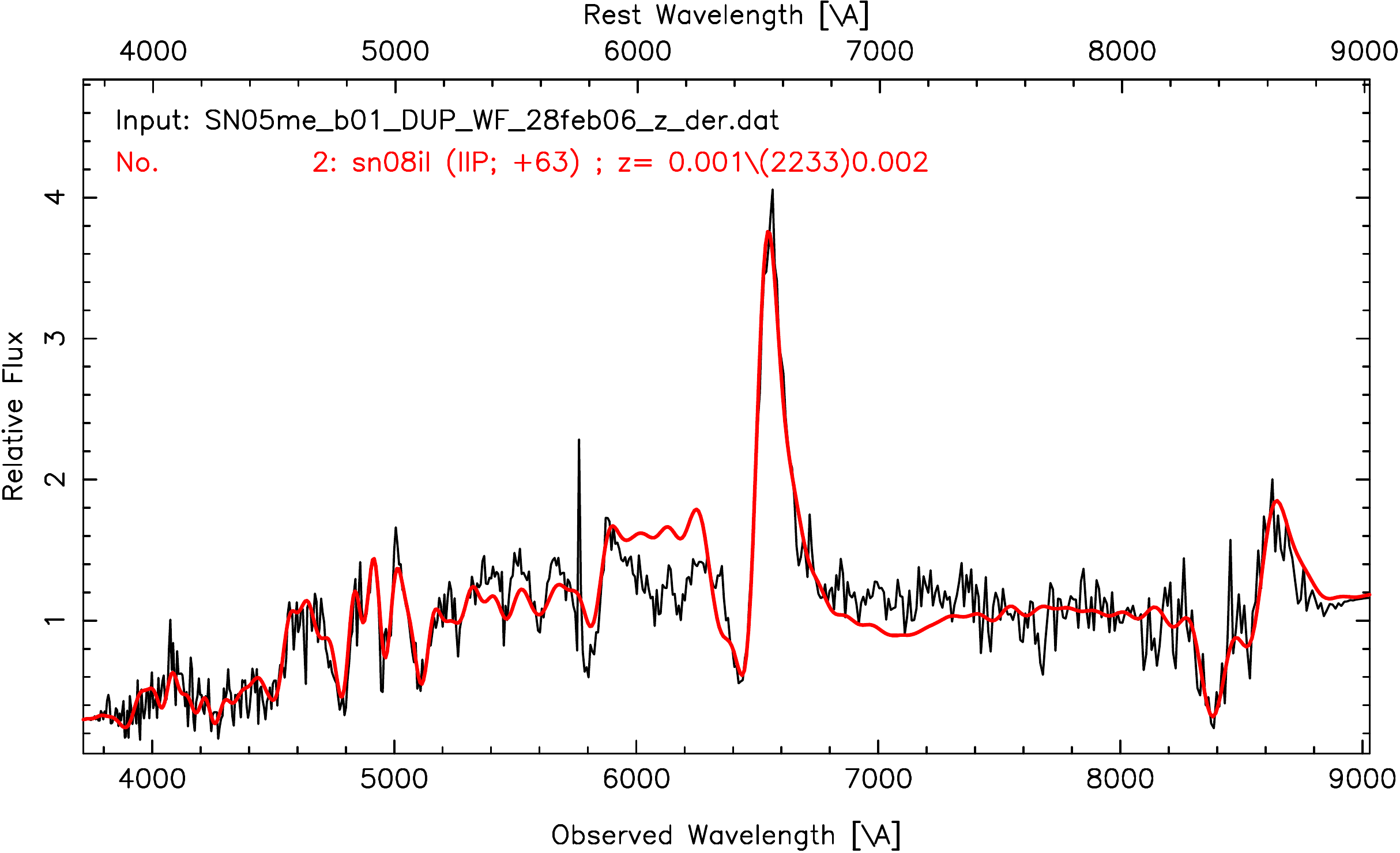}
\includegraphics[width=4.4cm]{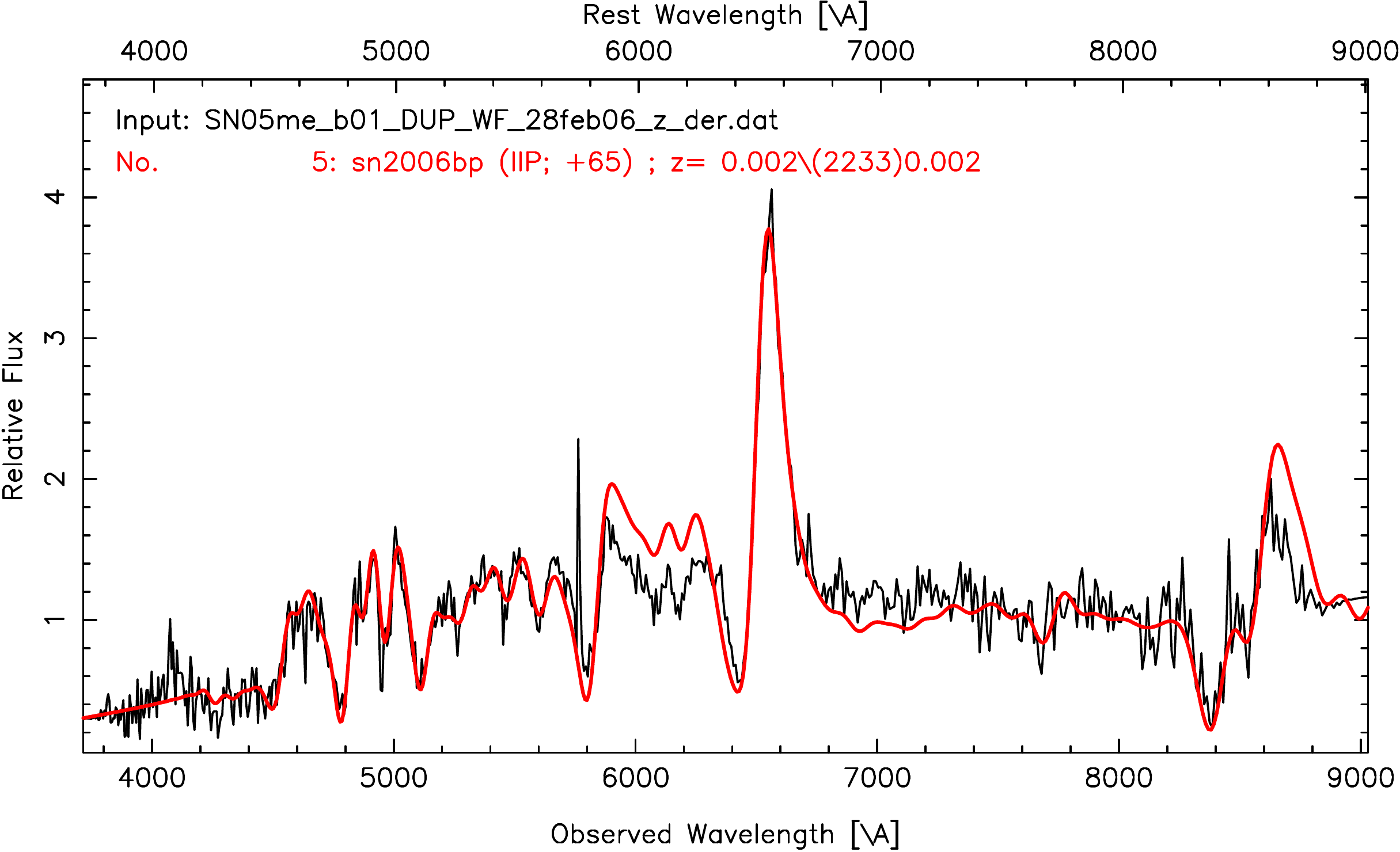}
\includegraphics[width=4.4cm]{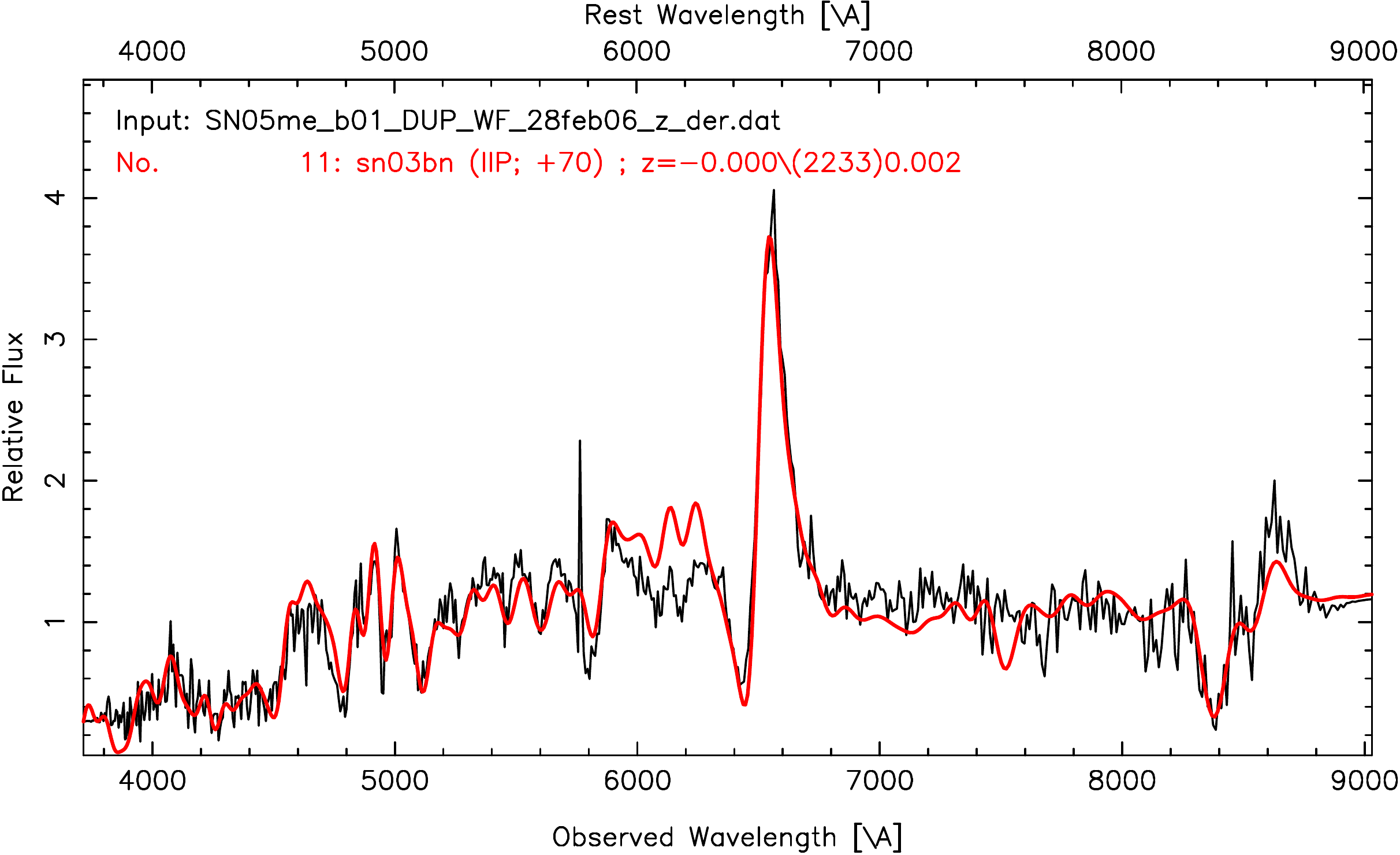}
\includegraphics[width=4.4cm]{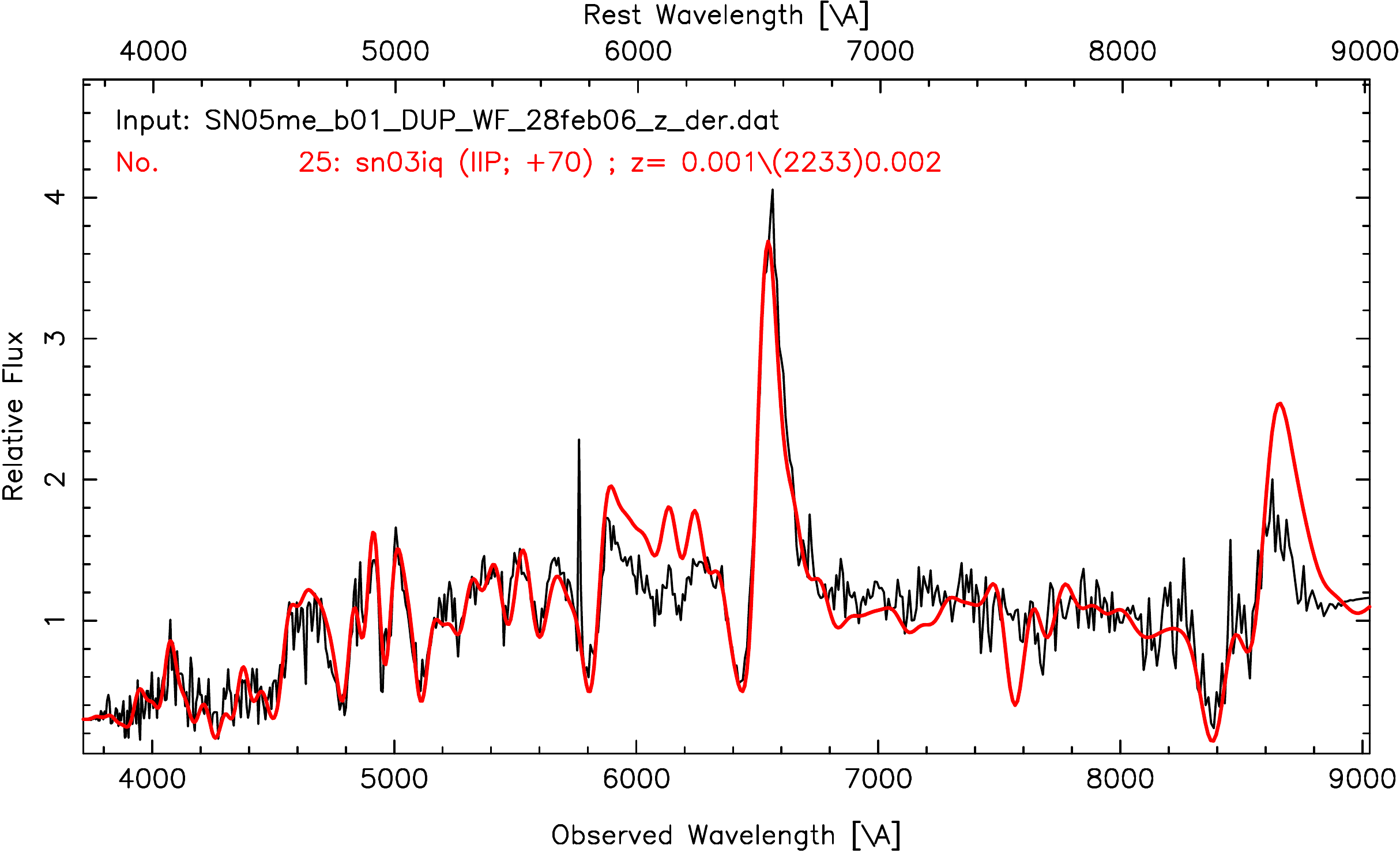}
\caption{Best spectral matching of SN~2005me using SNID. The plots show SN~2005me compared with 
SN~2004et, SN~2008il, SN~2006bp, SN~2003bn, and SN~2003iq at 74, 63, 74, 70 and 70 days from explosion.}
\end{figure}

\clearpage

\begin{figure}
\centering
\includegraphics[width=4.4cm]{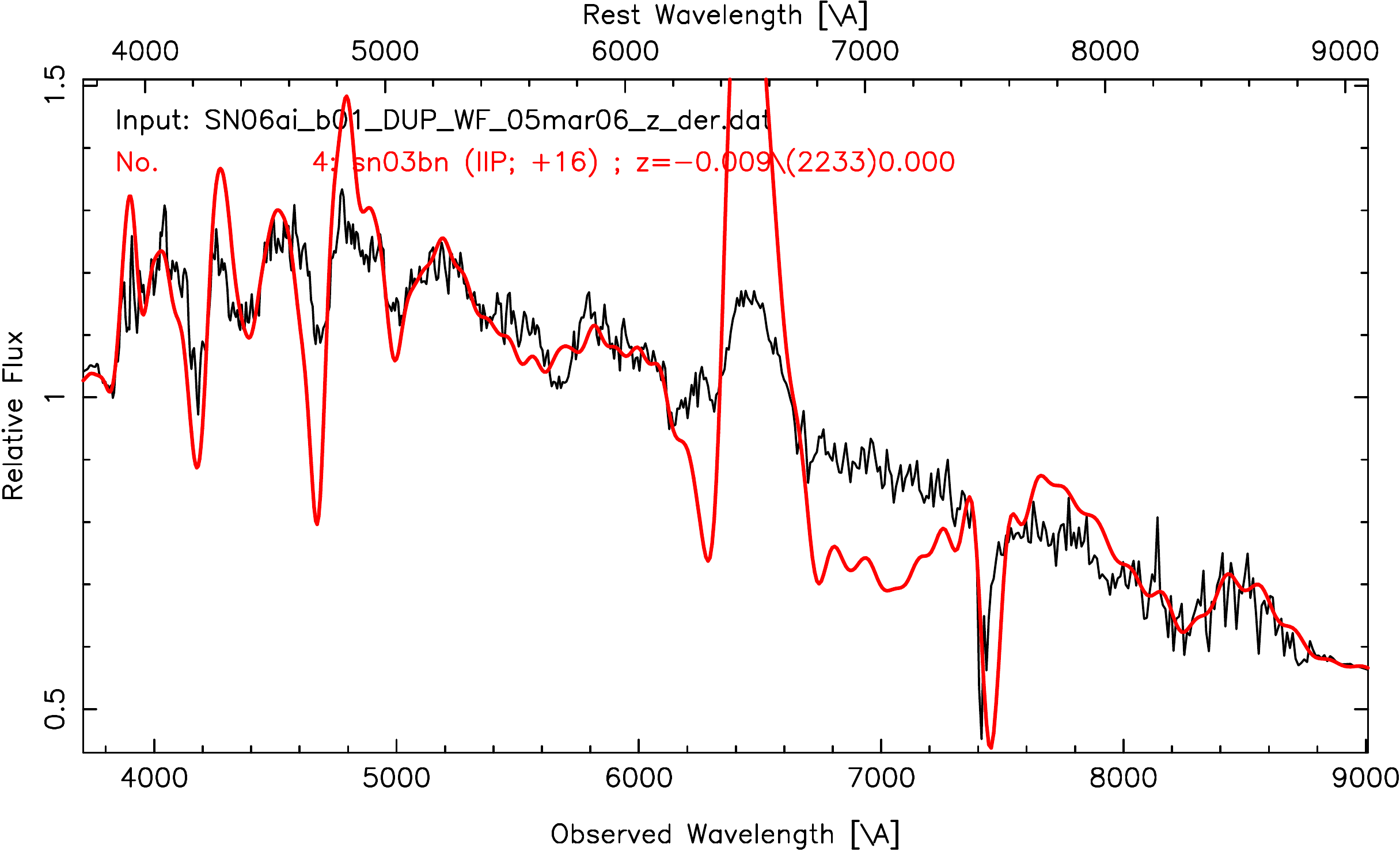}
\includegraphics[width=4.4cm]{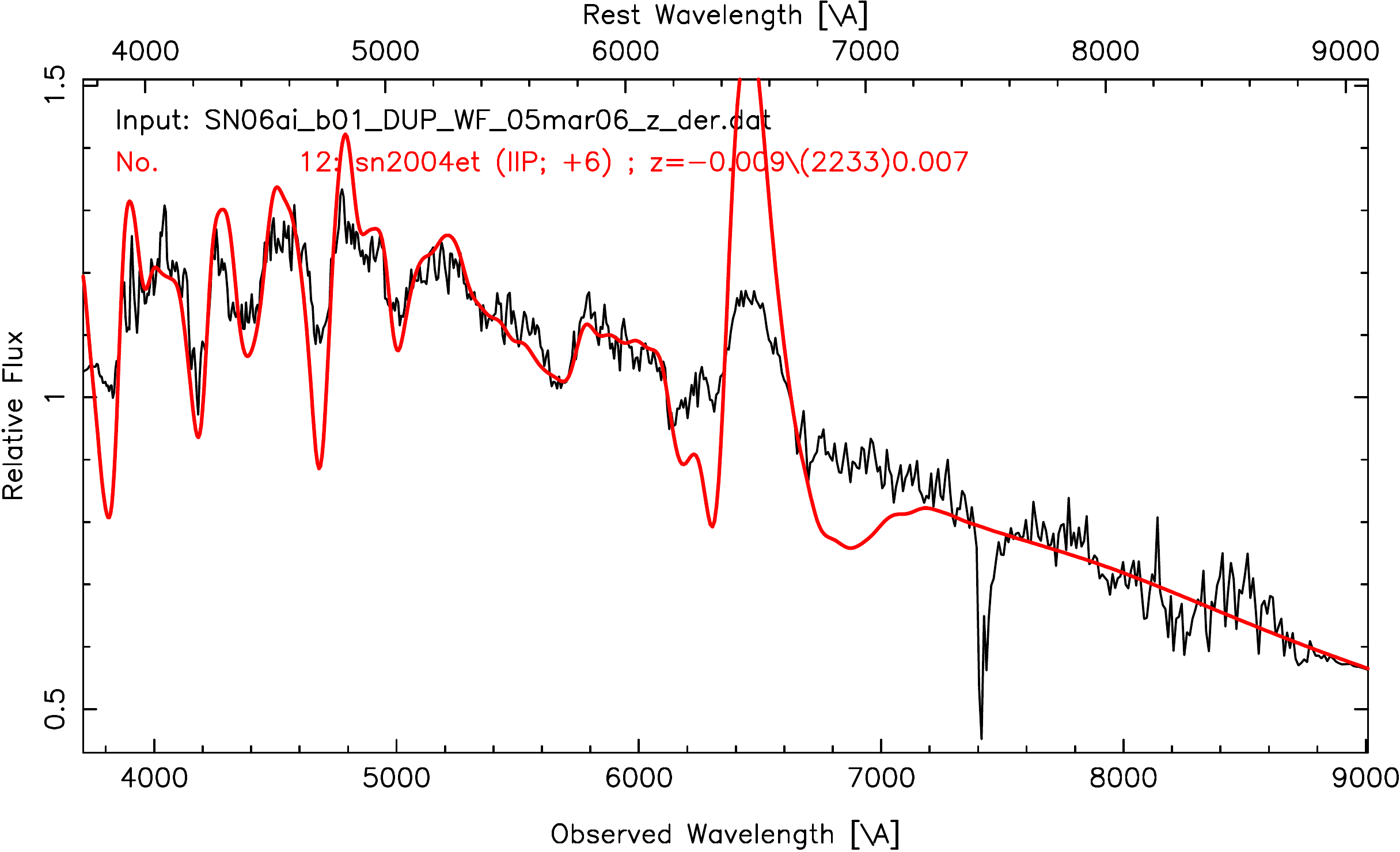}
\includegraphics[width=4.4cm]{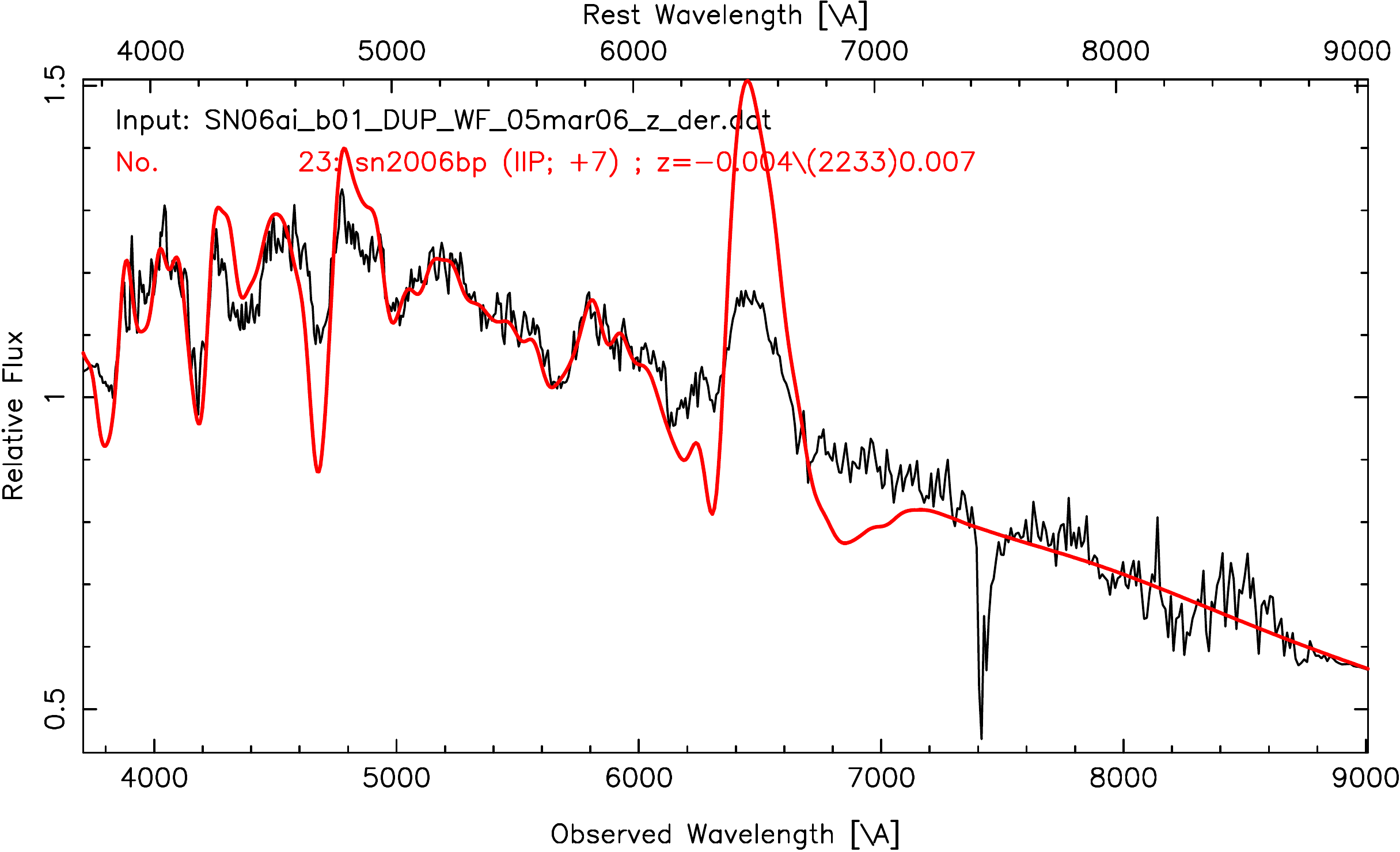}
\caption{Best spectral matching of SN~2006ai using SNID. The plots show SN~2006ai compared with 
SN~2003bn, SN~2004et, and SN~2006bp at 16, 22, and 16 days from explosion.}
\end{figure}

\begin{figure}
\centering
\includegraphics[width=4.4cm]{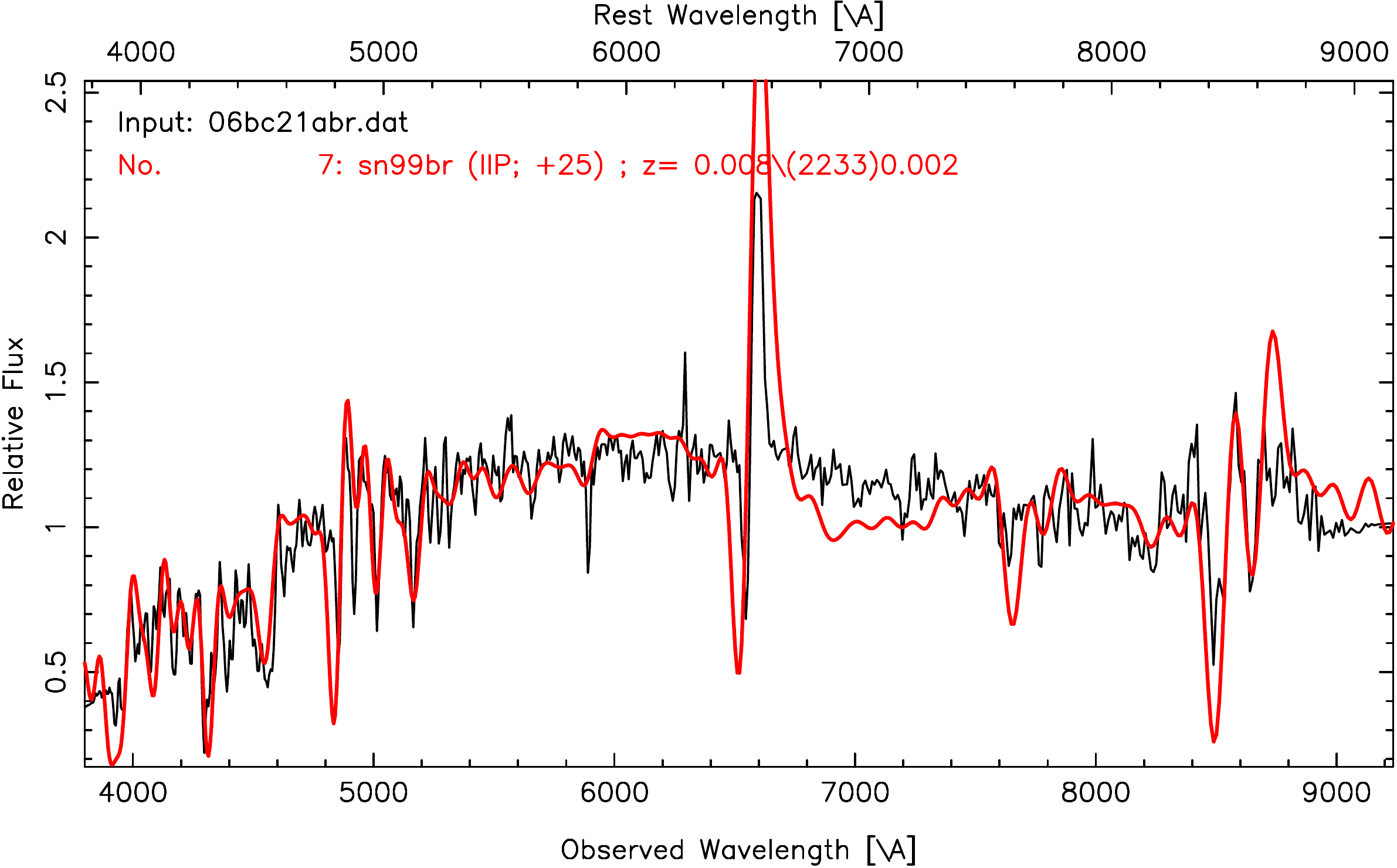}
\includegraphics[width=4.4cm]{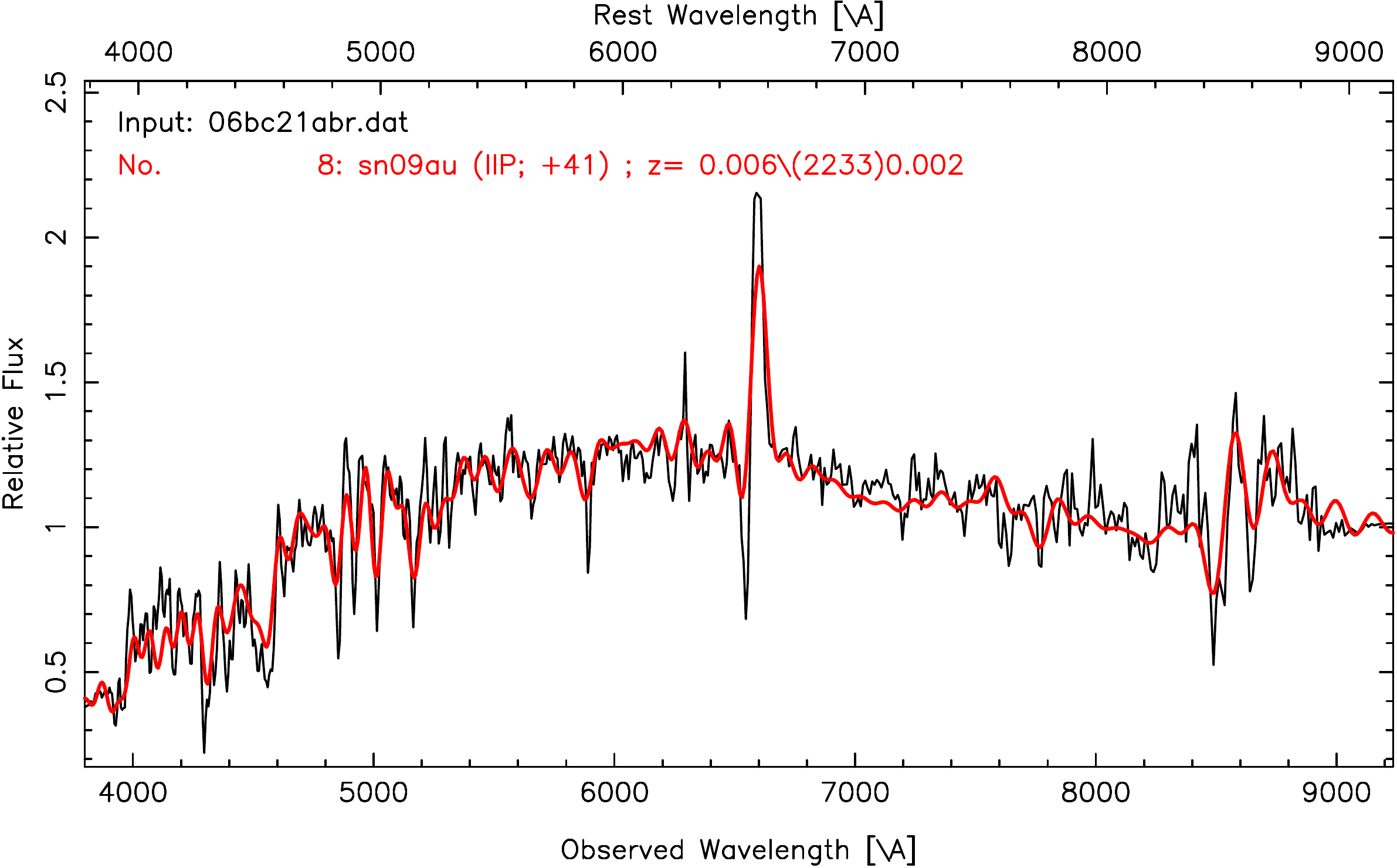}
\caption{Best spectral matching of SN~2006bc using SNID. The plots show SN~2006bc compared with 
SN~1999br and SN~2009au at 25 and 41 days from explosion.}
\end{figure}

\begin{figure}
\centering
\includegraphics[width=4.4cm]{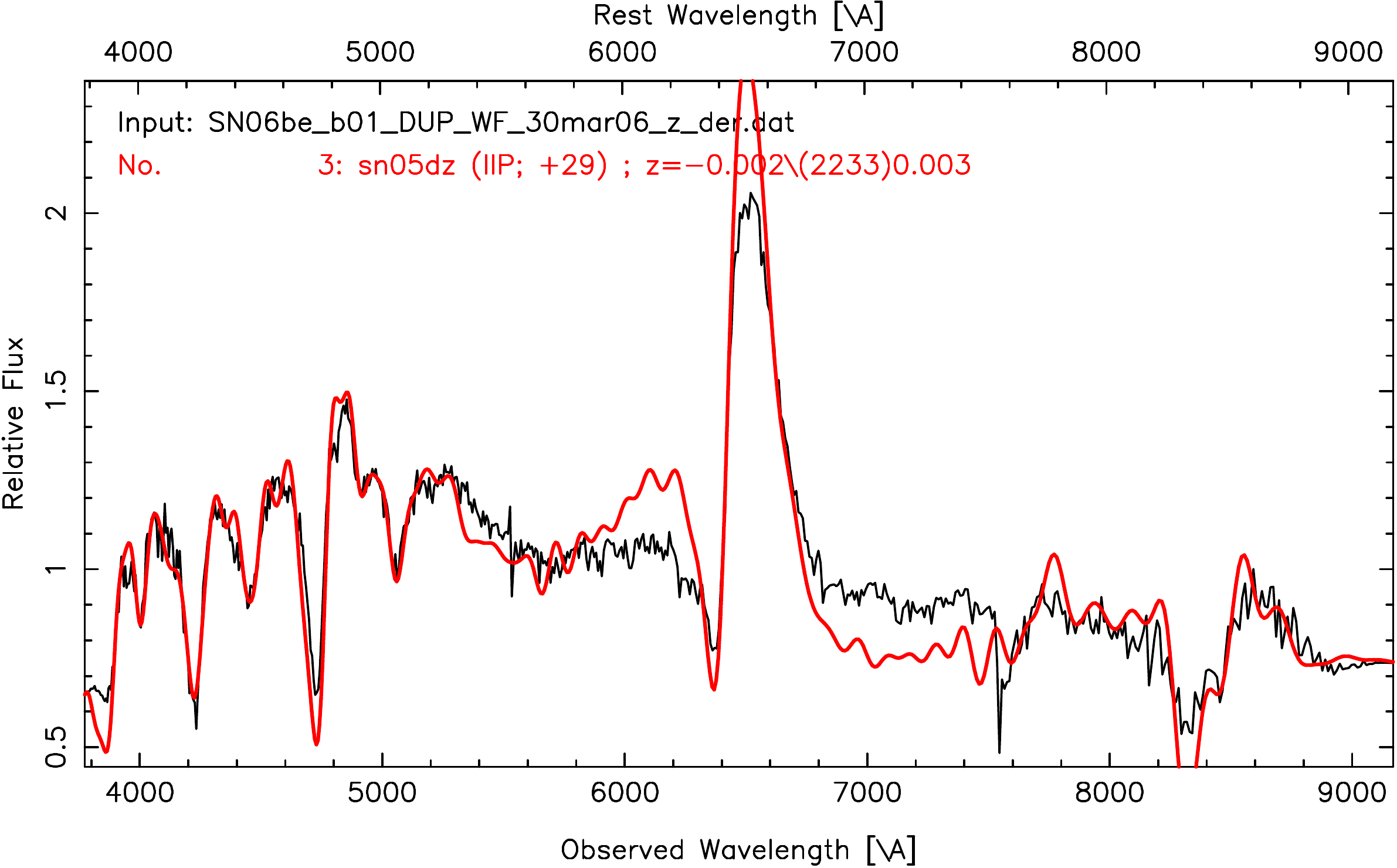}
\includegraphics[width=4.4cm]{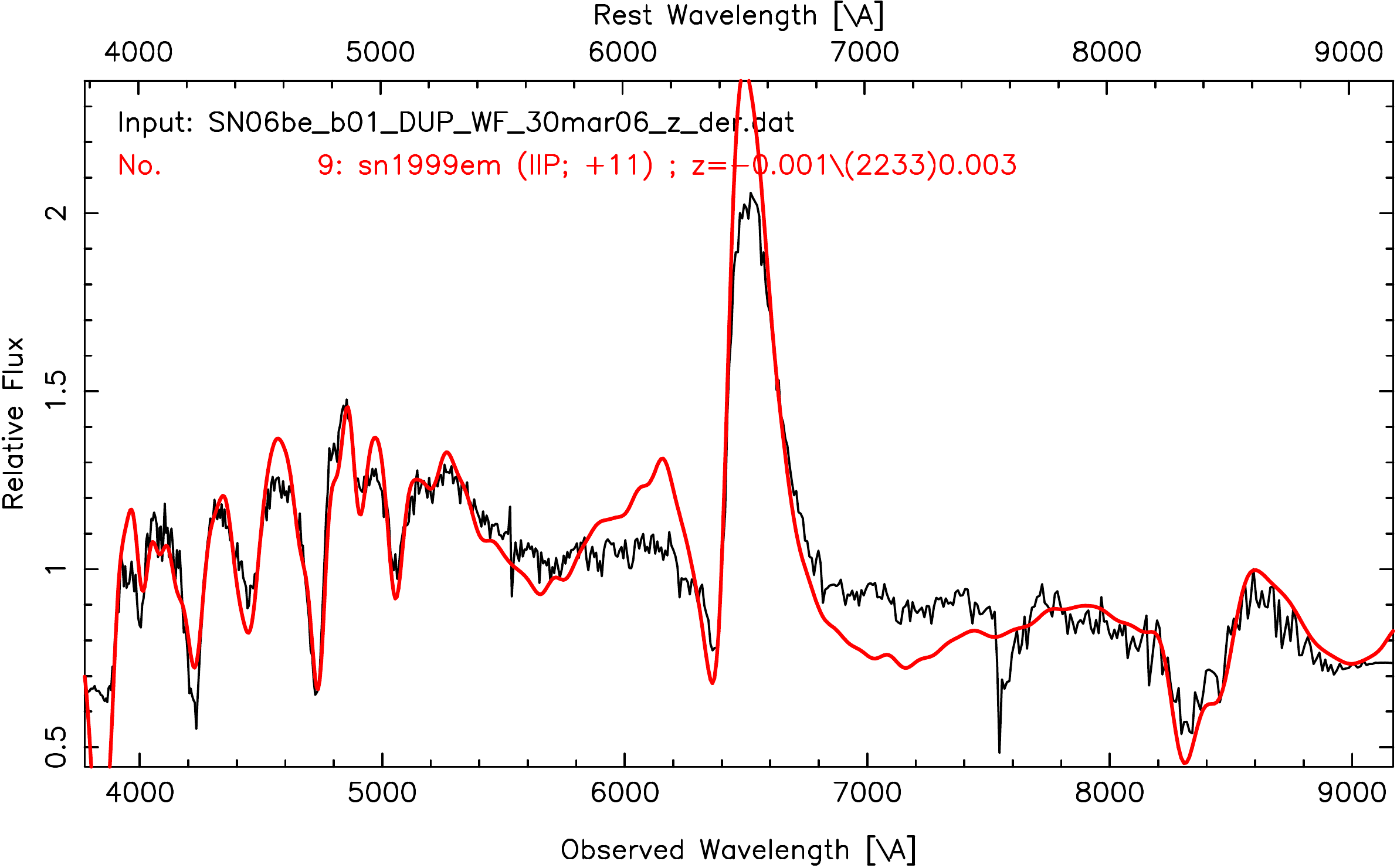}
\includegraphics[width=4.4cm]{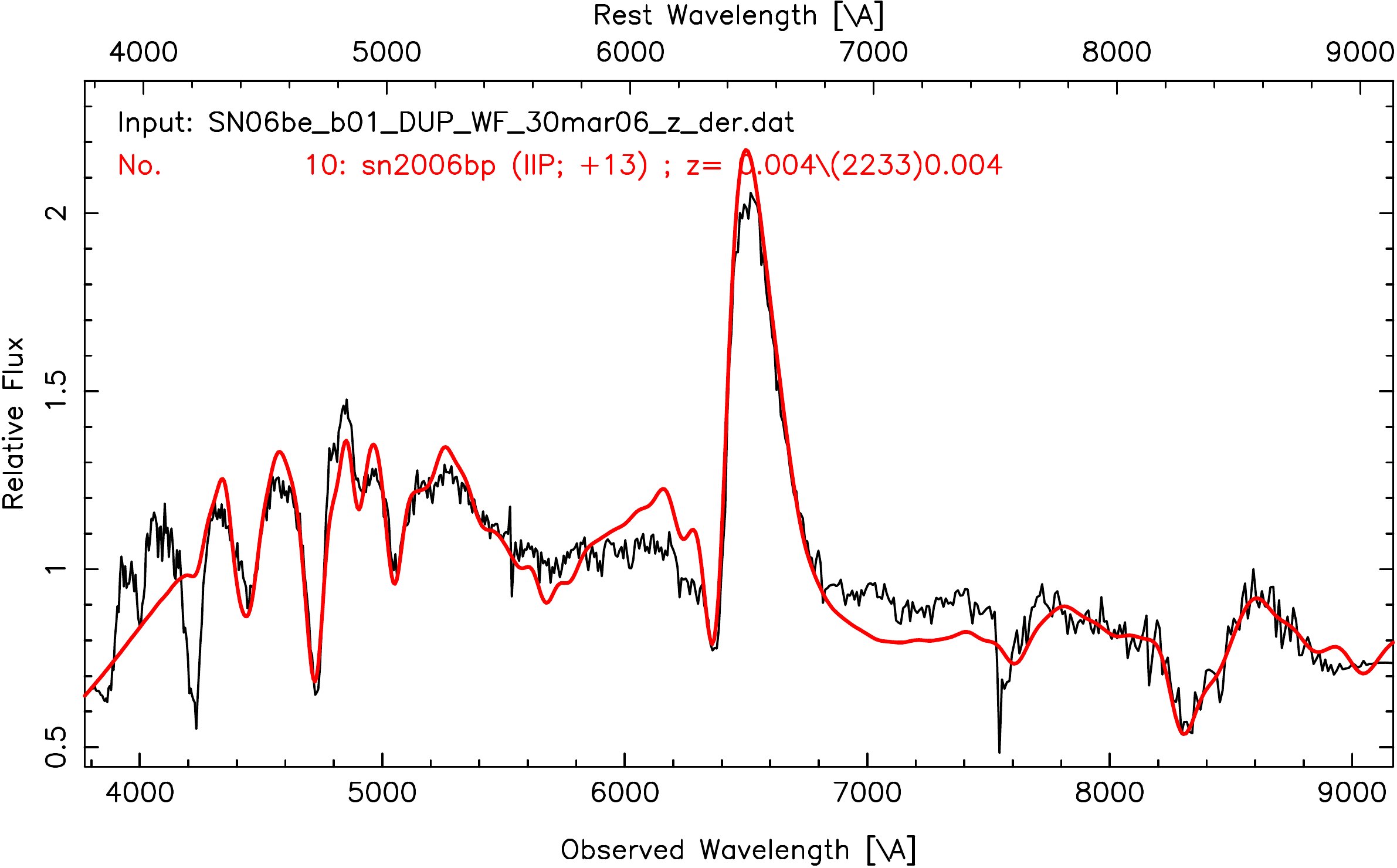}
\includegraphics[width=4.4cm]{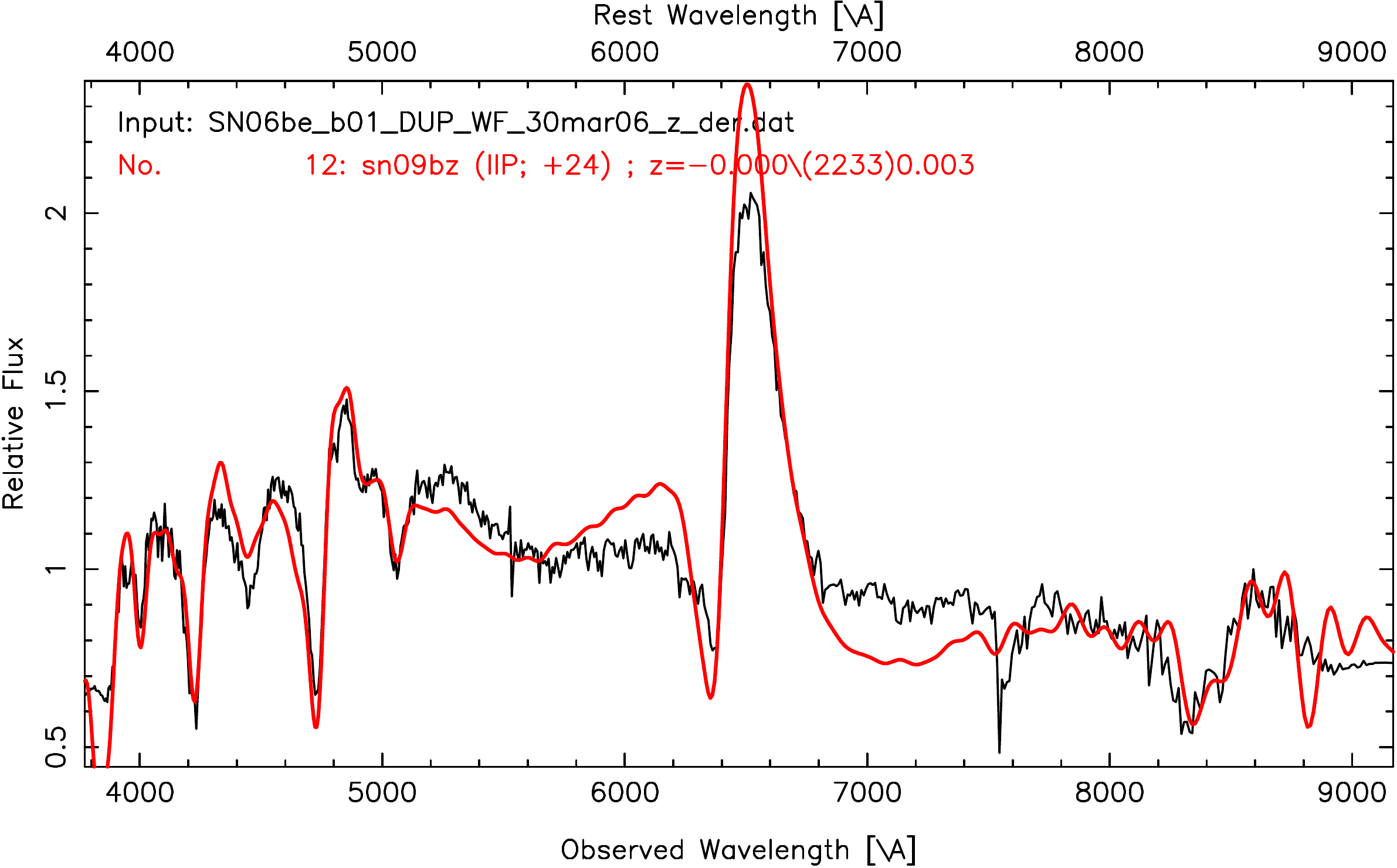}
\includegraphics[width=4.4cm]{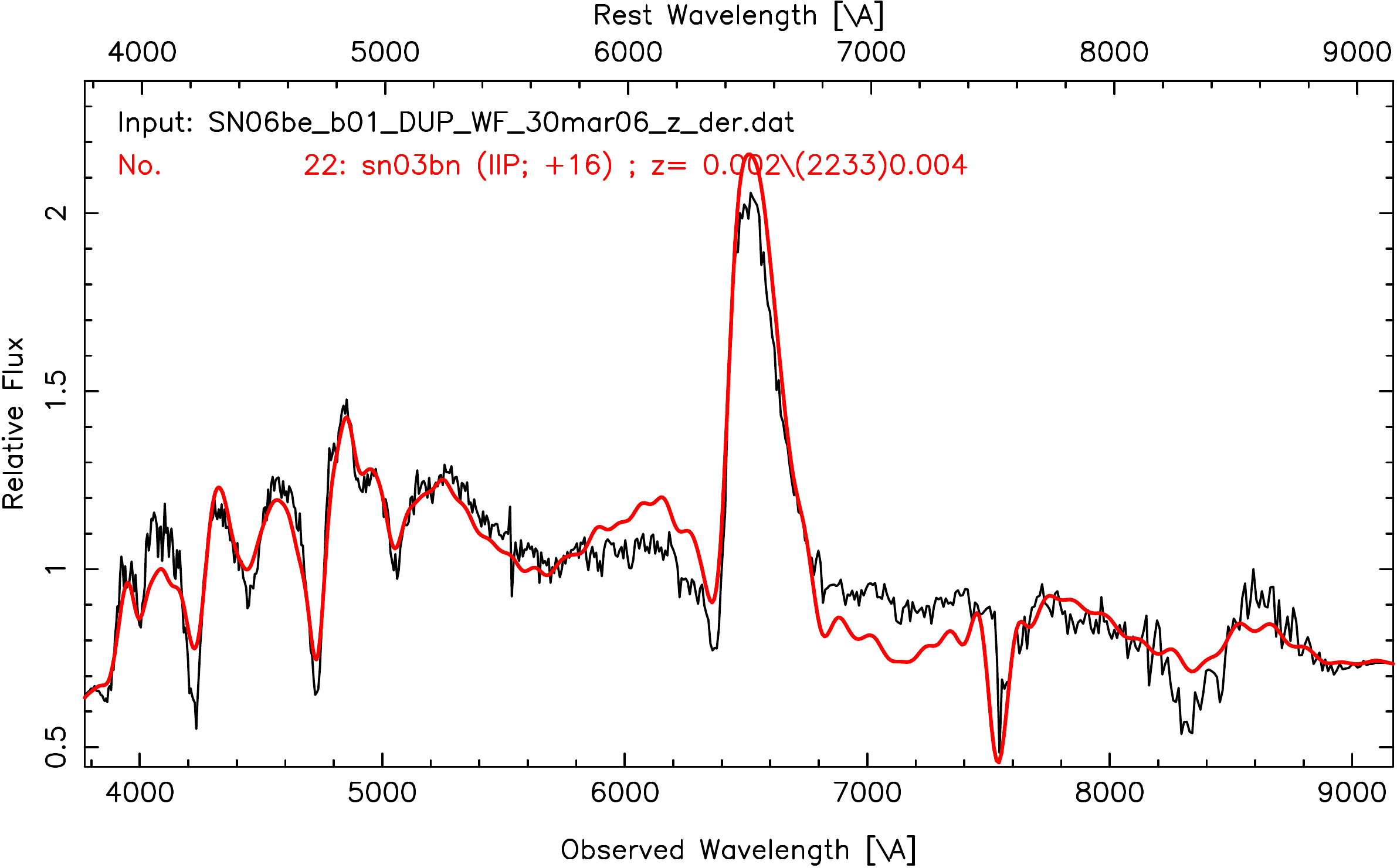}
\caption{Best spectral matching of SN~2006be using SNID. The plots show SN~2006be compared with 
SN~2005dz, SN~1999em, SN~2006bp, SN~2009bz, and SN~2003bn at 29, 21, 22 24, and 16 days from explosion.}
\end{figure}

\clearpage

\begin{figure}
\centering
\includegraphics[width=4.4cm]{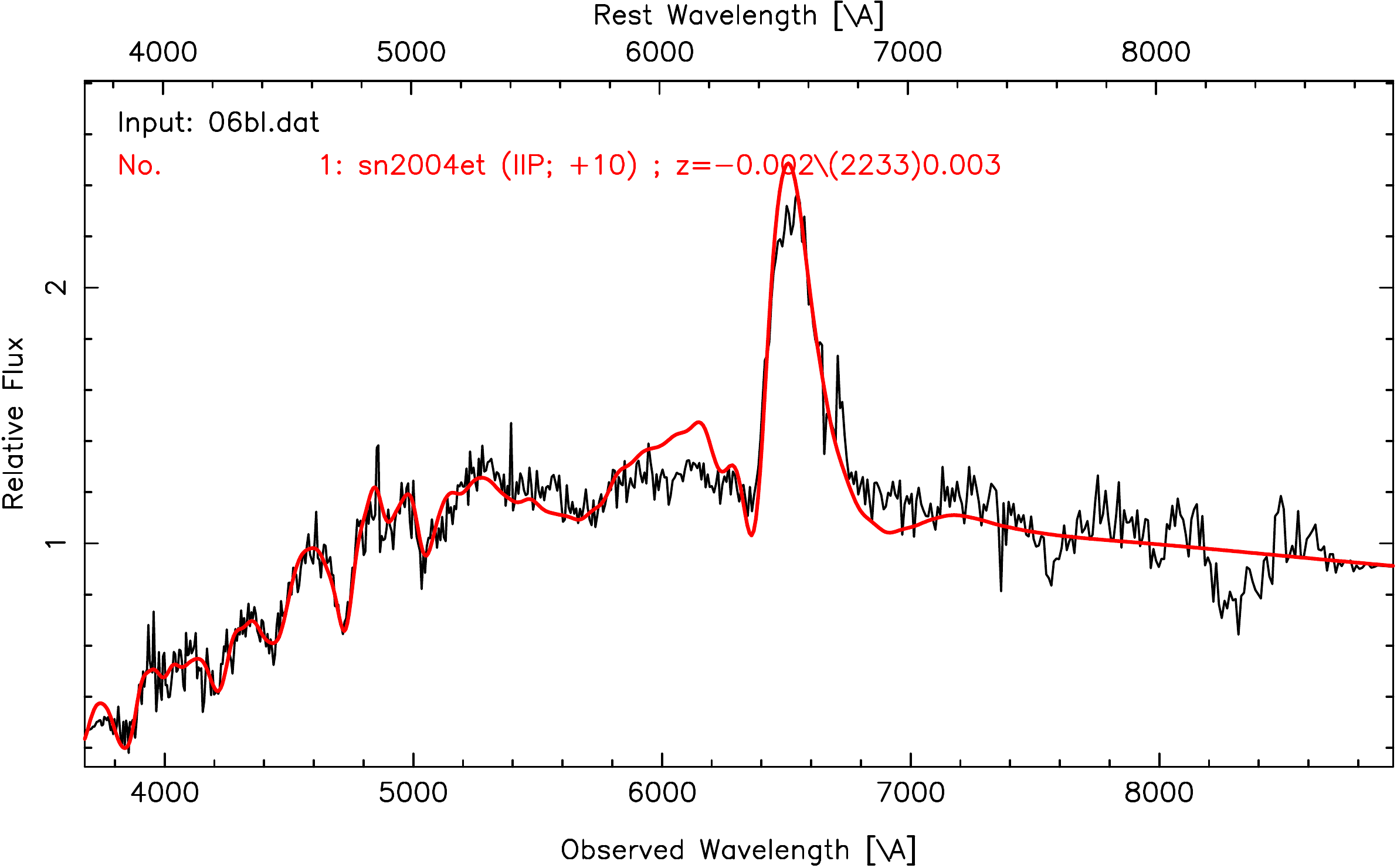}
\includegraphics[width=4.4cm]{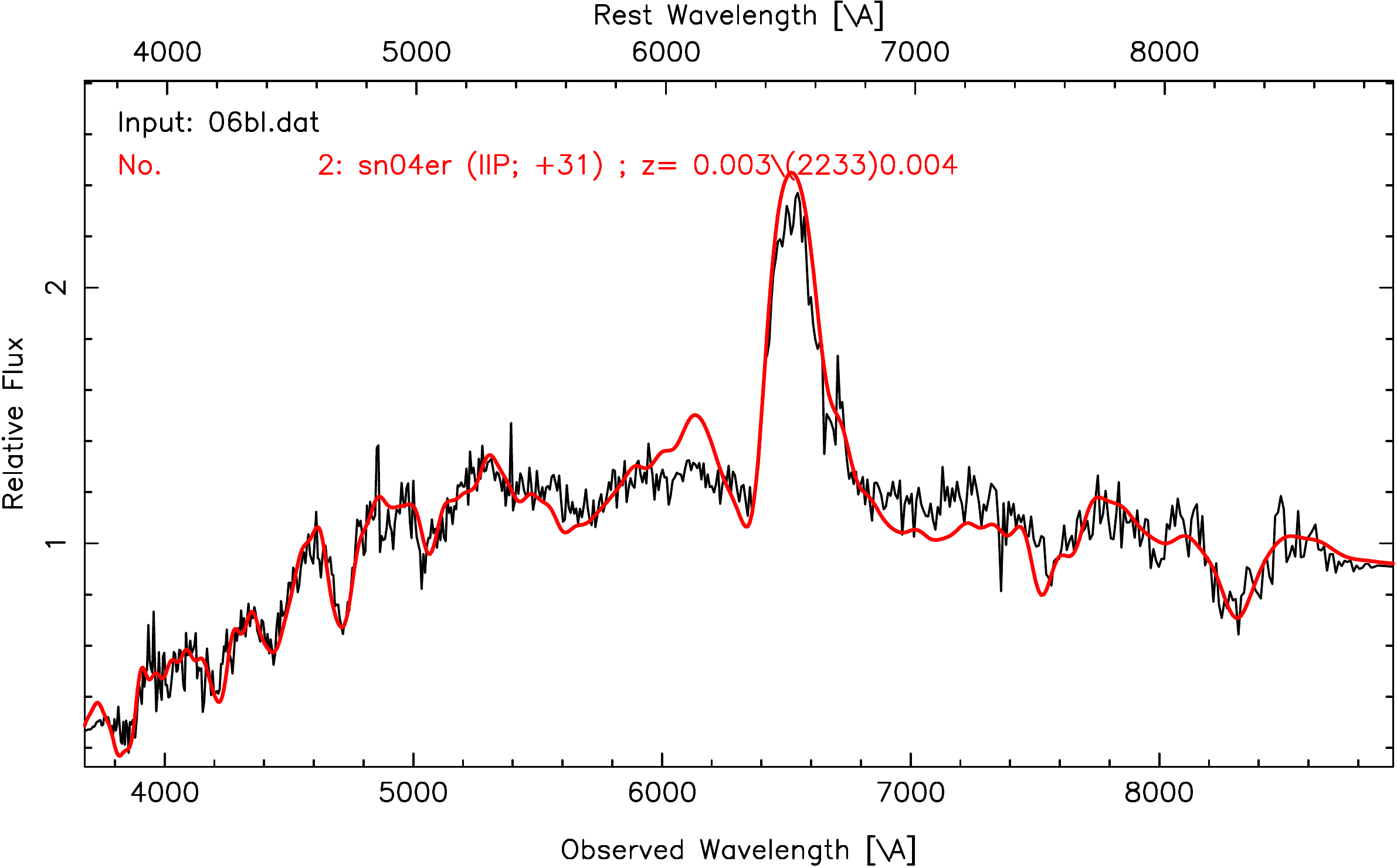}
\includegraphics[width=4.4cm]{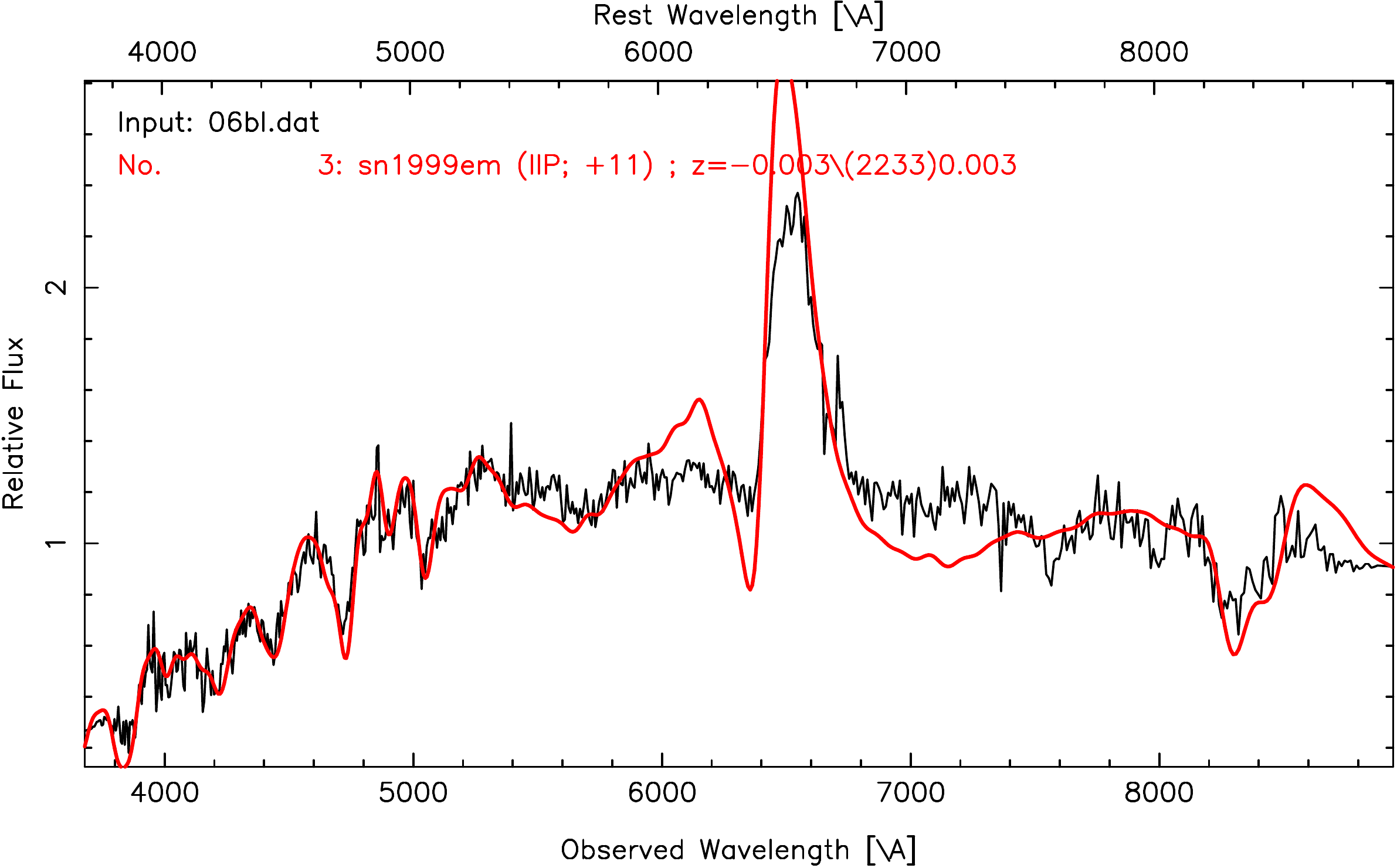}
\includegraphics[width=4.4cm]{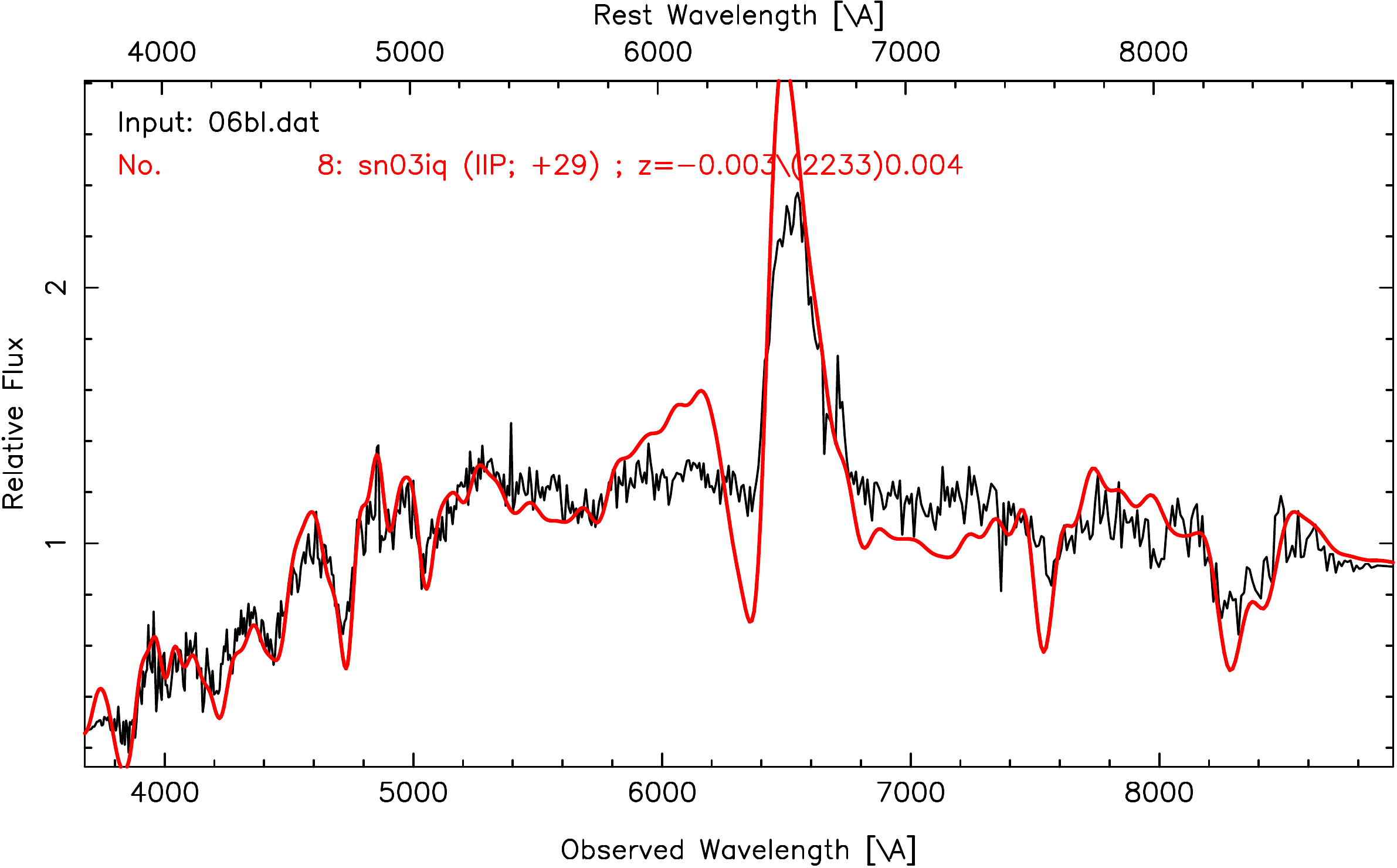}
\includegraphics[width=4.4cm]{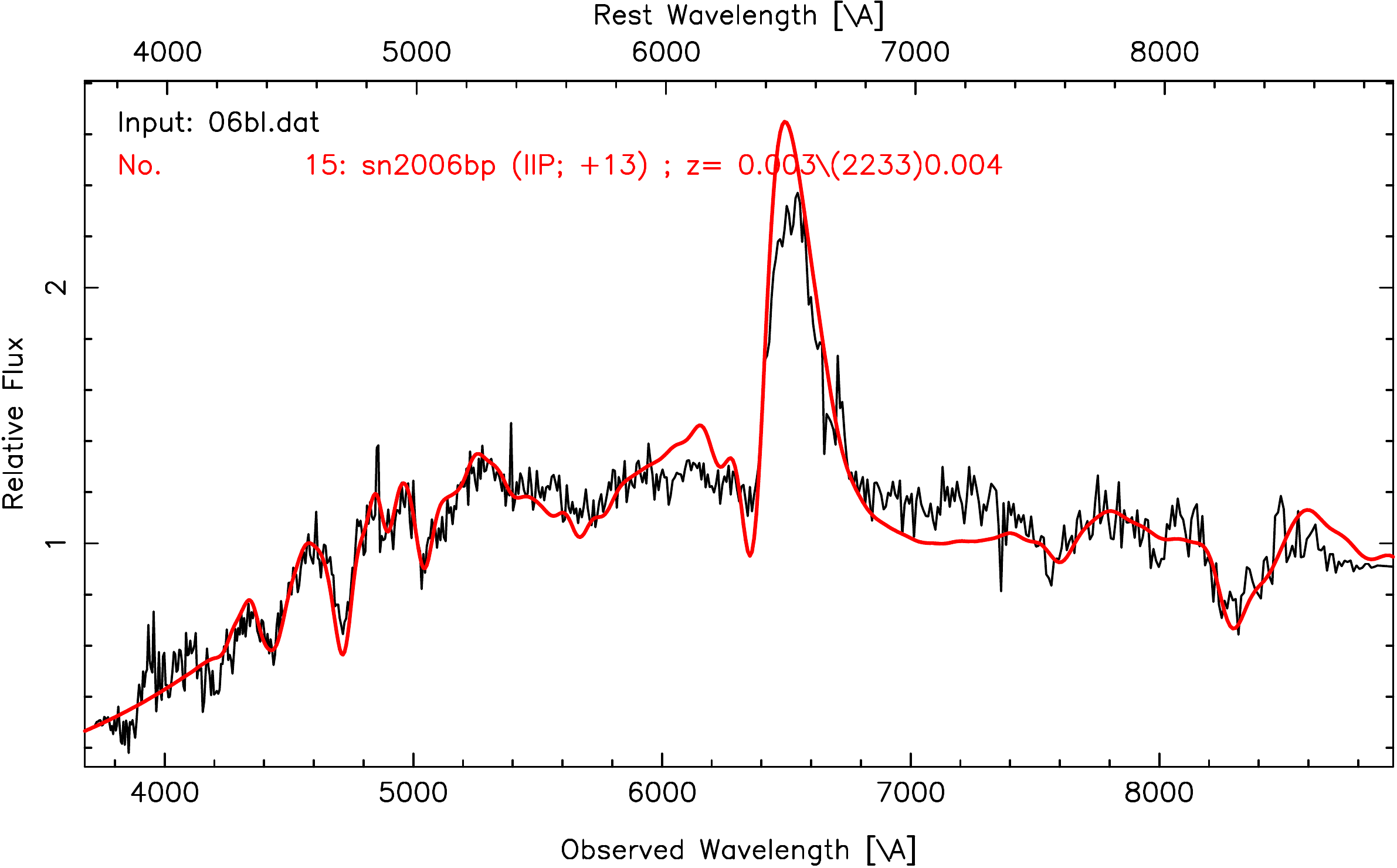}
\includegraphics[width=4.4cm]{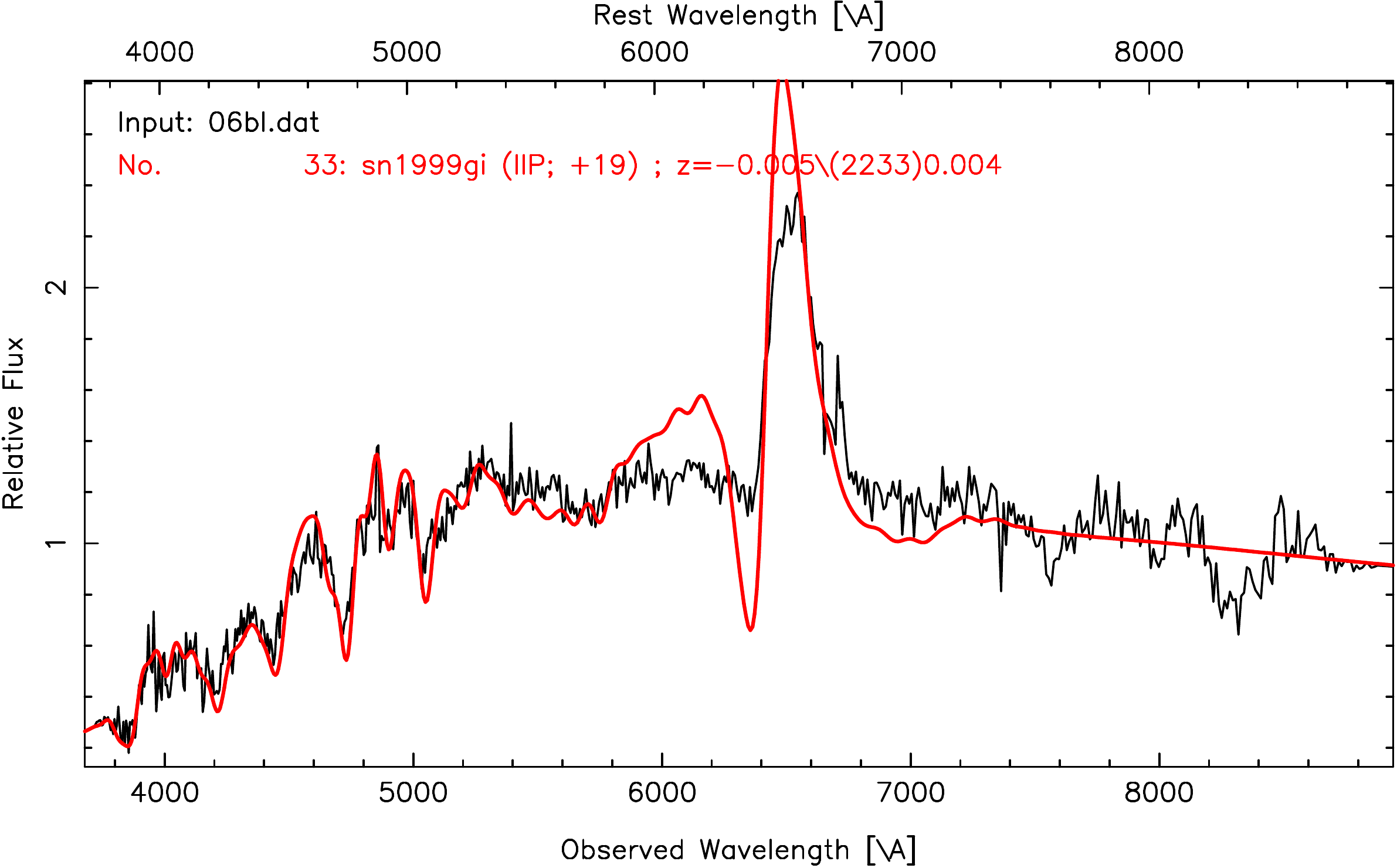}
\includegraphics[width=4.4cm]{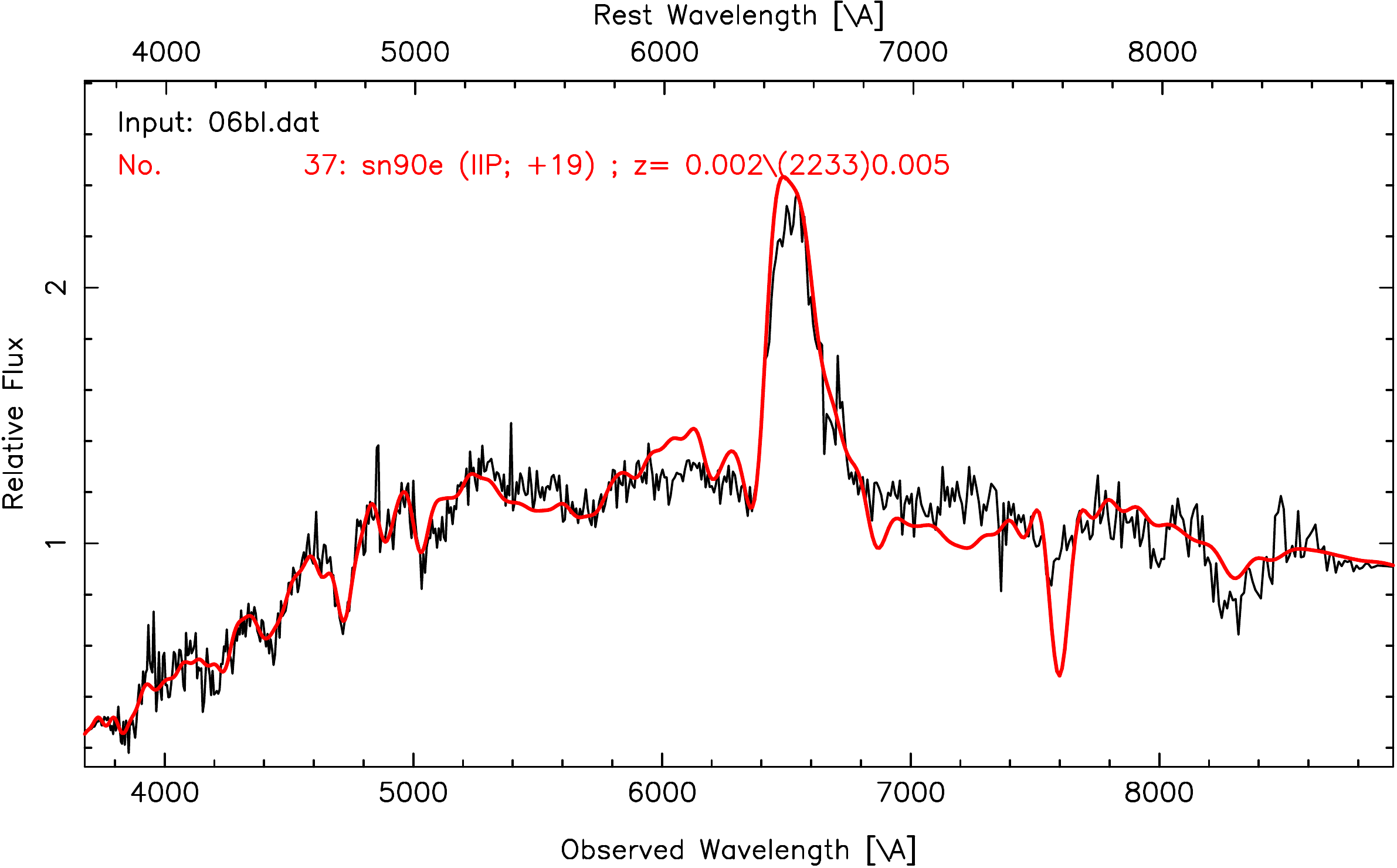}
\caption{Best spectral matching of SN~2006bl using SNID. The plots show SN~2006bl compared with 
SN~2004et, SN~2004er, SN~1999em, SN~2003iq, SN~2006bp, SN~1999gi, and SN~1990E at 26, 31, 21, 29, 22, 31 and 19 days from explosion.}
\end{figure}

\begin{figure}
\centering
\includegraphics[width=4.4cm]{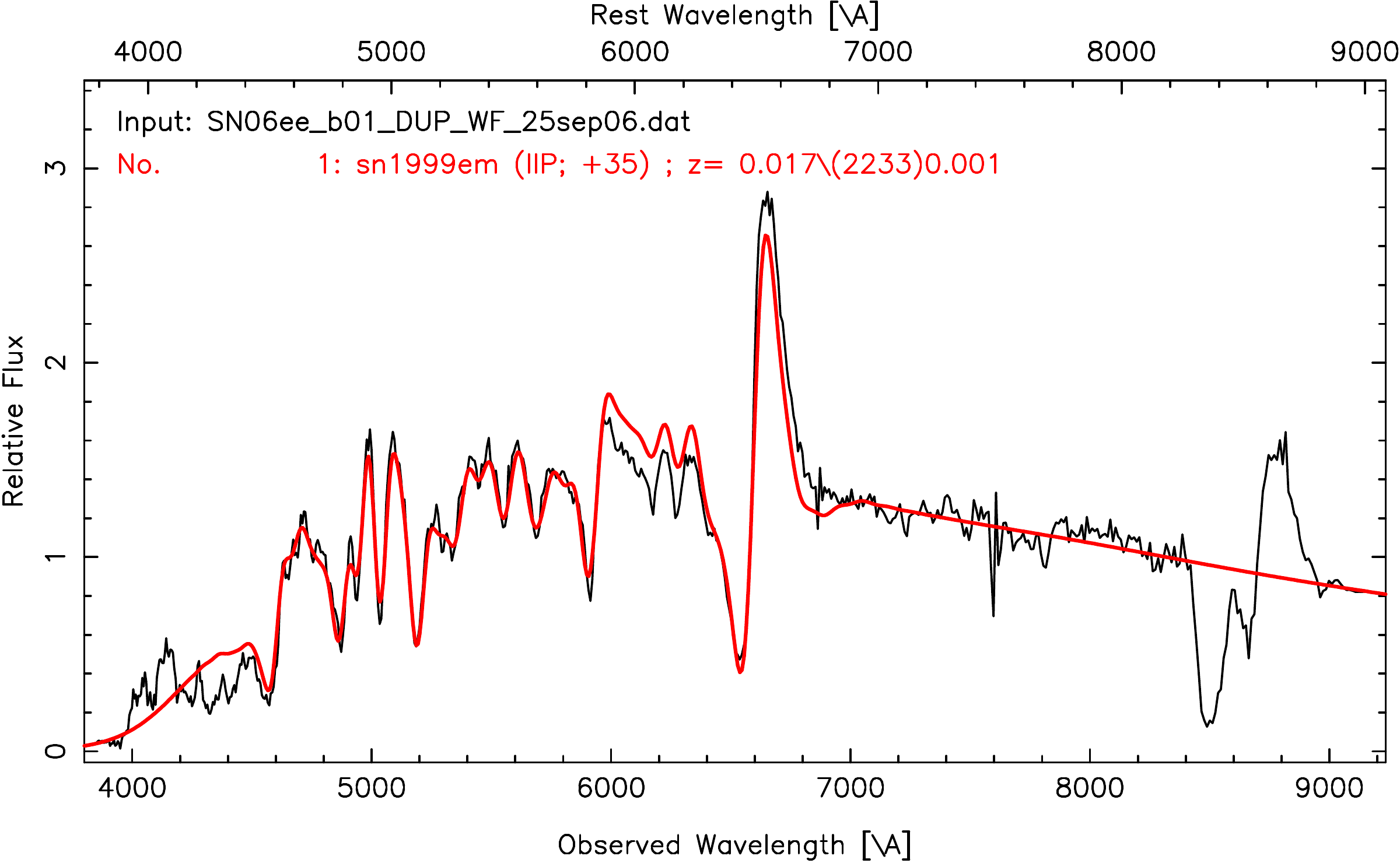}
\includegraphics[width=4.4cm]{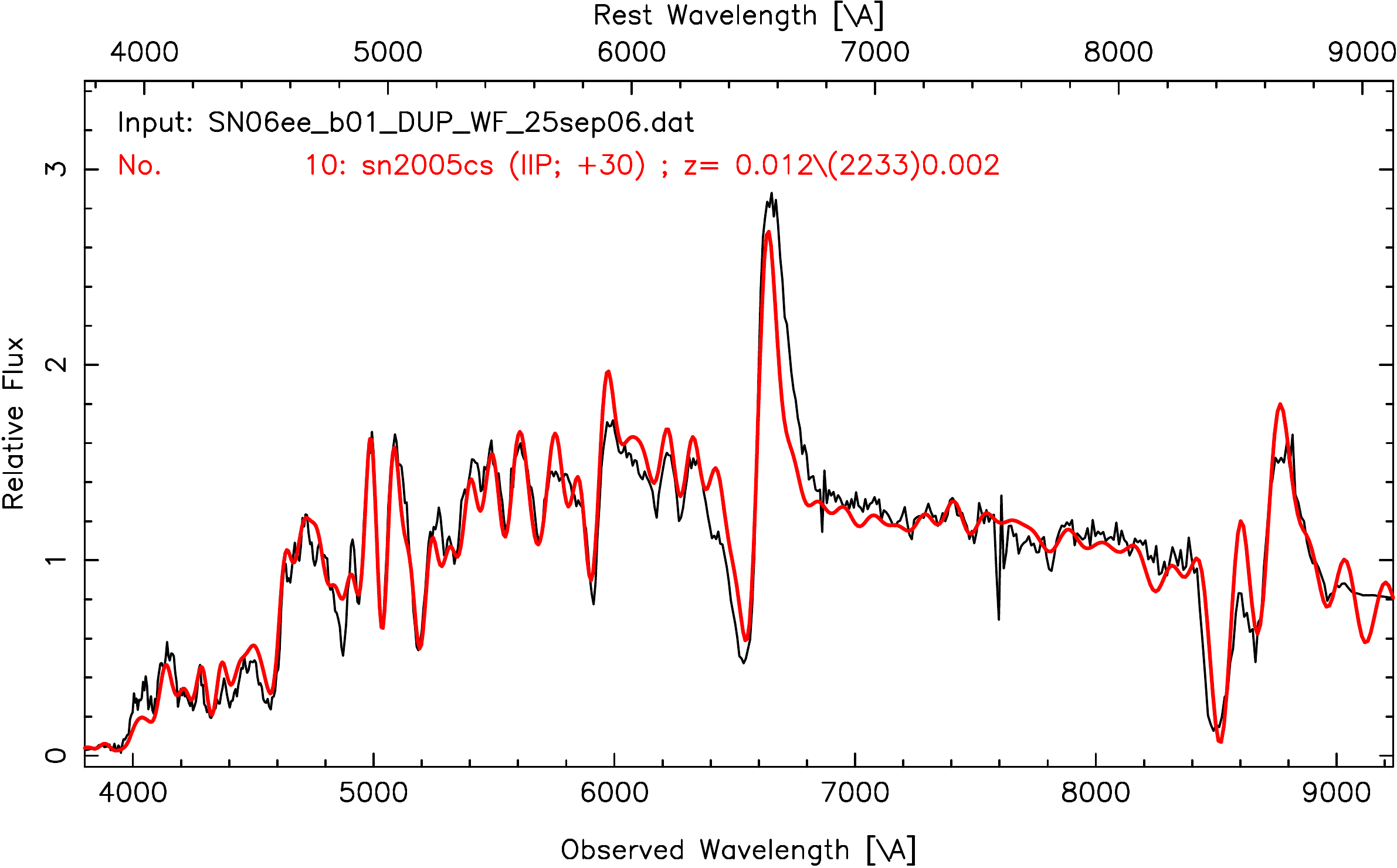}
\includegraphics[width=4.4cm]{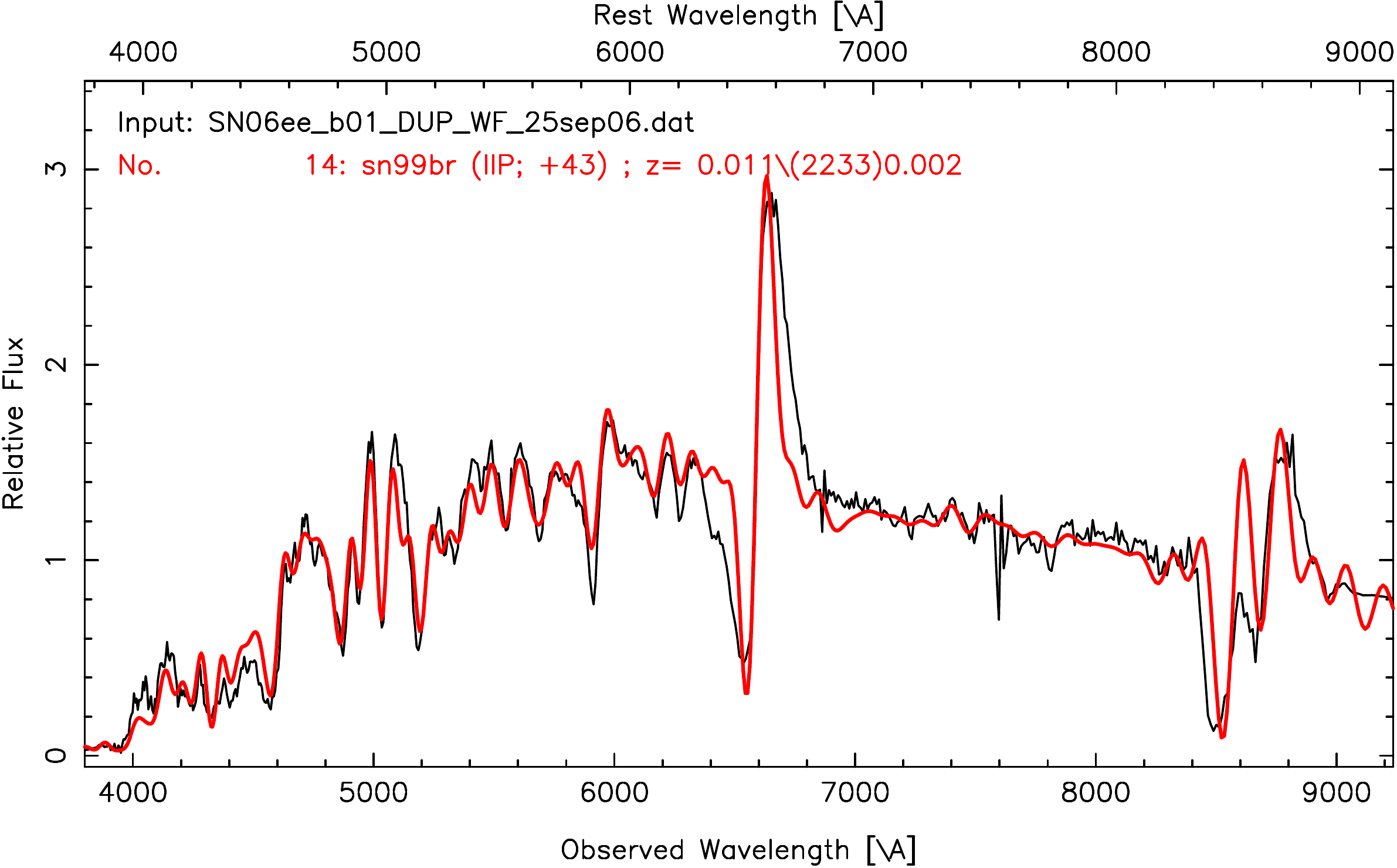}
\caption{Best spectral matching of SN~2006ee using SNID. The plots show SN~2006ee compared with 
SN~1999em, SN~2005cs, and SN~1999br at 45, 36, and 43 days from explosion.}
\end{figure}

\clearpage

\begin{figure}
\centering
\includegraphics[width=4.4cm]{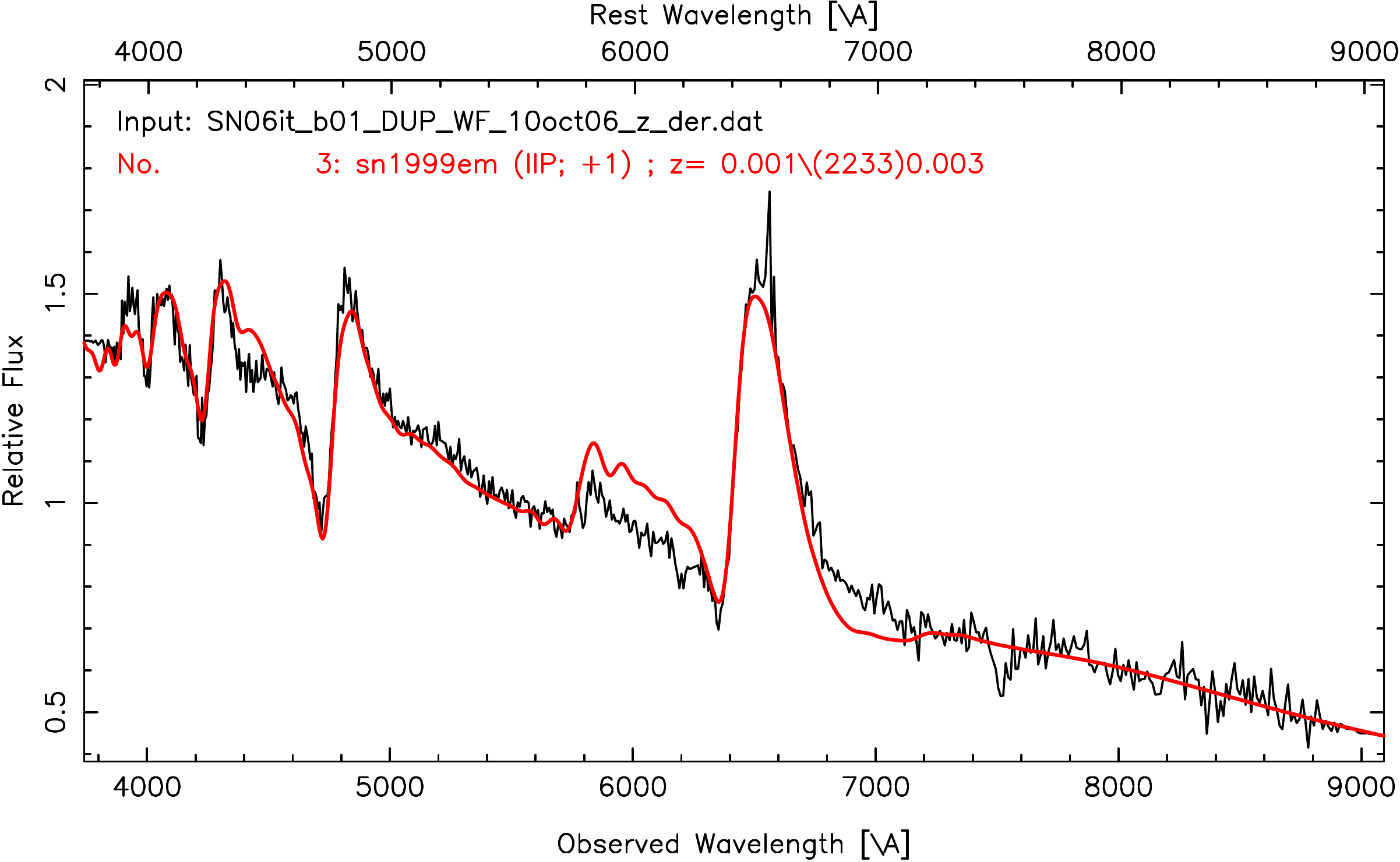}
\includegraphics[width=4.4cm]{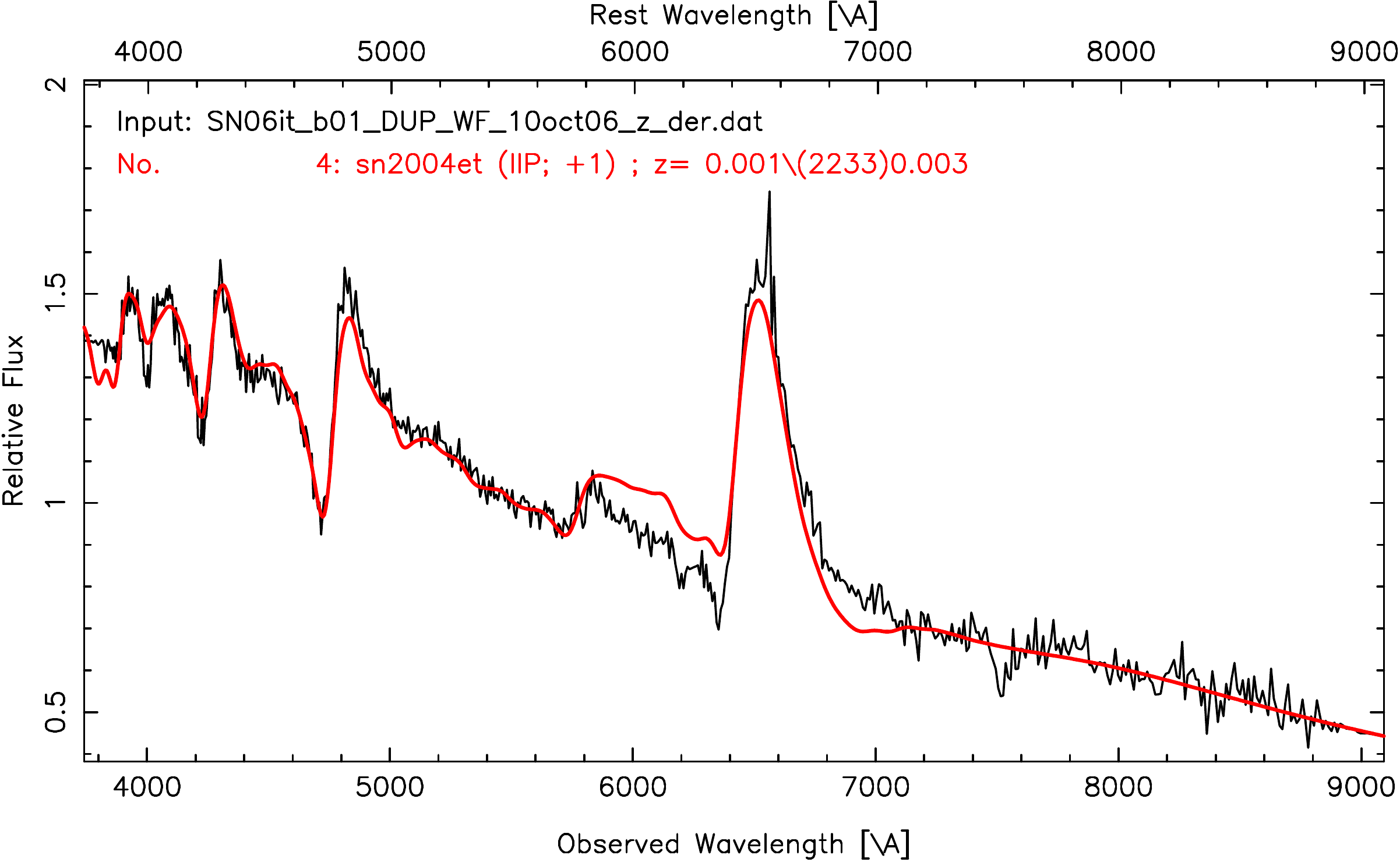}
\includegraphics[width=4.4cm]{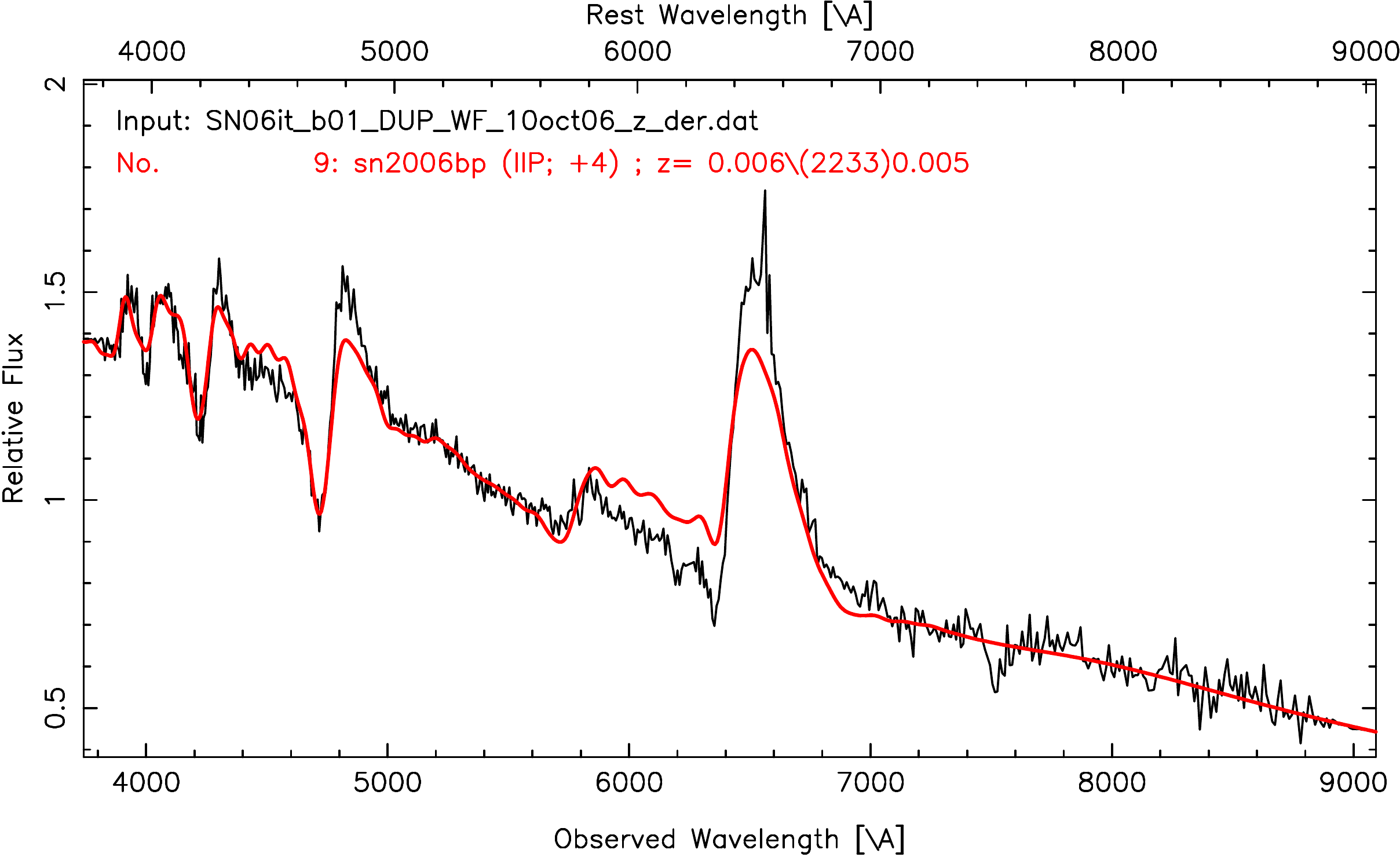}
\includegraphics[width=4.4cm]{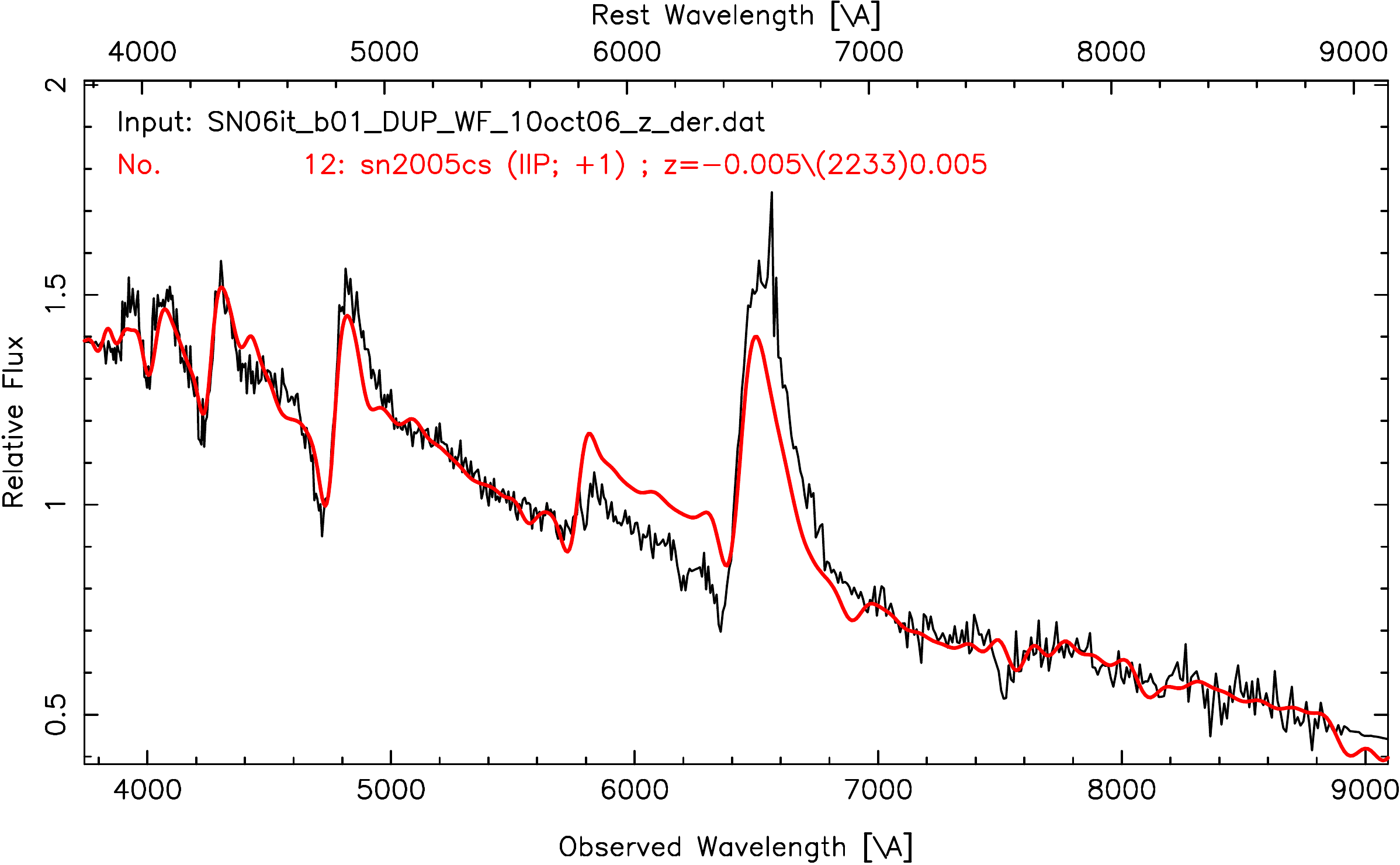}
\caption{Best spectral matching of SN~2006it using SNID. The plots show SN~2006it compared with 
SN~1999em, SN~2004et, SN~2006bp, and SN~2005cs at 12, 17, 13 and 7 days from explosion.}
\end{figure}

\begin{figure}
\centering
\includegraphics[width=4.4cm]{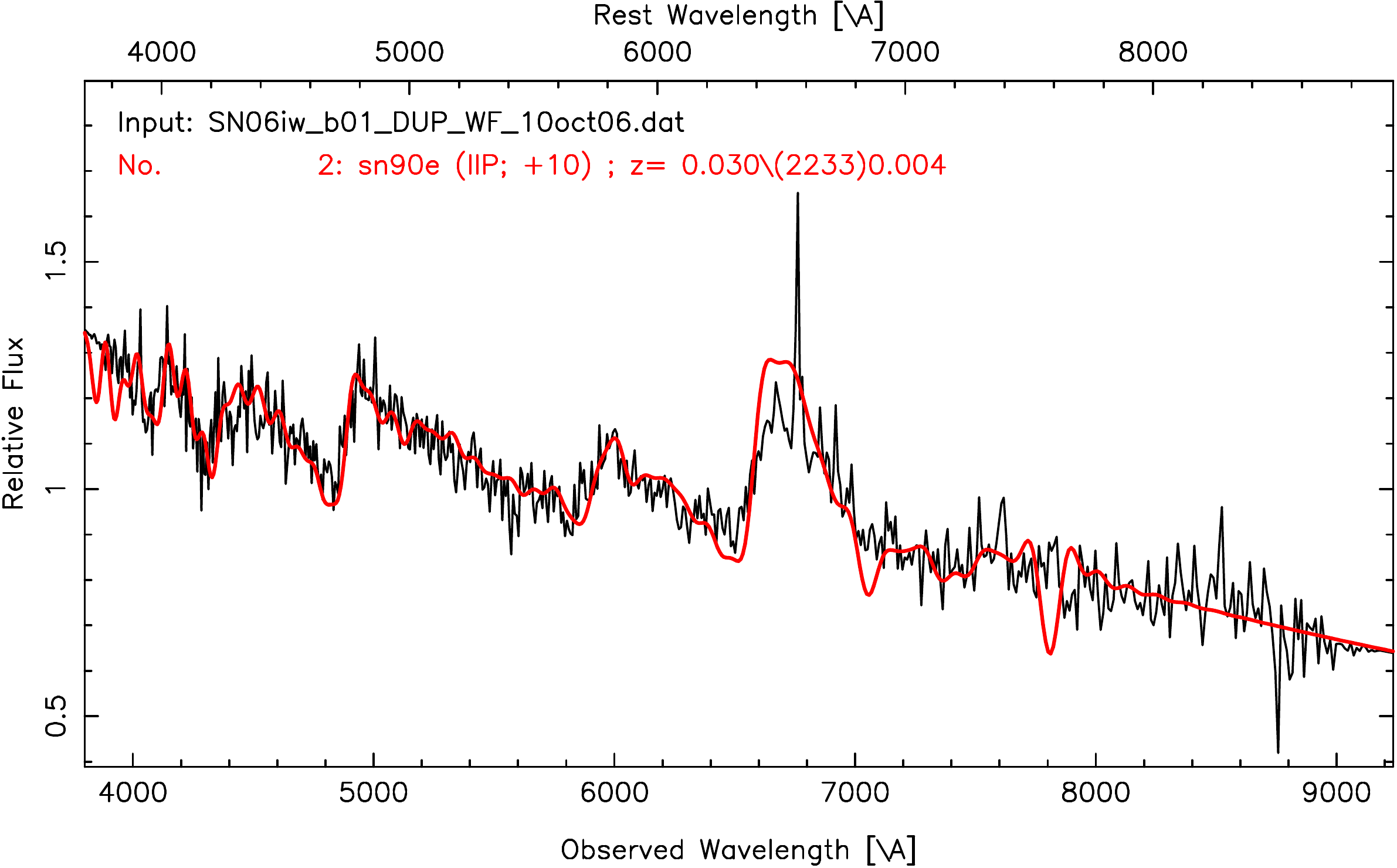}
\includegraphics[width=4.4cm]{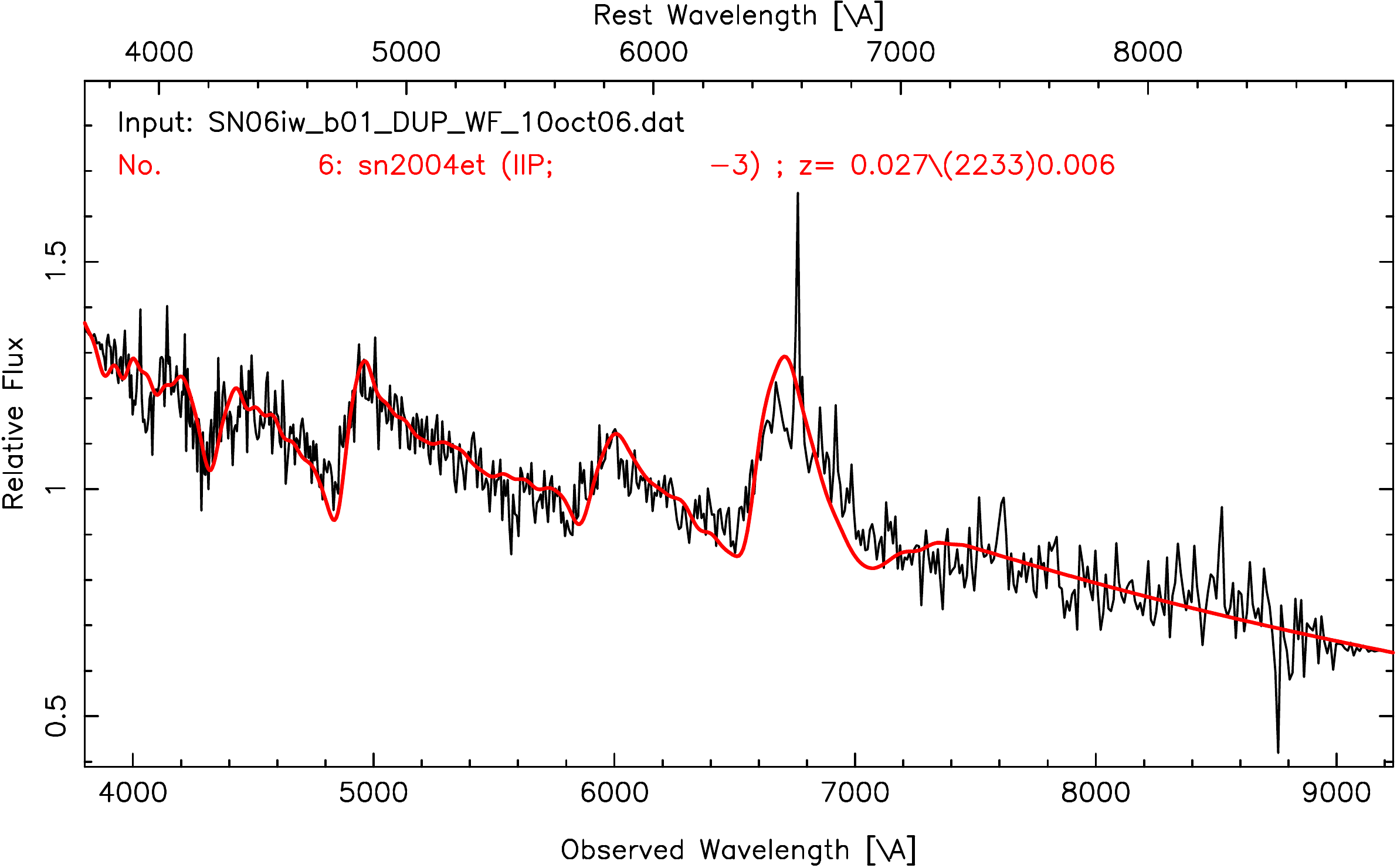}
\includegraphics[width=4.4cm]{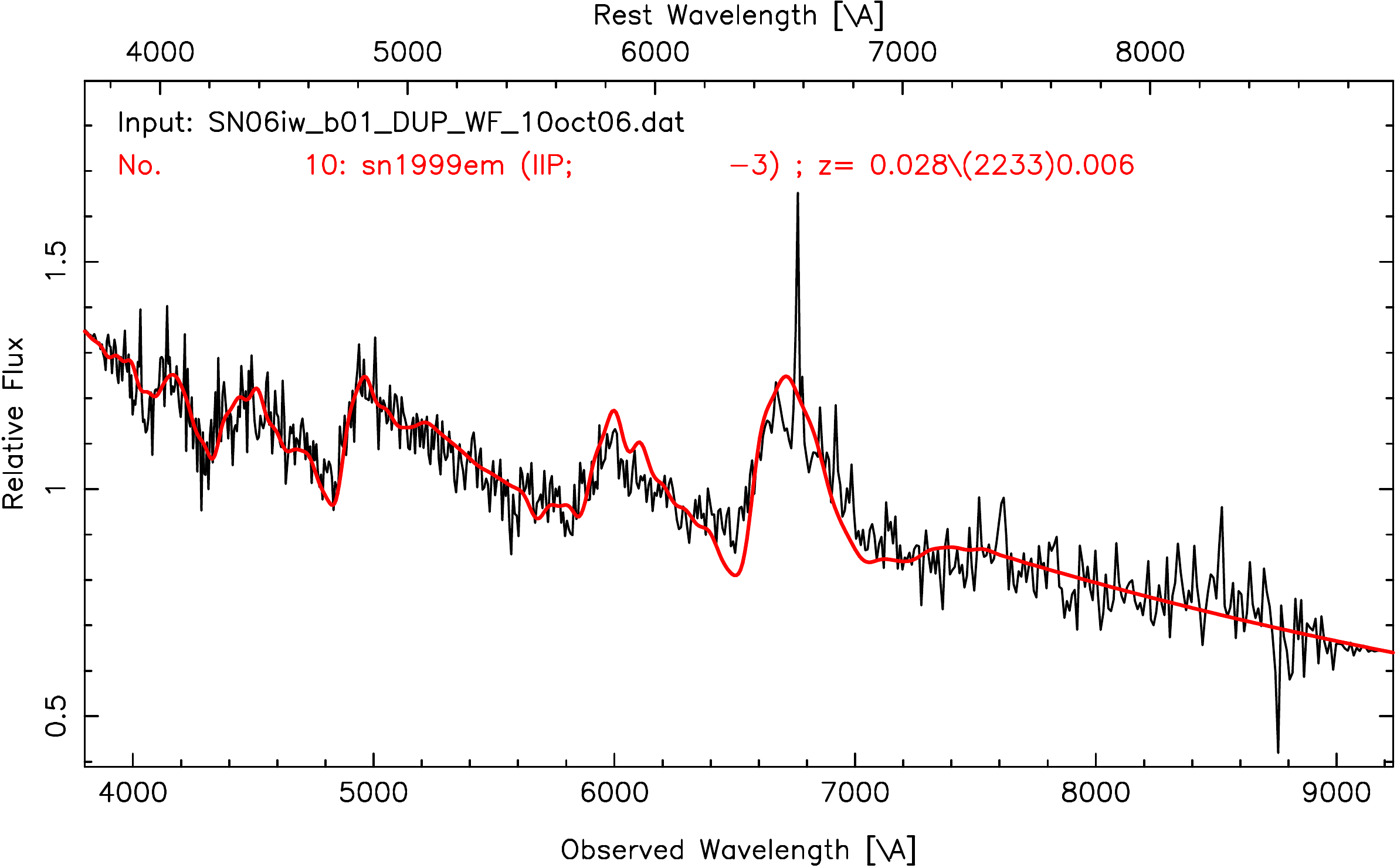}
\includegraphics[width=4.4cm]{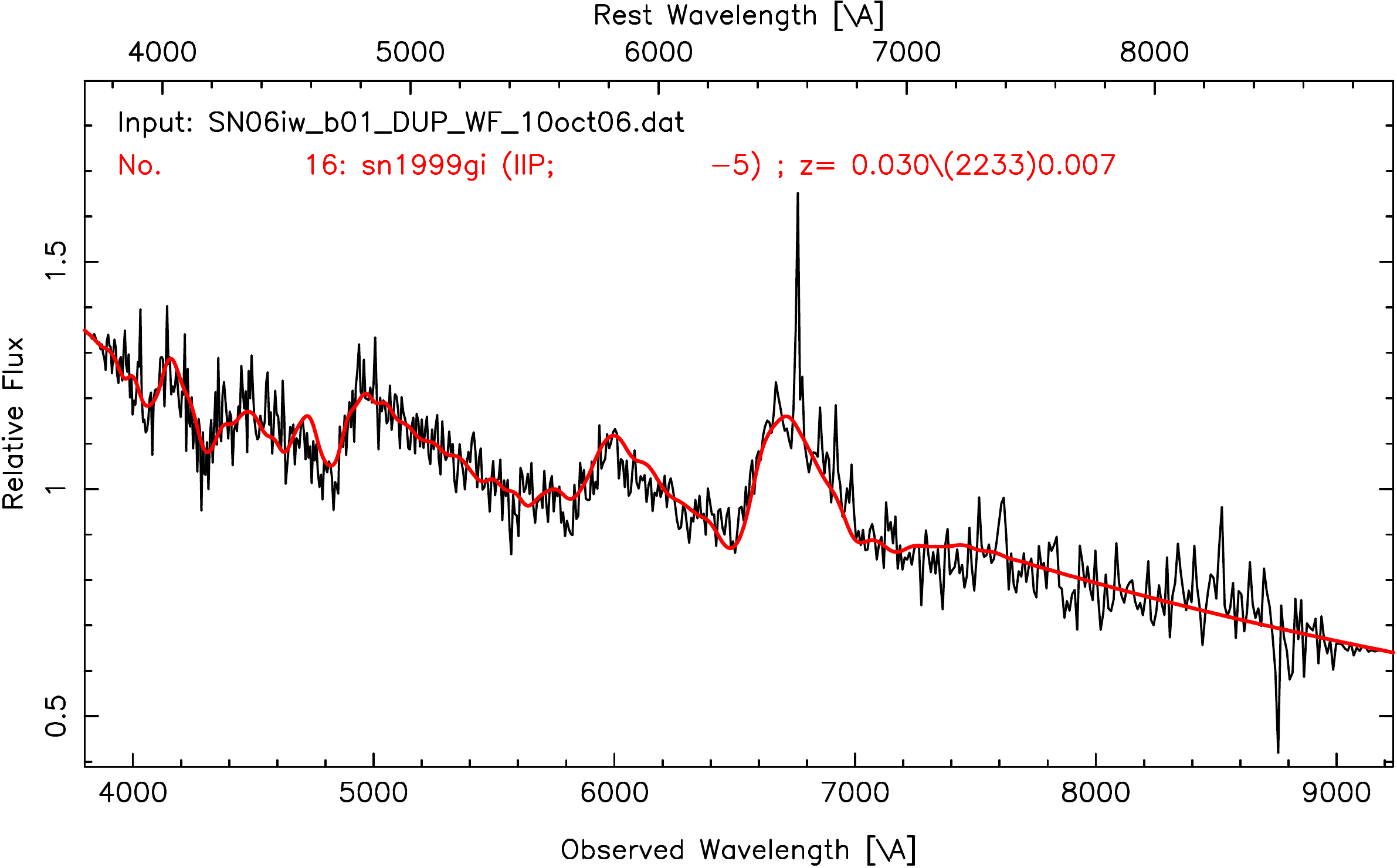}
\caption{Best spectral matching of SN~2006iw using SNID. The plots show SN~2006iw compared with 
SN~1990E, SN~2004et, SN~1999em, and SN~1999gi at 10, 13, 7, and 7 days from explosion.}
\end{figure}

\clearpage

\begin{figure}
\centering
\includegraphics[width=4.4cm]{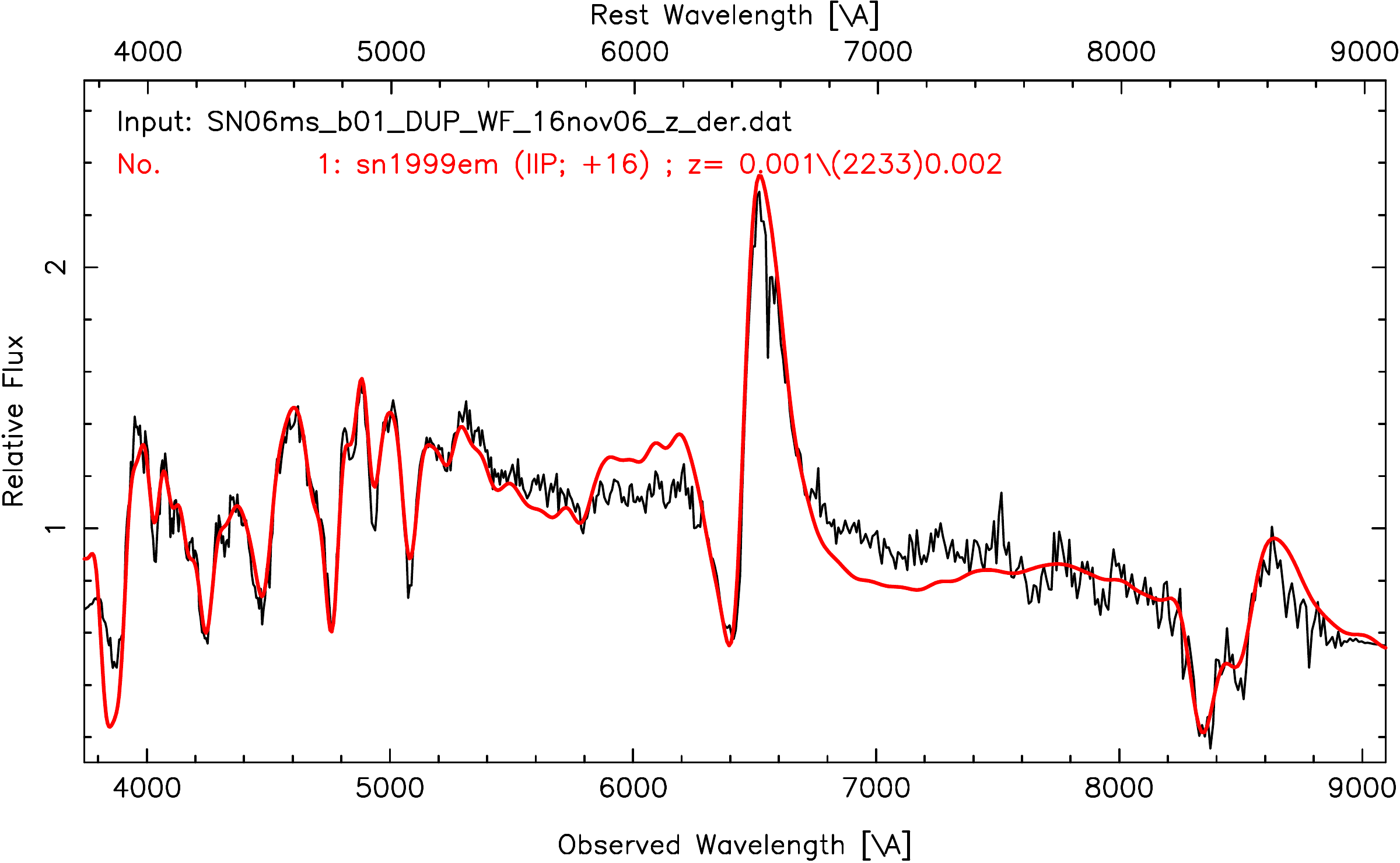}
\includegraphics[width=4.4cm]{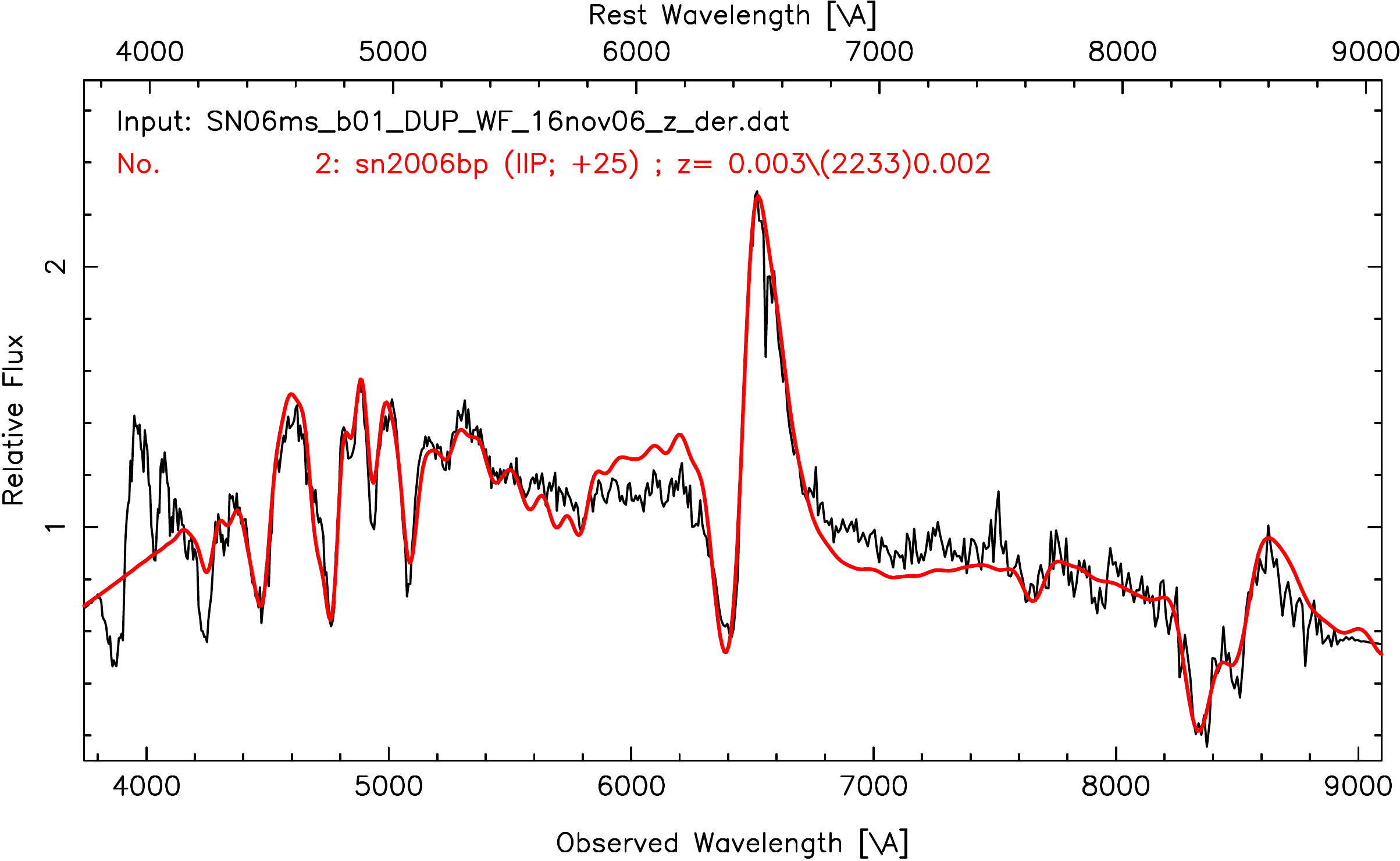}
\includegraphics[width=4.4cm]{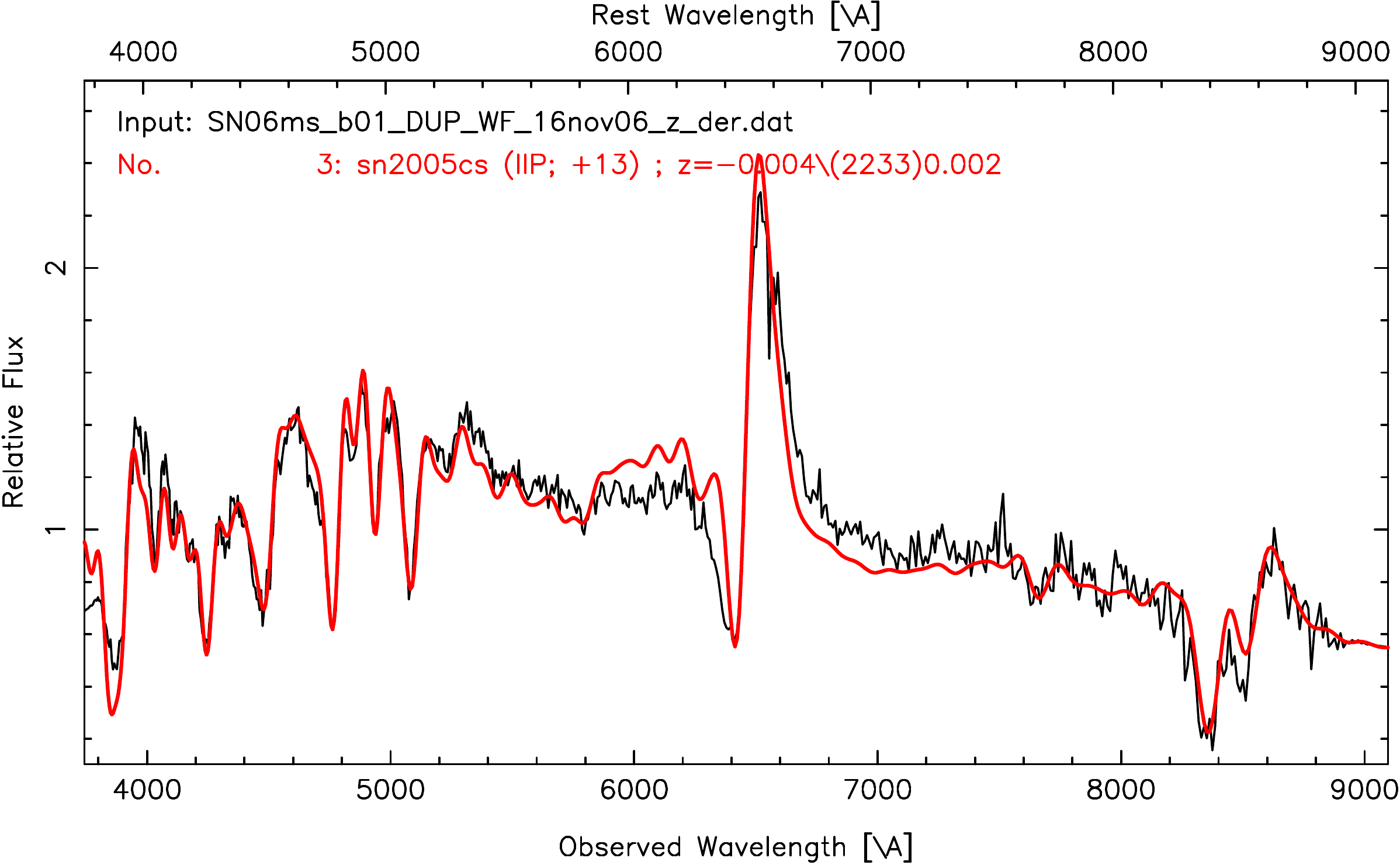}
\includegraphics[width=4.4cm]{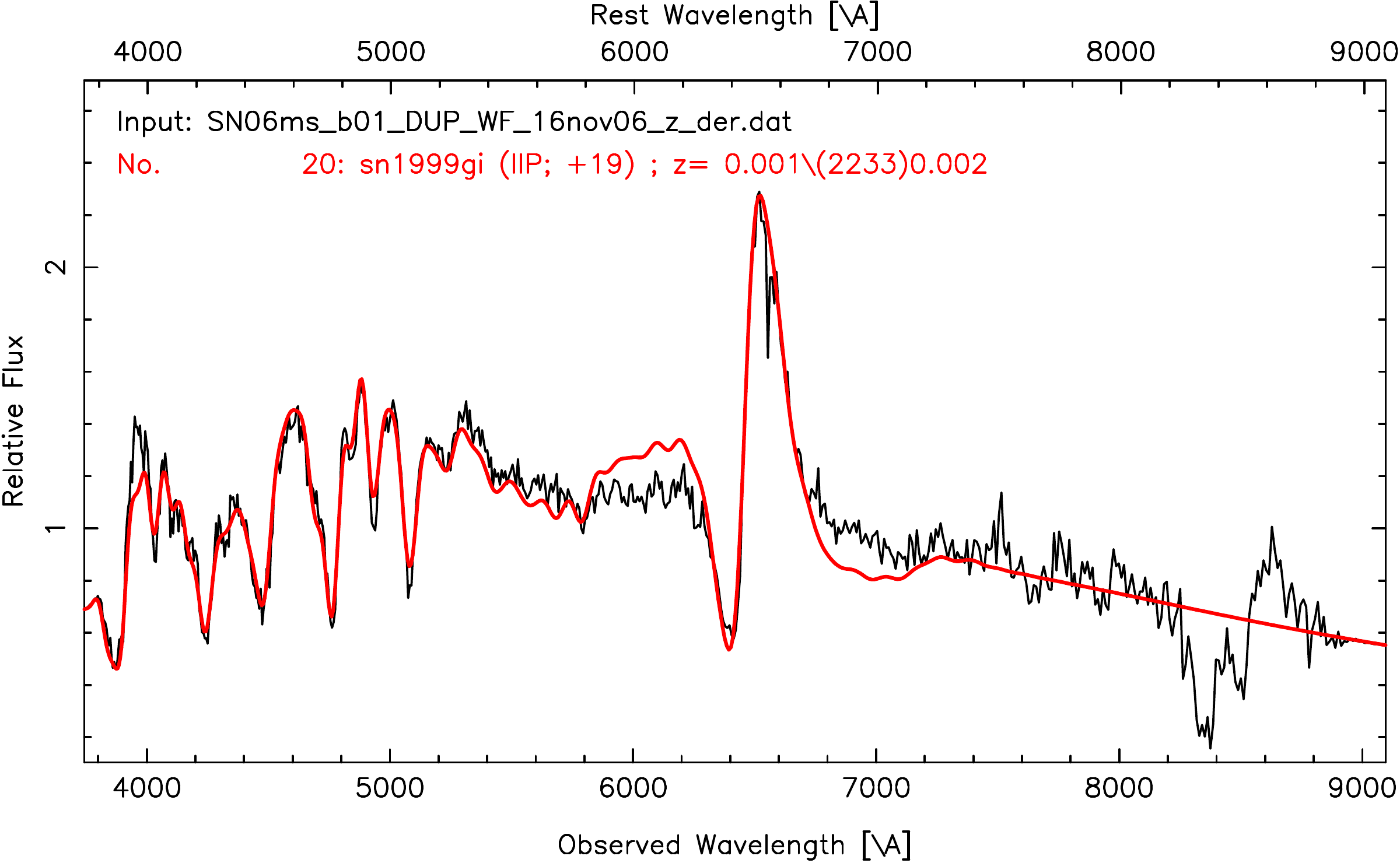}
\caption{Best spectral matching of SN~2006ms using SNID. The plots show SN~2006ms compared with 
SN~1999em, SN~2006bp, SN~2005cs, and SN~1999gi at 26, 34, 19, smf 31 days from explosion.}
\end{figure}

\begin{figure}
\centering
\includegraphics[width=4.4cm]{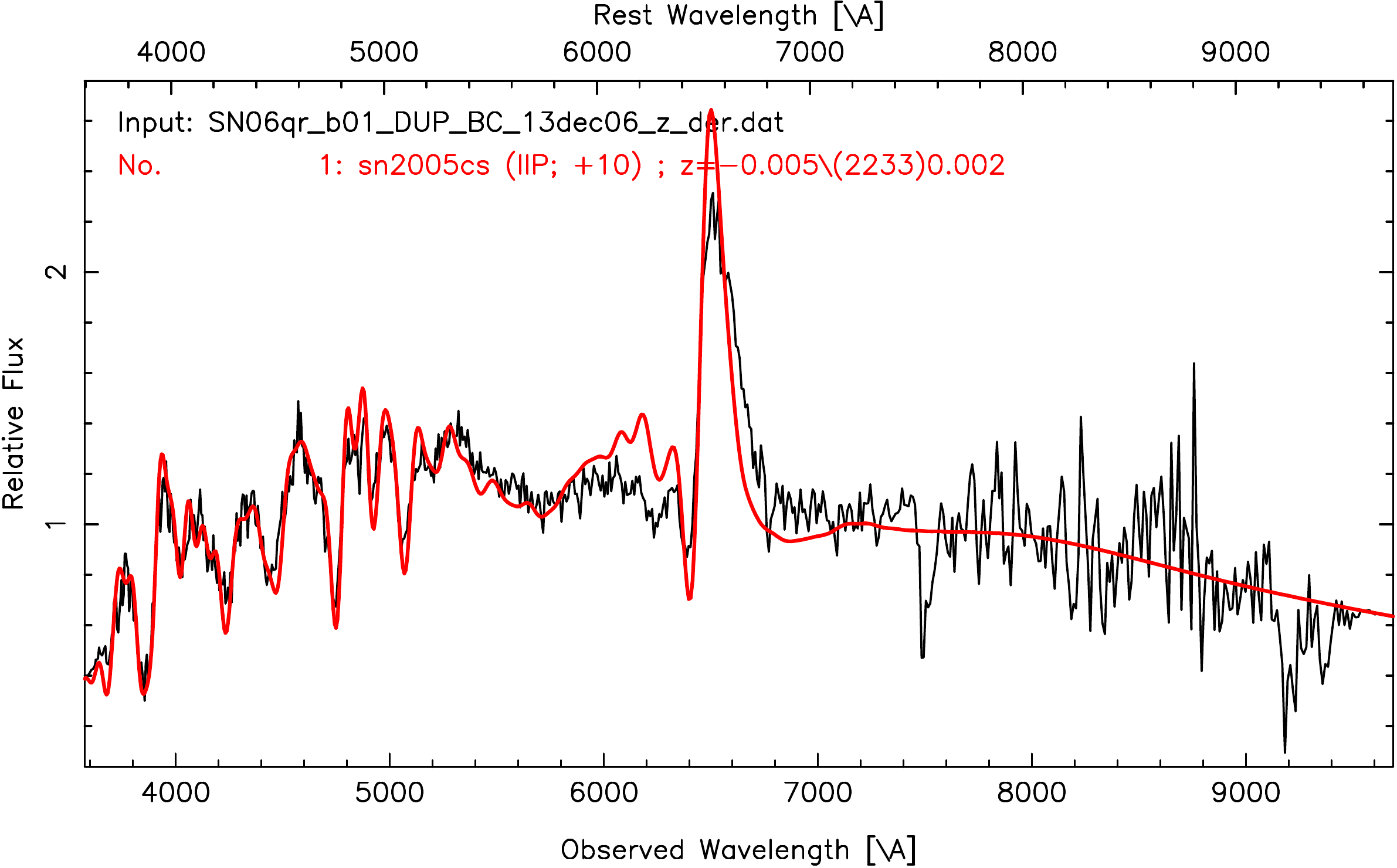}
\includegraphics[width=4.4cm]{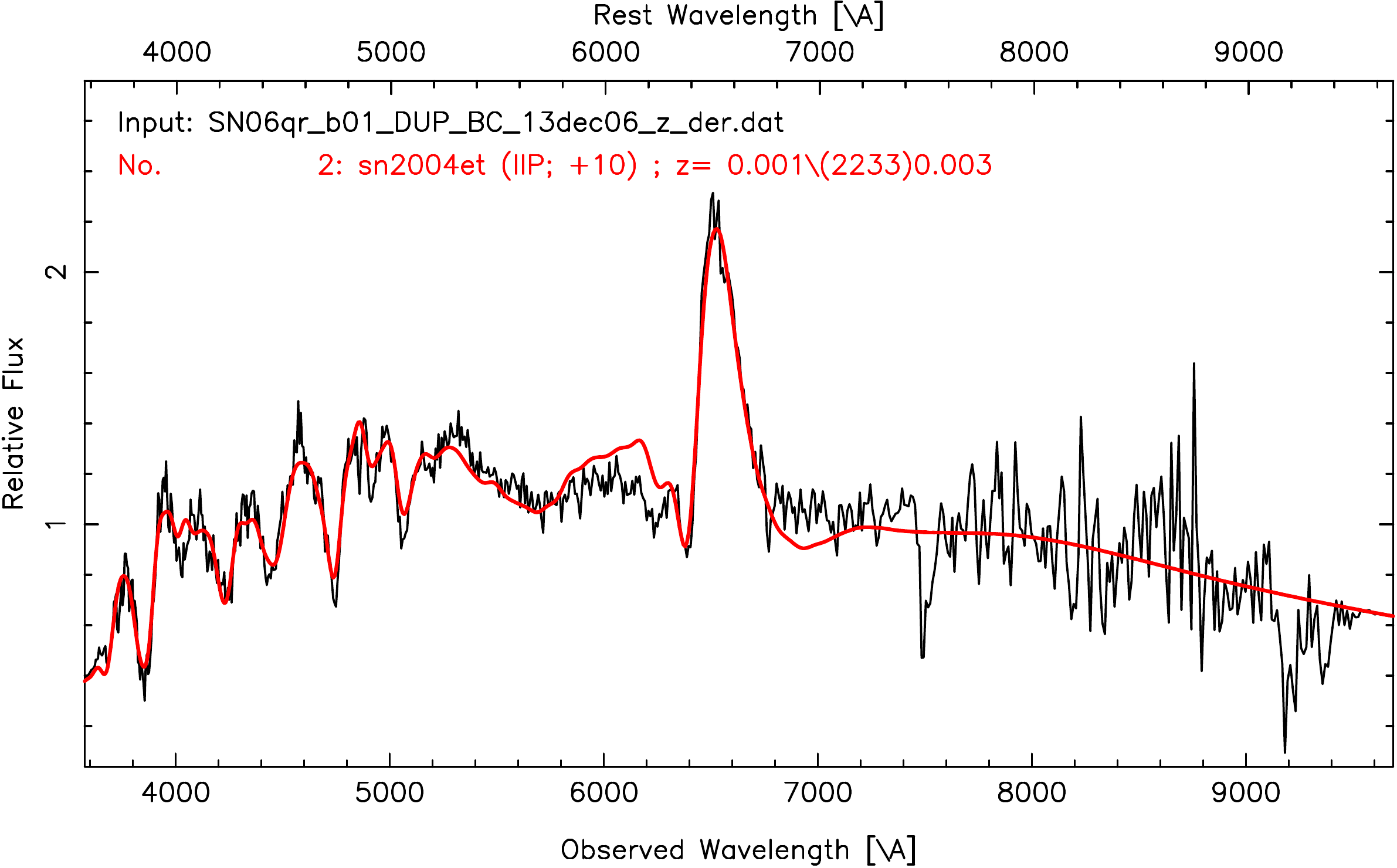}
\includegraphics[width=4.4cm]{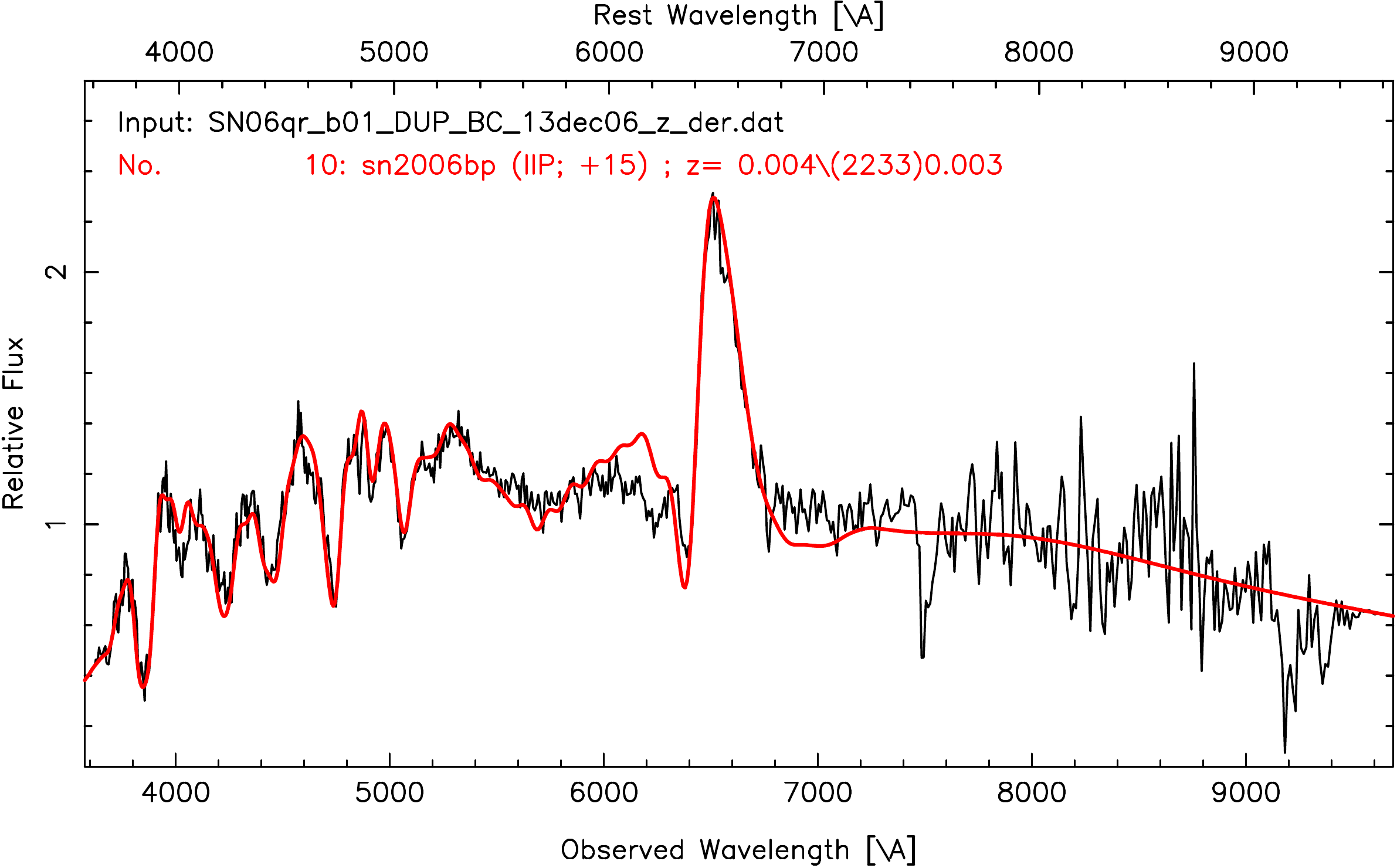}
\caption{Best spectral matching of SN~2006qr using SNID. The plots show SN~2006qr compared with 
SN~2005cs, SN~2004et, and SN~2006bp at 16, 26, and 24 days from explosion.}
\end{figure}

\begin{figure}
\centering
\includegraphics[width=4.4cm]{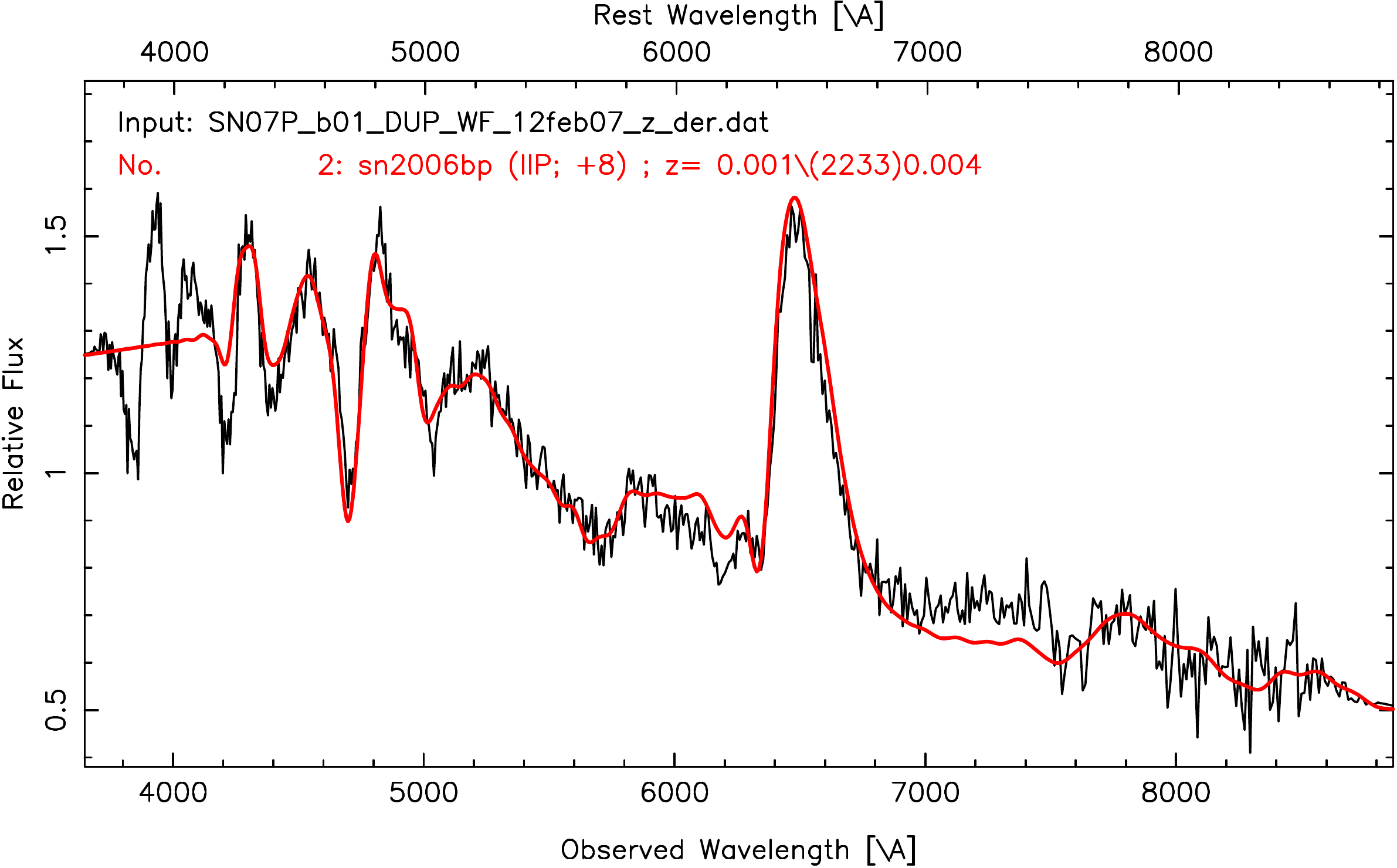}
\includegraphics[width=4.4cm]{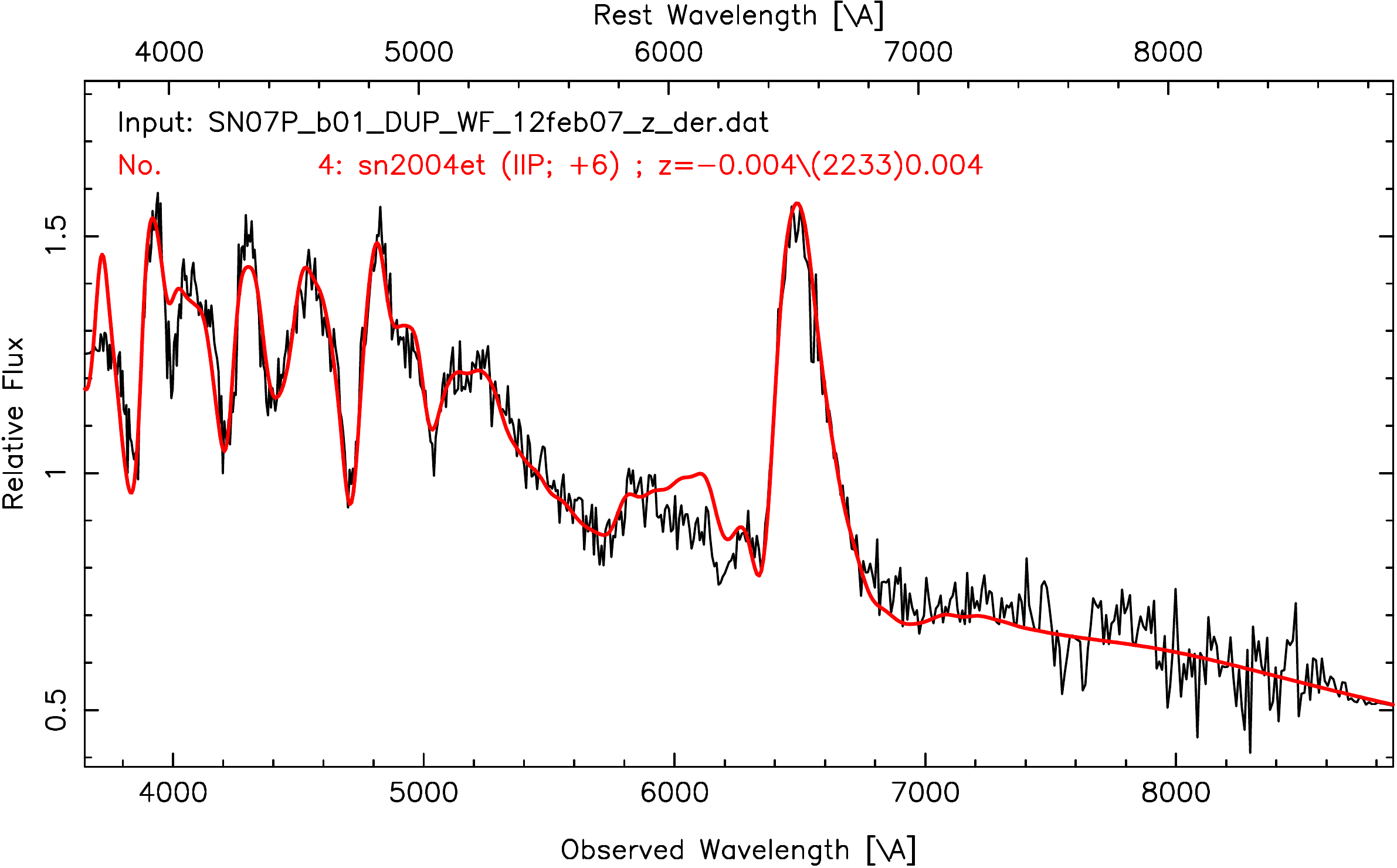}
\includegraphics[width=4.4cm]{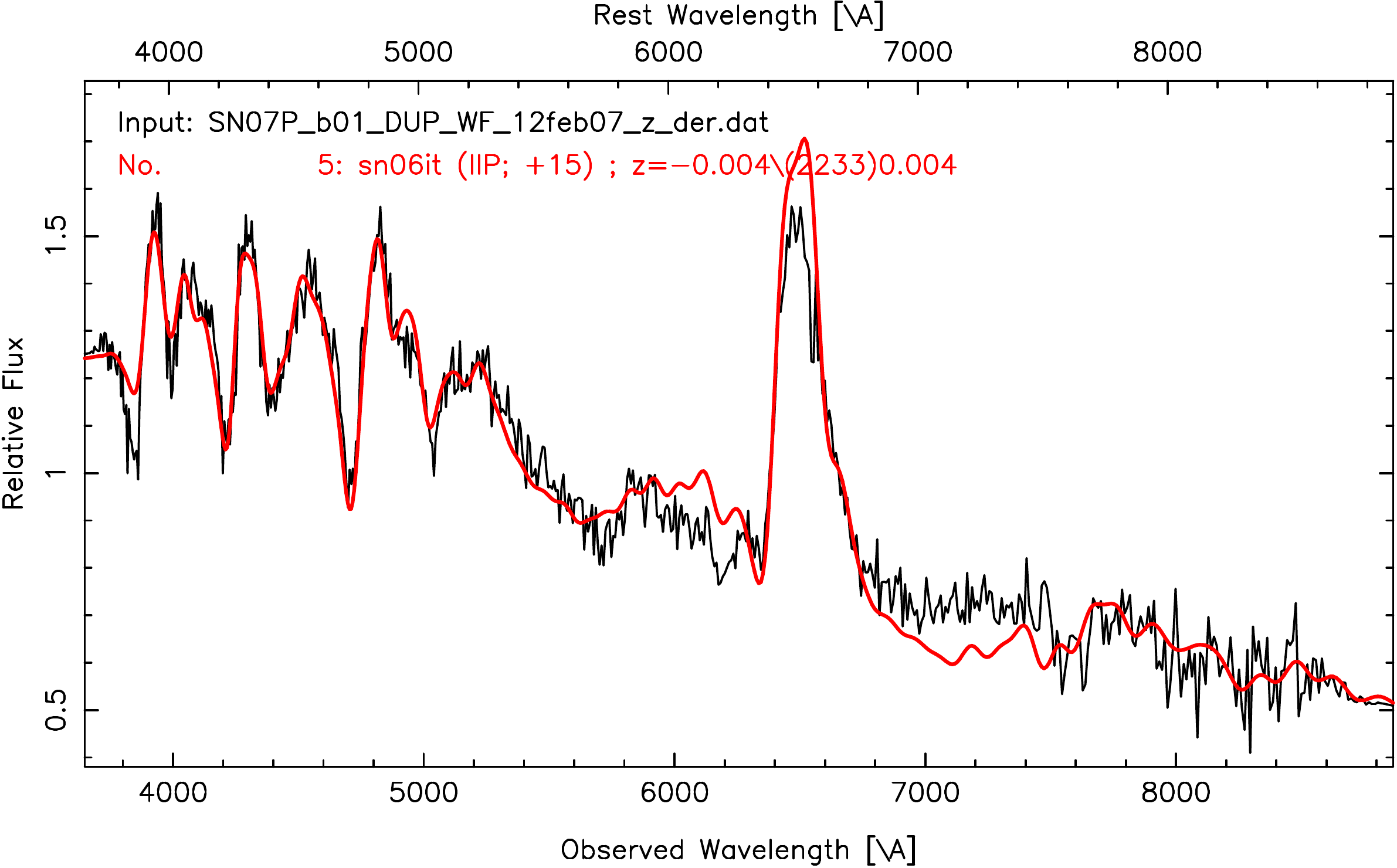}
\includegraphics[width=4.4cm]{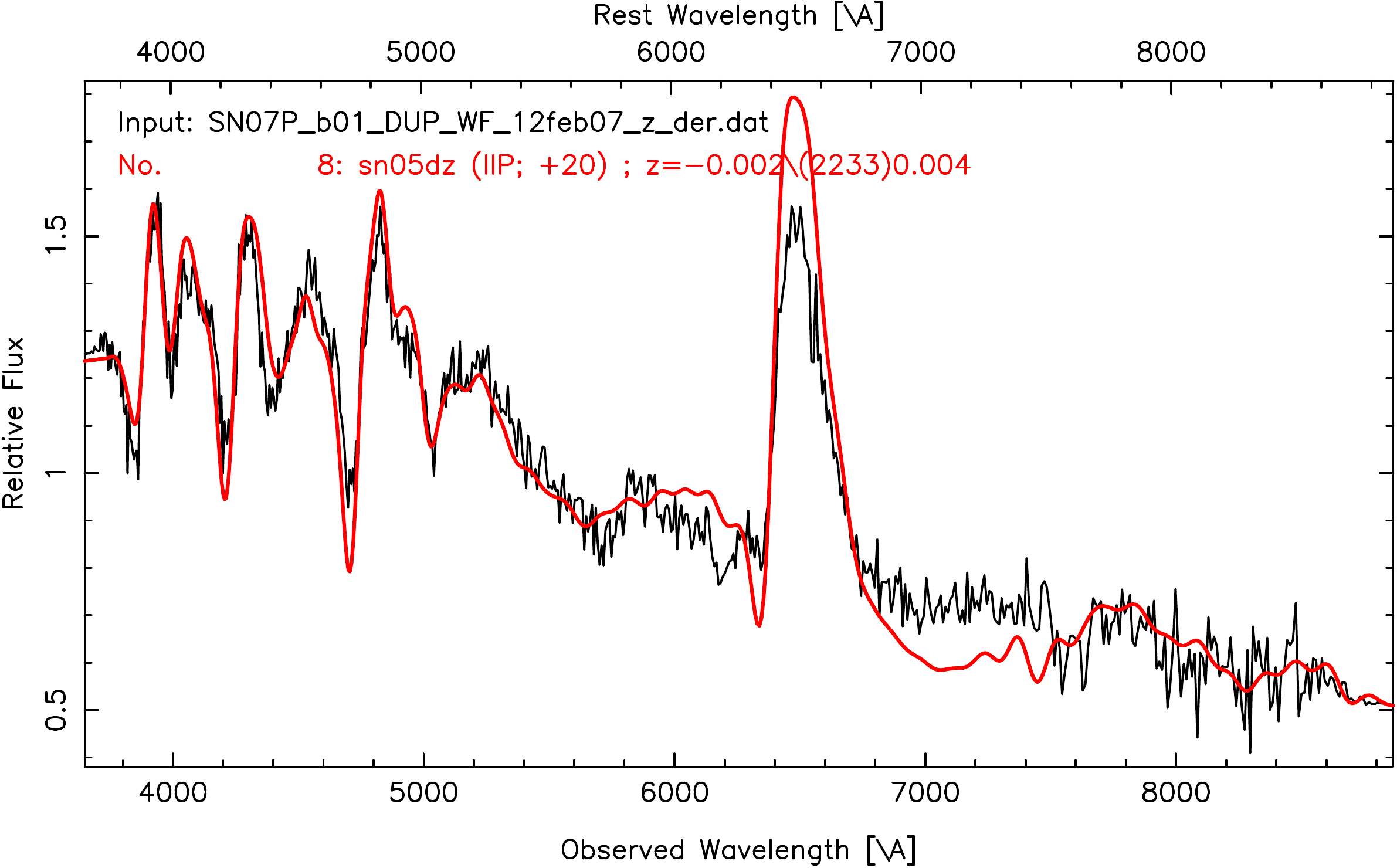}
\caption{Best spectral matching of SN~2007P using SNID. The plots show SN~2007P compared with 
SN~2006bp, SN~2004et, SN~2006it, and SN~2005dz at 17. 22, 15, and 20 days from explosion.}
\end{figure}

\clearpage

\begin{figure}
\centering
\includegraphics[width=4.4cm]{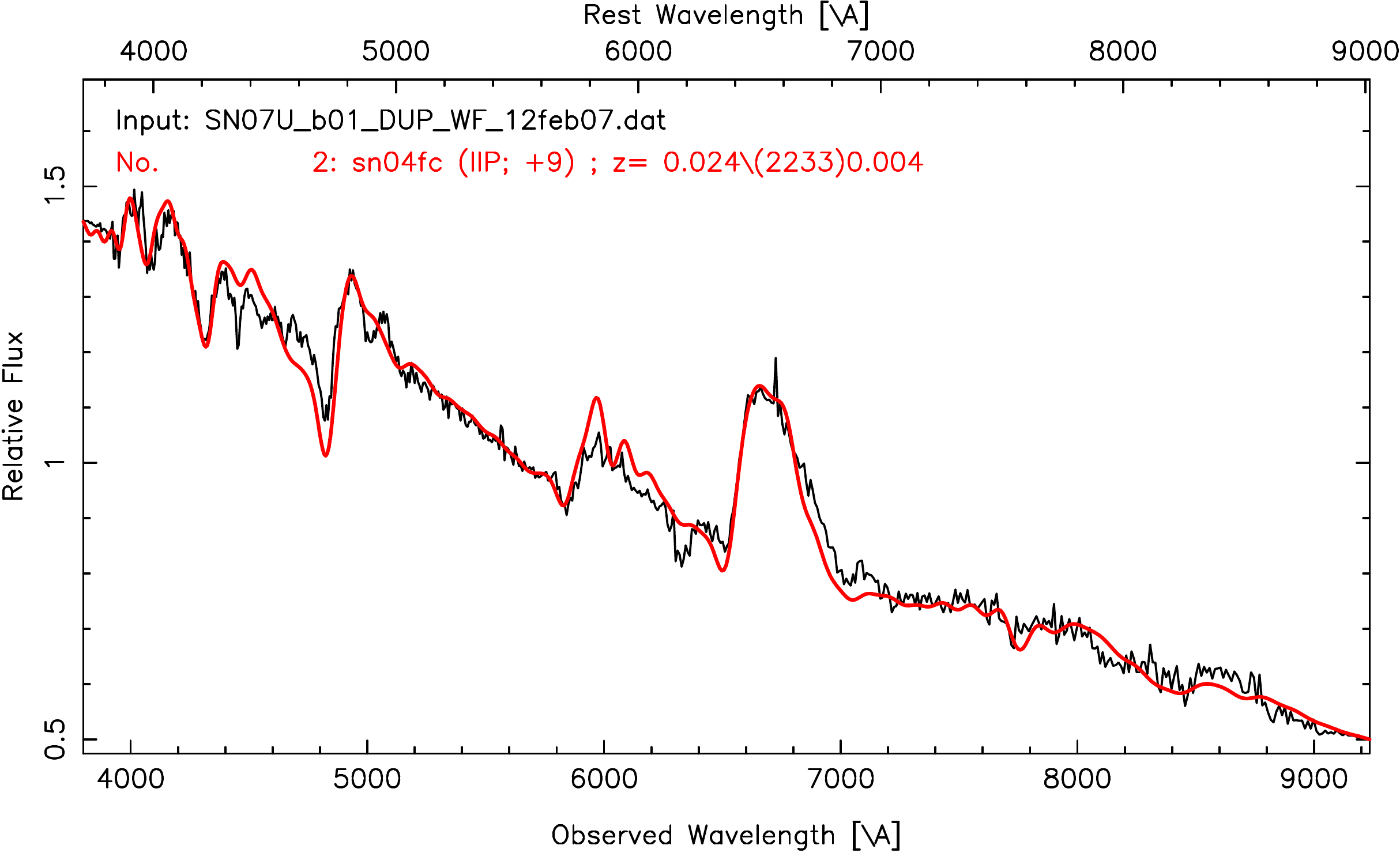}
\includegraphics[width=4.4cm]{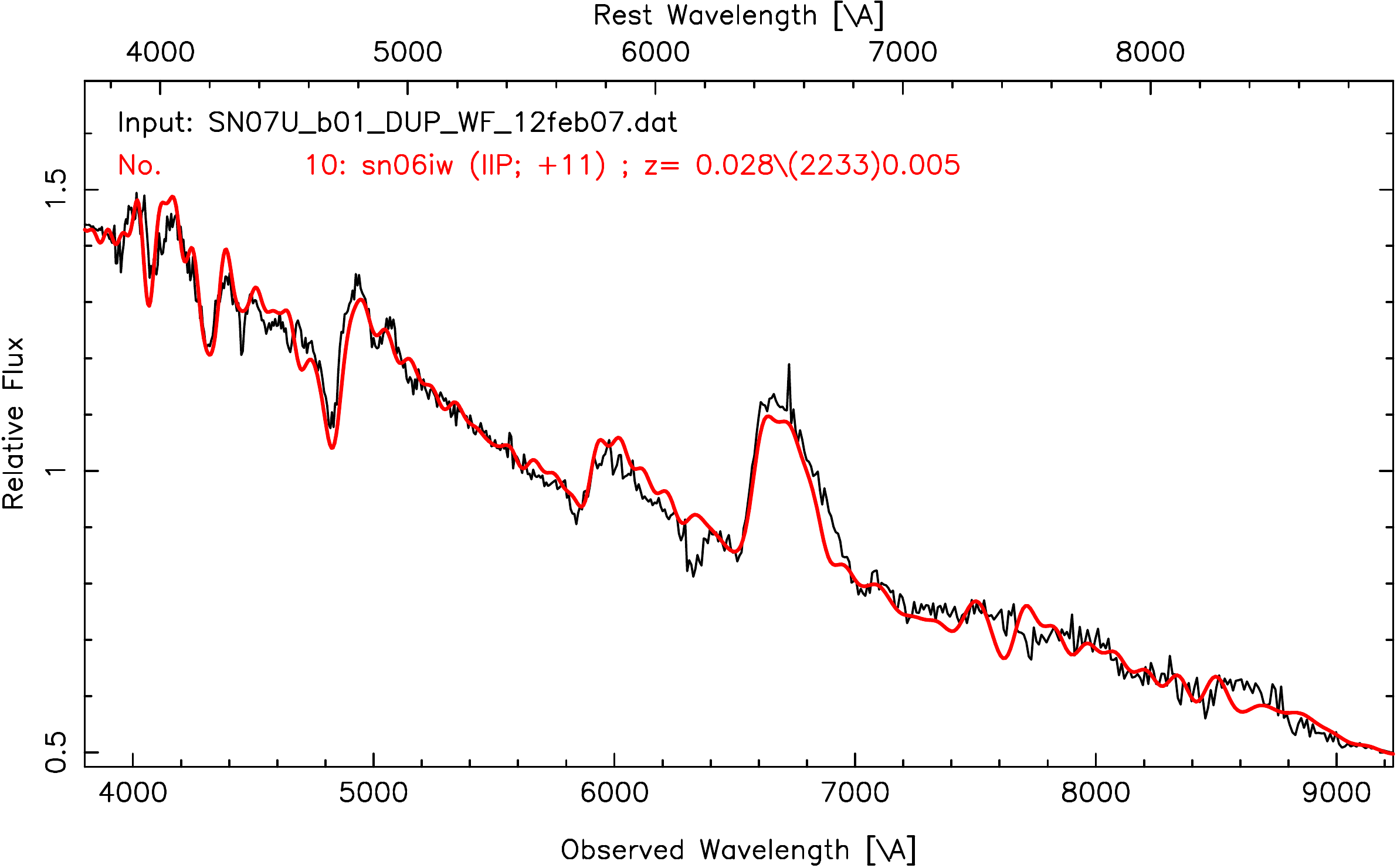}
\includegraphics[width=4.4cm]{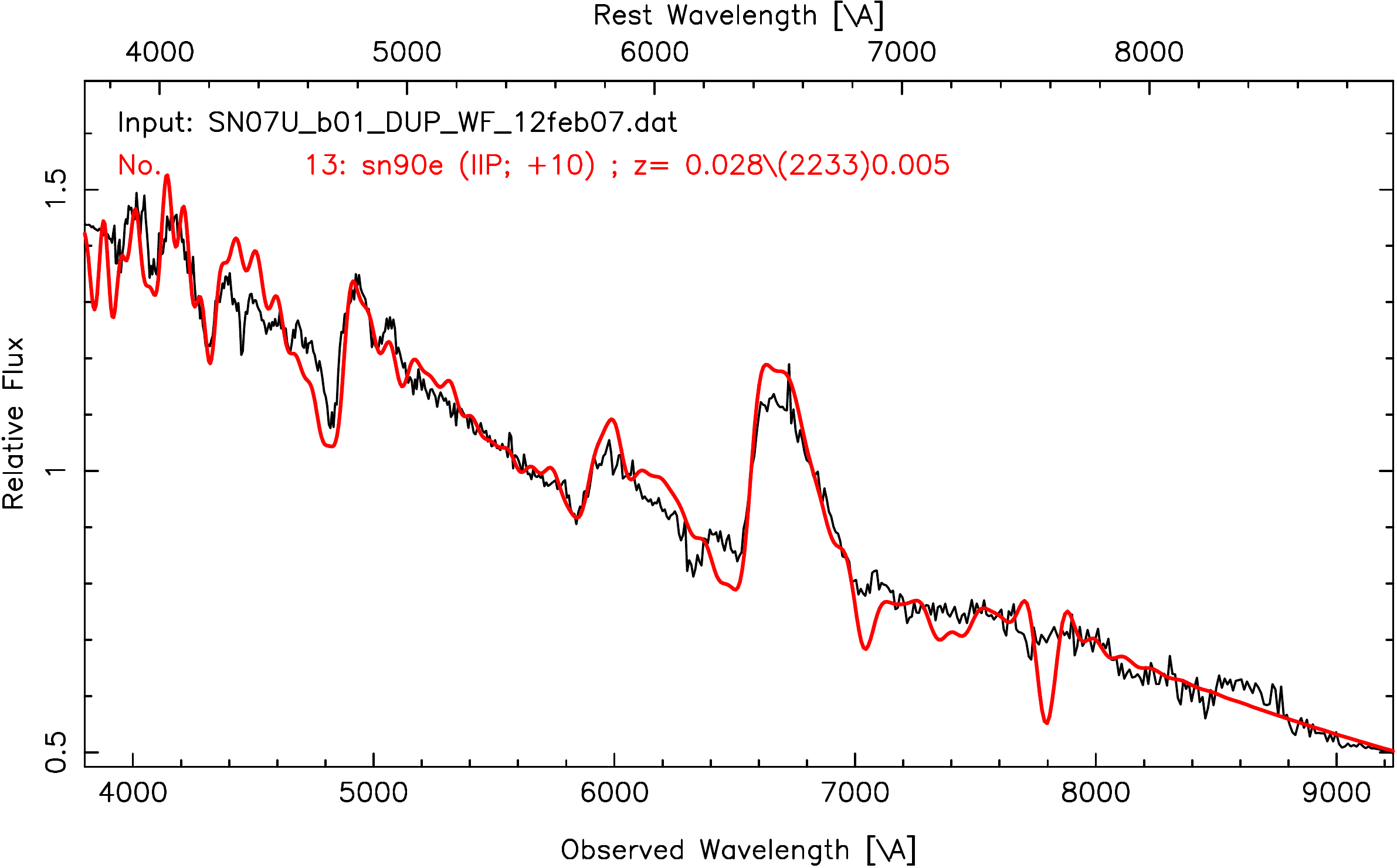}
\caption{Best spectral matching of SN~2007U using SNID. The plots show SN~2007U compared with 
SN~2004fc, SN~2006iw, and SN~1990E at 9, 11, and 10 days from explosion.}
\end{figure}

\begin{figure}
\centering
\includegraphics[width=4.4cm]{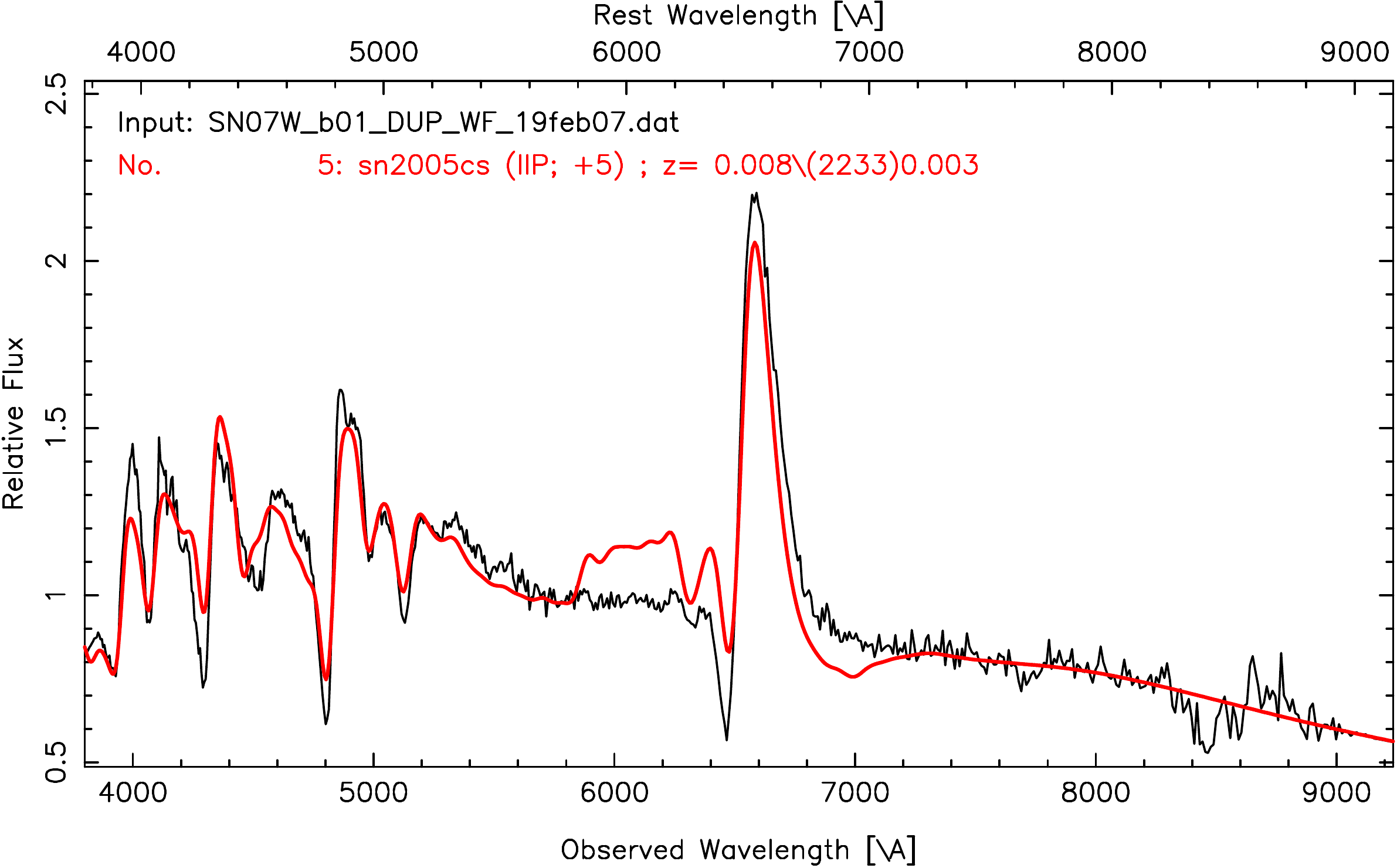}
\includegraphics[width=4.4cm]{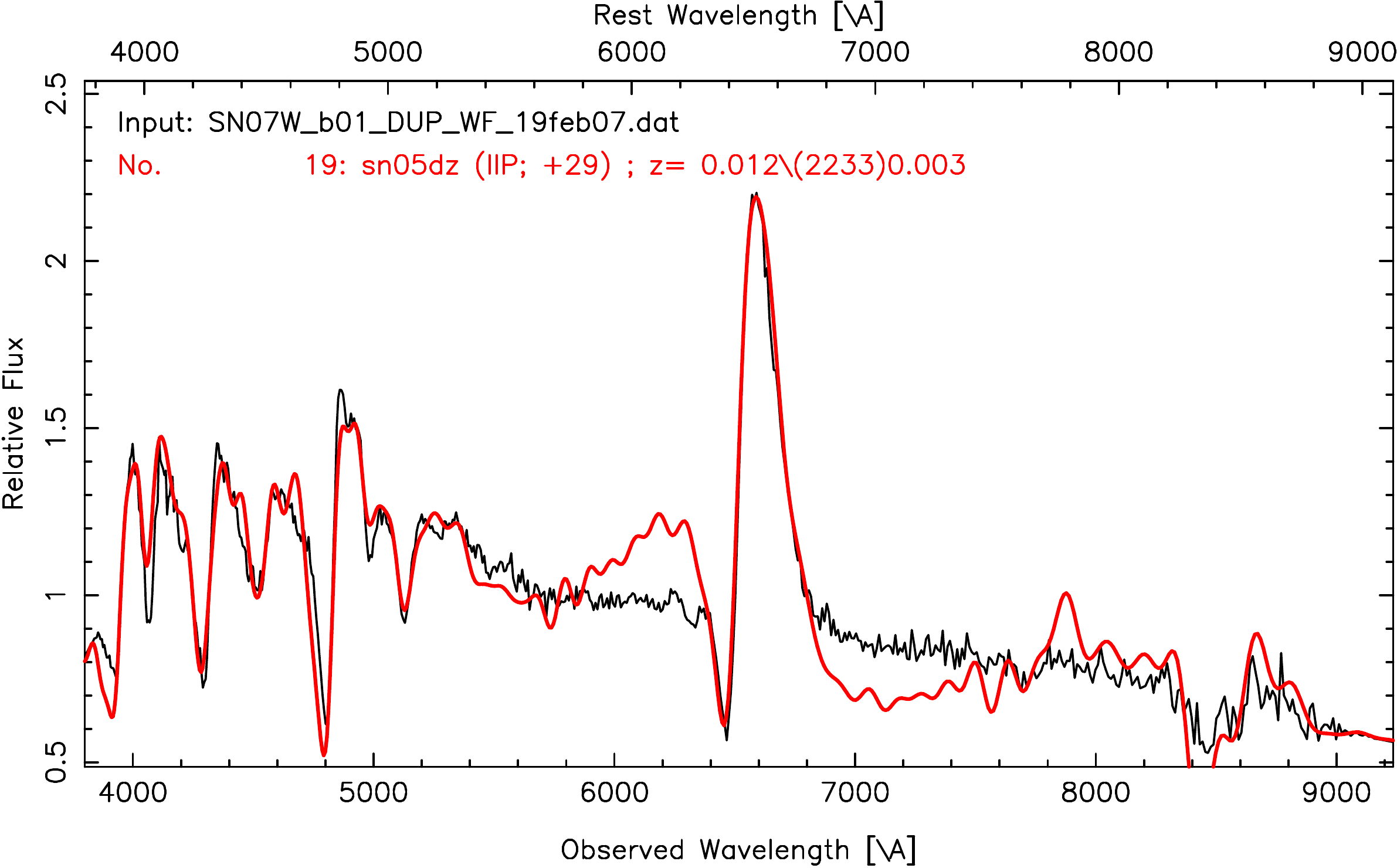}
\includegraphics[width=4.4cm]{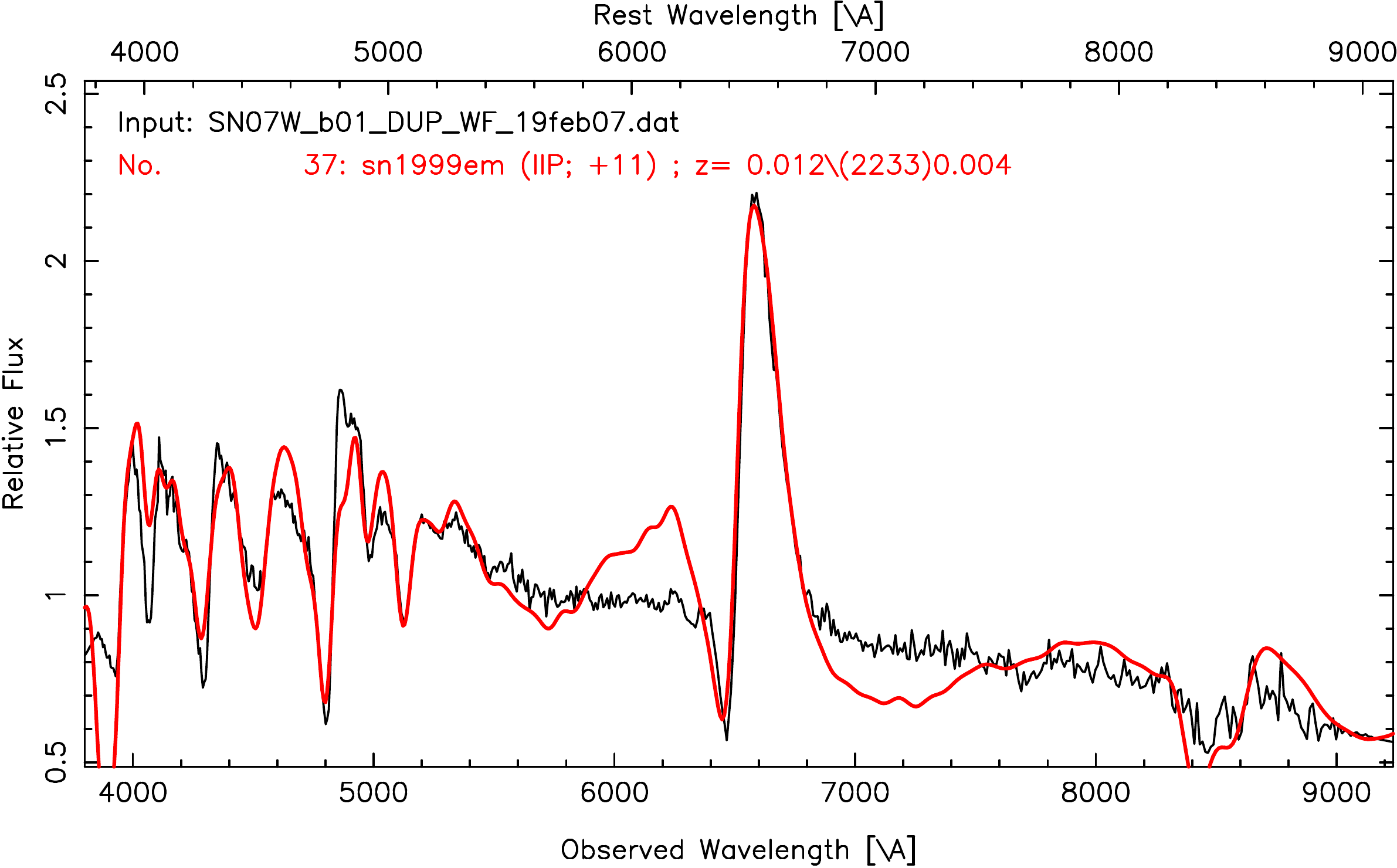}
\caption{Best spectral matching of SN~2007W using SNID. The plots show SN~2007W compared with 
SN~2005cs, SN~2005dz, and SN~1999em at 11, 29, and 21 days from explosion.}
\end{figure}

\begin{figure}
\centering
\includegraphics[width=4.4cm]{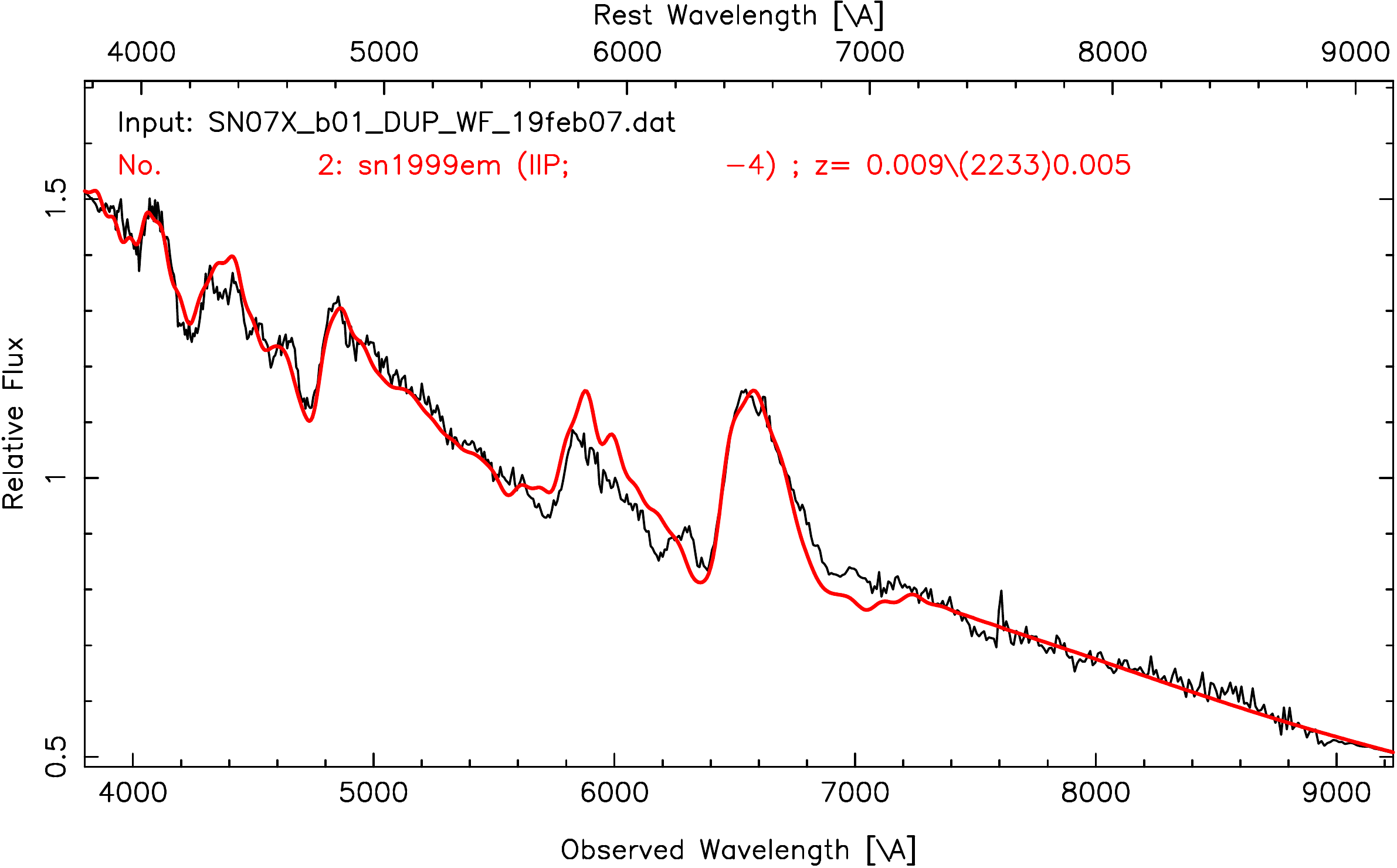}
\includegraphics[width=4.4cm]{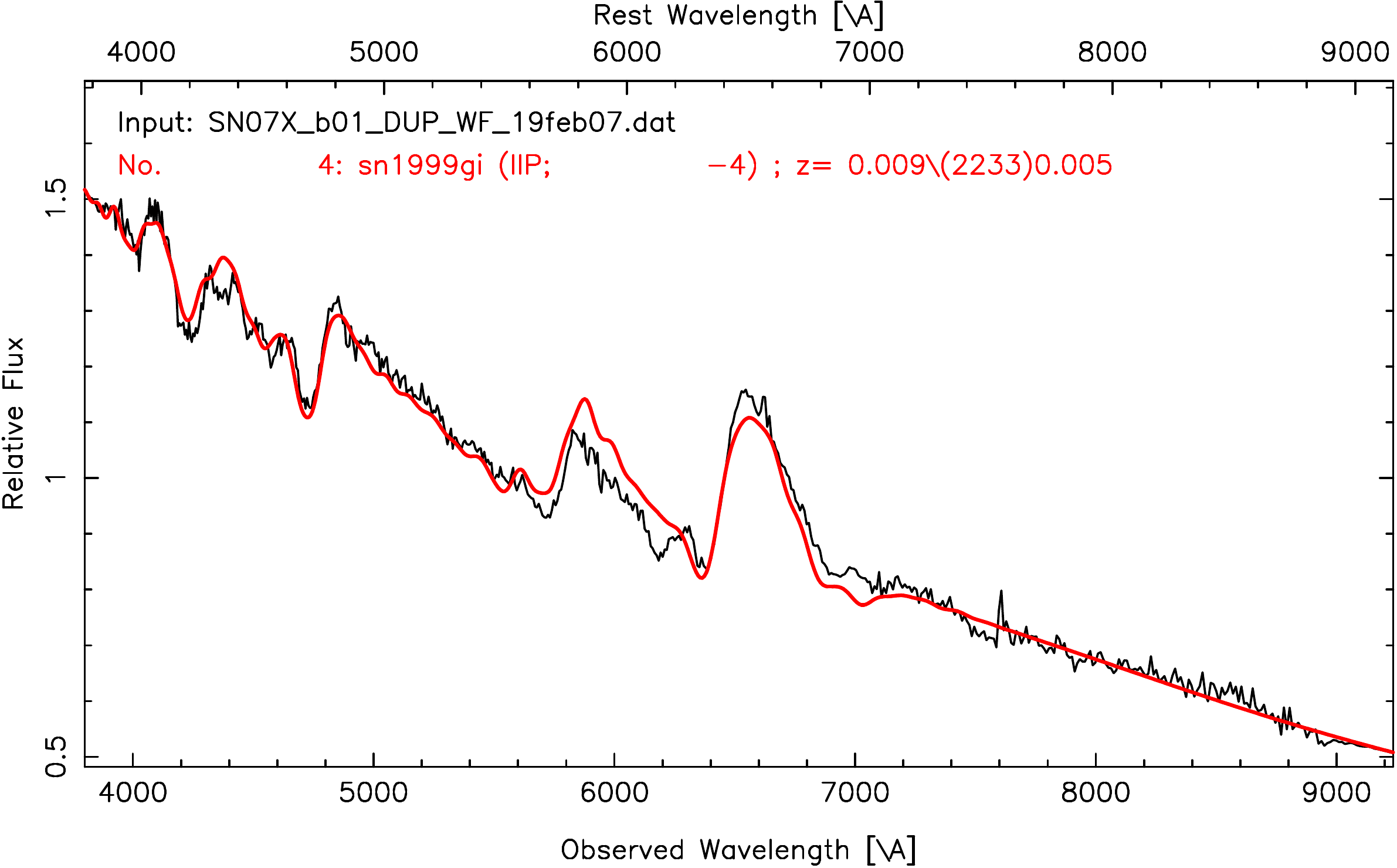}
\includegraphics[width=4.4cm]{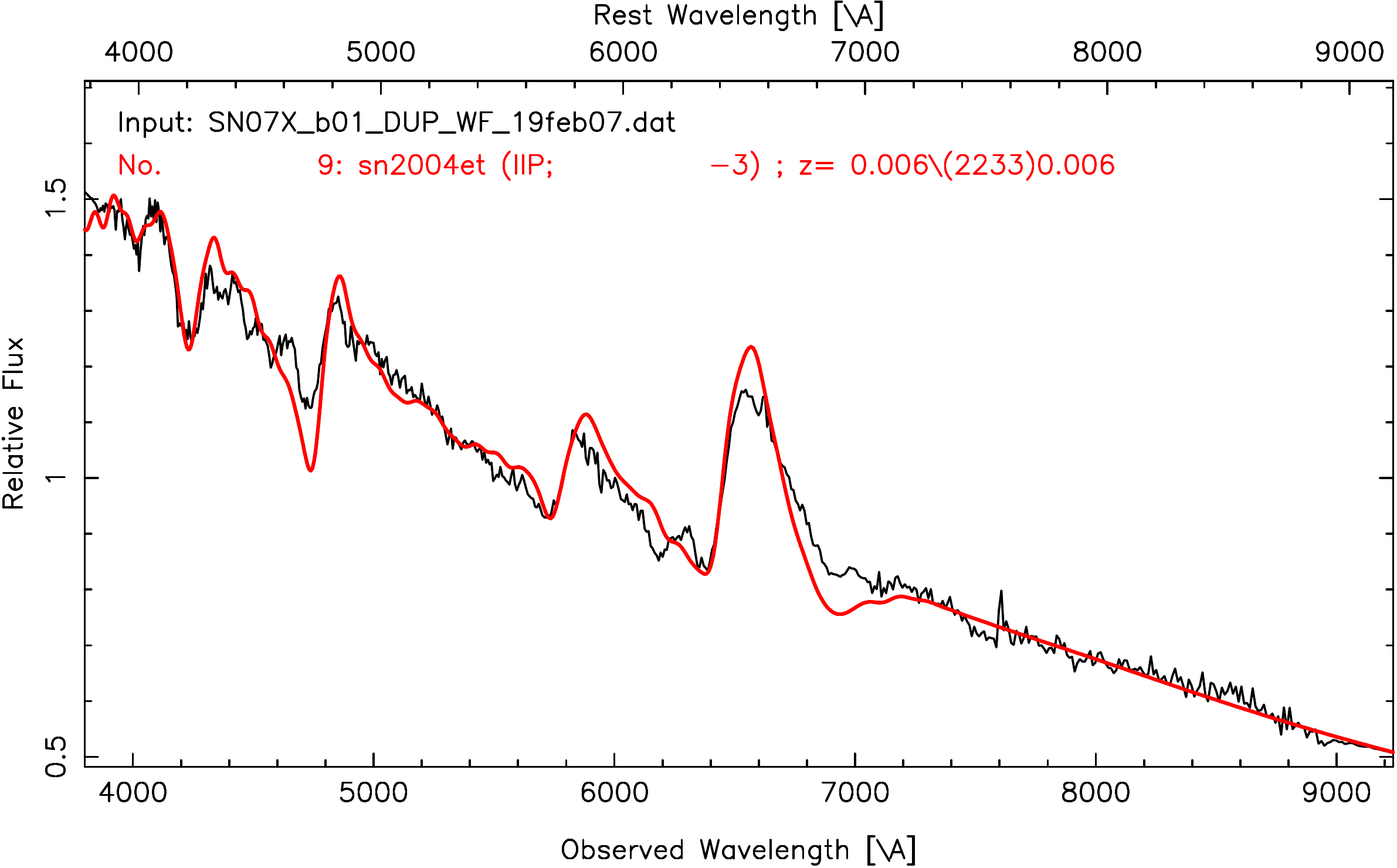}
\includegraphics[width=4.4cm]{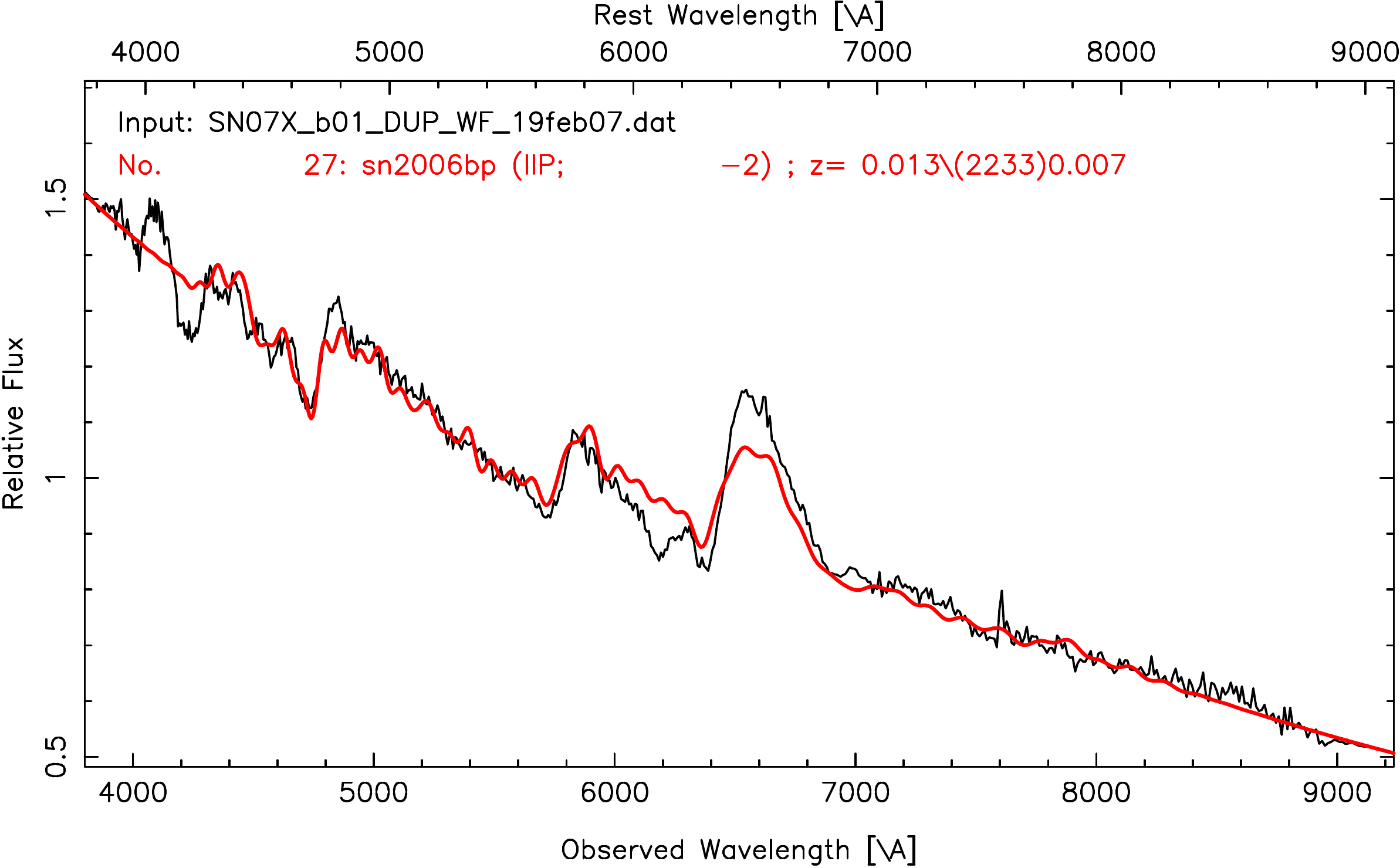}
\caption{Best spectral matching of SN~2007X using SNID. The plots show SN~2007X compared with 
SN~1999em, SN~1999gi, SN~2004et, and SN~2006bp at 6, 8, 13, and 7 days from explosion.}
\end{figure}

\clearpage

\begin{figure}
\centering
\includegraphics[width=4.4cm]{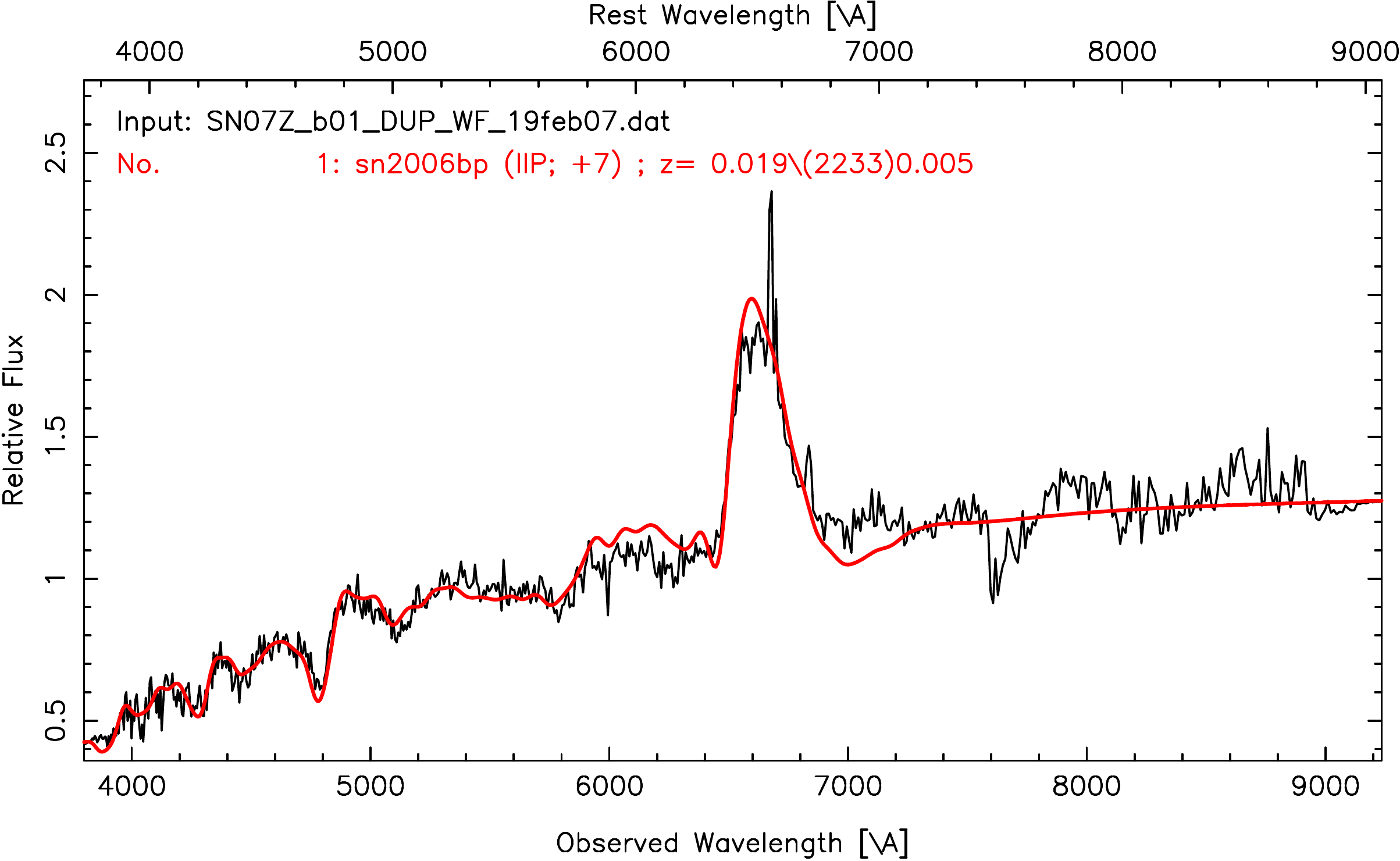}
\includegraphics[width=4.4cm]{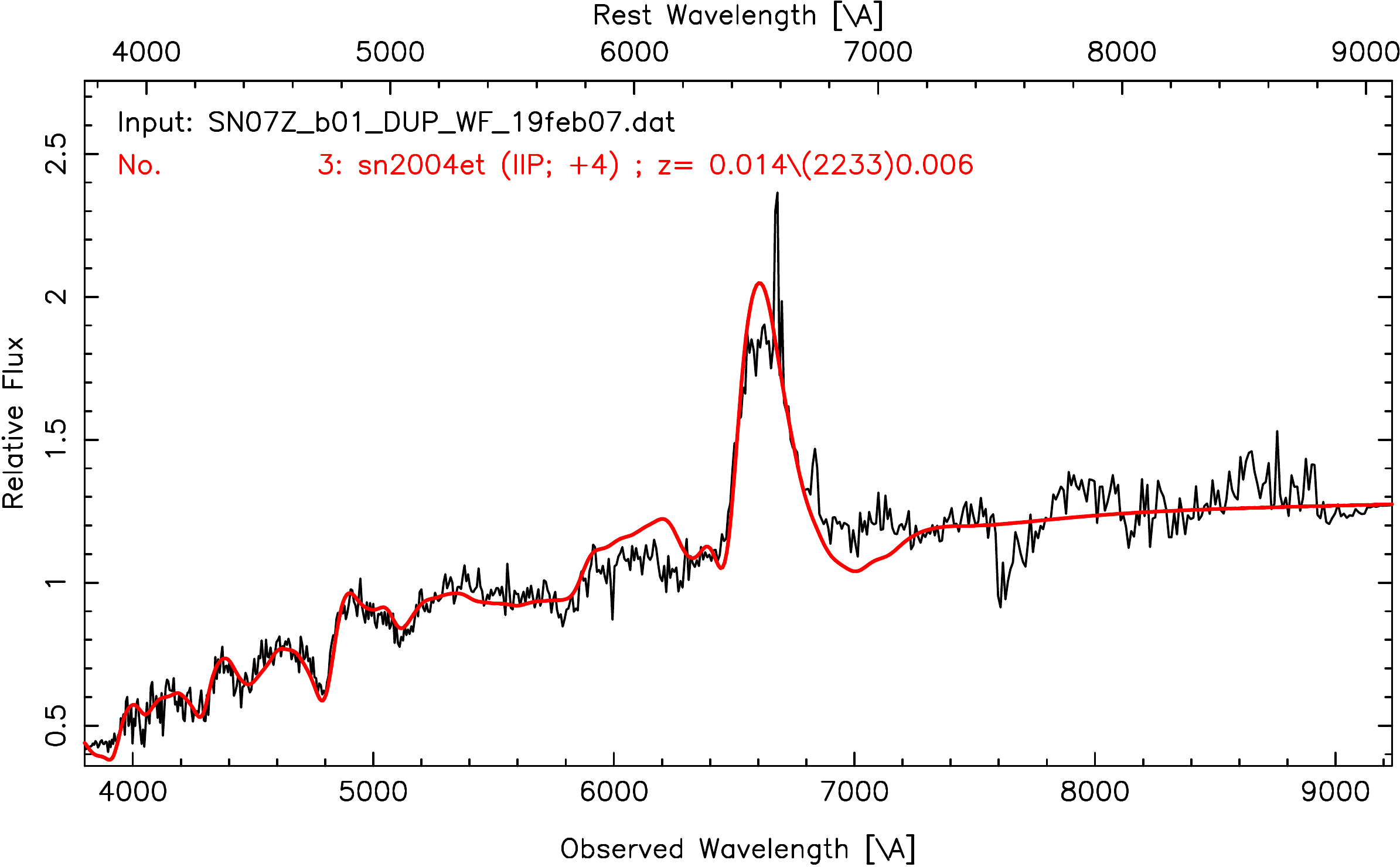}
\includegraphics[width=4.4cm]{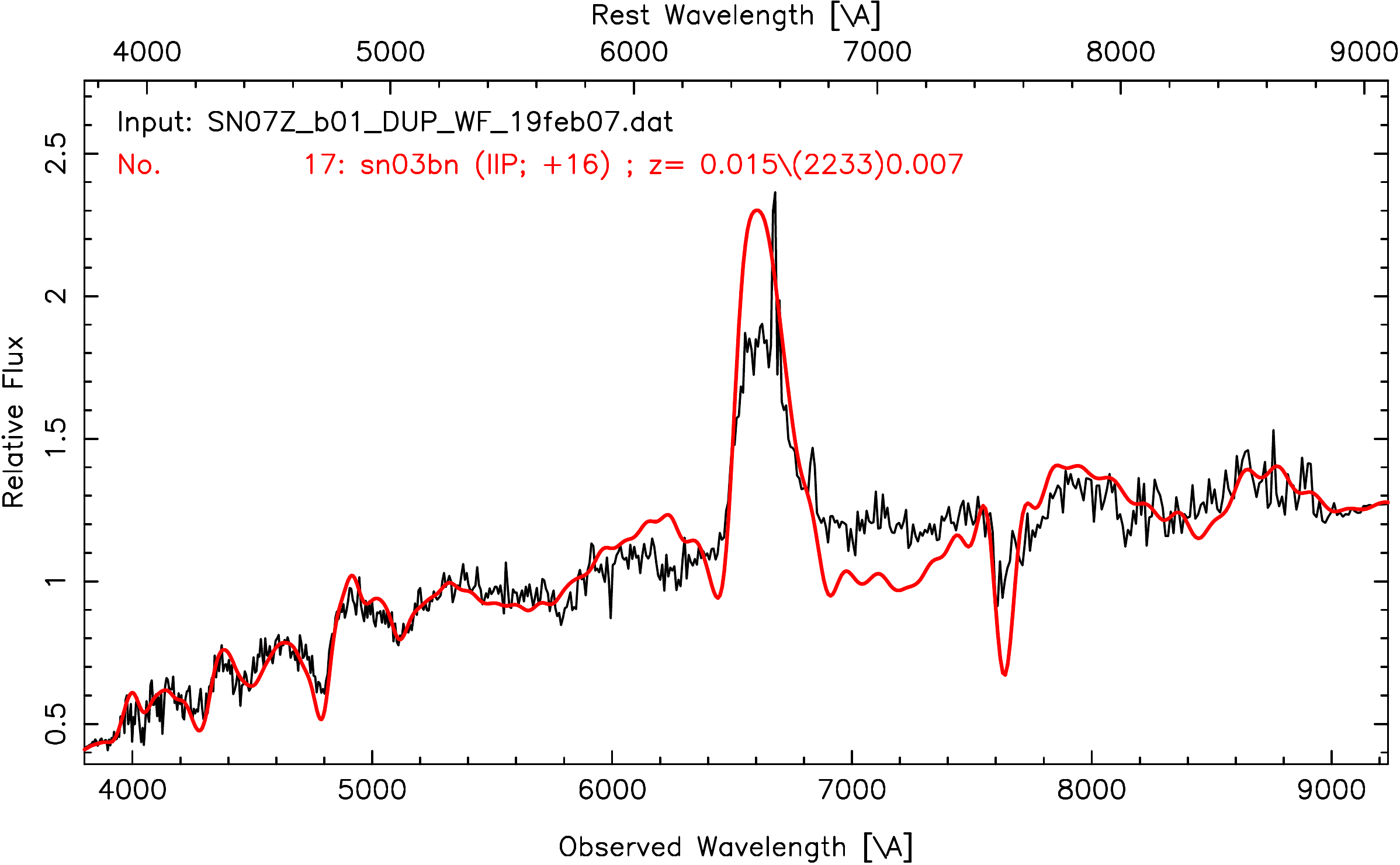}
\includegraphics[width=4.4cm]{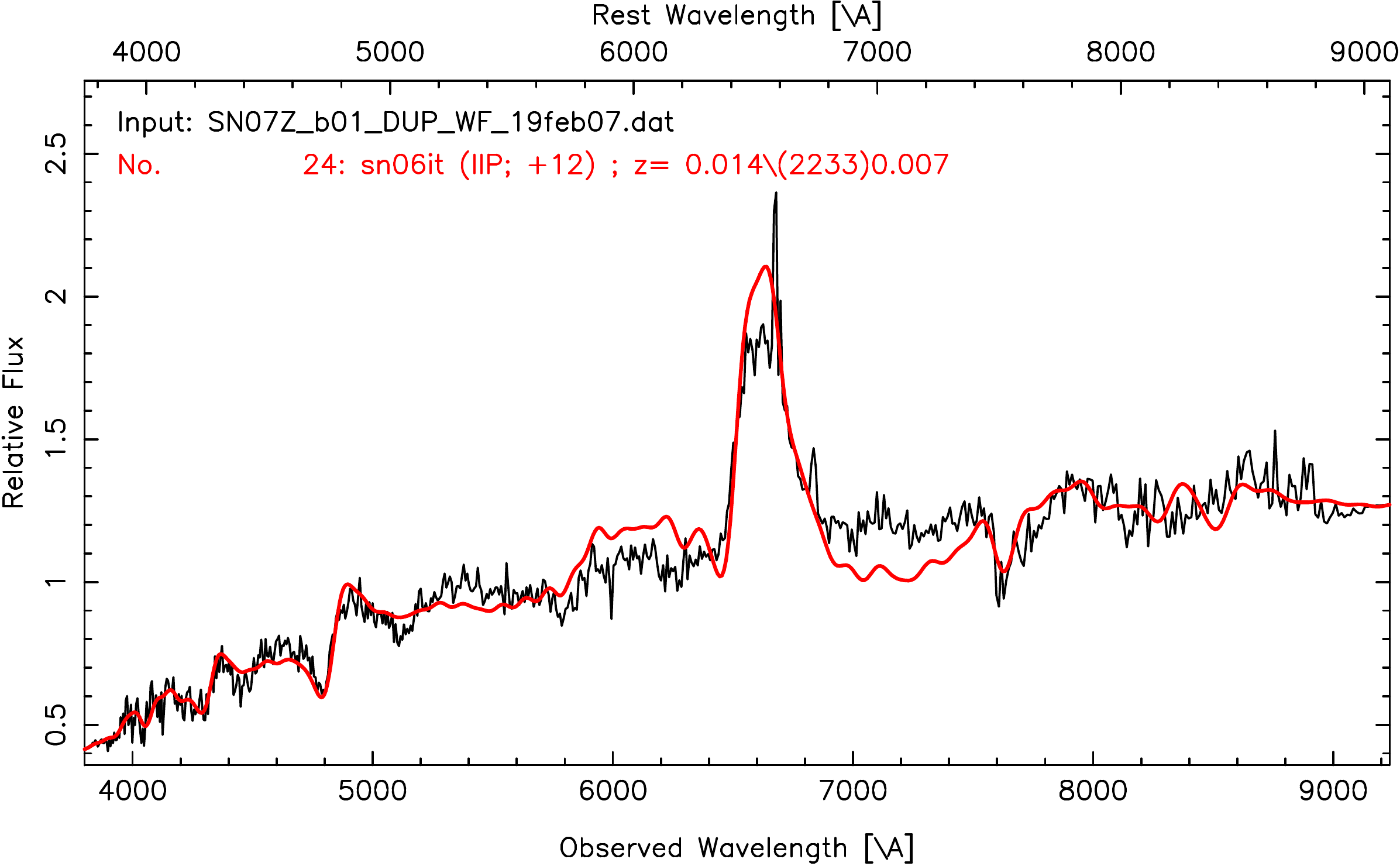}
\caption{Best spectral matching of SN~2007Z using SNID. The plots show SN~2007Z compared with 
SN~2006bp, SN~2004et, SN~2003bn, and SN~2006it at 16, 17, 16, and 12 days from explosion.}
\end{figure}

\begin{figure}
\centering
\includegraphics[width=4.4cm]{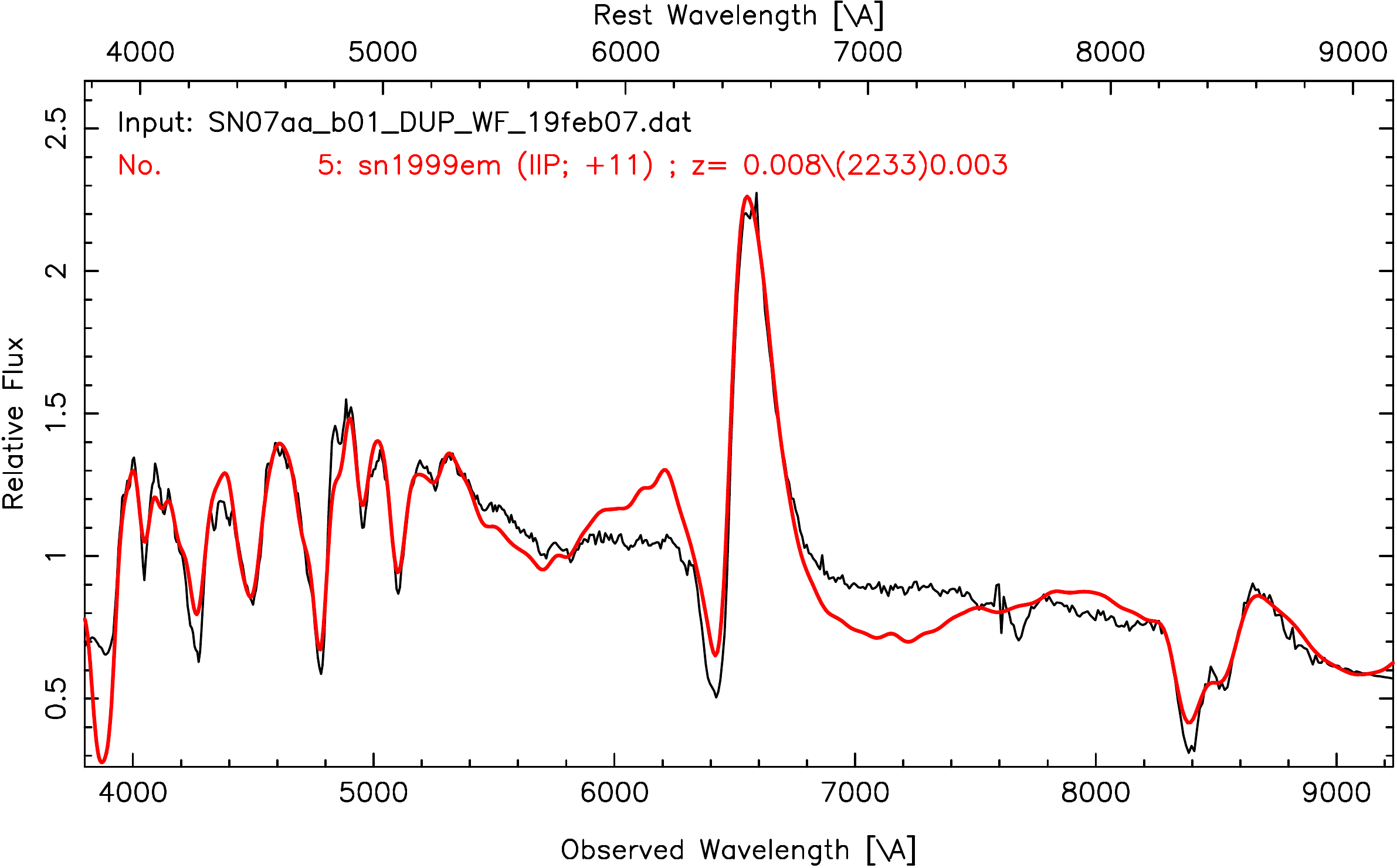}
\includegraphics[width=4.4cm]{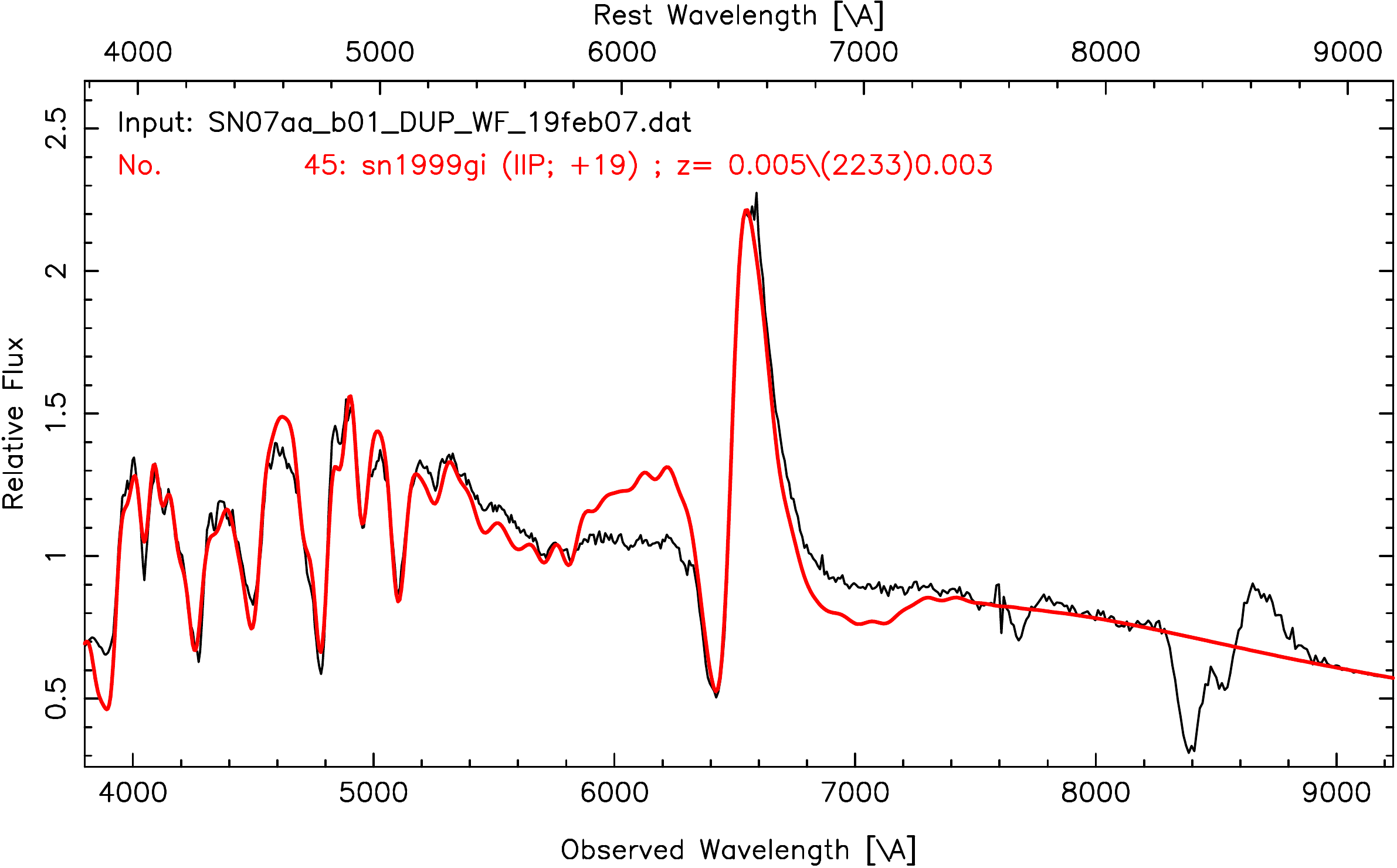}
\includegraphics[width=4.4cm]{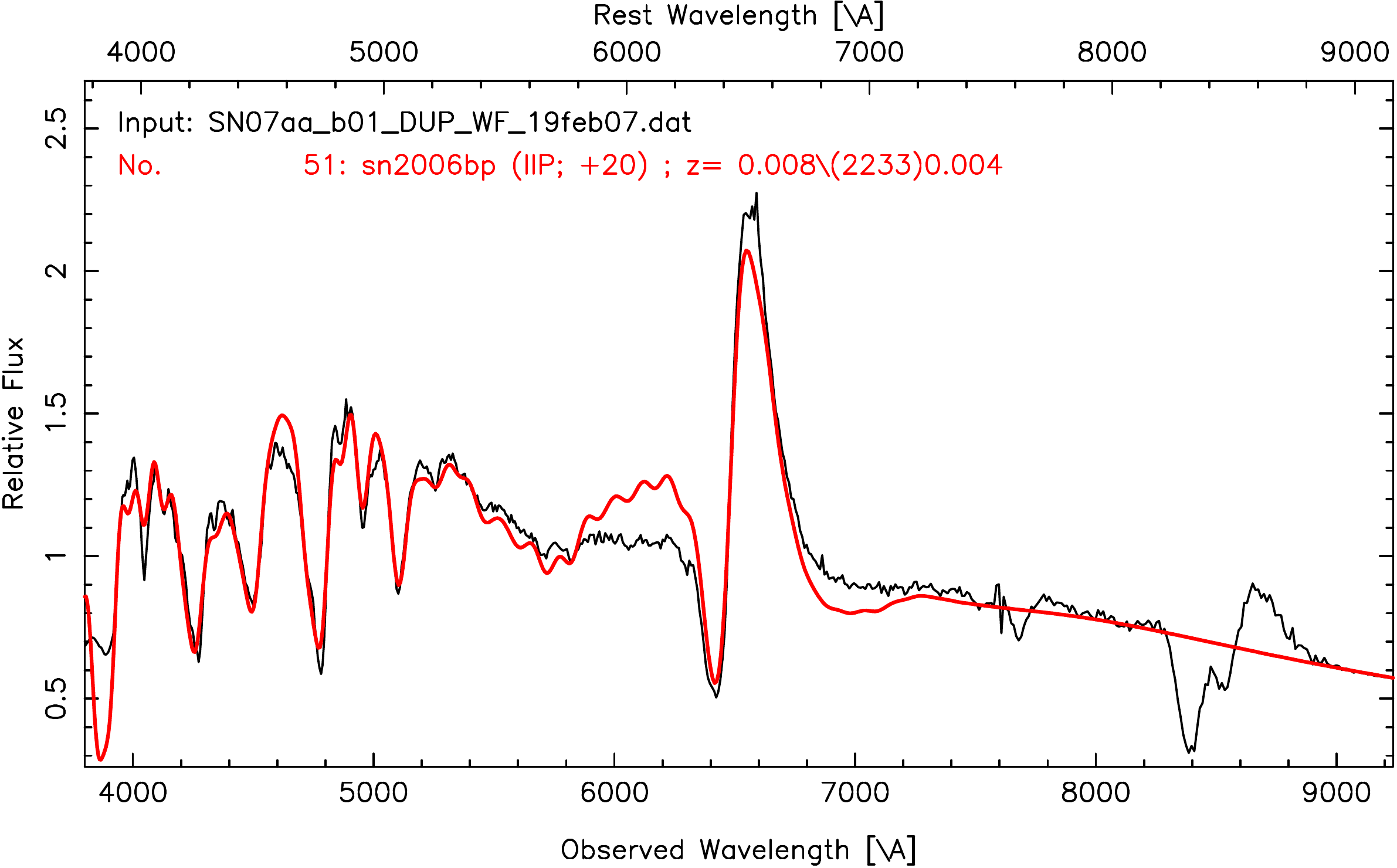}
\includegraphics[width=4.4cm]{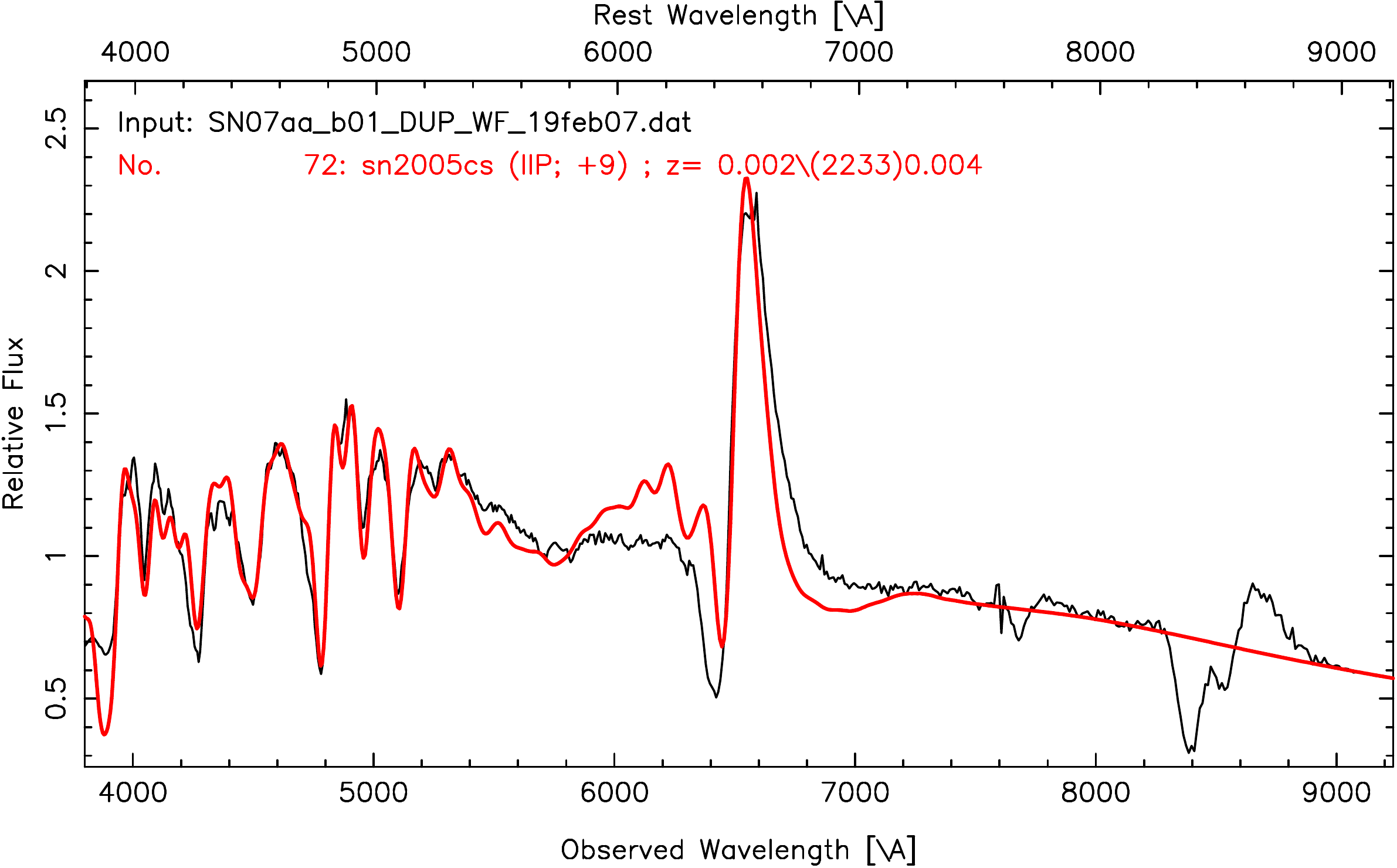}
\caption{Best spectral matching of SN~2007aa using SNID. The plots show SN~2007aa compared with 
SN~1999em, SN~1999gi, SN~2006bp, and SN~2005cs at 21, 31, 29, and 15 days from explosion.}
\end{figure}

\clearpage

\begin{figure}
\centering
\includegraphics[width=4.4cm]{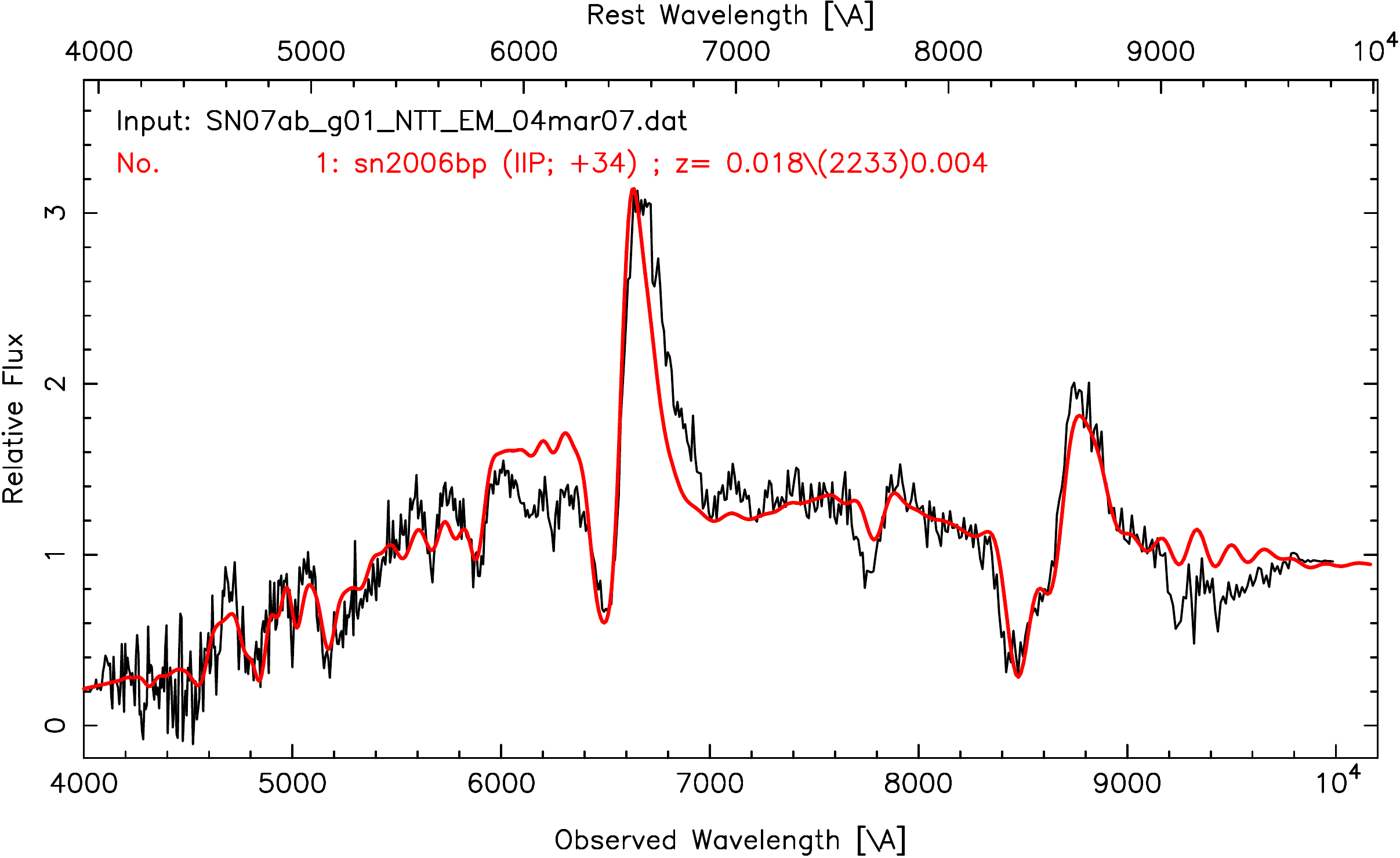}
\includegraphics[width=4.4cm]{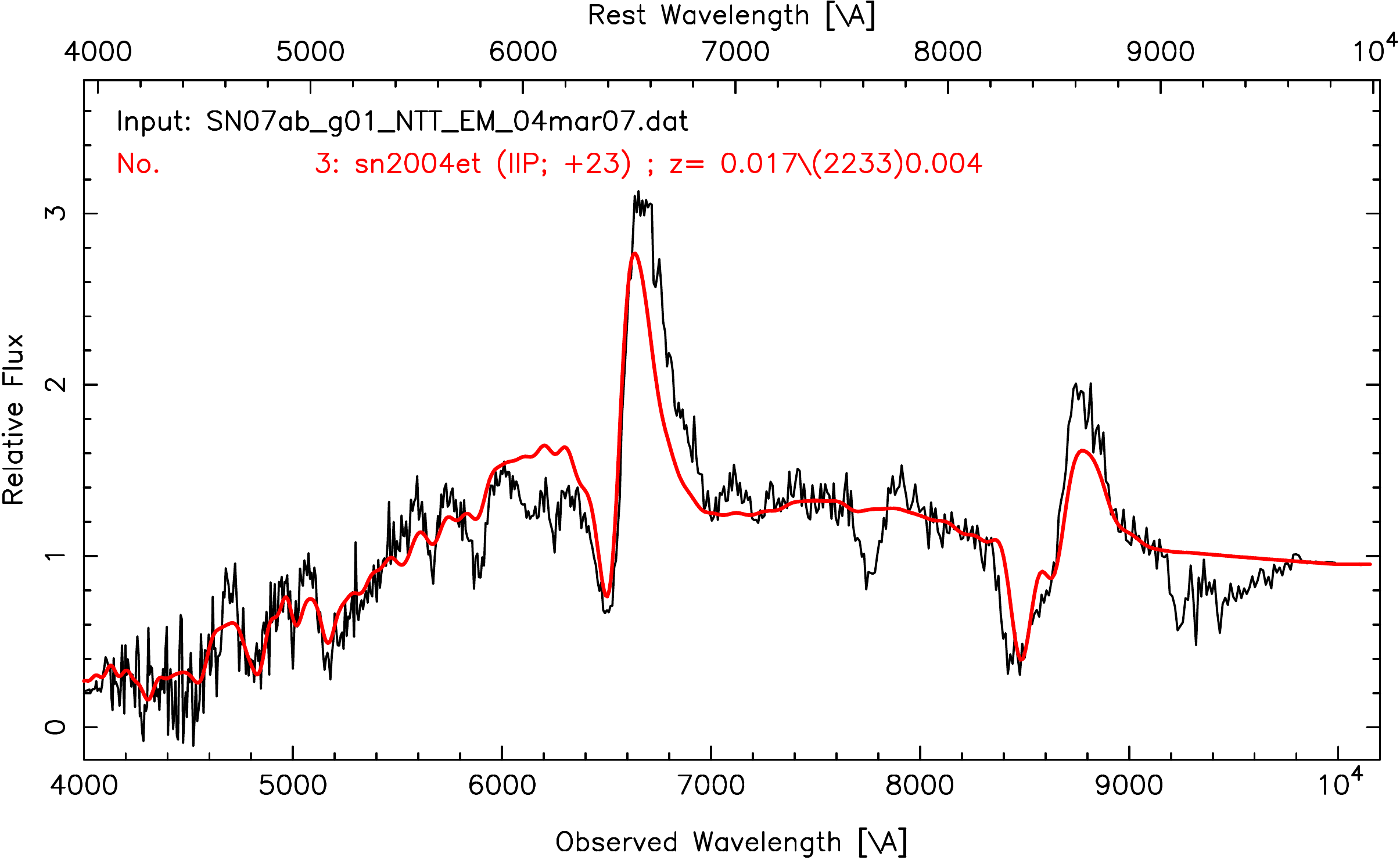}
\includegraphics[width=4.4cm]{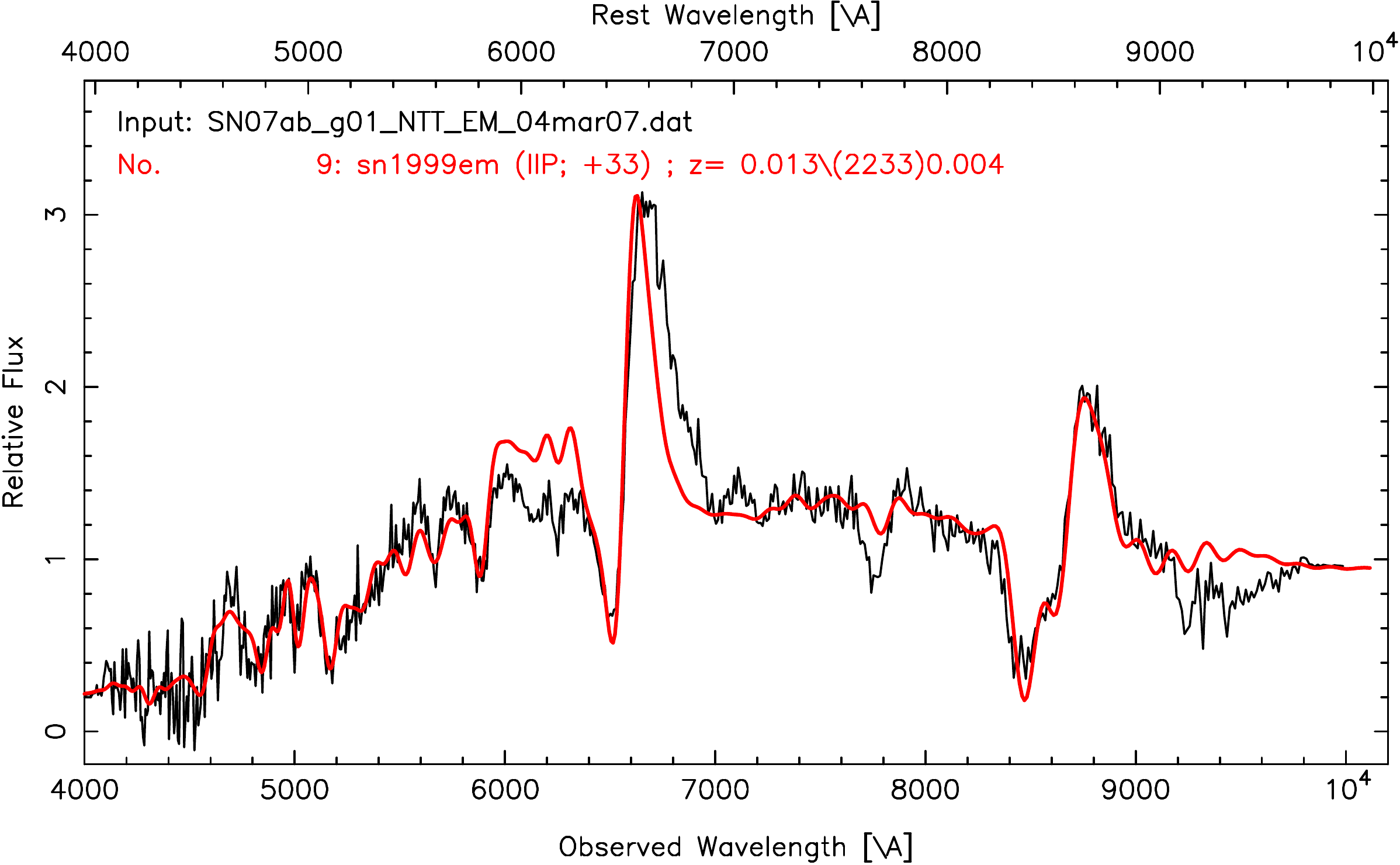}
\includegraphics[width=4.4cm]{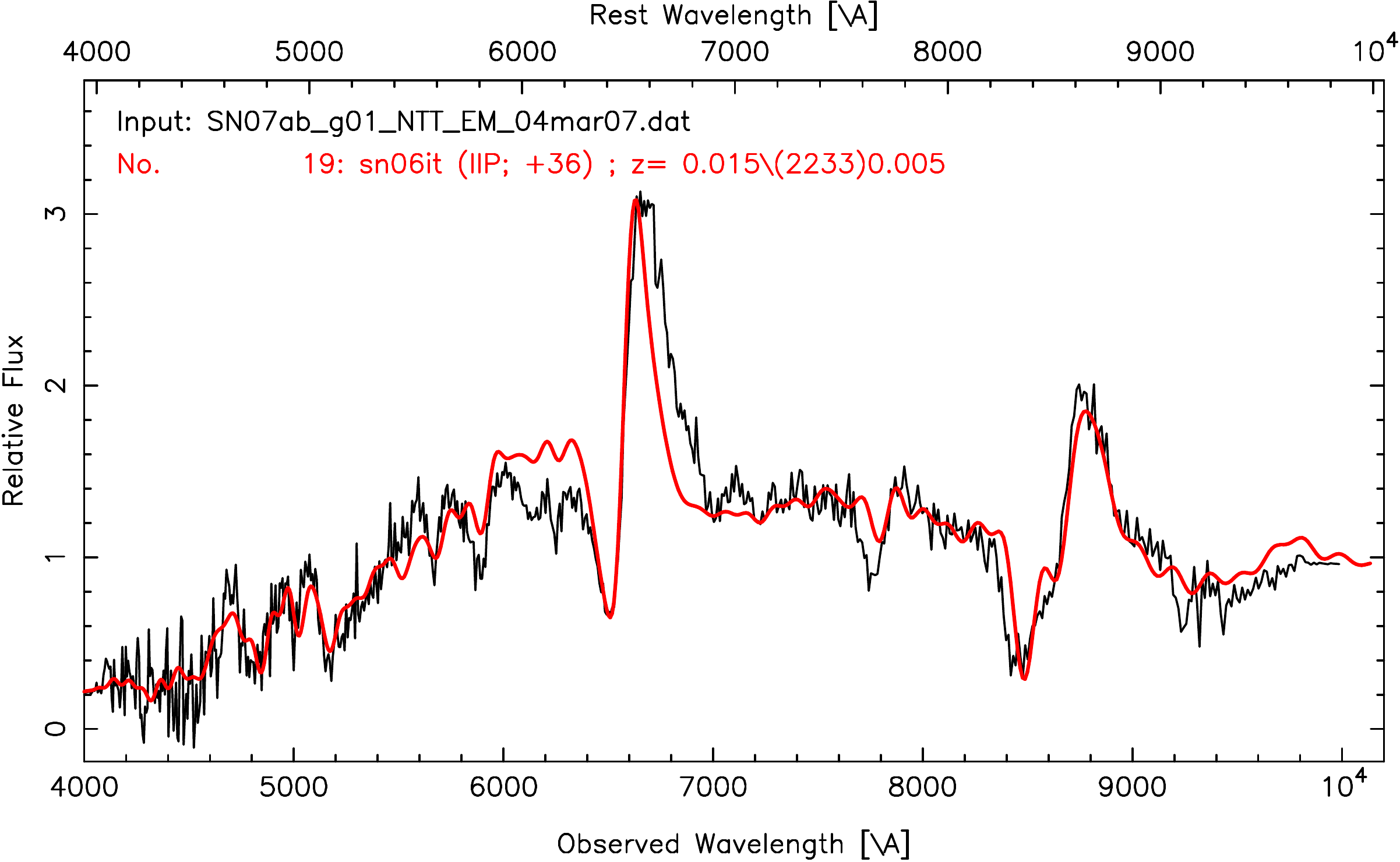}
\includegraphics[width=4.4cm]{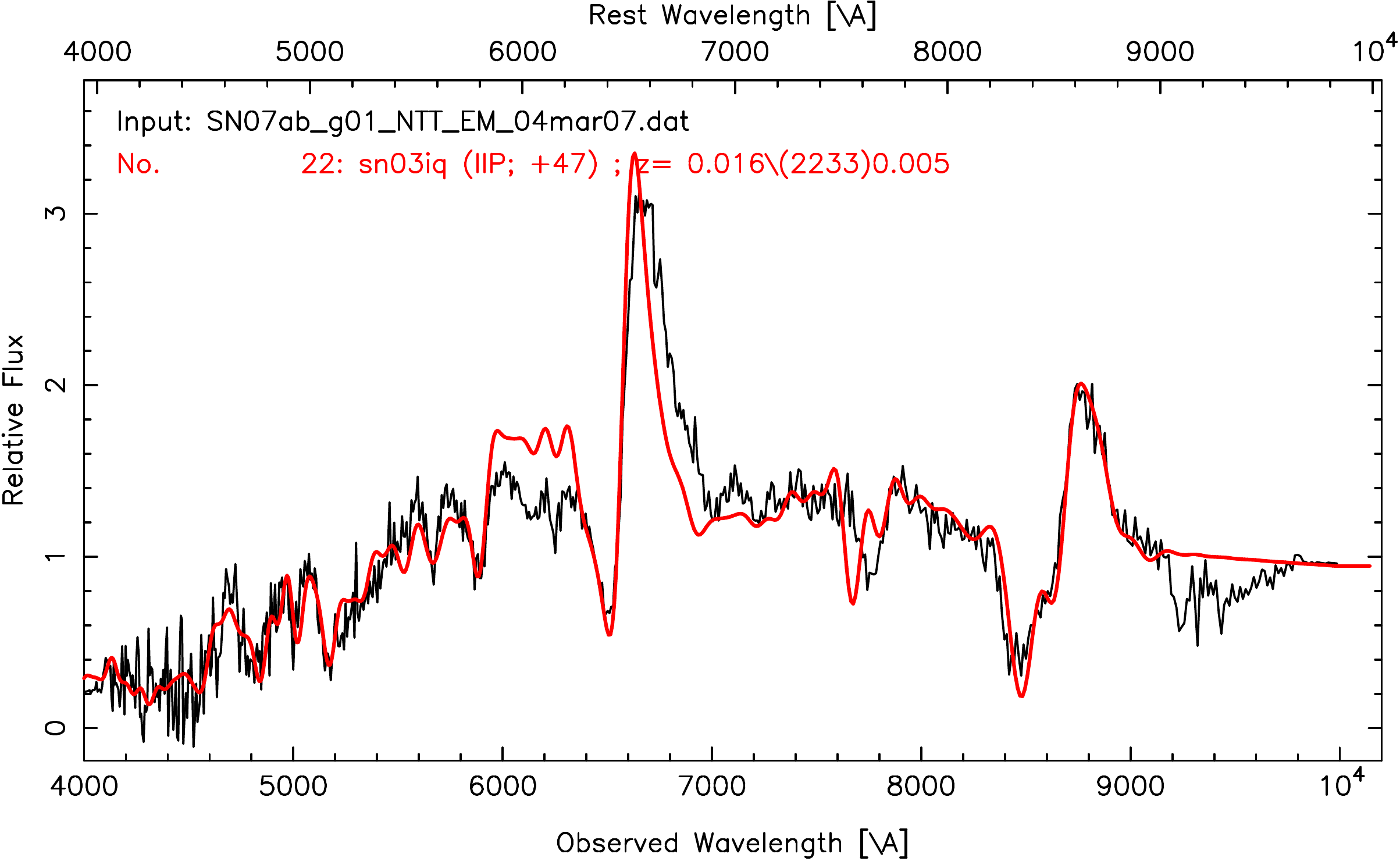}
\caption{Best spectral matching of SN~2007ab using SNID. The plots show SN~2007ab compared with 
SN~2006bp, SN~2004et, SN~1999em, SN~2006it and SN~2003iq at 43, 39, 43, 36, and 47 days from explosion.}
\end{figure}

\begin{figure}
\centering
\includegraphics[width=4.4cm]{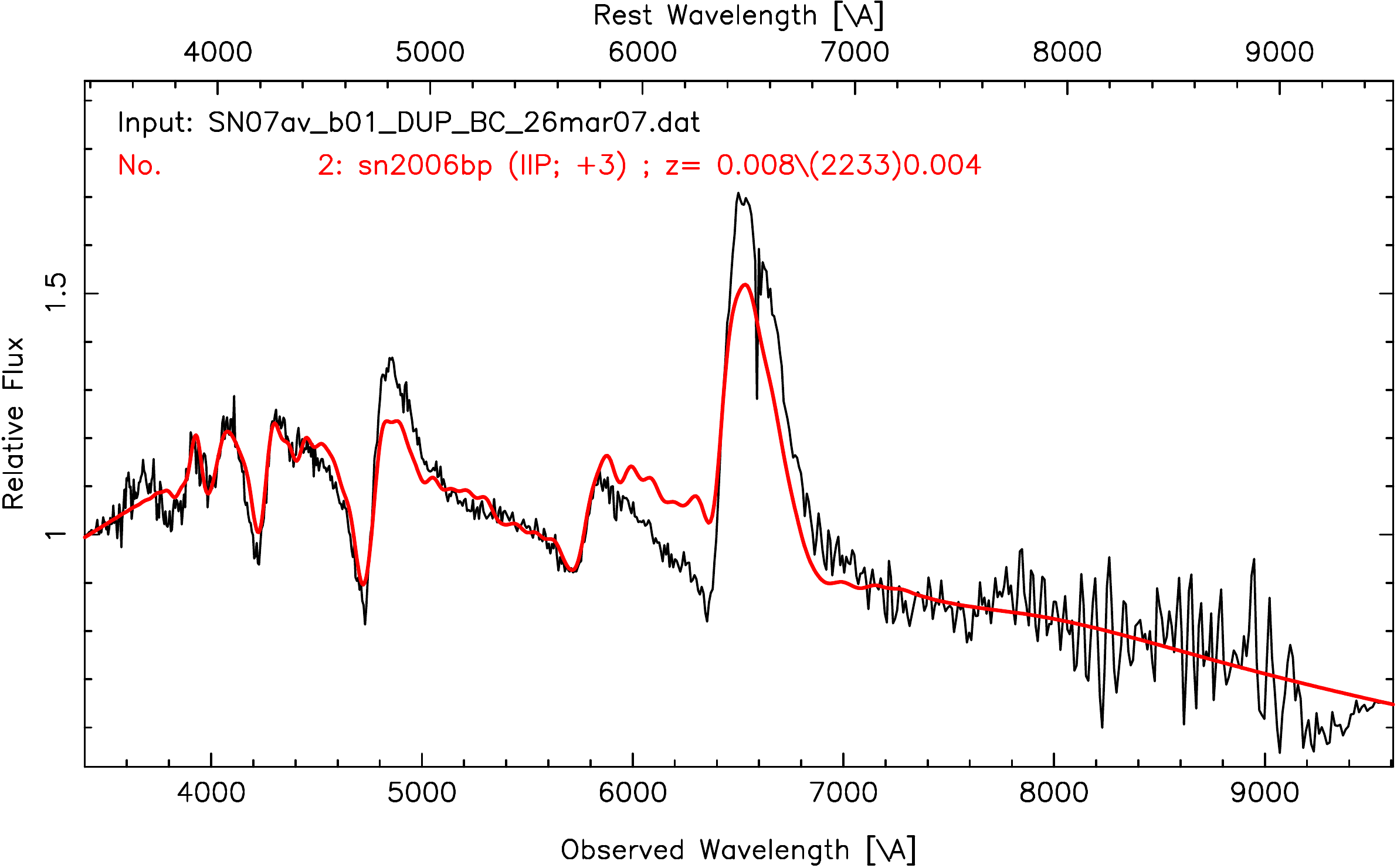}
\includegraphics[width=4.4cm]{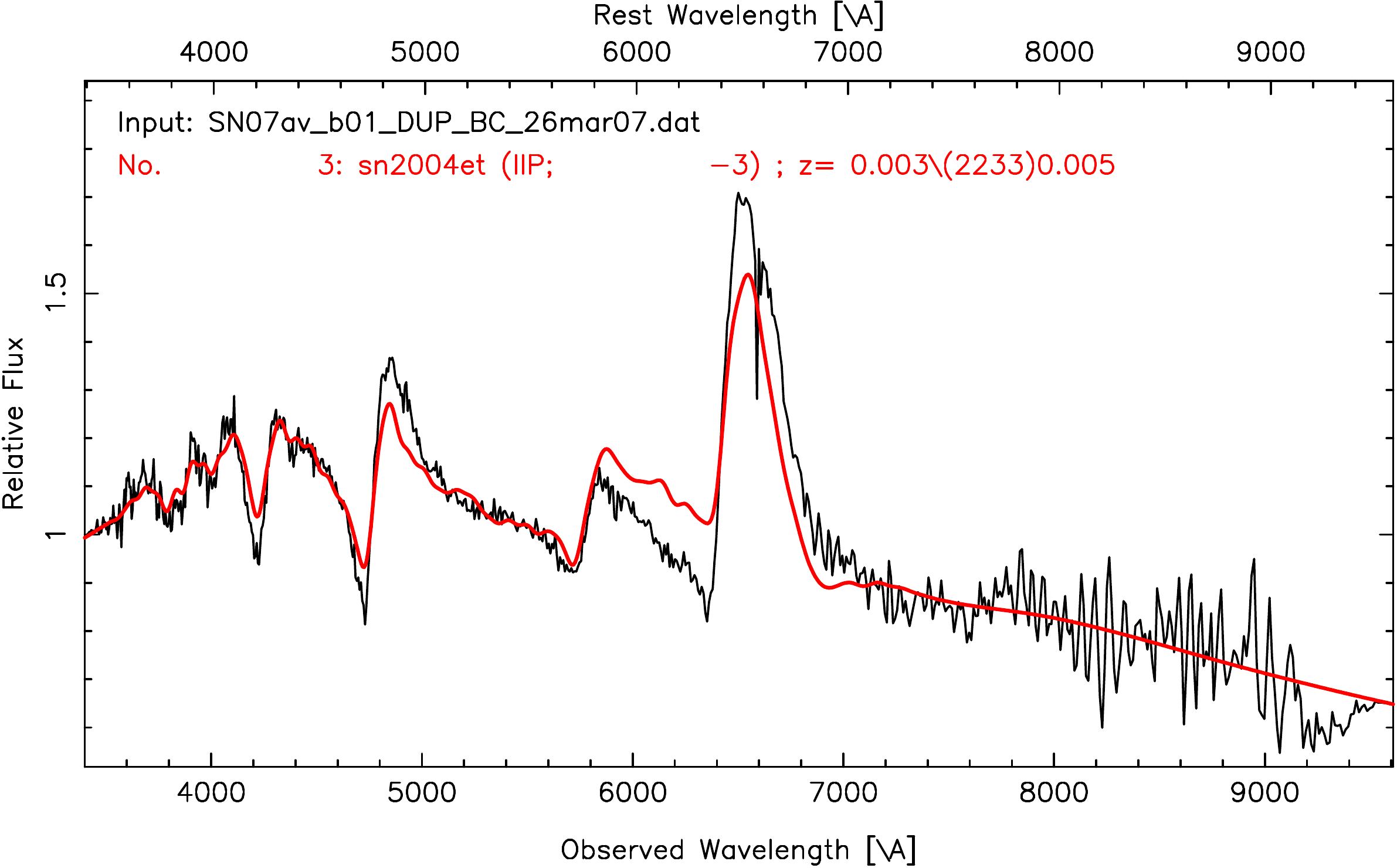}
\includegraphics[width=4.4cm]{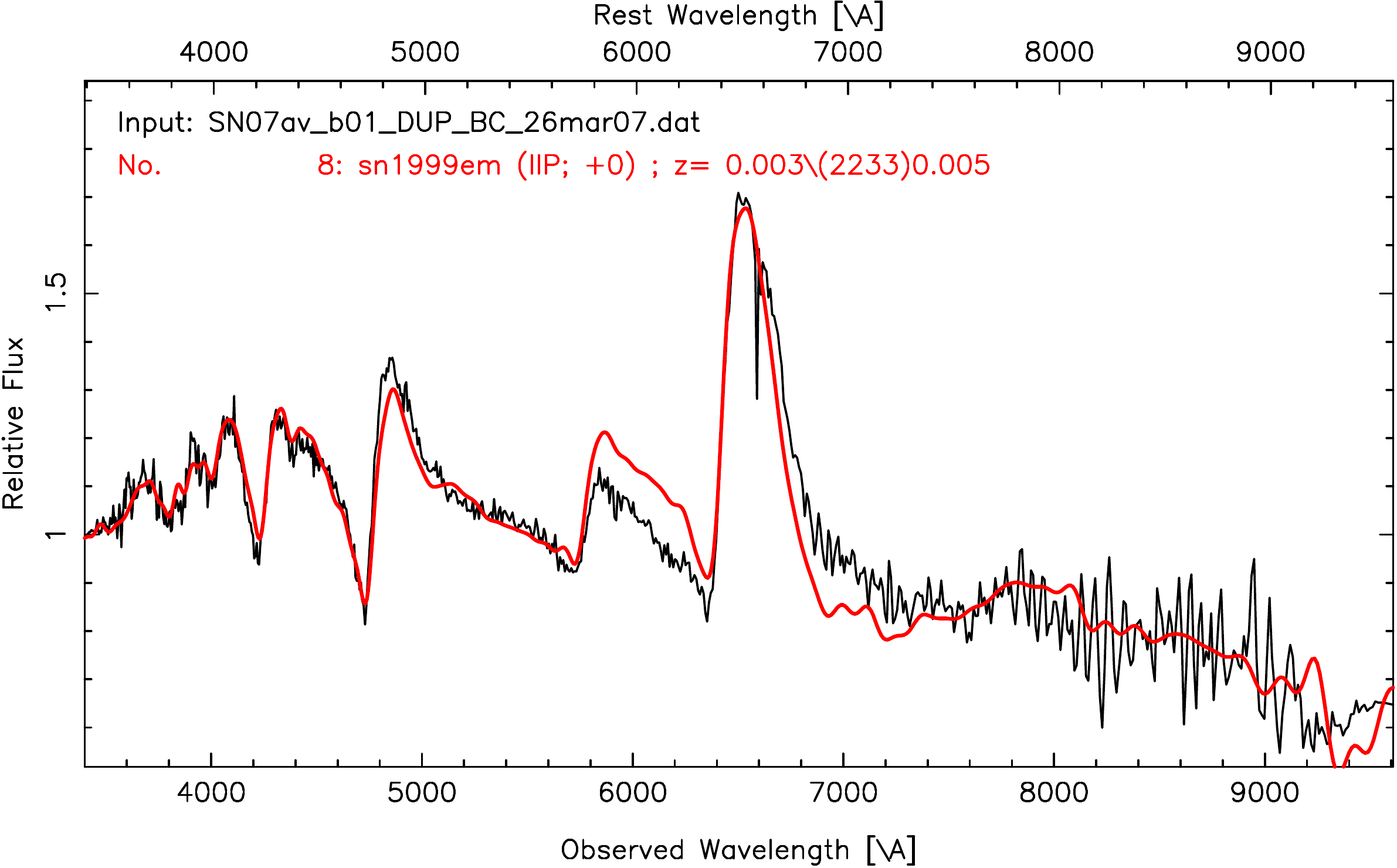}
\includegraphics[width=4.4cm]{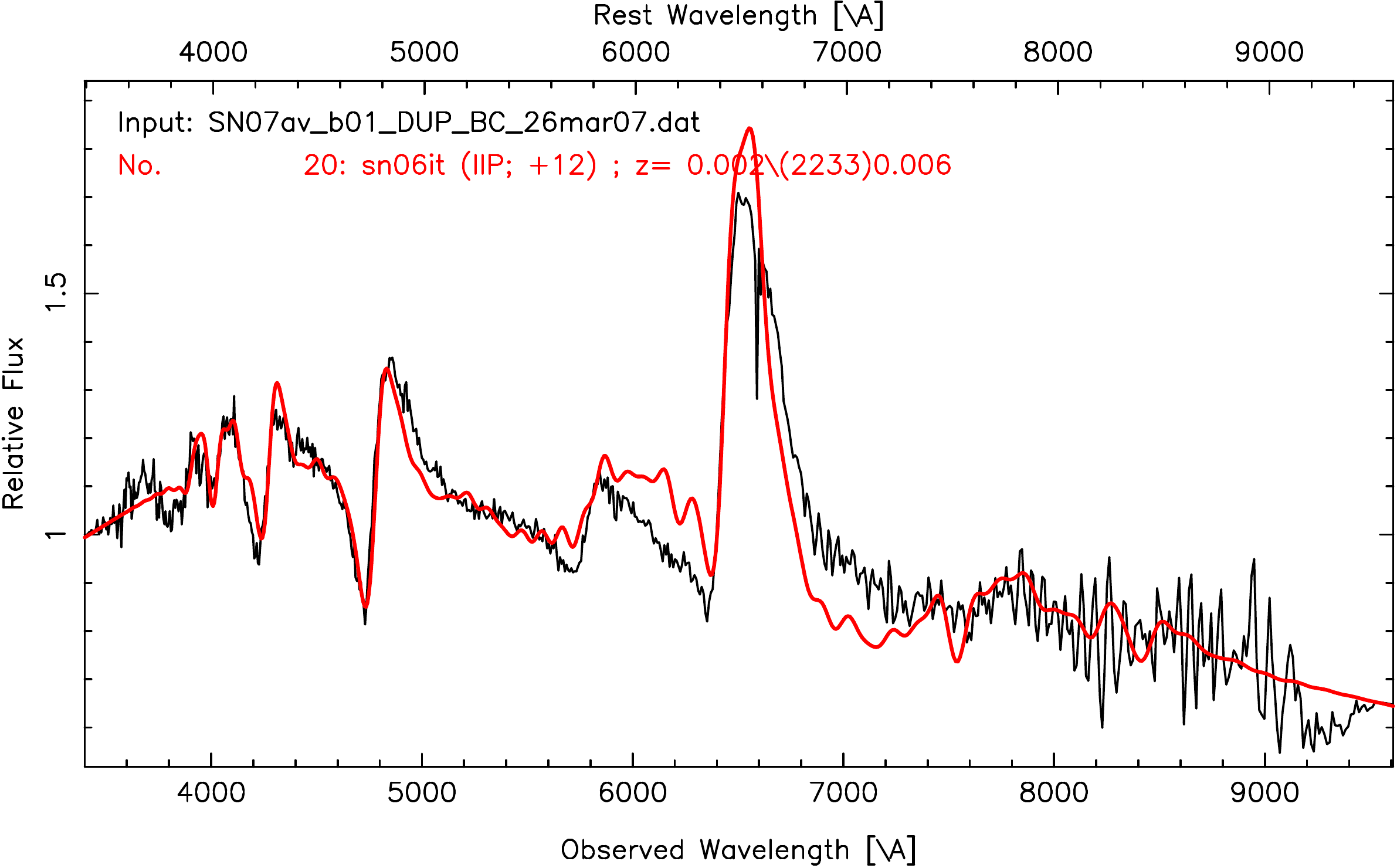}
\caption{Best spectral matching of SN~2007av using SNID. The plots show SN~2007av compared with 
SN~2006bp, SN~2004et, SN~1999em, and SN~2006it at 12, 13, 10, and 12 days from explosion.}
\end{figure}

\clearpage

\begin{figure}
\centering
\includegraphics[width=4.4cm]{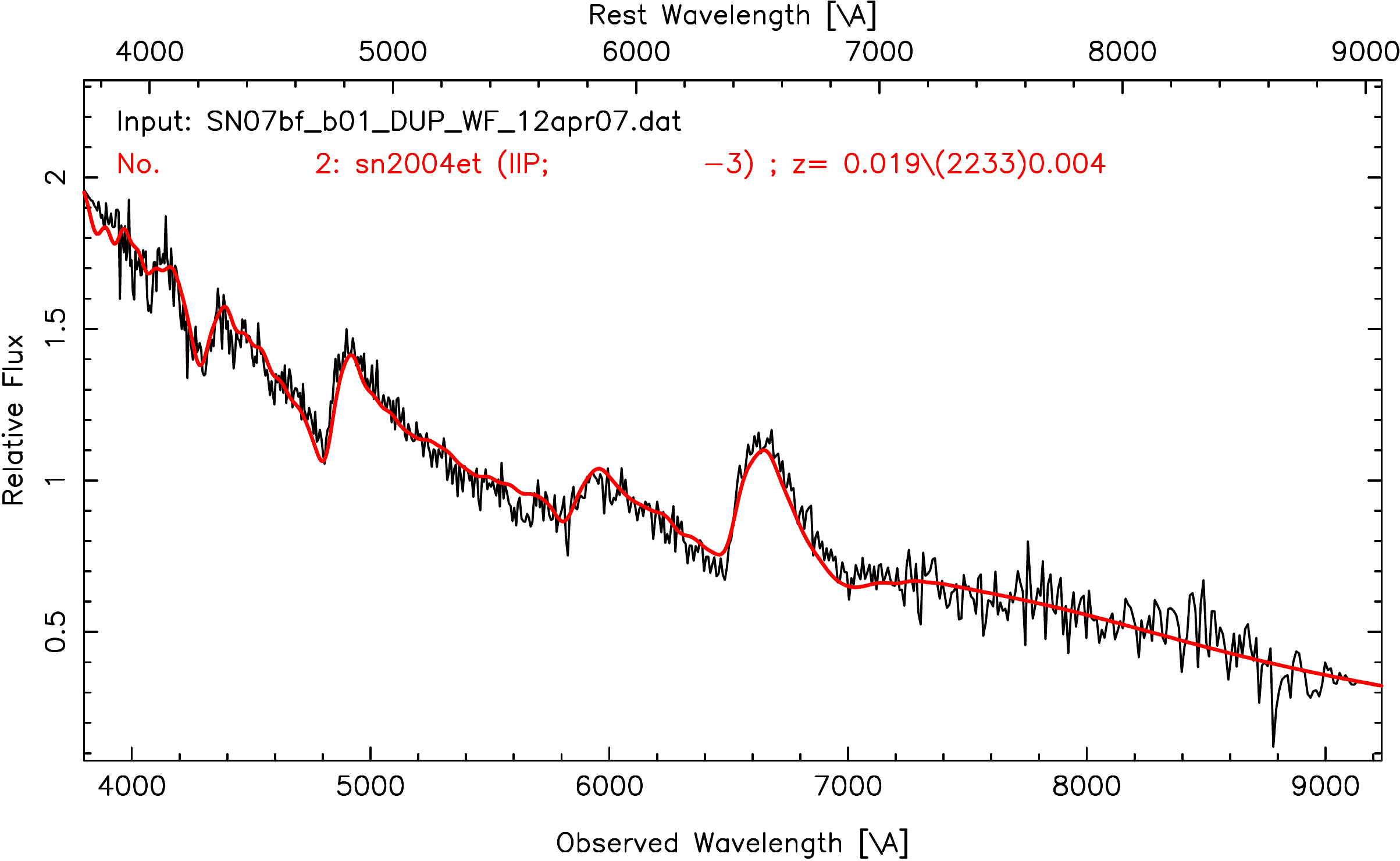}
\includegraphics[width=4.4cm]{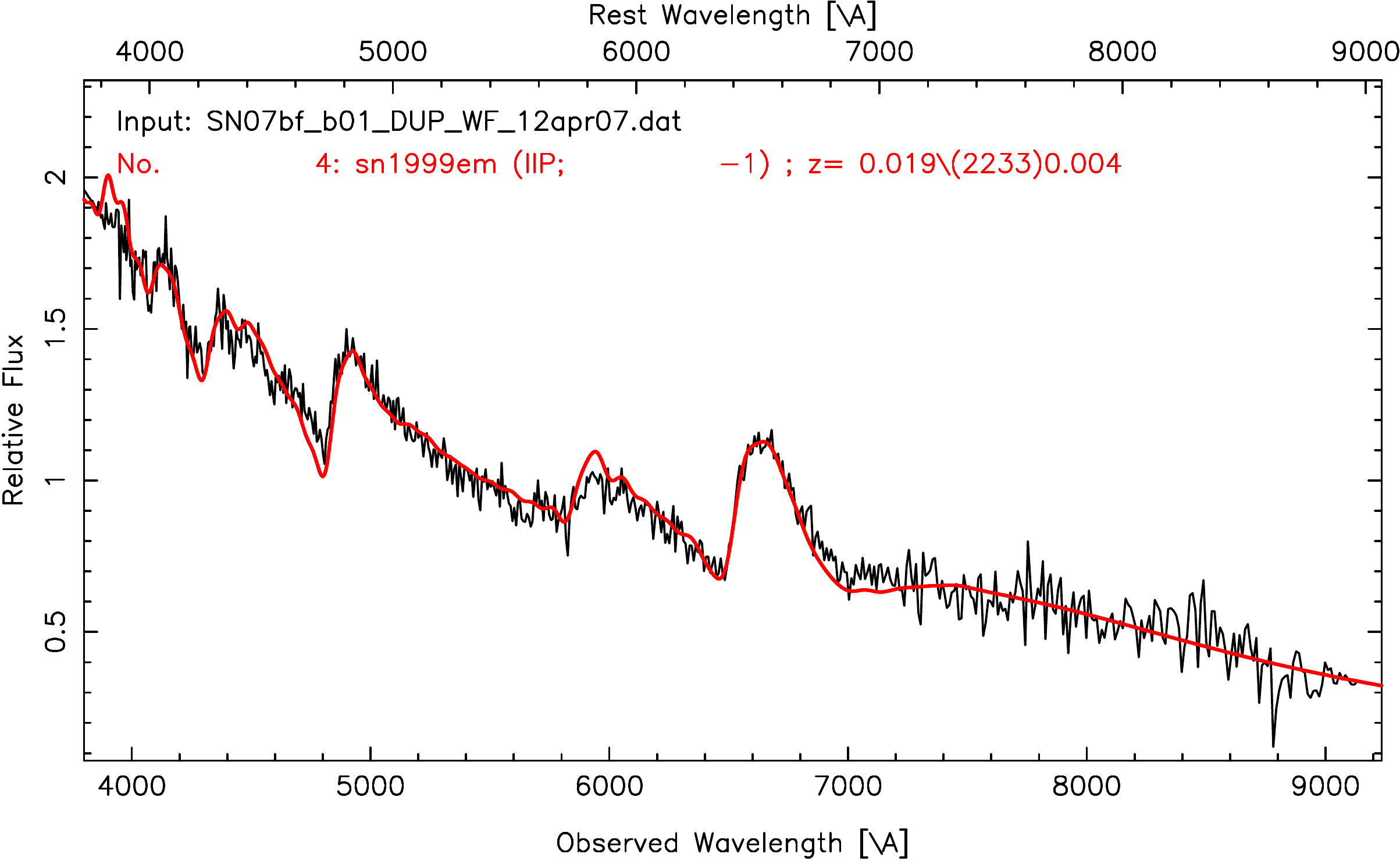}
\includegraphics[width=4.4cm]{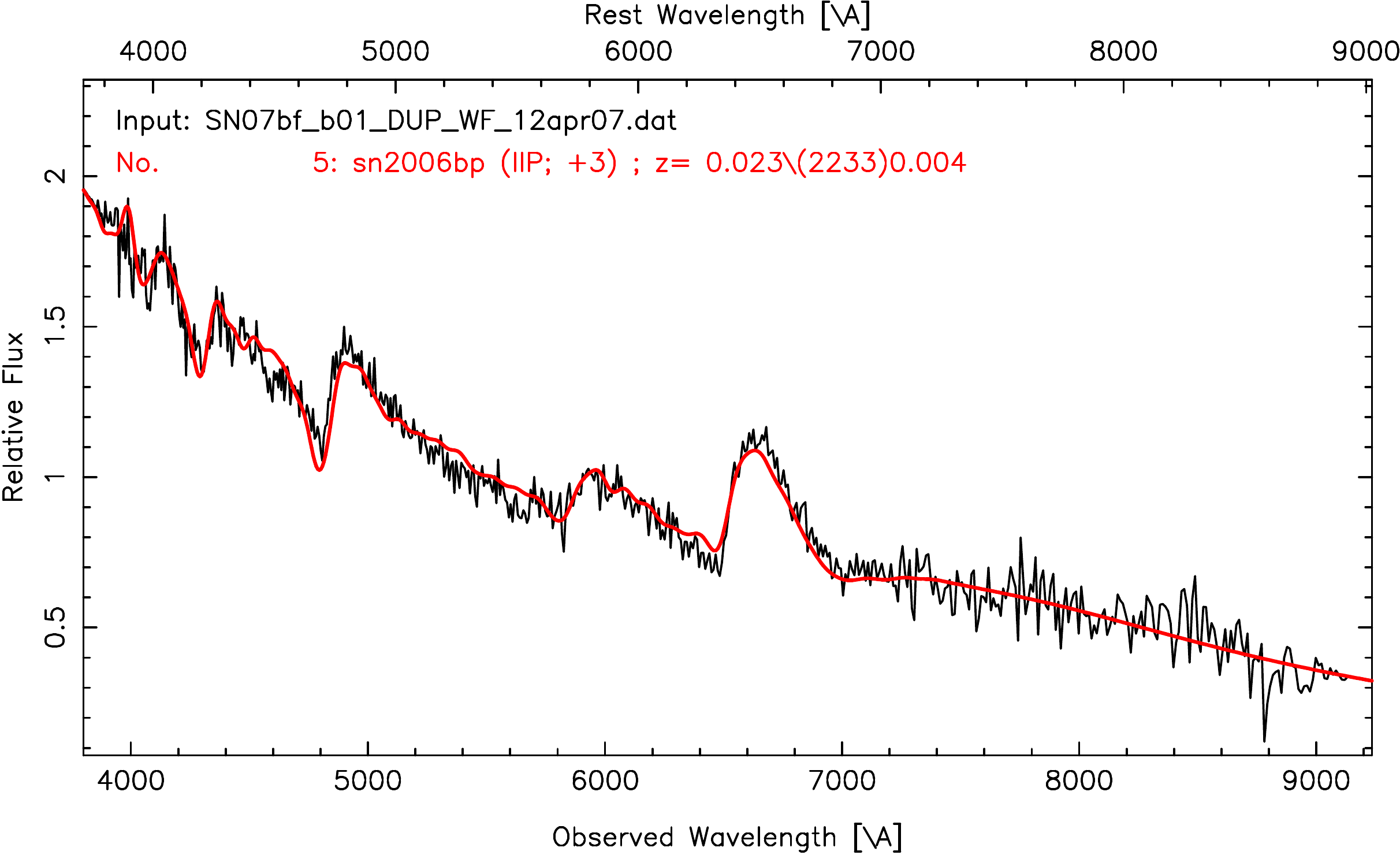}
\caption{Best spectral matching of SN~2007bf using SNID. The plots show SN~2007bf compared with 
SN~2004et, SN~1999em, and SN~2006bp at 13, 9, and 12 days from explosion.}
\end{figure}

\begin{figure}
\centering
\includegraphics[width=4.4cm]{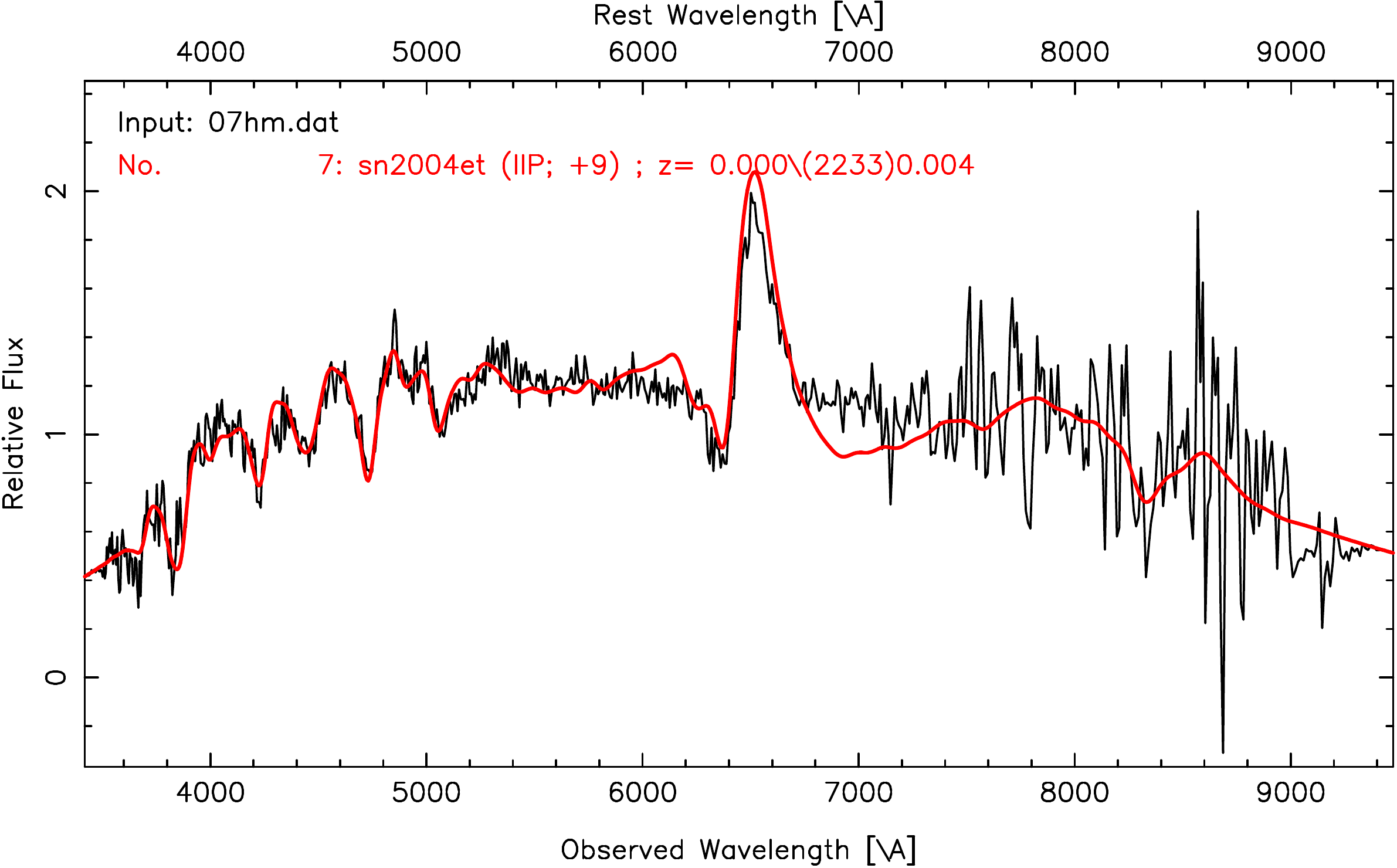}
\includegraphics[width=4.4cm]{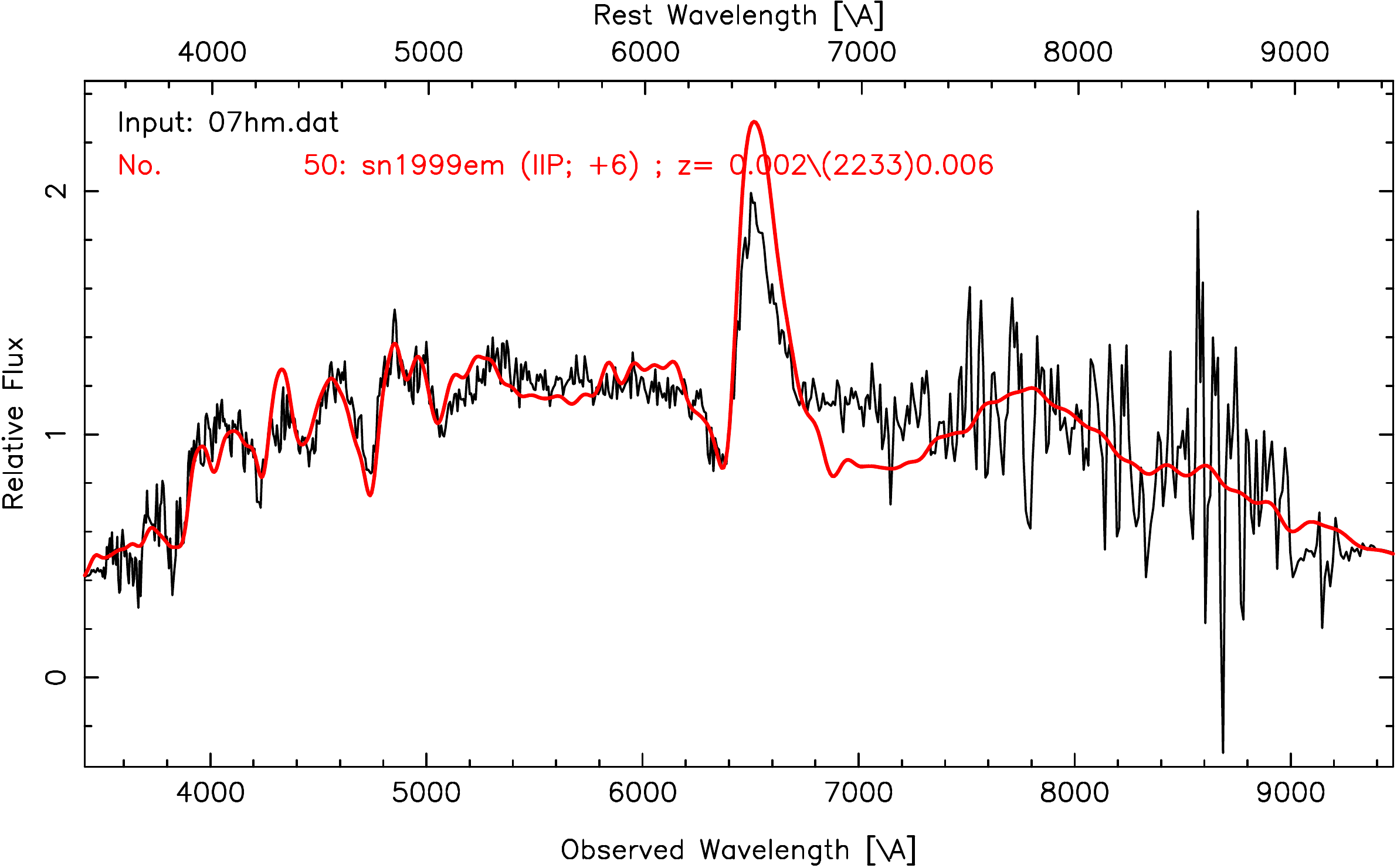}
\includegraphics[width=4.4cm]{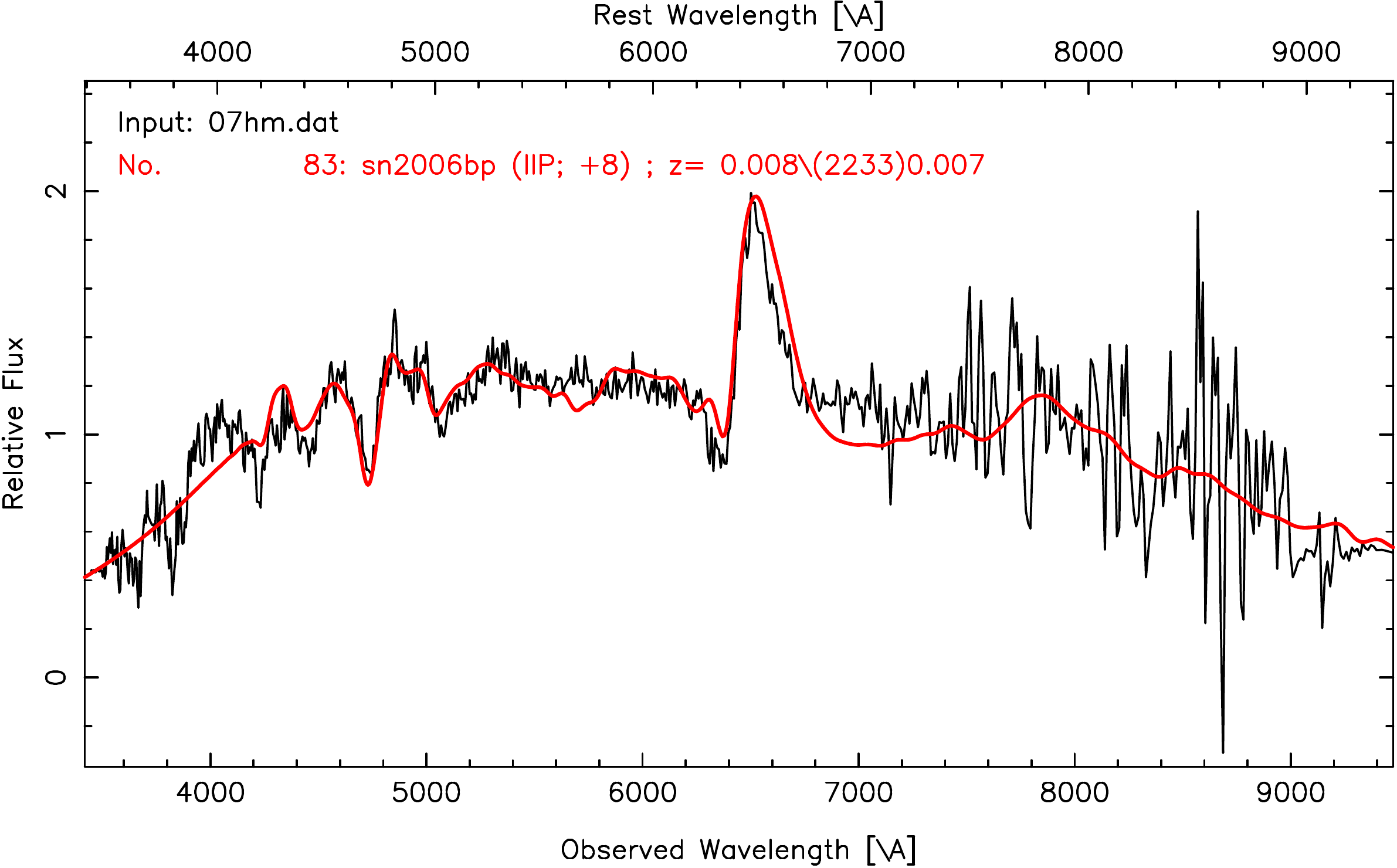}
\caption{Best spectral matching of SN~2007hm using SNID. The plots show SN~2007hm compared with 
SN~2004et, SN~1999em, and SN~2006bp at 25, 16, and 17 days from explosion.}
\end{figure}

\begin{figure}
\centering
\includegraphics[width=4.4cm]{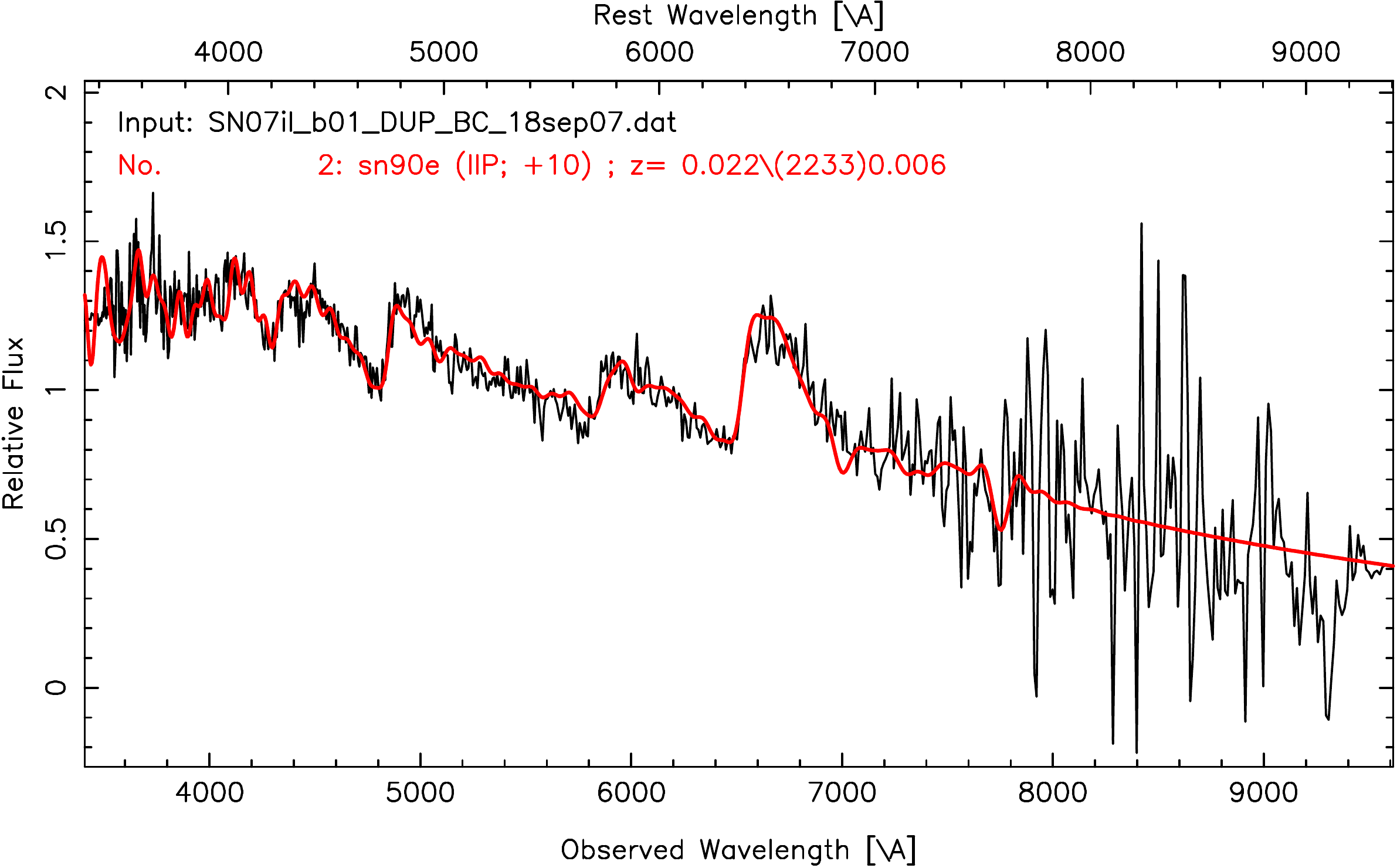}
\includegraphics[width=4.4cm]{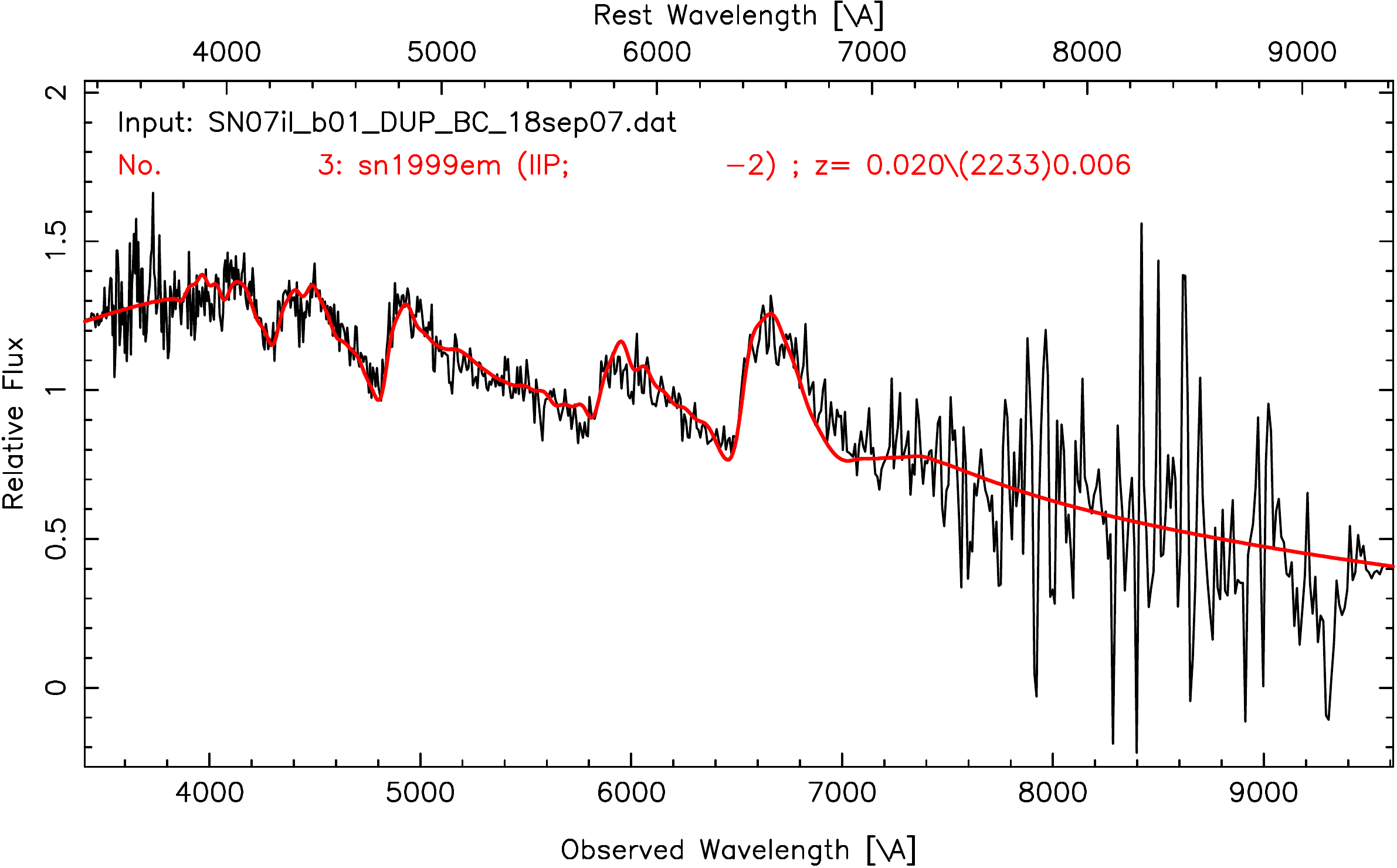}
\includegraphics[width=4.4cm]{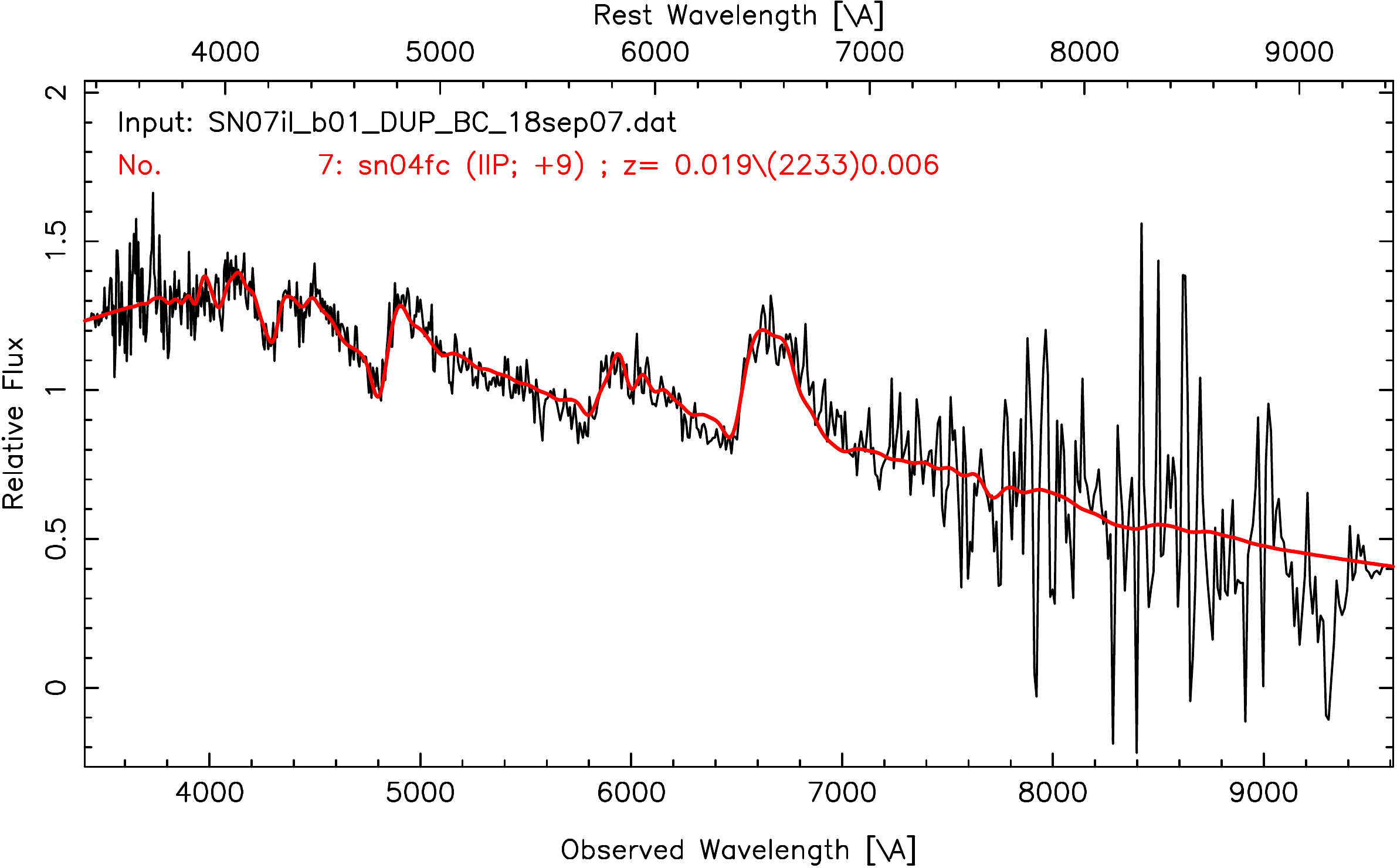}
\includegraphics[width=4.4cm]{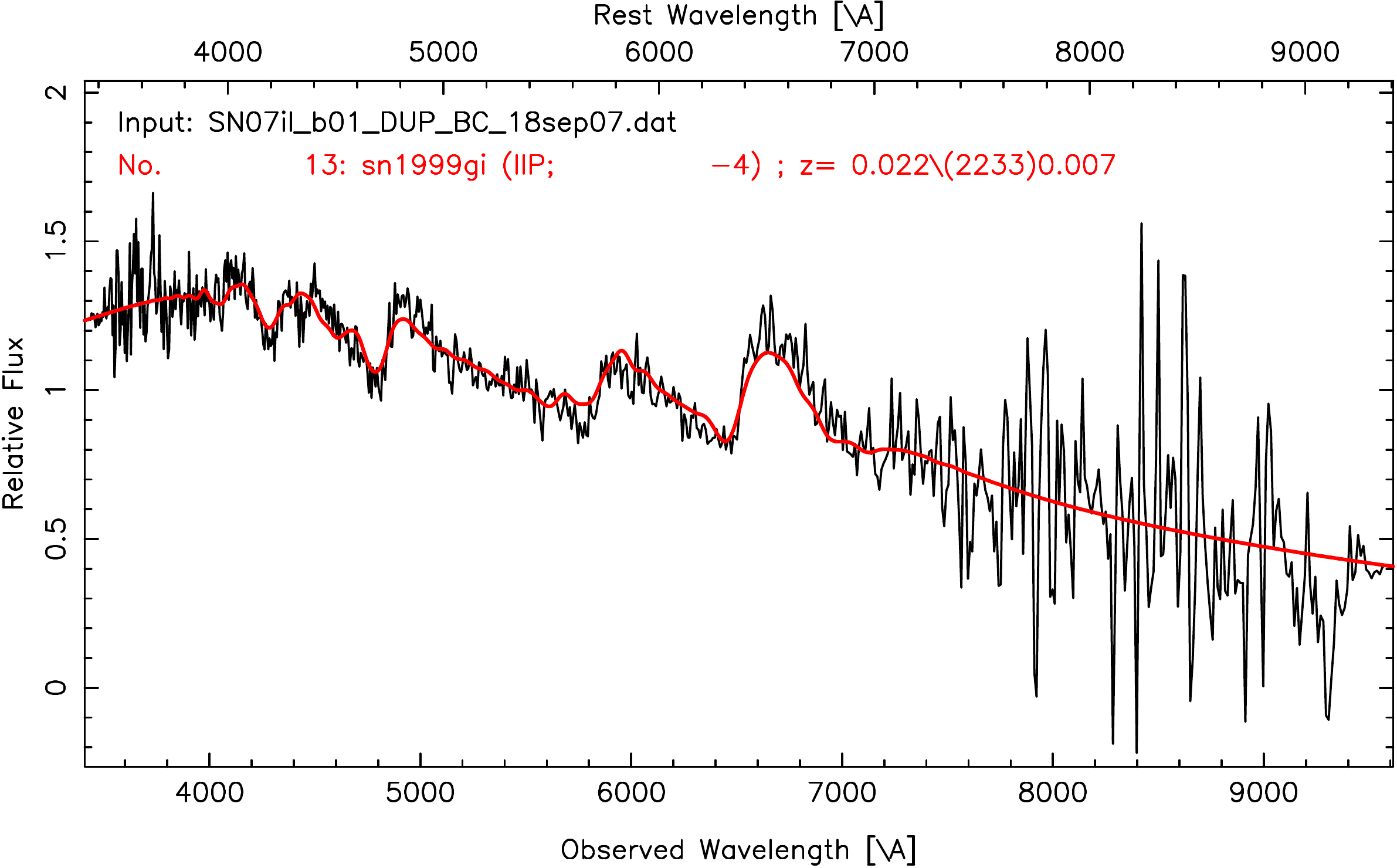}
\includegraphics[width=4.4cm]{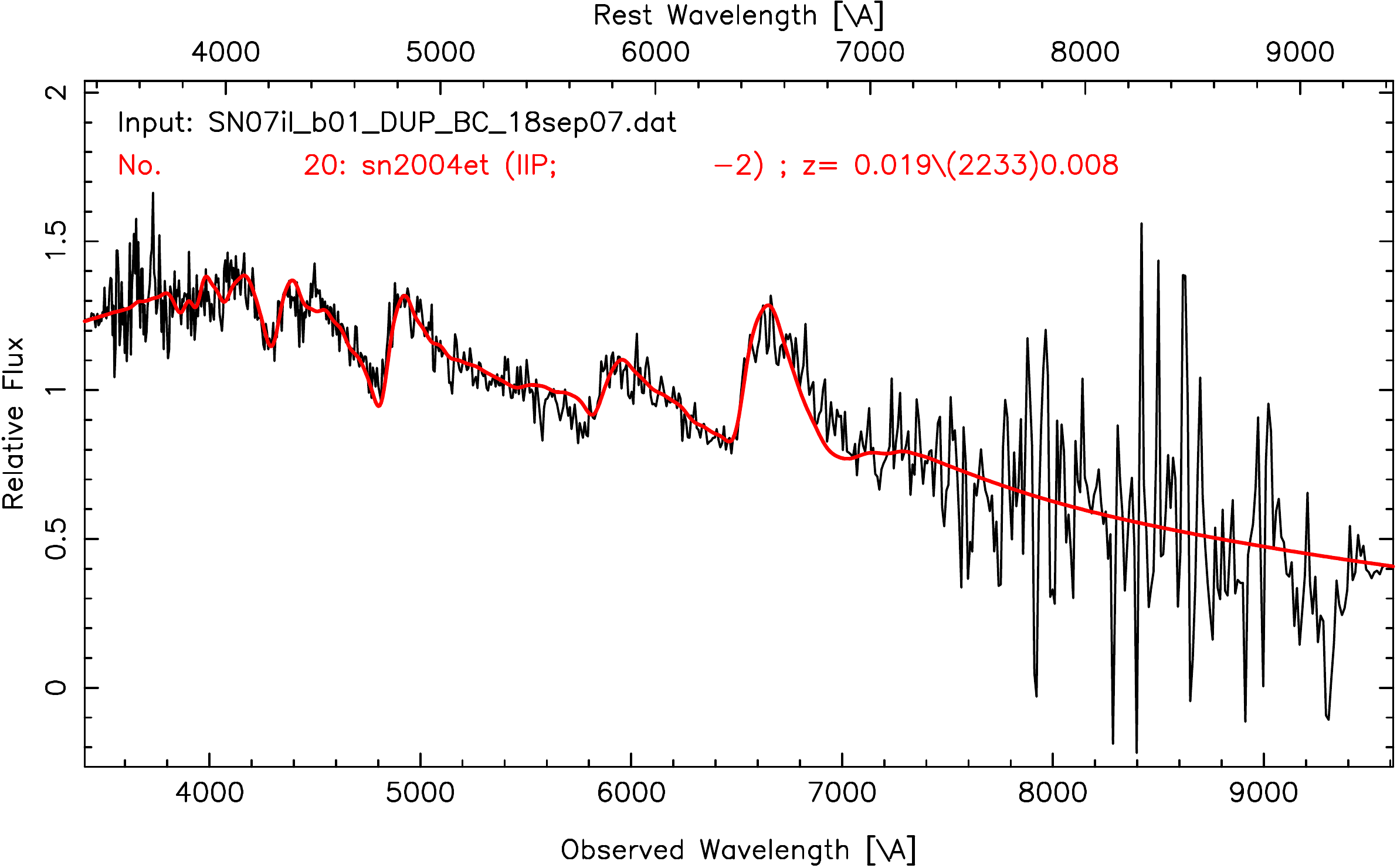}
\includegraphics[width=4.4cm]{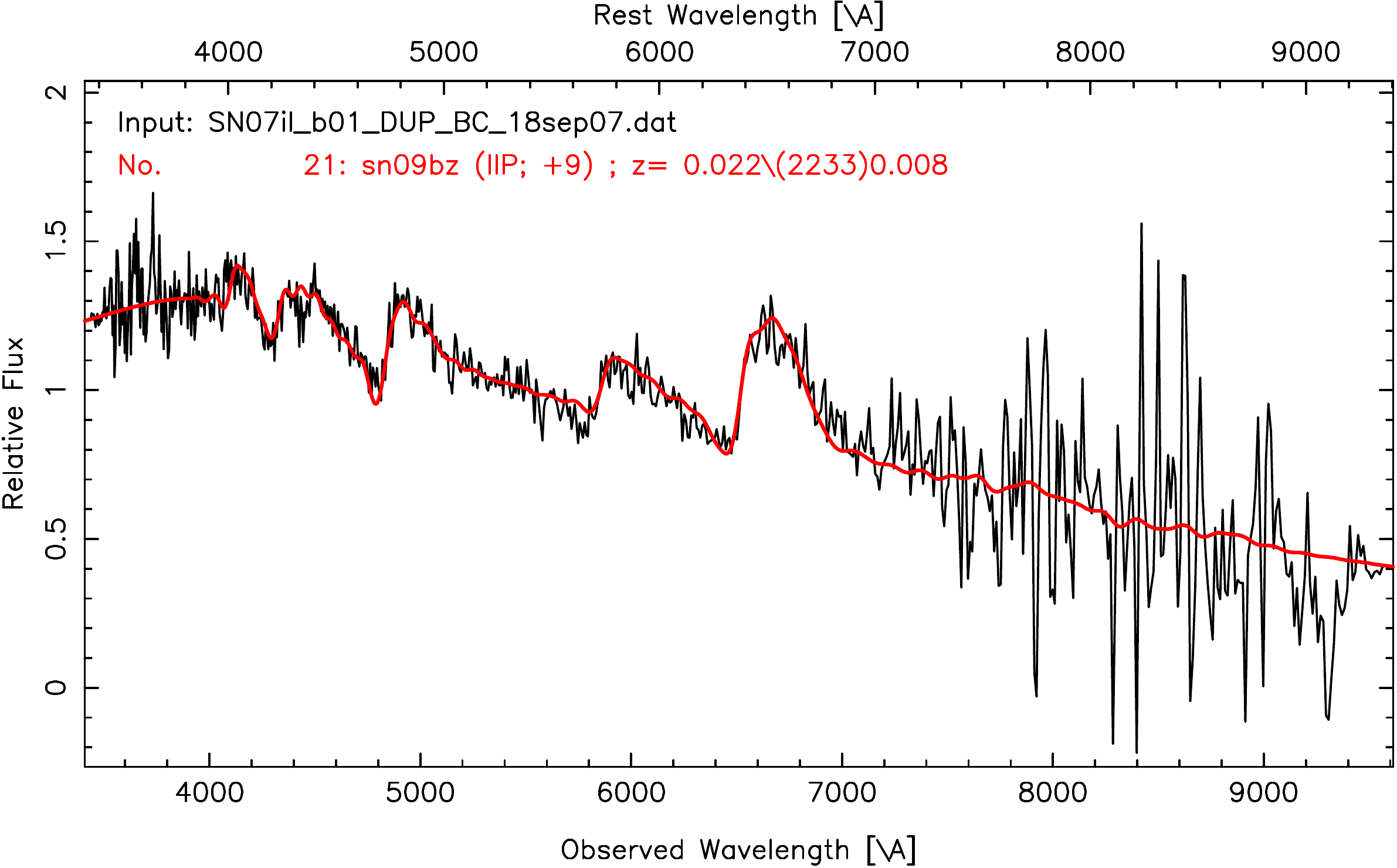}
\caption{Best spectral matching of SN~2007il using SNID. The plots show SN~2007il compared with 
SN~1990E, SN~1999em, SN~2004fc, SN~1999gi, SN~2004et, and SN~2009bz at 10, 8, 9, 8, 14, and 9 days from explosion.}
\end{figure}

\clearpage

\begin{figure}
\centering
\includegraphics[width=4.4cm]{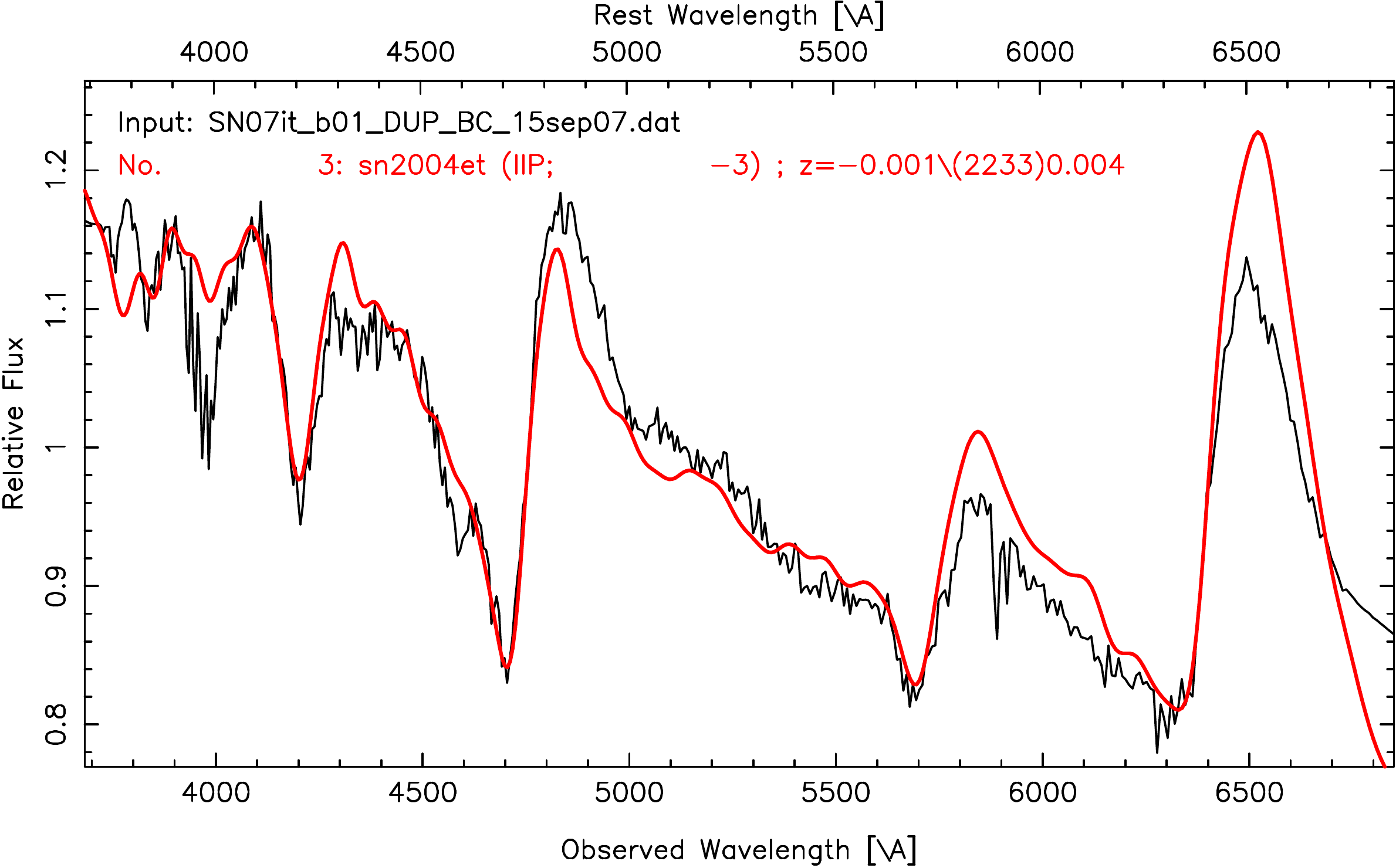}
\includegraphics[width=4.4cm]{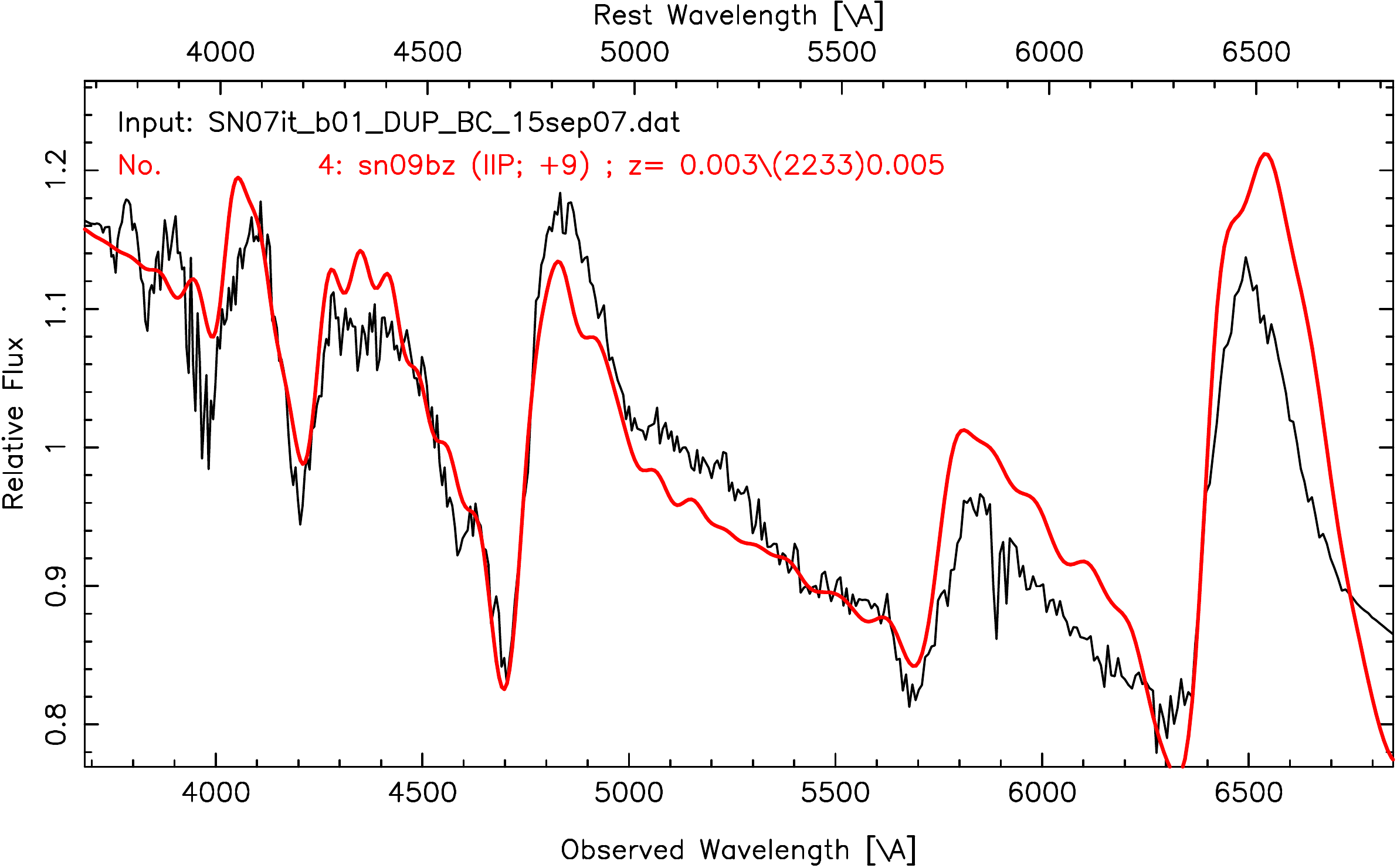}
\includegraphics[width=4.4cm]{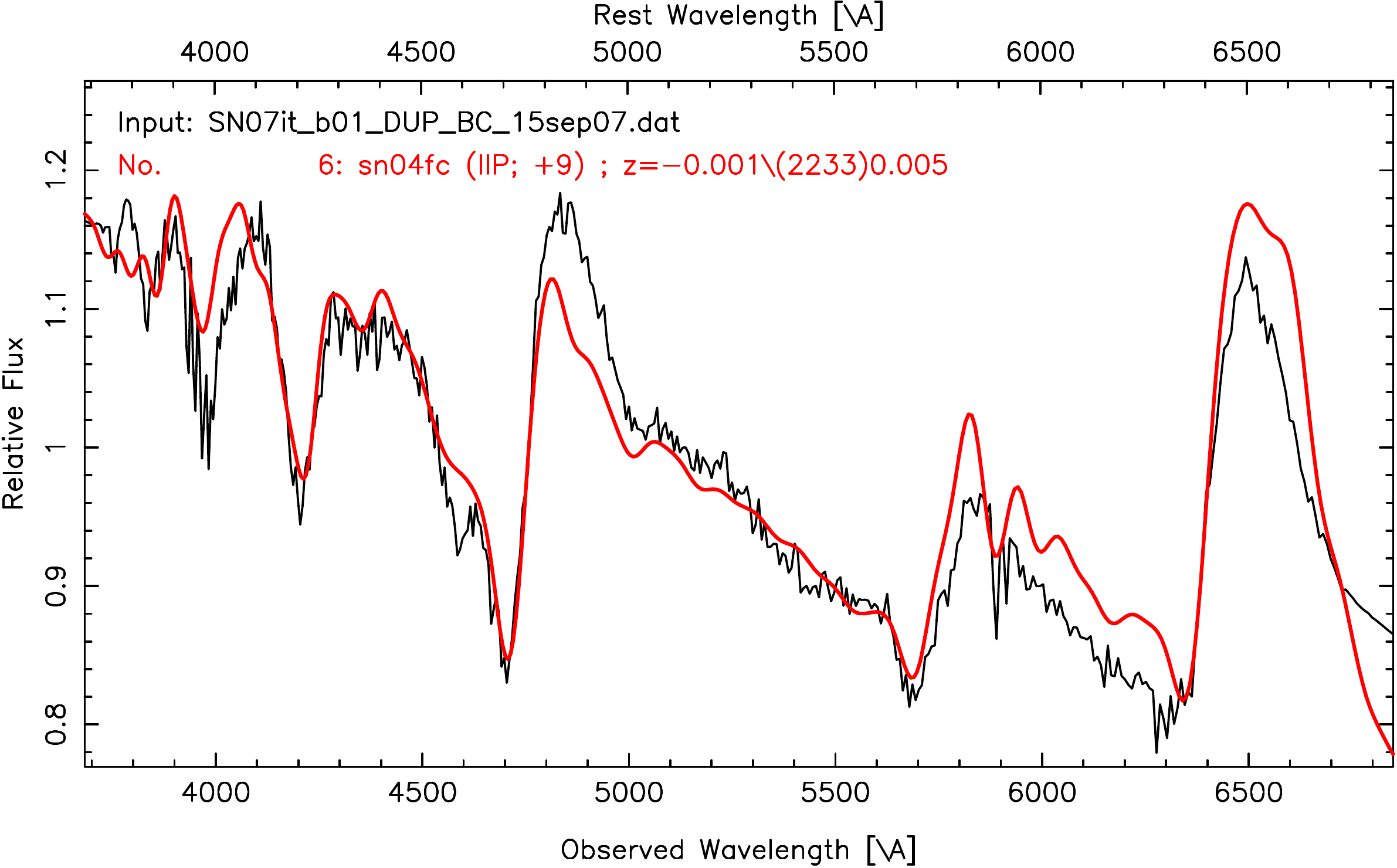}
\includegraphics[width=4.4cm]{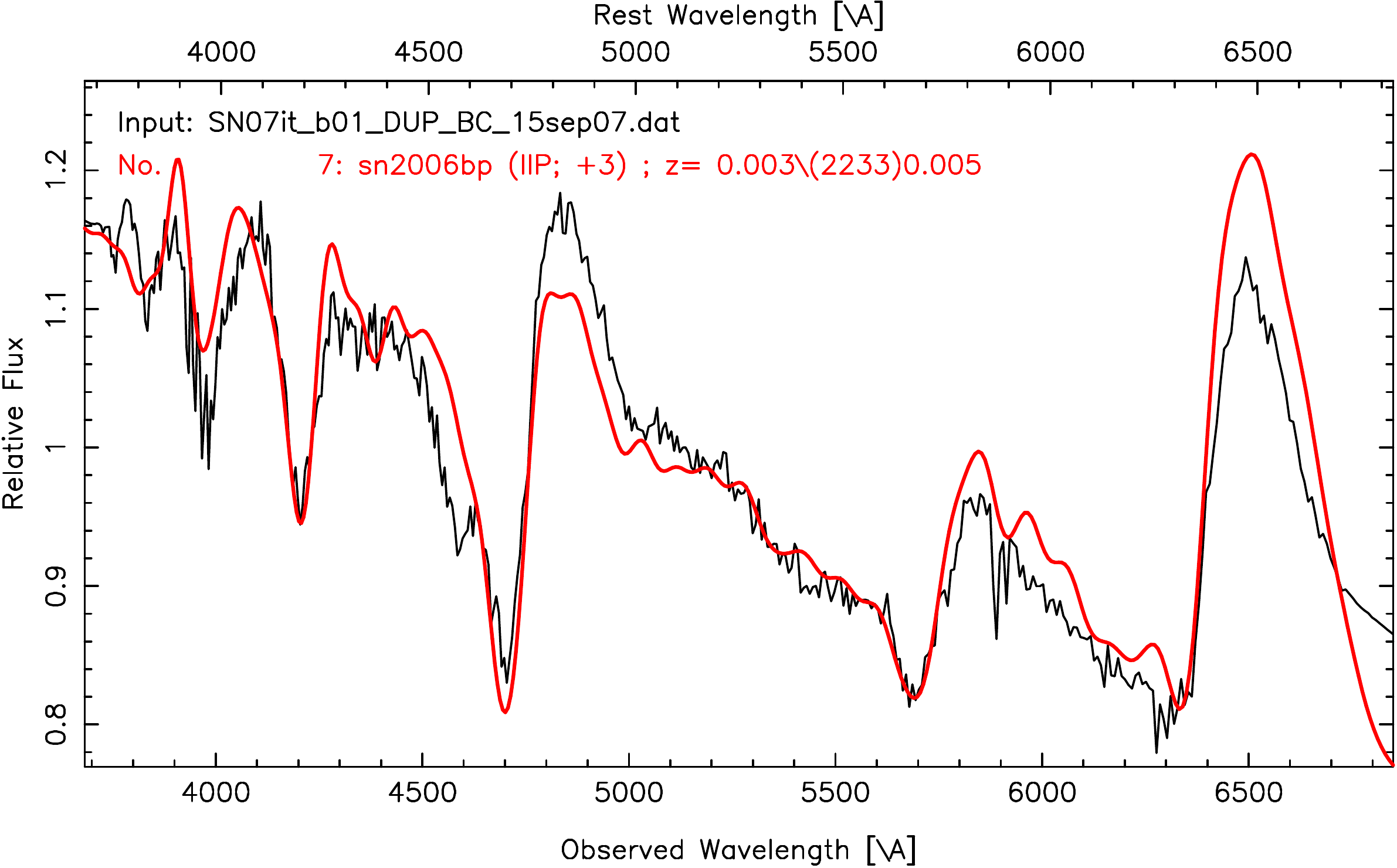}
\includegraphics[width=4.4cm]{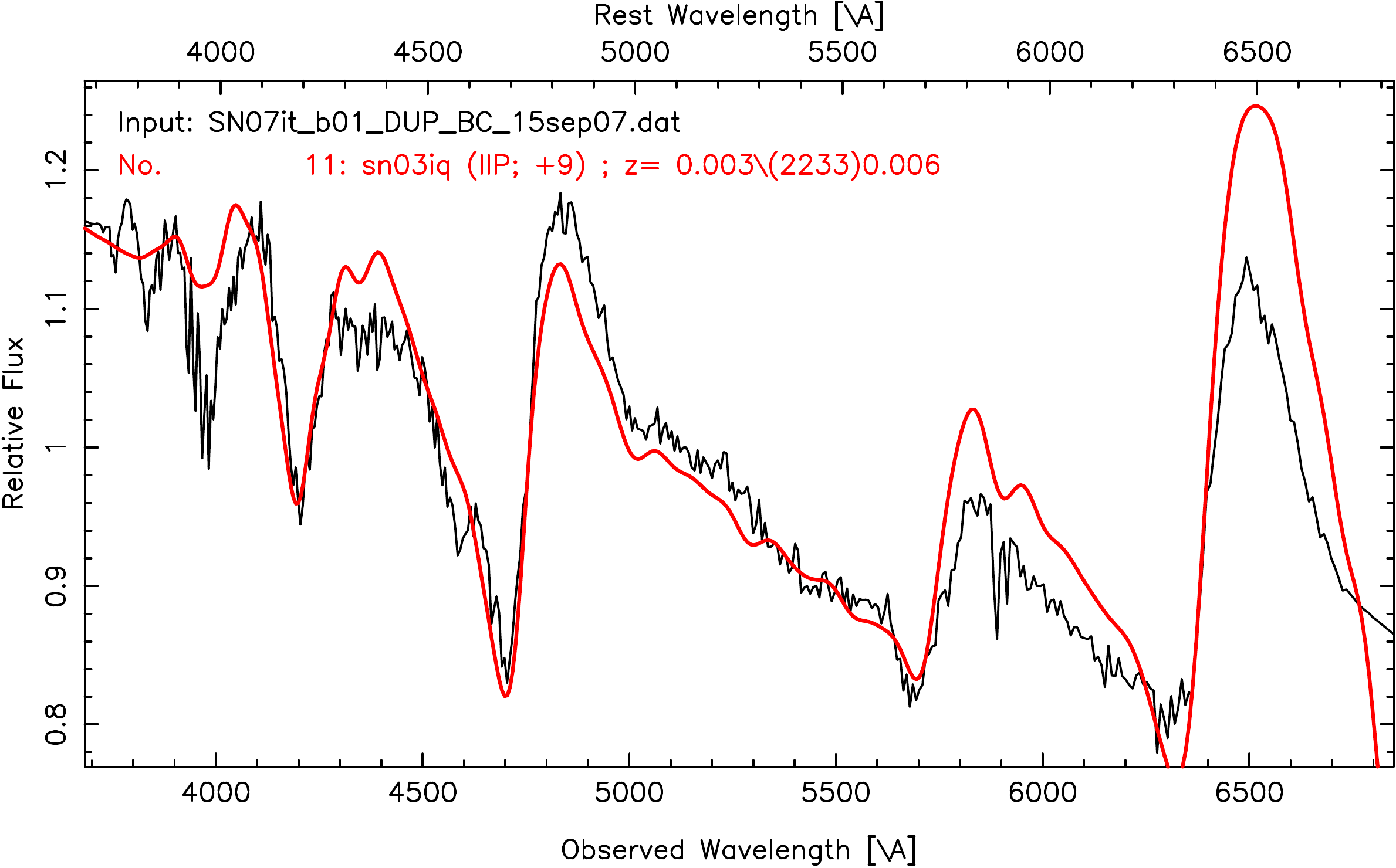}
\caption{Best spectral matching of SN~2007it using SNID. The plots show SN~2007it compared with 
SN~2004et, SN~2009bz, SN~2004fc, SN~2006bp, and SN~2003iq at 13, 9, 9, 12, and 9 days from explosion.}
\end{figure}

\begin{figure}
\centering
\includegraphics[width=4.4cm]{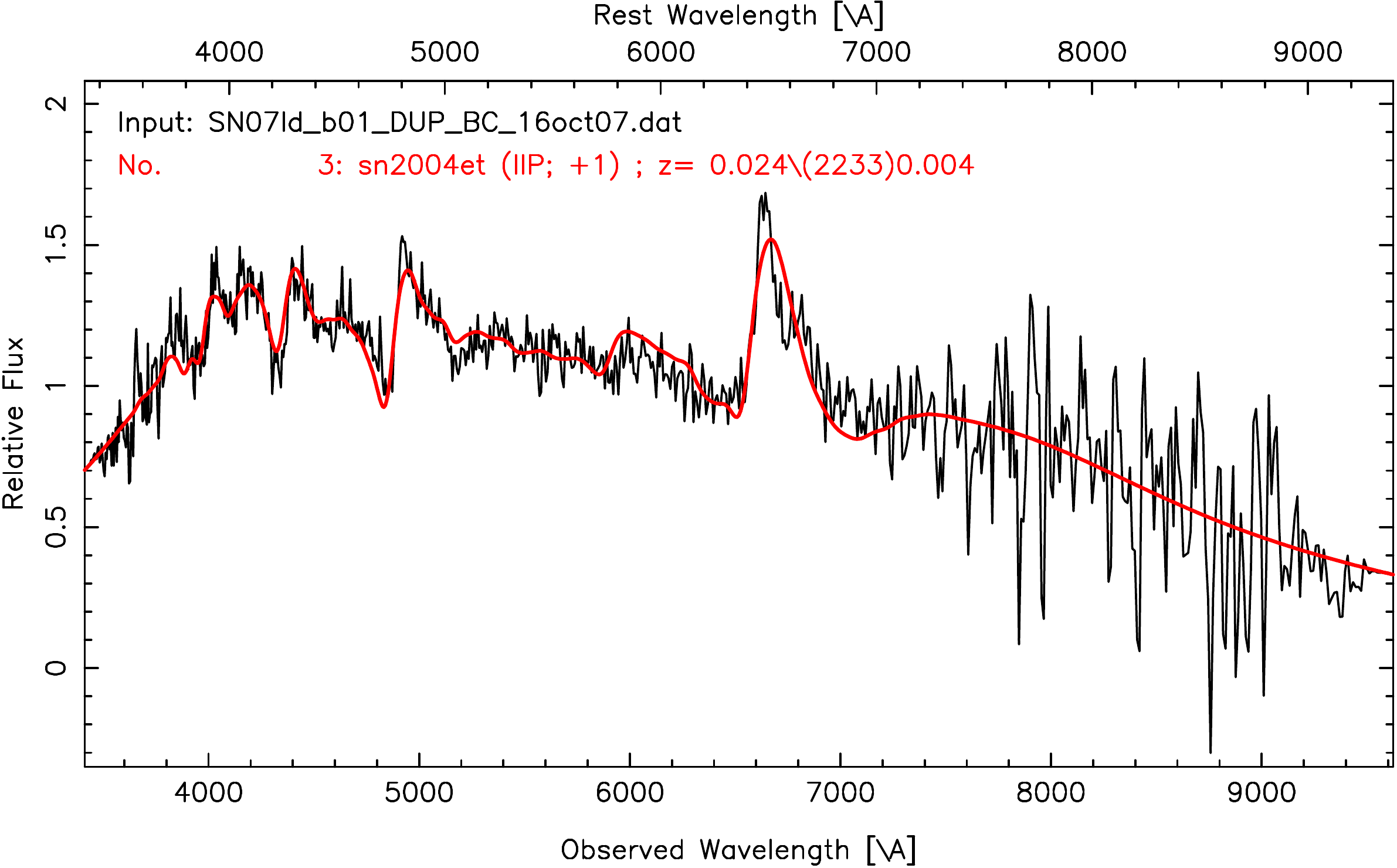}
\includegraphics[width=4.4cm]{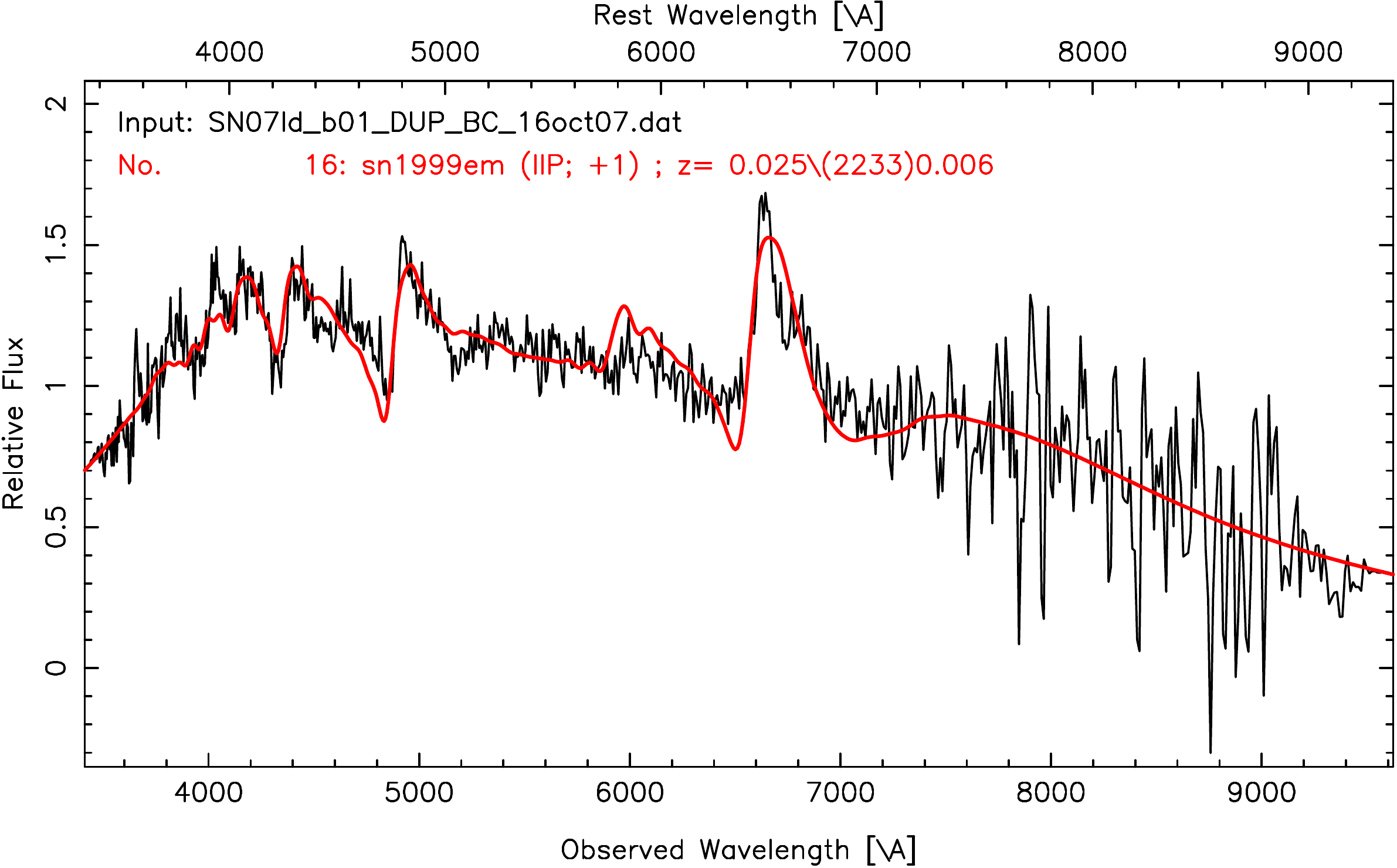}
\includegraphics[width=4.4cm]{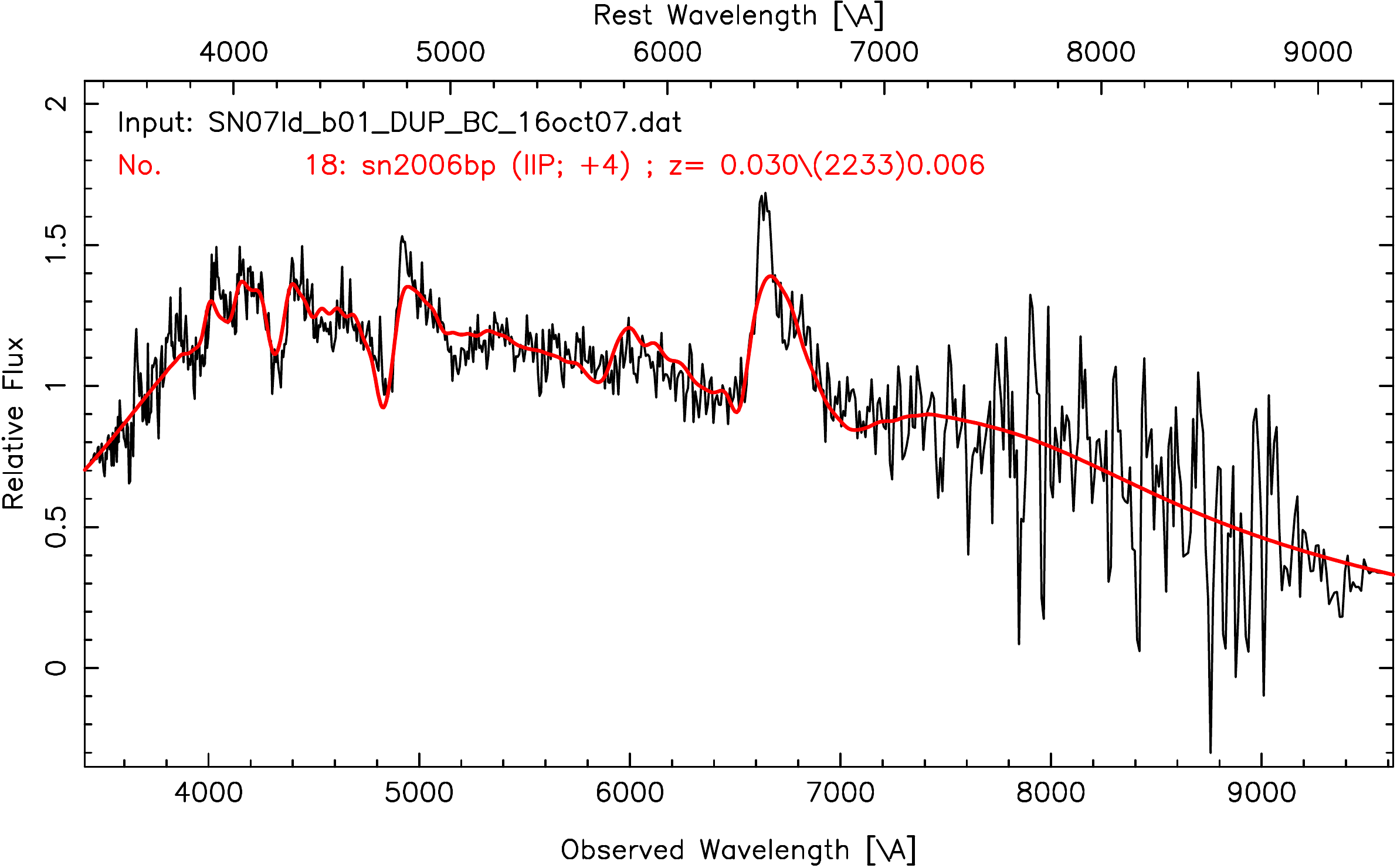}
\includegraphics[width=4.4cm]{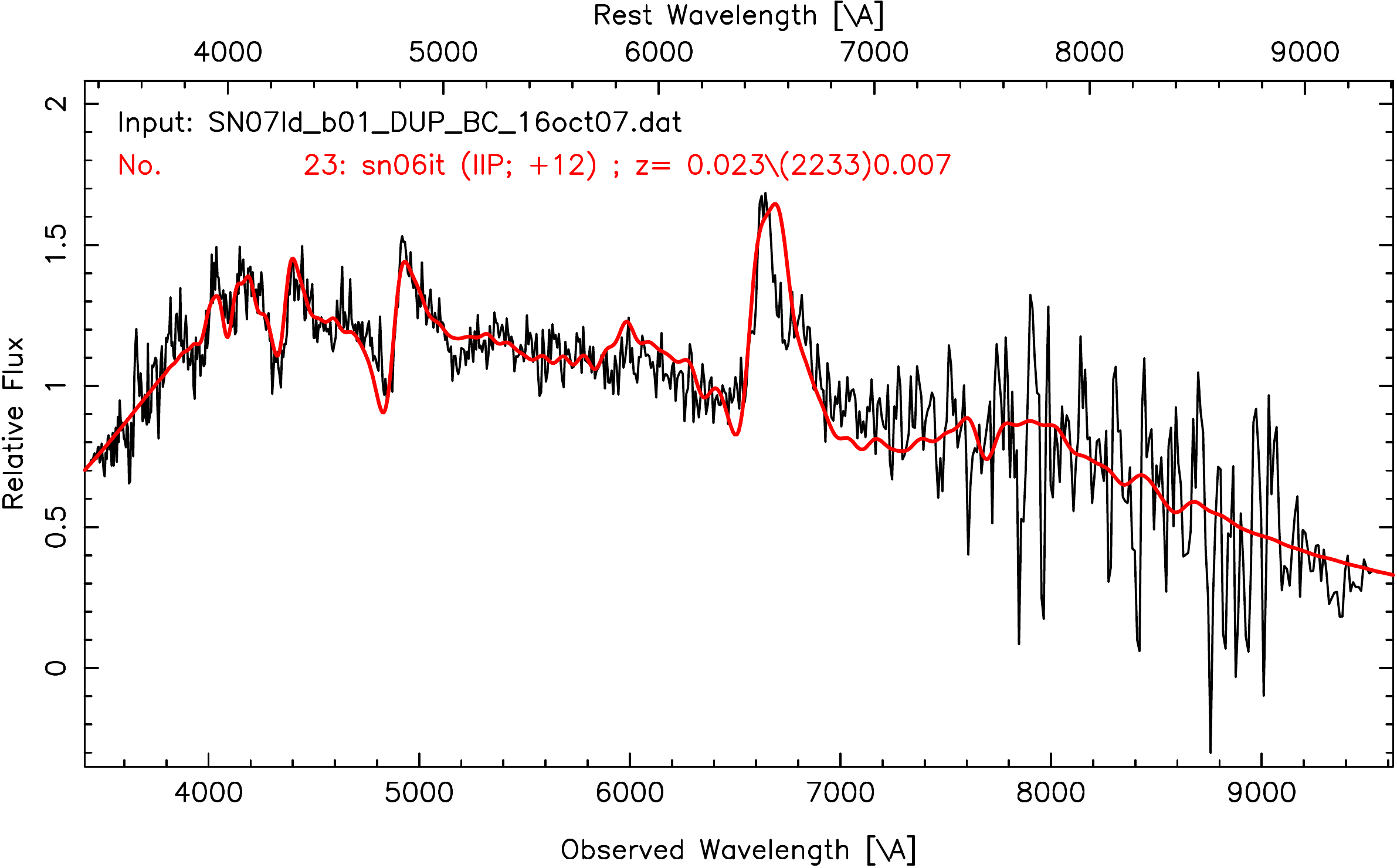}
\caption{Best spectral matching of SN~2007ld using SNID. The plots show SN~2007ld compared with 
SN~2004et, SN~1999em, SN~2006bp, and SN~2006it at 17, 11, 13m and 12 days from explosion.}
\end{figure}

\clearpage

\begin{figure}
\centering
\includegraphics[width=4.4cm]{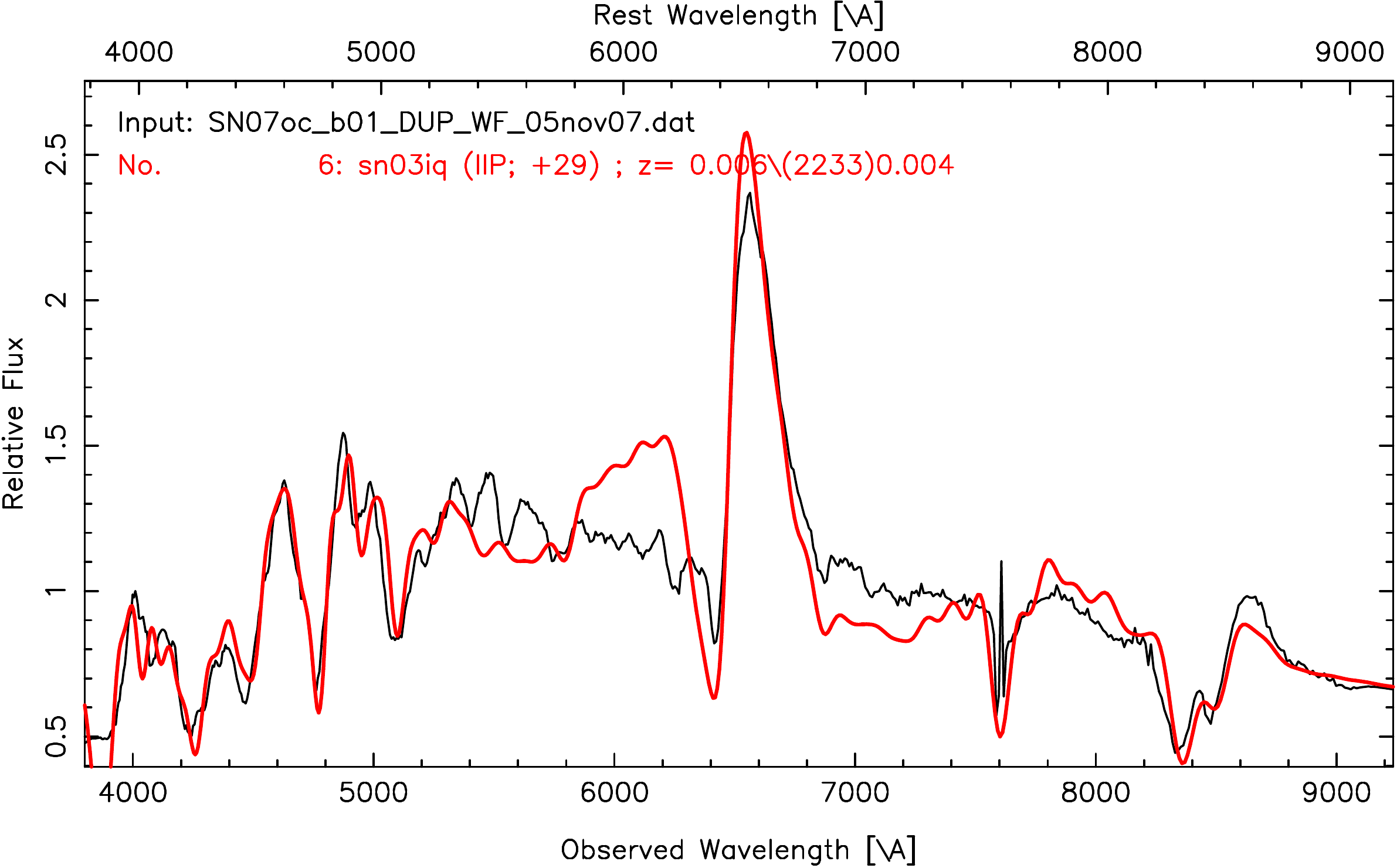}
\includegraphics[width=4.4cm]{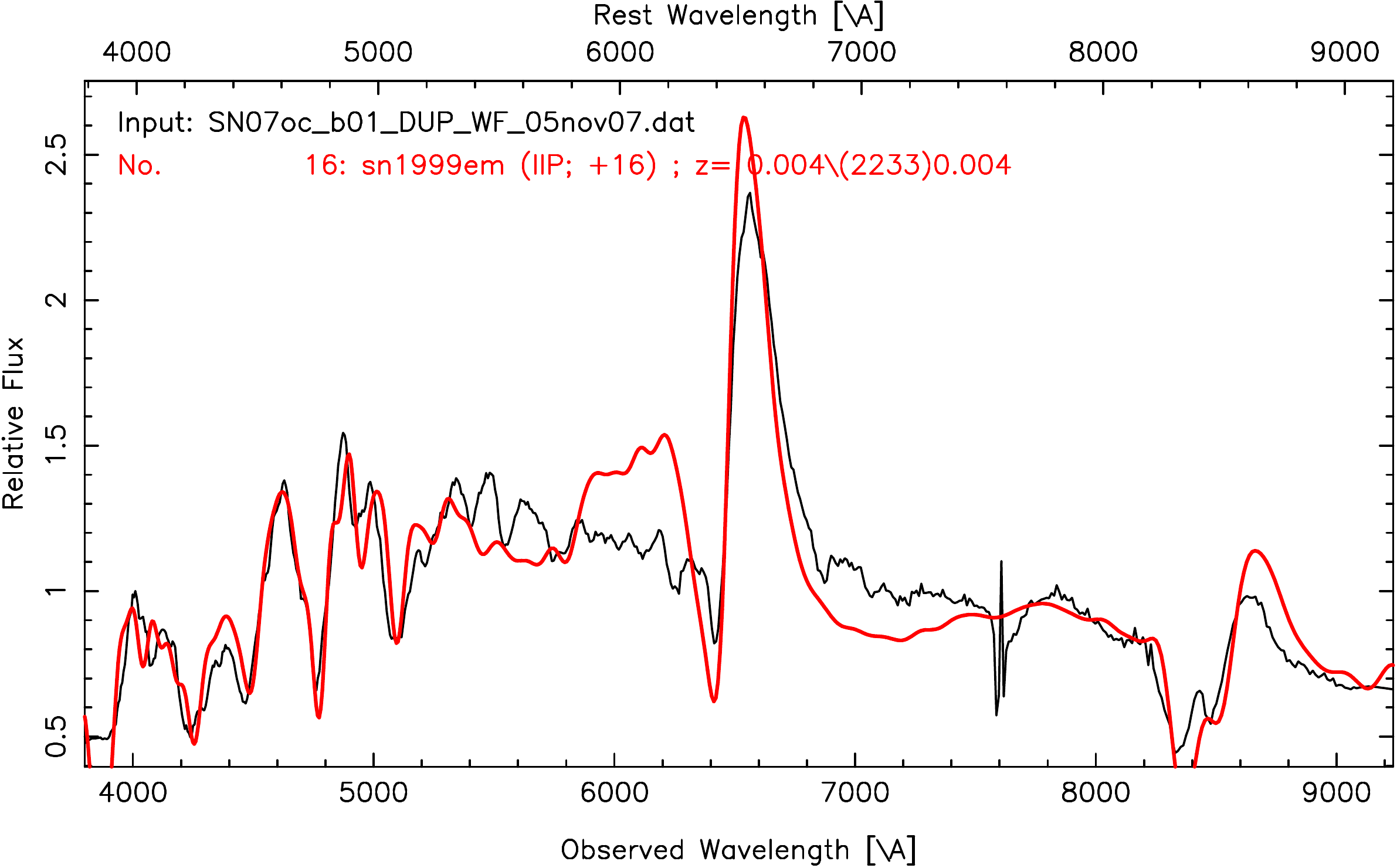}
\includegraphics[width=4.4cm]{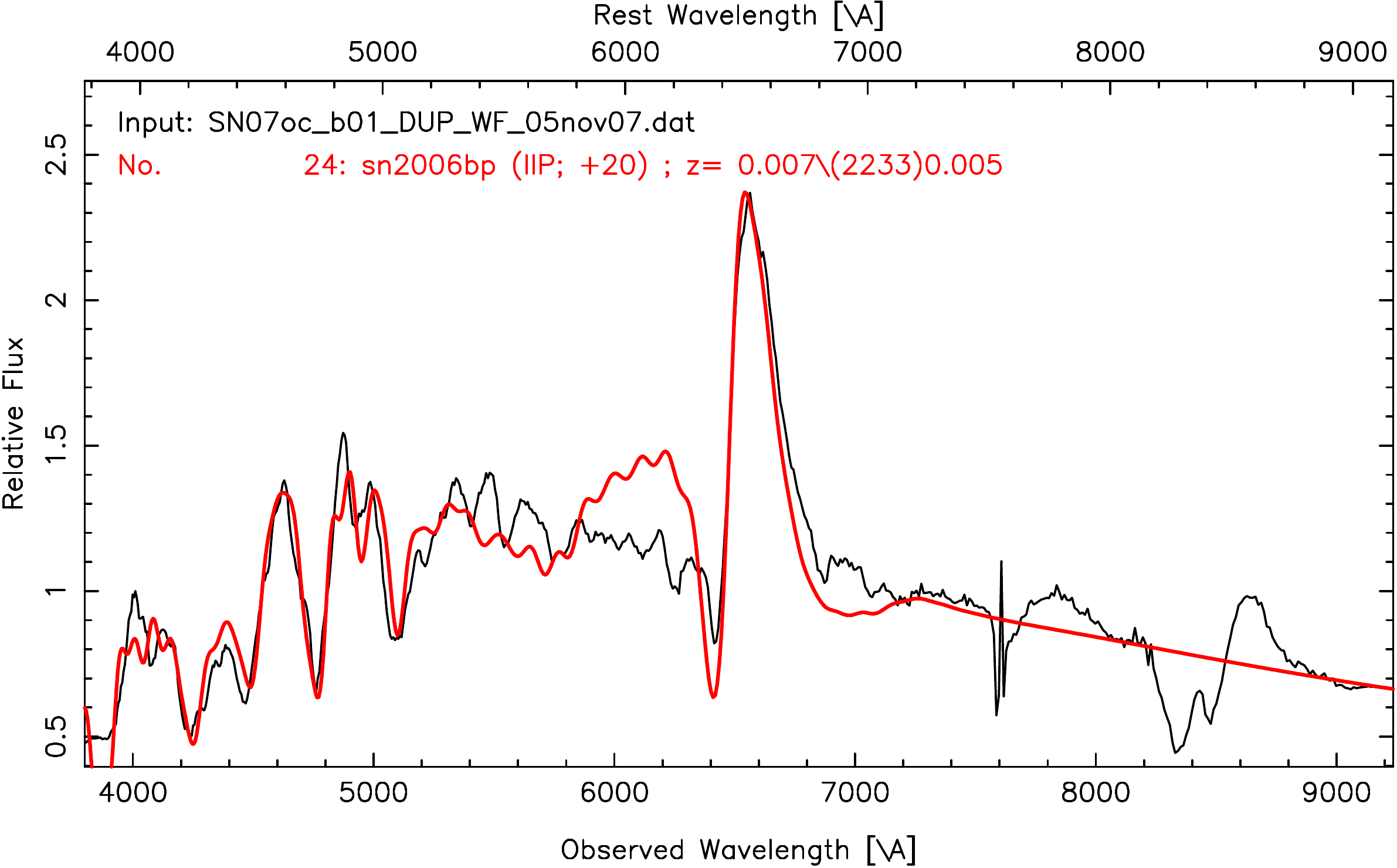}
\caption{Best spectral matching of SN~2007oc using SNID. The plots show SN~2007oc compared with 
SN~2003iq, SN~1999em, and SN~2006bp at 29, 26, and 29 days from explosion.}
\end{figure}

\begin{figure}
\centering
\includegraphics[width=4.4cm]{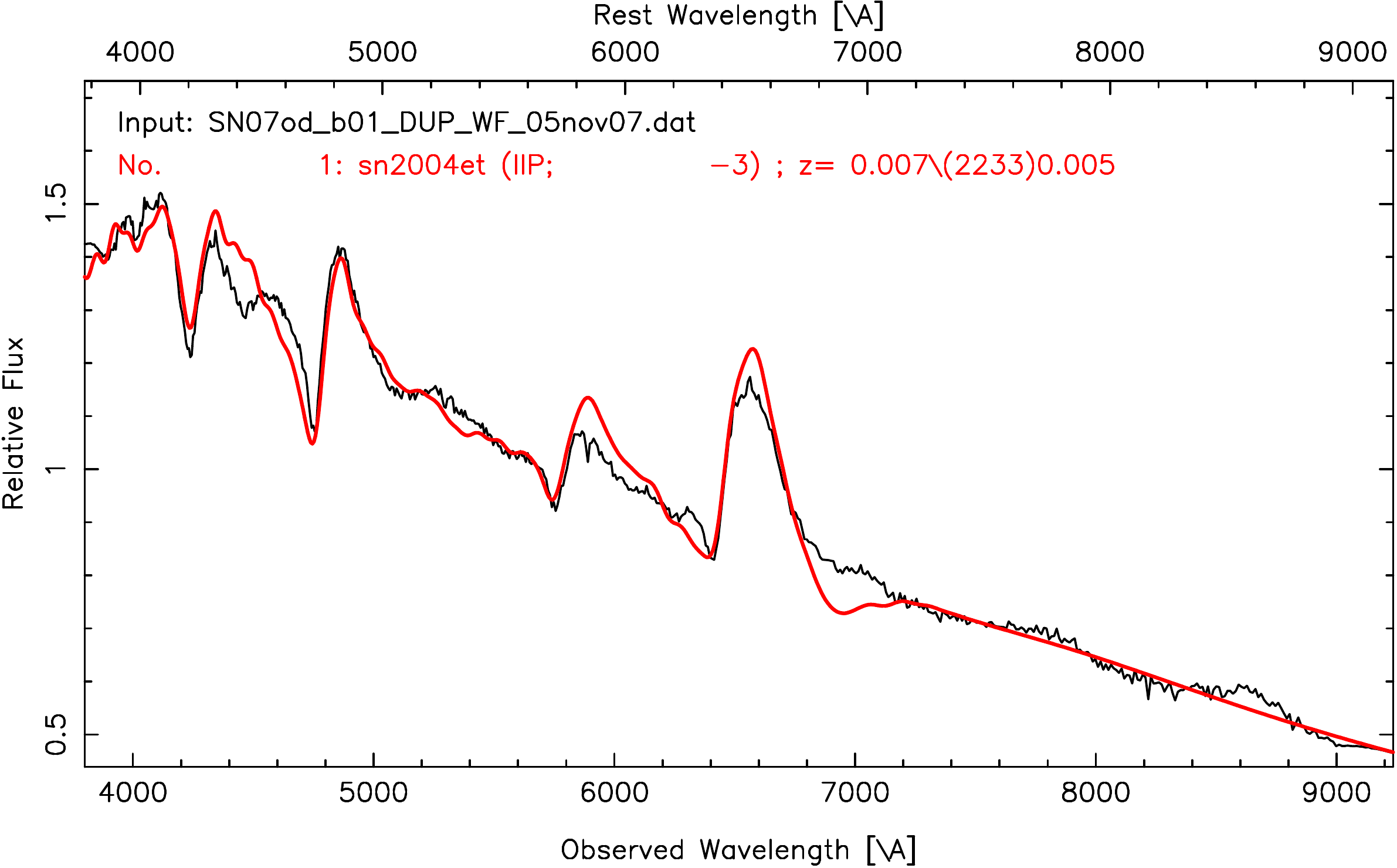}
\includegraphics[width=4.4cm]{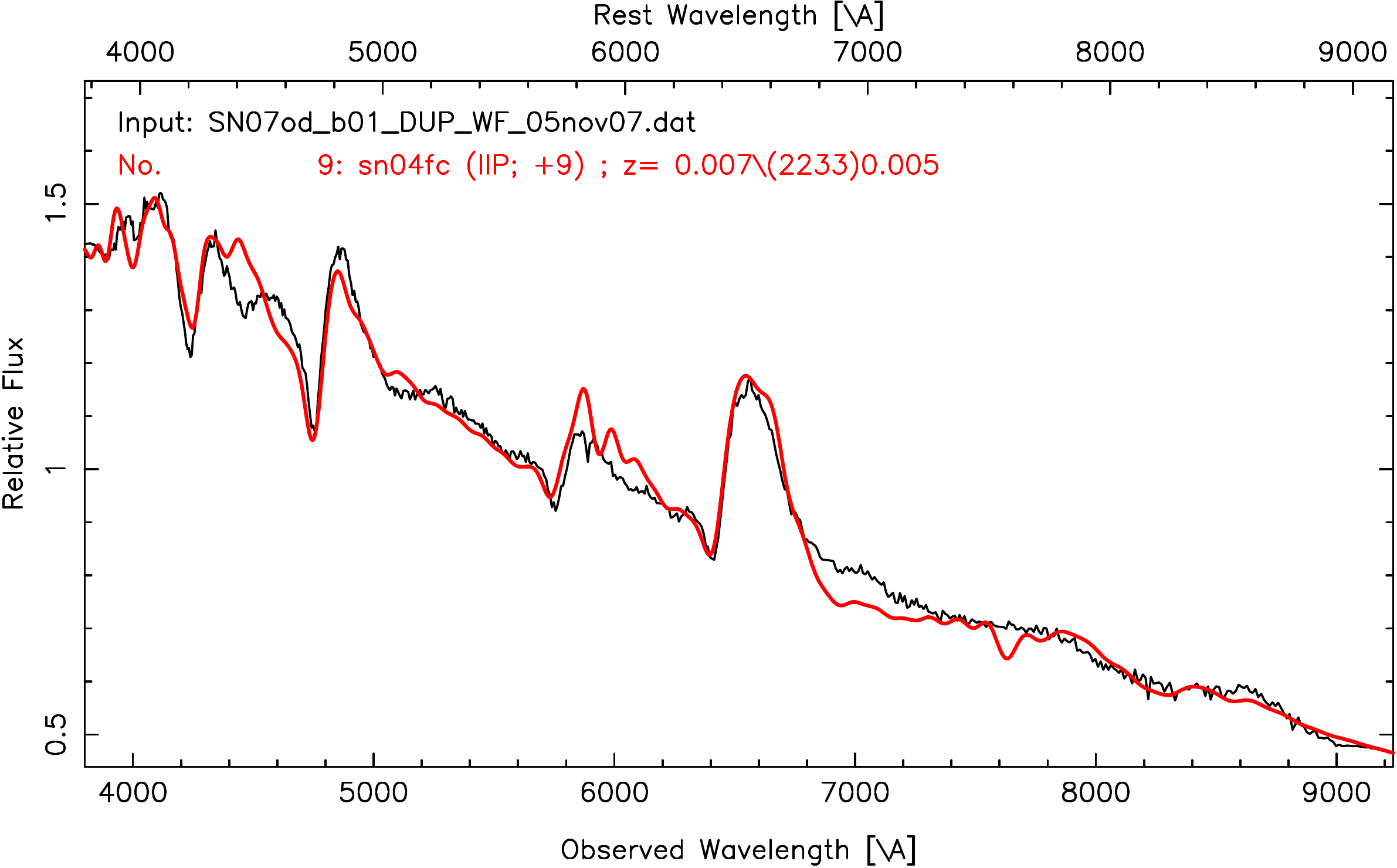}
\includegraphics[width=4.4cm]{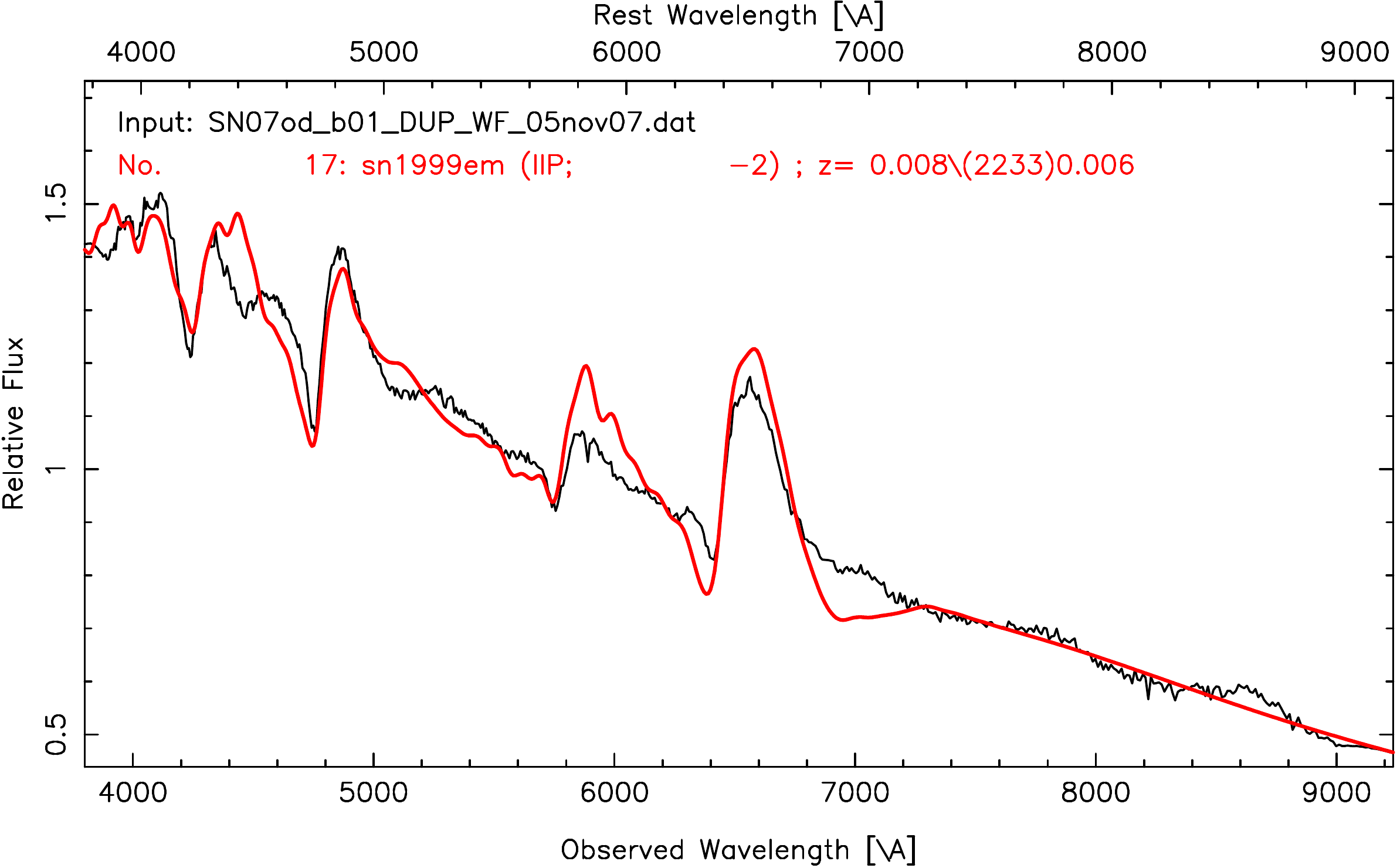}
\includegraphics[width=4.4cm]{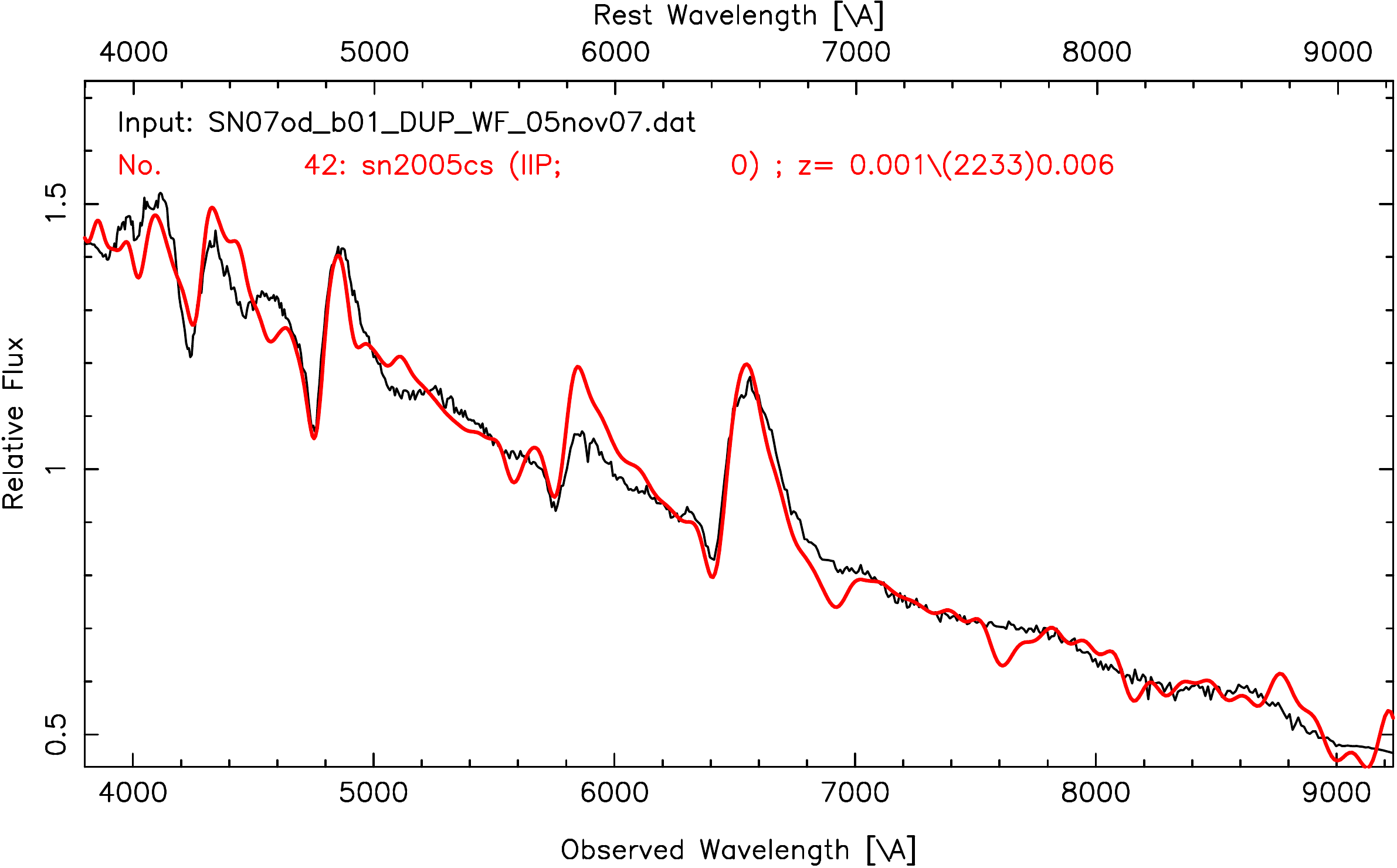}
\caption{Best spectral matching of SN~2007od using SNID. The plots show SN~2007od compared with 
SN~2004et, SN~2004fc, SN~1999em, and SN~2005cs at 13, 9, 8, and 6 days from explosion.}
\end{figure}

\clearpage

\begin{figure}
\centering
\includegraphics[width=4.4cm]{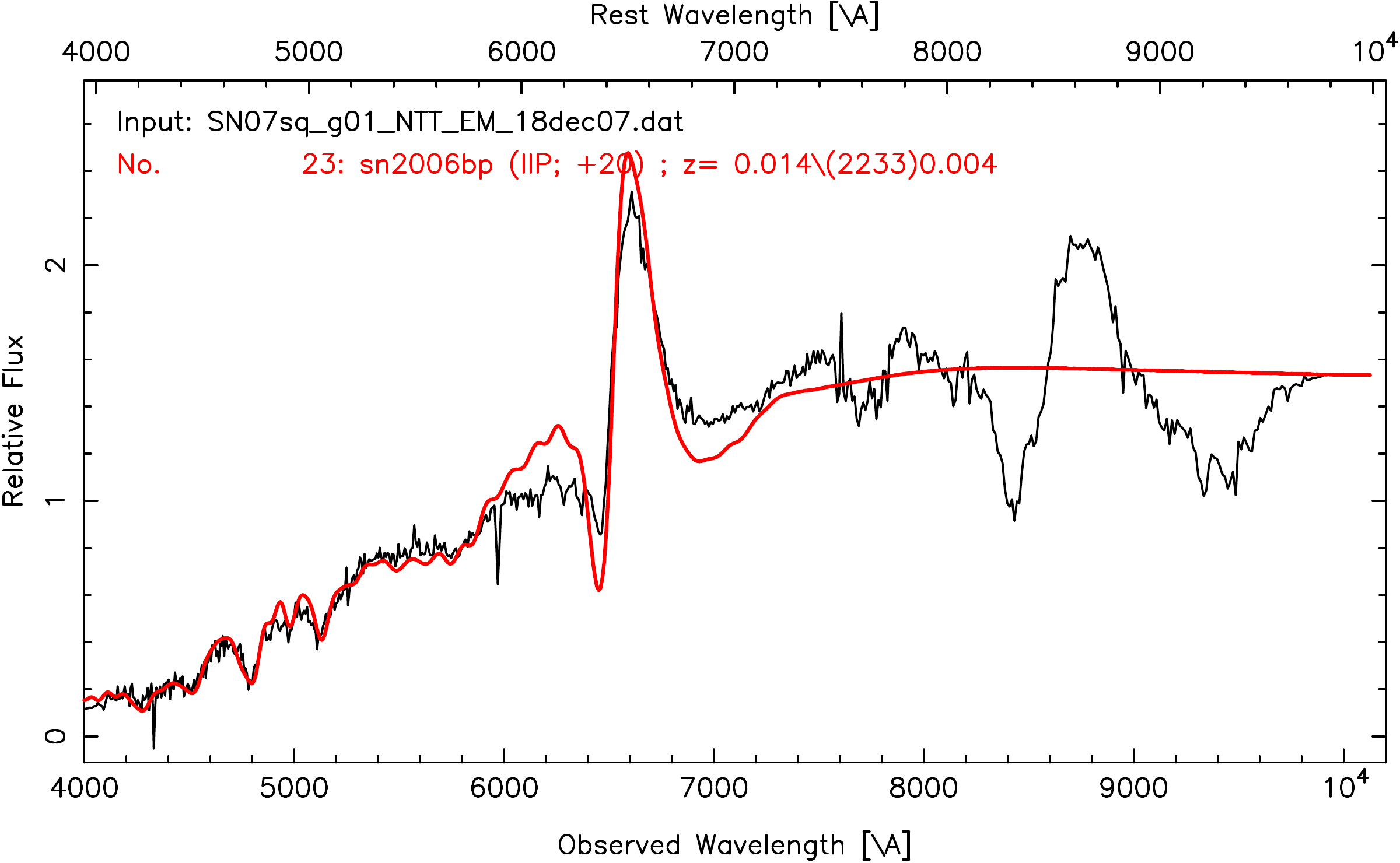}
\includegraphics[width=4.4cm]{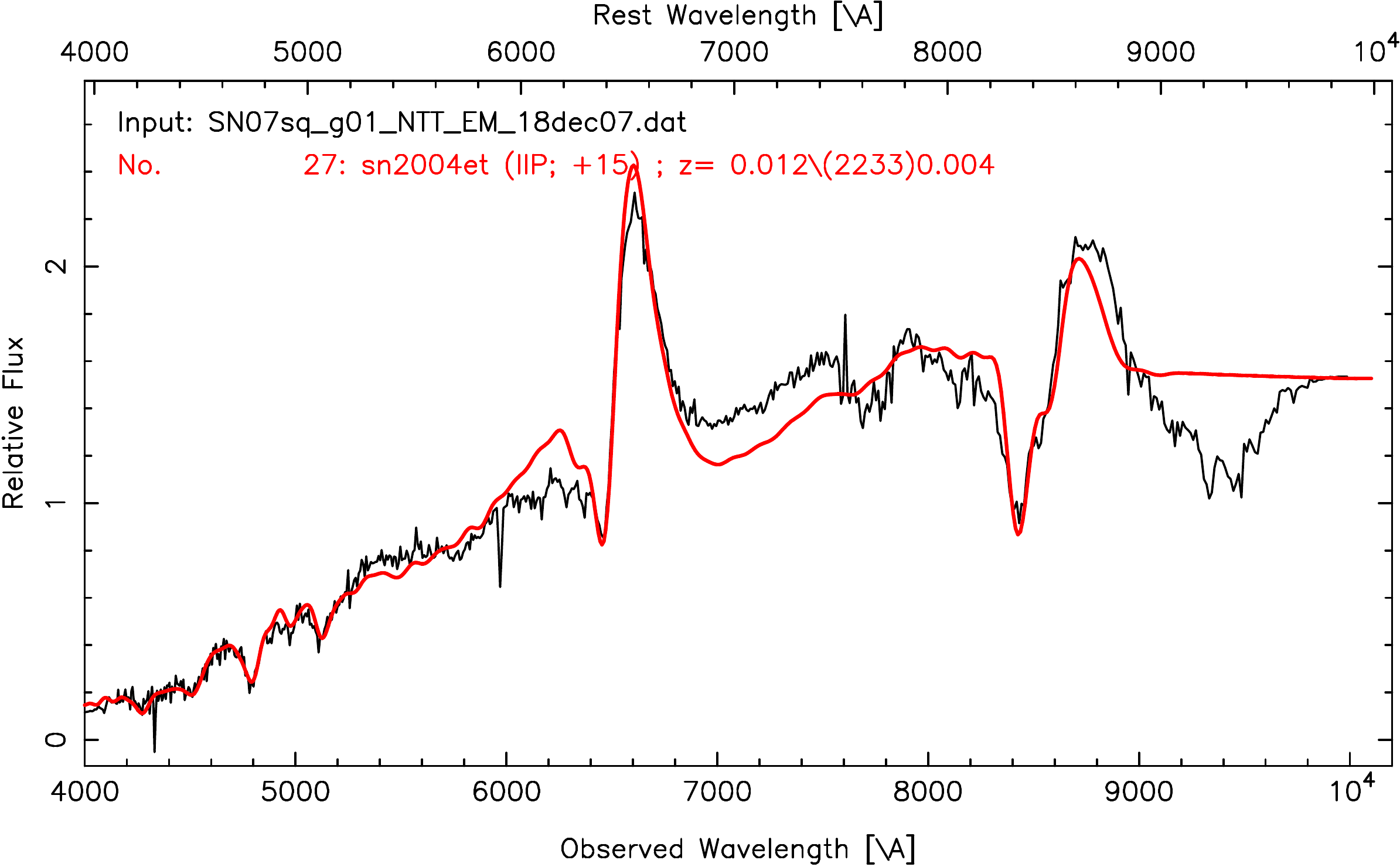}
\includegraphics[width=4.4cm]{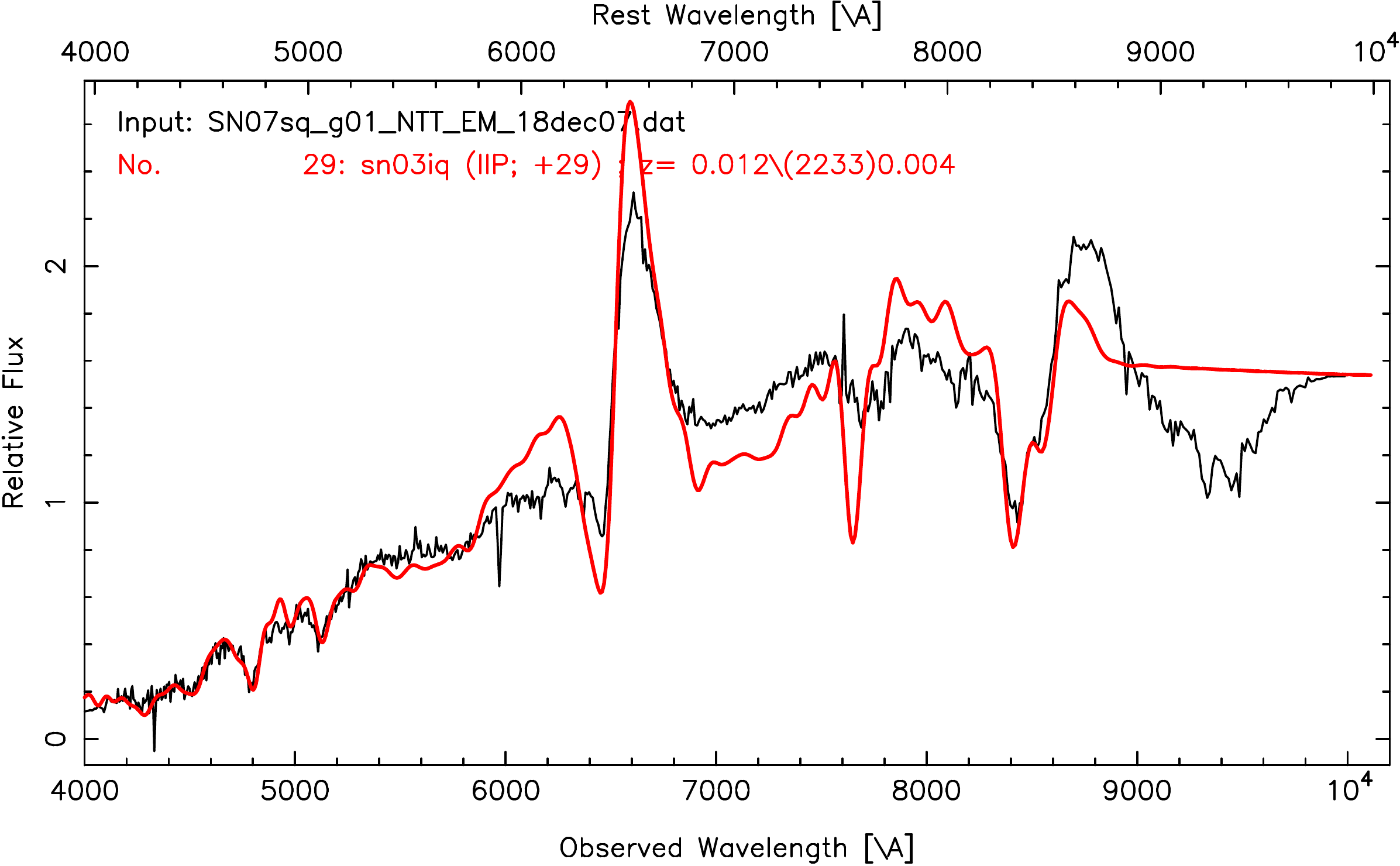}
\includegraphics[width=4.4cm]{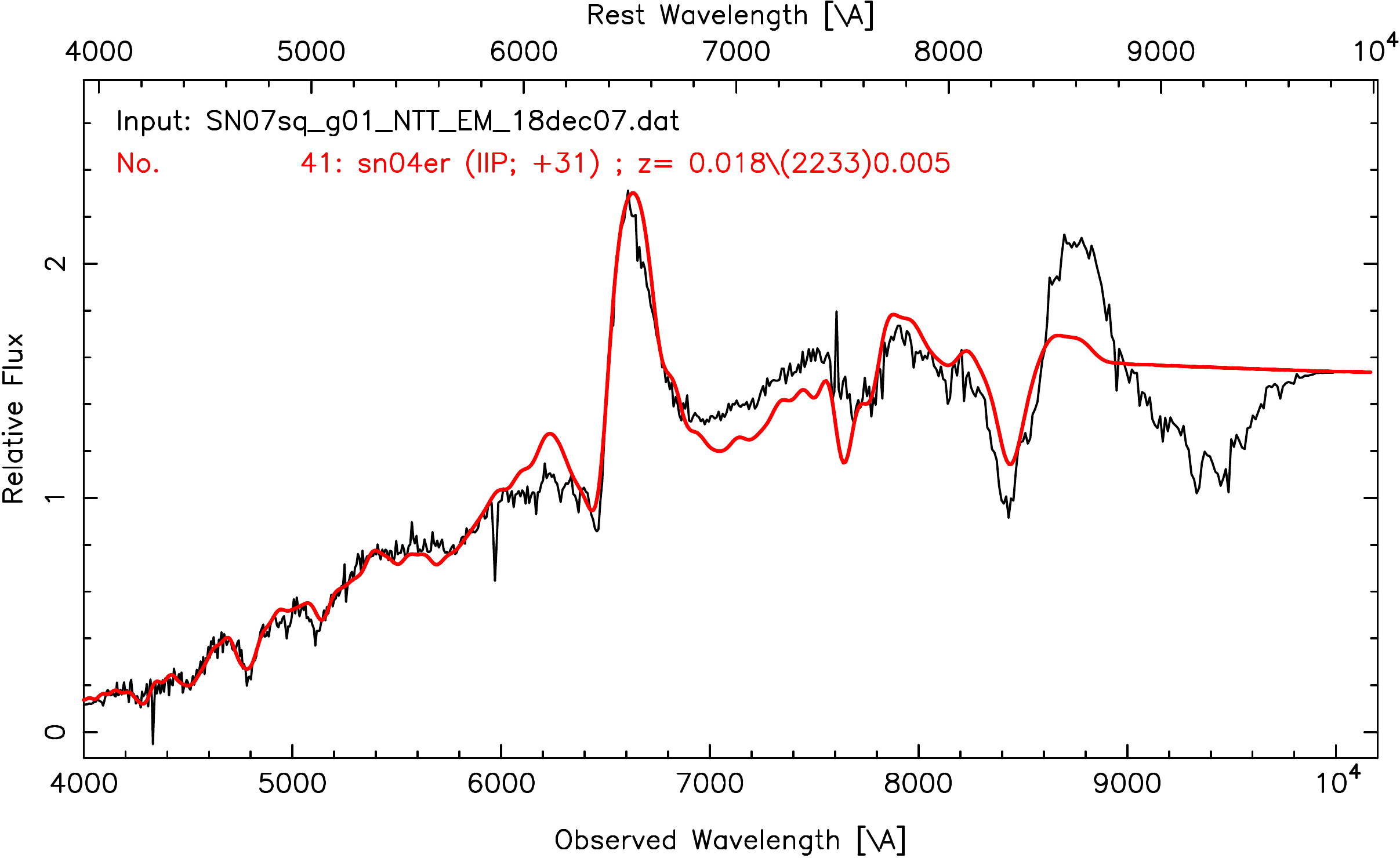}
\caption{Best spectral matching of SN~2007sq using SNID. The plots show SN~2007sq compared with 
SN~2006bp, SN~2004et, SN~2003iq, and SN~2004er at 29, 31, 29, and 31 days from explosion.}
\end{figure}

\begin{figure}
\centering
\includegraphics[width=4.4cm]{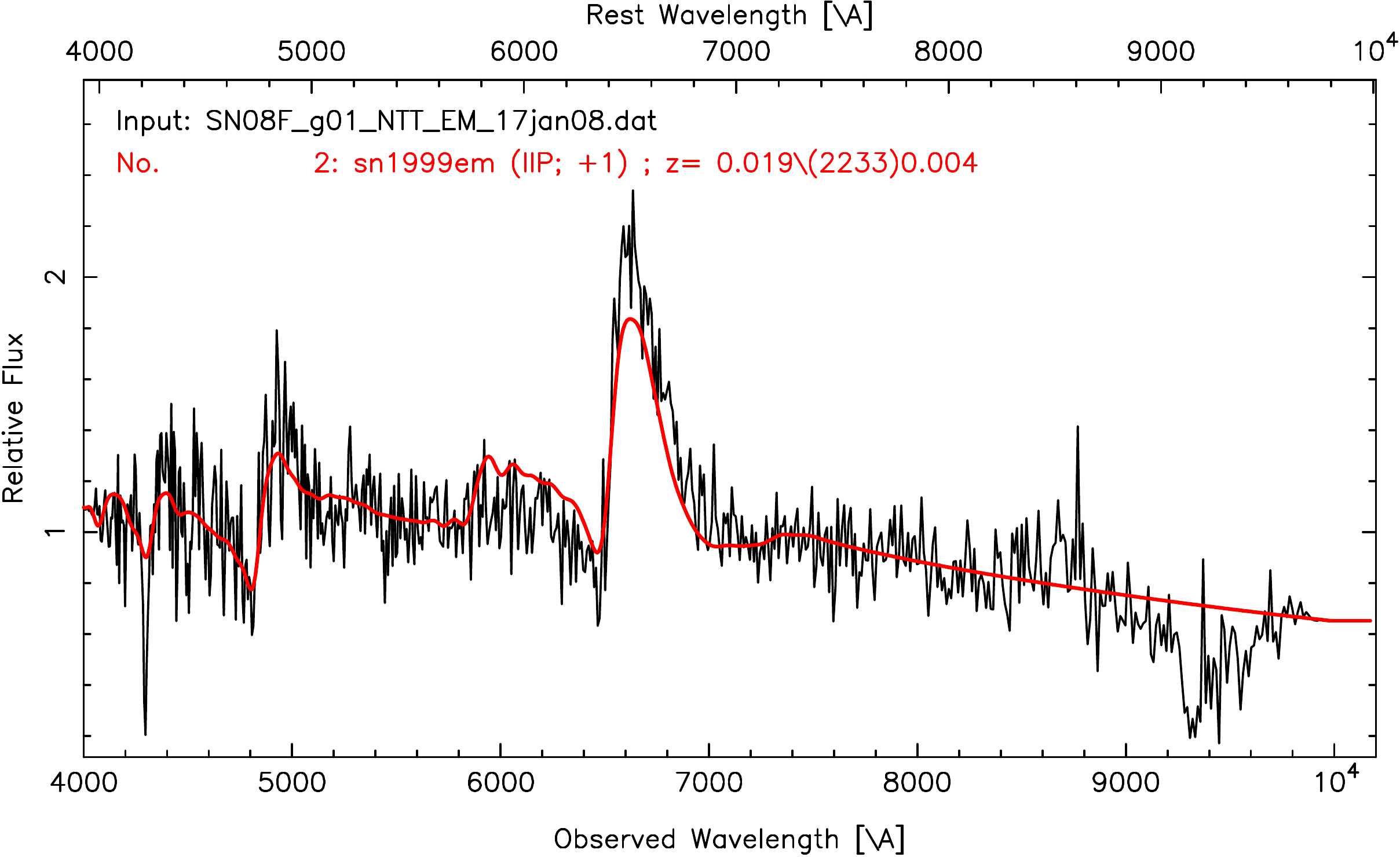}
\includegraphics[width=4.4cm]{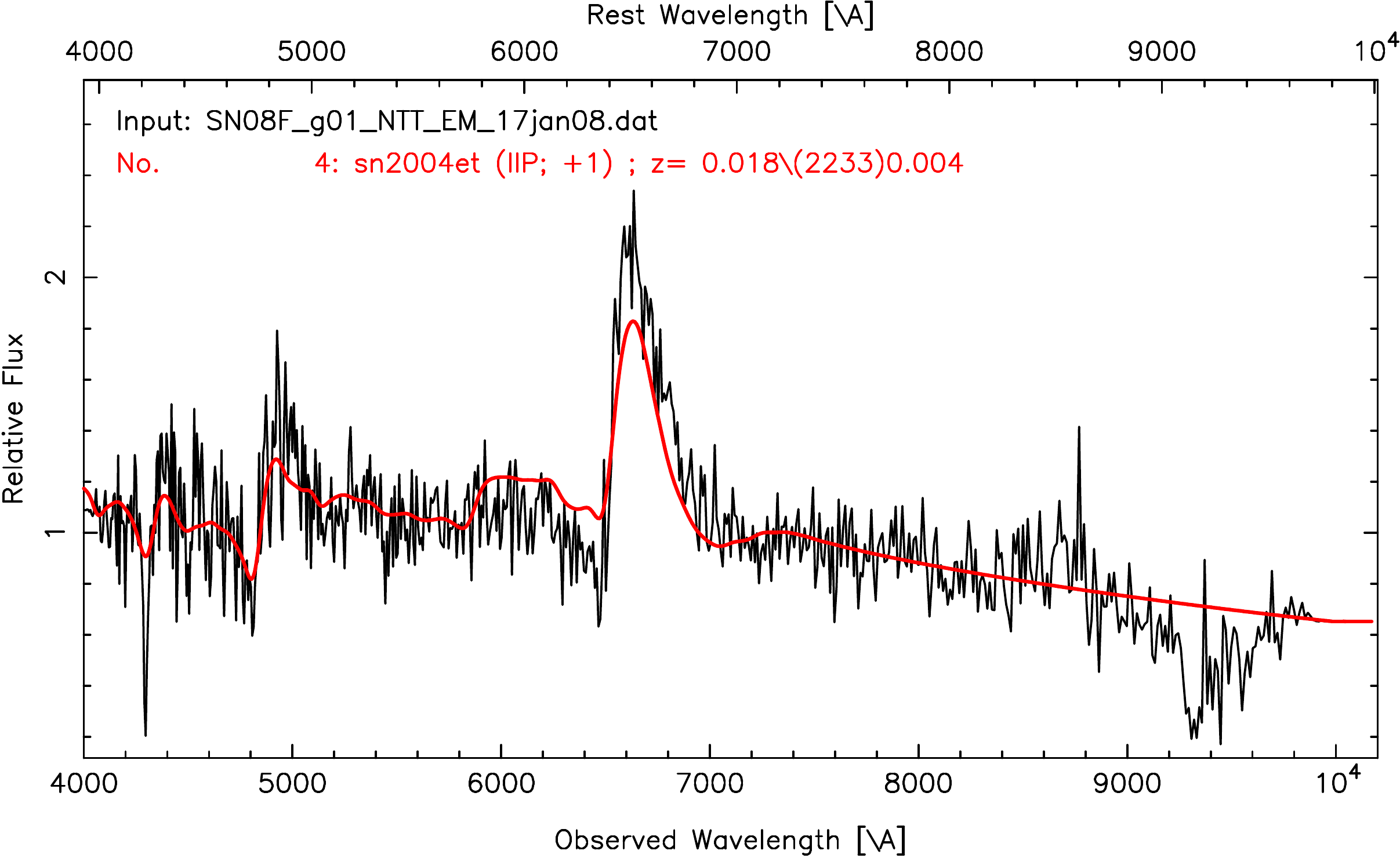}
\includegraphics[width=4.4cm]{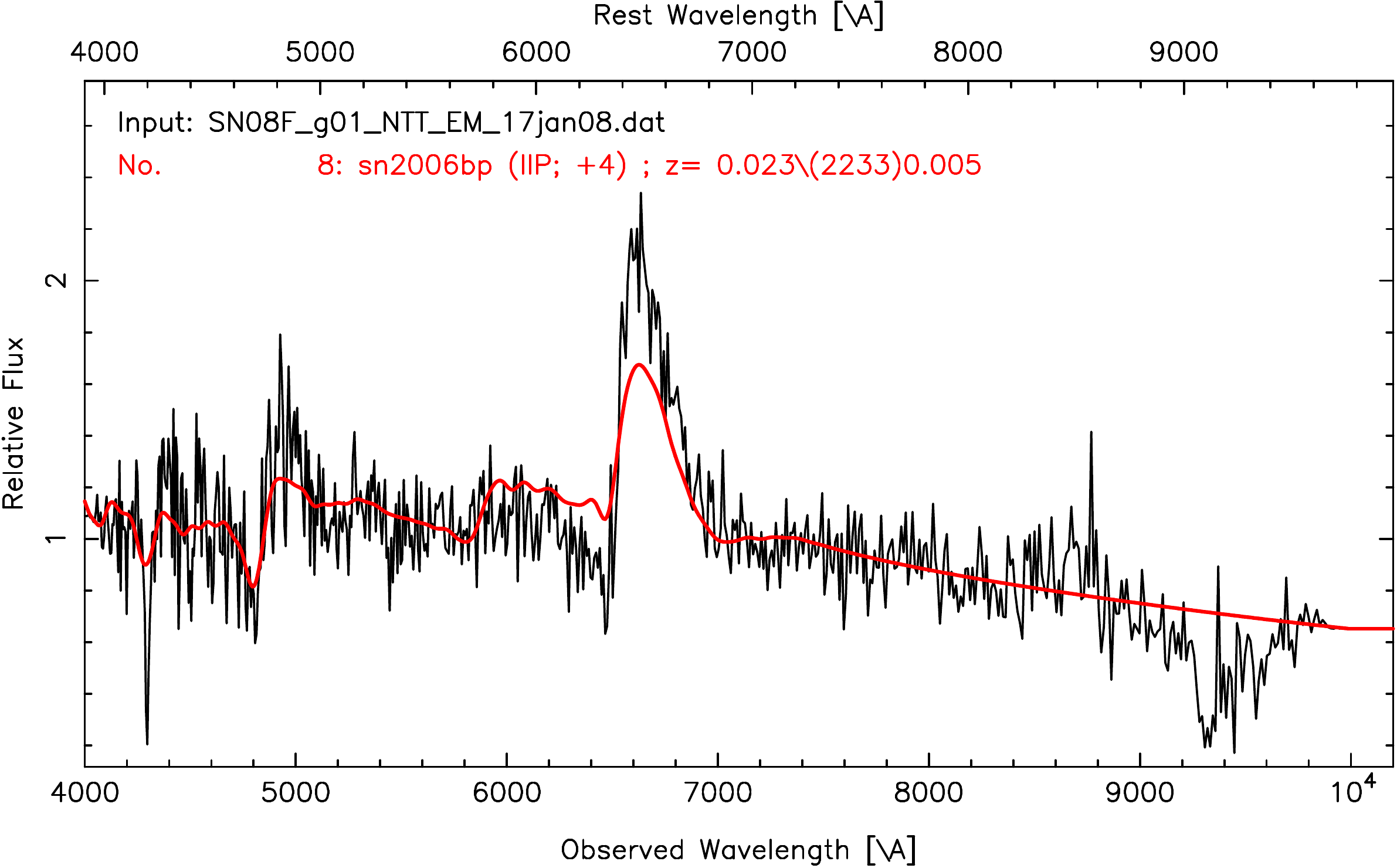}
\includegraphics[width=4.4cm]{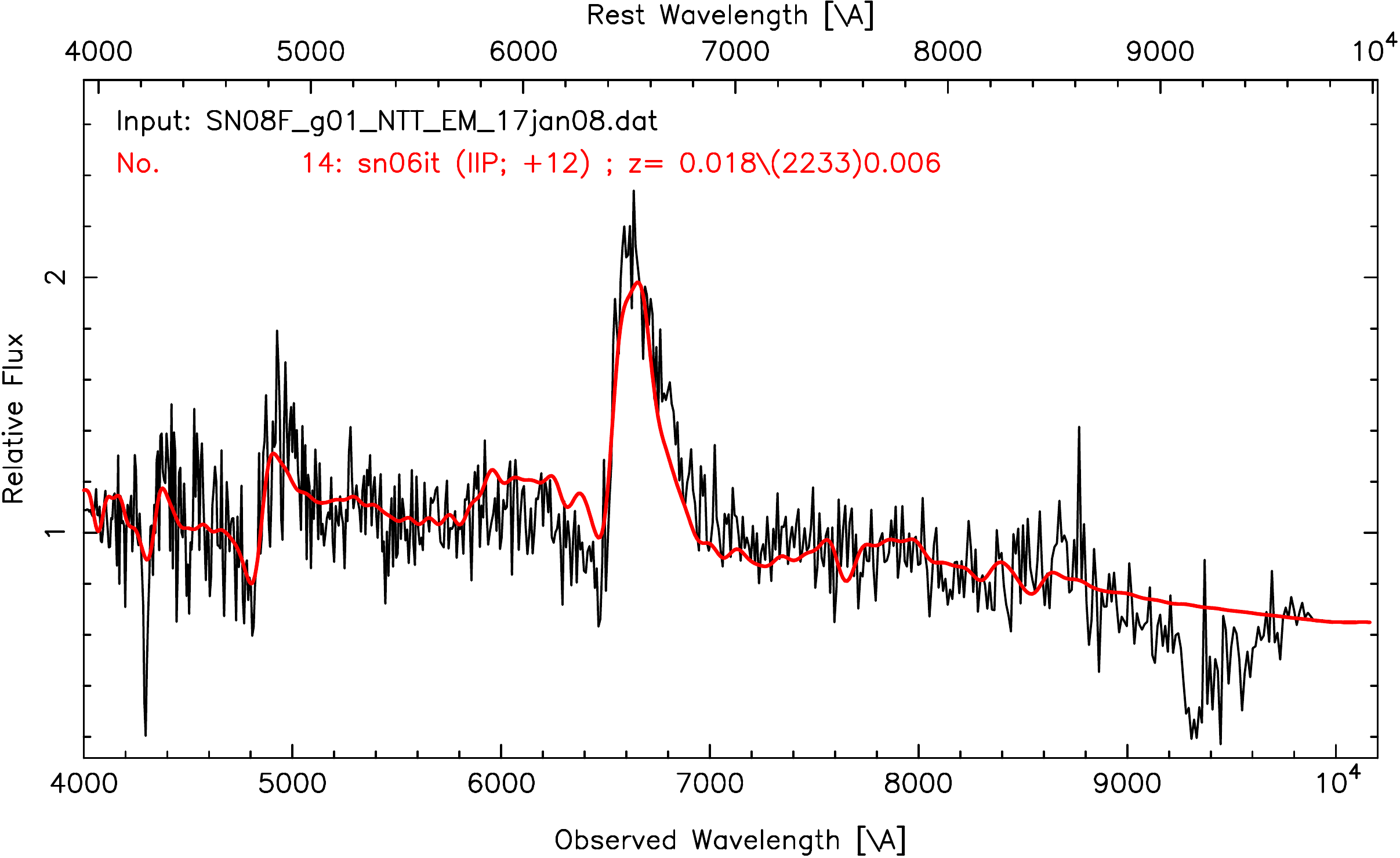}
\caption{Best spectral matching of SN~2008F using SNID. The plots show SN~2008F compared with 
SN~1999em, SN~2004et, SN~2006bp, and SN~2006it at 11, 17, 13, and 12 days from explosion.}
\end{figure}

\clearpage

\begin{figure}
\centering
\includegraphics[width=4.4cm]{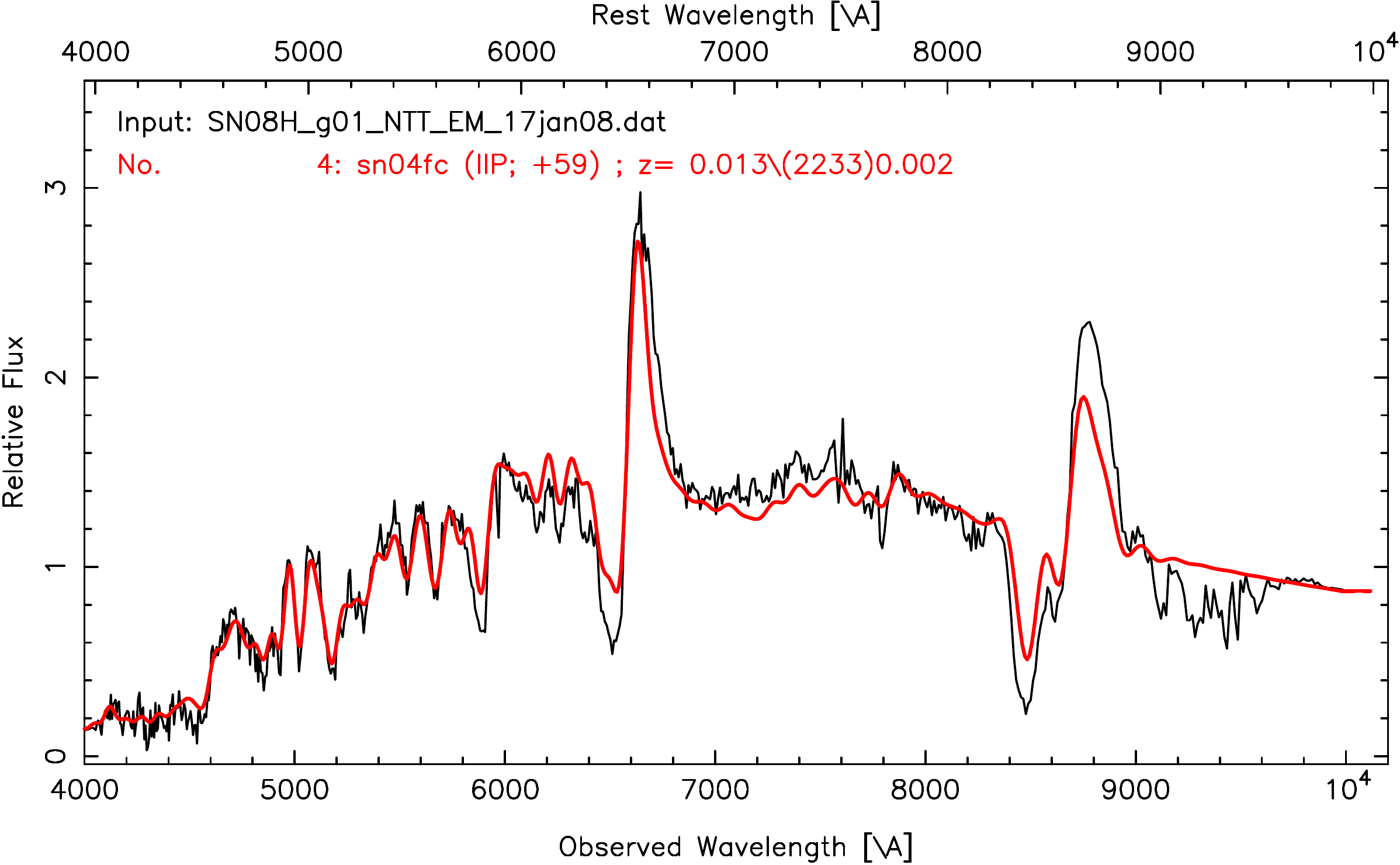}
\includegraphics[width=4.4cm]{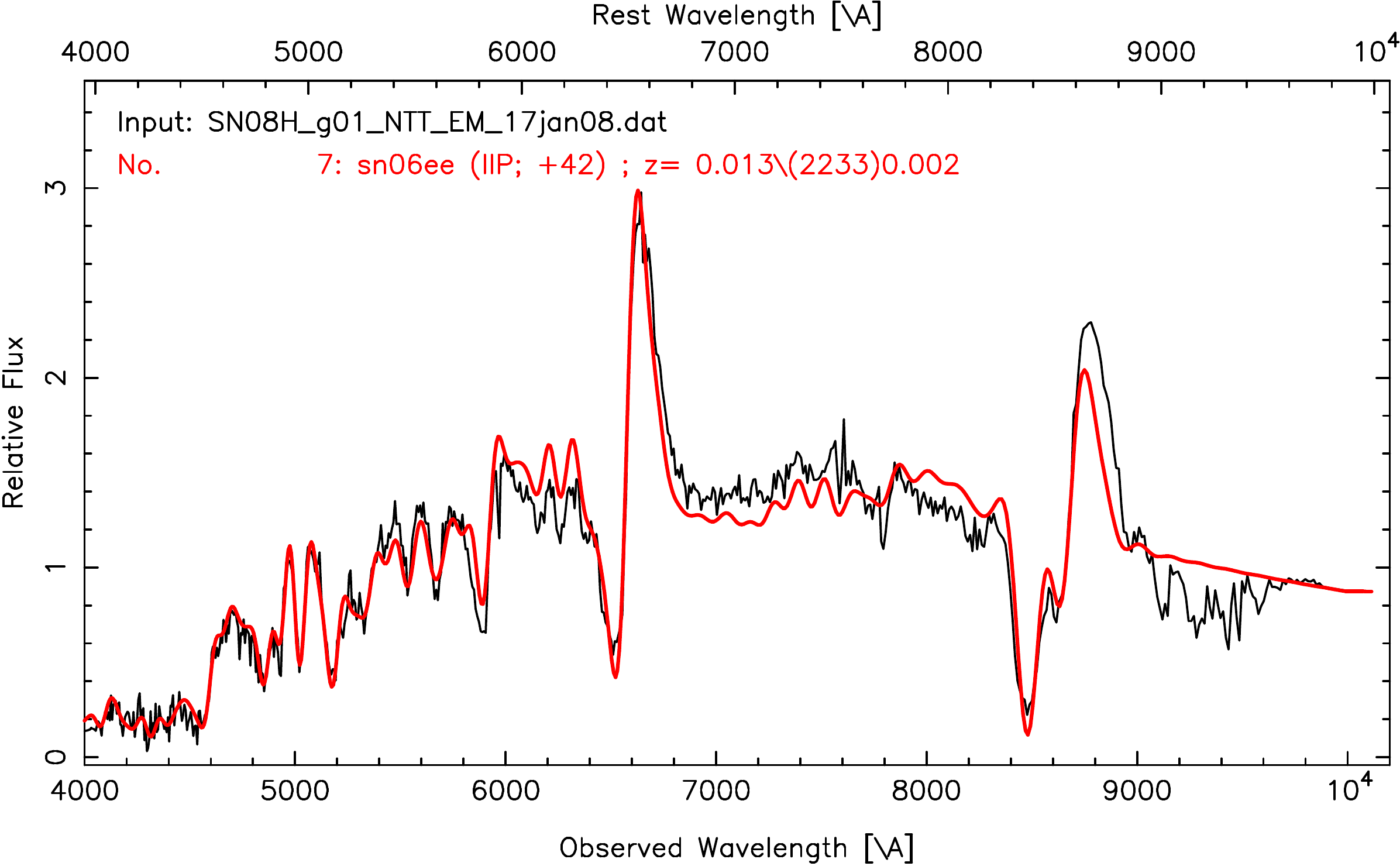}
\includegraphics[width=4.4cm]{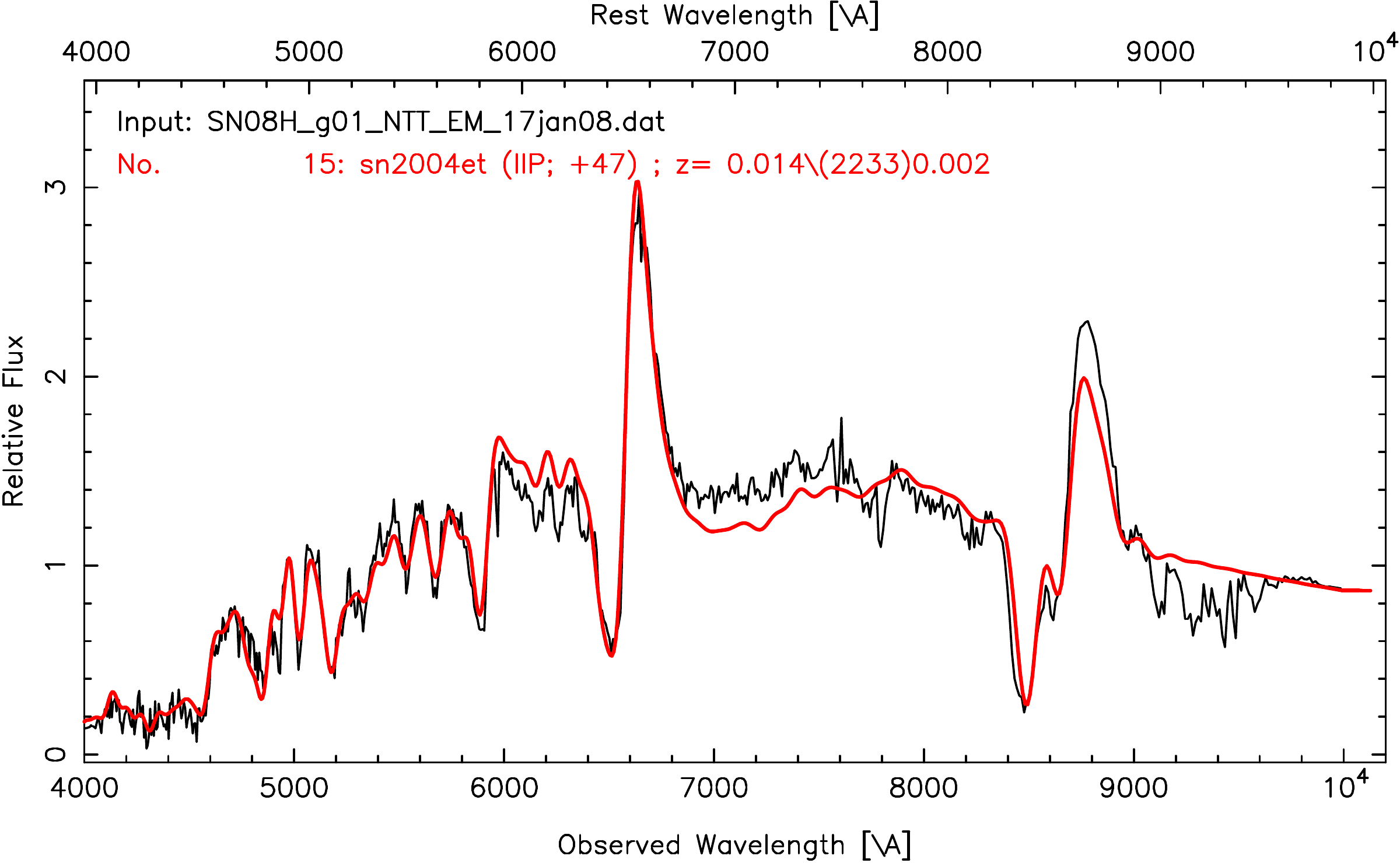}
\includegraphics[width=4.4cm]{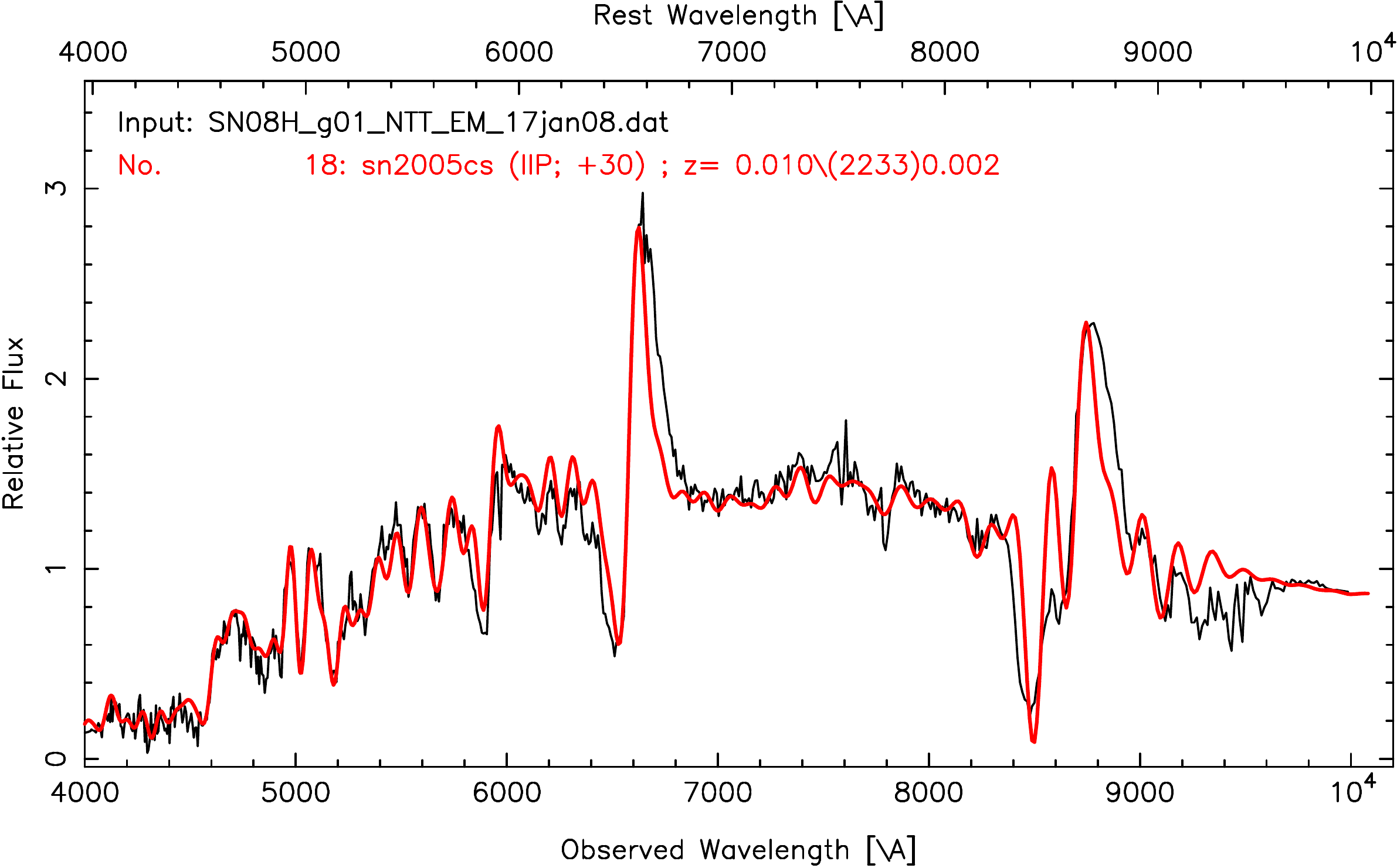}
\includegraphics[width=4.4cm]{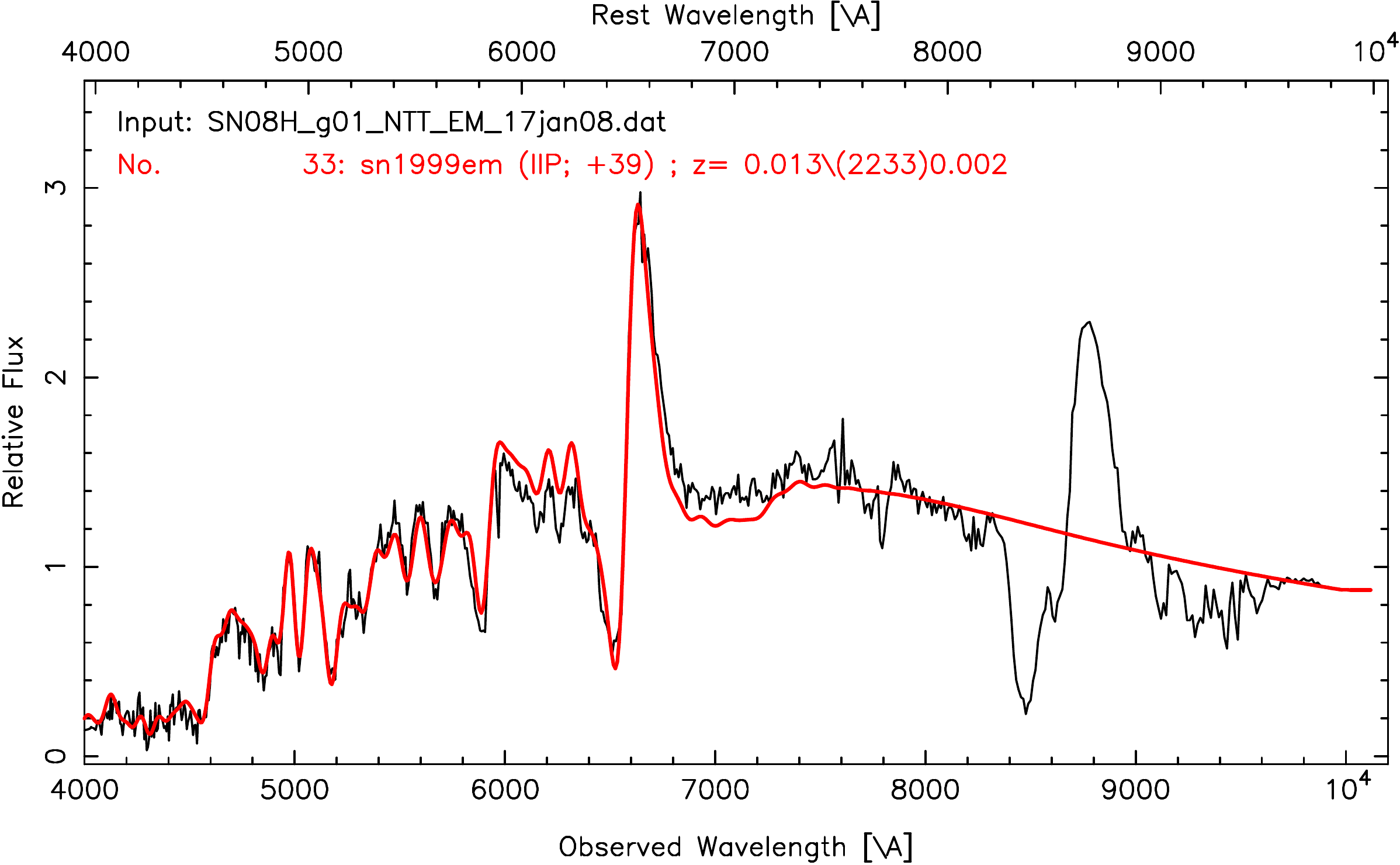}
\caption{Best spectral matching of SN~2008H using SNID. The plots show SN~2008H compared with 
SN~2004fc, SN~2006ee, SN~2004et, SN~2005cs, and SN~1999em at 59, 42, 63, 36, and 49 days from explosion.}
\end{figure}

\begin{figure}
\centering
\includegraphics[width=4.4cm]{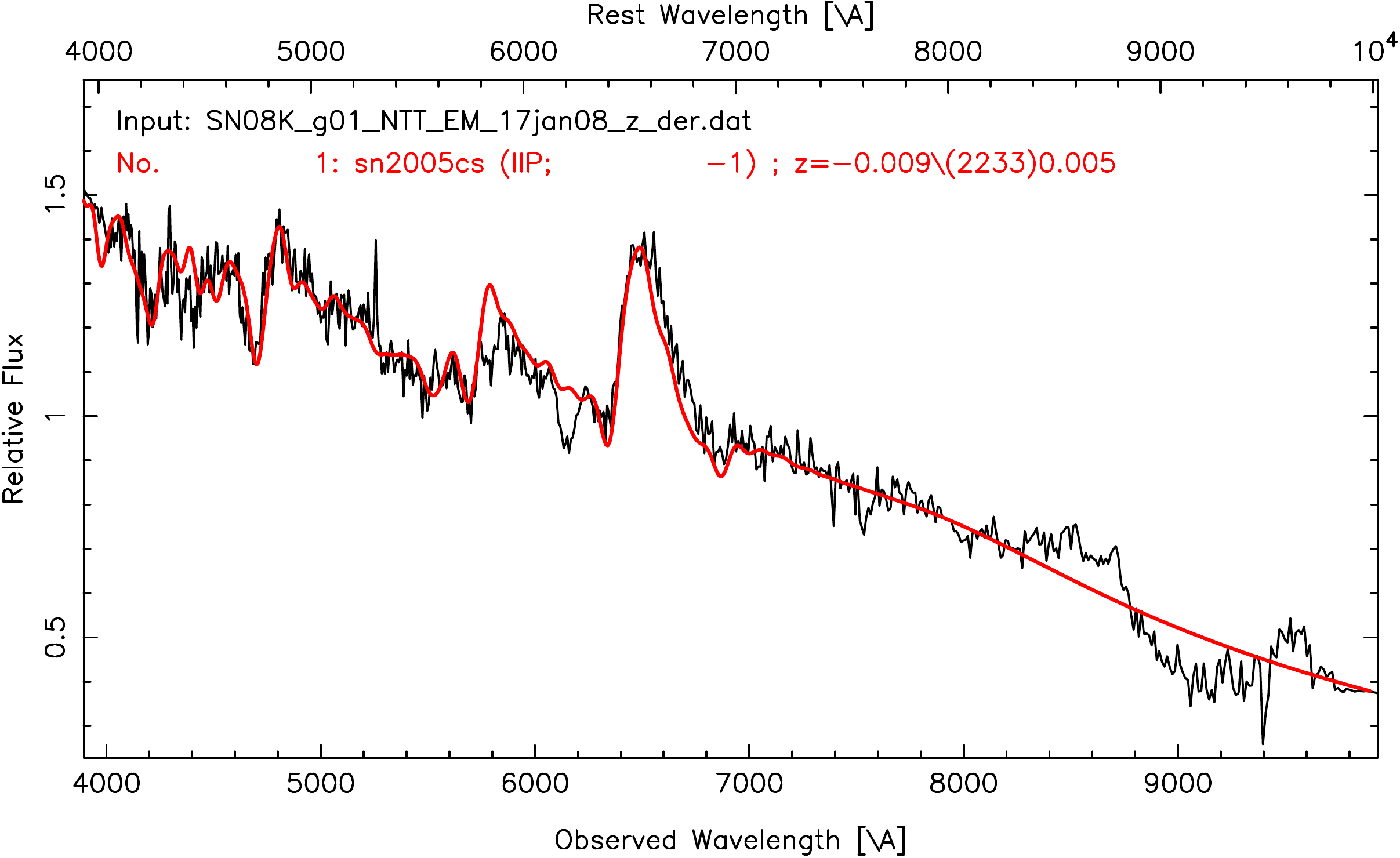}
\includegraphics[width=4.4cm]{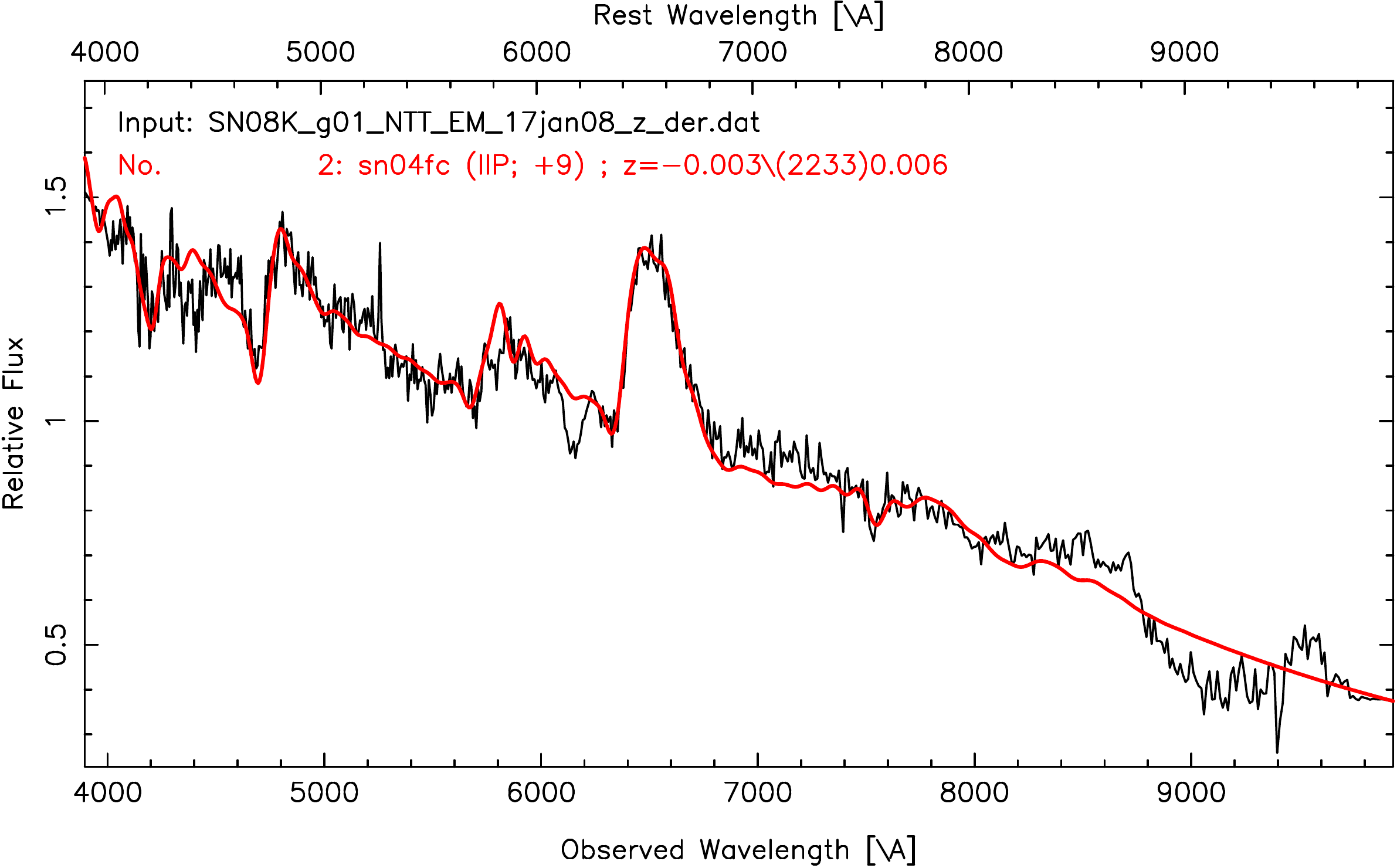}
\includegraphics[width=4.4cm]{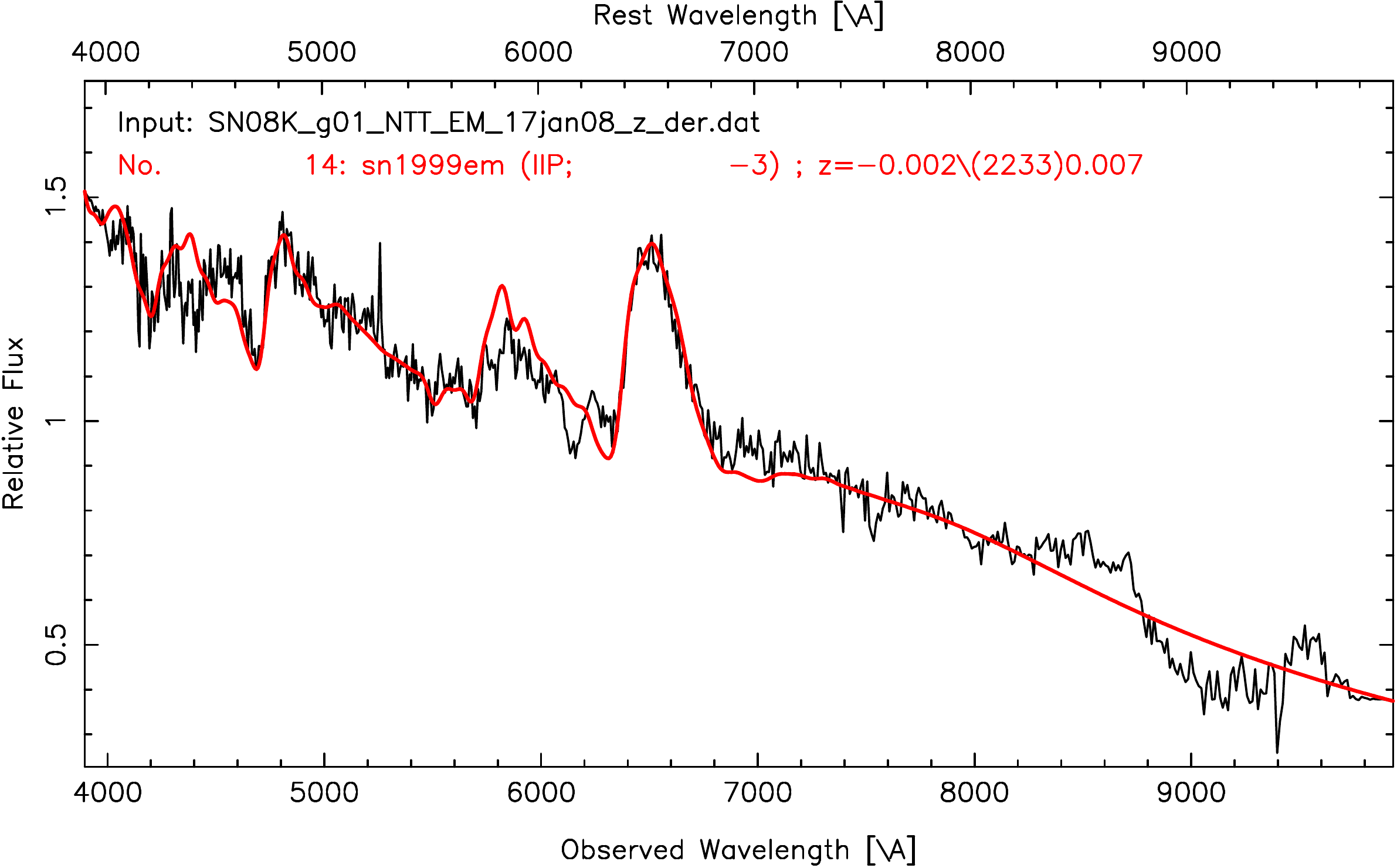}
\includegraphics[width=4.4cm]{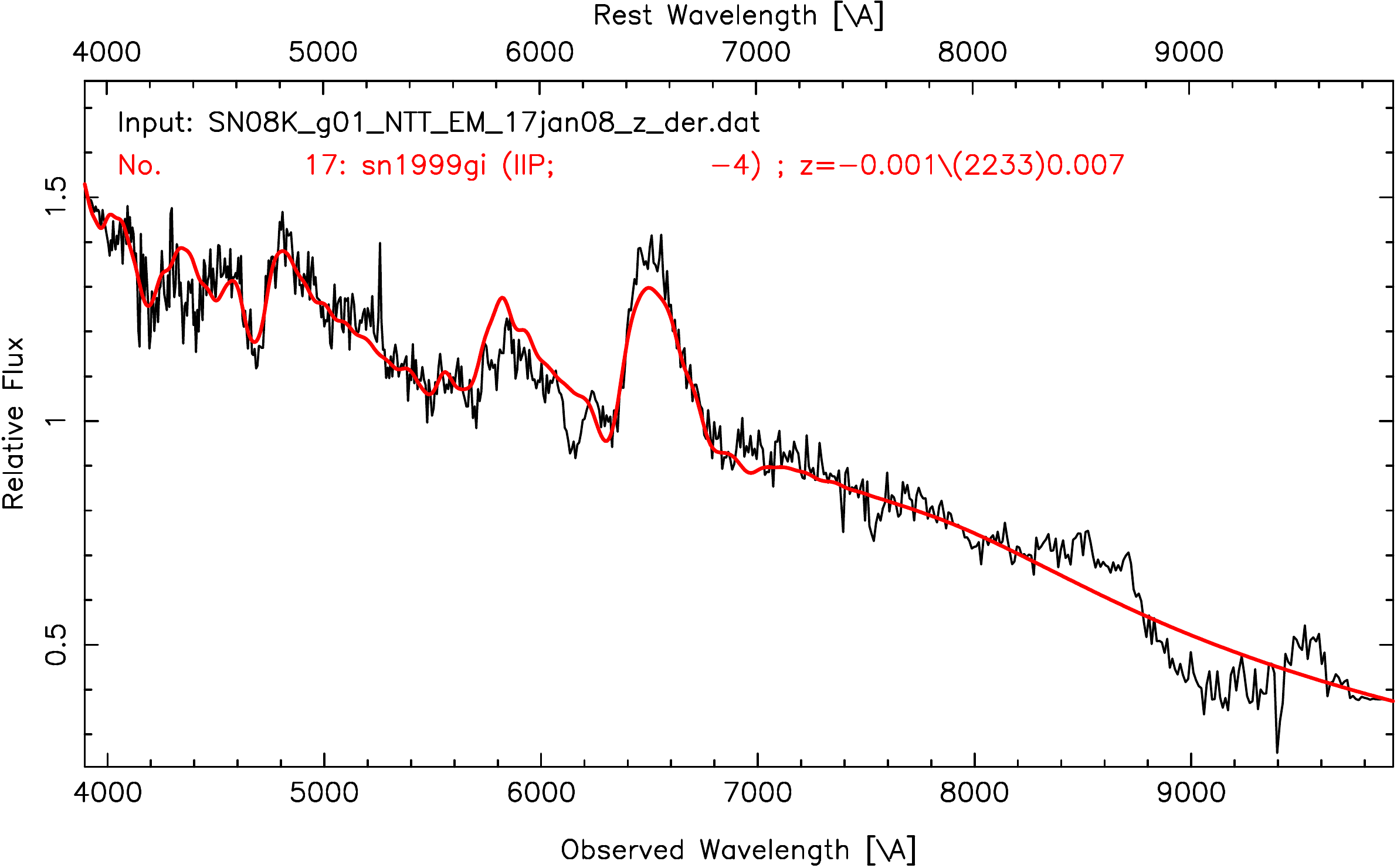}
\caption{Best spectral matching of SN~2008K using SNID. The plots show SN~2008K compared with 
SN~2005cs, SN~2004fc, SN~1999em, and SN~1999gi at 5, 9, 7 and 8 days from explosion.}
\end{figure}

\clearpage

\begin{figure}
\centering
\includegraphics[width=4.4cm]{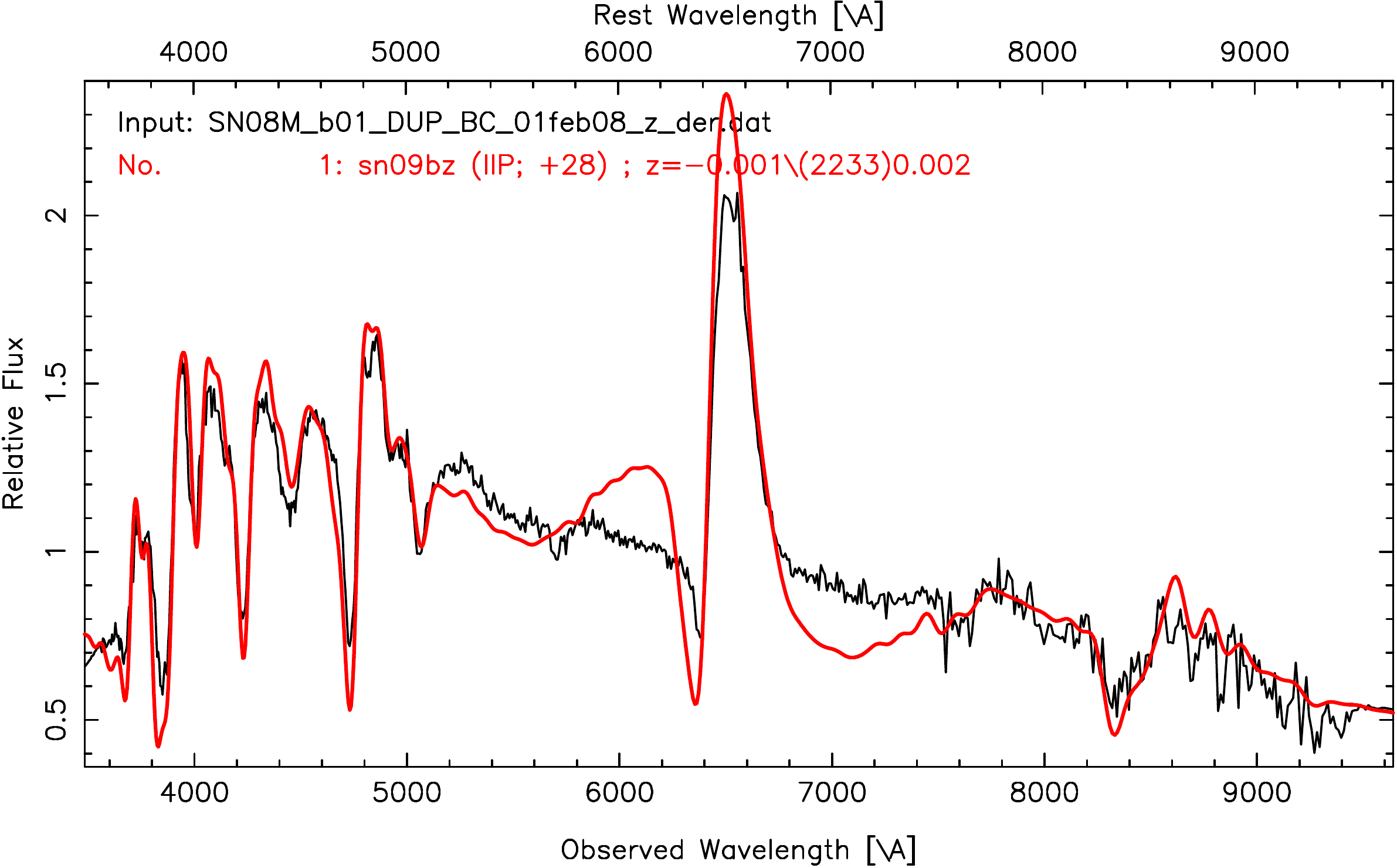}
\includegraphics[width=4.4cm]{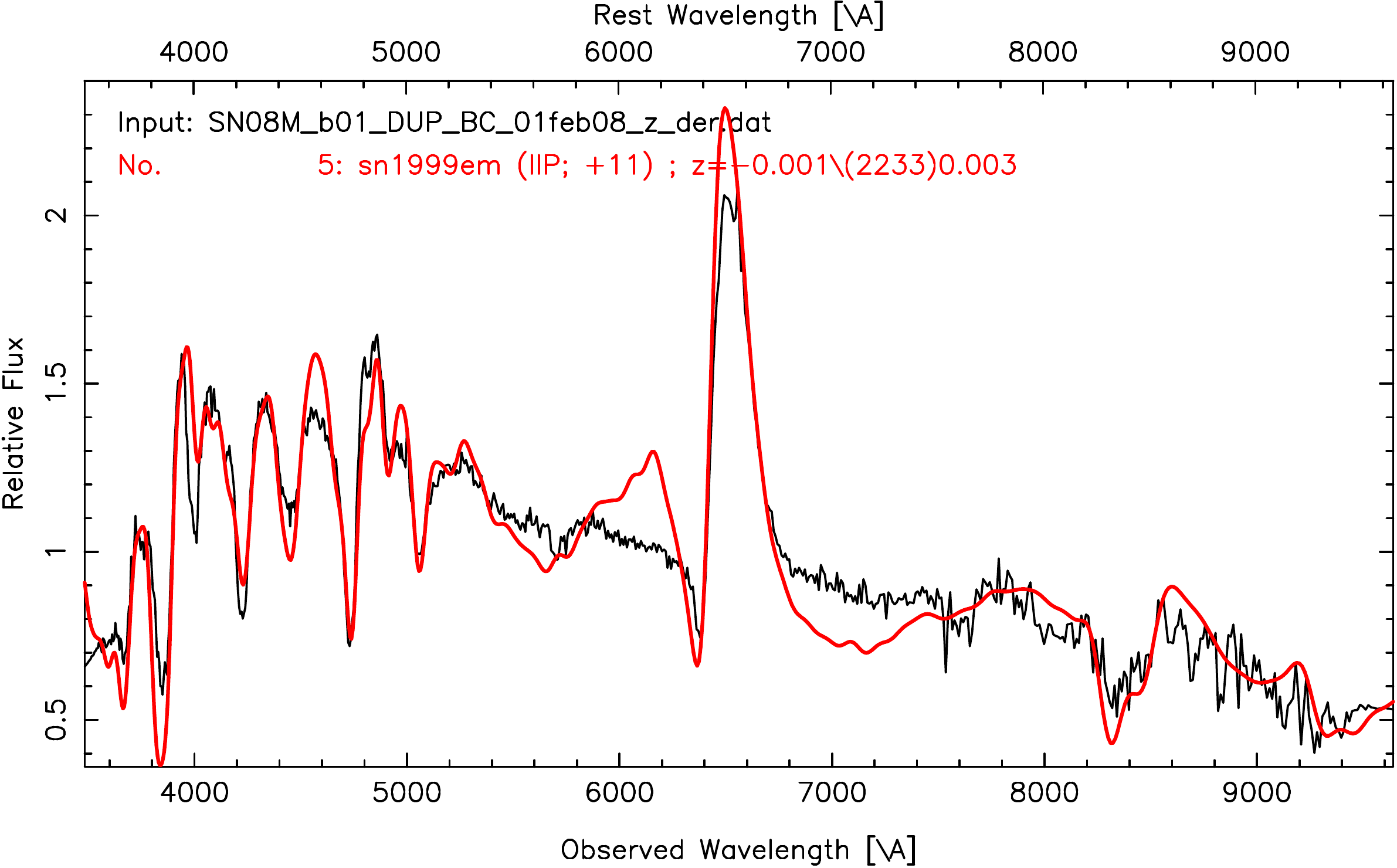}
\includegraphics[width=4.4cm]{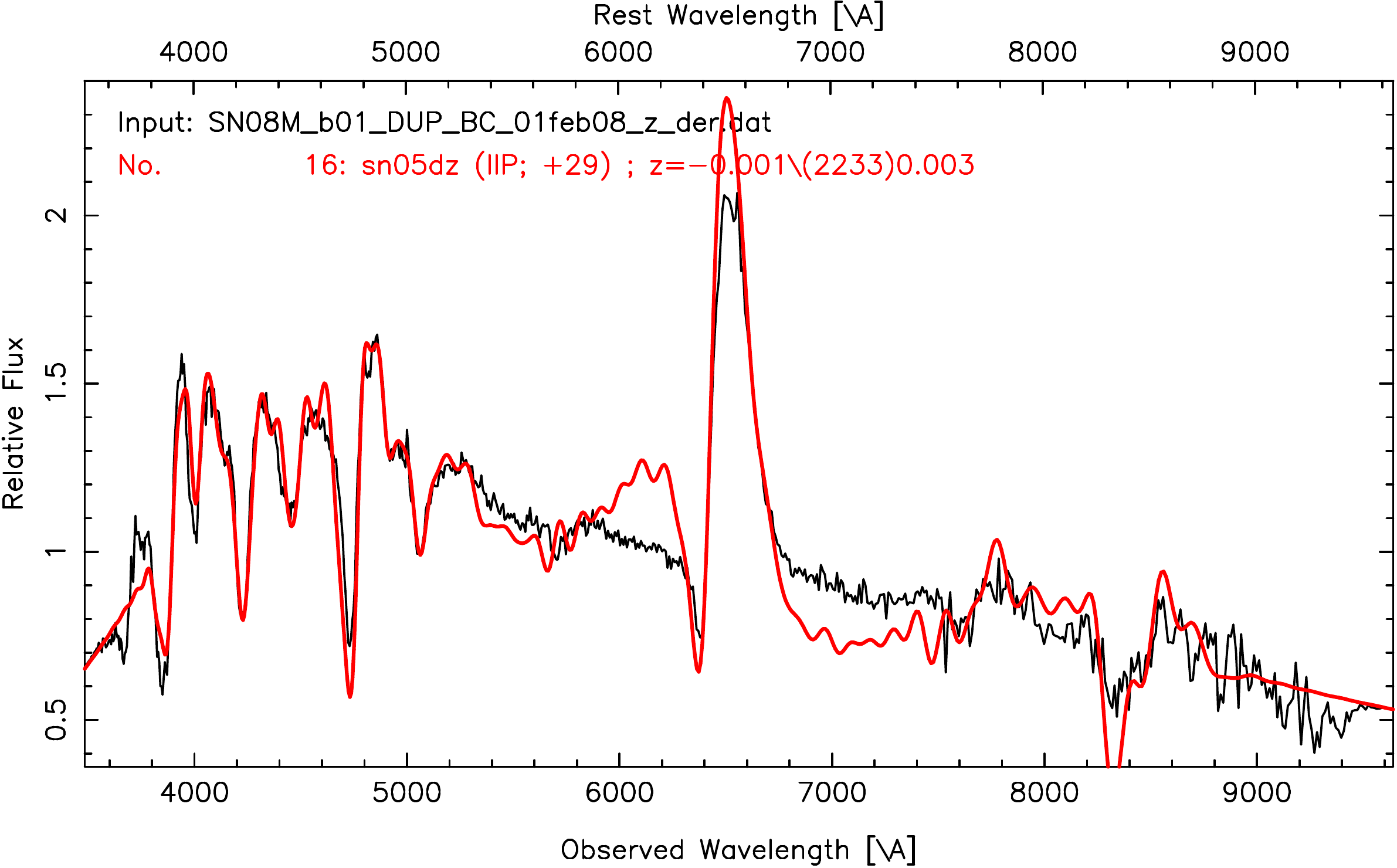}
\includegraphics[width=4.4cm]{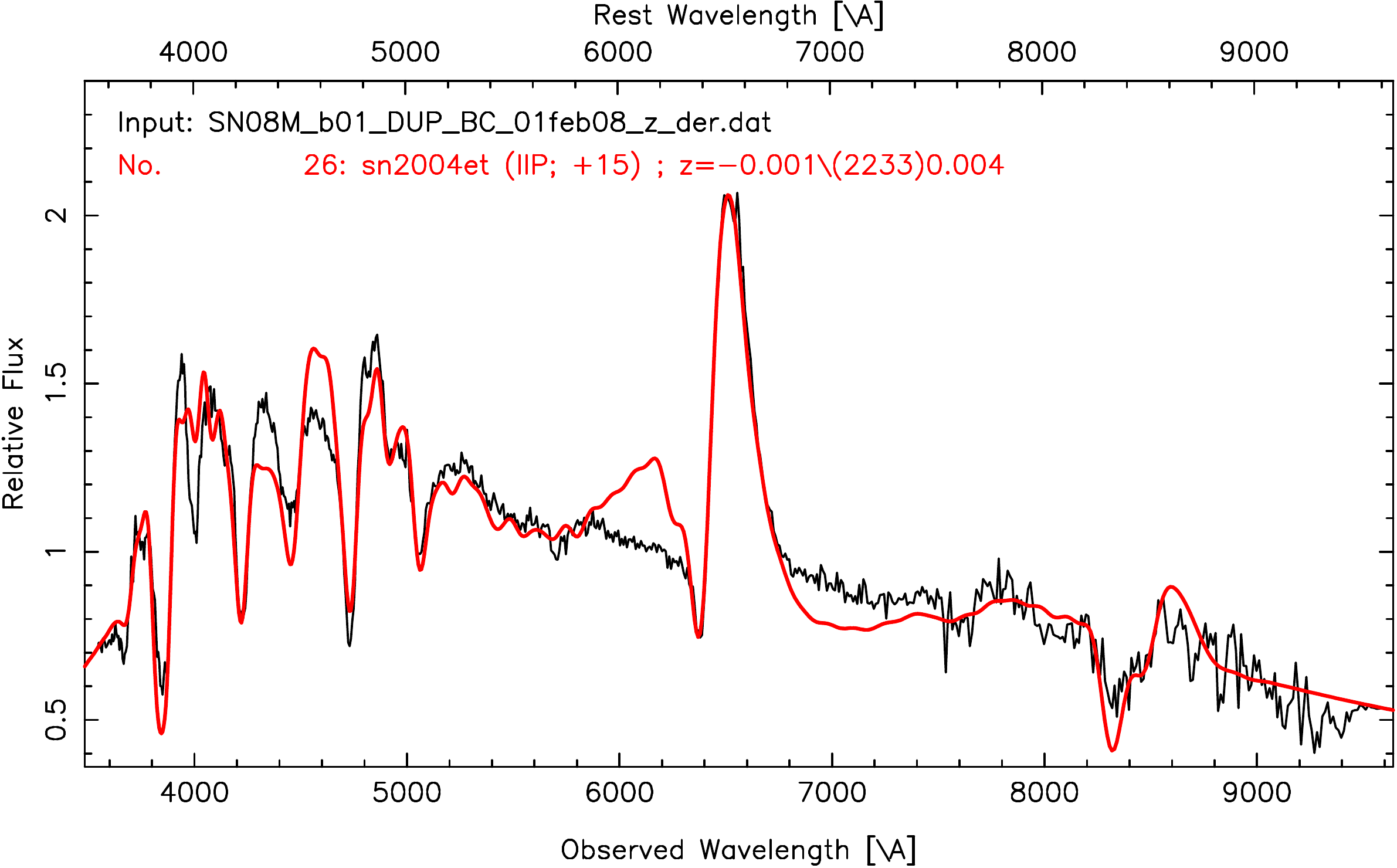}
\caption{Best spectral matching of SN~2008M using SNID. The plots show SN~2008M compared with 
SN~2009bz, SN~1999em, SN~2005dz, and SN~2004et at 28, 21, 24, and 26 days from explosion.}
\end{figure}

\begin{figure}
\centering
\includegraphics[width=4.4cm]{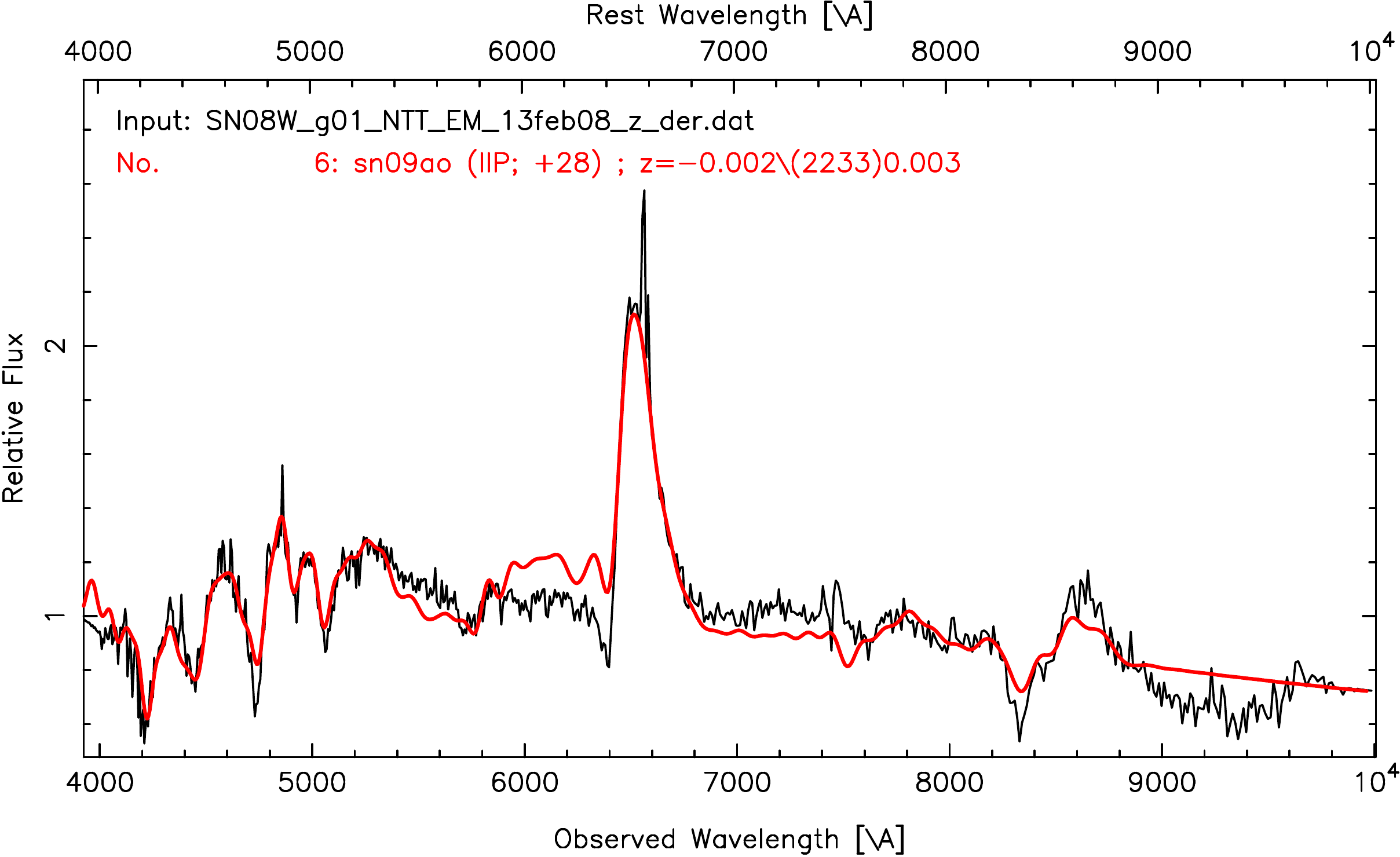}
\includegraphics[width=4.4cm]{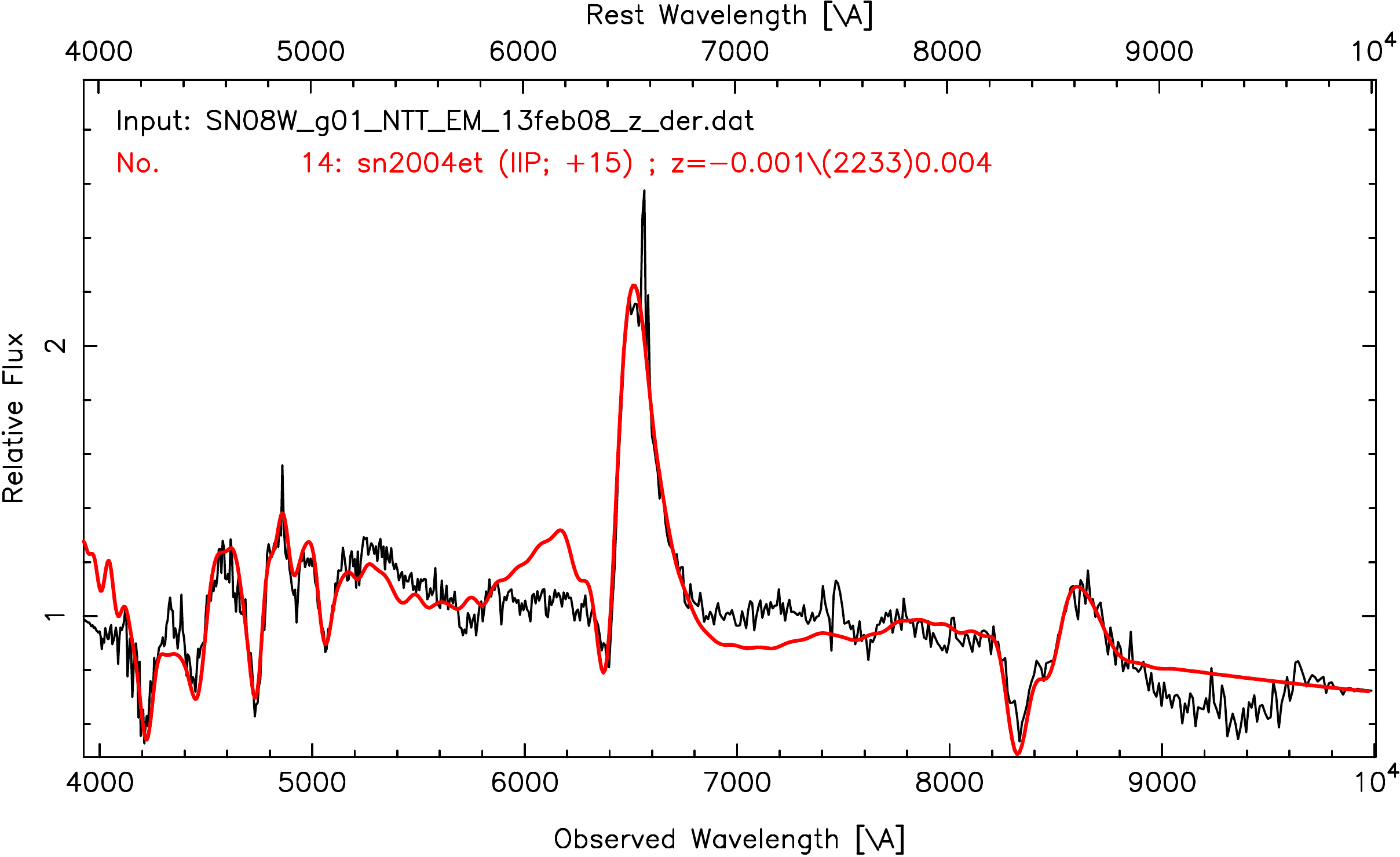}
\includegraphics[width=4.4cm]{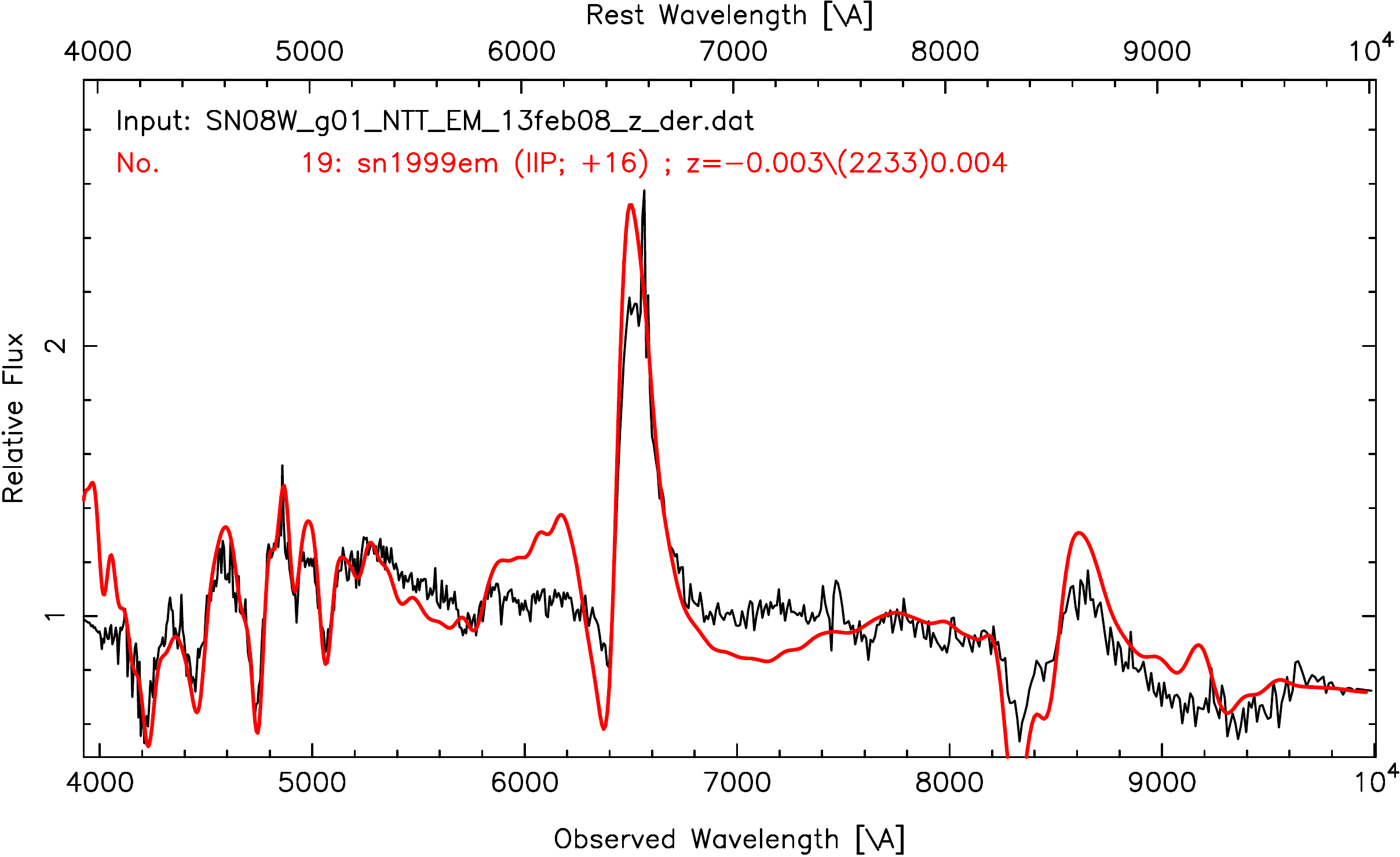}
\includegraphics[width=4.4cm]{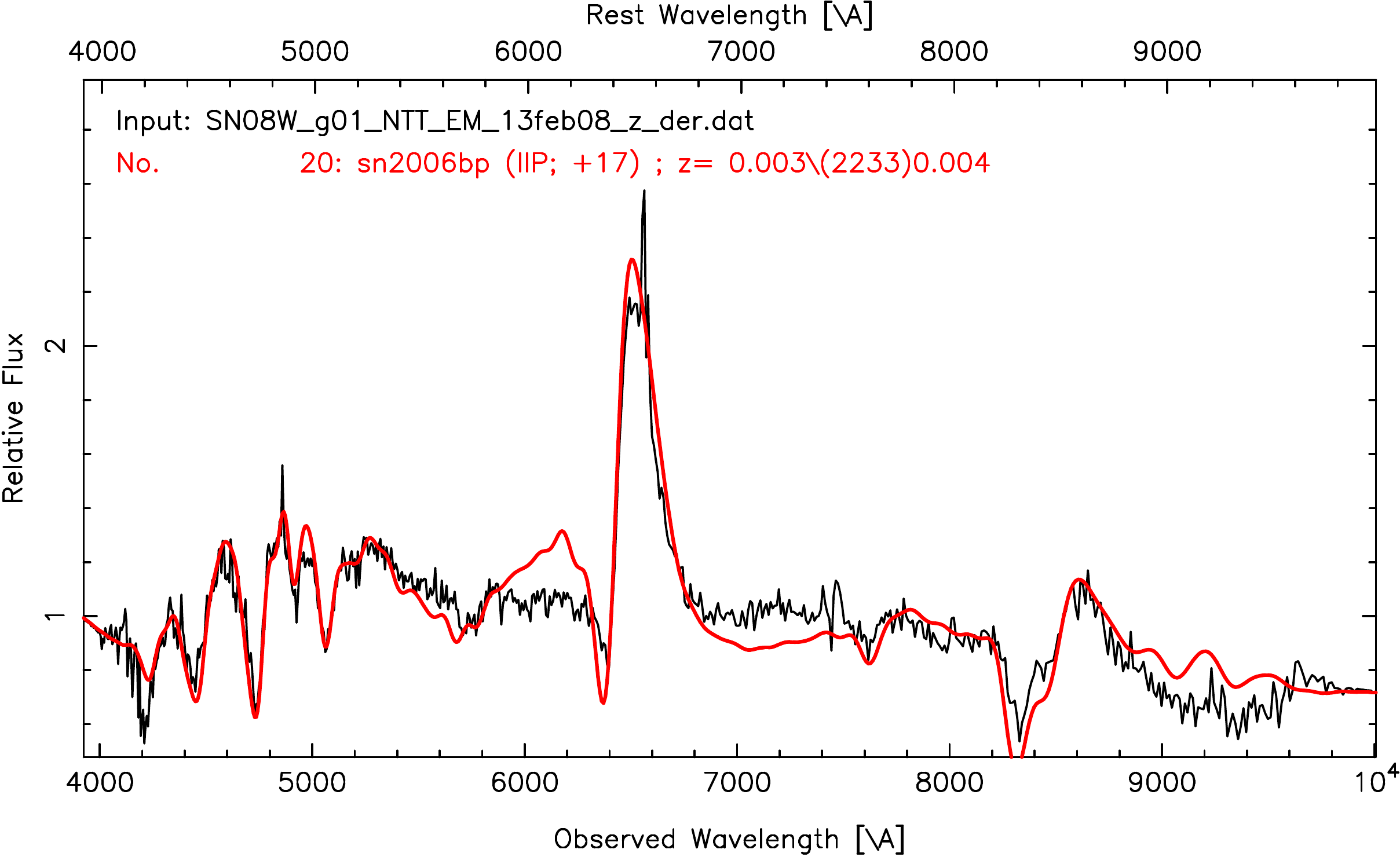}
\includegraphics[width=4.4cm]{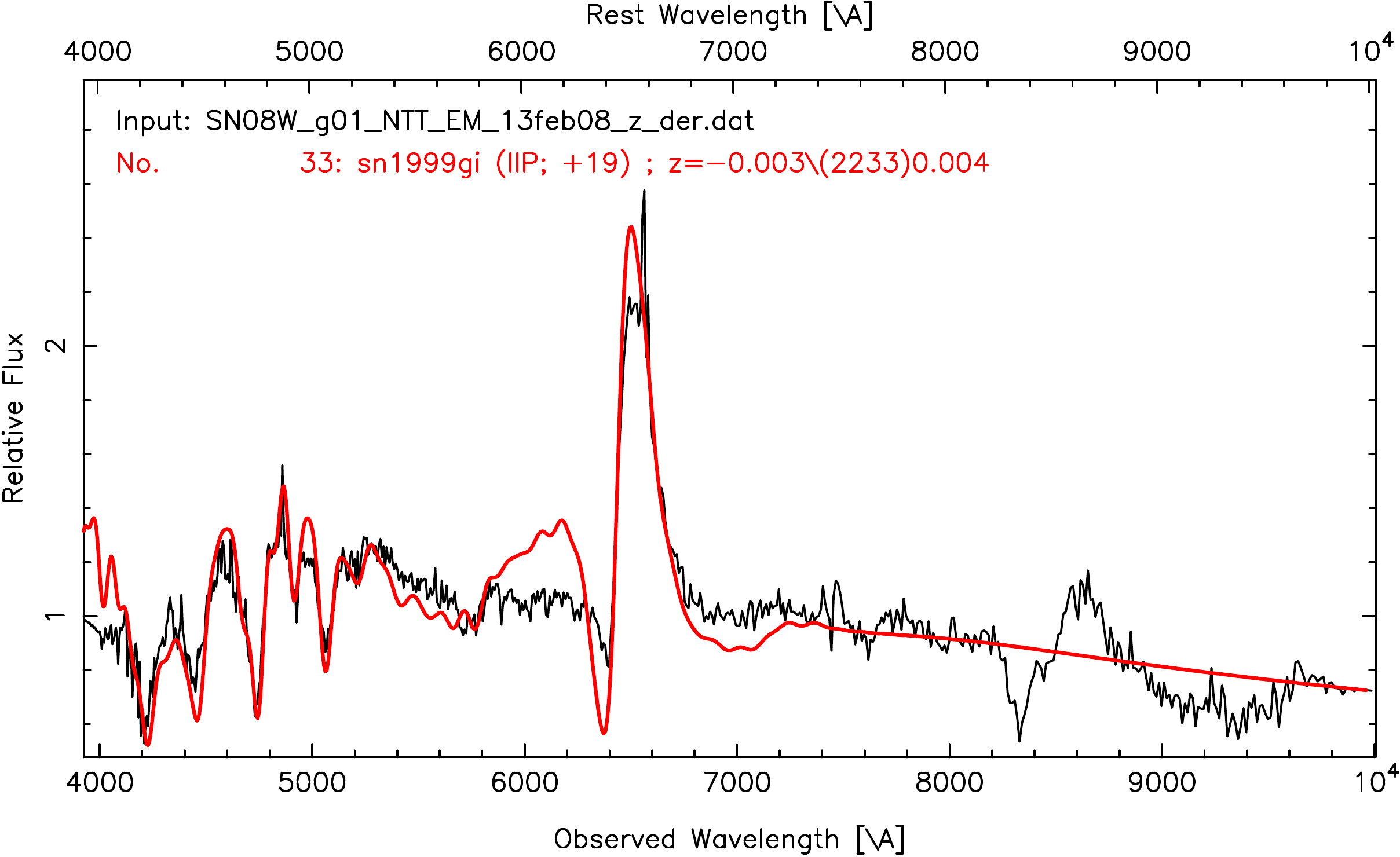}
\includegraphics[width=4.4cm]{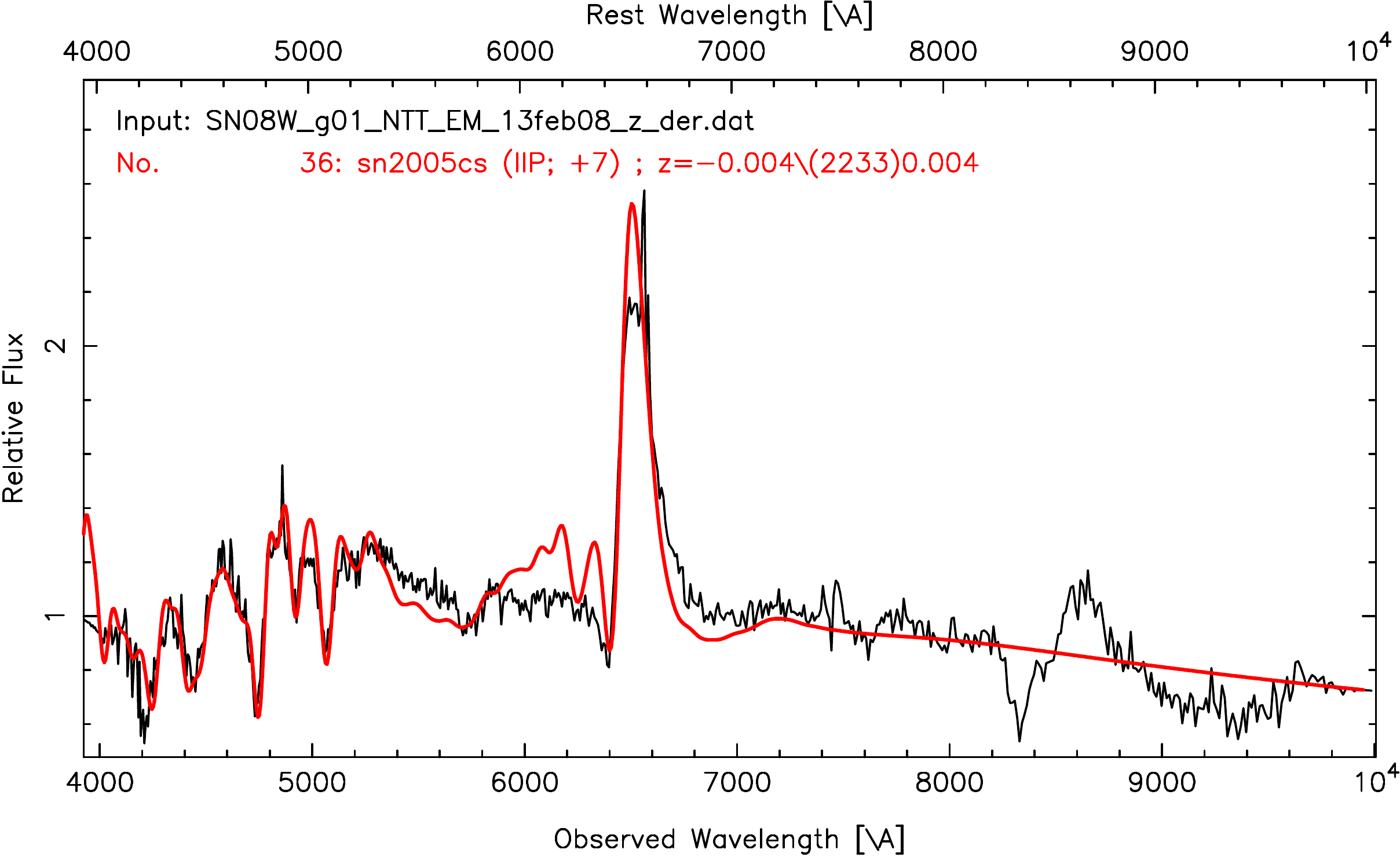}
\caption{Best spectral matching of SN~2008W using SNID. The plots show SN~2008W compared with 
SN~2009ao, SN~2004et, SN~1999em, SN~2006bp, SN~1999gi, and SN~2005cs at 28, 31, 26, 26, 31, and 17 days from explosion.}
\end{figure}

\clearpage

\begin{figure}
\centering
\includegraphics[width=4.4cm]{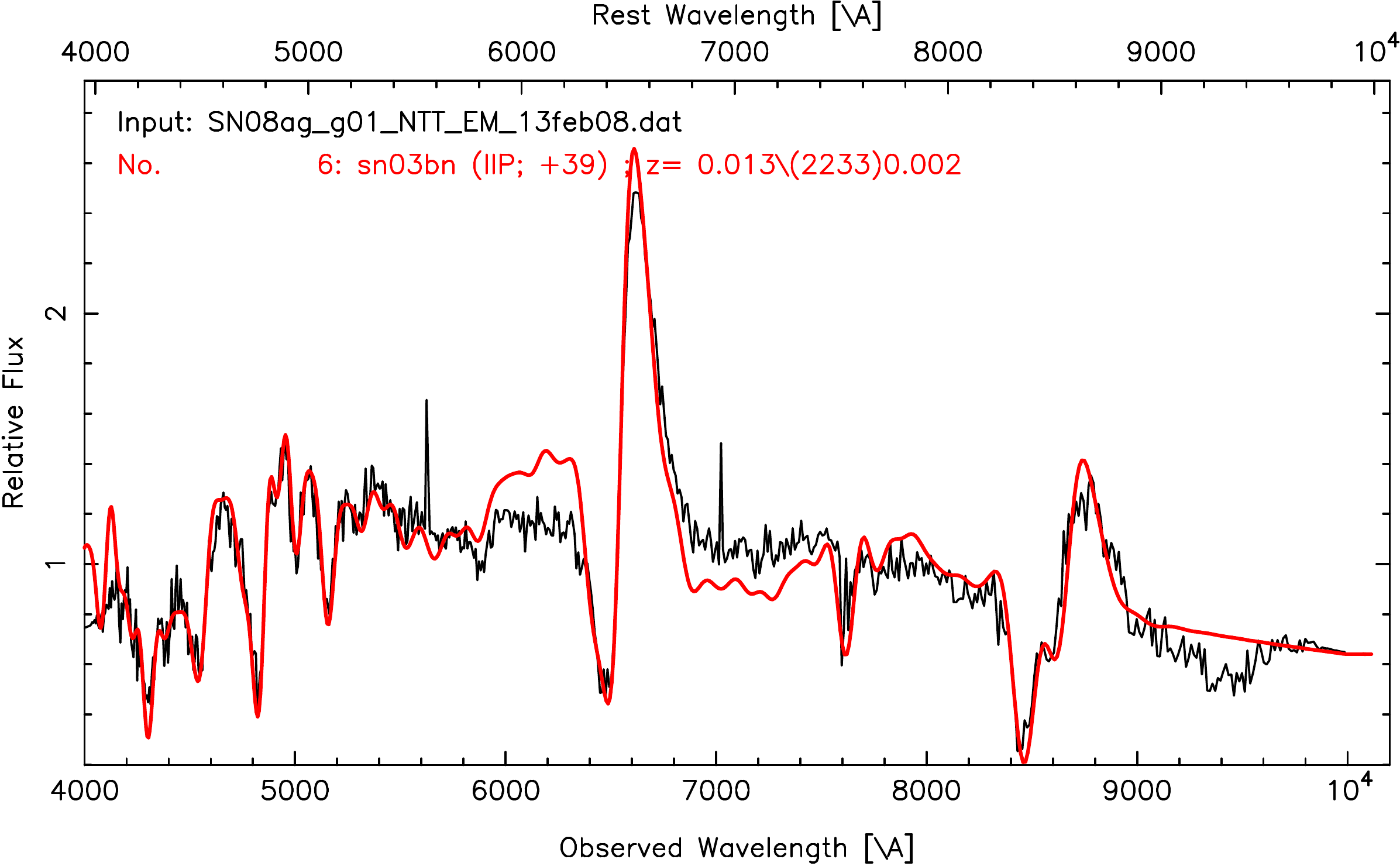}
\includegraphics[width=4.4cm]{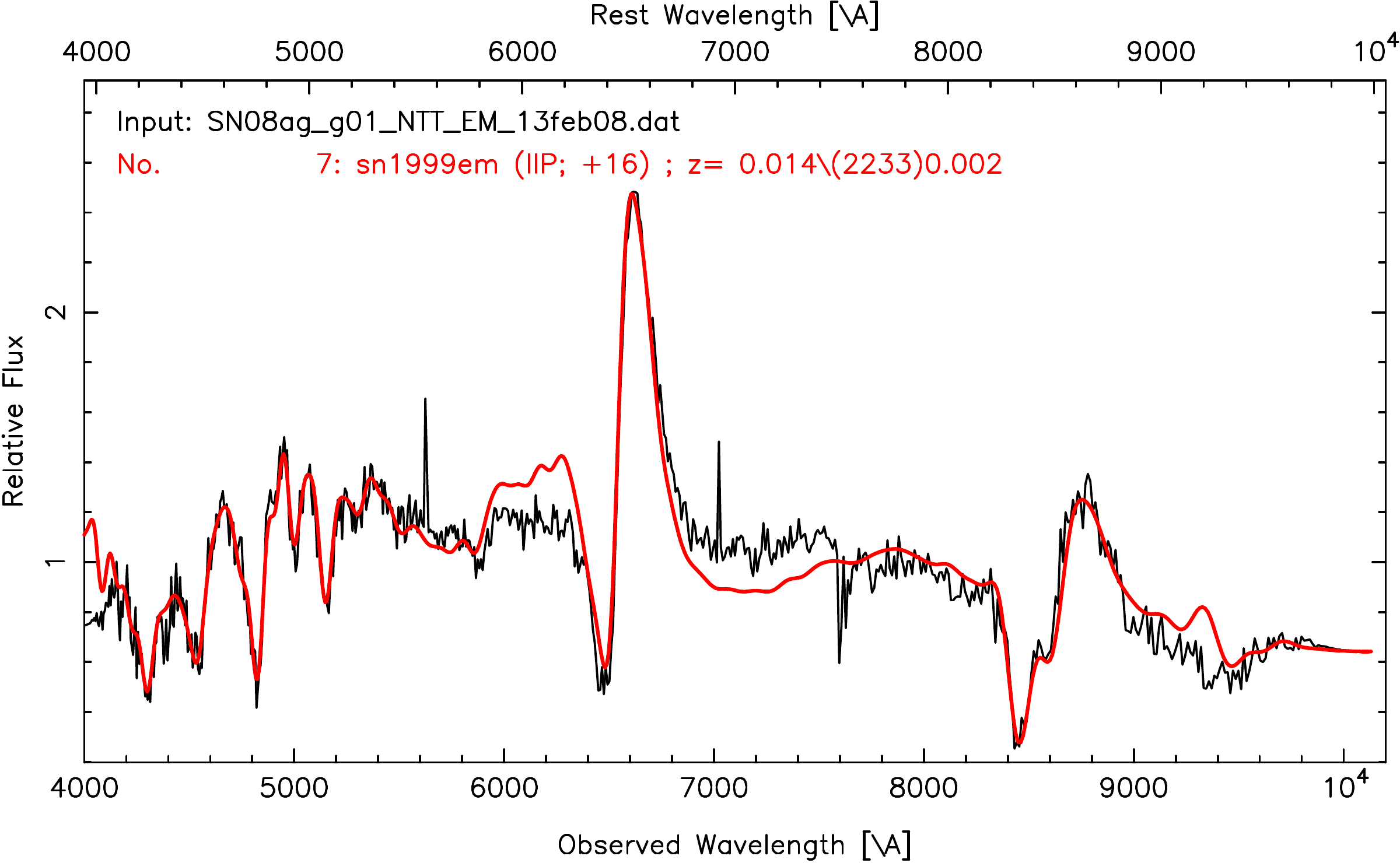}
\includegraphics[width=4.4cm]{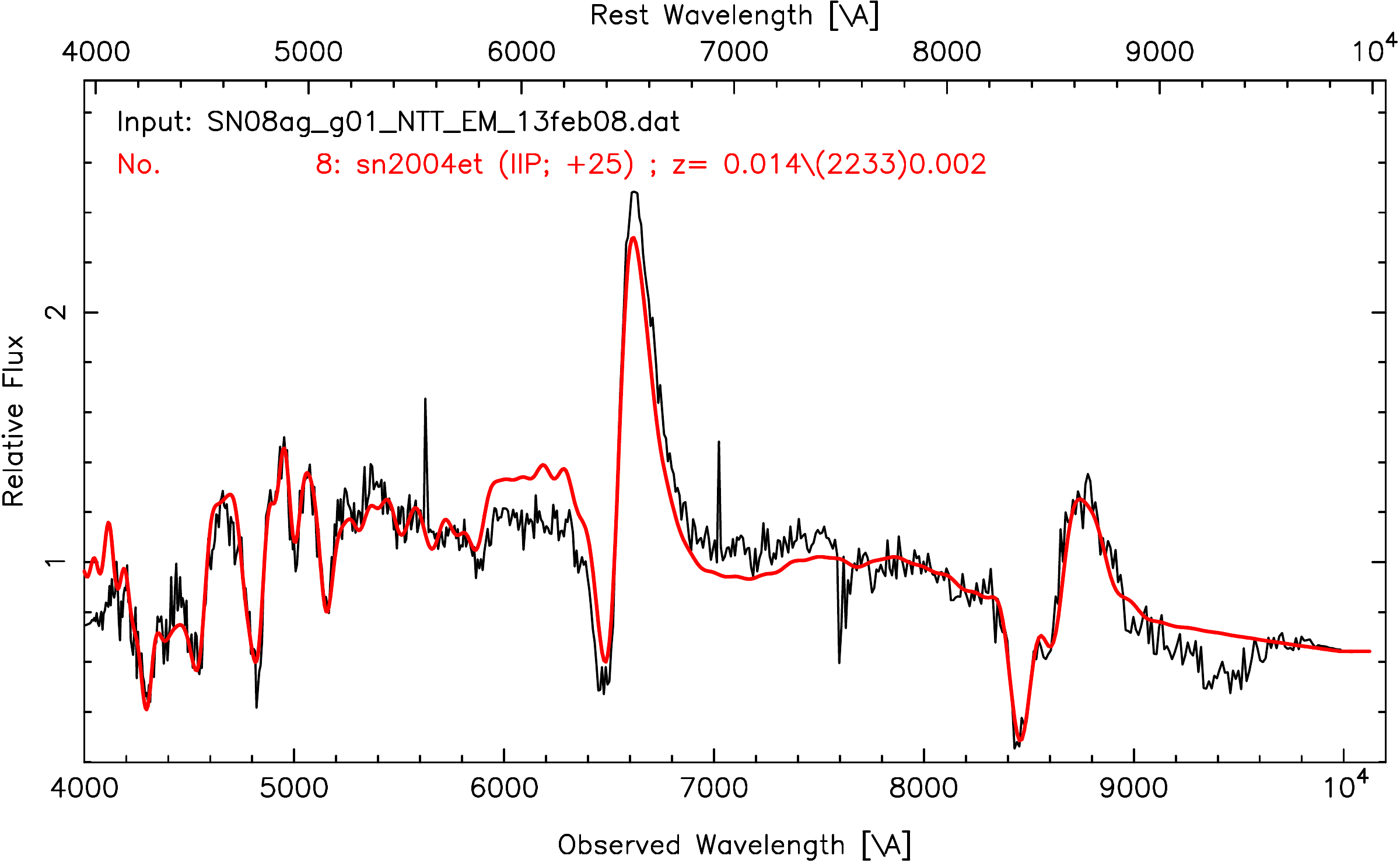}
\includegraphics[width=4.4cm]{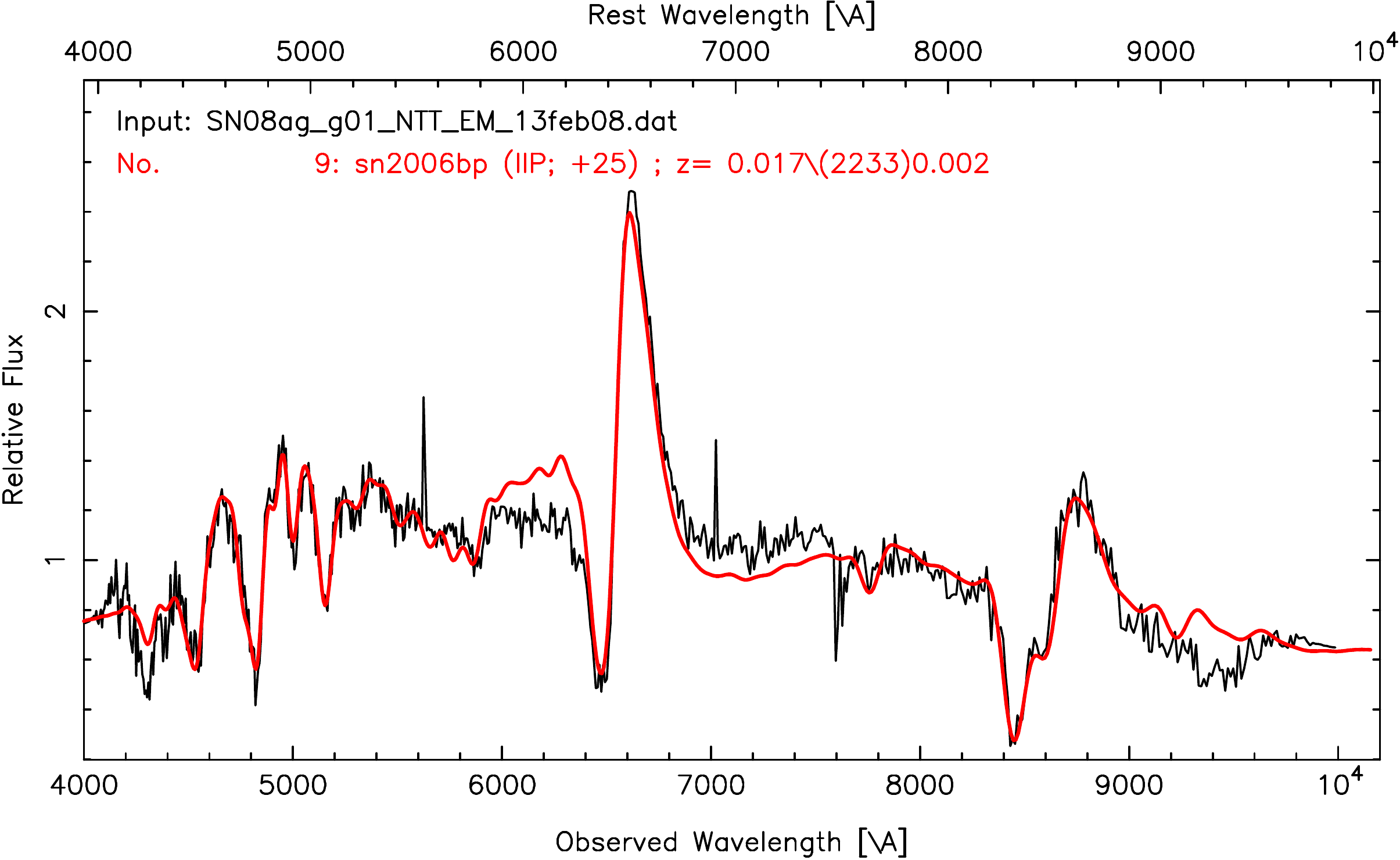}
\includegraphics[width=4.4cm]{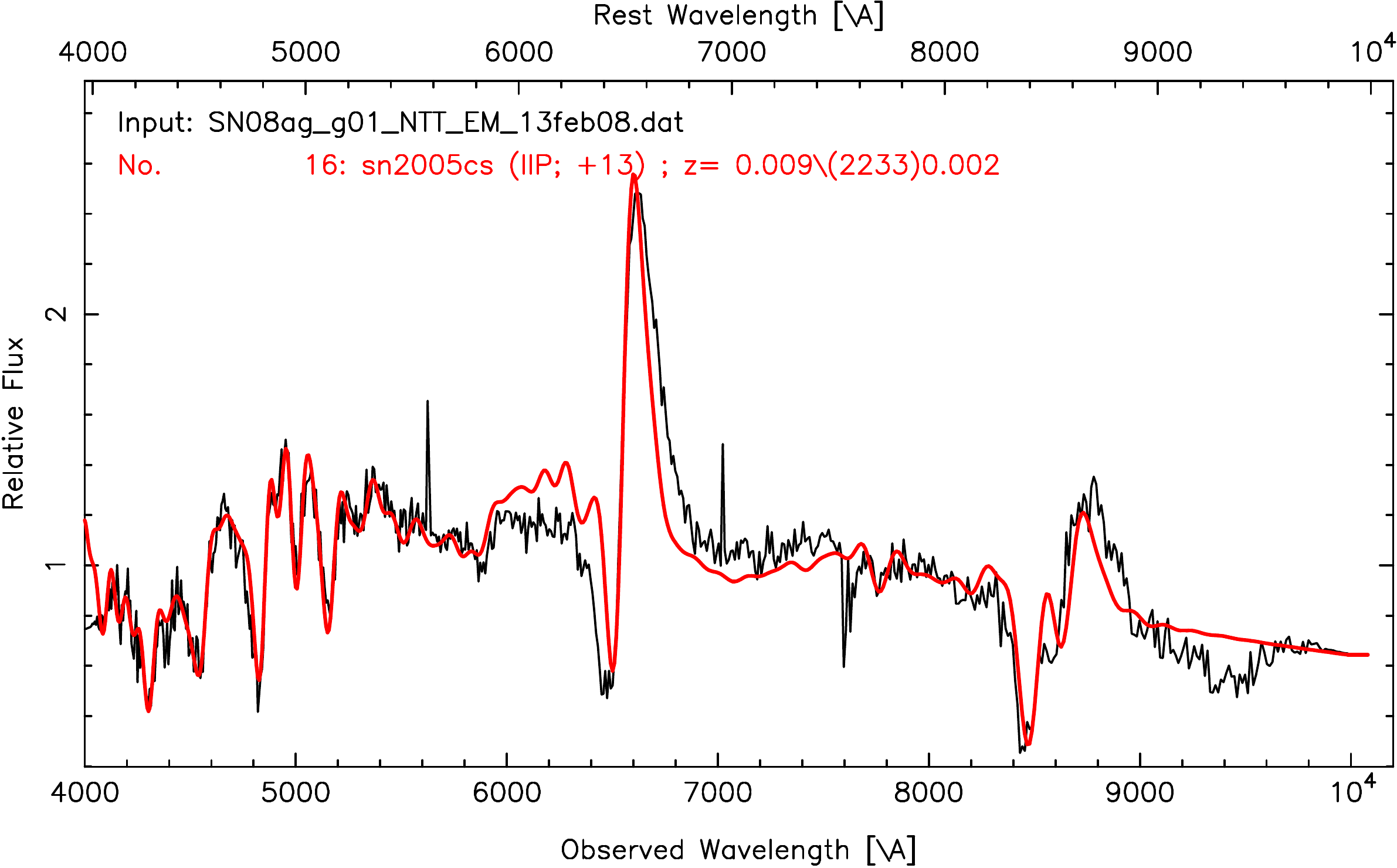}
\includegraphics[width=4.4cm]{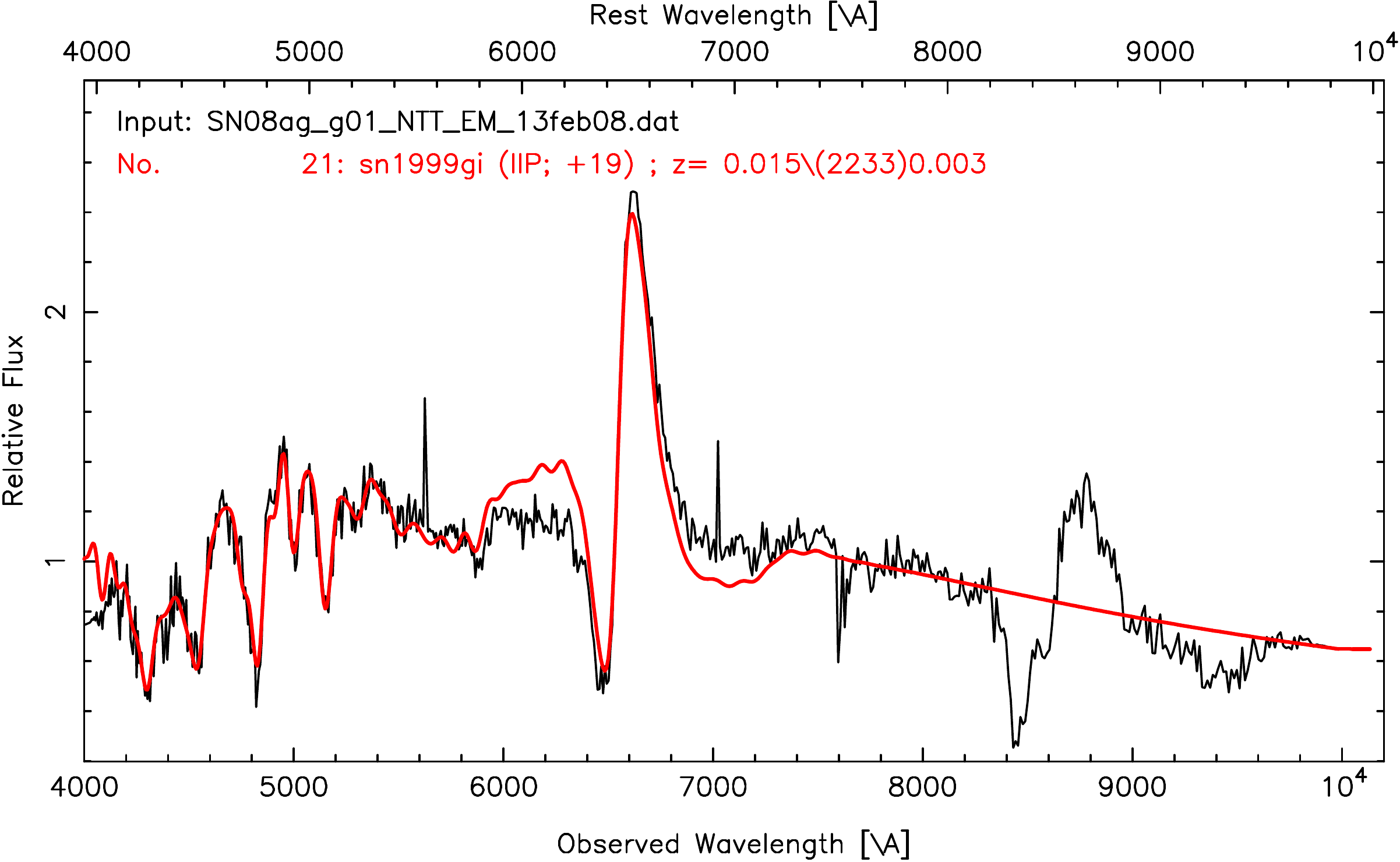}
\caption{Best spectral matching of SN~2008ag using SNID. The plots show SN~2008ag compared with 
SN~2003bn, SN~1999em, SN~2004et, SN~2006bp, SN~2005cs, and SN~1999gi at 39, 26, 41, 34, 19, and 31 days from explosion.}
\end{figure}

\begin{figure}
\centering
\includegraphics[width=4.4cm]{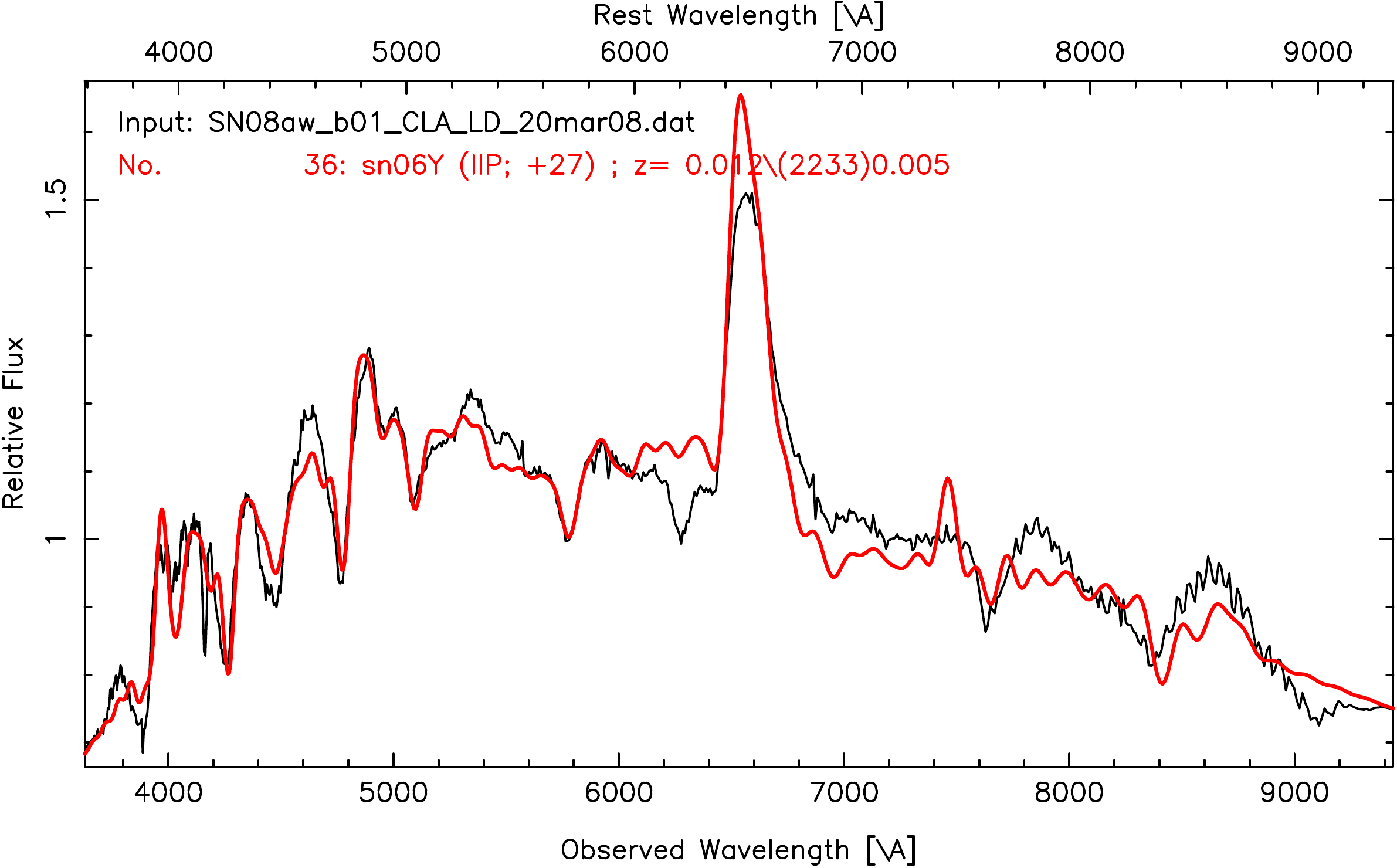}
\includegraphics[width=4.4cm]{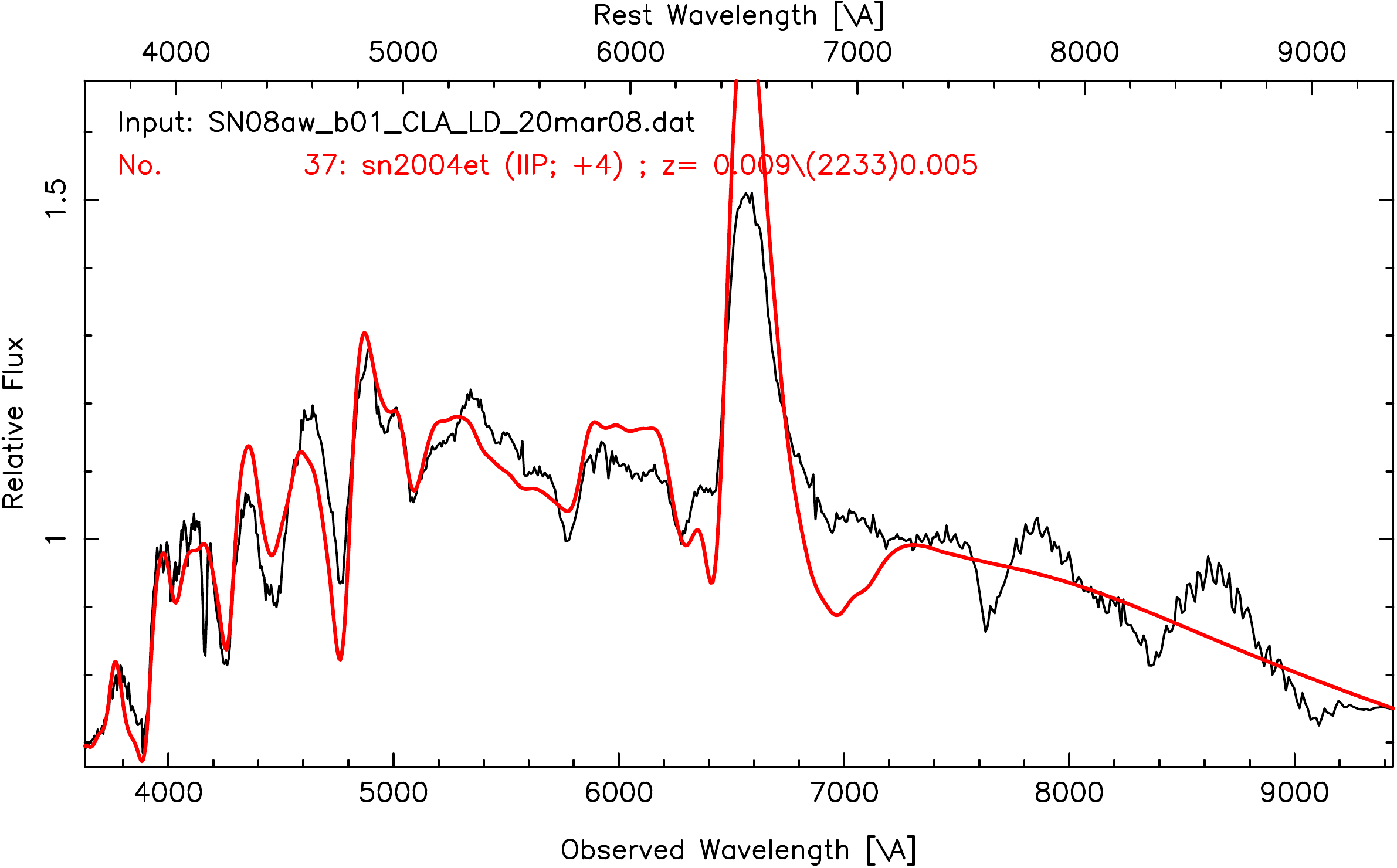}
\caption{Best spectral matching of SN~2008aw using SNID. The plots show SN~2008aw compared with 
SN~2006Y and SN~2004et at 27 and 20 days from explosion.}
\end{figure}

\begin{figure}
\centering
\includegraphics[width=4.4cm]{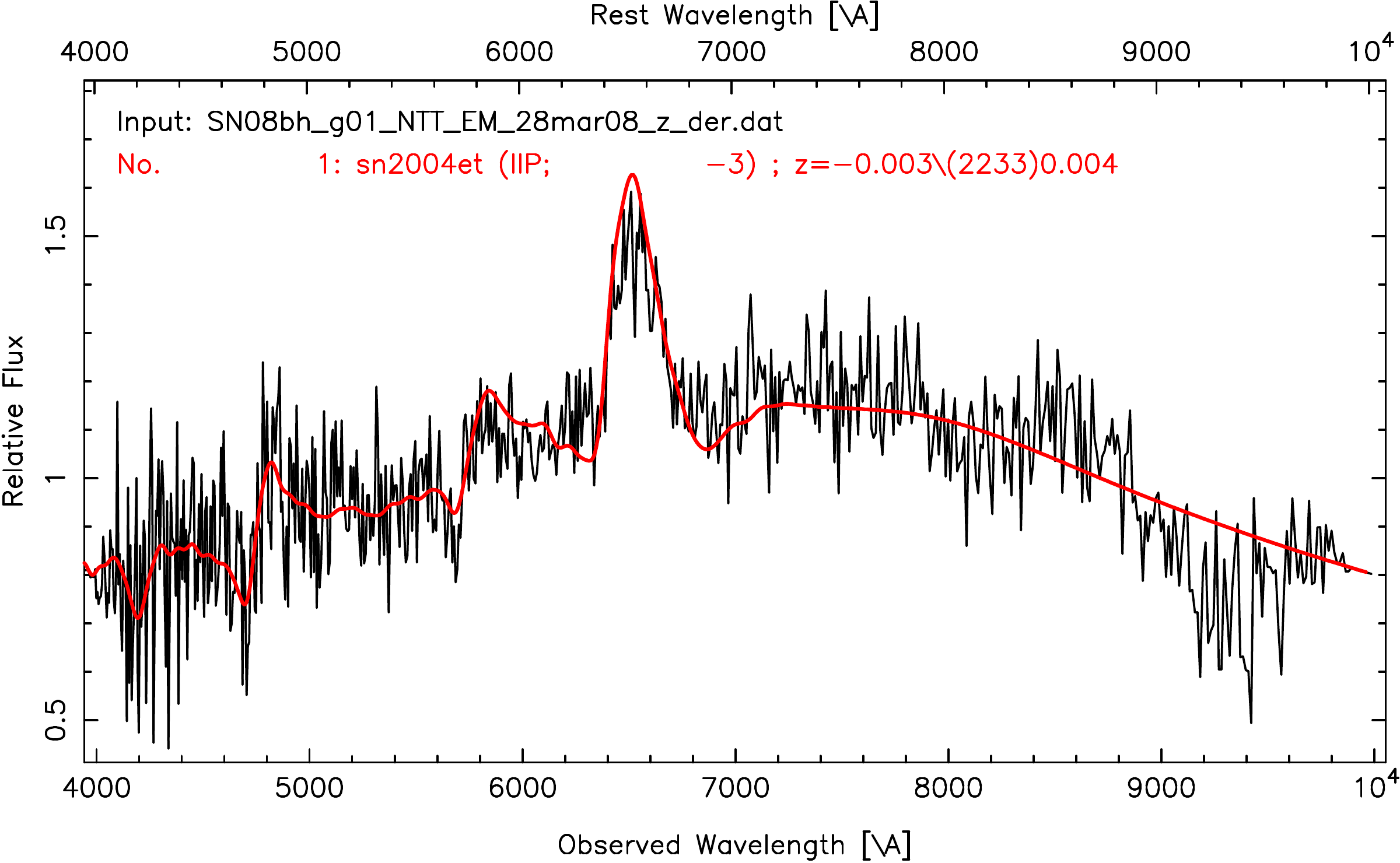}
\includegraphics[width=4.4cm]{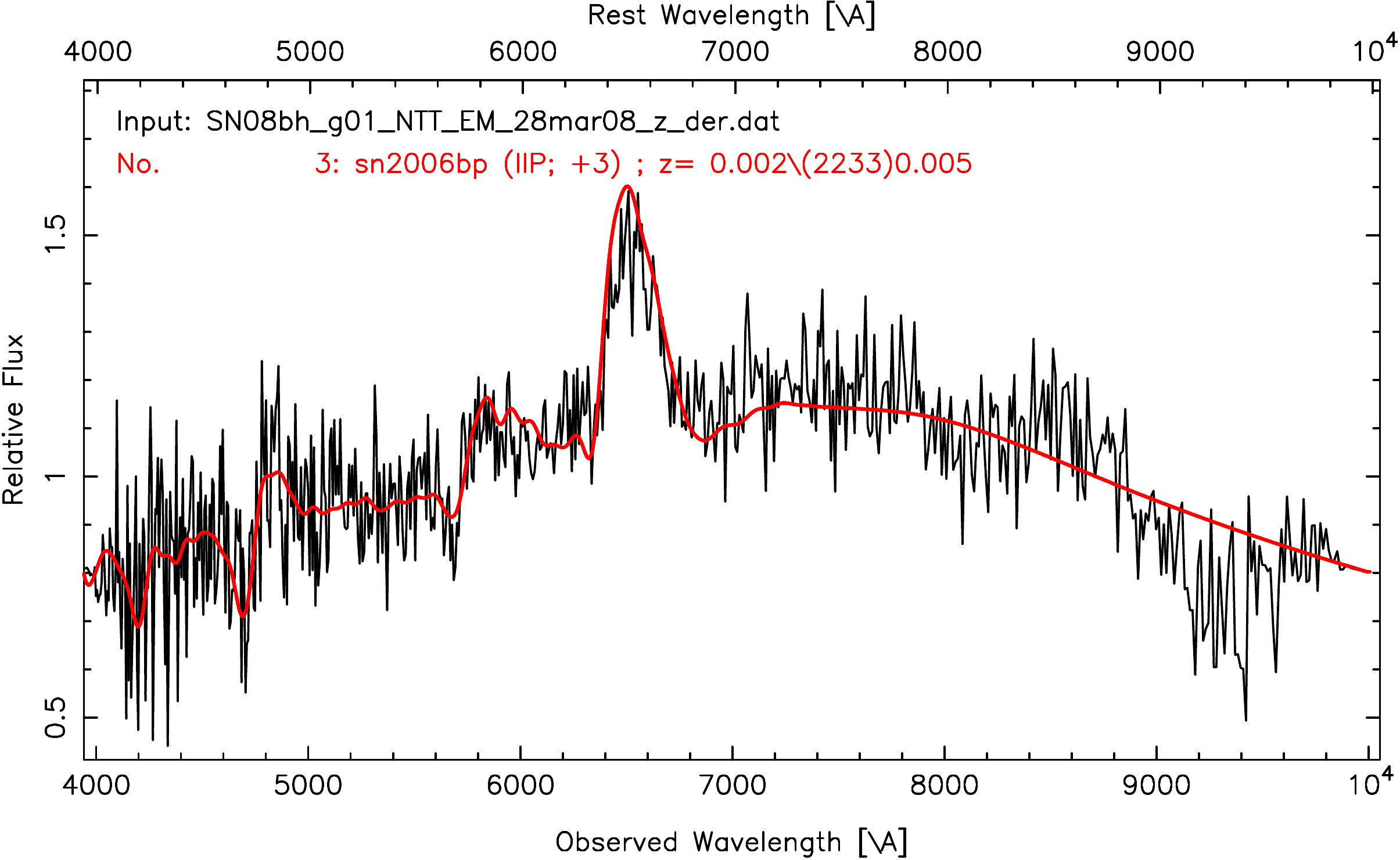}
\caption{Best spectral matching of SN~2008bh using SNID. The plots show SN~2008bh compared with 
SN~2004et and SN~2006bp at 13 and 12 days from explosion.}
\end{figure}

\clearpage

\begin{figure}
\centering
\includegraphics[width=4.4cm]{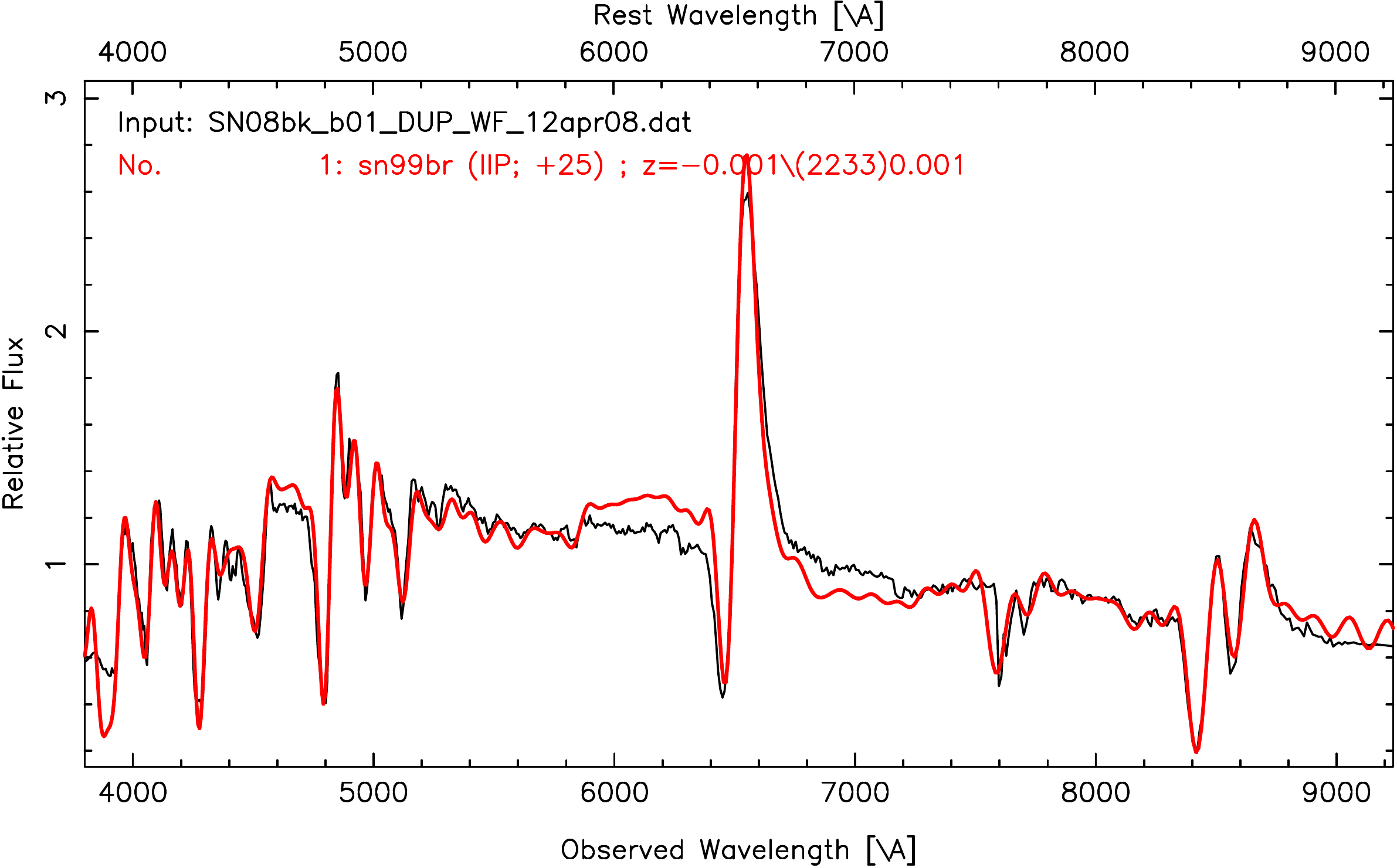}
\includegraphics[width=4.4cm]{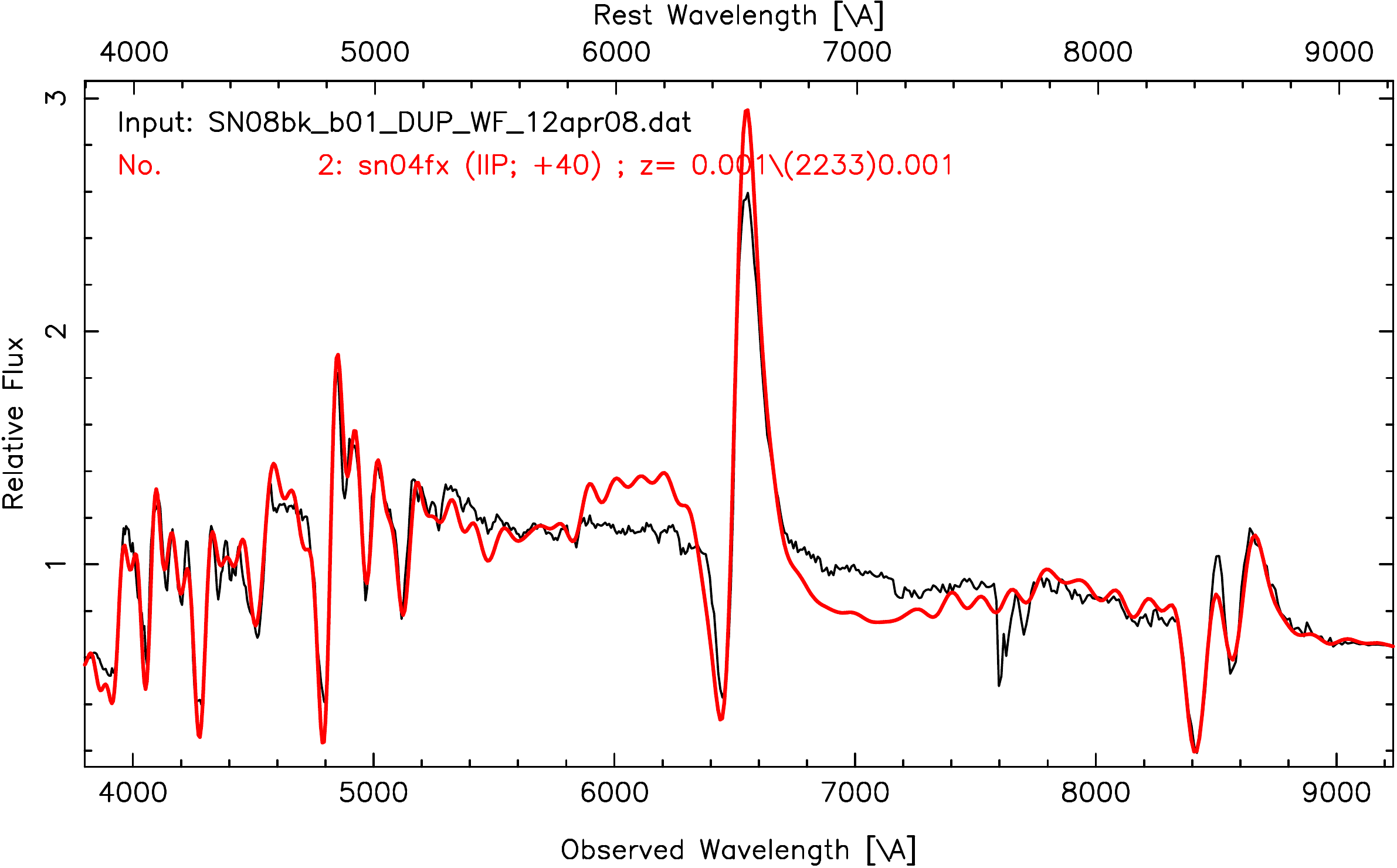}
\includegraphics[width=4.4cm]{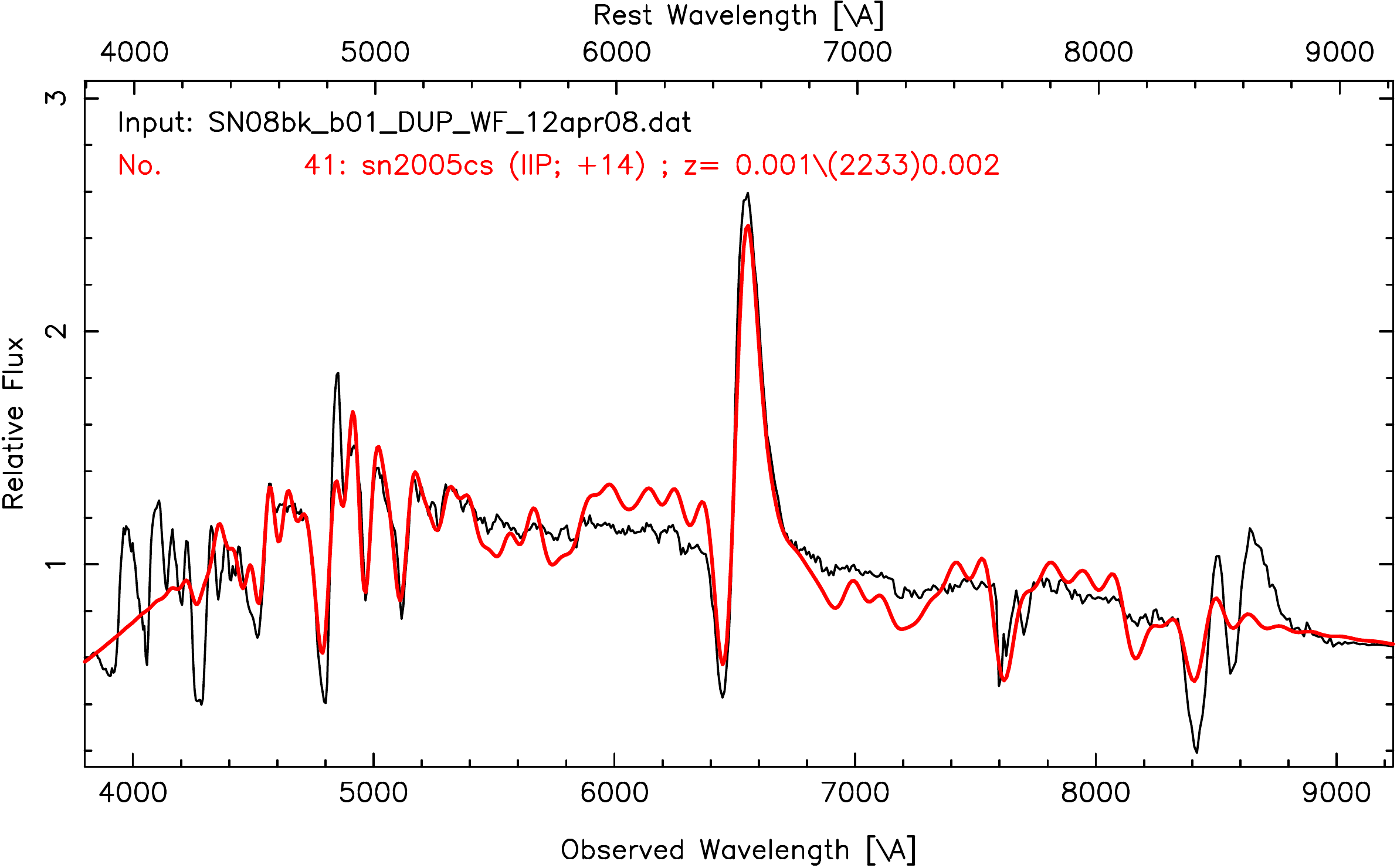}
\caption{Best spectral matching of SN~2008bk using SNID. The plots show SN~2008bk compared with 
SN~1999br, SN~2004fx, and SN~2005cs at 25, 40, and 20 days from explosion.}
\end{figure}

\begin{figure}
\centering
\includegraphics[width=4.4cm]{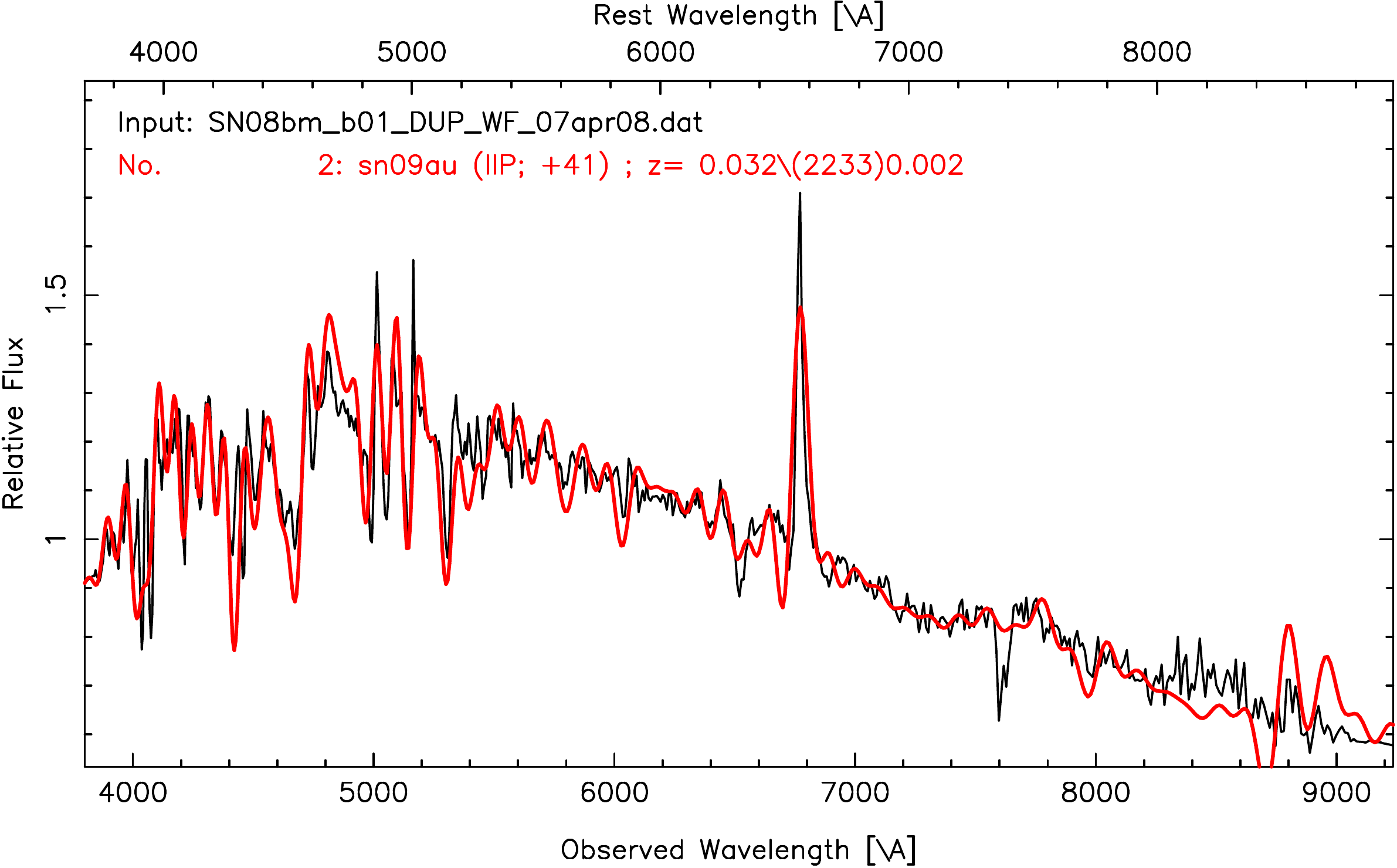}
\caption{Best spectral matching of SN~2008bm using SNID. The plots show SN~2008bm compared with 
SN~2009au at 41 days from explosion.}
\end{figure}

\begin{figure}
\centering
\includegraphics[width=4.4cm]{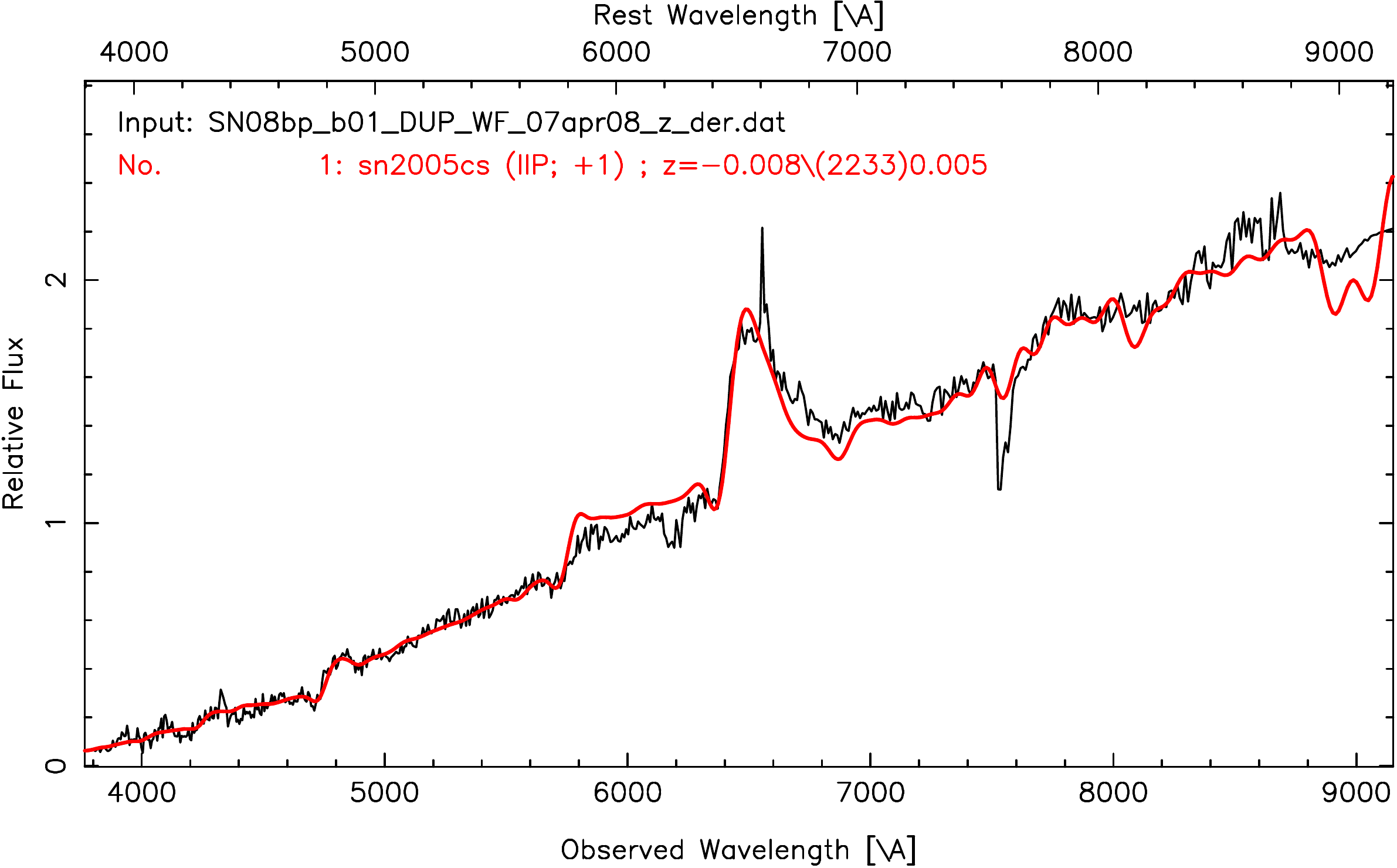}
\includegraphics[width=4.4cm]{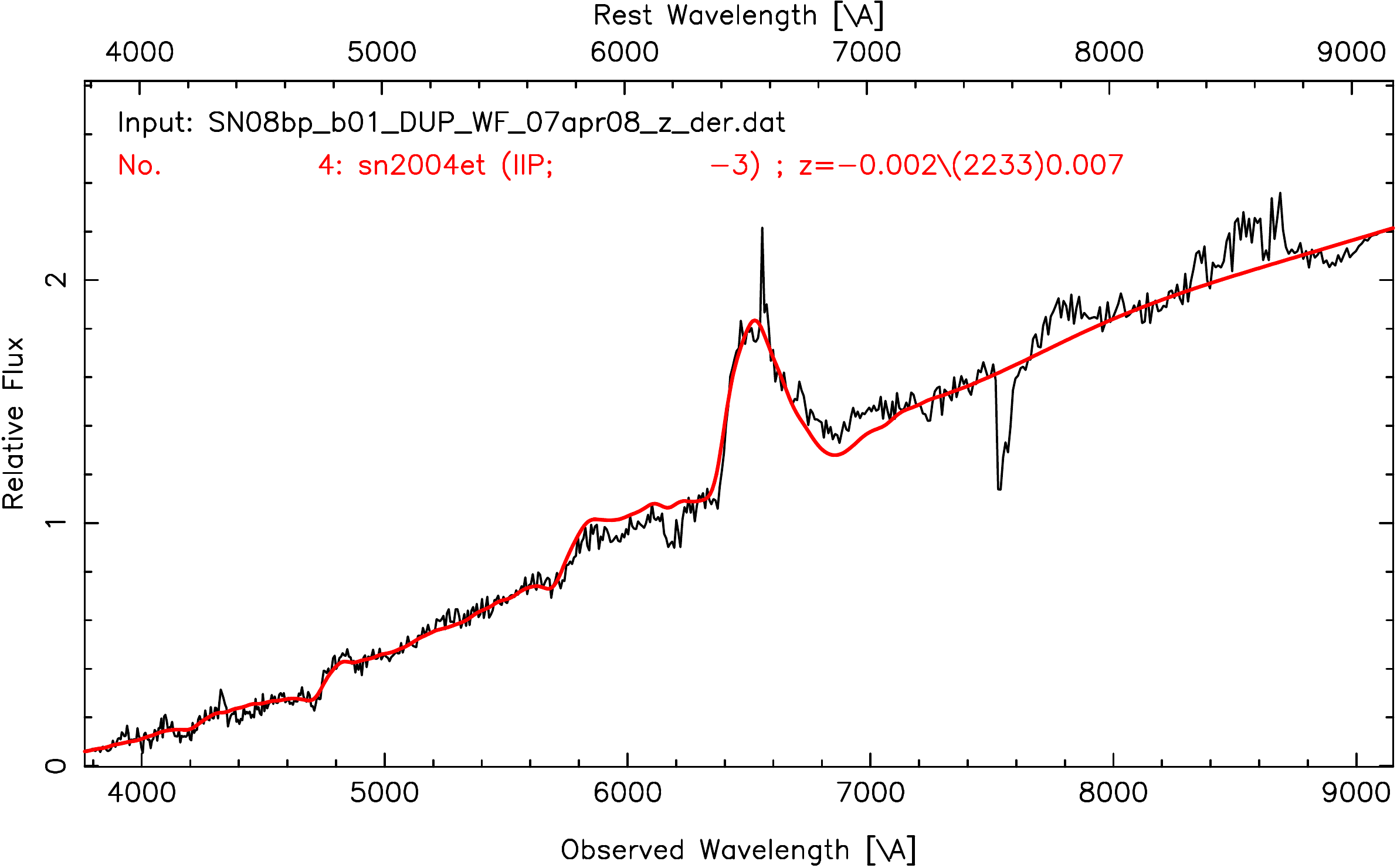}
\includegraphics[width=4.4cm]{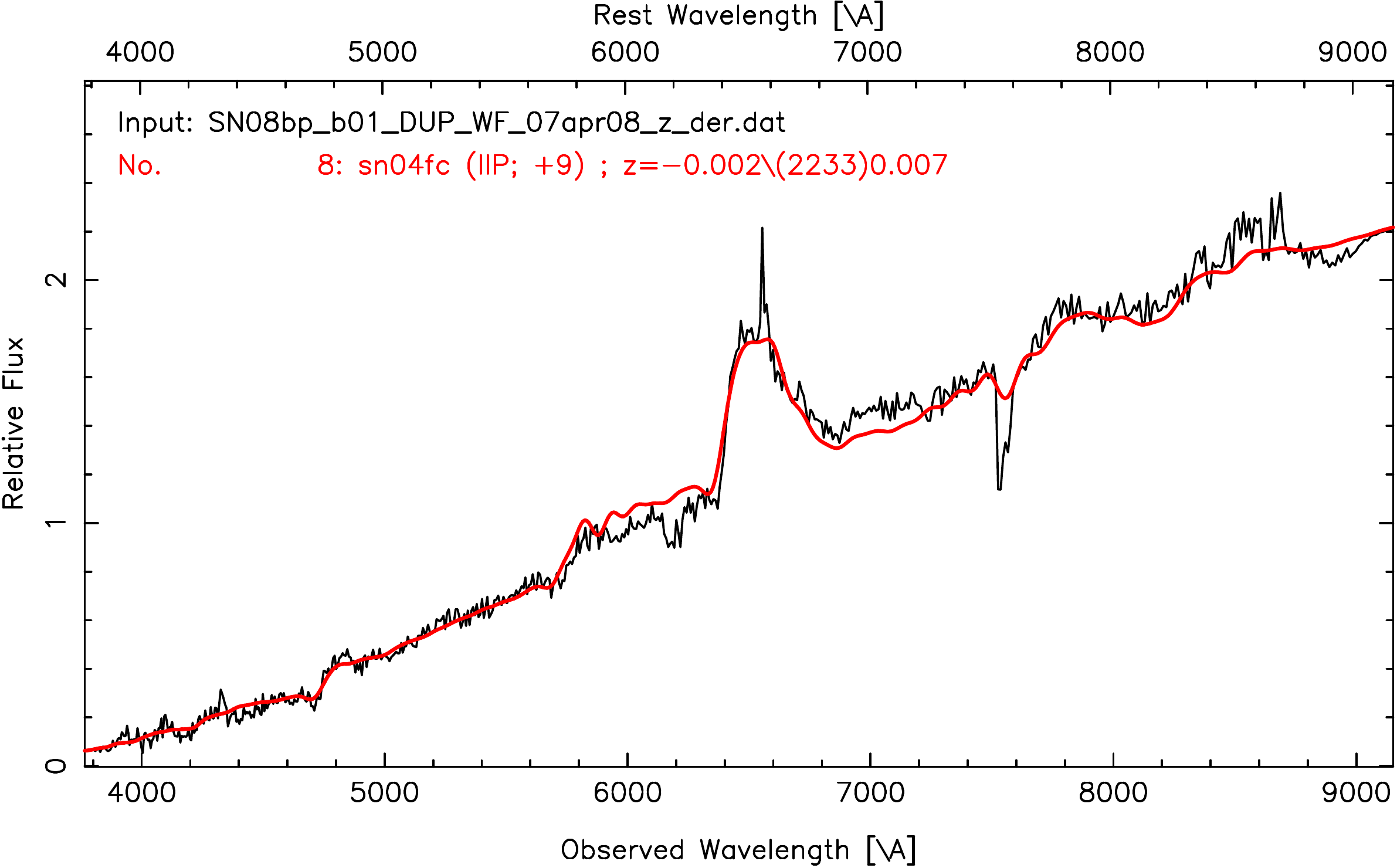}
\includegraphics[width=4.4cm]{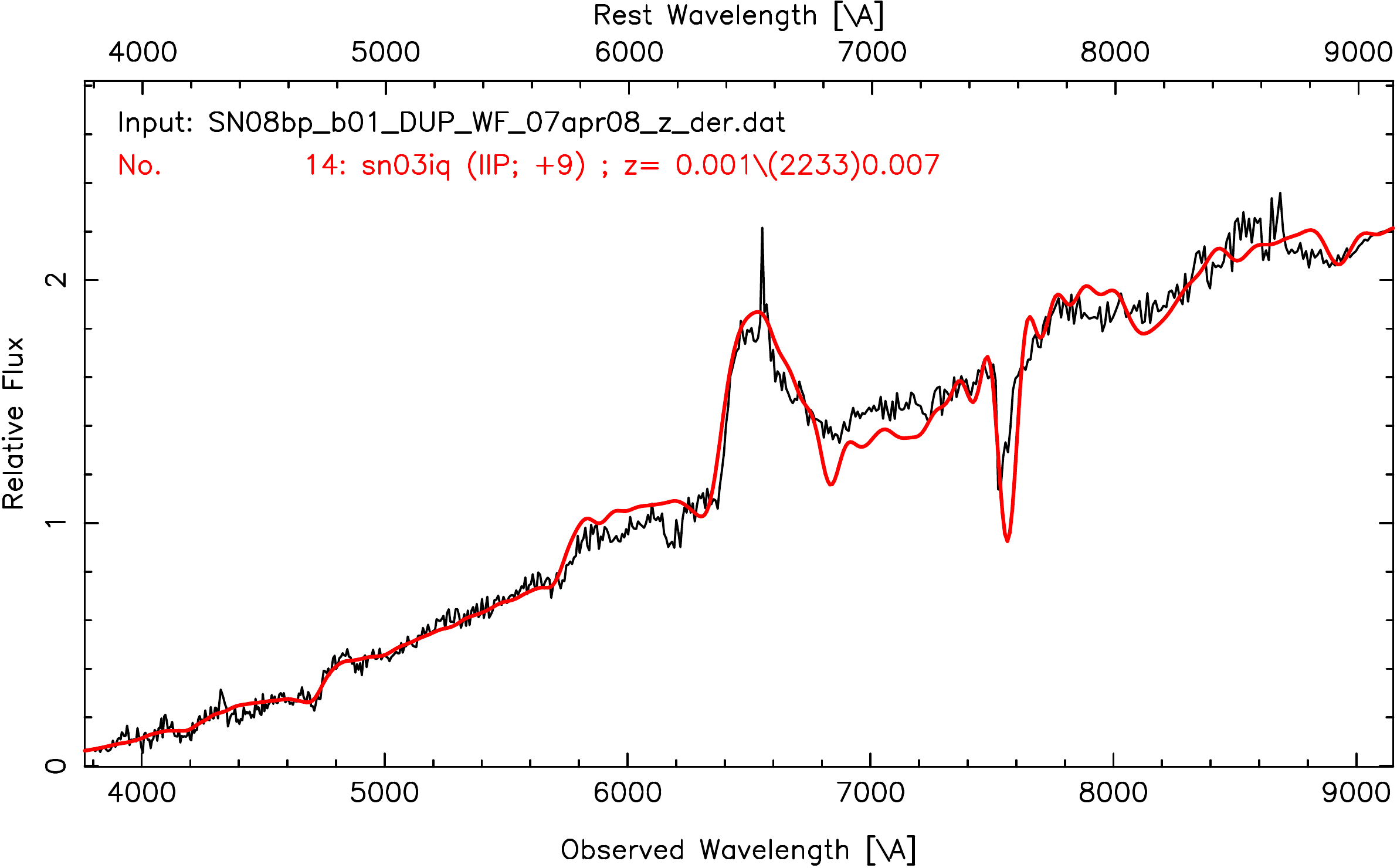}
\includegraphics[width=4.4cm]{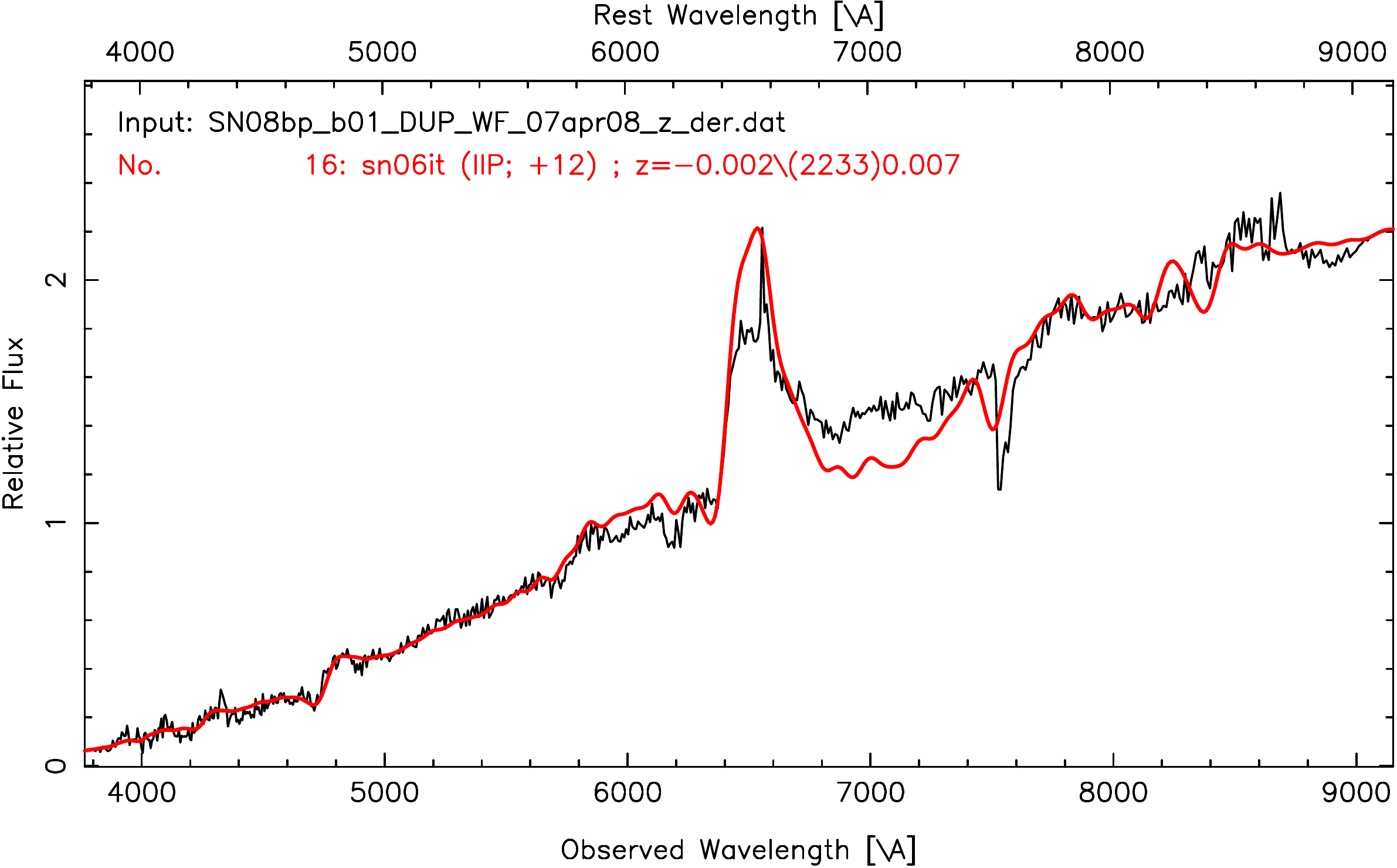}
\includegraphics[width=4.4cm]{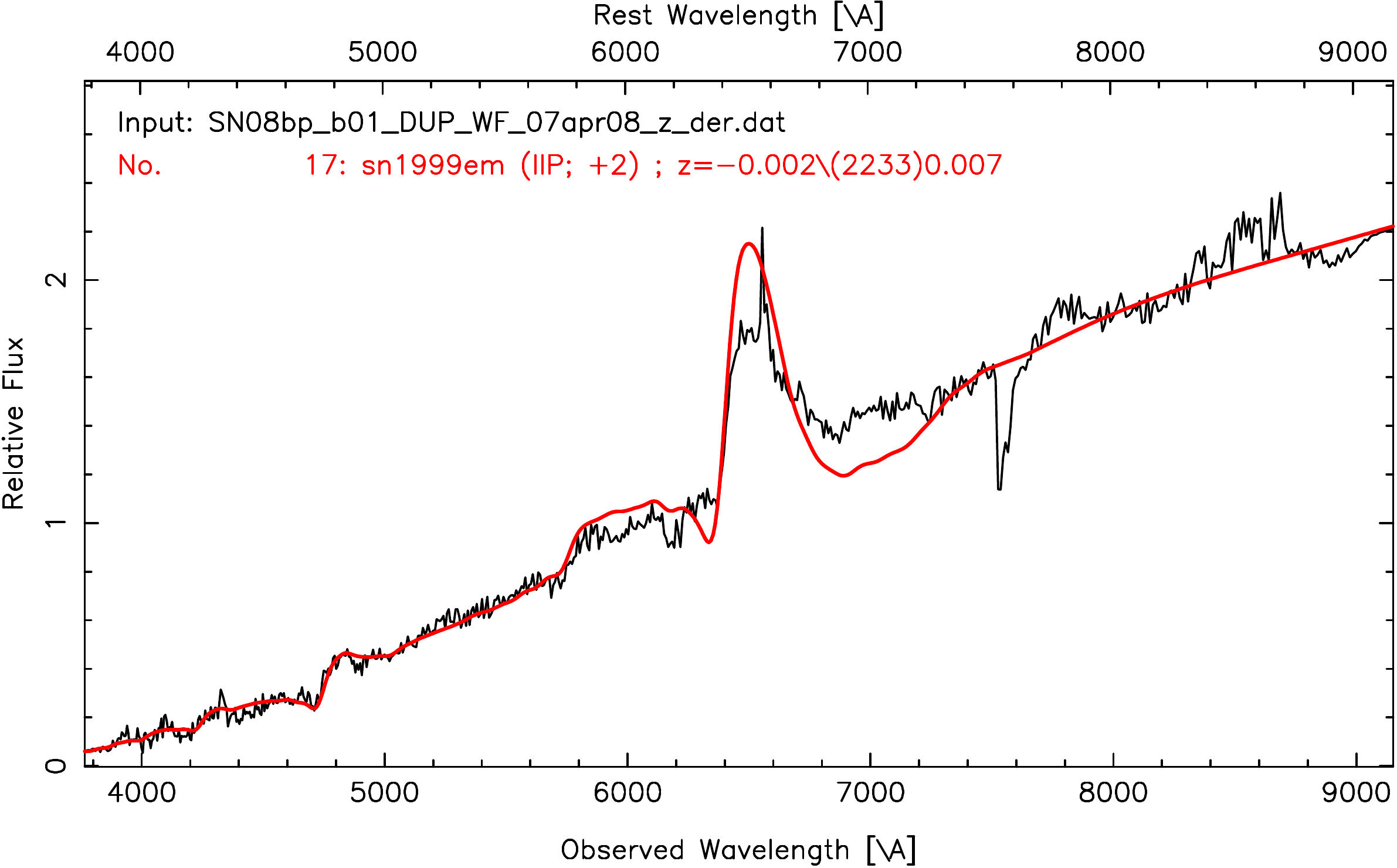}
\caption{Best spectral matching of SN~2008bp using SNID. The plots show SN~2008bp compared with 
SN~2005cs, SN~2004et, SN~2004fc, SN~2003iq, SN~2006it, and SN~1999em at 7, 13, 9, 9, 12, and 12 days from explosion.}
\end{figure}

\clearpage

\begin{figure}
\centering
\includegraphics[width=4.4cm]{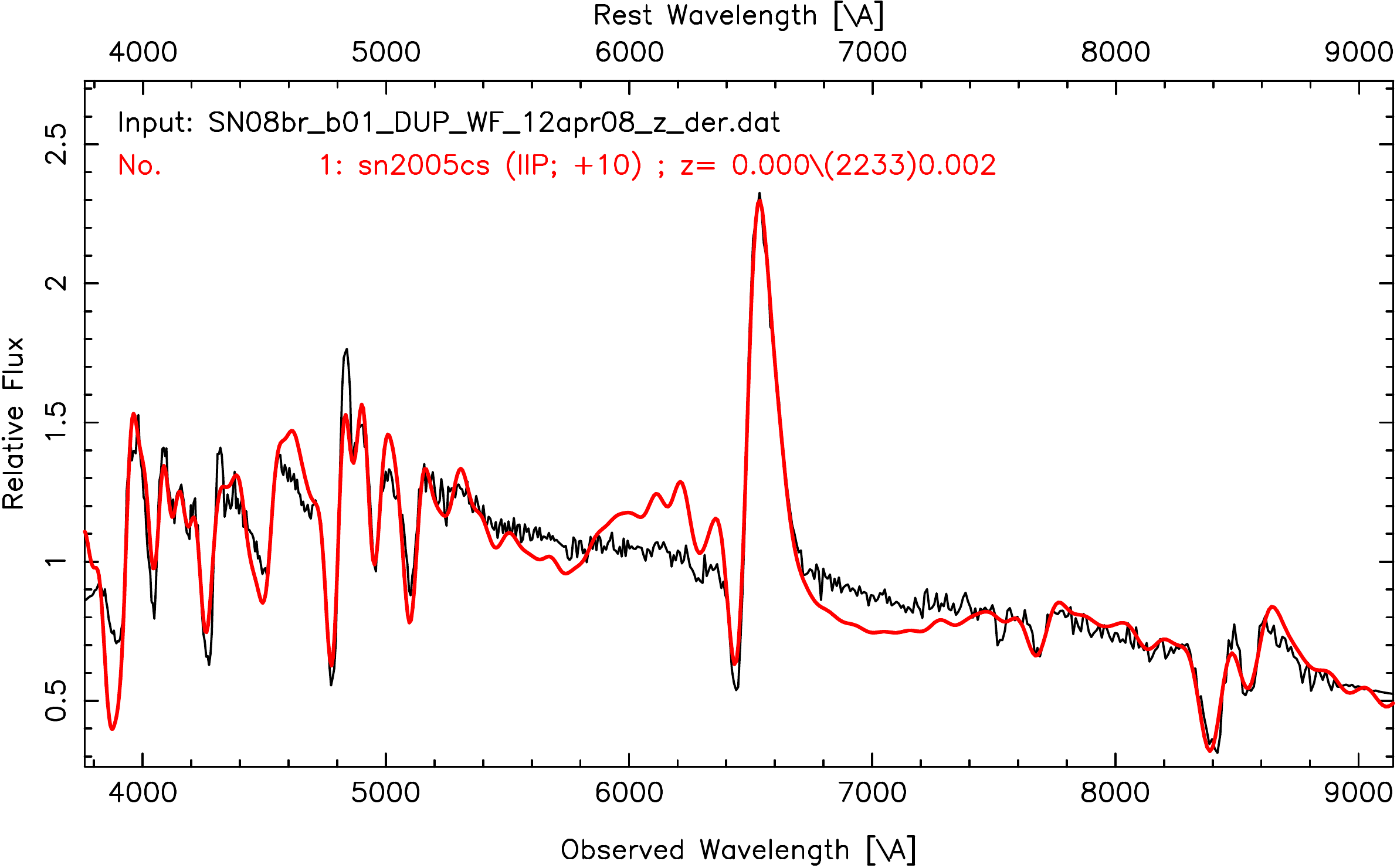}
\includegraphics[width=4.4cm]{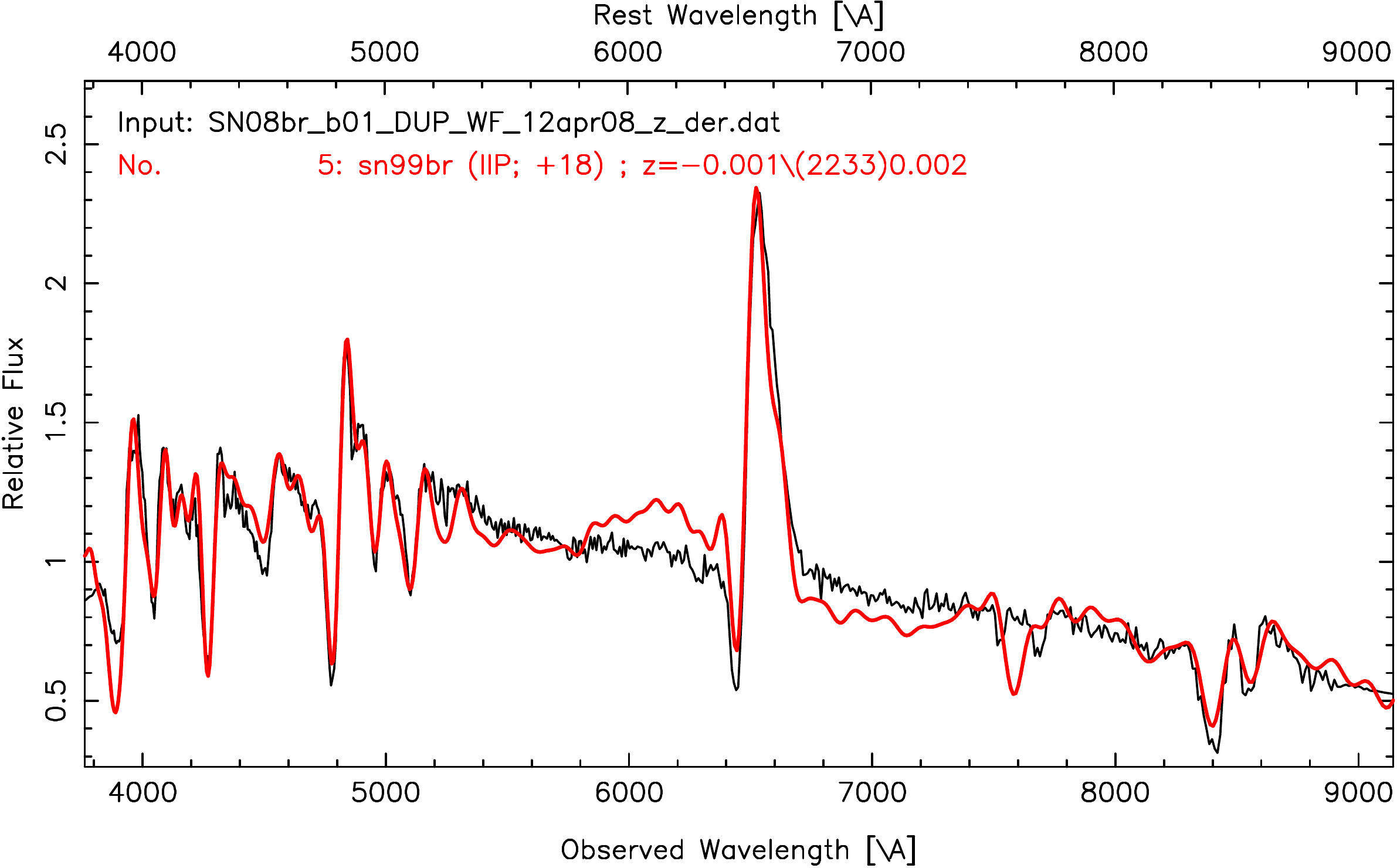}
\caption{Best spectral matching of SN~2008br using SNID. The plots show SN~2008br compared with 
SN~2005cs and SN~1999br at 16, 18 days from explosion.}
\end{figure}

\begin{figure}
\centering
\includegraphics[width=4.4cm]{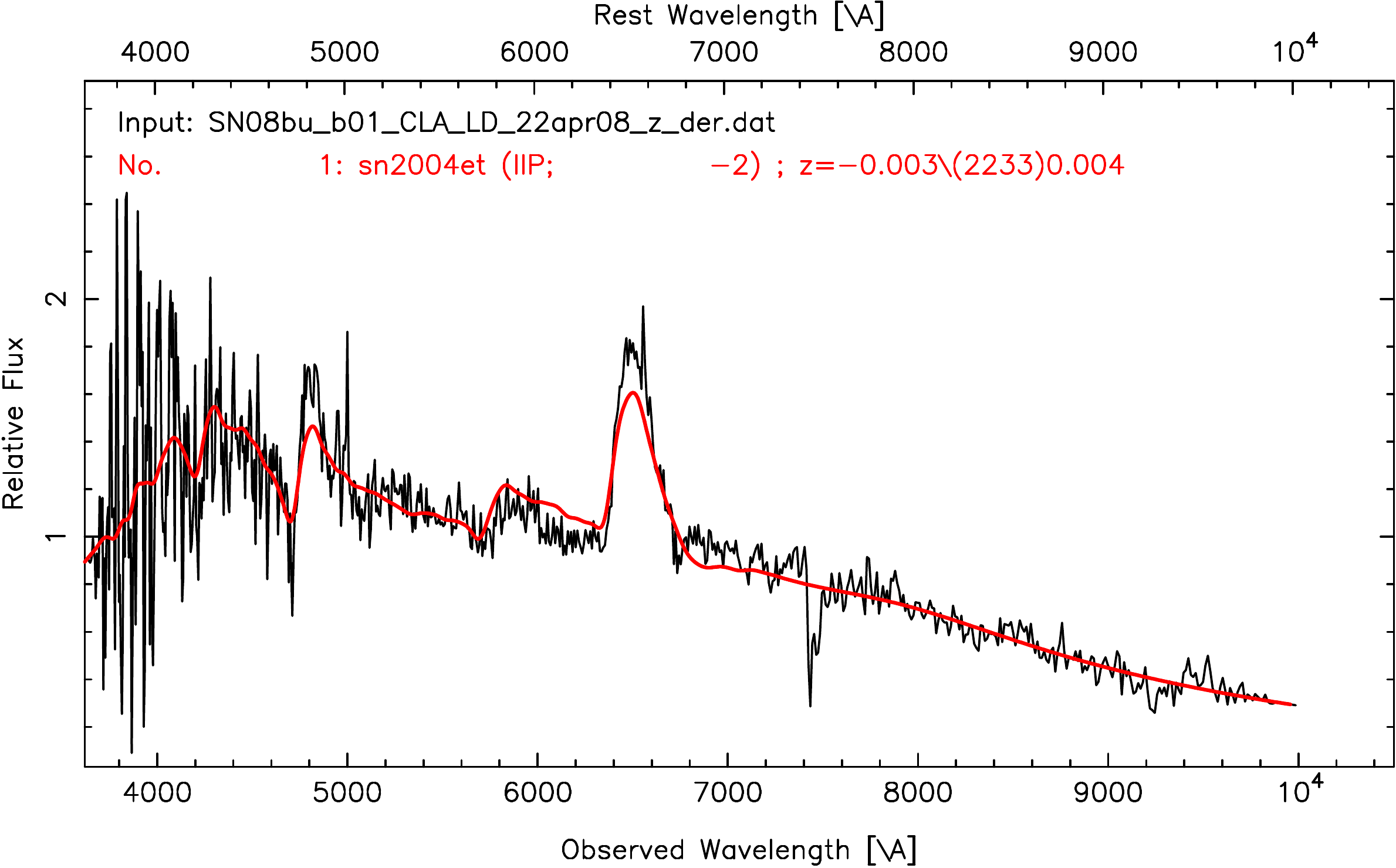}
\includegraphics[width=4.4cm]{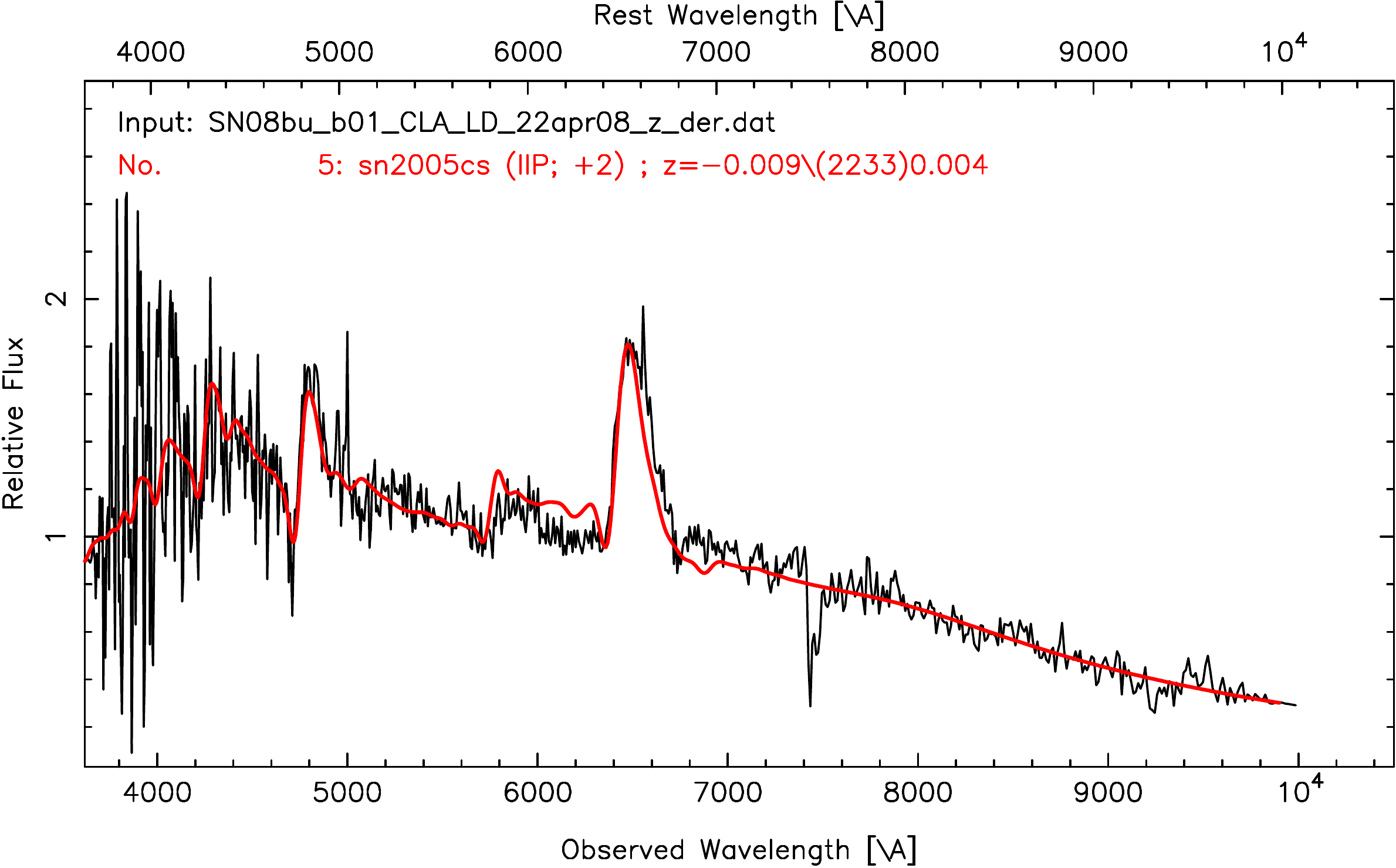}
\includegraphics[width=4.4cm]{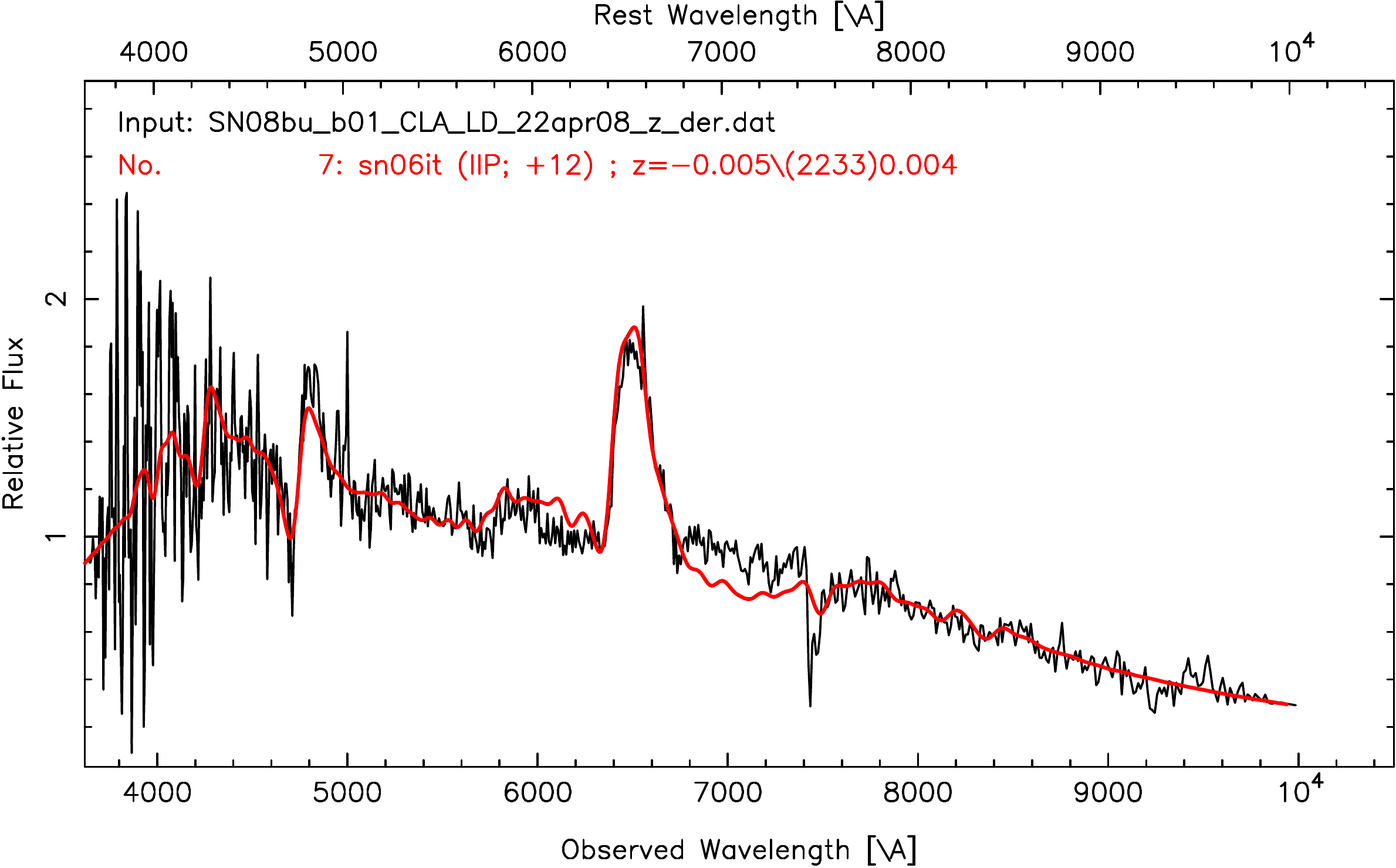}
\includegraphics[width=4.4cm]{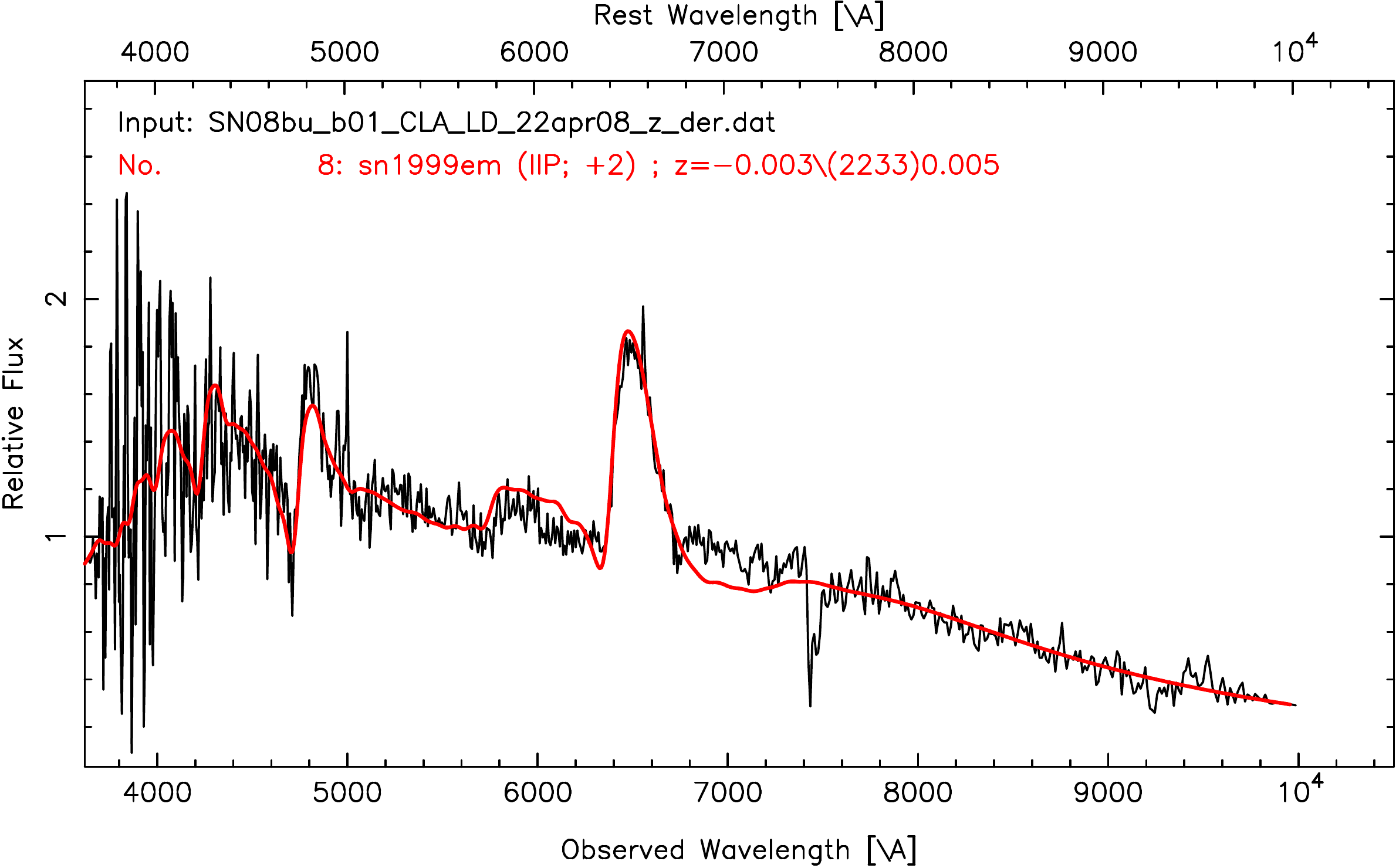}
\includegraphics[width=4.4cm]{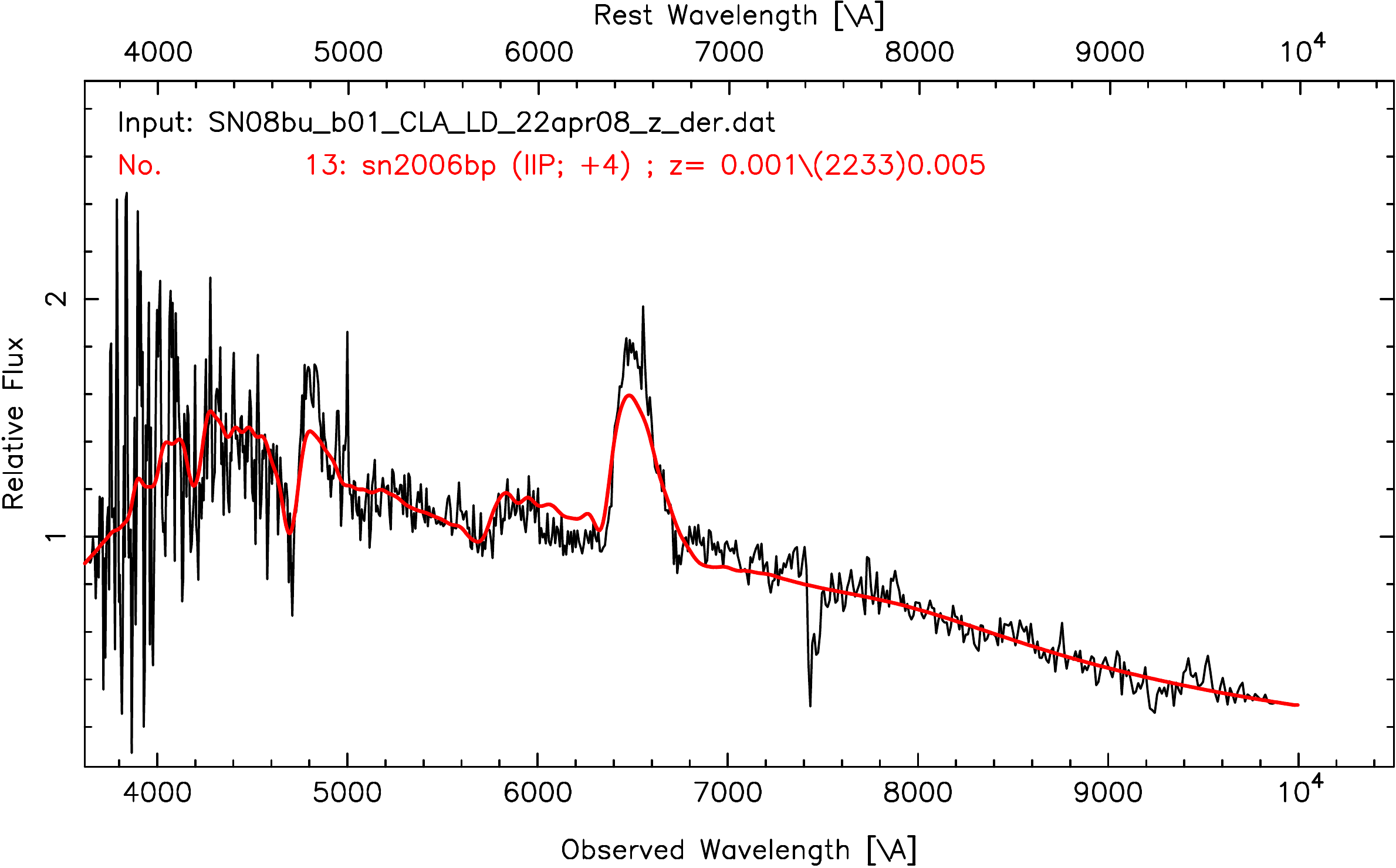}
\caption{Best spectral matching of SN~2008bu using SNID. The plots show SN~2008bu compared with 
SN~2004et, SN~2005cs, SN~2006it, SN~1999em, and SN~2006bp at 14, 8, 12, 12, and 13 days from explosion.}
\end{figure}

\begin{figure}
\centering
\includegraphics[width=4.4cm]{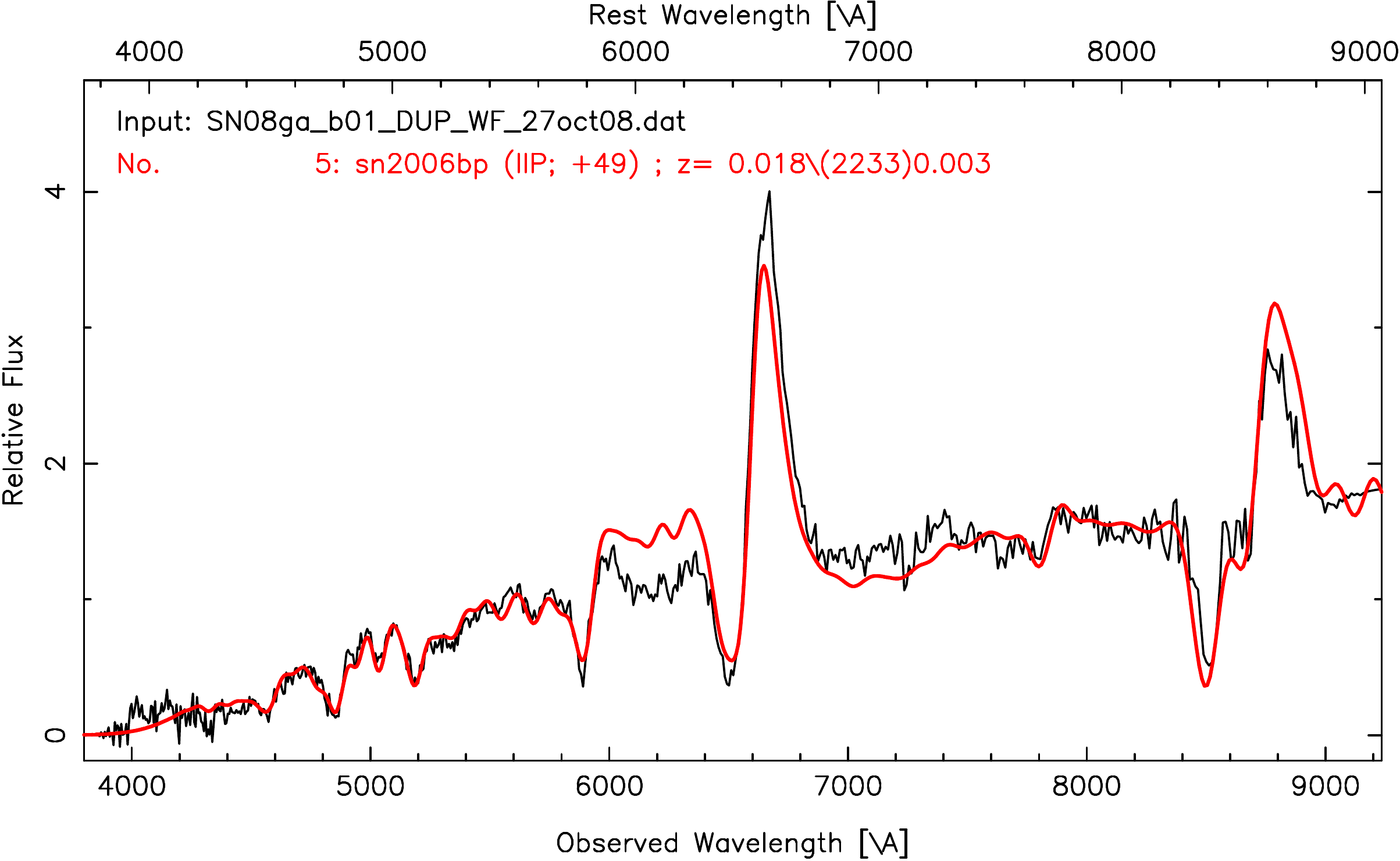}
\includegraphics[width=4.4cm]{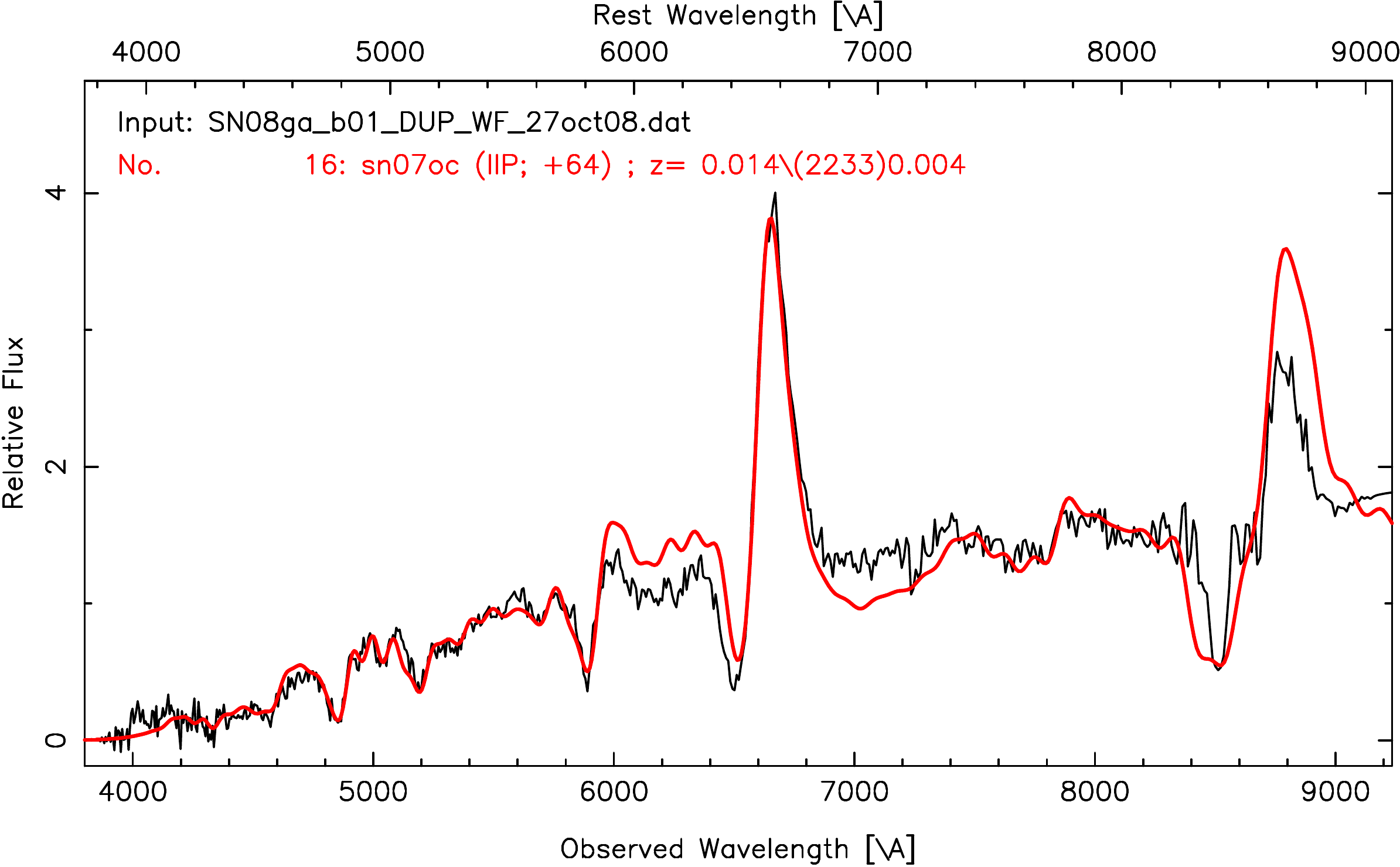}
\includegraphics[width=4.4cm]{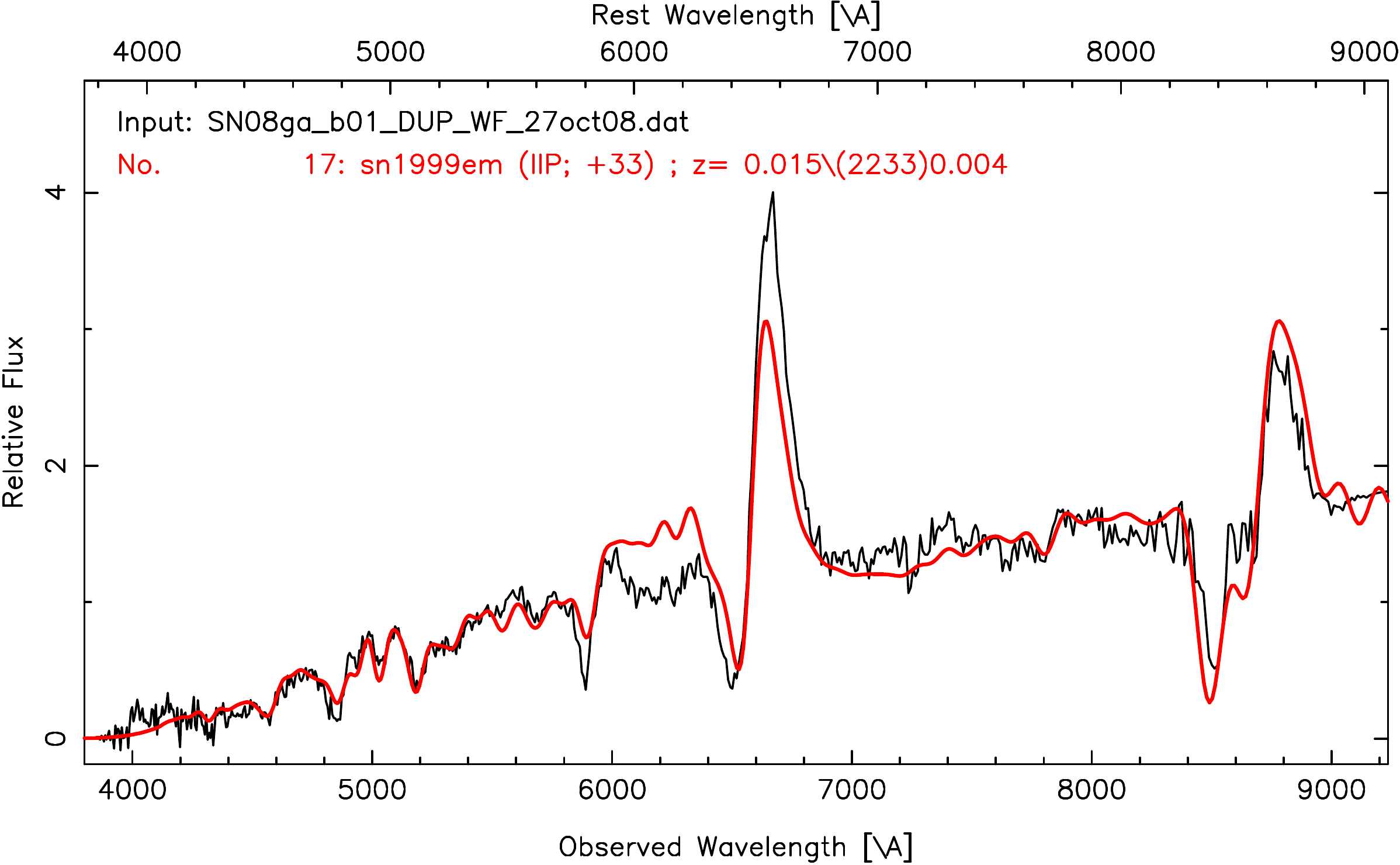}
\caption{Best spectral matching of SN~2008ga using SNID. The plots show SN~2008ga compared with 
SN~2006bp, SN~2007oc, and SN~1999em at 58, 64, and 43 days from explosion.}
\end{figure}

\clearpage

\begin{figure}
\centering
\includegraphics[width=4.4cm]{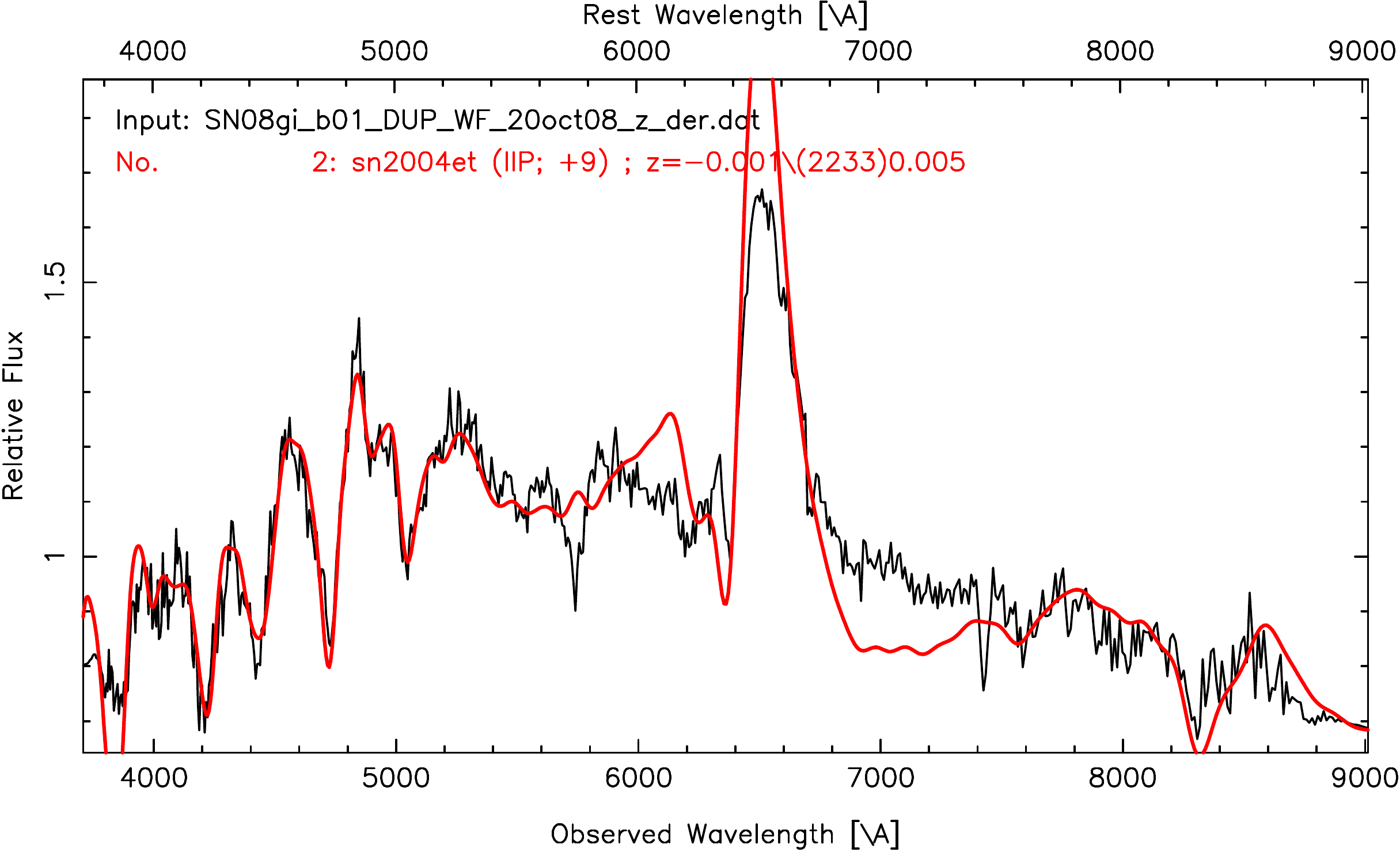}
\includegraphics[width=4.4cm]{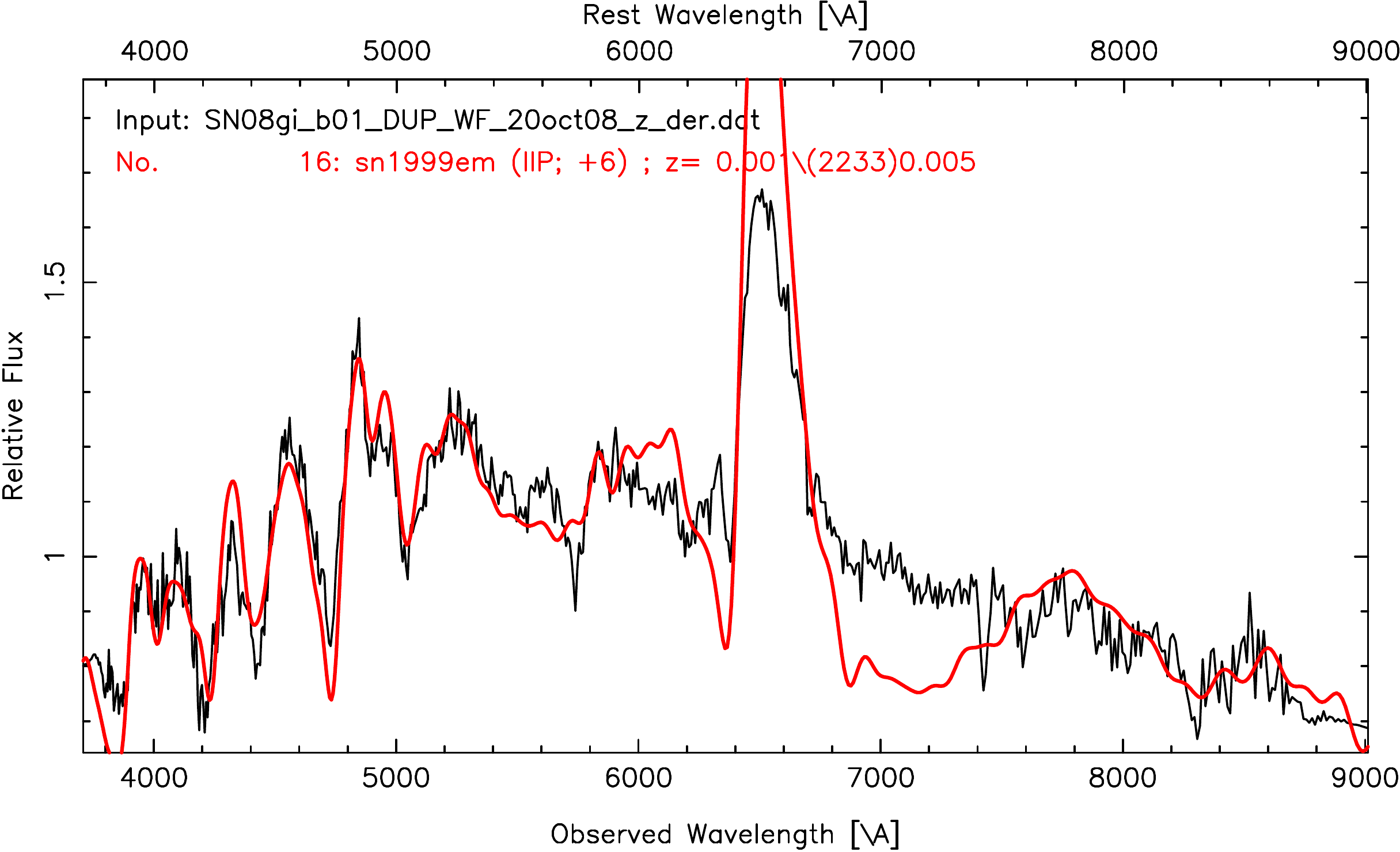}
\caption{Best spectral matching of SN~2008gi using SNID. The plots show SN~2008gi compared with 
SN~2004et and SN~1999em at 25 and 16 days from explosion.}
\end{figure}

\begin{figure}
\centering
\includegraphics[width=4.4cm]{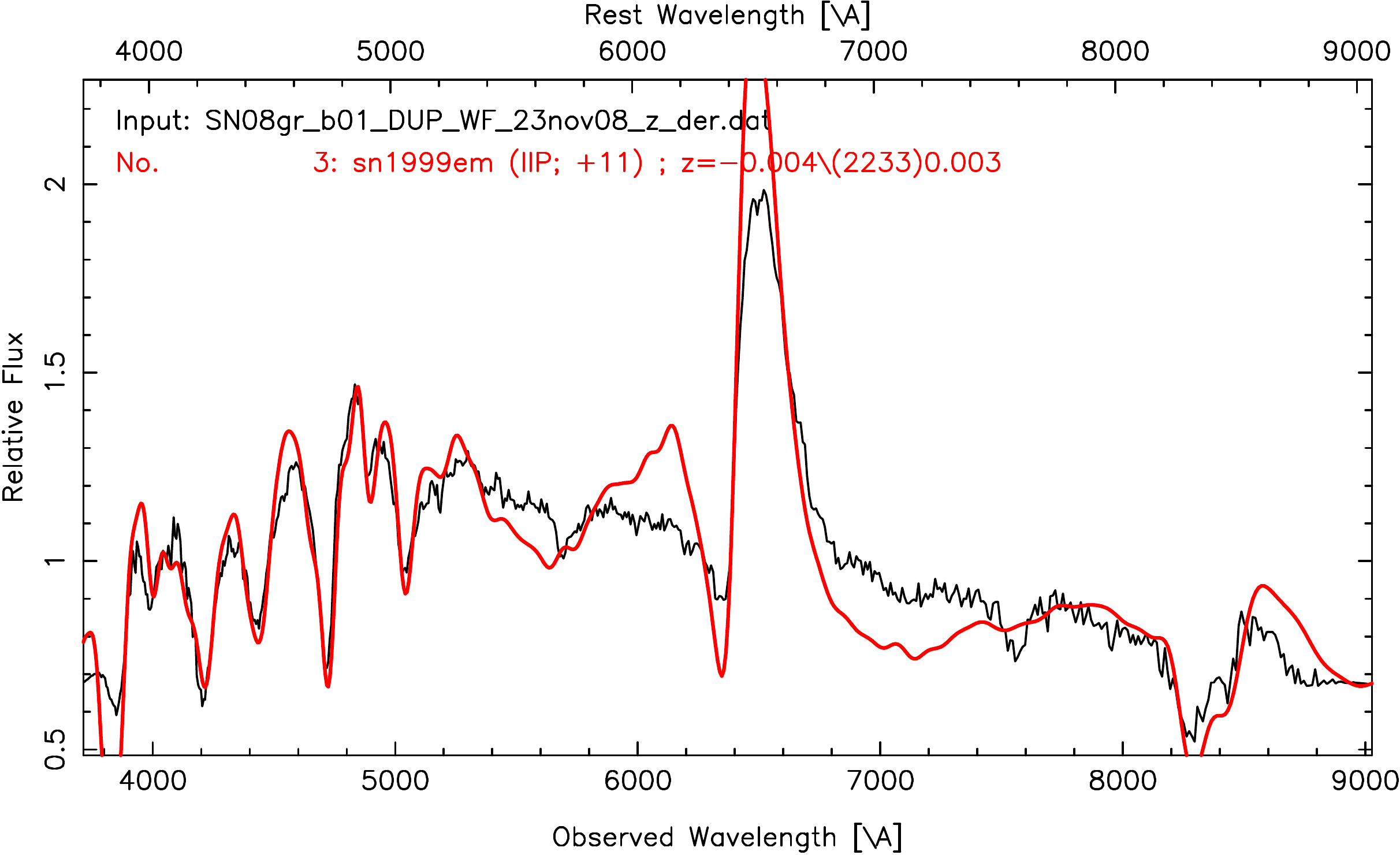}
\includegraphics[width=4.4cm]{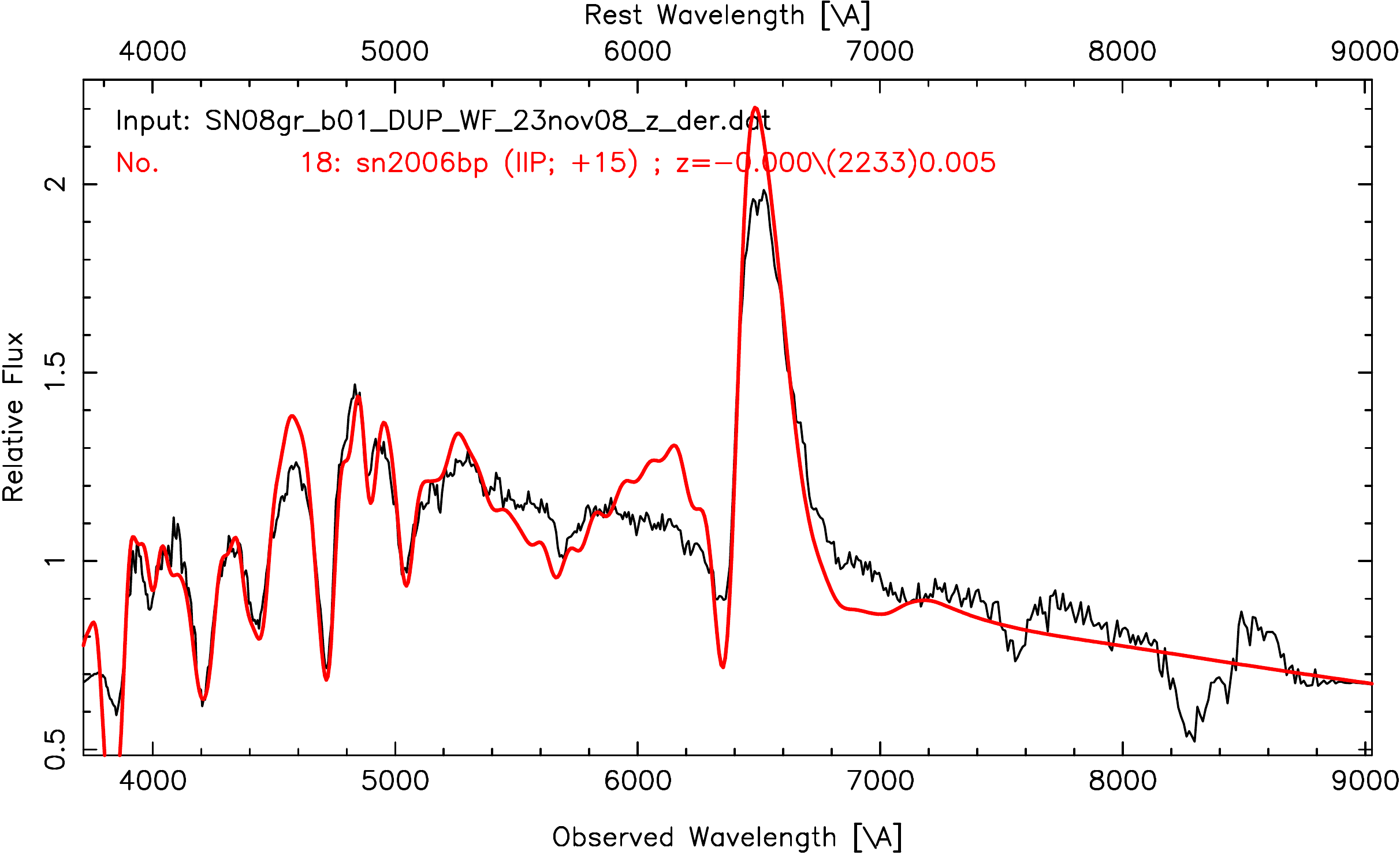}
\includegraphics[width=4.4cm]{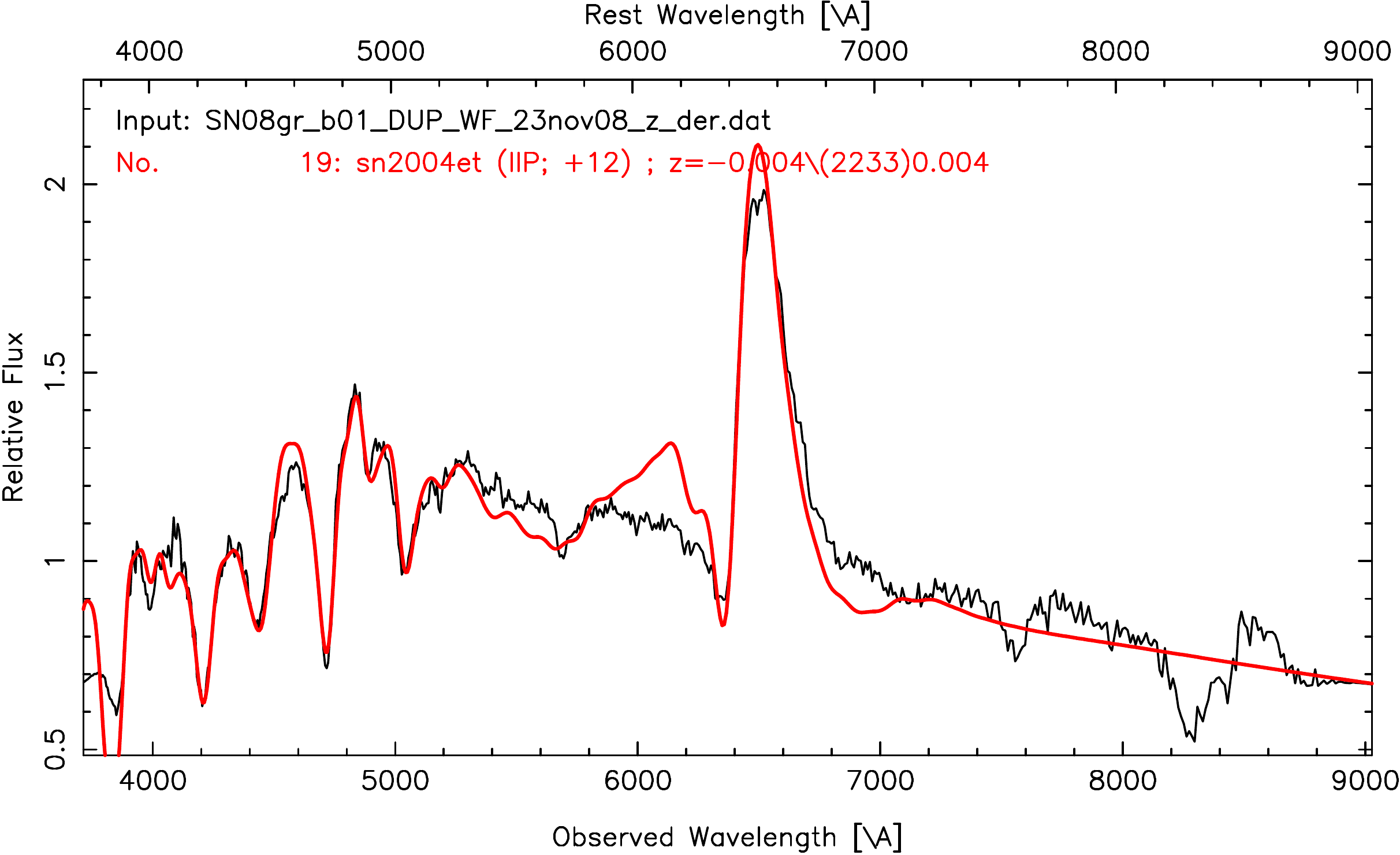}
\caption{Best spectral matching of SN~2008gr using SNID. The plots show SN~2008gr compared with 
SN~1999em, SN~2006bp, and SN~2004et at 21, 24, and 28 days from explosion.}
\end{figure}

\clearpage

\begin{figure}
\centering
\includegraphics[width=4.4cm]{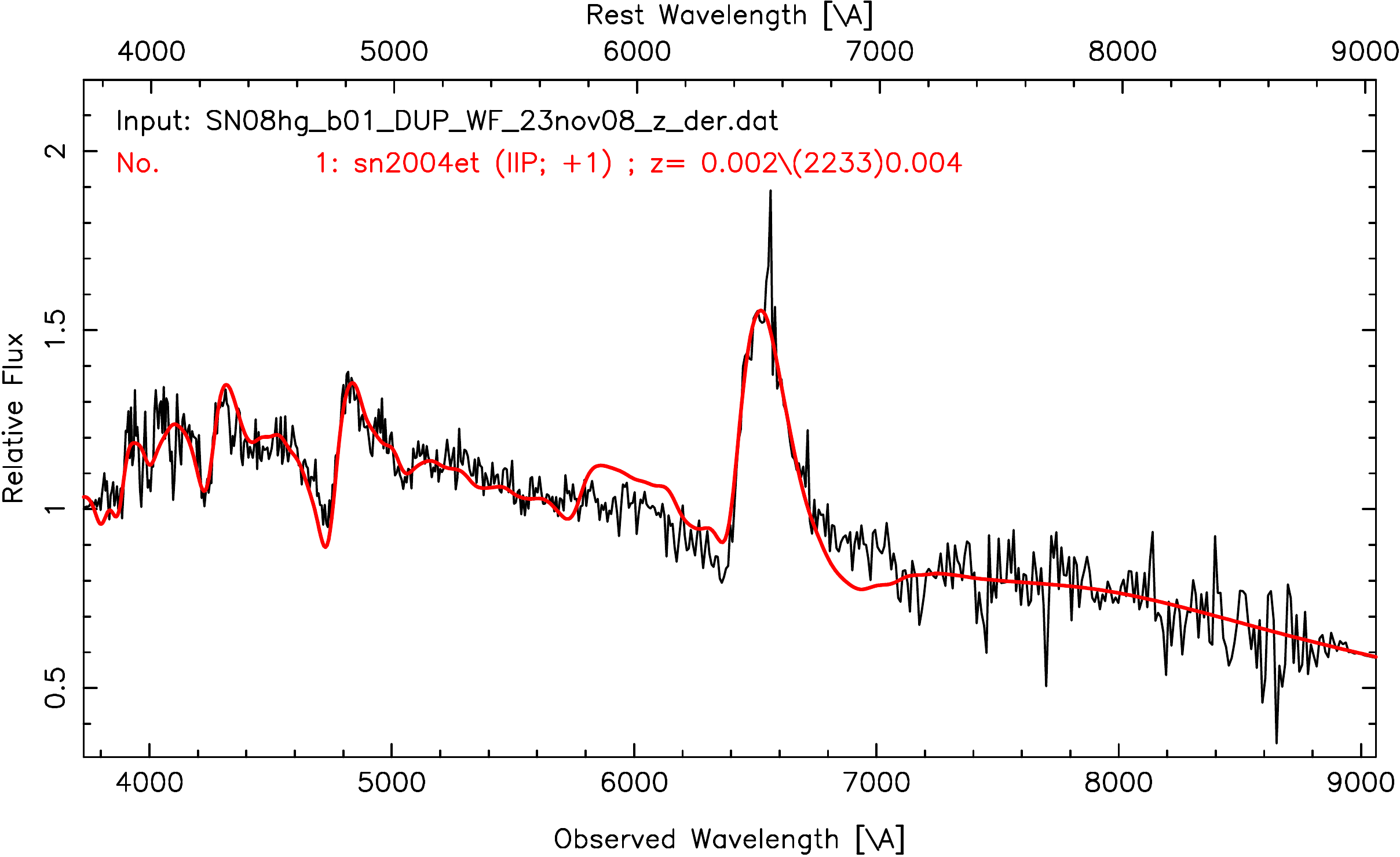}
\includegraphics[width=4.4cm]{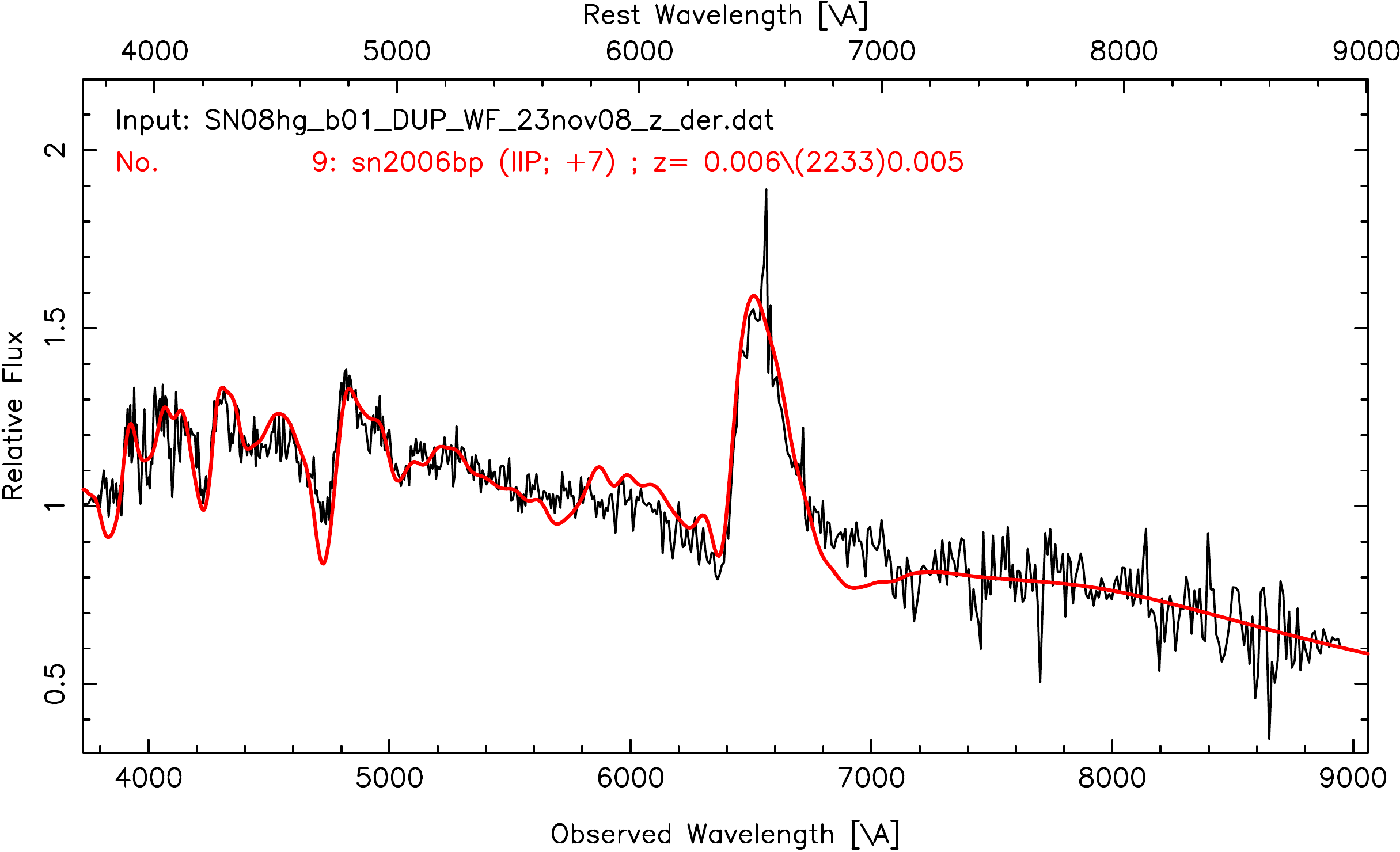}
\includegraphics[width=4.4cm]{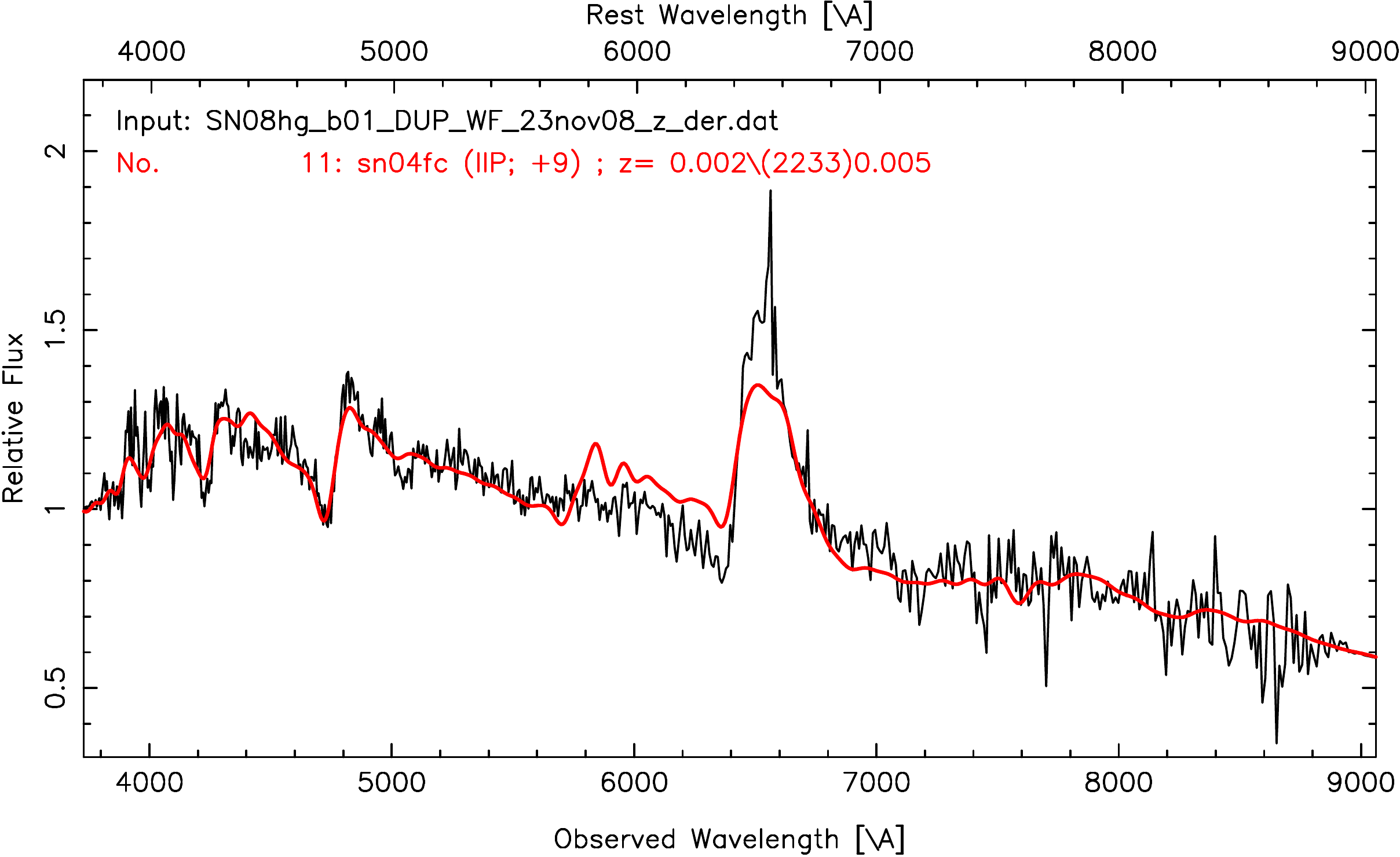}
\includegraphics[width=4.4cm]{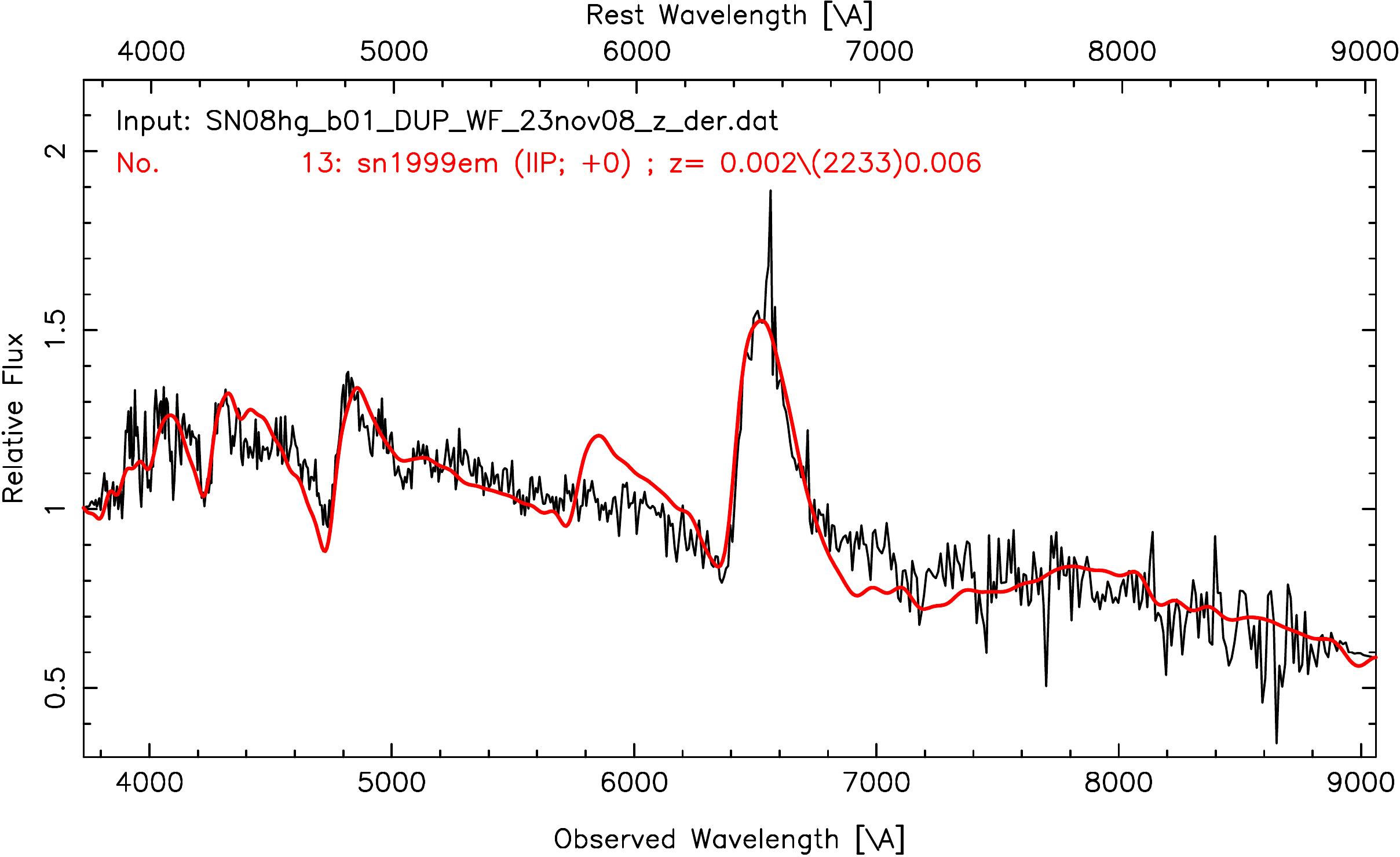}
\includegraphics[width=4.4cm]{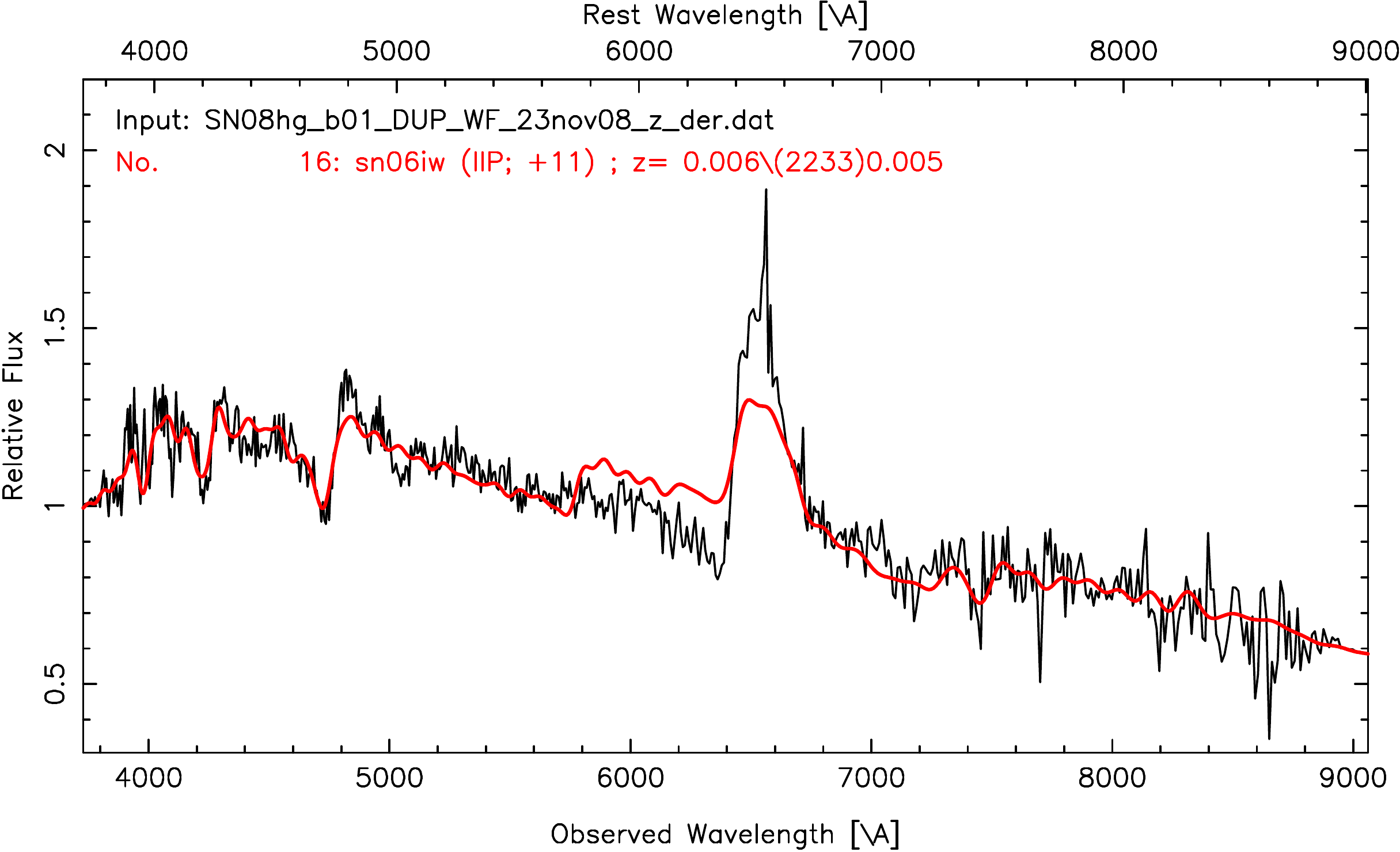}
\caption{Best spectral matching of SN~2008hg using SNID. The plots show SN~2008hg compared with 
SN~2004et, SN~2006bp, SN~2004fc, SN~1999em, and 2006iw at 17, 13, 9, 10, amd 11 days from explosion.}
\end{figure}

\begin{figure}
\centering
\includegraphics[width=4.4cm]{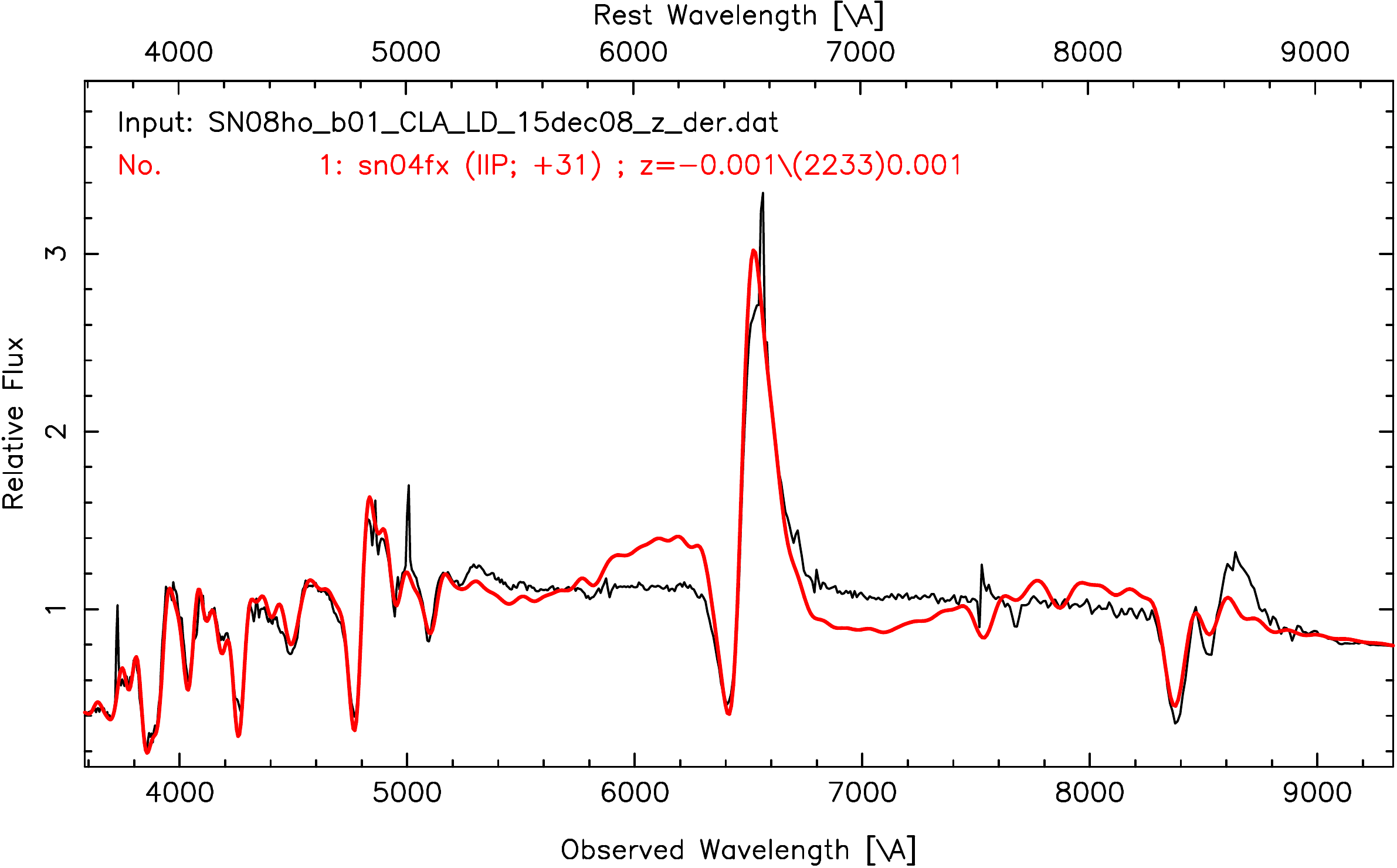}
\includegraphics[width=4.4cm]{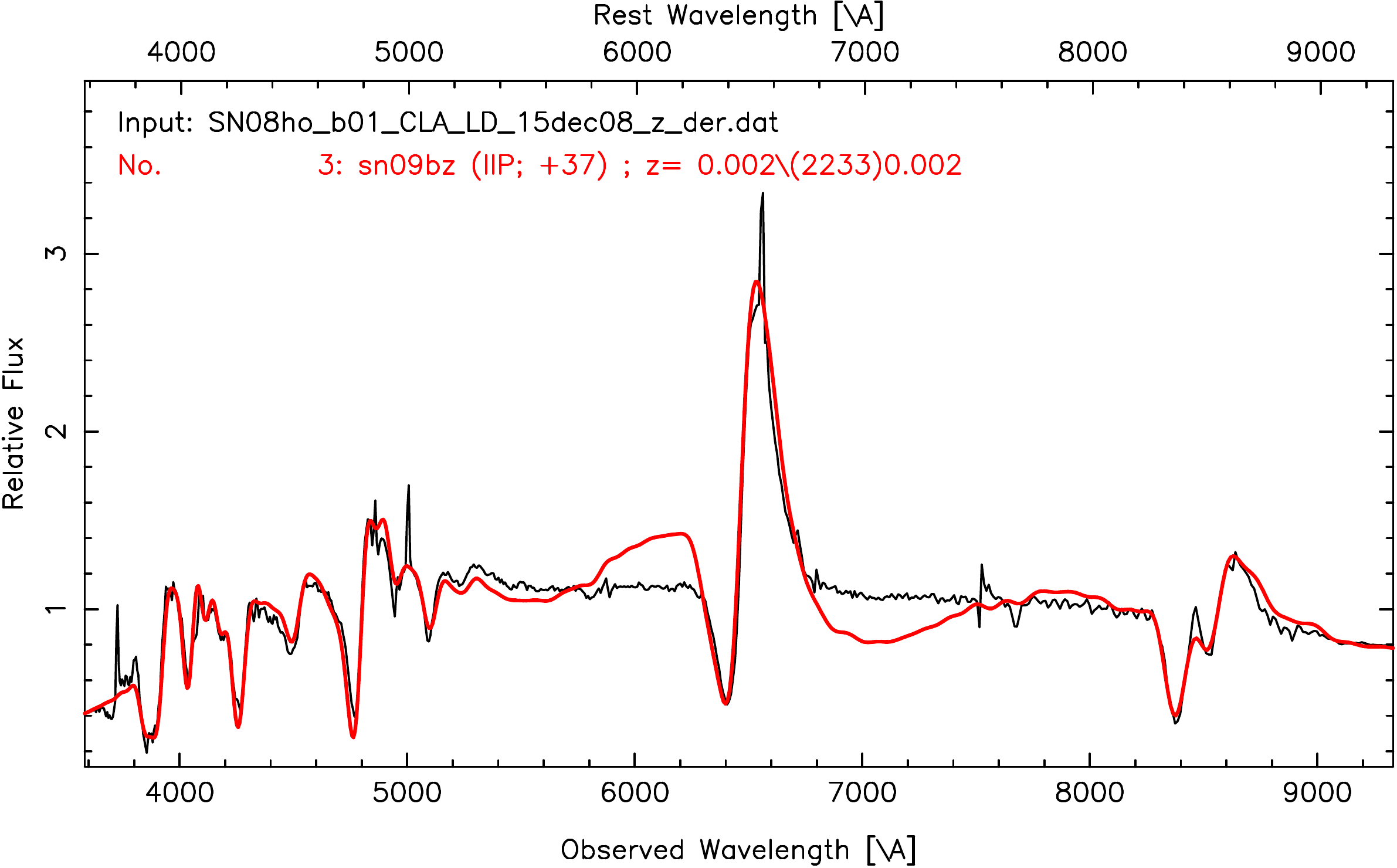}
\includegraphics[width=4.4cm]{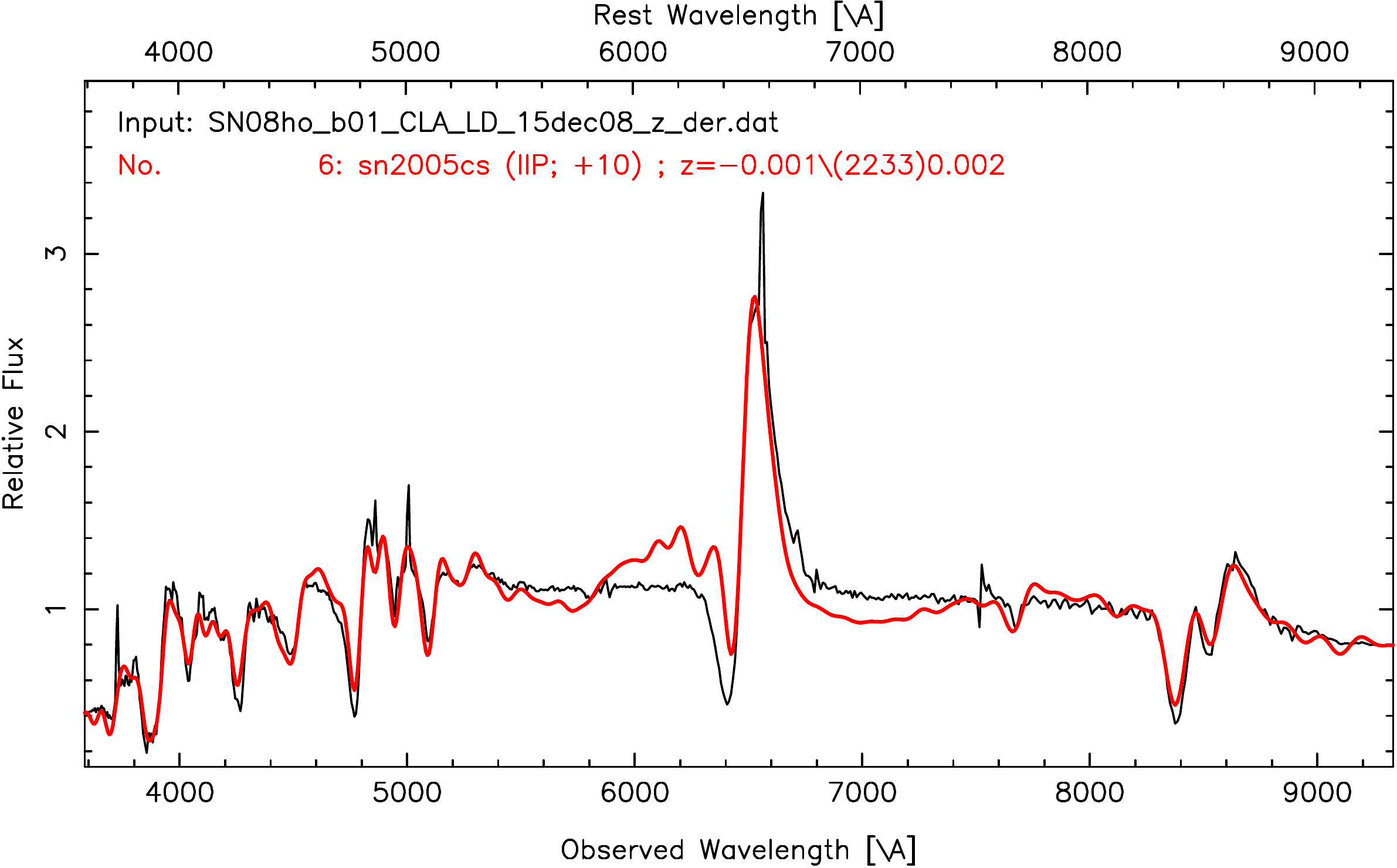}
\includegraphics[width=4.4cm]{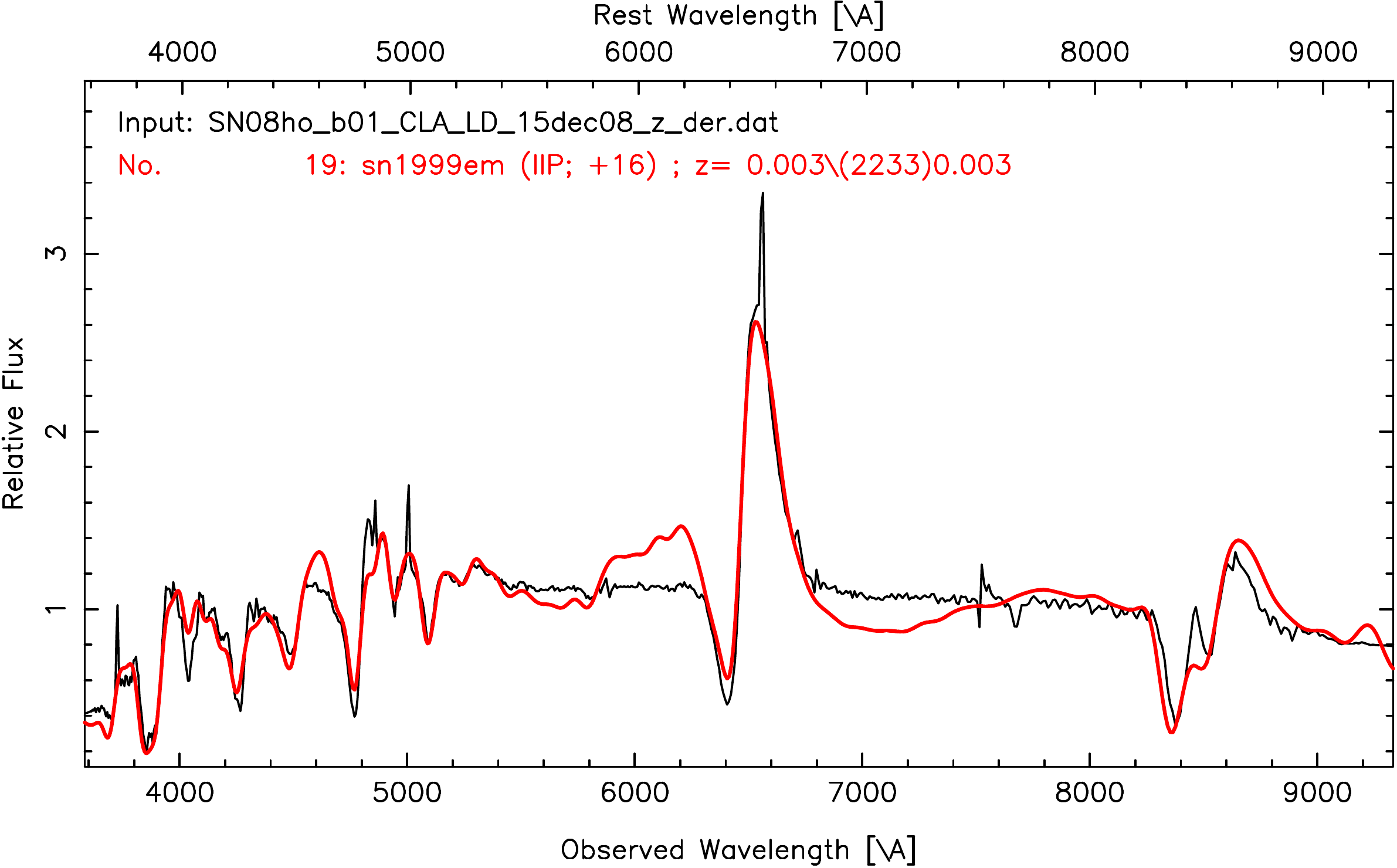}
\caption{Best spectral matching of SN~2008ho using SNID. The plots show SN~2008ho compared with 
SN~2004fx, SN~2009bz, SN~2005cs, and SN~1999em at 31, 37, 16, and 26 days from explosion.}
\end{figure}

\clearpage

\begin{figure}
\centering
\includegraphics[width=4.4cm]{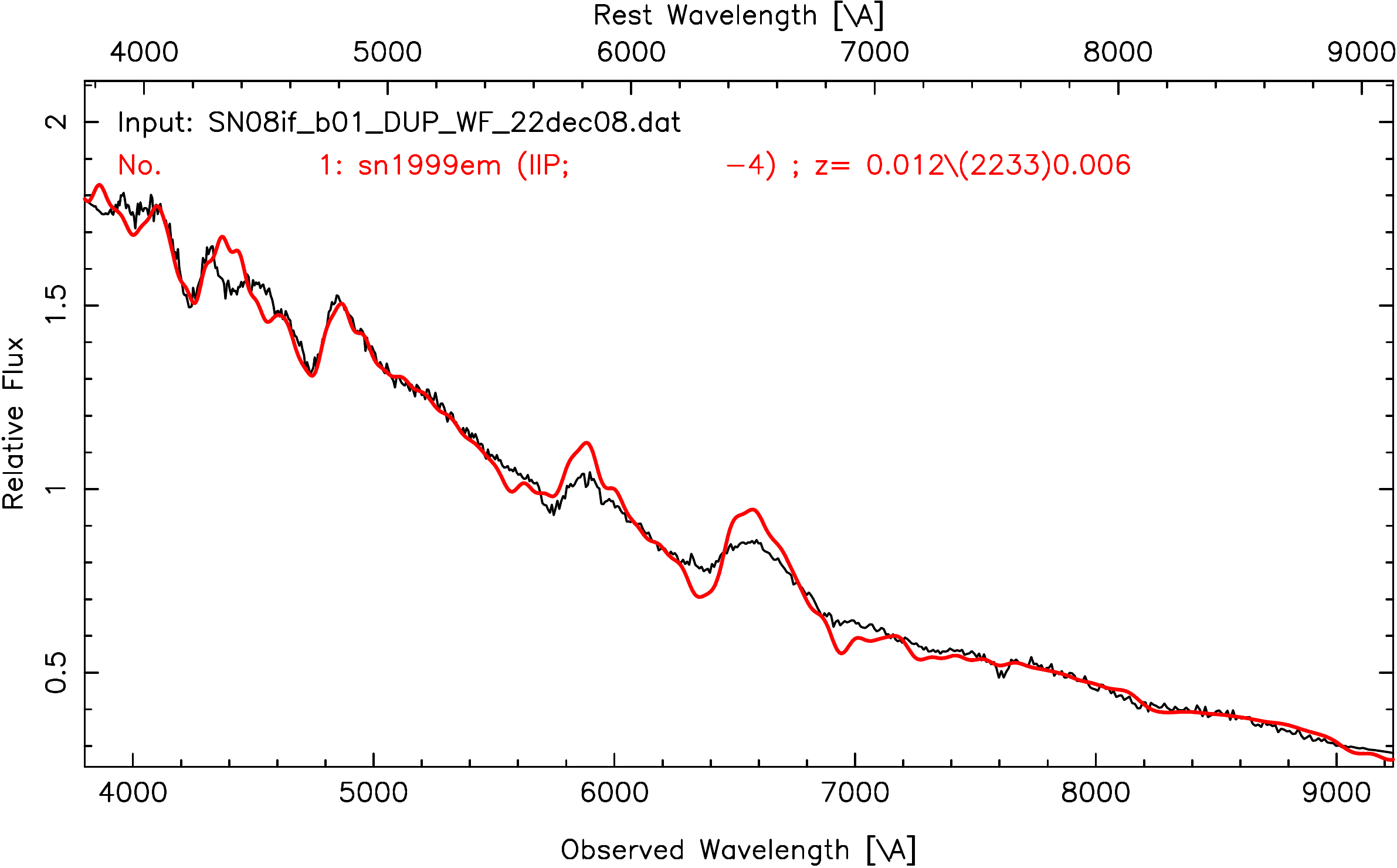}
\includegraphics[width=4.4cm]{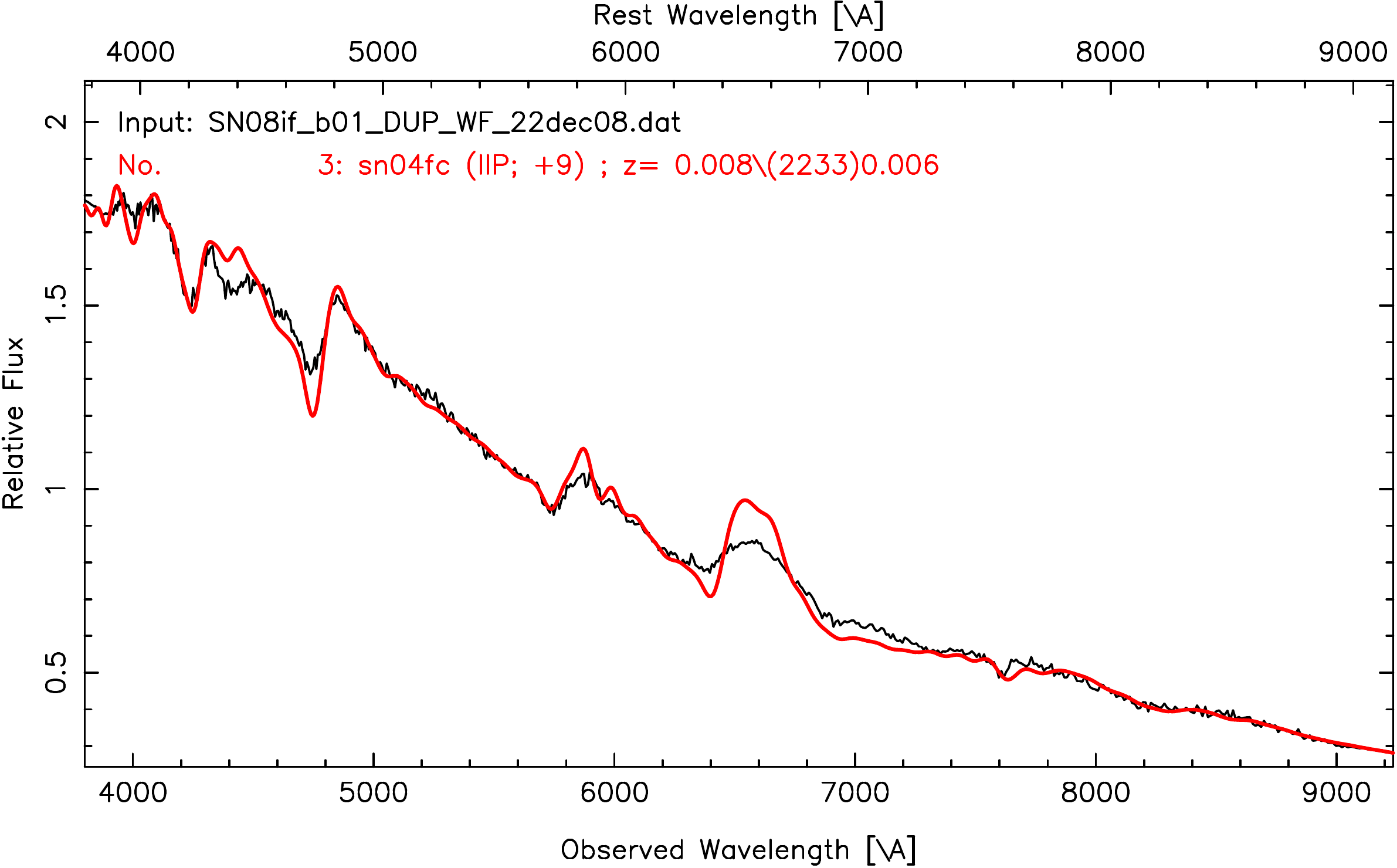}
\includegraphics[width=4.4cm]{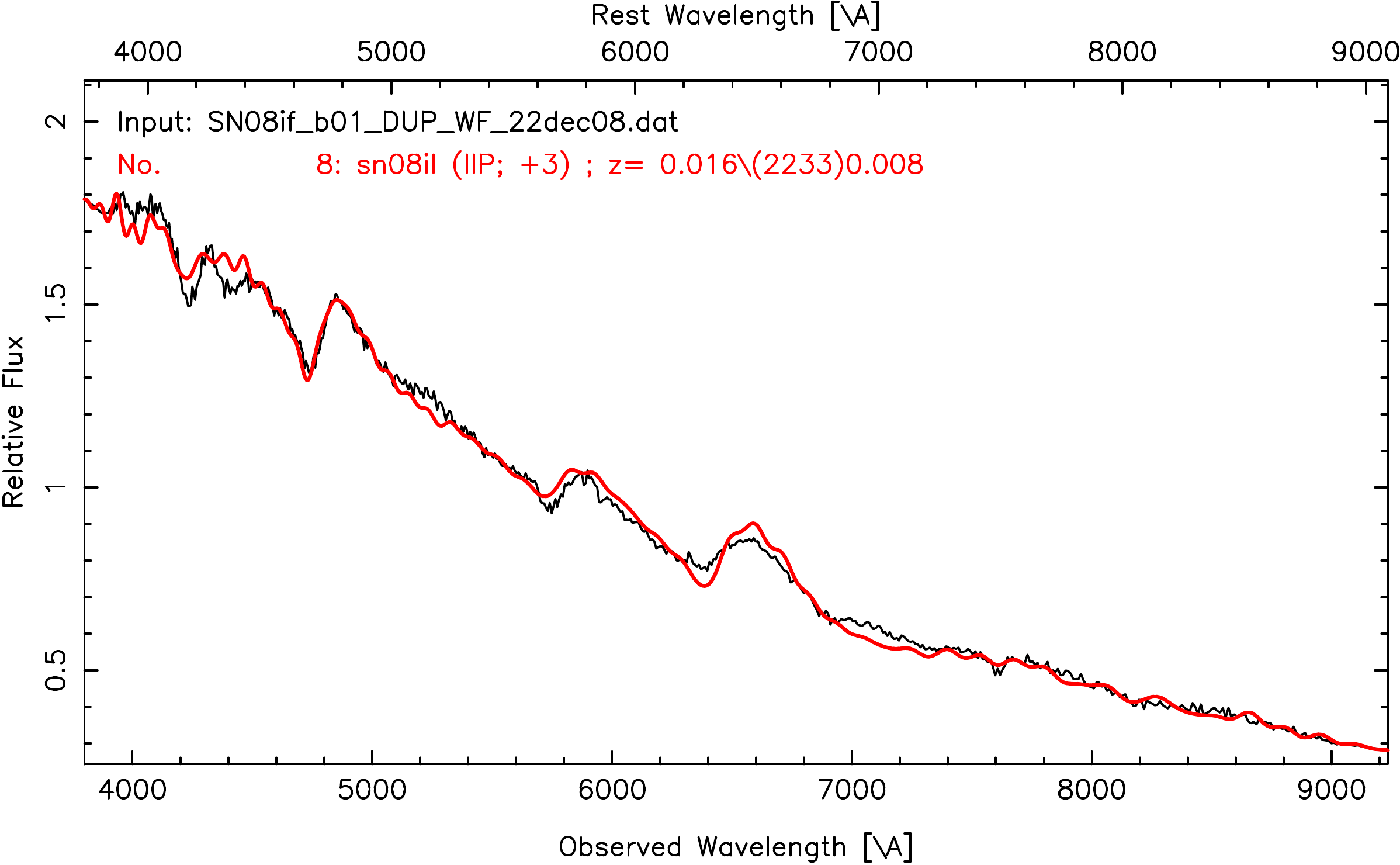}
\includegraphics[width=4.4cm]{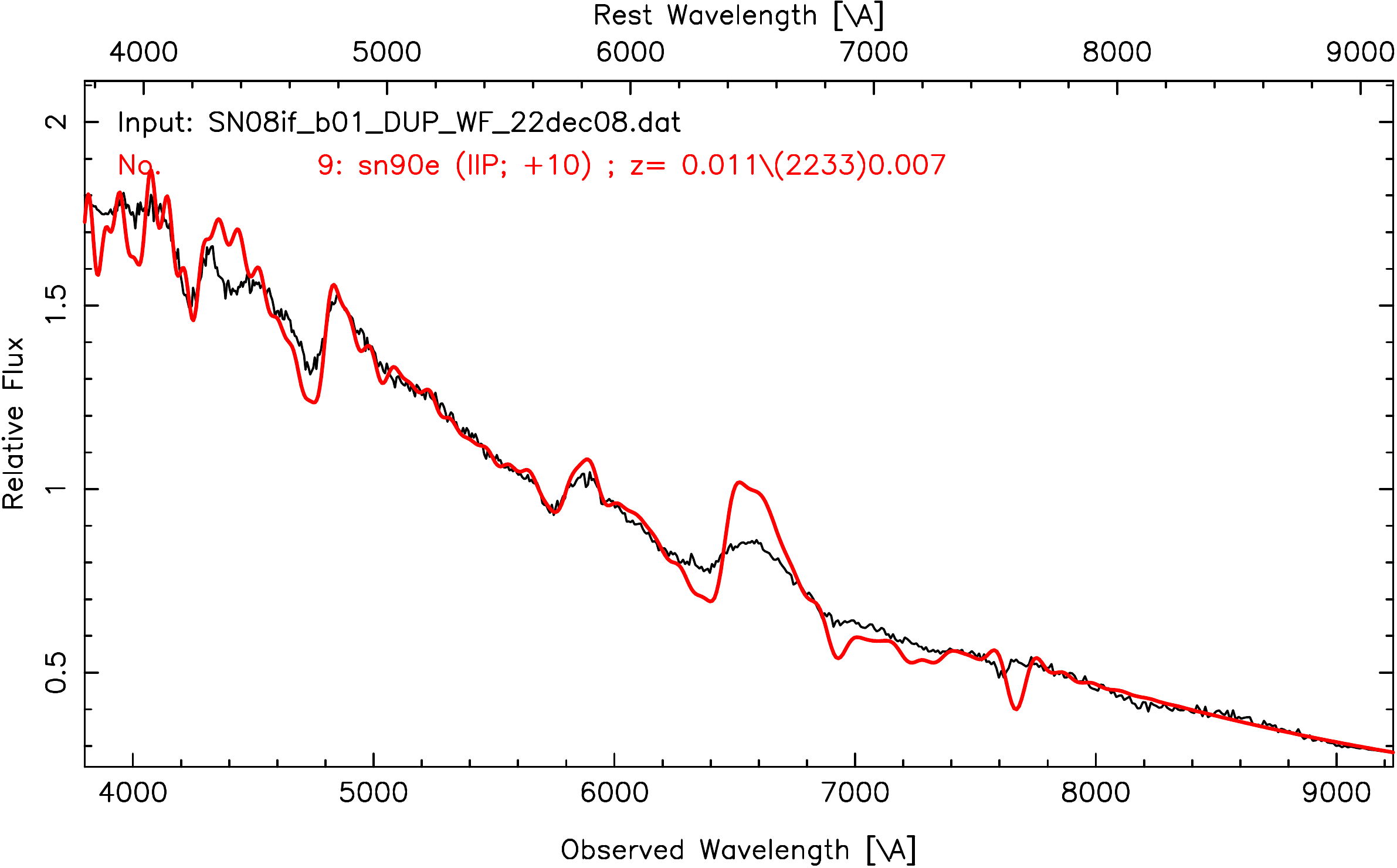}
\includegraphics[width=4.4cm]{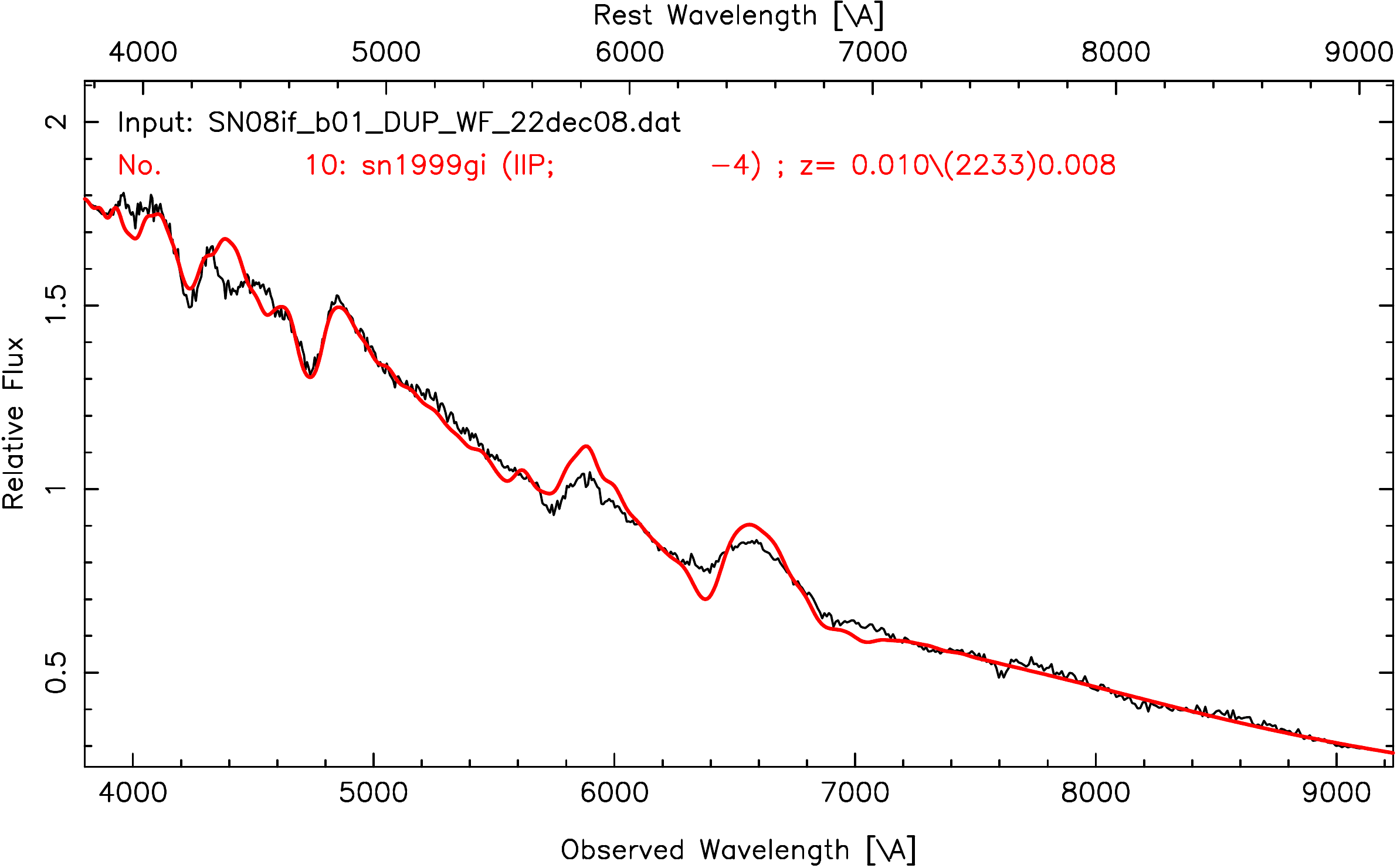}
\includegraphics[width=4.4cm]{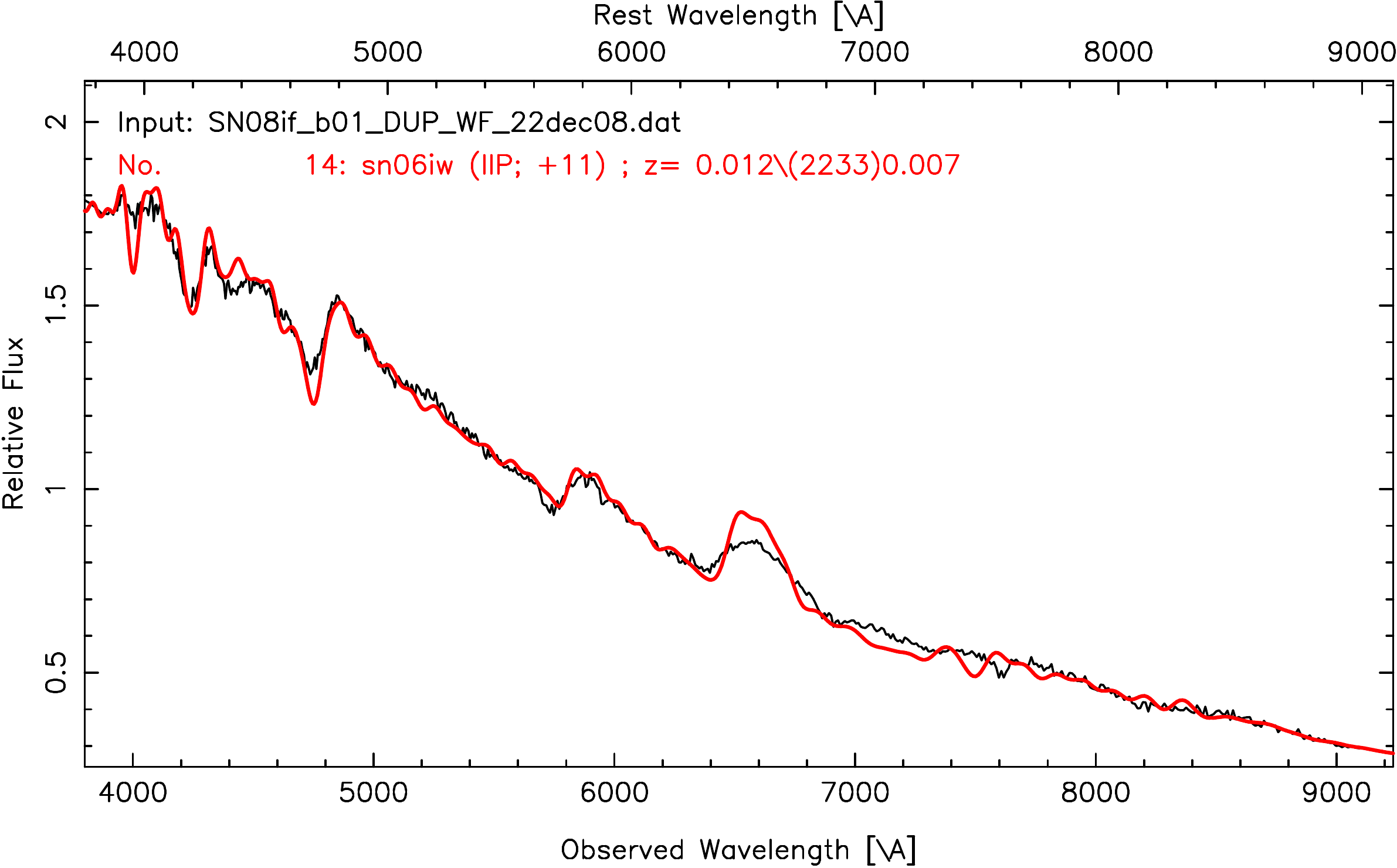}
\caption{Best spectral matching of SN~2008if using SNID. The plots show SN~2008if compared with 
SN~1999em, SN~2004fc, SN~2008il, SN~1990E, SN~1999gi, and SN~2006iw at 6, 9, 3, 10, 8, and 11 days from explosion.}
\end{figure}

\begin{figure}
\centering
\includegraphics[width=4.4cm]{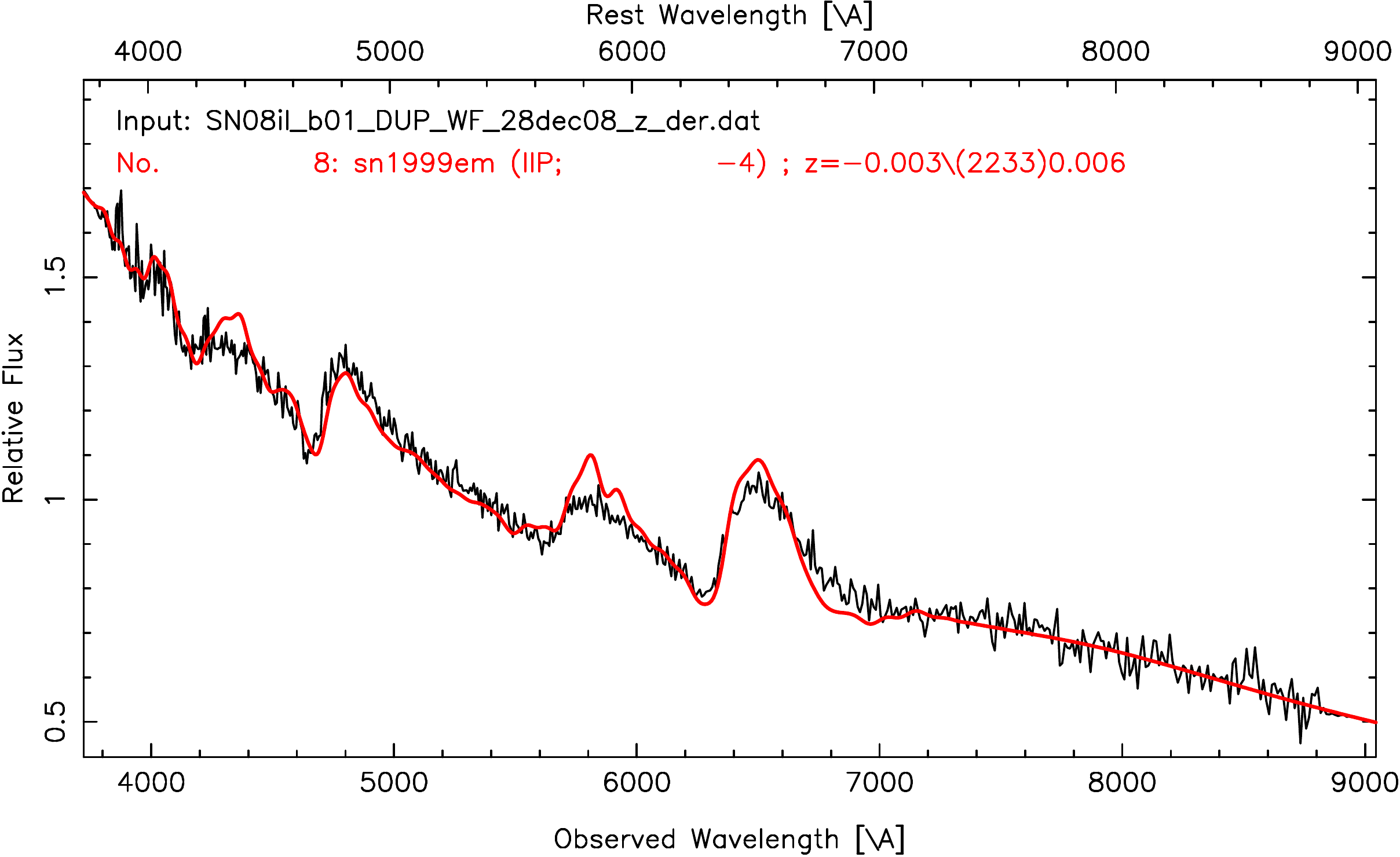}
\includegraphics[width=4.4cm]{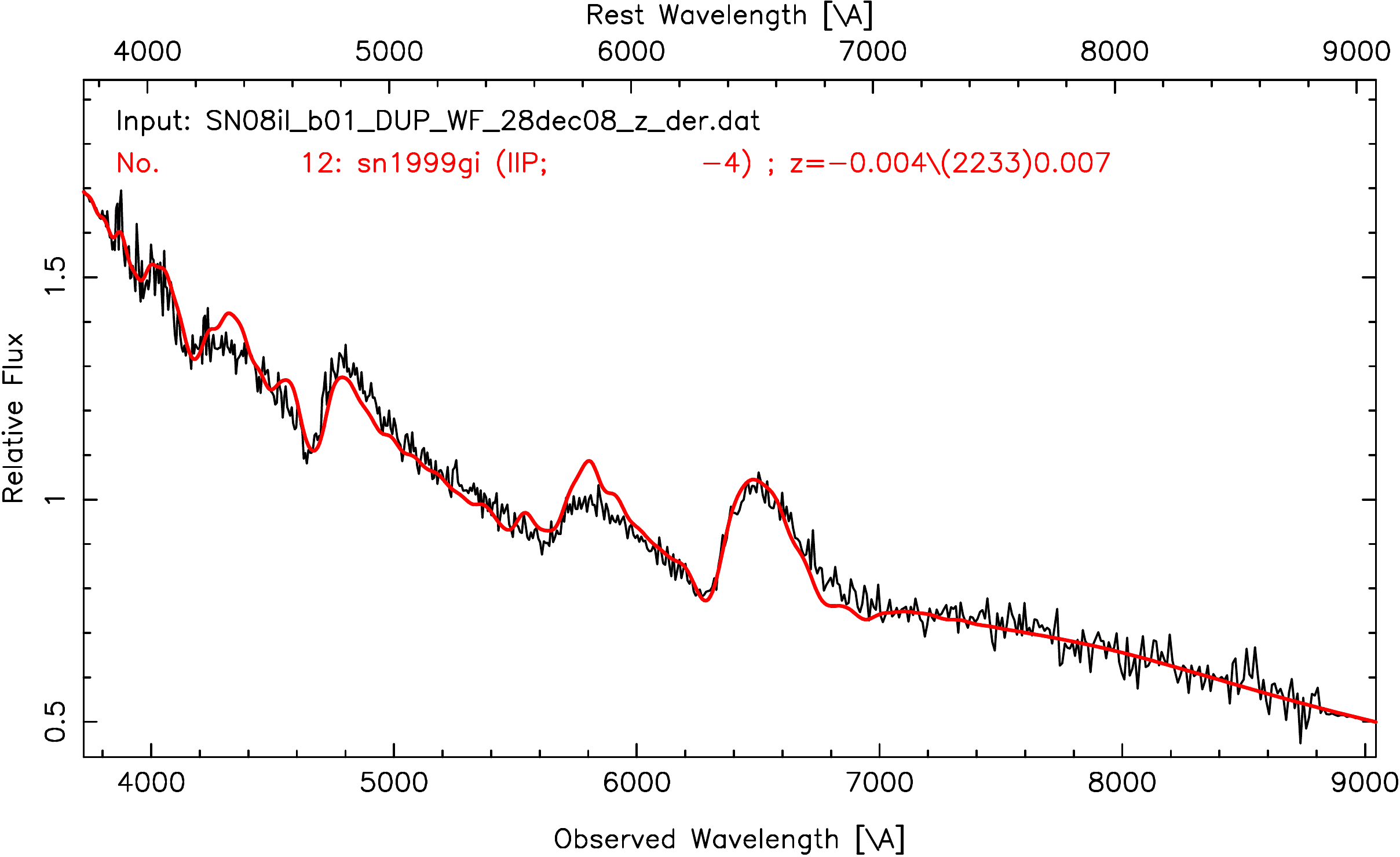}
\includegraphics[width=4.4cm]{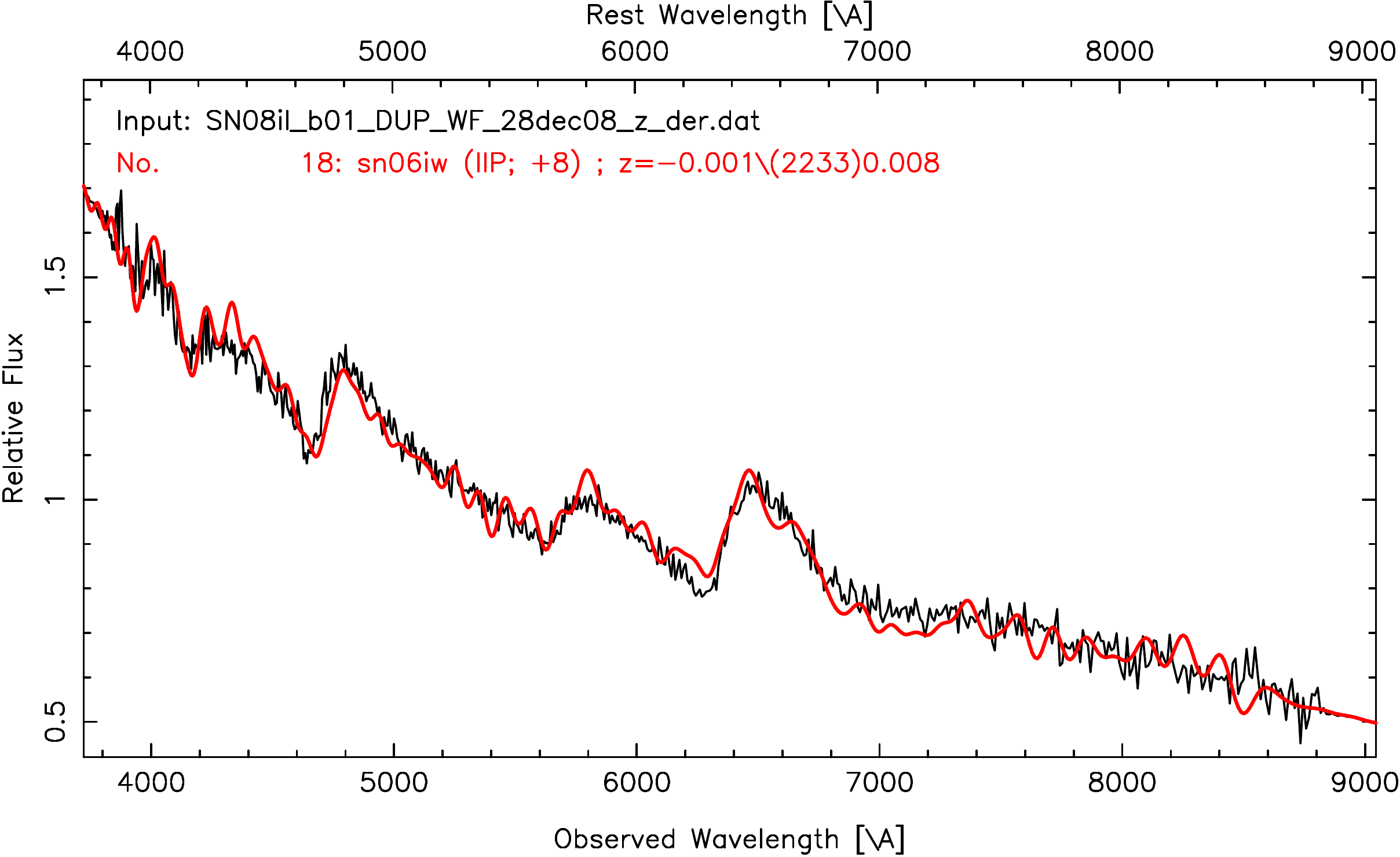}
\caption{Best spectral matching of SN~2008il using SNID. The plots show SN~2008il compared with 
SN~1999em, SN~1999gi, and SN~2006iw at 6, 8, and 8 days from explosion.}
\end{figure}

\clearpage

\begin{figure}
\centering
\includegraphics[width=4.4cm]{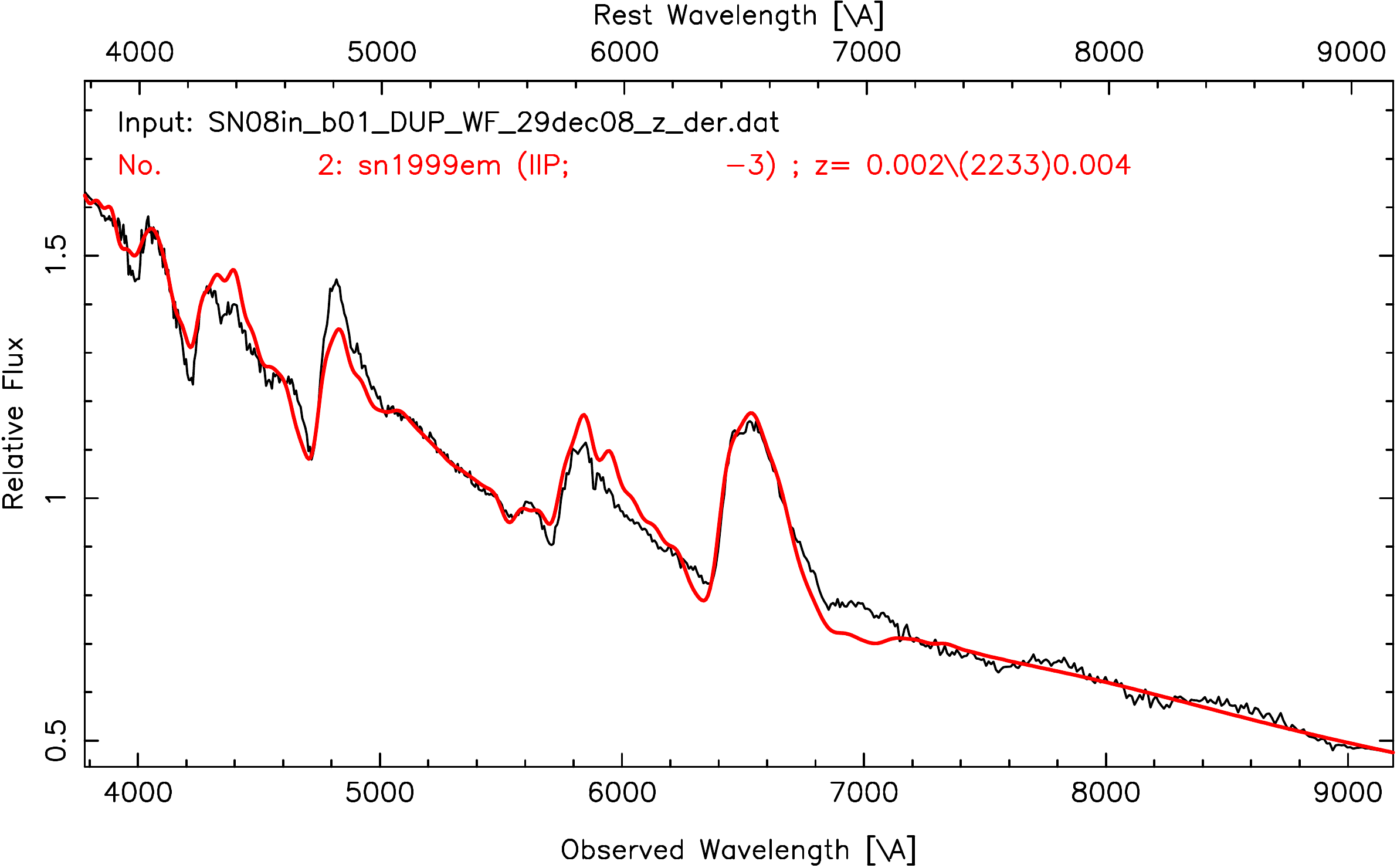}
\includegraphics[width=4.4cm]{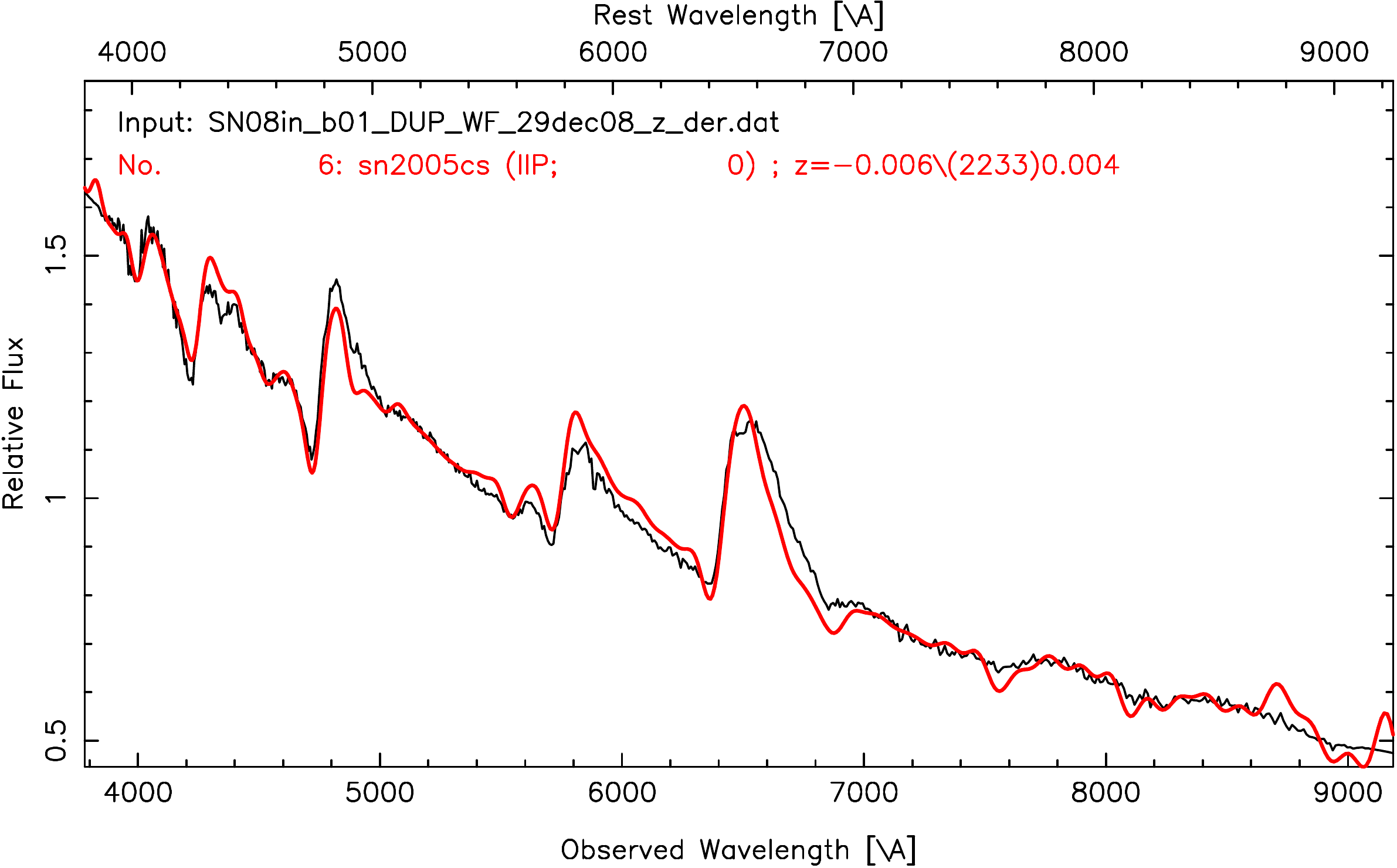}
\includegraphics[width=4.4cm]{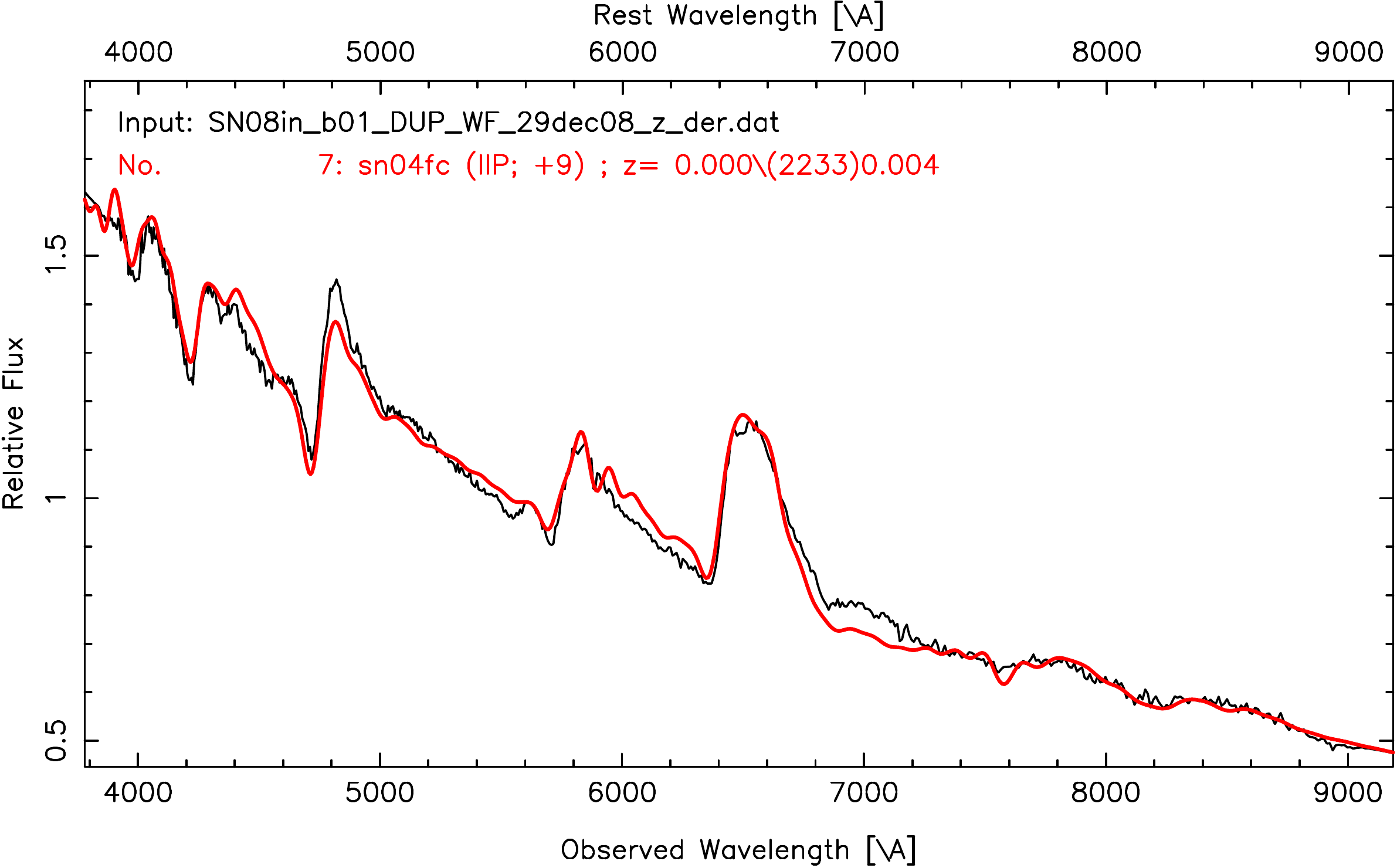}
\includegraphics[width=4.4cm]{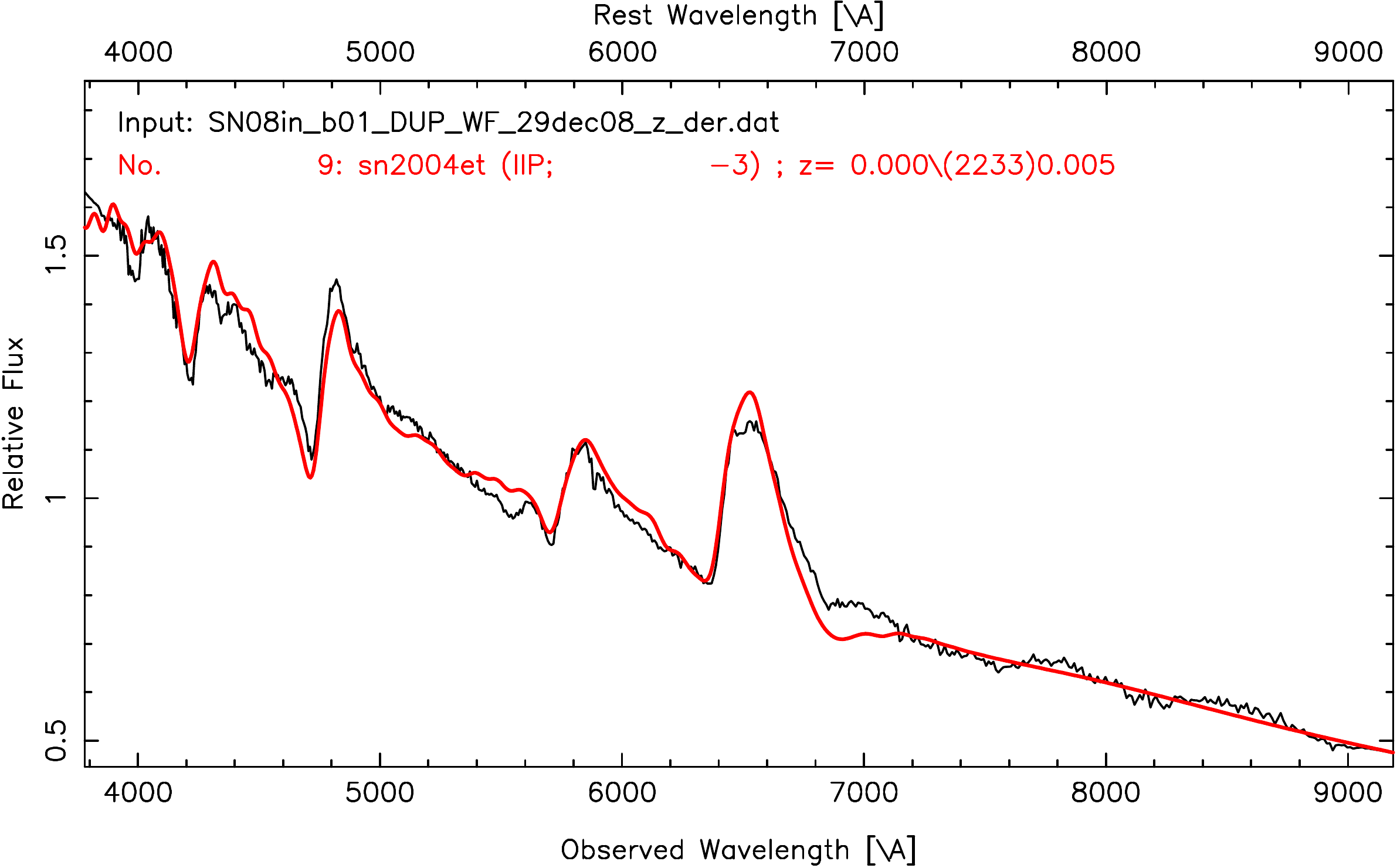}
\caption{Best spectral matching of SN~2008in using SNID. The plots show SN~2008in compared with 
SN~1999em, SN~2005cs, SN~2004fc, and SN~2004et at 7, 6, 9, and 13 days from explosion.}
\end{figure}

\begin{figure}
\centering
\includegraphics[width=4.4cm]{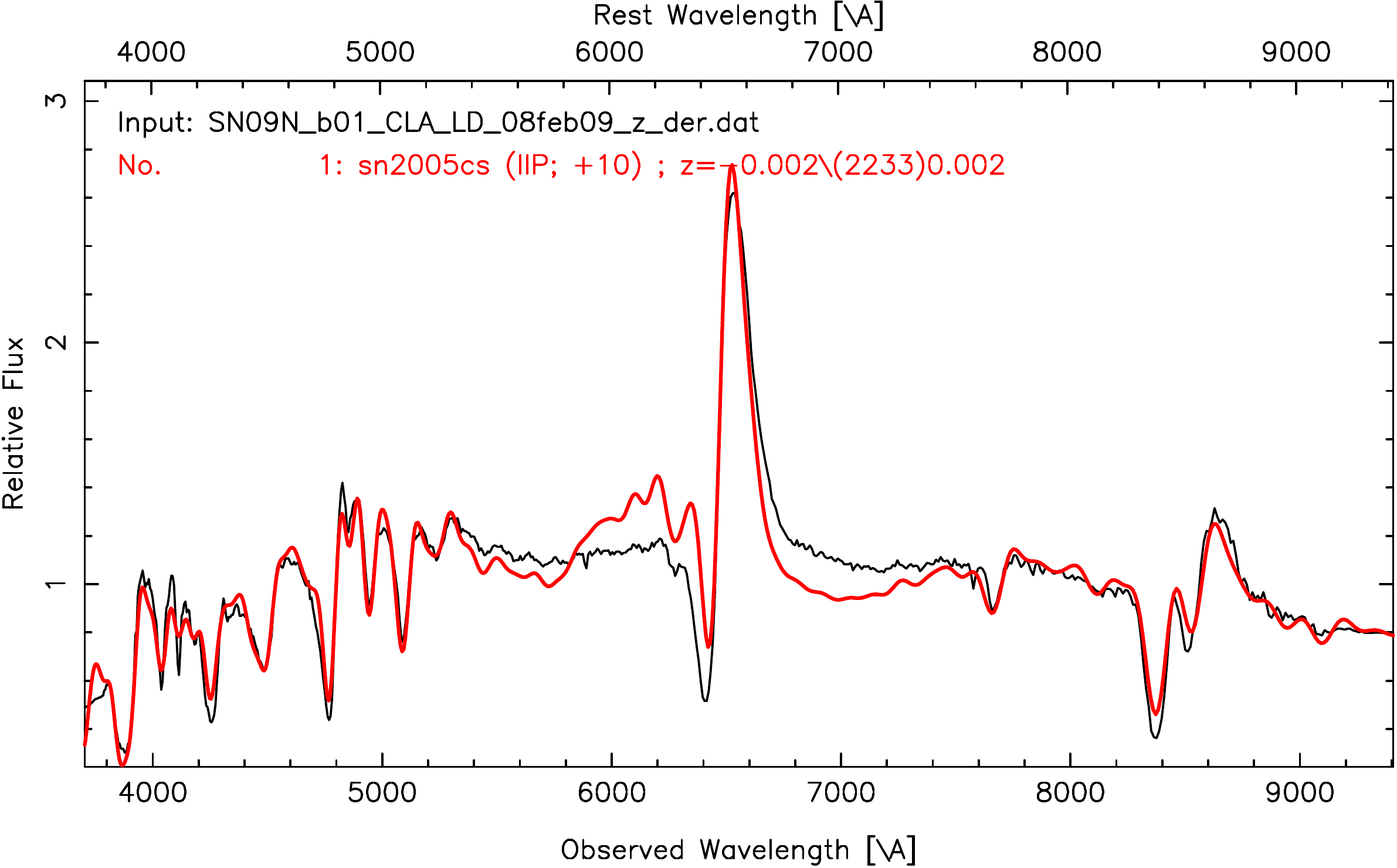}
\includegraphics[width=4.4cm]{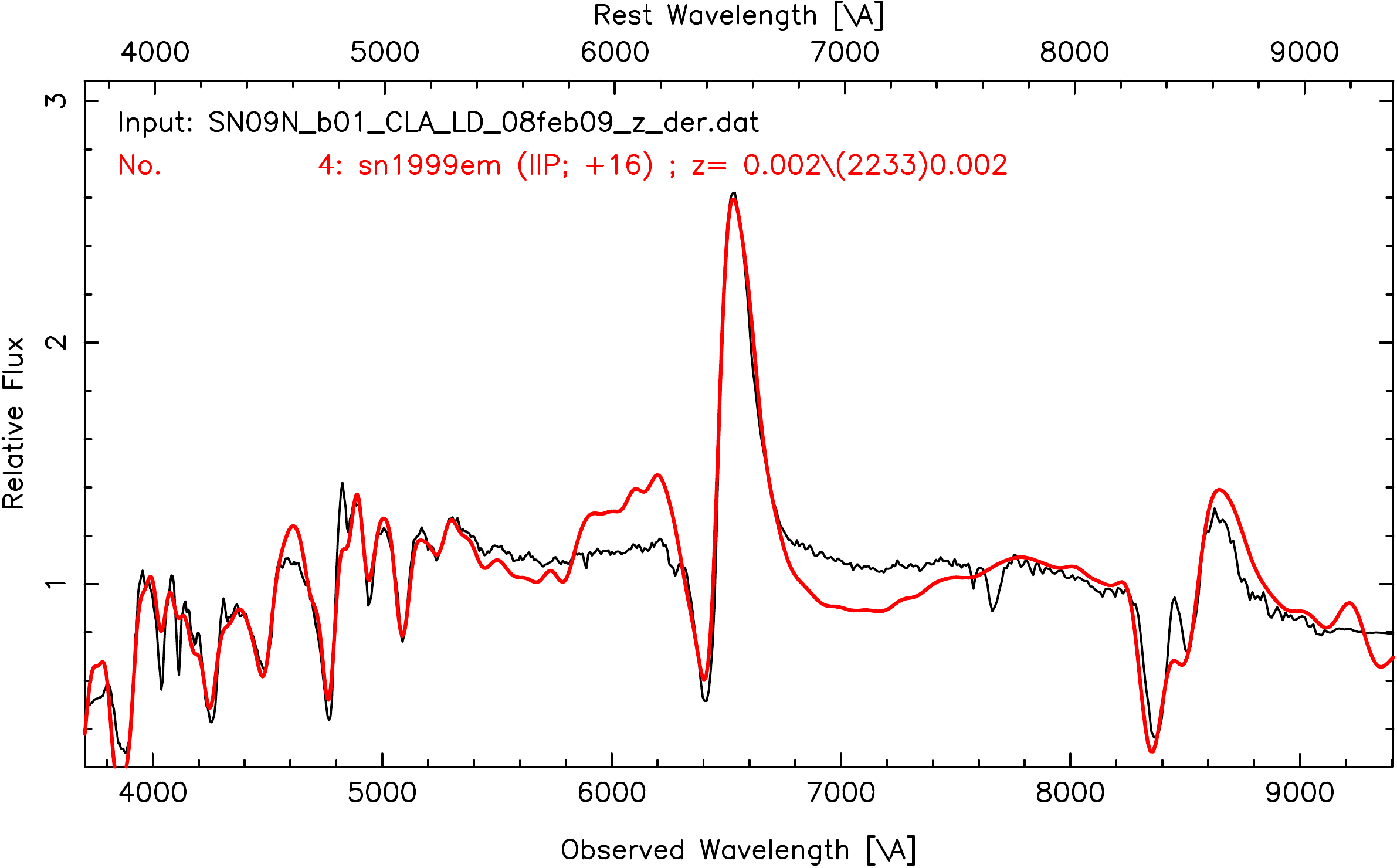}
\includegraphics[width=4.4cm]{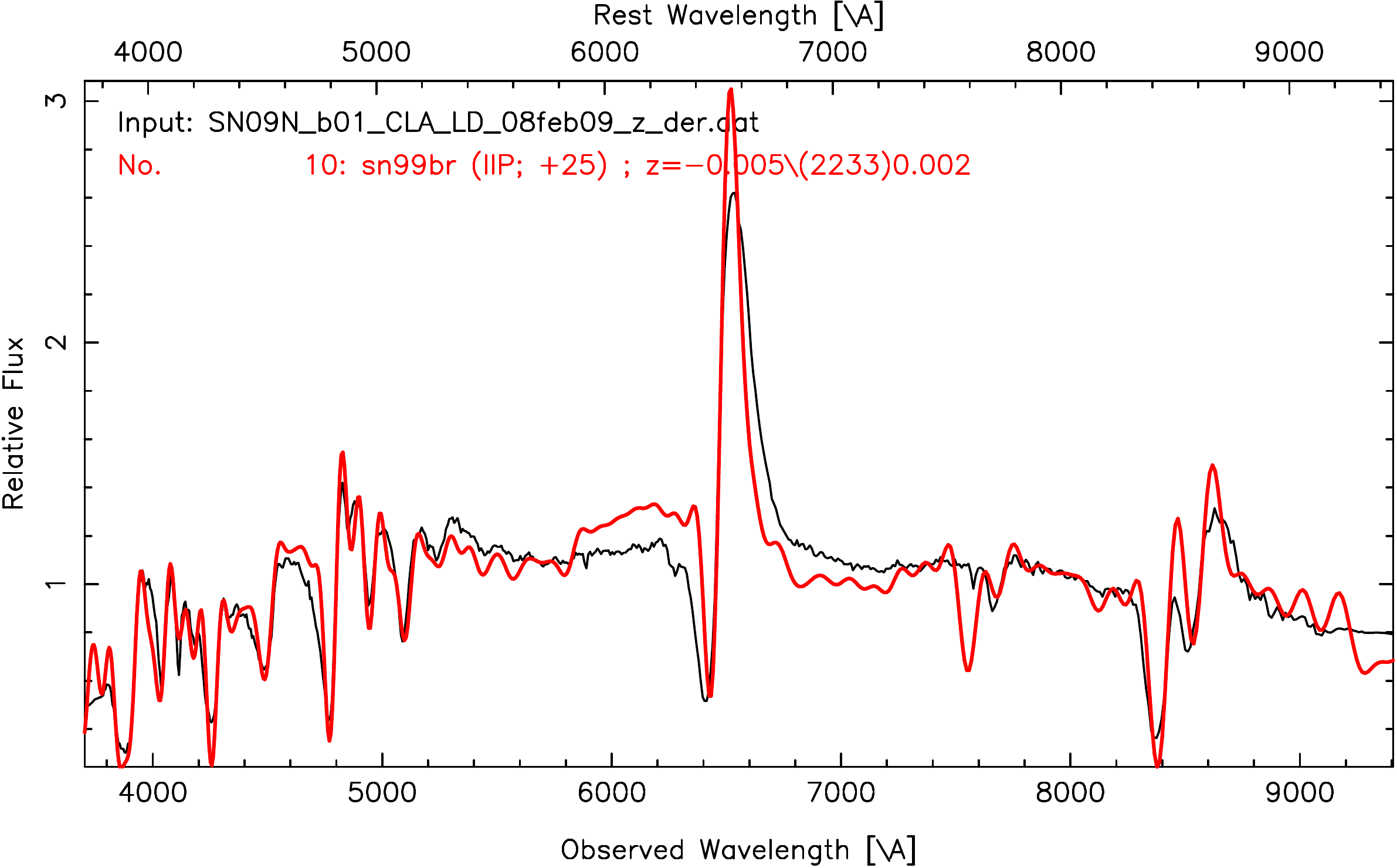}
\includegraphics[width=4.4cm]{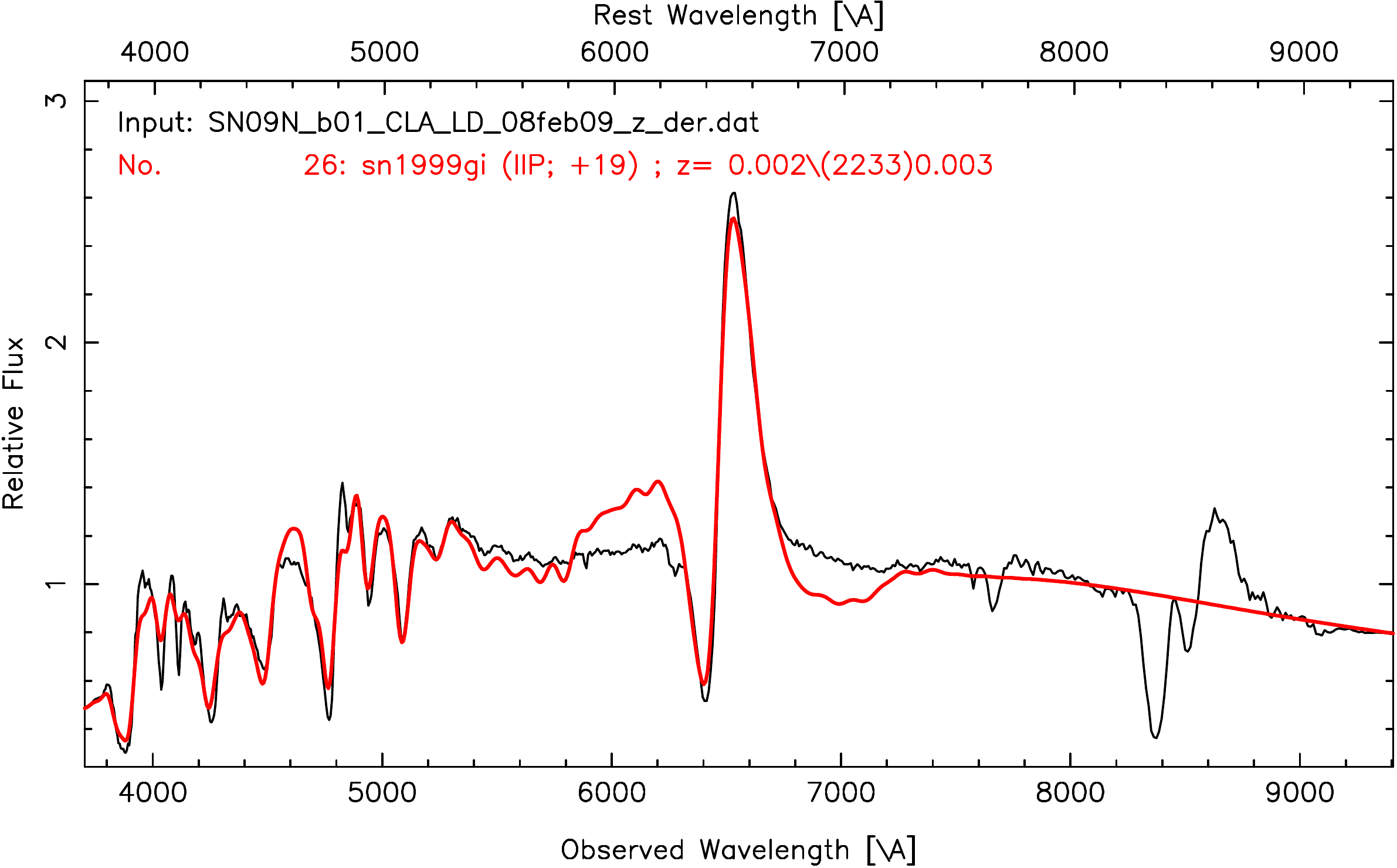}
\caption{Best spectral matching of SN~2009N using SNID. The plots show SN~2009N compared with 
SN~2005cs, SN~1999em, SN~1999br, and SN~1999gi at 22, 16, 25, 31 days from explosion.}
\end{figure}

\begin{figure}
\centering
\includegraphics[width=4.4cm]{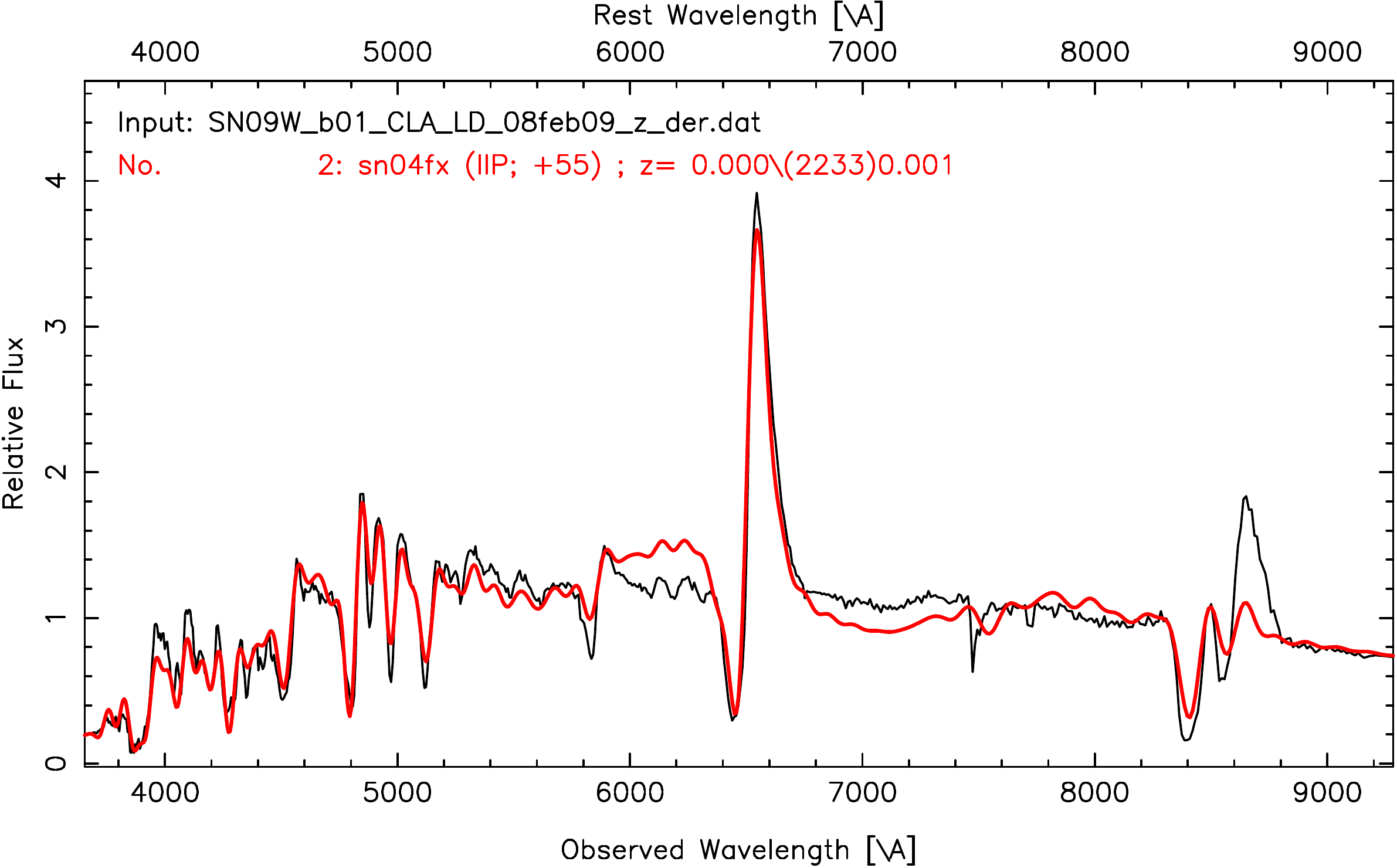}
\includegraphics[width=4.4cm]{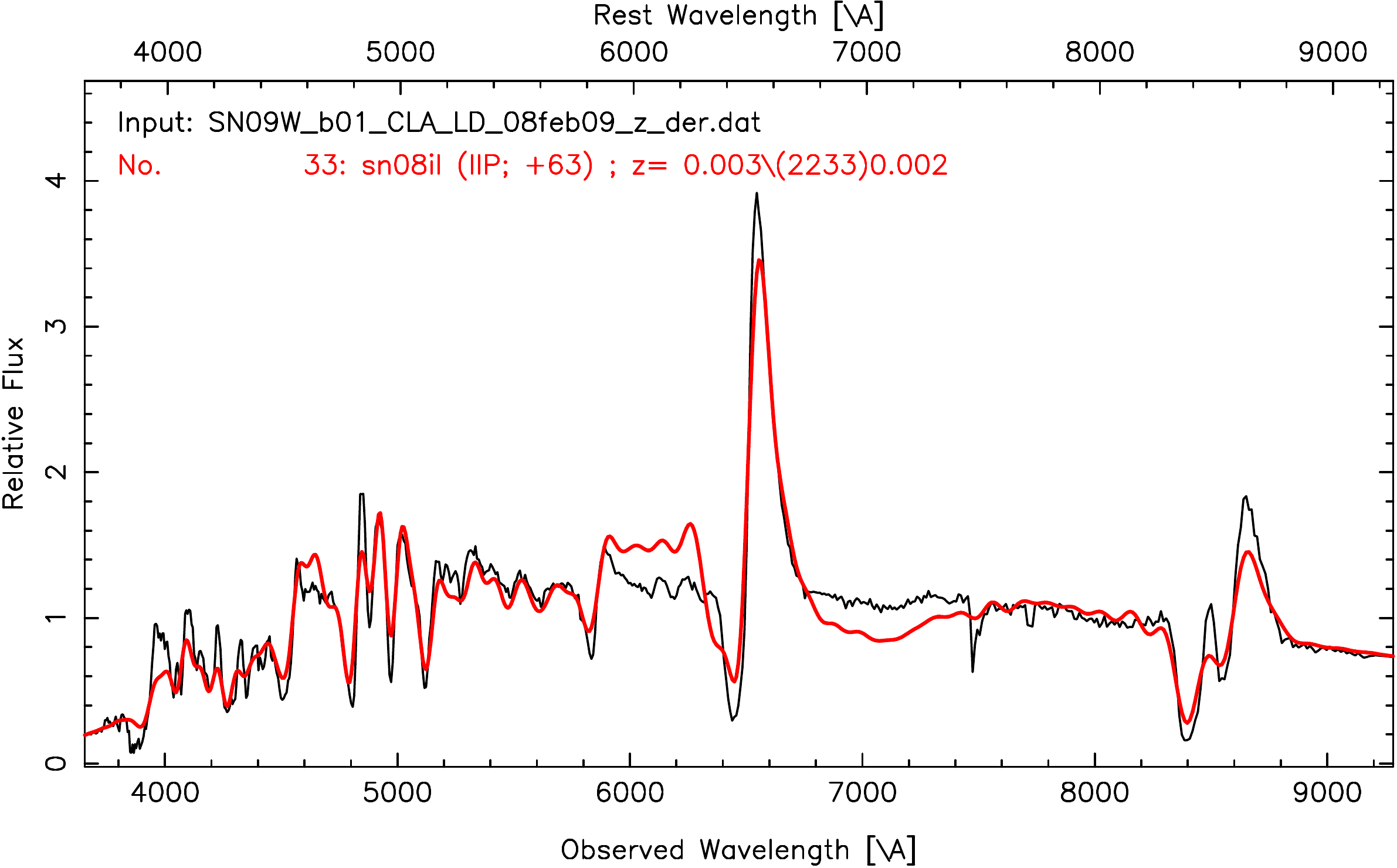}
\includegraphics[width=4.4cm]{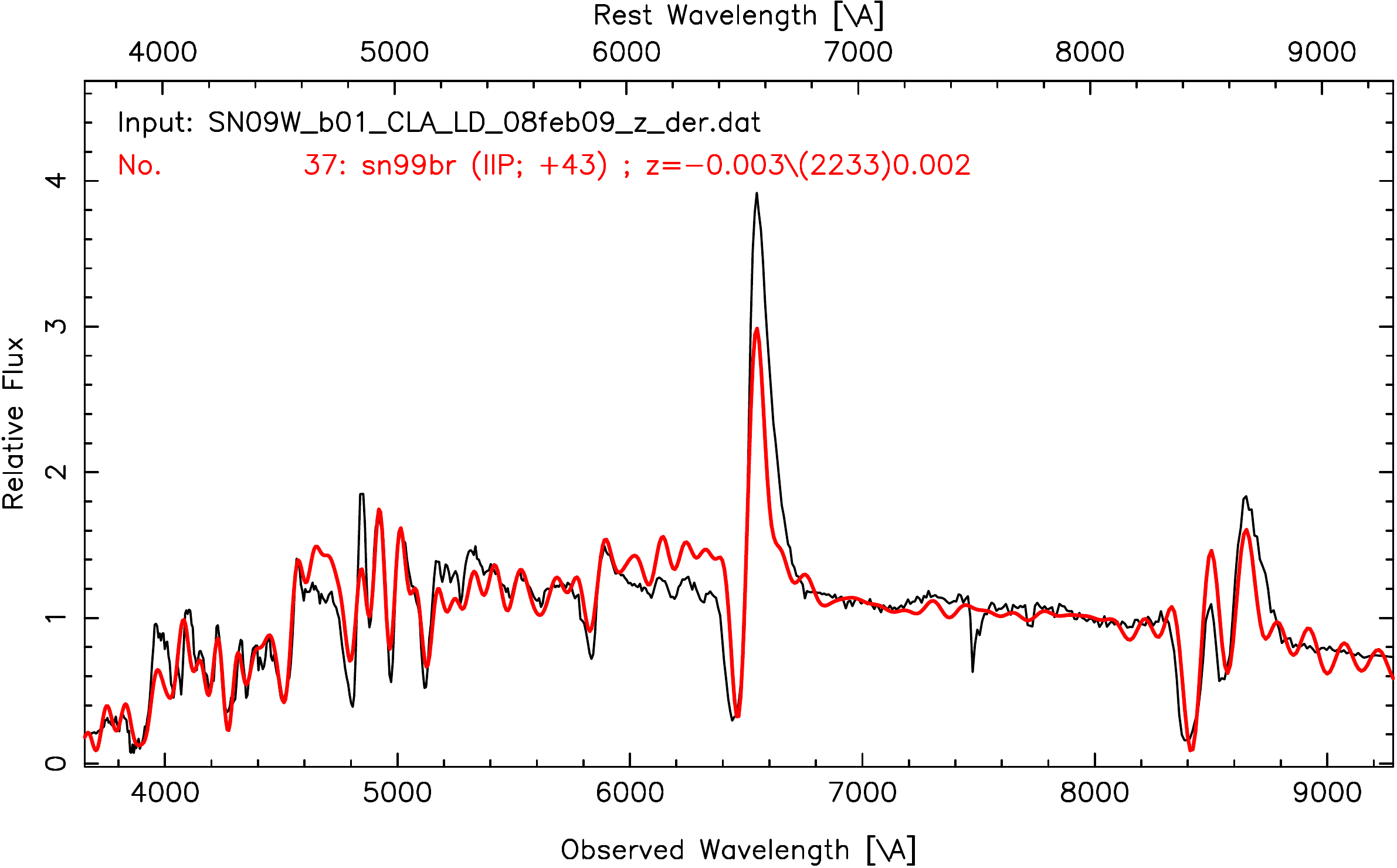}
\caption{Best spectral matching of SN~2009W using SNID. The plots show SN~2009W compared with 
SN~2004fx, SN~2008il, and SN~1999br at 55, 63, and 43 days from explosion.}
\end{figure}

\clearpage

\begin{figure}
\centering
\includegraphics[width=4.4cm]{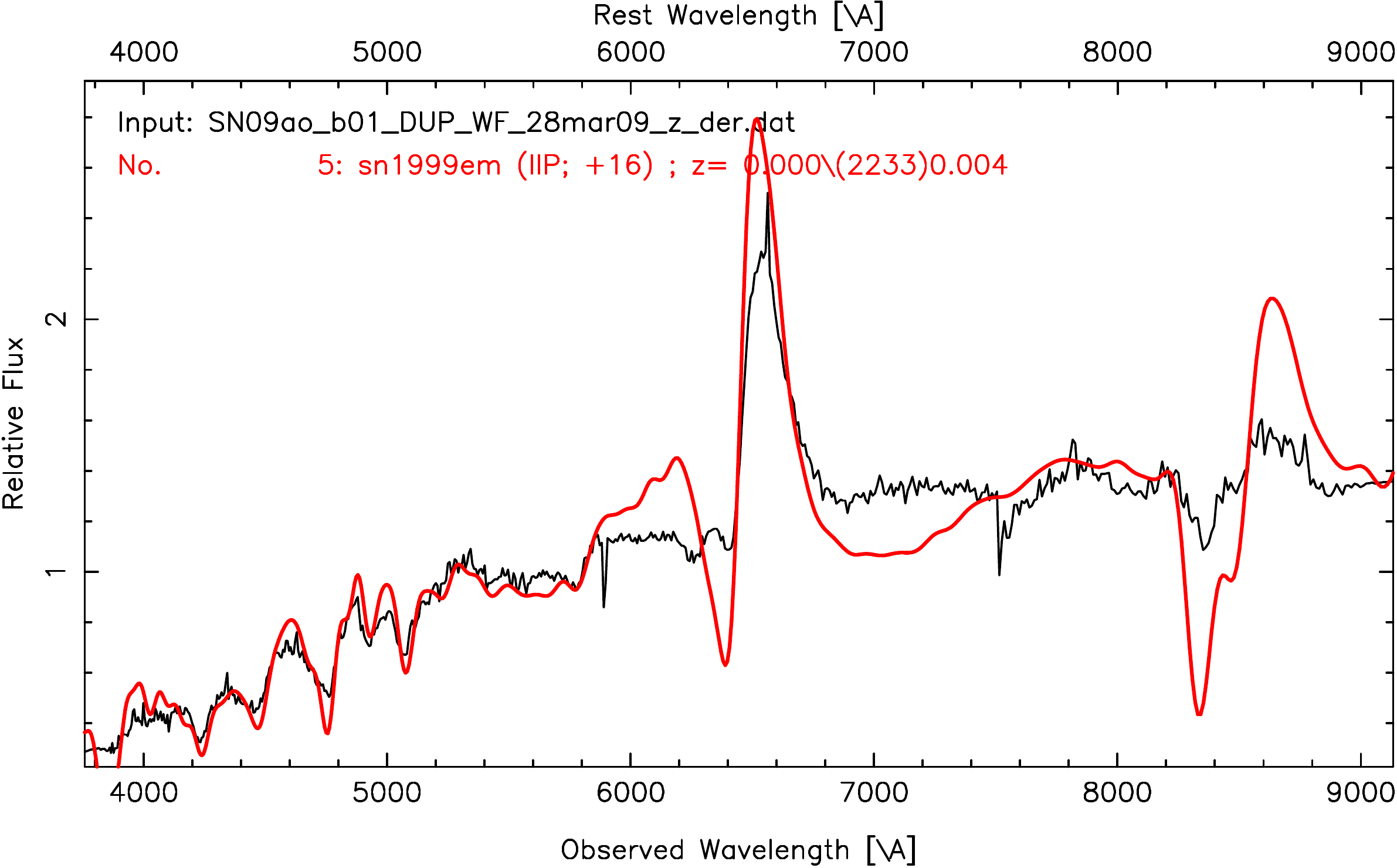}
\includegraphics[width=4.4cm]{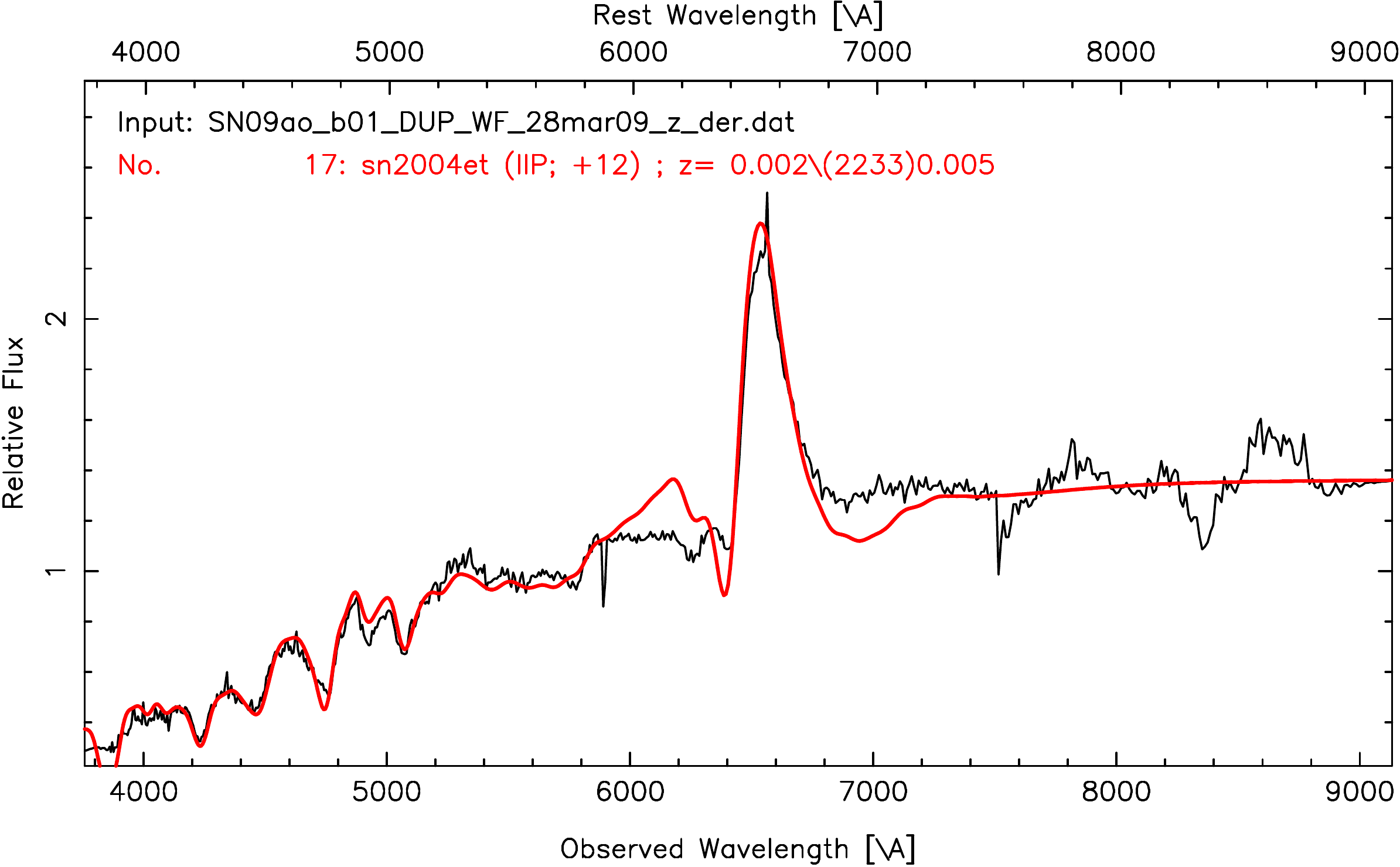}
\includegraphics[width=4.4cm]{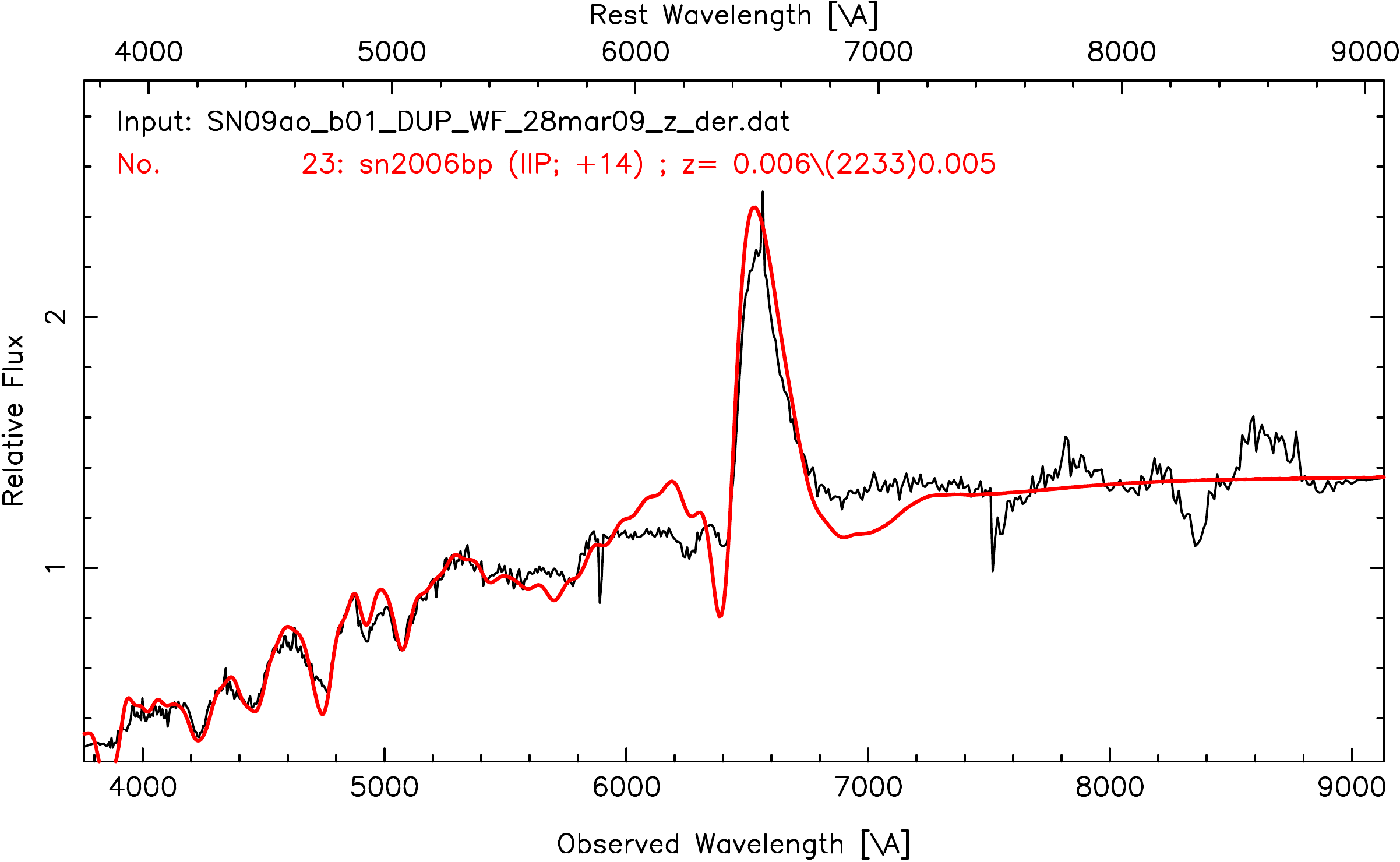}
\caption{Best spectral matching of SN~2009ao using SNID. The plots show SN~2009ao compared with 
SN~1999em, SN~2004et, and SN~2006bp at 26, 22, and 24 days from explosion.}
\end{figure}

\begin{figure}
\centering
\includegraphics[width=4.4cm]{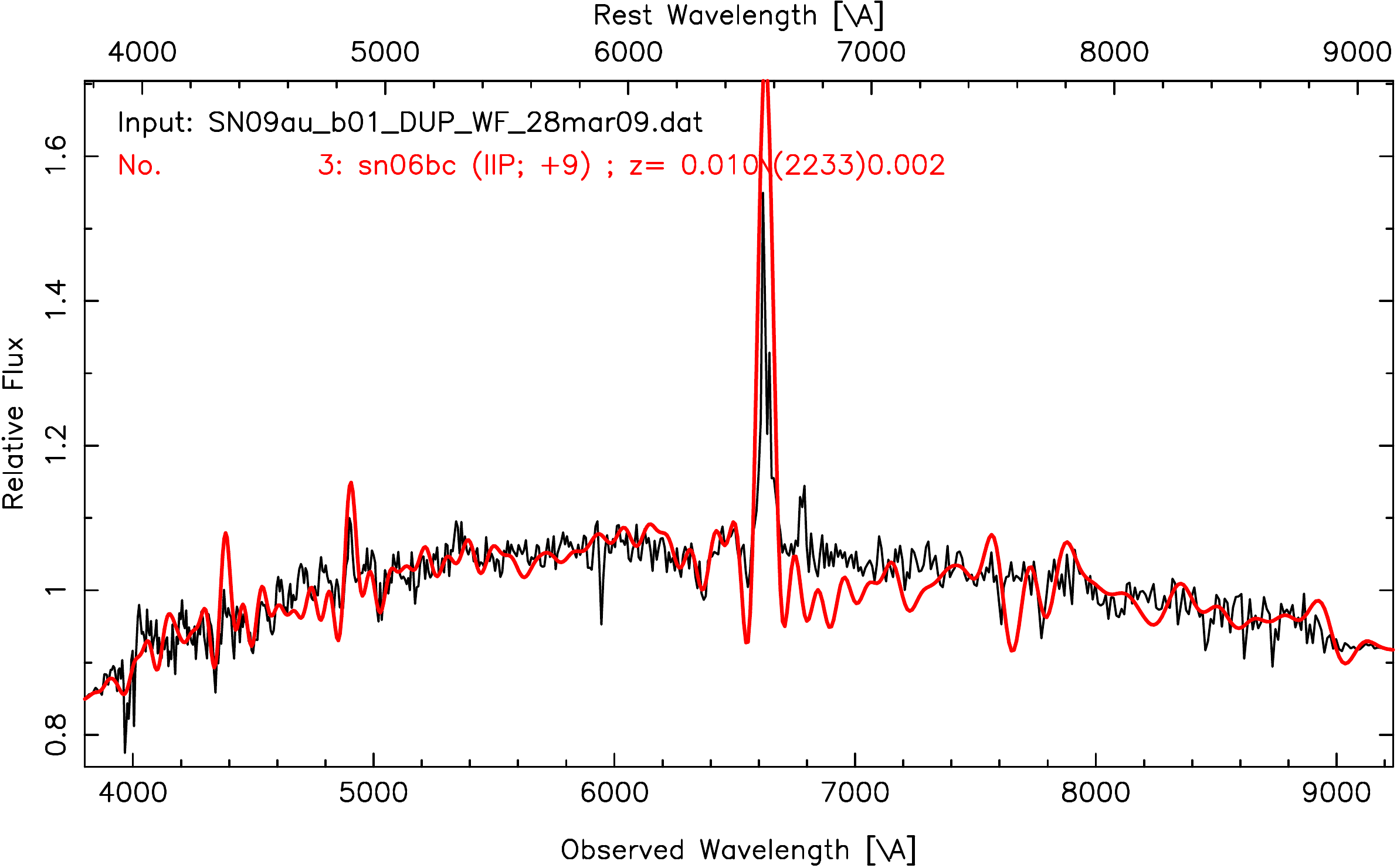}
\caption{Best spectral matching of SN~2009au using SNID. The plots show SN~2009au compared with 
SN~2006bc at 9 days from explosion.}
\end{figure}

\clearpage

\begin{figure}
\centering
\includegraphics[width=4.4cm]{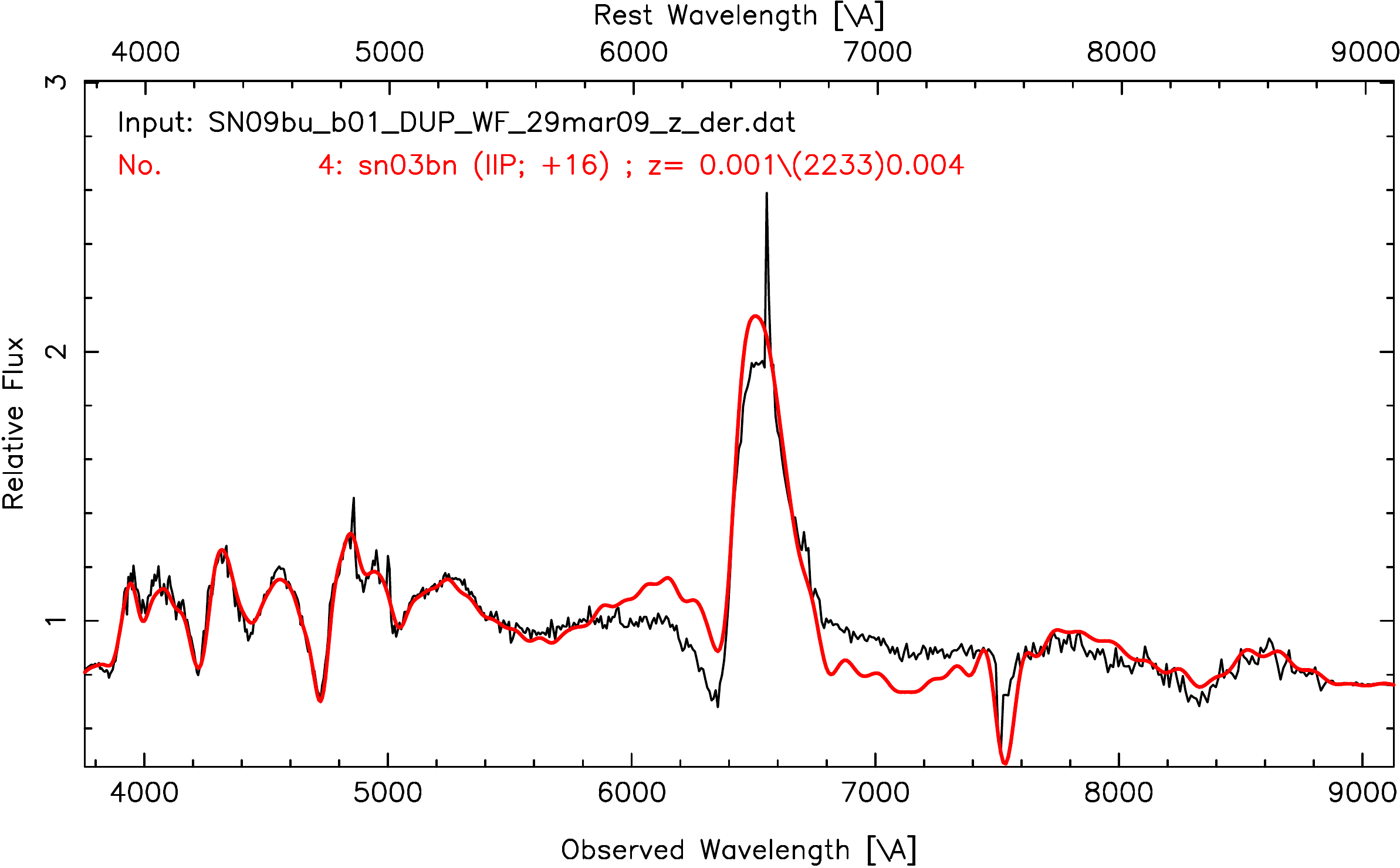}
\includegraphics[width=4.4cm]{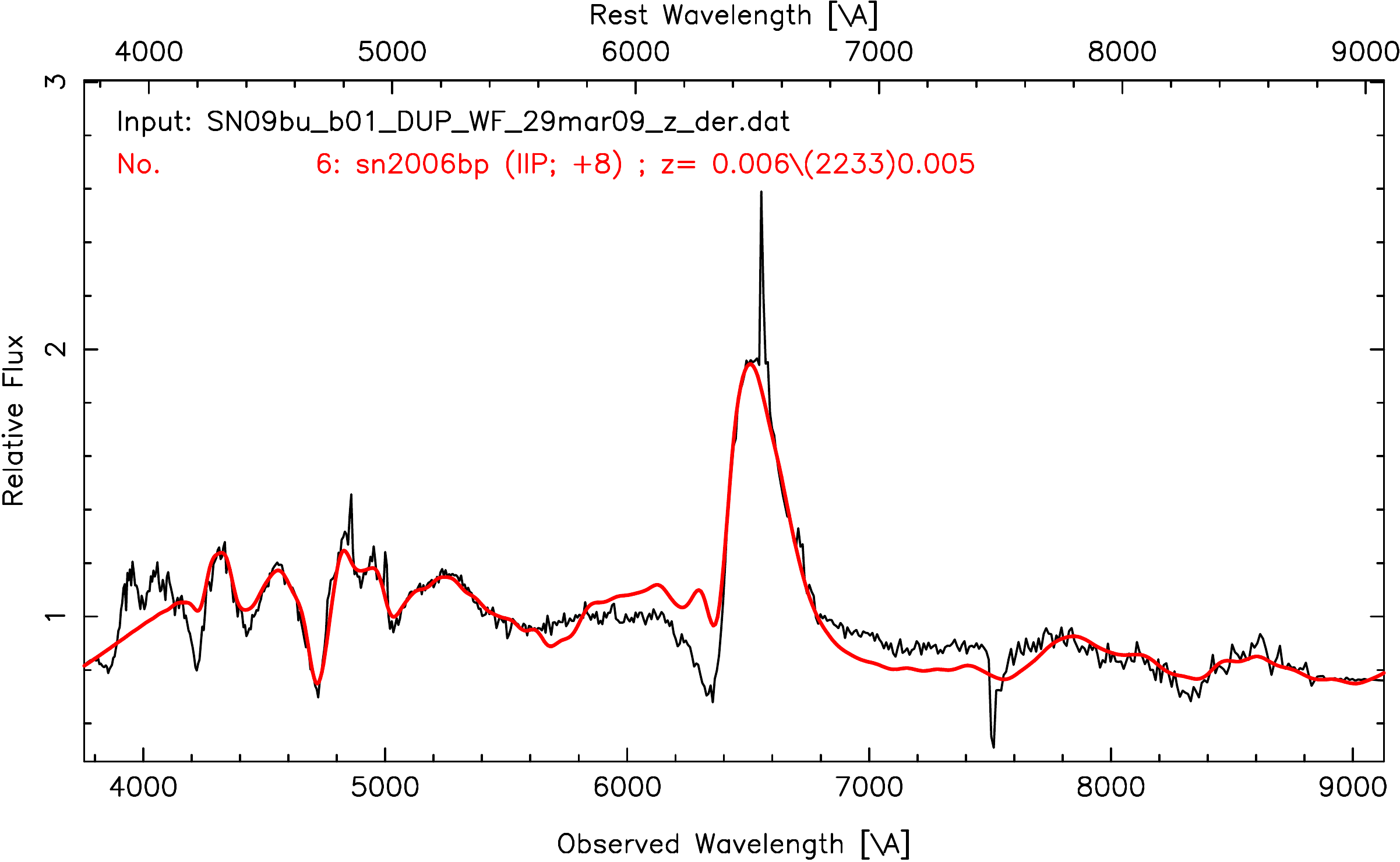}
\includegraphics[width=4.4cm]{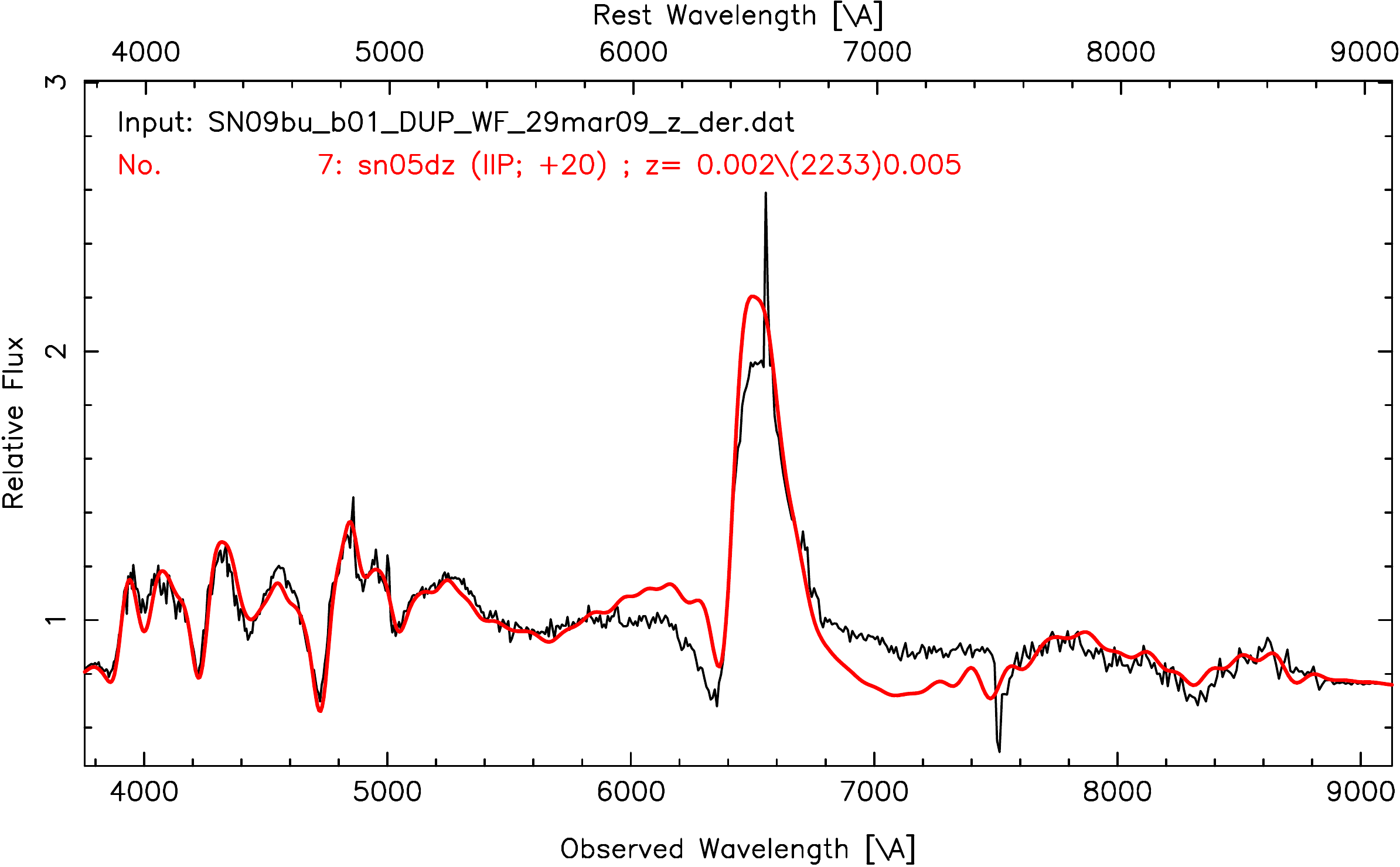}
\includegraphics[width=4.4cm]{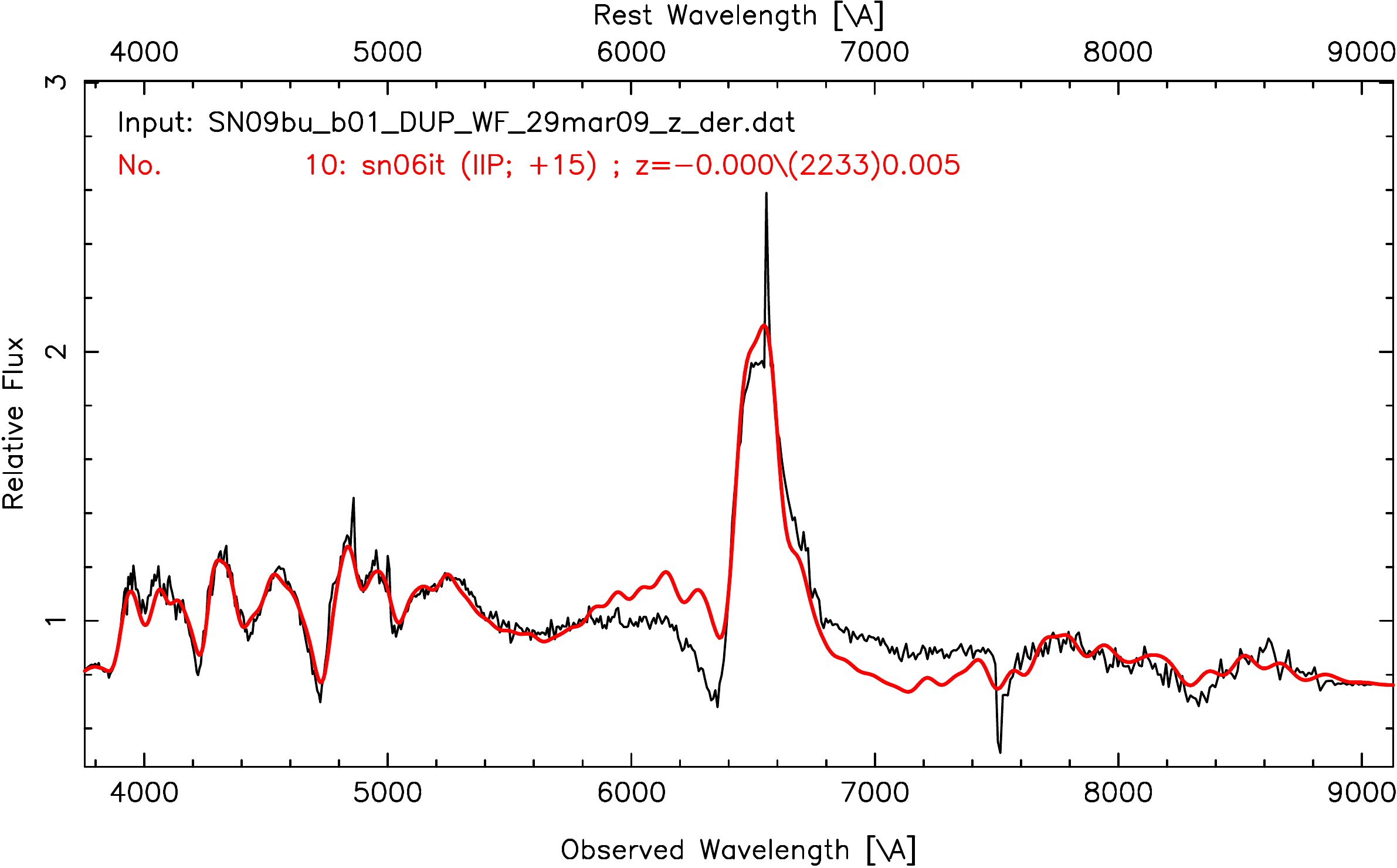}
\includegraphics[width=4.4cm]{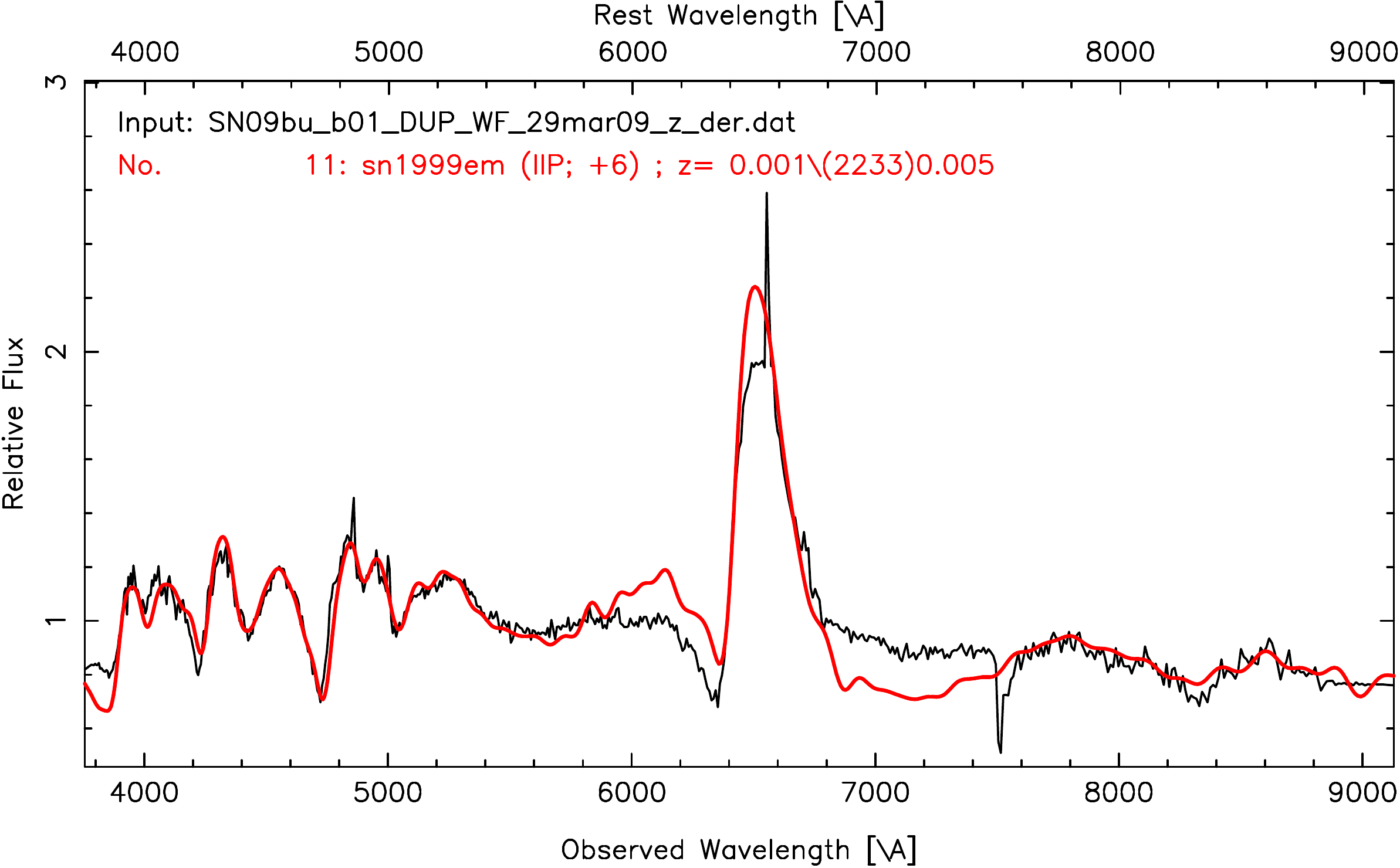}
\includegraphics[width=4.4cm]{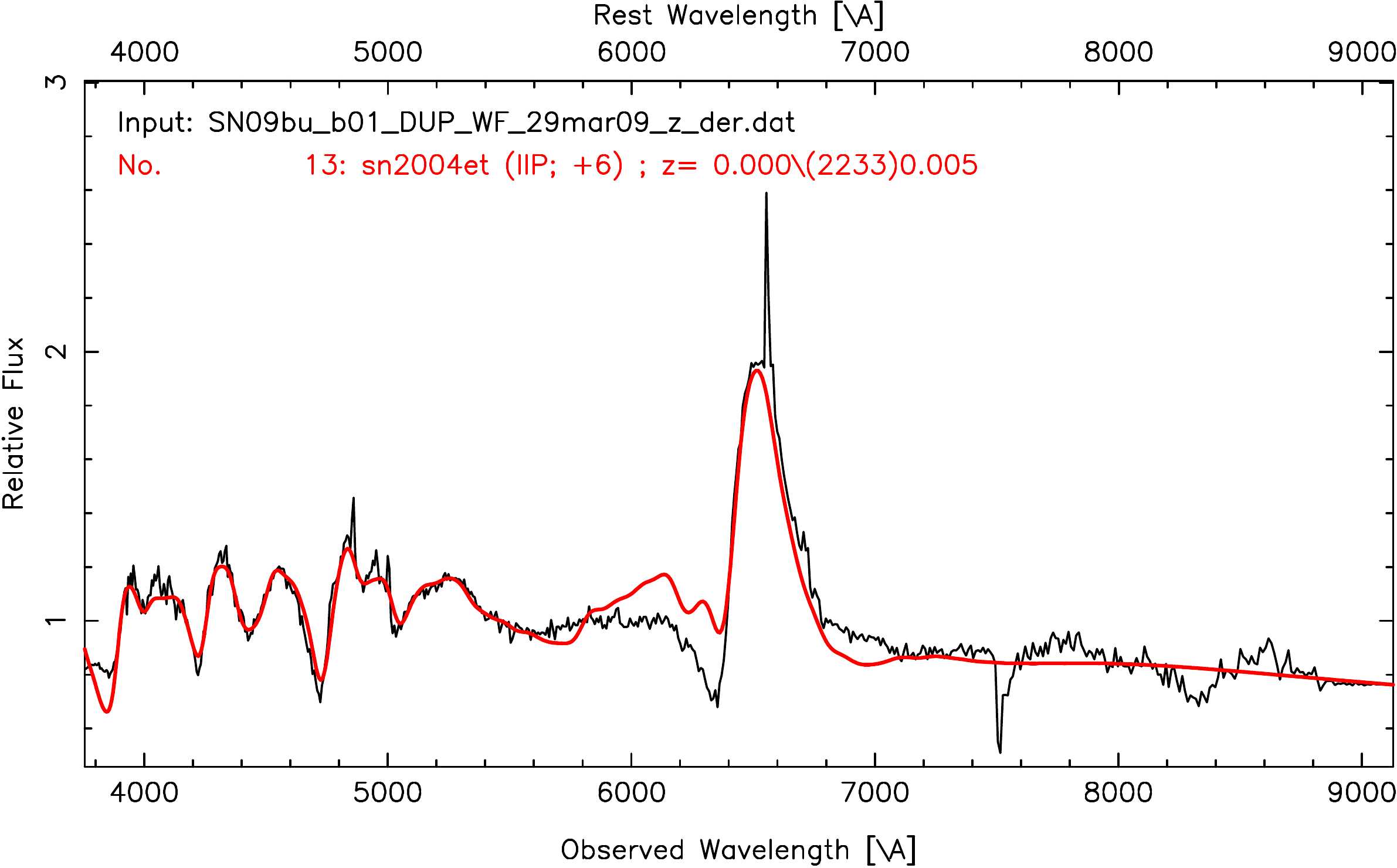}
\caption{Best spectral matching of SN~2009bu using SNID. The plots show SN~2009bu compared with 
SN~2003bn, SN~2006bp, SN~2005dz, SN~2006it, SN~1999em, and SN~2004et at 16, 17, 20, 15, 16, and 22 days from explosion.}
\end{figure}

\begin{figure}
\centering
\includegraphics[width=4.4cm]{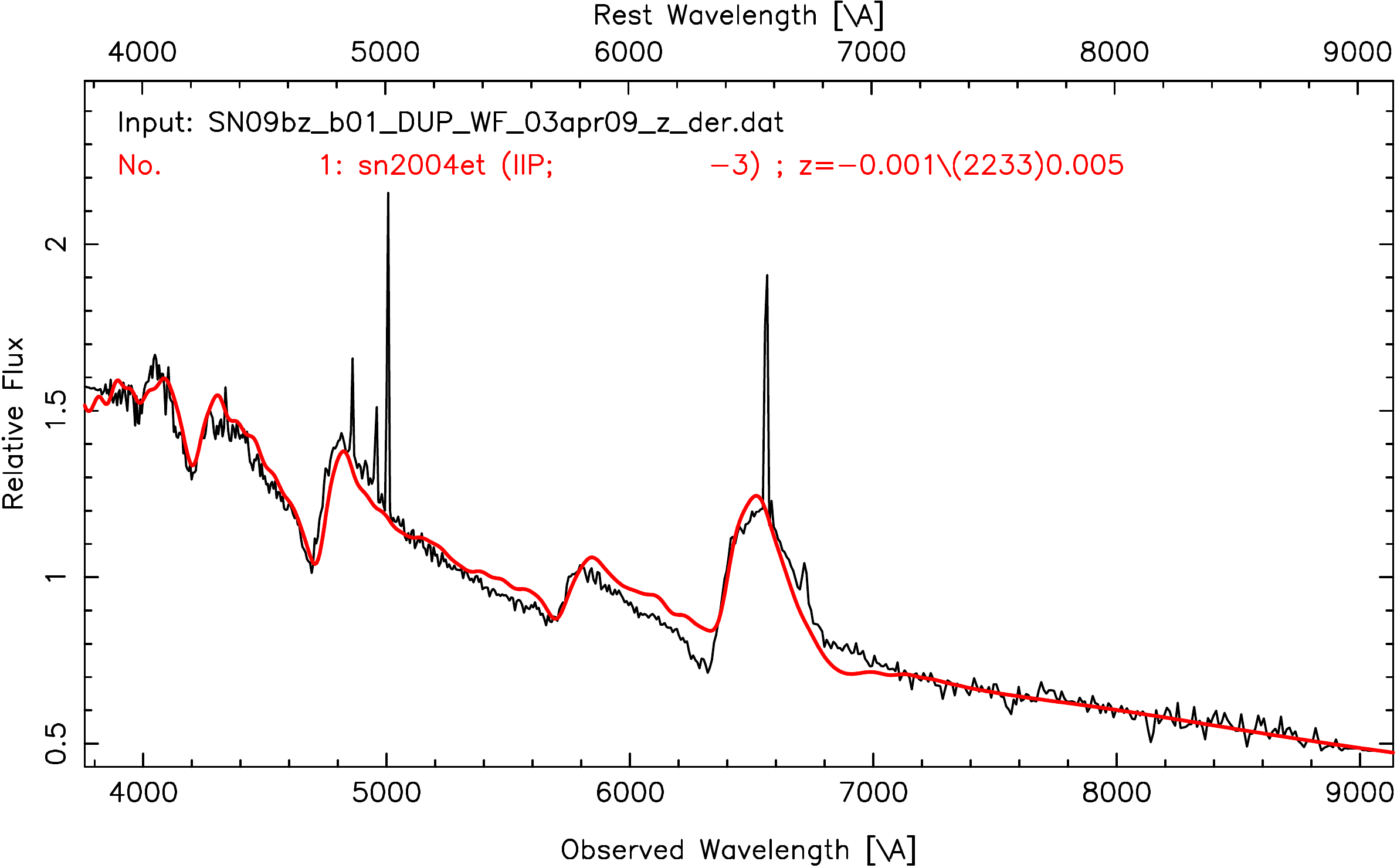}
\includegraphics[width=4.4cm]{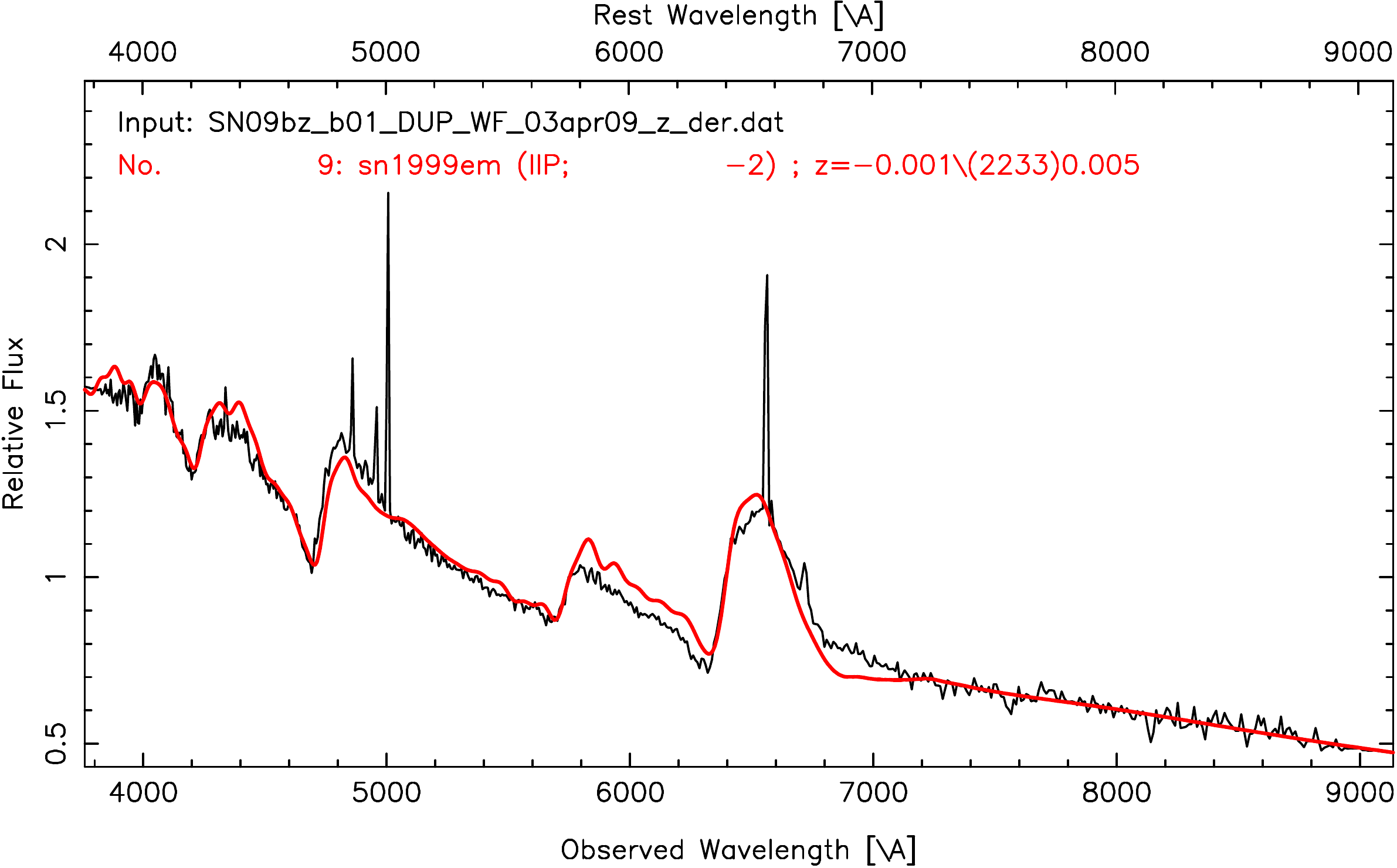}
\includegraphics[width=4.4cm]{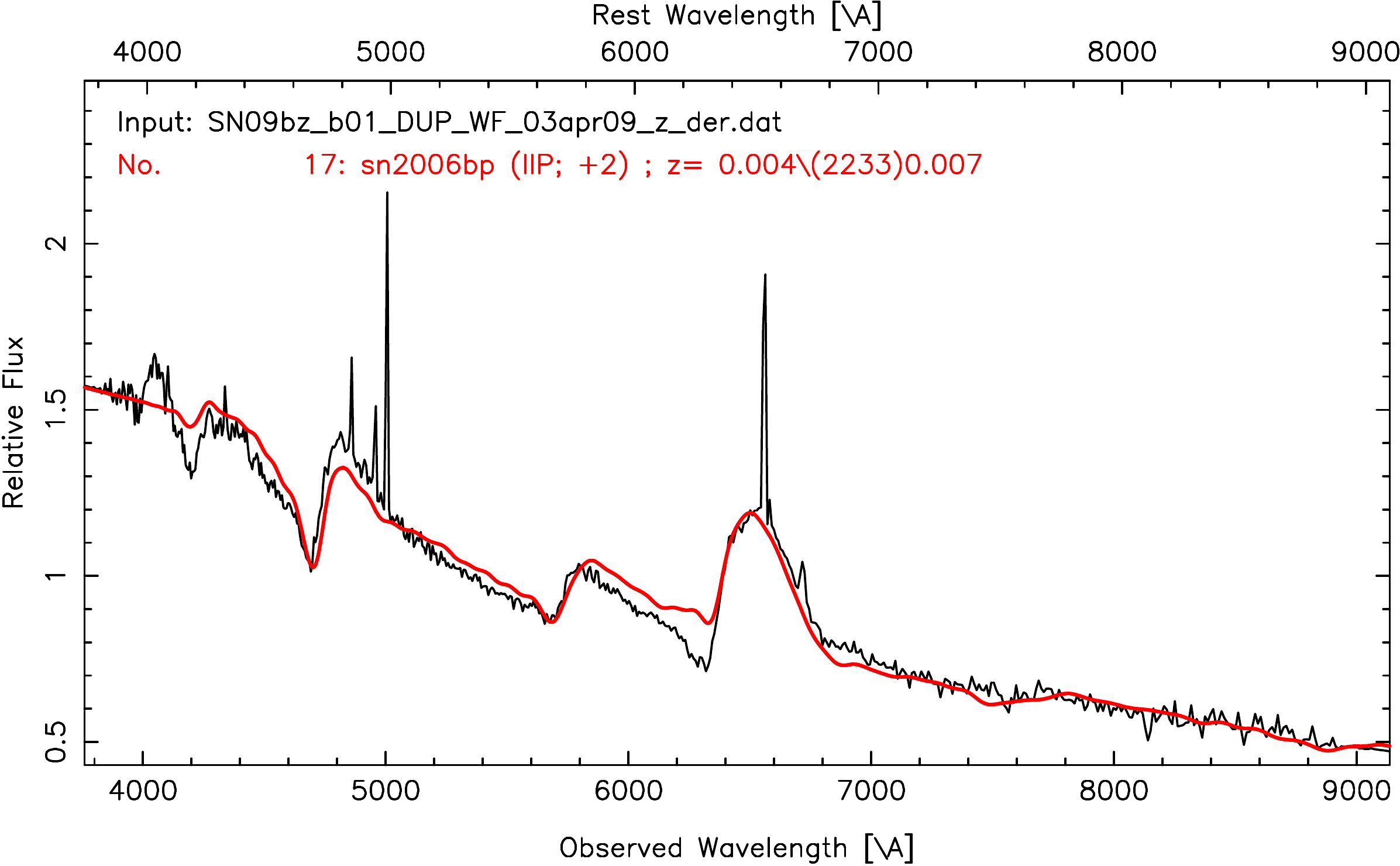}
\includegraphics[width=4.4cm]{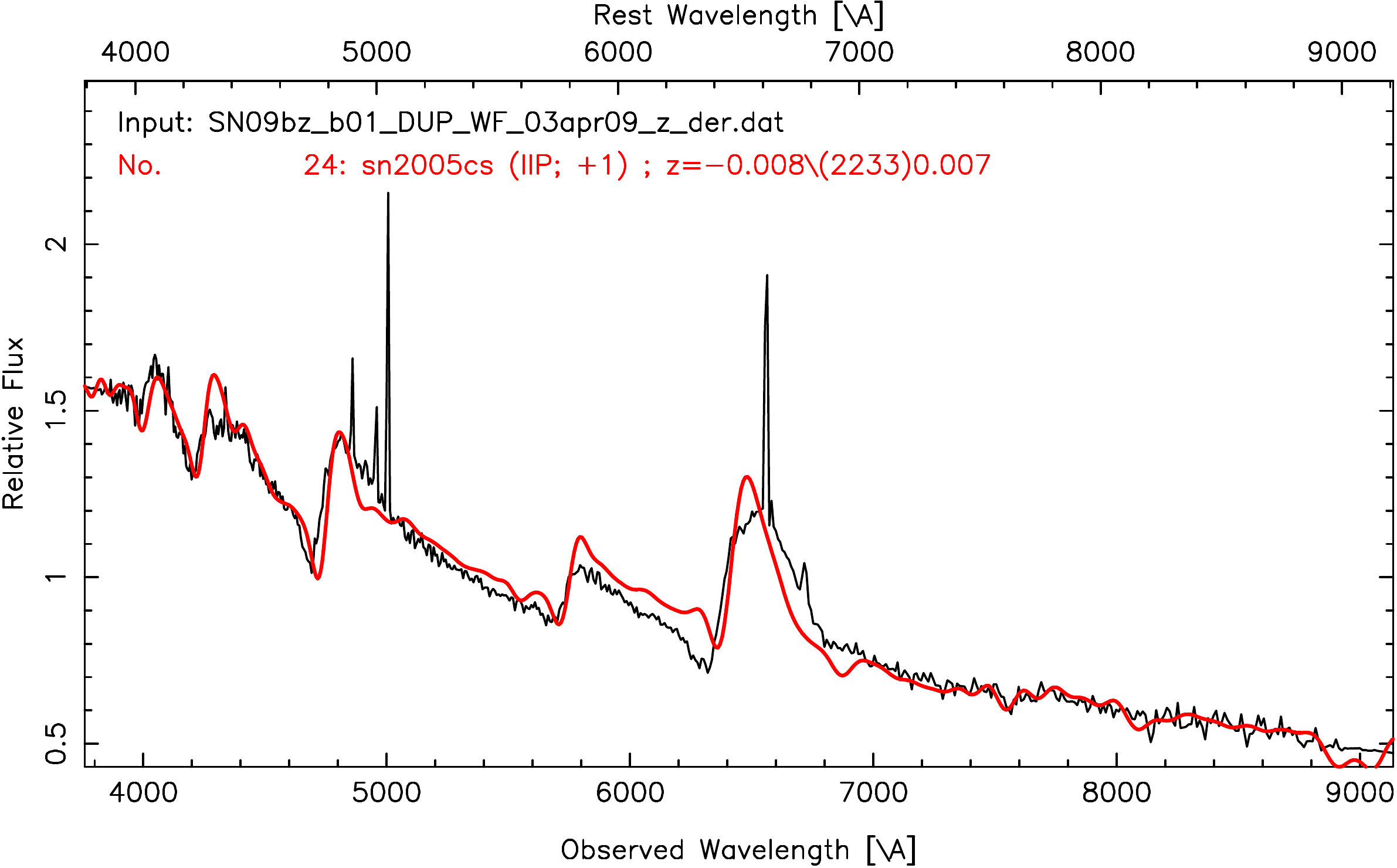}
\caption{Best spectral matching of SN~2009bz using SNID. The plots show SN~2009bz compared with 
SN~2004et, SN~1999em, SN~2006bp, and SN~2005cs at 13, 8, 11, and 7 days from explosion.}
\end{figure}

\newpage
\newpage
\end{document}